\documentclass[floatfix,prb,aps,nofootinbib,amssymb,twocolumn,superscriptaddress,10pt]{revtex4-2}

\usepackage[T1]{fontenc}
\usepackage{graphicx}
\usepackage{changepage}
\usepackage{xcolor}
\usepackage{dcolumn}
\usepackage{comment}
\usepackage{amsmath,amssymb,bbold,bm}
\usepackage{hyperref}
\usepackage{amsfonts}
\usepackage{dsfont}
\usepackage{listings}
\usepackage{braket}
\usepackage{placeins}
\usepackage{pgffor}
\usepackage{multirow}
\usepackage{tikz}
\usepackage{color}
\usepackage{xr}
\usepackage[normalem]{ulem}
\usepackage{cleveref}
\usepackage{physics}
\usepackage{blindtext}
\usepackage{tabularx}
\usepackage{ltablex}
\usepackage{array}
\usepackage{pifont}
\usepackage[caption=false]{subfig}
\usepackage{siunitx}
\usepackage{graphics}
\usepackage{pdfpages}
\usepackage{nicematrix}
\usepackage{ragged2e}

\makeatletter
\def\maketag@@@#1{\hbox{\m@th\normalfont\normalsize#1}}
\makeatother

\makeatletter     
\renewcommand\onecolumngrid{
\do@columngrid{one}{\@ne}%
\def\set@footnotewidth{\onecolumngrid}
\def\footnoterule{\kern-6pt\hrule width 1.5in\kern6pt}%
}

\newcommand\numberthis{\refstepcounter{equation}\tag{\theequation}}

\crefname{appendix}{Appendix}{Appendices}
\crefname{equation}{Eq.}{Eqs.}
\crefname{figure}{Fig.}{Figs.}
\crefname{table}{Table}{Tables}
\crefname{section}{Section}{Sections}
\crefname{enumi}{Case}{Cases}
\creflabelformat{appendix}{[#2#1#3]}
\crefrangelabelformat{figure}{#3#1#4--(#5\crefstripprefix{#1}{#2}#6}
\crefmultiformat{figure}{Figs.~#2#1\xdef\mycreffirstarg{#1}#3}%
 { and~#2#1#3}{, #2#1#3}{ and~#2#1#3}

\let\vec\mathbf

\newcolumntype{M}[1]{>{\centering\arraybackslash}m{#1}}
\newcolumntype{N}[1]{ >{\vbox to 3ex\bgroup\vfill\centering}p{#1} <{\egroup}}
\newcolumntype{C}{>{\begin{tiny}}c<{\end{tiny}}}
\newcolumntype{U}{>{\begin{small}}c<{\end{small}}}

\usepackage{etoolbox} 

\NiceMatrixOptions{code-for-last-col=\scriptstyle,code-for-first-row=\scriptstyle}

\allowdisplaybreaks

\makeatletter
	\let\oldoutpupage\@outputpage@head
	\patchcmd{\@outputpage@head}{\@ifx{\LS@rot\@undefined}{}{\LS@rot}}{}{}{}
\makeatother

\begin{document}
\title{TBG as Topological Heavy Fermion: II. Analytical approximations of the model parameters}
\author{Dumitru C\u{a}lug\u{a}ru}
\thanks{These authors contributed equally.}
\affiliation{Department of Physics, Princeton University, Princeton, New Jersey 08544, USA}

\author{Maksim Borovkov}
\thanks{These authors contributed equally.}
\affiliation{Department of Physics, Princeton University, Princeton, New Jersey 08544, USA}

\author{Liam L.H.~Lau}
\affiliation{Center for Materials Theory, Department of Physics and Astronomy, Rutgers University, 136 Frelinghuysen Rd., Piscataway, NJ 08854-8019, USA}

\author{Piers Coleman}
\affiliation{Center for Materials Theory, Department of Physics and Astronomy, Rutgers University, 136 Frelinghuysen Rd., Piscataway, NJ 08854-8019, USA}
\affiliation{Department of Physics, Royal Holloway, University of London, Egham, Surrey TW20 0EX, UK}

\author{Zhi-Da Song}
\affiliation{International Center for Quantum Materials, School of Physics, Peking University, Beijing 100871, China}

\author{B.~Andrei Bernevig}
\thanks{These authors contributed equally.}
\affiliation{Department of Physics, Princeton University, Princeton, New Jersey 08544, USA}
\affiliation{Donostia International Physics Center, P. Manuel de Lardizabal 4, 20018 Donostia-San Sebastian, Spain}
\affiliation{IKERBASQUE, Basque Foundation for Science, 48009 Bilbao, Spain}{}
\date{\today}

\begin{abstract}
The recently-introduced Topological Heavy Fermion (THF) model~\cite{SON22} of twisted bilayer graphene (TBG) aims to reconcile the quantum-dot-like electronic structure of the latter observed by scanning tunneling microscopy, with its electron delocalization seen in transport measurements. The THF model achieves this by coupling localized (heavy) fermions with anomalous conduction electrons. Originally, the parameters of the THF model were obtained numerically from the Bistritzer-Macdonald (BM) model of TBG~\cite{SON22}. In this work, we derive analytical expressions for the THF model parameters as a function of the twist angle, the ratio between the tunneling amplitudes at the $AA$ and $AB$ regions ($w_0 / w_1$), and the screening length of the interaction potential. By numerically computing the THF model parameters across an extensive experimentally-relevant parameter space, we show that the resulting approximations are remarkably good, {\it i.e.}{} within the 30\% relative error for almost the entire parameter space. At the single-particle level, the THF model accurately captures the energy spectrum of the BM model over a large phase space of angles and tunneling amplitude ratios. When interactions are included, we also show that the THF description of TBG is good around the magic angle for realistic values of the tunneling amplitude ratios ($0.6 \leq w_0/w_1 \leq 1.0$), for which the hybridization between the localized and conduction fermions $\gamma$ is smaller than the onsite repulsion of the heavy fermions $U_1$ ({\it i.e.}{} $\abs{\gamma} < U_1$).
\end{abstract}

\maketitle

\section{Introduction}\label{sec:introduction}

Constructing a unified theory of the strongly-correlated electronic phenomena in twisted bilayer graphene (TBG), observed in both transport~\cite{CAO18,CAO18a,LU19,YAN19,SHA19,SAI20,STE20,CAO20,CAO20a,SER20,CAO21,POL19,JIA19,ZON20,SAI21,DAS21,WU21a,ROZ21,LU21,SAI21,PAR21c,HES21,JAO22,LIU21e,STE21,SAI21a,DAS22} and spectroscopy~\cite{ARO20,XIE19,CHO19,KER19,WON20,NUC20,CHO21a,LIS21,GRO22} experiments, has become one of the major goals of contemporary condensed matter physics. The exotic physics of TBG near the magic angle~\cite{BIS11} arises from the interplay of strong electron-electron interactions and non-trivial topology of its flat bands. In overcoming these theoretical challenges, multiple approaches were pursued in order to understand the correlated insulating states~\cite{XIE21a,LED20,ABO20,REP20,SHE21,BUL20,LIA21,BER21b,CEA20,ZHA20,XIE21,ZHA21,HOF22,KAN19,VAF21,ZOU18,KOS18,XU18,BUL20,YUA18,PO18c,EFI18,RAD18,XU18,RAD19,WU19,THO18,CLA19,EUG20,REP20,FER20,WU20b,BUL20a,KWA21,KAN20a,KAN21,LIU21,PAD18,OCH18,VEN18,DOD18,SEO19,PO18,PIX19,XIE20a,VAF20,SOE20,LIU21a,DA21}, as well as the superconducting phases~\cite{GUI18,YOU19,LIA19,WU18,ISO18,LIU18,GON19,KHA21,KON20,CHR20,LEW21,KEN18,HUA19,GUO18,CHI20a} of TBG. One such approach consists of starting from the single-particle Bistritzer-Macdonald (BM) model~\cite{BIS11} and obtaining the correlated insulating states within a momentum space formalism~\cite{BUL20,LIA21,BER21b,CEA20,ZHA20,XIE21,ZHA21,HOF22}. Another way of tackling the many-body problem theoretically is through effective lattice models of TBG~\cite{KAN19,VAF21,RAD18,ZOU18,KOS18,XU18,RAD19,BUL20,YUA18,PO18c} to which Hubbard-like interactions are added. Owing to the topological obstruction of the TBG band structure~\cite{SON19,PO19,HEJ19,SON21,BOU19}, the construction of such lattice models is done, however, at the expense of certain symmetries of TBG, which are not preserved therein. Finally, some phenomenological models were also introduced~\cite{EFI18,XU18b,WU19,THO18,DA19,CLA19,EUG20,REP20,FER20,HUA20a}.

The topological heavy-fermion (THF) model~\cite{SON22} uses a completely different approach designed to naturally explain the coexistence of two seemingly contradictory electronic behaviors of TBG. On the one hand, scanning-tunnelling microscopy measurements~\cite{XIE19,WON20} have revealed quantum dot-like structures, pointing to a localized nature of the electronic states. On the other hand, various transport measurements clearly report a non-localized electronic behavior~\cite{CAO18,LU19,YAN19,SAI20,STE20}. By mapping the BM model to system of localized (heavy) lattice fermions hybridized with anomalous conduction (light) electrons, Ref.~\cite{SON22} solves the aforementioned contradiction without breaking any symmetries of TBG. 

Motivated by the prospect of applying pre-existing heavy-fermion machinery to the TBG problem~\cite{SI10,GEG08,COL84,DZE10,TSV83,WER06,LU13,WEN14,KOT06,EME92,FRE18,FUR94,CAS96,MAR97a,CHA95a,CHO22,SHI22a,HU23a,ZHO23,LAU23}, the purpose of this work is twofold. Firstly, by starting from the BM model~\cite{BIS11} and its approximation~\cite{BER21}, we derive analytical expressions for the THF model parameters in terms of the twist angle, interlayer tunneling amplitudes, and the electron-electron interaction potential. Secondly, we extend the analysis conducted in Ref.~\cite{SON22} at the magic angle and numerically obtain the parameters of the THF model across a large, experimentally-relevant phase space. The simple analytical approximations are seen to match the numerically obtained THF parameter to a relative error smaller than 30\% across the vast majority of the phase space we explore. Taken together, our complementary analytical and numerical analyses show the range of validity of the heavy-fermion model. We find that the flat and closest remote bands of TBG within the single-particle BM model Hamiltonian are extremely well-fitted by the THF model. However, the usefulness of the latter rests in the strict delineation of the local ($f$) and itinerant ($c$) fermion energy scales. This happens when the hybridization $\gamma$ between the two types of fermion species is smaller than the Hubbard onsite repulsion of the heavy fermions $U_1$ ({\it i.e.}{} $\abs{\gamma} < U_1$). This condition happens exactly around the magic angle and for realistic tunneling amplitude ratios ($0.6 \leq w_0 / w_1 \leq 1.0$).

This paper is organized as follows. We start by setting the notation and reviewing the BM model~\cite{BIS11}, its tripod and hexagon approximations~\cite{BER21}, and finally the THF model~\cite{SON22} in \cref{sec:model_review}. \cref{sec:analytics_single_particle} is devoted to obtaining analytical expressions for the parameters of the THF single-particle Hamiltonian from the BM model, while in \cref{sec:analytics_many_body}, we derive approximations for the THF interaction Hamiltonian parameters. Following the method derived in Ref.~\cite{SON22}, we numerically obtain the single-particle and interaction parameters of the THF model within a large phase space around the magic angle in \cref{sec:numerical_simulations}. We confirm the validity of our analytical expressions derived in \cref{sec:analytics_single_particle,sec:analytics_many_body} by comparing them against the numerical results. The main features of the THF model away from the magic angle, as well as its applicability as an effective model of TBG, are also discussed in \cref{sec:numerical_simulations}. In \cref{sec:inthamsym}, we derive the continuous symmetries of the THF interaction Hamiltonian arising under different limits. Finally, the overall conclusions of this work presented in \cref{sec:conclusions}.

\section{Models Review}\label{sec:model_review}
Our discussion starts with an overview of the BM model~\cite{BIS11} and its analytical approximations: the tripod and Hexagon models~\cite{BER21}. We further outline the THF model and the single-particle Hamiltonian in \cref{subsec:topological_HF}. Finally, we outline the projected Coulomb-interaction Hamiltonian in the THF basis. A unified review of the BM model, its approximations, and the THF model is provided in \cref{app:sec:BM_model,app:sec:HF_model,app:sec:HF_interaction}.

\subsection{The Bistritzer-Macdonald model and approximations}\label{subsec:BM_approx_models}

The BM model was originally introduced in Ref.~\cite{BIS11}. We will employ the same notation as the one used in Refs.~\cite{BER21,SON21,BER21b,BER21a,SON19,LIA21,XIE21}. For the single-layer graphene layer $l = \pm $, we denote by $\hat{c}^\dagger_{l,\mathbf{p},\alpha,s}$ the fermionic operator, which creates an electron of momentum $\mathbf{p}$, graphene sublattice $\alpha \in \{1,2\}$, and spin $s \in \{\uparrow,\downarrow\}$. The physics of TBG arises from the hybridization of the single-layer graphene electronic states in valleys $K$ and $K'$, which we denote by $\eta=+$ and $\eta=-$, respectively. The corresponding single-layer momenta, with the origin at the $\Gamma$ point of the single-layer graphene BZ, are given by $\mathbf{p} = \pm \mathbf{K}_l$, where $\mathbf{K}_l$ is the graphene $K$ point of layer $l$. We define vectors $\mathbf{q}_i = C_{3z}^{i-1}(\mathbf{K}_- - \mathbf{K}_+)$ for $i=1,2,3$, together with the moir\'e reciprocal lattice vectors $\mathbf{b}_{Mj} = \mathbf{q}_3 - \mathbf{q}_j$, $j=1,2$. Finally, we introduce the moir\'e reciprocal lattice $\mathcal{Q}_0 = \mathbb{Z}\mathbf{b}_{M1} + \mathbb{Z}\mathbf{b}_{M2}$, as well as the auxiliary lattices $\mathcal{Q}_{\pm}=\mathcal{Q}_0 \pm \mathbf{q}_1$. 

The BM model fermions are created by the operators $\hat{c}^\dagger_{\mathbf{k},\mathbf{Q}, \alpha, \eta, s} = \hat{c}^\dagger_{l, \eta \mathbf{K}_{l} + \mathbf{k} - \mathbf{Q}, \alpha, s}$~\cite{BIS11}, for valley $\eta$, momentum $\mathbf{k}$ [measured from the $\Gamma_M$ point of the moir\'e Brillouin Zone (MBZ)], and plane-wave $\mathbf{Q} \in \mathcal{Q}_{\pm}$. The BM model Hamiltonian in this basis reads as
\begin{equation}
    \hat{H}_{\textrm{BM}} = \sum_{\substack{\mathbf{k} \in \textrm{MBZ} \\ \mathbf{Q}, \mathbf{Q}^{\prime}}} \sum_{\substack{\eta, s \\ \alpha, \alpha^{\prime}}} h^{(\eta)}_{\mathbf{Q} \alpha, \mathbf{Q}^{\prime} \alpha^{\prime}} (\mathbf{k}) \hat{c}^\dagger_{\mathbf{k},\mathbf{Q},\alpha,\eta,s}\hat{c}_{\mathbf{k},\mathbf{Q}',\alpha',\eta,s},
\end{equation}
where the first-quantized Hamiltonian matrix $h^{(\eta)} (\mathbf{k})$ is given in \cref{eq:1VBM_hamiltonian}. The single-particle BM Hamiltonian $\hat{H}_{\textrm{BM}}$ depends on the single-layer graphene Dirac velocity $v_F$, as well as on the interlayer hopping amplitudes at the $AA$ and $AB$ stacking centers, denoted by $w_0$ and $w_1$, respectively. In general, as a result of lattice relaxation and corrugation effects, $0 \leq w_0 < w_1$~\cite{DAI16,JAI16,SON21,UCH14,WIJ15}. Unless mentioned otherwise, throughout this paper, we will rescale all the parameters according to~\cite{BER21}
\begin{equation}
    E \rightarrow \frac{E}{v_F k_{\theta}},\qquad \mathbf{k} \rightarrow \frac{\mathbf{k}}{k_{\theta}},
\end{equation}
where $k_{\theta} = |\mathbf{K}_+ - \mathbf{K}_-| = 2|\mathbf{K}_+|\sin{\frac{\theta}{2}}$. We employ $v_F = \SI{5.944}{\eV \angstrom}$, $\abs{\vec{K}}=\SI{1.703}{\angstrom^{-1}}$, and $w_1=\SI{110}{\milli\eV}$ in our numerical calculations.

Ref.~\cite{BER21} has argued that good approximations of the BM model eigenstates along the high-symmetry lines in the MBZ can be obtained by considering only a limited number of plane-wave states $\mathbf{Q}$. For example, the Tripod model considers only the four (ten for the two-shell approximation) closest plane-wave states around the $K_M$ point; similarly, the Hexagon model considers six plane-wave states around the $\Gamma_M$ point~\cite{BER21}. We will make extensive use of these two approximations~\cite{BER21} (see \cref{app:subsec:tripod,app:subsec:hexagon}) throughout this work in order to obtain analytical expressions for the THF parameters.

\subsection{The THF model}\label{subsec:topological_HF}
The single-particle THF model~\cite{SON22} was designed to capture the physics of the TBG low-energy bands around the charge neutrality point, while fully preserving the symmetries and topology of the BM model. It was first shown in Refs.~\cite{PO18c,SON19} that the active bands are subjected to a fragile topological obstruction~\cite{BRA17,BOU19}, while the whole model is anomalous~\cite{SON21} with particle-hole symmetry~\cite{SON19}. This, in turn, implies that no effective two-band lattice model of the active TBG bands that preserves their symmetry and topological properties can be constructed. Ref.~\cite{SON22} solves the topological obstruction and anomaly by introducing two types of electrons which are hybridized with one another: heavy $f$-electrons and itinerant (conducting) $c$-electrons (see \cref{app:sec:HF_model} for more details). The $f$-electrons are located at the $AA$-stacking sites of TBG, form a triangular lattice, and transform in the same way as $p_x \pm ip_y$ orbitals under the symmetry group of TBG. The $f$-electron states almost-completely span the BM active bands, except for a small region around the $\Gamma_M$ point, where they do not induce the correct irreducible representations (irreps)~\cite{SON22}. In order to match the irreps of the BM model, four conduction $c$-electron bands (in each valley and for each spin) are hybridized with the $f$-electrons near the $\Gamma_M$ point.

The fermion operators for the $f$-electron of the orbital $p_x + ip_y$ ($p_x - ip_y$) with the orbital quantum number $\alpha=1$ ($\alpha=2$), valley $\eta=\pm$, spin $s\in \{\uparrow,\downarrow\}$ at the lattice site $\mathbf{R}$ are denoted by $\hat{f}^\dagger_{\mathbf{R}, \alpha, \eta, s}$. Similarly, the $c$-electron of band $a \in \{1,2,3,4\}$, valley $\eta=\pm$, spin $s \in \{\uparrow,\downarrow\}$, and momentum $\mathbf{k}$ is given by $\hat{c}^\dagger_{\mathbf{k}, a, \eta, s}$. The BM model operators can be projected into the THF basis as~\cite{SON22}
\begin{align}
    \hat{c}^\dagger_{\mathbf{k},\mathbf{Q},\beta,\eta,s} &\approx \frac{1}{\sqrt{N}}\sum_{\alpha}\sum_{\mathbf{R}}e^{i\mathbf{k}\cdot\mathbf{R}} v^{(\eta)*}_{\mathbf{Q}\beta,\alpha}(\mathbf{k})\hat{f}^\dagger_{\mathbf{R},\alpha,\eta,s}  \nonumber \\ &+ \sum_a \tilde{u}^{(\eta)*}_{\mathbf{Q}\beta,a}(\mathbf{k})\hat{c}^\dagger_{\mathbf{k},a,\eta,s},
    \label{eq:main_projection_BM}
\end{align}
where $N$ is the number of moir\'e unit cells and the $f$-electron and $c$-electron wave functions are given by $v^{(\eta)}_{\mathbf{Q}\beta,\alpha}(\mathbf{k})$ and $\tilde{u}^{(\eta)}_{\mathbf{Q}\beta,a}(\mathbf{k})$, respectively. At the $\Gamma_M$ point, the $\hat{f}^\dagger_{\mathbf{k},\alpha,\eta,s}$ and $\hat{c}^\dagger_{\mathbf{k},a,\eta,s}$ electrons with $a = 1, 2$ transform according to two $\Gamma_3$ irreps, whereas the $\hat{c}^\dagger_{\mathbf{k},a,\eta,s}$ electrons with $a = 3, 4$ transform as the $\Gamma_1 \oplus \Gamma_2$ representation. The continuous real-space wave function of the $f$-electron orbital  $\alpha$ in layer $l=\pm$, sublattice $\beta=1,2$ and valley $\eta=\pm$ is denoted by $w_{l\beta,\alpha}^{(\eta)}(\vec{r})$ and is obtained by Fourier-transforming $v_{\mathbf{Q}\beta,\alpha}^{(\eta)}(\mathbf{k})$ according to \cref{eq:local_orbital_r2q}. For $l=+$, $\alpha = 1$, $\eta = +$ the $f$-electron wave functions in sublattice $\beta=1,2$ are given respectively by~\cite{SON22}
\begin{align}
    w^{(+)}_{+1,1}(\mathbf{r}) &= \frac{\alpha_1}{\sqrt{2}}\frac{1}{\sqrt{\pi\lambda_1^2}}e^{i\frac{\pi}{4}-\mathbf{r}^2/(2\lambda_1^2)}, \label{eq:main_wannier11}\\
    w^{(+)}_{+2,1}(\mathbf{r}) &= -\frac{\alpha_2}{\sqrt{2}}\frac{x+iy}{\lambda_2^2\sqrt{\pi}}e^{i\frac{\pi}{4}-\mathbf{r}^2/(2\lambda_2^2)}, \label{eq:main_wannier12}
\end{align}
with the other components being obtained from symmetry considerations~\cite{SON22}, as shown in \cref{app:subsec:local_orbitals}. In \cref{eq:main_wannier11,eq:main_wannier12}, the amplitudes $\alpha_1$ and $\alpha_2$ characterize the weight of the $f$-electrons in the two graphene sublattices, while $\lambda_1$ and $\lambda_2$ denote the corresponding wave function spreads~\cite{SON22}. 

\begin{figure}[t] 
\centering
\includegraphics[scale=1.0]{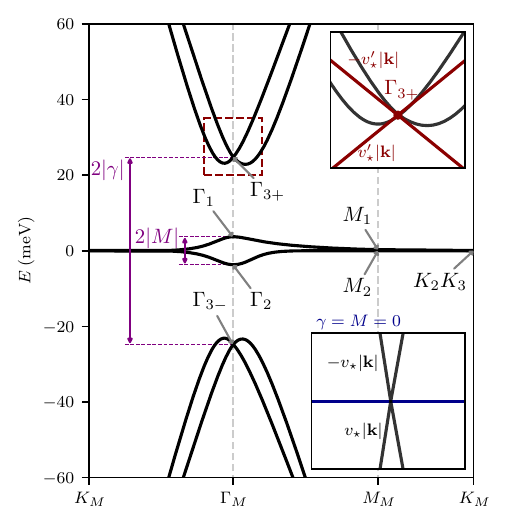}
\label{fig:main_1}
\caption{Schematic band structure of the THF model. The momentum-space irreps at the $\Gamma_M$, $M_M$, and $K_M$ points are indicated by gray arrows. The value $2\abs{\gamma}$ is the energetic splitting between the $\Gamma_{3\pm}$ irrep states at the $\Gamma_M$ point, respectively, while the value $2\abs{M}$ is the energetic splitting between the $\Gamma_{1,2}$ irrep states. We set the $f$-electron nearest-neighbor hopping ($t_0$) to zero. The top-right inset shows a zoomed-in plot of the dispersive bands around the $\Gamma_{3+}$ irrep (dashed square of the main plot). The corresponding Dirac cone has Fermi velocity $v_{\star}'$ (the linear dispersion of a Dirac cone with Fermi velocity $v_{\star}'$ is superimposed in red as guide to the eye). The bottom-right inset plot depicts the band structure of the THF model in the limit of $\gamma = M = 0$. In this limit, the doubly-degenerate Dirac cone with Dirac velocity $v_{\star}$ is formed by the remote bands. Away from the $\gamma = M = 0$ limit, $v_{\star}$ is also the group velocity of the conduction electrons away from charge neutrality.}
\end{figure}

The THF Hamiltonian can be written as~\cite{SON22}

{\small\begin{align}
    &\hat{H}_0 = \sum_{\eta,s}\biggl[\sum_{\left<\mathbf{R},\mathbf{R}'\right>}H_{\alpha,\alpha'}^{(f,\eta)}(\mathbf{R},\mathbf{R}')\hat{f}^\dagger_{\mathbf{R}',\alpha',\eta,s}\hat{f}_{\mathbf{R},\alpha,\eta,s} \nonumber \\
    &+ \sum_{a,\alpha}\sum_{\substack{|\mathbf{k}| < \Lambda_c \\ \mathbf{R}}}e^{i\mathbf{k}\cdot\mathbf{R}-|\mathbf{k}|^2\lambda^2/2}H^{(cf,\eta)}_{a,\alpha}(\mathbf{k})\hat{c}^\dagger_{\mathbf{k},a,\eta,s}\hat{f}_{\mathbf{R},\alpha,\eta,s} + \mathrm{h.c.} \nonumber \\
    &+ \sum_{a,a'}\sum_{|\mathbf{k}|<\Lambda_c}H^{(c,\eta)}_{a,a'}(\mathbf{k})\hat{c}^\dagger_{\mathbf{k},a,\eta,s}\hat{c}_{\mathbf{k},a',\eta, s}\biggr],
    \label{eq:main_HF_Hamiltonian}
\end{align}}where $\textrm{h.c.}$ denotes hermitian conjugation.
In \cref{eq:main_HF_Hamiltonian}, $\left<\mathbf{R},\mathbf{R}'\right>$ indicates that the sum runs over nearest-neighbor (NN) lattice sites; $\Lambda_c$ denotes the momentum cutoff for the conduction band electrons, for which the model only includes the low-energy states around the $\Gamma_M$ point~\cite{SON22}. The form of the valley-diagonal matrix elements $H_{\alpha,\alpha'}^{(f,\eta)}(\mathbf{R},\mathbf{R}')$, $H^{(c,\eta)}_{a,a'}(\mathbf{k})$, and $H^{(cf,\eta)}_{a,\alpha}(\mathbf{k})$ was obtained in Ref.~\cite{SON22} from symmetry considerations. The exponential suppression of the $f$-$c$ hybridization through the factor $e^{-|\mathbf{k}|^2\lambda^2 /2}$ was introduced empirically to account for the localized nature of the $f$-electrons, with $\lambda$ being related to the spreads $\lambda_{1,2}$ of the $f$-electrons from \cref{eq:main_wannier11,eq:main_wannier12} according to $\lambda = \sqrt{\lambda_1^2 + \lambda_2^2}$~\cite{SON22}.

The matrix elements describing the $f$-electron term were shown to be~\cite{SON22}
\begin{equation}
    H_{\alpha,\alpha'}^{(f,\eta)}(\mathbf{R},\mathbf{R}') = \begin{cases}
        -\mu\delta_{\alpha,\alpha'}, & \text{if }\mathbf{R}=\mathbf{R}' \\
        t_0\delta_{\bar{\alpha},\alpha'}, & \text{if } \mathbf{R} \text{ and } \mathbf{R}' \text{are NN.}
    \end{cases}
    \label{eq:main_f_hamiltonian}
\end{equation}
where $\bar{\alpha}=3-\alpha$. The parameter $t_0$ is the NN hopping amplitude of the local $f$-electrons and was set to zero in Ref.~\cite{SON22} (but could and will, in principle, differ from zero away from the magic angle), while $\mu$ is the chemical potential in the Grand-Canonical Ensemble.

The matrix elements governing the $c$-electrons are given by~\cite{SON22}
\begin{equation}
    H^{(c,+)}_{a,a^{\prime}}(\mathbf{k}) =
    \begin{pmatrix}
        0 & v_{\star}(k_x\sigma_0 + i k_y \sigma_z) \\
        v_{\star}(k_x\sigma_0 - i k_y \sigma_z) & M\sigma_x
    \end{pmatrix},
    \label{eq:main_c_hamiltonian}
\end{equation}
in valley $\eta=+$, while the matrix elements in valley $\eta=-$ can be obtained via time-reversal symmetry (see \cref{app:sec:HF_model}).

Finally, the $f$-$c$ hybridization term reads as~\cite{SON22}
\begin{equation}
    H^{(cf,\eta)}_{a,\alpha}(\mathbf{k}) =
    \begin{pmatrix}
    \gamma \sigma_0 + v^{\prime}_{\star}(\eta k_x \sigma_x  + k_y \sigma_y) \\
    v_{\star}^{\prime\prime}(\eta k_x \sigma_x  - k_y \sigma_y)
    \end{pmatrix}.
    \label{eq:main_fc_hamiltonian}
\end{equation}

The physical meaning of the parameters $M$, $\gamma$, $v_{\star}$, $v_{\star}'$ from \cref{eq:main_c_hamiltonian,eq:main_fc_hamiltonian} is outlined schematically in \cref{fig:main_1}. The bandwidth of the active bands is given by $2|M|$, while the splitting between the states that form the two $\Gamma_{3}$ irreps ({\it i.e.}{}, the splitting between the remote bands at the $\Gamma_M$ point) is given by $2|\gamma|$. In an abuse of notation, we will henceforth denote the $\Gamma_3$ irrep formed by the electron (hole) remote bands as $\Gamma_{3+}$ ($\Gamma_{3-}$). The Dirac velocity of the Dirac cone formed by the states which transform as  the $\Gamma_3$ irrep is given by $v_{\star}'$. Furthermore, in the limit $M=\gamma=0$ ({\it i.e.}{} in the so-called isotropic limit, $w_0 = w_1 = 1/\sqrt{13}$~\cite{BER21}), the dispersion of the higher bands becomes linear and the corresponding Dirac velocity is given by $v_{\star}$. The contribution of the NN $f$-electron hopping term $t_0$ to the dispersion of the active bands is very small around the magic-angle and thus, following Ref.~\cite{SON22}, is neglected in our analytical calculations, but not in our numerical analysis from \cref{sec:numerical_simulations}. Finally, the $v_{\star}''$ parameter, describing the coupling between the conduction $c$-electrons of the $a =3,4$ bands and the $f$-electrons is numerically computed to be smaller than $v_{\star}$ and $v_{\star}'$ (see \cref{app:sec:numerics}), and therefore is also neglected in our analytical calculations~\cite{SON22}.

\subsection{The interaction THF Hamiltonian}\label{subsec:Coulomb_HF}
The many-body THF model is obtained by adding the interaction term $\hat{H}_I$ to the single-particle Hamiltonian $\hat{H}_0$~\cite{SON22}: $\hat{H} = \hat{H}_0 + \hat{H}_I$. The interaction term $\hat{H}_I$ reads as~\cite{BER21a}
\begin{equation}
    \hat{H}_I = \frac{1}{2}\int d^2 \vec{r}_1 d^2 \vec{r}_2 V(\vec{r}_1 - \vec{r}_2) :\hat{\rho}(\vec{r}_1)::\hat{\rho}(\vec{r}_2):\;,
    \label{eq:main_interaction_general_hamiltonian}
\end{equation}
where the Fourier transformation of the Coulomb screened potential is given by~\cite{BER21a}
\begin{equation}
    V(\mathbf{q}) = (\pi U_{\xi} \xi^2) \frac{\tanh{(|\mathbf{q}|\xi/2)}}{|\mathbf{q}|\xi}.
    \label{eq:main_Coulomb}
\end{equation}
In \cref{eq:main_Coulomb}, $\xi$ is the screening length and $U_{\xi} = e^2/(\epsilon\xi)$ is the energy scale of the interaction (see \cref{app:subsec:Coulomb_interaction_general}).
The normal ordered density operator $:\hat{\rho}(\mathbf{r}):=\hat{\rho}(\mathbf{r}) - \bra{G_0}\hat{\rho}(\mathbf{r})\ket{G_0}$ is defined with respect to a state $\ket{G_0}$ at the charge neutrality point~\cite{SON22,BER21a}. In terms of the BM model real-space fermions $\hat{c}^\dagger_{l,\alpha,\eta,s}(\vec{r})$ (see \cref{app:sec:BM_model}), the density operator is naturally given by~\cite{SON22}
\begin{equation}
    \hat{\rho}(\mathbf{r}) = \sum_{l,\beta,\eta,s}\hat{c}^\dagger_{l,\alpha,\eta,s}(\vec{r})\hat{c}_{l,\alpha,\eta,s}(\vec{r}).
    \label{eq:main_density_definition}
\end{equation}
Taking the Fourier transformation of \cref{eq:main_density_definition} and projecting the BM model fermions into the THF model basis according to \cref{eq:main_projection_BM}, the interaction term $\hat{H}_I$ can be rewritten as the sum of the following contributions~\cite{SON22}
\begin{equation}
    \hat{H}_I = \hat{H}_U + \hat{H}_V + \hat{H}_W + \hat{H}_J + \hat{H}_{\tilde{J}} + \hat{H}_K.
    \label{eq:main_Coulomb_projected_in_BM}
\end{equation}
A full review of each of the contributions from \cref{eq:main_Coulomb_projected_in_BM} is provided in \cref{app:sec:HF_interaction}. Briefly, the $\hat{H}_U = \hat{H}_{U_1} + \hat{H}_{U_2}$ term comprises both the onsite $f$-electron orbital interaction $\hat{H}_{U_1}$, characterized by the $U_1$ parameter and the NN repulsion $\hat{H}_{U_2}$ contribution, whose energy scale is given by the $U_2$ parameter. Furthermore, the $\hat{H}_V$ term is the Coulomb repulsion between the conduction $c$-electrons governed by the interaction potential $V(\vec{q})$ given in \cref{eq:main_Coulomb}. The $\hat{H}_W$ term is the density-density interaction between the $f$- and $c$-electrons and is characterized by two parameters, $W_{1}$ and $W_{3}$. More specifically, the $W_1$ parameter quantifies the repulsion between the $f$-electrons and $c$-electrons forming the $\Gamma_3$ irrep, while $W_3$ corresponds to the interaction between the $f$-electrons and $\Gamma_1\oplus\Gamma_2$ $c$-electrons. $\hat{H}_J + \hat{H}_{\tilde{J}}$ is the sum of the $f$-$c$ exchange interaction and double hybridization ($f^{\dagger}f^{\dagger}cc$) terms. The strengths of both interactions are given by the $J$ parameter. Finally, the $\hat{H}_K$ term corresponds to the high-energy process of creating a particle (hole) in the $f$-electron bands and two holes (particles) and a particle (hole) in the conduction bands. As this is a high-energy process, we neglect the $\hat{H}_K$ term in what follows.

\section{Analytical Expressions: Single-Particle Hamiltonian}\label{sec:analytics_single_particle}

\begin{table*}[t]
    \centering
    \begin{tabular}{|l|p{6.5cm}|l|c|}
    \hline
    Parameter & Description & Approximation & Reference\\
    \hline
    \multirow{2}{*}{$\lambda_1$} &\multirow{2}{=}{\justifying  Spread of the $\hat{f}^\dagger_{\vec{R},1,+,\uparrow}$ orbital on the $\beta=1$ graphene sublattice} & $\lambda_1^{\text{1-shell}} = \sqrt{-2\ln{w_1}}$ & \cref{eq:main_lambda1_approx_final,eq:lambda1_alpha_ratio_1shell_approximation}\\
    \cline{3-4}
     & & $\lambda_1^{\textrm{2-shell}} = \sqrt{2\ln{\frac{1 - w_0^2}{w_1}}}$ & \cref{eq:tripod_lambda_2shell_approximation} \\
    \hline
    $\lambda_2$ & Spread of the $\hat{f}^\dagger_{\vec{R},1,+,\uparrow}$ orbital on the $\beta=2$ graphene sublattice & $\lambda_2 \approx \lambda_1$ & \cref{eq:main_lambda12_approx,eq:app_lambda12_approx}  \\
    \hline
    \multirow{2}{*}{$\frac{\alpha_1}{\alpha_2}$} & \multirow{2}{=}{\justifying Ratio of the amplitudes of the $f$-electron on the two graphene sublattices} & $\left(\frac{\alpha_1}{\alpha_2}\right)^{\text{1-shell}} = \frac{w_1}{w_0}\sqrt{-2\ln{w_1}}$ & \cref{eq:main_a1ra2_approx_final,eq:lambda1_alpha_ratio_1shell_approximation}\\
    \cline{3-4}
     & & $\left(\frac{\alpha_1}{\alpha_2}\right)^{\textrm{2-shell}} = \frac{w_1}{w_0} \sqrt{2\ln{\frac{1 - w_0^2}{w_1}}}$ & \cref{eq:tripod_lambda_2shell_approximation} \\
    \hline
    \multirow{2}{*}{$M$} & \multirow{2}{=}{Mass term of the conduction electrons} & $M^{\textrm{1-shell}} = \left|2w_1 - \sqrt{1 + w_0^2}\right|$ & \cref{eq:main_M_approximation,eq:M_app_approximation_one_shell}\\
    \cline{3-4}
     & & $M^{\textrm{2-shell}} = \left|\frac{-4 \sqrt{w_0^2+1} w_1+w_0^2+w_1^2+2}{2 \sqrt{w_0^2+1}}\right|$ & \cref{eq:M_approximation_two_shell,eq:main_M_approximation_two_shell} \\
    \hline
    $v_{\star}$ & Dirac velocity of the conduction electrons & $v_{\star}^{\textrm{hex}} = \sqrt{\frac{12}{13}}$ & \cref{eq:main_vstar_approx,eq:vstar_approx} \\
    \hline
    $\gamma$ & $f-c$ hybridization parameter & $\gamma^{\textrm{1-shell}} = \frac{1}{2}\left(\sqrt{4 + w_0^2} - \sqrt{9w_0^2 + 4w_1^2}\right)$ & \cref{eq:main_gamma_1shell,eq:app_gamma_1shell} \\
    \hline
    $v_{\star}^{\prime}$ & Dirac velocity of the $\Gamma_3$ states & $v_{\star}^{\prime \textrm{approx.}}$ & \cref{eq:approximation_vstarprime} \\
    \hline
    
    \end{tabular}
    \caption{Approximations of the single-particle parameters of the THF Hamiltonian. For each parameter, we list its description, its approximation(s), as well as the equation(s) where the corresponding approximations are derived. The expression for $v_{\star}^{\prime \textrm{approx.}}$ is rather cumbersome and is thus relegated to \cref{app:subsec:analytic_other_hexagon}.}
    \label{tab:summary_single_particle_params}
\end{table*}

In this section, we derive analytical expressions for the THF model parameters. The THF model was derived on general symmetry principles~\cite{SON22} and thus is expected to remain valid for a range of angles and tunneling amplitude ratios, up to a change in its parameters. To obtain the THF model parameters analytically, we match specific features of the TBG spectrum ({\it i.e.}{} either the eigenstate wave functions directly or the energy dispersion) within both the BM and THF models. More precisely, by approximating the $f$-electron wave functions $v_{\mathbf{Q}\beta,\alpha}^{(\eta)}$ from \cref{eq:main_projection_BM} around the $K_M$ point with the eigenstates of the Tripod model, we can directly compute the spreads $\lambda_1$, $\lambda_2$ and amplitudes (normalization prefactors) $\alpha_1$, $\alpha_2$ of the $f$-electron orbitals from \cref{eq:main_wannier11,eq:main_wannier12}. Next, we will use the Hexagon model states with the appropriate gauge-fixing conditions to approximate the $c$-electron wave functions $\tilde{u}_{\mathbf{Q}\beta,a}$ (see \cref{app:subsec:BM_Chern} for details on gauge fixing). This will enable us to obtain the conduction electron parameters $M$, $\gamma$, as well as the hybridization parameters $v_{\star}$, $v_{\star}'$ in terms of the $w_0$ and $w_1$ parameters of the BM model. The derivations are presented briefly here and summarized in \cref{tab:summary_single_particle_params}, while the detailed calculation are relegated to \cref{app:sec:analytic_single_particle}. The validity of these approximations will be assessed in \cref{sec:numerical_simulations} by comparing them with numerical results. Finally, in \cref{app:subsec:analytic_dirac_velocity} we also obtain an expression for the renormalized Dirac velocity $v_D^{\textrm{(THF)}}$ of the flat bands at the $K_M$ point starting from the THF model and compare to the expression derived from the BM model~\cite{BER21} in \cref{app:subsec:additional_numerics_single_particle}.

\subsection{Local \texorpdfstring{$f$}{f}-electrons and the Tripod model}\label{subsec:main_local_tripod}
We start our discussion by outlining the derivation of the wave function of the $f$-electron orbitals from the one-shell Tripod model approximation of the BM model~\cite{BER21}. A more detailed derivation together with the two-shell Tripod model approximation is provided in \cref{app:subsec:analytic_forbitals_tripod}.

As shown in \cref{eq:main_wannier11,eq:main_wannier12} the $f$-electron orbital wave functions are characterized by the spread parameters $\lambda_1$ and $\lambda_2$ and the sublattice amplitudes $\alpha_\beta$ (for sublattice $\beta=1,2$). By construction, the $f$-electron states are fully supported on the active TBG bands around the $K_M$ point~\cite{SON22}. As a result, the former are unitarily related to the latter. Under an appropriately fixed gauge (see \cref{app:subsec:analytic_forbitals_tripod} for the gauge fixing conditions), we can equate the $f$-electron wave function in momentum space and the active TBG band wave functions $U^{e_Y}_{\mathbf{Q}\beta,\eta}(\mathbf{k})$ expressed in the Chern band basis (see \cref{app:subsec:BM_Chern})~\cite{AHN19,HEJ21,BUL20,BER21,SON21},
\begin{equation}
    v^{(\eta)}_{\mathbf{Q}\beta,\alpha}(\mathbf{k}) \approx U^{e_Y}_{\mathbf{Q}\beta,\eta}(\mathbf{k}),
\end{equation}
where we require $(-1)^{\alpha+1}=\eta e_Y$ from symmetry. We obtain the approximate Chern band TBG wave functions $U^{e_Y}_{\mathbf{Q}\beta,\eta}(\mathbf{k})$ from the one-shell Tripod model, while the $f$-electron wave functions $v^{(\eta)}_{\mathbf{Q}\beta,\alpha}(\mathbf{k})$ are derived from the Fourier transformation of the real-space wave functions in \cref{eq:main_wannier11,eq:main_wannier12} (see \cref{app:subsec:analytic_forbitals_tripod}).
We focus on the momentum $\mathbf{k}=\mathbf{q}_1$ corresponding to the $K_M$ point and the plane wave states corresponding to $\mathbf{Q} = \mathbf{q}_1,2\mathbf{q}_1$, for which we derive in \cref{app:subsec:analytic_forbitals_tripod} that the ratio of wave function components is
\begin{align}
    -iw_1 &= \frac{U^{+1}_{2\mathbf{q}_1,1,+}(\mathbf{q}_1)}{U^{+1}_{\mathbf{q}_1,1,+}(\mathbf{q}_1)} = \frac{v^{(+)}_{2\mathbf{q}_1 1,1}(\mathbf{q}_1)}{v^{(+)}_{\mathbf{q}_1 1,1}(\mathbf{q}_1)} = -ie^{-\frac{\lambda_1^2}{2}}, \label{eq:main_w1_lambda1} \\
    iw_0 &= \frac{U^{+1}_{2\mathbf{q}_1,2,+}(\mathbf{q}_1)}{U^{+1}_{\mathbf{q}_1,1,+}(\mathbf{q}_1)} = \frac{v^{(+)}_{2\mathbf{q}_1 2,1}(\mathbf{q}_1)}{v^{(+)}_{\mathbf{q}_1 1,1}(\mathbf{q}_1)} = i\frac{\alpha_2\lambda_2^2}{\alpha_1\lambda_1}e^{-\frac{\lambda_2^2}{2}}.
    \label{eq:main_w0_HF}
\end{align}
We thus obtain two equations linking the renormalized BM model parameters $w_0$, $w_1$ and the $f$-electron parameters $\lambda_1$, $\lambda_2$, $\alpha_1$, $\alpha_2$. One other equation can be obtained from \cref{eq:main_w1_lambda1,eq:main_w0_HF}, for which the normalization condition reads as
\begin{equation}
    \alpha_1^2 + \alpha_2^2 = 1.
    \label{eq:main_normalization_amplitudes}
\end{equation}
Finally, the spreads of the $f$-electron wave functions in the two sublattices can be assumed to be almost equal, {\it i.e.}{}
\begin{equation}
    \lambda_1 \approx \lambda_2.
    \label{eq:main_lambda12_approx}
\end{equation}
This approximation is verified and confirmed to hold numerically within 20\% relative error in \cref{app:sec:numerics}. By solving \cref{eq:main_w0_HF,eq:main_w1_lambda1,eq:main_normalization_amplitudes,eq:main_lambda12_approx}, we obtain in the one-shell Tripod model approximation of the $f$-electron wave function parameters,
\begin{align}
    \lambda_1^{\text{1-shell}} &= \sqrt{-2\ln{w_1}}, \label{eq:main_lambda1_approx_final} \\ 
    \left(\frac{\alpha_1}{\alpha_2}\right)^{\text{1-shell}} &= \frac{w_1}{w_0}\sqrt{-2\ln{w_1}}, \label{eq:main_a1ra2_approx_final}
\end{align}
which are expressed in units of $1/k_{\theta}$. Note that the decay length $\lambda$ in the hybridization term of the THF Hamiltonian in \cref{eq:main_HF_Hamiltonian} is given by $\lambda = \sqrt{\lambda_1^2 + \lambda_2^2}$ and therefore can be approximated as
\begin{equation}
    \lambda^{\text{1-shell}} = 2\sqrt{-\ln{w_1}}.
\end{equation}
Including the second shell in the Tripod model~\cite{BER21}, we can obtain a further approximation $\lambda_1^{\textrm{2-shell}}$, as discussed in \cref{app:subsec:analytic_forbitals_tripod}. The expression $\lambda_1^{\textrm{2-shell}}$ is not defined in the entire parameter space $(\theta, w_0/w_1)$. Nevertheless, within the domain it \emph{is} defined, it approximates the numerically-obtained values within a 30\% relative error, as we demonstrate in \cref{app:subsec:additional_numerics_single_particle}.

\subsection{Conduction \texorpdfstring{$c$}{c}-electrons and the Hexagon model}\label{subsec:main_conduction_hexagon}
We now derive the parameters pertaining to the $c$-electron Hamiltonian $H^{(cf,\eta)}_{a,\alpha}(\mathbf{k})$. As can be seen from \cref{fig:main_1}, the energetic splitting within the THF model between the states transforming as the $\Gamma_{1}$ and $\Gamma_{2}$ irreps is $2\abs{M}$. According to Ref.~\cite{BER21}, the same energetic splitting can be obtained within the Hexagon model. In order to match the energy spectra within the BM and THF models, we must have

\begin{equation}
    M^{\textrm{1-shell}} = \left|2w_1-\sqrt{1+w_0^2}\right|.
    \label{eq:main_M_approximation}
\end{equation}
A better approximation can be obtained by considering the two-shell hexagonal model derived in Ref.~\cite{BER21}. Again, by matching the energy spectra at the $\Gamma_M$ point (see \cref{app:subsec:analytic_cbands_hexagon}), we obtain another approximation
\begin{equation}
    M^{\textrm{2-shell}} = \left|\frac{-4 \sqrt{w_0^2+1} w_1+w_0^2+w_1^2+2}{2 \sqrt{w_0^2+1}}\right|,
    \label{eq:main_M_approximation_two_shell}
\end{equation}
which will be used further in \cref{sec:numerical_simulations}.

In the absence of the $f$-$c$ electron hybridization ({\it i.e.}{} $\gamma=v_{\star}'=0$), the band structure has a parabolic band touching point at the $\Gamma_M$ point~\cite{SON22}. If, moreover, we set $M=0$, a Dirac cone emerges at the $\Gamma_M$ point whose Dirac velocity is given by $v_{\star}$ (see the inset in \cref{fig:main_1}). Analogously, within the hexagon approximation of the BM model, in the isotropic limit $w_0=w_1=\frac{1}{\sqrt{3}}$, the band structure develops a Dirac cone at the $\Gamma_M$ point, with the Dirac velocity obtained in Ref.~\cite{BER21}. Comparing the dispersion relation within the BM and THF models (see \cref{app:subsec:analytic_cbands_hexagon}), we can find an approximation of the $v_{\star}$ THF parameter
\begin{equation}
    \label{eq:main_vstar_approx}
    v_{\star}^{\textrm{hex}} = \sqrt{\frac{12}{13}}.
\end{equation}
The validity of this approximation will be assessed in \cref{sec:numerical_simulations}.

\subsection{Hybridization terms}\label{subsec:main_hybridization}
Finally, we derive the hybridization term parameters $\gamma$ and $v_{\star}'$ from the BM model. We note the energetic splitting between the four states forming the two $\Gamma_3$ irreps (which are related to each other by particle-hole symmetry and have opposite eigenenergies) is given by $2 \abs{\gamma}$ within the THF model. This energetic splitting can be also obtained from the one-shell Hexagon model~\cite{BER21}, allowing us to approximate $\gamma$ as
\begin{equation}
    \label{eq:main_gamma_1shell}
    \gamma^{\textrm{1-shell}} = \frac{1}{2}\left(\sqrt{4 + w_0^2} - \sqrt{9w_0^2 + 4w_1^2}\right).
\end{equation}

As seen in \cref{fig:main_1}, at the $\Gamma_M$ point, the states corresponding to the $\Gamma_3$ irreps form a Dirac cone, a fact that can be proved by performing a $\mathbf{k}\cdot\mathbf{p}$ expansion of the THF Hamiltonian~\cite{SON22}. The Dirac velocity of the $\Gamma_3$ states is given by the $v_{\star}'$ parameter, as shown in the top inset of \cref{fig:main_1} (without the $v_{\star}'$ parameter, the remote bands would be degenerate; $v_{\star}'$ gives the velocity of the ``Rashba''-like point). By performing the $\mathbf{k}\cdot\mathbf{p}$ expansion of the one-shell Hexagon model Hamiltonian (see \cref{app:subsec:analytic_other_hexagon}), one can obtain the Dirac velocity of the same bands from within the BM model, thus yielding an expression for $v_{\star}'$ in terms of $w_0$ and $w_1$. The resulting expression $v_{\star}'^{\textrm{approx.}}$ is rather bulky and is thus relegated to \cref{app:subsec:analytic_other_hexagon}.

\section{Analytical Expressions: Interaction Hamiltonian}\label{sec:analytics_many_body}

\begin{table*}[t]
    \centering
    \begin{tabular}{|l|p{4cm}|p{8.5 cm}|c|}
    \hline
    Parameter & Description & Approximation & Reference\\
    \hline
    $U_1$ & Density-density interaction of the $f$-electrons & $U_1^{\textrm{approx.}} = \frac{\xi U_{\xi}}{\lambda_1} \left[\frac{41}{48}\sqrt{\frac{\pi}{2}} - 2 \frac{\lambda_1}{\xi}\ln{2} + 2(1 + \alpha_2^2) \left(\frac{\lambda_1}{\xi}\right)^3\frac{3}{4}\zeta(3)\right]$ & \cref{eq:main_U1_final,eq:U1_final}  \\
    \hline
    \multirow{3}{*}{$W_1$, $W_3$} &\multirow{3}{=}{\justifying  Density-density interaction between the $c$-electrons and the $f$-electrons} & $W_1^{\textrm{1st approx.}} = \frac{2\pi}{\sqrt{3}} \left( \frac{\xi}{a_M} \right)^2$ & \multirow{2}{*}{\cref{eq:main_W1_W3_0th,eq:fc_interaction_strength_approx_13}}\\
    \cline{3-3}
    & & $W_3^{\textrm{1st approx.}} = \frac{2\pi}{\sqrt{3}} \left( \frac{\xi}{a_M} \right)^2$ & \\
    \cline{3-4}
    & & $W_1^{\textrm{2nd approx.}} = \frac{2\pi}{\sqrt{3}} \left( \frac{\xi}{a_M} \right)^2 U_{\xi} - \frac{n_f(\mathbf{b}_{M1})V(\mathbf{b}_{M1})}{\Omega_0}$ & \cref{eq:main_W1_W3_1st,eq:W1_W3_1st} \\
    \hline
    $V(\mathbf{q})$ & Density-density interaction of the $c$-electrons & $V^{\textrm{approx.}} = \frac{2\pi}{\sqrt{3}} \left( \frac{\xi}{a_M} \right)^2 U_{\xi}$ & \cref{eq:main_V,eq:interaction_cc_approximate_equality}  \\
    \hline
    \end{tabular}
    \caption{Approximations of the many-body parameters of the THF Hamiltonian. Similarly to \cref{tab:summary_single_particle_params}, we list the description of each parameter, its approximation(s), as well as the equation(s) where the corresponding approximations are derived.}
    \label{tab:summary_many_body_params}
\end{table*}

As discussed in \cref{subsec:Coulomb_HF}, the interaction terms of the THF model Hamiltonian are fully characterized by six parameters, namely $U_{1,2}$, $W_{1,3}$, $V$ and $J$. In the same spirit as in \cref{sec:analytics_single_particle}, we approximate the THF model wave functions for the $f$- and $c$-electrons and derive analytical expressions for the following interaction parameters: $U_{1}$, $W_{1,3}$ and $V$ (see \cref{app:sec:analytic_many-body} for the detailed derivation). We start with the onsite $f$-electron density-density interaction and use the Gaussian approximations of the orbital wave functions to compute the parameter $U_1$. Note that the parameter $U_2$ is obtained directly from $U_1$~\cite{SON22}, as outlined in \cref{app:subsec:analytic_interaction_ff}. Next, we use the conduction $c$-electron states to derive the $f$-$c$ density-density interaction parameters $W_{1,3}$. By approximating the Coulomb potential $V(\mathbf{q}) \approx V(\vec{0})$ we also obtain the effective $c$-$c$ density-density interaction parameter $V$. The $J$ parameter was not obtained analytically in this paper, as its expression is very cumbersome, however we outline the method in \cref{app:subsec:analytic_exchange_interaction}. Finally, analytical expressions for the THF form factors are derived and in \cref{app:sec:form_factors}.

\subsection{\texorpdfstring{$f$-$f$}{f-f} density-density interaction strength}
The $U_1$ parameter is the strength of the onsite $f$-electron repulsion. In momentum space, it can be written as (see \cref{app:subsec:analytic_interaction_ff})~\cite{SON22},
\begin{equation}
    U_1 = \int\frac{\dd^2{\mathbf{q}}}{(2\pi)^2}V(\mathbf{q})\abs{n_f(\mathbf{q})}^2,
    \label{eq:main_U1}
\end{equation}
where $V(\mathbf{q})$ is the Fourier transformation of the Coulomb screening potential from \cref{eq:main_Coulomb} and $n_f(\mathbf{q})$ is the Fourier transformation of the Wannier states density 
\begin{equation}
    n_f(\mathbf{r}) = \sum_{l,\beta}|w_{l\beta,\alpha}^{(\eta)}(\mathbf{r})|^2.
    \label{eq:main_Wannier_density}
\end{equation}
The explicit form of $n_f(\mathbf{r})$ is given in \cref{eq:wannier_density_full} and depends on the $\lambda_{1,2}$ and $\alpha_{1,2}$ parameters, but does not depend on the $\alpha$ index. Plugging the expression of $V(\mathbf{q})$ from \cref{eq:main_Coulomb} into \cref{eq:main_U1}, we can reduce the integral to the following expression
\begin{equation}
    U_1 = \xi U_{\xi}\int_0^{+\infty}\dd{q}\tanh{\left(\frac{\xi q}{2}\right)}\abs{n_f(q)}^2.
\end{equation}
We then use the series expansion of the hyperbolic tangent in order to express the integral as an infinite sum of Gaussian integrals and evaluate each term in the series separately. Taking the limit $\xi / \lambda_1 \gg 1$, which is justified for the typical experimental setup ($\xi \sim \SI{10}{\nano \meter}$, while $\lambda_1 \sim \SI{2.35}{\nano \meter}$), and only considering the first two leading contributions, we obtain
\begin{align}
    U_1 &\approx \frac{\xi U_{\xi}}{\lambda_1}\biggl[\frac{41}{48}\sqrt{\frac{\pi}{2}} - 2\sum_{k=1}^{\infty}(-1)^k\left(\frac{1}{a(k)} - \frac{1+\alpha_2^2}{a(k)^3}\right)\biggr].
    \label{eq:main_U1_approximation}
\end{align}

In \cref{eq:main_U1_approximation}, we also used the approximation $\lambda_1\approx\lambda_2$. The two series in \cref{eq:main_U1_approximation} are well-known convergent series and, by evaluating them, we eventually obtain
\begin{align}
    U_1^{\textrm{approx.}} &= \frac{\xi U_{\xi}}{\lambda_1}\biggl[\frac{41}{48}\sqrt{\frac{\pi}{2}} - 2 \frac{\lambda_1}{\xi}\ln{2} \nonumber \\ &+ (1+\alpha_2^2)\frac{3}{2} \left(\frac{\lambda_1}{\xi}\right)^3\zeta(3)\biggr],
    \label{eq:main_U1_final}
\end{align}
where $\zeta(x)$ is the Riemann zeta function and $\zeta(3) \approx 1.2$. \cref{eq:main_U1_final} depends on two single-particle parameters: $\lambda_1$ and $\alpha_2$. As will be shown in \cref{sec:numerical_simulations}, we find that using the one-shell tripod approximations $\lambda^{\textrm{1-shell}}$ and $(\alpha_1/\alpha_2)^{\textrm{1-shell}}$ from \cref{eq:main_lambda1_approx_final,eq:main_a1ra2_approx_final}, together with the normalization condition $\alpha_1^2 + \alpha_2^2 = 1$, \cref{eq:main_U1_final} gives an excellent agreement ({\it i.e.}{} within 30\% relative error for almost the entire parameter space we consider) with the numerically calculated $U_1$ value.

\subsection{\texorpdfstring{$f$-$c$}{f-c} density-density interaction strength}
\label{subsec:main_fc_interaction}
The $W_1$ and $W_3$ parameters describe the $f$-$c$ density-density interaction term $\hat{H}_W$, where $W_1$ ($W_3$) corresponds to the Coulomb interaction between the $\Gamma_3$ ($\Gamma_1\oplus\Gamma_2$) conduction $c$-electrons and the Wannier $f$-electrons (see \cref{app:subsec:review_fc_interaction})~\cite{SON22}. As such, the parameters could, in principle, be obtained analytically by calculating the overlap between the Wannier and conduction electron densities weighted by the Coulomb potential in momentum space. This entails a summation over reciprocal lattice vectors $\vec{G}$ of the form~\cite{SON22}:
\begin{equation}
    \small
    \frac{1}{\Omega_0} \sum_{l,\beta} \sum_{\mathbf{Q}\in \mathcal{Q}_{l\eta}} \sum_{\mathbf{G}} n_f(\mathbf{G}) V(\mathbf{G})   \tilde{u}^{(\eta)*}_{\mathbf{Q}\beta,a}(\vec{0}) \tilde{u}^{(\eta)}_{\mathbf{Q}+ \mathbf{G} \beta,a'}(\vec{0}),
    \label{eq:main_W1W3_calculation}
\end{equation}
where setting $a=a'=1,2$ gives the $W_1$ parameter and $a=a'=3,4$, the $W_3$ parameter. 

We know, however, that the Wannier states and conduction electron wave functions decay exponentially in momentum space~\cite{BER21} and, therefore, a good approximation would be to terminate the series at the $\vec{G} = 0$ term. In this way we obtain (see \cref{app:subsec:analytic_interaction_fc})
\begin{equation}
    \footnotesize
    W_1^{\textrm{1st approx.}} = W_3^{\textrm{1st approx.}} = W \equiv \frac{2\pi}{\sqrt{3}} \left( \frac{\xi}{a_M} \right)^2 U_{\xi}.
    \label{eq:main_W1_W3_0th}
\end{equation}

As will be shown numerically in \cref{sec:numerical_simulations}, across a large parameter space, $W_1$ differs from $W_3$ by a relatively small amount (see \cref{app:sec:tables}). In order to capture this difference analytically, for the $W_1$ parameter we take one further order in the summation over reciprocal lattice vectors $\vec{G}$, such that $|\vec{G}| \leq |\vec{b}_{M1}|$. By using the $C_{3z}$-symmetry properties of the conduction electron wave functions, we obtain
\begin{equation}
    \small
     W_1^{\textrm{2nd approx.}} = \frac{2\pi}{\sqrt{3}} \left( \frac{\xi}{a_M} \right)^2 U_{\xi} - \frac{n_f(\mathbf{b}_{M1})V(\mathbf{b}_{M1})}{\Omega_0},
    \label{eq:main_W1_W3_1st}
\end{equation}
where $n_f(\mathbf{q})$ is the Fourier transformation of the Wannier density from \cref{eq:main_Wannier_density} and $V(\mathbf{q})$ is the Fourier transformation of the Coulomb potential, as given by \cref{eq:main_Coulomb}.

\subsection{\texorpdfstring{$c$-$c$}{c-c} density-density interaction strength}
 The Coulomb interaction between the conduction electron densities $\hat{H}_V$ is governed by the interaction matrix elements proportional to $V(\mathbf{q})\delta_{a_1,a_1'}\delta_{a_2,a_2'}$ (see \cref{app:subsec:review_cc_interaction})~\cite{SON22}.
In \cref{eq:main_HF_Hamiltonian}, we impose a momentum cutoff for the $c$-electrons, as we are only interested in the low-energy physics of the system. As such, we only consider scattering processes with a small magnitude of the momentum transfer $\mathbf{q}$. For these processes, the Coulomb potential does not deviate from its value at zero momentum $V(\mathbf{q}) \approx V(\vec{0})$ and we could approximate the $c$-$c$ density-density interaction as being independent of the momentum transfer and being governed by a single parameter
\begin{equation}
    V^{\textrm{approx.}} = \frac{1}{\Omega_0}V(\vec{0}) = W = \frac{2\pi}{\sqrt{3}} \left( \frac{\xi}{a_M} \right)^2 U_{\xi},
    \label{eq:main_V}
\end{equation}
where $W$ was defined in \cref{eq:main_W1W3_calculation}. In \cref{app:subsec:analytic_interaction_cc}, we estimate the validity of this approximation.

\section{Numerical Simulations}\label{sec:numerical_simulations}

\begin{figure}
\centering
\includegraphics[scale=1.0]{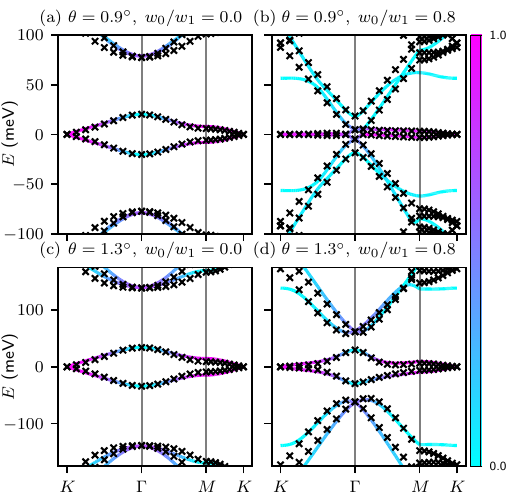}
\caption{Band structures of the BM and THF models at various angles and tunneling amplitude ratios around the magic angle. The BM and THF model band structures are shown by lines and crosses, respectively. Additionally, the BM model bands are colored according to the weight of the $f$-electrons on them. We indicate the angle and tunneling amplitude ratio above each panel.}
\label{fig:main_band_structures}
\end{figure}

\begin{figure*}[t!]
    \centering
    \includegraphics[width=\textwidth]{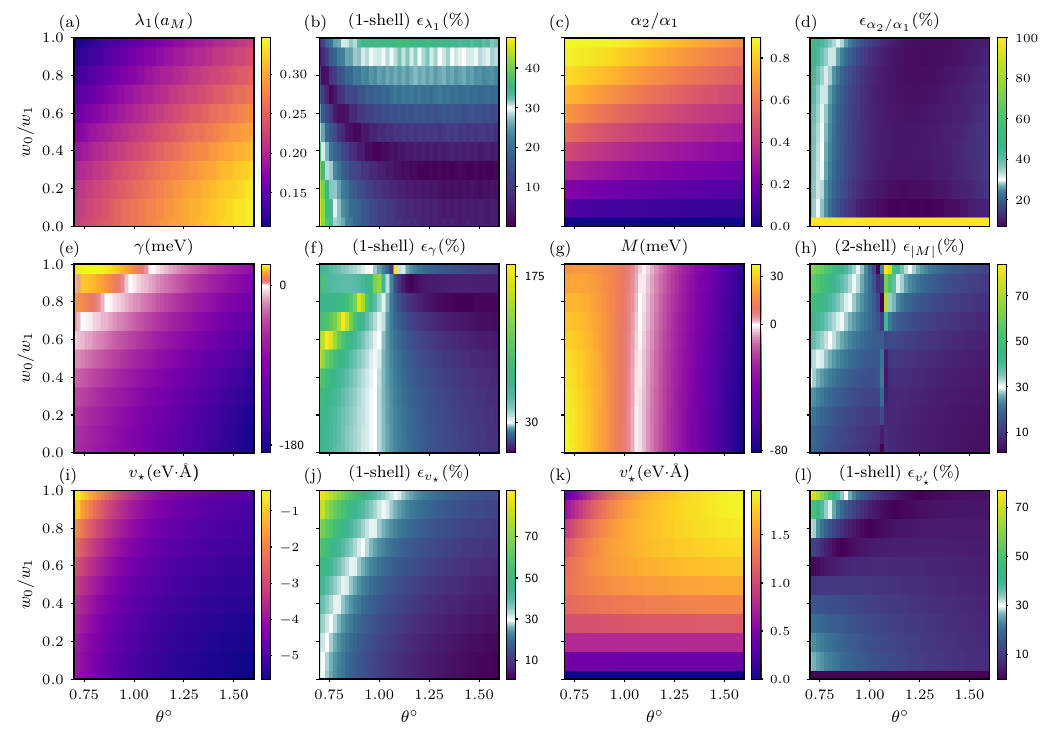}
    \subfloat{\label{fig:main_sp_params:a}}\subfloat{\label{fig:main_sp_params:b}}\subfloat{\label{fig:main_sp_params:c}}\subfloat{\label{fig:main_sp_params:d}}\subfloat{\label{fig:main_sp_params:e}}\subfloat{\label{fig:main_sp_params:f}}\subfloat{\label{fig:main_sp_params:g}}\subfloat{\label{fig:main_sp_params:h}}\subfloat{\label{fig:main_sp_params:i}}\subfloat{\label{fig:main_sp_params:j}}\subfloat{\label{fig:main_sp_params:k}}\subfloat{\label{fig:main_sp_params:l}}\caption{Numerical results for the THF single-particle parameters for different values of $0.0 \leq w_0/w_1 \leq 1.0$ and $\SI{0.7}{\degree} \leq \theta \leq \SI{1.6}{\degree}$ and the relative error of the analytic approximation of the corresponding parameter. (a), (c), (e), (g), (i), and (k) depict the simulated parameter in the units indicated on top of the corresponding panel. In (e) and (g), we indicate the zero-energy level in white. (b), (d), (f), (h), (j), and (l) depict the relative error between the numerical results and the analytical approximations of the THF parameters obtained within the one-shell tripod ($\lambda_1$, $\alpha_2/\alpha_1$), one-shell hexagon ($\gamma$, $v_{\star}$, $v_{\star}'$), and two-shell hexagon ($M$) approximations (see \cref{app:subsec:additional_numerics_single_particle} for other approximations). The color indicates the relative percentage error $\epsilon$. The 30\% relative error threshold is indicated by a white line in the color map. Note the divergence of the relative error when the parameter corresponding parameter approaches zero. The figure indicates that for a wide range of parameters, the agreement between the analytical approximations and the numerically obtained values is within $30\%$.}
    \label{fig:main_sp_params}
\end{figure*}

\begin{figure}
    \centering
    \includegraphics[]{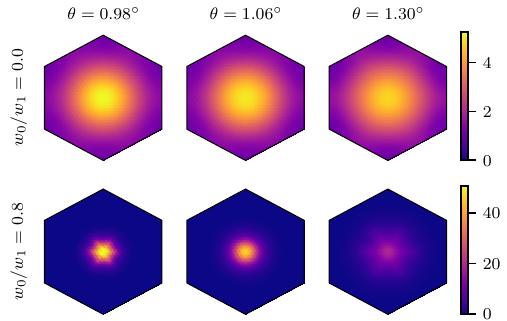}
    \caption{The Berry curvature of the THF Chern-$(+1)$ band for different twist angles and tunneling amplitude ratios. The parameters used are indicated at the top and left parts of the figure. The Berry curvature is plotted in the $\Gamma_M$-centered first Brillouin zone.}
    \label{fig:main_berry}
\end{figure}

In this section, we numerically compute the THF model single-particle and interaction parameters for different values of the twist angle ($\SI{0.7}{\degree} \leq \theta \leq \SI{1.6}{\degree}$), tunneling amplitude ratio ($0.0 \leq w_0/w_1 \leq 1.0$), and screening length ($\SI{2}{\nano\meter} \leq \xi \leq \SI{50}{\nano\meter}$). We follow the same procedure as in Ref.~\cite{SON22}, which relies on calculating the $f$-electron wave functions $v^{(\eta)}_{\mathbf{Q}\beta,\alpha}(\mathbf{k})$ by projecting Gaussian $p_x \pm i p_y$ orbitals located at the $1a$ Wyckoff position on the six bands of TBG near charge neutrality and building Maximally Localized Wannier Functions~\cite{MAR97b,SOU01a,PIZ20} using the Wannier90 software~\cite{PIZ20} (see \cref{app:subsec:numeric_details}). The conduction $c$-electron wave functions $\tilde{u}^{(\eta)}_{\mathbf{Q}\beta,a}(\mathbf{k})$ are then obtained by taking the orthogonal complement to the $f$-electron states at the $\Gamma_M$ point, and extended in its vicinity by approximating $\tilde{u}^{(\eta)}_{\mathbf{Q}\beta,a}(\mathbf{k}) \approx \tilde{u}^{(\eta)}_{\mathbf{Q}\beta,a}(\vec{0})$. As such, all the parameters of the single-particle THF model can be computed by calculating the corresponding expectation values of the BM-model Hamiltonian between the $f$- or $c$-electron states. The main results are presented in \cref{subsec:numerical_single_particle}, with additional details being relegated to \cref{app:subsec:additional_numerics_single_particle}. 

The interaction model parameters ($U_1$, $W_{1,3}$, $J$) are obtained by numerically evaluating the corresponding Coulomb integrals from the \cref{tab:Coulomb_summary}, using the same method employed by Ref.~\cite{SON22}. The interaction Hamiltonian parameters are presented in \cref{subsec:numerical_many_body} for $w_0/w_1 = 0.8$, with \cref{app:subsec:additional_numerics_many-body} containing the results at other tunneling amplitude ratios.

For both the single-particle and interaction parameters, we investigate the agreement between the analytical expressions obtained previously in \cref{sec:analytics_single_particle,sec:analytics_many_body} and the numerical result. To quantify the discrepancy, for any parameter $X$ for which $X_{\textrm{an}}$ denotes the analytically-obtained value and $X_{\textrm{num}}$ is the simulation result, we provide the relative error, $\epsilon_X = \Delta_X / \max{(X_{\textrm{num}}, X_{\textrm{an}})}$, where $\Delta_X = |X_{\textrm{num}} - X_{\textrm{an}}|$ is the absolute error. We note that $\epsilon_X$ is not always a reliable indicator of the agreement between the analytical expressions and the numerical results due to its divergence whenever the corresponding parameter $X$ vanishes. Finally, in light of our analytical and numerical results, we discuss the THF model parameters, their approximations, and the validity of the THF model itself away from the magic angle in \cref{subsec:discussion_away_ma}.

\subsection{Single-particle THF parameters}\label{subsec:numerical_single_particle}

The single-particle THF model was derived based on general symmetry principles~\cite{SON22}. As such, it is naturally expected to reproduce the BM model band structure at various twist angles and tunneling amplitude ratios, by appropriately changing the THF single-particle parameters. In \cref{fig:main_band_structures}, we compare the BM and THF model band structures at two angles ($\theta = \SI{0.90}{\degree}$ and $\theta = \SI{1.30}{\degree}$) around the magic angle, for $w_0/w_1 = 0.0$ ({\it i.e.}{} in the chiral limit~\cite{TAR19}), as well as for the realistic $w_0/w_1=0.8$. We find that, although initially devised for magic angle TBG~\cite{SON22}, the THF model captures the BM model band structure near charge neutrality remarkably well, even \emph{away} from the magic angle. This is also confirmed by our more detailed analysis in \cref{app:sec:tables}, where the band structures of the two models are compared for different twist angles across a larger interval. 

Having established that the THF model accurately reproduces the BM model band structure even away from the magic angle, we now proceed to show the numerically computed single-particle parameters $\lambda_1$, $\alpha_2/\alpha_1$, $\gamma$, $M$, $v_{\star}$ and $v_{\star}'$ as a function of the twist angle $\theta$ and the tunneling amplitude ratio $w_0/w_1$ in \cref{fig:main_sp_params}. For every parameter $X$, we also plot the relative error $\epsilon_X$ between the numerical value and its best analytical estimation. For simplicity, only the approximations presented in \cref{sec:analytics_single_particle} are compared here with the numerical results. The validity of the other analytical approximations derived in \cref{app:sec:analytic_single_particle} is assessed in \cref{app:subsec:additional_numerics_single_particle}.

Assuming $\lambda_1 \approx \lambda_2$ (see the discussion in \cref{app:subsec:additional_numerics_single_particle} for justification), we compare the numerical result for the $\lambda_1$ and $\alpha_2/\alpha_1$ parameters with the same quantities obtained from the one-shell Tripod model approximation, as given by \cref{app:subsec:analytic_forbitals_tripod}. For the $\gamma$, $v_{\star}$ and $v_{\star}'$ parameters, we compare the numerical results with the approximations obtained from one-shell Hexagon model (see \cref{app:subsec:analytic_other_hexagon,app:subsec:analytic_cbands_hexagon}). For the $M$ parameter, however, we use the two-shell hexagon approximation (derived in \cref{app:subsec:analytic_cbands_hexagon}), since it gives a better agreement. We refer the reader to \cref{app:sec:numerics} for a discussion of other numerical results that include the $f$-electron hopping amplitude $t_0$, the total weight of the Wannier states on the active bands $\mathcal{W}$, the spread $\lambda_2$ of the $\alpha=2$ orbital, and relative errors for the $\lambda^{\textrm{2-shell}}$ (derived in \cref{app:subsec:analytic_forbitals_tripod}) and $M^{\textrm{1-shell}}$ (derived in \cref{app:subsec:analytic_cbands_hexagon}) approximations. 

The behavior of the numerically obtained single-particle THF model parameters can be explained on general grounds based on the BM model. First, we note that the $f$-fermion localization length $\lambda_1$ (and also $\lambda_2$, shown in \cref{fig:G1:a} of \cref{app:subsec:additional_numerics_single_particle}) does not vary significantly over the phase space, increasing slightly towards the chiral limit. The $f$-fermion states' sublattice amplitude ratio $\alpha_2/\alpha_1$ approaches zero in the chiral limit. This is expected, as the $f$-electron wave functions become sublattice-polarized in the chiral limit~\cite{SON22} and also explains the divergence in the relative error $\epsilon_{(\alpha_2/\alpha_1)}$ in the chiral limit, as both the numerical and the analytical values of $\alpha_2/\alpha_1$ approach zero. 

As seen from \cref{fig:main_1}, the $\gamma$ ($M$) parameter is related to the gap between the $\Gamma_{3+}$ and $\Gamma_{3-}$ ($\Gamma_1$ and $\Gamma_2$) irrep states at the $\Gamma_M$ point. As such, we naturally expect $M$ to vanish at the magic angle $\theta \approx \SI{1.05}{\degree}$. Also at the magic angle, but in the isotropic limit ($w_0/w_1=1.0$), one also expects that $\gamma$ vanishes~\cite{BER21}. This behavior can indeed be observed in our simulations. Again, we note the divergence of the relative errors $\epsilon_{\gamma}$ and $\epsilon_{M}$ at the points where $\gamma$ and $M$, respectively, approach zero. 

The group velocity of the remote bands (parameterized by $v_{\star}$) is relatively large for most of the phase diagram shown in \cref{fig:main_sp_params:i}, being comparable with the single-layer graphene Fermi velocity $v_F = \SI{5.944}{\eV \angstrom}$~\cite{BER21}. In contrast, the $v_{\star}'$ parameter shown in \cref{fig:main_sp_params:k} parameter decreases from about one-third of $v_F$ away from the chiral limit to exactly zero in the chiral limit. This can be explained through our analytical approximation of $v_{\star}'$ derived in \cref{app:subsec:analytic_other_hexagon} and is a consequence of the exact intra-valley ``inversion'' symmetry of TBG in the chiral limit~\cite{WAN21a}. 

Finally, we note that similarly to the TBG flat bands~\cite{BER21, SON21,BUL20,HEJ21}, the active THF bands can be recombined into Chern-$(\pm 1)$ bands~\cite{SON22}. In addition to the so-called ideal droplet condition, the flatness of the Berry curvatures of the TBG Chern bands is key for realizing fractional Chern insulator phases in TBG~\cite{LED20,REP20,REP20a,SHE21}. In \cref{fig:main_berry}, we plot the Berry curvature of the THF Chern-$(+1)$ band for various twist angles and tunneling amplitude ratios. We find that the Berry curvature is spread more evenly across the Brillouin zone in the chiral limit ($w_0/w_1 = 0.0$), but becomes more concentrated near the $\Gamma_M$ point for realistic values of the tunneling amplitude ratio $w_0/w_1 = 0.8$.  

\subsection{Interaction THF parameters}\label{subsec:numerical_many_body}
\begin{figure*}[t!]
    \centering
    \includegraphics[width=\textwidth]{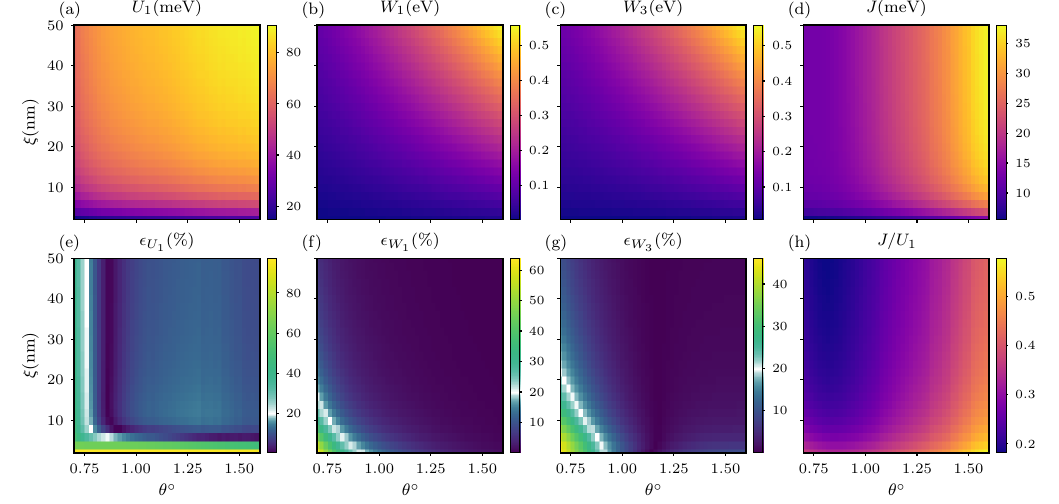}
    \subfloat{\label{fig:main_int_params:a}}\subfloat{\label{fig:main_int_params:b}}\subfloat{\label{fig:main_int_params:c}}\subfloat{\label{fig:main_int_params:d}}\subfloat{\label{fig:main_int_params:e}}\subfloat{\label{fig:main_int_params:f}}\subfloat{\label{fig:main_int_params:g}}\subfloat{\label{fig:main_int_params:h}}\caption{Numerical results for the THF interaction parameters as a function of the twist angle $\SI{0.7}{\degree} \leq \theta \leq \SI{1.6}{\degree}$ and the screening length $\SI{2}{\nano\meter} \leq \xi \leq \SI{50}{\nano\meter}$ at the fixed tunneling amplitude ratio $w_0/w_1=0.8$ and the relative error of the corresponding analytic approximation. In (a)-(d), we depict the interaction parameters $U_1$, $W_{1,3}$ and $J$ in the units indicated at the top of the panels.
    In (e)-(g), we plot the relative error of the analytically obtained parameters compared to the numerically computed values. The 20\% relative error threshold is indicated by a white line in the color map. (h) shows the numerically simulated ratio of the exchange interaction strength $J$ and the one-site interaction strength $U_1$.}
    \label{fig:main_int_params}
\end{figure*}
Having discussed the single-particle THF parameters, we now consider the interaction ones. In this case, the phase-space over which the parameters are obtained is three-dimensional, meaning that we can vary the twist angle $\theta$, the tunneling amplitude ratio $w_0/w_1$, and the screening length $\xi$ defined in \cref{eq:main_Coulomb}. As such, we focus on the experimentally-relevant 2 $w_0/w_1 = 0.8$~\cite{DAI16,JAI16,SON21,UCH14,WIJ15} (see \cref{app:subsec:additional_numerics_many-body} for additional tunneling amplitude ratios). In \cref{fig:main_int_params}, we plot the numerically computed parameters $U_1$, $W_{1,3}$ and $J$ as functions of the twist angle $\SI{0.7}{\degree} \leq \theta \leq \SI{1.6}{\degree}$ and the screening length $\SI{2}{\nano\meter} \leq \xi \leq \SI{50}{\nano\meter}$, where both parameters can in principle be tuned experimentally.

We also checked the validity of the approximations derived in \cref{sec:analytics_many_body} by comparing them with the numerical results. For the $U_1$ parameter expression from \cref{eq:main_U1_final} we employed the analytical expressions for $\alpha_2$ and $\lambda_1$ parameters obtained from the one-shell Tripod model. For the $W_1$ parameter we use the second approximation from \cref{eq:main_W1_W3_1st}, as it gives better agreement, while for the $W_3$ parameter we only consider first approximation from \cref{eq:main_W1_W3_0th}. The $V(\mathbf{q})$-parameter is not numerically computed, as it is just the Fourier transformation of the screened Coulomb potential. Similarly to \cref{fig:main_sp_params}, we show the relative errors for the $U_1$, $W_{1}$, and $W_{3}$ parameters in \cref{fig:main_int_params:e,fig:main_int_params:f,fig:main_int_params:g}.

Finally, in \cref{fig:main_int_params:h}, we plot the ratio of the exchange ($J$) and the onsite ($U_1$) interaction strengths. For the entire phase space, the exchange interaction is more than two-times smaller than the onsite interaction. We will leverage this fact in our discussion of the interaction THF Hamiltonian symmetries in \cref{sec:inthamsym}.

\subsection{The THF model away from magic angle}
\label{subsec:discussion_away_ma}

Our combined numerical and analytical analysis of the THF model around the magic angle reveals a series of general features concerning the variation of its parameters and its applicability as an effective model of TBG:
\begin{itemize}
    \item For most of the parameter space, our analytical expressions approximate the numerically-obtained THF parameters to an error of less than $30\%$. Remarkably, \cref{fig:discussion_thf_params:a} shows that \emph{all} single-particle parameters can be obtained analytically up to a $30\%$ error for any tunneling amplitude ratio $0 \leq w_0/w_1 \lesssim 0.8$ and twist angle $\SI{1.00}{\degree} \lesssim \theta \leq \SI{1.60}{\degree}$.
    
    \item The magic angle is seen in the small $M$ value for all $w_0/w_1$ and $\theta \approx \SI{1.05}{\degree}$.
    \item The $t_0$ parameter characterizing the hopping of $f$-electrons is also minimized for all $w_0/w_1$ around the magic angle.
    
    \item Although $M$ and $t_0$ are independent parameters, they are minimized (in the case of $t_0$) or vanish completely (in the case of $M$) around the same angle, as shown in \cref{fig:discussion_thf_params:b}. For the THF model, the magic angle can thus be defined as the angle for which $M$ vanishes and $t_0$ is minimized.
    
    \item The parameter $\gamma$ changes sign for $w_0/w_1= 0.9$ at the magic angle. As shown in \cref{fig:discussion_thf_params:b}, $\gamma$ also vanishes for some $w_0/w_1 \leq 1.0$ for any angle $\theta \lesssim \SI{1.05}{\degree}$. This coincides with the gap between the remote and flat bands vanishing. In the limit $\gamma=0$, the THF model is exactly solvable~\cite{HU23} and constitutes a good effective model of TBG.  
    
    \item As shown in \cref{fig:G2} of \cref{app:subsec:additional_numerics_many-body}, the onsite repulsion $U_1$ decreases for larger angles and/or for smaller amplitude ratios $w_0/w_1$ ({\it i.e.}{}, closer to the chiral limit).
    
    \item The group velocity of the THF remote bands ($v_{\star}$) is comparable to (but always smaller than) the single-layer graphene Dirac velocity for most of the phase diagram considered in \cref{fig:main_sp_params}.
    
    \item The BM model band structure near charge neutrality is very well-fitted by the THF model, even \emph{away} from the magic angle, showing that the latter is an \emph{excellent} model for the single-particle physics of TBG.
    
    \item The \emph{full} many-body THF model is only useful whenever there is a separation of energy scales ({\it i.e.}{}, the $f-c$ hybridization is smaller than the onsite interaction of the $f$-electrons). In \cref{fig:discussion_thf_params:c}, we see that around the magic angle and, more importantly, around the realistic tunneling amplitude ratio $w_0/w_1=0.8$, $\gamma / U_1$ is small, showing that the THF model is a good effective model of TBG. For $w_0/w_1=0$, \cref{fig:discussion_thf_params:c} shows that $\gamma/U_1$ is large, and hence the THF model cannot provide a great approximation of the TBG physics, due to the large $f-c$ mixing around the $\Gamma_M$ point. The decreased localization of the $f$-electron wave functions coupled with an increased gap between the active and remote TBG bands show that a projected, strongly-coupled, momentum-space description of the problem~\cite{LIA21} might be more suitable in the chiral limit ($w_0 / w_1 = 0$), as was confirmed by exact diagonalization studies~\cite{XIE21}. We note, however, that the tunneling amplitude ratio $w_0 / w_1 = 0$ is unrealistic and inconsistent with the experimentally-observed quantum-dot-like behaviour of TBG~\cite{XIE19,WON20}.

    \item \Cref{fig:discussion_thf_params:d} reveals that the $U_1$ and $W_{1}$ interaction parameters become approximately equal for a small region of the explored phase space. We analyze the emerging continuous symmetries of the interaction Hamiltonian in \cref{sec:inthamsym}.
    
\end{itemize}

\section{Symmetries of the interaction Hamiltonian}\label{sec:inthamsym}

\begin{figure}[t!]
    \centering
    \includegraphics[width=\columnwidth]{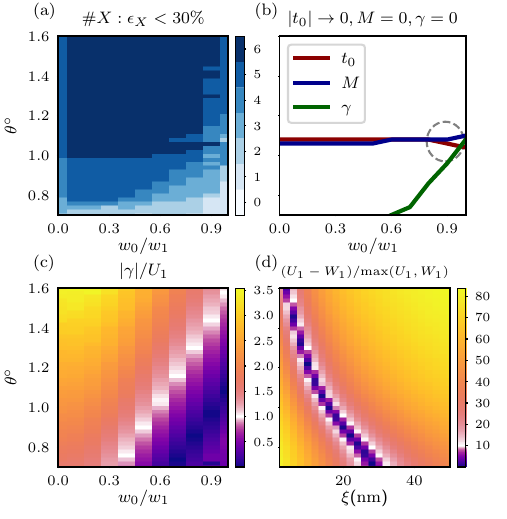}
    \subfloat{\label{fig:discussion_thf_params:a}}\subfloat{\label{fig:discussion_thf_params:b}}\subfloat{\label{fig:discussion_thf_params:c}}\subfloat{\label{fig:discussion_thf_params:d}}\caption{The relation between various THF parameters and their approximations. (a) shows how many of the single-particle THF parameters $\lambda_1$, $\alpha_2/\alpha_1$, $\gamma$, $v_{\star}$, $v_{\star}'$, and $M$ are approximated by the analytical formulae to an error smaller than $30\%$ across the explored phase space. In (b), we show the lines along which the $M$ parameter or the $f-c$ hybridization parameter $\gamma$ vanish, as well as the line along which the nearest-neighbor hopping amplitude $t_0$ of the $f$-electrons is minimized (and also approximately vanishes). $\abs{M}$ and $t_0$ are both minimized for all tunneling amplitude ratios around the magic angle. The dashed gray circle indicates the region near the magic angle for which $\gamma$ also vanishes and renders the THF model exactly solvable~\cite{HU23}. The ratio between the $f-c$ hybridization amplitude and the onsite interaction $\abs{\gamma}/\abs{U_1}$ is shown in (c) at a typical screening length $\xi = \SI{10}{\nano\meter}$. We also plot the relative error of the approximation $U_1 \approx W_1$ in (d) for $w_0/w_1 = 0.8$. The relative error is defined as $\epsilon = (U_1 - W_1) / \textrm{max}(U_1,W_1)$. A small region emerges where the approximation $U_1 \approx W_1$ holds to an error smaller than $10\%$. }
    \label{fig:discussion_thf_params}
\end{figure}

The interaction part of the THF Hamiltonian is close to a highly-symmetric point. In this section, we outline the continuous symmetries of the THF interaction Hamiltonian arising under different limits, and refer the reader to \cref{app:sec:symmetries} for the detailed deviations. We start by splitting the interaction Hamiltonian from \cref{eq:main_Coulomb_projected_in_BM} into two sums (neglecting the $\hat{H}_K$ term). The first one, dubbed as the density-density interaction, is given by $\hat{H}_I^{\textrm{dens.-dens.}} = \hat{H}_{U_1} + \hat{H}_{W} + \hat{H}_V$, while the second one consists of the exchange and double hybridization interactions, $\hat{H}_J + \hat{H}_{\tilde{J}}$. Keeping in mind that $J$ is smaller than $U_1$, as seen in \cref{fig:discussion_thf_params:a}, one can first consider the density-density interaction term $\hat{H}_I^{\textrm{dens.-dens.}}$ individually.

Within the approximation 
\begin{equation}
    U_1 \approx W_1 \approx W_3 = W = \frac{2\pi}{\sqrt{3}}\left(\frac{\xi}{a_M}\right)^2U_{\xi},
    \label{eq:main_interaction_approximation}
\end{equation} the density-density interaction term can be rewritten as
\begin{equation}
    \hat{H}_I^{\textrm{dens.-dens.}} \approx \frac{1}{2}W\sum_{\mathbf{R}}:\hat{\rho}_{\mathbf{R}}::\hat{\rho}_{\mathbf{R}}:,
    \label{eq:main_dens-dens_term_expanded}
\end{equation}
where the total density operator $\hat{\rho}_{\mathbf{R}}$ is the sum between the $f$- and $c$-electron density operators, $\hat{\rho}_{\mathbf{R}} = \hat{\rho}^f_{\mathbf{R}} + \hat{\rho}^c_{\mathbf{R}}$. The total density operator is thus the inner product of a 24-dimensional spinor (whose entries are the six -- four conduction and two heavy -- fermions for each spin and valley flavor) and its hermitian conjugate. As a result, the density-density interaction term remains invariant under unitary transformations of the 24-dimensional spinor and enjoys an enlarged $U(24)$ symmetry (see \cref{app:subsec:symmetry_density-density-interaction}). 

In \cref{fig:discussion_thf_params:b}, we compare the $U_1$ and $W_1$ interaction parameters. We remind the reader that the $W_3$ parameter is almost equal to $W_1 \approx W$, and that the screened Coulomb potential $V(\mathbf{q})$, governing the $c$-electron interaction, can also be approximated as $W$, as argued near \cref{eq:main_V}. Thus, it suffices to check only for the equality between $U_1$ and $W_1$. We find out that in the experimentally-relevant submanifold $w_0/w_1=0.8$, a small region emerges where the approximation $U_1 \approx W_1 \approx W_3 \approx V$ holds to an error smaller than 20\%, thus rendering the density-density interaction Hamiltonian $\hat{H}_I^{\textrm{dens.-dens.}}$ approximately $U(24)$-symmetric.

One can also independently consider the exchange and double hybridization interaction terms. After some algebra (see \cref{app:subsec:higher_symmetry}), these interaction terms can be rewritten as
\begin{equation}
    \hat{H}_{J} + \hat{H}_{\tilde{J}} = \frac{J}{4}\sum_{\mathbf{R},\eta, \alpha}\hat{D}_{\mathbf{R} \alpha \eta} \hat{D}^\dagger_{\mathbf{R}\alpha\eta} = \frac{J}{2}\{\hat{D}^\dagger_{\mathbf{R}1+},\hat{D}_{\mathbf{R}1+}\},
    \label{eq:main_U8_Drepresentation}
\end{equation}
with the $\hat{D}_{\mathbf{R}1+}$ operator being given by
\begin{align}
    \hat{D}_{\mathbf{R}1+} &= \sum_{s_1}\bigl(\hat{f}^\dagger_{\mathbf{R} 1+ s_1}\hat{c}_{\mathbf{R} 3+ s_1} + \hat{c}^\dagger_{\mathbf{R} 3 - s_1}\hat{f}_{\mathbf{R} 1 - s_1} \nonumber \\ &- \hat{f}^\dagger_{\mathbf{R} 2 - s_1}\hat{c}_{\mathbf{R} 4 - s_1} - \hat{c}^\dagger_{\mathbf{R} 4 + s_1}\hat{f}_{\mathbf{R} 2 + s_1}\bigr) \nonumber \\ &= \frac{1}{2} \hat{\Psi}^\dagger_{\mathbf{R}}(\sigma_z\tau_0\zeta_x s_0 + i\sigma_0\tau_z\zeta_y s_0 ) \hat{\Psi}_{\mathbf{R}}.
    \label{eq:main_DR1+_pauli_representation}
\end{align}
In \cref{eq:main_DR1+_pauli_representation}, we have introduced a 16-dimensional spinor $\hat{\Psi}_{\mathbf{R}} = (\hat{f}_{\mathbf{R}1\pm \uparrow\downarrow}, \hat{c}_{\mathbf{R}3\pm \uparrow\downarrow},\hat{f}_{\mathbf{R}2\pm \uparrow\downarrow}, \hat{c}_{\mathbf{R}4\pm \uparrow\downarrow})$, the Pauli matrices $\zeta_{0,x,y,z}$ acting in the $(f,c)$-type of fermion space, as well as the Pauli matrices $\sigma_{0,x,y,z}$, acting in the orbital space $\alpha={1,2}$ for $f$-electrons and $\alpha+2={3,4}$ for $c$-electrons. We also use $\tau_{0,x,y,z}$ and $s_{0,x,y,z}$ to denote the Pauli matrices acting in the valley $\eta=\pm$ and spin $s={\uparrow,\downarrow}$ spaces, respectively. We refer the reader to the full derivation provided in \cref{app:subsec:higher_symmetry}.

\Cref{eq:main_U8_Drepresentation} provides the means for deriving additional enlarged symmetries of the exchange and double hybridization interaction terms. Any Hermitian local quadratic operator $\hat{\Sigma}_{\mathbf{R}}$ that commutes with the $\hat{D}^\dagger_{\mathbf{R}1+}$ operators will generate a continuous symmetry of the exchange and double hybridization interaction terms (see \cref{app:subsec:higher_symmetry}). We find a set of 64 such local quadratic operators $\hat{\Sigma}_{\mathbf{R}}^{(1,\ldots,64)}$, which are isomorphic to the $U(8)$ Lie algebra generators. Additionally, a 65th local operator $\hat{\Sigma}_{\mathbf{R}}^{(65)}$ commutes with the exchange and double hybridization interaction terms $\hat{H}_{J} + \hat{H}_{\tilde{J}}$ and also commutes with the other 64 symmetry generators. All of these operators can be found in \cref{app:subsec:higher_symmetry}. As such, one can perform unitary rotations belonging to the $U(8)\times U(1)$ symmetry group within $\alpha=1,2$ $f$-electron orbitals and $a=3,4$ conduction $c$-electron bands. Since the exchange and double hybridization interaction terms do not involve the $a=1,2$ conduction $c$-electron bands, those can also be rotated independently, thus giving rise to another $U(8)$ symmetry of $\hat{H}_{J} + \hat{H}_{\tilde{J}}$. We conclude that the symmetry group of the exchange and hybridization interaction terms is close to a $U(8)\times U(8) \times U(1)$ group.

Treating the exchange and double hybridization interaction terms as a perturbation to the density-density interaction Hamiltonian, we can then say that the $\hat{H}_{J} + \hat{H}_{\tilde{J}}$ terms reduce the $U(24)$ symmetry of $\hat{H}_I^{\textrm{dens.-dens.}}$ to an approximate $U(8)\times U(8) \times U(1)$ symmetry of $\hat{H}_I$, which holds (for the interaction Hamiltonian only) in the regime depicted in \cref{fig:discussion_thf_params:b}. The single-particle term $\hat{H}_{0}$, however, further breaks this symmetry to the $U(4) \times U(4)$ group (in the THF chiral-flat limit with $M=0$ and $v_{\star} = 0$), the flat $U(4)$ group (in the THF flat limit with $M=0$ and $v_{\star} \neq 0$), the chiral $U(4)$ group (in the THF chiral limit with $M \neq 0$ and $v_{\star} = 0$), or the general $U(2) \times U(2)$ group (whenever $M \neq 0$ and $v_{\star} \neq 0$).

\section{Conclusions}\label{sec:conclusions}

In this article, we focused on deriving analytical expressions for the THF Hamiltonian parameters. This was achieved using a two-step process. Firstly, we matched various features of the TBG spectrum between the THF and BM models. Secondly, we employed the tripod and hexagon approximations of the BM model~\cite{BER21} to obtain simple, analytically-tractable, approximations for the single-particle and many-body parameters of the THF Hamiltonian.

Additionally, we have conducted a comprehensive numerical analysis of the THF Hamiltonian parameters across a large, experimentally-relevant, parameter space of various twist angles, tunneling amplitude ratios, and Coulomb potential screening lengths. By comparing them with the numerical results, we found that the analytical approximations we have derived in the first half of this work perform remarkably well across an extensive region of angles, tunneling amplitude ratios, and/or screening lengths. 

Finally, in addition to obtaining the variation of the THF Hamiltonian parameters around the magic angle, our combined analytical and numerical study has also allowed us to assess the applicability of the THF model as an effective model for the TBG physics. We found that the single-particle band structure of the BM model is very well-fitted by the THF model for almost the entire phase space of twist angles and amplitude ratios we have explored. On the other hand, the many-body THF model is only useful whenever the $f-c$ hybridization is smaller than the onsite repulsion of the $f$-electrons, a regime which holds at realistic values of the tunneling amplitude ratio.

\begin{acknowledgments}
The simulations presented in this work were performed using the Princeton Research Computing resources at Princeton University, which is a consortium of groups led by the Princeton Institute for Computational Science and Engineering (PICSciE) and Office of Information Technology's Research Computing. 
D.C. and B.A.B. were primarily supported by the DOE Grant No. DE-SC0016239, the Simons Investigator Grant No. 404513, the Gordon and Betty Moore Foundation through Grant No. GBMF8685 towards the Princeton theory program, and the Gordon and Betty Moore Foundation's EPiQS Initiative (Grant No. GBMF11070). 
D.C. acknowledges the hospitality of the Donostia International Physics Center, at which part of this work was carried out. 
L.L.H.L. and P.C. are supported by the Office of Basic Energy Sciences, Material Sciences and Engineering Division,  U.S. Department of Energy (DOE) under Contract  DE-FG02-99ER45790.
Z.-D. S. was supported by
National Natural Science Foundation of China (General Program No.\ 12274005), 
National Key Research and Development Program of China (No.\ 2021YFA1401900).
\end{acknowledgments} \vspace{-\baselineskip} \bibliographystyle{apsrev4-2}
\bibliography{TBG_1.bib,TBG_2.bib}
\renewcommand{\thetable}{S\arabic{table}}
\renewcommand{\thefigure}{S\arabic{figure}}

\appendix
\onecolumngrid
\newpage
\tableofcontents
\newpage

\section{The Bistritzer-Macdonald model}\label{app:sec:BM_model}
In this appendix, we provide a brief review of the Bistritzer-Macdonald (BM) model~\cite{BIS11}. The reader is pointed to Refs.~\cite{SON19,BER21,SON21,BER21a,LIA21,BER21b,XIE21} for a more systematic introduction. We start by reviewing the single-particle Hamiltonian and formalizing our notation (which is consistent with Refs.~\cite{SON19,BER21,SON21,BER21a}). We then outline two approximations of the BM model~\cite{BER21}: the Tripod model in \cref{app:subsec:tripod} and the Hexagon model in \cref{app:subsec:hexagon}. Finally, we summarize the symmetries of the model and the gauge-fixing conditions used in this paper.

\subsection{Single-particle Hamiltonian}\label{app:sec:BM_single_particle_Hamiltonian}
\begin{figure}
    \centering
    \includegraphics{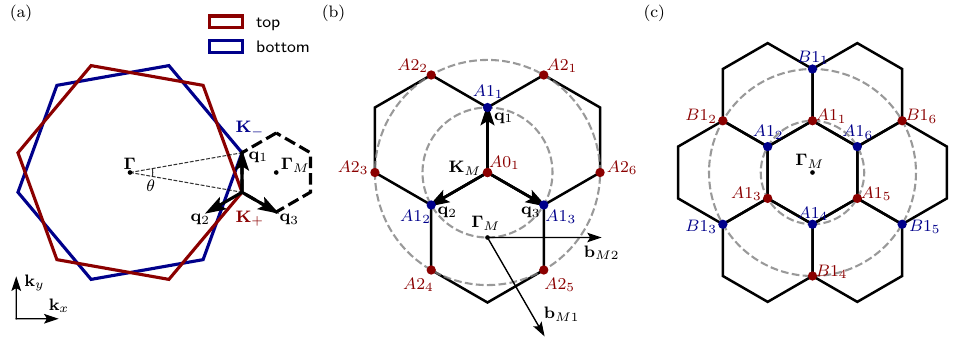}\subfloat{\label{fig:A1:a}}\subfloat{\label{fig:A1:b}}\subfloat{\label{fig:A1:c}}\caption{Moir\'e Brillouin Zone (MBZ), as well as the tripod and hexagonal model lattices~\cite{BER21}. (a) Two graphene single-layer Brillouin zones (top layer denoted by red, bottom layer by blue) are rotated by an angle $-\frac{\theta}{2}$ (top) and $+\frac{\theta}{2}$ (bottom) relative to the $x$-axis. The $K$ points of the top and bottom layers are labeled as $\mathbf{K}_{\pm}$ For small twist angles $\theta$ a translation symmetry emerges and gives rise to the Moir\'e Brillouin zone (black hexagon). (b) The Tripod model~\cite{BER21} with two shells consisting of the point labeled as $A0$ (the $K_M$ point of the MBZ), and the two shells shown as gray circles labeled as $A1$ and $A2$. Each $\mathbf{Q}$ point within a shell is indexed counterclockwise. The case when only the first shell ($A1$) and the $K_M$ point ({\it i.e.}{} the ``zeroth'' shell) are considered is the original Tripod model introduced by Ref.~\cite{BIS11}. (c) The two-shell Hexagon model. Shells are illustrated by the gray circles labeled as $A1$ and $B1$. Within each shell the $\mathbf{Q}$ points are indexed counterclockwise.}
    \label{fig:A1}
    
\end{figure}

Twisted Bilayer Graphene (TBG) consists of two stacked single-graphene layers, labeled by the index $l$ as top ($l=+$) and bottom ($l=-$). The two layers $l=\pm$ are rotated by the respective angle $\mp \frac{\theta}{2}$ relative to the $x$-axis. We denote by $\hat{c}^\dagger_{l,\mathbf{p},\alpha,s}$ the fermionic operator creating an electron of momentum $\mathbf{p}$ in layer $l$, sublattice $\alpha \in \{1,2\}$, and spin $s \in \{\uparrow,\downarrow\}$. For a small angle $\theta$, a moir\'e translation symmetry emerges. Letting $\mathbf{K}_l$ denote the graphene $K$ point of layer $l$, as shown in \cref{fig:A1:a}, we define $\vec{q}_1,\vec{q}_2,\vec{q}_3$ as
\begin{equation}
    \mathbf{q}_1 = (\mathbf{K}_- - \mathbf{K}_+) = k_{\theta}(0,1)^T,\qquad \mathbf{q}_2 = C_{3z}\mathbf{q}_1 = k_{\theta}(-\frac{\sqrt{3}}{2},-\frac{1}{2})^T,\qquad \mathbf{q}_3 = C_{3z}^2\mathbf{q}_1 = k_{\theta}(\frac{\sqrt{3}}{2},-\frac{1}{2})^T,
    \label{eq:q_vectors}
\end{equation}
where $k_{\theta} = |\mathbf{K}_+ - \mathbf{K}_-| = 2|\mathbf{K}_+|\sin{\frac{\theta}{2}}$ depends on the twist angle. The corresponding moir\'e reciprocal vectors read as
\begin{equation}
    \mathbf{b}_{M1} = \mathbf{q}_3 - \mathbf{q}_1,\qquad \mathbf{b}_{M2} = \mathbf{q}_3 - \mathbf{q}_2,
\end{equation}
and span the moir\'e reciprocal lattice $\mathcal{Q}_0 = \mathbb{Z}\mathbf{b}_{M1} + \mathbb{Z}\mathbf{b}_{M2}$. We can then define the real-space lattice vectors $\vec{a}_{M1}$ and $\vec{a}_{M2}$ according to the formula $\vec{a}_{Mi}\cdot\mathbf{b}_{Mj} = 2\pi\delta_{ij}$. This implies $\vec{a}_{M1} = \frac{2\pi}{3k_{\theta}}(\sqrt{3},1)$ and $\vec{a}_{M2} = \frac{2\pi}{3k_{\theta}}(-\sqrt{3},1)$. To describe the basis states of the BM model, we also introduce the $\vec{\mathcal{Q}}_{\pm}$ lattices as $\vec{\mathcal{Q}}_+ = \vec{\mathcal{Q}}_0 + \mathbf{q}_1$ and $\vec{\mathcal{Q}}_- = \vec{\mathcal{Q}}_0 - \mathbf{q}_1$, which together form a honeycomb lattice $\mathcal{Q} = \mathcal{Q}_{+}\oplus \mathcal{Q}_{-}$.

We now define the basis states of the BM model. The low energy states of the system in valley $\eta$ are given by~\cite{BER21}

\begin{equation}
    \hat{c}^\dagger_{\mathbf{k},\mathbf{Q}, \alpha, \eta, s} = \hat{c}^\dagger_{l, \eta \mathbf{K}_{l} + \mathbf{k} - \mathbf{Q}, \alpha, s},\qquad  \textrm{for} \qquad  \mathbf{Q} \in \mathcal{Q}_{\pm},
    \label{eq:BM_model_creation_operator}
\end{equation}
 where $\mathbf{k}$ is measured from the $\Gamma_M$ point of the moir\'e Brillouin zone (MBZ). In this basis, the Hamiltonian of the BM model reads as~\cite{BIS11,BER21}
\begin{equation}
    \hat{H}_{\textrm{BM}} = \sum_{\eta, s}\sum_{\mathbf{k} \in \textrm{MBZ}} \sum_{\alpha, \alpha^{\prime}} \sum_{\mathbf{Q}, \mathbf{Q}^{\prime}} h^{(\eta)}_{\mathbf{Q} \alpha, \mathbf{Q}^{\prime} \alpha^{\prime}} (\mathbf{k}) \hat{c}^\dagger_{\mathbf{k},\mathbf{Q},\alpha,\eta,s}\hat{c}_{\mathbf{k},\mathbf{Q}',\alpha',\eta,s},
    \label{eq:BM_model_hamiltonian}
\end{equation}
where the first-quantized Hamiltonian for the $\eta=+$ valley is given by
\begin{equation}
    h^{(+)}_{\mathbf{Q} \alpha, \mathbf{Q}^{\prime} \alpha^{\prime}} (\mathbf{k}) = v_F(\mathbf{k} - \mathbf{Q})\cdot\boldsymbol{\sigma}\delta_{\mathbf{Q},\mathbf{Q}^{\prime}} + \sum_{j=1}^3 \left[T_j\right]_{\alpha\alpha^{\prime}}\delta_{\mathbf{Q},\mathbf{Q}^{\prime}+\mathbf{q}_j},
    \label{eq:1VBM_hamiltonian}
\end{equation}
with the tunneling matrices
\begin{equation}
    T_j = w_0 \sigma_0 + w_1\sigma_x\cos{\frac{2\pi(j-1)}{3}} + w_1\sigma_y\sin{\frac{2\pi(j-1)}{3}}.
    \label{eq:tunneling_matrix}
\end{equation}
In \cref{eq:tunneling_matrix}, parameters $w_0,w_1$ are the interlayer couplings at the AA-stacking and AB-stacking regions, respectively. 
For the numerical calculations, we will use the same parameters as in Ref.~\cite{SON22}: $v_F = \SI{5.944}{\eV \cdot \angstrom}$, $|\mathbf{K}_{\pm}| = \SI{1.703}{\angstrom^{-1}}$, and $w_1 = \SI{110}{\meV}$. To simplify derivations, in what follows, we will employ dimensionless units by rescaling all the energies and momenta according to Ref.~\cite{BER21}
\begin{equation}
    \label{app:eqn:nondim_units}
    E \rightarrow \frac{E}{v_F k_{\theta}},\qquad \mathbf{k} \rightarrow \frac{\mathbf{k}}{k_{\theta}}.
\end{equation}
The single-particle Hamiltonian for the opposite valley $\eta = -$ is obtained from time-reversal symmetry (see \cref{app:subsec:BM_symmetries})
\begin{equation}
    h^{(-)}_{\mathbf{Q} \alpha, \mathbf{Q}^{\prime} \alpha^{\prime}} (\mathbf{k}) = h^{(+)*}_{-\mathbf{Q} \alpha, -\mathbf{Q}^{\prime} \alpha^{\prime}} (-\mathbf{k}).
\end{equation}

We define the energy band basis of the BM model Hamiltonian \cref{eq:BM_model_hamiltonian} as
\begin{equation}
    \hat{c}^\dagger_{\mathbf{k},n,\eta,s} = \sum_{\mathbf{Q}, \alpha}u_{\mathbf{Q}\alpha,n\eta}(\mathbf{k})\hat{c}^\dagger_{\mathbf{k},\mathbf{Q},\alpha,\eta,s},
\end{equation}
where $u_{\mathbf{Q}\alpha,n\eta}(\mathbf{k})$ is the wave function of the band $n$ and valley $\eta$ satisfying the Schr\"odinger equation for the single-particle Hamiltonian
\begin{equation}
    \sum_{\mathbf{Q}^{\prime},\alpha^{\prime}} h^{(\eta)}_{\mathbf{Q}^{\prime}\alpha^{\prime},\mathbf{Q}\alpha}(\mathbf{k}) u_{\mathbf{Q}^{\prime}\alpha^{\prime},n\eta}(\mathbf{k}) = \epsilon_{n\eta}(\mathbf{k})u_{\mathbf{Q}\alpha,n\eta}(\mathbf{k}).
\end{equation}
Here, the band with the label $n = +|n|$ ($n=-|n|$) indicates the $|n|$-th conduction (valence) band above (below) the charge neutrality point. The eigenstate wave functions $u_{\mathbf{Q}\alpha,n\eta}(\mathbf{k})$ satisfy the embedding relation for shifting momentum $\mathbf{k}$ by a moir\'e reciprocal vector $\vec{G}_M$~\cite{BER21a}
\begin{equation}
u_{\mathbf{Q} \beta,n \eta}(\mathbf{k} + \vec{G}_{M}) = u_{\mathbf{Q}-\vec{G}_M \beta ,n \eta}(\mathbf{k}).
\label{eq:embedding_condition}
\end{equation}

For later use in defining the THF model, we also introduce the real-space basis states as the Fourier transformation of the BM-model basis states from \cref{eq:BM_model_creation_operator}

\begin{equation}
    \hat{c}^\dagger_{l,\alpha,\eta,s}(\vec{r}) = \frac{1}{\sqrt{\Omega_{\textrm{tot}}}}\sum_{\mathbf{k}\in\textrm{MBZ}}\sum_{\mathbf{Q}\in\mathcal{Q}_{l\eta}}e^{-i(\mathbf{k}-\mathbf{Q})\cdot\vec{r}}\hat{c}^\dagger_{\mathbf{k},\mathbf{Q},\alpha,\eta,s},
    \label{eq:real_space_basis}
\end{equation}
where $\Omega_{\textrm{tot}} = N\Omega_0$ is the total area of the sample of $N$ moir\'e unit cells having an area $\Omega_0 = \vec{a}_{M1}\times \vec{a}_{M2} = \frac{8\pi^2}{3\sqrt{3}k_{\theta}}$. Under the translation operator $T_{\vec{R}}$, defined according to
\begin{equation}
    T_{\vec{R}}\hat{c}^\dagger_{\mathbf{k},\mathbf{Q},\alpha,\eta,s}T^{-1}_{\vec{R}} = e^{-i\mathbf{k}\cdot\vec{R}}\hat{c}^\dagger_{\mathbf{k},\mathbf{Q},\alpha,\eta,s},
\end{equation}
where $\vec{R} = n_1\vec{a}_{M1} + n_2\vec{a}_{M2}$, $n_1,n_2 \in \mathds{Z}$, the real-space basis states $\hat{c}^\dagger_{l,\alpha,\eta,s}(\vec{r})$ transform as
\begin{align}
    T_{\vec{R}}\hat{c}^\dagger_{l,\alpha,\eta,s}(\vec{r})T_{\vec{R}}^{-1}&=
    \frac{1}{\sqrt{\Omega_{\textrm{tot}}}}\sum_{\mathbf{k}\in\textrm{MBZ}}\sum_{\mathbf{Q}\in\mathcal{Q}_{\pm}}e^{-i(\mathbf{k}-\mathbf{Q})\cdot\vec{r}}T_{\vec{R}}\hat{c}^\dagger_{\mathbf{k},\mathbf{Q},\alpha,\eta,s}T_{\vec{R}}^{-1} \\
    &= \frac{1}{\sqrt{\Omega_{\textrm{tot}}}}\sum_{\mathbf{k}\in\textrm{MBZ}}\sum_{\mathbf{Q}\in\mathcal{Q}_{\pm}}e^{-i(\mathbf{k}-\mathbf{Q})\cdot\vec{r}}e^{-i\mathbf{k}\cdot\vec{R}}\hat{c}^\dagger_{\mathbf{k},\mathbf{Q},\alpha,\eta,s}.
\end{align}
With an auxiliary definition
\begin{equation}
    \Delta\mathbf{K}_{l} = \begin{cases}
        \mathbf{q}_2,\qquad l=+ \\
        -\mathbf{q}_3,\qquad l=-
    \end{cases},
\end{equation}
we notice that $\eta \Delta \mathbf{K}_{l} - \mathbf{Q} \in \mathcal{Q}_0\; (\textrm{for}\;\mathbf{Q}\in\mathcal{Q}_{\pm})$, which implies that $e^{i(\eta \Delta \mathbf{K}_{l} - \mathbf{Q})\cdot\vec{R}}=1$. This allows us to write
\begin{equation}
    T_{\vec{R}}\hat{c}^\dagger_{l,\alpha,\eta,s}(\vec{r})T_{\vec{R}}^{-1}=
     \frac{1}{\sqrt{\Omega_{\textrm{tot}}}}\sum_{\mathbf{k}\in\textrm{MBZ}}\sum_{\mathbf{Q}\in\mathcal{Q}_{\pm}}e^{-i(\mathbf{k}-\mathbf{Q})\cdot(\vec{r}+\vec{R})}e^{-i\eta \Delta \mathbf{K}_{l}\cdot\vec{R}}\hat{c}^\dagger_{\mathbf{k},\mathbf{Q},\alpha,\eta,s} = e^{-i\eta \Delta \mathbf{K}_{l}\cdot\vec{R}}\hat{c}^\dagger_{l,\alpha,\eta,s}(\vec{r} + \vec{R}).
     \label{eq:BM_real_space_transformation}
\end{equation}
\cref{eq:BM_real_space_transformation} will be useful for reviewing the Wannier states of the THF model in \cref{app:subsec:local_orbitals}. 

\subsection{The Tripod model}\label{app:subsec:tripod}
In this section, we review the Tripod model of TBG -- a $K_M$ point centered BM model approximation with a small number of plane-waves. This amounts to only including a small number of $\mathbf{Q}$ points in the $\mathcal{Q}$ lattice~\cite{BIS11,BER21}, as shown in the lattice depicted in \cref{fig:A1}(b). In the two-shell Tripod model we approximate the BM model $\mathcal{Q}$ lattice as
\begin{equation}
\mathcal{Q}^{(\textrm{tripod})} = \underbrace{A0_1}_{\textrm{0th-shell}} \oplus \underbrace{(A1_1, A1_2, A1_3)}_{\textrm{1st-shell}} \oplus \underbrace{(A2_1, A2_2, A2_3,A2_3,A2_4,A2_5,A2_6)}_{\textrm{2nd-shell}},
\label{eq:tripod_shells}
\end{equation}
while for the one-shell Tripod model ({\it i.e.}{} the original Tripod model of TBG derived by Ref.~\cite{BIS11}), we consider only the zeroth and the first shells. In both models, we focus only on the $\eta=+$ valley. 

We denote by $\delta \mathbf{k}$ the momentum deviation from the $K_M$ point, {\it i.e.}{} $\delta \mathbf{k} = \mathbf{k} - \mathbf{q}_1$. The single-particle states of the one-shell Tripod model can be written as
\begin{equation} \label{eq:tripod_spinor_definition}
    \ket{\Psi(\delta\mathbf{k})} = \sum_{\alpha}\left[\psi_{A0_1,\alpha}(\delta \mathbf{k})\hat{c}^\dagger_{\mathbf{q}_1 + \delta \mathbf{k}, \mathbf{q}_1, \alpha, +, s}  + \sum_{i=1}^3 \psi_{A1_i,\alpha}(\delta \mathbf{k})\hat{c}^\dagger_{\mathbf{q}_1 + \delta \mathbf{k}, \mathbf{q}_1 + \mathbf{q}_i, \alpha, +, s} \right] \ket{0}.
\end{equation}
In what follows, we will make the $\delta \mathbf{k}$-dependence  $\psi$ spinors implicit. The first-quantized Hamiltonian acting on the eight-dimensional spinor $\Psi^T = (\psi_{A0_1}^T,\psi_{A1_1}^T,\psi_{A1_2}^T,\psi_{A1_3}^T)$ is given by
\begin{equation}
    H^{\textrm{(1-shell tripod)}}(\delta \mathbf{k}, w_0, w_1) =
    \begin{pmatrix}
    \delta \mathbf{k} \cdot \boldsymbol{\sigma} & T_1 & T_2 & T_3 \\
    T_1 & (\delta \mathbf{k} - \mathbf{q}_1) \cdot \boldsymbol{\sigma} & 0 & 0 \\
    T_2 & 0 & (\delta \mathbf{k} - \mathbf{q}_2)\cdot \boldsymbol{\sigma} & 0 \\
    T_3 & 0 & 0 & (\delta \mathbf{k} - \mathbf{q}_3)\cdot \boldsymbol{\sigma}
    \end{pmatrix}
    \label{eq:tripod_hamiltonian}
\end{equation}
The Schr\"odinger equation $H^{\textrm{(1-shell tripod)}}(\delta \mathbf{k}, w_0, w_1) \Psi = E(\delta\mathbf{k}) \Psi$ can be rewritten as a system of linear equations

\begin{equation}
    \begin{cases}
    (\delta \mathbf{k} \cdot \boldsymbol{\sigma}) \psi_{A0_1} + \sum_{i=1}^3T_i\psi_{A1_i} = E(\delta\mathbf{k})\psi_{A0_1} \\
    T_i\psi_{A0_1} + \left[(\delta \mathbf{k} - \mathbf{q}_i)\cdot \boldsymbol{\sigma}\right]\psi_{A1_i} = E(\delta\mathbf{k})\psi_{A1_i},\qquad i=1,2,3
    \end{cases}.
    \label{eq:tripod_system}
\end{equation}
The second equation allows us to express $\psi_{A1_i}$ in terms of $\psi_{A0_1}$,
\begin{equation}
    \psi_{A1_i} =  \left[ E(\delta \mathbf{k}) - (\delta \mathbf{k} - \mathbf{q}_i)\cdot\boldsymbol{\sigma}\right]^{-1}T_i\psi_{A0_1}.
    \label{eq:tripod_A1i_from_A01}
\end{equation}
Plugging \cref{eq:tripod_A1i_from_A01} into \cref{eq:tripod_system}, we obtain an equation for $\psi_{A0_1}$,
\begin{equation}
    (\delta \mathbf{k} \cdot \boldsymbol{\sigma})\psi_{A0_1} + \sum_{i=1}^3 T_i\frac{E(\delta \mathbf{k}) + (\delta \mathbf{k} - \mathbf{q}_i)\cdot \boldsymbol{\sigma}}{E(\delta \mathbf{k})^2 - (\delta \mathbf{k} - \mathbf{q}_i)^2}T_i \psi_{A0_1} = E(\delta \mathbf{k}) \psi_{A0_1}.
    \label{eq:tripod_psiA01_full}
\end{equation}
We want to show the existence of a double-degenerate zero-energy state exactly at the $K_M$ point, where $\delta \mathbf{k} = \vec{0}$. Letting $E(\vec{0})=0$, \cref{eq:tripod_system} simplifies to
\begin{equation}
    \sum_j T_j (\mathbf{q}_j \cdot \boldsymbol{\sigma}) T_j \psi_{A0_1} = 0,\qquad 
     -(\mathbf{q}_j \cdot \boldsymbol{\sigma})T_j  \psi_{A0_1} = \psi_{A1_j}
     \label{eq:A01_A1j}
\end{equation}
We notice that $\sum_j T_j (\mathbf{q}_j \cdot \boldsymbol{\sigma}) T_j = 0$, and therefore a solution exists. As such, one concludes that $E(\delta\mathbf{k}) \rightarrow 0$ in the vicinity of $\abs{\delta \mathbf{k}} \rightarrow 0$ due to the continuity of the flat-bands. By expanding \cref{eq:tripod_psiA01_full} to linear order in $E(\delta\mathbf{k})$ and $\delta \mathbf{k}$ we obtain
\begin{equation}
    (1-3w_1^2)\delta \mathbf{k} \cdot\sigma\psi_{A0_1} = \left[ 1 + 3(w_0^2 + w_1^2) \right] E(\delta\mathbf{k})\psi_{A0_1}.
\end{equation}
Finally, we can introduce the renormalized Dirac velocity~\cite{BIS11,BER21},
\begin{equation}
    v_D^{\textrm{(Tripod)}} = \frac{1 - 3w_1^2}{1 + 3(w_0^2 + w_1^2)},
    \label{eq:tripod_dirac_velocity}
\end{equation}
and rewrite the equations for the flat-band eigenstates in the one-shell tripod approximation as
\begin{align}
    &v_D^{\textrm{(Tripod)}} (\delta \mathbf{k} \cdot \boldsymbol{\sigma}) \psi_{A0_1} = E(\delta\mathbf{k}) \psi_{A0_1}, \numberthis \label{eq:tripod_equations_1}\\
    &\psi_{A1_i} = -(\mathbf{q}_i \cdot \boldsymbol{\sigma}) T_i \psi_{A0_1} \numberthis \label{eq:tripod_equations_2}.
\end{align}
\cref{eq:tripod_dirac_velocity} implies that the magic angle condition is $w_1=1/\sqrt{3},\;\;\forall w_0$, at which the Dirac velocity $v_D^{\textrm{(Tripod)}}$ vanishes at the $\vec{K}_M$ point. \cref{eq:tripod_equations_1} can now be easily solved and the two eigenstates corresponding to the flat bands $n=\pm1$ can be found to be
\begin{align}
    \psi_{A0_1}^{(n=-1)} &= \frac{\alpha}{\sqrt{2}}(-e^{-i\phi(\delta \mathbf{k})},1),
    \label{eq:tripod_A01_solution1} \\
    \psi_{A0_1}^{(n=+1)} &= \frac{\beta}{\sqrt{2}}(e^{-i\phi(\delta \mathbf{k})},1),
    \label{eq:tripod_A01_solution2}
\end{align}
where we have introduced $\phi(\delta \mathbf{k}) = \arctan{\frac{\delta k_x}{\delta k_y}}$ and the phases $\alpha, \beta$ are to be fixed. The discussion of the gauge fixing will be postponed to \cref{app:subsec:analytic_forbitals_tripod}. We will approximate the THF Wannier states with the Tripod model states \cref{eq:tripod_A01_solution1,eq:tripod_A01_solution2}.

Finally, we discuss the two-shell Tripod model [see \cref{eq:tripod_shells}]. An eigenstate of this model is given by
\begin{equation}
    \ket{\Psi} = \sum_{\alpha}\left[\psi_{A0_1,\alpha}\hat{c}^\dagger_{\mathbf{q}_1 + \delta \mathbf{k}, \mathbf{q}_1, \alpha, +, s}  + \sum_{i=1}^3 \psi_{A1_i,\alpha}\hat{c}^\dagger_{\mathbf{q}_1 + \delta \mathbf{k}, \mathbf{q}_1 + \mathbf{q}_i, \alpha, +, s} + \sum_{j=1}^6 \psi_{A2_j,\alpha}\hat{c}^\dagger_{\mathbf{q}_1 + \delta \mathbf{k}, \mathbf{q}_1 + \vec{G}_j, \alpha, +, s}\right] \ket{0},
\end{equation}
where $\vec{G}_j = \hat{C}_{6z}^j (\mathbf{b}_{M2} - \mathbf{b}_{M1})$ for $1 \leq j \leq 6$ are the six nearest neighbor moir\'e lattice vectors. Note that we have again made the $\delta \mathbf{k}$-dependence implicit in the spinors $\psi$ and the state $\ket{\Psi}$ for the sake of brevity. The first-quantized Hamiltonian acting on the twenty-dimensional spinor $\Psi^T = (\psi_{A0_1}^T,\psi_{A1_1}^T,\psi_{A1_2}^T,\psi_{A1_3}^T, \psi_{A2_1}^T, \ldots, \psi_{A2_6}^T)$ is given by
\begin{equation}
    H^{\textrm{(2-shell tripod)}}(\delta \mathbf{k},w_0,w_1) = 
    \begin{pmatrix}
    H^{\textrm{(1-shell tripod)}}(\delta \mathbf{k},w_0,w_1) & \hat{T}(w_0,w_1) \\
    \hat{T}^T(w_0,w_1) & H^{\textrm{2nd-shell}}(\delta \mathbf{k})
    \end{pmatrix},
\end{equation}
where
\begin{equation}
    H^{\textrm{2nd-shell}}_{ij}(\delta\mathbf{k}) = \delta_{ij}(\delta \mathbf{k} - \vec{G}_j)\cdot\boldsymbol{\sigma},
    \qquad \hat{T} = 
    \begin{pmatrix}
    0 & 0 & 0 & 0 & 0 & 0 \\
    T_2 & T_3 & 0 & 0 & 0 & 0 \\
    0 & 0 & T_3 & T_1 & 0 & 0 \\
    0 & 0 & 0 & 0 & T_1 & T_2
    \end{pmatrix},
\end{equation}
and $H^{\textrm{(1-shell tripod)}}$ was given in \cref{eq:tripod_hamiltonian}. The corresponding Schr\"odinger equation can be rewritten as a system of linear equations
\begin{align}
    \left(\delta \mathbf{k} \cdot \boldsymbol{\sigma}\right) \psi_{A0_1} + \sum_{i=1}^3T_i\psi_{A1_i} &= E(\delta\mathbf{k})\psi_{A0_1} \\
    T_i\psi_{A0_1} + \left[(\delta \mathbf{k} - \mathbf{q}_i)\cdot \boldsymbol{\sigma}\right]\psi_{A1_i} + \sum_{j=1}^6\hat{T}_{i+1j}\psi_{A2_j} &= E(\delta\mathbf{k})\psi_{A1_i}, \quad i=1,2,3
    \label{eq:tripod_A01_A1i_A2j} \\
    \sum_{i^{\prime}=1}^3\hat{T}^T_{ji^{\prime}+1}\psi_{A1_{i^{\prime}}} + \left[(\delta \mathbf{k} - \vec{G}_j)\cdot \boldsymbol{\sigma}\right] \psi_{A2_j} &= E(\delta\mathbf{k}) \psi_{A2_j},\quad j=1,\ldots,6.
    \label{eq:tripod_A2j_A1i}
\end{align}
For later use in \cref{app:subsec:analytic_forbitals_tripod}, we will now derive the relation between $\psi_{A0_1}$ and $\psi_{A1_1}$ corresponding to the TBG flat bands at the $K_M$ point. To do so, we solve the above system letting $\delta \mathbf{k} = \vec{0}$ and $E(\delta\mathbf{k})=0$. From \cref{eq:tripod_A2j_A1i}, and using the fact that $|\vec{G}_j|^2 = 3$, we find
\begin{equation}
    \psi_{A2_j} = \frac{1}{3}(\vec{G}_j\cdot\boldsymbol{\sigma})\sum_{i^{\prime}=1}^3\hat{T}^T_{ji^{\prime}+1}\psi_{A1_{i^{\prime}}},
\end{equation}
which we then plug into \cref{eq:tripod_A01_A1i_A2j} (for $i = 1$) to afford
\begin{equation}
    T_1\psi_{A0_1} + (-\vec{q}_1\cdot\boldsymbol{\sigma})\psi_{A1_1} + \frac{1}{3}\sum_{j=1}^6\sum_{i^{\prime}=1}^3\hat{T}_{2j}(\vec{G}_j\cdot\boldsymbol{\sigma})\hat{T}^T_{ji^{\prime}+1}\psi_{A1_{i^{\prime}}} = 0.
    \label{eq:2ndshell_A01_A11}
\end{equation}
We notice that $\hat{T}_{2j}$ has non-zero elements only if $j=1,2$. This, in turn, implies that $\hat{T}^T_{ji^{\prime} + 1}$ would have non-zero elements only for $i^{\prime} = 1$, simplifying \cref{eq:2ndshell_A01_A11}, which we write as
\begin{equation}
    T_1\psi_{A0_1} + (-\vec{q}_1\cdot\boldsymbol{\sigma})\psi_{A1_1} + \frac{1}{3}\left[T_2(\vec{G}_1\cdot\boldsymbol{\sigma})T_2+ T_3(\vec{G}_2\cdot\boldsymbol{\sigma})T_3\right]\psi_{A1_1} = 0.
\end{equation}
One can show, through a straightforward calculation that $\frac{1}{3}\left[T_2(\vec{G}_1\cdot\boldsymbol{\sigma})T_2+ T_3(\vec{G}_2\cdot\boldsymbol{\sigma})T_3\right] = w_0^2\sigma_y$. Noticing that $\mathbf{q}_1 \cdot \boldsymbol{\sigma} = \sigma_y$, we can express $\psi_{A1_1}$ in terms of $\psi_{A0_1}$ as
\begin{equation}
    \psi_{A1_1} = \frac{1}{1-w_0^2}\sigma_yT_1\psi_{A0_1}
    \label{eq:2nd_shell_A11_A01}.
\end{equation}
We will use \cref{eq:2nd_shell_A11_A01} for obtaining the parameters of the local-fermion wave functions in \cref{app:subsec:analytic_forbitals_tripod}.

\subsection{The Hexagon model}\label{app:subsec:hexagon}
The Tripod model is useful for approximating the TBG eigenspectrum near the $K_M$ point. To obtain an approximation of the spectrum at the $\Gamma_M$ point, we now review the Hexagon model introduced in Ref.~\cite{BER21}. As shown in \cref{fig:A1:c}, in the hexagonal model one considers two shells, $A$ and $B$. In what follows, we will revise the hexagon single-shell model, and quote results for the two shell approximation. For the single-shell Hexagon model, similarly to the Tripod modeled discussed in \cref{app:subsec:tripod}, we consider a reduced amount of the plane-waves, corresponding to the $A1_1,A1_2,A1_3,A1_4,A1_5,A1_6$ lattice sites with the respective coordinates $\mathbf{Q}_1=\mathbf{q}_1,\;\mathbf{Q}_2=-\mathbf{q}_3,\;\mathbf{Q}_3=\mathbf{q}_2,\;\mathbf{Q}_4=-\mathbf{q}_1,\;\mathbf{Q}_5=\mathbf{q}_3,\;\mathbf{Q}_6=-\mathbf{q}_2$. An eigenstate in the $\eta=+$ valley can be written as
\begin{equation}
    \ket{\Psi(\mathbf{k})} = \sum_{\alpha}\sum_{j=1}^{6}\psi_{A1_j}(\mathbf{k})\hat{c}^\dagger_{\mathbf{k},\mathbf{Q}_j,\alpha,+,s} \ket{0}.
    \label{eq:hexagon_spinor_definition}
\end{equation}
The first-quantized Hamiltonian for the valley $\eta=+$, acting on the 12-dimensional spinor $\Psi = (\psi_{A1_1}^T,\dots, \psi_{A1_6}^T)^T$ is given by
\begin{equation}
    H^{\textrm{hex}}(\mathbf{k},w_0,w_1) = 
    \begin{pmatrix}
        (\mathbf{k}-\mathbf{q}_1)\cdot\boldsymbol{\sigma} & T_2 & 0 & 0 & 0 & T_3 \\
        T_2 & (\mathbf{k}+\mathbf{q}_3)\cdot\boldsymbol{\sigma} & T_1 & 0 & 0 & 0 \\
        0 & T_1 & (\mathbf{k}-\mathbf{q}_2)\cdot\boldsymbol{\sigma} & T_3 & 0 & 0 \\
        0 & 0 & T_3 & (\mathbf{k}+\mathbf{q}_1)\cdot\boldsymbol{\sigma} & T_2 & 0 \\
        0 & 0 & 0 & T_2 & (\mathbf{k}-\mathbf{q}_3)\cdot\boldsymbol{\sigma} & T_1 \\
        T_3 & 0 & 0 & 0 & T_1 & (\mathbf{k}+\mathbf{q}_2)\cdot\boldsymbol{\sigma}
    \end{pmatrix},
    \label{eq:hexagon_hamiltonian}
\end{equation}
where we suppressed the $\mathbf{k}$ dependence in the spinors $\Psi$ and $\psi$ for the sake of brevity. 

\begin{figure}
    \centering
    \includegraphics{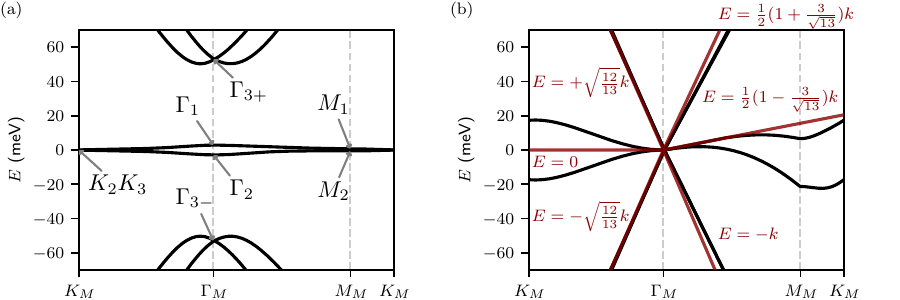}\subfloat{\label{fig:A3:a}}\subfloat{\label{fig:A3:b}}\caption{The BM, hexagon and the approximate six-band model band structures. (a) Schematics of the BM model band structure. The irreps at $K_M$, $\Gamma_M$ and $M_M$ points are indicated with the gray arrows. (b) The band structure (black) of the Hexagon model from \cref{eq:hexagon_hamiltonian} in the isotropic limit $w_0=w_1=1/\sqrt{3}$ and the dispersion (red) of the approximate six-band model from \cref{eq:Hexagon_6band_hamiltonian}. We label the dispersion relations for each of the red lines.}
    \label{fig:A3}
\end{figure}

The Hexagon model Hamiltonian \cref{eq:hexagon_hamiltonian} cannot be solved analytically for a general momentum $\mathbf{k}$ and general BM model parameters $w_0$, $w_1$. However, it is possible to obtain analytical expressions for the eigenenergies at the $\Gamma_M$ point for any  $w_0$, $w_1$. The results can be found in Table I of Ref.~\cite{BER21}. The six eigenstates at the $\Gamma_M$ point, closest to the charge-neutrality point, form $\Gamma_1\oplus\Gamma_2\oplus2\Gamma_3$ irreducible representations (irreps) of the little group of the $\Gamma_M$ point~\cite{SON19}. In an abuse of notation, we will refer to the specific eigenstates of the BM model at the $\Gamma_M$ point by their irreps. We indicate the corresponding eigenenergies in the band structure in \cref{fig:A3:a}, where we label the positive and negative energy $2\Gamma_3$ irreps as $\Gamma_{3+}$ and $\Gamma_{3-}$, respectively. From the one-shell hexagon approximation the analytical expressions for the $\Gamma_1,\Gamma_2$ and $\Gamma_{3\pm}$ eigenenergies are given by~\cite{BER21}
\begin{align}
    E_{\Gamma_{1}} &= + (2w_1 - \sqrt{1 + w_0^2}),
    \numberthis \label{eq:hexagon_gamma1} \\
    E_{\Gamma_{2}} &= - (2w_1 - \sqrt{1 + w_0^2}),
    \numberthis \label{eq:hexagon_gamma2} \\
    E_{\Gamma_{3-}} &= -\frac{1}{2}(\sqrt{4+w_0^2} - \sqrt{9w_0^2 + 4w_1^2}),
    \numberthis \label{eq:hexagon_gamma3-} \\
    E_{\Gamma_{3+}} &= +\frac{1}{2}(\sqrt{4+w_0^2} - \sqrt{9w_0^2 + 4w_1^2}).
    \numberthis \label{eq:hexagon_gamma3+}
\end{align}

The energetic splittings between $\Gamma_1$ and $\Gamma_2$ irreps, as well as between $\Gamma_{3\pm}$ irreps are well-captured by the THF model~\cite{SON22}. The discussion of how to relate the THF model parameters to the energetic splittings introduced in the hexagonal model is relegated to \cref{app:subsec:analytic_cbands_hexagon,app:subsec:analytic_other_hexagon}. In \cref{app:subsec:analytic_cbands_hexagon}, we will also employ a slightly better approximation of the energy splitting between the $\Gamma_1$ and $\Gamma_2$ irreps which was obtained within the two-band approximation of the Hexagon model in the so-called second magic manifold ($w_0\leq1/\sqrt{3},\;w_1 = \frac{1}{2}\sqrt{1+w_0^2}$)~\cite{BER21}. The result of this approximation reads~\cite{BER21}
\begin{equation}
    E_{\Gamma_1} - E_{\Gamma_2} = \left|\frac{-4 w_1\sqrt{w_0^2+1}+w_0^2+w_1^2+2}{2 \sqrt{w_0^2+1}}\right|.
    \label{eq:hexagon_2ndshell_bandwigth}
\end{equation}

To make a connection with the THF model parameters characterizing the conduction band electrons, defined in \cref{app:subsec:HF_conduction_bands,,app:subsec:HF_coupling}, we also discuss the isotropic limit of the Hexagon model in which $w_0=w_1=1/\sqrt{3}$. In the isotropic limit, a six-fold degenerate zero-energy state at the $\Gamma_M$ point emerges in the Hexagon model, as seen in \cref{fig:A3:b}. In order to obtain approximate analytical expressions for the energy dispersion near the $\Gamma_M$ point, we write an effective six-band model from the $\mathbf{k}\cdot\mathbf{p}$ expansion of the hexagon Hamiltonian given by \cref{eq:hexagon_hamiltonian}
\begin{equation}
    H^{\textrm{6-band}}_{ij}(\mathbf{k}) = \Psi_{i}^{\dagger}\left(\mathds{1}_{6\times6}\otimes\mathbf{k}\cdot\boldsymbol{\sigma}\right)\Psi_{j},
    \label{eq:Hexagon_6band_hamiltonian}
\end{equation}
where $\Psi_{i}$ (for $1 \leq i \leq 6$) are the 12-dimensional spinors denoting the eigenstates of the first-quantized Hexagon model Hamiltonian at $\mathbf{k}=\vec{0}$ from \cref{eq:hexagon_hamiltonian} corresponding to the six-degenerate subspace at zero energy and satisfying the eigenvalue equation
\begin{equation}
    H^{\textrm{hex}}(\vec{0},1/\sqrt{3},1/\sqrt{3})\Psi_{i} = 0.
\end{equation}
The explicit form of the eigenstates $\Psi_i$ and the six-band Hamiltonian $H^{\textrm{6-band}}_{ij}(\mathbf{k})$ are given in Ref.~\cite{BER21}. To extract the energy dispersion of the $6$-band $\mathbf{k}\cdot\mathbf{p}$ model, we find the roots of the characteristic polynomial of $H^{\textrm{6-band}} (\mathbf{k})$~\cite{BER21}
\begin{equation}
    \det(E -  H^{\textrm{6-band}}(\mathbf{k})) = 0 \implies \left[13E^2 - 12 (k_x^2 + k_y^2)E + k_x(k_x^2 - 3k_y^2)\right]^2 = 0.
    \label{eq:Hexagon_characteristic}
\end{equation}
Along the $\Gamma_M-K_M$ line [{\it i.e.}{}, $\mathbf{k} = k \left(0 , 1 \right)$], therefore, \cref{eq:Hexagon_characteristic} yields
\begin{equation}
    \Gamma_M-K_M:\qquad (13E^3 - 12k^2E)^2 = 0 \implies E = 0,\; \pm \sqrt{\frac{12}{13}}k.
\end{equation}
Along the $\Gamma_M-M_M$ line [{\it i.e.}{}, $\mathbf{k} = k \left(1, 0 \right)$], from \cref{eq:Hexagon_characteristic} we obtain
\begin{equation}
     \Gamma_M-M_M:\qquad (k+E)^2(k^2 - 13kE + 13 E^2)^2 = 0 \implies E = -k,\;\frac{1}{2}(1 \pm \frac{3}{\sqrt{13}})k.
\end{equation}
Denoting the dispersion around the $\Gamma_M$ point as $E(k) = v_{\star}k$, we obtain the following values for the Dirac velocity $v_{\star}$ of the Hexagon model Hamiltonian in the isotropic limit $w_0=w_1=1/\sqrt{3}$ along the $\Gamma_M-K_M$ and $\Gamma_M-M_M$ directions
\begin{equation}
    v_{\star} = \begin{cases}
    0, & \mathbf{k} \in \Gamma_M-K_M \\
    \pm \sqrt{\frac{12}{13}}v_F, & \mathbf{k} \in \Gamma_M-K_M \\
    -v_F, & \mathbf{k} \in \Gamma_M-M_M \\
    \frac{1}{2}(1 \pm \frac{3}{\sqrt{13}})v_F, & \mathbf{k} \in \Gamma_M-M_M
    \end{cases}.
\end{equation}
We plot the resulting dispersion in \cref{fig:A3:b}.

\subsection{Review of the BM model symmetries}\label{app:subsec:BM_symmetries}
The full crystalline symmetries of the BM model form the $P622$ space group generated by $C_{6z}$, $C_{2x}$, translations, and the time reversal symmetry $T$~\cite{SON19,BER21a}. Within each valley, the BM model is described by the magnetic space group $P6^{\prime}2^{\prime}2$, generated by $C_{2z}T$, $C_{2x}$, $C_{3z}$, and translations~\cite{SON19,BER21a}. We denote the action of a symmetry operator $g$ on the BM model basis state $\hat{c}^\dagger_{\mathbf{k}, \mathbf{Q}, \alpha, \eta, s}$ as

\begin{equation}
    \hat{g}\hat{c}^\dagger_{\mathbf{k},\mathbf{Q},\alpha,\eta,s}g^{-1} = \sum_{\mathbf{Q}^{\prime}, \alpha^{\prime}, \eta^{\prime}}
    \left[D(g)\right]_{\mathbf{Q}^{\prime}\alpha^{\prime} \eta^{\prime},\mathbf{Q}\alpha\eta}
    \hat{c}^\dagger_{g\mathbf{k},\mathbf{Q}^{\prime},\alpha^{\prime},\eta^{\prime},s}
\end{equation}
where $ \left[D(g)\right]_{\mathbf{Q}^{\prime}\alpha^{\prime} \eta^{\prime},\mathbf{Q}\alpha\eta}$ is the representation of the symmetry operator $g$. We note that in the absence of the spin-orbit coupling, the representation $D(g)$ does not act on the spin degrees of freedom. The representation matrices for the space group generators and time-reversal symmetries are defined as
\begin{align}
    \left[D(T)\right]_{\mathbf{Q}^{\prime}\alpha^{\prime} \eta^{\prime},\mathbf{Q}\alpha\eta} &= \delta_{\mathbf{Q}^{\prime},-\mathbf{Q}}\left[\sigma_0\right]_{\alpha^{\prime},\alpha}\left[\tau_x\right]_{\eta^{\prime},\eta}, \numberthis \label{eq:BM_symmetries_T} \\
    \left[D(C_{3z})\right]_{\mathbf{Q}^{\prime}\alpha^{\prime} \eta^{\prime},\mathbf{Q}\alpha\eta} &= \delta_{\mathbf{Q}^{\prime},C_{3z}\mathbf{Q}}\left[e^{i\frac{2\pi}{2}\sigma_z \tau_z}\right]_{\alpha^{\prime}\eta^{\prime},\alpha\eta}, \numberthis \label{eq:BM_symmetries_C3z} \\
    \left[D(C_{2x})\right]_{\mathbf{Q}^{\prime}\alpha^{\prime} \eta^{\prime},\mathbf{Q}\alpha\eta} &= \delta_{\mathbf{Q}^{\prime},-\mathbf{Q}}\left[\sigma_0\right]_{\alpha^{\prime},\alpha}\left[\tau_x\right]_{\eta^{\prime},\eta}, \numberthis \label{eq:BM_symmetries_C2x} \\ \left[D(C_{2z}T)\right]_{\mathbf{Q}^{\prime}\alpha^{\prime} \eta^{\prime},\mathbf{Q}\alpha\eta} &= \delta_{\mathbf{Q}^{\prime},\mathbf{Q}}\left[\sigma_x\right]_{\alpha^{\prime},\alpha}\left[\tau_0\right]_{\eta^{\prime},\eta}, \numberthis \label{eq:BM_symmetries_C2zT}
\end{align}
where $\sigma_{x,y,z} (\sigma_0)$ are the Pauli (identity) matrices acting in the sublattice space and $\tau_{x,y,z} (\tau_0)$ are the Pauli (identity) matrices acting in the valley space. The BM model also possesses a unitary particle-hole symmetry $P$~\cite{SON19,SON21} whose action is defined as
\begin{equation}
    P\hat{c}^\dagger_{\mathbf{k},\mathbf{Q},\eta,\alpha,s}P^{-1} = \sum_{\mathbf{Q}',\alpha',\eta'}\left[D(P)\right]_{\mathbf{Q}^{\prime}\alpha^{\prime} \eta^{\prime},\mathbf{Q}\alpha\eta}\hat{c}^\dagger_{-\mathbf{k},\mathbf{Q}',\eta',\alpha',s},
\end{equation}
and the representation matrix $D(P)$ is given by
\begin{equation}
     \left[D(P)\right]_{\mathbf{Q}^{\prime}\alpha^{\prime} \eta^{\prime},\mathbf{Q}\alpha\eta} = \zeta_{\mathbf{Q}} \delta_{\mathbf{Q}^{\prime},-\mathbf{Q}}\left[\sigma_0\right]_{\alpha^{\prime},\alpha}\left[\tau_z\right]_{\eta^{\prime},\eta},
    \label{eq:BM_PH_symmetry}
\end{equation}
where $\zeta_{\mathbf{Q}} = \pm1$ for $\mathbf{Q} \in \mathcal{Q}_{\pm}$. The unitary particle-hole symmetry anticommutes with the BM model Hamiltonian defined in \cref{eq:BM_model_hamiltonian}~\cite{SON19,SON21}
\begin{equation}
    \{\hat{H}_{\textrm{BM}}, P\} = 0.
\end{equation}
The particle-hole symmetry can be combined with $C_{2z}$ to obtain a $\mathbf{k}$-preserving symmetry whose representation matrix reads as
\begin{equation}
    \left[D(C_{2z}P)\right]_{\mathbf{Q}^{\prime}\alpha^{\prime} \eta^{\prime},\mathbf{Q}\alpha\eta} = -i\zeta_{\mathbf{Q}}\delta_{\mathbf{Q}^{\prime},\mathbf{Q}}[\sigma_x]_{\alpha^{\prime}\alpha}[\tau_y]_{\eta^{\prime}\eta}
    \label{eq:BM_c2zP_symm}
\end{equation}

One can also define the action of a symmetry $g$ in the real space basis
\begin{equation}
    \hat{g}\hat{c}^\dagger_{l,\alpha,\eta,s}(\vec{r})\hat{g}^{-1} = \sum_{l^{\prime},\alpha^{\prime},\eta^{\prime}}D_{l^{\prime}\alpha^{\prime}\eta^{\prime},l\alpha\eta}(g)\hat{c}^\dagger_{l^{\prime},\alpha^{\prime},\eta^{\prime},s}(g\vec{r}).
\end{equation}
From \cref{eq:real_space_basis} and the symmetry representations in momentum space given in \cref{eq:BM_symmetries_T,eq:BM_symmetries_C3z,eq:BM_symmetries_C2x,eq:BM_symmetries_C2zT}, the representation matrices in the real space basis are given by~\cite{SON22}
\begin{align}
    D_{l^{\prime}\alpha^{\prime}\eta^{\prime},l\alpha\eta}(T) &= [\rho_0]_{l^{\prime}l}[\sigma_0]_{\alpha^{\prime}\alpha}[\tau_x]_{\eta^{\prime}\eta} \label{eq:real_basis_symmetries_T}\\
    D_{l^{\prime}\alpha^{\prime}\eta^{\prime},l\alpha\eta}(C_{3z}) &= [\rho_0]_{l^{\prime}l}[e^{i\frac{2\pi}{3}\sigma_z\tau_z}]_{\alpha^{\prime}\eta^{\prime},\alpha\eta}, \label{eq:real_basis_symmetries_C3z}\\
    D_{l^{\prime}\alpha^{\prime}\eta^{\prime},l\alpha\eta}(C_{2x}) &= [\rho_x]_{l^{\prime}l}[\sigma_x]_{\alpha^{\prime}\alpha}[\tau_0]_{\eta^{\prime}\eta}, \label{eq:real_basis_symmetries_C2x}\\
    D_{l^{\prime}\alpha^{\prime}\eta^{\prime},l\alpha\eta}(C_{2z}T) &= [\rho_0]_{l^{\prime}l}[\sigma_x]_{\alpha^{\prime}\alpha}[\tau_0]_{\eta^{\prime}\eta},
    \label{eq:real_basis_symmetries_C2zT}\\
    D_{l^{\prime}\alpha^{\prime}\eta^{\prime},l\alpha\eta}(P) &= [-i\rho_y]_{l^{\prime}l}[\sigma_0]_{\alpha^{\prime}\alpha}[\tau_0]_{\eta^{\prime}\eta}, \label{eq:real_basis_symmetries_P}
\end{align}
where $\sigma_{0,x,y,z}$, $\tau_{0,x,y,z}$, $\rho_{0,x,y,z}$ are the identity and Pauli matrices in the sublattice, valley and layer spaces, respectively.

\subsection{Chern band basis and gauge fixing}\label{app:subsec:BM_Chern}
In this section, we briefly review the gauge-fixing conditions for the BM model eigenstates~\cite{BER21,SON21} and the Chern band basis~\cite{BER21, SON21,BUL20,HEJ21}. At the end of the section, we derive the Chern band basis transformation under BM model symmetries for further use in the analytical derivation of the THF model parameters in \cref{app:subsec:analytic_forbitals_tripod}.

When $g$ is a symmetry commuting (anticommuting) of the model, {\it i.e.}{} satisfying $[\hat{H}_0,g] = 0$ ($\{\hat{H}_0,g\}=0$), if $u_{\mathbf{Q}\alpha,n\eta}(\mathbf{k})$ is an eigenstate wave function of TBG, then $\sum_{\mathbf{Q}^{\prime},\alpha^{\prime},\eta^{\prime}}[D(g)]_{\mathbf{Q}^{\prime}\alpha^{\prime}\eta^{\prime},\mathbf{Q}\alpha\eta}u_{\mathbf{Q}^{\prime}\alpha^{\prime},n\eta^{\prime}}(\mathbf{k})$ is also an eigenstate at momentum $g\mathbf{k}$ with the same (opposite) energy, with an additional complex conjugation when $g$ is antiunitary. This allows us to introduce a sewing matrix $B^g(\mathbf{k})$~\cite{BER21a,LIA21}
\begin{equation}
    \sum_{\mathbf{Q}^{\prime},\alpha^{\prime}}[D(g)]_{\mathbf{Q}^{\prime}\alpha^{\prime}\eta^{\prime},\mathbf{Q}\alpha\eta}u_{\mathbf{Q}^{\prime}\alpha^{\prime},n\eta^{\prime}}(\mathbf{k}) = \sum_{m} [B^g(\mathbf{k})]_{m\eta^{\prime},n\eta}u_{\mathbf{Q}\alpha,m\eta}(g\mathbf{k})
    \label{eq:definition_sewing_matrix}
\end{equation}
In this work, we employ the following gauge-fixing conventions for the $C_{2z}P$ and $C_{2z}T$ symmetries~\cite{BER21a,LIA21} 
\begin{equation}
    [B^{C_{2z}T}(\mathbf{k})]_{m\eta^{\prime},n\eta} = \delta_{m,n}\delta_{\eta^{\prime},\eta},\qquad [B^{C_{2z}P}(\mathbf{k})]_{m\eta^{\prime},n\eta} = -\textrm{sgn}(n)\eta^{\prime}\delta_{-m,n}\delta_{\eta^{\prime},-\eta}.
    \label{eq:BM_gauge_fixing}
\end{equation}
Additionally, we fix the relative sign between the eigenstates corresponding to the electron and valence bands of TBG by imposing~\cite{BER21a,LIA21} 
\begin{equation}
    \lim_{\mathbf{q}\rightarrow0}\sum_{\mathbf{Q},\alpha}
    \left|\hat{u}^\dagger_{n\eta,\mathbf{Q}\alpha}(\mathbf{k}+\mathbf{q})u_{\mathbf{Q}\alpha,n\eta}(\mathbf{k}) - \hat{u}^\dagger_{-n\eta,\mathbf{Q}\alpha}(\mathbf{k}+\mathbf{q})u_{\mathbf{Q}\alpha,-n\eta}(\mathbf{k})\right| = 0.
\end{equation}
In this gauge choice, we introduce a Chern-band basis for the TBG active bands $n_B = \pm1$~\cite{AHN19,HEJ21,BUL20,BER21,SON21}
\begin{equation}
    d^{\dagger}_{\mathbf{k},e_Y,\eta,s} = \frac{\hat{c}^\dagger_{\mathbf{k},n_B,\eta,s} + i e_Y \hat{c}^\dagger_{\mathbf{k},-n_B,\eta,s}}{\sqrt{2}},\;\;(e_Y=\pm1).
\end{equation}
In terms of the wave functions, the Chern band basis wave function can be written as
\begin{equation}
    U^{e_Y}_{\mathbf{Q}\alpha,\eta} = \frac{1}{\sqrt{2}}\left(u_{\mathbf{Q}\alpha,+1\eta} + ie_Yu_{\mathbf{Q}\alpha,-1\eta}\right).
    \label{eq:chern_wave function}
\end{equation}
From \cref{eq:BM_gauge_fixing}, we derive the transformation of the Chern band basis under the action of the $C_{2z}T$ and $C_{2z}P$ symmetries
\begin{alignat}{3}
    C_{2z}T: && \quad U^{-e_Y*}_{\mathbf{Q}\bar{\alpha},\eta}  &= U^{e_Y}_{\mathbf{Q}\alpha,\eta},
    \label{eq:BM_chern_transform_C2zT} \\
    C_{2z}P: && \quad \zeta_{\mathbf{Q}}U^{e_Y}_{\mathbf{Q}\bar{\alpha},\eta} &= ie_YU^{e_Y}_{\mathbf{Q}\alpha,-\eta},
    \label{eq:BM_chern_transform_C2zP}
\end{alignat}
where $\bar{\alpha} = 3-\alpha$ for $\alpha=1,2$.

In \cref{app:subsec:analytic_forbitals_tripod}, we will additionally fix the gauge of the Chern band basis states at the $K_M$ point by invoking the $C_{3z}$ symmetry. In the final part of this section, we derive the form of the $C_{3z}$ symmetry sewing matrix $B^{C_{3z}}(\mathbf{k})$ at the $K_M$ point. Since the $C_{3z}$ symmetry does not exchange the valleys, as seen in \cref{eq:BM_symmetries_C3z}, the corresponding sewing matrix takes the following form
\begin{equation}
    \left[B^{C_{3z}}(\mathbf{k})\right]_{m\eta',n\eta} = \delta_{\eta',\eta}B_{mn}(\mathbf{k}),
    \label{eq:C3z_sewing_matrix_general}
\end{equation}
where the complex matrix $B_{mn}$ will be obtained below. For the $C_{3z}$ symmetry the following commutation relations hold~\cite{SON21,BER21a}
\begin{equation}
    \left[C_{3z},C_{2z}T\right]=0 \qquad \left[C_{3z},C_{2z}P\right]=0,
\end{equation}
which translate, respectively, using \cref{eq:definition_sewing_matrix}, into constraints on the sewing matrix
\begin{equation}
    B_{mn}(\mathbf{k}) = B_{mn}^{\star}(\mathbf{k}),\qquad \text{sgn}(m)B_{mn}(\mathbf{k}) = \text{sgn}(n)B_{-m,-n}(\mathbf{k}).
    \label{eq:C3z_constraints_sewing_matrix}
\end{equation}
As a unitary matrix, the sewing matrix $B^{C_{3z}}$ has determinant of modulus one. As such, the matrix $B_{mn}(\mathbf{k})$ satisfying the constraints from \cref{eq:C3z_constraints_sewing_matrix} can be generically written as
\begin{equation}
    B_{mn} (\mathbf{k}) = \begin{pmatrix}
        \cos{\theta(\mathbf{k})} & -\sin{\theta(\mathbf{k})} \\
        \sin{\theta(\mathbf{k})} & \cos{\theta(\mathbf{k})}
    \end{pmatrix}_{mn},
    \label{eq:sewing_matrix_C3z_bands}
\end{equation}
where the phase $\theta(\mathbf{k})\in \{0,\pi\}$ when the active TBG bands are non-degenerate at $\mathbf{k}$ ({\it i.e.}{} at a generic point away from $K_M$). On the contrary, when $\mathbf{k}$ is at the $K_M$ point, the phase $\theta(\mathbf{k})$ can assume any value $\theta(\mathbf{k})\in [0,2\pi)$ and has to be fixed further. From \cref{eq:definition_sewing_matrix,eq:BM_symmetries_C3z,eq:sewing_matrix_C3z_bands,eq:C3z_sewing_matrix_general}, we derive the way the Chern basis states transform under the $C_{3z}$ symmetry:
\begin{equation}
    e^{i\eta\frac{2\pi}{3}(-1)^{\beta+1}}U^{e_Y}_{C_{3z}\mathbf{Q}\beta,\eta}(C_{3z}\mathbf{k}) = e^{ie_Y\theta(\mathbf{k})} U^{e_Y}_{\mathbf{Q}\beta,\eta}(\mathbf{k}).
    \label{eq:Chern_C3z_transformation}
\end{equation}
The phase $\theta(\mathbf{k})$ will be fixed in \cref{app:subsec:analytic_forbitals_tripod}.

\section{The single-particle THF model}\label{app:sec:HF_model}
In this section, we review the single-particle THF model, first introduced in Ref.~\cite{SON22}. The THF model comprises two types of fermions, the so-called ``heavy'' $(f)$ and ``conduction'' $(c)$ electrons. The $f$-fermions represent electronic states localized at the TBG $AA$-sites, transforming as $p_x \pm i p_y$ orbitals under the TBG symmetry group, while the $c$-fermions correspond to semimetallic conduction band electronic states. Introducing two types of fermions enables one to resolve the stable topological obstruction of the entire continuum BM model~\cite{SON21,SON22}, in addition to the fragile topological obstruction of the TBG active bands~\cite{SON22,SON21,SON19,BUL20,AHN19,PO18c}. 

The goal of this section is to formalize the notation of the THF model~\cite{SON22} and define its parameters, which will be obtained analytically in \cref{app:sec:analytic_single_particle} and calculated numerically in \cref{app:sec:numerics}. We start by reviewing the $f$-electron states, discussing their symmetry properties, and providing the $f$-electron part of the single-particle THF Hamiltonian. In \cref{app:subsec:HF_conduction_bands}, we then proceed with the construction of the conduction band electronic states, review their symmetry properties, and provide the conduction band part of the single-particle Hamiltonian. In \cref{app:subsec:HF_coupling}, we discuss the coupling terms between the $f$- and $c$-electron states and finally summarize the full single-particle Hamiltonian in \cref{app:subsec:HF_single_summary}. 

\subsection{Local orbital electrons}\label{app:subsec:local_orbitals}
By construction, the local $f$-electron states transform as $p_x \pm ip_y$ orbitals within the symmetry group of TBG~\cite{SON22}. For brevity, we will find it useful to employ the first-quantized formalism. As such, we define the momentum space basis states as
\begin{equation}
    \ket{\mathbf{k},\mathbf{Q},\alpha,\eta,s} = \hat{c}^\dagger_{\mathbf{k},\mathbf{Q},\alpha,\eta,s}\ket{0}.
\end{equation}
Correspondingly, the continuous real-space basis states are given by 
\begin{equation}
    \ket{\vec{r},l,\alpha,\eta,s}=\hat{c}^\dagger_{l,\alpha,\eta,s}(\vec{r})\ket{0}= \frac{1}{\sqrt{\Omega_{\textrm{tot}}}}\sum_{\mathbf{k}\in\textrm{MBZ}}\sum_{\mathbf{Q}\in\mathcal{Q}_{l\eta}}e^{-i(\mathbf{k}-\mathbf{Q})\vec{r}}\ket{\mathbf{k},\mathbf{Q},\alpha,\eta,s},
\end{equation}
where $\mathbf{r}$ is a continuous variable.

The $f$-fermion wave functions are defined as Wannier states
\begin{equation}
    \ket{W_{\vec{R},\alpha,\eta,s}} = \frac{1}{\sqrt{N}}\sum_{l=\pm}\sum_{\mathbf{k}\in\textrm{MBZ}}\sum_{\beta}\sum_{\mathcal{Q}\in\mathcal{Q}_{l\eta}} \ket{\mathbf{k},\mathbf{Q},\beta,\eta,s}e^{-i\mathbf{k}\cdot\vec{R}}v^{(\eta)}_{\mathbf{Q}\beta,\alpha}(\mathbf{k})
    \label{eq:wannier_state}
\end{equation}
where $\alpha \in \{1,2\}$ denotes the orbital quantum number, such that $\alpha=1$ for the $p_x + ip_y$ orbital and $\alpha=2$ for the $p_x -ip_y$ orbital. In \cref{eq:wannier_state}, $N$ is the number of Moir\'e unit cells. The Fourier components $v^{(\eta)}_{\mathbf{Q}\beta,\alpha}$ are obtained either analytically or numerically. In the former case, $v^{(\eta)}_{\mathbf{Q}\beta,\alpha}$ are derived from an approximation of the BM model and expressed as a function of the BM model parameters. We discuss this approach in \cref{app:subsec:analytic_forbitals_tripod}. In the later case, the components $v^{(\eta)}_{\mathbf{Q}\beta,\alpha}$ are obtained through the disentanglement and Wannierization procedures, as was done in Ref.~\cite{SON22} and briefly reviewed in \cref{app:sec:numerics}.

We can write an expression for the $f$-electron wave functions in real space as
\begin{equation}
    \small
\bra{\vec{r},l,\beta,\eta,s}\ket{W_{\vec{R},\alpha,\eta,s}}=\bra{\vec{r},l,\beta,\eta,s}T_{\vec{R}}\ket{W_{0,\alpha,\eta,s}}=e^{-i\eta\Delta\mathbf{K}_{l}\cdot\vec{R}}\bra{\vec{r}-\vec{R},l,\beta,\eta,s}\ket{W_{0,\alpha,\eta,s}}=e^{-i\eta\Delta\mathbf{K}_l\cdot\vec{R}}w_{l\beta,\alpha}^{(\eta)}(\vec{r}-\vec{R}),
\end{equation}
where $w^{(\eta)}_{l\beta,\alpha}(\vec{r}-\vec{R}) = e^{i\eta\Delta\vec{K}_l\cdot\vec{R}}\bra{\vec{r}-\vec{R},l,\beta,\eta,s}\ket{W_{0,\alpha,\eta,s}}$ is the real-space Wannier function. From \cref{eq:wannier_state} we infer
\begin{equation}
    w^{(\eta)}_{l\beta,\alpha}(\vec{r}-\vec{R})=e^{i\eta\Delta\mathbf{K}_l\cdot\vec{R}}\bra{\vec{r},l,\beta,\eta,s}\ket{W_{\vec{R},\alpha,\eta,s}}=\frac{1}{N\Omega_{\textrm{tot}}}\sum_{\mathbf{k}\in\textrm{MBZ}}\sum_{\mathbf{Q}\in\mathcal{Q}_{l\eta}}e^{i(\mathbf{k}-\mathbf{Q})\cdot(\vec{r}-\vec{R})}v_{\mathbf{Q}\beta,\alpha}^{(\eta)}(\mathbf{k}),
    \label{eq:local_orbital_r2q}
\end{equation}
which can be inverted as
\begin{equation}
    v^{(\eta)}_{\mathbf{Q}\beta,\alpha}(\mathbf{k}) = \frac{1}{\sqrt{\Omega_0}}\int d^2 \vec{r}\; w^{(\eta)}_{l_{\mathbf{Q},
    \eta},\beta,\alpha}(\vec{r}-\vec{R})e^{-i(\mathbf{k}-\mathbf{Q})(\vec{r}-\vec{R})}.
    \label{eq:wannier_fourier}
\end{equation}
Here $l_{\mathbf{Q},\eta}=\zeta_{\mathbf{Q}}\eta$ is the graphene layer on which the basis state $\ket{\mathbf{k},\mathbf{Q},\beta,\eta,s}$ is supported.
Note that there are 32 components of the real space functions $w_{l\beta,\alpha}^{(\eta)}(\vec{r})$: two per layer $l=\pm$, two per sublattice $\beta=\pm$, two per valley $\eta=\pm$, two per orbital $\alpha=1,2$, and, finally, two per spin $s=\uparrow,\downarrow$. However, as we will show below, only two are independent, with the rest following from symmetry constraints. More specifically, Ref.~\cite{SON22} has argued that a first-order approximation of the two independent components labeled by $l=+,\beta=1,2,\alpha=1,\eta=+$ is given by
\begin{equation}
    w^{(+)}_{+1,1}(\vec{r})=\frac{\alpha_1}{\sqrt{2}}\frac{1}{\sqrt{\pi\lambda_1^2}}e^{i\frac{\pi}{4}-\vec{r}^2/(2\lambda_1^2)},\qquad w^{(+)}_{+2,1}(\vec{r})=-\frac{\alpha_2}{\sqrt{2}}\frac{x + iy}{\lambda_2^2\sqrt{\pi}}e^{i\frac{\pi}{4} - \vec{r}^2/(2\lambda_2^2)},
    \label{eq:wannier_real_space_functions}
\end{equation}
where the parameters $\lambda_1,\lambda_2,\alpha_1,\alpha_2$ are to be obtained either analytically, as will be done in \cref{app:subsec:analytic_forbitals_tripod}, or calculated numerically, as was done in~\cite{SON22} and will be done in  \cref{app:sec:numerics}.

For any symmetry operator $g$ (with an additional complex conjugation in case $g$ is anti-unitary) the action on the Wannier states reads as~\cite{SON22}
\begin{equation}
    \sum_{l^{\prime},\beta^{\prime}}w_{l^{\prime}\beta^{\prime},\alpha}^{(\eta)}(g\vec{r})[D(g)]_{l^{\prime}\beta^{\prime}\eta,l\beta\eta^{\prime}}=\sum_{\alpha^{\prime}}w_{l\beta,\alpha^{\prime}}^{(\eta^{\prime})}(\vec{r})[D^f(g)]_{\alpha^{\prime}\eta^{\prime},\alpha\eta},
    \label{eq:wannier_symmetry_transform}
\end{equation}
where $D^f(g)$ is the symmetry representation matrix in the basis of the Wannier states. The $D^f(g)$ matrices are obtained from the fact that $\ket{W_{\vec{R},\alpha,\eta,s}}$ have symmetry properties identical to $p_x \pm i p_y$ located orbitals at the $1a$ Wyckoff position~\cite{SON22}. The symmetry representation matrices therefore read as 
\begin{align}
    [D^f(T)]_{\alpha^{\prime}\eta^{\prime},\alpha\eta}&=[\sigma_0]_{\alpha^{\prime},\alpha}[\tau_x]_{\eta^{\prime}\eta},
    \label{eq:wannier_symmetries_T} \\
    [D^f(C_{3z})]_{\alpha^{\prime}\eta^{\prime},\alpha\eta}&=[e^{i\frac{2\pi}{3}\sigma_z\tau_z}]_{\alpha^{\prime}\eta^{\prime},\alpha\eta},
    \label{eq:wannier_symmetries_C3z} \\
    [D^f(C_{2x})]_{\alpha^{\prime}\eta^{\prime},\alpha\eta}&=[\sigma_x]_{\alpha^{\prime},\alpha}[\tau_0]_{\eta^{\prime},\eta},
    \label{eq:wannier_symmetries_C2x} \\
    [D^f(C_{2z}T)]_{\alpha^{\prime}\eta^{\prime},\alpha\eta}&=[\sigma_x]_{\alpha^{\prime},\alpha}[\tau_0]_{\eta^{\prime},\eta},
    \label{eq:wannier_symmetries_C2zT} \\
    [D^f(P)]_{\alpha^{\prime}\eta^{\prime},\alpha\eta}&=i[\sigma_z]_{\alpha^{\prime},\alpha}[\tau_z]_{\eta^{\prime},\eta},
    \label{eq:wannier_symmetries_P} \\
    [D^f(C_{2z}P)]_{\alpha^{\prime}\eta^{\prime},\alpha\eta}&=-i[\sigma_y]_{\alpha^{\prime},\alpha}[\tau_y]_{\eta^{\prime},\eta},
    \label{eq:wannier_symmetries_C2zP}
\end{align}
where $\sigma_{0,x,y,z}$ and $\tau_{0,x,y,z}$ denote the identity and Pauli matrices in the orbital $\alpha=1,2$ and valley $\eta=\pm$ degrees of freedom, respectively.

As a result of the time-reversal symmetry, \cref{eq:BM_symmetries_T,,eq:wannier_symmetries_T,,eq:wannier_symmetry_transform} impose
\begin{equation}
    w_{l\beta,\alpha}^{(\eta)} = w_{l\beta,\alpha}^{(-\eta)*},
    \label{eq:wannier_TRS}
\end{equation}
where the complex conjugation is stemming from the antiunitarity of the time-reversal operator $T$.
Similarly, as a consequence of the particle-hole symmetry, from \cref{eq:BM_PH_symmetry,eq:wannier_symmetries_P,eq:wannier_symmetry_transform} we derive 
\begin{equation}
    w^{(\eta)}_{-l\beta,\alpha}(-\vec{r})=il\eta(-1)^{\alpha}w^{(\eta)}_{l\beta,\alpha}(\vec{r}).
    \label{eq:wannier_P_constraint}
\end{equation}
Finally, the $C_{2z}T$ symmetry imposes, from \cref{eq:BM_symmetries_C2zT,eq:wannier_symmetries_C2zT,eq:wannier_symmetry_transform},
\begin{equation}
    w_{l\beta,\alpha}^{(\eta)}(\vec{r})=w_{{l\bar{\beta},\bar{\alpha}}}^{(\eta)*}(-\vec{r}).
    \label{eq:wannier_C2zT_constraint}
\end{equation}
Given the constraints listed in \cref{eq:wannier_TRS,,eq:wannier_C2zT_constraint,,eq:wannier_P_constraint}, we can write down all the components of the real-space Wannier functions
\begin{alignat}{4}
    & w_{l1,1}^{(\eta)}(\vec{r}) && = \frac{\alpha_1}{\sqrt{2}}\frac{1}{\sqrt{\pi\lambda_1^2}}e^{i\frac{\pi}{4}l\eta - \vec{r}^2/(2\lambda_1^2)}, \qquad &&
    w_{l2,1}^{(\eta)}(\vec{r}) &&=-l\frac{\alpha_2}{\sqrt{2}}\frac{x + i\eta y}{\lambda_2^2\sqrt{\pi}}e^{i\frac{\pi}{4}l\eta - \vec{r}^2/(2\lambda_2^2)},
    \label{eq:wannier_states_I} \\
    & w_{l1,2}^{(\eta)}(\vec{r}) &&= l\frac{\alpha_2}{\sqrt{2}}\frac{x - i\eta y}{\lambda_2^2\sqrt{\pi}}e^{-i\frac{\pi}{4}l\eta - \vec{r}^2/(2\lambda_2^2)},
    \qquad && w_{l2,2}^{(\eta)}(\vec{r}) &&= \frac{\alpha_1}{\sqrt{2}}\frac{1}{\sqrt{\pi\lambda_1^2}}e^{-i\frac{\pi}{4}l\eta - \vec{r}^2/(2\lambda_1^2)}.
    \label{eq:wannier_states_II}
\end{alignat} 
Fourier-transforming the real-space Wannier functions according to \cref{eq:wannier_fourier}, we can also obtain the momentum space $f$-electron wave functions
\begin{align} 
    \widetilde{v}_{\mathbf{Q} 1, 1}^{(\eta)}(\mathbf{k}) =& \alpha_{1} \sqrt{ \frac{2\pi \lambda_{1}^2}{\Omega_M \mathcal{N}_{f,\mathbf{k}}} } e^{ i\frac{\pi}4 \zeta_\mathbf{Q} - \frac12 (\mathbf{k}-\mathbf{Q})^2 \lambda_{1}^2}, \nonumber\\
    \widetilde{v}_{\mathbf{Q} 2, 1}^{(\eta)}(\mathbf{k}) =& \alpha_{2}   \sqrt{ \frac{2\pi \lambda_{2}^4}{\Omega_M \mathcal{N}_{f,\mathbf{k}}} }  \zeta_\mathbf{Q} \left[ i\eta (k_x-Q_x) -  (k_y-Q_y) \right] e^{ i\frac{\pi}4 \zeta_\mathbf{Q} - \frac12 (\mathbf{k}-\mathbf{Q})^2 \lambda_{2}^2} \label{eq:vQ-1}, \\
    \widetilde{v}_{\mathbf{Q} 1, 2}^{(\eta)}(\mathbf{k}) =& \alpha_{2}  \sqrt{ \frac{2\pi \lambda_{2}^4}{\Omega_M \mathcal{N}_{f,\mathbf{k}} } } \zeta_\mathbf{Q}  \left[-i\eta (k_x-Q_x) - (k_y-Q_y) \right] e^{-i\frac{\pi}4 \zeta_\mathbf{Q} - \frac12 (\mathbf{k}-\mathbf{Q})^2 \lambda_{2}^2} ,\nonumber\\
    \widetilde{v}_{\mathbf{Q} 2, 2}^{(\eta)}(\mathbf{k}) =&  \alpha_{1} \sqrt{ \frac{2\pi \lambda_{1}^2}{\Omega_M \mathcal{N}_{f,\mathbf{k}}} } e^{-i\frac{\pi}4 \zeta_\mathbf{Q} - \frac12 (\mathbf{k}-\mathbf{Q})^2 \lambda_{1}^2} \label{eq:vQ-2}, 
\end{align}
In \cref{eq:vQ-1,eq:vQ-2}, we have introduced the normalization factor $\mathcal{N}_{f,\mathbf{k}}$, which can be determined to be 
\begin{equation}
\mathcal{N}_{f,\mathbf{k}} = \alpha_1^2 \frac{2\pi \lambda_1^2}{\Omega_M} 
    \sum_{\mathbf{Q}} e^{-(\mathbf{k}-\mathbf{Q})^2\lambda_1^2} 
+ \alpha_2^2 \frac{2\pi \lambda_2^2}{\Omega_M} \sum_{\mathbf{Q}} 
    (\mathbf{k}-\mathbf{Q})^2 e^{-(\mathbf{k}-\mathbf{Q})^2 \lambda_2^2} \ . 
\end{equation}

The analytic computation of the parameters $\lambda_1,\lambda_2, \alpha_1,\alpha_2$ will rely on the symmetry properties of the Wannier states in the momentum-space $v^{(\eta)}_{\mathbf{Q}\beta,\alpha}(\mathbf{k})$ under $C_{3z}$, $C_{2z}T$, and $C_{2z}P$ transformations, which we will now review~\cite{SON22}. By definition,
\begin{equation}
    \bra{\mathbf{k}, \mathbf{Q}, \beta, \tilde{\eta},s}\ket{W_{0,\alpha,\eta,s}} = v^{(\eta)}_{\mathbf{Q}\beta,\alpha}(\mathbf{k}) \delta_{\eta,\tilde{\eta}},
\end{equation}
where $\alpha=1,2$ denotes the orbital component and $\beta = 1,2$ the graphene sublattice. The symmetry action  on the Wannier state and the Bloch state can be written in the first-quantized formalism as
\begin{align}
    g\ket{W_{0,\alpha, \eta, s}} &= \sum_{\alpha^{\prime},\eta^{\prime}} [D^f(g)]_{\alpha^{\prime}\eta^{\prime}, \alpha\eta}\ket{W_{0,\alpha^{\prime},\eta^{\prime},s}}, \\
    \bra{\mathbf{k},\mathbf{Q},\beta,\tilde{\eta},s}g &= \sum_{\mathbf{Q}^{\prime},\beta^{\prime},\eta^{\prime}} [D(g)]_{\mathbf{Q}^{\prime}\beta^{\prime}\eta^{\prime},\mathbf{Q}\beta\tilde{\eta}}\bra{g\mathbf{k}, \mathbf{Q}^{\prime},\beta^{\prime},\eta^{\prime},s},
\end{align}
where $D(g)$ and $D^f(g)$ are given by \crefrange{eq:BM_symmetries_T}{eq:BM_symmetries_C2zT} and \crefrange{eq:wannier_symmetries_T}{eq:wannier_symmetries_C2zP}, respectively. In this way, for each Wannier and Bloch states, we find
\begin{align}
    &\bra{\mathbf{k},\mathbf{Q},\beta,\tilde{\eta},s}g\ket{W_{0,\alpha, \eta, s}} = \nonumber \\
    =&\sum_{\alpha^{\prime},\eta^{\prime}} [D^f(g)]_{\alpha^{\prime}\eta^{\prime}, \alpha\eta}\bra{\mathbf{k},\mathbf{Q},\beta,\tilde{\eta},s}\ket{W_{0,\alpha^{\prime},\eta^{\prime},s}} = \sum_{\alpha^{\prime},\eta^{\prime}} [D^f(g)]_{\alpha^{\prime}\eta^{\prime}, \alpha\eta} v^{(\eta^{\prime})}_{\mathbf{Q}\beta,\alpha^{\prime}}(\mathbf{k})\delta_{\tilde{\eta},\eta^{\prime}},
    \label{eq:wannier_g_acts_wannier} \\
    &\bra{\mathbf{k},\mathbf{Q},\beta,\tilde{\eta},s}g\ket{W_{0,\alpha, \eta, s}} =  \nonumber \\
    =&\sum_{\mathbf{Q}^{\prime},\beta^{\prime},\eta^{\prime}} [D(g)]_{\mathbf{Q}^{\prime}\beta^{\prime}\eta^{\prime},\mathbf{Q}\beta\tilde{\eta}}\bra{g\mathbf{k}, \mathbf{Q}^{\prime},\beta^{\prime},\eta^{\prime},s}\ket{W_{0,\alpha, \eta, s}} = \sum_{\mathbf{Q}^{\prime},\beta^{\prime},\eta^{\prime}} [D(g)]_{\mathbf{Q}^{\prime}\beta^{\prime}\eta^{\prime},\mathbf{Q}\beta\tilde{\eta}} v^{(\eta)}_{\mathbf{Q}^{\prime}\beta^{\prime},\alpha}(g\mathbf{k})\delta_{\eta^{\prime},\eta}.
    \label{eq:wannier_g_acts_BM}
\end{align}
From \cref{eq:wannier_g_acts_wannier,eq:wannier_g_acts_BM} we can derive the symmetry transformation of the Wannier states in momentum space
\begin{equation}
    \sum_{\mathbf{Q}^{\prime},\beta^{\prime}} [D(g)]_{\mathbf{Q}^{\prime}\beta^{\prime}\eta,\mathbf{Q}\beta\tilde{\eta}} v^{(\eta)}_{\mathbf{Q}^{\prime}\beta^{\prime},\alpha}(g\mathbf{k}) = \sum_{\alpha^{\prime}} [D^f(g)]_{\alpha^{\prime}\tilde{\eta}, \alpha\eta} v^{(\tilde{\eta})}_{\mathbf{Q}\beta,\alpha^{\prime}}(\mathbf{k}).
    \label{eq:wannier_transform_momentum}
\end{equation}
Finally, we introduce the creation operators of the Wannier states in real-space as
\begin{equation}
    \hat{f}^\dagger_{\mathbf{R},\alpha,\eta,s} = \sum_{l,\beta}\int \dd^2{\mathbf{r}} \bra{\mathbf{r},l,\beta,\eta,s}\ket{W_{\mathbf{R},\alpha,\eta,s}} \hat{c}^\dagger_{l,\beta,\eta,s}(\mathbf{r}) = \sum_{l,\beta}e^{-i\eta\Delta\mathbf{K}_l\cdot\mathbf{R}}\int \dd^2{\mathbf{r}} w_{l\beta,\alpha}^{(\eta)}(\mathbf{r}-\mathbf{R})\hat{c}^\dagger_{l,\beta,\eta,s}(\mathbf{r}).
\end{equation}
The corresponding momentum space definition is therefore given by
\begin{equation}
    \hat{f}^\dagger_{\mathbf{k},\alpha,\eta,s} = \frac{1}{\sqrt{N}}\sum_{\mathbf{R}}e^{i\mathbf{k}\cdot\mathbf{R}}\hat{f}^\dagger_{\mathbf{R},\alpha,\eta,s} = \sum_{\mathbf{Q}, \beta}\hat{c}^\dagger_{\mathbf{k},\mathbf{Q},\beta,\eta,s}v_{\mathbf{Q}\beta,\alpha}^{(\eta)}(\mathbf{k}).
    \label{eq:heavy_fermion_operator_fourier}
\end{equation}
Due to the small overlap between neighboring Wannier orbitals~\cite{SON22}, in the THF model, we neglect any hopping between orbitals at different lattice sites in our analytical calculations from \cref{app:sec:analytic_single_particle}. Therefore, the local orbital part of the THF model Hamiltonian in the grand canonical ensemble is given simply by
\begin{equation}
    \hat{H}^{(f)}_0 = -\mu\sum_{\eta s}\sum_{\mathbf{R}}\hat{f}^\dagger_{\mathbf{R},\alpha,\eta,s}\hat{f}_{\mathbf{R},\alpha,\eta,s},
    \label{eq:local_orbitals_hamiltonian}
\end{equation}
where $\mu$ is the chemical potential.

\subsection{Conduction band electrons}\label{app:subsec:HF_conduction_bands}
At the $\Gamma_M$ point, the Wannier functions transform according to the $\Gamma_3$ irrep of the corresponding little group. In contrast, the active TBG bands transform according to the $\Gamma_1 \oplus \Gamma_2$ representation~\cite{SON22}. As such, in order to obtain the correct band structure at the $\Gamma_M$ point, in the THF model, the conduction electrons indexed by $1 \leq a \leq 4$ within each valley and spin flavors are added~\cite{SON22}. By construction, they form the $\Gamma_1 \oplus \Gamma_2 \oplus \Gamma_3$ representation at the $\Gamma_M$ point. Additionally, the conduction electrons have a large kinetic energy away from the $\Gamma_M$ point. Therefore, we consider only momenta below a certain cutoff $|\mathbf{k}| < \Lambda_c$. We denote the creation operator of the conduction electron of the momentum $\mathbf{k}$, band $a$, valley $\eta$, spin $s$ as $\hat{c}^\dagger_{\mathbf{k},a,\eta,s}$. The latter can be written in the Bloch wave basis as
\begin{equation}
    \hat{c}^\dagger_{\mathbf{k},a,\eta,s} = \sum_{\mathbf{Q},\beta}\tilde{u}_{\mathbf{Q}\beta,a}^{(\eta)}(\mathbf{k})\hat{c}^\dagger_{\mathbf{k},\mathbf{Q},\beta,\eta,s}.
    \label{eq:conduction_operator_definition}
\end{equation}
As the $c$-electrons are only relevant in the proximity of the $\Gamma_M$ point ({\it i.e.}{} their energy increases rapidly away from the $\Gamma_M$ point), hereafter, we will approximate $\tilde{u}_{\mathbf{Q}\beta,a}^{(\eta)}(\mathbf{k}) \approx \tilde{u}_{\mathbf{Q}\beta,a}^{(\eta)}(\vec{0})$~\cite{SON22}. We construct the wave functions $\tilde{u}_{\mathbf{Q}\beta,a}^{(\eta)}(\mathbf{k})$ through the same procedure as the one used in Ref.~\cite{SON22}. First, we define the projector into the six BM model bands $n=\pm3,\pm2,\pm1$ as
\begin{equation}
    P^{(\eta)}_{\mathbf{Q}^{\prime}\alpha^{\prime},\mathbf{Q}\alpha}(\mathbf{k}) = \sum_{n=\pm3,\pm2,\pm1}u^{(\eta)}_{\mathbf{Q}^{\prime}\alpha^{\prime},n}(\mathbf{k})u^{(\eta)*}_{\mathbf{Q}\alpha,n}(\mathbf{k}).
    \label{eq:projector_BM}
\end{equation}
Similarly, we define the projector into the $f$-electron bands
\begin{equation}
    Q^{(\eta)}_{\mathbf{Q}^{\prime}\beta^{\prime},\mathbf{Q}\beta}(\mathbf{k}) = \sum_{\alpha=1,2}v_{\mathbf{Q}^{\prime}\beta^{\prime},\alpha}^{(\eta)}(\mathbf{k})v_{\mathbf{Q}\beta,\alpha}^{(\eta)*}(\mathbf{k}).
    \label{eq:projector_local}
\end{equation}
The conduction electron states $\tilde{u}_{\mathbf{Q}\beta,a}^{(\eta)}$ are therefore the eigenstates of the operator $P^{(\eta)}(\mathbf{k}) - P^{(\eta)}(\mathbf{k})Q^{(\eta)}(\mathbf{k})P^{(\eta)}(\mathbf{k})$ with eigenvalue 1. We note that the Wannier states are only supported on the six bands ($n=\pm1,\pm2,\pm3$) near the charge neutrality point, $P^{(\eta)}(\mathbf{k})Q^{(\eta)}(\mathbf{k})P^{(\eta)}(\mathbf{k}) = Q^{(\eta)}(\mathbf{k})$. Therefore, due to the approximation $\tilde{u}_{\mathbf{Q}\beta,a}^{(\eta)}(\mathbf{k}) \approx \tilde{u}_{\mathbf{Q}\beta,a}^{(\eta)}(\vec{0})$ we have to solve the equation
\begin{equation}
    (P^{(\eta)}(\vec{0}) - Q^{(\eta)}(\vec{0}))\tilde{u}_{\mathbf{Q}\beta,a}^{(\eta)}(\vec{0}) = \tilde{u}_{\mathbf{Q}\beta,a}^{(\eta)}(\vec{0})
    \label{eq:conduction_projection_equation}
\end{equation}
either analytically, as will be discussed in \cref{app:subsec:analytic_cbands_hexagon}, or numerically as was done in Ref.~\cite{SON22} and will be done in \cref{app:sec:numerics}.
We label the states $\tilde{u}_{\mathbf{Q}\beta,a}^{(\eta)}$ with the index $a$, such that $\tilde{u}_{\mathbf{Q}\beta,a}^{(\eta)}$ for $a=1,2$ transform as the $\Gamma_3$ irrep and $\tilde{u}_{\mathbf{Q}\beta,a}^{(\eta)}$  for $a=3,4$ transform as the $\Gamma_1 \oplus \Gamma_2$ irreps. The action of a symmetry $g$ can be written in a form analogous to \cref{eq:wannier_transform_momentum}
\begin{equation}
     \sum_{\mathbf{Q}^{\prime},\beta^{\prime}} [D(g)]_{\mathbf{Q}^{\prime}\beta^{\prime}\eta,\mathbf{Q}\beta\tilde{\eta}} \tilde{u}^{(\eta)}_{\mathbf{Q}^{\prime}\beta^{\prime},a}(g\mathbf{k}) = \sum_{\alpha^{\prime}} [D^c(g)]_{a^{\prime}\tilde{\eta}, a\eta} \tilde{u}^{(\tilde{\eta})}_{\mathbf{Q}\beta,a^{\prime}}(\mathbf{k}),
    \label{eq:conduction_transform_momentum}
\end{equation}
in which we fix the representation matrices $D^c(g)$ according to Ref.~\cite{SON22}
\begin{align}
    [D^c(T)]_{a^{\prime}\eta^{\prime},a\eta} &= [\sigma_0\oplus\sigma_0]_{a^{\prime}a}[\tau_x]_{\eta^{\prime}\eta},
    \label{eq:conduction_representations_T} \\
    [D^c(C_{3z})]_{a^{\prime}\eta^{\prime},a\eta} &= [e^{i\eta\frac{2\pi}{3}\sigma_z} \oplus \sigma_0]_{a^{\prime}a}[\tau_0]_{\eta^{\prime}\eta},
    \label{eq:conduction_representations_C3z} \\
    [D^c(C_{2x})]_{a^{\prime}\eta^{\prime},a\eta} &= [\sigma_x\oplus\sigma_x]_{a^{\prime}a}[\tau_0][\tau_0]_{\eta^{\prime}\eta},
    \label{eq:conduction_representations_C2x} \\
    [D^c(C_{2z}T)]_{a^{\prime}\eta^{\prime},a\eta} &= [\sigma_x\oplus\sigma_x]_{a^{\prime}a}[\tau_0]_{\eta^{\prime}\eta},
    \label{eq:conduction_representations_C2zT} \\
    [D^c(P)]_{a^{\prime}\eta^{\prime},a\eta} &=  [-i\sigma_z\oplus -i\sigma_z]_{a^{\prime}a}[\tau_z]_{\eta^{\prime}\eta},
    \label{eq:conduction_representations_P} \\
    [D^c(C_{2z}P)]_{a^{\prime}\eta^{\prime},a\eta} &=  [i\sigma_y\oplus i\sigma_y]_{a^{\prime}a}[\tau_y]_{\eta^{\prime}\eta},
    \label{eq:conduction_representations_C2zP}
\end{align}
where $\sigma_{0,x,y,z}$ and $\tau_{0,x,y,z}$ are the identity, Pauli matrices in the $a=1,2$ or $a=3,4$ bands and valley subspaces respectively.
We can write down the Hamiltonian for the conduction electrons in the first-quantized formalism. As was shown in Ref.~\cite{SON21}, in the basis of the conduction electron bands $1 \leq a \leq 4$, the Hamiltonian for the $\eta=+$ valley reads as
\begin{equation}
    H^{(c,+)}(\mathbf{k}) =
    \begin{pmatrix}
        0 & v_{\star}(k_x\sigma_0 + i k_y \sigma_z) \\
        v_{\star}(k_x\sigma_0 - i k_y \sigma_z) & M\sigma_x
    \end{pmatrix},
    \label{eq:conduction_hamiltonian}
\end{equation}
with the opposite valley $(\eta=-)$ Hamiltonian following by time-reversal symmetry
\begin{equation}
    H^{(c,-)}(\mathbf{k}) = H^{(c,+)*}(-\mathbf{k}).
\end{equation}
The two parameters, $M$ and $v_{\star}$, are either obtained analytically from the Hexagon model (see \cref{app:subsec:analytic_cbands_hexagon}) or computed numerically in Ref.~\cite{SON22} and \cref{app:sec:numerics}).
The conduction electron Hamiltonian therefore reads as
\begin{equation}
    \hat{H}^{(c)}_0 = \sum_{\eta,s}\sum_{a,a^{\prime}}\sum_{|\mathbf{k}|<\Lambda_c}(H^{(c,\eta)}_{a,a^{\prime}}(\mathbf{k}) - \mu\delta_{aa^{\prime}})\hat{c}^\dagger_{\mathbf{k} a \eta s}\hat{c}_{\mathbf{k} a^{\prime} \eta s},
    \label{eq:conduction_band_hamiltonian}
\end{equation}
where $\mu$ is the chemical potential and $\Lambda_c$ is the momentum cutoff of the $c$-electrons, which are only relevant in the proximity of the $\Gamma_M$ point.

\subsection{Coupling Hamiltonian}\label{app:subsec:HF_coupling}
The local $f$- and conduction band $c$-fermions are coupled at the single-particle level according to~\cite{SON22}
\begin{equation}
    \hat{H}^{(cf)}_0 = \sum_{\eta, s}\sum_{a, \alpha}\sum_{|\mathbf{k}| < \Lambda_c}\sum_{\mathbf{R}}\bra{0}\hat{c}_{\mathbf{k},a,\eta,s}\hat{H}_{\textrm{BM}}\hat{f}^\dagger_{\mathbf{R},\alpha,\eta,s}\ket{0}\hat{c}^\dagger_{\mathbf{k},a,\eta,s}\hat{f}_{\mathbf{R},\alpha,\eta,s} + \mathrm{h.c.}
\end{equation}
In Ref.~\cite{SON22}, the overlap term $\bra{0}\hat{c}_{\mathbf{k},a,\eta,s}\hat{H}_{\textrm{BM}}\hat{f}^\dagger_{\mathbf{R},\alpha,\eta,s}\ket{0}$ was argued to be exponentially decaying in momentum space with the characteristic length $1/\lambda$, given by
\begin{equation}
    \lambda = \sqrt{\sum_{l\beta}\int \dd^2{\mathbf{r}}|w^{(\eta)}_{l\beta,\alpha}(\mathbf{r})|^2\mathbf{r}^2},
\end{equation}
where $w^{(\eta)}_{l\beta,\alpha}(\mathbf{r})$ are the Wannier states wave functions defined in \cref{eq:wannier_states_I,,eq:wannier_states_II}. Following this approximation, and applying the Fourier transform to the local orbitals, the coupling term can be rewritten as~\cite{SON22}
\begin{equation}
    \hat{H}^{(cf)}_0 = \sum_{\eta, s}\sum_{a,\alpha}\sum_{|\mathbf{k}| < \Lambda_c}e^{-|\mathbf{k}|^2\lambda^2/2}H^{(cf,\eta)}_{a,\alpha}(\mathbf{k})\hat{c}^\dagger_{\mathbf{k},a,\eta,s}\hat{f}_{\mathbf{k},\alpha,\eta,s} + \mathrm{h.c.},
    \label{eq:coupling_hamiltonian}
\end{equation}
and $H^{(cf,\eta)}_{a,\alpha}(\mathbf{k})$ is given by~\cite{SON22}
\begin{equation}
    \label{eq:coupling_hamiltonian_matrix}
    H^{(cf,\eta)}(\mathbf{k}) =
    \begin{pmatrix}
    \gamma \sigma_0 + v^{\prime}_{\star}(\eta k_x \sigma_x  + k_y \sigma_y) \\
    v_{\star}^{\prime\prime}(\eta k_x \sigma_x  - k_y \sigma_y)
    \end{pmatrix}.
\end{equation}
The coupling parameters $\gamma,\;v^{\prime}_{\star}$ are either obtained analytically as will be done in \cref{app:subsec:analytic_other_hexagon} or computed numerically as was done in Ref.~\cite{SON22} and will be done in \cref{app:sec:numerics}. Numerically, we also find that $v_{\star}^{\prime\prime}$ is small compared to $v_{\star}$ and $v^{\prime}_{\star}$ for the experimentally-relevant BM parameter region $\SI{0.8}{\degree} \leq \theta \leq \SI{1.6}{\degree}$ and $w_0/w_1 \gtrsim 0.1$. Therefore, following Ref.~\cite{SON22}, we will neglect $v_{\star}^{\prime\prime}$ in our analytical calculations from \cref{app:sec:analytic_single_particle}.

\subsection{Summary of the single-particle Hamiltonian}\label{app:subsec:HF_single_summary}
Collecting all the terms \cref{eq:local_orbitals_hamiltonian,,eq:conduction_band_hamiltonian,,eq:coupling_hamiltonian}, we obtain the single-particle Hamiltonian of the THF model~\cite{SON22}
\begin{align}
    \hat{H}_0 &= -\mu\sum_{\eta, s}\sum_{\mathbf{R}}\hat{f}^\dagger_{\mathbf{R},\alpha,\eta,s}\hat{f}_{\mathbf{R},\alpha,\eta,s} + \sum_{\eta,s}\sum_{a,a^{\prime}}\sum_{|\mathbf{k}|<\Lambda_c}(H^{(c,\eta)}_{a,a^{\prime}}(\mathbf{k}) - \mu\delta_{aa^{\prime}})\hat{c}^\dagger_{\mathbf{k} a \eta s}\hat{c}_{\mathbf{k} a^{\prime} \eta s} \nonumber \\
    &+ \sum_{\eta,s}\sum_{a,\alpha}\sum_{|\mathbf{k}| < \Lambda_c}e^{-|\mathbf{k}|^2\lambda^2/2}H^{(cf,\eta)}_{a,\alpha}(\mathbf{k})\hat{c}^\dagger_{\mathbf{k},a,\eta,s}\hat{f}_{\mathbf{k},\alpha,\eta,s} + \mathrm{h.c.}
    \label{eq:single_particle_hamiltonian}
\end{align}
Although the cutoff $\Lambda_c$ is set to be small, it is convenient for further calculations to extend it to infinity $\Lambda_c \rightarrow +\infty$, thus restoring the translational symmetry~\cite{SON22}.

\section{The Coulomb interaction in the THF model}\label{app:sec:HF_interaction}
In this appendix, we review the Coulomb interaction Hamiltonian of the THF model~\cite{SON22}. We start by projecting the Coulomb interaction Hamiltonian into the THF states reviewed in \cref{app:sec:HF_model}. We then summarize all possible terms of the projected interaction Coulomb Hamiltonian in \cref{tab:Coulomb_summary} and provide the detailed analytical expressions for each of the terms.

\subsection{Screened Coulomb interaction in TBG}\label{app:subsec:Coulomb_interaction_general}

Consider the screened Coulomb interaction Hamiltonian~\cite{BER21a}
\begin{equation}
    \hat{H}_I = \frac{1}{2}\int d^2 \vec{r}_1 d^2 \vec{r}_2 V(\vec{r}_1 - \vec{r}_2) :\hat{\rho}(\vec{r}_1)::\hat{\rho}(\vec{r}_2):.
    \label{eq:interaction_general_hamiltonian}
\end{equation}
In \cref{eq:interaction_general_hamiltonian}, $:\hat{\rho}(\mathbf{r}): = \hat{\rho}(\mathbf{r}) - \bra{G_0}\hat{\rho}(\mathbf{r})\ket{G_0}$ is the normal-ordered density operator shifted to have zero expectation for a state $\ket{G_0}$ at  at the charge neutrality point. The density operator $\hat{\rho}(\mathbf{r})$ is given by
\begin{equation}
     \hat{\rho}(\mathbf{r}) = \sum_{l, \beta, \eta, s} \hat{c}^\dagger_{l, \beta,\eta,s}(\mathbf{r})\hat{c}_{l, \beta,\eta,s}(\mathbf{r}),
     \label{eq:density_operator_definition}
\end{equation}
and the state $\ket{G_0}$ is defined such as~\cite{SON22}
\begin{equation}
    \bra{G_0}\hat{c}^\dagger_{\mathbf{k},\mathbf{Q},\alpha,\eta,s}\hat{c}_{\mathbf{k}^{\prime},\mathbf{Q}^{\prime},\alpha^{\prime},\eta^{\prime},s^{\prime}}\ket{G_0} = \frac{1}{2}\delta_{\mathbf{k},\mathbf{k}^{\prime}}\delta_{\mathbf{Q},\mathbf{Q}^{\prime}}\delta_{\alpha,\alpha^{\prime}}\delta_{\eta,\eta^{\prime}}\delta_{s,s^{\prime}}.
\end{equation}
We approximate the Coulomb interaction $V(\mathbf{r})$ in \cref{eq:interaction_general_hamiltonian} with the double-gate screened potential~\cite{BER21a}, given by
\begin{equation}
    V(\mathbf{r}) = U_{\xi}\sum_{n=-\infty}^{\infty}\frac{(-1)^n}{\sqrt{(\mathbf{r}/\xi)^2 + n^2}},
    \label{eq:Coulomb_interaction}
\end{equation}
In \cref{eq:Coulomb_interaction}, $\xi$ is the distance between the screening gates, while $U_{\xi} = e^2 / (\epsilon \xi)$ (where $\epsilon \approx 6$ is the dielectric constant) is the Coulomb interaction scale. In this work, we will investigate the dependence of the Coulomb interaction parameters on the screening length $\xi$, and hence $U_{\xi}$ will be rescaled accordingly. We note, however, that for the typical experimental setup~\cite{BER21a}, $\xi=\SI{10}{\nano \meter}$ and, therefore, $U_{\xi} = \SI{24}{\milli \eV}$.

For later use, we also define and compute the Fourier transformation of the Coulomb interaction from \cref{eq:Coulomb_interaction}
\begin{equation}
    V(\mathbf{q}) = \int \dd^2{\mathbf{r}}_1 V(\mathbf{r}_1)e^{-i\mathbf{q}\cdot\mathbf{r}_1} = (\pi U_{\xi} \xi^2) \frac{\tanh{(|\mathbf{q}|\xi/2)}}{|\mathbf{q}|\xi}.
    \label{eq:Coulomb_interaction_fourier}
\end{equation}

To project the density operator $\hat{\rho}(\mathbf{r})$ into the THF states, we recall from \cref{eq:heavy_fermion_operator_fourier,eq:conduction_operator_definition} that the BM model creation operator \cref{eq:BM_model_creation_operator} can be projected onto the basis of the THF model as
\begin{equation}
    \hat{c}^\dagger_{\mathbf{k},\mathbf{Q},\beta,\eta,s} \approx \sum_{\alpha} v^{(\eta)*}_{\mathbf{Q}\beta,\alpha}(\mathbf{k})\hat{f}^\dagger_{\mathbf{k},\alpha,\eta,s} + \sum_a \tilde{u}^{(\eta)*}_{\mathbf{Q}\beta,a}(\mathbf{k})\hat{c}^\dagger_{\mathbf{k},a,\eta,s}.
\end{equation}
Therefore, the real-space operators from \cref{eq:real_space_basis} can be obtained via the Fourier transformation
\begin{equation}
    \hat{c}^\dagger_{l,\beta,\eta,s}(\mathbf{r}) \approx \sum_{\mathbf{R}, \alpha} e^{i\eta\Delta\mathbf{K}_l\cdot\mathbf{R}}w_{l\beta,\alpha}^{(\eta)*}(\mathbf{r}-\mathbf{R})\hat{f}^\dagger_{\mathbf{R},\alpha,\eta,s} + \frac{1}{\sqrt{\Omega_{\textrm{tot}}}}\sum_{|\mathbf{k}|<\Lambda_c}\sum_a\sum_{\mathbf{Q} \in \mathcal{Q}_{l_{\eta}}}e^{-i(\mathbf{k}-\mathbf{Q})\cdot\mathbf{R}}\tilde{u}^{(\eta)*}_{\mathbf{Q}\beta,a}(\mathbf{k})\hat{c}^\dagger_{\mathbf{k},a,\eta,s}.
    \label{eq:interaction_real_space_projection}
\end{equation}
Plugging \cref{eq:interaction_real_space_projection} into the definition of the density operator in \cref{eq:density_operator_definition} and after some simplifications that the reader can find in Ref.~\cite{SON22}, we obtain for the normal ordered operator
\begin{equation}
    :\hat{\rho}(\mathbf{r}): = :\hat{\rho}_{ff}(\mathbf{r}): + :\hat{\rho}_{cc}(\mathbf{r}): + :\hat{\rho}_{fc}(\mathbf{r}): + :\hat{\rho}_{cf}(\mathbf{r}):,
    \label{eq:interaction_total_density}
\end{equation}
where $\hat{\rho}_{ff}(\mathbf{r})$ is the $f$-electron density
\begin{equation}
    :\hat{\rho}_{ff}(\mathbf{r}): = \sum_{\eta, s}\sum_{\mathbf{R}, \alpha}n_f(\mathbf{r}-\mathbf{R}):\hat{f}^\dagger_{\mathbf{R},\alpha,\eta,s}\hat{f}_{\mathbf{R},\alpha,\eta,s}:,
    \label{eq:interaction_ff_term}
\end{equation}
with
\begin{equation}
    n_f(\mathbf{r} - \mathbf{R}) = \sum_{l,\beta}|w_{l\beta,\alpha}^{(\eta)}(\mathbf{r}-\mathbf{R})|^2,
    \label{eq:heavy_fermion_density}
\end{equation}
being the probability density of the $f$-electrons written in terms of the Wannier functions $w_{l\beta,\alpha}^{(\eta)}(\mathbf{r}-\mathbf{R})$ from \cref{eq:local_orbital_r2q}. We note that due to the symmetry constraints from \cref{eq:wannier_P_constraint,,eq:wannier_C2zT_constraint}, the density $n_f(\mathbf{r})$ does not depend on the valley $\eta$ and sublattice $\alpha$. As such, we omit the $\eta$ and $\alpha$ indices for the density $n_f(\mathbf{r})$ on the left-hand side of \cref{eq:heavy_fermion_density} and in what follows. The $f$-orbital density $n_f(\mathbf{r})$ in \cref{eq:heavy_fermion_density} can also be written in the momentum space with the help of \cref{eq:local_orbital_r2q} as
\begin{equation}
    n_f(\mathbf{r}) = \sum_{l,\beta}|w_{l\beta,\alpha}^{(\eta)}(\mathbf{r})|^2 = \frac{1}{N^2\Omega_0}\sum_{\mathbf{k}, \mathbf{k}^{\prime}}\sum_{l,\beta}\sum_{\mathbf{Q}, \mathbf{Q}^{\prime}\in\mathcal{Q}_{l_{\eta}}}e^{i(\mathbf{k} - \mathbf{Q} - \mathbf{k}^{\prime} + \mathbf{Q}^{\prime})\cdot \mathbf{r}}v^{(\eta)*}_{\mathbf{Q}^{\prime}\beta,\alpha}(\mathbf{k}^{\prime})v^{(\eta)}_{\mathbf{Q}^{\prime}\beta,\alpha}(\mathbf{k}),
\end{equation}
and the Fourier transformation of the density is defined as
\begin{equation}
    n_f(\mathbf{r}) = \frac{1}{N\Omega_0} \sum_{\mathbf{q} \in \textrm{MBZ}} \sum_{\mathbf{G} \in \mathcal{Q}_0} n_f(\mathbf{q}+\mathbf{G}) e^{-i(\mathbf{q}+\mathbf{G})\cdot\mathbf{r}}, \qquad
n_f( \mathbf{q} + \mathbf{G} ) = \frac{1}{N} \sum_{\mathbf{k}} \sum_{\mathbf{Q},\beta}
    v^{(\eta)*}_{\mathbf{Q}-\mathbf{G} \beta,\alpha}(\mathbf{k}+\mathbf{q})
    v^{(\eta)}_{\mathbf{Q}\beta,\alpha}(\mathbf{k}).
    \label{eq:interaction_density_fourier_transform}
\end{equation}

Furthermore, in \cref{eq:interaction_total_density}, $\hat{\rho}_{cc}(\mathbf{r})$ denotes the conduction electron density
\begin{equation}
    :\hat{\rho}_{cc}(\mathbf{r}): = \frac{1}{\Omega_{\textrm{tot}}}\sum_{\eta, s}\sum \limits_{\substack{l\beta \\ a a^{\prime}}} \sum \limits_{\substack{|\mathbf{k}|<\Lambda_c \\ \mathbf{Q} \in \mathcal{Q}_{l_{\eta}}}} e^{-i(\mathbf{k} - \mathbf{Q} - \mathbf{k}^{\prime} + \mathbf{Q}^{\prime})\cdot \mathbf{r}}\tilde{u}_{\mathbf{Q}\beta,a}^{(\eta)*}(\mathbf{k})\tilde{u}^{(\eta)}_{\mathbf{Q}^{\prime}\beta,a^{\prime}}(\mathbf{k}^{\prime}):\hat{c}^\dagger_{\mathbf{k},a,\eta,s}\hat{c}_{\mathbf{k}^{\prime},a^{\prime},\eta,s}:,
    \label{eq:interaction_cc_term}
\end{equation}
and $\hat{\rho}_{fc}(\mathbf{r})$ and $\hat{\rho}_{cf}(\mathbf{r})$ are the complex-conjugated hybridization terms, with
\begin{equation}
    :\hat{\rho}_{fc}(\mathbf{r}): = \frac{1}{\sqrt{\Omega_{\textrm{tot}}}} \sum_{\eta, s}\sum \limits_{\substack{l\beta \\ a \alpha}} \sum \limits_{\substack{|\mathbf{k}|<\Lambda_c \\ \mathbf{Q} \in \mathcal{Q}_{l_{\eta}} \\ \mathbf{R}}} w_{l\beta,\alpha}^{(\eta)*}(\mathbf{r} - \mathbf{R})\tilde{u}_{\mathbf{Q}\beta,a}^{(\eta)}(\mathbf{k})e^{i\eta\Delta\mathbf{K}_l\cdot\mathbf{R} + i(\mathbf{k} - \mathbf{Q})\cdot \mathbf{r}}\hat{f}^\dagger_{\mathbf{R},\alpha,\eta,s}\hat{c}_{\mathbf{k},\alpha,\eta,s},
    \label{eq:interaction_fc_term}
\end{equation}
and
\begin{equation}
    :\hat{\rho}(\mathbf{r})_{cf}: = (:\hat{\rho}_{fc}(\mathbf{r}):)^{\dagger}.
    \label{eq:interaction_cf_term}
\end{equation}
Plugging \crefrange{eq:interaction_ff_term}{eq:interaction_cf_term} into \cref{eq:interaction_total_density}, we find that the interaction Hamiltonian in \cref{eq:interaction_general_hamiltonian} has ten terms in total~\cite{SON22}. Out of these ten terms, four are ``diagonal'' (of the form $:\hat{\rho}_{A}(\mathbf{r}_1):V(\mathbf{r}_1 - \mathbf{r}_2):\hat{\rho}_{A}(\mathbf{r}_2):$), and six are ``off-diagonal'' ({\it i.e.}{} of the form $:\hat{\rho}_{A}(\mathbf{r}_1):V(\mathbf{r}_1 - \mathbf{r}_2):\hat{\rho}_{B}(\mathbf{r}_2): + :\hat{\rho}_{B}(\mathbf{r}_1):V(\mathbf{r}_1 - \mathbf{r}_2):\hat{\rho}_{A}(\mathbf{r}_2):$, where $A\neq B$). To simplify the notation, we use the symbol $(\leftrightarrow)$ to denote the term $\hat{\rho}_{B}(\mathbf{r}_1)V(\mathbf{r}_1 - \mathbf{r}_2)\hat{\rho}_{A}(\mathbf{r}_2)$ obtained by swapping the $\hat{\rho}_{A}(\mathbf{r}_1)$ and $\hat{\rho}_{B}(\mathbf{r}_2)$ operators in the expression $\hat{\rho}_{A}(\mathbf{r}_1)V(\mathbf{r}_1 - \mathbf{r}_2)\hat{\rho}_{B}(\mathbf{r}_2)$.

We summarize the physical meaning of these terms and the THF model parameters characterizing the energetic scale the corresponding interaction strengths in \cref{tab:Coulomb_summary}. Each of the parameters is computed numerically for a large parameter space in \cref{app:sec:numerics}. We also derive approximate analytical expressions for the parameters $U$, $V$, $W$ in \crefrange{app:subsec:analytic_interaction_ff}{app:subsec:analytic_interaction_cc}. Finally, we discuss the enlarged continuous symmetries of the interaction Hamiltonian in \cref{app:sec:symmetries}. In the remaining part of this appendix, we review the interacting terms $\hat{H}_U$, $\hat{H}_V$, $\hat{H}_W$, $\hat{H}_{\tilde{J}+}$, and $\hat{H}_J$ listed in \cref{tab:Coulomb_summary} and derive their general expressions, which are later used either for analytic approximations or in order to derive the symmetries of the interacting Hamiltonian. We note that the terms $:\hat{\rho}_{ff}(\mathbf{r}_1):V(\mathbf{r}_1 - \mathbf{r}_2):\hat{\rho}_{fc}(\mathbf{r}_2): + (\leftrightarrow)$ and $:\hat{\rho}_{ff}(\mathbf{r}_1):V(\mathbf{r}_1 - \mathbf{r}_2):\hat{\rho}_{cf}(\mathbf{r}_2): + (\leftrightarrow)$ were proved to be zero in Ref.~\cite{SON22}, while the terms $\hat{H}_K$ and $\hat{H}_{K+}$ were shown to be an order of magnitude smaller compared to other interaction terms, and therefore will not be considered~\cite{SON22}.

\begin{table}[t]
    \centering
    \begin{tabular}{ |M{1.0cm}|M{5.5cm}|M{2cm}|m{8.5cm}|}
    \hline
       $\mathbf{\hat{H}_I}$ \textbf{term} & \textbf{THF projection}  & \textbf{Parameter} & \textbf{Physical meaning} \\ \hline
        \multicolumn{4}{|c|}{\textbf{Diagonal terms}} \\ \hline
        $\hat{H}_U$ & $:\hat{\rho}_{ff}(\mathbf{r}_1):V(\mathbf{r}_1 - \mathbf{r}_2):\hat{\rho}_{ff}(\mathbf{r}_2):$ & $U$ & Density-density interaction of the local $f$-orbitals \\ \hline
        $\hat{H}_V$ & $:\hat{\rho}_{cc}(\mathbf{r}_1):V(\mathbf{r}_1 - \mathbf{r}_2):\hat{\rho}_{cc}(\mathbf{r}_2):$ & $V$ & Density-density interaction of the conduction $c$-electrons \\ \hline
        $\hat{H}_{\tilde{J}+}$ & $:\hat{\rho}_{fc}(\mathbf{r}_1):V(\mathbf{r}_1 - \mathbf{r}_2):\hat{\rho}_{fc}(\mathbf{r}_2):$ & $J$ & High-energy process of creating a pair of particles in the $f$-electron orbitals and two holes in the conduction bands. \\ \hline
        $\hat{H}_{\tilde{J}+}^{\dagger}$ & $:\hat{\rho}_{cf}(\mathbf{r}_1):V(\mathbf{r}_1 - \mathbf{r}_2):\hat{\rho}_{cf}(\mathbf{r}_2):$ & $J$ & High-energy process of creating a pair of particles in the conduction bands and two holes in the $f$-electron bands. \\ \hline
        \multicolumn{4}{|c|}{\textbf{Off-diagonal terms}} \\
        \hline
         $\hat{H}_W$ & $:\hat{\rho}_{ff}(\mathbf{r}_1):V(\mathbf{r}_1 - \mathbf{r}_2):\hat{\rho}_{cc}(\mathbf{r}_2): + (\leftrightarrow)$ & $W$ & Density-density interaction between the $f$-electrons and conduction electrons \\ \hline
         NA & $:\hat{\rho}_{ff}(\mathbf{r}_1):V(\mathbf{r}_1 - \mathbf{r}_2):\hat{\rho}_{fc}(\mathbf{r}_2): + (\leftrightarrow)$ & NA & Process of creating two particles and a hole in the $f$-electron bands and a hole in the conduction bands. Due to the symmetries of TBG, this process has zero probability \\ \hline
         NA & $:\hat{\rho}_{ff}(\mathbf{r}_1):V(\mathbf{r}_1 - \mathbf{r}_2):\hat{\rho}_{cf}(\mathbf{r}_2): + (\leftrightarrow)$ & NA & Process of creating a particle and two holes in the $f$-electron bands and a particle in the conduction bands. Due to the symmetries of TBG, this process has zero probability \\ \hline
         $\hat{H}_{K+}$ & $:\hat{\rho}_{cc}(\mathbf{r}_1):V(\mathbf{r}_1 - \mathbf{r}_2):\hat{\rho}_{fc}(\mathbf{r}_2): + (\leftrightarrow)$ & $K$ & High-energy process of creating a particle in the local orbitals and two holes and a particle in the conduction bands \\ \hline
         $\hat{H}_{K+}^{\dagger}$ & $:\hat{\rho}_{cc}(\mathbf{r}_1):V(\mathbf{r}_1 - \mathbf{r}_2):\hat{\rho}_{cf}(\mathbf{r}_2): + (\leftrightarrow)$ & $K$ & High-energy process of creating a hole in the $f$-electron bands and two particles and a hole in the conduction bands \\ \hline
         $\hat{H}_{J}$ & $:\hat{\rho}_{fc}(\mathbf{r}_1):V(\mathbf{r}_1 - \mathbf{r}_2):\hat{\rho}_{cf}(\mathbf{r}_2): + (\leftrightarrow)$ & $J$ & Exchange interaction between the $f$-electrons and the conduction-band fermions \\ \hline
    \end{tabular}
    \caption{Summary of the interaction Hamiltonian terms within the THF model. The first column indicates the projected term in the interacting Hamiltonian $\hat{H}_I$. The second column lists the corresponding expression for each term in the first column. We denote the exchange of $:\hat{\rho}_{A}(\mathbf{r}_1):$ and $:\hat{\rho}_{B}(\mathbf{r}_2):$ operators in $:\hat{\rho}_{A}(\mathbf{r}_1):V(\mathbf{r}_1 - \mathbf{r}_2):\hat{\rho}_{B}(\mathbf{r}_1):$ with the $(\leftrightarrow)$ symbol. The third column provides the interaction strength parameter, which characterizes each projected term. Finally, the fourth column explains the physical meaning of the projected term in the first column. In this table, NA stands for not-applicable and implies that these terms are zero~\cite{SON22}.}
    \label{tab:Coulomb_summary}
\end{table}

\subsection{The $f$-$f$ density-density interaction}\label{app:subsec:review_ff_interaction}
The density-density interaction of the $f$-electrons, $\hat{H}_U$ (see \cref{tab:Coulomb_summary}), is obtained by plugging \cref{eq:interaction_ff_term} into \cref{eq:interaction_general_hamiltonian}~\cite{SON22}
\begin{equation}
    \hat{H}_{U} = \frac{1}{2}\sum_{\mathbf{R}, \mathbf{R}^{\prime}}\sum_{\alpha, \eta, s}\sum_{\alpha^{\prime},\eta^{\prime},s^{\prime}}U(\mathbf{R} - \mathbf{R}^{\prime}):\hat{f}^\dagger_{\mathbf{R}\alpha s} \hat{f}_{\mathbf{R} \alpha \eta s}::\hat{f}^\dagger_{\mathbf{R}^{\prime}\alpha^{\prime}\eta^{\prime}s^{\prime}}\hat{f}_{\mathbf{R}^{\prime}\alpha^{\prime}\eta^{\prime}s^{\prime}}:,
    \label{eq:original_ff_interaction}
\end{equation}
with
\begin{equation}
    U(\mathbf{R}) = \int d^2 \mathbf{r}_1 d^2 \mathbf{r}_2 V(\mathbf{r}_1 - \mathbf{r}_2 - \mathbf{R})n_f(\mathbf{r}_1)n_f(\mathbf{r}_2)
\end{equation}
being the interaction parameter we have to calculate. Because of the divergent nature of $V(\mathbf{r})$ at $\mathbf{r} = \vec{0}$, the integral is best evaluated in the momentum space. With the help of \cref{eq:Coulomb_interaction_fourier,eq:interaction_density_fourier_transform} the function $U(\mathbf{R})$ is given by
\begin{equation}
    U(\mathbf{R}) = \frac{1}{N\Omega_0}\sum_{\mathbf{q} \in \textrm{MBZ}} \sum_{\mathbf{G} \in \mathcal{Q}_0} V(\mathbf{q} + \mathbf{G})e^{-i\mathbf{q}\cdot\mathbf{R}}n_f(\mathbf{q} + \mathbf{G})n_f(-\mathbf{q} - \mathbf{G}),
    \label{eq:interaction_ff_formula}
\end{equation}
where $N$ is the number of the Moir\'e unit cells and $\Omega_0$ is the area of the single moir\'e unit cell. We find numerically that the onsite repulsion strength $U(\vec{0})$ is much larger than the one corresponding to the nearest neighbor $U(\vec{a}_{M})$. Therefore, the term $U(\mathbf{R})$ with $\mathbf{R} = 0$ dominates the physics of the $f$-electrons. 

Ref.~\cite{SON22} suggested that the expression for the onsite interaction term $\hat{H}_U$ in \cref{eq:original_ff_interaction} can be truncated at the nearest-neighbour level
\begin{equation} \label{eq:HI1}
{\small\hat{H}_{U,\textrm{n.n.}} = 
\frac{U_1}{2}\sum_{\mathbf{R}} \sum_{\substack{\alpha,\eta, s \\ \alpha',\eta',s'}}
:f_{\mathbf{R} \alpha \eta s}^\dagger f_{\mathbf{R} \alpha\eta s}:
    :f_{\mathbf{R} \alpha' \eta' s'}^\dagger f_{\mathbf{R} \alpha'\eta' s'}: + \frac{U_2}{2} \sum_{\langle \mathbf{R}\mathbf{R}' \rangle} \sum_{\substack{\alpha,\eta, s \\ \alpha',\eta',s'}} 
    :f_{\mathbf{R} \alpha \eta s}^\dagger f_{\mathbf{R} \alpha\eta s}:
    :f_{\mathbf{R}' \alpha' \eta' s'}^\dagger f_{\mathbf{R}' \alpha'\eta' s'}:,}
\end{equation}where $\langle \mathbf{R}\mathbf{R}' \rangle$ denotes the sum over NN lattice sites and $U_1$ and $U_2$ are the onsite and the effective NN interaction parameters respectively. The effective parameter $U_2$ is chosen to also take into account the \emph{all} Coulomb repulsion terms \emph{beyond} the onsite repulsion, not just the NN one. As such, instead of setting $U_2 = U(\mathbf{a}_M)$, the parameter $U_2$ is obtained by equating the expressions in \cref{eq:original_ff_interaction,eq:HI1} within the Hartree approximation. We briefly outline the procedure here. 

First, we require \normalsize
\begin{equation}
\sum_{\alpha,\eta, s}\langle:f_{\mathbf{R} \alpha \eta s}^\dagger f_{\mathbf{R} \alpha\eta s}:\rangle = N_f \qquad \textrm{for any } \mathbf{R},
\end{equation}
where $\left<...\right>$ denotes taking an expectation value for any Slater determinant state $\ket{G}$. As such, the Hamiltonian in \cref{eq:original_ff_interaction} in the Hartree approximation is given by~\cite{SON22}
\begin{align}
 \hat{H}_U^{\textbf{MF}}= \left(N_f \sum_{\mathbf{R}}  U(\mathbf{R})\right) \sum_{\alpha',\eta',s'} \sum_{\mathbf{R}'}  :f_{\mathbf{R}' \alpha' \eta' s'}^\dagger f_{\mathbf{R}' \alpha'\eta' s'}: \ ,
 \label{eq:interaction_U_MF}
\end{align}
where we dropped the constant term. In the Hartree approximation, the Hamiltonian from \cref{eq:HI1}  reads~\cite{SON22}
\begin{align}
\hat{H}_{U,\textrm{n.n.}}^{\textbf{MF}} = 
N_f (U_1 + 6 U_2)  \sum_{\alpha',\eta',s'}     \sum_{\mathbf{R}'}  :f_{\mathbf{R}' \alpha' \eta' s'}^\dagger f_{\mathbf{R}' \alpha'\eta' s'}: \ .
\label{eq:interaction_U_nnMF}
\end{align}
To obtain the value for $U_2$ we equate \cref{eq:interaction_U_MF,eq:interaction_U_nnMF}:
\begin{equation}
U_1 + 6 U_2 = \sum_{\mathbf{R}}  U(\mathbf{R}) = \frac{1}{\Omega_0} \sum_{\mathbf{G}} V(\mathbf{G}) n_f(\mathbf{G}) n_f(-\mathbf{G}),
\end{equation}
which yields us
\begin{equation}
    U_2 = \frac{1}{6}\left(-U_1 + \frac{1}{\Omega_0} \sum_{\mathbf{G}} V(\mathbf{G}) n_f(\mathbf{G}) n_f(-\mathbf{G})\right).
\end{equation}

\subsection{The $f$-$c$ density-density interaction}\label{app:subsec:review_fc_interaction}
The density-density interaction of the local $f$-orbitals and conduction band electrons, $\hat{H}_W$ (see \cref{tab:Coulomb_summary}), is obtained by substituting the expressions for the $f$-electron density $\hat{\rho}_{ff}$ from \cref{eq:interaction_ff_term} and for the conduction band electron density $\hat{\rho}_{cc}$ from \cref{eq:interaction_cc_term} into
\begin{equation}
    \hat{H}_W = \frac{1}{2}\int d^2 \vec{r}_1 d^2 \vec{r}_2 \left[:\hat{\rho}_{ff}(\vec{r}_1):V(\vec{r}_1 - \vec{r}_2) :\hat{\rho}_{cc}(\vec{r}_2): + :\hat{\rho}_{cc}(\vec{r}_1):V(\vec{r}_1 - \vec{r}_2) :\hat{\rho}_{ff}(\vec{r}_2):\right].
\end{equation}
The resulting term reads as
\begin{align}
\hat{H}_{W} 
=& \frac1{\Omega_{\textrm{tot}}} \int \dd^2{\mathbf{r}}_1 \dd^2{\mathbf{r}}_2 \sum_{\mathbf{R}, \alpha,\eta_1, s_1} 
    :f_{\mathbf{R} \alpha\eta_1 s_1}^\dagger f_{\mathbf{R} \alpha\eta_1 s_1}: n_f(\mathbf{r}_1-\mathbf{R}) \sum_{\eta_2,s_2, a,a^{\prime}} \sum_{l, \beta} \sum_{\mathbf{Q},\mathbf{Q}^{\prime}\in \mathcal{Q}_{l\eta_2} } \sum_{|\mathbf{k}|, |\mathbf{k}^{\prime}|<\Lambda_c } V(\mathbf{r}_1-\mathbf{r}_2) \nonumber\\
 & \times \tilde{u}^{(\eta_2)*}_{\mathbf{Q}\beta,a}(\mathbf{k}) \tilde{u}^{(\eta_2)}_{\mathbf{Q}'\beta,a'}(\mathbf{k}') e^{-i(\mathbf{k}-\mathbf{Q}-\mathbf{k}'+\mathbf{Q}')\cdot\mathbf{r}_2} 
    :c_{\mathbf{k} a\eta_2 s_2}^\dagger  c_{\mathbf{k}' a' \eta_2 s_2}:.
\end{align}
Applying the Fourier transformation from \cref{eq:Coulomb_interaction_fourier}
\begin{equation}
    \int \dd^2{\mathbf{r}}_2  V(\mathbf{r}_1-\mathbf{r}_2)  e^{-i(\mathbf{k}-\mathbf{Q}-\mathbf{k}'+\mathbf{Q}')\cdot\mathbf{r}_2} = V(\mathbf{k}-\mathbf{Q}-\mathbf{k}'+\mathbf{Q}') e^{-i(\mathbf{k}-\mathbf{Q}-\mathbf{k}'+\mathbf{Q}')\cdot\mathbf{r}_1},
\end{equation}
and the variable substitution $\mathbf{r}_1\to \mathbf{r}_1 + \mathbf{R}$ we can rewrite $\hat{H}_W$ as

\begin{align}
\hat{H}_W =& \frac{1}{\Omega_{\textrm{tot}}} \int \dd^2{\mathbf{r}_1} \sum_{\mathbf{R}, \alpha,\eta_1, s_1} 
:f_{\mathbf{R} \alpha\eta_1 s_1}^{\dagger} f_{\mathbf{R} \alpha\eta_1 s_1}: n_f(\mathbf{r}_1) 
    \sum_{|\mathbf{k}|, |\mathbf{k}'|<\Lambda_c}  \sum_{\eta_2,s_2, a,a', l } \sum_{\mathbf{Q},\mathbf{Q}'\in \mathcal{Q}_{l\eta_2} }  
    e^{-i(\mathbf{k}-\mathbf{Q}-\mathbf{k}'+\mathbf{Q}')\cdot(\mathbf{r}_1+\mathbf{R})}  \nonumber\\ 
& \times  V(\mathbf{k}-\mathbf{Q}-\mathbf{k}'+\mathbf{Q}') \tilde{u}^{(\eta_2)*}_{\mathbf{Q}\beta,a}(\mathbf{k}) \tilde{u}^{(\eta_2)}_{\mathbf{Q}'\beta,a'}(\mathbf{k}')
    :c_{\mathbf{k} a\eta_2 s_2}^{\dagger}  c_{\mathbf{k}' a' \eta_2 s_2}:.
\end{align}
 Introducing the integral
\begin{equation}
    X_{aa'}^{(\eta_2)} (\mathbf{k},\mathbf{k}')
= \frac{1}{\Omega_0} \int d^2\mathbf{r}\  n_f(\mathbf{r}) \sum_{l,\beta} \sum_{\mathbf{Q},\mathbf{Q}'\in \mathcal{Q}_{l\eta_2}} V(\mathbf{k}-\mathbf{Q}-\mathbf{k}'+\mathbf{Q}')  e^{-i(\mathbf{k}-\mathbf{Q}-\mathbf{k}'+\mathbf{Q}')\cdot\mathbf{r}}  \tilde{u}^{(\eta_2)*}_{\mathbf{Q}\beta,a}(\mathbf{k}) \tilde{u}^{(\eta_2)}_{\mathbf{Q}'\beta,a'}(\mathbf{k}'),
\label{eq:interaction_fc_density_Xintegral}
\end{equation}
and using the fact that  $e^{-i(\mathbf{Q} - \mathbf{Q}')\cdot \mathbf{R}} = 1$, which follows from $\mathbf{Q} - \mathbf{Q}' \in \mathcal{Q}_0$, $\hat{H}_W$ can be rewritten as
\begin{equation}
\hat{H}_{W} = \frac{1}{N}\sum_{\mathbf{R}, \alpha,\eta_1, s_1} \sum_{|\mathbf{k}|, |\mathbf{k}'|<\Lambda_c}  \sum_{\eta_2,s_2, a,a'} 
    X_{aa'}^{(\eta_2)} (\mathbf{k},\mathbf{k}') e^{-i(\mathbf{k}-\mathbf{k}')\cdot\mathbf{R}} 
    :f_{\mathbf{R} \alpha\eta_1 s_1}^\dagger f_{\mathbf{R} \alpha\eta_1 s_1}: 
    :c_{\mathbf{k} a\eta_s s_2}^\dagger  c_{\mathbf{k}' a' \eta_s s_2}:\;.
    \label{eq:interaction_fc_density}
\end{equation}
We are interested only in the low-energy states, implying that we can approximate the integral from \cref{eq:interaction_fc_density_Xintegral} as $X_{aa'}^{(\eta_2)} (\mathbf{k},\mathbf{k}') \approx X_{aa'}^{(\eta_2)} (\vec{0},\vec{0})$~\cite{SON22}, where
\begin{align}
X_{aa'}^{(\eta_2)} (\vec{0},\vec{0})
=& \frac{1}{\Omega_0} \int \dd^2{\mathbf{r}_1} n_f(\mathbf{r}) \sum_{l,\beta} \sum_{\mathbf{Q},\mathbf{Q}'\in \mathcal{Q}_{l\eta_2}} V(\mathbf{Q}-\mathbf{Q}') e^{i(\mathbf{Q}-\mathbf{Q}')\cdot\mathbf{r}}  \tilde{u}^{(\eta_2)*}_{\mathbf{Q}\beta,a}(\vec{0}) \tilde{u}^{(\eta_2)}_{\mathbf{Q}'\beta,a'}(\vec{0}) \nonumber\\
=& \frac1{\Omega_0 } \sum_{l,\beta} \sum_{\mathbf{Q},\mathbf{Q}'\in \mathcal{Q}_{l\eta_2}} n_f(\mathbf{Q}-\mathbf{Q}') V(\mathbf{Q}-\mathbf{Q}')   \tilde{u}^{(\eta_2)*}_{\mathbf{Q}\beta,a}(\vec{0}) \tilde{u}^{(\eta_2)}_{\mathbf{Q}'\beta,a'}(\vec{0})\ . 
\end{align}

In Ref.~\cite{SON22}, the form of the matrix $X_{aa'}^{(\eta)} \equiv X_{aa'}^{(\eta)} (\vec{0},\vec{0})$ was derived with the aid of Schur's lemma. In this work, we will find it insightful to re-derive the form of $X_{aa'}^{(\eta)}$ differently, without the aid of Schur's lemma. We start by obtaining the symmetry properties of the matrix $X_{aa'}^{(\eta)_2}$ from the symmetry representations formed by the conduction $c$-electrons. From \cref{eq:conduction_transform_momentum}, the matrix $X_{aa'}^{(\eta_2)}$ transforms under a symmetry operator $g$ as
{\small
\begin{align}
    &  \sum_{a,a'} [D^c(g)]^{\star}_{a\eta_2 b\eta} X_{aa'}^{\eta_2 } [D^c(g)]_{a'\eta_2 b' \eta'} = \nonumber \\ 
    &=  \frac{1}{\Omega_0} \sum_{l,\beta} \sum_{\mathbf{Q},\mathbf{Q}'\in \mathcal{Q}_{l\eta_2}} n_f(\mathbf{Q}-\mathbf{Q}') V(\mathbf{Q}-\mathbf{Q}') \sum_{a}  [D^c(g)]^{\star}_{a\eta_2 b\eta}(g)  \tilde{u}^{(\eta_2)*}_{\mathbf{Q}\beta,a}(\vec{0}) \sum_{a'} [D^c(g)]_{a'\eta_2 b' \eta'} \tilde{u}^{(\eta_2)}_{\mathbf{Q}'\beta,a'}(\vec{0})\  \nonumber \\ 
    & =  \frac{1}{\Omega_0} \sum_{l,\beta} \sum_{\mathbf{Q},\mathbf{Q}'\in \mathcal{Q}_{l\eta_2}} n_f(\mathbf{Q}-\mathbf{Q}') V(\mathbf{Q}-\mathbf{Q}')   \sum_{\mathbf{Q}_1,\beta_1} [D(g)]^{\star}_{\mathbf{Q}\beta \eta_2, \mathbf{Q}_1 \beta_1 \eta} \tilde{u}^{(\eta)*}_{\mathbf{Q}_1\beta_1,b}(\vec{0}) \sum_{\mathbf{Q}_2,\beta_2} [D(g)]_{\mathbf{Q}' \beta \eta_2, \mathbf{Q}_2 \beta_2 \eta'} \tilde{u}^{(\eta')}_{\mathbf{Q}_2\beta_2,b'}(\vec{0})\  \nonumber \\ 
    &=\frac{1}{\Omega_0 }\sum_l\sum_{\mathbf{Q}_1,\beta_1, \mathbf{Q}_2,\beta_2} \tilde{u}^{\eta \star}_{Q_1 \beta_1 b}(\vec{0}) \tilde{u}^{\eta'}_{\mathbf{Q}_2\beta_2 b'}(\vec{0})\sum_\beta \sum_{\mathbf{Q},\mathbf{Q}'\in \mathbf{Q}_{l\eta_2}}  n_f(\mathbf{Q}-\mathbf{Q}') V(\mathbf{Q}-\mathbf{Q}')  [D(g)]^{\star}_{\mathbf{Q}\beta \eta_2, \mathbf{Q}_1 \beta_1 \eta} [D(g)]_{\mathbf{Q}' \beta \eta_2, \mathbf{Q}_2 \beta_2 \eta'}\ ,
\end{align}}where $[D^c(g)]_{a\eta_2 b\eta}$ are the symmetry representation matrices of the conduction fermions listed in \crefrange{eq:conduction_representations_T}{eq:conduction_representations_C2zP} and $[D(g)]_{\mathbf{Q}\beta \eta_2, \mathbf{Q}_1 \beta_1 \eta}$ are given by \crefrange{eq:BM_symmetries_T}{eq:BM_PH_symmetry}. We note that the symmetry representation matrices $D[(g)]_{\mathbf{Q}' \beta \eta_2, \mathbf{Q}_2 \beta_2 \eta'}$ can be written in the form
\begin{equation}
    D[(g)]_{\mathbf{Q}' \beta \eta_2, \mathbf{Q}_2 \beta_2 \eta'} = f_{\mathbf{Q}}\delta_{\mathbf{Q}_2, g\mathbf{Q}'}[D^s(g)]_{\beta ,\beta_2 }[D^v(g)]_{ \eta_2, \eta'},
\end{equation}
where $f_{\mathbf{Q}} = \zeta_{\mathbf{Q}}$ for $P$ symmetry and $f_{\mathbf{Q}} = 1$ otherwise. The symmetry action thus factorizes into a matrix acting on the sublattice index and another one acting on the valley index independently. This enables us to trace out the $\mathbf{Q}_1$ and $\mathbf{Q}_2$ indices
\normalsize{
\begin{align}
   &\sum_{a,a'} [D^c(g)]^{\star}_{a\eta_2 b\eta} X_{aa'}^{\eta_2 } [D^c(g)]_{a'\eta_2 b' \eta'}  \nonumber \\ & = \frac{1}{\Omega_0 }\sum_l \sum_\beta \sum_{\mathbf{Q},\mathbf{Q}'\in \mathbf{Q}_{l\eta_2}}  n_f(\mathbf{Q}-\mathbf{Q}') V(\mathbf{Q}-\mathbf{Q}')  [D(g)]^{\star}_{\mathbf{Q}\beta \eta_2,\beta_1 \eta} [D(g)]_{\mathbf{Q}' \beta \eta_2,\beta_2 \eta'}   \tilde{u}^{\eta \star}_{g\mathbf{Q} \beta_1 b}(\vec{0}) \tilde{u}^{\eta'}_{g \mathbf{Q}'\beta_2 b'}(\vec{0})
   \label{eq:Winteraction_Xintermidiate}
\end{align}}
From the unitarity of the symmetry operators we find
\begin{equation}
    \sum_{\beta}[D(g)]^{\star}_{\beta ,\beta_1 }[D(g)]_{\beta,\beta_2} = \delta_{\beta_1,\beta_2},
\end{equation}
which allows us to trace out the $\beta,\; \beta_2$ indices in \cref{eq:Winteraction_Xintermidiate} and obtain
\begin{align}
    &\sum_{a,a'}[D^c(g)]^{\star}_{a\eta_2 b\eta} X_{aa'}^{\eta_2 } [D^c(g)]_{a'\eta_2 b' \eta'} 0 \nonumber \\ & =\frac{1}{\Omega_0 }\sum_{\beta_1, l} \sum_{\mathbf{Q},\mathbf{Q}'\in \mathbf{Q}_{l\eta_2}} \tilde{u}^{\eta \star}_{\mathbf{Q} \beta_1 b}(\vec{0}) \tilde{u}^{\eta'}_{\mathbf{Q}'\beta_2 b'}(\vec{0})  n_f(\mathbf{Q}-\mathbf{Q}') V(\mathbf{Q}-\mathbf{Q}') f_\mathbf{Q} f_{\mathbf{Q}'} [D(g)]^{\star}_{\eta_2 ,\eta }[D(g)]_{\eta_2,\eta'} \nonumber \\ &=\frac{1}{\Omega_0 }\sum_{\beta_1, l} \sum_{\mathbf{Q},\mathbf{Q}'\in \mathbf{Q}_{l\eta_2}} u^{\eta \star}_{\mathbf{Q} \beta_1 b}(\vec{0}) u^{\eta'}_{\mathbf{Q}'\beta_2 b'}(\vec{0})  n_f(\mathbf{Q}-\mathbf{Q}') V(\mathbf{Q}-\mathbf{Q}')  [D(g)]^{\star}_{\eta_2 ,\eta }[D(g)]_{\eta_2,\eta'},
\end{align}
where we have used the fact that $f_{\mathbf{Q}}f_{\mathbf{Q}'} = 1$ since both $\mathbf{Q}$ and $\mathbf{Q}'$ belong to the same layer $\mathcal{Q}_{l\eta_2}$. The valleys are related by $\eta = \eta' = g^{-1}\eta_2$, where $g\eta$ expresses the effect of the symmetry $g$ on the valley index $\eta$. With this shorthand notation, we can write
\begin{equation}
    \sum_{a,a'}[D^c(g)]^{\star}_{a(g\eta), b\eta} X_{aa'}^{(g\eta) } [D^c(g)]_{a'(g\eta), b' \eta'} = \frac{1}{\Omega_0 }\sum_{\beta_1, l} \sum_{\mathbf{Q},\mathbf{Q}'\in \mathbf{Q}_{l(g\eta)}} u^{\eta \star}_{\mathbf{Q} \beta_1 b}(\vec{0}) u^{\eta}_{\mathbf{Q}'\beta_2 b'}(\vec{0})  n_f(\mathbf{Q}-\mathbf{Q}') V(\mathbf{Q}-\mathbf{Q}') = X_{bb'}^{\eta },
    \label{eq:interaction_Xintegral_symmetries}
\end{equation}
{\it i.e.}{} implying that the matrix $X^{(\eta)}_{aa'}$ commutes with all valley conserving symmetry representations $D^c(g)$. Representing $X^{\eta}_{aa'}$ as a block-matrix
\begin{equation}
    X^{(\eta)} = 
    \begin{pmatrix}
    X_{\Gamma_3} & X_{\Gamma_3\Gamma_{12}} \\
    X_{\Gamma_{12}\Gamma_3} & X_{\Gamma_1\Gamma_2}
    \end{pmatrix},
\end{equation}
where $X_{\Gamma_3}$, $X_{\Gamma_3\Gamma_{12}}$, $X_{\Gamma_{12}\Gamma_3}$ and $X_{\Gamma_1\Gamma_2}$ are $2\times2$ blocks. We consider the valley-preserving crystalline symmetries, generated by $C_{2z}T$, $C_{2x}$ and $C_{3z}$. The constraints implied, according to \cref{eq:interaction_Xintegral_symmetries}, by the $C_{2z}T$, $C_{2x}$ and $C_{3z}$ symmetries, respectively, read as
\begin{align}
    (\sigma_0\otimes \sigma_x) X^{(\eta)\star}   (\sigma_0\otimes \sigma_x) &= X^{(\eta)},
    \label{eq:Xintegral_C2zTconstraint} \\
    (\sigma_0\otimes \sigma_x) X^{(\eta)}   (\sigma_0\otimes \sigma_x) &= X^{(\eta)},
    \label{eq:Xintegral_C2xconstraint} \\
    \left(\exp(- i \frac{2\pi}{3}\sigma_z ) \oplus \sigma_0 \right) X^{(\eta)} \left(\exp( i \frac{2\pi}{3}\sigma_z ) \oplus \sigma_0 \right)
 &= X^{(\eta)}.
 \label{eq:Xintegral_C3zconstraint}
\end{align}
From the combination of \cref{eq:Xintegral_C2zTconstraint,,eq:Xintegral_C2xconstraint} we infer that the $X^{(\eta)}$ matrix entries are real~\cite{SON22}. From \cref{eq:Xintegral_C2xconstraint}, it follows that each of the blocks of the $X^{(\eta)}$ matrix has diagonal elements equal to each other, as well as non-diagonal ones~\cite{SON22}. Finally, \cref{eq:Xintegral_C3zconstraint} implies $X_{\Gamma_{12}\Gamma_3} = X_{\Gamma_3\Gamma_{12}} = 0$ and $X_{\Gamma_3}$ is diagonal~\cite{SON22}. We note, that the crystalline symmetries do not put any constraints for the $\Gamma_1\Gamma_2$ block of the $X^{(\eta)}$ matrix. Invoking the particle-hole symmetry $P$ representation from \cref{eq:conduction_representations_P}, which preserves the valley, we find that non-diagonal elements of the $\Gamma_1\Gamma_2$ block must be zero~\cite{SON22}. Additionally, given that the matrix elements are real, the time-reversal symmetry implies, that $X^{(\eta)}$ does not depend on valley~\cite{SON22}. With this we can write~\cite{SON22}
\begin{equation}
    X^{\eta} =
    \begin{pmatrix}
    W_1 & 0 & 0 & 0 \\
    0 & W_1 & 0 & 0 \\
    0 & 0 & W_3 & 0 \\
    0 & 0 & 0 & W_3
    \end{pmatrix},
\end{equation}
where the real parameters $W_1$ and $W_3$ are either obtained numerically in \cref{app:sec:numerics} or calculated analytically in \cref{app:subsec:analytic_interaction_fc}.

\subsection{The $c$-$c$ density-density interaction}\label{app:subsec:review_cc_interaction}
The term in the interaction Hamiltonian corresponding to the density-density interaction of the conduction band electrons, $\hat{H}_V$ (see \cref{tab:Coulomb_summary}), is obtained by plugging the conduction band electron density $\hat{\rho}_{cc}$ from \cref{eq:interaction_cc_term} into
\begin{equation}
   \hat{H}_V = \frac{1}{2}\int d^2 \vec{r}_1 d^2 \vec{r}_2 :\hat{\rho}_{cc}(\vec{r}_1):V(\vec{r}_1 - \vec{r}_2) :\hat{\rho}_{cc}(\vec{r}_2):.
\end{equation}
The term reads
\begin{align}
    \hat{H}_V &= \frac{1}{2\Omega_{\textrm{tot}}^2}\sum_{\beta_1, l_1, \eta_1, s_1} \sum_{\beta_2, l_2, \eta_2, s_2} \sum_{\substack{|\mathbf{k}_1|,|\mathbf{k}_1'|<\Lambda_c \\ \mathbf{Q}_1\mathbf{Q}_1'\in\mathcal{Q}_{l_1\eta_1}}}\sum_{\substack{|\mathbf{k}_2|,|\mathbf{k}_2'|<\Lambda_c \\ \mathbf{Q}_2\mathbf{Q}_2'\in\mathcal{Q}_{l_2\eta_2}}} \int \dd^2{\mathbf{r}_1} \dd^2{\mathbf{r}_2} V(\mathbf{r}_1 - \mathbf{r}_2) \nonumber \\
    &\times e^{-i(\mathbf{k}_1 - \mathbf{Q}_1 - \mathbf{k}_1' + \mathbf{Q}_1')\cdot\mathbf{r}_1}\tilde{u}_{\mathbf{Q}_1\beta_1,a_1}^{(\eta_1)*}(\mathbf{k}_1)\tilde{u}_{\mathbf{Q}'_1\beta_1,a'_1}^{(\eta_1)}(\mathbf{k}_1')e^{-i(\mathbf{k}_2 - \mathbf{Q}_2 - \mathbf{k}_2' + \mathbf{Q}_2')\cdot\mathbf{r}_2}\tilde{u}_{\mathbf{Q}_2'\beta_2,a_2'}^{(\eta_2)*}(\mathbf{k}_2')\tilde{u}_{\mathbf{Q}_2\beta_2,a_2}^{(\eta_2)}(\mathbf{k}_2)  \nonumber \\
    & \times :\hat{c}^\dagger_{\mathbf{k}_1a_1\eta_1s_1}\hat{c}_{\mathbf{k}_1'a_1'\eta_1s_1}::\hat{c}^\dagger_{\mathbf{k}_2'a_2'\eta_2s_2}\hat{c}_{\mathbf{k}_2a_2\eta_2s_2}:
\end{align}
Integrating over $\mathbf{r}_1$ and $\mathbf{r}_2$ and using the Fourier transformation of the Coulomb potential from \cref{eq:Coulomb_interaction_fourier}, we derive the momentum conversation law $\mathbf{k}_1' - \mathbf{Q}_1' - \mathbf{k}_1 + \mathbf{Q}_1 = \mathbf{k}_2'-\mathbf{Q}_2'-\mathbf{k}_2+\mathbf{Q}_2$~\cite{SON21}. Since $\mathbf{k}_1',\mathbf{k}_1,\mathbf{k}_2,\mathbf{k}_2'$ are all small momenta around the $\Gamma_M$ point and $\mathbf{Q}_1 - \mathbf{Q}_1',\;\mathbf{Q}_2 - \mathbf{Q}_2'$ both belonging to the set of reciprocal lattice vectors $\mathcal{Q}_0$, we obtain $\mathbf{k}_1' - \mathbf{k}_1 = \mathbf{k}_2' - \mathbf{k}_2$ together with $- \mathbf{Q}_1' + \mathbf{Q}_1 =-\mathbf{Q}_2'+\mathbf{Q}_2$. Introducing 
the momentum $\mathbf{q} = \mathbf{k}_1'-\mathbf{k}_1 = \mathbf{k}_2' - \mathbf{k}_2$ and the reciprocal lattice vector $\mathbf{G} = -\mathbf{Q}_1' + \mathbf{Q}_1 = -\mathbf{Q}_2' + \mathbf{Q}_2$ the conduction band electron density-density interaction term can be rewritten as
{\small
\begin{align}
    & \hat{H}_V = \frac{1}{2\Omega_0N} \sum_{\beta_1,l_1,\eta_1,s_1}\sum_{\beta_2,l_2,\eta_2,s_2}\sum_{a_1,a_1',a_2,a_2'}\sum_{\substack{|\mathbf{k}_1|<\Lambda_c \\ \mathbf{Q}_1 \in \mathcal{Q}_{l_1\eta_1}}}\sum_{\substack{|\mathbf{k}_2|<\Lambda_c \\ \mathbf{Q}_2 \in \mathcal{Q}_{l_2\eta_2}}}\sum_{\mathbf{G}}\sum_{\substack{\mathbf{q} \\ |\mathbf{k}_1 + \mathbf{q}_1|,|\mathbf{k}_2+\mathbf{q}|<\Lambda_c}} V(\mathbf{q} + \mathbf{G}) \nonumber \\
    &\times \tilde{u}_{\mathbf{Q}_1\beta_1,a_1}^{(\eta_1)*}(\mathbf{k}_1)\tilde{u}_{\mathbf{Q}_1 - \mathbf{G}\beta_1,a'_1}^{(\eta_1)}(\mathbf{k}_1 + \mathbf{q})\tilde{u}_{\mathbf{Q}_2-\mathbf{G}\beta_2,a_2'}^{(\eta_2)*}(\mathbf{k}_2 + \mathbf{q})\tilde{u}_{\mathbf{Q}_2\beta_2,a_2}^{(\eta_2)}(\mathbf{k}_2):\hat{c}^\dagger_{\mathbf{k}_1a_1\eta_1s_1}\hat{c}_{\mathbf{k}_1+\mathbf{q} a_1'\eta_1s_1}::\hat{c}^\dagger_{\mathbf{k}_2+\mathbf{q} a_2'\eta_2s_2}\hat{c}_{\mathbf{k}_2 a_2\eta_2s_2}:.
\end{align}}From now on, we will use a simplified notation and denote the wave function $\tilde{u}^{(\eta)}_{\mathbf{Q}\alpha,a}(\mathbf{k})$ as a ket vector $\ket{\tilde{u}^{(\eta)}_{a}(\mathbf{k})}$. Using this simplification, we introduce the matrix 
\begin{equation}
X_{\eta_1 a_1 a_1', \eta_2 a_2 a_2'}(\mathbf{k}_1,\mathbf{k}_2; \mathbf{q}) = \frac1{\Omega_0} \sum_{\mathbf{G}} V(\mathbf{q}+\mathbf{G}) 
    \bra{\tilde{u}^{(\eta_1)}_{a_1}(\mathbf{k}_1)} \ket{\tilde{u}^{(\eta_1)}_{a_1'}(\mathbf{k}_1+\mathbf{q}+\mathbf{G})}
\bra{\tilde{u}^{(\eta_2)}_{a_2'}(\mathbf{k}_2+\mathbf{q}+\mathbf{G})} \ket{\tilde{u}^{(\eta_2)}_{a_2}(\mathbf{k}_2)},
\label{eq:interaction_cc_Xmatrix}
\end{equation}
such that the interaction term is rewritten in a simpler way as
\begin{align}
    \hat{H}_V = \frac{1}{2N} &\sum_{\eta_1,s_1,a_1,a_1'}\sum_{\eta_2,s_2,a_2,a_2'}\sum_{|\mathbf{k}_1|,|\mathbf{k}_2|<\Lambda_c}\sum_{\substack{\mathbf{q} \\ |\mathbf{k}_1+\mathbf{q}|,|\mathbf{k}_2+\mathbf{q}|<\Lambda_c}}X_{\eta_1 a_1 a_1', \eta_2 a_2 a_2'}(\mathbf{k}_1,\mathbf{k}_2; \mathbf{q}) \nonumber \\ &:\hat{c}^\dagger_{\mathbf{k}_1a_1\eta_1s_1}\hat{c}_{\mathbf{k}_1+\mathbf{q} a_1'\eta_1s_1}::\hat{c}^\dagger_{\mathbf{k}_2+\mathbf{q} a_2'\eta_2s_2}\hat{c}_{\mathbf{k}_2 a_2\eta_2s_2}:\;.
\end{align}
As we are interested in the low-energy physics, and the conduction band electrons are low-energy only in the vicinity of the $\Gamma_M$ point, we approximate the matrix in \cref{eq:interaction_cc_Xmatrix} as
\begin{equation}
    \small
    X_{\eta_1 a_1 a_1', \eta_2 a_2 a_2'}(\mathbf{k}_1,\mathbf{k}_2; \mathbf{q}) \approx X_{\eta_1 a_1 a_1', \eta_2 a_2 a_2'}(\vec{0},\vec{0};\mathbf{q})  =  \frac1{\Omega_0} \sum_{\mathbf{G}} V(\mathbf{q}+\mathbf{G}) 
  \bra{\tilde{u}^{(\eta_1)}_{a_1}(\vec{0})} \ket{\tilde{u}^{(\eta_1)}_{a_1'}(\mathbf{G})   }
   \bra{ \tilde{u}^{(\eta_2)}_{a_2'}(\mathbf{G})} \ket{ \tilde{u}^{(\eta_2)}_{a_2}(\vec{0})   }.
   \label{eq:interaction_cc_Xmatrix_final}
\end{equation}
We discuss the analytical approximation of the matrix \cref{eq:interaction_cc_Xmatrix_final} in \cref{app:subsec:analytic_interaction_cc}.

\subsection{The $f$-$c$ exchange interaction }\label{app:subsec:review_fc_exchange_interaction}
The exchange interaction term in the interaction Hamiltonian, $\hat{H}_J$ (see \cref{tab:Coulomb_summary}), is obtained by plugging \cref{eq:interaction_fc_term,,eq:interaction_cf_term} into
\begin{equation}
    \hat{H}_J = \frac{1}{2}\int d^2 \vec{r}_1 d^2 \vec{r}_2 \left[:\hat{\rho}_{fc}(\vec{r}_1):V(\vec{r}_1 - \vec{r}_2) :\hat{\rho}_{cf}(\vec{r}_2): + :\hat{\rho}_{cf}(\vec{r}_1):V(\vec{r}_1 - \vec{r}_2) :\hat{\rho}_{fc}(\vec{r}_2):\right].
\end{equation}
After multiple simplifications, discussed in details in Ref.~\cite{SON22}, the exchange interaction term reads \normalsize
\begin{align}
\hat{H}_{J} &= \frac1{2N} \sum_{\substack{\eta_1\alpha_1a_1 \\ {\eta_2\alpha_2 a_2}}} \sum_{\mathbf{R}} \sum_{|\mathbf{k}_1|,|\mathbf{k}_2|<\Lambda_c }  
    X_{\eta_1\alpha_1a_1, \eta_2\alpha_2 a_2}(\mathbf{k}_1,\mathbf{k}_2) e^{i(\mathbf{k}_1-\mathbf{k}_2)\cdot\mathbf{R}}   \nonumber \\
&\times
    \left(
    \hat{f}^\dagger_{\mathbf{R} \alpha_1\eta_1 s_1} \hat{c}_{ \mathbf{k}_1 a_1\eta_1 s_1} \hat{c}^\dagger_{\mathbf{k}_2 a_2 \eta_2 s_2} \hat{f}_{\mathbf{R} \alpha_2 \eta_2 s_2} 
    + \hat{c}^\dagger_{\mathbf{k}_2 a_2 \eta_2 s_2} \hat{f}_{\mathbf{R} \alpha_2 \eta_2 s_2} \hat{f}^\dagger_{\mathbf{R} \alpha_1\eta_1 s_1} \hat{c}_{ \mathbf{k}_1 a_1\eta_1 s_1}\right) \numberthis \label{eq:interaction_exchange_J_term} \;,
\end{align}
where we define the matrix
\begin{equation} 
X_{\eta_1 \alpha_1 a_1, \eta_2 \alpha_2 a_2}(\mathbf{k}_1,\mathbf{k}_2) = \int \frac{\dd^2{\mathbf{q}}}{(2\pi)^2} \ V(\mathbf{q}) 
\bra{v^{(\eta_1)}_{\alpha_1}(\mathbf{k}_1-\mathbf{q})} \ket{\tilde{u}^{(\eta_1)}_{a_1}(\mathbf{k}_1)}
\bra{\tilde{u}^{(\eta_2)}_{a_2}(\mathbf{k}_2)} \ket{v^{(\eta_2)}_{\alpha_2}(\mathbf{k}_2-\mathbf{q})}.
\label{eq:exchange_matrix}
\end{equation}
Note that similarly to \cref{eq:interaction_cc_Xmatrix}, we have written the wave function $v_{\mathbf{Q}\beta,\alpha}^{(\eta)}(\mathbf{k})$ as a ket vector $\ket{v_{\alpha}^{(\eta)}(\mathbf{k})}$.

The low-energy physics involves the conduction band electrons with momentum close to the $\Gamma_M$ point, {\it i.e.}{} $\mathbf{k}_{1,2} \sim \vec{0}$ in \cref{eq:exchange_matrix}. We therefore approximate the exchange matrix as
\begin{equation}
    {\mathcal{J}}_{\eta_1 \alpha_1 a_1, \eta_2 \alpha_2 a_2} = 
X_{\eta_1 \alpha_1 a_1, \eta_2 \alpha_2 a_2}(\vec{0},\vec{0}) = \int \frac{\dd^2{\mathbf{q}}}{(2\pi)^2} \ V(\mathbf{q}) 
\bra{v^{(\eta_1)}_{\alpha_1}(-\mathbf{q})} \ket{\tilde{u}^{(\eta_1)}_{a_1}(\vec{0})}
\bra{\tilde{u}^{(\eta_2)}_{a_2}(\vec{0})} \ket{v^{(\eta_2)}_{\alpha_2}(-\mathbf{q})}.
\label{eq:exchange_matrix_approximate}
\end{equation}
After rearranging the creation and annihilation operators in \cref{eq:interaction_exchange_J_term} and using the approximation given in \cref{eq:exchange_matrix_approximate}, the $\hat{H}_J$ term can be written as~\cite{SON22}
\begin{equation}
\hat{H}_{J} = - \frac1{N} \sum_{ \substack{\eta_1\alpha_1a_1 \\ {\eta_2\alpha_2 a_2} } } \sum_{\mathbf{R}} \sum_{|\mathbf{k}_1|,|\mathbf{k}_2|<\Lambda_c }  
    {\mathcal{J}}_{\eta_1\alpha_1a_1, \eta_2\alpha_2 a_2} e^{i(\mathbf{k}_1-\mathbf{k}_2)\cdot\mathbf{R}}  
    :\hat{f}^\dagger_{\mathbf{R} \alpha_1\eta_1 s_1}\hat{f}_{\mathbf{R} \alpha_2 \eta_2 s_2} :
    :\hat{c}^\dagger_{\mathbf{k}_2 a_2 \eta_2 s_2} \hat{c}_{ \mathbf{k}_1 a_1\eta_1 s_1} :\ .
    \label{eq:exchange_term_J_hamiltonian_final1}
\end{equation}
We note that $\hat{H}_{J}$ corresponds to a ferromagnetic exchange interaction, [the minus sign in \cref{eq:exchange_term_J_hamiltonian_final1} follows from commuting the fermion operators $f^\dagger c c^\dagger f = -f^\dagger f c^\dagger c$].

We now review the symmetry constraints imposed on the exchange matrix ${\mathcal{J}}_{\eta_1 \alpha_1 a_1, \eta_2 \alpha_2 a_2}$ \cite{SON22}. Recall the transformation rules of the $f$- and $c$-electron states under a symmetry $g$, \cref{eq:wannier_transform_momentum,,eq:conduction_transform_momentum}, which in the simplified Dirac notation for $\ket{\tilde{u}_a^{(\eta)}(\mathbf{k})}$ and $\ket{v_{\alpha}^{(\eta)}(\mathbf{k})}$ read as
\begin{align}
     g \ket{\tilde{u}^{(\eta)}_{a}(g\mathbf{k})} &= \sum_{a^{\prime}} [D^c(g)]_{a^{\prime}\tilde{\eta}, a\eta} \ket{\tilde{u}^{(\tilde{\eta})}_{a^{\prime}}(\mathbf{k})}, \\
     g \ket{v^{(\eta)}_{\alpha}(g\mathbf{k})} &= \sum_{\alpha^{\prime}} [D^f(g)]_{\alpha^{\prime}\tilde{\eta}, \alpha\eta} \ket{v^{(\tilde{\eta})}_{\alpha^{\prime}}(\mathbf{k})},
\end{align}
where the representation matrices $[D^{f,c}(g)]_{a^{\prime}\tilde{\eta}, a\eta}$ for the $f$- and $c$-electrons are listed in \crefrange{eq:wannier_symmetries_T}{eq:wannier_symmetries_C2zP} and \crefrange{eq:conduction_representations_T}{eq:conduction_representations_C2zP}, respectively. Inserting the identity $e = g^{-1}g$ into the inner products in \cref{eq:exchange_matrix_approximate}, we obtain the constraints imposed by a symmetry $g$
\begin{align}
    &{\mathcal{J}}_{\eta_1 \alpha_1 a_1, \eta_2 \alpha_2 a_2} = \sum_{\alpha_1',a_1'}\sum_{\alpha_2',a_2'}[D^{f\dagger}(g)]_{\alpha_1'\tilde{\eta}_1, \alpha_1\eta_1}[D^c(g)]_{a_1'\tilde{\eta}_1, a_1\eta_1} {\mathcal{J}}_{\tilde{\eta}_1 \alpha_1' a_1', \tilde{\eta}_2 \alpha_2' a_2'}[D^{c\dagger}(g)]_{a_2'\tilde{\eta}_2, a_2\eta_2}[D^{f}(g)]_{\alpha_2'\tilde{\eta}_2,\alpha_2\eta_2} \nonumber \\
    &= \sum_{\alpha_1',a_1'}[D^{f\dagger}(g)]_{\alpha_1'\tilde{\eta}_1, \alpha_1\eta_1}[D^c(g)]_{a_1'\tilde{\eta}_1, a_1\eta_1} {\mathcal{J}}_{\tilde{\eta}_1 \alpha_1' a_1', \eta_2 \alpha_2 a_2} = \sum_{\alpha_2',a_2'}{\mathcal{J}}_{\eta_1 \alpha_1 a_1, \tilde{\eta}_2 \alpha_2' a_2'}[D^{c\dagger}(g)]_{a_2'\tilde{\eta}_2, a_2\eta_2}[D^{f}(g)]_{\alpha_2'\tilde{\eta}_2,\alpha_2\eta_2}.
    \label{eq:exchange_J_constraint}
\end{align}
For the $C_{2z}P$ symmetry from \cref{eq:wannier_symmetries_C2zP,,eq:conduction_representations_C2zP,,eq:exchange_J_constraint} we obtain~\cite{SON22}
\begin{equation}
{\mathcal{J}}_{\eta_1 \alpha_1 a_1, \eta_2 \alpha_2 a_2} = - e^{i\pi(\alpha_2 - a_2)} {\mathcal{J}}_{\eta_1 \alpha_1 a_1, -\eta_2 \bar{\alpha}_2 \bar{a}_2} = - e^{i\pi (a_1 - \alpha_1)} {\mathcal{J}}_{-\eta_1 \bar{\alpha}_1 \bar{a}_1, \eta_2 \alpha_2 a_2},
\label{eq:exchange_J_C2zP}
\end{equation}
where $\bar{\alpha} = 2,1$ for $\alpha=1,2$ and $\bar{a} = 2,1,4,3$ for $a = 1,2,3,4$.
Invoking the $C_{2x}$ symmetry, \cref{eq:wannier_symmetries_C2x,,eq:conduction_representations_C2x,,eq:exchange_J_constraint} yield~\cite{SON22}
\begin{equation}
{\mathcal{J}}_{\eta_1 \alpha_1 a_1, \eta_2 \alpha_2 a_2} = {\mathcal{J}}_{\eta_1 \bar{\alpha}_1 \bar{a}_1, \eta_2 \bar{\alpha}_2 \bar{a}_2},
\label{eq:exchange_J_C2x}
\end{equation}
while the $C_{2y} = C_{2x}C_{2z}$ symmetry, following \cref{eq:wannier_symmetries_C2x,,eq:wannier_symmetries_C2zT,,eq:wannier_symmetries_T,,eq:conduction_representations_C2x,,eq:conduction_representations_C2zT,,eq:conduction_representations_T,,eq:exchange_J_constraint}, implies~\cite{SON22}
\begin{equation}
{\mathcal{J}}_{\eta_1 \alpha_1 a_1, \eta_2 \alpha_2 a_2} = {\mathcal{J}}_{-\eta_1 \alpha_1 a_1, -\eta_2 \alpha_2 a_2}\ .
\label{eq:exchange_J_C2y}
\end{equation}
The $C_{2z}T$ symmetry together with \cref{eq:wannier_symmetries_C2zT,,eq:conduction_representations_C2zT,,eq:exchange_J_constraint} requires the ${\mathcal{J}}$ matrix to satisfy~\cite{SON22}
\begin{equation}
{\mathcal{J}}^*_{\eta_1 \alpha_1 a_1, \eta_2 \alpha_2 a_2} = {\mathcal{J}}_{\eta_1 \bar{\alpha}_1 \bar{a}_1, \eta_2 \bar{\alpha}_2 \bar{a}_2},
\label{eq:exchange_J_C2zT}
\end{equation}
which coupled with \cref{eq:exchange_J_C2x} implies that the ${\mathcal{J}}$ matrix is real. Finally, for the $C_{3z}$ symmetry, using \cref{eq:wannier_symmetries_C3z,,eq:conduction_representations_C3z,,eq:exchange_J_constraint}, we obtain~\cite{SON22}
\begin{equation}
    {\mathcal{J}}_{\eta_1 \alpha_1 a_1, \eta_2 \alpha_2 a_2} = \zeta_{\eta_1\alpha_1}^*\zeta_{\eta_1 a_1}\zeta^*_{\eta_2 a_2}\zeta_{\eta_2\alpha_2}{\mathcal{J}}_{\eta_1 \alpha_1 a_1, \eta_2 \alpha_2 a_2},
    \label{eq:exchange_J_C3z}
\end{equation}
where $\zeta_{\eta a}$ and $\zeta_{\eta \alpha}$ are the $C_{3z}$ eigenvalues given by the matrix elements $[D^c(C_{3z})]_{a\eta,a\eta}$ from the \cref{eq:conduction_representations_C3z} and $[D^f(C_{3z})]_{\alpha\eta,\alpha\eta}$ from the \cref{eq:wannier_symmetries_C3z}, respectively.

\begin{figure}
    \centering
    \includegraphics[]{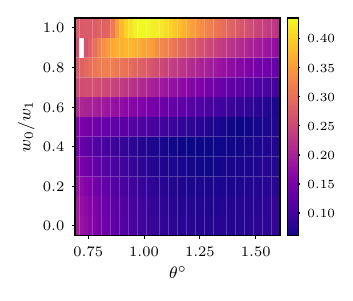}
    \caption{The ratio between the leading matrix element of ${\mathcal{J}}$ ({\it i.e.}{} $J$) and the largest matrix element of ${\mathcal{J}}$ we ignore as a function of the twist angle $\theta$ and the tunneling amplitude ratio $w_0/w_1$. The matrix elements of ${\mathcal{J}}$ are computed numerically from \cref{eq:exchange_matrix_approximate} for the typical screening length $\xi = \SI{10}{\nano\meter}$ and for the same parameters as the ones employed in \cref{app:sec:tables}. The isolated white point corresponds to a point where $J$ is not the leading matrix element of ${\mathcal{J}}$.}
    \label{fig:app_ratio_of_j}
\end{figure}

The constraints in \cref{eq:exchange_J_C3z,eq:exchange_J_C2zT,eq:exchange_J_C2x,eq:exchange_J_C2zP} limit the number of independent \emph{real} components of ${\mathcal{J}}$ to just six, namely ${\mathcal{J}}_{-11,-11}$, ${\mathcal{J}}_{-11,-22}$, ${\mathcal{J}}_{-12,-12}$, ${\mathcal{J}}_{-12,-13}$, ${\mathcal{J}}_{-13,-13}$, and ${\mathcal{J}}_{-14,-14}$. Ref.~\cite{SON22} has argued that a good approximation is to keep only the largest matrix element of ${\mathcal{J}}$
\begin{equation}
    {\mathcal{J}}_{\eta 1 3,\eta 1 3} = {\mathcal{J}}_{\eta 2 4,\eta 2 4} = -{\mathcal{J}}_{\eta 1 3,-\eta 2 4} = -{\mathcal{J}}_{\eta 2 4,-\eta 1 3} = J.
    \label{eq:exchange_J_parameter}
\end{equation}
In \cref{fig:app_ratio_of_j}, we plot the ratio between the leading matrix element of ${\mathcal{J}}$ ({\it i.e.}{} $J$) and the largest matrix element of ${\mathcal{J}}$ we ignore. The approximation of only keeping the leading contribution of ${\mathcal{J}}$ holds across the entire phase space we consider at least as well as it does at the magic angle~\cite{SON22}: the leading matrix element of ${\mathcal{J}}$ is at least four-times larger than the next contribution (except for the region around the magic angle and non-realistic equal tunneling amplitudes $w_0/w_1 = 1.0$). In what follows, we will follow Ref.~\cite{SON22} and ignore all the matrix elements of ${\mathcal{J}}$ except $J$. 

The exchange interaction strength parameter $J$ was calculated in~\cite{SON22} and will be calculated for a larger phase space in \cref{app:sec:numerics}. An analytic expression for $J $ can also be obtained, as we discuss in \cref{app:subsec:analytic_exchange_interaction}. The exchange interaction term from \cref{eq:exchange_term_J_hamiltonian_final1} can be simplified in a more compact form as~\cite{SON22}
\begin{equation}
\hat{H}_J = - \frac{J}{2N} \sum_{\mathbf{R}, s_1, s_2} \sum_{\alpha,\alpha',\eta,\eta'} \sum_{|\mathbf{k}_1|,|\mathbf{k}_2|<\Lambda_c }   
     e^{i( \mathbf{k}_1- \mathbf{k}_2 )\cdot\mathbf{R} } 
     ( \eta\eta' + (-1)^{\alpha+\alpha'} )
     :\hat{f}^\dagger_{\mathbf{R} \alpha \eta s_1} f_{\mathbf{R} \alpha' \eta' s_2}:  :\hat{c}^\dagger_{\mathbf{k}_2, \alpha'+2, \eta' s_2} \hat{c}_{ \mathbf{k}_1, \alpha+2, \eta s_1}:.
     \label{eq:HJ_explicit}
\end{equation}
We note that the coefficient $( \eta\eta' + (-1)^{\alpha+\alpha'} )$ is nonzero only for $\eta= \eta'$ and $\alpha = \alpha '$ or for $\bar{\eta}= \eta'$ and $\bar{ \alpha} = \alpha '$. 

\subsection{The double hybridization terms $f^{\dagger}f^{\dagger}cc$}\label{app:subsec:review_ffcc_interaction}
The double hybridization term in the interaction Hamiltonian, $\hat{H}_{\tilde{J}+}$ (see \cref{tab:Coulomb_summary}), is obtained by plugging \cref{eq:interaction_cf_term,,eq:interaction_fc_term} into
\begin{equation}
    \hat{H}_{\tilde{J}+} = \frac{1}{2}\int d^2 \vec{r}_1 d^2 \vec{r}_2 \left[:\hat{\rho}_{fc}(\vec{r}_1):V(\vec{r}_1 - \vec{r}_2) :\hat{\rho}_{fc}(\vec{r}_2):\right],
    \label{eq:interaction_Hjplus}
\end{equation}
while the corresponding term $:\hat{\rho}_{cf}:V:\hat{\rho}_{cf}:$ can be obtained by the hermitian conjugation of \cref{eq:interaction_Hjplus}. Following~\cite{SON22} and inserting the Fourier transform of the Coulomb interaction \cref{eq:Coulomb_interaction_fourier}, we write
{\small
\begin{align}
\hat{H}_{\tilde{J} +} =&  \frac1{\Omega_{\textrm{tot}}} \int \frac{\dd^2{\mathbf{q}}}{(2\pi)^2} \int \dd^2{\mathbf{r}_1} \dd^2{\mathbf{r}_2} V(\mathbf{q}) e^{-i\mathbf{q}\cdot(\mathbf{r}_1-\mathbf{r}_2)}
    \sum_{ \substack{ \beta_1, l_1,\eta_1, s_1 \\ \beta_2, l_2, \eta_2, s_2}  } 
    \sum_{ \substack{ \mathbf{R}, \alpha_1, \alpha_2 \\ a_1 a_2} }  
    \sum_{ \substack{ |\mathbf{k}_1|<\Lambda_c \\ |\mathbf{k}_2|<\Lambda_c }}
    \sum_{\substack{\mathbf{Q}_1 \in \mathcal{Q}_{l_1\eta_1} \\ \mathbf{Q}_2\in \mathcal{Q}_{l_2\eta_2}}}
    w^{(\eta_1)*}_{l_1 \beta_1, \alpha_1}(\mathbf{r}_1-\mathbf{R}) 
    \tilde{u}^{(\eta_1)}_{\mathbf{Q}_1\beta_1,a_1}(\mathbf{k}_1) e^{i( \mathbf{k}_1-\mathbf{Q}_1)\cdot\mathbf{r}_1} \nonumber\\
& \times w^{(\eta_2)*}_{l_2 \beta_2, \alpha_2}(\mathbf{r}_2-\mathbf{R}) 
        \tilde{u}^{(\eta_2)}_{\mathbf{Q}_2\beta_2,a_2}(\mathbf{k}_2) e^{i(\mathbf{k}_2-\mathbf{Q}_2)\cdot\mathbf{r}_2}
        e^{i(\eta_1\Delta\mathbf{K}_{l_1} + \eta_2\Delta\mathbf{K}_{l_2})\cdot\mathbf{R}} 
        \hat{f}^\dagger_{\mathbf{R} \alpha_1 \eta_1 s_1} \hat{c}_{\mathbf{k}_1 a_1 \eta_1 s_1} 
        \hat{f}^\dagger_{\mathbf{R} \alpha_2 \eta_2 s_2} \hat{c}_{\mathbf{k}_2 a_2 \eta_2 s_2}
    \label{eq:double_exchange_J_I}
\end{align}}Using \cref{eq:wannier_fourier}, we simplify \cref{eq:double_exchange_J_I} as
\begin{align}
\hat{H}_{\tilde{J} +} =&  \frac1{N} \int \frac{\dd^2{\mathbf{q}}}{(2\pi)^2}  V(\mathbf{q}) 
    \sum_{ \substack{ \beta_1, l_1,\eta_1, s_1 \\ \beta_2, l_2, \eta_2, s_2}  } 
    \sum_{ \substack{ \mathbf{R}, \alpha_1, \alpha_2 \\ a_1, a_2} }  
    \sum_{ \substack{ |\mathbf{k}_1|<\Lambda_c \\ |\mathbf{k}_2|<\Lambda_c }}
    \sum_{\substack{\mathbf{Q}_1 \in \mathcal{Q}_{l_1\eta_1} \\ \mathbf{Q}_2\in \mathcal{Q}_{l_2\eta_2}}}
    v^{(\eta_1)*}_{\mathbf{Q}_1\beta_1, \alpha_1}(\mathbf{k}_1-\mathbf{q}) \tilde{u}^{(\eta_1)}_{\mathbf{Q}_1\beta_1,a_1}(\mathbf{k}_1)
    v^{(\eta_2)*}_{\mathbf{Q}_2\beta_2, \alpha_2}(\mathbf{k}_2+\mathbf{q}) \tilde{u}^{(\eta_2)}_{\mathbf{Q}_2\beta_2,a_2}(\mathbf{k}_2)
     \nonumber \\
&\times e^{i(\eta_1 \Delta\mathbf{K}_{l_1}  + \mathbf{k}_1-\mathbf{Q}_1-\mathbf{q} + \eta_2 \Delta\mathbf{K}_{l_2} + \mathbf{k}_2 - \mathbf{Q}_2 +\mathbf{q})\cdot\mathbf{R}}
        \hat{f}^\dagger_{\mathbf{R} \alpha_1 \eta_1 s_1} \hat{c}_{\mathbf{k}_1 a_1 \eta_1 s_1} 
        \hat{f}^\dagger_{\mathbf{R} \alpha_2 \eta_2 s_2} \hat{c}_{\mathbf{k}_2 a_2 \eta_2 s_2}  \ .
        \label{eq:double_exchange_J_II}
\end{align}
We notice that $e^{i(\eta_1 \Delta\mathbf{K}_{l_1} -\mathbf{Q}_1 + \eta_2 \Delta\mathbf{K}_{l_2} - \mathbf{Q}_2)\cdot\mathbf{R}}=1$ since $ \eta_1 \Delta\mathbf{K}_{l_1} -\mathbf{Q}_1 + \eta_2 \Delta\mathbf{K}_{l_2} - \mathbf{Q}_2$ belongs to the moir\'e reciprocal lattice. This enables us to further simplify \cref{eq:double_exchange_J_II} \normalsize
\begin{equation}
\hat{H}_{\tilde{J} +} =  \frac{1}{N} 
    \sum_{  \eta_1, s_1,  \eta_2, s_2  } 
    \sum_{ \substack{ \mathbf{R}, \alpha_1, \alpha_2 \\ a_1, a_2} }  
    \sum_{ \substack{ |\mathbf{k}_1|<\Lambda_c \\ |\mathbf{k}_2|<\Lambda_c }}
    \tilde{X}_{\eta_1\alpha_1 a_1, \eta_2\alpha_2 a_2}(\mathbf{k}_1,\mathbf{k}_2)
     e^{i(\mathbf{k}_1+ \mathbf{k}_2)\cdot\mathbf{R}}
     \hat{f}^\dagger_{\mathbf{R} \alpha_1 \eta_1 s_1} \hat{f}^\dagger_{\mathbf{R} \alpha_2 \eta_2 s_2}
      \hat{c}_{\mathbf{k}_2 a_2 \eta_2 s_2} \hat{c}_{\mathbf{k}_1 a_1 \eta_1 s_1} \ ,
\end{equation}
where we introduced the matrix
\begin{equation}
\tilde{X}_{\eta_1\alpha_1 a_1, \eta_2\alpha_2 a_2}(\mathbf{k}_1,\mathbf{k}_2) = \int \frac{\dd^2{\mathbf{q}}}{(2\pi)^2}  V(\mathbf{q}) 
    \bra{v^{(\eta_1)}_{\alpha_1}(\mathbf{k}_1-\mathbf{q})}\ket{\tilde{u}^{(\eta_1)}_{a_1}(\mathbf{k}_1)}
   \bra{v^{(\eta_2)}_{\alpha_2}(\mathbf{k}_2+\mathbf{q})} \ket{\tilde{u}^{(\eta_2)}_{a_2}(\mathbf{k}_2)}.
\end{equation}
Since we consider only the low-energy physics, we can approximate $\mathbf{k} \approx \vec{0}$ and therefore introduce the matrix
\begin{equation}
\tilde{J}_{\eta_1\alpha_1 a_1, \eta_2\alpha_2 a_2} = \tilde{X}_{\eta_1\alpha_1 a_1, \eta_2\alpha_2 a_2}(\vec{0},\vec{0}) = \int \frac{\dd^2{\mathbf{q}}}{(2\pi)^2}  V(\mathbf{q}) 
    \bra{v^{(\eta_1)}_{\alpha_1}(-\mathbf{q})}\ket{\tilde{u}^{(\eta_1)}_{a_1}(\vec{0})}
   \bra{v^{(\eta_2)}_{\alpha_2}(\mathbf{q})} \ket{\tilde{u}^{(\eta_2)}_{a_2}(\vec{0})}.
\end{equation}
The matrix $\tilde{J}_{\eta_1\alpha_1 a_1, \eta_2\alpha_2 a_2}$ transforms under the symmetries of TBG analogously to \cref{eq:exchange_J_constraint}~\cite{SON22}. Moreover, the time-reversal symmetry $T$, \cref{eq:wannier_symmetries_T,,eq:conduction_representations_T} implies
\begin{equation}
\tilde{J}_{\eta_1\alpha_1 a_1, \eta_2\alpha_2 a_2}=J_{\eta_1\alpha_1 a_1,- \eta_2\alpha_2 a_2},
\end{equation} and hence relates parameters $\tilde{J}$ and $J$, introduced in \cref{eq:exchange_J_parameter}~\cite{SON22}. Following Ref.~\cite{SON22}, as well as our discussion surrounding \cref{eq:exchange_J_parameter}, we keep only the leading terms and set the others to zero. Therefore, $\tilde{J}_{\eta_1\alpha_1 a_1, \eta_2\alpha_2 a_2}$ can be written more compactly as~\cite{SON22}
\begin{equation}
\tilde{J}_{\eta_1\alpha_1 a_1, \eta_2\alpha_2 a_2} = \tilde{J}_{\eta \alpha \alpha+2, \eta' \alpha' \alpha'+2} = \frac{J}{2}(-\eta \eta' +(-1)^{\alpha+ \alpha'}).
\end{equation}
As such, the double exchange term in \cref{eq:interaction_Hjplus} reads as
\begin{align}
\hat{H}_{\tilde{J}+} & =  \frac{J}{4N} 
    \sum_{ \mathbf{R}, s_1, s_2 } \sum_{\alpha,\alpha',\eta,\eta'}  
    \sum_{  |\mathbf{k}_1|, |\mathbf{k}_2|<\Lambda_c } 
     e^{i(\mathbf{k}_1+ \mathbf{k}_2)\cdot\mathbf{R}}
    ( -\eta\eta' + (-1)^{\alpha+\alpha'} )
    \hat{f}^\dagger_{\mathbf{R} \alpha \eta s_1} \hat{f}^\dagger_{\mathbf{R} \alpha' \eta' s_2}
    \hat{c}_{\mathbf{k}_2, \alpha'+2, \eta' s_2} \hat{c}_{\mathbf{k}_1, \alpha+2, \eta s_1}  \nonumber \\ &=  \frac{J}{4N} 
    \sum_{ \mathbf{R}, s_1, s_2 } \sum_{\alpha,\alpha',\eta,\eta'}  
    \sum_{  |\mathbf{k}_1|, |\mathbf{k}_2|<\Lambda_c } 
     e^{i(\mathbf{k}_1+ \mathbf{k}_2)\cdot\mathbf{R}}
    ( \eta\eta' + (-1)^{\alpha+\alpha'} )
    \hat{f}^\dagger_{\mathbf{R} \alpha \eta s_1} \hat{f}^\dagger_{\mathbf{R} \alpha' -\eta' s_2}
    \hat{c}_{\mathbf{k}_2, \alpha'+2, -\eta' s_2} \hat{c}_{\mathbf{k}_1, \alpha+2, \eta s_1}.
\end{align}
We introduce the total double exchange term $\hat{H}_{\tilde{J}}$ as the sum of $\hat{H}_{\tilde{J}+}$ and its hermitian conjugate
\begin{equation} \label{eq:HJstar_two_term}
    \hat{H}_{\tilde{J}} = \hat{H}_{\tilde{J} +} + \hat{H}_{\tilde{J} +}^\dagger \ ,
\end{equation}
and discuss the enlarged continuous symmetries of the exchange and double hybridization interaction terms from \cref{eq:HJ_explicit} and \cref{eq:HJstar_two_term}, respectively, in \cref{app:subsec:higher_symmetry}.

\section{THF single-particle parameters from the Bistritzer-Macdonald model}\label{app:sec:analytic_single_particle}
In this appendix, we provide detailed derivations of the analytical expressions for the THF model single-particle parameters. In \crefrange{app:subsec:analytic_forbitals_tripod}{app:subsec:analytic_other_hexagon}, we discuss the parameters that characterize the THF single-particle Hamiltonian, the summary of which can be found in \cref{app:subsec:HF_single_summary}. In particular, we derive expressions for the local $f$-fermion parameters $\lambda_1$, $\lambda_2$, $\alpha_1$, $\alpha_2$ from the Tripod model (see \cref{app:subsec:tripod}), expressions for the parameters $M$ and $v_{\star}$ that characterize the conduction $c$-fermions from the Hexagon model (see \cref{app:subsec:tripod}), and finally, expressions for parameters $\gamma$ and $v_{\star}'$ of the hybridization between $f$- and $c$-fermions also from the Hexagon model. Additionally, in \cref{app:subsec:analytic_dirac_velocity}, we derive the renormalized Dirac velocity $v_D$ of the THF $f$-electron bands at the $K_M$ point.

\subsection{Local fermion orbital parameters from the Tripod model}\label{app:subsec:analytic_forbitals_tripod}
In this section, we derive the analytical expressions for the Wannier states real space function parameters $\lambda_1$, $\lambda_2$, $\alpha_1$, $\alpha_2$ introduced in \cref{eq:wannier_real_space_functions}. Analytically, it is more convenient to work in momentum space and consider the representation of the Wannier states in the Bloch basis $v^{(\eta)}_{\mathbf{Q}\beta,\alpha}(\mathbf{k})$. Our strategy will be to approximate the functions from \cref{eq:wannier_state} as a linear combination of the BM model eigenstates $u_{\mathbf{Q}\beta,n\eta}(\mathbf{k})$. This approximation is almost exact at the $K_M$ point, where, by construction, the THF states are fully supported by the active TBG bands and, hence, are linear combinations of the latter. In turn, we employ the Tripod model reviewed in \cref{app:subsec:tripod} to find analytical expressions of the TBG active band states $u_{\mathbf{Q}\beta,n\eta}(\mathbf{k})$ near the $K_M$ point.

From \cref{eq:wannier_transform_momentum} we find that the THF states $v_{\mathbf{Q}\beta,\alpha}^{(\eta)}(\mathbf{k})$ satisfy the following gauge conditions for $C_{2z}P$, $C_{2z}T$ symmetries respectively
\begin{align}
    \zeta_{\mathbf{Q}}v_{\mathbf{Q}\beta,\alpha}^{(\eta)}(\mathbf{k}) &= i \zeta_\alpha v_{\mathbf{Q}\bar{\beta},\bar{\alpha}}^{(-\eta)}(\mathbf{k}),
    \label{eq:Wannier_repr_C2zP} \\
    v_{\mathbf{Q}\beta,\alpha}^{(\eta)}(\mathbf{k}) &= v^{(\eta)*}_{\mathbf{Q}\bar{\beta},\bar{\alpha}}(\mathbf{k}),
     \label{eq:Wannier_repr_C2zT}
\end{align}

We see that under the $C_{2z}P$ and $C_{2z}T$ symmetries, the Wannier states transform in the same way as the Chern band basis states [see \cref{eq:BM_chern_transform_C2zP,,eq:BM_chern_transform_C2zT}] with the gauge fixed according to \cref{eq:BM_gauge_fixing} for a general momentum $\mathbf{k}$. As such, and since at the MBZ boundary the $f$-fermion states are linear combinations of the TBG active bands, we can approximately identify the Wannier states with the Chern basis states
\begin{equation}
    v^{(\eta)}_{\mathbf{Q}\beta,\alpha}(\mathbf{k}) \approx U^{e_Y}_{\mathbf{Q}\beta,\eta}(\mathbf{k}), \quad
    \text{for} \quad (-1)^{\alpha+1} = \eta e_Y \quad \text{and } \mathbf{k} \text{ at the edge of the MBZ}.
    \label{eq:local_fermion_tripod_approximation}
\end{equation}
The symmetry transformation rules of the Wannier states in \cref{eq:Wannier_repr_C2zP,eq:Wannier_repr_C2zT} and those of the Chern basis states, listed in \cref{eq:BM_chern_transform_C2zP,eq:BM_chern_transform_C2zT}, are thus matched.

In the vicinity of $K_M$ point ($\mathbf{k} = \mathbf{q}_1 + \delta \mathbf{k}$), in \cref{app:subsec:tripod}, we derived the eigenstates of the BM model in \cref{eq:tripod_A01_solution1,,eq:tripod_A01_solution2}, [expressed in the spinor basis of \cref{eq:tripod_spinor_definition}]
\begin{equation}
        \psi_{A0_1}^{(n=-1)}(\delta\mathbf{k}) = \frac{\alpha}{\sqrt{2}}(-e^{-i\phi(\delta \mathbf{k})},1)^T,\qquad  \psi_{A0_1}^{(n=+1)}(\delta\mathbf{k}) = \frac{\beta}{\sqrt{2}}(e^{-i\phi(\delta \mathbf{k})},1)^{T},
\end{equation}
where the phases $\alpha,\beta$ are fixed via the $C_{2z}T$ gauge condition in \cref{eq:BM_gauge_fixing} as
\begin{equation}
    \sigma_x \psi_{A0_1}^{(n=\pm1)*} = \psi_{A0_1}^{(n=\pm1)} \implies \alpha^2 = -e^{-i\phi(\delta \mathbf{k})},\;\; \beta^2 = e^{i\phi(\delta \mathbf{k})}.
\end{equation}
With this, we can construct the Chern band wave function for the $A0_1$ site [$\mathbf{Q}=\mathbf{q}_1$, see \cref{fig:A1:b}] in the valley $\eta=+$ as
\begin{equation}
    U_{\mathbf{q}_1\beta,+}^{e_Y}(\mathbf{q}_1 + \delta\mathbf{k}) = \frac{1}{\sqrt{2}}\left(\psi_{A0_1}^{(n=-1)}(\delta\mathbf{k}) + i e_Y \psi_{A0_1}^{(n=+1)}(\delta\mathbf{k})\right) = \frac{i}{2}\left((-1 + e_Y)e^{-i\phi(\delta\mathbf{k})/2}, (1 + e_Y)e^{i\phi(\delta\mathbf{k})/2}\right)^{T}.
\end{equation}
With the approximation \cref{eq:local_fermion_tripod_approximation} we relate
\begin{equation}
    v_{\mathbf{q}_1\beta,1}^{+}(\mathbf{q}_1 + \delta\mathbf{k}) = U^{+1}_{\mathbf{q}_1\beta,+}(\mathbf{q}_1 + \delta\mathbf{k}),\qquad  v_{\mathbf{q}_1\beta,2}^{+}(\mathbf{q}_1 + \delta\mathbf{k}) = U^{-1}_{\mathbf{q}_1\beta,+}(\mathbf{q}_1 + \delta\mathbf{k}),
\end{equation}
and find that exactly at the $K_M$ point ($\delta \mathbf{k} = 0$) the phase $\phi(\delta\mathbf{k})$ is not defined. Hence, this strategy, while applicable for general momentum $\mathbf{k}$, is not valid at exactly the $K_M$ point. We note, however, that while $C_{2z}T$ and $C_{2z}P$ fix the gauge freedom, at the $K_M$ point, the BM model has two degenerate eigenstates at zero energy and, therefore, a $SU(2)$ gauge freedom, which we will fix by invoking the $C_{3z}$ symmetry. We recall that from \cref{eq:Chern_C3z_transformation}, the Chern band wave function $U^{e_Y}_{\mathbf{Q}\beta,\eta}(\mathbf{k})$ transforms under the $C_{3z}$ symmetry according to 
\begin{equation}
    e^{i\eta\frac{2\pi}{3}(-1)^{\beta+1}}U^{e_Y}_{C_{3z}\mathbf{Q}\beta,\eta}(C_{3z}\mathbf{k}) = e^{ie_Y\theta(\mathbf{k})} U^{e_Y}_{\mathbf{Q}\beta,\eta}(\mathbf{k}),
\end{equation}
for a phase $\theta(\mathbf{k})$ that is to be determined for $\mathbf{k} = \mathbf{q}_1$. At the $K_M$ point and for $\mathbf{Q}=\mathbf{q}_1$ we obtain
\begin{equation}
    U^{e_Y}_{C_{3z}\mathbf{q}_1\beta,\eta}(C_{3z}\mathbf{q}_1) = U^{e_Y}_{\mathbf{q}_1\beta,\eta}(\mathbf{q}_1) = e^{ie_Y\theta(\mathbf{q}_1)} e^{-i\eta\frac{2\pi}{3}(-1)^{\beta+1}} U^{e_Y}_{\mathbf{q}_1\beta,\eta}(\mathbf{q}_1),
    \label{eq:chern_basis_C3z}
\end{equation}
where we used \cref{eq:embedding_condition} in the first equality. To move forward, we recall from \cref{eq:BM_symmetries_C3z,eq:wannier_transform_momentum,eq:wannier_symmetries_C3z} that under $C_{3z}$ symmetry, the Wannier states transform as
\begin{equation}
    v_{C_{3z}\mathbf{Q}\beta,\alpha}^{(\eta)}(C_{3z}\mathbf{k})e^{i\eta\frac{2\pi}{3}(-1)^{\beta+1}} = v_{\mathbf{Q}\beta,\alpha}^{(\eta)}(\mathbf{k})e^{i\eta\frac{2\pi}{3}(-1)^{\alpha+1}},
     \label{eq:Wannier_repr_C3z}
\end{equation}
which for $\mathbf{k} = \mathbf{Q} = \mathbf{q}_1$ can be rewritten as
\begin{equation}
    v_{\mathbf{q}_1\beta,\alpha}^{(\eta)}(\mathbf{q}_1)e^{i\eta\frac{2\pi}{3}(-1)^{\beta+1}} = v_{\mathbf{Q}\beta,\alpha}^{(\eta)}(\mathbf{q}_1)e^{i\eta\frac{2\pi}{3}(-1)^{\alpha+1}}.
    \label{eq:wannier_C3z_transformation_q1}
\end{equation}
Here we used that $v^{(\eta)}_{\mathbf{q}_1 + \vec{G}_M \beta,\alpha}(\mathbf{q}_1 + \vec{G}_M) = v^{(\eta)}_{\mathbf{q}_1\beta,\alpha}(\mathbf{q}_1)$ (for a moir\'e reciprocal vector $\vec{G}_M$), which immediately follows from \cref{eq:wannier_fourier}. From \cref{eq:wannier_C3z_transformation_q1} and the convention $(-1)^{\alpha+1} = e_Y\eta$, it follows, that the amplitude $v^{(\eta)}_{\mathbf{q}_1\beta,\alpha}(\mathbf{q}_1)$ is non-zero only for $e_Y = \eta(-1)^{\beta+1}$. Since the Wannier states have to transform in the same way as the Chern basis, the same expression $e_Y = \eta(-1)^{\beta+1}$ has to hold for the Chern basis \cref{eq:chern_basis_C3z}. Plugging it in, we find $\theta(\mathbf{q}_1) = \frac{2\pi}{3}$. Therefore, we obtain up to a constant $c$
\begin{equation}
    U^{e_Y}_{\mathbf{q}_1\beta,\eta}(\mathbf{q}_1) = c\delta_{e_Y,\eta(-1)^{\beta+1}},
\end{equation}
We use the Tripod model \cref{eq:tripod_equations_2} to obtain the Chern states for the $A1_1$ lattice site ($\mathbf{Q}=2\mathbf{q}_1$) at the $K_M$ point. We find
\begin{equation}
    U^{+1}_{2\mathbf{q}_1\beta,+}(\mathbf{q}_1) = (\mathbf{q}_1\cdot\boldsymbol{\sigma})T_1U^{+1}_{\mathbf{q}_1\beta,+}(\mathbf{q}_1) = (-iw_1,iw_0)_{\beta}.
\end{equation}
Using the approximation from \cref{eq:local_fermion_tripod_approximation}, we then have the following ratios
\begin{equation}
    -iw_1 = \frac{U^{+1}_{2\mathbf{q}_11,+}(\mathbf{q}_1)}{U^{+1}_{\mathbf{q}_11,+}(\mathbf{q}_1)} = \frac{v^{(+)}_{2\mathbf{q}_11,1}(\mathbf{q}_1)}{v^{(+)}_{\mathbf{q}_11,1}(\mathbf{q}_1)},\qquad iw_0 = \frac{U^{+1}_{2\mathbf{q}_12,+}(\mathbf{q}_1)}{U^{+1}_{\mathbf{q}_11,+}(\mathbf{q}_1)} = \frac{v^{(+)}_{2\mathbf{q}_12,1}(\mathbf{q}_1)}{v^{(+)}_{\mathbf{q}_11,1}(\mathbf{q}_1)}.
    \label{eq:local_lambda_expressions}
\end{equation}
We calculate the Wannier states $v^{(+)}_{\mathbf{q}_11,1}(\mathbf{q}_1),\;v^{(+)}_{2\mathbf{q}_11,1}(\mathbf{q}_1),\;v^{(+)}_{2\mathbf{q}_12,1}(\mathbf{q}_1)$ in momentum space, by plugging the corresponding real-space function \cref{eq:wannier_states_I} into \cref{eq:wannier_fourier}. The resulting expressions are given by
\begin{align}
    v^{(+)}_{\mathbf{q}_11,1}(\mathbf{q}_1) &= \frac{\alpha_1\sqrt{2\pi}e^{i\frac{\pi}{4}}}{\lambda_1\sqrt{\Omega_0}}\int_0^{+\infty}re^{-\frac{r^2}{2\lambda_1^2}}dr =  \frac{\alpha_1\lambda_1\sqrt{2\pi}}{\sqrt{\Omega_0}}e^{i\frac{\pi}{4}},
    \label{eq:local_tripod_v111} \\
    v^{(+)}_{2\mathbf{q}_11,1}(\mathbf{q}_1) &= \frac{\alpha_1 \sqrt{2\pi}e^{-i\frac{\pi}{4}}}{\lambda_1\sqrt{\Omega_0}} \int_0^{+\infty} r e^{-\frac{r^2}{2\lambda_1^2}} J_{0}(q_1 r) dr = \frac{\alpha_1 \lambda_1 \sqrt{2\pi}}{\sqrt{\Omega_0}}e^{-i\frac{\pi}{4}}e^{-\frac{\lambda_1^2}{2}},
     \label{eq:local_tripod_v211} \\
    v^{(+)}_{2\mathbf{q}_11,1}(\mathbf{q}_1) &= -\frac{\alpha_2\sqrt{2\pi}e^{-i\frac{\pi}{4}}}{\lambda_2^2\sqrt{\Omega_0}}\int_0^{+\infty}r^2e^{-\frac{r^2}{2\lambda_2^2}}J_1(q_1 r)dr =  -\frac{\alpha_2\lambda_2^2\sqrt{2\pi}}{\sqrt{\Omega_0}}e^{-i\frac{\pi}{4}},
    \label{eq:local_tripod_v221}
\end{align}
where $q_1 = |\mathbf{q}_1| = 1$ and $J_{0,1}(z)$ are the Bessel functions of the first kind. Plugging \cref{eq:local_tripod_v111,,eq:local_tripod_v211,,eq:local_tripod_v221} in \cref{eq:local_lambda_expressions}, we obtain
\begin{equation}
    w_1 = e^{-\frac{\lambda_1^2}{2}},\qquad w_0=\frac{\alpha_2\lambda_2^2}{\alpha_1\lambda_1}e^{-\frac{\lambda_2^2}{2}}.
\end{equation}
We could find another relation for $\lambda_1$, $\lambda_2$ by considering further plane-wave states. However, as we will show later in \cref{app:sec:numerics}, an approximation $\lambda_1\approx\lambda_2$ is valid up to a 20\% error for a significant part of the BM model parameter space. Together with the normalization condition this, gives us the other two equations required to close the system
\begin{equation}
    \label{eq:app_lambda12_approx}
    \lambda_1=\lambda_2,\qquad \alpha_1^2 + \alpha_2^2 = 1.
\end{equation}
Solving the system, we find
\begin{equation}
    \lambda_1^{\textrm{1-shell}} = \sqrt{-2\ln{w_1}},\qquad \left(\frac{\alpha_1}{\alpha_2}\right)^{\textrm{1-shell}} = \frac{w_1}{w_0}\sqrt{-2\ln{w_1}},
    \label{eq:lambda1_alpha_ratio_1shell_approximation}
\end{equation}
where we note that $w_0,w_1$ are dimensionless and $\lambda_1$ is in units of $1/k_{\theta}$ as discussed at the end of \cref{app:sec:BM_model}.

In the end of this section, we also consider the two-shell tripod approximation of the BM model, discussed in \cref{app:subsec:tripod} in detail. From the definition of the Chern basis and \cref{eq:2nd_shell_A11_A01} we obtain
\begin{equation}
     U^{+1}_{2\mathbf{q}_1\beta,+}(\mathbf{q}_1) = \frac{(\mathbf{q}_1\cdot\boldsymbol{\sigma})}{1-w_0^2}T_1U^{+1}_{\mathbf{q}_1\beta,+}(\mathbf{q}_1) = \left(\frac{-iw_1}{1-w_0^2},\frac{iw_0}{1-w_0^2}\right)_{\beta},
\end{equation}
and plug it in \cref{eq:local_lambda_expressions}. Analogously to the previous calculation, we obtain a better approximation for $\lambda_1$ and $\alpha_1 / \alpha_2$
\begin{equation}
    \lambda_1^{\textrm{2-shell}} = \sqrt{2\ln{\frac{1 - w_0^2}{w_1}}},\qquad 
    \left(\frac{\alpha_1}{\alpha_2}\right)^{\textrm{2-shell}} = \frac{w_1}{w_0} \sqrt{2\ln{\frac{1 - w_0^2}{w_1}}}.
    \label{eq:tripod_lambda_2shell_approximation}
\end{equation}
We compare analytical approximations for the THF parameters $\lambda_1,\;\lambda_2,\;\alpha_1/\alpha_2$ to the numerical values in \cref{fig:main_sp_params} of the main text and in \cref{fig:G1} in \cref{app:subsec:additional_numerics_single_particle}.

\subsection{Conduction band parameters from the Hexagon model}\label{app:subsec:analytic_cbands_hexagon}
In this section, we map the conduction band electron states with the wave functions $\tilde{u}_{\mathbf{Q}\beta,a}^{(\eta)}(\mathbf{k}),\;a=3,4$ which form the $\Gamma_1\oplus\Gamma_2$ representation, to the eigenstates of the Hexagon model reviewed in \cref{app:subsec:hexagon}. As the result, we provide analytical expressions for parameters $M$ and $v_{\star}$ which characterize the conduction band part of the THF model Hamiltonian in \cref{eq:conduction_band_hamiltonian}. Analogously to  \cref{app:subsec:HF_conduction_bands}, here we focus on the $\Gamma_M$ point, setting $\mathbf{k}=\vec{0}$, and omitting the momentum dependence in what follows. We consider the lattice sites $A1_{1,\dots,6}$ from \cref{fig:A1:c} with the corresponding wave-vectors $\mathbf{Q} = \pm \mathbf{q}_i$, $i=1,2,3$. From the $C_{2x}$, $T$, $C_{2z}T$ and $C_{2z}P$ symmetries and \crefrange{eq:conduction_transform_momentum}{eq:conduction_representations_C2zP} we obtain for $a=3,4$ and $\mathbf{Q} = l\mathbf{q}_1$, $l\in \{\pm1\}$,
\begin{equation}
    \tilde{u}_{l\mathbf{q}_1,\bar{\beta},\bar{a}}^{(\eta)} = \tilde{u}^{(\eta)}_{-l\mathbf{q}_1,\beta,a},\qquad \tilde{u}^{(-\eta)*}_{l\mathbf{q}_1,\beta,a}=\tilde{u}^{(\eta)}_{-l\mathbf{q}_1,\beta,a},\qquad \tilde{u}^{(\eta)*}_{l\mathbf{q}_1,\bar{\beta},\bar{a}} = \tilde{u}^{(\eta)}_{l\mathbf{q}_1,\beta,a},\qquad l\tilde{u}^{(\eta)}_{l\mathbf{q}_1,\bar{\beta},\bar{a}}=i(-1)^{a+1}\tilde{u}_{l\mathbf{q}_1,\beta,a}^{(-\eta)}.
    \label{eq:conduction_wave function_constraints}
\end{equation}
We rewrite the wave function $\tilde{u}$ as the product of the amplitude $X$ and phase $\phi$
\begin{equation}
    \tilde{u}_{l\mathbf{q}_1,\beta,a}^{(\eta)} = X^{(\eta)}_{l\mathbf{q}_1,\beta,a}e^{i\phi^{(\eta)}_{l\mathbf{q}_1,\beta,a}},\;\;X^{(\eta)}_{l\mathbf{q}_1,\beta,a} \geq 0.
\end{equation}
The set of constraints in \cref{eq:conduction_wave function_constraints} implies that
\begin{equation}
    X_{l\mathbf{q}_1,\bar{\beta},\bar{a}}^{(\eta)}=X_{-l\mathbf{q}_1,\beta,a}^{(\eta)},\qquad X_{l\mathbf{q}_1,\beta,a}^{(-\eta)} = X_{-l\mathbf{q}_1,\beta,a}^{(\eta)},\qquad  X_{l\mathbf{q}_1,\bar{\beta},\bar{a}}^{(\eta)}=X_{l\mathbf{q}_1,\beta,a}^{(\eta)},\qquad  X_{l\mathbf{q}_1,\bar{\beta},\bar{a}}^{(-\eta)}=X_{l\mathbf{q}_1,\beta,a}^{(\eta)},
    \label{eq:conduction_amplitudes_constraints}
\end{equation}
for the amplitudes and similarly for the phases,
\begin{alignat}{2}
    \phi_{l\mathbf{q}_1,\bar{\beta},\bar{a}}^{(\eta)}&=\phi_{-l\mathbf{q}_1,\beta,a}^{(\eta)},\qquad 
    &&\phi_{l\mathbf{q}_1,\beta,a}^{(-\eta)} = -  \phi_{-l\mathbf{q}_1,\beta,a}^{(\eta)},\label{eq:conduction_phases_constraints_1} \\ \phi^{(\eta)}_{l\mathbf{q}_1,\beta,a}&=-\phi^{(\eta)}_{l\mathbf{q}_1,\bar{\beta},\bar{a}},\qquad 
    &&\frac{\pi}{2}(1 + l) + \phi^{(\eta)}_{l\mathbf{q}_1,\beta,a} = \frac{\pi}{2}(-1)^{a+1} + \phi^{(-\eta)}_{l\mathbf{q}_1,\bar{\beta},\bar{a}},
    \label{eq:conduction_phases_constraints_2}
\end{alignat}
where all the equalities are to be understood modulo $2\pi$ and $\bar{a} = 7 - a$.
From \cref{eq:conduction_amplitudes_constraints}, we infer that $X_{-l\mathbf{q}_1,\beta,a}^{(\eta)} = X_{-l\mathbf{q}_1,\bar{\beta},\bar{a}}^{(\eta)} = X_{l\mathbf{q}_1,\beta,a}^{(\eta)} \equiv X_{\beta,a}^{(\eta)}$, which means that the amplitude does not depend on $l$. Similarly, $\phi_{l\mathbf{q}_1,\beta,a}^{(-\eta)} = -\phi_{-l\mathbf{q}_1,\beta,a}^{(\eta)} = - \phi_{l\mathbf{q}_1,\bar{\beta},\bar{a}}^{(\eta)} = \phi_{l\mathbf{q}_1,\beta,a}^{(\eta)} \equiv \phi_{l\mathbf{q}_1,\beta,a}$, meaning that the phase does not depend on the valley. Given this constraint and using the $C_{2z}P$ symmetry condition in \cref{eq:conduction_phases_constraints_1,eq:conduction_phases_constraints_2}, we express the phase as
\begin{equation}
    \phi_{l\mathbf{q}_1,\beta,a} = -\frac{\pi}{4}(1 + l) + \frac{\pi}{4}(-1)^{a+1} + \pi \xi_{l,a},
\end{equation}
where we introduced a factor $\xi_{l,a}$ that depends on indices $l$ and $a$, and appears due to the fact that we divided by 2 while working modulo $2\pi$. We choose $\pi \xi_{l,a}$ to be consistent with the constraints \cref{eq:conduction_phases_constraints_1,eq:conduction_phases_constraints_2} and find that it is non-zero only for $l=1,\; a=4$. As such, we conclude that $\xi_{l,a} = \delta_{l,1}\delta_{4,a}$ and, therefore, the conduction electron state is given by
\begin{equation}
    \tilde{u}_{l\mathbf{q}_1,\beta,a}^{(\eta)} = X^{(\eta)}_{\beta,a}e^{i\frac{\pi}{4}l(2a-7)},\; \textrm{for}\;a=3,4,\;\;l=\pm1,\;\;X^{(\eta)}_{\beta,a} \geq 0
    \label{eq:state_lpm1}
\end{equation}
We can invoke the $C_{3z}$ symmetry to relate the wave function $\tilde{u}_{\mathbf{Q}\beta,a}^{(\eta)}$ at  $\mathbf{Q}=\pm\mathbf{q}_3,\pm\mathbf{q}_2$ to $\tilde{u}_{\pm\mathbf{q}_1\beta,a}^{(\eta)}$. From \cref{eq:conduction_transform_momentum} we obtain
\begin{equation}
    \tilde{u}_{l\mathbf{q}_3,\beta,a}^{(\eta)}=e^{-i\eta\frac{2\pi}{3}(-1)^{\beta+1}}\tilde{u}_{l\mathbf{q}_1,\beta,a}^{(\eta)},\qquad \tilde{u}_{l\mathbf{q}_2,\beta,a}^{(\eta)}=e^{-i\eta\frac{4\pi}{3}(-1)^{\beta+1}}\tilde{u}_{l\mathbf{q}_2,\beta,a}^{(\eta)},\qquad \textrm{for}\;l=\pm1.
    \label{eq:conduction_C3z}
\end{equation}
Furthermore, the normalization condition implies
\begin{equation}
    \sum_{p=1,2,3}\sum_{l=\pm1}\sum_{\beta=1,2}\left|u_{l\mathbf{q}_p,\beta,a}^{(\eta)}\right|^2 = 1,
    \label{eq:conduction_normalization}
\end{equation}
which, in turn, leads to
\begin{equation}
    (X^{(\eta)}_{1,a})^2 + (X^{(\eta)}_{2,a})^2 = \frac{1}{6},\; \textrm{for}\;a=3,4.
\end{equation}
Recall from \cref{eq:conduction_amplitudes_constraints} that $X^{(+)}_{\beta,a} = X^{(-)}_{\beta,a}$. This entails that there are only four real positive numbers $X^{(+)}_{1,a},\; X^{(+)}_{2,a}$ for $a=3,4$ that characterize the conduction electrons in the Hexagon model. To construct these states, we observe that the conduction electron wave functions are orthogonal to the Wannier states. Indeed, for a symmetry operator $g$ that does not change the valley it is straightforward to show from \cref{eq:wannier_transform_momentum,eq:conduction_transform_momentum} that
\begin{equation}
    \sum_{\mathbf{Q},\beta}\sum_{\alpha^{\prime}}[D^f(g)]_{\alpha\alpha^{\prime},\eta\eta}^{\dagger}\tilde{u}_{\mathbf{Q}\beta,a}^{(\eta)}v^{(\eta)*}_{\mathbf{Q}\beta,\alpha^{\prime}} = \sum_{\mathbf{Q},\beta}\sum_{a^{\prime}}[D^c(g)]_{a^{\prime}a,\eta\eta}\tilde{u}_{\mathbf{Q}\beta,a^{\prime}}^{(\eta)}v^{(\eta)*}_{\mathbf{Q}\beta,\alpha}.
\end{equation}
Therefore, for the $C_{3z}$ symmetry and the symmetry representations given by \cref{eq:wannier_symmetries_C3z,eq:conduction_representations_C3z}, we have
\begin{equation}
    \sum_{\mathbf{Q},\beta}\left(1 - e^{-i\eta\frac{2\pi}{3}(-1)^{\alpha+1}}\right)\tilde{u}_{\mathbf{Q}\beta,a}^{(\eta)}v^{(\eta)*}_{\mathbf{Q}\beta,\alpha} = 0,
\end{equation}
which implies that
\begin{equation}
    \sum_{\mathbf{Q},\beta}\tilde{u}_{\mathbf{Q}\beta,a}^{(\eta)}v^{(\eta)*}_{\mathbf{Q}\beta,\alpha} = 0,\;\textrm{for}\;a=3,4\;\textrm{and}\,\alpha=1,2.
    \label{eq:conduction_orthogonality}
\end{equation}
As discussed in \cref{app:subsec:HF_conduction_bands}, the conduction electron wave functions are given as the eigenvalue-one eigenstates of the projector $P^{(\eta)}(\vec{0}) - Q^{(\eta)}(\vec{0})$ from \cref{eq:conduction_projection_equation}. Together with the orthogonality condition \cref{eq:conduction_orthogonality}, we obtain
\begin{equation}
    \sum_{\mathbf{Q}',\beta'}(P^{(\eta)}_{\mathbf{Q}'\beta',\mathbf{Q}\beta}(\vec{0}) - Q^{(\eta)}_{\mathbf{Q}'\beta',\mathbf{Q}\beta}(\vec{0}))\tilde{u}^{(\eta)}_{\mathbf{Q}'\beta',a} = \sum_{\mathbf{Q}',\beta'}P^{(\eta)}_{\mathbf{Q}'\beta',\mathbf{Q}\beta}(\vec{0})\tilde{u}^{(\eta)}_{\mathbf{Q}'\beta',a} = \tilde{u}^{(\eta)}_{\mathbf{Q}\beta,a},\;\textrm{for}\;a=3,4.
\end{equation}
Since the projector $P^{(\eta)}(\vec{0})$ into the six BM model bands at the $\Gamma_M$ point involves a projector into the two $\Gamma_3$ representations and a projector into $\Gamma_1\oplus\Gamma_2$ representation, it follows that the conduction electron states $\tilde{u}^{(\eta)}_{\mathbf{Q}\beta,a},\;a=3,4$ are linear combinations of the BM bands $u_{\mathbf{Q}\beta,\eta\pm1}(\vec{0})$. As such, we can identify the conduction electron states and the BM model $\Gamma_1$, $\Gamma_2$ states. Written in the Hexagon model spinor basis from \cref{eq:hexagon_spinor_definition}, where the 12-dimensional spinor reads $\Psi = (\psi_{A1_1}^T,\dots, \psi_{A1_6}^T)^T$, we find these states to be
\begin{align}
\psi_{\Gamma_1} &= e^{-i \frac{\pi}{4}}S( i \alpha^\star, \alpha,  e^{ \frac{ \pi i}{3}} \alpha, 
 e^{ \frac{ \pi i}{6}} \alpha^\star,  e^{ \frac{2\pi i}{3}} i
   \alpha^\star, -e^{  \frac{\pi i}{3}} \alpha , -  \alpha, 
 -i \alpha^\star, -e^{ \frac{5 \pi i}{6}}  \alpha^\star, e^{ \frac{2\pi i}{3}} \alpha, 
 -i e^{ \frac{\pi i}{6}} \alpha, e^{\frac{5 \pi  i}{6}} \alpha^\star  ), \label{eq:G1_hex_state}\\  \psi_{\Gamma_2} &= e^{-i \frac{\pi}{4}} S  ( -i  \alpha, \alpha^\star,  -e^{ \frac{ \pi i}{3}} \alpha^\star, 
 e^{ \frac{ \pi i}{6}} \alpha,  e^{\frac{\pi i}{6}}
   \alpha, -e^{  \frac{\pi i}{3}} \alpha^\star ,    \alpha^\star,
-i \alpha, e^{ \frac{5\pi i}{6}}  \alpha, e^{ \frac{2\pi i}{3}} \alpha^\star, 
 i e^{ \frac{ \pi i}{6}} \alpha^\star,  e^{\frac{5\pi  i}{6}} \alpha  ), \label{eq:G2_hex_state}
\end{align}
where
\begin{equation}
S=\frac{1}{\sqrt{ 24 \sqrt{1+w_0^2}(\sqrt{1+w_0^2}+1)} }, \;\;\; \alpha = 1 + i w_0 + \sqrt{1+ w_0^2}.
\end{equation}
The states from \cref{eq:G1_hex_state,eq:G1_hex_state} form the same representation of the $C_{3z}$ symmetry as the conduction electron states $u^{(+)}_{\mathbf{Q},\beta, a}(\vec{0})$, $a=3,4$ eigenstates [see \cref{eq:conduction_representations_C3z}]. In order to ensure that the $C_{2z}T$ and $C_{2z}P$ symmetry representations given in \cref{eq:conduction_representations_C2zT,eq:conduction_representations_C2zP} are also satisfied, we take the linear combination of \cref{eq:G1_hex_state,eq:G2_hex_state}
\begin{equation}
u^{(+)}_{\mathbf{Q},\beta 3}(\vec{0})= \frac{1}{\sqrt{2}} (\psi_{\Gamma_1}+ \psi_{\Gamma_2}), \;\;\;u^{(+)}_{\mathbf{Q},\beta 4}(\vec{0})= \frac{1}{\sqrt{2}} (\psi_{\Gamma_1}- \psi_{\Gamma_2}).
\end{equation} 
The $u^{(+)}_{\mathbf{Q},\beta a}(\vec{0})$ states, therefore, are given by
{\small \begin{align}
u^{(+)}_{\mathbf{Q},\beta 3}(\vec{0}) &= S \sqrt{2}   e^{-i\frac{\pi}{4}}(w_0 ,   \theta,  -e^{-i \frac{2\pi}{3} } i  w_0   , - i e^{i \frac{2\pi}{3}}  \theta  , e^{i \frac{2\pi}{3} }  w_0,  e^{-i \frac{2\pi}{3}} \theta,-i  w_0,- i  \theta , e^{-i \frac{2\pi}{3} }  w_0 ,  e^{i \frac{2\pi}{3}} \theta,-e^{i \frac{2\pi}{3} } i  w_0 ,- i e^{-i \frac{2\pi}{3}}  \theta ) \label{eq:U3_eigenstate} \\   u^{(+)}_{\mathbf{Q},\beta 4}(\vec{0}) &= S \sqrt{2}   e^{-i\frac{\pi}{4}}(i \theta ,i  w_0,- e^{-i \frac{2\pi}{3} } \theta,-e^{i \frac{2\pi}{3} }  w_0 ,i e^{i \frac{2\pi}{3} } \theta ,i e^{-i \frac{2\pi}{3} }w_0,- \theta ,-  w_0,i e^{-i \frac{2\pi}{3} } \theta , i e^{i \frac{2\pi}{3} } w_0,- e^{i \frac{2\pi}{3} }\theta , -e^{-i \frac{2\pi}{3} }w_0 ), \label{eq:U4_eigenstate}
\end{align}}where $\theta= 1+\sqrt{1+w_0^2}$. The $\Gamma_1 \oplus \Gamma_2$ eigenstates can be used to analytically compute the $f$-$c$ exchange interaction parameter $J$, as explained in \cref{app:subsec:analytic_exchange_interaction}.

The THF model parameters for the conduction electron Hamiltonian from \cref{eq:conduction_hamiltonian} can be easily identified by comparing the spectra of the THF and BM models near the $\Gamma_M$ point. Firstly, the mass $M$ is the energetic splitting between the conduction electron bands $a=3,4$~\cite{BER21}, which, as we just showed with \cref{eq:U3_eigenstate,eq:U4_eigenstate}, correspond to the active BM model bands. Secondly, within the Hexagon model, we can approximate the splitting between the $\Gamma_1$ and $\Gamma_2$ irreps states from \cref{eq:hexagon_gamma1,eq:hexagon_gamma2}. This entails that $M$ can be approximated as
\begin{equation}
    \label{eq:M_app_approximation_one_shell}
    M^{\textrm{1-shell}} = \left|2w_1 - \sqrt{1 + w_0^2}\right|.
\end{equation}
Using the more accurate TBG bandwidth estimation from the two-band approximation of the Hexagon model in the second magic manifold~\cite{BER21}, a better approximation of the THF mass $M$ can be obtained from \cref{eq:hexagon_2ndshell_bandwigth}
\begin{equation}
    M^{\textrm{2-shell}} = \left|\frac{-4 \sqrt{w_0^2+1} w_1+w_0^2+w_1^2+2}{2 \sqrt{w_0^2+1}}\right|.
    \label{eq:M_approximation_two_shell}
\end{equation}

In the absence of the mass term ({\it i.e.}{} $M = 0$), the dispersion of the conduction band Hamiltonian from \cref{eq:conduction_hamiltonian} becomes linear, with the corresponding Dirac velocity being given by $v_{\star}$~\cite{SON22}. Moreover, the states transforming as the representation $\Gamma_1\oplus\Gamma_2\oplus\Gamma_3$ form a degenerate four-dimensional subspace at zero-energy. The condition $M=0$ is satisfied within the isotropic limit $w_0 = w_1 = \frac{1}{\sqrt{3}}$. Therefore, we can map the $M=0$ THF dispersion to the dispersion of the six-band model approximation of the hexagonal model in the isotropic limit, discussed in \cref{app:subsec:hexagon}. As shown in \cref{fig:A3:b}, along the $\Gamma_M$-$K_M$ path, the band structure of the six-band approximation of the Hexagon model in the isotropic limit is identical to the band structure of the THF model without hybridization ($\gamma=0$) and in the mass-less limit ($M=0$), featuring two degenerate flat bands and two degenerate Dirac cones. By matching the Dirac velocity of the dispersive bands in the THF and BM models, we obtain (see \cref{fig:A3:b})
\begin{equation}
    \label{eq:vstar_approx}
    v_{\star}^{\text{hex}} = \sqrt{\frac{12}{13}}.
\end{equation}
We compare the analytical approximations for the THF parameters $M$ and $v_{\star}$ with the numerical results in \cref{fig:main_sp_params} of the main text.

\subsection{Single-particle coupling parameters from the Hexagon model }\label{app:subsec:analytic_other_hexagon}
As shown in \cref{eq:coupling_hamiltonian_matrix}, for the THF model, the parameter $\gamma$ describes the splitting between the states forming the two $\Gamma_{3\pm}$ irreps (see also \cref{fig:main_1} from the main text). On the other hand, the splitting between the $\Gamma_{3\pm}$ irreps can be obtained within the one-shell approximation of the Hexagon model, with the respective energies being given by \cref{eq:hexagon_gamma3-,,eq:hexagon_gamma3+}. With this, one can approximate the $\gamma$ parameter according to
\begin{equation}
    \label{eq:app_gamma_1shell}
    \gamma^{\textrm{1-shell}} = \frac{1}{2} \left( \sqrt{4+w_0^2} - \sqrt{9w_0^2 + 4w_1^2} \right).
\end{equation}

The $\vec{k}$-linear hybridization term between the $f$- and $c$-electrons that transform as the $\Gamma_{3}$ irrep is parameterized by $v_{\star}'$. As shown in \cref{fig:main_1} from the main text, $v_{\star}'$ is also the Dirac velocity of the two Dirac cones formed by the remote bands. To extract $v_{\star}'$, we can therefore perform a $\mathbf{k}\cdot\mathbf{p}$ expansion of the hexagonal model Hamiltonian $H^{\textrm{hex}}(\mathbf{k},w_0,w_1)$ from \cref{eq:hexagon_hamiltonian} for the $\Gamma_{3}$-irrep states around the $\Gamma_M$ point, as will be described below.

For small $\mathbf{k} = k \textbf{q}_2$ (which, without loss of generality, we have chosen to be the in $\Gamma_M-K_M$ direction along the $\mathbf{q}_2$ vector), the hexagonal model Hamiltonian $H^{\textrm{hex}}(\mathbf{k},w_0,w_1)$ can be expanded as
\begin{equation}
	H^{\textrm{hex}}(k\mathbf{q}_2,w_0,w_1) = H^{\textrm{hex}}\left(\vec{0},w_0,w_1\right) + k V^{\text{hex}}  + \mathcal{O}\left(k^2\right),
\end{equation}
where the perturbation matrix $V^{\text{hex}}$ is given by
\begin{equation}
V^{\text{hex}}  = \frac{\partial H^{\textrm{hex}}\left(k \textbf{q}_2, w_0, w_1\right)}{\partial k} \biggr\rvert_{k=0}.
\end{equation}
We then let $\Phi_{\Gamma_3}^{1,2}$ denote the two (orthonormal) positive-energy eigenstates of $H^{\textrm{hex}}\left(\vec{0},w_0,w_1\right)$ corresponding to the $\Gamma_3$ irrep such that 
\begin{equation}
	H^{\textrm{hex}}\left(\vec{0},w_0,w_1\right) \Phi_{\Gamma_3}^{1,2} = E_{\Gamma_3} \Phi_{\Gamma_3}^{1,2},\qquad \Phi^{i\dagger}_{\Gamma_3}\Phi^{j}_{\Gamma_3} = \delta_{i,j},
	\label{eq:G3_eigenvectors}
\end{equation}
with $E_{\Gamma_3} = \gamma^{(\text{1-shell})}$. To find the dispersion of the positive-energy $\Gamma_3$-irrep states around the $\Gamma_M$ point along the $\Gamma_M-K_M$ direction, we employ first-order perturbation theory to obtain
\begin{equation}
    \Phi_{\Gamma_3}^{\dagger i} H^{\textrm{hex}}(k\mathbf{q}_2,w_0,w_1) \Phi_{\Gamma_3}^{j} = E_{\Gamma_3} \delta_{i,j} + k V_{\Gamma_3,ij}^{\text{hex}} + \mathcal{O}\left(k^2\right),
    \label{eq:projected_perturbation_scheme}
\end{equation}
where $V_{\Gamma_3}^{\text{hex}}$ is the perturbation matrix $V^{\text{hex}}$ projected into the $\Phi_{\Gamma_3}^{1,2}$ subspace
\begin{equation}
	V_{\Gamma_3,ij}^{\text{hex}} =  \Phi_{\Gamma_3}^{i\dagger} V^{\text{hex}}\Phi_{\Gamma_3}^j, \quad \text{for} \quad i,j=1,2.
	\label{eq:vstarprime_small_perturbation_matrix}
\end{equation}
The Dirac velocity of the Dirac cone formed by the $\Gamma_3$-irrep states at energy $E_{\Gamma_3} = \gamma^{(\text{1-shell})}$ (given by $v_{\star}'$ in the THF model) can then be inferred from the projected perturbation matrix $V_{\Gamma_3}^{\text{hex}}$: the eigenvalues of $V_{\Gamma_3}^{\text{hex}}$ are $\pm v_{\star}^{\prime \textrm{approx.}}$, where $v_{\star}^{\prime \textrm{approx.}}$ is an analytical approximation of $v_{\star}'$. 

For general values of the tunneling amplitudes $w_0$ and $w_1$, the eigenvalues of the Hexagon model Hamiltonian at the $\Gamma_M$ point $H^{\textrm{hex}}(\vec{0},w_0,w_1)$ have been obtained analytically~\cite{BER21}. To also derive the corresponding eigenvectors necessary for computing $V_{\Gamma_3}^{\text{hex}}$, we will first employ the $C_{3z}$ and $P$ symmetries (see \cref{app:subsec:BM_symmetries}) to block-diagonalize the $H^{\textrm{hex}}(\vec{0},w_0,w_1)$ Hamiltonian (as will be explained below). 

The $C_{3z}$ and $P$ symmetries of TBG commute ($[C_{3z},P] = 0$)~\cite{BER21a}, and thus admit a common eigenbasis. To find it explicitly, we first note that the representation matrices of $C_{3z}$ and $P$ from \cref{eq:BM_symmetries_C3z,eq:BM_PH_symmetry} expressed in the basis of $H^{\textrm{hex}}(\vec{0},w_0,w_1)$ from \cref{eq:hexagon_hamiltonian} are given by
\begin{align}
    D^{\text{hex}}(C_{3z}) &= \begin{pmatrix}
        0 & 0 & 1 \\
        1 & 0 & 0 \\
        0 & 1 & 0
    \end{pmatrix}
    \otimes
    \begin{pmatrix}
        1 & 0 \\
        0 & 1 
    \end{pmatrix}
     \otimes
    e^{i \frac{2 \pi}{3} \sigma_z}, \\
    D^{\text{hex}}(P) &= 
    \left[
        -\begin{pmatrix}
            0 & 0 & 1 \\
            1 & 0 & 0 \\
            0 & 1 & 0
        \end{pmatrix}
    \otimes
        \begin{pmatrix}
            0 & 0\\
            1 & 0 
        \end{pmatrix}
        +\begin{pmatrix}
            0 & 0 & 1 \\
            1 & 0 & 0 \\
            0 & 1 & 0
        \end{pmatrix}^{t}
    \otimes
        \begin{pmatrix}
            0 & 0\\
            1 & 0 
        \end{pmatrix}^{t}
    \right]
    \otimes
    \sigma_0,
\end{align}
where $\otimes$ denotes the matrix Kronecker product. Defining $\omega = e^{i \frac{2 \pi}{3}}$, a common orthonormal eigenbasis for $D^{\text{hex}}(C_{3z})$ and $D^{\text{hex}}(P)$ can be chosen to be
\begin{equation}
    R = 
    \frac{1}{\sqrt{6}}\left[
        \begin{pmatrix}
            1 & 1 & 1 \\
            1 & \omega & \omega^* \\
            1 & \omega^* & \omega
        \end{pmatrix}
    \otimes
        \begin{pmatrix}
            1 & 1\\
            0 & 0 
        \end{pmatrix}
        +\begin{pmatrix}
            1 & \omega^* & \omega \\
            1 & 1 & 1 \\
            1 & \omega & \omega^*
        \end{pmatrix}
    \otimes
        \begin{pmatrix}
            0 & 0\\
            i & -i 
        \end{pmatrix}
    \right]
    \otimes
    \sigma_0,
\end{equation}
such that
\begin{align}
    R^{\dagger} D^{\text{hex}}(C_{3z}) R &= \begin{pmatrix}
        1 & 0 & 0 \\
        0 & \omega^* & 0 \\
        0 & 0 & \omega
    \end{pmatrix}
    \otimes
    \begin{pmatrix}
        1 & 0 \\
        0 & 1 
    \end{pmatrix}
     \otimes
    e^{i \frac{2 \pi}{3} \sigma_z}, \\
    R^{\dagger} D^{\text{hex}}(P) R &= \begin{pmatrix}
        1 & 0 & 0 \\
        0 & 1 & 0 \\
        0 & 0 & 1
    \end{pmatrix}
    \otimes
    \begin{pmatrix}
        i & 0 \\
        0 & -i 
    \end{pmatrix}
     \otimes
    \sigma_0
\end{align}
are manifestly diagonal. Since $C_{3z}$ is a commuting symmetry of TBG ({\it i.e.}{} $\left[H^{\textrm{hex}}\left(\vec{0},w_0,w_1\right) , D^{\text{hex}}(C_{3z}) \right] = 0$), the Hexagon model Hamiltonian will block-diagonalize according to $C_{3z}$ eigenvalues. We are interested in the positive-energy eigenstates of $H^{\textrm{hex}}\left(\vec{0},w_0,w_1\right)$ that transform according to the two-dimensional $\Gamma_{3}$ irrep (whose $C_{3z}$ eigenvalues are $\omega$ and $\omega^*$). To find $\Phi_{\Gamma_3}^{1,2}$, we can therefore project $H^{\textrm{hex}}\left(\vec{0},w_0,w_1\right)$ into the subspace spanned by the $C_{3z}$ eigenvectors with eigenvalues 
$\omega$ and $\omega^*$. The corresponding projector
\begin{equation}
    \tilde{R}=\frac{1}{\sqrt{6}}\begin{pmatrix}
         1 & 0 & 0 & 1 & 1 & 0 & 0 & 1 \\
         0 & 1 & 1 & 0 & 0 & 1 & 1 & 0 \\
         i & 0 & 0 & i e^{\frac{2 i \pi }{3}} & -i & 0 & 0 & -i e^{\frac{2 i \pi}{3}} \\
         0 & i & i e^{-\frac{2 i \pi }{3}} & 0 & 0 & -i & -i e^{-\frac{2 i \pi }{3}} & 0 \\
         1 & 0 & 0 & e^{-\frac{2 i \pi }{3}} & 1 & 0 & 0 & e^{-\frac{2 i \pi }{3}} \\
         0 & 1 & e^{\frac{2 i \pi }{3}} & 0 & 0 & 1 & e^{\frac{2 i \pi }{3}} & 0 \\
         i & 0 & 0 & i & -i & 0 & 0 & -i \\
         0 & i & i & 0 & 0 & -i & -i & 0 \\
         1 & 0 & 0 & e^{\frac{2 i \pi }{3}} & 1 & 0 & 0 & e^{\frac{2 i \pi }{3}} \\
         0 & 1 & e^{-\frac{2 i \pi }{3}} & 0 & 0 & 1 & e^{-\frac{2 i \pi }{3}} & 0 \\
         i & 0 & 0 & i e^{-\frac{2 i \pi }{3}} & -i & 0 & 0 & -i e^{-\frac{2 i \pi}{3}} \\
         0 & i & i e^{\frac{2 i \pi }{3}} & 0 & 0 & -i & -i e^{\frac{2 i \pi }{3}} & 0 \\
    \end{pmatrix}
\end{equation}
is obtained from the columns of $R$ with indices $i=1,2,6,9,3,4,8,\,\text{and}\,11$ (in this order). In the basis spanned by the columns of $\tilde{R}$, the representation matrices of $C_{3z}$ and $P$ are given respectively by
\begin{align}
    \tilde{R}^{\dagger} D^{\text{hex}}(C_{3z}) \tilde{R} &= 
    \begin{pmatrix}
        1 & 0 & 0 & 0 \\
        0 & 1 & 0 & 0 \\
        0 & 0 & 1 & 0 \\
        0 & 0 & 0 & 1 \\
    \end{pmatrix} \otimes
    \begin{pmatrix}
        \omega & 0 \\
        0 & \omega^* \\
    \end{pmatrix}, \\
    \tilde{R}^{\dagger} D^{\text{hex}}(P) \tilde{R} &= \begin{pmatrix}
        i & 0 \\
        0 & -i \\
    \end{pmatrix} \otimes
    \begin{pmatrix}
        1 & 0 & 0 & 0 \\
        0 & 1 & 0 & 0 \\
        0 & 0 & 1 & 0 \\
        0 & 0 & 0 & 1 \\
    \end{pmatrix}. \label{app:eqn:proj_p_sym}
\end{align}

The particle-hole transformation $P$ is an anticommuting symmetry of TBG~\cite{SON19,BER21a}, with $\left\lbrace H^{\textrm{hex}}\left(\vec{0},w_0,w_1\right) , D^{\text{hex}}(P) \right\rbrace = 0$. In the space spanned by the columns of $\tilde{R}$, this implies that 
\begin{equation}
     \left\lbrace \tilde{R}^{\dagger} H^{\textrm{hex}}\left(\vec{0},w_0,w_1\right) \tilde{R} , \tilde{R}^{\dagger} D^{\text{hex}}(P) \tilde{R} \right\rbrace = 0, \label{app:eqn:proj_p_h_anticomm}
\end{equation}
which, owing to the form of $\tilde{R}^{\dagger} D^{\text{hex}}(P) \tilde{R}$ from \cref{app:eqn:proj_p_sym}, implies that the projected Hexagon model Hamiltonian has an off-diagonal block structure
\begin{equation}
     \tilde{R}^{\dagger} H^{\textrm{hex}}\left(\vec{0},w_0,w_1\right) \tilde{R} = \begin{pNiceMatrix}[last-col=3,first-row]
		4 & 4 \\
		\mathbb{0} & S & 4 \\
		S^{\dagger} & \mathbb{0}  & 4 \\
	\end{pNiceMatrix}.
    \label{app:eqn:off_diag_hex_ham}
\end{equation}
In \cref{app:eqn:off_diag_hex_ham}, the dimensions of each block have been indicated outside the matrix, while the $4 \times 4$-dimensional matrix $S$ is given by
\begin{equation}
    S = \begin{pmatrix}
        -2 i w_0 & 0 & i \left(w_1+1\right) & 0 \\
        0 & -2 i w_0 & 0 & i \left(w_1-1\right) \\
        i \left(w_1-1\right) & 0 & i w_0 & 0 \\
        0 & i \left(w_1+1\right) & 0 & i w_0 \\
    \end{pmatrix}.
    \label{eq:S_block}
\end{equation}
The projected Hexagon model Hamiltonian can therefore be diagonalized through a singular value decomposition of $S$
\begin{equation}
    S = W\Sigma U^{\dagger},
\end{equation}
where $U$, $W$ are unitary $4\times4$ matrices whose columns are the left and right singular eigenvectors of $S$, while $\Sigma$ is a diagonal $4\times4$ matrix with real positive entries. Both $\Sigma$ and the singular eigenvectors can be obtained analytically in terms of the $w_0$ and $w_1$. Their expressions are, however, very cumbersome and will not be given explicitly here. Finally, by conjugating the projected Hexagon model Hamiltonian with the unitary
\begin{equation}
    \mathcal{U} = \frac{1}{\sqrt{2}} \begin{pNiceMatrix}[last-col=3,first-row]
		4 & 4 \\
		W & W & 4 \\
		U & -U  & 4 \\
	\end{pNiceMatrix},
\end{equation}
$\tilde{R}^{\dagger} H^{\textrm{hex}}\left(\vec{0},w_0,w_1\right) \tilde{R}$ can be brought to an explicitly diagonal form
\begin{equation}
    \mathcal{U}^{\dagger}  \tilde{R}^{\dagger} H^{\textrm{hex}}\left(\vec{0},w_0,w_1\right) \tilde{R} \mathcal{U} = \begin{pNiceMatrix}[last-col=3,first-row]
		4 & 4 \\
		\Sigma & \mathbb{0} & 4 \\
		\mathbb{0} & -\Sigma  & 4 \\
	\end{pNiceMatrix}.
\end{equation}
The columns of the $12 \times 8$ matrix $\tilde{R} \mathcal{U}$ are precisely the orthonormal eigenvectors of $H^{\textrm{hex}}\left(\vec{0},w_0,w_1\right)$ transforming according to the $\Gamma_3$ irrep of the little group of $\Gamma_M$. The analytical expressions for the eigenvectors $\Phi_{\Gamma_3}^{1,2}$ in terms of $w_0$ and $w_1$ can be found from the columns of $\tilde{R} \mathcal{U}$ that correspond to the eigenvalue $E_{\Gamma_3} = \gamma^{(\text{1-shell})}$.

With the analytical expressions of the $\Phi_{\Gamma_3}^{1,2}$ eigenvectors at hand (which will not be given explicitly here for the sake of brevity), we can construct and diagonalize the projected perturbation matrix $V_{\Gamma_3}^{\text{hex}}$ defined in \cref{eq:vstarprime_small_perturbation_matrix} and obtain the analytical approximation of the $v_{\star}^{\prime}$ parameter 
\begin{equation}
    \label{eq:approximation_vstarprime}
    v_{\star}^{\prime \textrm{approx.}} = \frac{1}{4}w_0\left(\frac{3}{\sqrt{9w_0^2 + 4w_1^2}} - \frac{1}{\sqrt{4 + w_0^2}}\right) + \frac{2B^2w_0}{\sqrt{(4 B^2 w_0^2 + C_1^2)(4 B^2 w_0^2 + C_2^2)}},
\end{equation}
where we have introduced auxiliary coefficients to simplify the notation
\begin{align}
    B &= 3 + 3w_0^2 + w_1^2 - \sqrt{(4 + w_0^2)(9w_0^2 + 4w_1^2)}, \\
    C_1 &= (1 + w_1)(-4w_1 + \sqrt{(4 + w_0^2)(9w_0^2 + 4w_1^2)}) - w_0^2(9 + w_1), \\
    C_2 &= (-1 + w_1)(4w_1 + \sqrt{(4 + w_0^2)(9w_0^2 + 4w_1^2)}) - w_0^2(-9 + w_1).
\end{align}
It is worth noting that, by direct substitution, $v_{\star}^{\prime \textrm{approx.}} = 0$ for $w_0 = 0$. This is not just an artifact of the present approximation, but is a consequence of a more general property of TBG in the chiral limit. In the chiral limit, the TBG Hamiltonian additionally anticommutes with the chiral symmetry operator $C$, and hence \emph{commutes} with $C P$ ({\it i.e.}{} the product between the chiral symmetry and unitary particle-hole operators). The system thus features a intra-valley ``inversion'' symmetry~\cite{WAN21a}. Since $[C_{3z}, C P] = 0$, one finds that the eigenstates transforming as the $\Gamma_3$ irrep must carry identical ``inversion'' eigenvalues under $C P$. As a consequence, the linear term in the $\vec{k} \cdot \vec{p}$ expansion from \cref{eq:projected_perturbation_scheme} will necessarily vanish as a consequence of the $C P$ symmetry.

We compare the analytical approximations for the THF parameters $\gamma,\;v^{\prime}_{\star}$ of the conduction electrons to the numerical calculations in \cref{fig:main_sp_params} of the main text.

\subsection{Analytic calculation of the renormalized Dirac velocity from the THF model}\label{app:subsec:analytic_dirac_velocity}
In this section, we compute the renormalized Dirac velocity $v_D$ of the Dirac cone formed by the two active bands at the moire $K_M$ and $K'_M$ points for both valleys of the THF model. We start by plugging the expression of the Fourier-transformed $f$-electron operators from \cref{eq:heavy_fermion_operator_fourier} into the single-particle THF Hamiltonian from \cref{eq:single_particle_hamiltonian}
\begin{equation}\label{eq:fourierf}
\hat{H}_{0}=\sum_{|\mathbf{k}| < \Lambda_c}\sum_{a,
a', \eta, s} H_{a,a'}^{\left(c, \eta\right)} \left(\mathbf{k}\right)
\hat{c}^\dagger_{\mathbf{k} a \eta s} \hat{c}_{\mathbf{k} a' \eta s} + 
\frac{1}{\sqrt{N}}\sum_{\mathbf{R}}\sum_{|\textbf{k}|<\Lambda_c} \sum_{\mathbf{k}' \in \text{MBZ}} \sum_{\alpha, a, \eta, s} 
\big( 
	e^{i(\mathbf{k} - \mathbf{k}') \cdot\mathbf{R}}V_{\alpha
a}^{\left(\eta\right)}
\left(\textbf{k}\right)\hat{f}^\dagger_{\mathbf{k}' \alpha \eta
s}\hat{c}_{\textbf{k} a \eta s} + \mathrm{h.c.} \big),
\end{equation}
where the chemical potential has been set to zero ($\mu = 0$) and we have defined the hybridization coefficient as
\begin{equation}
    V_{\alpha a}^{\left(\eta\right)}\left(\mathbf{k}\right) \equiv e^{-|\mathbf{k}|^2 \lambda^2/2} H_{\alpha a}^{\left(cf,\eta\right)}\left(\mathbf{k}\right).
\end{equation} 
Using the identity $\frac{1}{N} \sum_{\mathbf{R}} e^{i(\mathbf{k} - \mathbf{k}')\cdot\mathbf{R}} = \sum_{\vec{G}\in\mathbf{Q}_0} \delta_{\mathbf{k}, \mathbf{k}' - \vec{G}}$, we rewrite \cref{eq:fourierf} as
\begin{equation}
	\hat{H}_{0}=\sum_{\vec{G}\in\mathbf{Q}_0} \sum_{\mathbf{k} \in \text{MBZ}} \sum_{a, \eta, s} \Big[\sum_{a'} H_{a,a'}^{\left(c, \eta\right)} \left(\mathbf{k}-\vec{G} \right)
\hat{c}^\dagger_{\mathbf{k}-\vec{G}, a \eta s} \hat{c}_{\mathbf{k}-\vec{G}, a' \eta s} +
	\sum_{\alpha} 
\big( 
	V_{\alpha
a}^{\left(\eta\right)}
\left(\mathbf{k}-\vec{G}\right)\hat{f}^\dagger_{\mathbf{k} \alpha \eta
	s}\hat{c}_{\mathbf{k}-\vec{G},  a \eta s} + \mathrm{h.c.} \big)\Big].
\label{eq:startingpoint}
\end{equation}

To find $v_D$, we focus on the low-energy physics of $\hat{H}_0$ for momenta $\mathbf{k}$ around the $K_M$ point, such that $\mathbf{k} = -\mathbf{q}_2 + \delta\mathbf{k}$, where $|\delta \mathbf{k}| \ll 1$. Since $V_{\alpha a}^{\left(\eta\right)} \left(\mathbf{k}\right)$ decays exponentially on a scale $1/\lambda \leq |\mathbf{b}_{M1}|$ and the difference in energy between the $\hat{c}^\dagger_{\mathbf{k}-\vec{G},  a \eta s}$ and $\hat{f}^\dagger_{\mathbf{k} \alpha \eta s}$ electrons increases linearly with $\abs{\mathbf{k}-\vec{G}}$, a good approximation is to keep only several $\mathbf{G}$ terms in the summation in \cref{eq:startingpoint}. We choose to keep $\vec{G}_0 = \textbf{0}$, $\vec{G}_1 =  \mathbf{b}_{M2} - \textbf{b}_{M1}$ and $\vec{G}_2 = \mathbf{b}_{M2}$, such that the $C_{3z}$ symmetry of $\hat{H}_{0}$ is preserved (as will be explained below). Letting the eigenstates of the reduced model be
\begin{align}
    \footnotesize
    &\ket{\Psi(\delta \mathbf{k})} = \left(\sum_{\alpha}\psi_{\alpha}^{\left(f\right)}(\delta\mathbf{k})\hat{f}^\dagger_{\mathbf{k}\alpha,\eta s} + \sum_{a}\big\{\psi_{a}^{\left(0c\right)}(\delta \mathbf{k})\hat{c}^\dagger_{\mathbf{k} a,\eta s} + \psi_{a}^{\left(1c\right)}(\delta\mathbf{k})\hat{c}^\dagger_{\mathbf{k}-\vec{G}_1 a,\eta s} + \psi_{a}^{\left(2c\right)}(\delta\mathbf{k})\hat{c}^\dagger_{\mathbf{k} - \vec{G}_2 a,\eta s}\big\}\right)\biggr\rvert_{\mathbf{k}=-\mathbf{q}_2 + \delta\mathbf{k}} \ket{0} \nonumber \\
    &= \left(\sum_{\alpha}\psi_{\alpha}^{\left(f\right)}(\delta\mathbf{k})\hat{f}^\dagger_{\delta \mathbf{k} -\mathbf{q}_2 \alpha,\eta s} + \sum_{a}\big\{\psi_{a}^{\left(0c\right)}(\delta \mathbf{k})\hat{c}^\dagger_{\delta \mathbf{k} -\mathbf{q}_2 a,\eta s} + \psi_{a}^{\left(1c\right)}(\delta\mathbf{k})\hat{c}^\dagger_{\delta \mathbf{k} -\mathbf{q}_1 a,\eta s} + \psi_{a}^{\left(2c\right)}(\delta\mathbf{k})\hat{c}^\dagger_{\delta \mathbf{k} -\mathbf{q}_3 a,\eta s}\big\}\right) \ket{0},
\end{align}
we can rewrite the Schr\"odinger equation $\hat{H}_0\ket{\Psi(\delta\mathbf{k})} = E (\delta\mathbf{k}) \ket{\Psi(\delta \mathbf{k})}$ in the first-quantized formalism  as
\begin{equation}
    H^{(\eta)}\left(\delta \mathbf{k}\right) \Psi \left( \delta \mathbf{k} \right) = E (\delta\mathbf{k}) \Psi \left( \delta \mathbf{k} \right),
    \label{app:eqn:scrodineger_thf}
\end{equation}
with the Hamiltonian matrix
\begin{equation}
	\footnotesize H^{(\eta)}\left(\delta \mathbf{k}\right) = \begin{pmatrix} 0 & V^{ (\eta)}(-\mathbf{q}_2+\delta \mathbf{k})  & V^{(\eta)}(-\mathbf{q}_2 - \vec{G}_1 +\delta \mathbf{k})  & V^{(\eta)}(-\mathbf{q}_2 - \vec{G}_2  +\delta \mathbf{k}) \\ V^{\dagger (\eta)}(-\mathbf{q}_2  +\delta \mathbf{k}) & H^{(c, \eta)}\left(-\mathbf{q}_2+\delta \mathbf{k} \right) & 0 & 0 \\ V^{\dagger (\eta)}(-\mathbf{q}_2 - \vec{G}_1 +\delta \mathbf{k})  & 0 & H^{(c, \eta)}(-\mathbf{q}_2-\vec{G}_1 +\delta \mathbf{k}) & 0 \\ V^{\dagger (\eta)}(-\mathbf{q}_2-\vec{G}_2 + \delta \mathbf{k}) & 0 & 0 & H^{(c, \eta)}(-\mathbf{q}_2 - \vec{G}_2 + \delta \mathbf{k}) \end{pmatrix}
    \label{app:eqn:hamiltonian_THF_tripod}
\end{equation}
acting on the fourteen-dimensional spinor 
\begin{equation}
\Psi \left( \delta\mathbf{k} \right) = \left( \psi^{(f)} \left( \delta\mathbf{k} \right), \psi^{(0c)} \left( \delta\mathbf{k} \right) , \psi^{(1c)} \left( \delta\mathbf{k} \right), \psi^{(2c)} \left( \delta\mathbf{k} \right) \right).
\end{equation}

The Hamiltonian of the reduced model from \cref{app:eqn:hamiltonian_THF_tripod} preserves the $C_{3z}$ symmetry of TBG. To see this explicitly, we first note that the representation matrices of the $C_{3z}$ symmetry from \cref{eq:wannier_symmetries_C3z,eq:conduction_representations_C3z} expressed in the basis of the reduced Hamiltonian $H^{(\eta)}\left(\delta \mathbf{k}\right)$ read as 
\begin{equation}
    \mathcal{D}^{(\eta)} (C_{3z}) = \begin{pmatrix}
        D^{(\eta)}_{f} (C_{3z}) & 0 & 0 & 0 \\
        0 & 0 & 0 & D^{(\eta)}_{c} (C_{3z}) \\
        0 & D^{(\eta)}_{c} (C_{3z}) & 0 & 0 \\
        0 & 0 & D^{(\eta)}_{c} (C_{3z}) & 0 \\
    \end{pmatrix},
\end{equation}
where the $2 \times 2$ and $4 \times 4$ matrix blocks $D^{(\eta)}_{f} (C_{3z})$ and $D^{(\eta)}_{c} (C_{3z})$ are given, respectively, by
\begin{equation}
    D^{(\eta)}_{f} (C_{3z}) = e^{i\eta \frac{2\pi}{3}\sigma_z} 
    \quad \text{and} \quad
    D^{(\eta)}_{c} (C_{3z}) = e^{i\eta\frac{2\pi}{3}\sigma_z} \oplus \sigma_0.
\end{equation}
The Hamiltonian of the reduced model from \cref{app:eqn:hamiltonian_THF_tripod} can then be shown to obey
\begin{equation}
    \mathcal{D}^{(\eta)} (C_{3z}) H^{(\eta)}\left(\delta \mathbf{k}\right) \mathcal{D}^{\dagger (\eta)} (C_{3z}) = H^{(\eta)}\left(C_{3z}\delta \mathbf{k}\right),
\end{equation}
being symmetric under $C_{3z}$ rotations.

To obtain the low-energy spectrum of $H^{(\eta)}\left(\delta \mathbf{k}\right)$ we start by re-writing \cref{app:eqn:scrodineger_thf} in component form
\begin{align}
	&\sum_{i=0}^2 \sum_{a} V^{(\eta)}_{\alpha a}\left(-\mathbf{q}_2 - \vec{G}_i + \delta \mathbf{k}\right) \psi_{a}^{\left(ic\right)} = E \psi_{\alpha}^{\left(f\right)}, \label{app:eqn:scrodineger_thf_1_1}\\
	& \sum_{\alpha}V^{\dagger (\eta)}_{a \alpha}\left(-\mathbf{q}_2 - \vec{G}_i+\delta \mathbf{k}\right) \psi_{\alpha}^{\left(f\right)}+ \sum_{a'} H^{(c, \eta)}_{a a'}\left(-\mathbf{q}_2 - \vec{G}_i + \delta \mathbf{k} \right) \psi_{a'}^{\left(ic\right)} = E \psi_{a}^{\left(ic\right)},\quad \text{for}\quad i=0,1,2, \label{app:eqn:scrodineger_thf_1_2}
\end{align}
where, for brevity, the $\delta \mathbf{k}$ dependence has been made implicit. Rearranging \cref{app:eqn:scrodineger_thf_1_2}, we obtain 
\begin{equation}
\psi_{a}^{\left(ic\right)} = \sum_{a' \alpha} \left[E - H^{(c, \eta)}\left(-\mathbf{q}_2 - \vec{G}_i + \delta \mathbf{k}\right)\right]^{-1}_{a a'} V^{\dagger (\eta)}_{a' \alpha}\left(-\mathbf{q}_2 - \vec{G}_i + \delta \mathbf{k}\right) \psi_{\alpha}^{\left(f\right)}. \label{app:eqn:scrodineger_thf_2}
\end{equation}
Substituting the expression of $\psi_{a}^{\left(ic\right)}$ from \cref{app:eqn:scrodineger_thf_2} into \cref{app:eqn:scrodineger_thf_1_1}, we obtain
\begin{equation}\label{eq:felectronenergy}
	E \psi^{\left(f\right)} = \sum_{i=0,1,2} V^{ (\eta)}\left(-\mathbf{q}_2 - \vec{G}_i + \delta \mathbf{k}\right)\left[E - H^{(c, \eta)}\left(-\mathbf{q}_2 - \vec{G}_i + \delta \mathbf{k}\right)\right]^{-1} V^{\dagger (\eta)}\left(-\mathbf{q}_2 - \vec{G}_i + \delta \mathbf{k}\right) \psi^{\left(f\right)}.
\end{equation}
where the $\alpha$ and $a$ indices have been suppressed. \cref{eq:felectronenergy} is reminiscent of \cref{eq:tripod_psiA01_full} for the Tripod model, and a similar approach can be applied: since $E \rightarrow 0$ for $\delta \mathbf{k} \rightarrow 0$ (as expected from the BM model), we can find the low-energy physics from \cref{eq:felectronenergy} by working to linear order in $\delta \mathbf{k}$ and $E$. Using the identity $\eta^2 = +1$, we can compute the inverse matrix in \cref{eq:felectronenergy} to be
\begin{align}
	\nonumber  &\left[E - H^{(c, \eta)}\left(\mathbf{k}\right)\right]^{-1} = 
	\\ &\frac{1}{v_{\star}^4|\mathbf{k}|^4}
    \resizebox{.8\hsize}{!}{$\begin{pmatrix} (-M^2 -v_{\star}^2 |\mathbf{k}|^2)E \sigma_0 - M v_{\star}^2 \left[(k_y^2 -k_x^2) \sigma_x + 2 k_x k_y \eta \sigma_y\right] & - v_{\star}^3 |\mathbf{k}|^2 (\eta k_x \sigma_0 + i k_y \sigma_z) + M v_{\star}E (\eta k_x \sigma_x - k_y \sigma_y) \\
	- v_{\star}^3 |\mathbf{k}|^2 (\eta k_x \sigma_0 - i k_y \sigma_z) + M v_{\star}E (\eta k_x \sigma_x - k_y \sigma_y) & -v_{\star}^2|\mathbf{k}|^2 E \sigma_0
	\end{pmatrix} $}\label{eq:inverseneglectesquared}.
\end{align}
Plugging \cref{eq:inverseneglectesquared} in \cref{eq:felectronenergy} and expanding to the first order in $\delta \mathbf{k}$ and $E$, we obtain the low-energy dispersion near the $K_M$ point in the valley $\eta = +1$ to be
\begin{equation}
	v_D^{\textrm{(THF)}} \; \left(\sigma_x \delta k_y - \sigma_y \delta k_x\right)  \psi^{\left(f\right)}(\delta \mathbf{k})=   E (\delta \mathbf{k})\psi^{\left(f\right)}(\delta \mathbf{k}),
\end{equation}
where the corresponding Dirac velocity is given by
\begin{equation}
     v_D^{\textrm{(THF)}} = \frac{3 M v_{\star}^2 \left(-2(v_{\star}'^2 + \gamma^2) +  \lambda^2(v_{\star}'-\gamma)(v_{\star}'+\gamma) \right)}{e^{\lambda^2}v_{\star}^4 + 3(M^2 + v_{\star}^2)(v_{\star}'^2 + \gamma^2)}.
	\label{eq:THF_Dirac_velocity_final}
\end{equation}

Repeating the calculation for the $K'_M$ point in the $\eta=+1$ valley results in $-v_D^{\textrm{(THF)}} \; \left(\sigma_x \delta k_y - \sigma_y \delta k_x\right) \psi^{\left(f\right)} = E \psi^{\left(f\right)}$, which has the same chirality as the Dirac cone at $K_M$ in the same valley. The Dirac cone structure of the THF model for the $K_M$ and $K'_M$ points in the $\eta=-1$ valley is $\pm v_D^{\textrm{(THF)}} \; \left( \sigma_x \delta k_y + \sigma_y \delta k_x\right)$, respectively. Note that the chirality of the Dirac cones in the $\eta=-1$ valley is opposite to the Dirac cones in the $\eta=+1$ valley.

Finally, we note that renormalized Dirac velocity of the THF $f$-electron bands can vanish in three different limits (note that we are using the non-dimensional units of \cref{app:eqn:nondim_units}).
\begin{itemize}
    \item $M=0$, which corresponds to the flat band limit~\cite{SON22}.
    \item $\gamma = 0$ and $\lambda = \sqrt{2}$ . In this case, the Dirac velocity of the $f$-electron bands vanishes, but the active TBG bands can still be dispersive due to a nonzero $M$ term.
    \item $v_{\star}' = \pm \gamma \sqrt{\frac{2+\lambda^2}{\lambda^2-2}}$ and $\lambda > \sqrt{2}$. Similarly to the previous case, the vanishing of $v_D^{\textrm{(THF)}}$ does not necessarily imply flat active TBG bands, as the latter can still disperse as a result of a nonzero $M$ term.
\end{itemize}
For the phase space we explore in \cref{app:subsec:additional_numerics_single_particle}, we find that $v_D^{\textrm{(THF)}}$ only vanishes whenever $M = 0$ ({\it i.e.}{}, the other two conditions are never satisfied). 

\section{THF interaction strength parameters}\label{app:sec:analytic_many-body}
In this appendix, we provide detailed calculations of the interaction strength parameters $U_1$, $W_{1,3}$, and $V$ and briefly discuss a strategy for obtaining an analytical approximation of the $J$ parameter. The physical meaning of these parameters was outlined in \cref{tab:Coulomb_summary}. For each of the parameters, we use the general expressions reviewed in \cref{app:sec:HF_interaction} and derive analytical approximations. We start with the $f$-$f$ density-density interaction in \cref{app:subsec:analytic_interaction_ff} and compute the $U_1$ parameter. In \cref{app:subsec:analytic_interaction_fc}, we derive an analytical approximation of the $f$-$c$ density-density interaction strength parameter $W_{1,3}$. We then calculate the $c$-$c$ density-density interaction parameter $V$ in \cref{app:subsec:analytic_interaction_cc} and, finally, we outline the calculation for the $f$-$c$ exchange interaction parameter $J$ in \cref{app:subsec:analytic_exchange_interaction}.

\subsection{Analytical calculation of the $f$-$f$ density-density interaction strength}\label{app:subsec:analytic_interaction_ff}

In this section, we compute the $f$-$f$ density-density interaction strength parameter (first term in \cref{tab:Coulomb_summary}). We start from \cref{eq:interaction_ff_formula} obtained in \cref{app:subsec:review_ff_interaction}. The $f$-orbital density can be calculated from the Gaussian analytic expressions of the Wannier states from  \cref{eq:wannier_states_I,,eq:wannier_states_II}
\begin{equation}
    n_f(\mathbf{r})= \sum_{l,\beta}|w_{l\beta,\alpha}^{(\eta)}(\mathbf{r})|^2 = \frac{1}{\pi}\left(\frac{\alpha_1^2}{\lambda_1^2}e^{-\frac{r^2}{\lambda_1^2}} + \frac{\alpha_2^2 r^2}{\lambda_2^4}e^{-\frac{r^2}{\lambda_2^2}}  \right).
\end{equation}
With the definition of the Fourier transformation of the density $n_f(\mathbf{r})$ from \cref{eq:interaction_density_fourier_transform}, we obtain the density function in the momentum space
\begin{align}
    n_f(\mathbf{q}) &= \int \dd^2{\mathbf{r}} e^{i\mathbf{q}\cdot\mathbf{r}}n_f(\mathbf{r}) = 2\pi \int_0^{+\infty}n_f(r)rJ_0(qr)\dd{r} = \alpha_1^2  e^{-\frac{ q^2\lambda_1^2}{4}} + \alpha_2^2 \left(1 - \frac{q^2 \lambda_2^2}{4} \right) e^{- \frac{q^2\lambda_2^2}{4}},
    \label{eq:wannier_density_full}
\end{align}
where $q=|\mathbf{q}|$ and $J_0(z)$ is the Bessel function of the first kind. We can use the approximation $\lambda_1 \approx \lambda_2$ from the \cref{app:subsec:analytic_forbitals_tripod} and the normalization condition $\alpha_1^2 + \alpha_2^2 = 1$, which gives
\begin{equation}
    n_f(\mathbf{q}) \approx \left(1 - \alpha_2^2\frac{q^2\lambda_1^2}{4}\right)e^{-\frac{q^2\lambda_1^2}{4}}.
    \label{eq:wannier_density_approx}
\end{equation}

With the help of \cref{eq:Coulomb_interaction_fourier}, we can compute the onsite interaction strength $U(\mathbf{R})$ from \cref{eq:interaction_ff_formula} at the zeroth site $\mathbf{R} = 0$, which we dubbed as $U_1$ in \cref{app:subsec:review_ff_interaction}
\begin{equation}
    U_1 = \frac{1}{N\Omega_0}\sum_{\mathbf{q} \in \textrm{MBZ}} \sum_{\mathbf{G} \in \mathcal{Q}_0} V(\mathbf{q} + \mathbf{G})n_f(\mathbf{q} + \mathbf{G})n_f(-\mathbf{q} - \mathbf{G}) = \frac{1}{N\Omega_0}\int_{\substack{\textrm{entire} \\ \textrm{space}}} \frac{\Omega_{\textrm{tot}}\dd^2{\mathbf{q}}}{(2\pi)^2}V(\mathbf{q})n_f^2(\mathbf{q}),
    \label{eq:ff_interaction_U0}
\end{equation}
where we have used
\begin{equation}
    \sum_{\mathbf{q} \in \textrm{MBZ}} \sum_{\mathbf{G} \in \mathcal{Q}_0} \rightarrow \int_{\substack{\textrm{entire} \\ \textrm{space}}} \frac{\Omega_{\textrm{tot}}}{(2\pi)^2}\dd^2{\mathbf{q}},\qquad \Omega_{\textrm{tot}} = N\Omega_0.
\end{equation}
Plugging the expressions of the Fourier transformed Wannier state density from \cref{eq:interaction_density_fourier_transform} and of the Coulomb interaction potential from \cref{eq:Coulomb_interaction_fourier}, we obtain an integral that can be evaluated analytically
\begin{equation}
    U_1 = \xi U_{\xi}\int_0^{+\infty}\dd{q} \tanh{\left(\frac{\xi q}{2}\right)}n_f^2(q).
    \label{eq:ff_interaction_U0_integral}
\end{equation}
Expanding the hyperbolic tangent
\begin{equation}
    \tanh{x} = 1 + 2\sum_{k=1}^{\infty}(-1)^k e^{-2kx},
\end{equation}
and plugging it together with an approximated density expression from \cref{eq:wannier_density_approx} into \cref{eq:ff_interaction_U0_integral}, we obtain
\begin{equation}
    U_1 = \frac{\xi U_{\xi}}{\lambda_1}\left[\int_0^{+\infty}\dd{x}e^{-\frac{x^2}{2}}\left(1 - \alpha_2^2\frac{x^2}{4}\right)^2 + 2 \sum_{k=1}^{\infty}(-1)^k\int_0^{+\infty}\dd{x}e^{-\frac{x^2}{2}-a(k)x}\left(1 - \alpha_2^2\frac{x^2}{4}\right)^2\right]
    \label{eq:interaction_integral_series}
\end{equation}
where we substituted $a(k) = \frac{k\xi}{\lambda_1}$ and $x = q \lambda_1$. Evaluating each integral in \cref{eq:interaction_integral_series} and taking the limit $a(k)\rightarrow +\infty$ for \textit{each} $k$ (an approximation which will be justified below), we find the approximation
\begin{equation}
    U_1 \approx \frac{\xi U_{\xi}}{\lambda_1}\left[\frac{41}{48}\sqrt{\frac{\pi}{2}} - 2\sum_{k=1}^{\infty}(-1)^k\left(\frac{1}{a(k)} - \frac{1 + \alpha^2_2}{a(k)^3}\right)\right].
    \label{eq:U1_before_taking_limit_ainf}
\end{equation}
Taking the limit $a(k) \rightarrow \infty$ for each $k$ is justified by analyzing the order of magnitude for the quantities involved. From the numerical calculations~\cite{SON22} (see also \cref{app:sec:tables}) we can assume $\lambda_1 \sim 0.185 a_M = \SI{2.37}{\nano \meter}$, where $a_M = \frac{4\pi}{\sqrt{3}G_1} = \SI{13}{\nano \meter}$. We can estimate by the order of magnitude $\xi \sim \SI{10}{\nano \meter}$. Therefore $a(k) = \frac{k\xi}{\lambda_1} \sim 3 k$. This implies that in the sum over $k$ in \cref{eq:U1_before_taking_limit_ainf}, the neglected term of the order $\mathcal{O}(\frac{1}{k^5})$ is at least $\sim 10k^2$ times smaller than the terms that we kept. Generally, we expect the approximation to work better for larger $\xi$ and does not give a good agreement for small $\xi$, which we indeed observe numerically in \cref{app:sec:numerics}. Plugging $a(k)$ back and evaluating the series in $k$, we find
\begin{equation}
    \label{eq:U1_final}
    U_1^{\textrm{approx.}} = \frac{\xi U_{\xi}}{\lambda_1}\left[\frac{41}{48}\sqrt{\frac{\pi}{2}} - 2 \frac{\lambda_1}{\xi}\ln{2} + 2(1 + \alpha_2^2) \left(\frac{\lambda_1}{\xi}\right)^3\frac{3}{4}\zeta(3)\right],
\end{equation}
where $\zeta(x)$ is the Riemann zeta function and $\zeta(3) \approx 1.2$. In practice, we use the approximations for the spread $\lambda_1$ and the normalization factors ratio $\alpha_1/\alpha_2$ given by \cref{eq:lambda1_alpha_ratio_1shell_approximation} together with the normalization condition $\alpha_1^2 + \alpha_2^2 = 1$. The comparison with the numerical result was given in \cref{fig:main_int_params} of the main text.

\subsection{Analytical calculation of the $f$-$c$ density-density interaction }\label{app:subsec:analytic_interaction_fc}
In this section, we analytically compute the $f$-$c$ density-density interaction strength parameter ({\it i.e.}{}, the fifth term in \cref{tab:Coulomb_summary}). As discussed in \cref{app:subsec:review_fc_interaction}, this interaction term is described by two parameters $W_1$ and $W_3$, which are the diagonal entries of the matrix $X^{\eta}_{aa'}$ introduced in \cref{eq:interaction_fc_density_Xintegral}. From \cref{eq:interaction_fc_density_Xintegral}, we have that
\begin{equation}
    X^{\eta}_{aa'}(\vec{0},\vec{0}) = \frac{1}{\Omega_0 } \sum_{l,\beta} \sum_{\mathbf{Q},\mathbf{Q}'\in \mathcal{Q}_{l\eta_2}} n_f(\mathbf{Q}-\mathbf{Q}') V(\mathbf{Q}-\mathbf{Q}')   \tilde{u}^{(\eta_2)*}_{\mathbf{Q}\beta,a}(\vec{0}) \tilde{u}^{(\eta_2)}_{\mathbf{Q}'\beta,a'}(\vec{0}),
\end{equation}
which we can rewrite by noting that $\mathbf{Q}-\mathbf{Q}' = \mathbf{G} \in \mathcal{Q}_0$
\begin{equation}
    X^{\eta}_{aa'}(\vec{0},\vec{0}) = \frac{1}{\Omega_0} \sum_{l,\beta} \sum_{\mathbf{Q}\in \mathcal{Q}_{l\eta}} \sum_{\mathbf{G}} n_f(\mathbf{G}) V(\mathbf{G})   \tilde{u}^{(\eta)*}_{\mathbf{Q}\beta,a}(\vec{0}) \tilde{u}^{(\eta)}_{\mathbf{Q}+ \mathbf{G} \beta,a'}(\vec{0}).
\end{equation}
We can explicitly split the sum into the terms with $\mathbf{G} = \vec{0}$ and $\mathbf{G} \neq \vec{0}$
\begin{align}
    X^{\eta}_{aa'}(\vec{0},\vec{0}) &=  \frac{1}{\Omega_0 } \sum_{l,\beta} \sum_{\mathbf{Q}\in \mathcal{Q}_{l\eta}}  n_f(\vec{0}) V(\vec{0})   \tilde{u}^{(\eta)*}_{\mathbf{Q}\beta,a}(\vec{0}) \tilde{u}^{(\eta)}_{\mathbf{Q} \beta,a'}(\vec{0})\ +  \frac1{ \Omega_0 } \sum_{l,\beta} \sum_{\mathbf{Q}\in \mathcal{Q}_{l\eta}} \sum_{\mathbf{G} \ne 0} n_f(\mathbf{G}) V(\mathbf{G})   \tilde{u}^{(\eta)*}_{\mathbf{Q}\beta,a}(\vec{0}) \tilde{u}^{(\eta)}_{\mathbf{Q} + \mathbf{G} \beta,a'}(\vec{0})\ \nonumber  \\
    &= \frac{2\pi}{\sqrt{3}} \left(  \frac{\xi}{a_M} \right)^2 U_\xi \delta_{a,a'} +  \frac{1}{ \Omega_0 } \sum_{l,\beta} \sum_{\mathbf{Q}\in \mathcal{Q}_{l\eta}} \sum_{\mathbf{G} \ne 0} n_f(\mathbf{G}) V(\mathbf{G})   \tilde{u}^{(\eta)*}_{\mathbf{Q}\beta,a}(\vec{0}) \tilde{u}^{(\eta)}_{\mathbf{Q}+ \mathbf{G} \beta,a'}(\vec{0}),
    \label{eq:interaction_fc_Xintegral_final}
\end{align}
where $a_M = \frac{2\pi}{3}(\sqrt{3},1)$ is a real-space moir\'e lattice vector. In \cref{eq:interaction_fc_Xintegral_final}, we have used the fact that $\Omega_0 = (2\pi)^2/\Omega_{\textrm{BZ}}$ and $\Omega_{\textrm{BZ}} = \sqrt{3}|\mathbf{b}_{M1}|^2/2$. Given the exponential decay in $\vec{\mathbf{G}}$ of the wave function $\tilde{u}_{\mathbf{Q}+\mathbf{G}\beta,a}^{(\eta)}(\vec{0})$, we can truncate the expression at the first term and obtain
\begin{equation}
    W_1^{\textrm{1st approx.}} = W_3 = W \equiv \frac{2\pi}{\sqrt{3}} \left( \frac{\xi}{a_M} \right)^2 U_{\xi}.
    \label{eq:fc_interaction_strength_approx_13}
\end{equation}
Numerically, we find that the $W_1$ and $W_3$ interaction strength parameters are slightly different (see \cref{subsec:numerical_many_body}). In order to capture this difference analytically, for the $W_1$ parameter, we will truncate \cref{eq:interaction_fc_Xintegral_final} at the smallest $|\mathbf{G}| = |\mathbf{b}_{M1}|$ for the conduction electron states forming the $\Gamma_1\oplus\Gamma_2$ representations. In the hexagon approximation, described in \cref{app:subsec:analytic_cbands_hexagon}, we can write from the $\mathbf{G} \neq \vec{0}$ terms of the expansion in \cref{eq:interaction_fc_Xintegral_final}
{\small
\begin{align}
    &\frac{1}{\Omega_0 } \sum_{l,\beta} \sum_{\mathbf{Q}\in \mathcal{Q}_{l\eta}} \sum_{\mathbf{G} \ne \vec{0}} n_f(\mathbf{G}) V(\mathbf{G})   \tilde{u}^{(\eta)*}_{\mathbf{Q}\beta,a}(\vec{0}) \tilde{u}^{(\eta)}_{\mathbf{Q}+ \mathbf{G} \beta,a'}(\vec{0}) =  n_f(\mathbf{b}_{M1}) V(\mathbf{b}_{M1}) \frac{1}{N \Omega_0 } \sum_{l,\beta} \sum_{\mathbf{Q}\in \mathcal{Q}_{l\eta}} \sum_{|\mathbf{G}| = |\mathbf{b}_{M1}|} \tilde{u}^{(\eta)*}_{\mathbf{Q}\beta,a}(\vec{0}) \tilde{u}^{(\eta)}_{\mathbf{Q}+ \mathbf{G} \beta,a'}(\vec{0})   \nonumber \\
    =& n_f(\mathbf{b}_{M1}) V(\mathbf{b}_{M1}) \frac{\delta_{a,a'}}{\Omega_0 } \sum_{\beta} [\tilde{u}_{\mathbf{q}_1\beta,a}^{(\eta)*}\tilde{u}_{\mathbf{q}_2\beta,a}^{(\eta)} + \tilde{u}_{-\mathbf{q}_3\beta,a}^{(\eta)*}\tilde{u}_{-\mathbf{q}_1\beta,a}^{(\eta)} + \tilde{u}_{\mathbf{q}_2\beta,a}^{(\eta)*}\tilde{u}_{\mathbf{q}_3\beta,a}^{(\eta)} + \tilde{u}_{-\mathbf{q}_1\beta,a}^{(\eta)*}\tilde{u}_{-\mathbf{q}_2\beta,a}^{(\eta)} +  
    \tilde{u}_{\mathbf{q}_3\beta,a}^{(\eta)*}\tilde{u}_{\mathbf{q}_1\beta,a}^{(\eta)} + \mathrm{h.c.}],
\end{align}}The sum in the parenthesis can be evaluated from the symmetry considerations. We obtain

\begin{align}
    & \sum_{\beta}\left[\tilde{u}_{\mathbf{q}_1\beta,a}^{(\eta)*}\tilde{u}_{\mathbf{q}_2\beta,a}^{(\eta)} + \tilde{u}_{-\mathbf{q}_3\beta,a}^{(\eta)*}\tilde{u}_{-\mathbf{q}_1\beta,a}^{(\eta)} + \tilde{u}_{\mathbf{q}_2\beta,a}^{(\eta)*}\tilde{u}_{\mathbf{q}_3\beta,a}^{(\eta)} + \tilde{u}_{-\mathbf{q}_1\beta,a}^{(\eta)*}\tilde{u}_{-\mathbf{q}_2\beta,a}^{(\eta)} +  
     \tilde{u}_{\mathbf{q}_3\beta,a}^{(\eta)*}\tilde{u}_{\mathbf{q}_1\beta,a}^{(\eta)} + \mathrm{h.c.}\right]  \nonumber \\
    =&  \sum_{\beta}\sum_{l=\pm} \left[\tilde{u}_{l\mathbf{q}_1\beta,a}^{(\eta)*}(\tilde{u}_{l\mathbf{q}_2\beta,a}^{(\eta)} + \tilde{u}_{l\mathbf{q}_3\beta,a}^{(\eta)}) +  \tilde{u}_{l\mathbf{q}_2\beta,a}^{(\eta)*}(\tilde{u}_{l\mathbf{q}_1\beta,a}^{(\eta)} + \tilde{u}_{l\mathbf{q}_3\beta,a}^{(\eta)}) + 
    \tilde{u}_{l\mathbf{q}_3\beta,a}^{(\eta)*}(\tilde{u}_{l\mathbf{q}_1\beta,a}^{(\eta)} + \tilde{u}_{l\mathbf{q}_2\beta,a}^{(\eta)})\right]  \nonumber \\
    =& - \sum_{p=1,2,3}\sum_{\beta}\sum_{l=\pm} \tilde{u}_{l\mathbf{q}_p\beta,a}^{(\eta)*} \tilde{u}_{l\mathbf{q}_p\beta,a}^{(\eta)} = -1,
\end{align}
where we used \cref{eq:conduction_C3z} and, consequently, the relation
\begin{equation}
    \tilde{u}_{\pm\mathbf{q}_1\beta,a}^{(\eta)} + \tilde{u}_{\pm\mathbf{q}_2\beta,a}^{(\eta)} + \tilde{u}_{\pm\mathbf{q}_3\beta,a}^{(\eta)} = 0,
\end{equation}
as well as the normalization condition in \cref{eq:conduction_normalization}. Thus, the interaction strength parameter $W_3$ reads
\begin{equation}
    \label{eq:W1_W3_1st}
    W_1^{\textrm{2nd approx.}} = \frac{2\pi}{\sqrt{3}} \left( \frac{\xi}{a_M} \right)^2 U_{\xi} - \frac{1}{\Omega_0}n_f(\mathbf{b}_{M1})V(\mathbf{b}_{M1}),
\end{equation}
where the expression for $n_f(\mathbf{G})$ is given by \cref{eq:wannier_density_approx}. We note that the parameter $W_1$ in the second approximation is slightly smaller than $W_3$, which gives a good approximation of the numerical simulations, see main text and \cref{app:sec:numerics,app:subsec:additional_numerics_many-body}.

\subsection{Analytical calculation of the $c$-$c$ density-density interaction }\label{app:subsec:analytic_interaction_cc}
From \cref{app:subsec:review_cc_interaction}, we recall the interaction matrix $X_{\eta_1 a_1 a_1', \eta_2 a_2 a_2'}(\mathbf{k}_1,\mathbf{k}_2; \mathbf{q})$, which is given by \cref{eq:interaction_cc_Xmatrix_final}. Ref.~\cite{SON22} has shown numerically that only the $\mathbf{G}=\vec{0}$  term dominates in the summation from \cref{eq:interaction_cc_Xmatrix_final}. Therefore, a good analytic approximation would be to keep only the $\mathbf{G}=\vec{0}$ term and write the interaction term as
\begin{equation} \label{eq:HV-explicit}
\hat{H}_{V} \approx \frac{1}{2\Omega_0 N} \sum_{\eta_1, s_1, a_1} \sum_{\eta_2, s_2, a_2} 
    \sum_{|\mathbf{k}_1|,|\mathbf{k}_2|<\Lambda_c} \sum_{\substack{\mathbf{q}\\ |\mathbf{k}_1+\mathbf{q}|, |\mathbf{k}_2+\mathbf{q}|<\Lambda_c } } 
    V(\mathbf{q}) 
    :\hat{c}^\dagger_{\mathbf{k}_1 a_1\eta_1 s_1} \hat{c}_{\mathbf{k}_1+\mathbf{q} a_1 \eta_1 s_1}:
    :\hat{c}^\dagger_{\mathbf{k}_2+\mathbf{q} a_2 \eta_2 s_2} \hat{c}_{\mathbf{k}_2 a_2 \eta_2 s_2}:\ ,
\end{equation}
where according to \cref{eq:Coulomb_interaction_fourier}
\begin{equation}
    \frac{V(\mathbf{q})}{V(\vec{0})} = \frac{\tanh{\xi q/2}}{\xi q/2}.
    \label{eq:Coulomb_fourier_formula_to_etimate}
\end{equation}
Since we are interested in low-energy physics, the range of $\mathbf{q}$ is small and can be approximated as $|\mathbf{q}| \lesssim 0.1|\mathbf{b}_{M1}|$. Given this range, we can estimate the deviation of $V(\mathbf{q})$ from $V(\vec{0})$. Plugging the values in \cref{eq:Coulomb_fourier_formula_to_etimate}, we find $V(0.1|\mathbf{b}_{M1}|)/V(\vec{0}) \sim 0.97$ and hence we can further approximate \cref{eq:HV-explicit}

\begin{equation} \label{eq:HV-explicit2}
\hat{H}_{V} \approx \frac{1}{2\Omega_0 N} V(\vec{0}) \sum_{\eta_1, s_1, a_1} \sum_{\eta_2, s_2, a_2} 
    \sum_{|\mathbf{k}_1|,|\mathbf{k}_2|<\Lambda_c} \sum_{\substack{\mathbf{q}\\ |\mathbf{k}_1+\mathbf{q}|, |\mathbf{k}_2+\mathbf{q}|<\Lambda_c } } 
    :\hat{c}^\dagger_{\mathbf{k}_1 a_1\eta_1 s_1} \hat{c}_{\mathbf{k}_1+\mathbf{q} a_1 \eta_1 s_1}:
    :\hat{c}^\dagger_{\mathbf{k}_2+\mathbf{q} a_2 \eta_2 s_2} \hat{c}_{\mathbf{k}_2 a_2 \eta_2 s_2}:\ .
\end{equation}
In this way, comparing to \cref{eq:interaction_fc_Xintegral_final}, we find
\begin{equation}
    \frac{1}{\Omega_0}V(\vec{0}) \approx W_1 \approx W_3 \approx W = \frac{2\pi}{\sqrt{3}} \left( \frac{\xi}{a_M} \right)^2 U_{\xi},
    \label{eq:interaction_cc_approximate_equality}
\end{equation}
where the second approximation becomes an equality if we truncate the calculation for $W_3$ at the same term as $W_1$ (see \cref{app:subsec:analytic_interaction_fc}). The approximate equality of the interaction strengths in \cref{eq:interaction_cc_approximate_equality} leads to the emergence of higher symmetries, which will be discussed in \cref{app:sec:symmetries}.

\subsection{Analytic calculation of the $f$-$c$ exchange interaction}\label{app:subsec:analytic_exchange_interaction}
In this section, we outline a calculation for the $f$-$c$ exchange interaction parameter $J$, which per \cref{app:subsec:review_fc_exchange_interaction} is given by the matrix element in the \cref{eq:exchange_matrix_approximate}
{\small\begin{align}
    J = &{\mathcal{J}}_{\eta 1 3,\eta 1 3} = \int \frac{\dd^2{\mathbf{q}}}{(2\pi)^2} \ V(\mathbf{q}) 
\bra{v^{(\eta)}_{1}(-\mathbf{q})} \ket{\tilde{u}^{(\eta)}_{3}(\vec{0})}
\bra{\tilde{u}^{(\eta)}_{3}(\vec{0})} \ket{v^{(\eta)}_{1}(-\mathbf{q})} = \int \frac{\dd^2{\mathbf{q}}}{(2\pi)^2} \ V(\mathbf{q}) 
\left|\sum_{\mathbf{Q},\beta}v^{(+)*}_{\mathbf{Q}\beta,1}(-\mathbf{q})\tilde{u}^{(+)}_{\mathbf{Q}\beta,3}(\vec{0})\right|^2
\label{eq:analytic_J_formula}
\end{align}}where $V(\mathbf{q})$ is given by \cref{eq:Coulomb_interaction_fourier} and we have set $\eta=+$ without loss of generality. The momentum-space wave functions of the $f$-electrons were obtained in \cref{eq:vQ-1,eq:vQ-2}. On the other hand, the analytical expression of the $c$-electron wave function $\tilde{u}^{(+)}_{\mathbf{Q}\beta,3}(\vec{0})$ was derived in \cref{eq:U3_eigenstate}. By plugging \cref{eq:vQ-1,eq:vQ-2,eq:U3_eigenstate} into \cref{eq:analytic_J_formula}, one can obtain the analytic expression for the $f$-$c$ exchange interaction parameter $J$. The resulting formula however is cumbersome and not particularly illuminating, and is beyond the scope of the present work.

\section{Analytical approximations to the density form factors}\label{app:sec:form_factors}

As reviewed in \cref{app:sec:HF_interaction}, the THF interaction Hamiltonian is obtained by projecting the Coulomb density-density interaction of TBG into the THF electron bands. In this appendix, we also derive an expression for the TBG electron density operator in terms of the $f$- and $c$-electron operators via the corresponding form factors (which we introduce below). By approximating \emph{both} the $f$- and the $c$-electron wave functions with appropriately-chosen Gaussian profiles, we also obtain analytical expressions for the THF form factors.

\subsection{Gaussian wave functions of \texorpdfstring{$f$- and $c$-}{f- and c-}electrons} 

\begin{figure}[t]
    \centering
    \includegraphics[width=0.9\linewidth]{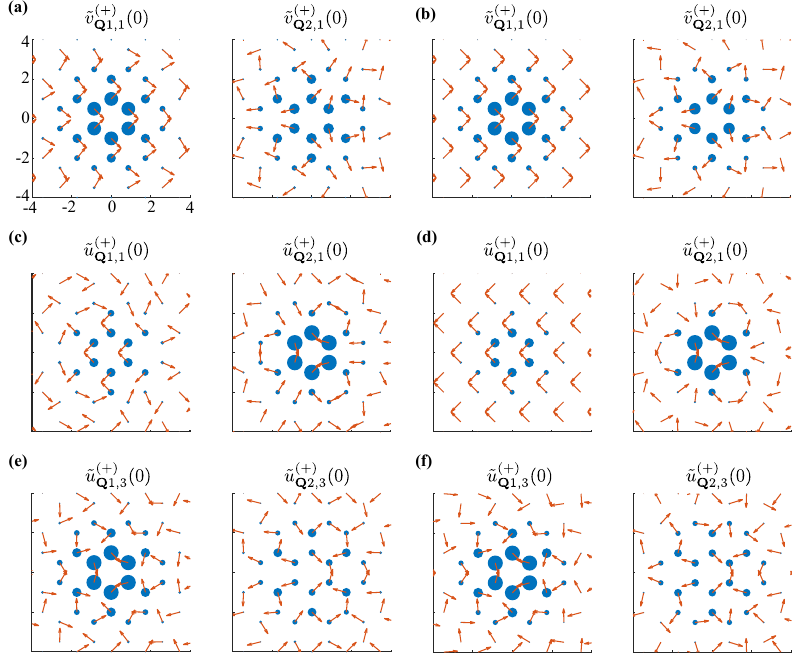}
    \subfloat{\label{fig:wave_fc:a}}\subfloat{\label{fig:wave_fc:b}}\subfloat{\label{fig:wave_fc:c}}\subfloat{\label{fig:wave_fc:d}}\subfloat{\label{fig:wave_fc:e}}\subfloat{\label{fig:wave_fc:f}}\caption{The numerically-obtained wave functions of the THF model and their Gaussian approximations. We employ $w_0/w_1=0.8$ and $\theta=\SI{1.05}{\degree}$. 
    The $f$-electron wave function $\widetilde{v}_{\mathbf{Q} \beta,\alpha=1}^{(+)}(0)$ is shown in (a) and (b), while the $c$-electron wave functions $\widetilde{u}_{\mathbf{Q} \beta,a=1}^{(+)}(0)$
    and $\widetilde{u}_{\mathbf{Q} \beta,a=3}^{(+)}(0)$ are shown, respectively, in (c) and (d), as well as (e) and (f). The numerically-obtained wave functions are plotted in (a), (c), and (e), with the corresponding Gaussian approximations shown in (b), (d), and (f).}
\label{fig:wave_fc}
\end{figure}

We use the analytical Gaussian wave functions from \cref{eq:wannier_states_I,eq:wannier_states_II} to approximate the Wannier functions of the $f$-electrons. As shown in \cref{fig:wave_fc:a,fig:wave_fc:b}, the momentum space Gaussian wave functions from \cref{eq:vQ-1,eq:vQ-2} agree with the wave functions obtained numerically from the continuum BM model. 

We also introduce the following analytical expressions for the $c$-electron wave functions at $\mathbf{k}=0$
{\footnotesize\begin{alignat}{4}
    &\widetilde{u}_{\mathbf{Q} 1, 1}^{(\eta)}(0)
&&= - \alpha_{c1} \sqrt{ \frac{2\pi \lambda_{c1}^2}{\Omega_M \mathcal{N}_{c1} } } e^{- i \frac{\pi}4 \zeta_\mathbf{Q} - \frac12 \mathbf{Q}^2 \lambda_{c1}^2 } ,\qquad &&
    \widetilde{u}_{\mathbf{Q} 2, 1}^{(\eta)}(0)
&&= \alpha_{c2} \sqrt{ \frac{\pi \lambda_{c2}^6 }{\Omega_M \mathcal{N}_{c1}} } (i\eta Q_x + Q_y)^2 e^{- i \frac{\pi}4 \zeta_\mathbf{Q} - \frac12 \mathbf{Q}^2 \lambda_{c2}^2 }, \label{app:eqn:anal_func_c_1} \\
    &\widetilde{u}_{\mathbf{Q} 1, 2}^{(\eta)}(0)
&&= \alpha_{c2} \sqrt{ \frac{\pi \lambda_{c2}^6 }{\Omega_M \mathcal{N}_{c2}} } (-i\eta Q_x + Q_y)^2 e^{i \frac{\pi}4 \zeta_\mathbf{Q} - \frac12 \mathbf{Q}^2 \lambda_{c2}^2 } ,\qquad &&
    \widetilde{u}_{\mathbf{Q} 2, 2}^{(\eta)}(0)
&&= - \alpha_{c1} \sqrt{ \frac{2\pi \lambda_{c1}^2}{\Omega_M \mathcal{N}_{c2}} } e^{i \frac{\pi}4 \zeta_\mathbf{Q} - \frac12 \mathbf{Q}^2 \lambda_{c1}^2 }, \label{app:eqn:anal_func_c_2} \\
    &\widetilde{u}_{\mathbf{Q} 1, 3}^{(\eta)}(0)
&&= \alpha_{c3} \sqrt{ \frac{2\pi \lambda_{c3}^4 }{\Omega_M \mathcal{N}_{c3}} } \zeta_\mathbf{Q} (-i \eta Q_x + Q_y) e^{ - i \frac{\pi}4 \zeta_\mathbf{Q} - \frac12 \mathbf{Q}^2 \lambda_{c3}^2 },\qquad &&
    \widetilde{u}_{\mathbf{Q} 2, 3}^{(\eta)}(0)
&&= \alpha_{c4} \sqrt{ \frac{\pi \lambda_{c4}^6 }{\Omega_M \mathcal{N}_{c3}} } (-i\eta Q_x +  Q_y)^2 e^{ - i \frac{\pi}4 \zeta_\mathbf{Q} - \frac12 \mathbf{Q}^2 \lambda_{c4}^2 }, \label{app:eqn:anal_func_c_3}\\
    &\widetilde{u}_{\mathbf{Q} 1, 4}^{(\eta)}(0) &&= \alpha_{c4} \sqrt{ \frac{\pi \lambda_{c4}^6 }{\Omega_M \mathcal{N}_{c4}} } (i\eta Q_x +  Q_y)^2 e^{ i \frac{\pi}4 \zeta_\mathbf{Q} - \frac12 \mathbf{Q}^2 \lambda_{c4}^2 },\qquad &&
    \widetilde{u}_{\mathbf{Q} 2, 4}^{(\eta)}(0)
&&= \alpha_{c3} \sqrt{ \frac{2\pi \lambda_{c3}^4 }{\Omega_M \mathcal{N}_{c4}} } \zeta_\mathbf{Q} (i \eta Q_x + Q_y) e^{ i \frac{\pi}4 \zeta_\mathbf{Q} - \frac12 \mathbf{Q}^2 \lambda_{c3}^2 }. \label{app:eqn:anal_func_c_4}
\end{alignat}}where $\mathcal{N}_{c1}$, $\mathcal{N}_{c2}$, $\mathcal{N}_{c3}$, $\mathcal{N}_{c4}$ are normalization factors. \Cref{app:eqn:anal_func_c_1,app:eqn:anal_func_c_2,app:eqn:anal_func_c_3,app:eqn:anal_func_c_4} are not derived from any approximation of the solutions of the continuum BM Hamiltonian. Instead, we find that they simply match well with the numerical solutions. 
For $w_0/w_1=0.8$ and $\theta=1.05^\circ$, we fit the parameters in \cref{app:eqn:anal_func_c_1,app:eqn:anal_func_c_2,app:eqn:anal_func_c_3,app:eqn:anal_func_c_4} to the numerical solutions to obtain
\begin{alignat}{5}
    & \lambda_{c1} = 0.2194 a_M,\qquad 
    && \lambda_{c2} = 0.3299 a_M,\qquad 
    && \alpha_{c1} = 0.3958,\qquad 
    && \alpha_{c2} = 0.9183, \qquad 
    &&\mathcal{N}_{c1} = \mathcal{N}_{c2} = 1.2905, \\ 
    &\lambda_{c3} = 0.2430 a_M,\qquad 
    && \lambda_{c4} = 0.2241 a_M,\qquad 
    && \alpha_{c3} = 0.9257,\qquad 
    && \alpha_{c4} = 0.3783, \qquad 
    &&\mathcal{N}_{c3} = \mathcal{N}_{c4} = 1.1102.
\end{alignat} 
The comparison between the $c$-electron Gaussian wave functions and the corresponding numerical wave functions obtained from the continuum model is shown in \cref{fig:wave_fc:c,fig:wave_fc:d,fig:wave_fc:e,fig:wave_fc:f}. Similarly to the $f$-electrons, the numerical $c$-electron wave functions are very well-approximated by a Gaussian profile. Finally, we note that in order to obtain $\widetilde{u}_{\mathbf{Q} \beta,a}^{(\eta)}(\mathbf{k})$ at some small (but nonzero) momentum $\mathbf{k}$, we use the following approximation 
\begin{equation}
    \widetilde{u}_{\mathbf{Q} \beta,a}^{(\eta)}(\mathbf{k}) \approx 
    \widetilde{u}_{\mathbf{Q}-\mathbf{k} \beta,a}^{(\eta)}(0),\qquad \text{for}\qquad \mathbf{k}<\Lambda_{c0} \ . 
\end{equation}

\subsection{The density operator}

The TBG density operator can be expressed in terms of the $f$- and $c$-electron operators as 
\begin{align} \label{eq:rho-fc-r}
\hat{\rho} (\mathbf{r})
=& \sum_{\beta l\eta s}\left[ \sum_{\substack{\mathbf{R} \alpha \\ \mathbf{R}' \alpha'} }  
    e^{i\eta\Delta\mathbf{K}_l\cdot(\mathbf{R}-\mathbf{R}')}
    w^{(\eta)*}_{l \beta, \alpha}(\mathbf{r}-\mathbf{R})  w^{(\eta)}_{l\beta, \alpha'}(\mathbf{r}-\mathbf{R}') f_{\mathbf{R} \alpha\eta s}^\dagger f_{\mathbf{R}' \alpha' \eta s}  \right. \nonumber\\
    + & \frac1{N\Omega_{M}}\sum_{\substack{|\mathbf{k}|,|\mathbf{k}'|<\Lambda_{c0} \\a a'\\ \mathbf{Q}\mathbf{Q}'\in \mathcal{Q}_{l\eta} }}  e^{-i(\mathbf{k}-\mathbf{Q}-\mathbf{k}'+\mathbf{Q}')\cdot\mathbf{r}} 
    \widetilde{u}^{(\eta)*}_{\mathbf{Q}\beta,a}(\mathbf{k}) \widetilde{u}^{(\eta)}_{\mathbf{Q}'\beta,a'}(\mathbf{k}') c_{\mathbf{k} a\eta s}^\dagger c_{\mathbf{k}' a' \eta s} \nonumber\\
    + & \left. \frac1{\sqrt{N\Omega_{M}}} \sum_{\mathbf{R} \alpha a} \sum_{\substack{|\mathbf{k}|<\Lambda_{c0} \\ \mathbf{Q}\in \mathcal{Q}_{l\eta} }}  \left( 
    w^{(\eta)*}_{l \beta, \alpha}(\mathbf{r}-\mathbf{R}) \widetilde{u}^{(\eta)}_{\mathbf{Q}\beta,a}(\mathbf{k}) e^{i\eta\Delta\mathbf{K}_l\cdot\mathbf{R} + i(\mathbf{k}-\mathbf{Q})\cdot\mathbf{r}}
    f_{\mathbf{R} \alpha\eta s}^\dagger c_{ \mathbf{k} a\eta s}
    + \mathrm{h.c.}  \right)  \right]
\end{align}
where $\Lambda_{c0}$ is the cutoff at which the actual Bloch wave functions of passive bands are significantly different from those of the THF model. Applying the Fourier transformation
\begin{equation}
    \hat{\rho}_{\mathbf{q}+\mathbf{G}} = \int d^2\mathbf{r}\ e^{i\mathbf{q}\cdot\mathbf{r}} \hat{\rho}(\mathbf{r}),
\end{equation}
we obtain 
{\small
\begin{align} \label{eq:rho-fc}
\rho_{\mathbf{q}+\mathbf{G}}
= & \sum_{\mathbf{k}\in \mathrm{MBZ}} \sum_{\eta s } \bigg( \sum_{\alpha\beta} \mathcal{M}_{\alpha,\beta}^{(f,\eta)}(\mathbf{k},\mathbf{q}+\mathbf{G}) f_{\mathbf{p} \alpha \eta s}^\dagger f_{\mathbf{k} \beta \eta s}
+ \sum_{ab} \theta(\Lambda_{c0} - |\mathbf{k}|)\theta(\Lambda_{c0} - |\mathbf{p}|) \mathcal{M}_{\alpha,\beta}^{(c,\eta)}(\mathbf{k},\mathbf{q}+\mathbf{G}) c_{\mathbf{p} a \eta s}^\dagger c_{\mathbf{k} b \eta s} \nonumber\\
& + \sum_{\alpha a} \left( \theta(\Lambda_{c0}-|\mathbf{k}|) \mathcal{M}_{\alpha,a}^{(fc,\eta)}(\mathbf{k},\mathbf{q}+\mathbf{G}) 
    f_{\mathbf{p} \alpha\eta s}^\dagger c_{\mathbf{k} a \eta s}  \right) 
+ \sum_{\alpha a} \left( \theta(\Lambda_{c0}-|\mathbf{p}|) \mathcal{M}_{a,\alpha}^{(cf,\eta)}(\mathbf{k},\mathbf{q}+\mathbf{G}) 
    c_{\mathbf{p} a \eta s}^\dagger f_{\mathbf{k} \alpha\eta s}  \right)  \bigg),
\end{align}}where $\mathbf{p}$ is the image of $\mathbf{k}+\mathbf{q}+\mathbf{G}$ in the first MBZ,  and $\mathcal{M}$ are the form factors 
\begin{align}
    \mathcal{M}^{(f,\eta)}_{\alpha,\beta}(\mathbf{k},\mathbf{q}+\mathbf{G}) &= \sum_{\mathbf{Q} \alpha'} \widetilde{v}^{(\eta)*}_{\mathbf{Q}-\mathbf{G} \alpha', \alpha}(\mathbf{k}+\mathbf{q}) \widetilde{v}^{(\eta)}_{\mathbf{Q} \alpha',\beta}(\mathbf{k}) \\
    \mathcal{M}^{(c,\eta)}_{a,b}(\mathbf{k},\mathbf{q}+\mathbf{G}) &= \sum_{\mathbf{Q} \alpha'} \widetilde{u}^{(\eta)*}_{\mathbf{Q}-\mathbf{G} \alpha', a}(\mathbf{k}+\mathbf{q}) \widetilde{u}^{(\eta)}_{\mathbf{Q} \alpha',b}(\mathbf{k}) \\
    \mathcal{M}^{(fc,\eta)}_{\alpha,a}(\mathbf{k},\mathbf{q}+\mathbf{G}) &= \sum_{\mathbf{Q} \alpha'} \widetilde{v}^{(\eta)*}_{\mathbf{Q}-\mathbf{G} \alpha',\alpha}(\mathbf{k}+\mathbf{q}) \widetilde{u}^{(\eta)}_{\mathbf{Q}\alpha',a}(\mathbf{k}) \\ 
    \mathcal{M}^{(cf,\eta)}_{a,\alpha}(\mathbf{k},\mathbf{q}+\mathbf{G}) &= \sum_{\mathbf{Q} \alpha'} \widetilde{u}^{(\eta)*}_{\mathbf{Q}-\mathbf{G}\alpha',a}(\mathbf{k}+\mathbf{q}) \widetilde{v}^{(\eta)}_{\mathbf{Q} \alpha',\alpha}(\mathbf{k}) = \mathcal{M}^{(fc,\eta)*}_{\alpha,a}(\mathbf{k}+\mathbf{q}, - \mathbf{q} - \mathbf{G}) .  
\end{align}
Note that because $f$-electrons are well-localized in real space, their momentum-space wave functions are smooth and periodic, {\it i.e.}{} $\widetilde{v}_{\mathbf{Q} \alpha', \alpha}^{(\eta)}(\mathbf{k}+\mathbf{G}) = \widetilde{v}_{\mathbf{Q}-\mathbf{G} \alpha', \alpha}^{(\eta)}(\mathbf{k})$, implying that $f_{\mathbf{k}+\mathbf{G} \alpha\eta s} = f_{\mathbf{k} \alpha\eta s}$.
Therefore, we can also write $f_{\mathbf{p}\alpha\eta s}$ as $f_{\mathbf{k}+\mathbf{q} \alpha\eta s}$. However, these properties do not apply to $c$-electrons: suppose we could also construct Wannier functions for $c$-electrons, then one would have $\widetilde{u}_{\mathbf{Q} \alpha', a}^{(\eta)}(\mathbf{k}+\mathbf{G}) = \widetilde{u}_{\mathbf{Q}-\mathbf{G} \alpha', a}^{(\eta)}(\mathbf{k})$ and hence $c_{\mathbf{k}+\mathbf{G} a\eta s} = c_{\mathbf{k} a\eta s}$. The second term in \cref{eq:rho-fc} would then have a similar behavior as the first term, and we would not need to truncate the momenta of $c$-electrons. However, such a (smooth) $\widetilde{u}_{\mathbf{Q}-\mathbf{G} \alpha', a}^{(\eta)}(\mathbf{k})$ does not exist due to the topology of the model. Thus, we have to introduce a cutoff $\Lambda_{c0}$ beyond which $\widetilde{u}_{\mathbf{Q}-\mathbf{G} \alpha', a}^{(\eta)}(\mathbf{k})$ can not be smoothly defined. It is worth mentioning that the second term in \cref{eq:rho-fc} allows Umklapp processes, for which $\mathbf{k}$ ($|\mathbf{k}|<\Lambda_{c0}$) is in the first MBZ and $\mathbf{k}+\mathbf{q}+\mathbf{G}$ is in another MBZ. 

We can equivalently write the density operator as 
\begin{align} \label{eq:rho-fc1}
\rho_{\mathbf{q}+\mathbf{G}}
= & \sum_{\mathbf{k}\in \mathrm{MBZ}} \sum_{\eta s } \bigg( \sum_{\alpha\beta} \mathcal{M}_{\alpha,\beta}^{(f,\eta)}(\mathbf{k},\mathbf{q}+\mathbf{G}) f_{\mathbf{k}+\mathbf{q} \alpha \eta s}^\dagger f_{\mathbf{k} \beta \eta s} \nonumber\\
+ & \sum_{ab} \sum_{\mathbf{P} \mathbf{P}'} g(\mathbf{k}+\mathbf{q}+\mathbf{P}') g(\mathbf{k}+\mathbf{P}) \mathcal{M}_{\alpha,\beta}^{(c,\eta)}(\mathbf{k},\mathbf{q}+\mathbf{G}) c_{\mathbf{k}+\mathbf{q}+\mathbf{P}', a \eta s}^\dagger c_{\mathbf{k}+\mathbf{P} b \eta s} \nonumber\\
+ & \sum_{\alpha a} \sum_{\mathbf{P}} g(\mathbf{k}+\mathbf{P}) \mathcal{M}_{\alpha,a}^{(fc,\eta)}(\mathbf{k},\mathbf{q}+\mathbf{G}) 
    f_{\mathbf{k}+\mathbf{q} \alpha\eta s}^\dagger c_{\mathbf{k}+\mathbf{P} a \eta s} \nonumber\\
+ & \sum_{\alpha a} \sum_{\mathbf{P}} g(\mathbf{k}+\mathbf{q}+\mathbf{P}) \mathcal{M}_{a \alpha}^{(cf,\eta)}(\mathbf{k},\mathbf{q}+\mathbf{G}) 
    c_{\mathbf{k}+\mathbf{q}+\mathbf{P} a \eta s}^\dagger f_{\mathbf{k} \alpha\eta s} \bigg) 
\end{align}
where $g(\mathbf{k}) $ can be chosen as $\theta(  \Lambda_{c0} - |\mathbf{k}| )$ to reproduce \cref{eq:rho-fc}. 
In practice, we choose $g(\mathbf{k})$ to be a soft truncation function that quickly decays to zero when $\abs{\mathbf{k}}$ exceeds $\Lambda_{c0}$. Hereafter, we choose $g(\mathbf{k})$ to be 
\begin{equation}
    g(\mathbf{k}) = \prod_{j=0}^{5} \frac12 \mathrm{erfc} \left(  \frac{\kappa}{2\pi} \mathbf{k} \cdot C^{j}_{6z} \mathbf{a}_{M1} - \frac{\kappa}2  \right)\
\end{equation}
The function $\mathrm{erfc}(x)$ quickly decays to zero when $x>0$ and quickly approaches two when $x<0$. The lines defined by $\frac{1}{2\pi}  \mathbf{k} \cdot  C^{j}_{6z} \mathbf{a}_{M1} - \frac{1}2=0$ for $0 \leq j < 6$ enclose the first MBZ. Thus $g(\mathbf{k})$ approaches one when $\mathbf{k}$ is small (in the first MBZ) and approaches zero when $\mathbf{k}$ is large (outside the first MBZ). $\kappa$ is a parameter tuning the sharpness of $g(\mathbf{k})$. In practice, we choose $\kappa=10$. 

\subsection{Gaussian form factors}

In this section, we derive analytical expressions of the form factors. In order to obtain analytical results, we will make two approximations. First, we will replace the summation over $\mathbf{Q}$ by an integral, {\it i.e.} 
\begin{equation} \label{eq:form-app1}
    \sum_{\mathbf{Q}} \to \frac{2\Omega_M}{(2\pi)^2} \int d^2\mathbf{Q}\ ,
\end{equation}
where $ \frac{2\Omega_M}{(2\pi)^2} $ is the inverse of the average area of each $\mathbf{Q}$ point in the momentum space. 
This approximation will be justified if $k_\theta \lambda_{1,2}\ll 1$.
Second, we will assume that
\begin{equation} \label{eq:form-app2}
     \mathcal{N}_\mathbf{k}=     \mathcal{N}_{c1}=\mathcal{N}_{c2}=\mathcal{N}_{c3}=\mathcal{N}_{c4}=1\ . 
\end{equation}
As we will see, the second approximation is consistent with the first one. 

\subsubsection{Gaussian approximation of \texorpdfstring{$\mathcal{M}^{(f,\eta)}$}{Mf}}

We first consider the diagonal elements of $\mathcal{M}^{(f,\eta)}$. 
Replacing the summation by an integral, we obtain
{\small
\begin{align}
    \mathcal{M}^{(f,\eta)}_{1,1}(\mathbf{k},\mathbf{q}) = \mathcal{M}^{(f,\eta)}_{2,2}(\mathbf{k},\mathbf{q}) &=  \alpha_{1}^2 \frac{2\pi \lambda_{1}^2}{\Omega_M} \sum_{\mathbf{Q}} e^{-\frac12(\mathbf{k}-\mathbf{Q})^2 \lambda_{1}^2 - \frac12 (\mathbf{k}-\mathbf{Q}+\mathbf{q})^2 \lambda_{1}^2}  + \alpha_{2}^2 \frac{2\pi \lambda_{1}^4}{\Omega_M} \sum_{\mathbf{Q}} e^{-\frac12 (\mathbf{k}-\mathbf{Q})^2 \lambda_{2}^2 -\frac12 (\mathbf{k}-\mathbf{Q}+\mathbf{q})^2 \lambda_{2}^2} \nonumber\\
    & \quad \times  \left(  -i\eta(k_x-Q_x + q_x ) -  (k_y-Q_y + q_y )  \right)
    \left(  i\eta(k_x-Q_x) - (k_y-Q_y)  \right) \nonumber\\
    &\approx  \alpha_{1}^2 \frac{\lambda_{1}^2}{\pi} \int d^2\mathbf{Q}\ e^{-\frac12 \mathbf{Q}^2 \lambda_{1}^2 - \frac12 (-\mathbf{q}+\mathbf{Q})^2 \lambda_{1}^2 }  \nonumber\\
    & \quad + \alpha_{2}^2 \frac{\lambda_{2}^2}{\pi} \int d^2\mathbf{Q}\  e^{- \frac12 \mathbf{Q}^2 \lambda_{2}^2 - \frac12 (-\mathbf{q}+\mathbf{Q})^2 \lambda_{2}^2} \left(  i\eta(Q_x - q_x ) +  (Q_y -q_y )  \right)
   \left( -i\eta Q_x  +Q_y  \right) \nonumber\\
   &\approx  \alpha_{1}^2 \exp\left(  - \frac14 \mathbf{q}^2 \lambda_{1}^2  \right)
    + \alpha_{2}^2 \exp\left(  - \frac14 \mathbf{q}^2 \lambda_{2}^2  \right)
    \left( 1 - \frac14 \mathbf{q}^2 \lambda_{2}^2  \right)\ . \label{app:form_fact_f} 
\end{align}}The above result also extends to $\mathcal{M}^{(f,\eta)}_{1,1}(\mathbf{k},\mathbf{q}+\mathbf{G})$, with $\mathbf{G}$ being a nonzero reciprocal lattice vector. When $\mathbf{q}+\mathbf{G}=0$, \cref{app:form_fact_f} is simply $\mathcal{M}^{(f,\eta)}_{1,1}(\mathbf{k},0) = 1$, which is nothing but the normalization condition of the Bloch states. 
Thus, our second approximation, $\mathcal{N}_\mathbf{k}=1$, is consistent with this analytical expression. As we replace the summation over $\mathbf{Q}$ by integral, the form factors, which depend on $\mathbf{k}$ through the $\mathbf{k}-\mathbf{Q}$ terms, must become independent of $\mathbf{k}$.  

One can also argue the form of $\mathcal{M}^{(f,\eta)}_{\alpha,\beta}(\mathbf{k},\mathbf{q}+\mathbf{G})$ from a real space picture. 
The density profile of the Wannier function from \cref{eq:wannier_states_I} is 
\begin{equation}
n(\mathbf{r}) = \frac{\alpha_{1}^2}{\pi \lambda_{1}^2} e^{- \mathbf{r}^2 /\lambda_{1}^2} + 
    \frac{\alpha_{2}^2}{\pi \lambda_{2}^4} \mathbf{r}^2 e^{-\mathbf{r}^2/\lambda_{2}^2} 
\end{equation}
Its Fourier transformation is 
\begin{equation}
n(\mathbf{q}) = \int d^2 \mathbf{r}\ e^{i\mathbf{q}\cdot\mathbf{r}} n(\mathbf{r}) 
    = \alpha_{1}^2 \exp\left(  - \frac14 \mathbf{q}^2 \lambda_{1}^2  \right)
    + \alpha_{2}^2 \exp\left(  - \frac14 \mathbf{q}^2 \lambda_{2}^2  \right)
    \left( 1 - \frac14 \mathbf{q}^2 \lambda_{2}^2  \right)\ . 
\end{equation}
\Cref{app:form_fact_f} is precisely the Fourier transformation of the density profile of the Gaussian Wannier functions. The contribution from $f$-electrons belonging to different sites are omitted here. Thus, we also refer to \cref{eq:form-app1,eq:form-app2} as the one-center approximation. 

Ref.~\cite{SON22} has shown that if only the one-center integrals of the real space Wannier functions are kept, the off-diagonal terms $\mathcal{M}^{(f,\eta)}_{1,2}(\mathbf{q})$, $\mathcal{M}^{(f,\eta)}_{2,1}(\mathbf{q})$ are zero due to the emergent particle-hole symmetry. Here, we give a justification in momentum space. We denote the anti-unitary particle-hole symmetry as $\mathcal{P} = P\cdot C_{2z}T$. 
The wave functions satisfy $\mathcal{P} \ket{\widetilde{v}_{1}(\mathbf{k})} = \ket{\widetilde{v}_{2}(-\mathbf{k})} $ and $\mathcal{P} \ket{\widetilde{v}_{2}(\mathbf{k})} =-\ket{\widetilde{v}_{1}(-\mathbf{k})} $, where $\ket{\widetilde{v}_{\alpha}(\mathbf{k})}$ is a shorthand for the vector $\widetilde{v}_{\mathbf{Q}\alpha',\alpha}(\mathbf{k})$. As a result, one must have
\begin{equation}
\mathcal{M}^{(f,\eta)}_{1,2}(\mathbf{k},\mathbf{q}) = 
\langle  \widetilde{v}_{1}(\mathbf{k}+\mathbf{q}) | \widetilde{v}_{2}(\mathbf{k})  \rangle =  \langle  \mathcal{P} \widetilde{v}_{1}(\mathbf{k}+\mathbf{q}) |  \mathcal{P} \widetilde{v}_{2}(\mathbf{k})  \rangle^* = - \langle  \widetilde{v}_{1}(-\mathbf{k}) | \widetilde{v}_{2}(-\mathbf{k}-\mathbf{q})   \rangle
= - \mathcal{M}^{(f,\eta)}_{1,2}(-\mathbf{k}-\mathbf{q},\mathbf{q}). \label{eq:Mf12=0} 
\end{equation}
\Cref{eq:Mf12=0} also extends to $\mathcal{M}^{(f,\eta)}_{1,2}(\mathbf{k},\mathbf{q}+\mathbf{G})$, for $\mathbf{G}$ being a nonzero reciprocal lattice. 
Under the one-center approximation from \cref{eq:form-app1}, $\mathcal{M}$ is $\mathbf{k}$-independent, and so it follows that $\mathcal{M}^{(f,\eta)}_{1,2}(\mathbf{q}) = 0 $. Similarly, $  \mathcal{M}^{(f,\eta)}_{2,1}(\mathbf{q}) = 0 $ under the one-center approximation.

\subsubsection{Gaussian approximation of \texorpdfstring{$\mathcal{M}^{(c,\eta)}$}{Mc}}

We now calculate the form factors of the $c$-electrons under the approximation in \cref{eq:form-app1}. Similarly to the $f$-electrons, the normalization approximation in \cref{eq:form-app2} is consistent with treating $\mathbf{Q}$ as a continuous variable according to \cref{eq:form-app1}. The diagonal elements of the $\mathcal{M}^{(c,\eta)}$ form factor for the $a=1,2$ $c$-electrons are
{\small
\begin{align}
& \mathcal{M}^{(c,\eta)}_{1,1}(\mathbf{q}) = \mathcal{M}^{(c,\eta)}_{2,2}(\mathbf{q}) \nonumber\\
=&  \alpha_{c1}^2 \frac{2\pi \lambda_{c1}^2}{\Omega_M} \sum_{\mathbf{Q}} e^{-\frac12\mathbf{Q}^2 \lambda_{c1}^2 - \frac12 (\mathbf{Q}-\mathbf{q})^2 \lambda_{c1}^2} 
+ \alpha_{c2}^2 \frac{\pi \lambda_{c2}^6}{\Omega_M} \sum_{\mathbf{Q}} e^{-\frac12\mathbf{Q}^2 \lambda_{c1}^2 - \frac12 (\mathbf{Q}-\mathbf{q})^2 \lambda_{c1}^2} (-i\eta(Q_x-q_x)+Q_y-q_y)^2 (i\eta Q_x + Q_y)^2 
\nonumber\\
 \approx & \alpha_{c1}^2 \frac{\lambda_{c1}^2}{\pi} 
    \int d^2\mathbf{Q}\ e^{-\frac12\mathbf{Q}^2\lambda_{c1}^2 - \frac12(\mathbf{Q}-\mathbf{q})^2\lambda_{c1}^2 } 
     + \alpha_{c2}^2 \frac{\lambda_{c2}^6}{2\pi}
  \int d^2\mathbf{Q}\ e^{-\frac12\mathbf{Q}^2\lambda_{c2}^2 - \frac12(\mathbf{Q}-\mathbf{q})^2\lambda_{c2}^2 } 
  (-i\eta(Q_x-q_x) + Q_y-q_y)^2 (i\eta Q_x + Q_y)^2 \nonumber\\
\approx & \alpha_{c1}^2 \exp\left( -\frac14 \mathbf{q}^2 \lambda_{c1}^2 \right)
    + \alpha_{c2}^2 \exp\left( -\frac14 \mathbf{q}^2 \lambda_{c2}^2 \right)
        \left(   1 - \frac12 \mathbf{q}^2\lambda_{c2}^2 + \frac1{32} \mathbf{q}^4\lambda_{c2}^4  \right)\ . 
\end{align}}The above result also extends to $ \mathcal{M}^{(c,\eta)}_{1,1}(\mathbf{q}+\mathbf{G})$ for a nonzero reciprocal lattice $\mathbf{G}$. For the same reason as in \cref{eq:Mf12=0}, the particle-hole symmetry requires that $M^{(c,\eta)}_{1,2}(\mathbf{q}) = M^{(c,\eta)}_{2,1}(\mathbf{q})=0$. 

The diagonal elements of the $\mathcal{M}^{(c,\eta)}$ form factor for the $a=3,4$ $c$-electrons read as
\begin{align}
 \mathcal{M}^{(c,\eta)}_{3,3}(\mathbf{q}) &= \mathcal{M}^{(c,\eta)}_{4,4}(\mathbf{q}) 
    = \alpha_{c3}^2 \frac{2\pi \lambda_{c3}^4}{\Omega_M} \sum_{\mathbf{Q}} 
        e^{-\frac12\mathbf{Q}^2 \lambda_{c3}^2- \frac12(\mathbf{Q}-\mathbf{q})^2 \lambda_{c3}^2}
        (i\eta(Q_x-q_x)+Q_y-q_y)(-i\eta Q_x + Q_y) \nonumber\\
&\quad + \alpha_{c4}^2 \frac{\pi \lambda_{c4}^6}{\Omega_M} \sum_{\mathbf{Q}} 
e^{-\frac12\mathbf{Q}^2 \lambda_{c4}^2- \frac12(\mathbf{Q}-\mathbf{q})^2 \lambda_{c4}^2}
(i\eta(Q_x-q_x)+Q_y-q_y)^2(-i\eta Q_x + Q_y) ^2 \nonumber\\
&\approx  \alpha_{c3}^2 \frac{\lambda_{c3}^4}{\pi} \int d^2\mathbf{Q}\  
    e^{-\frac12\mathbf{Q}^2 \lambda_{c3}^2- \frac12(\mathbf{Q}-\mathbf{q})^2 \lambda_{c3}^2}
    (i\eta(Q_x-q_x)+Q_y-q_y)(-i\eta Q_x + Q_y) \nonumber\\
&\quad + \alpha_{c4}^2 \frac{ \lambda_{c4}^6}{2\pi} \int d^2\mathbf{Q}\ 
    e^{-\frac12\mathbf{Q}^2 \lambda_{c4}^2- \frac12(\mathbf{Q}-\mathbf{q})^2 \lambda_{c4}^2}
    (i\eta(Q_x-q_x)+Q_y-q_y)^2(-i\eta Q_x + Q_y) ^2 \nonumber\\
& \approx 
    \alpha_{c3}^2 \exp\left(  - \frac14 \mathbf{q}^2 \lambda_{c3}^2  \right)
    \left( 1 - \frac14 \mathbf{q}^2 \lambda_{c3}^2  \right) + 
    \alpha_{c4}^2 \exp\left( -\frac14 \mathbf{q}^2 \lambda_{c4}^2 \right)
        \left(   1 - \frac12 \mathbf{q}^2\lambda_{c4}^2 + \frac1{32} \mathbf{q}^4\lambda_{c4}^4  \right), 
\end{align}
which also extends to $ M^{(c,\eta)}_{3,3}(\mathbf{q}+\mathbf{G})$ for a nonzero reciprocal lattice $\mathbf{G}$. Similarly to \cref{eq:Mf12=0}, $M^{(c,\eta)}_{3,4}(\mathbf{q}) = M^{(c,\eta)}_{4,3}(\mathbf{q})=0$ due to the particle-hole symmetry. 

The matrix elements of the $\mathcal{M}^{(c,\eta)}$ form factor between $a=1,2$ and $a=3,4$ $c$-electrons are given by 
\begin{align}
    \mathcal{M}^{(c,\eta)}_{1,3}(\mathbf{q}) &=  \mathcal{M}^{(c,\eta)*}_{2,4}(\mathbf{q}) =  -\alpha_{c1}\alpha_{c3} \frac{2\pi \lambda_{c1} \lambda_{c3}^2}{\Omega_M}
    \sum_{\mathbf{Q}} e^{- \frac12 (\mathbf{Q}-\mathbf{q})^2 \lambda_{c1}^2 - \frac12 \mathbf{Q}^2 \lambda_{c3}^2 }
    \zeta_\mathbf{Q} (-i\eta Q_x + Q_y) \nonumber\\
    & + \alpha_{c2} \alpha_{c4} \frac{\pi \lambda_{c2}^3 \lambda_{c4}^3}{\Omega_M} \sum_\mathbf{Q}
    e^{- \frac12 (\mathbf{Q}-\mathbf{q})^2 \lambda_{c2}^2 - \frac12 \mathbf{Q}^2 \lambda_{c4}^2 }
    (-i\eta(Q_x-q_x) + Q_y-q_y)^2 (-i\eta Q_x + Q_y)^2 ,\\
    \mathcal{M}^{(c,\eta)}_{1,4}(\mathbf{q}) &= \mathcal{M}^{(c,\eta)*}_{2,3}(\mathbf{q}) = 
    - \alpha_{c1}\alpha_{c4} \frac{\sqrt2 \pi \lambda_{c1}^2 \lambda_{c4}^3 }{\Omega_M}
    \sum_{\mathbf{Q}} e^{- \frac12 (\mathbf{Q}-\mathbf{q})^2 \lambda_{c1}^2 - \frac12 \mathbf{Q}^2 \lambda_{c4}^2}
    i \zeta_\mathbf{Q} (i\eta Q_x + Q_y)^2  \nonumber\\
    & + \alpha_{c2} \alpha_{c3} \frac{\sqrt2 \pi \lambda_{c2}^3 \lambda_{c3}^2}{\Omega_M}
    \sum_\mathbf{Q} e^{-\frac12 (\mathbf{Q}-\mathbf{q})^2 \lambda_{c2}^2 - \frac12 \mathbf{Q}^2 \lambda_{c3}^2 }
    i (-i\eta (Q_x-q_x)^2 + Q_y-q_y)^2 (i\eta Q_x + Q_y) \ .
\end{align}
Note that the first term of $\mathcal{M}^{(c,\eta)}_{1,3}$ and the first term of $\mathcal{M}^{(c,\eta)}_{1,4}$ contain the sublattice $\zeta_\mathbf{Q}$ factor, which equals to $+1$ for $\mathbf{Q}\in \mathcal{Q}_+$ and $-1$ for $\mathbf{Q}\in \mathcal{Q}_-$. Because $\zeta_\mathbf{Q}$ is not a continuous function of $\mathbf{Q}$, we cannot apply \cref{eq:form-app1} to the terms which contain it. 
However, we argue that one can neglect these terms due to the fast oscillation of $\zeta_\mathbf{Q}$. We then have 
{\small
\begin{align}
 \mathcal{M}^{(c,\eta)}_{1,3}(\mathbf{q}) =&  \mathcal{M}^{(c,\eta)*}_{2,4}(\mathbf{q}) 
    \approx   \alpha_{c2}\alpha_{c4} \frac{\lambda_{c2}^3 \lambda_{c4}^3 }{2\pi}
    \int d^2\mathbf{Q} \ 
    e^{-\frac12(\mathbf{Q}-\mathbf{q})^2\lambda_{c2}^2 - \frac12\mathbf{Q}^2\lambda_{c4}^2 } 
    (-i\eta(Q_x-q_x) + Q_y-q_y)^2 (-i\eta Q_x + Q_y)^2  \nonumber\\
\approx & \alpha_{c2} \alpha_{c4} \exp\left( - \frac{\lambda_{c2}^2 \lambda_{c4}^2}{2(\lambda_{c2}^2+\lambda_{c4}^2)} \mathbf{q}^2 \right) 
    \frac{\lambda_{c2}^7 \lambda_{c4}^7}{(\lambda_{c2}^2+\lambda_{c4}^2)^5} (q_x + i\eta q_y)^4 ,\\
\mathcal{M}^{(c,\eta)}_{1,4}(\mathbf{q}) =& \mathcal{M}^{(c,\eta)*}_{2,3}(\mathbf{q}) \approx  
\alpha_{c2} \alpha_{c3} \frac{\lambda_{c2}^3 \lambda_{c3}^2}{\sqrt2 \pi }
    \int d^2\mathbf{Q}\ e^{-\frac12 (\mathbf{Q}-\mathbf{q})^2 \lambda_{c2}^2 - \frac12 \mathbf{Q}^2 \lambda_{c3}^2 }
    i (-i\eta (Q_x-q_x)^2 + Q_y-q_y)^2 (i\eta Q_x + Q_y)
\nonumber\\
\approx & \alpha_{c2} \alpha_{c3} \exp\left( - \frac{\lambda_{c2}^2 \lambda_{c3}^2}{2(\lambda_{c2}^2+\lambda_{c3}^2)} \mathbf{q}^2 \right) 
   \frac{\lambda_{c2}^3 \lambda_{c3}^4 }{ (\lambda_{c2}^2+\lambda_{c3}^2)^4 } 
   (q_x + i\eta q_y)
   \left(  -4\sqrt2 (\lambda_{c2}^2 + \lambda_{c3}^2) + \sqrt2 \lambda_{c2}^2\lambda_{c3}^2 \mathbf{q}^2  \right)\ . 
\end{align}}

\subsubsection{Gaussian approximation of \texorpdfstring{$\mathcal{M}^{(fc,\eta)}$}{Mfc} and \texorpdfstring{$\mathcal{M}^{(cf,\eta)}$}{Mcf}}

Finally, we compute the form factor $\mathcal{M}^{(fc,\eta)}$ under the approximations of \cref{eq:form-app1,eq:form-app2}. 
In addition, we also approximate $\mathcal{M}^{(fc,\eta)}_{\alpha,a}(\mathbf{k},\mathbf{q}+\mathbf{G}) = \langle \widetilde{v}^{(\eta)}_{\alpha}(\mathbf{k}+\mathbf{q}+\mathbf{G}) | \widetilde{u}^{(\eta)}_{a}(\mathbf{k})  \rangle$ as $\mathcal{M}^{(fc,\eta)}_{\alpha,a}(\mathbf{q}+\mathbf{G}) = \langle \widetilde{v}^{(\eta)}_{\alpha}(\mathbf{q}+\mathbf{G}) | \widetilde{u}^{(\eta)}_{a}(0)  \rangle$, which is reasonable because the momentum $\mathbf{k}$ of the $c$-electrons is limited to a small region around $\mathbf{k}=0$. 
Following the calculations for $\mathcal{M}^{(f,\eta)}$ and $\mathcal{M}^{(c,\eta)}$, we obtain
{\small
\begin{align}
\mathcal{M}_{1,1}^{(fc,\eta)}(\mathbf{q}) &=  
\mathcal{M}_{2,2}^{(fc,\eta)*}(\mathbf{q}) = 
    -\alpha_{1} \alpha_{c1} \frac{2\pi \lambda_{1} \lambda_{c1} }{\Omega_M} \sum_{\mathbf{Q}}  
    -i\zeta_\mathbf{Q} e^{- \frac12 (\mathbf{Q}-\mathbf{q})^2 \lambda_{1}^2 - \frac12 \mathbf{Q}^2 \lambda_{c1}^2 } \nonumber\\
& \quad + \alpha_{2} \alpha_{c2}  \frac{\sqrt2 \pi \lambda_{2}^2 \lambda_{c2}^3}{\Omega_M}
    \sum_\mathbf{Q} e^{-\frac12 (\mathbf{Q}-\mathbf{q})^2\lambda_{2}^2 - \frac12 \mathbf{Q}^2 \lambda_{c2}^2 } (-i) ( -i\eta(q_x-Q_x) - (q_y-Q_y) ) (i\eta Q_x + Q_y)^2 \nonumber\\
&\approx \alpha_{2} \alpha_{c2}  \frac{\lambda_{2}^2 \lambda_{c2}^3}{\sqrt2 \pi}
    \int d^2\mathbf{Q}\ e^{-\frac12 (\mathbf{Q}-\mathbf{q})^2\lambda_{2}^2 - \frac12 \mathbf{Q}^2 \lambda_{c2}^2 }(-i) ( -i\eta(q_x-Q_x) - (q_y-Q_y) ) (i\eta Q_x + Q_y)^2 \nonumber\\
&\approx \alpha_{2} \alpha_{c2} \exp\left(  - \frac{\lambda_{2}^2\lambda_{c2}^2}{2(\lambda_{2}^2+\lambda_{c2}^2)} \mathbf{q}^2  \right) \frac{\sqrt2 \lambda_{2}^6 \lambda_{c2}^5}{(\lambda_{2}^2+\lambda_{c2}^2)^4}  (q_x - i\eta q_y)^3\ , \\
 \mathcal{M}_{1,2}^{(fc,\eta)}(\mathbf{q}) &=  
\mathcal{M}_{2,1}^{(fc,\eta)*}(\mathbf{q}) = 
\alpha_{1} \alpha_{c2} \frac{\sqrt2 \pi \lambda_{1} \lambda_{c2}^3}{\Omega_M}
    \sum_\mathbf{Q} e^{-\frac12 (\mathbf{Q}-\mathbf{q})^2 \lambda_{1}^2 -\frac12 \mathbf{Q}^2 \lambda_{c2}^2}
    (-i\eta Q_x + Q_y)^2  \nonumber\\
& \quad - \alpha_{2} \alpha_{c1} \frac{2\pi \lambda_{2}^2 \lambda_{c1}}{\Omega_M}
    \sum_\mathbf{Q} e^{-\frac12(\mathbf{Q}-\mathbf{q})^2 \lambda_{2}^2 - \frac12\mathbf{Q}^2 \lambda_{c1}^2 }
    \zeta_\mathbf{Q} (-i\eta (q_x-Q_x)-(q_y-Q_y)) \nonumber\\
& \approx \alpha_{1} \alpha_{c2} \frac{\lambda_{1} \lambda_{c2}^3}{\sqrt2 \pi}
    \int d^2\mathbf{Q}\ e^{-\frac12 (\mathbf{Q}-\mathbf{q})^2 \lambda_{1}^2 -\frac12 \mathbf{Q}^2 \lambda_{c2}^2}
    (-i\eta Q_x + Q_y)^2  \nonumber\\
& \approx - 
\alpha_{1} \alpha_{c2} \exp\left(  - \frac{\lambda_{1}^2\lambda_{c2}^2}{2(\lambda_{1}^2+\lambda_{c2}^2)} \mathbf{q}^2  \right) \frac{\sqrt2 \lambda_{1}^5 \lambda_{c2}^3}{(\lambda_{1}^2+\lambda_{c2}^2)^3}
(q_x +i \eta q_y)^2\ , \\
 \mathcal{M}_{1,3}^{(fc,\eta)}(\mathbf{q})
&= \mathcal{M}_{2,4}^{(fc,\eta)*}(\mathbf{q})
= \alpha_{1} \alpha_{c3} \frac{2\pi \lambda_{1}\lambda_{c3}^2 }{\Omega_M} \sum_{\mathbf{Q}} 
    e^{-\frac12 (\mathbf{Q}-\mathbf{q})^2 \lambda_{1}^2  - \frac12 \mathbf{Q}^2 \lambda_{c3}^2 }
    (-i) (-i\eta Q_x + Q_y) \nonumber\\
& \quad + \alpha_{2} \alpha_{c4} \frac{\sqrt2 \pi \lambda_{2}^2 \lambda_{c4}^3 }{\Omega_M}
    \sum_\mathbf{Q}  e^{-\frac12(\mathbf{Q}-\mathbf{q})^2 \lambda_{2}^2 - \frac12 \mathbf{Q}^2 \lambda_{c4}^2 } (-i) (  -i\eta (q_x - Q_x) - (q_y-Q_y) ) ( -i\eta Q_x + Q_y )^2 \nonumber\\
&\approx \alpha_{1} \alpha_{c3} \frac{\lambda_{1}\lambda_{c3}^2 }{\pi} \int d^2\mathbf{Q}\ 
    e^{-\frac12 (\mathbf{Q}-\mathbf{q})^2 \lambda_{1}^2  - \frac12 \mathbf{Q}^2 \lambda_{c3}^2 }
    (-i) (-i\eta Q_x + Q_y) \nonumber\\
& \quad + \alpha_{2} \alpha_{c4} \frac{\lambda_{2}^2 \lambda_{c4}^3 }{\sqrt2 \pi}
    \int d^2\mathbf{Q} \   e^{-\frac12(\mathbf{Q}-\mathbf{q})^2 \lambda_{2}^2 - \frac12 \mathbf{Q}^2 \lambda_{c4}^2 } (-i) (  -i\eta (q_x - Q_x) - (q_y-Q_y) ) ( -i\eta Q_x + Q_y )^2 \nonumber\\
&\approx -  \alpha_{1} \alpha_{c3}  \exp \left(  - \frac{\lambda_{1}^2\lambda_{c3}^2}{2(\lambda_{1}^2+\lambda_{c3}^2)} \mathbf{q}^2  \right) \frac{2\lambda_{1}^3 \lambda_{c3}^2}{(\lambda_{1}^2+\lambda_{c3}^2)^2} (q_x + i\eta q_y) \nonumber\\
& \quad+  \alpha_{2} \alpha_{c4}  \exp \left(  - \frac{\lambda_{2}^2\lambda_{c4}^2}{2(\lambda_{2}^2+\lambda_{c4}^2)} \mathbf{q}^2  \right)  \frac{ \lambda_{2}^4 \lambda_{c4}^3}{(\lambda_{2}^2+\lambda_{c4}^2)^4} (q_x + i\eta q_y) \left(   -4\sqrt2 (\lambda_{2}^2 + \lambda_{c4}^2) + \sqrt2 \lambda_{2}^2\lambda_{c4}^2 \mathbf{q}^2  \right)\ , \\
 \mathcal{M}_{1,4}^{(fc,\eta)}(\mathbf{q}) &= \mathcal{M}_{2,3}^{(fc,\eta)*}(\mathbf{q}) 
    = \alpha_{1} \alpha_{c4} \frac{\sqrt2 \pi \lambda_{1} \lambda_{c4}^3}{\Omega_M} \sum_\mathbf{Q}
    e^{-\frac12 (\mathbf{Q}-\mathbf{q})^2 \lambda_{1}^2 - \frac12 \mathbf{Q}^2 \lambda_{c4}^2} (i\eta Q_x + Q_y)^2 \nonumber\\
& \quad + \alpha_{2} \alpha_{c3} \frac{2\pi \lambda_{2}^2 \lambda_{c3}^2}{\Omega_M} \sum_\mathbf{Q}
    e^{-\frac12 (\mathbf{Q}-\mathbf{q})^2 \lambda_{2}^2 - \frac12 \mathbf{Q}^2 \lambda_{c3}^2} 
    (-i\eta (q_x-Q_x) - (q_y-Q_y) ) (i\eta Q_x + Q_y) \nonumber\\
& \approx  \alpha_{1} \alpha_{c4} \frac{ \lambda_{1} \lambda_{c4}^3}{\sqrt2 \pi} \int d^2\mathbf{Q}\ 
    e^{-\frac12 (\mathbf{Q}-\mathbf{q})^2 \lambda_{1}^2 - \frac12 \mathbf{Q}^2 \lambda_{c4}^2} (i\eta Q_x + Q_y)^2 \nonumber\\
& \quad + \alpha_{2} \alpha_{c3} \frac{\lambda_{2}^2 \lambda_{c3}^2}{\pi} \int d^2\mathbf{Q}\ 
    e^{-\frac12 (\mathbf{Q}-\mathbf{q})^2 \lambda_{2}^2 - \frac12 \mathbf{Q}^2 \lambda_{c3}^2} 
    (-i\eta (q_x-Q_x) - (q_y-Q_y) ) (i\eta Q_x + Q_y) \nonumber\\
&\approx  - \alpha_{1} \alpha_{c4} \exp\left(  - \frac{\lambda_{1}^2\lambda_{c4}^2}{2(\lambda_{1}^2+\lambda_{c4}^2)} \mathbf{q}^2  \right) \frac{\sqrt2 \lambda_{1}^5 \lambda_{c4}^3}{(\lambda_{1}^2+\lambda_{c4}^2)^3} (q_x  - i \eta q_y)^2 \nonumber \\
& \quad + \alpha_{2} \alpha_{c3} \exp\left(  - \frac{\lambda_{2}^2\lambda_{c3}^2}{2(\lambda_{2}^2+\lambda_{c3}^2)} \mathbf{q}^2  \right) \frac{2 \lambda_{2}^4 \lambda_{c3}^4}{(\lambda_{2}^2+\lambda_{c3}^2)^3} (q_x  - i \eta q_y)^2\ . 
\end{align}}

Due to the Hermiticity of the density operator, the form factor $\mathcal{M}^{(cf,\eta)}$ is given by 
\begin{equation}
    \mathcal{M}^{(cf,\eta)}_{a,\alpha}(\mathbf{q}) = \mathcal{M}^{(fc,\eta)*}_{\alpha,a}(-\mathbf{q})\ . 
\end{equation}

\section{Higher symmetries of the interaction Hamiltonian}\label{app:sec:symmetries}
In this appendix, we explore the higher symmetries of the interaction Hamiltonian that arise in several limits of parameters $U_1$, $W_{1,3}$, $V$ and $J$. We start obtaining the symmetry group of the density-density interaction terms of $\hat{H}_I$ (namely $\hat{H}_U + \hat{H}_W + \hat{H}_V$). We then derive the symmetry group of the exchange and double hybridization interaction terms of $\hat{H}_I$ (given by $\hat{H}_J + \hat{H}_{\tilde{J}}$). Finally, we discuss the symmetries of the entire interaction Hamiltonian. We note that these symmetries hold in the absence of the kinetic term. The latter reduces the symmetry of the THF model to the continuous symmetries derived in Ref.~\cite{SON22}. 

\subsection{Density-density interactions and the $U(24)$ symmetry}\label{app:subsec:symmetry_density-density-interaction}
In this section, we consider the terms of the interaction Hamiltonian containing only the density-density interaction terms, {\it i.e.}{} the first, second and fifth terms in \cref{tab:Coulomb_summary}.
\begin{equation}
    \hat{H}_I^{(\textrm{dens-dens})} = \hat{H}_U + \hat{H}_W + \hat{H}_V,
    \label{eq:U24_interaction_Hamiltonian}
\end{equation}
where $\hat{H}_U$ is the $f$-electron density-density interaction, given by \cref{eq:interaction_ff_formula},  $\hat{H}_W$ is the $f$-$c$-electron density-density interaction from \cref{eq:interaction_fc_density} and $\hat{H}_V$ is the $c$-electron density-density interaction given by \cref{eq:HV-explicit2} in the approximation set by \cref{eq:interaction_cc_approximate_equality}.

Although the $c$-electrons do not have a localized real-space representation, formally we can define the Fourier transformation of the density operator $\hat{\rho}_{\mathbf{q}}^c$
\begin{equation}
    \hat{\rho}_{\mathbf{q}}^c = \sum_{\eta, s, a}\sum_{|\mathbf{k}|,|\mathbf{k}+\mathbf{q}|<\Lambda_c}\hat{c}^\dagger_{\mathbf{k} + \mathbf{q} a \eta s}\hat{c}_{\mathbf{k} a \eta s};\qquad \hat{\rho}_{\mathbf{R}}^c=\frac{1}{N}\sum_{\mathbf{q}<\Lambda_c}e^{-i\mathbf{q}\cdot\mathbf{R}}\hat{\rho}_{\mathbf{q}}^c,
    \label{eq:U24symmetry_Fourier_conduction}
\end{equation}
where the cutoff $\Lambda_c = \partial \text{BZ}$ is taken at the boundary of the BZ, such that the number of quantum states $N$ is matched between the real and momentum spaces.

Plugging \cref{eq:U24symmetry_Fourier_conduction} in \cref{eq:U24_interaction_Hamiltonian} and neglecting the NN interaction term $U_2$ in \cref{eq:interaction_ff_formula}, we obtain
\begin{equation}
    \small 
    \hat{H}_I^{(\textrm{dens-dens})} \approx  \frac{U_1}{2}\sum_{\alpha,\eta, s}\sum_{\alpha',\eta', s'}:\hat{f}^\dagger_{\mathbf{R} \alpha \eta s}\hat{f}_{\mathbf{R} \alpha\eta s}:
    :\hat{f}^\dagger_{\mathbf{R} \alpha' \eta' s'} \hat{f}_{\mathbf{R} \alpha'\eta' s'}: + W\sum_{\mathbf{R}}\sum_{\alpha, \eta, s}\hat{f}^\dagger_{\mathbf{R} \alpha \eta s} \hat{f}_{\mathbf{R} \alpha\eta s}::\hat{\rho}_{\mathbf{R}}^c: + \frac{W}{2}\sum_{R}:\hat{\rho}_{\mathbf{R}}^c::\hat{\rho}_{\mathbf{R}}^c:,
    \label{eq:U24_interaction_frho}
\end{equation}
where we approximated the $f$-$c$ interaction strength parameter with $W_1 = W_3 = W$, as argued in \cref{eq:fc_interaction_strength_approx_13}.
A higher symmetry emerges when $U_1 = W$. To see why this is so, we first introduce a shorthand notation $\hat{\rho}_{\mathbf{R}}^f = \sum_{\alpha, \eta, s} f_{\mathbf{R} \alpha \eta s}^\dagger f_{\mathbf{R} \alpha\eta s}$, and rewrite \cref{eq:U24_interaction_frho} as
\begin{equation}
    \hat{H}_I^{(\textrm{dens-dens})} =
    \frac{1}{2}W\sum_{\mathbf{R}}:\hat{\rho}_{\mathbf{R}}^f + \hat{\rho}_{\mathbf{R}}^c::\hat{\rho}_{\mathbf{R}}^f + \hat{\rho}_{\mathbf{R}}^c: = \frac{1}{2}W\sum_{\mathbf{R}}:\hat{\rho}_{\mathbf{R}}::\hat{\rho}_{\mathbf{R}}:,
    \label{eq:U24_interaction_symmetric}
\end{equation}
where we substituted $\hat{\rho}_{\mathbf{R}} = \hat{\rho}_{\mathbf{R}}^f + \hat{\rho}_{\mathbf{R}}^c$. We notice that the density $\hat{\rho}_{\mathbf{R}}$ can be written as the inner product of a 24-dimensional spinor, {\it i.e.}{} $\hat{\rho}_{\mathbf{R}} = \psi^{\dagger}_{\mathbf{R}}\psi_{\mathbf{R}}$, where
\begin{equation}
    \psi_{\mathbf{R}} = (\hat{f}_{\mathbf{R},1},\hat{f}_{\mathbf{R},2},\ldots,\hat{f}_{\mathbf{R},8}, \hat{c}_{\mathbf{R},1},\hat{c}_{\mathbf{R},2},\ldots,\hat{c}_{\mathbf{R},16}),
    \label{eq:U24_interaction_spinor}
\end{equation}
and indices $1 \leq i \leq 8$ for the $f$-electrons and $1 \leq j \leq 16$ for the $c$-electrons encode the triplets $(\alpha, \eta, s)$ and $(a,\eta,s)$ correspondingly. Here we introduced an auxiliary Fourier transformation
\begin{equation}
    \hat{c}^\dagger_{\mathbf{k},j} = \sum_{\mathbf{R}}\hat{c}^\dagger_{\mathbf{R},j}e^{i\mathbf{k}\cdot\mathbf{R}},\qquad \hat{c}^\dagger_{\mathbf{R},j} = \frac{1}{\sqrt{N}}\sum_{|\mathbf{k}|<\Lambda_c}e^{-i\mathbf{k}\cdot\mathbf{R}}\hat{c}^\dagger_{\mathbf{k},j}.
    \label{eq:c_cre_Fourier}
\end{equation}
From \cref{eq:U24_interaction_spinor} we see that a transformation $\psi_{\mathbf{R}} \rightarrow U\psi_{\mathbf{R}}$, where $U\in U(24)$ is a $24\times24$ unitary matrix, leaves the interaction Hamiltonian given by \cref{eq:U24_interaction_symmetric} invariant. We conclude that in the limit $U_1=W_1=W_2=V$, the density-density interaction Hamiltonian from \cref{eq:U24_interaction_Hamiltonian} enjoys an enlarged $U(24)$ symmetry.

\subsection{Interaction Hamiltonian and the $U(8)\times U(8) \times U(1)$ symmetry}\label{app:subsec:higher_symmetry}
In this section, we consider the exchange interaction term and the double hybridization term given by \cref{eq:HJ_explicit,eq:HJstar_two_term} respectively. Invoking the auxiliary Fourier transform for the conduction band electrons creation and annihilation operators introduced in \cref{eq:c_cre_Fourier}, the $\hat{H}_J$ term reads
\begin{equation}
    \hat{H}_J = - \frac{J}{2} \sum_{\mathbf{R}, s_1, s_2} \sum_{\alpha,\alpha',\eta,\eta'}( \eta\eta' + (-1)^{\alpha+\alpha'} )
     :\hat{f}^\dagger_{\mathbf{R} \alpha \eta s_1} f_{\mathbf{R} \alpha' \eta' s_2}:  :\hat{c}^\dagger_{\mathbf{R}, \alpha'+2, \eta' s_2} \hat{c}_{ \mathbf{R}, \alpha+2, \eta s_1}:\ ,
     \label{eq:HJ_Fourier}
\end{equation}
and the $\hat{H}_{\tilde{J}}$ term reads
\begin{align}
    \hat{H}_{\tilde{J}} &= 
    \frac{J}{4} 
    \sum_{ \mathbf{R}, s_1, s_2 } \sum_{\alpha,\alpha',\eta,\eta'}  
    ( \eta\eta' + (-1)^{\alpha+\alpha'} ) \biggl(
    \hat{f}^\dagger_{\mathbf{R} \alpha \eta s_1} \hat{f}^\dagger_{\mathbf{R} \alpha' -\eta' s_2} 
    \hat{c}_{\mathbf{R}, \alpha'+2, -\eta' s_2} \hat{c}_{\mathbf{R}, \alpha+2, \eta s_1}  \nonumber \\  &+\hat{c}^\dagger_{\mathbf{R}, \alpha'+2, -\eta' s_2} \hat{c}^\dagger_{\mathbf{R}, \alpha+2, \eta s_1}\hat{f}_{\mathbf{R} \alpha \eta s_1} \hat{f}_{\mathbf{R} \alpha' -\eta' s_2} \biggr),
    \label{eq:HJ_tilde_Fourier}
\end{align}
After a straightforward calculation, we can represent the sum of the terms in \cref{eq:HJ_Fourier,,eq:HJ_tilde_Fourier} as a product
\begin{align}
    \hat{H}_{J} + \hat{H}_{\tilde{J}} &= -\frac{J}{2} \sum_{ \mathbf{R}, s_1, s_2 } \sum_{\alpha,\alpha',\eta,\eta'}  
    ( \eta\eta' + (-1)^{\alpha+\alpha'} ) \nonumber \\ & \times \left(\hat{f}^\dagger_{\mathbf{R} \alpha \eta s_1}\hat{c}_{\mathbf{R},\alpha+2, \eta s_1}  + \hat{c}^\dagger_{\mathbf{R} \alpha+2, -\eta s_1} \hat{f}_{\mathbf{R}\alpha,-\eta, s_1}  \right) \left(\hat{f}_{\mathbf{R} \alpha' \eta' s_2}\hat{c}^\dagger_{\mathbf{R},\alpha'+2, \eta' s_2} +\hat{c}_{\mathbf{R} \alpha'+2, -\eta' s_2} \hat{f}^\dagger_{\mathbf{R}\alpha',-\eta', s_2}  \right).
    \label{eq:sum_J_Jtilde}
\end{align}
Introducing an operator
\begin{equation}
    \hat{O}_{\mathbf{R}\alpha \eta} = \sum_{s_1} \hat{f}^\dagger_{\mathbf{R} \alpha \eta s_1}\hat{c}_{\mathbf{R}\alpha+2, \eta s_1}  + \hat{c}^\dagger_{\mathbf{R} \alpha+2, -\eta s_1} \hat{f}_{\mathbf{R}\alpha,-\eta, s_1} ,\qquad \hat{O}^\dagger_{\mathbf{R}\alpha \eta} = \hat{O}_{\mathbf{R}\alpha -\eta}, 
\end{equation}
we can rewrite \cref{eq:sum_J_Jtilde} as 
\begin{equation}
    \hat{H}_{J} + \hat{H}_{\tilde{J}} = \frac{J}{2}\sum_{\mathbf{R}}\sum_{\eta, \alpha } \left(\hat{O}_{\mathbf{R} \alpha \eta } \hat{O}_{\mathbf{R} \alpha -\eta} - \hat{O}_{\mathbf{R} \alpha \eta } \hat{O}_{\mathbf{R} \bar{\alpha} \eta}  \right),
\end{equation}
where $\bar{\alpha} = 2,1$ for $\alpha=1,2$. We also define the operator
\begin{equation}
    \hat{D}_{\mathbf{R} \alpha \eta }= \hat{O}_{\mathbf{R}\alpha \eta} - \hat{O}_{\mathbf{R}\bar{\alpha} -\eta},
    \label{eq:Doperator_definition}
\end{equation}
which obeys
\begin{equation}
     \hat{D}^\dagger_{\mathbf{R}\alpha\eta} = \hat{D}_{\mathbf{R}\alpha,-\eta},\qquad \hat{D}_{\mathbf{R}\bar{\alpha},-\eta}=-\hat{D}_{\mathbf{R}\alpha\eta} ,
     \label{eq:operator_D_constraints}
\end{equation} 
and using the definition in \cref{eq:Doperator_definition}, we arrive to a very simple expression for \cref{eq:sum_J_Jtilde}
\begin{equation}
    \hat{H}_{J} + \hat{H}_{\tilde{J}} = \frac{J}{4}\sum_{\mathbf{R},\eta, \alpha}\hat{D}_{\mathbf{R} \alpha \eta} \hat{D}^\dagger_{\mathbf{R}\alpha\eta} = \frac{J}{2}\{\hat{D}^\dagger_{\mathbf{R}1+},\hat{D}_{\mathbf{R}1+}\},
    \label{eq:U8_Drepresentation}
\end{equation}
where in the last equality, we used the properties from \cref{eq:operator_D_constraints}.

We are looking for quadratic local symmetry operators which commute with the interaction terms from \cref{eq:U8_Drepresentation}. To simplify our search, we note that for a certain $\hat{\Sigma}_{\mathbf{R}}$ operator:
\begin{align}
    \left[\hat{\Sigma}_{\mathbf{R}}, \left\{\hat{D}^\dagger_{\mathbf{R}1+},\hat{D}_{\mathbf{R}1+}\right\}\right] = \left\{\left[\hat{\Sigma}_{\mathbf{R}},\hat{D}^\dagger_{\mathbf{R}1+}\right],\hat{D}_{\mathbf{R}1+}\right\} + \left\{\left[\hat{\Sigma}_{\mathbf{R}},\hat{D}_{\mathbf{R}1+}\right],\hat{D}^\dagger_{\mathbf{R}1+}\right\},
\end{align}
from which we conclude that $\hat{\Sigma}_{\mathbf{R}}$ is the symmetry of the interacting terms in \cref{eq:U8_Drepresentation} when it commutes with the $\hat{D}_{\mathbf{R}1+}$ operator. For the analysis of the \cref{eq:U8_Drepresentation} symmetries, we find it useful to introduce the Pauli matrices $\zeta_{0,x,y,z}$, which act in the $(f,c)$-type of fermion space, as well as the matrices $\sigma_{0,x,y,z}$, acting in the orbital space $\alpha={1,2}$ for $f$-electrons and $\alpha+2={3,4}$ for $c$-electrons. As in the previous sections, we also use $\tau_{0,x,y,z}$ and $s_{0,x,y,z}$ to denote the Pauli matrices acting in the valley $\eta=\pm$ and spin $s={\uparrow,\downarrow}$ spaces respectively. The action of these Pauli matrices in their respective subspaces will be exemplified below. We define the basis of the $f$-electrons and $\Gamma_1\oplus\Gamma_2$ electrons with the 16-dimensional spinor operator $\hat{\Psi}_{\mathbf{R}}$ as
{\tiny
\begin{equation}
    \hat{\Psi}_{\mathbf{R}} = \biggl(\overbrace{\underbrace{\hat{f}_{\mathbf{R}1+\uparrow},\hat{f}_{\mathbf{R}1+\downarrow},\hat{c}_{\mathbf{R}3+\uparrow},\hat{c}_{\mathbf{R}3+\uparrow}}_{\eta=+},\underbrace{\hat{f}_{\mathbf{R}1-\uparrow},\hat{f}_{\mathbf{R}1-\downarrow},\hat{c}_{\mathbf{R}3-\uparrow},\hat{c}_{\mathbf{R}3-\uparrow}}_{\eta=-}}^{\alpha=1},\overbrace{\underbrace{\hat{f}_{\mathbf{R}2+\uparrow},\hat{f}_{\mathbf{R}2+\downarrow},\hat{c}_{\mathbf{R}4+\uparrow},\hat{c}_{\mathbf{R}4+\uparrow}}_{\eta=+},\underbrace{\hat{f}_{\mathbf{R}2-\uparrow},\hat{f}_{\mathbf{R}2-\downarrow},\hat{c}_{\mathbf{R}4-\uparrow},\hat{c}_{\mathbf{R}4-\uparrow}}_{\eta=-}}^{\alpha=2}\biggr)^T.
\end{equation}}To exemplify the action of the various Pauli matrices in their respective spaces, we note that the 16-dimensional matrix corresponding to the simultaneous action of the $\sigma_x$, $\tau_0$, $\zeta_y$, and $s_z$ Pauli matrices in the orbital, valley, $(f,c)$-type of fermion, and spin spaces, respectively, is denoted by $\sigma_x \tau_0 \zeta_y s_z$ and is given explicitly by the Kronecker product $\sigma_x \otimes \tau_0 \otimes \zeta_y \otimes s_z$ in the basis of the $\hat{\Psi}_{\mathbf{R}}$.

Within this basis, the $\hat{D}_{\mathbf{R}1+}$ operator can be written as
\begin{equation}
    \small
    \hat{D}_{\mathbf{R}1+} = \sum_{s_1}\left(\hat{f}^\dagger_{\mathbf{R} 1+ s_1}\hat{c}_{\mathbf{R} 3+ s_1} + \hat{c}^\dagger_{\mathbf{R} 3 - s_1}\hat{f}_{\mathbf{R} 1 - s_1} - \hat{f}^\dagger_{\mathbf{R} 2 - s_1}\hat{c}_{\mathbf{R} 4 - s_1} - \hat{c}^\dagger_{\mathbf{R} 4 + s_1}\hat{f}_{\mathbf{R} 2 + s_1}\right) = \frac{1}{2} \hat{\Psi}^\dagger_{\mathbf{R}}(\sigma_z\tau_0\zeta_x s_0 + i\sigma_0\tau_z\zeta_y s_0 ) \hat{\Psi}_{\mathbf{R}}.
    \label{eq:DR1+_pauli_representation}
\end{equation}
We find 64 local symmetry generators $\hat{\Sigma}_{\mathbf{R}}^{\mu\nu}$ which commute with the $\hat{D}_{\mathbf{R}1+}$ operator
\begin{equation}
    \left[\hat{\Sigma}_{\mathbf{R}}^{\mu\nu}, \hat{D}_{\mathbf{R}1+}\right] = 0, \qquad \mu=1,\ldots16,\;\nu=1,\ldots 4,
    \label{eq:U8_symmetry_commutation}
\end{equation}
where $\hat{\Sigma}_{\mathbf{R}}^{\mu\nu}$ reads
\begin{equation}
    \hat{\Sigma}_{\mathbf{R}}^{\mu\nu} = \hat{\Psi}^\dagger_{\mathbf{R}}S_{\mu}\otimes s_{\nu}\hat{\Psi}_{\mathbf{R}},
\end{equation}
and $s_{\nu}$ are the spin Pauli matrices, while the components of $S_{\mu}$ are listed below:
\begin{alignat}{4}
    &\sigma_0\tau_0\zeta_0 \qquad && \sigma_x\tau_0\zeta_y \qquad && \sigma_y\tau_0\zeta_y \qquad && \sigma_z\tau_0\zeta_0, \nonumber \\
    &\sigma_0\tau_x\zeta_x \qquad && \sigma_x\tau_x\zeta_z \qquad && \sigma_y\tau_x\zeta_z \qquad && \sigma_z\tau_x\zeta_x, \nonumber \\
    &\sigma_0\tau_y\zeta_x \qquad && \sigma_x\tau_y\zeta_z \qquad && \sigma_y\tau_y\zeta_z \qquad && \sigma_z\tau_y\zeta_x, \nonumber \\
    &\sigma_0\tau_z\zeta_0 \qquad && \sigma_x\tau_z\zeta_y \qquad && \sigma_y\tau_z\zeta_y \qquad && \sigma_z\tau_z\zeta_0.
    \label{eq:U8_generators}
\end{alignat}
The matrices $S_{\mu}\otimes s_{\nu}$ generate a Lie algebra isomorphic to the $U(8)$ algebra, whose conventional generators have the form $g^a\otimes g^b \otimes g^c$, where $g^{a,b,c}$ ($a,b,c \in \{0,x,y,z\}$) are the Pauli matrices. We construct an explicit isomorphism by assigning $s_c \rightarrow g^c$, and $\sigma_a\tau_b\zeta_{b'} \rightarrow g^{f(a)}g^{f(b)}$ according to the following rules:
\begin{align}
    \sigma_0\tau_0\zeta_0 \rightarrow +g^0g^0 &&
    \sigma_x\tau_0\zeta_y \rightarrow +g^0g^x &&
    \sigma_y\tau_0\zeta_y \rightarrow -g^0g^y &&
    \sigma_z\tau_0\zeta_0 \rightarrow -g^0g^z\nonumber \\
    \sigma_0\tau_x\zeta_x \rightarrow -g^xg^z &&
    \sigma_x\tau_x\zeta_z \rightarrow -g^xg^y &&
    \sigma_y\tau_x\zeta_z \rightarrow -g^xg^x &&
    \sigma_z\tau_x\zeta_x \rightarrow -g^xg^0 \nonumber \\
    \sigma_0\tau_y\zeta_x \rightarrow -g^yg^z &&
    \sigma_x\tau_y\zeta_z \rightarrow -g^yg^y &&
    \sigma_y\tau_y\zeta_z \rightarrow -g^yg^x &&
    \sigma_z\tau_y\zeta_x \rightarrow -g^yg^0 \nonumber \\
    \sigma_0\tau_z\zeta_0 \rightarrow +g^zg^0 &&
    \sigma_x\tau_z\zeta_y \rightarrow -g^zg^x &&
    \sigma_y\tau_z\zeta_y \rightarrow +g^zg^y &&
    \sigma_z\tau_z\zeta_0 \rightarrow +g^zg^z
    \label{eq:U8_isomorphism}
\end{align}
A direct calculation demonstrates that the Lie bracket operation is preserved. We note that the symmetry generators $\hat{\Sigma}_{\mathbf{R}}^{\mu\nu}$ also commute with the $\hat{D}^\dagger_{\mathbf{R}1+}$ operator, which follows from \cref{eq:U8_symmetry_commutation} and the hermiticity of the Pauli matrices, which entails that $\hat{H}_{J} + \hat{H}_{\tilde{J}}$ also commutes with all the symmetry generators. 
Additionally, we find a 65-th generator $\hat{\Sigma}_{\mathbf{R}}^{65}$ given by
\begin{equation}
    \hat{\Sigma}_{\mathbf{R}}^{65} = \hat{\Psi}^\dagger_{\mathbf{R}}\sigma_z\tau_z\zeta_z s_0 \hat{\Psi}_{\mathbf{R}},
\end{equation}
which commutes with all the $U(8)$-symmetry generators $\hat{\Sigma}_{\mathbf{R}}^{\mu\nu}$ and satisfies
\begin{equation}
    \frac{1}{2}\left[\hat{\Sigma}_{\mathbf{R}}^{65}, \hat{D}_{\mathbf{R}1+}\right] =  \hat{D}_{\mathbf{R}1+}, \qquad \frac{1}{2}\left[\hat{\Sigma}_{\mathbf{R}}^{65}, \hat{D}^\dagger_{\mathbf{R}1+}\right] =  -\hat{D}^\dagger_{\mathbf{R}1+}.
\end{equation}
This implies that $\hat{\Sigma}_{\mathbf{R}}^{65}$ is also a symmetry of $\hat{H}_{J} + \hat{H}_{\tilde{J}}$
\begin{equation}
 \left[\hat{\Sigma}_{\mathbf{R}}^{65}, \hat{H}_{J} + \hat{H}_{\tilde{J}}\right] = 0.
\end{equation}
Introducing matrices $\Sigma_{\mu\nu} = S_{\mu}\otimes s_{\nu}$ and $\Sigma^{65} = \sigma_z\tau_z\zeta_z s_0$ we can rewrite the terms $\hat{H}_{J} + \hat{H}_{\tilde{J}}$ as
\begin{equation}
    \hat{H}_{J} + \hat{H}_{\tilde{J}} = -\frac{J}{16}\sum_{\mathbf{R}}\sum_{\mu\nu}\left(\hat{\Psi}^\dagger_{\mathbf{R}}\Sigma^{65}\Sigma_{\mu\nu}\hat{\Psi}_{\mathbf{R}}\hat{\Psi}^\dagger_{\mathbf{R}}\Sigma^{65}\Sigma_{\mu\nu}\hat{\Psi}_{\mathbf{R}} - \hat{\Psi}^\dagger_{\mathbf{R}}\Sigma_{\mu\nu}\hat{\Psi}_{\mathbf{R}}\hat{\Psi}^\dagger_{\mathbf{R}}\Sigma_{\mu\nu}\hat{\Psi}_{\mathbf{R}}\right) + \frac{J}{2}\sum_{\mathbf{R}}\hat{\Psi}^\dagger_{\mathbf{R}}\hat{\Psi}_{\mathbf{R}}.
\end{equation}
Finally, we note that the terms $\hat{H}_{J} + \hat{H}_{\tilde{J}}$ do not contain the $\hat{c}_{\mathbf{R},a=1,2,\eta,s}$ operators, corresponding to the $\Gamma_3$ irreps. Therefore, $\hat{H}_{J} + \hat{H}_{\tilde{J}}$ is symmetric under spin-valley-flavor rotations within the $\Gamma_3$ irrep subspace. This implies that the resulting symmetry of the $\hat{H}_{J} + \hat{H}_{\tilde{J}}$ interacting terms is
\begin{equation}
    U(8)\times U(8) \times U(1).
\end{equation}

As shown in \cref{app:subsec:symmetry_density-density-interaction}, in the case $U_1 = W_1 = W_3$, an enlarged $U(24)$ symmetry of the density-density term \cref{eq:U24_interaction_Hamiltonian} emerges. Since the $U(8)\times U(8) \times U(1)$ group is a subgroup of $U(24)$, we conclude that with the condition $U_1 = W_1 = W_3$, the total interacting Hamiltonian
\begin{equation}
    \hat{H}_I = \hat{H}_U + \hat{H}_W + \hat{H}_V + \hat{H}_{J} + \hat{H}_{\tilde{J}}
\end{equation}
possesses the $U(8)\times U(8) \times U(1)$ symmetry. When the condition $U_1 = W_1 = W_3$ is not satisfied, the symmetry is only approximate.

\section{Additional numerical results}\label{app:sec:numerics}
In this appendix, we outline the method used for numerically obtaining the THF model parameters across a large parameter space. We start by reviewing the computational procedure employed by Ref.~\cite{SON22}, on which our method is also based and which uses the Wannier90 software~\cite{MAR97b,SOU01a,PIZ20} to obtain the $f$-electron wave functions and the THF parameters. We then explain how the former is adapted to work seamlessly across an extensive parameter space. We then present additional numerical results concerning the THF parameters and/or their analytical approximations, which were not discussed in \cref{sec:numerical_simulations}.

\subsection{Details of numerical calculation}\label{app:subsec:numeric_details}

\begin{figure}
    \centering
    \includegraphics{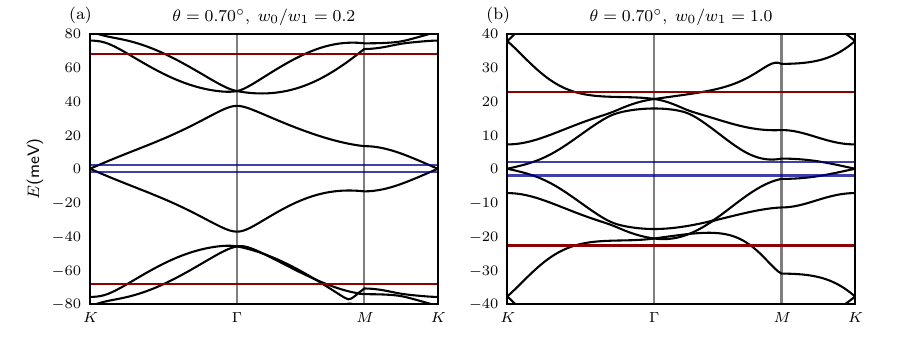}\subfloat{\label{fig:setting_windows:a}}\subfloat{\label{fig:setting_windows:b}}\caption{The BM band structure for the two different cases discussed around \cref{app:eqn:position_of_en_win}, illustrating the choice of the frozen (blue) and disentanglement (red) windows in Wannier90. In (a), the 
    disentanglement window is set to include the $\Gamma_3$ irreps of the four closest remote bands, but exclude the $K_M$ and $M_M$ points of the latter. The band structure in (a) constitutes the typical band structure across the parameter space we explore in this work. This prescription fails for small values of the twist angle $\theta \gtrsim \SI{0.75}{\degree}$ and large values of $w_0 / w_1 \gtrsim 0.9$ when the $\Gamma_3$ irreps of the four closest remote bands are higher in energy than the $K_M$ or $M_M$ points of the remote bands. A typical band structure corresponding to this case is shown in (b). In this case, we set the disentanglement window to include the $\Gamma_3$ irreps. In both cases, the frozen energy window is set to include the $K_M$ points of the active bands.}
    \label{fig:setting_windows}
\end{figure}

Following Ref.~\cite{SON22}, we build trial Wannier functions for the $f$-electrons transforming as  $p_x \pm i p_y$ orbitals at the $1a$ Wyckoff position under the TBG symmetry group. Denoting by $\alpha=1$ for the $p_x + ip_i$ ($\alpha=2$ for $p_x - ip_y$) orbital component of the Wannier state at the lattice site $\mathbf{R}$, valley $\eta$ and spin $s$, we write for the trial functions~\cite{SON22}
\begin{equation}
    \ket{W'_{\mathbf{R},\alpha,\eta,s}} = \sqrt{\frac{2\pi\lambda_0^2}{\Omega_{\textrm{tot}}}}\sum_{l=\pm}\sum_{\mathbf{k}\in\textrm{MBZ}}\sum_{\mathbf{Q}\in\mathcal{Q}_{l\eta}}e^{(-1)^{\alpha + 1}i\frac{\pi}{4}l\eta - i\mathbf{k}\cdot\mathbf{R} - \frac{1}{2}\lambda_0^2(\mathbf{k} - \mathbf{Q})^2}\ket{\mathbf{k},\mathbf{Q},\alpha,\eta,s},
    \label{eq:trial_wave_function}
\end{equation}
where $\ket{\mathbf{k},\mathbf{Q},\alpha,\eta,s}$ is the BM model basis states of the sublattice $\alpha\in\{1,2\}$. In \cref{eq:trial_wave_function} $\lambda_0$ is the trial value for the spread of the Wannier state. Following Ref.~\cite{SON22}, we set $\lambda_0 = 0.1|\mathbf{a}_{M1}|$.

We project the Wannier orbitals on the 16 bands $\ket{\mathbf{k},n,\eta,s} = \hat{c}^\dagger_{\mathbf{k},n,\eta,s}\ket{0}$ of the BM model closest to charge neutrality, thus obtaining the trial overlap functions $A^{\eta}_{n,\alpha(k)} = \bra{\mathbf{k},n,\eta,s}\ket{W'_{\mathbf{R},\alpha,\eta,s}} (n=\{\pm1, \ldots, \pm8\}, \alpha=1,2,\ \eta=\pm)$. The trial overlap function are then fed into the Wannier90 software~\cite{MAR97b,SOU01a,PIZ20} to build maximally-localized Wannier using an almost identical procedure as the one in Ref.~\cite{SON22}.

The only difference between our approach and the one used in Ref.~\cite{SON22} concerns the choice of the disentanglement and frozen energy windows\footnote{We remind the reader that the Wannier90 package will only project the Wannier functions on the bands within the \emph{disentanglement} energy window. This means that the overlap between the bands within the disentanglement window and the Wannier functions can assume any value between zero and one. In contrast, the bands within the \emph{frozen} energy window will have an overlap of one with the Wannier states. In other words, the bands within the disentanglement window will be only partially supported by the Wannier functions, whereas the bands within the frozen window will be \emph{fully} supported by the Wannier functions.} for the disentanglement and wannierization steps of the Wannier90 package~\cite{PIZ20}. In contrast to Ref.~\cite{SON22}, which employs the same energy windows for different tunneling amplitude ratios ($w_0/w_1$), the much larger parameter space that we consider here requires us to adjust the disentanglement energy window for each value of the twist angle and tunneling amplitude ratio, using a procedure which will be described below. Firstly, we note that the $p_x \pm i p_y$ orbitals located at the $1a$ Wyckoff position ($p_x,p_y @ 1a$ orbitals) induce a $\Gamma_3$ irrep at the $\Gamma_M$ point of the moir\'e Brillouin zone. In order to ensure that the maximally localized Wannier functions obtained from Wannier90 obey the appropriate symmetries corresponding to $p_x,p_y @ 1a$ orbitals, we must fix the disentanglement energy window such that it includes the $\Gamma_3$ irreps of the BM model near charge neutrality (which are contributed by the remote bands). Secondly, the $f$-electrons must be supported entirely on the active bands at the $K_M$ and $M_M$ points. To achieve this, we fix the \emph{frozen} ({\it i.e.}{}, not the disentanglement) energy window to be located between $\pm \SI{0.0005}{\milli\eV}$ (thus ensuring that the $f$-electrons span the two Dirac cones at the $K_M$ and $K'_M$ points of the active bands). Additionally, we always exclude the $M_M$, $K_M$, and $K'_M$ points of the remote bands from the disentanglement window, thus ensuring that the $f$-electrons are only supported on the active band states at these points of the moir\'e Brillouin zone. 
Finally, to achieve a smoothly defined set of Wannier functions over the entire parameter space and to ensure that the correct irreps are included or excluded from the disentanglement energy window (as explained above), we set the disentanglement energy window to be located between $E^{-}_{\textrm{dis}}$ and $E^{+}_{\textrm{dis}}$, as shown in \cref{fig:setting_windows:a}, with
\begin{equation}
    E^{\pm}_{\textrm{dis}} = \pm x\min\left[ \abs{\epsilon_{ \pm 2,+}\left( K_M \right)}, \abs{\epsilon_{ \pm 2,+} \left(M_M \right)} \right] + (1-x) E_{\Gamma_3,\pm},
    \label{app:eqn:position_of_en_win}
\end{equation}
where $E_{\Gamma_3,+}$ ($E_{\Gamma_3,-}$) represents the energy of the lowest (highest) $\Gamma_3$ irrep located above (below) charge neutrality and we set $x=0.89$. It is worth noting, however, that for small values of the twist angle $\theta \sim \SI{0.75}{\degree}$ and large values of $w_0 / w_1 \sim 0.9$, $\abs{E_{\Gamma_3,\pm}} > \abs{\epsilon_{ \pm 2,+}\left( K_M \right)}$ and $\abs{E_{\Gamma_3,\pm}} > \abs{\epsilon_{ \pm 2,+}\left( M_M \right)}$. In this case, which is illustrated in \cref{fig:setting_windows:b}, we set the window according to $E^{\pm}_{\textrm{dis}} = \pm 1.2 \times |E_{\Gamma_3}|$, in order to include the correct states.

\subsection{Additional numerical results: single-particle parameters}\label{app:subsec:additional_numerics_single_particle}

\begin{figure}[t!]
    \includegraphics[width=\textwidth]{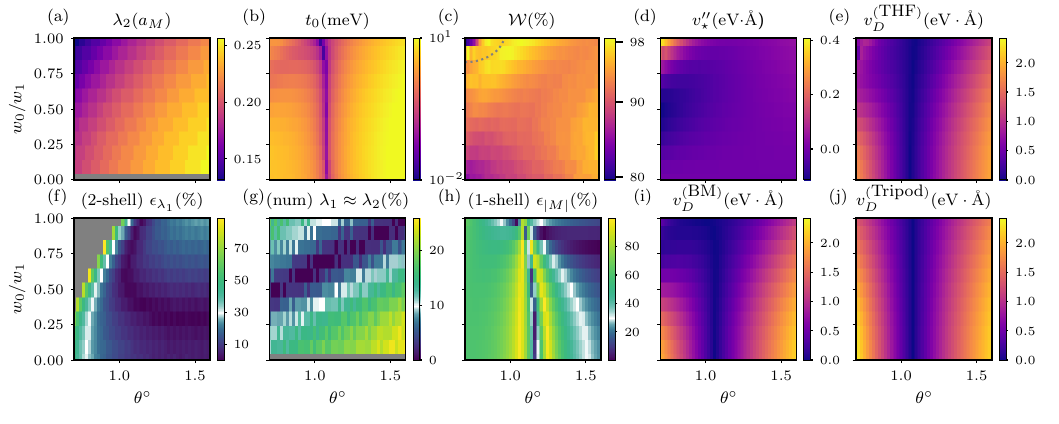}
    \subfloat{\label{fig:G1:a}}\subfloat{\label{fig:G1:b}}\subfloat{\label{fig:G1:c}}\subfloat{\label{fig:G1:d}}\subfloat{\label{fig:G1:e}}\subfloat{\label{fig:G1:f}}\subfloat{\label{fig:G1:g}}\subfloat{\label{fig:G1:h}}\subfloat{\label{fig:G1:i}}\subfloat{\label{fig:G1:j}}\caption{Additional numerical results concerning the single-particle parameters of the THF model as a function of the ratio $w_0/w_1$ and twist angle $\theta$. We employ $v_F = \SI{5.944}{\eV \angstrom}$, $\abs{\vec{K}}=\SI{1.703}{\angstrom^{-1}}$, and $w_1=\SI{110}{\milli\eV}$. The numerically obtained values of $\lambda_2$, $t_0$ (the nearest neighbor hopping amplitude of the THF $f$-electrons), $\mathcal{W}$ (the total weight of the $f$-electrons on the active TBG bands), and $v_{\star}^{\prime\prime}$ are shown in (a)-(d), respectively. Note that $\lambda_2$ is not defined for $w_0/w_1 = 0.0$, as the weight of the $\alpha=1$ Wannier orbital on the $\beta=2$ graphene sublattice vanishes in the chiral limit~\cite{SON22}. As a result, we represent the corresponding region by a gray line. In (c), the gray arc serves as a rough guide for the region where $\abs{E_{\Gamma_3,\pm}} > \abs{\epsilon_{ \pm 2,+}\left( K_M \right)}, \abs{\epsilon_{ \pm 2,+}\left( M_M \right)}$ [see \cref{fig:setting_windows:b} for a typical band structure in this region]. In this region, the $f$-electrons are also contributed by the remote TBG bands at the $M_M$ and $K_M$  points, causing a relative decrease in $\mathcal{W}$. We show the Dirac velocity of the active TBG bands at the $K_M$ point obtained from the THF model through \cref{eq:THF_Dirac_velocity_final} in (e). The bottom rows assess the validity of various THF single-particle parameter approximations. The relative error of the two-shell Tripod model approximation of the $\lambda_1$ parameter and  one-shell Hexagon model approximation of the $M$ parameter are shown in (e) and (g), respectively. The gray area in (e) denotes the region where the expression in \cref{eq:tripod_lambda_2shell_approximation} is undefined. The relative difference between the numerically-computed $f$-electron spread parameters, $|\lambda_1 - \lambda_2| / \text{max}(\lambda_1,\lambda_2)$ is shown in (f). As $\lambda_2$ is undefined in the chiral limit, the corresponding region is represented by a gray line in (f). Finally,
    we plot the Dirac velocity of the active TBG bands at the $K_M$ point obtained from the BM model numerically and its one-shell Tripod approximation in (i) and (j), respectively.}
    \label{fig:G1}
\end{figure}

In \cref{fig:G1}, we provide numerical results on the single-particle THF model that were not discussed in the main text. In the upper row of \cref{fig:G1}, we plot the numerically obtained values of the spread of the $\alpha = 1$ Wannier orbital on the $\beta = 2$ graphene sublattice $\lambda_2$, the hopping parameter $t_0$, the weight of the $f$-electron states on the active bands (denoted by $\mathcal{W}$), the $f$-$c$ electron hybridization parameter $v_{\star}''$, as well as the Dirac velocity $v_D^{\textrm(THF)}$ of the TBG active bands obtained from the THF model according to \cref{eq:THF_Dirac_velocity_final}.

First, we note that in the chiral limit ($w_0/w_1 = 0$), the heavy fermion wave functions become graphene-sublattice polarized~\cite{SON22} and, as a result, the $\alpha = 2$ amplitude of the $\beta = 1$ $f$-electron orbital vanishes. In this case [marked by a gray line in \cref{fig:G1:a}], one cannot define the spread $\lambda_2$. Second, we clearly observe that the hopping parameter $t_0$ goes to zero in the vicinity of the magic-angle, which indicates the validity of the model in this parameter region. In general, the $f$-electrons show a large overlap with the active TBG bands across the entire parameter space. One exception is the region corresponding to the upper-left corner of the $\mathcal{W}$ plot from \cref{fig:G1:c}, where, as discussed in \cref{app:subsec:numeric_details}, the energy of the $\Gamma_3$ irrep gets larger than the corresponding energies of the active bands at the $K_M$ and $M_M$ points. As such, the $f$-electrons are also supported on the remote bands, leading to a drop in the weight parameter $\mathcal{W}$. For the $v_{\star}''$ parameter, we confirm that it is generally smaller than $v_{\star}$ and $v_{\star}'$ and thus it is a good approximation to neglect it. Finally, $v_D^{\textrm(THF)}$ vanishes at the magic angle for any tunneling amplitude ratio.

In the lower row of \cref{fig:G1}, we assess the validity of various analytical approximations of the THF single-particle parameters. First, in \cref{fig:G1:f}, we plot the relative error of approximation \cref{eq:tripod_lambda_2shell_approximation} for the spread parameter $\lambda_1^{(\textrm{2-shell})}$ obtained in \cref{app:subsec:analytic_forbitals_tripod} within the two-shell Tripod model. This approximation works well across a large region of the phase space we consider, except for small angles $\theta \sim \SI{0.75}{\degree}$ and tunneling amplitude ratios close to unity $w_0/w_1 \sim 1$, for which the expression in \cref{eq:tripod_lambda_2shell_approximation} is invalid. Next, we assess the validity of the assumption $\lambda_1 \approx \lambda_2$ used in deriving the approximations in \cref{sec:analytics_single_particle}. In \cref{fig:G1:g} we plot the relative difference between the two spread parameters, {\it i.e.}{} $|\lambda_1 - \lambda_2| / \text{max}(\lambda_1,\lambda_2)$, where $\lambda_1$ and $\lambda_2$ are both obtained numerically. We see that the approximation $\lambda_1 \approx \lambda_2$ works remarkably well, with the two spread parameters being within $30\%$ of one another across the entire phase diagram. It is worth noting that in the heavy fermion model proposed by Ref.~\cite{SHI22a}, the corresponding local orbitals also feature equal spreads within the two graphene sublattices.  

In \cref{fig:G1:h}, we plot the relative error of the one-shell Hexagon model approximation $M^{\textrm{1-shell}}$ from \cref{eq:M_app_approximation_one_shell}. Comparing \cref{fig:G1:h} with \cref{fig:main_sp_params:h}, we see that the two-shell Hexagon model approximation $M^{\textrm{2-shell}}$ from \cref{eq:M_approximation_two_shell} agrees much better with the numerical results. Finally, we plot the Dirac velocity of the TBG active bands obtained from the BM model and its one-shell Tripod approximation in \cref{fig:G1:i,fig:G1:j}, respectively. We see that the THF model approximates this Dirac velocity much better than the one-shell Tripod model (particularly at lower angles and large tunneling amplitude ratio).

\subsection{Additional numerical results: interaction parameters}\label{app:subsec:additional_numerics_many-body}

\begin{figure}[t!]
    \includegraphics[width=4.5 in]{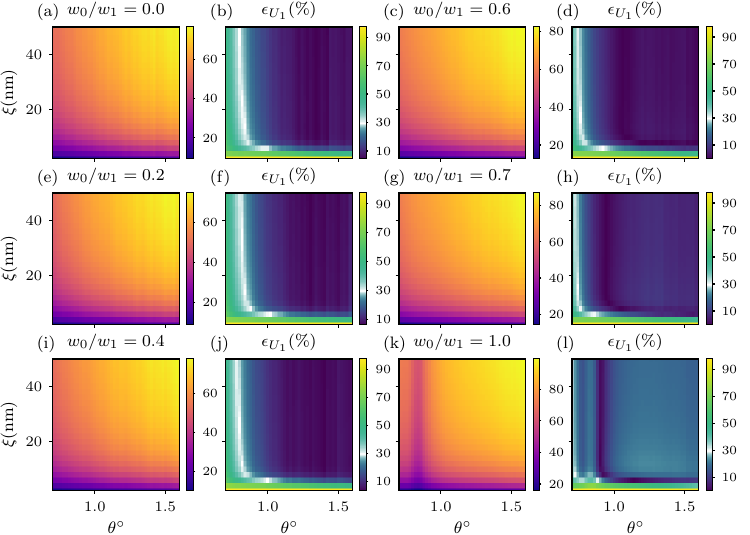}
    \subfloat{\label{fig:G3:a}}\subfloat{\label{fig:G3:b}}\subfloat{\label{fig:G3:c}}\subfloat{\label{fig:G3:d}}\subfloat{\label{fig:G3:e}}\subfloat{\label{fig:G3:f}}\subfloat{\label{fig:G3:g}}\subfloat{\label{fig:G3:h}}\subfloat{\label{fig:G3:i}}\subfloat{\label{fig:G3:j}}\subfloat{\label{fig:G3:k}}\subfloat{\label{fig:G3:l}}\caption{Additional numerical results on the $U_1$ interaction parameter of the THF model as a function of the twist angle $\theta$ and screening length $\xi$ for different values of the tunneling amplitude ratio $w_0/w_1 \in \{0.0,0.2,0.4,0.6,0.7,1.0\}$. We employ $v_F = \SI{5.944}{\eV \angstrom}$, $\abs{\vec{K}}=\SI{1.703}{\angstrom^{-1}}$, and $w_1=\SI{110}{\milli\eV}$. The Coulomb interaction scale $U_{\xi}$ is chosen such that $U_{\xi} = \SI{24}{\milli \eV}$ at $\xi = \SI{10}{\nano\meter}$. We plot the numerically calculated values of $U_1$ in (a), (c), (e), (g), (i), and (k), while in (b), (d), (f), (h), (j), and (l) we present the relative error of the approximation $U_1^{\textrm{approx.}}$ from \cref{eq:U1_final}.} 
    \label{fig:G3}
\end{figure}

\begin{figure}[b!]
    \includegraphics[width=4.5 in]{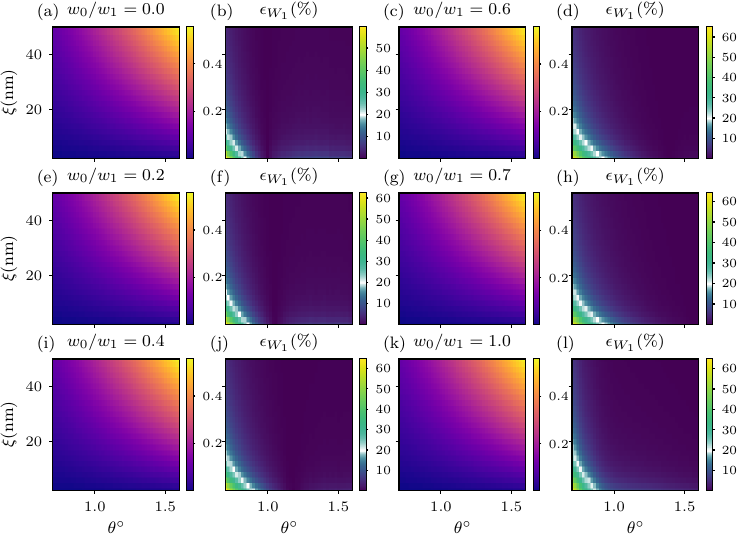}
    \subfloat{\label{fig:G4:a}}\subfloat{\label{fig:G4:b}}\subfloat{\label{fig:G4:c}}\subfloat{\label{fig:G4:d}}\subfloat{\label{fig:G4:e}}\subfloat{\label{fig:G4:f}}\subfloat{\label{fig:G4:g}}\subfloat{\label{fig:G4:h}}\subfloat{\label{fig:G4:i}}\subfloat{\label{fig:G4:j}}\subfloat{\label{fig:G4:k}}\subfloat{\label{fig:G4:l}}\caption{Additional numerical results concerning the $W_1$ interaction parameter of the THF model as a function of the twist angle $\theta$ and screening length $\xi$ for different values of the tunneling amplitude ratio $w_0/w_1 \in \{0.0,0.2,0.4,0.6,0.7,1.0\}$. We employ $v_F = \SI{5.944}{\eV \angstrom}$, $\abs{\vec{K}}=\SI{1.703}{\angstrom^{-1}}$, and $w_1=\SI{110}{\milli\eV}$. The Coulomb interaction scale $U_{\xi}$ is chosen such that $U_{\xi} = \SI{24}{\milli \eV}$ at $\xi = \SI{10}{\nano\meter}$. We plot the numerically calculated values of $W_1$ in panels (a), (c), (e), (g), (i), and (k), while in panels (b), (d), (f), (h), (j), and (l) we present the relative error of the approximation $W_1^{\textrm{2nd approx.}}$ from \cref{eq:W1_W3_1st}} 
    \label{fig:G4}
\end{figure}

\begin{figure}[b!]
    \includegraphics[width=4.5 in]{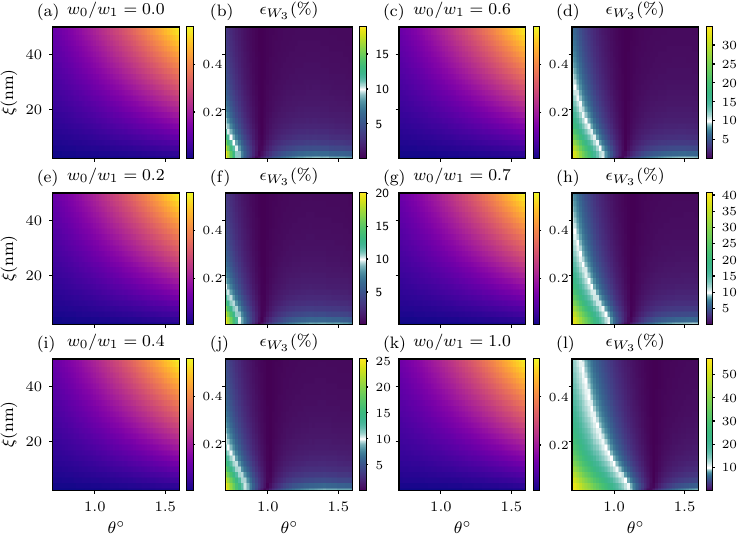}
    \subfloat{\label{fig:G5:a}}\subfloat{\label{fig:G5:b}}\subfloat{\label{fig:G5:c}}\subfloat{\label{fig:G5:d}}\subfloat{\label{fig:G5:e}}\subfloat{\label{fig:G5:f}}\subfloat{\label{fig:G5:g}}\subfloat{\label{fig:G5:h}}\subfloat{\label{fig:G5:i}}\subfloat{\label{fig:G5:j}}\subfloat{\label{fig:G5:k}}\subfloat{\label{fig:G5:l}}\caption{Additional numerical results concerning the $W_3$ interaction parameter of the THF model as a function of the twist angle $\theta$ and screening length $\xi$ for different values of the tunneling amplitude ratio $w_0/w_1 \in \{0.0,0.2,0.4,0.6,0.7,1.0\}$. We employ $v_F = \SI{5.944}{\eV \angstrom}$, $\abs{\vec{K}}=\SI{1.703}{\angstrom^{-1}}$, and $w_1=\SI{110}{\milli\eV}$. The Coulomb interaction scale $U_{\xi}$ is chosen such that $U_{\xi} = \SI{24}{\milli \eV}$ at $\xi = \SI{10}{\nano\meter}$. We plot the numerically calculated values of $W_1$ in panels (a), (c), (e), (g), (i), and (k), while in panels (b), (d), (f), (h), (j), and (l) we present the relative error of the approximation $W_3^{\textrm{1st approx.}}$ from \cref{eq:fc_interaction_strength_approx_13}} 
    \label{fig:G5}
\end{figure}

\begin{figure}[t!]
    \includegraphics[width=6.354 in]{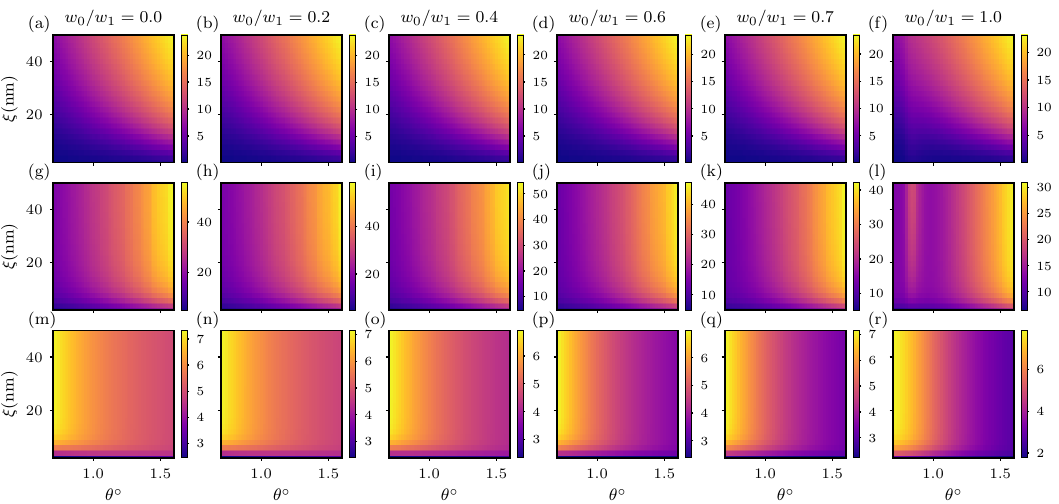}
    \subfloat{\label{fig:G6:a}}\subfloat{\label{fig:G6:b}}\subfloat{\label{fig:G6:c}}\subfloat{\label{fig:G6:d}}\subfloat{\label{fig:G6:e}}\subfloat{\label{fig:G6:f}}\subfloat{\label{fig:G6:g}}\subfloat{\label{fig:G6:h}}\subfloat{\label{fig:G6:i}}\subfloat{\label{fig:G6:j}}\subfloat{\label{fig:G6:k}}\subfloat{\label{fig:G6:l}}\subfloat{\label{fig:G6:m}}\subfloat{\label{fig:G6:n}}\subfloat{\label{fig:G6:o}}\subfloat{\label{fig:G6:p}}\subfloat{\label{fig:G6:q}}\subfloat{\label{fig:G6:r}}\caption{Additional numerical results concerning the $U_2$, $J$ and $K$ interaction parameters of the THF model as a function of the twist angle $\theta$ and screening length $\xi$ for different values of the tunneling amplitude ratio $w_0/w_1 \in \{0.0,0.2,0.4,0.6,0.7,1.0\}$. We employ $v_F = \SI{5.944}{\eV \angstrom}$, $\abs{\vec{K}}=\SI{1.703}{\angstrom^{-1}}$, and $w_1=\SI{110}{\milli\eV}$. The Coulomb interaction scale $U_{\xi}$ is chosen such that $U_{\xi} = \SI{24}{\milli \eV}$ at $\xi = \SI{10}{\nano\meter}$. In (a)-(f) we plot the $U_2$ parameter, in (g)-(l), the $J$ parameter and in (m)-(r), the $K$ parameter.} 
    \label{fig:G6}
\end{figure}

In this section, we supplement our analysis of the THF interaction Hamiltonian parameters from \cref{subsec:numerical_many_body} with additional numerical results obtained at other tunneling amplitude ratios $w_0/w_1 \in \left\lbrace 0.0, 0.2, 0.4, 0.6, 0.7, 1.0 \right\rbrace$. In \cref{fig:G3}, we plot the dependence of the onsite repulsion strength $U_1$ on the twist angle $\theta$ and the screening length $\xi$, and assess the validity of the approximation from \cref{eq:U1_final}, for which we used the $\lambda_1^{\textrm{1-shell}}$ and $(\alpha_1/\alpha_2)^{\textrm{1-shell}}$ approximations from \cref{eq:lambda1_alpha_ratio_1shell_approximation}. The variation of the $W_1$ and $W_3$ interaction parameters across the same parameter space is shown in \cref{fig:G4} and \cref{fig:G5}, respectively, together with the relative errors of the corresponding analytical approximations derived in \cref{app:sec:analytic_many-body}. For completeness, we also plot the variation of the NN repulsion strength ($U_2$), exchange interaction strength ($J$), and density-hybridization repulsion strength ($K$)~\cite{SON22} in the three rows of \cref{fig:G6}. 

\begin{figure}[h!]
    \includegraphics[width=3.708 in]{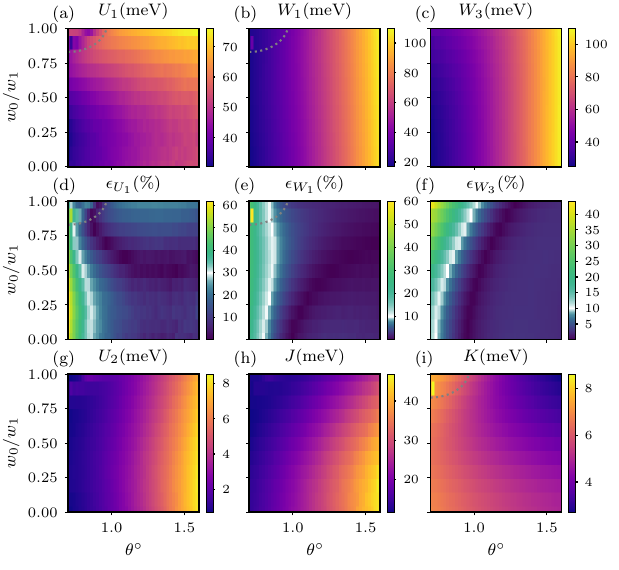}
    \subfloat{\label{fig:G2:a}}\subfloat{\label{fig:G2:b}}\subfloat{\label{fig:G2:c}}\subfloat{\label{fig:G2:d}}\subfloat{\label{fig:G2:e}}\subfloat{\label{fig:G2:f}}\subfloat{\label{fig:G2:g}}\subfloat{\label{fig:G2:h}}\subfloat{\label{fig:G2:i}}\caption{Additional numerical results concerning the interaction parameters of the THF model as a function of the twist angle $\theta$ and tunneling amplitude ratio $w_0/w_1$. We employ $v_F = \SI{5.944}{\eV \angstrom}$, $\abs{\vec{K}}=\SI{1.703}{\angstrom^{-1}}$, and $w_1=\SI{110}{\milli\eV}$. The Coulomb interaction scale $U_{\xi}$ is chosen such that $U_{\xi} = \SI{24}{\milli \eV}$ at $\xi = \SI{10}{\nano\meter}$. In (a)-(c) and (g)-(i) we plot the numerically obtained values for the $U_1$, $W_1$, $W_3$, $U_2$, $J$ and $K$ parameters, while in (d)-(f) we assess the validity of the THF interaction parameter approximations. The relative errors are provided for the $U_1$, $W_1$ and $W_3$ parameters using the approximations from \cref{eq:U1_final,eq:fc_interaction_strength_approx_13,eq:W1_W3_1st}. In (b), (d), (e) and (i), the white arc serves as a rough guide for the region where $\abs{E_{\Gamma_3,\pm}} > \abs{\epsilon_{ \pm 2,+}\left( K_M \right)}, \abs{\epsilon_{ \pm 2,+}\left( M_M \right)}$, and, as a consequence, the $f$-electrons are also contributed by the remote bands at the $M_M$ and $K_M$  points.} 
    \label{fig:G2}
\end{figure}

In \cref{fig:main_int_params} of the main text, we plotted the interaction parameters $U_1$, $W_{1,3}$, $J$ and their corresponding relative errors as the function of the twist angle $\theta$ and the screening length $\xi$ at a fixed tunneling amplitude ratio $w_0/w_1 = 0.8$. In \cref{fig:G2}, we fix the screening length to a typical value $\xi = \SI{10}{\nano\meter}$ instead and plot the $U_1$, $W_1$, $W_3$, $U_2$, $J$, and $K$ parameters as the function of the twist angle $\theta$ and the amplitude tunneling ratio $w_0/w_1$. For $U_1$, $W_1$ and $W_3$, we also show the validity of the corresponding analytical approximations from \cref{eq:U1_final,eq:fc_interaction_strength_approx_13,eq:W1_W3_1st}. We find that the all the interacting strengths increase with the twist angle $\theta$, except for the density-hybridization repulsion strength $K$, which decreases.

\FloatBarrier
\newpage
\clearpage
\clearpage
\section{Tables with numerical results and band-structure comparison}\label{app:sec:tables}
This appendix provides a detailed and complete picture of all the numerically-obtained single-particle and interaction THF parameters for different values of the twist angle $\SI{0.70}{\degree} \leq \theta \leq \SI{1.60}{\degree}$, tunneling amplitude ratio $ 0.0 \leq w_0/w_1 \leq 1.0$, and screening length $ \SI{2}{\nano\meter} \leq \xi \leq \SI{50}{\nano\meter}$. 

We employ the method devised by Ref.~\cite{SON22}, which was adapted as explained in \cref{app:subsec:numeric_details} to work across the larger parameter space that we consider in this work. For the BM model used in obtaining the TBG bands used in the Wannierization stage, we employ $v_F = \SI{5.944}{\eV \angstrom}$, $\abs{\vec{K}}=\SI{1.703}{\angstrom^{-1}}$, and $w_1=\SI{110}{\milli\eV}$ (see \cref{app:sec:BM_model}), while changing the twist angle $\theta$ and the tunneling amplitude ratio $w_0/w_1$. 

We also assume that the electron-electron interaction is given by the double-gated screened Coulomb potential from \cref{eq:Coulomb_interaction}, and obtain the THF interaction parameters at different screening lengths $\xi$. The Coulomb interaction scale $U_{\xi}$ is chosen such that $U_{\xi} = \SI{24}{\milli \eV}$ at $\xi = \SI{10}{\nano\meter}$, and scaled accordingly at different screening lengths (see \cref{app:sec:HF_interaction}).

For convenience, the format of the tables and figures for different twist angles $\theta$ is the same. For each angle $\theta$, the results are summarized on two pages. On the top-left part of the first page, a table summarizes the single-particle parameters of the THF model at different tunneling ratios $w_0/w_1$. For each tunneling ratio, we also report the ratio between $\gamma$ (the gap between the active and remote TBG bands) and the onsite ($U_1$) and NN ($U_2$) repulsion parameters. A plot of the BM and THF band structures at $w_0/w_1=0.8$ is shown on the top-right part of the page. Finally, the parameters of the THF interaction Hamiltonian are tabulated for different tunneling ratios and screening lengths on the bottom of the first page and on the second page.



\section*{Supplementary material}

Total weight of the heavy-fermions on the active bands after wannierisation: 93.4%

FIG. S18. Band structures of the BM and THF models near charge neutrality for $w_0/w_1 = 0.8$, depicted by lines and crosses, respectively. The BM bands are colored according to the weight of the $f$-electron wave function on them. We use the same BM parameters as in Table S4.

| $w_0/w_1$ | $\alpha_1$ | $\alpha_2$ | $\lambda_1$ | $\lambda_2$ | $\lambda$ | $\mathcal{W}$ | $t_0$ | $\gamma$ | $M$ | $v_\star$ | $v_\star'$ | $v_\star''$ | $\gamma/U_1$ | $\gamma/U_2$ |
|---|---|---|---|---|---|---|---|---|---|---|---|---|---|---|
| 1.00 | 0.745 | 0.668 | 0.107 | 0.131 | 0.394 | 0.796 | 1.797 | 20.663 | 17.854 | -0.436 | 0.293 | 0.4078 | 0.39 | 30.29 |
| 0.90 | 0.753 | 0.658 | 0.119 | 0.137 | 0.531 | 0.931 | 1.443 | 18.255 | 21.792 | -1.087 | 0.592 | 0.2367 | 0.45 | 14.64 |
| 0.80 | 0.766 | 0.643 | 0.131 | 0.143 | 0.451 | 0.934 | 0.864 | 12.123 | 25.441 | -1.731 | 0.873 | 0.0788 | 0.28 | 12.60 |
| 0.70 | 0.790 | 0.613 | 0.143 | 0.155 | 0.356 | 0.944 | 0.887 | 3.238 | 28.632 | -2.288 | 1.085 | -0.0327 | 0.07 | 4.40 |
| 0.60 | 0.824 | 0.566 | 0.155 | 0.161 | 0.346 | 0.925 | 1.500 | -7.226 | 31.307 | -2.740 | 1.211 | -0.0940 | -0.18 | -9.89 |
| 0.50 | 0.865 | 0.503 | 0.167 | 0.173 | 0.350 | 0.919 | 2.335 | -18.227 | 33.470 | -3.098 | 1.250 | -0.1141 | -0.50 | -22.96 |
| 0.40 | 0.905 | 0.425 | 0.185 | 0.179 | 0.358 | 0.916 | 3.135 | -28.886 | 35.157 | -3.379 | 1.197 | -0.1042 | -0.86 | -34.11 |
| 0.30 | 0.943 | 0.334 | 0.197 | 0.185 | 0.373 | 0.911 | 3.885 | -38.410 | 36.414 | -3.596 | 1.045 | -0.0755 | -1.24 | -42.21 |
| 0.20 | 0.972 | 0.233 | 0.209 | 0.185 | 0.368 | 0.894 | 4.224 | -46.025 | 37.279 | -3.756 | 0.786 | -0.0402 | -1.51 | -52.37 |
| 0.10 | 0.993 | 0.121 | 0.215 | 0.191 | 0.357 | 0.877 | 4.364 | -50.990 | 37.785 | -3.858 | 0.426 | -0.0113 | -1.66 | -62.86 |
| 0.00 | 1.000 | 0.000 | 0.221 | 0.000 | 0.337 | 0.855 | 4.129 | -52.722 | 37.952 | -3.893 | 0.000 | -0.0000 | -1.64 | -76.02 |

TABLE S4. Parameters of the $f$-electron Wannier functions and the THF single-particle Hamiltonian for different values of the tunneling ratio $w_0/w_1$. $\mathcal{W}$ denotes the total weight of the THF $f$-electrons on the active bands. In computing the ratios $\gamma/U_1$ and $\gamma/U_2$, we employ the on-site and nearest-neighbor repulsion parameters $U_1$ and $U_2$ obtained numerically for $\xi = 10$ nm, as given in Table S5. We employ $v_F = 5.944$ eV·Å, $|\mathbf{K}| = 1.703$ Å$^{-1}$, $w_1 = 110$ meV, and $\theta = 0.70°$ for the BM model.

| $w_0/w_1$ | $\xi$/nm | 2 | 4 | 6 | 8 | 10 | 12 | 14 | 16 | 18 | 20 | 22 | 24 | 26 | 28 | 30 | 32 | 34 | 36 | 38 | 40 | 42 | 44 | 46 | 48 | 50 |
|---|---|---|---|---|---|---|---|---|---|---|---|---|---|---|---|---|---|---|---|---|---|---|---|---|---|---|
| 1.0 | $U_1$ | 18.79 | 32.48 | 41.67 | 47.98 | 52.51 | 55.89 | 58.50 | 60.58 | 62.27 | 63.68 | 64.86 | 65.87 | 66.75 | 67.51 | 68.19 | 68.78 | 69.32 | 69.80 | 70.23 | 70.63 | 70.98 | 71.31 | 71.61 | 71.89 | 72.15 |
| | $U_2$ | 0.09 | 0.21 | 0.34 | 0.49 | 0.68 | 0.93 | 1.22 | 1.56 | 1.91 | 2.29 | 2.66 | 3.03 | 3.40 | 3.75 | 4.08 | 4.40 | 4.71 | 5.00 | 5.27 | 5.53 | 5.77 | 6.00 | 6.22 | 6.42 | 6.61 |
| | $W_{1,2}$ | 2.96 | 6.62 | 10.73 | 14.98 | 19.27 | 23.57 | 27.87 | 32.16 | 36.46 | 40.76 | 45.05 | 49.35 | 53.64 | 57.94 | 62.24 | 66.53 | 70.83 | 75.12 | 79.42 | 83.72 | 88.01 | 92.31 | 96.60 | 100.90 | 105.20 |
| | $W_{3,4}$ | 9.94 | 19.21 | 26.88 | 33.11 | 38.44 | 43.25 | 47.81 | 52.23 | 56.59 | 60.91 | 65.23 | 69.53 | 73.83 | 78.13 | 82.42 | 86.72 | 91.01 | 95.31 | 99.61 | 103.90 | 108.20 | 112.50 | 116.79 | 121.09 | 125.38 |
| | $J$ | 6.47 | 9.86 | 11.36 | 12.06 | 12.39 | 12.56 | 12.65 | 12.70 | 12.73 | 12.74 | 12.75 | 12.76 | 12.76 | 12.76 | 12.76 | 12.76 | 12.76 | 12.77 | 12.77 | 12.77 | 12.77 | 12.77 | 12.77 | 12.77 | 12.77 |
| | $K$ | 5.32 | 6.57 | 7.22 | 7.55 | 7.71 | 7.79 | 7.82 | 7.84 | 7.85 | 7.85 | 7.86 | 7.86 | 7.86 | 7.86 | 7.86 | 7.86 | 7.86 | 7.86 | 7.86 | 7.86 | 7.86 | 7.86 | 7.86 | 7.86 | 7.86 |
| 0.9 | $U_1$ | 14.27 | 24.91 | 32.21 | 37.30 | 41.01 | 43.80 | 45.99 | 47.76 | 49.21 | 50.42 | 51.46 | 52.35 | 53.13 | 53.82 | 54.43 | 54.97 | 55.46 | 55.90 | 56.30 | 56.67 | 57.00 | 57.31 | 57.60 | 57.86 | 58.10 |
| | $U_2$ | 0.26 | 0.52 | 0.76 | 0.99 | 1.25 | 1.53 | 1.85 | 2.19 | 2.55 | 2.91 | 3.27 | 3.63 | 3.98 | 4.31 | 4.63 | 4.94 | 5.23 | 5.51 | 5.77 | 6.01 | 6.24 | 6.47 | 6.67 | 6.87 | 7.06 |
| | $W_{1,2}$ | 2.95 | 6.45 | 10.38 | 14.51 | 18.72 | 22.98 | 27.26 | 31.55 | 35.84 | 40.14 | 44.43 | 48.73 | 53.02 | 57.32 | 61.61 | 65.91 | 70.21 | 74.50 | 78.80 | 83.09 | 87.39 | 91.69 | 95.98 | 100.28 | 104.57 |
| | $W_{3,4}$ | 8.72 | 16.92 | 23.89 | 29.74 | 34.87 | 39.58 | 44.09 | 48.49 | 52.83 | 57.15 | 61.46 | 65.76 | 70.06 | 74.36 | 78.65 | 82.95 | 87.25 | 91.54 | 95.84 | 100.14 | 104.43 | 108.73 | 113.02 | 117.32 | 121.61 |
| | $J$ | 5.98 | 9.37 | 11.12 | 12.06 | 12.61 | 12.96 | 13.20 | 13.39 | 13.53 | 13.65 | 13.75 | 13.84 | 13.92 | 13.99 | 14.05 | 14.11 | 14.16 | 14.21 | 14.25 | 14.29 | 14.32 | 14.35 | 14.38 | 14.41 | 14.43 |
| | $K$ | 2.89 | 5.03 | 6.29 | 6.96 | 7.30 | 7.48 | 7.56 | 7.60 | 7.62 | 7.63 | 7.64 | 7.64 | 7.64 | 7.64 | 7.64 | 7.64 | 7.64 | 7.64 | 7.64 | 7.64 | 7.64 | 7.64 | 7.64 | 7.64 | 7.64 |
| 0.8 | $U_1$ | 14.80 | 26.09 | 34.00 | 39.60 | 43.60 | 46.78 | 49.21 | 51.15 | 52.74 | 54.07 | 55.20 | 56.16 | 57.00 | 57.74 | 58.39 | 58.96 | 59.48 | 59.95 | 60.37 | 60.75 | 61.10 | 61.42 | 61.72 | 61.99 | 62.24 |
| | $U_2$ | 0.27 | 0.55 | 0.84 | 1.14 | 1.46 | 1.80 | 2.15 | 2.52 | 2.89 | 3.26 | 3.63 | 3.99 | 4.35 | 4.69 | 5.02 | 5.34 | 5.64 | 5.93 | 6.20 | 6.46 | 6.71 | 6.94 | 7.16 | 7.37 | 7.57 |
| | $W_{1,2}$ | 2.89 | 6.27 | 10.09 | 14.14 | 18.32 | 22.55 | 26.82 | 31.10 | 35.39 | 39.69 | 43.98 | 48.27 | 52.57 | 56.87 | 61.16 | 65.46 | 69.75 | 74.05 | 78.35 | 82.64 | 86.94 | 91.23 | 95.53 | 99.83 | 104.12 |
| | $W_{3,4}$ | 7.97 | 15.56 | 22.14 | 27.78 | 32.80 | 37.46 | 41.94 | 46.32 | 50.66 | 54.98 | 59.29 | 63.59 | 67.89 | 72.18 | 76.48 | 80.78 | 85.07 | 89.37 | 93.66 | 97.96 | 102.25 | 106.55 | 110.85 | 115.14 | 119.44 |
| | $J$ | 5.54 | 8.86 | 10.58 | 11.48 | 11.98 | 12.26 | 12.44 | 12.55 | 12.63 | 12.68 | 12.72 | 12.75 | 12.78 | 12.80 | 12.81 | 12.83 | 12.84 | 12.85 | 12.86 | 12.87 | 12.87 | 12.88 | 12.89 | 12.89 | 12.90 |
| | $K$ | 2.70 | 4.75 | 6.00 | 6.69 | 7.04 | 7.22 | 7.31 | 7.35 | 7.37 | 7.38 | 7.39 | 7.39 | 7.39 | 7.39 | 7.39 | 7.39 | 7.39 | 7.39 | 7.39 | 7.39 | 7.39 | 7.39 | 7.39 | 7.39 | 7.39 |
| 0.7 | $U_1$ | 14.46 | 25.74 | 33.82 | 39.63 | 43.93 | 47.20 | 49.77 | 51.83 | 53.52 | 54.93 | 56.12 | 57.14 | 58.03 | 58.80 | 59.48 | 60.08 | 60.62 | 61.11 | 61.55 | 61.94 | 62.31 | 62.64 | 62.94 | 63.22 | 63.48 |
| | $U_2$ | 0.12 | 0.25 | 0.38 | 0.54 | 0.74 | 0.99 | 1.29 | 1.63 | 1.99 | 2.37 | 2.75 | 3.13 | 3.49 | 3.85 | 4.19 | 4.51 | 4.82 | 5.11 | 5.38 | 5.64 | 5.89 | 6.12 | 6.34 | 6.55 | 6.74 |
| | $W_{1,2}$ | 2.92 | 6.28 | 10.05 | 14.06 | 18.22 | 22.44 | 26.71 | 30.99 | 35.27 | 39.56 | 43.86 | 48.15 | 52.45 | 56.75 | 61.04 | 65.34 | 69.63 | 73.93 | 78.22 | 82.52 | 86.82 | 91.11 | 95.41 | 99.70 | 104.00 |
| | $W_{3,4}$ | 7.25 | 14.23 | 20.42 | 25.84 | 30.74 | 35.34 | 39.79 | 44.16 | 48.49 | 52.81 | 57.11 | 61.41 | 65.71 | 70.01 | 74.30 | 78.60 | 82.90 | 87.19 | 91.49 | 95.78 | 100.08 | 104.38 | 108.67 | 112.97 | 117.26 |
| | $J$ | 5.21 | 8.46 | 10.20 | 11.13 | 11.64 | 11.94 | 12.12 | 12.23 | 12.30 | 12.35 | 12.38 | 12.41 | 12.42 | 12.43 | 12.43 | 12.43 | 12.43 | 12.43 | 12.43 | 12.44 | 12.44 | 12.44 | 12.44 | 12.44 | 12.44 |
| | $K$ | 2.55 | 4.52 | 5.75 | 6.44 | 6.80 | 6.98 | 7.07 | 7.11 | 7.14 | 7.15 | 7.15 | 7.16 | 7.16 | 7.16 | 7.16 | 7.16 | 7.16 | 7.16 | 7.16 | 7.16 | 7.16 | 7.16 | 7.16 | 7.16 | 7.16 |

| $u_0/u_1$ | $\xi$/nm | 2 | 4 | 6 | 8 | 10 | 12 | 14 | 16 | 18 | 20 | 22 | 24 | 26 | 28 | 30 | 32 | 34 | 36 | 38 | 40 | 42 | 44 | 46 | 48 | 50 |
|---|---|---|---|---|---|---|---|---|---|---|---|---|---|---|---|---|---|---|---|---|---|---|---|---|---|---|
| 0.6 | $U_1$ | 12.87 | 23.12 | 30.63 | 36.13 | 40.24 | 43.42 | 45.92 | 47.95 | 49.62 | 51.01 | 52.20 | 53.22 | 54.10 | 54.87 | 55.55 | 56.15 | 56.69 | 57.18 | 57.62 | 58.01 | 58.38 | 58.71 | 59.01 | 59.30 | 59.55 |
| | $U_2$ | 0.11 | 0.23 | 0.36 | 0.52 | 0.73 | 0.99 | 1.30 | 1.64 | 2.01 | 2.40 | 2.78 | 3.16 | 3.53 | 3.88 | 4.22 | 4.55 | 4.85 | 5.15 | 5.42 | 5.68 | 5.93 | 6.16 | 6.38 | 6.59 | 6.78 |
| | $W_{1,2}$ | 2.99 | 6.36 | 10.00 | 14.11 | 18.25 | 22.47 | 26.73 | 31.01 | 35.29 | 39.59 | 43.88 | 48.17 | 52.47 | 56.77 | 61.06 | 65.36 | 69.65 | 73.95 | 78.24 | 82.54 | 86.84 | 91.13 | 95.43 | 99.72 | 104.02 |
| | $W_{3,4}$ | 6.59 | 13.00 | 18.81 | 24.00 | 28.75 | 33.33 | 37.75 | 42.10 | 46.43 | 50.74 | 55.04 | 59.34 | 63.64 | 67.94 | 72.23 | 76.53 | 80.83 | 85.12 | 89.42 | 93.71 | 98.01 | 102.31 | 106.60 | 110.90 | 115.19 |
| | $J$ | 4.95 | 8.19 | 10.04 | 11.09 | 11.73 | 12.13 | 12.40 | 12.58 | 12.71 | 12.80 | 12.87 | 12.92 | 12.96 | 13.00 | 13.02 | 13.04 | 13.06 | 13.07 | 13.08 | 13.09 | 13.09 | 13.10 | 13.10 | 13.11 | 13.11 |
| | $K$ | 2.43 | 4.35 | 5.57 | 6.26 | 6.62 | 6.80 | 6.90 | 6.94 | 6.97 | 6.98 | 6.98 | 6.98 | 6.98 | 6.98 | 6.98 | 6.98 | 6.98 | 6.98 | 6.98 | 6.98 | 6.98 | 6.98 | 6.98 | 6.98 | 6.98 |
| 0.5 | $U_1$ | 11.35 | 20.57 | 27.46 | 32.58 | 36.48 | 39.51 | 41.92 | 43.89 | 45.51 | 46.88 | 48.04 | 49.04 | 49.91 | 50.68 | 51.35 | 51.95 | 52.49 | 52.97 | 53.40 | 53.80 | 54.16 | 54.49 | 54.79 | 55.07 | 55.33 |
| | $U_2$ | 0.12 | 0.25 | 0.40 | 0.57 | 0.79 | 1.07 | 1.39 | 1.74 | 2.11 | 2.50 | 2.88 | 3.26 | 3.63 | 3.99 | 4.33 | 4.65 | 4.96 | 5.25 | 5.53 | 5.79 | 6.03 | 6.26 | 6.48 | 6.69 | 6.89 |
| | $W_{1,2}$ | 3.00 | 6.35 | 10.08 | 14.06 | 18.19 | 22.41 | 26.66 | 30.94 | 35.23 | 39.52 | 43.81 | 48.10 | 52.40 | 56.70 | 60.99 | 65.29 | 69.58 | 73.88 | 78.18 | 82.47 | 86.77 | 91.06 | 95.36 | 99.65 | 103.95 |
| | $W_{3,4}$ | 6.11 | 12.08 | 17.59 | 22.60 | 27.29 | 31.79 | 36.18 | 40.53 | 44.85 | 49.16 | 53.46 | 57.76 | 62.05 | 66.35 | 70.65 | 74.94 | 79.24 | 83.54 | 87.83 | 92.13 | 96.42 | 100.72 | 105.02 | 109.31 | 113.61 |
| | $J$ | 4.81 | 8.12 | 10.13 | 11.36 | 12.16 | 12.71 | 13.10 | 13.38 | 13.60 | 13.76 | 13.89 | 14.00 | 14.08 | 14.15 | 14.20 | 14.25 | 14.28 | 14.32 | 14.34 | 14.36 | 14.38 | 14.38 | 14.41 | 14.42 | 14.43 |
| | $K$ | 2.37 | 4.25 | 5.47 | 6.17 | 6.53 | 6.72 | 6.82 | 6.87 | 6.89 | 6.89 | 6.90 | 6.90 | 6.90 | 6.90 | 6.90 | 6.90 | 6.90 | 6.90 | 6.90 | 6.90 | 6.90 | 6.90 | 6.90 | 6.90 | 6.90 |
| 0.4 | $U_1$ | 10.18 | 18.63 | 25.03 | 29.86 | 33.56 | 36.47 | 38.81 | 40.72 | 42.31 | 43.65 | 44.80 | 45.78 | 46.64 | 47.39 | 48.06 | 48.66 | 49.19 | 49.67 | 50.10 | 50.49 | 50.85 | 51.18 | 51.48 | 51.76 | 52.02 |
| | $U_2$ | 0.13 | 0.27 | 0.43 | 0.61 | 0.85 | 1.13 | 1.46 | 1.81 | 2.19 | 2.58 | 2.97 | 3.35 | 3.72 | 4.07 | 4.41 | 4.74 | 5.04 | 5.33 | 5.61 | 5.87 | 6.11 | 6.34 | 6.56 | 6.77 | 6.96 |
| | $W_{1,2}$ | 2.96 | 6.26 | 9.94 | 13.90 | 18.02 | 22.22 | 26.47 | 30.75 | 35.03 | 39.33 | 43.62 | 47.91 | 52.21 | 56.50 | 60.80 | 65.09 | 69.39 | 73.69 | 77.98 | 82.28 | 86.57 | 90.87 | 95.17 | 99.46 | 103.76 |
| | $W_{3,4}$ | 5.76 | 11.42 | 16.71 | 21.60 | 26.22 | 30.68 | 35.06 | 39.39 | 43.71 | 48.02 | 52.32 | 56.61 | 60.91 | 65.21 | 69.50 | 73.80 | 78.10 | 82.39 | 86.69 | 90.98 | 95.28 | 99.58 | 103.87 | 108.17 | 112.46 |
| | $J$ | 4.74 | 8.16 | 10.32 | 11.72 | 12.67 | 13.35 | 13.85 | 14.23 | 14.52 | 14.75 | 14.93 | 15.08 | 15.20 | 15.30 | 15.38 | 15.44 | 15.50 | 15.54 | 15.58 | 15.62 | 15.65 | 15.67 | 15.69 | 15.71 | 15.72 |
| | $K$ | 2.35 | 4.23 | 5.46 | 6.17 | 6.54 | 6.73 | 6.82 | 6.87 | 6.89 | 6.90 | 6.91 | 6.91 | 6.91 | 6.91 | 6.91 | 6.91 | 6.91 | 6.91 | 6.91 | 6.91 | 6.91 | 6.91 | 6.91 | 6.91 | 6.91 |
| 0.3 | $U_1$ | 9.25 | 16.96 | 22.91 | 27.45 | 30.98 | 33.76 | 36.02 | 37.87 | 39.41 | 40.72 | 41.84 | 42.81 | 43.65 | 44.39 | 45.05 | 45.64 | 46.16 | 46.64 | 47.07 | 47.46 | 47.81 | 48.14 | 48.44 | 48.72 | 48.97 |
| | $U_2$ | 0.14 | 0.29 | 0.46 | 0.66 | 0.91 | 1.20 | 1.54 | 1.90 | 2.28 | 2.67 | 3.03 | 3.41 | 3.78 | 4.13 | 4.47 | 4.79 | 5.10 | 5.39 | 5.66 | 5.92 | 6.17 | 6.40 | 6.61 | 6.84 | 7.01 |
| | $W_{1,2}$ | 2.90 | 6.13 | 9.77 | 13.69 | 17.79 | 21.99 | 26.24 | 30.51 | 34.80 | 39.09 | 43.38 | 47.67 | 51.97 | 56.26 | 60.56 | 64.85 | 69.15 | 73.45 | 77.74 | 82.04 | 86.33 | 90.63 | 94.93 | 99.22 | 103.52 |
| | $W_{3,4}$ | 5.52 | 10.95 | 16.08 | 20.68 | 25.04 | 29.88 | 34.24 | 38.58 | 42.89 | 47.19 | 51.49 | 55.79 | 60.09 | 64.38 | 68.68 | 72.97 | 77.27 | 81.57 | 85.86 | 90.16 | 94.45 | 98.75 | 103.05 | 107.34 | 111.64 |
| | $J$ | 4.74 | 8.23 | 10.54 | 12.13 | 13.21 | 14.05 | 14.66 | 15.13 | 15.49 | 15.77 | 15.99 | 16.16 | 16.30 | 16.42 | 16.51 | 16.59 | 16.66 | 16.72 | 16.77 | 16.81 | 16.85 | 16.89 | 16.92 | 16.96 | 17.03 |
| | $K$ | 2.34 | 4.27 | 5.52 | 6.24 | 6.61 | 6.80 | 6.88 | 6.90 | 6.91 | 6.91 | 6.91 | 6.91 | 6.91 | 6.91 | 6.91 | 6.91 | 6.91 | 6.91 | 6.91 | 6.91 | 6.91 | 6.91 | 6.91 | 6.91 | 6.91 |
| 0.2 | $U_1$ | 8.98 | 16.54 | 22.41 | 26.92 | 30.43 | 33.22 | 35.47 | 37.33 | 38.88 | 40.19 | 41.31 | 42.29 | 43.13 | 43.88 | 44.54 | 45.13 | 45.65 | 46.13 | 46.56 | 46.95 | 47.31 | 47.63 | 47.93 | 48.21 | 48.47 |
| | $U_2$ | 0.13 | 0.28 | 0.44 | 0.64 | 0.88 | 1.17 | 1.50 | 1.86 | 2.25 | 2.64 | 3.03 | 3.41 | 3.78 | 4.13 | 4.47 | 4.79 | 5.10 | 5.39 | 5.66 | 5.92 | 6.17 | 6.40 | 6.61 | 6.82 | 7.01 |
| | $W_{1,2}$ | 2.78 | 5.92 | 9.49 | 13.37 | 17.40 | 21.64 | 25.89 | 30.15 | 34.44 | 38.73 | 43.02 | 47.31 | 51.61 | 55.91 | 60.20 | 64.50 | 68.79 | 73.09 | 77.38 | 81.68 | 85.98 | 90.27 | 94.57 | 98.86 | 103.16 |
| | $W_{3,4}$ | 5.38 | 10.69 | 15.73 | 20.47 | 25.01 | 29.43 | 33.79 | 38.12 | 42.43 | 46.73 | 51.03 | 55.33 | 59.62 | 63.92 | 68.22 | 72.51 | 76.81 | 81.10 | 85.40 | 89.70 | 93.99 | 98.29 | 102.58 | 106.88 | 111.18 |
| | $J$ | 4.78 | 8.36 | 10.77 | 12.41 | 13.58 | 14.41 | 15.07 | 15.60 | 16.03 | 16.32 | 16.57 | 16.77 | 16.94 | 17.08 | 17.19 | 17.29 | 17.37 | 17.43 | 17.49 | 17.54 | 17.58 | 17.61 | 17.64 | 17.67 | 17.69 |
| | $K$ | 2.43 | 4.34 | 5.62 | 6.36 | 6.75 | 6.95 | 7.05 | 7.10 | 7.12 | 7.14 | 7.15 | 7.15 | 7.15 | 7.15 | 7.15 | 7.15 | 7.15 | 7.15 | 7.15 | 7.15 | 7.15 | 7.15 | 7.15 | 7.15 | 7.15 |
| 0.1 | $U_1$ | 9.03 | 16.65 | 22.60 | 27.18 | 30.76 | 33.59 | 35.88 | 37.76 | 39.35 | 40.68 | 41.81 | 42.78 | 43.64 | 44.39 | 45.06 | 45.65 | 46.18 | 46.66 | 47.09 | 47.48 | 47.84 | 48.17 | 48.47 | 48.75 | 49.01 |
| | $U_2$ | 0.12 | 0.24 | 0.39 | 0.58 | 0.81 | 1.10 | 1.43 | 1.79 | 2.17 | 2.56 | 2.95 | 3.33 | 3.70 | 4.05 | 4.39 | 4.72 | 5.03 | 5.32 | 5.59 | 5.85 | 6.09 | 6.33 | 6.54 | 6.75 | 6.94 |
| | $W_{1,2}$ | 2.67 | 5.72 | 9.23 | 13.08 | 17.14 | 21.32 | 25.56 | 29.83 | 34.11 | 38.40 | 42.69 | 46.99 | 51.28 | 55.58 | 59.87 | 64.17 | 68.46 | 72.76 | 77.06 | 81.35 | 85.65 | 89.94 | 94.24 | 98.54 | 102.83 |
| | $W_{3,4}$ | 5.30 | 10.55 | 15.55 | 20.27 | 24.79 | 29.20 | 33.56 | 37.89 | 42.20 | 46.50 | 50.80 | 55.09 | 59.39 | 63.69 | 67.98 | 72.28 | 76.57 | 80.87 | 85.17 | 89.46 | 93.76 | 98.06 | 102.35 | 106.65 | 110.94 |
| | $J$ | 4.83 | 8.47 | 10.94 | 12.63 | 13.83 | 14.71 | 15.39 | 15.92 | 16.35 | 16.67 | 16.92 | 17.12 | 17.29 | 17.42 | 17.52 | 17.59 | 17.67 | 17.74 | 17.78 | 17.81 | 17.83 | 17.84 | 17.86 | 17.87 | 17.88 |
| | $K$ | 2.43 | 4.41 | 5.72 | 6.47 | 6.87 | 7.07 | 7.18 | 7.23 | 7.25 | 7.26 | 7.27 | 7.27 | 7.27 | 7.27 | 7.27 | 7.27 | 7.27 | 7.27 | 7.27 | 7.27 | 7.27 | 7.27 | 7.27 | 7.27 | 7.27 |
| 0.0 | $U_1$ | 9.09 | 17.45 | 23.69 | 28.48 | 32.20 | 35.15 | 37.52 | 39.46 | 41.08 | 42.44 | 43.61 | 44.61 | 45.48 | 46.25 | 46.92 | 47.52 | 48.06 | 48.55 | 48.98 | 49.38 | 49.74 | 50.07 | 50.38 | 50.66 | 50.92 |
| | $U_2$ | 0.09 | 0.19 | 0.31 | 0.48 | 0.69 | 0.97 | 1.29 | 1.64 | 2.02 | 2.41 | 2.80 | 3.18 | 3.55 | 3.91 | 4.25 | 4.58 | 4.88 | 5.18 | 5.45 | 5.71 | 5.96 | 6.19 | 6.41 | 6.62 | 6.81 |
| | $W_{1,2}$ | 2.45 | 5.57 | 9.04 | 12.86 | 16.91 | 21.08 | 25.32 | 29.58 | 33.87 | 38.15 | 42.45 | 46.74 | 51.04 | 55.33 | 59.63 | 63.92 | 68.22 | 72.51 | 76.81 | 81.11 | 85.40 | 89.70 | 93.99 | 98.29 | 102.59 |
| | $W_{3,4}$ | 5.30 | 10.54 | 15.54 | 20.26 | 24.79 | 29.20 | 33.56 | 37.88 | 42.20 | 46.50 | 50.80 | 55.09 | 59.39 | 63.69 | 67.98 | 72.28 | 76.57 | 80.87 | 85.17 | 89.46 | 93.76 | 98.06 | 102.35 | 106.65 | 110.94 |
| | $J$ | 4.90 | 8.54 | 11.01 | 12.67 | 13.83 | 14.67 | 15.28 | 15.75 | 16.10 | 16.38 | 16.59 | 16.76 | 16.90 | 17.01 | 17.10 | 17.17 | 17.23 | 17.28 | 17.32 | 17.35 | 17.38 | 17.41 | 17.43 | 17.45 | 17.46 |
| | $K$ | 2.45 | 4.44 | 5.76 | 6.52 | 6.92 | 7.12 | 7.23 | 7.28 | 7.30 | 7.31 | 7.32 | 7.32 | 7.32 | 7.32 | 7.32 | 7.32 | 7.32 | 7.32 | 7.32 | 7.32 | 7.32 | 7.32 | 7.32 | 7.32 | 7.32 |

TABLE S5: Parameters of the THF interaction Hamiltonian for different ratios $u_0/u_1$ and screening lengths $\xi$ of the double-gate interaction potential ($U_\xi = 24$ meV for $\xi = 10$ nm). We use the same BM model parameters as in Table S4.

Total weight of the heavy-fermions on the active bands after wannierisation: 93.7%

Wannier orbitals support — 0.0 to 1.0

Energy (meV) — $K$ $\Gamma$ $M$ $K$

FIG. S19. Band structures of the BM and THF models near charge neutrality for $w_0/w_1 = 0.8$, depicted by lines and crosses, respectively. The BM bands are colored according to the weight of the $f$-electron wave function on them. We use the same BM parameters as in Table S6.

TABLE S6. Parameters of the $f$-electron Wannier functions and the THF single-particle Hamiltonian for different values of the tunneling ratio $w_0/w_1$. $\mathcal{W}$ denotes the total weight of the THF $f$-electrons on the active bands. In computing the ratios $\gamma/U_1$ and $\gamma/U_2$, we employ the on-site and nearest-neighbor repulsion parameters $U_1$ and $U_2$ obtained numerically for $\xi = 10$nm, as given in Table S7. We employ $v_F = 5.944$eV$\cdot$Å, $|\mathbf{K}| = 1.703$Å$^{-1}$, $w_1 = 110$meV, and $\theta = 0.72°$ for the BM model.

| $w_0/w_1$ | $\alpha_1$ | $\alpha_2$ | $\lambda$ | $\lambda_2$ | $\lambda$ | $\mathcal{W}$ | $t_0$ | $\gamma$ | $M$ | $v_\star$ | $v''_\star$ | $\gamma/U_1$ | $\gamma/U_2$ |
|---|---|---|---|---|---|---|---|---|---|---|---|---|---|
| 1.00 | 0.667 | 0.113 | 0.407 | 0.816 | 1.790 | 21.778 | 18.594 | -0.674 | 0.371 | 0.3721 | 0.41 | 29.73 |
| 0.90 | 0.751 | 0.661 | 0.143 | 0.711 | 0.985 | 1.237 | 22.275 | -1.014 | 0.676 | 0.1527 | 0.07 | 1.47 |
| 0.80 | 0.768 | 0.640 | 0.149 | 0.423 | 0.937 | 0.789 | 25.610 | -1.974 | 0.945 | 0.0570 | 0.25 | 11.51 |
| 0.70 | 0.794 | 0.608 | 0.155 | 0.353 | 0.931 | 0.956 | 28.481 | -2.504 | 1.139 | -0.0413 | 0.04 | 2.11 |
| 0.60 | 0.829 | 0.560 | 0.167 | 0.342 | 0.934 | 1.564 | 30.862 | -2.929 | 1.249 | -0.0934 | -0.23 | -11.87 |
| 0.50 | 0.869 | 0.495 | 0.173 | 0.350 | 0.924 | 2.377 | 32.774 | -3.264 | 1.273 | -0.1085 | -0.55 | -24.10 |
| 0.40 | 0.909 | 0.418 | 0.185 | 0.179 | 0.916 | 3.102 | 34.258 | -3.528 | 1.210 | -0.0972 | -0.90 | -35.24 |
| 0.30 | 0.945 | 0.328 | 0.203 | 0.185 | 0.905 | 3.740 | 35.359 | -3.732 | 1.050 | -0.0697 | -1.26 | -44.24 |
| 0.20 | 0.974 | 0.228 | 0.215 | 0.191 | 0.889 | 4.128 | 36.117 | -3.884 | 0.786 | -0.0369 | -1.53 | -53.67 |
| 0.10 | 0.993 | 0.118 | 0.221 | 0.191 | 0.864 | 4.090 | 36.559 | -3.979 | 0.424 | -0.0103 | -1.63 | -68.11 |
| 0.00 | 1.000 | 0.000 | 0.221 | 0.000 | 0.851 | 3.977 | 36.705 | -4.013 | 0.000 | -0.0000 | -1.65 | -76.91 |

| $w_0/w_1$ | $\xi$/nm | 2 | 4 | 6 | 8 | 10 | 12 | 14 | 16 | 18 | 20 | 22 | 24 | 26 | 28 | 30 | 32 | 34 | 36 | 38 | 40 | 42 | 44 | 46 | 48 | 50 |
|---|---|---|---|---|---|---|---|---|---|---|---|---|---|---|---|---|---|---|---|---|---|---|---|---|---|---|
| 1.0 | $U_1$ | 18.94 | 32.69 | 41.90 | 48.22 | 52.75 | 56.13 | 58.74 | 60.81 | 62.50 | 63.91 | 65.09 | 66.10 | 66.98 | 67.74 | 68.41 | 69.01 | 69.54 | 70.02 | 70.45 | 70.85 | 71.21 | 71.53 | 71.83 | 72.11 | 72.37 |
| | $U_2$ | 0.10 | 0.22 | 0.36 | 0.52 | 0.73 | 1.00 | 1.32 | 1.68 | 2.06 | 2.46 | 2.85 | 3.24 | 3.62 | 3.98 | 4.33 | 4.66 | 4.97 | 5.27 | 5.55 | 5.81 | 6.06 | 6.29 | 6.51 | 6.72 | 6.92 |
| | $W_{1,2}$ | 3.00 | 6.73 | 10.94 | 15.36 | 19.85 | 24.37 | 28.90 | 33.44 | 37.99 | 42.53 | 47.07 | 51.62 | 56.16 | 60.71 | 65.25 | 69.80 | 74.34 | 78.89 | 83.43 | 87.98 | 92.52 | 97.07 | 101.61 | 106.16 | 110.70 |
| | $W_{3,4}$ | 9.98 | 19.29 | 27.01 | 33.35 | 38.82 | 43.83 | 48.60 | 53.25 | 57.84 | 62.41 | 66.97 | 71.52 | 76.07 | 80.62 | 85.16 | 89.71 | 94.25 | 98.80 | 103.34 | 107.89 | 112.43 | 116.98 | 121.52 | 126.07 | 130.61 |
| | $J$ | 6.51 | 9.92 | 11.43 | 12.13 | 12.46 | 12.63 | 12.72 | 12.77 | 12.80 | 12.82 | 12.83 | 12.84 | 12.84 | 12.84 | 12.84 | 12.85 | 12.85 | 12.85 | 12.85 | 12.85 | 12.85 | 12.85 | 12.85 | 12.85 | 12.85 |
| | $K$ | 5.28 | 6.52 | 7.16 | 7.47 | 7.62 | 7.70 | 7.73 | 7.75 | 7.75 | 7.76 | 7.76 | 7.76 | 7.76 | 7.76 | 7.76 | 7.76 | 7.76 | 7.76 | 7.76 | 7.76 | 7.76 | 7.76 | 7.76 | 7.76 | 7.76 |
| 0.9 | $U_1$ | 11.80 | 20.61 | 26.65 | 30.89 | 33.99 | 36.35 | 38.21 | 39.71 | 40.97 | 42.02 | 42.93 | 43.72 | 44.42 | 45.03 | 45.58 | 46.07 | 46.52 | 46.92 | 47.29 | 47.63 | 47.94 | 48.23 | 48.49 | 48.73 | 48.96 |
| | $U_2$ | 0.41 | 0.78 | 1.10 | 1.40 | 1.71 | 2.02 | 2.37 | 2.72 | 3.09 | 3.45 | 3.81 | 4.16 | 4.50 | 4.83 | 5.14 | 5.44 | 5.72 | 5.98 | 6.23 | 6.47 | 6.69 | 6.91 | 7.11 | 7.30 | 7.48 |
| | $W_{1,2}$ | 9.65 | 17.74 | 24.36 | 30.01 | 35.11 | 39.93 | 44.60 | 49.21 | 53.79 | 58.34 | 62.90 | 67.44 | 71.99 | 76.54 | 81.08 | 85.63 | 90.17 | 94.72 | 99.26 | 103.81 | 108.35 | 112.90 | 117.44 | 121.99 | 126.53 |
| | $W_{3,4}$ | 8.58 | 16.65 | 23.57 | 29.47 | 34.71 | 39.61 | 44.32 | 48.95 | 53.53 | 58.10 | 62.65 | 67.20 | 71.75 | 76.29 | 80.84 | 85.38 | 89.93 | 94.47 | 99.02 | 103.56 | 108.11 | 112.65 | 117.20 | 121.74 | 126.29 |
| | $J$ | 5.66 | 8.97 | 10.68 | 11.62 | 12.17 | 12.53 | 12.78 | 12.97 | 13.11 | 13.23 | 13.33 | 13.42 | 13.48 | 13.55 | 13.61 | 13.66 | 13.71 | 13.75 | 13.78 | 13.81 | 13.84 | 13.86 | 13.89 | 13.91 | 13.93 |
| | $K$ | 3.62 | 6.09 | 7.50 | 8.24 | 8.62 | 8.80 | 8.89 | 8.93 | 8.95 | 8.96 | 8.96 | 8.96 | 8.96 | 8.96 | 8.97 | 8.97 | 8.97 | 8.97 | 8.97 | 8.97 | 8.97 | 8.97 | 8.97 | 8.97 | 8.97 |
| 0.8 | $U_1$ | 15.43 | 27.18 | 35.39 | 41.19 | 45.42 | 48.62 | 51.12 | 53.12 | 54.76 | 56.13 | 57.29 | 58.28 | 59.14 | 59.89 | 60.55 | 61.14 | 61.67 | 62.14 | 62.57 | 62.96 | 63.32 | 63.64 | 63.94 | 64.22 | 64.47 |
| | $U_2$ | 0.18 | 0.36 | 0.55 | 0.74 | 0.97 | 1.25 | 1.58 | 1.95 | 2.33 | 2.73 | 3.12 | 3.51 | 3.89 | 4.25 | 4.60 | 4.93 | 5.24 | 5.53 | 5.81 | 6.07 | 6.32 | 6.55 | 6.77 | 6.98 | 7.17 |
| | $W_{1,2}$ | 3.02 | 6.55 | 10.55 | 14.81 | 19.22 | 23.70 | 28.21 | 32.74 | 37.28 | 41.82 | 46.37 | 50.91 | 55.45 | 60.00 | 64.54 | 69.09 | 73.63 | 78.18 | 82.72 | 87.27 | 91.81 | 96.36 | 100.90 | 105.45 | 109.99 |
| | $W_{3,4}$ | 8.18 | 15.80 | 22.51 | 28.30 | 33.49 | 38.36 | 43.06 | 47.68 | 52.26 | 56.82 | 61.38 | 65.93 | 70.47 | 75.02 | 79.56 | 84.11 | 88.65 | 93.20 | 97.74 | 102.29 | 106.83 | 111.38 | 115.92 | 120.47 | 125.01 |
| | $J$ | 5.62 | 8.97 | 10.68 | 11.57 | 12.06 | 12.34 | 12.50 | 12.61 | 12.67 | 12.72 | 12.75 | 12.78 | 12.80 | 12.81 | 12.83 | 12.84 | 12.85 | 12.86 | 12.86 | 12.87 | 12.88 | 12.88 | 12.89 | 12.89 | 12.89 |
| | $K$ | 2.69 | 4.72 | 5.95 | 6.61 | 6.95 | 7.11 | 7.19 | 7.23 | 7.25 | 7.25 | 7.26 | 7.26 | 7.26 | 7.26 | 7.26 | 7.26 | 7.26 | 7.26 | 7.26 | 7.26 | 7.26 | 7.26 | 7.26 | 7.26 | 7.26 |
| 0.7 | $U_1$ | 14.75 | 26.23 | 34.44 | 40.32 | 44.67 | 47.99 | 50.58 | 52.67 | 54.38 | 55.80 | 57.00 | 58.03 | 58.92 | 59.68 | 60.39 | 61.00 | 61.54 | 62.03 | 62.47 | 62.87 | 63.23 | 63.56 | 63.87 | 64.15 | 64.41 |
| | $U_2$ | 0.12 | 0.25 | 0.39 | 0.56 | 0.77 | 1.05 | 1.38 | 1.74 | 2.13 | 2.53 | 2.94 | 3.33 | 3.72 | 4.09 | 4.44 | 4.77 | 5.09 | 5.39 | 5.67 | 5.94 | 6.19 | 6.42 | 6.65 | 6.86 | 7.06 |
| | $W_{1,2}$ | 3.09 | 6.62 | 10.59 | 14.84 | 19.23 | 23.70 | 28.21 | 32.74 | 37.28 | 41.82 | 46.36 | 50.91 | 55.45 | 59.99 | 64.54 | 69.08 | 73.63 | 78.17 | 82.72 | 87.26 | 91.81 | 96.35 | 100.90 | 105.44 | 109.99 |
| | $W_{3,4}$ | 7.35 | 14.43 | 20.74 | 26.30 | 31.38 | 36.19 | 40.87 | 45.48 | 50.05 | 54.61 | 59.16 | 63.71 | 68.26 | 72.80 | 77.35 | 81.89 | 86.44 | 90.98 | 95.53 | 100.07 | 104.62 | 109.16 | 113.71 | 118.25 | 122.80 |
| | $J$ | 5.28 | 8.57 | 10.32 | 11.26 | 11.78 | 12.09 | 12.27 | 12.39 | 12.46 | 12.51 | 12.54 | 12.58 | 12.59 | 12.60 | 12.61 | 12.61 | 12.61 | 12.61 | 12.61 | 12.61 | 12.61 | 12.61 | 12.61 | 12.61 | 12.61 |
| | $K$ | 2.54 | 4.49 | 5.70 | 6.36 | 6.70 | 6.87 | 6.95 | 6.99 | 7.01 | 7.02 | 7.02 | 7.02 | 7.02 | 7.02 | 7.02 | 7.02 | 7.02 | 7.02 | 7.02 | 7.02 | 7.02 | 7.02 | 7.02 | 7.02 | 7.02 |

TABLE S7. Parameters of the THF interaction Hamiltonian for different ratios $u_0/u_1$ and screening lengths $\xi$ of the double-gate interaction potential ($U_\xi = 24$ meV for $\xi = 10$ nm). We use the same BM model parameters as in Table S6.

| $u_0/u_1$ | $\xi$/nm | 2 | 4 | 6 | 8 | 10 | 12 | 14 | 16 | 18 | 20 | 22 | 24 | 26 | 28 | 30 | 32 | 34 | 36 | 38 | 40 | 42 | 44 | 46 | 48 | 50 |
|---|---|---|---|---|---|---|---|---|---|---|---|---|---|---|---|---|---|---|---|---|---|---|---|---|---|---|
| 0.6 | $U_1$ | 13.17 | 23.63 | 31.28 | 36.86 | 41.04 | 44.26 | 46.80 | 48.85 | 50.54 | 51.95 | 53.15 | 54.18 | 55.07 | 55.84 | 56.53 | 57.14 | 57.68 | 58.17 | 58.61 | 59.01 | 59.37 | 59.71 | 60.01 | 60.30 | 60.56 |
|  | $U_2$ | 0.12 | 0.24 | 0.38 | 0.55 | 0.78 | 1.06 | 1.40 | 1.77 | 2.17 | 2.57 | 2.98 | 3.38 | 3.76 | 4.13 | 4.49 | 4.82 | 5.14 | 5.44 | 5.72 | 5.99 | 6.24 | 6.48 | 6.70 | 6.91 | 7.11 |
|  | $W_{1,2}$ | 3.15 | 6.70 | 10.66 | 14.90 | 19.28 | 23.75 | 28.26 | 32.78 | 37.32 | 41.86 | 46.41 | 50.95 | 55.49 | 60.04 | 64.58 | 69.13 | 73.67 | 78.22 | 82.76 | 87.31 | 91.85 | 96.40 | 100.94 | 105.49 | 110.03 |
|  | $W_{3,4}$ | 6.72 | 13.25 | 19.20 | 24.56 | 29.53 | 34.29 | 38.94 | 43.53 | 48.10 | 52.66 | 57.21 | 61.76 | 66.30 | 70.85 | 75.39 | 79.94 | 84.48 | 89.03 | 93.57 | 98.12 | 102.66 | 107.21 | 111.75 | 116.30 | 120.84 |
|  | $J$ | 5.06 | 8.37 | 10.26 | 11.35 | 12.02 | 12.44 | 12.73 | 12.93 | 13.07 | 13.18 | 13.26 | 13.32 | 13.37 | 13.41 | 13.44 | 13.46 | 13.48 | 13.50 | 13.51 | 13.52 | 13.53 | 13.54 | 13.55 | 13.55 | 13.56 |
|  | $K$ | 2.43 | 4.33 | 5.53 | 6.19 | 6.53 | 6.70 | 6.78 | 6.82 | 6.85 | 6.85 | 6.86 | 6.86 | 6.86 | 6.86 | 6.86 | 6.86 | 6.86 | 6.86 | 6.86 | 6.79 | 6.79 | 6.79 | 6.79 | 6.79 | 6.79 |
| 0.5 | $U_1$ | 11.63 | 21.04 | 28.06 | 33.27 | 37.22 | 40.29 | 42.73 | 44.72 | 46.37 | 47.75 | 48.92 | 49.93 | 50.81 | 51.58 | 52.25 | 52.86 | 53.40 | 53.88 | 54.32 | 54.72 | 55.08 | 55.41 | 55.72 | 56.00 | 56.26 |
|  | $U_2$ | 0.13 | 0.27 | 0.42 | 0.61 | 0.85 | 1.15 | 1.49 | 1.87 | 2.27 | 2.68 | 3.09 | 3.49 | 3.87 | 4.25 | 4.60 | 4.93 | 5.25 | 5.55 | 5.83 | 6.10 | 6.35 | 6.57 | 6.81 | 7.02 | 7.22 |
|  | $W_{1,2}$ | 3.16 | 6.70 | 10.64 | 14.85 | 19.23 | 23.69 | 28.20 | 32.73 | 37.26 | 41.80 | 46.34 | 50.89 | 55.43 | 59.98 | 64.52 | 69.07 | 73.61 | 78.16 | 82.70 | 87.25 | 91.79 | 96.34 | 100.88 | 105.43 | 109.97 |
|  | $W_{3,4}$ | 6.24 | 12.36 | 18.02 | 23.22 | 28.11 | 32.82 | 37.45 | 42.04 | 46.60 | 51.16 | 55.71 | 60.25 | 64.80 | 69.34 | 73.89 | 78.43 | 82.98 | 87.52 | 92.07 | 96.61 | 101.16 | 105.70 | 110.25 | 114.79 | 119.34 |
|  | $J$ | 4.94 | 8.34 | 10.40 | 11.68 | 12.52 | 13.10 | 13.51 | 13.82 | 14.05 | 14.23 | 14.37 | 14.49 | 14.58 | 14.65 | 14.71 | 14.76 | 14.80 | 14.83 | 14.86 | 14.88 | 14.90 | 14.92 | 14.94 | 14.95 | 14.96 |
|  | $K$ | 2.38 | 4.25 | 5.45 | 6.11 | 6.46 | 6.63 | 6.71 | 6.75 | 6.77 | 6.78 | 6.78 | 6.78 | 6.78 | 6.78 | 6.79 | 6.79 | 6.79 | 6.79 | 6.79 | 6.79 | 6.79 | 6.79 | 6.79 | 6.79 | 6.79 |
| 0.4 | $U_1$ | 10.58 | 19.26 | 25.85 | 30.80 | 34.60 | 37.57 | 39.95 | 41.90 | 43.51 | 44.87 | 46.03 | 47.03 | 47.90 | 48.66 | 49.34 | 49.94 | 50.49 | 50.97 | 51.39 | 51.79 | 52.15 | 52.48 | 52.78 | 53.06 | 53.32 |
|  | $U_2$ | 0.13 | 0.28 | 0.44 | 0.64 | 0.89 | 1.19 | 1.54 | 1.93 | 2.33 | 2.74 | 3.15 | 3.55 | 3.94 | 4.31 | 4.66 | 5.00 | 5.31 | 5.61 | 5.89 | 6.16 | 6.41 | 6.65 | 6.87 | 7.08 | 7.28 |
|  | $W_{1,2}$ | 3.11 | 6.59 | 10.48 | 14.67 | 19.04 | 23.49 | 27.99 | 32.52 | 37.05 | 41.59 | 46.13 | 50.68 | 55.22 | 59.77 | 64.31 | 68.86 | 73.40 | 77.95 | 82.49 | 87.04 | 91.58 | 96.13 | 100.67 | 105.22 | 109.76 |
|  | $W_{3,4}$ | 5.90 | 11.74 | 17.20 | 22.28 | 27.10 | 31.79 | 36.40 | 40.98 | 45.54 | 50.10 | 54.64 | 59.19 | 63.74 | 68.28 | 72.83 | 77.37 | 81.92 | 86.46 | 91.01 | 95.55 | 100.10 | 104.64 | 109.19 | 113.73 | 117.52 |
|  | $J$ | 4.91 | 8.40 | 10.63 | 12.08 | 13.00 | 13.67 | 14.14 | 14.47 | 14.71 | 14.88 | 15.01 | 15.11 | 15.19 | 15.25 | 15.30 | 15.34 | 15.37 | 15.40 | 15.42 | 16.09 | 16.12 | 16.14 | 16.16 | 16.18 | 16.20 |
|  | $K$ | 2.36 | 4.24 | 5.45 | 6.12 | 6.47 | 6.65 | 6.73 | 6.77 | 6.79 | 6.80 | 6.80 | 6.81 | 6.81 | 6.81 | 6.81 | 6.81 | 6.81 | 6.81 | 6.81 | 6.81 | 6.81 | 6.81 | 6.81 | 6.81 | 6.81 |
| 0.3 | $U_1$ | 9.75 | 17.85 | 24.08 | 28.81 | 32.47 | 35.35 | 37.68 | 39.58 | 41.17 | 42.51 | 43.65 | 44.64 | 45.50 | 46.25 | 46.92 | 47.51 | 48.04 | 48.52 | 48.96 | 49.35 | 49.71 | 50.04 | 50.34 | 50.62 | 50.88 |
|  | $U_2$ | 0.13 | 0.29 | 0.46 | 0.66 | 0.92 | 1.24 | 1.59 | 1.98 | 2.39 | 2.80 | 3.21 | 3.61 | 3.99 | 4.36 | 4.71 | 5.03 | 5.35 | 5.65 | 5.93 | 6.19 | 6.44 | 6.68 | 6.92 | 7.13 | 7.31 |
|  | $W_{1,2}$ | 3.03 | 6.43 | 10.27 | 14.43 | 18.78 | 23.22 | 27.72 | 32.24 | 36.78 | 41.32 | 45.86 | 50.40 | 54.95 | 59.49 | 64.04 | 68.58 | 73.13 | 77.67 | 82.22 | 86.76 | 91.31 | 95.85 | 100.40 | 104.94 | 109.49 |
|  | $W_{3,4}$ | 5.69 | 11.30 | 16.61 | 21.61 | 26.39 | 31.06 | 35.66 | 40.23 | 44.79 | 49.34 | 53.89 | 58.44 | 62.98 | 67.53 | 72.07 | 76.62 | 81.16 | 85.71 | 90.25 | 94.80 | 99.34 | 103.89 | 108.43 | 112.98 | 117.52 |
|  | $J$ | 4.92 | 8.50 | 10.87 | 12.46 | 13.56 | 14.28 | 14.73 | 15.00 | 15.21 | 15.35 | 15.47 | 15.56 | 15.63 | 15.68 | 15.72 | 15.75 | 15.78 | 15.80 | 15.81 | 15.83 | 17.24 | 17.29 | 17.31 | 17.33 | 17.33 |
|  | $K$ | 2.38 | 4.28 | 5.52 | 6.21 | 6.56 | 6.74 | 6.81 | 6.85 | 6.88 | 6.89 | 6.90 | 6.90 | 6.91 | 6.91 | 6.91 | 6.91 | 6.91 | 6.91 | 6.91 | 6.91 | 6.91 | 6.91 | 6.91 | 6.91 | 6.91 |
| 0.2 | $U_1$ | 9.08 | 17.28 | 23.38 | 28.05 | 31.68 | 34.55 | 36.86 | 38.76 | 40.34 | 41.68 | 42.82 | 43.81 | 44.67 | 45.43 | 46.09 | 46.69 | 47.22 | 47.70 | 48.14 | 48.53 | 48.89 | 49.22 | 49.52 | 49.80 | 50.06 |
|  | $U_2$ | 0.13 | 0.27 | 0.44 | 0.64 | 0.90 | 1.22 | 1.57 | 1.96 | 2.37 | 2.78 | 3.19 | 3.59 | 3.98 | 4.35 | 4.70 | 5.03 | 5.36 | 5.66 | 5.93 | 6.19 | 6.44 | 6.68 | 6.90 | 7.11 | 7.31 |
|  | $W_{1,2}$ | 2.91 | 6.22 | 10.00 | 14.12 | 18.45 | 22.89 | 27.38 | 31.90 | 36.44 | 40.97 | 45.52 | 50.06 | 54.60 | 59.15 | 63.69 | 68.24 | 72.78 | 77.33 | 81.87 | 86.42 | 90.96 | 95.51 | 100.05 | 104.60 | 109.14 |
|  | $W_{3,4}$ | 5.60 | 11.03 | 16.26 | 21.20 | 25.96 | 30.61 | 35.21 | 39.78 | 44.34 | 48.89 | 53.43 | 57.98 | 62.53 | 67.07 | 71.62 | 76.16 | 80.71 | 85.25 | 89.80 | 94.34 | 98.86 | 103.43 | 107.97 | 112.52 | 117.00 |
|  | $J$ | 4.94 | 8.65 | 11.10 | 12.79 | 13.98 | 14.86 | 15.47 | 15.90 | 16.22 | 16.46 | 16.64 | 16.77 | 16.87 | 16.94 | 17.00 | 17.05 | 17.09 | 17.12 | 17.16 | 17.20 | 17.24 | 17.83 | 17.88 | 17.97 | 18.02 |
|  | $K$ | 2.42 | 4.36 | 5.62 | 6.33 | 6.69 | 6.88 | 6.96 | 7.00 | 7.03 | 7.04 | 7.04 | 7.04 | 7.05 | 7.05 | 7.05 | 7.05 | 7.05 | 7.05 | 7.05 | 7.05 | 7.05 | 7.05 | 7.05 | 7.05 | 7.05 |
| 0.1 | $U_1$ | 9.68 | 17.82 | 24.14 | 29.08 | 32.74 | 35.70 | 38.09 | 40.04 | 41.66 | 43.03 | 44.20 | 45.20 | 46.08 | 46.84 | 47.52 | 48.12 | 48.66 | 49.15 | 49.59 | 49.99 | 50.35 | 50.68 | 50.99 | 51.27 | 51.53 |
|  | $U_2$ | 0.08 | 0.22 | 0.38 | 0.57 | 0.78 | 1.09 | 1.44 | 1.82 | 2.22 | 2.64 | 3.05 | 3.45 | 3.83 | 4.21 | 4.56 | 4.90 | 5.21 | 5.51 | 5.80 | 6.06 | 6.31 | 6.55 | 6.77 | 6.98 | 7.18 |
|  | $W_{1,2}$ | 2.77 | 5.97 | 9.68 | 13.76 | 18.07 | 22.50 | 26.99 | 31.50 | 36.03 | 40.57 | 45.12 | 49.66 | 54.20 | 58.75 | 63.29 | 67.84 | 72.38 | 76.93 | 81.47 | 86.02 | 90.56 | 95.11 | 99.65 | 104.20 | 108.74 |
|  | $W_{3,4}$ | 5.48 | 11.26 | 16.50 | 21.04 | 25.79 | 30.43 | 35.03 | 39.60 | 44.15 | 48.70 | 53.25 | 57.80 | 62.34 | 66.89 | 71.43 | 75.98 | 80.52 | 85.07 | 89.61 | 94.16 | 98.70 | 103.25 | 107.79 | 112.34 | 116.88 |
|  | $J$ | 4.63 | 8.75 | 11.26 | 12.97 | 14.16 | 14.76 | 15.12 | 15.38 | 15.60 | 15.78 | 15.93 | 16.05 | 16.15 | 16.50 | 17.17 | 17.33 | 17.57 | 17.61 | 17.67 | 17.72 | 17.76 | 17.83 | 17.87 | 17.91 | 17.91 |
|  | $K$ | 2.46 | 4.43 | 5.71 | 6.44 | 6.81 | 7.00 | 7.09 | 7.13 | 7.15 | 7.16 | 7.16 | 7.16 | 7.17 | 7.17 | 7.17 | 7.17 | 7.17 | 7.17 | 7.17 | 7.17 | 7.17 | 7.17 | 7.17 | 7.17 | 7.17 |
| 0.0 | $U_1$ | 9.23 | 16.23 | 24.70 | 29.65 | 33.48 | 36.51 | 38.94 | 40.92 | 42.57 | 43.96 | 45.14 | 46.15 | 47.04 | 47.81 | 48.49 | 49.09 | 49.64 | 50.13 | 50.57 | 50.97 | 51.34 | 51.67 | 51.98 | 52.26 | 52.52 |
|  | $U_2$ | 0.09 | 0.19 | 0.31 | 0.48 | 0.72 | 1.01 | 1.36 | 1.74 | 2.14 | 2.55 | 2.96 | 3.36 | 3.75 | 4.12 | 4.48 | 4.82 | 5.13 | 5.44 | 5.72 | 5.99 | 6.24 | 6.48 | 6.70 | 6.91 | 7.22 |
|  | $W_{1,2}$ | 2.77 | 5.86 | 9.53 | 13.59 | 17.89 | 22.31 | 26.80 | 31.32 | 35.85 | 40.38 | 44.93 | 49.47 | 54.01 | 58.56 | 63.10 | 67.65 | 72.19 | 76.74 | 81.28 | 85.83 | 90.37 | 94.92 | 99.46 | 104.01 | 108.55 |
|  | $W_{3,4}$ | 5.47 | 10.89 | 16.08 | 21.00 | 25.75 | 30.39 | 34.99 | 39.56 | 44.11 | 48.66 | 53.21 | 57.76 | 62.30 | 66.85 | 71.39 | 75.94 | 80.48 | 85.03 | 89.57 | 94.12 | 98.66 | 103.21 | 107.75 | 112.30 | 116.84 |
|  | $J$ | 5.04 | 8.81 | 11.34 | 13.04 | 14.22 | 15.06 | 15.68 | 16.14 | 16.49 | 16.76 | 16.97 | 17.14 | 17.27 | 17.37 | 17.46 | 17.53 | 17.58 | 17.63 | 17.67 | 17.70 | 17.73 | 17.75 | 17.77 | 17.79 | 17.80 |
|  | $K$ | 2.47 | 4.46 | 5.75 | 6.48 | 6.81 | 7.04 | 7.14 | 7.18 | 7.20 | 7.21 | 7.21 | 7.21 | 7.17 | 7.17 | 7.17 | 7.17 | 7.17 | 7.17 | 7.17 | 7.22 | 7.17 | 7.22 | 7.22 | 7.22 | 7.22 |

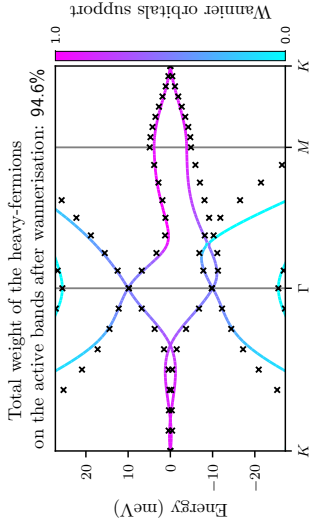

Total weight of the heavy-fermions on the active bands after wannierisation: 94.6%

FIG. S20. Band structures of the BM and THF models near charge neutrality for $w_0/w_1 = 0.8$, depicted by lines and crosses, respectively. The BM bands are colored according to the weight of the $f$-electron wave function on them. We use the same BM parameters as in Table S8.

TABLE S8. Parameters of the $f$-electron Wannier functions and the THF single-particle Hamiltonian for different values of the tunneling ratio $w_0/w_1$. $\mathcal{W}$ denotes the total weight of the THF $f$-electrons on the active bands. In computing the ratios $\gamma/U_1$ and $\gamma/U_2$, we employ the on-site and nearest-neighbor repulsion parameters $U_1$ and $U_2$ obtained numerically for $\xi = 10$ nm, as given in Table S9. We employ $v_F = 5.944$ eV·Å, $|\mathbf{K}| = 1.703$ Å$^{-1}$, $w_1 = 110$ meV, and $\theta = 0.74°$ for the BM model.

| $w_0/w_1$ | $\alpha_1$ | $\alpha_2$ | $\lambda_1$ | $\lambda_2$ | $\lambda$ | $\mathcal{W}$ | $t_0$ | $\gamma$ | $M$ | $v_\star$ | $v'_\star$ | $v''_\star$ | $\gamma/U_1$ | $\gamma/U_2$ |
|---|---|---|---|---|---|---|---|---|---|---|---|---|---|---|
| 1.00 | 0.745 | 0.667 | 0.113 | 0.137 | 0.434 | 0.833 | 1.779 | 22.608 | 19.173 | -0.925 | 0.455 | 0.3344 | 0.43 | 28.12 |
| 0.90 | 0.754 | 0.656 | 0.125 | 0.143 | 0.526 | 0.954 | 1.120 | 17.980 | 22.564 | -1.600 | 0.759 | 0.1701 | 0.42 | 12.27 |
| 0.80 | 0.770 | 0.638 | 0.137 | 0.149 | 0.394 | 0.946 | 0.729 | 10.030 | 25.573 | -2.207 | 1.013 | 0.0378 | 0.21 | 10.27 |
| 0.70 | 0.797 | 0.604 | 0.149 | 0.161 | 0.346 | 0.961 | 0.974 | -0.119 | 28.128 | -2.707 | 1.190 | -0.0482 | -0.00 | -0.15 |
| 0.60 | 0.833 | 0.553 | 0.161 | 0.167 | 0.340 | 0.936 | 1.620 | -11.380 | 30.227 | -3.105 | 1.283 | -0.0921 | -0.27 | -13.64 |
| 0.50 | 0.873 | 0.488 | 0.173 | 0.173 | 0.351 | 0.928 | 2.408 | -22.845 | 31.901 | -3.419 | 1.295 | -0.1031 | -0.60 | -24.88 |
| 0.40 | 0.912 | 0.411 | 0.191 | 0.185 | 0.351 | 0.913 | 3.053 | -33.741 | 33.195 | -3.666 | 1.221 | -0.0908 | -0.95 | -36.21 |
| 0.30 | 0.947 | 0.322 | 0.203 | 0.185 | 0.358 | 0.900 | 3.616 | -43.350 | 34.153 | -3.859 | 1.054 | -0.0644 | -1.29 | -45.23 |
| 0.20 | 0.975 | 0.224 | 0.215 | 0.191 | 0.356 | 0.884 | 3.957 | -50.959 | 34.810 | -4.002 | 0.785 | -0.0338 | -1.55 | -54.72 |
| 0.10 | 0.993 | 0.117 | 0.221 | 0.191 | 0.342 | 0.863 | 4.010 | -55.885 | 35.194 | -4.093 | 0.422 | -0.0094 | -1.66 | -67.23 |
| 0.00 | 1.000 | 0.000 | 0.227 | 0.190 | 0.342 | 0.860 | 4.107 | -57.596 | 35.321 | -4.124 | -0.000 | -0.0000 | -1.72 | -69.73 |

| $w_0/w_1$ | $\xi$/nm | 2 | 4 | 6 | 8 | 10 | 12 | 14 | 16 | 18 | 20 | 22 | 24 | 26 | 28 | 30 | 32 | 34 | 36 | 38 | 40 | 42 | 44 | 46 | 48 | 50 |
|---|---|---|---|---|---|---|---|---|---|---|---|---|---|---|---|---|---|---|---|---|---|---|---|---|---|---|
| 1.0 | $U_1$ | 18.85 | 32.50 | 41.63 | 47.88 | 52.36 | 55.70 | 58.28 | 60.33 | 62.01 | 63.39 | 64.57 | 65.57 | 66.43 | 67.19 | 67.86 | 68.45 | 68.98 | 69.46 | 69.89 | 70.28 | 70.63 | 70.96 | 71.26 | 71.54 | 71.79 |
| | $U_2$ | 0.11 | 0.24 | 0.39 | 0.57 | 0.80 | 1.10 | 1.44 | 1.83 | 2.23 | 2.65 | 3.06 | 3.47 | 3.86 | 4.23 | 4.59 | 4.93 | 5.24 | 5.55 | 5.83 | 6.10 | 6.35 | 6.59 | 6.81 | 7.02 | 7.22 |
| | $W_{1,2}$ | 3.09 | 6.90 | 11.26 | 15.86 | 20.57 | 25.33 | 30.12 | 34.91 | 39.70 | 44.50 | 49.30 | 54.10 | 58.90 | 63.71 | 68.51 | 73.31 | 78.11 | 82.91 | 87.71 | 92.51 | 97.31 | 102.11 | 106.91 | 111.71 | 116.52 |
| | $W_{3,4}$ | 9.99 | 19.31 | 27.08 | 33.52 | 39.15 | 44.36 | 49.35 | 54.24 | 59.09 | 63.91 | 68.72 | 73.52 | 78.32 | 83.13 | 87.93 | 92.73 | 97.53 | 102.33 | 107.13 | 111.93 | 116.73 | 121.53 | 126.34 | 131.14 | 135.94 |
| | $J$ | 6.52 | 9.03 | 11.45 | 12.16 | 12.50 | 12.67 | 12.77 | 12.82 | 12.85 | 12.87 | 12.88 | 12.89 | 12.90 | 12.90 | 12.91 | 12.91 | 12.91 | 12.92 | 12.92 | 12.92 | 12.92 | 12.92 | 12.93 | 12.93 | 12.93 |
| | $K$ | 5.24 | 6.46 | 7.09 | 7.39 | 7.53 | 7.60 | 7.63 | 7.64 | 7.65 | 7.65 | 7.66 | 7.66 | 7.66 | 7.66 | 7.66 | 7.66 | 7.66 | 7.66 | 7.66 | 7.66 | 7.66 | 7.66 | 7.66 | 7.66 | 7.66 |
| 0.9 | $U_1$ | 14.91 | 25.98 | 33.53 | 38.79 | 42.61 | 45.50 | 47.75 | 49.57 | 51.06 | 52.31 | 53.38 | 54.30 | 55.10 | 55.80 | 56.42 | 56.98 | 57.48 | 57.93 | 58.34 | 58.71 | 59.05 | 59.36 | 59.65 | 59.92 | 60.16 |
| | $U_2$ | 0.31 | 0.62 | 0.90 | 1.17 | 1.46 | 1.80 | 2.17 | 2.56 | 2.97 | 3.38 | 3.78 | 4.18 | 4.56 | 4.92 | 5.26 | 5.50 | 5.90 | 6.19 | 6.47 | 6.73 | 6.97 | 7.20 | 7.42 | 7.62 | 7.82 |
| | $W_{1,2}$ | 3.17 | 6.94 | 11.20 | 15.73 | 20.41 | 25.15 | 29.92 | 34.71 | 39.50 | 44.30 | 49.10 | 53.90 | 58.70 | 63.50 | 68.30 | 73.11 | 77.91 | 82.71 | 87.51 | 92.31 | 97.11 | 101.91 | 106.71 | 111.51 | 116.31 |
| | $W_{3,4}$ | 8.95 | 17.37 | 24.57 | 30.70 | 36.18 | 41.31 | 46.26 | 51.14 | 55.97 | 60.79 | 65.60 | 70.40 | 75.21 | 80.01 | 84.81 | 89.61 | 94.41 | 99.21 | 104.01 | 108.81 | 113.61 | 118.42 | 123.22 | 128.02 | 132.82 |
| | $J$ | 9.68 | 11.52 | 12.53 | 13.12 | 13.51 | 13.78 | 13.99 | 14.16 | 14.29 | 14.41 | 14.51 | 14.60 | 14.68 | 14.75 | 14.81 | 14.87 | 14.92 | 14.97 | 15.01 | 15.05 | 15.08 | 15.11 | 15.14 | 15.15 | 15.17 |
| | $K$ | 2.86 | 4.95 | 6.17 | 6.80 | 7.12 | 7.27 | 7.34 | 7.37 | 7.39 | 7.40 | 7.40 | 7.40 | 7.40 | 7.40 | 7.40 | 7.40 | 7.40 | 7.40 | 7.40 | 7.40 | 7.40 | 7.40 | 7.40 | 7.40 | 7.40 |
| 0.8 | $U_1$ | 16.07 | 28.27 | 36.77 | 42.77 | 47.14 | 50.45 | 53.03 | 55.09 | 56.78 | 58.18 | 59.37 | 60.38 | 61.26 | 62.03 | 62.70 | 63.31 | 63.84 | 64.32 | 64.76 | 65.15 | 65.52 | 65.85 | 66.16 | 66.43 | 66.69 |
| | $U_2$ | 0.17 | 0.35 | 0.53 | 0.73 | 0.98 | 1.28 | 1.64 | 2.03 | 2.44 | 2.86 | 3.28 | 3.69 | 4.09 | 4.47 | 4.83 | 5.17 | 5.49 | 5.80 | 6.08 | 6.35 | 6.61 | 6.85 | 7.07 | 7.28 | 7.48 |
| | $W_{1,2}$ | 3.17 | 6.87 | 11.07 | 15.56 | 20.22 | 24.95 | 29.71 | 34.50 | 39.29 | 44.09 | 48.89 | 53.69 | 58.49 | 63.29 | 68.09 | 72.89 | 77.69 | 82.49 | 87.30 | 92.10 | 96.90 | 101.70 | 106.50 | 111.30 | 116.10 |
| | $W_{3,4}$ | 8.20 | 16.03 | 22.86 | 28.81 | 34.19 | 39.28 | 44.21 | 49.08 | 53.91 | 58.72 | 63.53 | 68.33 | 73.14 | 77.94 | 82.74 | 87.54 | 92.34 | 97.14 | 101.94 | 106.74 | 111.54 | 116.35 | 121.15 | 125.95 | 130.75 |
| | $J$ | 5.69 | 9.07 | 10.78 | 11.67 | 12.14 | 12.41 | 12.56 | 12.66 | 12.72 | 12.76 | 12.78 | 12.80 | 12.82 | 12.83 | 12.84 | 12.85 | 12.85 | 12.86 | 12.86 | 12.87 | 12.87 | 12.88 | 12.88 | 12.88 | 12.89 |
| | $K$ | 2.67 | 4.68 | 5.88 | 6.52 | 6.84 | 6.99 | 7.06 | 7.09 | 7.11 | 7.12 | 7.12 | 7.12 | 7.12 | 7.12 | 7.12 | 7.12 | 7.12 | 7.12 | 7.12 | 7.12 | 7.12 | 7.12 | 7.12 | 7.12 | 7.12 |
| 0.7 | $U_1$ | 15.07 | 26.78 | 35.13 | 41.11 | 45.53 | 48.89 | 51.52 | 53.63 | 55.36 | 56.80 | 58.02 | 59.06 | 59.96 | 60.74 | 61.43 | 62.03 | 62.59 | 63.08 | 63.53 | 63.93 | 64.29 | 64.63 | 64.94 | 65.22 | 65.48 |
| | $U_2$ | 0.13 | 0.25 | 0.40 | 0.58 | 0.81 | 1.11 | 1.47 | 1.86 | 2.28 | 2.71 | 3.13 | 3.55 | 3.95 | 4.33 | 4.70 | 5.04 | 5.37 | 5.68 | 5.97 | 6.24 | 6.50 | 6.74 | 6.97 | 7.18 | 7.38 |
| | $W_{1,2}$ | 3.16 | 6.99 | 11.18 | 15.67 | 20.31 | 25.04 | 29.80 | 34.59 | 39.38 | 44.18 | 48.98 | 53.78 | 58.58 | 63.38 | 68.18 | 72.98 | 77.78 | 82.58 | 87.38 | 92.18 | 96.98 | 101.78 | 106.58 | 111.39 | 116.19 |
| | $W_{3,4}$ | 7.45 | 14.64 | 21.08 | 26.81 | 32.08 | 37.11 | 42.02 | 46.87 | 51.70 | 56.51 | 61.32 | 66.12 | 70.92 | 75.73 | 80.53 | 85.33 | 90.13 | 94.93 | 99.73 | 104.53 | 109.33 | 114.13 | 118.93 | 123.74 | 128.54 |
| | $J$ | 5.37 | 8.70 | 10.47 | 11.42 | 11.96 | 12.28 | 12.47 | 12.59 | 12.67 | 12.72 | 12.76 | 12.78 | 12.80 | 12.81 | 12.82 | 12.83 | 12.84 | 12.84 | 12.85 | 12.84 | 12.85 | 12.85 | 12.85 | 12.85 | 12.85 |
| | $K$ | 2.53 | 4.46 | 5.64 | 6.28 | 6.60 | 6.75 | 6.82 | 6.86 | 6.87 | 6.88 | 6.88 | 6.88 | 6.89 | 6.89 | 6.89 | 6.89 | 6.89 | 6.89 | 6.89 | 6.89 | 6.89 | 6.89 | 6.89 | 6.89 | 6.89 |

| $u_0/u_1$ | $\xi$/nm | 2 | 4 | 6 | 8 | 10 | 12 | 14 | 16 | 18 | 20 | 22 | 24 | 26 | 28 | 30 | 32 | 34 | 36 | 38 | 40 | 42 | 44 | 46 | 48 | 50 |
|---|---|---|---|---|---|---|---|---|---|---|---|---|---|---|---|---|---|---|---|---|---|---|---|---|---|---|
| 0.6 | $U_1$ | 13.45 | 24.11 | 31.88 | 37.54 | 41.78 | 45.03 | 47.60 | 49.68 | 51.38 | 52.81 | 54.01 | 55.05 | 55.95 | 56.73 | 57.42 | 58.03 | 58.58 | 59.07 | 59.51 | 59.91 | 60.28 | 60.61 | 60.92 | 61.20 | 61.47 |
|  | $U_2$ | 0.12 | 0.25 | 0.40 | 0.59 | 0.83 | 1.14 | 1.51 | 1.91 | 2.33 | 2.76 | 3.19 | 3.61 | 4.01 | 4.39 | 4.76 | 5.11 | 5.43 | 5.74 | 6.03 | 6.31 | 6.56 | 6.81 | 7.03 | 7.25 | 7.45 |
|  | $W_{1,2}$ | 3.33 | 7.07 | 11.26 | 15.74 | 20.38 | 25.10 | 29.87 | 34.65 | 39.44 | 44.24 | 49.04 | 53.84 | 58.64 | 63.44 | 68.24 | 73.04 | 77.84 | 82.64 | 87.44 | 92.25 | 97.05 | 101.85 | 106.65 | 111.45 | 116.25 |
|  | $W_{3,4}$ | 6.84 | 13.51 | 19.60 | 25.13 | 30.31 | 35.29 | 40.18 | 45.03 | 49.85 | 54.66 | 59.46 | 64.26 | 69.07 | 73.87 | 78.67 | 83.47 | 88.27 | 93.07 | 97.87 | 102.67 | 107.47 | 112.27 | 117.08 | 121.88 | 126.68 |
|  | $J$ | 5.17 | 8.56 | 10.50 | 11.63 | 12.33 | 12.78 | 13.09 | 13.31 | 13.47 | 13.59 | 13.68 | 13.75 | 13.81 | 13.85 | 13.89 | 13.92 | 13.94 | 13.96 | 13.98 | 13.99 | 14.00 | 14.01 | 14.02 | 14.03 | 14.03 |
|  | $K$ | 2.43 | 4.32 | 5.49 | 6.12 | 6.44 | 6.59 | 6.66 | 6.72 | 6.73 | 6.73 | 6.73 | 6.73 | 6.73 | 6.73 | 6.73 | 6.73 | 6.73 | 6.73 | 6.73 | 6.73 | 6.73 | 6.73 | 6.73 | 6.73 | 6.73 |
| 0.5 | $U_1$ | 11.86 | 21.44 | 28.56 | 33.84 | 37.84 | 40.94 | 43.41 | 45.42 | 47.08 | 48.47 | 49.65 | 50.67 | 51.55 | 52.32 | 53.00 | 53.61 | 54.15 | 54.63 | 55.07 | 55.47 | 55.84 | 56.17 | 56.48 | 56.76 | 57.02 |
|  | $U_2$ | 0.14 | 0.28 | 0.45 | 0.65 | 0.92 | 1.24 | 1.62 | 2.02 | 2.45 | 2.88 | 3.31 | 3.73 | 4.13 | 4.52 | 4.88 | 5.23 | 5.55 | 5.86 | 6.15 | 6.42 | 6.68 | 6.93 | 7.15 | 7.36 | 7.56 |
|  | $W_{1,2}$ | 3.34 | 7.07 | 11.24 | 15.70 | 20.33 | 25.05 | 29.81 | 34.60 | 39.39 | 44.18 | 48.98 | 53.78 | 58.58 | 63.39 | 68.19 | 72.99 | 77.79 | 82.59 | 87.39 | 92.19 | 97.00 | 101.79 | 106.59 | 111.39 | 116.20 |
|  | $W_{3,4}$ | 6.39 | 12.65 | 18.47 | 23.85 | 28.95 | 33.90 | 38.77 | 43.61 | 48.42 | 53.23 | 58.04 | 62.84 | 67.64 | 72.44 | 77.24 | 82.04 | 86.84 | 91.65 | 96.45 | 101.25 | 106.05 | 110.85 | 115.65 | 120.45 | 125.25 |
|  | $J$ | 5.07 | 8.56 | 10.69 | 12.02 | 12.82 | 13.52 | 13.96 | 14.30 | 14.55 | 14.75 | 14.90 | 15.02 | 15.12 | 15.21 | 15.27 | 15.33 | 15.37 | 15.41 | 15.44 | 15.47 | 15.49 | 15.51 | 15.53 | 15.54 | 15.55 |
|  | $K$ | 2.38 | 4.25 | 5.42 | 6.05 | 6.38 | 6.53 | 6.60 | 6.64 | 6.66 | 6.67 | 6.67 | 6.67 | 6.67 | 6.67 | 6.67 | 6.67 | 6.67 | 6.67 | 6.67 | 6.67 | 6.67 | 6.67 | 6.67 | 6.67 | 6.67 |
| 0.4 | $U_1$ | 10.93 | 19.89 | 26.65 | 31.73 | 35.61 | 38.64 | 41.07 | 43.05 | 44.68 | 46.06 | 47.24 | 48.25 | 49.12 | 49.89 | 50.57 | 51.18 | 51.72 | 52.21 | 52.64 | 53.04 | 53.41 | 53.74 | 54.04 | 54.32 | 54.58 |
|  | $U_2$ | 0.13 | 0.28 | 0.45 | 0.66 | 0.93 | 1.26 | 1.64 | 2.05 | 2.48 | 2.91 | 3.34 | 3.76 | 4.17 | 4.55 | 4.92 | 5.26 | 5.59 | 5.89 | 6.18 | 6.46 | 6.71 | 6.95 | 7.18 | 7.39 | 7.60 |
|  | $W_{1,2}$ | 3.27 | 6.94 | 11.05 | 15.49 | 20.11 | 24.82 | 29.58 | 34.36 | 39.15 | 43.95 | 48.75 | 53.55 | 58.35 | 63.15 | 67.95 | 72.75 | 77.55 | 82.35 | 87.15 | 91.95 | 96.75 | 101.56 | 106.36 | 111.16 | 115.96 |
|  | $W_{3,4}$ | 6.07 | 12.06 | 17.70 | 22.98 | 28.02 | 32.94 | 37.80 | 42.63 | 47.45 | 52.25 | 57.06 | 61.86 | 66.66 | 71.46 | 76.26 | 81.06 | 85.86 | 90.66 | 95.47 | 100.27 | 105.07 | 109.87 | 114.67 | 119.47 | 124.27 |
|  | $J$ | 5.07 | 8.65 | 10.95 | 12.44 | 13.40 | 14.18 | 14.72 | 15.12 | 15.43 | 15.67 | 15.86 | 16.01 | 16.14 | 16.24 | 16.32 | 16.38 | 16.44 | 16.48 | 16.52 | 16.55 | 16.58 | 16.61 | 16.63 | 16.64 | 16.66 |
|  | $K$ | 2.37 | 4.25 | 5.43 | 6.08 | 6.40 | 6.56 | 6.63 | 6.67 | 6.69 | 6.70 | 6.70 | 6.70 | 6.70 | 6.70 | 6.70 | 6.70 | 6.70 | 6.70 | 6.70 | 6.70 | 6.70 | 6.70 | 6.70 | 6.70 | 6.70 |
| 0.3 | $U_1$ | 10.16 | 18.58 | 24.91 | 29.91 | 33.67 | 36.63 | 39.01 | 40.95 | 42.57 | 43.93 | 45.09 | 46.09 | 46.96 | 47.72 | 48.40 | 49.00 | 49.54 | 50.02 | 50.46 | 50.85 | 51.22 | 51.55 | 51.85 | 52.13 | 52.39 |
|  | $U_2$ | 0.13 | 0.29 | 0.46 | 0.68 | 0.96 | 1.29 | 1.68 | 2.09 | 2.52 | 2.96 | 3.39 | 3.80 | 4.21 | 4.59 | 4.96 | 5.30 | 5.63 | 5.93 | 6.22 | 6.49 | 6.75 | 6.99 | 7.22 | 7.43 | 7.63 |
|  | $W_{1,2}$ | 3.17 | 6.76 | 10.82 | 15.22 | 19.83 | 24.53 | 29.29 | 34.07 | 38.86 | 43.65 | 48.45 | 53.25 | 58.05 | 62.85 | 67.65 | 72.46 | 77.26 | 82.06 | 86.86 | 91.66 | 96.46 | 101.26 | 106.06 | 110.86 | 115.66 |
|  | $W_{3,4}$ | 5.86 | 11.39 | 17.15 | 22.35 | 27.36 | 32.26 | 37.11 | 41.94 | 46.75 | 51.56 | 56.36 | 61.16 | 65.96 | 70.76 | 75.56 | 80.37 | 85.17 | 89.97 | 94.77 | 99.57 | 104.37 | 109.17 | 113.97 | 118.77 | 123.57 |
|  | $J$ | 5.11 | 8.90 | 11.44 | 13.17 | 14.38 | 15.27 | 15.93 | 16.44 | 16.82 | 17.13 | 17.37 | 17.56 | 17.72 | 17.84 | 17.94 | 18.03 | 18.10 | 18.15 | 18.20 | 18.24 | 18.28 | 18.31 | 18.33 | 18.35 | 18.37 |
|  | $K$ | 2.44 | 4.38 | 5.61 | 6.29 | 6.63 | 6.80 | 6.88 | 6.91 | 6.93 | 6.94 | 6.94 | 6.94 | 6.94 | 6.94 | 6.94 | 6.95 | 6.95 | 6.95 | 6.95 | 6.95 | 6.95 | 6.95 | 6.95 | 6.95 | 6.95 |
| 0.2 | $U_1$ | 10.02 | 18.42 | 24.91 | 29.87 | 33.70 | 36.72 | 39.15 | 41.13 | 42.78 | 44.16 | 45.34 | 46.36 | 47.24 | 48.01 | 48.69 | 49.30 | 49.84 | 50.33 | 50.77 | 51.17 | 51.54 | 51.87 | 52.18 | 52.46 | 52.72 |
|  | $U_2$ | 0.13 | 0.27 | 0.44 | 0.66 | 0.93 | 1.27 | 1.65 | 2.06 | 2.50 | 2.93 | 3.36 | 3.78 | 4.18 | 4.57 | 4.94 | 5.28 | 5.60 | 5.91 | 6.20 | 6.47 | 6.73 | 6.97 | 7.19 | 7.41 | 7.61 |
|  | $W_{1,2}$ | 2.92 | 6.30 | 10.23 | 14.56 | 19.13 | 23.82 | 28.56 | 33.34 | 38.13 | 42.92 | 47.72 | 52.52 | 57.32 | 62.12 | 66.92 | 71.72 | 76.52 | 81.33 | 86.13 | 90.93 | 95.73 | 100.53 | 105.33 | 110.13 | 114.93 |
|  | $W_{3,4}$ | 5.65 | 11.27 | 16.65 | 21.79 | 26.77 | 31.65 | 36.49 | 41.31 | 46.12 | 50.93 | 55.73 | 60.53 | 65.34 | 70.14 | 74.94 | 79.74 | 84.54 | 89.34 | 94.14 | 98.94 | 103.74 | 108.54 | 113.34 | 118.15 | 122.95 |
|  | $J$ | 5.11 | 8.93 | 11.51 | 13.28 | 14.50 | 15.41 | 16.09 | 16.57 | 16.94 | 17.23 | 17.47 | 17.65 | 17.82 | 17.96 | 18.07 | 18.17 | 18.24 | 18.30 | 18.35 | 18.39 | 18.43 | 18.46 | 18.48 | 18.51 | 18.54 |
|  | $K$ | 2.48 | 4.45 | 5.71 | 6.40 | 6.75 | 6.92 | 7.00 | 7.03 | 7.05 | 7.06 | 7.06 | 7.07 | 7.07 | 7.07 | 7.07 | 7.07 | 7.07 | 7.07 | 7.07 | 7.07 | 7.07 | 7.07 | 7.07 | 7.07 | 7.07 |
| 0.1 | $U_1$ | 9.95 | 18.30 | 24.76 | 29.70 | 33.53 | 36.54 | 38.97 | 40.95 | 42.59 | 43.98 | 45.16 | 46.17 | 47.05 | 47.83 | 48.51 | 49.12 | 49.66 | 50.15 | 50.59 | 50.99 | 51.35 | 51.68 | 51.99 | 52.27 | 52.53 |
|  | $U_2$ | 0.10 | 0.22 | 0.37 | 0.56 | 0.83 | 1.16 | 1.54 | 1.95 | 2.38 | 2.81 | 3.24 | 3.66 | 4.06 | 4.45 | 4.82 | 5.16 | 5.49 | 5.80 | 6.09 | 6.36 | 6.62 | 6.86 | 7.08 | 7.30 | 7.50 |
|  | $W_{1,2}$ | 2.88 | 6.24 | 10.23 | 14.47 | 19.03 | 23.72 | 28.47 | 33.24 | 38.03 | 42.82 | 47.62 | 52.42 | 57.22 | 62.02 | 66.82 | 71.62 | 76.43 | 81.23 | 86.03 | 90.83 | 95.63 | 100.43 | 105.23 | 110.03 | 114.83 |
|  | $W_{3,4}$ | 5.63 | 11.22 | 16.59 | 21.71 | 26.68 | 31.57 | 36.41 | 41.23 | 46.04 | 50.84 | 55.65 | 60.45 | 65.25 | 70.05 | 74.85 | 79.65 | 84.45 | 89.26 | 94.06 | 98.86 | 103.66 | 108.46 | 113.26 | 118.06 | 122.86 |
|  | $J$ | 5.18 | 9.06 | 11.68 | 13.45 | 14.69 | 15.59 | 16.25 | 16.74 | 17.12 | 17.41 | 17.64 | 17.82 | 17.96 | 18.07 | 18.17 | 18.24 | 18.30 | 18.35 | 18.39 | 18.43 | 18.46 | 18.48 | 18.51 | 18.52 | 18.54 |
|  | $K$ | 2.49 | 4.48 | 5.74 | 6.44 | 6.79 | 6.96 | 7.04 | 7.08 | 7.10 | 7.11 | 7.11 | 7.12 | 7.12 | 7.12 | 7.12 | 7.12 | 7.12 | 7.12 | 7.12 | 7.12 | 7.12 | 7.12 | 7.12 | 7.12 | 7.12 |
| 0.0 | $U_1$ | 9.95 | 18.30 | 24.76 | 29.70 | 33.53 | 36.54 | 38.97 | 40.95 | 42.59 | 43.98 | 45.16 | 46.17 | 47.05 | 47.83 | 48.51 | 49.12 | 49.66 | 50.15 | 50.59 | 50.99 | 51.35 | 51.68 | 51.99 | 52.27 | 52.53 |
|  | $U_2$ | 0.10 | 0.22 | 0.37 | 0.56 | 0.83 | 1.16 | 1.54 | 1.95 | 2.38 | 2.81 | 3.24 | 3.66 | 4.06 | 4.45 | 4.82 | 5.16 | 5.49 | 5.80 | 6.09 | 6.36 | 6.62 | 6.86 | 7.08 | 7.30 | 7.50 |
|  | $W_{1,2}$ | 2.88 | 6.24 | 10.23 | 14.47 | 19.03 | 23.72 | 28.47 | 33.24 | 38.03 | 42.82 | 47.62 | 52.42 | 57.22 | 62.02 | 66.82 | 71.62 | 76.43 | 81.23 | 86.03 | 90.83 | 95.63 | 100.43 | 105.23 | 110.03 | 114.83 |
|  | $W_{3,4}$ | 5.63 | 11.22 | 16.59 | 21.71 | 26.68 | 31.57 | 36.41 | 41.23 | 46.04 | 50.84 | 55.65 | 60.45 | 65.25 | 70.05 | 74.85 | 79.65 | 84.45 | 89.26 | 94.06 | 98.86 | 103.66 | 108.46 | 113.26 | 118.06 | 122.86 |
|  | $J$ | 5.18 | 9.06 | 11.68 | 13.45 | 14.69 | 15.59 | 16.25 | 16.74 | 17.12 | 17.41 | 17.64 | 17.82 | 17.96 | 18.07 | 18.17 | 18.24 | 18.30 | 18.35 | 18.39 | 18.43 | 18.46 | 18.48 | 18.51 | 18.52 | 18.54 |
|  | $K$ | 2.49 | 4.48 | 5.74 | 6.44 | 6.79 | 6.96 | 7.04 | 7.08 | 7.10 | 7.11 | 7.11 | 7.12 | 7.12 | 7.12 | 7.12 | 7.12 | 7.12 | 7.12 | 7.12 | 7.12 | 7.12 | 7.12 | 7.12 | 7.12 | 7.12 |

TABLE S9: Parameters of the THF interaction Hamiltonian for different ratios $u_0/u_1$ and screening lengths $\xi$ of the double-gate interaction potential ($U_\xi = 24$ meV for $\xi = 10$ nm). We use the same BM model parameters as in Table S8.

FIG. S21. Band structures of the BM and THF models near charge neutrality for $w_0/w_1 = 0.8$, depicted by lines and crosses, respectively. The BM bands are colored according to the weight of the $f$-electron wave function on them. We use the same BM parameters as in Table S10.

TABLE S10. Parameters of the $f$-electron Wannier functions and the THF single-particle Hamiltonian for different values of the tunneling ratio $w_0/w_1$. $\mathcal{W}$ denotes the total weight of the THF $f$-electrons on the active bands. In computing the ratios $\gamma/U_1$ and $\gamma/U_2$, we employ the on-site and nearest-neighbor repulsion parameters $U_1$ and $U_2$ obtained numerically for $\xi = 10$ nm, as given in Table S11. We employ $v_F = 5.944$ eV·Å, $|\mathbf{K}| = 1.703$ Å$^{-1}$, $w_1 = 110$ meV, and $\theta = 0.76°$ for the BM model.

| $w_0/w_1$ | $\alpha_1$ | $\alpha_2$ | $\lambda_1$ | $\lambda_2$ | $\lambda$ | $\mathcal{W}$ | $t_0$ | $\gamma$ | $M$ | $v_\star$ | $v_\star'$ | $v_\star''$ | $\gamma/U_1$ | $\gamma/U_2$ |
|---|---|---|---|---|---|---|---|---|---|---|---|---|---|---|
| 1.00 | 0.746 | 0.666 | 0.119 | 0.137 | 0.451 | 0.854 | 1.693 | 23.142 | 19.572 | -1.182 | 0.542 | 0.2964 | 0.44 | 25.98 |
| 0.90 | 0.756 | 0.655 | 0.125 | 0.149 | 0.596 | 0.969 | 1.013 | 17.447 | 22.650 | -1.849 | 0.840 | 0.1405 | 0.42 | 11.32 |
| 0.80 | 0.773 | 0.635 | 0.137 | 0.155 | 0.381 | 0.951 | 0.703 | 8.686 | 25.332 | -2.428 | 1.078 | 0.0213 | 0.18 | 8.72 |
| 0.70 | 0.801 | 0.599 | 0.149 | 0.161 | 0.341 | 0.959 | 1.018 | -2.006 | 27.580 | -2.897 | 1.236 | -0.0537 | -0.04 | -2.34 |
| 0.60 | 0.837 | 0.547 | 0.167 | 0.173 | 0.340 | 0.939 | 1.669 | -13.604 | 29.410 | -3.269 | 1.315 | -0.0903 | -0.32 | -15.09 |
| 0.50 | 0.876 | 0.482 | 0.179 | 0.179 | 0.351 | 0.933 | 2.412 | -25.262 | 30.862 | -3.563 | 1.315 | -0.0978 | -0.66 | -25.58 |
| 0.40 | 0.914 | 0.405 | 0.191 | 0.185 | 0.350 | 0.912 | 2.989 | -36.255 | 31.980 | -3.796 | 1.232 | -0.0848 | -0.99 | -36.59 |
| 0.30 | 0.948 | 0.317 | 0.209 | 0.191 | 0.356 | 0.897 | 3.518 | -45.897 | 32.804 | -3.978 | 1.057 | -0.0596 | -1.32 | -45.77 |
| 0.20 | 0.975 | 0.220 | 0.221 | 0.191 | 0.356 | 0.884 | 3.833 | -53.501 | 33.370 | -4.113 | 0.785 | -0.0311 | -1.58 | -53.90 |
| 0.10 | 0.994 | 0.114 | 0.227 | 0.197 | 0.345 | 0.866 | 3.959 | -58.409 | 33.699 | -4.199 | 0.421 | -0.0086 | -1.70 | -64.34 |
| 0.00 | 1.000 | 0.000 | 0.227 | 0.000 | 0.340 | 0.859 | 3.928 | -60.111 | 33.808 | -4.229 | -0.000 | -0.0000 | -1.74 | -68.63 |

| $w_0/w_1$ | $\xi$/nm | 2 | 4 | 6 | 8 | 10 | 12 | 14 | 16 | 18 | 20 | 22 | 24 | 26 | 28 | 30 | 32 | 34 | 36 | 38 | 40 | 42 | 44 | 46 | 48 | 50 |
|---|---|---|---|---|---|---|---|---|---|---|---|---|---|---|---|---|---|---|---|---|---|---|---|---|---|---|
| 1.0 | $U_1$ | 18.89 | 32.53 | 41.64 | 47.86 | 52.32 | 55.64 | 58.21 | 60.26 | 61.93 | 63.31 | 64.48 | 65.48 | 66.34 | 67.09 | 67.76 | 68.35 | 68.88 | 69.35 | 69.78 | 70.17 | 70.53 | 70.85 | 71.15 | 71.43 | 71.68 |
| | $U_2$ | 0.13 | 0.27 | 0.44 | 0.63 | 0.89 | 1.21 | 1.58 | 1.99 | 2.42 | 2.86 | 3.29 | 3.71 | 4.12 | 4.50 | 4.87 | 5.22 | 5.55 | 5.85 | 6.14 | 6.41 | 6.67 | 6.91 | 7.13 | 7.34 | 7.55 |
| | $W_{1,2}$ | 3.19 | 7.14 | 11.66 | 16.47 | 21.42 | 26.43 | 31.47 | 36.52 | 41.58 | 46.64 | 51.70 | 56.77 | 61.83 | 66.90 | 71.96 | 77.02 | 82.09 | 87.15 | 92.21 | 97.28 | 102.34 | 107.41 | 112.47 | 117.53 | 122.60 |
| | $W_{3,4}$ | 10.03 | 19.39 | 27.24 | 33.80 | 39.61 | 45.03 | 50.26 | 55.40 | 60.64 | 65.75 | 70.65 | 75.71 | 80.78 | 85.84 | 90.97 | 95.97 | 101.03 | 106.10 | 111.16 | 116.23 | 121.29 | 126.35 | 131.42 | 136.48 | 141.55 |
| | $J$ | 6.55 | 9.99 | 11.53 | 12.25 | 12.89 | 12.78 | 12.89 | 12.95 | 12.99 | 13.01 | 13.03 | 13.05 | 13.06 | 13.07 | 13.08 | 13.08 | 13.09 | 13.09 | 13.10 | 13.10 | 13.10 | 13.11 | 13.10 | 13.11 | 13.11 |
| | $K$ | 5.20 | 6.20 | 6.77 | 7.00 | 7.29 | 7.43 | 7.49 | 7.52 | 7.53 | 7.54 | 7.54 | 7.54 | 7.54 | 7.54 | 7.54 | 7.54 | 7.54 | 7.54 | 7.54 | 7.54 | 7.54 | 7.54 | 7.54 | 7.54 | 7.54 |
| 0.9 | $U_1$ | 14.65 | 23.76 | 32.90 | 38.04 | 41.78 | 44.60 | 46.80 | 48.58 | 50.04 | 51.26 | 52.31 | 53.21 | 53.99 | 54.68 | 55.29 | 55.83 | 56.32 | 56.77 | 57.17 | 57.53 | 57.87 | 58.18 | 58.46 | 58.72 | 58.97 |
| | $U_2$ | 0.33 | 0.65 | 0.94 | 1.23 | 1.54 | 1.89 | 2.28 | 2.69 | 3.11 | 3.53 | 3.95 | 4.35 | 4.74 | 5.11 | 5.46 | 5.79 | 6.10 | 6.39 | 6.67 | 6.93 | 7.17 | 7.40 | 7.62 | 7.83 | 8.02 |
| | $W_{1,2}$ | 3.36 | 7.31 | 11.78 | 16.56 | 21.48 | 26.48 | 31.52 | 36.57 | 41.63 | 46.69 | 51.75 | 56.81 | 61.88 | 66.94 | 72.00 | 77.07 | 82.13 | 87.20 | 92.26 | 97.32 | 102.39 | 107.45 | 112.52 | 117.58 | 122.64 |
| | $W_{3,4}$ | 8.98 | 17.43 | 24.70 | 30.96 | 36.62 | 41.97 | 47.16 | 52.29 | 57.38 | 62.40 | 67.53 | 72.59 | 77.66 | 82.72 | 87.79 | 92.85 | 97.91 | 102.98 | 108.04 | 113.11 | 118.17 | 123.23 | 128.30 | 133.36 | 138.43 |
| | $J$ | 9.63 | 11.46 | 12.46 | 13.05 | 13.44 | 13.72 | 13.93 | 14.10 | 14.24 | 14.36 | 14.47 | 14.57 | 14.66 | 14.73 | 14.80 | 14.87 | 14.92 | 14.98 | 15.02 | 15.07 | 15.10 | 15.14 | 15.15 | 15.14 | 15.20 |
| | $K$ | 2.84 | 4.90 | 6.10 | 6.71 | 7.00 | 7.14 | 7.21 | 7.24 | 7.25 | 7.26 | 7.26 | 7.26 | 7.26 | 7.26 | 7.26 | 7.26 | 7.26 | 7.26 | 7.26 | 7.26 | 7.26 | 7.26 | 7.26 | 7.26 | 7.26 |
| 0.8 | $U_1$ | 16.52 | 29.03 | 37.73 | 43.86 | 48.33 | 51.70 | 54.33 | 56.43 | 58.14 | 59.57 | 60.77 | 61.80 | 62.69 | 63.47 | 64.15 | 64.76 | 65.30 | 65.79 | 66.23 | 66.63 | 66.99 | 67.32 | 67.63 | 67.91 | 68.17 |
| | $U_2$ | 0.17 | 0.35 | 0.52 | 0.73 | 1.00 | 1.32 | 1.71 | 2.13 | 2.57 | 3.01 | 3.46 | 3.88 | 4.30 | 4.69 | 5.07 | 5.42 | 5.75 | 6.05 | 6.33 | 6.60 | 6.86 | 7.14 | 7.37 | 7.55 | 7.75 |
| | $W_{1,2}$ | 3.35 | 7.25 | 11.67 | 16.41 | 21.32 | 26.31 | 31.34 | 36.39 | 41.44 | 46.50 | 51.57 | 56.63 | 61.69 | 66.76 | 71.82 | 76.88 | 81.95 | 87.01 | 92.08 | 97.14 | 102.20 | 107.27 | 112.33 | 117.40 | 122.46 |
| | $W_{3,4}$ | 8.30 | 16.22 | 23.18 | 29.28 | 34.87 | 40.18 | 45.35 | 50.47 | 55.56 | 60.64 | 65.70 | 70.77 | 75.84 | 80.90 | 85.96 | 91.03 | 96.09 | 101.16 | 106.22 | 111.28 | 116.35 | 121.41 | 126.48 | 131.54 | 136.60 |
| | $J$ | 5.75 | 9.14 | 10.86 | 11.73 | 12.20 | 12.45 | 12.60 | 12.69 | 12.74 | 12.78 | 12.80 | 12.81 | 12.83 | 12.83 | 12.84 | 12.85 | 12.85 | 12.85 | 12.86 | 12.86 | 12.86 | 12.87 | 12.87 | 12.87 | 12.87 |
| | $K$ | 2.65 | 4.64 | 5.81 | 6.42 | 6.72 | 6.86 | 6.92 | 6.95 | 6.96 | 6.97 | 6.97 | 6.97 | 6.98 | 6.98 | 6.98 | 6.98 | 6.98 | 6.98 | 6.98 | 6.98 | 6.98 | 6.98 | 6.98 | 6.98 | 6.98 |
| 0.7 | $U_1$ | 15.39 | 27.31 | 35.79 | 41.86 | 46.33 | 49.73 | 52.40 | 54.53 | 56.28 | 57.73 | 58.96 | 60.01 | 60.92 | 61.71 | 62.41 | 63.02 | 63.57 | 64.07 | 64.51 | 64.92 | 65.29 | 65.62 | 65.93 | 66.22 | 66.48 |
| | $U_2$ | 0.13 | 0.26 | 0.41 | 0.60 | 0.86 | 1.18 | 1.57 | 1.99 | 2.44 | 2.89 | 3.33 | 3.77 | 4.19 | 4.59 | 4.96 | 5.32 | 5.66 | 5.97 | 6.27 | 6.55 | 6.81 | 7.06 | 7.29 | 7.51 | 7.71 |
| | $W_{1,2}$ | 3.45 | 7.39 | 11.82 | 16.55 | 21.46 | 26.45 | 31.48 | 36.53 | 41.58 | 46.64 | 51.70 | 56.77 | 61.83 | 66.90 | 71.96 | 77.02 | 82.09 | 87.15 | 92.22 | 97.28 | 102.34 | 107.41 | 112.47 | 117.53 | 122.60 |
| | $W_{3,4}$ | 7.56 | 14.87 | 21.44 | 27.33 | 32.81 | 38.08 | 43.23 | 48.34 | 53.42 | 58.50 | 63.56 | 68.63 | 73.69 | 78.76 | 83.82 | 88.89 | 93.95 | 99.02 | 104.08 | 109.14 | 114.21 | 119.27 | 124.33 | 129.40 | 134.46 |
| | $J$ | 5.45 | 8.84 | 10.64 | 11.62 | 12.17 | 12.50 | 12.70 | 12.83 | 12.93 | 12.99 | 13.03 | 13.06 | 13.09 | 13.10 | 13.12 | 13.13 | 13.14 | 13.15 | 13.15 | 13.15 | 13.16 | 13.16 | 13.16 | 13.16 | 13.16 |
| | $K$ | 2.52 | 4.43 | 5.58 | 6.19 | 6.48 | 6.62 | 6.69 | 6.72 | 6.73 | 6.74 | 6.74 | 6.74 | 6.74 | 6.74 | 6.74 | 6.74 | 6.74 | 6.74 | 6.74 | 6.74 | 6.74 | 6.74 | 6.74 | 6.74 | 6.74 |

| $u_0/u_1$ | $\xi$/nm | 2 | 4 | 6 | 8 | 10 | 12 | 14 | 16 | 18 | 20 | 22 | 24 | 26 | 28 | 30 | 32 | 34 | 36 | 38 | 40 | 42 | 44 | 46 | 48 | 50 |
|---|---|---|---|---|---|---|---|---|---|---|---|---|---|---|---|---|---|---|---|---|---|---|---|---|---|---|
| 0.6 | $U_1$ | 13.67 | 24.49 | 32.35 | 38.08 | 42.35 | 45.64 | 48.23 | 50.32 | 52.04 | 53.47 | 54.69 | 55.73 | 56.63 | 57.41 | 58.11 | 58.72 | 59.27 | 59.76 | 60.21 | 60.61 | 60.98 | 61.31 | 61.62 | 61.91 | 62.17 |
| | $U_2$ | 0.13 | 0.27 | 0.43 | 0.63 | 0.90 | 1.24 | 1.63 | 2.06 | 2.51 | 2.97 | 3.41 | 3.85 | 4.27 | 4.67 | 5.05 | 5.40 | 5.74 | 6.06 | 6.35 | 6.63 | 6.89 | 7.14 | 7.37 | 7.59 | 7.79 |
| | $W_{1,12}$ | 3.52 | 7.48 | 11.90 | 16.64 | 21.54 | 26.53 | 31.55 | 36.60 | 41.66 | 46.72 | 51.78 | 56.85 | 61.91 | 66.97 | 72.04 | 77.10 | 82.16 | 87.23 | 92.29 | 97.36 | 102.42 | 107.48 | 112.55 | 117.61 | 122.68 |
| | $W_{3,4}$ | 6.97 | 13.77 | 20.01 | 25.73 | 31.12 | 36.34 | 41.48 | 46.58 | 51.66 | 56.73 | 61.80 | 66.86 | 71.93 | 76.99 | 82.06 | 87.12 | 92.18 | 97.25 | 102.31 | 107.37 | 112.44 | 117.50 | 122.57 | 127.63 | 132.69 |
| | $J$ | 5.29 | 8.75 | 10.75 | 11.92 | 12.66 | 13.14 | 13.48 | 13.72 | 13.90 | 14.04 | 14.14 | 14.22 | 14.29 | 14.34 | 14.38 | 14.42 | 14.45 | 14.47 | 14.49 | 14.51 | 14.52 | 14.53 | 14.54 | 14.55 | 14.56 |
| | $K$ | 2.43 | 4.30 | 5.44 | 6.04 | 6.34 | 6.48 | 6.55 | 6.58 | 6.59 | 6.60 | 6.60 | 6.60 | 6.60 | 6.60 | 6.60 | 6.60 | 6.60 | 6.60 | 6.60 | 6.60 | 6.60 | 6.60 | 6.60 | 6.60 | 6.60 |
| 0.5 | $U_1$ | 12.12 | 21.88 | 29.11 | 34.46 | 38.51 | 41.65 | 44.15 | 46.18 | 47.85 | 49.25 | 50.44 | 51.47 | 52.35 | 53.13 | 53.82 | 54.42 | 54.97 | 55.46 | 55.90 | 56.30 | 56.66 | 57.00 | 57.30 | 57.59 | 57.85 |
| | $U_2$ | 0.14 | 0.30 | 0.48 | 0.70 | 0.99 | 1.34 | 1.74 | 2.18 | 2.63 | 3.09 | 3.54 | 3.97 | 4.39 | 4.79 | 5.17 | 5.52 | 5.86 | 6.17 | 6.47 | 6.75 | 7.01 | 7.26 | 7.48 | 7.70 | 7.91 |
| | $W_{1,12}$ | 3.52 | 7.46 | 11.86 | 16.58 | 21.48 | 26.46 | 31.49 | 36.54 | 41.59 | 46.65 | 51.72 | 56.78 | 61.84 | 66.91 | 71.97 | 77.03 | 82.10 | 87.16 | 92.23 | 97.29 | 102.35 | 107.42 | 112.48 | 117.55 | 122.61 |
| | $W_{3,4}$ | 6.53 | 12.95 | 18.95 | 24.63 | 29.85 | 35.04 | 40.16 | 45.25 | 50.33 | 55.40 | 60.47 | 65.53 | 70.59 | 75.66 | 80.72 | 85.79 | 90.85 | 95.91 | 100.98 | 106.04 | 111.11 | 116.17 | 121.23 | 126.30 | 131.36 |
| | $J$ | 5.21 | 8.79 | 10.99 | 12.38 | 13.31 | 13.96 | 14.43 | 14.79 | 15.06 | 15.27 | 15.44 | 15.58 | 15.68 | 15.77 | 15.84 | 15.90 | 15.95 | 15.99 | 16.03 | 16.06 | 16.08 | 16.10 | 16.12 | 16.14 | 16.15 |
| | $K$ | 2.39 | 4.24 | 5.38 | 5.99 | 6.29 | 6.43 | 6.50 | 6.53 | 6.54 | 6.55 | 6.55 | 6.55 | 6.55 | 6.55 | 6.55 | 6.55 | 6.55 | 6.55 | 6.55 | 6.55 | 6.55 | 6.55 | 6.55 | 6.55 | 6.55 |
| 0.4 | $U_1$ | 11.25 | 20.43 | 27.35 | 32.52 | 36.47 | 39.56 | 42.02 | 44.03 | 45.68 | 47.08 | 48.26 | 49.28 | 50.17 | 50.94 | 51.63 | 52.23 | 52.78 | 53.26 | 53.70 | 54.11 | 54.47 | 54.80 | 55.11 | 55.39 | 55.65 |
| | $U_2$ | 0.14 | 0.29 | 0.47 | 0.70 | 0.99 | 1.34 | 1.75 | 2.19 | 2.65 | 3.11 | 3.56 | 3.99 | 4.41 | 4.81 | 5.19 | 5.54 | 5.88 | 6.19 | 6.49 | 6.77 | 7.03 | 7.28 | 7.51 | 7.72 | 7.93 |
| | $W_{1,12}$ | 3.44 | 7.31 | 11.66 | 16.35 | 21.24 | 26.21 | 31.24 | 36.28 | 41.34 | 46.40 | 51.46 | 56.52 | 61.59 | 66.65 | 71.72 | 76.78 | 81.84 | 86.91 | 91.97 | 97.04 | 102.10 | 107.16 | 112.23 | 117.29 | 122.36 |
| | $W_{3,4}$ | 6.24 | 12.39 | 18.21 | 23.70 | 28.98 | 34.15 | 39.26 | 44.34 | 49.42 | 54.49 | 59.55 | 64.62 | 69.68 | 74.75 | 79.81 | 84.87 | 89.94 | 95.00 | 100.07 | 105.13 | 110.19 | 115.26 | 120.32 | 125.38 | 130.45 |
| | $J$ | 5.20 | 8.91 | 11.27 | 12.81 | 13.87 | 14.63 | 15.18 | 15.60 | 15.92 | 16.17 | 16.37 | 16.53 | 16.65 | 16.76 | 16.84 | 16.91 | 16.96 | 17.01 | 17.05 | 17.08 | 17.11 | 17.13 | 17.16 | 17.17 | 17.19 |
| | $K$ | 2.39 | 4.25 | 5.41 | 6.02 | 6.33 | 6.47 | 6.54 | 6.57 | 6.58 | 6.59 | 6.60 | 6.60 | 6.60 | 6.60 | 6.60 | 6.60 | 6.60 | 6.60 | 6.60 | 6.60 | 6.60 | 6.60 | 6.60 | 6.60 | 6.60 |
| 0.3 | $U_1$ | 10.53 | 19.23 | 25.85 | 30.86 | 34.71 | 37.73 | 40.15 | 42.13 | 43.77 | 45.15 | 46.33 | 47.34 | 48.22 | 48.99 | 49.67 | 50.27 | 50.81 | 51.30 | 51.74 | 52.14 | 52.50 | 52.84 | 53.14 | 53.42 | 53.68 |
| | $U_2$ | 0.14 | 0.29 | 0.47 | 0.71 | 1.00 | 1.36 | 1.77 | 2.22 | 2.67 | 3.13 | 3.58 | 4.01 | 4.44 | 4.83 | 5.21 | 5.56 | 5.90 | 6.22 | 6.51 | 6.79 | 7.05 | 7.29 | 7.53 | 7.74 | 7.95 |
| | $W_{1,12}$ | 3.33 | 7.12 | 11.41 | 16.07 | 20.94 | 25.91 | 30.93 | 35.98 | 41.03 | 46.09 | 51.15 | 56.22 | 61.28 | 66.34 | 71.41 | 76.47 | 81.53 | 86.60 | 91.66 | 96.73 | 101.79 | 106.85 | 111.92 | 116.98 | 122.05 |
| | $W_{3,4}$ | 6.03 | 12.00 | 17.69 | 23.11 | 28.36 | 33.51 | 38.61 | 43.69 | 48.77 | 53.83 | 58.90 | 63.96 | 69.03 | 74.09 | 79.16 | 84.22 | 89.28 | 94.35 | 99.41 | 104.48 | 109.54 | 114.60 | 119.67 | 124.73 | 129.80 |
| | $J$ | 5.22 | 9.03 | 11.54 | 13.12 | 14.39 | 15.24 | 15.87 | 16.33 | 16.72 | 17.00 | 17.23 | 17.41 | 17.55 | 17.67 | 17.76 | 17.84 | 17.90 | 17.95 | 18.01 | 18.04 | 18.06 | 18.09 | 18.11 | 18.13 | 18.15 |
| | $K$ | 2.42 | 4.31 | 5.49 | 6.12 | 6.43 | 6.58 | 6.65 | 6.68 | 6.69 | 6.70 | 6.70 | 6.70 | 6.70 | 6.71 | 6.71 | 6.71 | 6.71 | 6.71 | 6.71 | 6.71 | 6.71 | 6.71 | 6.71 | 6.71 | 6.71 |
| 0.2 | $U_1$ | 10.25 | 18.60 | 25.03 | 30.02 | 33.83 | 36.83 | 39.23 | 41.20 | 42.84 | 44.21 | 45.39 | 46.40 | 47.28 | 48.05 | 48.73 | 49.33 | 49.87 | 50.36 | 50.80 | 51.19 | 51.56 | 51.89 | 52.20 | 52.47 | 52.74 |
| | $U_2$ | 0.13 | 0.28 | 0.46 | 0.69 | 0.99 | 1.35 | 1.76 | 2.21 | 2.67 | 3.12 | 3.57 | 4.01 | 4.43 | 4.83 | 5.21 | 5.56 | 5.90 | 6.21 | 6.51 | 6.78 | 7.04 | 7.29 | 7.52 | 7.74 | 7.94 |
| | $W_{1,12}$ | 3.21 | 6.89 | 11.15 | 15.76 | 20.61 | 25.57 | 30.59 | 35.63 | 40.68 | 45.74 | 50.80 | 55.87 | 60.93 | 65.99 | 71.06 | 76.12 | 81.18 | 86.25 | 91.31 | 96.38 | 101.44 | 106.50 | 111.57 | 116.63 | 121.70 |
| | $W_{3,4}$ | 5.83 | 11.62 | 17.17 | 22.56 | 27.77 | 32.91 | 38.01 | 43.08 | 48.16 | 53.22 | 58.29 | 63.35 | 68.42 | 73.48 | 78.55 | 83.61 | 88.67 | 93.74 | 98.80 | 103.87 | 108.93 | 114.00 | 119.06 | 124.12 | 129.18 |
| | $J$ | 5.27 | 9.17 | 11.75 | 13.58 | 14.84 | 15.95 | 16.62 | 17.51 | 17.81 | 18.04 | 18.22 | 18.37 | 18.48 | 18.57 | 18.65 | 18.71 | 18.75 | 18.79 | 18.81 | 18.82 | 18.83 | 18.85 | 18.85 | 18.87 | 18.92 |
| | $K$ | 2.46 | 4.39 | 5.60 | 6.25 | 6.57 | 6.72 | 6.79 | 6.82 | 6.84 | 6.85 | 6.85 | 6.85 | 6.85 | 6.85 | 6.85 | 6.85 | 6.85 | 6.85 | 6.85 | 6.85 | 6.85 | 6.85 | 6.85 | 6.85 | 6.85 |
| 0.1 | $U_1$ | 10.31 | 18.81 | 25.41 | 30.43 | 34.31 | 37.37 | 39.81 | 41.81 | 43.47 | 44.86 | 46.05 | 47.07 | 47.96 | 48.73 | 49.42 | 50.03 | 50.57 | 51.06 | 51.50 | 51.90 | 52.27 | 52.61 | 52.91 | 53.19 | 53.46 |
| | $U_2$ | 0.11 | 0.24 | 0.40 | 0.62 | 0.91 | 1.26 | 1.67 | 2.11 | 2.57 | 3.03 | 3.48 | 3.92 | 4.34 | 4.73 | 5.11 | 5.47 | 5.81 | 6.12 | 6.42 | 6.70 | 6.97 | 7.20 | 7.43 | 7.65 | 7.86 |
| | $W_{1,12}$ | 3.08 | 6.67 | 10.84 | 15.44 | 20.27 | 25.23 | 30.24 | 35.28 | 40.33 | 45.39 | 50.45 | 55.52 | 60.58 | 65.64 | 70.71 | 75.77 | 80.83 | 85.90 | 90.96 | 96.03 | 101.09 | 106.15 | 111.22 | 116.28 | 121.35 |
| | $W_{3,4}$ | 5.63 | 11.59 | 17.16 | 22.51 | 27.72 | 32.85 | 37.95 | 43.03 | 48.10 | 53.16 | 58.23 | 63.29 | 68.36 | 73.42 | 78.49 | 83.55 | 88.61 | 93.68 | 98.74 | 103.81 | 108.87 | 113.93 | 119.00 | 124.06 | 129.13 |
| | $J$ | 5.35 | 9.29 | 11.99 | 13.77 | 15.04 | 15.95 | 16.73 | 17.23 | 17.62 | 17.91 | 18.14 | 18.32 | 18.45 | 18.57 | 18.65 | 18.71 | 18.75 | 18.80 | 18.83 | 18.85 | 18.89 | 18.92 | 18.94 | 18.96 | 18.96 |
| | $K$ | 2.50 | 4.47 | 5.70 | 6.36 | 6.68 | 6.84 | 6.91 | 6.94 | 6.96 | 6.97 | 6.97 | 6.97 | 6.97 | 6.97 | 6.97 | 6.97 | 6.97 | 6.97 | 6.97 | 6.97 | 6.97 | 6.97 | 6.97 | 6.97 | 6.97 |
| 0.0 | $U_1$ | 10.31 | 18.93 | 25.59 | 30.65 | 34.56 | 37.64 | 40.10 | 42.12 | 43.78 | 45.18 | 46.38 | 47.40 | 48.29 | 49.07 | 49.76 | 50.37 | 50.91 | 51.40 | 51.85 | 52.25 | 52.62 | 52.95 | 53.26 | 53.54 | 53.80 |
| | $U_2$ | 0.10 | 0.23 | 0.38 | 0.59 | 0.88 | 1.23 | 1.63 | 2.07 | 2.53 | 2.99 | 3.44 | 3.88 | 4.30 | 4.70 | 5.08 | 5.43 | 5.77 | 6.09 | 6.38 | 6.66 | 6.92 | 7.17 | 7.40 | 7.62 | 7.82 |
| | $W_{1,12}$ | 3.03 | 6.58 | 10.72 | 15.31 | 20.13 | 25.08 | 30.10 | 35.14 | 40.19 | 45.25 | 50.31 | 55.37 | 60.44 | 65.50 | 70.56 | 75.63 | 80.69 | 85.76 | 90.82 | 95.88 | 100.95 | 106.01 | 111.07 | 116.14 | 121.20 |
| | $W_{3,4}$ | 5.81 | 11.59 | 17.12 | 22.51 | 27.72 | 32.85 | 37.95 | 43.03 | 48.10 | 53.16 | 58.23 | 63.29 | 68.36 | 73.42 | 78.49 | 83.55 | 88.61 | 93.68 | 98.74 | 103.81 | 108.87 | 113.93 | 119.00 | 124.06 | 129.13 |
| | $J$ | 5.35 | 9.35 | 12.04 | 13.87 | 15.14 | 16.06 | 16.73 | 17.23 | 17.62 | 17.91 | 18.14 | 18.32 | 18.45 | 18.57 | 18.67 | 18.74 | 18.80 | 18.85 | 18.89 | 18.93 | 18.96 | 18.98 | 19.00 | 19.02 | 19.03 |
| | $K$ | 2.54 | 4.50 | 5.73 | 6.40 | 6.73 | 6.88 | 6.96 | 6.99 | 7.01 | 7.02 | 7.02 | 7.02 | 7.02 | 7.02 | 7.02 | 7.02 | 7.02 | 7.02 | 7.02 | 7.02 | 7.02 | 7.02 | 7.02 | 7.02 | 7.02 |

TABLE S11. Parameters of the THF interaction Hamiltonian for different ratios $u_0/u_1$ and screening lengths $\xi$ of the double-gate interaction potential ($U_\xi = 24$ meV for $\xi = 10$nm). We use the same BM model parameters as in Table S10.

## Figure S22

Total weight of the heavy-fermions on the active bands after wannierisation: 95.4%

Wannier orbitals support: 1.0 – 0.0

Energy (meV), axis ticks: 30, 20, 10, 0, −10, −20, −30; horizontal axis labels: $K$, $M$, $\Gamma$, $M$, $K$

FIG. S22. Band structures of the BM and THF models near charge neutrality for $w_0/w_1 = 0.8$, depicted by lines and crosses, respectively. The BM bands are colored according to the weight of the $f$-electron wave function on them. We use the same BM parameters as in Table S12.

TABLE S12. Parameters of the $f$-electron Wannier functions and the THF single-particle Hamiltonian for different values of the tunneling ratio $w_0/w_1$. $\mathcal{W}$ denotes the total weight of the THF $f$-electrons on the active bands. In computing the ratios $\gamma/U_1$ and $\gamma/U_2$, we employ the on-site and nearest-neighbor repulsion parameters $U_1$ and $U_2$ obtained numerically for $\xi = 10$ nm, as given in Table S13. We employ $v_F = 5.944$ eV·Å, $|\mathbf{K}| = 1.703$ Å$^{-1}$, $w_1 = 110$ meV, and $\theta = 0.78°$ for the BM model.

| $w_0/w_1$ | $\alpha_1$ | $\alpha_2$ | $\lambda_1$ | $\lambda_2$ | $\lambda$ | $\mathcal{W}$ | $t_0$ | $\gamma$ | $M$ | $v_\star$ | $v'_\star$ | $v''_\star$ | $\gamma/U_1$ | $\gamma/U_2$ |
|---|---|---|---|---|---|---|---|---|---|---|---|---|---|---|
| 1.00 | 0.746 | 0.666 | 0.119 | 0.143 | 0.494 | 0.891 | 1.562 | 23.376 | 19.779 | −1.440 | 0.629 | 0.2593 | 0.46 | 21.82 |
| 0.90 | 0.756 | 0.654 | 0.131 | 0.149 | 0.505 | 0.961 | 0.856 | 16.666 | 22.532 | −2.089 | 0.917 | 0.1137 | 0.37 | 11.07 |
| 0.80 | 0.775 | 0.632 | 0.143 | 0.155 | 0.365 | 0.954 | 0.681 | 7.164 | 24.890 | −2.635 | 1.138 | 0.0073 | 0.14 | 7.08 |
| 0.70 | 0.805 | 0.594 | 0.155 | 0.167 | 0.338 | 0.960 | 1.057 | −4.016 | 26.844 | −3.074 | 1.280 | −0.0577 | −0.09 | −4.39 |
| 0.60 | 0.841 | 0.541 | 0.167 | 0.173 | 0.340 | 0.941 | 1.703 | −15.914 | 28.421 | −3.421 | 1.345 | −0.0881 | −0.37 | −16.40 |
| 0.50 | 0.880 | 0.475 | 0.185 | 0.179 | 0.350 | 0.933 | 2.388 | −27.744 | 29.666 | −3.696 | 1.333 | −0.0926 | −0.71 | −26.45 |
| 0.40 | 0.917 | 0.398 | 0.197 | 0.185 | 0.351 | 0.910 | 2.916 | −38.822 | 30.620 | −3.915 | 1.242 | −0.0791 | −1.04 | −36.78 |
| 0.30 | 0.950 | 0.312 | 0.209 | 0.191 | 0.356 | 0.895 | 3.403 | −48.491 | 31.322 | −4.088 | 1.061 | −0.0551 | −1.36 | −45.48 |
| 0.20 | 0.976 | 0.216 | 0.221 | 0.197 | 0.356 | 0.883 | 3.688 | −56.089 | 31.803 | −4.216 | 0.784 | −0.0286 | −1.62 | −53.13 |
| 0.10 | 0.994 | 0.110 | 0.227 | 0.197 | 0.348 | 0.868 | 3.877 | −60.980 | 32.083 | −4.298 | 0.420 | −0.0079 | −1.75 | −61.77 |
| 0.00 | 1.000 | 0.000 | 0.233 | 0.000 | 0.344 | 0.863 | 3.873 | −62.673 | 32.174 | −4.326 | 0.000 | −0.0000 | −1.79 | −65.15 |

| $w_0/w_1$ | $\xi$/nm | 2 | 4 | 6 | 8 | 10 | 12 | 14 | 16 | 18 | 20 | 22 | 24 | 26 | 28 | 30 | 32 | 34 | 36 | 38 | 40 | 42 | 44 | 46 | 48 | 50 |
|---|---|---|---|---|---|---|---|---|---|---|---|---|---|---|---|---|---|---|---|---|---|---|---|---|---|---|
| 1.0 | $U_1$ | 18.42 | 31.70 | 40.54 | 46.59 | 50.92 | 54.15 | 56.66 | 58.65 | 60.28 | 61.63 | 62.77 | 63.75 | 64.60 | 65.34 | 65.96 | 66.58 | 67.10 | 67.67 | 67.99 | 68.38 | 68.73 | 69.05 | 69.35 | 69.62 | 69.88 |
|  | $U_2$ | 0.18 | 0.36 | 0.56 | 0.78 | 1.07 | 1.42 | 1.82 | 2.26 | 2.71 | 3.16 | 3.61 | 4.04 | 4.44 | 4.85 | 5.22 | 5.58 | 5.93 | 6.22 | 6.51 | 6.78 | 7.04 | 7.28 | 7.51 | 7.73 | 7.93 |
|  | $W_{1,2}$ | 3.35 | 7.46 | 12.18 | 17.23 | 22.43 | 27.71 | 33.01 | 38.33 | 43.66 | 48.99 | 54.33 | 59.60 | 64.99 | 70.33 | 75.66 | 81.00 | 86.33 | 91.66 | 97.03 | 102.37 | 107.70 | 113.03 | 118.36 | 123.70 | 129.00 |
|  | $W_{3,4}$ | 6.57 | 10.05 | 11.64 | 12.40 | 12.80 | 13.02 | 13.16 | 13.25 | 13.32 | 13.37 | 13.41 | 13.44 | 13.47 | 13.49 | 13.51 | 13.53 | 13.55 | 13.56 | 13.58 | 13.59 | 13.60 | 13.61 | 13.62 | 13.63 | 13.63 |
|  | $J$ | 6.20 | 6.91 | 7.19 | 7.31 | 7.37 | 7.39 | 7.40 | 7.41 | 7.41 | 7.41 | 7.41 | 7.41 | 7.41 | 7.41 | 7.41 | 7.41 | 7.41 | 7.41 | 7.41 | 7.41 | 7.41 | 7.41 | 7.41 | 7.41 | 7.41 |
|  | $K$ | 6.20 | 5.73 | 6.33 | 6.59 | 6.72 | 6.78 | 6.80 | 6.81 | 6.82 | 6.82 | 6.82 | 6.82 | 6.82 | 6.82 | 6.82 | 6.82 | 6.82 | 6.82 | 6.82 | 6.82 | 6.82 | 6.82 | 6.82 | 6.82 | 6.82 |
| 0.9 | $U_1$ | 16.04 | 27.88 | 35.93 | 41.53 | 45.47 | 48.63 | 51.01 | 52.91 | 54.48 | 55.79 | 56.90 | 57.85 | 58.69 | 59.41 | 60.06 | 60.63 | 61.14 | 61.60 | 62.02 | 62.40 | 62.75 | 63.07 | 63.37 | 63.64 | 63.89 |
|  | $U_2$ | 0.31 | 0.61 | 0.89 | 1.18 | 1.51 | 1.88 | 2.30 | 2.74 | 3.20 | 3.66 | 4.10 | 4.53 | 4.95 | 5.34 | 5.71 | 6.06 | 6.39 | 6.70 | 6.99 | 7.26 | 7.52 | 7.76 | 7.99 | 8.20 | 8.40 |
|  | $W_{1,2}$ | 3.48 | 7.59 | 12.28 | 17.29 | 22.48 | 27.74 | 33.05 | 38.37 | 43.69 | 49.03 | 54.36 | 59.69 | 65.02 | 70.36 | 75.69 | 81.03 | 86.36 | 91.69 | 97.03 | 102.37 | 107.70 | 113.03 | 118.36 | 123.70 | 129.03 |
|  | $W_{3,4}$ | 9.16 | 17.79 | 25.23 | 31.69 | 37.57 | 43.15 | 48.60 | 53.99 | 59.34 | 64.69 | 70.03 | 75.36 | 80.70 | 86.03 | 91.37 | 96.70 | 102.03 | 107.37 | 112.70 | 118.04 | 123.37 | 128.70 | 134.04 | 139.37 | 144.71 |
|  | $J$ | 9.76 | 11.57 | 12.53 | 13.08 | 13.43 | 13.67 | 13.85 | 13.99 | 14.11 | 14.21 | 14.29 | 14.37 | 14.44 | 14.50 | 14.55 | 14.60 | 14.64 | 14.68 | 14.72 | 14.75 | 14.78 | 14.81 | 14.83 | 14.85 | 14.85 |
|  | $K$ | 2.81 | 4.85 | 5.73 | 6.02 | 6.33 | 6.59 | 6.80 | 6.91 | 6.99 | 7.05 | 7.08 | 7.10 | 7.10 | 7.11 | 7.11 | 7.11 | 7.11 | 7.11 | 7.11 | 7.11 | 7.11 | 7.11 | 7.11 | 7.11 | 7.11 |
| 0.8 | $U_1$ | 17.00 | 29.86 | 38.78 | 45.05 | 49.62 | 53.07 | 55.75 | 57.90 | 59.64 | 61.09 | 62.31 | 63.36 | 64.26 | 65.05 | 65.74 | 66.36 | 66.90 | 67.40 | 67.84 | 68.24 | 68.61 | 68.94 | 69.25 | 69.54 | 69.80 |
|  | $U_2$ | 0.23 | 0.51 | 0.73 | 1.01 | 1.36 | 1.78 | 2.22 | 2.69 | 3.16 | 3.63 | 4.08 | 4.51 | 4.92 | 5.31 | 5.67 | 6.01 | 6.33 | 6.64 | 6.92 | 7.18 | 7.43 | 7.67 | 7.89 | 8.09 | 8.09 |
|  | $W_{1,2}$ | 3.54 | 7.65 | 12.31 | 17.31 | 22.49 | 27.75 | 33.05 | 38.37 | 43.70 | 49.03 | 54.36 | 59.69 | 65.03 | 70.36 | 75.70 | 81.03 | 86.36 | 91.70 | 97.03 | 102.37 | 107.70 | 113.03 | 118.37 | 123.70 | 129.04 |
|  | $W_{3,4}$ | 8.39 | 16.42 | 23.51 | 29.79 | 35.58 | 41.13 | 46.56 | 51.94 | 57.29 | 62.63 | 67.97 | 73.31 | 78.64 | 83.98 | 89.31 | 94.65 | 99.98 | 105.31 | 110.65 | 115.98 | 121.32 | 126.65 | 131.98 | 137.32 | 142.65 |
|  | $J$ | 5.81 | 9.23 | 10.94 | 11.81 | 12.27 | 12.52 | 12.66 | 12.75 | 12.79 | 12.82 | 12.84 | 12.85 | 12.86 | 12.87 | 12.87 | 12.87 | 12.88 | 12.88 | 12.88 | 12.88 | 12.88 | 12.88 | 12.88 | 12.88 | 12.88 |
|  | $K$ | 2.64 | 4.59 | 5.73 | 6.31 | 6.59 | 6.72 | 6.78 | 6.80 | 6.81 | 6.82 | 6.82 | 6.82 | 6.82 | 6.82 | 6.82 | 6.82 | 6.82 | 6.82 | 6.82 | 6.82 | 6.82 | 6.82 | 6.82 | 6.82 | 6.82 |
| 0.7 | $U_1$ | 15.66 | 27.76 | 36.35 | 42.02 | 46.36 | 50.46 | 53.15 | 55.30 | 57.07 | 58.53 | 59.77 | 60.83 | 61.74 | 62.54 | 63.24 | 63.86 | 64.41 | 64.91 | 65.35 | 65.76 | 66.13 | 66.47 | 66.78 | 67.06 | 67.32 |
|  | $U_2$ | 0.13 | 0.27 | 0.43 | 0.63 | 0.91 | 1.27 | 1.68 | 2.14 | 2.61 | 3.08 | 3.55 | 4.01 | 4.44 | 4.85 | 5.24 | 5.61 | 5.96 | 6.28 | 6.58 | 6.87 | 7.13 | 7.39 | 7.62 | 7.84 | 8.05 |
|  | $W_{1,2}$ | 3.54 | 7.82 | 12.52 | 17.50 | 22.67 | 27.93 | 33.23 | 38.55 | 43.88 | 49.21 | 54.54 | 59.88 | 65.21 | 70.55 | 75.88 | 81.21 | 86.55 | 91.88 | 97.22 | 102.55 | 107.88 | 113.22 | 118.55 | 123.88 | 129.22 |
|  | $W_{3,4}$ | 7.67 | 15.09 | 21.81 | 27.89 | 33.59 | 39.09 | 44.50 | 49.87 | 55.22 | 60.56 | 65.90 | 71.24 | 76.57 | 81.91 | 87.24 | 92.57 | 97.91 | 103.24 | 108.58 | 113.91 | 119.24 | 124.58 | 129.91 | 135.24 | 140.58 |
|  | $J$ | 5.55 | 8.98 | 10.82 | 11.83 | 12.40 | 12.75 | 12.97 | 13.12 | 13.22 | 13.29 | 13.34 | 13.38 | 13.41 | 13.43 | 13.45 | 13.46 | 13.47 | 13.48 | 13.48 | 13.49 | 13.49 | 13.50 | 13.50 | 13.51 | 13.52 |
|  | $K$ | 2.51 | 4.40 | 5.52 | 6.09 | 6.37 | 6.50 | 6.55 | 6.58 | 6.59 | 6.60 | 6.60 | 6.60 | 6.60 | 6.60 | 6.60 | 6.60 | 6.60 | 6.60 | 6.60 | 6.60 | 6.60 | 6.60 | 6.60 | 6.60 | 6.60 |

| $u_0/u_1$ | $\xi$/nm | 2 | 4 | 6 | 8 | 10 | 12 | 14 | 16 | 18 | 20 | 22 | 24 | 26 | 28 | 30 | 32 | 34 | 36 | 38 | 40 | 42 | 44 | 46 | 48 | 50 |
|---|---|---|---|---|---|---|---|---|---|---|---|---|---|---|---|---|---|---|---|---|---|---|---|---|---|---|
| 0.6 | $U_1$ | 13.92 | 24.00 | 32.86 | 38.65 | 42.97 | 46.29 | 48.90 | 51.01 | 52.74 | 54.19 | 55.41 | 56.45 | 57.36 | 58.15 | 58.85 | 59.46 | 60.01 | 60.51 | 60.96 | 61.36 | 61.73 | 62.06 | 62.37 | 62.66 | 62.92 |
| | $U_2$ | 0.14 | 0.28 | 0.45 | 0.68 | 0.97 | 1.34 | 1.76 | 2.22 | 2.70 | 3.17 | 3.64 | 4.10 | 4.53 | 4.95 | 5.34 | 5.70 | 6.05 | 6.37 | 6.67 | 6.96 | 7.23 | 7.48 | 7.71 | 7.93 | 8.14 |
| | $W_{1,2}$ | 3.72 | 7.60 | 12.58 | 17.58 | 22.75 | 28.01 | 33.31 | 38.63 | 43.96 | 49.29 | 54.62 | 59.95 | 65.29 | 70.62 | 75.96 | 81.29 | 86.62 | 91.96 | 97.29 | 102.63 | 107.96 | 113.29 | 118.63 | 123.96 | 129.30 |
| | $W_{3,4}$ | 7.10 | 14.04 | 20.45 | 26.37 | 31.99 | 37.45 | 42.85 | 48.21 | 53.56 | 58.90 | 64.23 | 69.57 | 74.90 | 80.24 | 85.57 | 90.90 | 96.24 | 101.57 | 106.91 | 112.24 | 117.57 | 122.91 | 128.24 | 133.58 | 138.91 |
| | $J$ | 5.41 | 8.95 | 11.01 | 12.23 | 13.00 | 13.51 | 13.87 | 14.14 | 14.33 | 14.48 | 14.60 | 14.69 | 14.76 | 14.82 | 14.86 | 14.90 | 14.93 | 14.96 | 14.98 | 15.00 | 15.02 | 15.03 | 15.04 | 15.05 | 15.06 |
| | $K$ | 2.43 | 4.28 | 5.39 | 5.96 | 6.24 | 6.37 | 6.42 | 6.45 | 6.46 | 6.47 | 6.47 | 6.47 | 6.47 | 6.47 | 6.47 | 6.47 | 6.47 | 6.47 | 6.47 | 6.47 | 6.47 | 6.47 | 6.47 | 6.47 | 6.47 |
| 0.5 | $U_1$ | 12.42 | 22.39 | 29.76 | 35.20 | 39.32 | 42.50 | 45.03 | 47.08 | 48.77 | 50.19 | 51.39 | 52.42 | 53.31 | 54.10 | 54.79 | 55.40 | 55.94 | 56.44 | 56.88 | 57.28 | 57.65 | 57.98 | 58.29 | 58.57 | 58.84 |
| | $U_2$ | 0.15 | 0.31 | 0.50 | 0.74 | 1.05 | 1.43 | 1.86 | 2.33 | 2.81 | 3.28 | 3.76 | 4.21 | 4.64 | 5.06 | 5.44 | 5.81 | 6.15 | 6.48 | 6.80 | 7.06 | 7.33 | 7.58 | 7.81 | 8.03 | 8.24 |
| | $W_{1,2}$ | 3.71 | 7.87 | 12.52 | 17.51 | 22.67 | 27.93 | 33.23 | 38.55 | 43.88 | 49.21 | 54.54 | 59.87 | 65.21 | 70.54 | 75.87 | 81.21 | 86.54 | 91.88 | 97.21 | 102.54 | 107.88 | 113.21 | 118.55 | 123.88 | 129.21 |
| | $W_{3,4}$ | 6.69 | 13.27 | 19.44 | 25.24 | 30.80 | 36.23 | 41.62 | 46.97 | 52.32 | 57.65 | 62.99 | 68.33 | 73.66 | 78.99 | 84.33 | 89.66 | 95.00 | 100.33 | 105.66 | 111.00 | 116.33 | 121.67 | 127.00 | 132.33 | 137.67 |
| | $J$ | 5.35 | 9.04 | 11.30 | 12.74 | 13.71 | 14.39 | 14.89 | 15.26 | 15.55 | 15.77 | 15.94 | 16.08 | 16.19 | 16.28 | 16.36 | 16.42 | 16.47 | 16.51 | 16.55 | 16.58 | 16.60 | 16.62 | 16.64 | 16.66 | 16.67 |
| | $K$ | 2.39 | 4.23 | 5.35 | 5.92 | 6.20 | 6.33 | 6.39 | 6.42 | 6.43 | 6.44 | 6.44 | 6.44 | 6.44 | 6.44 | 6.44 | 6.44 | 6.44 | 6.44 | 6.44 | 6.44 | 6.44 | 6.44 | 6.44 | 6.44 | 6.44 |
| 0.4 | $U_1$ | 11.54 | 20.94 | 27.99 | 33.26 | 37.27 | 40.40 | 42.89 | 44.92 | 46.59 | 48.00 | 49.20 | 50.22 | 51.11 | 51.89 | 52.58 | 53.19 | 53.74 | 54.23 | 54.67 | 55.07 | 55.44 | 55.77 | 56.08 | 56.36 | 56.63 |
| | $U_2$ | 0.15 | 0.31 | 0.49 | 0.74 | 1.06 | 1.44 | 1.88 | 2.34 | 2.83 | 3.31 | 3.78 | 4.23 | 4.67 | 5.08 | 5.47 | 5.83 | 6.18 | 6.50 | 6.80 | 7.09 | 7.35 | 7.60 | 7.84 | 8.06 | 8.26 |
| | $W_{1,2}$ | 3.62 | 7.71 | 12.31 | 17.27 | 22.42 | 27.67 | 32.97 | 38.29 | 43.61 | 48.94 | 54.27 | 59.61 | 64.94 | 70.28 | 75.61 | 80.94 | 86.28 | 91.61 | 96.95 | 102.28 | 107.61 | 112.95 | 118.28 | 123.61 | 128.95 |
| | $W_{3,4}$ | 6.40 | 12.73 | 18.75 | 24.46 | 29.97 | 35.39 | 40.77 | 46.12 | 51.46 | 56.80 | 62.13 | 67.47 | 72.80 | 78.14 | 83.47 | 88.80 | 94.14 | 99.47 | 104.81 | 110.14 | 115.47 | 120.81 | 126.14 | 131.48 | 136.81 |
| | $J$ | 5.36 | 9.16 | 11.60 | 13.20 | 14.29 | 15.08 | 15.65 | 16.09 | 16.42 | 16.68 | 16.88 | 17.04 | 17.17 | 17.28 | 17.36 | 17.43 | 17.49 | 17.54 | 17.58 | 17.61 | 17.64 | 17.66 | 17.68 | 17.70 | 17.72 |
| | $K$ | 2.40 | 4.25 | 5.38 | 5.97 | 6.25 | 6.38 | 6.44 | 6.47 | 6.48 | 6.49 | 6.49 | 6.49 | 6.49 | 6.49 | 6.49 | 6.49 | 6.49 | 6.49 | 6.49 | 6.49 | 6.49 | 6.49 | 6.49 | 6.49 | 6.49 |
| 0.3 | $U_1$ | 10.84 | 19.76 | 26.54 | 31.65 | 35.57 | 38.63 | 41.09 | 43.09 | 44.75 | 46.15 | 47.33 | 48.35 | 49.24 | 50.01 | 50.70 | 51.31 | 51.85 | 52.34 | 52.78 | 53.18 | 53.55 | 53.88 | 54.19 | 54.47 | 54.73 |
| | $U_2$ | 0.15 | 0.30 | 0.48 | 0.73 | 1.06 | 1.46 | 1.89 | 2.36 | 2.85 | 3.33 | 3.80 | 4.25 | 4.68 | 5.10 | 5.49 | 5.85 | 6.20 | 6.52 | 6.82 | 7.10 | 7.36 | 7.62 | 7.85 | 8.07 | 8.27 |
| | $W_{1,2}$ | 3.51 | 7.50 | 12.05 | 16.97 | 22.11 | 27.36 | 32.65 | 37.97 | 43.29 | 48.62 | 53.96 | 59.29 | 64.62 | 69.96 | 75.29 | 80.62 | 85.96 | 91.29 | 96.63 | 101.96 | 107.29 | 112.63 | 117.96 | 123.30 | 128.63 |
| | $W_{3,4}$ | 6.20 | 12.36 | 18.26 | 23.90 | 29.39 | 34.80 | 40.16 | 45.51 | 50.85 | 56.19 | 61.52 | 66.86 | 72.19 | 77.53 | 82.86 | 88.19 | 93.53 | 98.86 | 104.20 | 109.53 | 114.86 | 120.20 | 125.53 | 130.87 | 136.20 |
| | $J$ | 5.43 | 9.46 | 12.16 | 13.99 | 15.29 | 16.23 | 16.92 | 17.45 | 17.86 | 18.18 | 18.42 | 18.62 | 18.78 | 18.91 | 19.01 | 19.09 | 19.16 | 19.22 | 19.27 | 19.31 | 19.34 | 19.37 | 19.40 | 19.42 | 19.44 |
| | $K$ | 2.43 | 4.32 | 5.47 | 6.07 | 6.36 | 6.50 | 6.56 | 6.59 | 6.60 | 6.61 | 6.61 | 6.61 | 6.61 | 6.61 | 6.61 | 6.61 | 6.61 | 6.61 | 6.61 | 6.61 | 6.61 | 6.61 | 6.61 | 6.61 | 6.61 |
| 0.2 | $U_1$ | 10.49 | 19.22 | 25.92 | 31.02 | 34.94 | 38.03 | 40.49 | 42.19 | 43.85 | 45.59 | 46.78 | 47.81 | 48.70 | 49.48 | 50.16 | 50.77 | 51.32 | 51.81 | 52.25 | 52.66 | 53.02 | 53.36 | 53.67 | 53.95 | 54.21 |
| | $U_2$ | 0.14 | 0.29 | 0.48 | 0.73 | 1.07 | 1.46 | 1.89 | 2.36 | 2.76 | 3.25 | 3.72 | 4.17 | 4.61 | 5.02 | 5.41 | 5.78 | 6.12 | 6.44 | 6.75 | 7.03 | 7.30 | 7.55 | 7.78 | 8.00 | 8.21 |
| | $W_{1,2}$ | 3.38 | 7.27 | 11.75 | 16.65 | 21.77 | 27.01 | 32.30 | 37.61 | 42.94 | 48.27 | 53.60 | 58.93 | 64.27 | 69.60 | 74.93 | 80.27 | 85.60 | 90.93 | 96.27 | 101.60 | 106.94 | 112.27 | 117.61 | 122.94 | 128.27 |
| | $W_{3,4}$ | 6.08 | 12.12 | 17.95 | 23.56 | 29.02 | 34.42 | 39.78 | 45.13 | 50.47 | 55.81 | 61.14 | 66.48 | 71.81 | 77.14 | 82.48 | 87.81 | 93.15 | 98.48 | 103.81 | 109.15 | 114.48 | 119.82 | 125.15 | 130.48 | 135.82 |
| | $J$ | 5.48 | 9.58 | 12.34 | 14.21 | 15.53 | 16.48 | 17.17 | 17.70 | 18.10 | 18.40 | 18.64 | 18.83 | 18.98 | 19.11 | 19.19 | 19.27 | 19.34 | 19.39 | 19.43 | 19.47 | 19.50 | 19.52 | 19.54 | 19.56 | 19.58 |
| | $K$ | 2.52 | 4.48 | 5.68 | 6.31 | 6.62 | 6.76 | 6.82 | 6.85 | 6.86 | 6.87 | 6.87 | 6.87 | 6.87 | 6.87 | 6.87 | 6.87 | 6.87 | 6.87 | 6.87 | 6.87 | 6.87 | 6.87 | 6.87 | 6.87 | 6.87 |
| 0.1 | $U_1$ | 10.52 | 19.29 | 26.03 | 31.16 | 35.11 | 38.21 | 40.69 | 42.72 | 44.40 | 45.80 | 46.99 | 48.03 | 48.92 | 49.71 | 50.40 | 51.01 | 51.56 | 52.05 | 52.49 | 52.89 | 53.26 | 53.60 | 53.90 | 54.19 | 54.45 |
| | $U_2$ | 0.11 | 0.25 | 0.42 | 0.65 | 0.96 | 1.37 | 1.81 | 2.28 | 2.76 | 3.25 | 3.72 | 4.17 | 4.61 | 5.02 | 5.41 | 5.78 | 6.12 | 6.44 | 6.75 | 7.03 | 7.27 | 7.52 | 7.75 | 7.98 | 8.18 |
| | $W_{1,2}$ | 3.26 | 7.06 | 11.48 | 16.35 | 21.46 | 26.69 | 31.98 | 37.29 | 42.61 | 47.94 | 53.28 | 58.61 | 63.94 | 69.28 | 74.61 | 79.94 | 85.28 | 90.61 | 95.95 | 101.28 | 106.61 | 111.95 | 117.28 | 122.62 | 127.95 |
| | $W_{3,4}$ | 5.99 | 11.90 | 17.74 | 23.32 | 28.77 | 34.16 | 39.52 | 44.87 | 50.21 | 55.54 | 60.88 | 66.21 | 71.55 | 76.88 | 82.22 | 87.55 | 92.88 | 98.22 | 103.55 | 108.89 | 114.22 | 119.55 | 124.89 | 130.22 | 135.56 |
| | $J$ | 5.52 | 9.64 | 12.42 | 14.31 | 15.64 | 16.59 | 17.29 | 17.82 | 18.22 | 18.52 | 18.76 | 18.95 | 19.10 | 19.21 | 19.31 | 19.38 | 19.45 | 19.50 | 19.54 | 19.58 | 19.61 | 19.63 | 19.65 | 19.67 | 19.69 |
| | $K$ | 2.64 | 4.51 | 5.72 | 6.36 | 6.66 | 6.80 | 6.87 | 6.90 | 6.91 | 6.92 | 6.92 | 6.92 | 6.92 | 6.92 | 6.92 | 6.92 | 6.92 | 6.92 | 6.92 | 6.92 | 6.92 | 6.92 | 6.92 | 6.92 | 6.92 |
| 0.0 | $U_1$ | 10.52 | 19.29 | 26.03 | 31.16 | 35.11 | 38.21 | 40.69 | 42.72 | 44.40 | 45.80 | 47.00 | 48.03 | 48.92 | 49.71 | 50.40 | 51.01 | 51.56 | 52.05 | 52.49 | 52.89 | 53.26 | 53.60 | 53.90 | 54.19 | 54.45 |
| | $U_2$ | 0.11 | 0.25 | 0.42 | 0.65 | 0.96 | 1.35 | 1.78 | 2.25 | 2.73 | 3.22 | 3.69 | 4.14 | 4.58 | 4.99 | 5.38 | 5.75 | 6.09 | 6.42 | 6.72 | 7.00 | 7.27 | 7.52 | 7.75 | 7.98 | 8.18 |
| | $W_{1,2}$ | 3.21 | 6.97 | 11.37 | 16.22 | 21.33 | 26.56 | 31.84 | 37.15 | 42.48 | 47.81 | 53.14 | 58.47 | 63.81 | 69.14 | 74.47 | 79.81 | 85.14 | 90.48 | 95.81 | 101.14 | 106.48 | 111.81 | 117.14 | 122.48 | 127.81 |
| | $W_{3,4}$ | 5.98 | 11.90 | 17.74 | 23.32 | 28.77 | 34.16 | 39.52 | 44.87 | 50.21 | 55.54 | 60.88 | 66.21 | 71.55 | 76.88 | 82.22 | 87.55 | 92.88 | 98.22 | 103.55 | 108.89 | 114.22 | 119.55 | 124.89 | 130.22 | 135.56 |
| | $J$ | 5.52 | 9.64 | 12.42 | 14.31 | 15.64 | 16.59 | 17.29 | 17.82 | 18.22 | 18.52 | 18.76 | 18.95 | 19.10 | 19.21 | 19.31 | 19.38 | 19.45 | 19.50 | 19.54 | 19.58 | 19.61 | 19.63 | 19.65 | 19.67 | 19.69 |
| | $K$ | 2.64 | 4.51 | 5.72 | 6.36 | 6.66 | 6.80 | 6.87 | 6.90 | 6.91 | 6.92 | 6.92 | 6.92 | 6.92 | 6.92 | 6.92 | 6.92 | 6.92 | 6.92 | 6.92 | 6.92 | 6.92 | 6.92 | 6.92 | 6.92 | 6.92 |

TABLE S13: Parameters of the THF interaction Hamiltonian for different ratios $u_0/u_1$ and screening lengths $\xi$ of the double-gate interaction potential ($U_\xi = 24$ meV for $\xi = 10$nm). We use the same BM model parameters as in Table S12.

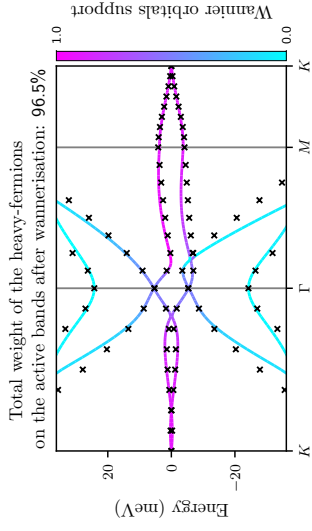

FIG. S23. Band structures of the BM and THF models near charge neutrality for $w_0/w_1 = 0.8$, depicted by lines and crosses, respectively. The BM bands are colored according to the weight of the $f$-electron wave function on them. We use the same BM parameters as in Table S14.

*(Figure labels: Total weight of the heavy-fermions on the active bands after wannierisation: 96.5%; vertical axis "Energy (meV)" with ticks 20, 0, −20; horizontal axis labels K, M, K; legend "Wannier orbitals support" from 1.0 to 0.0.)*

TABLE S14. Parameters of the $f$-electron Wannier functions and the THF single-particle Hamiltonian for different values of the tunneling ratio $w_0/w_1$. $\mathcal{W}$ denotes the total weight of the THF $f$-electrons on the active bands. In computing the ratios $\gamma/U_1$ and $v_\star'/U_1$, and $\gamma/U_2$, we employ the on-site and nearest-neighbor repulsion parameters $U_1$ and $U_2$ obtained numerically for $\xi = 10$ nm, as given in Table S15. We employ $v_F = 5.944$ eV·Å, $|\mathbf{K}| = 1.703$ Å$^{-1}$, $w_1 = 110$ meV, and $\theta = 0.80°$ for the BM model.

| $w_0/w_1$ | $\alpha_1$ | $\alpha_2$ | $\lambda_1$ | $\lambda_2$ | $\lambda$ | $\mathcal{W}$ | $t_0$ | $\gamma$ | $M$ | $v_\star$ | $v_\star'$ | $v_\star''$ | $\gamma/U_1$ | $v_\star'/U_1$ | $\gamma/U_2$ |
|---|---|---|---|---|---|---|---|---|---|---|---|---|---|---|---|
| 1.00 | 0.746 | 0.666 | 0.125 | 0.143 | 0.524 | 0.943 | 1.255 | 23.316 | 19.786 | −1.693 | 0.717 | 0.2240 | 0.11 | | 14.63 |
| 0.90 | 0.757 | 0.653 | 0.131 | 0.149 | 0.466 | 0.959 | 0.740 | 15.653 | 22.209 | −2.316 | 0.991 | 0.0900 | 0.33 | | 10.57 |
| 0.80 | 0.778 | 0.628 | 0.143 | 0.161 | 0.356 | 0.965 | 0.678 | 5.477 | 24.253 | −2.830 | 1.194 | 0.0044 | 0.11 | | 5.26 |
| 0.70 | 0.809 | 0.588 | 0.155 | 0.167 | 0.332 | 0.953 | 1.096 | −6.139 | 25.928 | −3.239 | 1.319 | −0.0604 | −0.13 | | −6.32 |
| 0.60 | 0.845 | 0.535 | 0.173 | 0.179 | 0.339 | 0.946 | 1.715 | −18.305 | 27.269 | −3.563 | 1.371 | −0.0855 | −0.42 | | −17.51 |
| 0.50 | 0.883 | 0.469 | 0.185 | 0.185 | 0.349 | 0.930 | 2.342 | −30.287 | 28.321 | −3.820 | 1.351 | −0.0876 | −0.75 | | −27.27 |
| 0.40 | 0.920 | 0.392 | 0.203 | 0.191 | 0.357 | 0.913 | 2.898 | −41.439 | 29.125 | −4.027 | 1.251 | −0.0738 | −1.10 | | −35.93 |
| 0.30 | 0.952 | 0.306 | 0.215 | 0.191 | 0.357 | 0.895 | 3.291 | −51.131 | 29.715 | −4.190 | 1.064 | −0.0510 | −1.41 | | −44.93 |
| 0.20 | 0.977 | 0.212 | 0.227 | 0.197 | 0.362 | 0.887 | 3.634 | −58.721 | 30.118 | −4.312 | 0.784 | −0.0263 | −1.67 | | −50.95 |
| 0.10 | 0.994 | 0.109 | 0.233 | 0.197 | 0.352 | 0.871 | 3.772 | −63.596 | 30.352 | −4.390 | 0.419 | −0.0072 | −1.79 | | −59.16 |
| 0.00 | 1.000 | 0.000 | 0.233 | 0.000 | 0.344 | 0.862 | 3.672 | −65.281 | 30.429 | −4.417 | 0.000 | −0.0000 | −1.81 | | −63.78 |

TABLE S15.

| $w_0/w_1$ | $\xi$/nm | 2 | 4 | 6 | 8 | 10 | 12 | 14 | 16 | 18 | 20 | 22 | 24 | 26 | 28 | 30 | 32 | 34 | 36 | 38 | 40 | 42 | 44 | 46 | 48 | 50 |
|---|---|---|---|---|---|---|---|---|---|---|---|---|---|---|---|---|---|---|---|---|---|---|---|---|---|---|
| 1.0 | $U_1$ | 17.16 | 29.50 | 37.72 | 43.34 | 47.38 | 50.42 | 52.78 | 54.67 | 56.23 | 57.52 | 58.62 | 59.57 | 60.40 | 61.12 | 61.76 | 62.33 | 62.84 | 63.30 | 63.71 | 64.09 | 64.44 | 64.76 | 65.05 | 65.32 | 65.57 |
| | $U_2$ | 0.33 | 0.64 | 0.94 | 1.24 | 1.59 | 2.00 | 2.44 | 2.91 | 3.39 | 3.86 | 4.32 | 4.77 | 5.19 | 5.60 | 5.98 | 6.33 | 6.67 | 6.98 | 7.28 | 7.55 | 7.81 | 8.05 | 8.28 | 8.50 | 8.70 |
| | $W_{1,2}$ | 3.55 | 7.86 | 12.81 | 18.11 | 23.58 | 29.12 | 34.71 | 40.30 | 45.91 | 51.52 | 57.13 | 62.74 | 68.35 | 73.96 | 79.57 | 85.18 | 90.79 | 96.41 | 102.02 | 107.63 | 113.24 | 118.85 | 124.46 | 130.07 | 135.68 |
| | $W_{3,4}$ | 10.15 | 19.61 | 27.41 | 34.45 | 40.64 | 46.52 | 52.25 | 57.91 | 63.55 | 69.17 | 74.78 | 80.40 | 86.01 | 91.62 | 97.23 | 102.84 | 108.46 | 114.07 | 119.68 | 125.28 | 130.90 | 136.51 | 142.12 | 147.73 | 153.34 |
| | $J$ | 6.72 | 10.38 | 12.17 | 13.12 | 13.68 | 14.04 | 14.31 | 14.51 | 14.67 | 14.81 | 14.93 | 15.04 | 15.13 | 15.21 | 15.27 | 15.34 | 15.39 | 15.44 | 15.48 | 15.52 | 15.55 | 15.58 | 15.61 | 15.63 | 15.65 |
| | $K$ | 2.99 | 5.09 | 6.25 | 6.81 | 7.07 | 7.18 | 7.23 | 7.26 | 7.27 | 7.27 | 7.27 | 7.27 | 7.27 | 7.27 | 7.27 | 7.27 | 7.27 | 7.27 | 7.27 | 7.27 | 7.27 | 7.27 | 7.27 | 7.27 | 7.27 |
| 0.9 | $U_1$ | 16.96 | 29.45 | 37.93 | 43.81 | 48.05 | 51.25 | 53.74 | 55.73 | 57.36 | 58.72 | 59.87 | 60.85 | 61.71 | 62.46 | 63.12 | 63.71 | 64.23 | 64.71 | 65.13 | 65.52 | 65.88 | 66.20 | 66.50 | 66.78 | 67.03 |
| | $U_2$ | 0.29 | 0.58 | 0.85 | 1.14 | 1.48 | 1.88 | 2.33 | 2.80 | 3.29 | 3.77 | 4.24 | 4.70 | 5.13 | 5.54 | 5.93 | 6.29 | 6.63 | 6.95 | 7.25 | 7.53 | 7.80 | 8.04 | 8.28 | 8.49 | 8.70 |
| | $W_{1,2}$ | 3.65 | 7.97 | 12.88 | 18.16 | 23.62 | 29.16 | 34.74 | 40.34 | 45.94 | 51.55 | 57.16 | 62.77 | 68.38 | 73.99 | 79.60 | 85.21 | 90.83 | 96.44 | 102.05 | 107.66 | 113.27 | 118.88 | 124.49 | 130.10 | 135.71 |
| | $W_{3,4}$ | 9.28 | 18.03 | 25.63 | 32.27 | 38.37 | 44.20 | 49.91 | 55.56 | 61.19 | 66.81 | 72.43 | 78.04 | 83.65 | 89.26 | 94.88 | 100.49 | 106.10 | 111.71 | 117.32 | 122.93 | 128.54 | 134.15 | 139.76 | 145.37 | 150.99 |
| | $J$ | 6.26 | 9.81 | 11.50 | 12.51 | 13.03 | 13.34 | 13.55 | 13.70 | 13.82 | 13.91 | 13.99 | 14.05 | 14.11 | 14.16 | 14.21 | 14.25 | 14.28 | 14.32 | 14.35 | 14.38 | 14.40 | 14.42 | 14.44 | 14.46 | 14.48 |
| | $K$ | 2.78 | 4.79 | 5.93 | 6.49 | 6.75 | 6.86 | 6.92 | 6.94 | 6.95 | 6.95 | 6.96 | 6.96 | 6.96 | 6.96 | 6.96 | 6.96 | 6.96 | 6.96 | 6.96 | 6.96 | 6.96 | 6.96 | 6.96 | 6.96 | 6.96 |
| 0.8 | $U_1$ | 17.40 | 30.52 | 39.62 | 46.06 | 50.64 | 54.15 | 56.87 | 59.04 | 60.81 | 62.28 | 63.52 | 64.57 | 65.49 | 66.28 | 66.98 | 67.60 | 68.15 | 68.65 | 69.10 | 69.50 | 69.87 | 70.21 | 70.52 | 70.80 | 71.06 |
| | $U_2$ | 0.33 | 0.51 | 0.74 | 1.04 | 1.42 | 1.86 | 2.34 | 2.84 | 3.33 | 3.82 | 4.29 | 4.73 | 5.16 | 5.56 | 5.93 | 6.28 | 6.61 | 6.92 | 7.21 | 7.48 | 7.74 | 8.00 | 8.20 | 8.41 | |
| | $W_{1,2}$ | 3.76 | 8.10 | 13.02 | 18.29 | 23.74 | 29.28 | 34.86 | 40.46 | 46.07 | 51.67 | 57.28 | 62.89 | 68.50 | 74.12 | 79.73 | 85.34 | 90.95 | 96.56 | 102.17 | 107.78 | 113.39 | 119.00 | 124.61 | 130.23 | 135.84 |
| | $W_{3,4}$ | 8.48 | 16.62 | 23.81 | 30.31 | 36.32 | 42.12 | 47.81 | 53.46 | 59.08 | 64.70 | 70.32 | 75.93 | 81.54 | 87.15 | 92.76 | 98.37 | 103.98 | 109.60 | 115.21 | 120.82 | 126.43 | 132.04 | 137.65 | 143.26 | 148.87 |
| | $J$ | 5.88 | 9.32 | 11.04 | 11.91 | 12.37 | 12.62 | 12.76 | 12.84 | 12.89 | 12.92 | 12.94 | 12.95 | 12.96 | 12.96 | 12.96 | 12.97 | 12.97 | 12.97 | 12.97 | 12.97 | 12.97 | 12.97 | 12.97 | 12.97 | 12.97 |
| | $K$ | 2.61 | 4.54 | 5.65 | 6.20 | 6.46 | 6.58 | 6.63 | 6.65 | 6.66 | 6.67 | 6.67 | 6.67 | 6.67 | 6.67 | 6.67 | 6.67 | 6.67 | 6.67 | 6.67 | 6.67 | 6.67 | 6.67 | 6.67 | 6.67 | 6.67 |
| 0.7 | $U_1$ | 15.96 | 28.28 | 37.00 | 43.22 | 47.80 | 51.29 | 54.01 | 56.18 | 57.96 | 59.44 | 60.69 | 61.76 | 62.68 | 63.48 | 64.18 | 64.81 | 65.37 | 65.86 | 66.31 | 66.72 | 67.09 | 67.43 | 67.74 | 68.03 | 68.29 |
| | $U_2$ | 0.13 | 0.28 | 0.44 | 0.67 | 0.97 | 1.36 | 1.80 | 2.29 | 2.79 | 3.29 | 3.78 | 4.25 | 4.70 | 5.13 | 5.53 | 5.90 | 6.26 | 6.59 | 6.90 | 7.19 | 7.46 | 7.72 | 7.96 | 8.18 | 8.39 |
| | $W_{1,2}$ | 3.88 | 8.27 | 13.21 | 18.48 | 23.94 | 29.48 | 35.06 | 40.66 | 46.26 | 51.87 | 57.48 | 63.09 | 68.70 | 74.31 | 79.92 | 85.54 | 91.15 | 96.76 | 102.37 | 107.98 | 113.59 | 119.20 | 124.81 | 130.43 | 136.03 |
| | $W_{3,4}$ | 7.78 | 15.34 | 22.22 | 28.40 | 34.42 | 40.18 | 45.85 | 51.49 | 57.11 | 62.73 | 68.34 | 73.96 | 79.57 | 85.18 | 90.79 | 96.40 | 102.01 | 107.62 | 113.23 | 118.85 | 124.46 | 130.07 | 135.68 | 141.29 | 146.90 |
| | $J$ | 5.66 | 9.16 | 11.05 | 12.09 | 12.70 | 13.07 | 13.32 | 13.48 | 13.60 | 13.69 | 13.75 | 13.80 | 13.83 | 13.86 | 13.89 | 13.91 | 13.93 | 13.93 | 13.96 | 13.97 | 13.97 | 13.98 | 13.98 | 13.98 | 13.98 |
| | $K$ | 2.49 | 4.36 | 5.45 | 6.00 | 6.25 | 6.37 | 6.42 | 6.44 | 6.45 | 6.46 | 6.46 | 6.46 | 6.46 | 6.46 | 6.46 | 6.46 | 6.46 | 6.46 | 6.46 | 6.46 | 6.46 | 6.46 | 6.46 | 6.46 | 6.46 |

TABLE S15. Parameters of the THF interaction Hamiltonian for different ratios $u_0/u_1$ and screening lengths $\xi$ of the double-gate interaction potential ($U_\xi = 24$ meV for $\xi = 10$ nm). We use the same BM model parameters as in Table S14.

| $u_0/u_1$ | $\xi$/nm | 2 | 4 | 6 | 8 | 10 | 12 | 14 | 16 | 18 | 20 | 22 | 24 | 26 | 28 | 30 | 32 | 34 | 36 | 38 | 40 | 42 | 44 | 46 | 48 | 50 |
|---|---|---|---|---|---|---|---|---|---|---|---|---|---|---|---|---|---|---|---|---|---|---|---|---|---|---|
| 0.6 | $U_1$ | 14.17 | 25.32 | 33.39 | 39.25 | 43.61 | 46.96 | 49.60 | 51.73 | 53.47 | 54.92 | 56.15 | 57.21 | 58.12 | 58.91 | 59.61 | 60.23 | 60.78 | 61.28 | 61.73 | 62.13 | 62.50 | 62.84 | 63.15 | 63.44 | 63.70 |
| | $U_2$ | 0.14 | 0.30 | 0.48 | 0.72 | 1.05 | 1.44 | 1.90 | 2.39 | 2.89 | 3.39 | 3.88 | 4.36 | 4.81 | 5.23 | 5.63 | 6.01 | 6.36 | 6.69 | 7.00 | 7.29 | 7.56 | 7.82 | 8.06 | 8.28 | 8.49 |
| | $W_{1,2}$ | 3.93 | 8.30 | 13.29 | 18.56 | 24.01 | 29.55 | 35.13 | 40.73 | 46.33 | 51.93 | 57.53 | 63.13 | 68.73 | 74.33 | 79.93 | 85.53 | 91.12 | 96.73 | 102.44 | 108.05 | 113.66 | 119.27 | 124.88 | 130.50 | 136.11 |
| | $W_{3,4}$ | 7.24 | 14.33 | 20.92 | 27.05 | 32.90 | 38.63 | 44.29 | 49.92 | 55.54 | 61.16 | 66.77 | 72.38 | 77.99 | 83.61 | 89.22 | 94.83 | 100.44 | 106.05 | 111.66 | 117.27 | 122.88 | 128.49 | 134.10 | 139.72 | 145.33 |
| | $J$ | 5.54 | 9.18 | 11.29 | 12.56 | 13.38 | 13.93 | 14.32 | 14.60 | 14.82 | 14.98 | 15.11 | 15.21 | 15.29 | 15.35 | 15.40 | 15.45 | 15.48 | 15.51 | 15.54 | 15.56 | 15.57 | 15.59 | 15.60 | 15.61 | 15.62 |
| | $K$ | 2.42 | 4.26 | 5.33 | 5.88 | 6.14 | 6.25 | 6.30 | 6.33 | 6.34 | 6.34 | 6.34 | 6.34 | 6.34 | 6.34 | 6.34 | 6.34 | 6.34 | 6.34 | 6.34 | 6.34 | 6.34 | 6.34 | 6.34 | 6.34 | 6.34 |
| 0.5 | $U_1$ | 12.73 | 22.92 | 30.44 | 35.98 | 40.15 | 43.38 | 45.94 | 48.02 | 49.73 | 51.16 | 52.37 | 53.41 | 54.31 | 55.09 | 55.79 | 56.40 | 56.95 | 57.45 | 57.89 | 58.30 | 58.66 | 59.00 | 59.31 | 59.59 | 59.86 |
| | $U_2$ | 0.32 | 0.52 | 0.77 | 1.11 | 1.52 | 1.98 | 2.48 | 2.98 | 3.48 | 3.98 | 4.48 | 4.96 | 5.41 | 5.77 | 6.10 | 6.45 | 6.78 | 7.09 | 7.38 | 7.65 | 7.90 | 8.14 | 8.36 | 8.57 | 8.76 |
| | $W_{1,2}$ | 3.91 | 8.30 | 13.21 | 18.47 | 23.92 | 29.46 | 35.03 | 40.63 | 46.23 | 51.84 | 57.45 | 63.06 | 68.67 | 74.28 | 79.90 | 85.51 | 91.12 | 96.73 | 102.34 | 107.95 | 113.56 | 119.17 | 124.78 | 130.39 | 136.01 |
| | $W_{3,4}$ | 6.84 | 13.60 | 19.97 | 25.65 | 31.79 | 37.49 | 43.14 | 48.76 | 54.38 | 60.00 | 65.61 | 71.22 | 76.83 | 82.44 | 88.05 | 93.67 | 99.28 | 104.89 | 110.50 | 116.11 | 121.72 | 127.33 | 132.94 | 138.55 | 144.16 |
| | $J$ | 5.50 | 9.28 | 11.62 | 13.10 | 14.10 | 14.81 | 15.32 | 15.71 | 16.00 | 16.23 | 16.41 | 16.55 | 16.66 | 16.75 | 16.83 | 16.89 | 16.94 | 16.98 | 17.02 | 17.05 | 17.07 | 17.09 | 17.11 | 17.13 | 17.14 |
| | $K$ | 2.40 | 4.22 | 5.31 | 5.86 | 6.11 | 6.23 | 6.28 | 6.31 | 6.32 | 6.32 | 6.32 | 6.32 | 6.32 | 6.32 | 6.32 | 6.32 | 6.32 | 6.32 | 6.32 | 6.32 | 6.32 | 6.32 | 6.32 | 6.32 | 6.32 |
| 0.4 | $U_1$ | 11.68 | 21.16 | 28.27 | 33.56 | 37.59 | 40.74 | 43.24 | 45.27 | 46.95 | 48.36 | 49.56 | 50.59 | 51.48 | 52.26 | 52.95 | 53.57 | 54.13 | 54.60 | 55.05 | 55.45 | 55.81 | 56.15 | 56.46 | 56.74 | 57.00 |
| | $U_2$ | 0.16 | 0.33 | 0.54 | 0.81 | 1.15 | 1.57 | 2.04 | 2.53 | 3.04 | 3.55 | 4.04 | 4.51 | 4.96 | 5.38 | 5.78 | 6.16 | 6.51 | 6.84 | 7.15 | 7.43 | 7.70 | 7.96 | 8.19 | 8.42 | 8.63 |
| | $W_{1,2}$ | 3.83 | 8.15 | 13.08 | 18.27 | 23.69 | 29.22 | 34.80 | 40.46 | 46.00 | 51.60 | 57.21 | 62.82 | 68.43 | 74.05 | 79.66 | 85.27 | 90.88 | 96.49 | 102.10 | 107.71 | 113.32 | 118.93 | 124.54 | 130.16 | 135.77 |
| | $W_{3,4}$ | 6.39 | 12.73 | 18.84 | 24.72 | 30.46 | 36.13 | 41.77 | 47.39 | 53.01 | 58.62 | 64.23 | 69.85 | 75.46 | 81.07 | 86.68 | 92.29 | 97.90 | 103.51 | 109.12 | 114.73 | 120.34 | 125.96 | 131.57 | 137.18 | 142.79 |
| | $J$ | 5.51 | 9.43 | 11.96 | 13.62 | 14.78 | 15.61 | 16.22 | 16.69 | 17.05 | 17.32 | 17.54 | 17.72 | 17.86 | 17.97 | 18.06 | 18.14 | 18.20 | 18.25 | 18.29 | 18.33 | 18.36 | 18.39 | 18.41 | 18.43 | 18.45 |
| | $K$ | 2.41 | 4.25 | 5.35 | 5.91 | 6.17 | 6.29 | 6.35 | 6.37 | 6.38 | 6.39 | 6.39 | 6.39 | 6.39 | 6.39 | 6.39 | 6.39 | 6.39 | 6.39 | 6.39 | 6.39 | 6.39 | 6.39 | 6.39 | 6.39 | 6.39 |
| 0.3 | $U_1$ | 11.12 | 20.25 | 27.16 | 32.36 | 36.34 | 39.45 | 41.94 | 43.96 | 45.63 | 47.04 | 48.24 | 49.26 | 50.13 | 50.93 | 51.62 | 52.23 | 52.78 | 53.27 | 53.71 | 54.12 | 54.48 | 54.82 | 55.13 | 55.41 | 55.67 |
| | $U_2$ | 0.32 | 0.52 | 0.80 | 1.15 | 1.56 | 2.03 | 2.55 | 3.05 | 3.54 | 4.03 | 4.52 | 5.01 | 5.40 | 5.79 | 6.17 | 6.50 | 6.83 | 7.14 | 7.45 | 7.71 | 7.97 | 8.21 | 8.43 | 8.64 | 8.84 |
| | $W_{1,2}$ | 3.69 | 7.91 | 12.72 | 17.92 | 23.34 | 28.87 | 34.44 | 40.04 | 45.64 | 51.25 | 56.86 | 62.47 | 68.08 | 73.69 | 79.30 | 84.91 | 90.52 | 96.13 | 101.74 | 107.36 | 112.97 | 118.58 | 124.19 | 129.80 | 135.41 |
| | $W_{3,4}$ | 6.38 | 12.56 | 18.52 | 24.32 | 30.03 | 35.68 | 41.29 | 46.89 | 52.49 | 58.09 | 63.69 | 69.30 | 74.90 | 80.51 | 86.11 | 91.72 | 97.32 | 102.93 | 108.53 | 114.14 | 119.74 | 125.35 | 130.95 | 136.56 | 142.42 |
| | $J$ | 5.45 | 9.76 | 12.52 | 14.32 | 15.62 | 16.59 | 17.29 | 17.82 | 18.22 | 18.54 | 18.78 | 18.97 | 19.11 | 19.18 | 19.19 | 19.21 | 19.22 | 19.23 | 19.24 | 19.25 | 19.25 | 19.26 | 19.26 | 19.27 | 19.27 |
| | $K$ | 2.45 | 4.33 | 5.45 | 6.02 | 6.29 | 6.41 | 6.47 | 6.50 | 6.51 | 6.51 | 6.51 | 6.51 | 6.51 | 6.51 | 6.51 | 6.51 | 6.51 | 6.51 | 6.51 | 6.51 | 6.51 | 6.51 | 6.51 | 6.51 | 6.51 |
| 0.2 | $U_1$ | 10.71 | 19.43 | 26.40 | 31.22 | 35.13 | 38.20 | 40.66 | 42.66 | 44.32 | 45.72 | 46.91 | 47.93 | 48.82 | 49.59 | 50.28 | 50.90 | 51.46 | 51.96 | 52.52 | 52.95 | 53.35 | 53.72 | 54.05 | 54.36 | 54.91 |
| | $U_2$ | 0.15 | 0.32 | 0.53 | 0.80 | 1.15 | 1.57 | 2.05 | 2.55 | 3.06 | 3.56 | 4.05 | 4.52 | 4.97 | 5.40 | 5.79 | 6.17 | 6.44 | 6.77 | 7.08 | 7.37 | 7.64 | 7.89 | 8.13 | 8.35 | 8.56 |
| | $W_{1,2}$ | 3.57 | 7.70 | 12.45 | 17.62 | 23.03 | 28.55 | 34.12 | 39.71 | 45.32 | 50.92 | 56.53 | 62.14 | 67.75 | 73.37 | 78.98 | 84.59 | 90.20 | 95.81 | 101.42 | 107.03 | 112.64 | 118.25 | 123.86 | 129.47 | 135.09 |
| | $W_{3,4}$ | 5.61 | 12.50 | 18.54 | 24.38 | 30.11 | 35.77 | 41.40 | 47.02 | 52.64 | 58.25 | 63.86 | 69.48 | 75.09 | 80.70 | 86.31 | 91.92 | 97.53 | 103.14 | 108.75 | 114.36 | 119.97 | 125.59 | 131.20 | 136.81 | 142.42 |
| | $J$ | 5.61 | 9.76 | 12.55 | 14.44 | 15.82 | 16.80 | 17.54 | 18.09 | 18.52 | 18.85 | 19.11 | 19.32 | 19.48 | 19.61 | 19.72 | 19.81 | 19.88 | 19.94 | 19.99 | 20.04 | 20.07 | 20.10 | 20.13 | 20.15 | 20.17 |
| | $K$ | 2.50 | 4.42 | 5.57 | 6.16 | 6.43 | 6.56 | 6.61 | 6.64 | 6.65 | 6.66 | 6.66 | 6.66 | 6.66 | 6.66 | 6.66 | 6.66 | 6.66 | 6.66 | 6.66 | 6.66 | 6.66 | 6.66 | 6.66 | 6.66 | 6.66 |
| 0.1 | $U_1$ | 10.87 | 19.90 | 26.81 | 31.95 | 35.85 | 38.90 | 41.34 | 43.33 | 44.96 | 46.35 | 47.54 | 48.55 | 49.43 | 50.20 | 50.85 | 51.51 | 52.06 | 52.55 | 53.00 | 53.40 | 53.77 | 54.11 | 54.42 | 54.76 | 55.35 |
| | $U_2$ | 0.12 | 0.28 | 0.47 | 0.73 | 1.08 | 1.49 | 1.96 | 2.45 | 2.97 | 3.47 | 3.97 | 4.44 | 4.89 | 5.31 | 5.71 | 6.09 | 6.44 | 6.77 | 7.08 | 7.37 | 7.64 | 7.89 | 8.13 | 8.35 | 8.56 |
| | $W_{1,2}$ | 3.37 | 7.35 | 12.01 | 17.31 | 22.70 | 28.21 | 33.78 | 39.37 | 44.97 | 50.58 | 56.19 | 61.80 | 67.41 | 73.02 | 78.63 | 84.24 | 89.86 | 95.47 | 101.08 | 106.69 | 112.30 | 117.91 | 123.52 | 129.13 | 134.74 |
| | $W_{3,4}$ | 6.19 | 12.38 | 18.39 | 24.21 | 29.83 | 35.54 | 41.22 | 46.84 | 52.45 | 58.07 | 63.68 | 69.29 | 74.90 | 80.51 | 86.12 | 91.73 | 97.35 | 102.96 | 108.57 | 114.18 | 119.79 | 125.40 | 131.01 | 136.62 | 142.23 |
| | $J$ | 5.70 | 9.88 | 12.73 | 14.63 | 16.03 | 17.01 | 17.74 | 18.28 | 18.70 | 19.02 | 19.26 | 19.45 | 19.59 | 19.71 | 19.79 | 19.83 | 19.88 | 19.92 | 19.95 | 19.98 | 20.00 | 20.01 | 20.02 | 20.02 | 20.03 |
| | $K$ | 2.56 | 4.50 | 5.67 | 6.27 | 6.55 | 6.68 | 6.73 | 6.76 | 6.77 | 6.78 | 6.78 | 6.78 | 6.78 | 6.78 | 6.78 | 6.78 | 6.78 | 6.78 | 6.78 | 6.78 | 6.78 | 6.78 | 6.78 | 6.78 | 6.78 |
| 0.0 | $U_1$ | 19.00 | 26.40 | 31.26 | 34.74 | 37.35 | 39.49 | 41.22 | 42.66 | 43.88 | 44.89 | 45.71 | 46.44 | 47.07 | 47.62 | 48.10 | 48.48 | 48.83 | 49.16 | 49.46 | 49.73 | 49.98 | 50.16 | 50.07 | 55.07 | 55.62 |
| | $U_2$ | 0.12 | 0.25 | 0.43 | 0.69 | 1.02 | 1.44 | 1.90 | 2.40 | 2.91 | 3.41 | 3.91 | 4.38 | 4.83 | 5.26 | 5.66 | 6.04 | 6.39 | 6.72 | 7.03 | 7.32 | 7.59 | 7.84 | 8.08 | 8.30 | 8.51 |
| | $W_{1,2}$ | 5.71 | 12.01 | 17.44 | 22.53 | 28.04 | 33.60 | 39.20 | 44.80 | 50.41 | 56.01 | 61.62 | 67.24 | 72.85 | 78.46 | 84.07 | 89.68 | 95.29 | 100.90 | 106.51 | 112.12 | 117.73 | 123.35 | 128.96 | 134.57 | 142.19 |
| | $W_{3,4}$ | 8.35 | 14.10 | 19.88 | 25.69 | 31.00 | 36.58 | 42.19 | 47.52 | 53.13 | 58.74 | 64.35 | 69.96 | 75.57 | 81.18 | 86.79 | 92.40 | 98.01 | 103.62 | 109.23 | 114.84 | 120.45 | 126.06 | 131.67 | 137.28 | 142.19 |
| | $J$ | 5.71 | 9.94 | 12.80 | 14.74 | 16.10 | 17.08 | 17.79 | 18.33 | 18.73 | 19.04 | 19.28 | 19.46 | 19.61 | 19.73 | 19.82 | 19.90 | 19.96 | 20.01 | 20.05 | 20.09 | 20.12 | 20.14 | 20.16 | 20.18 | 20.20 |
| | $K$ | 2.56 | 4.53 | 5.71 | 6.31 | 6.59 | 6.72 | 6.78 | 6.81 | 6.82 | 6.83 | 6.83 | 6.83 | 6.83 | 6.83 | 6.83 | 6.83 | 6.83 | 6.83 | 6.83 | 6.83 | 6.83 | 6.83 | 6.83 | 6.83 | 6.83 |

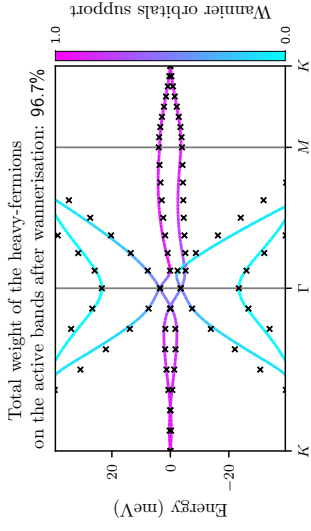

Total weight of the heavy-fermions
on the active bands after wannierisation: 96.7%

Wannier orbitals support

Energy (meV)

FIG. S24. Band structures of the BM and THF models near charge neutrality for $w_0/w_1 = 0.8$, depicted by lines and crosses, respectively. The BM bands are colored according to the weight of the $f$-electron wave function on them. We use the same BM parameters as in Table S16.

TABLE S16. Parameters of the $f$-electron Wannier functions and the THF single-particle Hamiltonian for different values of the tuning ratio $w_0/w_1$. $\mathcal{W}$ denotes the total weight of the THF $f$-electrons on the active bands. In computing the ratios $\gamma/U_1$ and $\gamma/U_2$, we employ the on-site and nearest-neighbor repulsion parameters $U_1$ and $U_2$ obtained numerically for $\xi = 10$ nm, as given in Table S17. We employ $v_F = 5.944\,\text{eV·Å}$, $|\mathbf{K}| = 1.703\,\text{Å}^{-1}$, $w_1 = 110$ meV, and $\theta = 0.82°$ for the BM model.

| $w_0/w_1$ | $\alpha_1$ | $\alpha_2$ | $\lambda$ | $\lambda_2$ | $\lambda$ | $\mathcal{W}$ | $t_0$ | $\gamma$ | $M$ | $v_\star$ | $v'_\star$ | $v''_\star$ | $\gamma/U_1$ | $\gamma/U_2$ |
|---|---|---|---|---|---|---|---|---|---|---|---|---|---|---|
| 1.00 | 0.747 | 0.665 | 0.125 | 0.149 | 0.550 | 0.965 | 1.037 | 22.972 | 19.590 | -1.938 | 0.803 | 0.1913 | 0.52 | 11.22 |
| 0.90 | 0.759 | 0.651 | 0.137 | 0.155 | 0.429 | 0.963 | 0.641 | 14.426 | 21.687 | -2.530 | 1.061 | 0.0692 | 0.29 | 9.97 |
| 0.80 | 0.781 | 0.624 | 0.149 | 0.161 | 0.346 | 0.967 | 0.682 | 3.641 | 23.430 | -3.011 | 1.246 | -0.0140 | 0.07 | 3.38 |
| 0.70 | 0.812 | 0.583 | 0.161 | 0.173 | 0.331 | 0.960 | 1.119 | -8.367 | 24.842 | -3.391 | 1.356 | -0.0621 | -0.17 | -8.03 |
| 0.60 | 0.849 | 0.529 | 0.173 | 0.179 | 0.341 | 0.947 | 1.710 | -20.770 | 25.964 | -3.693 | 1.397 | -0.0826 | -0.47 | -18.42 |
| 0.50 | 0.887 | 0.462 | 0.191 | 0.185 | 0.354 | 0.931 | 2.302 | -32.886 | 26.839 | -3.935 | 1.366 | -0.0827 | -0.81 | -27.27 |
| 0.40 | 0.922 | 0.386 | 0.203 | 0.191 | 0.358 | 0.912 | 2.792 | -44.103 | 27.505 | -4.130 | 1.259 | -0.0688 | -1.15 | -35.89 |
| 0.30 | 0.953 | 0.302 | 0.215 | 0.197 | 0.357 | 0.894 | 3.142 | -53.813 | 27.991 | -4.285 | 1.067 | -0.0472 | -1.45 | -44.48 |
| 0.20 | 0.978 | 0.208 | 0.227 | 0.197 | 0.364 | 0.888 | 3.500 | -61.395 | 28.323 | -4.401 | 0.784 | -0.0242 | -1.72 | -49.59 |
| 0.10 | 0.994 | 0.107 | 0.233 | 0.203 | 0.351 | 0.870 | 3.578 | -66.254 | 28.516 | -4.475 | 0.418 | -0.0066 | -1.82 | -57.91 |
| 0.00 | 1.000 | 0.000 | 0.239 | 0.000 | 0.349 | 0.867 | 3.590 | -67.932 | 28.579 | -4.501 | 0.000 | -0.0000 | -1.86 | -60.23 |

| $w_0/w_1$ | $\xi/\text{nm}$ | 2 | 4 | 6 | 8 | 10 | 12 | 14 | 16 | 18 | 20 | 22 | 24 | 26 | 28 | 30 | 32 | 34 | 36 | 38 | 40 | 42 | 44 | 46 | 48 | 50 |
|---|---|---|---|---|---|---|---|---|---|---|---|---|---|---|---|---|---|---|---|---|---|---|---|---|---|---|
| 1.0 | $U_1$ | 16.15 | 27.75 | 35.40 | 40.75 | 44.56 | 47.43 | 49.67 | 51.48 | 52.97 | 54.22 | 55.29 | 56.21 | 57.01 | 57.71 | 58.34 | 58.90 | 59.40 | 59.85 | 60.26 | 60.64 | 60.98 | 61.29 | 61.58 | 61.85 | 62.10 |
| | $U_2$ | 0.46 | 0.89 | 1.27 | 1.64 | 2.05 | 2.50 | 2.98 | 3.47 | 3.98 | 4.47 | 4.95 | 5.44 | 5.84 | 6.25 | 6.63 | 6.99 | 7.33 | 7.65 | 7.94 | 8.22 | 8.48 | 8.73 | 8.96 | 9.17 | 9.38 |
| | $W_{1,2}$ | 3.77 | 8.31 | 13.50 | 19.07 | 24.81 | 30.64 | 36.51 | 42.39 | 48.28 | 54.18 | 60.07 | 65.97 | 71.86 | 77.76 | 83.65 | 89.55 | 95.44 | 101.34 | 107.23 | 113.13 | 119.02 | 124.92 | 130.81 | 136.71 | 142.60 |
| | $W_{3,4}$ | 10.26 | 19.80 | 27.91 | 34.92 | 41.33 | 47.45 | 53.45 | 59.38 | 65.30 | 71.20 | 77.10 | 83.00 | 88.90 | 94.79 | 100.68 | 106.58 | 112.47 | 118.37 | 124.26 | 130.16 | 136.05 | 141.95 | 147.84 | 153.74 | 159.63 |
| | $J$ | 6.88 | 10.74 | 12.73 | 13.87 | 14.60 | 15.12 | 15.52 | 15.86 | 16.14 | 16.39 | 16.61 | 16.80 | 16.97 | 17.13 | 17.27 | 17.39 | 17.51 | 17.61 | 17.70 | 17.78 | 17.86 | 17.92 | 17.98 | 18.04 | 18.09 |
| | $K$ | 2.96 | 5.03 | 6.16 | 6.69 | 7.04 | 7.09 | 7.11 | 7.12 | 7.12 | 7.12 | 7.12 | 7.12 | 7.12 | 7.12 | 7.12 | 7.12 | 7.12 | 7.12 | 7.12 | 7.12 | 7.12 | 7.12 | 7.12 | 7.12 | 7.12 |
| 0.9 | $U_1$ | 17.85 | 30.98 | 39.87 | 46.02 | 50.46 | 53.79 | 56.38 | 58.45 | 60.14 | 61.55 | 62.73 | 63.75 | 64.63 | 65.40 | 66.07 | 66.68 | 67.21 | 67.70 | 68.13 | 68.53 | 68.89 | 69.22 | 69.52 | 69.80 | 70.06 |
| | $U_2$ | 0.27 | 0.54 | 0.80 | 1.09 | 1.45 | 1.87 | 2.35 | 2.85 | 3.37 | 3.88 | 4.37 | 4.85 | 5.30 | 5.73 | 6.13 | 6.51 | 6.86 | 7.19 | 7.50 | 7.79 | 8.06 | 8.32 | 8.55 | 8.78 | 8.99 |
| | $W_{1,2}$ | 3.85 | 8.39 | 13.56 | 19.11 | 24.85 | 30.68 | 36.54 | 42.42 | 48.31 | 54.21 | 60.10 | 66.00 | 71.89 | 77.79 | 83.68 | 89.58 | 95.47 | 101.37 | 107.26 | 113.16 | 119.05 | 124.95 | 130.84 | 136.74 | 142.63 |
| | $W_{3,4}$ | 9.38 | 18.26 | 26.01 | 32.84 | 39.17 | 45.26 | 51.24 | 57.17 | 63.08 | 68.98 | 74.88 | 80.78 | 86.67 | 92.57 | 98.46 | 104.36 | 110.25 | 116.15 | 122.04 | 127.94 | 133.83 | 139.73 | 145.62 | 151.52 | 157.41 |
| | $J$ | 9.85 | 11.59 | 12.47 | 12.94 | 13.22 | 13.39 | 13.50 | 13.59 | 13.65 | 13.70 | 13.74 | 13.78 | 13.81 | 13.83 | 13.86 | 13.89 | 13.92 | 13.93 | 13.92 | 13.92 | 13.95 | 13.96 | 14.39 | 14.39 | 14.40 |
| | $K$ | 2.75 | 4.73 | 5.83 | 6.37 | 6.61 | 6.71 | 6.76 | 6.78 | 6.79 | 6.79 | 6.79 | 6.79 | 6.80 | 6.80 | 6.80 | 6.80 | 6.80 | 6.80 | 6.80 | 6.80 | 6.80 | 6.80 | 6.80 | 6.80 | 6.80 |
| 0.8 | $U_1$ | 17.79 | 33.19 | 40.45 | 46.94 | 51.66 | 55.22 | 57.98 | 60.18 | 61.98 | 63.47 | 64.72 | 65.79 | 66.71 | 67.51 | 68.22 | 68.84 | 69.40 | 69.90 | 70.35 | 70.75 | 71.12 | 71.46 | 71.77 | 72.06 | 72.32 |
| | $U_2$ | 0.16 | 0.32 | 0.51 | 0.75 | 1.08 | 1.49 | 1.96 | 2.47 | 2.99 | 3.51 | 4.02 | 4.51 | 4.97 | 5.41 | 5.82 | 6.21 | 6.57 | 6.90 | 7.22 | 7.51 | 7.79 | 8.05 | 8.29 | 8.52 | 8.73 |
| | $W_{1,2}$ | 3.99 | 8.58 | 13.77 | 19.32 | 25.06 | 30.89 | 36.76 | 42.64 | 48.53 | 54.42 | 60.32 | 66.21 | 72.11 | 78.00 | 83.90 | 89.79 | 95.69 | 101.58 | 107.48 | 113.37 | 119.27 | 125.16 | 131.06 | 136.95 | 142.85 |
| | $W_{3,4}$ | 8.58 | 16.83 | 24.21 | 30.87 | 37.11 | 43.17 | 49.13 | 55.05 | 60.96 | 66.86 | 72.76 | 78.66 | 84.55 | 90.45 | 96.34 | 102.24 | 108.13 | 114.03 | 119.92 | 125.82 | 131.71 | 137.61 | 143.50 | 149.40 | 155.29 |
| | $J$ | 5.05 | 9.43 | 11.17 | 12.05 | 12.51 | 12.77 | 12.92 | 13.00 | 13.05 | 13.09 | 13.11 | 13.12 | 13.13 | 13.13 | 13.14 | 13.15 | 13.15 | 13.15 | 13.15 | 13.15 | 13.15 | 13.15 | 13.15 | 13.15 | 13.15 |
| | $K$ | 2.59 | 4.49 | 5.56 | 6.09 | 6.34 | 6.43 | 6.48 | 6.50 | 6.51 | 6.51 | 6.51 | 6.51 | 6.51 | 6.51 | 6.51 | 6.51 | 6.51 | 6.51 | 6.51 | 6.51 | 6.51 | 6.51 | 6.51 | 6.51 | 6.51 |
| 0.7 | $U_1$ | 16.20 | 28.67 | 37.48 | 43.76 | 48.39 | 51.89 | 54.63 | 56.83 | 58.62 | 60.11 | 61.36 | 62.43 | 63.36 | 64.16 | 64.87 | 65.49 | 66.06 | 66.56 | 67.02 | 67.42 | 67.79 | 68.13 | 68.44 | 68.73 | 68.99 |
| | $U_2$ | 0.14 | 0.29 | 0.47 | 0.71 | 1.04 | 1.46 | 1.94 | 2.45 | 2.98 | 3.50 | 4.01 | 4.50 | 4.97 | 5.41 | 5.82 | 6.21 | 6.57 | 6.91 | 7.22 | 7.52 | 7.80 | 8.05 | 8.30 | 8.52 | 8.74 |
| | $W_{1,2}$ | 4.11 | 8.76 | 13.97 | 19.53 | 25.27 | 31.10 | 36.97 | 42.85 | 48.74 | 54.63 | 60.53 | 66.42 | 72.32 | 78.21 | 84.11 | 90.00 | 95.90 | 101.79 | 107.69 | 113.58 | 119.48 | 125.37 | 131.27 | 137.16 | 143.06 |
| | $W_{3,4}$ | 7.90 | 15.59 | 22.64 | 29.12 | 35.29 | 41.31 | 47.26 | 53.17 | 59.08 | 64.98 | 70.88 | 76.77 | 82.67 | 88.56 | 94.46 | 100.35 | 106.25 | 112.14 | 118.04 | 123.93 | 129.83 | 135.72 | 141.62 | 147.51 | 153.41 |
| | $J$ | 5.76 | 9.33 | 11.27 | 12.34 | 12.98 | 13.38 | 13.65 | 13.83 | 13.96 | 14.05 | 14.13 | 14.19 | 14.23 | 14.27 | 14.29 | 14.33 | 14.35 | 14.36 | 14.38 | 14.39 | 14.40 | 14.41 | 14.42 | 14.39 | 14.40 |
| | $K$ | 2.48 | 4.32 | 5.38 | 5.90 | 6.13 | 6.24 | 6.28 | 6.30 | 6.31 | 6.32 | 6.32 | 6.32 | 6.32 | 6.32 | 6.32 | 6.32 | 6.32 | 6.32 | 6.32 | 6.32 | 6.32 | 6.32 | 6.32 | 6.32 | 6.32 |

| $u_0/u_1$ | $\xi$/nm | 2 | 4 | 6 | 8 | 10 | 12 | 14 | 16 | 18 | 20 | 22 | 24 | 26 | 28 | 30 | 32 | 34 | 36 | 38 | 40 | 42 | 44 | 46 | 48 | 50 |
|---|---|---|---|---|---|---|---|---|---|---|---|---|---|---|---|---|---|---|---|---|---|---|---|---|---|---|
| 0.6 | $U_1$ | 14.38 | 25.67 | 33.82 | 39.73 | 44.13 | 47.50 | 50.16 | 52.29 | 54.05 | 55.51 | 56.74 | 57.80 | 58.72 | 59.51 | 60.22 | 60.84 | 61.39 | 61.89 | 62.34 | 62.75 | 63.12 | 63.45 | 63.76 | 64.05 | 64.31 |
|  | $U_2$ | 0.15 | 0.32 | 0.51 | 0.66 | 0.84 | 1.13 | 1.56 | 2.04 | 2.56 | 3.09 | 3.62 | 4.13 | 4.62 | 5.08 | 5.52 | 5.93 | 6.32 | 6.68 | 7.02 | 7.33 | 7.63 | 7.90 | 8.16 | 8.40 | 8.63 |
|  | $W_{1,2}$ | 4.15 | 8.92 | 14.03 | 19.50 | 25.34 | 31.16 | 37.03 | 42.91 | 48.80 | 54.69 | 60.40 | 66.48 | 72.38 | 78.27 | 84.17 | 90.06 | 95.96 | 101.85 | 107.75 | 113.64 | 119.54 | 125.43 | 131.33 | 137.22 | 143.12 |
|  | $W_{3,4}$ | 7.38 | 14.64 | 21.41 | 27.76 | 33.86 | 39.85 | 45.79 | 51.70 | 57.60 | 63.50 | 69.40 | 75.29 | 81.19 | 87.08 | 92.98 | 98.87 | 104.77 | 110.66 | 116.56 | 122.45 | 128.35 | 134.24 | 140.14 | 146.04 | 151.93 |
|  | $J$ | 5.67 | 9.40 | 11.58 | 12.91 | 13.76 | 14.35 | 14.77 | 15.07 | 15.30 | 15.48 | 15.62 | 15.73 | 15.82 | 15.89 | 15.95 | 15.99 | 16.03 | 16.06 | 16.09 | 16.11 | 16.13 | 16.15 | 16.16 | 16.18 | 16.19 |
|  | $K$ | 2.42 | 4.23 | 5.28 | 5.80 | 6.03 | 6.14 | 6.18 | 6.20 | 6.21 | 6.21 | 6.21 | 6.21 | 6.22 | 6.22 | 6.22 | 6.22 | 6.22 | 6.22 | 6.22 | 6.22 | 6.22 | 6.22 | 6.22 | 6.22 | 6.22 |
| 0.5 | $U_1$ | 12.90 | 23.20 | 30.78 | 36.30 | 40.56 | 43.81 | 46.38 | 48.47 | 50.18 | 51.62 | 52.83 | 53.87 | 54.78 | 55.57 | 56.26 | 56.88 | 57.43 | 57.92 | 58.37 | 58.77 | 59.14 | 59.48 | 59.79 | 60.07 | 60.34 |
|  | $U_2$ | 0.16 | 0.34 | 0.56 | 0.84 | 1.21 | 1.65 | 2.14 | 2.66 | 3.20 | 3.72 | 4.23 | 4.72 | 5.19 | 5.62 | 6.04 | 6.42 | 6.78 | 7.12 | 7.43 | 7.72 | 8.00 | 8.26 | 8.50 | 8.72 | 8.94 |
|  | $W_{1,2}$ | 4.13 | 8.76 | 13.95 | 19.50 | 25.23 | 31.06 | 36.92 | 42.80 | 48.69 | 54.59 | 60.48 | 66.37 | 72.27 | 78.16 | 84.06 | 89.95 | 95.85 | 101.74 | 107.64 | 113.54 | 119.43 | 125.33 | 131.22 | 137.12 | 143.01 |
|  | $W_{3,4}$ | 7.01 | 13.94 | 20.51 | 26.75 | 32.81 | 38.78 | 44.71 | 50.62 | 56.52 | 62.41 | 68.31 | 74.21 | 80.10 | 86.00 | 91.89 | 97.79 | 103.68 | 109.58 | 115.47 | 121.37 | 127.26 | 133.16 | 139.05 | 144.95 | 150.84 |
|  | $J$ | 5.65 | 9.54 | 11.96 | 13.51 | 14.57 | 15.32 | 15.87 | 16.28 | 16.60 | 16.85 | 17.04 | 17.19 | 17.32 | 17.42 | 17.50 | 17.57 | 17.62 | 17.67 | 17.71 | 17.74 | 17.77 | 17.79 | 17.81 | 17.83 | 17.84 |
|  | $K$ | 2.40 | 4.21 | 5.26 | 5.79 | 6.03 | 6.13 | 6.18 | 6.20 | 6.21 | 6.21 | 6.21 | 6.21 | 6.21 | 6.21 | 6.21 | 6.21 | 6.21 | 6.21 | 6.21 | 6.21 | 6.21 | 6.21 | 6.21 | 6.21 | 6.21 |
| 0.4 | $U_1$ | 11.43 | 20.78 | 27.84 | 33.13 | 37.18 | 40.33 | 42.85 | 44.90 | 46.60 | 48.01 | 49.21 | 50.25 | 51.14 | 51.93 | 52.62 | 53.23 | 53.78 | 54.28 | 54.72 | 55.12 | 55.49 | 55.83 | 56.14 | 56.42 | 56.68 |
|  | $U_2$ | 0.16 | 0.35 | 0.57 | 0.86 | 1.23 | 1.68 | 2.17 | 2.70 | 3.23 | 3.76 | 4.27 | 4.76 | 5.23 | 5.66 | 6.07 | 6.46 | 6.82 | 7.15 | 7.47 | 7.76 | 8.02 | 8.29 | 8.52 | 8.75 | 8.97 |
|  | $W_{1,2}$ | 4.03 | 8.58 | 13.72 | 19.25 | 24.97 | 30.79 | 36.66 | 42.54 | 48.43 | 54.32 | 60.21 | 66.11 | 72.00 | 77.90 | 83.79 | 89.69 | 95.58 | 101.48 | 107.37 | 113.27 | 119.16 | 125.06 | 130.95 | 136.85 | 142.74 |
|  | $W_{3,4}$ | 6.74 | 13.45 | 19.45 | 25.58 | 31.58 | 37.53 | 43.44 | 49.35 | 55.25 | 61.14 | 67.04 | 72.93 | 78.83 | 84.72 | 90.62 | 96.51 | 102.41 | 108.30 | 114.20 | 120.09 | 125.99 | 131.88 | 137.78 | 143.67 | 149.57 |
|  | $J$ | 5.57 | 9.89 | 12.63 | 14.48 | 15.77 | 16.74 | 17.39 | 17.91 | 18.31 | 18.61 | 18.85 | 19.04 | 19.19 | 19.31 | 19.41 | 19.49 | 19.55 | 19.61 | 19.65 | 19.69 | 19.72 | 19.75 | 19.77 | 19.79 | 19.81 |
|  | $K$ | 2.47 | 4.35 | 5.43 | 5.97 | 6.22 | 6.33 | 6.38 | 6.40 | 6.41 | 6.42 | 6.42 | 6.42 | 6.42 | 6.42 | 6.42 | 6.42 | 6.42 | 6.42 | 6.42 | 6.42 | 6.42 | 6.42 | 6.42 | 6.42 | 6.42 |
| 0.3 | $U_1$ | 10.88 | 19.85 | 26.88 | 31.83 | 35.79 | 38.89 | 41.37 | 43.39 | 45.07 | 46.48 | 47.67 | 48.70 | 49.55 | 50.37 | 51.06 | 51.67 | 52.22 | 52.71 | 53.15 | 53.55 | 53.92 | 54.26 | 54.56 | 54.84 | 55.11 |
|  | $U_2$ | 0.16 | 0.33 | 0.55 | 0.83 | 1.21 | 1.66 | 2.16 | 2.68 | 3.22 | 3.75 | 4.26 | 4.75 | 5.23 | 5.65 | 6.06 | 6.45 | 6.80 | 7.14 | 7.45 | 7.75 | 8.02 | 8.28 | 8.52 | 8.75 | 8.96 |
|  | $W_{1,2}$ | 3.77 | 8.13 | 13.15 | 18.62 | 24.32 | 30.12 | 35.98 | 41.86 | 47.75 | 53.64 | 59.54 | 65.43 | 71.32 | 77.22 | 83.11 | 89.00 | 94.90 | 100.80 | 106.69 | 112.59 | 118.48 | 124.38 | 130.27 | 136.16 | 142.06 |
|  | $W_{3,4}$ | 6.45 | 12.89 | 19.16 | 25.26 | 31.24 | 37.18 | 43.09 | 49.00 | 54.90 | 60.79 | 66.69 | 72.58 | 78.48 | 84.37 | 90.27 | 96.16 | 102.06 | 107.95 | 113.85 | 119.74 | 125.64 | 131.53 | 137.43 | 143.32 | 149.22 |
|  | $J$ | 5.79 | 10.07 | 12.96 | 14.94 | 16.34 | 17.36 | 18.12 | 18.70 | 19.14 | 19.48 | 19.75 | 19.96 | 20.13 | 20.27 | 20.37 | 20.46 | 20.53 | 20.60 | 20.65 | 20.69 | 20.73 | 20.76 | 20.78 | 20.80 | 20.82 |
|  | $K$ | 2.52 | 4.43 | 5.55 | 6.11 | 6.37 | 6.48 | 6.53 | 6.55 | 6.56 | 6.57 | 6.57 | 6.57 | 6.57 | 6.57 | 6.57 | 6.57 | 6.57 | 6.57 | 6.57 | 6.57 | 6.57 | 6.57 | 6.57 | 6.57 | 6.57 |
| 0.2 | $U_1$ | 11.04 | 20.17 | 27.14 | 32.40 | 36.44 | 39.60 | 42.12 | 44.18 | 45.88 | 47.30 | 48.51 | 49.55 | 50.45 | 51.24 | 51.93 | 52.55 | 53.10 | 53.59 | 54.04 | 54.44 | 54.80 | 55.15 | 55.46 | 55.74 | 56.00 |
|  | $U_2$ | 0.13 | 0.29 | 0.49 | 0.77 | 1.14 | 1.59 | 2.09 | 2.62 | 3.15 | 3.68 | 4.19 | 4.68 | 5.15 | 5.58 | 6.00 | 6.38 | 6.74 | 7.08 | 7.39 | 7.69 | 7.96 | 8.22 | 8.46 | 8.69 | 8.90 |
|  | $W_{1,2}$ | 3.62 | 7.87 | 12.83 | 18.27 | 23.96 | 29.76 | 35.61 | 41.49 | 47.38 | 53.27 | 59.16 | 65.06 | 70.95 | 76.85 | 82.74 | 88.64 | 94.53 | 100.43 | 106.32 | 112.22 | 118.11 | 124.01 | 129.90 | 135.80 | 141.69 |
|  | $W_{3,4}$ | 6.39 | 12.78 | 19.02 | 25.10 | 31.08 | 37.01 | 42.93 | 48.83 | 54.73 | 60.62 | 66.52 | 72.41 | 78.31 | 84.20 | 90.10 | 96.00 | 101.89 | 107.79 | 113.69 | 119.58 | 125.48 | 131.37 | 137.27 | 143.16 | 149.05 |
|  | $J$ | 5.79 | 10.19 | 13.12 | 15.11 | 16.51 | 17.51 | 18.25 | 18.81 | 19.22 | 19.54 | 19.79 | 19.99 | 20.14 | 20.26 | 20.36 | 20.44 | 20.50 | 20.55 | 20.59 | 20.63 | 20.66 | 20.68 | 20.70 | 20.72 | 20.75 |
|  | $K$ | 2.56 | 4.51 | 5.65 | 6.22 | 6.48 | 6.60 | 6.65 | 6.67 | 6.68 | 6.69 | 6.69 | 6.69 | 6.69 | 6.69 | 6.69 | 6.69 | 6.69 | 6.69 | 6.69 | 6.69 | 6.69 | 6.69 | 6.69 | 6.69 | 6.69 |
| 0.1 | $U_1$ | 11.04 | 20.18 | 27.16 | 32.44 | 36.49 | 39.65 | 42.19 | 44.25 | 45.95 | 47.37 | 48.59 | 49.63 | 50.53 | 51.32 | 52.01 | 52.63 | 53.18 | 53.67 | 54.12 | 54.53 | 54.89 | 55.23 | 55.54 | 55.81 | 56.09 |
|  | $U_2$ | 0.13 | 0.28 | 0.48 | 0.76 | 1.13 | 1.57 | 2.07 | 2.60 | 3.14 | 3.66 | 4.18 | 4.67 | 5.13 | 5.57 | 5.98 | 6.37 | 6.74 | 7.06 | 7.38 | 7.67 | 7.95 | 8.21 | 8.45 | 8.67 | 8.89 |
|  | $W_{1,2}$ | 3.57 | 7.79 | 12.73 | 18.16 | 23.84 | 29.64 | 35.49 | 41.37 | 47.25 | 53.15 | 59.04 | 64.93 | 70.83 | 76.72 | 82.62 | 88.51 | 94.41 | 100.30 | 106.20 | 112.09 | 117.99 | 123.88 | 129.78 | 135.67 | 141.57 |
|  | $W_{3,4}$ | 6.38 | 12.75 | 18.98 | 25.03 | 30.66 | 36.96 | 42.87 | 48.77 | 54.67 | 60.57 | 66.46 | 72.36 | 78.25 | 84.15 | 90.04 | 95.94 | 101.83 | 107.73 | 113.62 | 119.52 | 125.41 | 131.31 | 137.20 | 143.10 | 148.99 |
|  | $J$ | 5.88 | 10.25 | 13.21 | 15.23 | 16.67 | 17.66 | 18.41 | 18.97 | 19.39 | 19.71 | 19.96 | 20.16 | 20.31 | 20.43 | 20.53 | 20.60 | 20.68 | 20.73 | 20.77 | 20.80 | 20.84 | 20.87 | 20.89 | 20.91 | 20.93 |
|  | $K$ | 2.58 | 4.54 | 5.69 | 6.26 | 6.53 | 6.65 | 6.70 | 6.72 | 6.73 | 6.74 | 6.74 | 6.74 | 6.74 | 6.74 | 6.74 | 6.74 | 6.74 | 6.74 | 6.74 | 6.74 | 6.74 | 6.74 | 6.74 | 6.74 | 6.74 |
| 0.0 | $U_1$ | 11.04 | 20.18 | 27.16 | 32.44 | 36.49 | 39.65 | 42.19 | 44.25 | 45.95 | 47.37 | 48.59 | 49.63 | 50.53 | 51.32 | 52.01 | 52.63 | 53.18 | 53.67 | 54.12 | 54.53 | 54.89 | 55.23 | 55.54 | 55.81 | 56.09 |
|  | $U_2$ | 0.13 | 0.28 | 0.48 | 0.76 | 1.13 | 1.57 | 2.07 | 2.60 | 3.14 | 3.66 | 4.18 | 4.67 | 5.13 | 5.57 | 5.98 | 6.37 | 6.74 | 7.06 | 7.38 | 7.67 | 7.95 | 8.21 | 8.45 | 8.67 | 8.89 |
|  | $W_{1,2}$ | 3.57 | 7.79 | 12.73 | 18.16 | 23.84 | 29.64 | 35.49 | 41.37 | 47.25 | 53.15 | 59.04 | 64.93 | 70.83 | 76.72 | 82.62 | 88.51 | 94.41 | 100.30 | 106.20 | 112.09 | 117.99 | 123.88 | 129.78 | 135.67 | 141.57 |
|  | $W_{3,4}$ | 6.38 | 12.75 | 18.98 | 25.03 | 30.66 | 36.96 | 42.87 | 48.77 | 54.67 | 60.57 | 66.46 | 72.36 | 78.25 | 84.15 | 90.04 | 95.94 | 101.83 | 107.73 | 113.62 | 119.52 | 125.41 | 131.31 | 137.20 | 143.10 | 148.99 |
|  | $J$ | 5.88 | 10.25 | 13.21 | 15.23 | 16.67 | 17.66 | 18.41 | 18.97 | 19.39 | 19.71 | 19.96 | 20.16 | 20.31 | 20.43 | 20.53 | 20.60 | 20.68 | 20.73 | 20.77 | 20.80 | 20.84 | 20.87 | 20.89 | 20.91 | 20.93 |
|  | $K$ | 2.58 | 4.54 | 5.69 | 6.26 | 6.53 | 6.65 | 6.70 | 6.72 | 6.73 | 6.74 | 6.74 | 6.74 | 6.74 | 6.74 | 6.74 | 6.74 | 6.74 | 6.74 | 6.74 | 6.74 | 6.74 | 6.74 | 6.74 | 6.74 | 6.74 |

TABLE S17. Parameters of the THF interaction Hamiltonian for different ratios $u_0/u_1$ and screening lengths $\xi$ of the double-gate interaction potential ($U_\xi = 24$ meV for $\xi = 10$nm). We use the same BM model parameters as in Table S16.

FIG. S25. Band structures of the BM and THF models near charge neutrality for $u_0/u_1 = 0.8$, depicted by lines and crosses, respectively. The BM bands are colored according to the weight of the $f$-electron wave function on them. We use the same BM parameters as in Table S18.

*(figure annotations: "Total weight of the heavy-fermions on the active bands after wannierisation: 97.4%"; colorbar "Wannier orbitals support" 0.0–1.0; vertical axis "Energy (meV)"; horizontal axis ticks $K$, $\Gamma$, $M$, $K$)*

TABLE S18. Parameters of the $f$-electron Wannier functions and the THF single-particle Hamiltonian for different values of the tunneling ratio $u_0/u_1$. $W$ denotes the total weight of the THF $f$-electrons on the active bands. In computing the ratios $\gamma/U_1$ and $\gamma/U_2$, we employ the on-site and nearest-neighbor repulsion parameters $U_1$ and $U_2$ obtained numerically for $\xi = 10$ nm, as given in Table S19. We employ $v_F = 5.944\,\mathrm{eV\,\text{Å}}$, $|\mathbf{K}| = 1.703\,\text{Å}^{-1}$, $w_1 = 110\,\mathrm{meV}$, and $\theta = 0.84°$ for the BM model.

| $u_0/u_1$ | $\alpha_1$ | $\alpha_2$ | $\lambda_1$ | $\lambda_2$ | $\lambda$ | $W$ | $t_0$ | $\gamma$ | $M$ | $v_\star$ | $v_\star'$ | $v_\star''$ | $\gamma/U_1$ | $\gamma/U_2$ |
|---|---|---|---|---|---|---|---|---|---|---|---|---|---|---|
| 1.00 | 0.747 | 0.665 | 0.131 | 0.149 | 0.585 | 0.980 | 0.899 | 22.359 | 19.192 | -2.173 | 0.885 | 0.1615 | 0.51 | 10.00 |
| 0.90 | 0.760 | 0.649 | 0.137 | 0.155 | 0.395 | 0.962 | 0.561 | 13.002 | 20.972 | -2.731 | 1.127 | 0.0514 | 0.25 | 9.20 |
| 0.80 | 0.784 | 0.621 | 0.149 | 0.167 | 0.339 | 0.974 | 0.696 | 1.668 | 22.429 | -3.178 | 1.294 | -0.0217 | 0.03 | 1.48 |
| 0.70 | 0.816 | 0.578 | 0.161 | 0.173 | 0.329 | 0.960 | 1.133 | -10.688 | 23.596 | -3.532 | 1.390 | -0.0628 | -0.22 | -0.56 |
| 0.60 | 0.853 | 0.523 | 0.179 | 0.179 | 0.343 | 0.947 | 1.685 | -23.303 | 24.516 | -3.814 | 1.419 | -0.0793 | -0.52 | -19.30 |
| 0.50 | 0.890 | 0.456 | 0.191 | 0.185 | 0.355 | 0.929 | 2.225 | -35.538 | 25.229 | -4.041 | 1.381 | -0.0778 | -0.78 | -27.56 |
| 0.40 | 0.925 | 0.381 | 0.209 | 0.191 | 0.359 | 0.911 | 2.664 | -46.811 | 25.767 | -4.225 | 1.267 | -0.0641 | -0.86 | -35.81 |
| 0.30 | 0.955 | 0.296 | 0.221 | 0.197 | 0.364 | 0.898 | 3.077 | -56.535 | 26.160 | -4.373 | 1.070 | -0.0437 | -1.20 | -42.69 |
| 0.20 | 0.979 | 0.204 | 0.233 | 0.203 | 0.364 | 0.887 | 3.301 | -64.108 | 26.426 | -4.484 | 0.784 | -0.0223 | -1.51 | -48.75 |
| 0.10 | 0.994 | 0.105 | 0.239 | 0.203 | 0.357 | 0.875 | 3.463 | -68.952 | 26.581 | -4.555 | 0.417 | -0.0061 | -1.87 | -54.87 |
| 0.00 | 1.000 | 0.000 | 0.239 | 0.000 | 0.358 | 0.874 | 3.566 | -70.622 | 26.631 | -4.580 | -0.000 | -0.0000 | -1.93 | -56.26 |

| $u_0/u_1$ | $\xi/\mathrm{nm}$ | 2 | 4 | 6 | 8 | 10 | 12 | 14 | 16 | 18 | 20 | 22 | 24 | 26 | 28 | 30 | 32 | 34 | 36 | 38 | 40 | 42 | 44 | 46 | 48 | 50 |
|---|---|---|---|---|---|---|---|---|---|---|---|---|---|---|---|---|---|---|---|---|---|---|---|---|---|---|---|
| 1.0 | $U_1$ | 15.88 | 27.27 | 34.83 | 40.01 | 43.75 | 46.58 | 48.79 | 50.57 | 52.04 | 53.28 | 54.34 | 55.25 | 56.04 | 56.74 | 57.36 | 57.92 | 58.42 | 58.87 | 59.28 | 59.65 | 59.99 | 60.30 | 60.59 | 60.86 | 61.10 |
| | $U_2$ | 0.51 | 1.39 | 1.80 | 2.24 | 2.71 | 3.22 | 3.74 | 4.27 | 4.78 | 5.27 | 5.74 | 6.18 | 6.60 | 6.99 | 7.35 | 7.70 | 8.02 | 8.32 | 8.60 | 8.86 | 9.11 | 9.34 | 9.56 | 9.76 | 9.96 |
| | $W_{1,2}$ | 4.02 | 8.80 | 14.26 | 20.11 | 26.15 | 32.28 | 38.44 | 44.61 | 50.79 | 56.98 | 63.16 | 69.35 | 75.53 | 81.72 | 87.91 | 94.09 | 100.28 | 106.46 | 112.65 | 118.84 | 125.02 | 131.21 | 137.39 | 143.58 | 149.77 |
| | $W_{3,4}$ | 10.31 | 19.90 | 28.11 | 35.30 | 41.93 | 48.32 | 54.59 | 60.81 | 67.01 | 73.20 | 79.39 | 85.58 | 91.76 | 97.95 | 104.14 | 110.32 | 116.51 | 122.69 | 128.88 | 135.07 | 141.25 | 147.44 | 153.62 | 159.81 | 166.00 |
| | $J$ | 6.92 | 10.83 | 12.89 | 14.08 | 14.86 | 15.43 | 15.88 | 16.20 | 16.58 | 16.87 | 17.13 | 17.36 | 17.57 | 17.76 | 17.94 | 18.09 | 18.23 | 18.36 | 18.48 | 18.59 | 18.69 | 18.78 | 18.86 | 18.94 | 19.01 |
| | $K$ | 2.92 | 4.96 | 6.05 | 6.78 | 6.89 | 6.93 | 6.96 | 6.95 | 6.95 | 6.96 | 6.96 | 6.96 | 6.96 | 6.96 | 6.96 | 6.96 | 6.96 | 6.96 | 6.96 | 6.96 | 6.96 | 6.96 | 6.96 | 6.96 | 6.96 |
| 0.9 | $U_1$ | 18.72 | 32.45 | 41.73 | 48.14 | 52.76 | 56.23 | 58.91 | 61.05 | 62.81 | 64.24 | 65.47 | 66.51 | 67.41 | 68.20 | 68.89 | 69.50 | 70.05 | 70.54 | 70.98 | 71.38 | 71.75 | 72.09 | 72.39 | 72.68 | 72.94 |
| | $U_2$ | 0.25 | 0.49 | 0.74 | 1.04 | 1.41 | 1.86 | 2.37 | 2.90 | 3.45 | 3.99 | 4.51 | 5.01 | 5.48 | 5.92 | 6.34 | 6.73 | 7.09 | 7.43 | 7.75 | 8.05 | 8.32 | 8.58 | 8.83 | 9.05 | 9.27 |
| | $W_{1,2}$ | 4.07 | 8.85 | 14.30 | 20.13 | 26.17 | 32.29 | 38.45 | 44.62 | 50.80 | 56.98 | 63.17 | 69.36 | 75.54 | 81.73 | 87.91 | 94.10 | 100.29 | 106.47 | 112.66 | 118.84 | 125.03 | 131.22 | 137.40 | 143.59 | 149.78 |
| | $W_{3,4}$ | 9.48 | 18.49 | 26.39 | 33.43 | 40.00 | 46.35 | 52.61 | 58.82 | 65.02 | 71.22 | 77.40 | 83.59 | 89.78 | 95.96 | 102.15 | 108.34 | 114.52 | 120.71 | 126.89 | 133.08 | 139.27 | 145.45 | 151.64 | 157.82 | 164.01 |
| | $J$ | 6.36 | 9.90 | 11.61 | 12.45 | 12.88 | 13.12 | 13.26 | 13.34 | 13.44 | 13.49 | 13.51 | 13.53 | 13.54 | 13.55 | 13.56 | 13.57 | 13.58 | 13.58 | 13.58 | 13.58 | 13.59 | 13.59 | 13.59 | 13.60 | 13.61 |
| | $K$ | 2.72 | 4.67 | 5.73 | 6.24 | 6.46 | 6.56 | 6.60 | 6.62 | 6.62 | 6.63 | 6.63 | 6.63 | 6.63 | 6.63 | 6.63 | 6.63 | 6.63 | 6.63 | 6.63 | 6.63 | 6.63 | 6.63 | 6.63 | 6.63 | 6.63 |
| 0.8 | $U_1$ | 18.13 | 31.74 | 41.14 | 47.72 | 52.50 | 56.10 | 58.90 | 61.12 | 62.94 | 64.45 | 65.70 | 66.77 | 67.71 | 68.51 | 69.22 | 69.85 | 70.41 | 70.91 | 71.37 | 71.77 | 72.15 | 72.49 | 72.80 | 73.09 | 73.35 |
| | $U_2$ | 0.16 | 0.32 | 0.51 | 0.77 | 1.12 | 1.57 | 2.07 | 2.61 | 3.16 | 3.71 | 4.23 | 4.74 | 5.22 | 5.67 | 6.09 | 6.49 | 6.86 | 7.20 | 7.52 | 7.82 | 8.10 | 8.37 | 8.61 | 8.84 | 9.06 |
| | $W_{1,2}$ | 4.23 | 9.09 | 14.57 | 20.42 | 26.45 | 32.57 | 38.73 | 44.91 | 51.09 | 57.27 | 63.46 | 69.64 | 75.83 | 82.02 | 88.20 | 94.39 | 100.57 | 106.76 | 112.95 | 119.13 | 125.32 | 131.51 | 137.69 | 143.88 | 150.06 |
| | $W_{3,4}$ | 8.68 | 17.06 | 24.59 | 31.46 | 37.95 | 44.27 | 50.51 | 56.72 | 62.92 | 69.11 | 75.30 | 81.48 | 87.67 | 93.86 | 100.04 | 106.23 | 112.41 | 118.60 | 124.79 | 130.97 | 137.16 | 143.34 | 149.53 | 155.72 | 161.90 |
| | $J$ | 6.03 | 9.55 | 11.32 | 12.21 | 12.69 | 12.96 | 13.12 | 13.22 | 13.28 | 13.32 | 13.34 | 13.36 | 13.38 | 13.39 | 13.39 | 13.40 | 13.40 | 13.41 | 13.41 | 13.41 | 13.41 | 13.41 | 13.41 | 13.42 | 13.42 |
| | $K$ | 2.57 | 4.44 | 5.47 | 5.97 | 6.19 | 6.29 | 6.33 | 6.35 | 6.35 | 6.36 | 6.36 | 6.36 | 6.36 | 6.36 | 6.36 | 6.36 | 6.36 | 6.36 | 6.36 | 6.36 | 6.36 | 6.36 | 6.36 | 6.36 | 6.36 |
| 0.7 | $U_1$ | 16.45 | 29.08 | 37.99 | 44.34 | 49.00 | 52.54 | 55.30 | 57.51 | 59.31 | 60.81 | 62.07 | 63.15 | 64.08 | 64.89 | 65.60 | 66.23 | 66.79 | 67.29 | 67.75 | 68.16 | 68.53 | 68.87 | 69.18 | 69.47 | 69.73 |
| | $U_2$ | 0.14 | 0.30 | 0.49 | 0.76 | 1.12 | 1.57 | 2.08 | 2.62 | 3.18 | 3.72 | 4.26 | 4.76 | 5.24 | 5.70 | 6.12 | 6.52 | 6.89 | 7.23 | 7.55 | 7.85 | 8.13 | 8.40 | 8.64 | 8.87 | 9.09 |
| | $W_{1,2}$ | 4.35 | 9.26 | 14.76 | 20.66 | 26.66 | 32.78 | 38.94 | 45.11 | 51.29 | 57.48 | 63.66 | 69.85 | 76.04 | 82.22 | 88.41 | 94.59 | 100.78 | 106.97 | 113.15 | 119.34 | 125.52 | 131.71 | 137.90 | 144.08 | 150.27 |
| | $W_{3,4}$ | 8.03 | 15.87 | 23.09 | 29.80 | 36.22 | 42.51 | 48.74 | 54.94 | 61.14 | 67.33 | 73.52 | 79.70 | 85.89 | 92.07 | 98.26 | 104.45 | 110.63 | 116.82 | 123.01 | 129.19 | 135.38 | 141.56 | 147.75 | 153.94 | 160.12 |
| | $J$ | 5.88 | 9.53 | 11.52 | 12.65 | 13.32 | 13.75 | 14.05 | 14.25 | 14.40 | 14.51 | 14.60 | 14.66 | 14.71 | 14.75 | 14.79 | 14.81 | 14.83 | 14.85 | 14.87 | 14.89 | 14.90 | 14.91 | 14.92 | 14.92 | 14.92 |
| | $K$ | 2.47 | 4.28 | 5.30 | 5.79 | 6.01 | 6.11 | 6.15 | 6.17 | 6.17 | 6.18 | 6.18 | 6.18 | 6.18 | 6.18 | 6.18 | 6.18 | 6.18 | 6.18 | 6.18 | 6.18 | 6.18 | 6.18 | 6.18 | 6.18 | 6.18 |

| $u_0/u_1$ | $\xi$/nm | 2 | 4 | 6 | 8 | 10 | 12 | 14 | 16 | 18 | 20 | 22 | 24 | 26 | 28 | 30 | 32 | 34 | 36 | 38 | 40 | 42 | 44 | 46 | 48 | 50 |
|---|---|---|---|---|---|---|---|---|---|---|---|---|---|---|---|---|---|---|---|---|---|---|---|---|---|---|
| 0.6 | $U_1$ | 14.61 | 26.05 | 34.30 | 40.27 | 44.71 | 48.11 | 50.79 | 52.94 | 54.70 | 56.17 | 57.41 | 58.48 | 59.39 | 60.19 | 60.90 | 61.52 | 62.08 | 62.58 | 63.03 | 63.43 | 63.81 | 64.15 | 64.46 | 64.74 | 65.01 |
| | $U_2$ | 0.16 | 0.33 | 0.54 | 0.83 | 1.21 | 1.67 | 2.19 | 2.74 | 3.29 | 3.84 | 4.37 | 4.88 | 5.36 | 5.81 | 6.23 | 6.63 | 7.00 | 7.34 | 7.66 | 7.96 | 8.24 | 8.50 | 8.75 | 8.98 | 9.19 |
| | $W_{1,2}$ | 4.39 | 9.31 | 14.81 | 20.67 | 26.70 | 32.83 | 38.89 | 45.16 | 51.34 | 57.53 | 63.71 | 69.90 | 76.08 | 82.27 | 88.46 | 94.64 | 100.83 | 107.01 | 113.20 | 119.39 | 125.57 | 131.76 | 137.94 | 144.13 | 150.32 |
| | $W_{3,4}$ | 7.53 | 14.96 | 21.92 | 28.51 | 34.87 | 41.14 | 47.36 | 53.56 | 59.75 | 65.94 | 72.13 | 78.31 | 84.50 | 90.69 | 96.87 | 103.06 | 109.24 | 115.43 | 121.62 | 127.80 | 133.99 | 140.18 | 146.36 | 152.55 | 158.73 |
| | $J$ | 5.81 | 9.63 | 11.88 | 13.25 | 14.15 | 14.76 | 15.20 | 15.53 | 15.77 | 15.96 | 16.11 | 16.22 | 16.31 | 16.39 | 16.45 | 16.50 | 16.54 | 16.57 | 16.60 | 16.62 | 16.64 | 16.66 | 16.68 | 16.69 | 16.70 |
| | $K$ | 2.42 | 4.21 | 5.22 | 5.71 | 5.93 | 6.03 | 6.08 | 6.09 | 6.10 | 6.10 | 6.10 | 6.10 | 6.10 | 6.10 | 6.10 | 6.10 | 6.10 | 6.10 | 6.10 | 6.10 | 6.10 | 6.10 | 6.10 | 6.10 | 6.10 |
| 0.5 | $U_1$ | 13.15 | 23.62 | 31.30 | 36.95 | 41.20 | 44.48 | 47.07 | 49.17 | 50.90 | 52.34 | 53.57 | 54.61 | 55.52 | 56.31 | 57.01 | 57.63 | 58.18 | 58.68 | 59.13 | 59.53 | 59.90 | 60.24 | 60.55 | 60.83 | 61.10 |
| | $U_2$ | 0.17 | 0.36 | 0.59 | 0.89 | 1.29 | 1.76 | 2.29 | 2.84 | 3.40 | 3.95 | 4.49 | 4.99 | 5.46 | 5.91 | 6.34 | 6.73 | 7.10 | 7.44 | 7.76 | 8.06 | 8.34 | 8.60 | 8.84 | 9.07 | 9.28 |
| | $W_{1,2}$ | 4.35 | 9.24 | 14.71 | 20.55 | 26.58 | 32.70 | 38.86 | 45.04 | 51.22 | 57.40 | 63.59 | 69.77 | 75.96 | 82.14 | 88.33 | 94.52 | 100.70 | 106.89 | 113.07 | 119.26 | 125.45 | 131.63 | 137.82 | 144.01 | 150.19 |
| | $W_{3,4}$ | 7.18 | 14.29 | 21.08 | 27.57 | 33.89 | 40.14 | 46.35 | 52.55 | 58.74 | 64.93 | 71.12 | 77.30 | 83.49 | 89.67 | 95.86 | 102.05 | 108.23 | 114.42 | 120.61 | 126.79 | 132.98 | 139.16 | 145.35 | 151.54 | 157.72 |
| | $J$ | 5.80 | 9.60 | 12.30 | 13.91 | 15.01 | 15.79 | 16.37 | 16.80 | 17.13 | 17.38 | 17.58 | 17.74 | 17.87 | 17.98 | 18.06 | 18.13 | 18.19 | 18.23 | 18.27 | 18.31 | 18.34 | 18.36 | 18.38 | 18.40 | 18.41 |
| | $K$ | 2.41 | 4.20 | 5.22 | 5.72 | 5.94 | 6.03 | 6.07 | 6.09 | 6.10 | 6.10 | 6.10 | 6.10 | 6.10 | 6.10 | 6.10 | 6.10 | 6.10 | 6.10 | 6.10 | 6.10 | 6.10 | 6.10 | 6.10 | 6.10 | 6.10 |
| 0.4 | $U_1$ | 12.24 | 22.13 | 29.49 | 34.96 | 39.10 | 42.32 | 44.88 | 46.96 | 48.67 | 50.10 | 51.32 | 52.36 | 53.26 | 54.05 | 54.75 | 55.36 | 55.91 | 56.41 | 56.86 | 57.26 | 57.63 | 57.97 | 58.28 | 58.56 | 58.82 |
| | $U_2$ | 0.17 | 0.36 | 0.59 | 0.91 | 1.31 | 1.79 | 2.32 | 2.87 | 3.43 | 3.98 | 4.51 | 5.02 | 5.49 | 5.95 | 6.37 | 6.76 | 7.13 | 7.47 | 7.79 | 8.09 | 8.36 | 8.63 | 8.87 | 9.10 | 9.31 |
| | $W_{1,2}$ | 4.24 | 9.04 | 14.46 | 20.28 | 26.31 | 32.42 | 38.58 | 44.75 | 50.93 | 57.12 | 63.30 | 69.49 | 75.67 | 81.86 | 88.04 | 94.23 | 100.42 | 106.60 | 112.79 | 118.97 | 125.16 | 131.35 | 137.53 | 143.72 | 149.91 |
| | $W_{3,4}$ | 6.93 | 13.83 | 20.40 | 26.92 | 33.21 | 39.45 | 45.65 | 51.85 | 58.04 | 64.22 | 70.41 | 76.60 | 82.78 | 88.97 | 95.16 | 101.34 | 107.53 | 113.71 | 119.90 | 126.09 | 132.27 | 138.46 | 144.65 | 150.83 | 157.02 |
| | $J$ | 5.84 | 9.99 | 12.68 | 14.40 | 15.70 | 16.59 | 17.24 | 17.74 | 18.11 | 18.41 | 18.62 | 18.82 | 18.96 | 19.08 | 19.17 | 19.25 | 19.31 | 19.36 | 19.41 | 19.45 | 19.48 | 19.50 | 19.53 | 19.55 | 19.56 |
| | $K$ | 2.43 | 4.25 | 5.29 | 5.79 | 6.02 | 6.12 | 6.16 | 6.18 | 6.19 | 6.19 | 6.19 | 6.19 | 6.19 | 6.19 | 6.19 | 6.19 | 6.19 | 6.19 | 6.19 | 6.19 | 6.19 | 6.19 | 6.19 | 6.19 | 6.19 |
| 0.3 | $U_1$ | 11.55 | 20.39 | 27.36 | 33.38 | 37.43 | 40.60 | 43.12 | 45.17 | 46.87 | 48.29 | 49.49 | 50.53 | 51.43 | 52.21 | 52.91 | 53.52 | 54.07 | 54.56 | 55.01 | 55.41 | 55.78 | 56.12 | 56.43 | 56.71 | 56.97 |
| | $U_2$ | 0.17 | 0.36 | 0.60 | 0.92 | 1.32 | 1.81 | 2.34 | 2.89 | 3.45 | 4.00 | 4.53 | 5.04 | 5.52 | 5.97 | 6.39 | 6.78 | 7.15 | 7.49 | 7.81 | 8.10 | 8.38 | 8.64 | 8.89 | 9.11 | 9.33 |
| | $W_{1,2}$ | 4.11 | 8.82 | 14.18 | 19.97 | 25.98 | 32.08 | 38.25 | 44.42 | 50.60 | 56.78 | 62.97 | 69.15 | 75.34 | 81.53 | 87.71 | 93.90 | 100.08 | 106.27 | 112.46 | 118.64 | 124.83 | 131.01 | 137.20 | 143.39 | 149.57 |
| | $W_{3,4}$ | 6.76 | 13.51 | 20.05 | 26.46 | 32.73 | 38.96 | 45.16 | 51.34 | 57.53 | 63.73 | 69.92 | 76.10 | 82.29 | 88.48 | 94.66 | 100.85 | 107.03 | 113.22 | 119.41 | 125.59 | 131.78 | 137.97 | 144.15 | 150.34 | 156.52 |
| | $J$ | 5.97 | 10.39 | 13.31 | 15.20 | 16.59 | 17.52 | 18.18 | 18.66 | 19.02 | 19.30 | 19.51 | 19.69 | 19.81 | 19.93 | 20.03 | 20.20 | 20.30 | 20.36 | 20.46 | 20.51 | 20.55 | 20.59 | 20.62 | 20.62 | 20.63 |
| | $K$ | 2.48 | 4.34 | 5.40 | 5.92 | 6.15 | 6.25 | 6.29 | 6.31 | 6.32 | 6.33 | 6.33 | 6.33 | 6.33 | 6.33 | 6.33 | 6.33 | 6.33 | 6.33 | 6.33 | 6.33 | 6.33 | 6.33 | 6.33 | 6.33 | 6.33 |
| 0.2 | $U_1$ | 11.19 | 20.43 | 27.45 | 32.75 | 36.81 | 39.98 | 42.52 | 44.58 | 46.28 | 47.71 | 48.92 | 49.96 | 50.86 | 51.65 | 52.35 | 52.96 | 53.51 | 54.01 | 54.46 | 54.86 | 55.23 | 55.57 | 55.88 | 56.16 | 56.14 |
| | $U_2$ | 0.16 | 0.35 | 0.59 | 0.91 | 1.32 | 1.80 | 2.33 | 2.89 | 3.45 | 4.00 | 4.53 | 5.04 | 5.52 | 5.97 | 6.39 | 6.78 | 7.09 | 7.49 | 7.75 | 8.05 | 8.33 | 8.59 | 8.83 | 9.11 | 9.33 |
| | $W_{1,2}$ | 3.96 | 8.56 | 13.87 | 19.63 | 25.63 | 31.73 | 37.88 | 44.06 | 50.24 | 56.42 | 62.60 | 68.79 | 74.98 | 81.16 | 87.35 | 93.53 | 99.72 | 105.91 | 112.09 | 118.28 | 124.46 | 130.65 | 136.84 | 143.03 | 149.21 |
| | $W_{3,4}$ | 6.65 | 13.31 | 19.67 | 26.01 | 32.27 | 38.48 | 44.68 | 50.88 | 57.06 | 63.25 | 69.44 | 75.62 | 81.81 | 88.00 | 94.18 | 100.37 | 106.55 | 112.74 | 118.93 | 125.11 | 131.30 | 137.49 | 143.67 | 149.86 | 156.04 |
| | $J$ | 5.97 | 10.52 | 13.55 | 15.62 | 17.08 | 18.13 | 18.91 | 19.48 | 19.93 | 20.27 | 20.53 | 20.73 | 20.89 | 21.02 | 21.13 | 21.21 | 21.28 | 21.21 | 21.25 | 21.28 | 21.31 | 21.33 | 21.35 | 21.36 | 21.38 |
| | $K$ | 2.54 | 4.44 | 5.53 | 6.06 | 6.30 | 6.40 | 6.44 | 6.47 | 6.48 | 6.48 | 6.48 | 6.48 | 6.48 | 6.48 | 6.48 | 6.48 | 6.48 | 6.48 | 6.48 | 6.48 | 6.48 | 6.48 | 6.48 | 6.48 | 6.48 |
| 0.1 | $U_1$ | 11.20 | 20.43 | 27.45 | 32.75 | 36.81 | 39.98 | 42.52 | 44.58 | 46.26 | 47.71 | 48.92 | 49.96 | 50.86 | 51.65 | 52.35 | 52.96 | 53.51 | 54.01 | 54.46 | 54.86 | 55.23 | 55.57 | 55.88 | 56.16 | 56.42 |
| | $U_2$ | 0.15 | 0.35 | 0.59 | 0.91 | 1.32 | 1.80 | 2.33 | 2.89 | 3.45 | 4.00 | 4.53 | 5.04 | 5.46 | 5.91 | 6.33 | 6.72 | 7.09 | 7.43 | 7.75 | 8.05 | 8.33 | 8.59 | 8.83 | 9.06 | 9.28 |
| | $W_{1,2}$ | 3.83 | 8.34 | 13.59 | 19.33 | 25.31 | 31.41 | 37.56 | 43.73 | 49.51 | 56.10 | 62.28 | 68.47 | 74.65 | 80.84 | 87.02 | 93.21 | 99.40 | 105.58 | 111.77 | 117.95 | 124.14 | 130.33 | 136.51 | 142.70 | 148.88 |
| | $W_{3,4}$ | 6.59 | 13.20 | 19.67 | 26.00 | 32.26 | 38.48 | 44.62 | 50.82 | 57.00 | 63.25 | 69.44 | 75.62 | 81.81 | 88.00 | 94.12 | 100.31 | 106.49 | 112.68 | 118.87 | 125.05 | 131.24 | 137.42 | 143.61 | 149.86 | 156.04 |
| | $J$ | 6.00 | 10.52 | 13.55 | 15.62 | 17.17 | 18.19 | 19.03 | 19.56 | 20.07 | 20.36 | 20.69 | 20.83 | 21.00 | 21.13 | 21.21 | 21.28 | 21.34 | 21.38 | 21.45 | 21.48 | 21.50 | 21.51 | 21.52 | 21.53 | 21.54 |
| | $K$ | 2.58 | 4.52 | 5.63 | 6.17 | 6.42 | 6.52 | 6.57 | 6.59 | 6.60 | 6.60 | 6.60 | 6.60 | 6.60 | 6.60 | 6.60 | 6.60 | 6.60 | 6.60 | 6.60 | 6.60 | 6.60 | 6.60 | 6.60 | 6.60 | 6.60 |
| 0.0 | $U_1$ | 11.11 | 20.28 | 27.27 | 32.55 | 36.60 | 39.77 | 42.30 | 44.35 | 46.06 | 47.48 | 48.69 | 49.73 | 50.64 | 51.42 | 52.12 | 52.73 | 53.28 | 53.78 | 54.23 | 54.63 | 55.00 | 55.34 | 55.65 | 55.93 | 56.19 |
| | $U_2$ | 0.15 | 0.32 | 0.54 | 0.85 | 1.26 | 1.74 | 2.27 | 2.83 | 3.39 | 3.94 | 4.47 | 4.98 | 5.46 | 5.91 | 6.33 | 6.72 | 7.09 | 7.43 | 7.75 | 8.05 | 8.33 | 8.59 | 8.83 | 9.06 | 9.28 |
| | $W_{1,2}$ | 3.80 | 8.28 | 13.51 | 19.24 | 25.22 | 31.32 | 37.47 | 43.64 | 49.82 | 56.00 | 62.19 | 68.38 | 74.56 | 80.75 | 86.93 | 93.12 | 99.31 | 105.49 | 111.68 | 117.86 | 124.05 | 130.24 | 136.42 | 142.61 | 148.79 |
| | $W_{3,4}$ | 6.56 | 13.16 | 19.62 | 25.96 | 32.21 | 38.42 | 44.62 | 50.82 | 57.01 | 63.19 | 69.38 | 75.56 | 81.75 | 87.94 | 94.12 | 100.31 | 106.49 | 112.68 | 118.87 | 125.05 | 131.24 | 137.42 | 143.61 | 149.80 | 155.98 |
| | $J$ | 6.06 | 10.58 | 13.64 | 15.75 | 17.23 | 18.30 | 19.09 | 19.68 | 20.13 | 20.47 | 20.73 | 20.94 | 21.10 | 21.24 | 21.34 | 21.42 | 21.49 | 21.55 | 21.59 | 21.63 | 21.66 | 21.69 | 21.71 | 21.74 | 21.76 |
| | $K$ | 2.60 | 4.55 | 5.67 | 6.22 | 6.46 | 6.57 | 6.61 | 6.63 | 6.64 | 6.65 | 6.65 | 6.65 | 6.65 | 6.65 | 6.65 | 6.65 | 6.65 | 6.65 | 6.65 | 6.65 | 6.65 | 6.65 | 6.65 | 6.65 | 6.65 |

TABLE S19. Parameters of the THF interaction Hamiltonian for different ratios $u_0/u_1$ and screening lengths $\xi$ of the double-gate interaction potential ($U_\xi = 24$ meV for $\xi = 10$nm). We use the same BM model parameters as in Table S18.

| $w_0/w_1$ | $\alpha_1$ | $\alpha_2$ | $\lambda_1$ | $\lambda_2$ | $\lambda$ | $\mathcal{W}$ | $t_0$ | $\gamma$ | $M$ | $v_\star$ | $v_\star'$ | $v_\star''$ | $\gamma/U_1$ | $\gamma/U_2$ |
|---|---|---|---|---|---|---|---|---|---|---|---|---|---|---|
| 1.00 | 0.748 | 0.664 | 0.131 | 0.149 | 0.597 | 0.986 | 0.759 | 21.494 | 18.596 | -2.394 | 0.964 | 0.1347 | 0.48 | 9.63 |
| 0.90 | 0.762 | 0.647 | 0.143 | 0.161 | 0.378 | 0.965 | 0.512 | 11.397 | 20.071 | -2.917 | 1.187 | 0.0362 | 0.21 | 8.10 |
| 0.80 | 0.787 | 0.617 | 0.155 | 0.167 | 0.336 | 0.976 | 0.698 | -0.429 | 21.260 | -3.333 | 1.338 | -0.0276 | -0.01 | -0.36 |
| 0.70 | 0.820 | 0.573 | 0.167 | 0.173 | 0.331 | 0.960 | 1.127 | -13.097 | 22.201 | -3.662 | 1.421 | -0.0627 | -0.26 | -10.88 |
| 0.60 | 0.856 | 0.517 | 0.179 | 0.185 | 0.344 | 0.944 | 1.641 | -25.899 | 22.935 | -3.926 | 1.441 | -0.0759 | -0.57 | -19.96 |
| 0.50 | 0.893 | 0.450 | 0.197 | 0.191 | 0.354 | 0.926 | 2.131 | -38.239 | 23.498 | -4.139 | 1.394 | -0.0731 | -0.91 | -28.15 |
| 0.40 | 0.927 | 0.375 | 0.209 | 0.197 | 0.359 | 0.909 | 2.518 | -49.560 | 23.921 | -4.314 | 1.274 | -0.0596 | -1.24 | -35.65 |
| 0.30 | 0.956 | 0.292 | 0.221 | 0.203 | 0.364 | 0.897 | 2.893 | -59.295 | 24.227 | -4.454 | 1.073 | -0.0404 | -1.55 | -42.18 |
| 0.20 | 0.980 | 0.201 | 0.233 | 0.203 | 0.364 | 0.887 | 3.088 | -66.859 | 24.434 | -4.561 | 0.784 | -0.0205 | -1.78 | -47.86 |
| 0.10 | 0.995 | 0.104 | 0.239 | 0.203 | 0.357 | 0.875 | 3.243 | -71.687 | 24.554 | -4.629 | 0.416 | -0.0056 | -1.90 | -53.65 |
| 0.00 | 1.000 | 0.000 | 0.245 | 0.000 | 0.357 | 0.873 | 3.312 | -73.351 | 24.593 | -4.653 | 0.000 | 0.0000 | -1.95 | -55.07 |

TABLE S20. Parameters of the $f$-electron Wannier functions and the THF single-particle Hamiltonian for different values of the tunneling ratio $w_0/w_1$. $\mathcal{W}$ denotes the total weight of the THF $f$-electrons on the active bands. In computing the ratios $\gamma/U_1$ and $\gamma/U_2$, we employ the on-site and nearest-neighbor repulsion parameters $U_1$ and $U_2$ obtained numerically for $\xi = 10$ nm, as given in Table S21. We employ $v_F = 5.944\,\text{eV·Å}$, $|\mathbf{K}| = 1.703\,\text{Å}^{-1}$, $w_1 = 110$ meV, and $\theta = 0.86°$ for the BM model.



Total weight of the heavy-fermions on the active bands after wannierisation: 97.6%

FIG. S26. Band structures of the BM and THF models near charge neutrality for $w_0/w_1 = 0.8$, depicted by lines and crosses, respectively. The BM bands are colored according to the weight of the $f$-electron wave function on them. We use the same BM parameters as in Table S20.

| $w_0/w_1$ | $\xi/\text{nm}$ | 2 | 4 | 6 | 8 | 10 | 12 | 14 | 16 | 18 | 20 | 22 | 24 | 26 | 28 | 30 | 32 | 34 | 36 | 38 | 40 | 42 | 44 | 46 | 48 | 50 |
|---|---|---|---|---|---|---|---|---|---|---|---|---|---|---|---|---|---|---|---|---|---|---|---|---|---|---|
| 1.0 | $U_1$ | 16.39 | 28.12 | 35.89 | 41.21 | 45.05 | 47.94 | 50.20 | 52.02 | 53.51 | 54.77 | 55.84 | 56.77 | 57.57 | 58.28 | 58.90 | 59.46 | 59.97 | 60.42 | 60.83 | 61.20 | 61.55 | 61.86 | 62.15 | 62.42 | 62.67 |
| | $U_2$ | 0.49 | 0.95 | 1.36 | 1.78 | 2.23 | 2.73 | 3.26 | 3.81 | 4.35 | 4.88 | 5.39 | 5.88 | 6.33 | 6.76 | 7.16 | 7.53 | 7.88 | 8.21 | 8.51 | 8.80 | 9.06 | 9.31 | 9.55 | 9.77 | 9.97 |
| | $W_{1,2}$ | 4.25 | 9.29 | 15.03 | 21.18 | 27.52 | 33.94 | 40.40 | 46.87 | 53.35 | 59.83 | 66.32 | 72.80 | 79.29 | 85.77 | 92.25 | 98.74 | 105.22 | 111.71 | 118.19 | 124.67 | 131.16 | 137.64 | 144.13 | 150.61 | 157.10 |
| | $W_{3,4}$ | 10.38 | 20.07 | 28.42 | 35.81 | 42.69 | 49.35 | 55.90 | 62.41 | 68.91 | 75.40 | 81.89 | 88.37 | 94.86 | 101.34 | 107.82 | 114.31 | 120.79 | 127.28 | 133.76 | 140.24 | 146.73 | 153.21 | 159.70 | 166.18 | 172.67 |
| | $J$ | 6.91 | 10.79 | 12.80 | 13.96 | 14.71 | 15.25 | 15.68 | 16.04 | 16.36 | 16.64 | 16.89 | 17.12 | 17.33 | 17.52 | 17.70 | 17.86 | 18.00 | 18.14 | 18.27 | 18.38 | 18.49 | 18.59 | 18.68 | 18.76 | 18.84 |
| | $K$ | 2.88 | 4.89 | 5.94 | 6.43 | 6.64 | 6.77 | 6.78 | 6.79 | 6.79 | 6.79 | 6.79 | 6.79 | 6.79 | 6.79 | 6.79 | 6.79 | 6.79 | 6.79 | 6.79 | 6.79 | 6.79 | 6.79 | 6.79 | 6.79 | 6.79 |
| 0.9 | $U_1$ | 19.33 | 33.48 | 43.09 | 49.61 | 54.35 | 57.96 | 60.84 | 62.83 | 64.61 | 66.08 | 67.33 | 68.39 | 69.30 | 70.10 | 70.80 | 71.42 | 71.97 | 72.47 | 72.91 | 73.32 | 73.69 | 74.03 | 74.33 | 74.62 | 74.88 |
| | $U_2$ | 0.23 | 0.46 | 0.71 | 1.01 | 1.41 | 1.89 | 2.42 | 2.99 | 3.56 | 4.12 | 4.67 | 5.19 | 5.67 | 6.13 | 6.56 | 6.96 | 7.33 | 7.68 | 8.01 | 8.31 | 8.59 | 8.85 | 9.10 | 9.33 | 9.55 |
| | $W_{1,2}$ | 4.32 | 9.37 | 15.11 | 21.25 | 27.59 | 34.01 | 40.46 | 46.94 | 53.42 | 59.90 | 66.38 | 72.87 | 79.35 | 85.84 | 92.32 | 98.80 | 105.29 | 111.77 | 118.26 | 124.74 | 131.23 | 137.71 | 144.19 | 150.68 | 157.16 |
| | $W_{3,4}$ | 9.57 | 18.69 | 26.75 | 34.00 | 40.82 | 47.45 | 53.99 | 60.50 | 66.99 | 73.48 | 79.97 | 86.45 | 92.94 | 99.42 | 105.91 | 112.39 | 118.87 | 125.36 | 131.84 | 138.33 | 144.81 | 151.30 | 157.78 | 164.26 | 170.75 |
| | $J$ | 6.40 | 9.95 | 11.63 | 12.44 | 12.85 | 13.07 | 13.19 | 13.26 | 13.31 | 13.33 | 13.35 | 13.37 | 13.38 | 13.38 | 13.39 | 13.39 | 13.40 | 13.40 | 13.40 | 13.40 | 13.41 | 13.41 | 13.41 | 13.41 | 13.41 |
| | $K$ | 2.69 | 4.60 | 5.63 | 6.10 | 6.31 | 6.40 | 6.44 | 6.45 | 6.46 | 6.46 | 6.46 | 6.46 | 6.46 | 6.46 | 6.46 | 6.46 | 6.46 | 6.46 | 6.46 | 6.46 | 6.46 | 6.46 | 6.46 | 6.46 | 6.46 |
| 0.8 | $U_1$ | 18.41 | 33.20 | 41.70 | 48.36 | 53.18 | 56.82 | 59.64 | 61.88 | 63.71 | 65.22 | 66.49 | 67.57 | 68.51 | 69.32 | 70.03 | 70.66 | 71.23 | 71.73 | 72.18 | 72.59 | 72.97 | 73.31 | 73.62 | 73.91 | 74.17 |
| | $U_2$ | 0.16 | 0.33 | 0.52 | 0.81 | 1.19 | 1.66 | 2.20 | 2.77 | 3.35 | 3.92 | 4.47 | 4.99 | 5.48 | 5.95 | 6.38 | 6.79 | 7.16 | 7.51 | 7.84 | 8.14 | 8.43 | 8.69 | 8.94 | 9.17 | 9.39 |
| | $W_{1,2}$ | 4.50 | 9.64 | 15.41 | 21.56 | 27.90 | 34.33 | 40.79 | 47.26 | 53.74 | 60.22 | 66.71 | 73.19 | 79.67 | 86.16 | 92.64 | 99.13 | 105.61 | 112.10 | 118.58 | 125.06 | 131.55 | 138.03 | 144.52 | 151.00 | 157.48 |
| | $W_{3,4}$ | 8.79 | 17.29 | 25.00 | 32.08 | 38.83 | 45.43 | 51.96 | 58.46 | 64.96 | 71.44 | 77.93 | 84.41 | 90.90 | 97.38 | 103.87 | 110.35 | 116.84 | 123.32 | 129.80 | 136.29 | 142.77 | 149.26 | 155.74 | 162.23 | 168.71 |
| | $J$ | 6.12 | 9.69 | 11.48 | 12.40 | 12.90 | 13.19 | 13.36 | 13.47 | 13.54 | 13.59 | 13.62 | 13.65 | 13.67 | 13.68 | 13.69 | 13.70 | 13.71 | 13.71 | 13.72 | 13.72 | 13.72 | 13.73 | 13.73 | 13.73 | 13.73 |
| | $K$ | 2.55 | 4.38 | 5.38 | 5.85 | 6.06 | 6.14 | 6.18 | 6.19 | 6.20 | 6.20 | 6.20 | 6.20 | 6.20 | 6.20 | 6.20 | 6.20 | 6.20 | 6.20 | 6.20 | 6.20 | 6.20 | 6.20 | 6.20 | 6.20 | 6.20 |
| 0.7 | $U_1$ | 16.65 | 29.40 | 38.38 | 44.77 | 49.46 | 53.02 | 55.80 | 58.02 | 59.83 | 61.34 | 62.60 | 63.68 | 64.62 | 65.43 | 66.14 | 66.77 | 67.33 | 67.84 | 68.29 | 68.70 | 69.07 | 69.42 | 69.73 | 70.02 | 70.28 |
| | $U_2$ | 0.15 | 0.32 | 0.52 | 0.81 | 1.20 | 1.68 | 2.23 | 2.81 | 3.39 | 3.96 | 4.51 | 5.04 | 5.53 | 6.00 | 6.43 | 6.83 | 7.21 | 7.56 | 7.89 | 8.19 | 8.48 | 8.74 | 8.99 | 9.23 | 9.44 |
| | $W_{1,2}$ | 4.60 | 9.69 | 15.32 | 21.75 | 28.10 | 34.52 | 40.98 | 47.46 | 53.94 | 60.42 | 66.90 | 73.39 | 79.87 | 86.36 | 92.84 | 99.32 | 105.81 | 112.29 | 118.77 | 125.26 | 131.74 | 138.23 | 144.71 | 151.20 | 157.68 |
| | $W_{3,4}$ | 8.17 | 16.16 | 23.57 | 30.51 | 37.20 | 43.77 | 50.29 | 56.79 | 63.28 | 69.76 | 76.25 | 82.73 | 89.22 | 95.70 | 102.19 | 108.67 | 115.16 | 121.64 | 128.12 | 134.61 | 141.09 | 147.58 | 154.06 | 160.54 | 167.03 |
| | $J$ | 6.00 | 9.73 | 11.77 | 12.94 | 13.65 | 14.11 | 14.43 | 14.65 | 14.81 | 14.94 | 15.03 | 15.10 | 15.15 | 15.20 | 15.23 | 15.25 | 15.28 | 15.30 | 15.32 | 15.34 | 15.35 | 15.37 | 15.38 | 15.39 | 15.40 |
| | $K$ | 2.46 | 4.24 | 5.23 | 5.69 | 5.90 | 5.98 | 6.02 | 6.03 | 6.04 | 6.04 | 6.04 | 6.04 | 6.04 | 6.04 | 6.04 | 6.04 | 6.04 | 6.04 | 6.04 | 6.04 | 6.04 | 6.04 | 6.04 | 6.04 | 6.04 |

| $u_0/u_1$ | $\xi$/nm | 2 | 4 | 6 | 8 | 10 | 12 | 14 | 16 | 18 | 20 | 22 | 24 | 26 | 28 | 30 | 32 | 34 | 36 | 38 | 40 | 42 | 44 | 46 | 48 | 50 |
|---|---|---|---|---|---|---|---|---|---|---|---|---|---|---|---|---|---|---|---|---|---|---|---|---|---|---|
| 0.6 | $U_1$ | 14.83 | 26.42 | 34.76 | 40.78 | 45.26 | 48.69 | 51.38 | 53.54 | 55.32 | 56.79 | 58.04 | 59.11 | 60.03 | 60.83 | 61.54 | 62.17 | 62.72 | 63.22 | 63.68 | 64.08 | 64.46 | 64.80 | 65.11 | 65.39 | 65.66 |
| | $U_2$ | 0.17 | 0.35 | 0.58 | 0.89 | 1.30 | 1.79 | 2.34 | 2.92 | 3.51 | 4.08 | 4.63 | 5.15 | 5.65 | 6.11 | 6.54 | 6.95 | 7.32 | 7.67 | 8.00 | 8.30 | 8.59 | 8.85 | 9.10 | 9.33 | 9.55 |
| | $W_{1,2}$ | 4.63 | 9.83 | 15.62 | 21.78 | 28.12 | 34.55 | 41.01 | 47.48 | 53.96 | 60.44 | 66.92 | 73.41 | 79.89 | 86.38 | 92.86 | 99.35 | 105.83 | 112.31 | 118.80 | 125.28 | 131.77 | 138.25 | 144.73 | 151.22 | 157.70 |
| | $W_{3,4}$ | 7.69 | 15.29 | 22.47 | 29.30 | 35.94 | 42.49 | 49.00 | 55.50 | 61.99 | 68.47 | 74.96 | 81.44 | 87.93 | 94.41 | 100.89 | 107.38 | 113.86 | 120.35 | 126.83 | 133.32 | 139.80 | 146.28 | 152.77 | 159.25 | 165.74 |
| | $J$ | 5.95 | 9.88 | 12.20 | 13.64 | 14.58 | 15.23 | 15.70 | 16.05 | 16.31 | 16.51 | 16.67 | 16.79 | 16.89 | 16.97 | 17.04 | 17.09 | 17.13 | 17.17 | 17.20 | 17.22 | 17.25 | 17.26 | 17.28 | 17.29 | 17.30 |
| | $K$ | 2.41 | 4.18 | 5.16 | 5.63 | 5.83 | 5.92 | 5.97 | 5.98 | 5.98 | 6.00 | 5.98 | 5.98 | 5.98 | 5.98 | 5.98 | 5.98 | 5.98 | 5.98 | 5.98 | 5.98 | 5.98 | 5.98 | 5.98 | 5.98 | 5.98 |
| 0.5 | $U_1$ | 13.47 | 24.17 | 32.01 | 37.75 | 42.06 | 45.38 | 48.01 | 50.13 | 51.88 | 53.34 | 54.57 | 55.62 | 56.54 | 57.33 | 58.04 | 58.66 | 59.21 | 59.71 | 60.16 | 60.57 | 60.94 | 61.27 | 61.58 | 61.87 | 62.13 |
| | $U_2$ | 0.17 | 0.37 | 0.61 | 0.93 | 1.36 | 1.86 | 2.42 | 3.00 | 3.59 | 4.16 | 4.71 | 5.23 | 5.73 | 6.19 | 6.62 | 7.02 | 7.40 | 7.75 | 8.07 | 8.37 | 8.66 | 8.92 | 9.17 | 9.40 | 9.61 |
| | $W_{1,2}$ | 4.57 | 9.73 | 15.49 | 21.68 | 27.97 | 34.39 | 40.85 | 47.32 | 53.80 | 60.29 | 66.77 | 73.25 | 79.74 | 86.22 | 92.71 | 99.19 | 105.67 | 112.16 | 118.64 | 125.13 | 131.61 | 138.10 | 144.58 | 151.07 | 157.55 |
| | $W_{3,4}$ | 7.36 | 14.67 | 21.68 | 28.43 | 35.03 | 41.56 | 48.07 | 54.56 | 61.05 | 67.54 | 74.02 | 80.50 | 86.99 | 93.47 | 99.96 | 106.44 | 112.93 | 119.41 | 125.89 | 132.38 | 138.86 | 145.35 | 151.83 | 158.32 | 164.80 |
| | $J$ | 5.97 | 10.07 | 12.64 | 14.29 | 15.42 | 16.22 | 16.81 | 17.24 | 17.58 | 17.84 | 18.04 | 18.20 | 18.32 | 18.43 | 18.51 | 18.58 | 18.63 | 18.68 | 18.72 | 18.75 | 18.78 | 18.80 | 18.82 | 18.84 | 18.85 |
| | $K$ | 2.41 | 4.19 | 5.18 | 5.64 | 5.85 | 5.94 | 5.97 | 5.99 | 6.00 | 6.00 | 6.00 | 6.00 | 6.00 | 6.00 | 6.00 | 6.00 | 6.00 | 6.00 | 6.00 | 6.00 | 6.00 | 6.00 | 6.00 | 6.00 | 6.00 |
| 0.4 | $U_1$ | 12.53 | 22.62 | 30.11 | 35.66 | 39.86 | 43.12 | 45.70 | 47.80 | 49.52 | 50.97 | 52.19 | 53.24 | 54.15 | 54.94 | 55.64 | 56.26 | 56.81 | 57.31 | 57.75 | 58.16 | 58.53 | 58.87 | 59.18 | 59.46 | 59.73 |
| | $U_2$ | 0.18 | 0.37 | 0.62 | 0.96 | 1.39 | 1.90 | 2.46 | 3.05 | 3.63 | 4.21 | 4.76 | 5.28 | 5.77 | 6.23 | 6.66 | 7.07 | 7.44 | 7.79 | 8.11 | 8.42 | 8.70 | 8.96 | 9.21 | 9.44 | 9.66 |
| | $W_{1,2}$ | 4.46 | 9.52 | 15.24 | 21.36 | 27.68 | 34.10 | 40.56 | 47.03 | 53.51 | 59.99 | 66.48 | 72.96 | 79.44 | 85.92 | 92.41 | 98.90 | 105.38 | 111.87 | 118.35 | 124.83 | 131.32 | 137.80 | 144.29 | 150.77 | 157.25 |
| | $W_{3,4}$ | 7.12 | 14.24 | 21.13 | 27.82 | 34.39 | 40.91 | 47.42 | 53.91 | 60.39 | 66.87 | 73.36 | 79.85 | 86.33 | 92.82 | 99.30 | 105.78 | 112.27 | 118.75 | 125.24 | 131.72 | 138.21 | 144.69 | 151.17 | 157.66 | 164.14 |
| | $J$ | 6.02 | 10.29 | 13.06 | 14.89 | 16.17 | 17.08 | 17.76 | 18.26 | 18.65 | 18.95 | 19.18 | 19.36 | 19.51 | 19.63 | 19.72 | 19.80 | 19.86 | 19.92 | 19.96 | 20.00 | 20.03 | 20.06 | 20.08 | 20.10 | 20.11 |
| | $K$ | 2.44 | 4.25 | 5.25 | 5.73 | 5.94 | 6.03 | 6.07 | 6.08 | 6.09 | 6.09 | 6.09 | 6.09 | 6.09 | 6.09 | 6.09 | 6.09 | 6.09 | 6.09 | 6.09 | 6.09 | 6.09 | 6.09 | 6.09 | 6.09 | 6.09 |
| 0.3 | $U_1$ | 11.85 | 21.49 | 28.73 | 34.14 | 38.26 | 41.46 | 44.02 | 46.09 | 47.80 | 49.23 | 50.45 | 51.49 | 52.39 | 53.18 | 53.88 | 54.50 | 55.05 | 55.54 | 55.99 | 56.39 | 56.76 | 57.10 | 57.41 | 57.70 | 57.96 |
| | $U_2$ | 0.17 | 0.37 | 0.63 | 0.97 | 1.41 | 1.92 | 2.49 | 3.07 | 3.65 | 4.23 | 4.78 | 5.30 | 5.79 | 6.26 | 6.69 | 7.09 | 7.46 | 7.81 | 8.13 | 8.44 | 8.72 | 8.98 | 9.23 | 9.46 | 9.68 |
| | $W_{1,2}$ | 4.32 | 9.28 | 14.94 | 21.03 | 27.35 | 33.76 | 40.22 | 46.69 | 53.17 | 59.65 | 66.13 | 72.62 | 79.10 | 85.58 | 92.07 | 98.55 | 105.04 | 111.52 | 118.00 | 124.49 | 130.97 | 137.46 | 143.94 | 150.43 | 156.91 |
| | $W_{3,4}$ | 6.96 | 13.93 | 20.74 | 27.39 | 33.95 | 40.46 | 46.96 | 53.45 | 59.94 | 66.42 | 72.91 | 79.39 | 85.87 | 92.36 | 98.84 | 105.33 | 111.81 | 118.29 | 124.78 | 131.26 | 137.75 | 144.23 | 150.72 | 157.20 | 163.68 |
| | $J$ | 6.09 | 10.57 | 13.45 | 15.35 | 16.68 | 17.62 | 18.32 | 18.83 | 19.23 | 19.53 | 19.77 | 19.96 | 20.11 | 20.23 | 20.33 | 20.41 | 20.48 | 20.53 | 20.58 | 20.62 | 20.65 | 20.68 | 20.71 | 20.73 | 21.19 |
| | $K$ | 2.50 | 4.34 | 5.37 | 5.87 | 6.08 | 6.17 | 6.21 | 6.23 | 6.24 | 6.24 | 6.24 | 6.24 | 6.24 | 6.24 | 6.24 | 6.24 | 6.24 | 6.24 | 6.24 | 6.24 | 6.24 | 6.24 | 6.24 | 6.24 | 6.24 |
| 0.2 | $U_1$ | 11.50 | 20.93 | 28.05 | 33.40 | 37.48 | 40.67 | 43.22 | 45.28 | 46.99 | 48.42 | 49.64 | 50.68 | 51.58 | 52.37 | 53.07 | 53.68 | 54.24 | 54.73 | 55.18 | 55.58 | 55.95 | 56.29 | 56.60 | 56.88 | 57.15 |
| | $U_2$ | 0.17 | 0.36 | 0.62 | 0.96 | 1.40 | 1.91 | 2.48 | 3.07 | 3.65 | 4.23 | 4.78 | 5.30 | 5.79 | 6.26 | 6.68 | 7.08 | 7.46 | 7.80 | 8.13 | 8.43 | 8.71 | 8.98 | 9.23 | 9.46 | 9.67 |
| | $W_{1,2}$ | 4.17 | 9.01 | 14.62 | 20.69 | 26.99 | 33.39 | 39.85 | 46.32 | 52.80 | 59.28 | 65.76 | 72.25 | 78.73 | 85.21 | 91.70 | 98.18 | 104.67 | 111.15 | 117.63 | 124.12 | 130.60 | 137.09 | 143.57 | 150.06 | 156.54 |
| | $W_{3,4}$ | 6.85 | 13.74 | 20.50 | 27.12 | 33.67 | 40.02 | 46.51 | 53.00 | 59.49 | 65.97 | 72.46 | 78.94 | 85.42 | 91.91 | 98.39 | 104.88 | 111.36 | 117.85 | 124.33 | 130.82 | 137.30 | 143.79 | 150.27 | 156.76 | 163.24 |
| | $J$ | 6.17 | 10.71 | 13.77 | 15.87 | 17.35 | 18.42 | 19.21 | 19.80 | 20.25 | 20.60 | 20.87 | 21.08 | 21.25 | 21.39 | 21.49 | 21.58 | 21.65 | 21.72 | 21.77 | 21.81 | 21.84 | 21.87 | 21.90 | 21.92 | 21.94 |
| | $K$ | 2.56 | 4.45 | 5.51 | 6.01 | 6.23 | 6.33 | 6.37 | 6.38 | 6.39 | 6.39 | 6.39 | 6.39 | 6.39 | 6.39 | 6.39 | 6.39 | 6.39 | 6.39 | 6.39 | 6.39 | 6.39 | 6.39 | 6.39 | 6.39 | 6.39 |
| 0.1 | $U_1$ | 11.52 | 20.98 | 28.16 | 33.56 | 37.68 | 40.90 | 43.47 | 45.55 | 47.27 | 48.71 | 49.93 | 50.98 | 51.89 | 52.68 | 53.38 | 54.00 | 54.55 | 55.05 | 55.50 | 55.90 | 56.27 | 56.61 | 56.92 | 57.21 | 57.47 |
| | $U_2$ | 0.15 | 0.33 | 0.57 | 0.90 | 1.34 | 1.85 | 2.41 | 3.00 | 3.58 | 4.16 | 4.71 | 5.24 | 5.73 | 6.19 | 6.62 | 7.02 | 7.40 | 7.75 | 8.07 | 8.38 | 8.66 | 8.92 | 9.17 | 9.40 | 9.62 |
| | $W_{1,2}$ | 4.03 | 8.79 | 14.33 | 20.36 | 26.67 | 33.07 | 39.52 | 45.99 | 52.47 | 58.95 | 65.43 | 71.92 | 78.40 | 84.88 | 91.37 | 97.85 | 104.34 | 110.82 | 117.31 | 123.79 | 130.27 | 136.76 | 143.24 | 149.73 | 156.21 |
| | $W_{3,4}$ | 6.79 | 13.63 | 20.36 | 26.97 | 33.51 | 40.02 | 46.51 | 53.00 | 59.49 | 65.97 | 72.46 | 78.94 | 85.43 | 91.91 | 98.40 | 104.88 | 111.36 | 117.85 | 124.28 | 130.76 | 137.24 | 143.73 | 150.21 | 156.70 | 163.18 |
| | $J$ | 6.23 | 10.84 | 13.99 | 16.09 | 17.58 | 18.66 | 19.44 | 20.03 | 20.47 | 20.81 | 21.07 | 21.28 | 21.44 | 21.58 | 21.67 | 21.75 | 21.82 | 21.87 | 21.92 | 21.96 | 21.99 | 22.02 | 22.04 | 22.06 | 22.08 |
| | $K$ | 2.61 | 4.53 | 5.61 | 6.13 | 6.35 | 6.45 | 6.49 | 6.51 | 6.52 | 6.52 | 6.52 | 6.52 | 6.52 | 6.52 | 6.52 | 6.52 | 6.52 | 6.52 | 6.52 | 6.52 | 6.52 | 6.52 | 6.52 | 6.52 | 6.52 |
| 0.0 | $U_1$ | 11.45 | 20.88 | 28.03 | 33.42 | 37.54 | 40.76 | 43.32 | 45.40 | 47.12 | 48.56 | 49.78 | 50.83 | 51.74 | 52.53 | 53.23 | 53.85 | 54.40 | 54.90 | 55.34 | 55.75 | 56.12 | 56.46 | 56.77 | 57.05 | 57.32 |
| | $U_2$ | 0.15 | 0.33 | 0.56 | 0.90 | 1.33 | 1.85 | 2.41 | 3.00 | 3.58 | 4.16 | 4.71 | 5.23 | 5.73 | 6.19 | 6.62 | 7.02 | 7.40 | 7.74 | 8.07 | 8.37 | 8.65 | 8.92 | 9.17 | 9.40 | 9.61 |
| | $W_{1,2}$ | 3.99 | 8.72 | 14.25 | 20.28 | 26.57 | 32.97 | 39.42 | 45.89 | 52.37 | 58.85 | 65.33 | 71.82 | 78.30 | 84.79 | 91.27 | 97.75 | 104.24 | 110.72 | 117.21 | 123.69 | 130.17 | 136.66 | 143.14 | 149.63 | 156.11 |
| | $W_{3,4}$ | 6.77 | 13.60 | 20.31 | 26.92 | 33.46 | 39.96 | 46.46 | 52.95 | 59.43 | 65.92 | 72.40 | 78.89 | 85.37 | 91.86 | 98.34 | 104.82 | 111.31 | 117.79 | 124.28 | 130.76 | 137.24 | 143.73 | 150.21 | 156.70 | 163.18 |
| | $J$ | 6.26 | 10.91 | 14.06 | 16.22 | 17.74 | 18.83 | 19.63 | 20.22 | 20.67 | 21.01 | 21.28 | 21.48 | 21.65 | 21.77 | 21.88 | 21.96 | 22.03 | 22.08 | 22.13 | 22.17 | 22.19 | 22.22 | 22.24 | 22.26 | 22.29 |
| | $K$ | 2.62 | 4.56 | 5.65 | 6.17 | 6.40 | 6.50 | 6.54 | 6.55 | 6.56 | 6.56 | 6.56 | 6.56 | 6.56 | 6.56 | 6.56 | 6.56 | 6.56 | 6.56 | 6.56 | 6.56 | 6.56 | 6.56 | 6.56 | 6.56 | 6.56 |

TABLE S21: Parameters of the THF interaction Hamiltonian for different ratios $u_0/u_1$ and screening lengths $\xi$ of the double-gate interaction potential ($U_\xi = 24$ meV for $\xi = 10$nm). We use the same BM model parameters as in Table S20.

| $w_0/w_1$ | $\alpha_1$ | $\alpha_2$ | $\lambda_1$ | $\lambda_2$ | $\lambda$ | $\mathcal{W}$ | $t_0$ | $\gamma$ | $M$ | $v_\star$ | $v_\star'$ | $v_\star''$ | $\gamma/U_1$ | $\gamma/U_2$ |
|---|---|---|---|---|---|---|---|---|---|---|---|---|---|---|
| 1.00 | 0.748 | 0.664 | 0.137 | 0.155 | 0.509 | 0.974 | 0.607 | 20.395 | 17.809 | -2.602 | 1.038 | 0.1112 | 0.41 | 9.66 |
| 0.90 | 0.764 | 0.645 | 0.143 | 0.161 | 0.366 | 0.969 | 0.475 | 9.628 | 18.995 | -3.089 | 1.244 | 0.0235 | 0.17 | 6.76 |
| 0.80 | 0.791 | 0.612 | 0.155 | 0.173 | 0.330 | 0.971 | 0.696 | -2.639 | 19.935 | -3.475 | 1.379 | -0.0321 | -0.05 | -2.11 |
| 0.70 | 0.824 | 0.567 | 0.173 | 0.179 | 0.335 | 0.961 | 1.121 | -15.584 | 20.667 | -3.781 | 1.450 | -0.0619 | -0.31 | -11.96 |
| 0.60 | 0.860 | 0.511 | 0.185 | 0.185 | 0.346 | 0.943 | 1.577 | -28.553 | 21.230 | -4.028 | 1.460 | -0.0723 | -0.62 | -20.53 |
| 0.50 | 0.896 | 0.444 | 0.197 | 0.191 | 0.355 | 0.924 | 2.016 | -40.986 | 21.657 | -4.230 | 1.406 | -0.0686 | -0.96 | -28.28 |
| 0.40 | 0.929 | 0.370 | 0.215 | 0.197 | 0.366 | 0.912 | 2.397 | -52.348 | 21.974 | -4.396 | 1.281 | -0.0554 | -1.30 | -34.65 |
| 0.30 | 0.958 | 0.287 | 0.227 | 0.203 | 0.365 | 0.897 | 2.705 | -62.091 | 22.201 | -4.530 | 1.075 | -0.0373 | -1.59 | -41.50 |
| 0.20 | 0.980 | 0.197 | 0.239 | 0.203 | 0.372 | 0.892 | 2.964 | -69.644 | 22.354 | -4.633 | 0.784 | -0.0189 | -1.85 | -45.62 |
| 0.10 | 0.995 | 0.102 | 0.245 | 0.209 | 0.366 | 0.881 | 3.095 | -74.457 | 22.442 | -4.698 | 0.416 | -0.0051 | -1.97 | -50.49 |
| 0.00 | 1.000 | 0.000 | 0.245 | 0.000 | 0.357 | 0.872 | 3.044 | -76.115 | 22.471 | -4.721 | 0.000 | -0.0000 | -1.98 | -53.85 |

TABLE S22. Parameters of the $f$-electron Wannier functions and the THF single-particle Hamiltonian for different values of the tunneling ratio $w_0/w_1$. $\mathcal{W}$ denotes the total weight of the THF $f$-electrons on the active bands. In computing the ratios $\gamma/U_1$ and $\gamma/U_2$, we employ the on-site and nearest-neighbor repulsion parameters $U_1$ and $U_2$ obtained numerically for $\xi = 10$ nm, as given in Table S23. We employ $v_F = 5.944$ eV·Å, $|\mathbf{K}| = 1.703$ Å$^{-1}$, $w_1 = 110$ meV, and $\theta = 0.88°$ for the BM model.

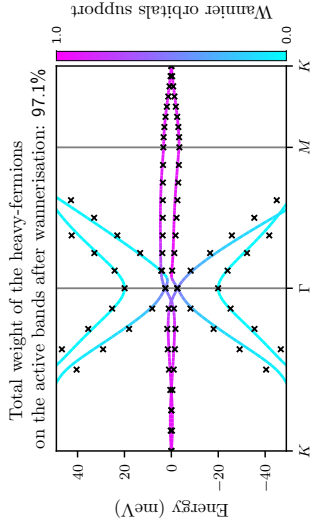

Total weight of the heavy-fermions on the active bands after wannierisation: 97.1%

Wannier orbitals support

FIG. S27. Band structures of the BM and THF models near charge neutrality for $w_0/w_1 = 0.8$, depicted by lines and crosses, respectively. The BM bands are colored according to the weight of the $f$-electron wave function on them. We use the same BM parameters as in Table S22.

| $w_0/w_1$ | $\xi/$nm | 2 | 4 | 6 | 8 | 10 | 12 | 14 | 16 | 18 | 20 | 22 | 24 | 26 | 28 | 30 | 32 | 34 | 36 | 38 | 40 | 42 | 44 | 46 | 48 | 50 |
|---|---|---|---|---|---|---|---|---|---|---|---|---|---|---|---|---|---|---|---|---|---|---|---|---|---|---|
| 1.0 | $U_1$ | 18.11 | 31.04 | 39.59 | 45.43 | 49.62 | 52.77 | 55.22 | 57.18 | 58.79 | 60.13 | 61.27 | 62.25 | 63.10 | 63.85 | 64.50 | 65.09 | 65.61 | 66.08 | 66.51 | 66.90 | 67.25 | 67.58 | 67.88 | 68.15 | 68.41 |
| | $U_2$ | 0.44 | 0.85 | 1.23 | 1.64 | 2.11 | 2.64 | 3.21 | 3.79 | 4.37 | 4.94 | 5.51 | 6.08 | 6.48 | 6.93 | 7.35 | 7.74 | 8.11 | 8.45 | 8.77 | 9.07 | 9.34 | 9.60 | 9.84 | 10.07 | 10.28 |
| | $W_{1,2}$ | 4.45 | 9.73 | 15.75 | 22.20 | 28.85 | 35.58 | 42.34 | 49.12 | 55.91 | 62.69 | 69.48 | 76.27 | 83.06 | 89.85 | 96.64 | 103.43 | 110.22 | 117.01 | 123.80 | 130.58 | 137.37 | 144.16 | 150.95 | 157.74 | 164.53 |
| | $W_{3,4}$ | 10.51 | 20.36 | 28.91 | 36.53 | 43.67 | 50.61 | 57.46 | 64.27 | 71.07 | 77.87 | 84.66 | 91.45 | 98.24 | 105.03 | 111.81 | 118.60 | 125.39 | 132.18 | 138.97 | 145.76 | 152.55 | 159.34 | 166.13 | 172.92 | 179.71 |
| | $J$ | 6.90 | 10.69 | 12.56 | 13.56 | 14.16 | 14.57 | 14.87 | 15.11 | 15.31 | 15.48 | 15.63 | 15.76 | 15.88 | 15.98 | 16.07 | 16.16 | 16.23 | 16.30 | 16.36 | 16.41 | 16.46 | 16.50 | 16.54 | 16.58 | 16.61 |
| | $K$ | 2.85 | 4.81 | 5.83 | 6.29 | 6.49 | 6.57 | 6.60 | 6.61 | 6.62 | 6.62 | 6.62 | 6.62 | 6.62 | 6.62 | 6.62 | 6.62 | 6.62 | 6.62 | 6.62 | 6.62 | 6.62 | 6.62 | 6.62 | 6.62 | 6.62 |
| 0.9 | $U_1$ | 19.83 | 34.32 | 44.08 | 50.81 | 55.64 | 59.26 | 62.05 | 64.27 | 66.08 | 67.58 | 68.83 | 69.90 | 70.83 | 71.64 | 72.34 | 72.97 | 73.53 | 74.03 | 74.48 | 74.88 | 75.26 | 75.60 | 75.91 | 76.19 | 76.46 |
| | $U_2$ | 0.22 | 0.44 | 0.69 | 1.00 | 1.42 | 1.93 | 2.50 | 3.10 | 3.70 | 4.29 | 4.85 | 5.39 | 5.89 | 6.36 | 6.80 | 7.21 | 7.59 | 7.95 | 8.28 | 8.58 | 8.87 | 9.14 | 9.39 | 9.62 | 9.84 |
| | $W_{1,2}$ | 4.59 | 9.93 | 15.98 | 22.43 | 29.08 | 35.81 | 42.57 | 49.35 | 56.14 | 62.93 | 69.72 | 76.50 | 83.29 | 90.08 | 96.87 | 103.66 | 110.45 | 117.24 | 124.03 | 130.82 | 137.61 | 144.40 | 151.19 | 157.98 | 164.77 |
| | $W_{3,4}$ | 9.67 | 18.90 | 27.13 | 34.60 | 41.68 | 48.59 | 55.43 | 62.24 | 69.04 | 75.83 | 82.62 | 89.41 | 96.20 | 102.99 | 109.78 | 116.57 | 123.36 | 130.15 | 136.94 | 143.73 | 150.52 | 157.31 | 164.09 | 170.88 | 177.67 |
| | $J$ | 6.45 | 10.01 | 11.69 | 12.49 | 12.89 | 13.09 | 13.20 | 13.27 | 13.34 | 13.35 | 13.34 | 13.35 | 13.35 | 13.36 | 13.36 | 13.36 | 13.36 | 13.36 | 13.36 | 13.36 | 13.36 | 13.36 | 13.36 | 13.37 | 13.37 |
| | $K$ | 2.66 | 4.53 | 5.52 | 5.97 | 6.16 | 6.24 | 6.27 | 6.28 | 6.29 | 6.29 | 6.29 | 6.29 | 6.29 | 6.29 | 6.29 | 6.29 | 6.29 | 6.29 | 6.29 | 6.29 | 6.29 | 6.29 | 6.29 | 6.29 | 6.29 |
| 0.8 | $U_1$ | 18.73 | 32.74 | 42.38 | 49.12 | 54.01 | 57.68 | 60.53 | 62.80 | 64.64 | 66.17 | 67.45 | 68.54 | 69.48 | 70.30 | 71.02 | 71.65 | 72.22 | 72.72 | 73.18 | 73.59 | 73.96 | 74.31 | 74.62 | 74.91 | 75.18 |
| | $U_2$ | 0.16 | 0.33 | 0.54 | 0.84 | 1.25 | 1.76 | 2.33 | 2.93 | 3.54 | 4.13 | 4.70 | 5.24 | 5.75 | 6.23 | 6.67 | 7.08 | 7.47 | 7.82 | 8.16 | 8.46 | 8.75 | 9.02 | 9.28 | 9.51 | 9.73 |
| | $W_{1,2}$ | 4.77 | 10.20 | 16.28 | 22.75 | 29.40 | 36.14 | 42.90 | 49.68 | 56.47 | 63.26 | 70.04 | 76.83 | 83.62 | 90.41 | 97.20 | 103.99 | 110.78 | 117.57 | 124.36 | 131.15 | 137.94 | 144.73 | 151.51 | 158.30 | 165.09 |
| | $W_{3,4}$ | 8.91 | 17.55 | 25.45 | 32.76 | 39.78 | 46.67 | 53.48 | 60.30 | 67.09 | 73.89 | 80.68 | 87.47 | 94.26 | 101.05 | 107.83 | 114.62 | 121.41 | 128.20 | 134.99 | 141.78 | 148.57 | 155.36 | 162.15 | 168.94 | 175.73 |
| | $J$ | 6.23 | 9.86 | 11.70 | 12.66 | 13.19 | 13.50 | 13.69 | 13.82 | 13.91 | 13.97 | 14.01 | 14.04 | 14.07 | 14.09 | 14.10 | 14.11 | 14.12 | 14.13 | 14.14 | 14.14 | 14.15 | 14.15 | 14.16 | 14.16 | 14.16 |
| | $K$ | 2.53 | 4.33 | 5.29 | 5.73 | 5.92 | 6.00 | 6.03 | 6.04 | 6.05 | 6.05 | 6.05 | 6.05 | 6.05 | 6.05 | 6.05 | 6.05 | 6.05 | 6.05 | 6.05 | 6.05 | 6.05 | 6.05 | 6.05 | 6.05 | 6.05 |
| 0.7 | $U_1$ | 16.80 | 29.63 | 38.66 | 45.08 | 49.79 | 53.36 | 56.15 | 58.37 | 60.19 | 61.70 | 62.97 | 64.05 | 64.99 | 65.80 | 66.51 | 67.14 | 67.71 | 68.21 | 68.67 | 69.08 | 69.45 | 69.79 | 70.11 | 70.39 | 70.66 |
| | $U_2$ | 0.16 | 0.36 | 0.56 | 0.88 | 1.30 | 1.82 | 2.40 | 3.00 | 3.61 | 4.21 | 4.78 | 5.32 | 5.83 | 6.30 | 6.75 | 7.16 | 7.54 | 7.90 | 8.23 | 8.54 | 8.83 | 9.10 | 9.35 | 9.58 | 9.80 |
| | $W_{1,2}$ | 4.87 | 10.36 | 16.46 | 22.94 | 29.59 | 36.33 | 43.09 | 49.87 | 56.66 | 63.45 | 70.24 | 77.02 | 83.81 | 90.60 | 97.39 | 104.18 | 110.97 | 117.76 | 124.55 | 131.34 | 138.13 | 144.92 | 151.71 | 158.50 | 165.28 |
| | $W_{3,4}$ | 8.31 | 16.46 | 24.08 | 31.27 | 38.23 | 45.09 | 51.91 | 58.71 | 65.50 | 72.29 | 79.08 | 85.87 | 92.66 | 99.45 | 106.24 | 113.03 | 119.82 | 126.61 | 133.40 | 140.19 | 146.98 | 153.77 | 160.56 | 167.34 | 174.13 |
| | $J$ | 6.12 | 9.94 | 12.00 | 13.29 | 14.05 | 14.55 | 14.90 | 15.15 | 15.33 | 15.47 | 15.58 | 15.66 | 15.73 | 15.79 | 15.83 | 15.87 | 15.89 | 15.92 | 15.94 | 15.95 | 15.97 | 15.98 | 15.99 | 16.00 | 16.01 |
| | $K$ | 2.45 | 4.20 | 5.15 | 5.59 | 5.78 | 5.86 | 5.89 | 5.90 | 5.91 | 5.91 | 5.91 | 5.91 | 5.91 | 5.91 | 5.91 | 5.91 | 5.91 | 5.91 | 5.91 | 5.91 | 5.91 | 5.91 | 5.91 | 5.91 | 5.91 |

| $u_0/u_1$ | $\xi$/nm | 2 | 4 | 6 | 8 | 10 | 12 | 14 | 16 | 18 | 20 | 22 | 24 | 26 | 28 | 30 | 32 | 34 | 36 | 38 | 40 | 42 | 44 | 46 | 48 | 50 |
|---|---|---|---|---|---|---|---|---|---|---|---|---|---|---|---|---|---|---|---|---|---|---|---|---|---|---|
| 0.6 | $U_1$ | 15.05 | 26.78 | 35.20 | 41.27 | 45.78 | 49.24 | 51.95 | 54.12 | 55.91 | 57.39 | 58.64 | 59.71 | 60.64 | 61.44 | 62.15 | 62.78 | 63.34 | 63.84 | 64.29 | 64.70 | 65.07 | 65.41 | 65.73 | 66.01 | 66.28 |
|  | $U_2$ | 0.17 | 0.37 | 0.61 | 0.95 | 1.39 | 1.92 | 2.51 | 3.12 | 3.73 | 4.32 | 4.89 | 5.43 | 5.94 | 6.41 | 6.85 | 7.27 | 7.65 | 8.00 | 8.34 | 8.64 | 8.93 | 9.20 | 9.45 | 9.68 | 9.90 |
|  | $W_{1,2}$ | 4.88 | 10.36 | 16.46 | 22.93 | 29.59 | 36.32 | 43.09 | 49.87 | 56.65 | 63.44 | 70.23 | 77.02 | 83.81 | 90.60 | 97.38 | 104.17 | 110.96 | 117.75 | 124.54 | 131.33 | 138.12 | 144.91 | 151.70 | 158.49 | 165.28 |
|  | $W_{3,4}$ | 7.86 | 15.65 | 23.05 | 30.14 | 37.06 | 43.90 | 50.71 | 57.51 | 64.31 | 71.10 | 77.89 | 84.68 | 91.46 | 98.25 | 105.04 | 111.83 | 118.62 | 125.41 | 132.20 | 138.99 | 145.78 | 152.57 | 159.36 | 166.15 | 172.94 |
|  | $J$ | 6.10 | 10.13 | 12.53 | 14.02 | 15.00 | 15.69 | 16.18 | 16.54 | 16.82 | 17.03 | 17.20 | 17.33 | 17.43 | 17.51 | 17.58 | 17.64 | 17.68 | 17.72 | 17.75 | 17.78 | 17.80 | 17.82 | 17.84 | 17.85 | 17.86 |
|  | $K$ | 2.41 | 4.15 | 5.10 | 5.54 | 5.73 | 5.81 | 5.84 | 5.85 | 5.86 | 5.86 | 5.86 | 5.86 | 5.86 | 5.86 | 5.86 | 5.86 | 5.86 | 5.86 | 5.86 | 5.86 | 5.86 | 5.86 | 5.86 | 5.86 | 5.86 |
| 0.5 | $U_1$ | 13.73 | 24.60 | 32.55 | 38.36 | 42.72 | 46.07 | 48.72 | 50.86 | 52.62 | 54.08 | 55.32 | 56.38 | 57.30 | 58.10 | 58.80 | 59.43 | 59.98 | 60.48 | 60.93 | 61.34 | 61.71 | 62.05 | 62.36 | 62.64 | 62.91 |
|  | $U_2$ | 0.18 | 0.38 | 0.64 | 0.99 | 1.45 | 1.99 | 2.58 | 3.19 | 3.80 | 4.40 | 4.97 | 5.51 | 6.03 | 6.49 | 6.93 | 7.34 | 7.72 | 8.07 | 8.40 | 8.71 | 9.00 | 9.27 | 9.52 | 9.75 | 9.97 |
|  | $W_{1,2}$ | 4.81 | 10.24 | 16.31 | 22.76 | 29.41 | 36.14 | 43.04 | 49.69 | 56.47 | 63.26 | 70.05 | 76.84 | 83.63 | 90.42 | 97.21 | 104.00 | 110.79 | 117.57 | 124.36 | 131.15 | 137.94 | 144.73 | 151.52 | 158.31 | 165.10 |
|  | $W_{3,4}$ | 7.54 | 15.06 | 22.31 | 29.33 | 36.21 | 43.04 | 49.85 | 56.64 | 63.44 | 70.23 | 77.02 | 83.81 | 90.60 | 97.39 | 104.17 | 110.96 | 117.75 | 124.54 | 131.33 | 138.12 | 144.91 | 151.70 | 158.49 | 165.28 | 172.07 |
|  | $J$ | 6.13 | 10.36 | 13.00 | 14.72 | 15.89 | 16.72 | 17.33 | 17.78 | 18.13 | 18.39 | 18.60 | 18.76 | 18.89 | 19.00 | 19.08 | 19.15 | 19.21 | 19.25 | 19.29 | 19.32 | 19.35 | 19.38 | 19.40 | 19.41 | 19.43 |
|  | $K$ | 2.42 | 4.17 | 5.13 | 5.57 | 5.76 | 5.84 | 5.87 | 5.89 | 5.89 | 5.90 | 5.90 | 5.90 | 5.90 | 5.90 | 5.90 | 5.90 | 5.90 | 5.90 | 5.90 | 5.90 | 5.90 | 5.90 | 5.90 | 5.90 | 5.90 |
| 0.4 | $U_1$ | 12.66 | 22.82 | 30.35 | 35.92 | 40.14 | 43.41 | 46.00 | 48.10 | 49.83 | 51.27 | 52.50 | 53.55 | 54.46 | 55.25 | 55.95 | 56.57 | 57.12 | 57.62 | 58.07 | 58.47 | 58.84 | 59.18 | 59.50 | 59.80 | 60.04 |
|  | $U_2$ | 0.19 | 0.41 | 0.68 | 1.04 | 1.51 | 2.06 | 2.65 | 3.27 | 3.88 | 4.47 | 5.04 | 5.58 | 6.09 | 6.56 | 7.00 | 7.41 | 7.79 | 8.14 | 8.47 | 8.78 | 9.06 | 9.33 | 9.58 | 9.81 | 10.03 |
|  | $W_{1,2}$ | 4.70 | 10.05 | 16.07 | 22.50 | 29.14 | 35.87 | 42.63 | 49.41 | 56.20 | 62.99 | 69.78 | 76.56 | 83.35 | 90.14 | 96.93 | 103.72 | 110.51 | 117.30 | 124.09 | 130.88 | 137.67 | 144.45 | 151.24 | 158.03 | 164.82 |
|  | $W_{3,4}$ | 7.32 | 14.65 | 21.79 | 28.75 | 35.61 | 42.43 | 49.23 | 56.03 | 62.82 | 69.61 | 76.40 | 83.19 | 89.98 | 96.77 | 103.56 | 110.35 | 117.13 | 123.92 | 130.71 | 137.50 | 144.29 | 151.08 | 157.87 | 164.66 | 171.45 |
|  | $J$ | 6.20 | 10.61 | 13.48 | 15.40 | 16.74 | 17.71 | 18.42 | 18.96 | 19.37 | 19.69 | 19.93 | 20.13 | 20.29 | 20.41 | 20.51 | 20.60 | 20.67 | 20.72 | 20.77 | 20.81 | 20.85 | 20.87 | 20.90 | 20.92 | 20.94 |
|  | $K$ | 2.45 | 4.24 | 5.22 | 5.67 | 5.87 | 5.95 | 5.98 | 5.99 | 6.00 | 6.00 | 6.00 | 6.00 | 6.00 | 6.00 | 6.00 | 6.00 | 6.00 | 6.00 | 6.00 | 6.00 | 6.00 | 6.00 | 6.00 | 6.00 | 6.00 |
| 0.3 | $U_1$ | 11.61 | 21.08 | 29.34 | 34.83 | 39.00 | 42.24 | 44.82 | 46.91 | 48.64 | 50.08 | 51.30 | 52.35 | 53.26 | 54.06 | 54.75 | 55.37 | 55.93 | 56.42 | 56.87 | 57.28 | 57.65 | 57.99 | 58.30 | 58.58 | 58.85 |
|  | $U_2$ | 0.18 | 0.39 | 0.66 | 1.03 | 1.50 | 2.05 | 2.64 | 3.26 | 3.87 | 4.47 | 5.04 | 5.61 | 6.11 | 6.55 | 7.00 | 7.40 | 7.78 | 8.13 | 8.46 | 8.77 | 9.06 | 9.32 | 9.57 | 9.81 | 10.02 |
|  | $W_{1,2}$ | 4.54 | 9.77 | 15.73 | 22.14 | 28.77 | 35.49 | 42.25 | 49.03 | 55.81 | 62.60 | 69.39 | 76.18 | 82.97 | 89.76 | 96.55 | 103.34 | 110.13 | 116.92 | 123.71 | 130.49 | 137.28 | 144.07 | 150.86 | 157.65 | 164.44 |
|  | $W_{3,4}$ | 7.17 | 14.37 | 21.43 | 28.35 | 35.20 | 42.02 | 48.82 | 55.61 | 62.40 | 69.19 | 75.98 | 82.77 | 89.56 | 96.35 | 103.14 | 109.93 | 116.72 | 123.51 | 130.30 | 137.08 | 143.87 | 150.66 | 157.45 | 164.24 | 171.03 |
|  | $J$ | 6.28 | 10.84 | 13.84 | 15.89 | 17.36 | 18.40 | 19.16 | 19.73 | 20.17 | 20.50 | 20.76 | 20.97 | 21.13 | 21.26 | 21.37 | 21.45 | 21.52 | 21.58 | 21.63 | 21.67 | 21.70 | 21.73 | 21.76 | 21.78 | 21.80 |
|  | $K$ | 2.51 | 4.34 | 5.35 | 5.81 | 6.01 | 6.10 | 6.13 | 6.14 | 6.15 | 6.15 | 6.15 | 6.15 | 6.15 | 6.15 | 6.15 | 6.15 | 6.15 | 6.15 | 6.15 | 6.15 | 6.15 | 6.15 | 6.15 | 6.15 | 6.15 |
| 0.2 | $U_1$ | 11.61 | 21.11 | 28.31 | 33.71 | 37.84 | 41.06 | 43.62 | 45.71 | 47.43 | 48.86 | 50.09 | 51.13 | 52.04 | 52.83 | 53.53 | 54.15 | 54.70 | 55.20 | 55.65 | 56.05 | 56.42 | 56.76 | 57.07 | 57.36 | 57.62 |
|  | $U_2$ | 0.19 | 0.40 | 0.68 | 1.05 | 1.53 | 2.08 | 2.68 | 3.29 | 3.91 | 4.50 | 5.07 | 5.61 | 6.13 | 6.58 | 7.02 | 7.43 | 7.81 | 8.17 | 8.49 | 8.80 | 9.09 | 9.35 | 9.60 | 9.84 | 10.05 |
|  | $W_{1,2}$ | 4.41 | 9.51 | 15.45 | 21.84 | 28.45 | 35.17 | 41.93 | 48.71 | 55.49 | 62.28 | 69.07 | 75.86 | 82.64 | 89.43 | 96.22 | 103.01 | 109.80 | 116.59 | 123.38 | 130.16 | 136.95 | 143.74 | 150.53 | 157.33 | 164.12 |
|  | $W_{3,4}$ | 7.07 | 14.19 | 21.28 | 28.10 | 34.94 | 41.75 | 48.54 | 55.34 | 62.13 | 68.92 | 75.71 | 82.50 | 89.28 | 96.07 | 102.86 | 109.65 | 116.44 | 123.23 | 130.02 | 136.81 | 143.60 | 150.39 | 157.18 | 163.97 | 170.76 |
|  | $J$ | 6.36 | 11.06 | 14.14 | 16.43 | 17.97 | 19.10 | 19.93 | 20.56 | 21.03 | 21.40 | 21.69 | 21.91 | 22.09 | 22.24 | 22.35 | 22.45 | 22.53 | 22.59 | 22.64 | 22.69 | 22.73 | 22.76 | 22.78 | 22.81 | 22.83 |
|  | $K$ | 2.58 | 4.45 | 5.49 | 5.96 | 6.17 | 6.25 | 6.29 | 6.30 | 6.31 | 6.31 | 6.31 | 6.31 | 6.31 | 6.31 | 6.31 | 6.31 | 6.31 | 6.31 | 6.31 | 6.31 | 6.31 | 6.31 | 6.31 | 6.31 | 6.31 |
| 0.1 | $U_1$ | 11.60 | 21.11 | 28.31 | 33.71 | 37.84 | 41.06 | 43.62 | 45.71 | 47.43 | 48.86 | 50.09 | 51.13 | 52.04 | 52.83 | 53.53 | 54.15 | 54.70 | 55.20 | 55.65 | 56.05 | 56.42 | 56.76 | 57.07 | 57.36 | 57.62 |
|  | $U_2$ | 0.18 | 0.37 | 0.64 | 1.00 | 1.47 | 2.03 | 2.62 | 3.24 | 3.85 | 4.45 | 5.02 | 5.56 | 6.06 | 6.53 | 6.97 | 7.38 | 7.76 | 8.12 | 8.45 | 8.75 | 9.04 | 9.31 | 9.56 | 9.79 | 9.96 |
|  | $W_{1,2}$ | 4.27 | 9.31 | 15.17 | 21.53 | 28.14 | 34.79 | 41.60 | 48.39 | 55.19 | 61.98 | 68.77 | 75.56 | 82.35 | 89.13 | 95.92 | 102.71 | 109.48 | 116.27 | 123.06 | 129.85 | 136.64 | 143.43 | 150.22 | 157.01 | 163.80 |
|  | $W_{3,4}$ | 7.01 | 14.08 | 21.07 | 27.96 | 34.79 | 41.60 | 48.39 | 55.19 | 61.98 | 68.77 | 75.56 | 82.35 | 89.13 | 95.92 | 102.71 | 109.48 | 116.27 | 123.06 | 129.85 | 136.64 | 143.43 | 150.24 | 157.03 | 163.82 | 170.56 |
|  | $J$ | 6.43 | 11.20 | 14.44 | 16.66 | 18.23 | 19.36 | 20.20 | 20.82 | 21.29 | 21.65 | 21.93 | 22.15 | 22.32 | 22.46 | 22.57 | 22.66 | 22.73 | 22.79 | 22.84 | 22.88 | 22.92 | 22.95 | 22.97 | 23.00 | 23.01 |
|  | $K$ | 2.63 | 4.54 | 5.59 | 6.08 | 6.29 | 6.38 | 6.41 | 6.43 | 6.43 | 6.44 | 6.44 | 6.44 | 6.44 | 6.44 | 6.44 | 6.44 | 6.44 | 6.44 | 6.44 | 6.44 | 6.44 | 6.44 | 6.44 | 6.44 | 6.44 |
| 0.0 | $U_1$ | 11.80 | 21.47 | 28.79 | 34.28 | 38.47 | 41.73 | 44.32 | 46.43 | 48.16 | 49.62 | 50.85 | 51.90 | 52.81 | 53.61 | 54.31 | 54.93 | 55.49 | 55.99 | 56.44 | 56.84 | 57.21 | 57.55 | 57.86 | 58.14 | 58.42 |
|  | $U_2$ | 0.15 | 0.34 | 0.59 | 0.95 | 1.41 | 1.96 | 2.56 | 3.17 | 3.79 | 4.38 | 4.95 | 5.49 | 6.00 | 6.47 | 6.91 | 7.32 | 7.71 | 8.06 | 8.39 | 8.70 | 8.98 | 9.25 | 9.50 | 9.74 | 9.96 |
|  | $W_{1,2}$ | 4.20 | 9.18 | 15.01 | 21.36 | 27.96 | 34.67 | 41.43 | 48.21 | 54.99 | 61.78 | 68.56 | 75.35 | 82.14 | 88.93 | 95.72 | 102.51 | 109.30 | 116.09 | 122.88 | 129.67 | 136.46 | 143.25 | 150.04 | 156.82 | 163.61 |
|  | $W_{3,4}$ | 6.99 | 14.05 | 21.03 | 27.92 | 34.75 | 41.55 | 48.35 | 55.14 | 61.93 | 68.72 | 75.51 | 82.30 | 89.09 | 95.88 | 102.67 | 109.46 | 116.25 | 123.04 | 129.83 | 136.62 | 143.41 | 150.20 | 156.98 | 163.77 | 170.56 |
|  | $J$ | 6.46 | 11.25 | 14.49 | 16.70 | 18.25 | 19.36 | 20.17 | 20.77 | 21.22 | 21.56 | 21.82 | 22.03 | 22.19 | 22.32 | 22.42 | 22.50 | 22.57 | 22.62 | 22.67 | 22.70 | 22.73 | 22.76 | 22.78 | 22.80 | 22.82 |
|  | $K$ | 2.64 | 4.57 | 5.63 | 6.12 | 6.34 | 6.43 | 6.46 | 6.47 | 6.48 | 6.48 | 6.48 | 6.48 | 6.48 | 6.48 | 6.48 | 6.48 | 6.48 | 6.48 | 6.48 | 6.48 | 6.48 | 6.48 | 6.48 | 6.48 | 6.48 |

TABLE S23: Parameters of the THF interaction Hamiltonian for different ratios $u_0/u_1$ and screening lengths $\xi$ of the double-gate interaction potential ($U_\xi = 24$ meV for $\xi = 10$nm). We use the same BM model parameters as in Table S22.

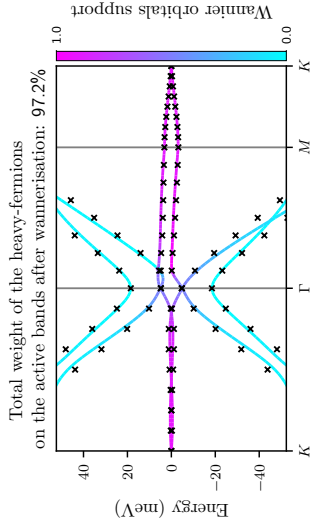

Total weight of the heavy-fermions on the active bands after wannierisation: 97.2%

Wannier orbitals support

Energy (meV)

**TABLE S24.** Parameters of the $f$-electron Wannier functions for different values of the tunneling ratio $w_0/w_1$. $\mathcal{W}$ denotes the total weight of the THF $f$-electrons on the active bands. In computing the ratios $\gamma/U_1$ and $\gamma/U_2$, we employ the on-site and nearest-neighbor repulsion parameters $U_1$ and $U_2$ obtained numerically for $\xi = 10$ nm, as given in Table S25. We employ $v_F = 5.944\,\mathrm{eV\cdot Å}$, $|\mathbf{K}| = 1.703\,\mathrm{Å}^{-1}$, $w_1 = 110$ meV, and $\theta = 0.90°$ for the BM model.

**FIG. S28.** Band structures of the BM and THF models near charge neutrality for $w_0/w_1 = 0.8$, depicted by lines and crosses, respectively. The BM bands are colored according to the weight of the $f$-electron wave function on them. We use the same BM parameters as in Table S24.

| $w_0/w_1$ | $\alpha_1$ | $\alpha_2$ | $\lambda_1$ | $\lambda_2$ | $\lambda$ | $\mathcal{W}$ | $t_0$ | $\gamma$ | $M$ | $v_\star$ | $v_\star'$ | $v_\star''$ | $\gamma/U_1$ | $\gamma/U_2$ |
|---|---|---|---|---|---|---|---|---|---|---|---|---|---|---|
| 1.00 | 0.749 | 0.663 | 0.137 | 0.155 | 0.471 | 0.971 | 0.491 | 19.081 | 16.837 | -2.795 | 1.108 | 0.0907 | 0.36 | 9.36 |
| 0.90 | 0.767 | 0.642 | 0.149 | 0.167 | 0.349 | 0.970 | 0.437 | 7.709 | 17.754 | -3.247 | 1.296 | 0.0130 | 0.13 | 5.37 |
| 0.80 | 0.794 | 0.608 | 0.161 | 0.173 | 0.329 | 0.972 | 0.685 | -4.951 | 18.462 | -3.606 | 1.417 | -0.0353 | -0.09 | -3.72 |
| 0.70 | 0.827 | 0.562 | 0.173 | 0.179 | 0.336 | 0.961 | 1.084 | -18.144 | 19.002 | -3.892 | 1.477 | -0.0606 | -0.36 | -13.01 |
| 0.60 | 0.863 | 0.505 | 0.185 | 0.185 | 0.348 | 0.940 | 1.493 | -31.261 | 19.409 | -4.123 | 1.479 | -0.0686 | -0.67 | -20.99 |
| 0.50 | 0.899 | 0.439 | 0.203 | 0.197 | 0.356 | 0.922 | 1.882 | -43.774 | 19.712 | -4.314 | 1.418 | -0.0641 | -1.01 | -28.37 |
| 0.40 | 0.931 | 0.365 | 0.215 | 0.197 | 0.367 | 0.911 | 2.216 | -55.172 | 19.933 | -4.472 | 1.287 | -0.0514 | -1.35 | -34.39 |
| 0.30 | 0.959 | 0.283 | 0.233 | 0.203 | 0.372 | 0.900 | 2.525 | -64.920 | 20.089 | -4.601 | 1.078 | -0.0344 | -1.65 | -39.98 |
| 0.20 | 0.981 | 0.195 | 0.239 | 0.209 | 0.372 | 0.891 | 2.718 | -72.462 | 20.192 | -4.699 | 0.784 | -0.0173 | -1.88 | -44.79 |
| 0.10 | 0.995 | 0.100 | 0.251 | 0.209 | 0.367 | 0.881 | 2.864 | -77.261 | 20.251 | -4.762 | 0.415 | -0.0047 | -2.00 | -49.16 |
| 0.00 | 1.000 | 0.000 | 0.251 | 0.000 | 0.363 | 0.877 | 2.891 | -78.912 | 20.270 | -4.784 | -0.000 | -0.0000 | -2.04 | -51.04 |

| $w_0/w_1$ | $\xi$/nm | 2 | 4 | 6 | 8 | 10 | 12 | 14 | 16 | 18 | 20 | 22 | 24 | 26 | 28 | 30 | 32 | 34 | 36 | 38 | 40 | 42 | 44 | 46 | 48 | 50 |
|---|---|---|---|---|---|---|---|---|---|---|---|---|---|---|---|---|---|---|---|---|---|---|---|---|---|---|
| 1.0 | $U_1$ | 19.24 | 32.95 | 41.09 | 48.16 | 52.58 | 55.88 | 58.45 | 60.50 | 62.18 | 63.57 | 64.75 | 65.76 | 66.64 | 67.40 | 68.08 | 68.68 | 69.21 | 69.69 | 70.13 | 70.52 | 70.88 | 71.21 | 71.52 | 71.80 | 72.05 |
| | $U_2$ | 0.39 | 0.77 | 1.14 | 1.55 | 2.04 | 2.59 | 3.19 | 3.81 | 4.42 | 5.02 | 5.58 | 6.12 | 6.62 | 7.09 | 7.53 | 7.93 | 8.31 | 8.66 | 8.98 | 9.29 | 9.57 | 9.84 | 10.08 | 10.31 | 10.53 |
| | $W_{1,2}$ | 4.70 | 10.26 | 16.59 | 23.36 | 30.32 | 37.36 | 44.44 | 51.53 | 58.63 | 65.73 | 72.83 | 79.93 | 87.04 | 94.14 | 101.24 | 108.34 | 115.44 | 122.54 | 129.64 | 136.74 | 143.85 | 150.95 | 158.05 | 165.15 | 172.25 |
| | $W_{3,4}$ | 10.61 | 20.09 | 29.32 | 37.17 | 44.59 | 51.82 | 58.97 | 66.09 | 73.20 | 80.31 | 87.41 | 94.51 | 101.62 | 108.72 | 115.82 | 122.92 | 130.02 | 137.12 | 144.22 | 151.32 | 158.43 | 165.53 | 172.63 | 179.73 | 186.83 |
| | $J$ | 6.91 | 10.63 | 12.42 | 13.33 | 13.85 | 14.18 | 14.40 | 14.58 | 14.72 | 14.83 | 14.93 | 15.02 | 15.10 | 15.16 | 15.22 | 15.27 | 15.32 | 15.36 | 15.39 | 15.43 | 15.45 | 15.48 | 15.50 | 15.52 | 15.54 |
| | $K$ | 2.80 | 4.72 | 5.71 | 6.14 | 6.32 | 6.40 | 6.42 | 6.44 | 6.44 | 6.44 | 6.44 | 6.44 | 6.44 | 6.44 | 6.44 | 6.44 | 6.44 | 6.44 | 6.44 | 6.44 | 6.44 | 6.44 | 6.44 | 6.44 | 6.44 |
| 0.9 | $U_1$ | 20.41 | 35.29 | 45.30 | 52.19 | 57.13 | 60.83 | 63.68 | 65.95 | 67.78 | 69.30 | 70.58 | 71.67 | 72.60 | 73.42 | 74.13 | 74.77 | 75.33 | 75.83 | 76.29 | 76.70 | 77.08 | 77.41 | 77.73 | 78.01 | 78.28 |
| | $U_2$ | 0.20 | 0.41 | 0.66 | 0.99 | 1.44 | 1.98 | 2.58 | 3.21 | 3.84 | 4.45 | 5.04 | 5.59 | 6.11 | 6.60 | 7.05 | 7.47 | 7.86 | 8.22 | 8.55 | 8.87 | 9.16 | 9.43 | 9.68 | 9.92 | 10.14 |
| | $W_{1,2}$ | 4.88 | 10.52 | 16.88 | 23.66 | 30.62 | 37.67 | 44.75 | 51.84 | 58.94 | 66.04 | 73.14 | 80.24 | 87.34 | 94.45 | 101.55 | 108.65 | 115.75 | 122.85 | 129.95 | 137.05 | 144.16 | 151.26 | 158.36 | 165.46 | 172.56 |
| | $W_{3,4}$ | 9.77 | 19.14 | 27.55 | 35.25 | 42.61 | 49.81 | 56.96 | 64.08 | 71.18 | 78.29 | 85.39 | 92.49 | 99.59 | 106.69 | 113.80 | 120.90 | 128.00 | 135.10 | 142.20 | 149.30 | 156.40 | 163.51 | 170.61 | 177.71 | 184.81 |
| | $J$ | 10.11 | 11.79 | 12.59 | 12.99 | 13.19 | 13.36 | 13.42 | 13.43 | 13.44 | 13.44 | 13.45 | 13.45 | 13.45 | 13.45 | 13.45 | 13.45 | 13.45 | 13.45 | 13.45 | 13.45 | 13.45 | 13.45 | 13.45 | 13.45 | 13.45 |
| | $K$ | 2.63 | 4.46 | 5.41 | 5.83 | 6.01 | 6.08 | 6.11 | 6.12 | 6.13 | 6.13 | 6.13 | 6.13 | 6.13 | 6.13 | 6.13 | 6.13 | 6.13 | 6.13 | 6.13 | 6.13 | 6.13 | 6.13 | 6.13 | 6.13 | 6.13 |
| 0.8 | $U_1$ | 18.96 | 33.10 | 42.83 | 49.62 | 54.54 | 58.24 | 61.10 | 63.38 | 65.23 | 66.77 | 68.05 | 69.15 | 70.09 | 70.91 | 71.63 | 72.27 | 72.84 | 73.34 | 73.80 | 74.21 | 74.59 | 74.93 | 75.25 | 75.54 | 75.80 |
| | $U_2$ | 0.16 | 0.34 | 0.56 | 0.89 | 1.33 | 1.88 | 2.48 | 3.11 | 3.75 | 4.36 | 4.95 | 5.51 | 6.03 | 6.52 | 6.97 | 7.40 | 7.79 | 8.15 | 8.49 | 8.80 | 9.09 | 9.37 | 9.62 | 9.86 | 10.08 |
| | $W_{1,2}$ | 5.06 | 10.79 | 17.20 | 23.99 | 30.96 | 38.01 | 45.09 | 52.18 | 59.28 | 66.38 | 73.48 | 80.58 | 87.69 | 94.79 | 101.89 | 108.99 | 116.09 | 123.19 | 130.29 | 137.40 | 144.50 | 151.60 | 158.70 | 165.80 | 172.90 |
| | $W_{3,4}$ | 9.03 | 17.83 | 25.92 | 33.69 | 40.78 | 47.90 | 55.09 | 62.21 | 69.31 | 76.42 | 83.52 | 90.62 | 97.72 | 104.82 | 111.93 | 119.03 | 126.13 | 133.23 | 140.33 | 147.43 | 154.53 | 161.64 | 168.74 | 175.84 | 182.94 |
| | $J$ | 6.33 | 10.03 | 11.91 | 12.91 | 13.47 | 13.80 | 14.01 | 14.16 | 14.25 | 14.33 | 14.38 | 14.42 | 14.45 | 14.47 | 14.49 | 14.51 | 14.52 | 14.53 | 14.54 | 14.54 | 14.55 | 14.55 | 14.56 | 14.56 | 14.57 |
| | $K$ | 2.50 | 4.27 | 5.20 | 5.61 | 5.79 | 5.86 | 5.89 | 5.90 | 5.90 | 5.90 | 5.90 | 5.90 | 5.90 | 5.90 | 5.90 | 5.90 | 5.90 | 5.90 | 5.90 | 5.90 | 5.90 | 5.90 | 5.90 | 5.90 | 5.90 |
| 0.7 | $U_1$ | 17.01 | 29.98 | 39.09 | 45.56 | 50.30 | 53.80 | 56.70 | 58.94 | 60.76 | 62.28 | 63.55 | 64.64 | 65.57 | 66.39 | 67.11 | 67.74 | 68.30 | 68.81 | 69.28 | 69.68 | 70.05 | 70.39 | 70.71 | 70.99 | 71.26 |
| | $U_2$ | 0.17 | 0.35 | 0.59 | 0.93 | 1.39 | 1.95 | 2.56 | 3.20 | 3.83 | 4.45 | 5.04 | 5.60 | 6.12 | 6.61 | 7.06 | 7.48 | 7.87 | 8.23 | 8.57 | 8.88 | 9.17 | 9.45 | 9.70 | 9.94 | 10.16 |
| | $W_{1,2}$ | 5.15 | 10.99 | 17.54 | 24.15 | 31.12 | 38.17 | 45.26 | 52.35 | 59.45 | 66.55 | 73.65 | 80.75 | 87.85 | 94.95 | 102.05 | 109.16 | 116.26 | 123.36 | 130.46 | 137.56 | 144.66 | 151.76 | 158.86 | 165.97 | 173.07 |
| | $W_{3,4}$ | 8.46 | 16.79 | 24.63 | 32.07 | 39.32 | 46.48 | 53.61 | 60.72 | 67.82 | 74.93 | 82.03 | 89.13 | 96.23 | 103.33 | 110.43 | 117.54 | 124.64 | 131.74 | 138.84 | 145.94 | 153.04 | 160.14 | 167.25 | 174.35 | 181.45 |
| | $J$ | 6.25 | 10.19 | 12.34 | 13.61 | 14.41 | 14.94 | 15.31 | 15.57 | 15.77 | 15.92 | 16.04 | 16.13 | 16.20 | 16.25 | 16.30 | 16.34 | 16.37 | 16.39 | 16.41 | 16.43 | 16.45 | 16.46 | 16.48 | 16.49 | 16.49 |
| | $K$ | 2.43 | 4.16 | 5.08 | 5.49 | 5.66 | 5.73 | 5.76 | 5.77 | 5.78 | 5.78 | 5.78 | 5.78 | 5.78 | 5.78 | 5.78 | 5.78 | 5.78 | 5.78 | 5.78 | 5.78 | 5.78 | 5.78 | 5.78 | 5.78 | 5.78 |

| $u_0/u_1$ | $\xi$/nm | 2 | 4 | 6 | 8 | 10 | 12 | 14 | 16 | 18 | 20 | 22 | 24 | 26 | 28 | 30 | 32 | 34 | 36 | 38 | 40 | 42 | 44 | 46 | 48 | 50 |
|---|---|---|---|---|---|---|---|---|---|---|---|---|---|---|---|---|---|---|---|---|---|---|---|---|---|---|
| 0.6 | $U_1$ | 15.27 | 27.14 | 33.64 | 41.77 | 46.31 | 49.79 | 52.52 | 54.71 | 56.50 | 57.99 | 59.25 | 60.32 | 61.25 | 62.06 | 62.77 | 63.40 | 63.96 | 64.46 | 64.91 | 65.32 | 65.70 | 66.04 | 66.35 | 66.64 | 66.90 |
|  | $U_2$ | 0.18 | 0.39 | 0.65 | 1.01 | 1.49 | 2.05 | 2.67 | 3.31 | 3.95 | 4.57 | 5.16 | 5.71 | 6.23 | 6.72 | 7.17 | 7.59 | 7.98 | 8.34 | 8.68 | 8.99 | 9.28 | 9.55 | 9.80 | 10.04 | 10.26 |
|  | $W_{1,2}$ | 5.14 | 10.92 | 17.33 | 24.12 | 31.10 | 38.14 | 45.23 | 52.32 | 59.42 | 66.52 | 73.62 | 80.72 | 87.82 | 94.92 | 102.02 | 109.12 | 116.23 | 123.33 | 130.43 | 137.53 | 144.63 | 151.73 | 158.83 | 165.94 | 173.04 |
|  | $W_{3,4}$ | 8.04 | 16.03 | 23.67 | 31.02 | 38.23 | 45.38 | 52.50 | 59.61 | 66.71 | 73.81 | 80.91 | 88.02 | 95.12 | 102.22 | 109.32 | 116.42 | 123.53 | 130.63 | 137.73 | 144.83 | 151.93 | 159.03 | 166.13 | 173.23 | 180.33 |
|  | $J$ | 6.26 | 10.40 | 12.87 | 14.42 | 15.45 | 16.16 | 16.68 | 17.06 | 17.35 | 17.58 | 17.75 | 17.89 | 17.99 | 18.05 | 18.15 | 18.21 | 18.26 | 18.29 | 18.33 | 18.35 | 18.38 | 18.40 | 18.41 | 18.43 | 18.44 |
|  | $K$ | 2.41 | 4.13 | 5.04 | 5.46 | 5.63 | 5.70 | 5.73 | 5.74 | 5.75 | 5.75 | 5.75 | 5.75 | 5.75 | 5.75 | 5.80 | 5.80 | 5.75 | 5.75 | 5.80 | 5.80 | 5.75 | 5.80 | 5.75 | 5.75 | 5.75 |
| 0.5 | $U_1$ | 13.99 | 25.04 | 33.09 | 38.97 | 43.38 | 46.77 | 49.44 | 51.59 | 53.36 | 54.83 | 56.08 | 57.14 | 58.06 | 58.87 | 59.57 | 60.20 | 60.76 | 61.26 | 61.71 | 62.12 | 62.49 | 62.83 | 63.14 | 63.43 | 63.69 |
|  | $U_2$ | 0.18 | 0.39 | 0.68 | 1.05 | 1.54 | 2.12 | 2.74 | 3.38 | 4.02 | 4.64 | 5.23 | 5.78 | 6.30 | 6.78 | 7.24 | 7.65 | 8.04 | 8.40 | 8.74 | 9.05 | 9.34 | 9.61 | 9.86 | 10.10 | 10.32 |
|  | $W_{1,2}$ | 5.06 | 10.77 | 17.15 | 23.93 | 30.90 | 37.95 | 45.03 | 52.12 | 59.22 | 66.32 | 73.42 | 80.52 | 87.62 | 94.72 | 101.82 | 108.92 | 116.03 | 123.13 | 130.23 | 137.33 | 144.43 | 151.53 | 158.63 | 165.74 | 172.84 |
|  | $W_{3,4}$ | 7.74 | 15.48 | 22.97 | 30.26 | 37.45 | 44.58 | 51.69 | 58.80 | 65.90 | 73.00 | 80.11 | 87.21 | 94.31 | 101.41 | 108.51 | 115.62 | 122.72 | 129.82 | 136.92 | 144.02 | 151.12 | 158.22 | 165.33 | 172.43 | 179.53 |
|  | $J$ | 6.31 | 10.65 | 13.38 | 15.15 | 16.36 | 17.22 | 17.85 | 18.32 | 18.67 | 18.94 | 19.15 | 19.32 | 19.45 | 19.56 | 19.65 | 19.72 | 19.77 | 19.82 | 19.86 | 19.89 | 19.92 | 19.94 | 19.96 | 19.98 | 20.00 |
|  | $K$ | 2.42 | 4.16 | 5.08 | 5.50 | 5.68 | 5.75 | 5.78 | 5.79 | 5.80 | 5.80 | 5.80 | 5.80 | 5.80 | 5.80 | 5.80 | 5.80 | 5.80 | 5.80 | 5.80 | 5.80 | 5.80 | 5.80 | 5.80 | 5.80 | 5.80 |
| 0.4 | $U_1$ | 12.94 | 23.30 | 30.95 | 36.60 | 40.87 | 44.17 | 46.77 | 48.98 | 50.76 | 52.11 | 53.34 | 54.39 | 55.31 | 56.10 | 56.81 | 57.43 | 57.98 | 58.48 | 58.93 | 59.34 | 59.71 | 60.05 | 60.36 | 60.64 | 60.91 |
|  | $U_2$ | 0.20 | 0.42 | 0.71 | 1.10 | 1.60 | 2.19 | 2.81 | 3.46 | 4.09 | 4.71 | 5.30 | 5.85 | 6.37 | 6.85 | 7.30 | 7.72 | 8.11 | 8.47 | 8.80 | 9.11 | 9.40 | 9.67 | 9.92 | 10.16 | 10.38 |
|  | $W_{1,2}$ | 4.94 | 10.57 | 16.90 | 23.66 | 30.61 | 37.66 | 44.71 | 51.83 | 58.93 | 66.03 | 73.13 | 80.23 | 87.33 | 94.43 | 101.53 | 108.63 | 115.74 | 122.84 | 129.94 | 137.04 | 144.14 | 151.24 | 158.34 | 165.44 | 172.55 |
|  | $W_{3,4}$ | 7.53 | 15.09 | 22.47 | 29.72 | 36.89 | 44.01 | 51.12 | 58.23 | 65.33 | 72.43 | 79.54 | 86.64 | 93.74 | 100.84 | 107.94 | 115.04 | 122.14 | 129.25 | 136.35 | 143.45 | 150.55 | 157.65 | 164.75 | 171.85 | 178.96 |
|  | $J$ | 6.38 | 10.92 | 13.88 | 15.86 | 17.24 | 18.23 | 18.96 | 19.51 | 19.93 | 20.25 | 20.50 | 20.70 | 20.86 | 20.99 | 21.09 | 21.17 | 21.24 | 21.30 | 21.35 | 21.39 | 21.42 | 21.45 | 21.47 | 21.49 | 21.51 |
|  | $K$ | 2.47 | 4.24 | 5.18 | 5.61 | 5.79 | 5.86 | 5.89 | 5.90 | 5.91 | 5.91 | 5.91 | 5.91 | 5.91 | 5.91 | 5.91 | 5.91 | 5.91 | 5.91 | 5.91 | 5.91 | 5.91 | 5.91 | 5.91 | 5.91 | 5.91 |
| 0.3 | $U_1$ | 12.27 | 22.20 | 29.61 | 35.13 | 39.31 | 42.57 | 45.16 | 47.25 | 48.98 | 50.43 | 51.65 | 52.70 | 53.61 | 54.41 | 55.10 | 55.73 | 56.28 | 56.78 | 57.22 | 57.63 | 58.00 | 58.34 | 58.65 | 58.93 | 59.20 |
|  | $U_2$ | 0.20 | 0.42 | 0.72 | 1.12 | 1.62 | 2.21 | 2.84 | 3.48 | 4.12 | 4.74 | 5.33 | 5.88 | 6.39 | 6.88 | 7.33 | 7.74 | 8.13 | 8.49 | 8.82 | 9.13 | 9.42 | 9.69 | 9.95 | 10.18 | 10.40 |
|  | $W_{1,2}$ | 4.79 | 10.31 | 16.59 | 23.32 | 30.27 | 37.30 | 44.38 | 51.47 | 58.57 | 65.67 | 72.77 | 79.87 | 86.97 | 94.07 | 101.18 | 108.28 | 115.38 | 122.48 | 129.58 | 136.68 | 143.78 | 150.89 | 157.99 | 165.09 | 172.19 |
|  | $W_{3,4}$ | 7.38 | 14.82 | 22.15 | 29.36 | 36.50 | 43.62 | 50.73 | 57.84 | 64.94 | 72.04 | 79.14 | 86.25 | 93.35 | 100.45 | 107.55 | 114.65 | 121.75 | 128.85 | 135.96 | 143.06 | 150.16 | 157.26 | 164.36 | 171.46 | 178.56 |
|  | $J$ | 6.48 | 11.11 | 14.68 | 16.92 | 18.51 | 19.66 | 20.46 | 21.13 | 21.61 | 21.98 | 22.26 | 22.49 | 22.67 | 22.81 | 22.93 | 23.02 | 23.10 | 23.16 | 23.21 | 23.26 | 23.29 | 23.32 | 23.35 | 23.37 | 23.39 |
|  | $K$ | 2.53 | 4.35 | 5.32 | 5.76 | 5.94 | 6.02 | 6.05 | 6.06 | 6.07 | 6.07 | 6.07 | 6.07 | 6.07 | 6.07 | 6.07 | 6.07 | 6.07 | 6.07 | 6.07 | 6.07 | 6.07 | 6.07 | 6.07 | 6.07 | 6.07 |
| 0.2 | $U_1$ | 11.92 | 21.63 | 28.93 | 34.29 | 38.55 | 41.78 | 44.45 | 46.55 | 48.28 | 49.73 | 50.96 | 52.01 | 52.92 | 53.72 | 54.42 | 55.04 | 55.60 | 56.10 | 56.54 | 56.95 | 57.32 | 57.66 | 57.97 | 58.26 | 58.52 |
|  | $U_2$ | 0.19 | 0.42 | 0.71 | 1.11 | 1.62 | 2.20 | 2.84 | 3.48 | 4.12 | 4.74 | 5.32 | 5.88 | 6.39 | 6.88 | 7.33 | 7.74 | 8.13 | 8.49 | 8.82 | 9.13 | 9.42 | 9.69 | 9.94 | 10.18 | 10.40 |
|  | $W_{1,2}$ | 4.63 | 10.04 | 16.24 | 22.97 | 29.90 | 36.94 | 44.02 | 51.10 | 58.20 | 65.30 | 72.40 | 79.50 | 86.60 | 93.70 | 100.81 | 107.91 | 115.01 | 122.11 | 129.21 | 136.31 | 143.41 | 150.52 | 157.62 | 164.72 | 171.82 |
|  | $W_{3,4}$ | 7.29 | 14.65 | 21.93 | 29.12 | 36.28 | 43.38 | 50.48 | 57.59 | 64.69 | 71.79 | 78.89 | 86.00 | 93.10 | 100.20 | 107.30 | 114.40 | 121.50 | 128.60 | 135.71 | 142.81 | 149.91 | 157.01 | 164.11 | 171.21 | 178.31 |
|  | $J$ | 6.57 | 11.41 | 14.68 | 17.17 | 18.78 | 19.95 | 20.79 | 21.42 | 21.89 | 22.26 | 22.54 | 22.76 | 22.93 | 23.07 | 23.18 | 23.27 | 23.34 | 23.40 | 23.45 | 23.49 | 23.53 | 23.56 | 23.58 | 23.60 | 23.62 |
|  | $K$ | 2.59 | 4.46 | 5.45 | 5.90 | 6.10 | 6.18 | 6.21 | 6.22 | 6.23 | 6.23 | 6.23 | 6.23 | 6.23 | 6.23 | 6.23 | 6.23 | 6.23 | 6.23 | 6.23 | 6.23 | 6.23 | 6.23 | 6.23 | 6.23 | 6.23 |
| 0.1 | $U_1$ | 11.89 | 21.60 | 28.89 | 34.22 | 38.60 | 41.84 | 44.51 | 46.61 | 48.34 | 49.79 | 51.02 | 52.07 | 52.99 | 53.79 | 54.48 | 55.11 | 55.66 | 56.16 | 56.60 | 57.01 | 57.38 | 57.72 | 58.03 | 58.31 | 58.58 |
|  | $U_2$ | 0.18 | 0.39 | 0.68 | 1.07 | 1.57 | 2.16 | 2.79 | 3.43 | 4.07 | 4.69 | 5.28 | 5.83 | 6.35 | 6.83 | 7.28 | 7.70 | 8.09 | 8.45 | 8.78 | 9.09 | 9.38 | 9.65 | 9.90 | 10.14 | 10.36 |
|  | $W_{1,2}$ | 4.50 | 9.81 | 15.98 | 22.67 | 29.59 | 36.63 | 43.70 | 50.79 | 57.89 | 64.99 | 72.09 | 79.19 | 86.29 | 93.39 | 100.49 | 107.59 | 114.69 | 121.80 | 128.90 | 136.00 | 143.10 | 150.20 | 157.30 | 164.40 | 171.51 |
|  | $W_{3,4}$ | 7.23 | 14.55 | 21.81 | 28.99 | 36.12 | 43.24 | 50.35 | 57.45 | 64.55 | 71.65 | 78.75 | 85.85 | 92.96 | 100.06 | 107.16 | 114.26 | 121.36 | 128.47 | 135.57 | 142.67 | 149.77 | 156.87 | 163.97 | 171.07 | 178.17 |
|  | $J$ | 6.64 | 11.55 | 14.89 | 17.17 | 18.88 | 20.04 | 20.89 | 21.51 | 21.99 | 22.34 | 22.62 | 22.83 | 23.00 | 23.14 | 23.24 | 23.33 | 23.40 | 23.46 | 23.51 | 23.55 | 23.58 | 23.61 | 23.63 | 23.65 | 23.67 |
|  | $K$ | 2.65 | 4.55 | 5.57 | 6.03 | 6.23 | 6.31 | 6.34 | 6.35 | 6.36 | 6.36 | 6.36 | 6.36 | 6.36 | 6.36 | 6.36 | 6.36 | 6.36 | 6.36 | 6.36 | 6.36 | 6.36 | 6.36 | 6.36 | 6.36 | 6.36 |
| 0.0 | $U_1$ | 11.93 | 21.68 | 29.04 | 34.55 | 38.75 | 42.03 | 44.63 | 46.74 | 48.47 | 49.93 | 51.16 | 52.21 | 53.13 | 53.93 | 54.63 | 55.25 | 55.81 | 56.30 | 56.75 | 57.16 | 57.53 | 57.87 | 58.18 | 58.47 | 58.73 |
|  | $U_2$ | 0.17 | 0.37 | 0.65 | 1.04 | 1.55 | 2.13 | 2.76 | 3.41 | 4.05 | 4.66 | 5.25 | 5.81 | 6.33 | 6.81 | 7.26 | 7.68 | 8.06 | 8.42 | 8.76 | 9.07 | 9.36 | 9.63 | 9.88 | 10.12 | 10.34 |
|  | $W_{1,2}$ | 4.44 | 9.72 | 15.87 | 22.54 | 29.47 | 36.50 | 43.57 | 50.66 | 57.76 | 64.85 | 71.96 | 79.06 | 86.16 | 93.26 | 100.36 | 107.46 | 114.56 | 121.66 | 128.77 | 135.87 | 142.97 | 150.07 | 157.17 | 164.27 | 171.37 |
|  | $W_{3,4}$ | 7.22 | 14.52 | 21.77 | 28.94 | 36.08 | 43.19 | 50.30 | 57.40 | 64.50 | 71.61 | 78.71 | 85.81 | 92.91 | 100.01 | 107.11 | 114.21 | 121.32 | 128.42 | 135.52 | 142.62 | 149.72 | 156.82 | 163.92 | 171.03 | 178.13 |
|  | $J$ | 6.67 | 11.62 | 14.97 | 17.27 | 18.88 | 20.04 | 20.89 | 21.51 | 21.99 | 22.34 | 22.62 | 22.83 | 23.00 | 23.14 | 23.24 | 23.33 | 23.40 | 23.46 | 23.51 | 23.55 | 23.58 | 23.61 | 23.63 | 23.65 | 23.67 |
|  | $K$ | 2.67 | 4.58 | 5.61 | 6.08 | 6.27 | 6.35 | 6.38 | 6.40 | 6.40 | 6.40 | 6.40 | 6.40 | 6.40 | 6.40 | 6.40 | 6.40 | 6.40 | 6.40 | 6.40 | 6.40 | 6.40 | 6.40 | 6.40 | 6.40 | 6.40 |

TABLE S25. Parameters of the THF interaction Hamiltonian for different ratios $u_0/u_1$ and screening lengths $\xi$ of the double-gate interaction potential ($U_\xi = 24$ meV for $\xi = 10$nm). We use the same BM model parameters as in Table S24.

| $w_0/w_1$ | $\alpha_1$ | $\alpha_2$ | $\lambda_1$ | $\lambda_2$ | $\lambda$ | $t_0$ | $\mathcal{W}$ | $\gamma$ | $M$ | $v_\star$ | $v_\star'$ | $v_\star''$ | $\gamma/U_1$ | $\gamma/U_2$ |
|---|---|---|---|---|---|---|---|---|---|---|---|---|---|---|
| 1.00 | 0.750 | 0.662 | 0.137 | 0.161 | 0.436 | 0.969 | 0.392 | 17.571 | 15.693 | -2.974 | 1.173 | 0.0730 | 0.32 | 8.89 |
| 0.90 | 0.769 | 0.639 | 0.149 | 0.167 | 0.344 | 0.973 | 0.409 | 5.656 | 16.357 | -3.393 | 1.344 | 0.0045 | 0.10 | 3.78 |
| 0.80 | 0.797 | 0.604 | 0.161 | 0.173 | 0.329 | 0.972 | 0.664 | -7.357 | 16.853 | -3.726 | 1.452 | -0.0374 | -0.13 | -5.18 |
| 0.70 | 0.831 | 0.557 | 0.179 | 0.185 | 0.338 | 0.957 | 1.032 | -20.769 | 17.218 | -3.993 | 1.502 | -0.0587 | -0.41 | -13.89 |
| 0.60 | 0.867 | 0.499 | 0.191 | 0.191 | 0.353 | 0.940 | 1.410 | -34.018 | 17.482 | -4.211 | 1.495 | -0.0648 | -0.73 | -21.20 |
| 0.50 | 0.901 | 0.433 | 0.203 | 0.197 | 0.358 | 0.921 | 1.728 | -46.602 | 17.672 | -4.391 | 1.429 | -0.0598 | -1.06 | -28.39 |
| 0.40 | 0.933 | 0.359 | 0.221 | 0.203 | 0.368 | 0.910 | 2.027 | -58.029 | 17.804 | -4.542 | 1.293 | -0.0476 | -1.40 | -34.02 |
| 0.30 | 0.961 | 0.278 | 0.233 | 0.209 | 0.377 | 0.903 | 2.278 | -67.781 | 17.895 | -4.666 | 1.080 | -0.0317 | -1.71 | -38.61 |
| 0.20 | 0.981 | 0.192 | 0.245 | 0.209 | 0.372 | 0.890 | 2.458 | -75.311 | 17.953 | -4.761 | 0.784 | -0.0159 | -1.91 | -43.94 |
| 0.10 | 0.995 | 0.098 | 0.251 | 0.209 | 0.373 | 0.886 | 2.588 | -80.097 | 17.986 | -4.822 | 0.415 | -0.0043 | -2.06 | -46.74 |
| 0.00 | 1.000 | 0.000 | 0.251 | 0.000 | 0.363 | 0.876 | 2.596 | -81.742 | 17.996 | -4.843 | 0.000 | -0.0000 | -2.06 | -49.88 |

TABLE S26. Parameters of the $f$-electron Wannier functions and the THF single-particle Hamiltonian for different values of the tunneling ratio $w_0/w_1$. $\mathcal{W}$ denotes the total weight of the THF $f$-electrons on the active bands. In computing the ratios $\gamma'/U_1$ and $\gamma'/U_2$, we employ the on-site and nearest-neighbor repulsion parameters $U_1$ and $U_2$ obtained numerically for $\xi = 10$ nm, as given in Table S27. We employ $v_F = 5.944$ eV Å, $|\mathbf{K}| = 1.703$ Å$^{-1}$, $w_1 = 110$ meV, and $\theta = 0.92°$ for the BM model.

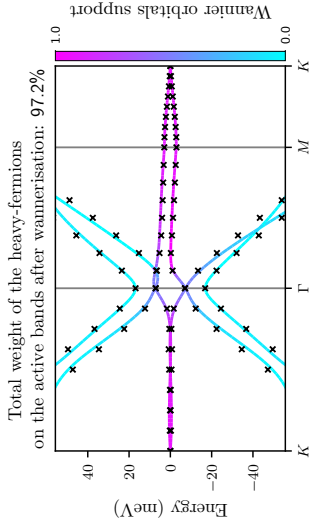

Total weight of the heavy-fermions on the active bands after wannierisation: 97.2%

Energy (meV) — Wannier orbitals support: 0.0 — 1.0

FIG. S29. Band structures of the BM and THF models near charge neutrality for $w_0/w_1 = 0.8$, depicted by lines and crosses, respectively. The BM bands are colored according to the weight of the $f$-electron wave function on them. We use the same BM parameters as in Table S26.

| $w_0/w_1$ | $\xi$/nm | 2 | 4 | 6 | 8 | 10 | 12 | 14 | 16 | 18 | 20 | 22 | 24 | 26 | 28 | 30 | 32 | 34 | 36 | 38 | 40 | 42 | 44 | 46 | 48 | 50 |
|---|---|---|---|---|---|---|---|---|---|---|---|---|---|---|---|---|---|---|---|---|---|---|---|---|---|---|
| 1.0 | $U_1$ | 20.30 | 34.72 | 44.22 | 50.69 | 55.31 | 58.76 | 61.43 | 63.56 | 65.30 | 66.74 | 67.95 | 68.99 | 69.89 | 70.67 | 71.36 | 71.97 | 72.52 | 73.01 | 73.45 | 73.85 | 74.22 | 74.55 | 74.86 | 75.14 | 75.40 |
|  | $U_2$ | 0.36 | 0.70 | 1.05 | 1.47 | 1.98 | 2.56 | 3.20 | 3.85 | 4.49 | 5.11 | 5.70 | 6.27 | 6.78 | 7.26 | 7.71 | 8.13 | 8.52 | 8.88 | 9.21 | 9.52 | 9.81 | 10.08 | 10.33 | 10.57 | 10.80 |
|  | $W_{1,2}$ | 4.98 | 10.84 | 17.49 | 24.59 | 31.87 | 39.24 | 46.64 | 54.05 | 61.47 | 68.89 | 76.31 | 83.73 | 91.15 | 98.57 | 105.99 | 113.41 | 120.83 | 128.25 | 135.67 | 143.09 | 150.51 | 157.93 | 165.35 | 172.78 | 180.20 |
|  | $W_{3,4}$ | 10.71 | 20.89 | 29.75 | 37.84 | 45.54 | 53.07 | 60.54 | 67.97 | 75.40 | 82.82 | 90.24 | 97.66 | 105.08 | 112.51 | 119.93 | 127.35 | 134.77 | 142.19 | 149.61 | 157.03 | 164.45 | 171.87 | 179.29 | 186.71 | 194.13 |
|  | $J$ | 6.93 | 10.61 | 12.32 | 13.16 | 13.61 | 13.87 | 14.03 | 14.15 | 14.24 | 14.32 | 14.38 | 14.43 | 14.47 | 14.51 | 14.54 | 14.57 | 14.59 | 14.61 | 14.63 | 14.65 | 14.66 | 14.67 | 14.68 | 14.69 | 14.70 |
|  | $K$ | 2.76 | 4.64 | 5.58 | 5.99 | 6.16 | 6.22 | 6.25 | 6.26 | 6.26 | 6.27 | 6.27 | 6.27 | 6.27 | 6.27 | 6.27 | 6.27 | 6.27 | 6.27 | 6.27 | 6.27 | 6.27 | 6.27 | 6.27 | 6.27 | 6.27 |
| 0.9 | $U_1$ | 20.75 | 35.84 | 45.98 | 52.96 | 57.96 | 61.69 | 64.57 | 66.86 | 68.71 | 70.25 | 71.53 | 72.63 | 73.57 | 74.39 | 75.10 | 75.74 | 76.31 | 76.81 | 77.27 | 77.68 | 78.06 | 78.41 | 78.73 | 79.03 | 79.27 |
|  | $U_2$ | 0.20 | 0.41 | 0.66 | 1.02 | 1.49 | 2.07 | 2.71 | 3.36 | 4.02 | 4.66 | 5.26 | 5.84 | 6.37 | 6.87 | 7.33 | 7.76 | 8.15 | 8.52 | 8.86 | 9.18 | 9.47 | 9.75 | 10.00 | 10.24 | 10.47 |
|  | $W_{1,2}$ | 5.19 | 11.15 | 17.85 | 24.95 | 32.25 | 39.62 | 47.02 | 54.43 | 61.85 | 69.27 | 76.69 | 84.11 | 91.53 | 98.95 | 106.37 | 113.79 | 121.21 | 128.63 | 136.05 | 143.47 | 150.89 | 158.31 | 165.73 | 173.15 | 180.57 |
|  | $W_{3,4}$ | 9.87 | 19.38 | 27.98 | 35.93 | 43.57 | 51.08 | 58.54 | 65.97 | 73.40 | 80.82 | 88.24 | 95.66 | 103.08 | 110.50 | 117.92 | 125.34 | 132.76 | 140.18 | 147.60 | 155.03 | 162.45 | 169.87 | 177.29 | 184.71 | 192.13 |
|  | $J$ | 6.60 | 10.22 | 11.91 | 12.72 | 13.13 | 13.34 | 13.45 | 13.51 | 13.55 | 13.57 | 13.59 | 13.60 | 13.60 | 13.61 | 13.61 | 13.61 | 13.61 | 13.61 | 13.61 | 13.61 | 13.61 | 13.61 | 13.61 | 13.61 | 13.61 |
|  | $K$ | 2.59 | 4.38 | 5.30 | 5.70 | 5.86 | 5.92 | 5.95 | 5.96 | 5.96 | 5.96 | 5.96 | 5.96 | 5.96 | 5.96 | 5.96 | 5.96 | 5.96 | 5.96 | 5.96 | 5.96 | 5.96 | 5.96 | 5.96 | 5.96 | 5.96 |
| 0.8 | $U_1$ | 19.18 | 33.46 | 43.26 | 50.10 | 55.05 | 58.78 | 61.66 | 63.95 | 65.81 | 67.35 | 68.64 | 69.74 | 70.69 | 71.52 | 72.24 | 72.88 | 73.44 | 73.95 | 74.41 | 74.82 | 75.20 | 75.54 | 75.86 | 76.15 | 76.41 |
|  | $U_2$ | 0.17 | 0.35 | 0.59 | 0.94 | 1.42 | 2.00 | 2.64 | 3.30 | 3.96 | 4.60 | 5.21 | 5.79 | 6.32 | 6.82 | 7.28 | 7.71 | 8.11 | 8.46 | 8.82 | 9.14 | 9.44 | 9.71 | 9.97 | 10.21 | 10.43 |
|  | $W_{1,2}$ | 5.36 | 11.41 | 18.15 | 25.27 | 32.57 | 39.94 | 47.34 | 54.75 | 62.17 | 69.59 | 77.01 | 84.43 | 91.85 | 99.27 | 106.69 | 114.11 | 121.53 | 128.95 | 136.38 | 143.80 | 151.22 | 158.64 | 166.06 | 173.48 | 180.90 |
|  | $W_{3,4}$ | 9.17 | 18.12 | 26.43 | 34.25 | 41.84 | 49.33 | 56.77 | 64.20 | 71.63 | 79.05 | 86.47 | 93.89 | 101.31 | 108.73 | 116.15 | 123.57 | 130.99 | 138.41 | 145.84 | 153.26 | 160.68 | 168.10 | 175.52 | 182.94 | 190.36 |
|  | $J$ | 6.44 | 10.22 | 12.16 | 13.19 | 13.78 | 14.15 | 14.38 | 14.54 | 14.65 | 14.74 | 14.80 | 14.85 | 14.88 | 14.91 | 14.94 | 14.95 | 14.97 | 14.98 | 14.99 | 15.00 | 15.01 | 15.01 | 15.02 | 15.02 | 15.03 |
|  | $K$ | 2.48 | 4.21 | 5.11 | 5.50 | 5.66 | 5.72 | 5.74 | 5.75 | 5.75 | 5.76 | 5.76 | 5.76 | 5.76 | 5.76 | 5.76 | 5.76 | 5.76 | 5.76 | 5.76 | 5.76 | 5.76 | 5.76 | 5.76 | 5.76 | 5.76 |
| 0.7 | $U_1$ | 17.21 | 30.19 | 39.49 | 46.00 | 50.77 | 54.38 | 57.20 | 59.45 | 61.28 | 62.80 | 64.08 | 65.17 | 66.11 | 66.93 | 67.65 | 68.28 | 68.85 | 69.35 | 69.81 | 70.22 | 70.60 | 70.94 | 71.25 | 71.54 | 71.81 |
|  | $U_2$ | 0.18 | 0.37 | 0.63 | 1.00 | 1.50 | 2.09 | 2.73 | 3.40 | 4.06 | 4.70 | 5.31 | 5.88 | 6.42 | 6.92 | 7.38 | 7.81 | 8.20 | 8.57 | 8.91 | 9.23 | 9.52 | 9.80 | 10.05 | 10.29 | 10.52 |
|  | $W_{1,2}$ | 5.43 | 11.52 | 18.27 | 25.40 | 32.70 | 40.08 | 47.48 | 54.89 | 62.31 | 69.73 | 77.15 | 84.57 | 91.99 | 99.41 | 106.83 | 114.25 | 121.67 | 129.09 | 136.51 | 143.93 | 151.35 | 158.77 | 166.19 | 173.62 | 181.04 |
|  | $W_{3,4}$ | 8.62 | 17.15 | 25.21 | 32.93 | 40.47 | 47.94 | 55.38 | 62.81 | 70.23 | 77.66 | 85.08 | 92.50 | 99.92 | 107.34 | 114.76 | 122.18 | 129.60 | 137.02 | 144.44 | 151.86 | 159.28 | 166.70 | 174.12 | 181.54 | 188.96 |
|  | $J$ | 6.40 | 10.41 | 12.58 | 13.99 | 14.83 | 15.39 | 15.79 | 16.08 | 16.29 | 16.45 | 16.58 | 16.68 | 16.75 | 16.81 | 16.86 | 16.90 | 16.94 | 16.97 | 17.00 | 17.02 | 17.04 | 17.05 | 17.06 | 17.06 | 17.07 |
|  | $K$ | 2.42 | 4.12 | 5.00 | 5.39 | 5.55 | 5.61 | 5.64 | 5.65 | 5.65 | 5.65 | 5.65 | 5.65 | 5.65 | 5.65 | 5.65 | 5.65 | 5.65 | 5.65 | 5.65 | 5.65 | 5.65 | 5.65 | 5.65 | 5.65 | 5.65 |

| $u_0/u_1$ | | 2 | 4 | 6 | 8 | 10 | 12 | 14 | 16 | 18 | 20 | 22 | 24 | 26 | 28 | 30 | 32 | 34 | 36 | 38 | 40 | 42 | 44 | 46 | 48 | 50 |
|---|---|---|---|---|---|---|---|---|---|---|---|---|---|---|---|---|---|---|---|---|---|---|---|---|---|---|
| 0.6 | $U_1$ | 15.41 | 27.37 | 35.91 | 42.07 | 46.63 | 50.12 | 52.86 | 55.05 | 56.85 | 58.34 | 59.60 | 60.68 | 61.61 | 62.42 | 63.13 | 63.76 | 64.32 | 64.82 | 65.28 | 65.69 | 66.06 | 66.40 | 66.71 | 67.00 | 67.27 |
| | $U_2$ | 0.19 | 0.41 | 0.70 | 1.09 | 1.60 | 2.21 | 2.86 | 3.53 | 4.19 | 4.83 | 5.44 | 6.01 | 6.54 | 7.04 | 7.50 | 7.92 | 8.32 | 8.69 | 9.03 | 9.34 | 9.64 | 9.91 | 10.17 | 10.40 | 10.63 |
| | $W_{1,2}$ | 5.42 | 11.50 | 18.24 | 25.36 | 32.66 | 40.03 | 47.43 | 54.85 | 62.27 | 69.69 | 77.11 | 84.53 | 91.95 | 99.37 | 106.79 | 114.21 | 121.63 | 129.05 | 136.47 | 143.89 | 151.31 | 158.73 | 166.15 | 173.57 | 180.99 |
| | $W_{3,4}$ | 8.23 | 16.42 | 24.31 | 31.94 | 39.45 | 46.91 | 54.34 | 61.77 | 69.19 | 76.61 | 84.04 | 91.46 | 98.88 | 106.30 | 113.72 | 121.14 | 128.56 | 135.98 | 143.40 | 150.82 | 158.24 | 165.66 | 173.08 | 180.50 | 187.92 |
| | $J$ | 6.42 | 10.68 | 13.25 | 14.86 | 15.95 | 16.70 | 17.25 | 17.66 | 17.97 | 18.21 | 18.39 | 18.54 | 18.66 | 18.75 | 18.82 | 18.89 | 18.94 | 18.98 | 19.01 | 19.04 | 19.07 | 19.09 | 19.11 | 19.12 | 19.13 |
| | $K$ | 2.41 | 4.10 | 4.99 | 5.38 | 5.53 | 5.60 | 5.62 | 5.64 | 5.64 | 5.64 | 5.64 | 5.64 | 5.64 | 5.64 | 5.64 | 5.64 | 5.64 | 5.64 | 5.64 | 5.64 | 5.64 | 5.64 | 5.64 | 5.64 | 5.64 |
| 0.5 | $U_1$ | 14.25 | 25.48 | 33.65 | 39.60 | 44.04 | 47.46 | 50.16 | 52.32 | 54.10 | 55.59 | 56.84 | 57.91 | 58.83 | 59.64 | 60.35 | 60.98 | 61.53 | 62.04 | 62.49 | 62.90 | 63.27 | 63.61 | 63.92 | 64.21 | 64.48 |
| | $U_2$ | 0.20 | 0.42 | 0.71 | 1.12 | 1.64 | 2.25 | 2.91 | 3.58 | 4.24 | 4.88 | 5.49 | 6.06 | 6.59 | 7.09 | 7.55 | 7.97 | 8.37 | 8.73 | 9.06 | 9.39 | 9.68 | 9.96 | 10.21 | 10.45 | 10.67 |
| | $W_{1,2}$ | 5.32 | 11.33 | 18.03 | 25.14 | 32.43 | 39.80 | 47.20 | 54.61 | 62.03 | 69.45 | 76.87 | 84.29 | 91.71 | 99.13 | 106.55 | 113.97 | 121.39 | 128.81 | 136.23 | 143.65 | 151.07 | 158.49 | 165.92 | 173.34 | 180.76 |
| | $W_{3,4}$ | 7.94 | 15.91 | 23.66 | 31.24 | 38.73 | 46.18 | 53.61 | 61.03 | 68.45 | 75.87 | 83.29 | 90.72 | 98.14 | 105.56 | 112.98 | 120.40 | 127.82 | 135.24 | 142.66 | 150.08 | 157.50 | 164.92 | 172.34 | 179.76 | 187.18 |
| | $J$ | 6.40 | 10.95 | 13.76 | 15.59 | 16.84 | 17.73 | 18.37 | 18.85 | 19.21 | 19.49 | 19.71 | 19.88 | 20.01 | 20.12 | 20.20 | 20.27 | 20.33 | 20.38 | 20.42 | 20.45 | 20.48 | 20.50 | 20.52 | 20.54 | 20.56 |
| | $K$ | 2.43 | 4.14 | 5.04 | 5.43 | 5.60 | 5.66 | 5.70 | 5.70 | 5.70 | 5.70 | 5.70 | 5.70 | 5.70 | 5.70 | 5.70 | 5.70 | 5.70 | 5.70 | 5.70 | 5.70 | 5.70 | 5.70 | 5.70 | 5.70 | 5.70 |
| 0.4 | $U_1$ | 13.20 | 23.74 | 31.51 | 37.23 | 41.55 | 44.88 | 47.52 | 49.66 | 51.44 | 52.87 | 54.11 | 55.17 | 56.09 | 56.89 | 57.60 | 58.22 | 58.78 | 59.28 | 59.73 | 60.13 | 60.51 | 60.84 | 61.16 | 61.44 | 61.71 |
| | $U_2$ | 0.21 | 0.44 | 0.75 | 1.17 | 1.71 | 2.32 | 2.98 | 3.66 | 4.32 | 4.96 | 5.56 | 6.13 | 6.66 | 7.16 | 7.62 | 8.04 | 8.43 | 8.80 | 9.14 | 9.45 | 9.75 | 10.02 | 10.27 | 10.51 | 10.74 |
| | $W_{1,2}$ | 5.19 | 11.11 | 17.76 | 24.85 | 32.13 | 39.50 | 46.90 | 54.31 | 61.73 | 69.15 | 76.57 | 83.99 | 91.41 | 98.83 | 106.25 | 113.67 | 121.09 | 128.51 | 135.93 | 143.35 | 150.77 | 158.19 | 165.62 | 173.04 | 180.46 |
| | $W_{3,4}$ | 7.74 | 15.54 | 23.17 | 30.70 | 38.21 | 45.65 | 53.08 | 60.50 | 67.92 | 75.34 | 82.76 | 90.19 | 97.61 | 105.03 | 112.45 | 119.87 | 127.29 | 134.71 | 142.13 | 149.55 | 156.97 | 164.39 | 171.81 | 179.23 | 186.65 |
| | $J$ | 6.58 | 11.26 | 14.31 | 16.35 | 17.77 | 18.80 | 19.54 | 20.11 | 20.54 | 20.86 | 21.12 | 21.32 | 21.48 | 21.61 | 21.71 | 21.79 | 21.86 | 21.92 | 21.97 | 22.01 | 22.04 | 22.07 | 22.09 | 22.11 | 22.13 |
| | $K$ | 2.48 | 4.23 | 5.15 | 5.55 | 5.72 | 5.78 | 5.81 | 5.82 | 5.83 | 5.83 | 5.83 | 5.83 | 5.83 | 5.83 | 5.83 | 5.83 | 5.83 | 5.83 | 5.83 | 5.83 | 5.83 | 5.83 | 5.83 | 5.83 | 5.83 |
| 0.3 | $U_1$ | 12.23 | 22.17 | 29.63 | 35.16 | 39.38 | 42.97 | 45.56 | 47.67 | 49.40 | 50.85 | 52.08 | 53.13 | 54.04 | 54.84 | 55.54 | 56.16 | 56.72 | 57.21 | 57.66 | 58.07 | 58.44 | 58.78 | 59.09 | 59.37 | 59.64 |
| | $U_2$ | 0.20 | 0.43 | 0.74 | 1.21 | 1.76 | 2.38 | 3.04 | 3.72 | 4.38 | 5.02 | 5.62 | 6.15 | 6.68 | 7.17 | 7.63 | 8.06 | 8.45 | 8.81 | 9.15 | 9.47 | 9.76 | 10.03 | 10.29 | 10.53 | 10.78 |
| | $W_{1,2}$ | 5.04 | 10.86 | 17.46 | 24.52 | 31.80 | 39.16 | 46.56 | 53.97 | 61.39 | 68.81 | 76.23 | 83.65 | 91.07 | 98.49 | 105.91 | 113.33 | 120.75 | 128.17 | 135.59 | 143.01 | 150.43 | 157.85 | 165.27 | 172.69 | 180.11 |
| | $W_{3,4}$ | 7.61 | 15.29 | 22.89 | 30.40 | 37.86 | 45.29 | 52.72 | 60.14 | 67.33 | 74.98 | 82.40 | 89.82 | 97.24 | 104.66 | 112.08 | 119.51 | 126.93 | 134.35 | 141.77 | 149.19 | 156.61 | 164.03 | 171.45 | 178.87 | 186.29 |
| | $J$ | 6.69 | 11.51 | 14.81 | 17.01 | 18.61 | 19.73 | 20.54 | 21.14 | 21.58 | 21.90 | 22.16 | 22.36 | 22.52 | 22.65 | 22.75 | 22.84 | 22.91 | 22.97 | 23.01 | 23.05 | 23.08 | 23.11 | 23.13 | 23.15 | 23.17 |
| | $K$ | 2.54 | 4.35 | 5.29 | 5.71 | 5.88 | 5.95 | 5.98 | 5.99 | 5.99 | 5.99 | 5.99 | 5.99 | 5.99 | 5.99 | 5.99 | 5.99 | 5.99 | 5.99 | 5.99 | 5.99 | 5.99 | 5.99 | 5.99 | 5.99 | 5.99 |
| 0.2 | $U_1$ | 12.01 | 21.81 | 29.17 | 34.68 | 38.88 | 42.14 | 44.74 | 46.84 | 48.58 | 50.03 | 51.26 | 52.32 | 53.23 | 54.02 | 54.73 | 55.35 | 55.90 | 56.40 | 56.85 | 57.26 | 57.63 | 57.97 | 58.28 | 58.56 | 58.83 |
| | $U_2$ | 0.20 | 0.43 | 0.74 | 1.17 | 1.71 | 2.34 | 3.00 | 3.67 | 4.34 | 4.98 | 5.58 | 6.15 | 6.68 | 7.18 | 7.63 | 8.06 | 8.45 | 8.82 | 9.16 | 9.47 | 9.76 | 10.04 | 10.29 | 10.53 | 10.75 |
| | $W_{1,2}$ | 4.87 | 10.56 | 17.10 | 24.13 | 31.40 | 38.76 | 46.15 | 53.56 | 60.98 | 68.40 | 75.82 | 83.24 | 90.66 | 98.08 | 105.50 | 112.92 | 120.34 | 127.76 | 135.18 | 142.60 | 150.02 | 157.44 | 164.86 | 172.28 | 179.71 |
| | $W_{3,4}$ | 7.52 | 15.13 | 22.69 | 30.11 | 37.63 | 45.06 | 52.49 | 59.91 | 67.33 | 74.75 | 82.17 | 89.59 | 97.01 | 104.43 | 111.85 | 119.27 | 126.69 | 134.11 | 141.53 | 148.96 | 156.38 | 163.80 | 171.22 | 178.64 | 186.06 |
| | $J$ | 6.78 | 11.76 | 15.12 | 17.42 | 19.04 | 20.20 | 21.06 | 21.70 | 22.18 | 22.55 | 22.84 | 23.06 | 23.23 | 23.38 | 23.50 | 23.59 | 23.67 | 23.73 | 23.78 | 23.82 | 23.86 | 23.89 | 23.92 | 23.94 | 23.96 |
| | $K$ | 2.61 | 4.47 | 5.44 | 5.87 | 6.04 | 6.11 | 6.14 | 6.15 | 6.16 | 6.16 | 6.16 | 6.16 | 6.16 | 6.16 | 6.16 | 6.16 | 6.16 | 6.16 | 6.16 | 6.16 | 6.16 | 6.16 | 6.16 | 6.16 | 6.16 |
| 0.1 | $U_1$ | 12.06 | 22.01 | 29.46 | 35.00 | 39.25 | 42.83 | 45.42 | 47.52 | 49.26 | 50.71 | 51.95 | 53.01 | 53.92 | 54.72 | 55.43 | 56.05 | 56.60 | 57.11 | 57.56 | 57.97 | 58.34 | 58.68 | 58.99 | 59.27 | 59.54 |
| | $U_2$ | 0.20 | 0.43 | 0.74 | 1.17 | 1.71 | 2.34 | 3.00 | 3.67 | 4.34 | 4.98 | 5.58 | 6.15 | 6.68 | 7.18 | 7.63 | 8.06 | 8.45 | 8.82 | 9.16 | 9.47 | 9.76 | 10.04 | 10.29 | 10.53 | 10.75 |
| | $W_{1,2}$ | 4.75 | 10.37 | 16.95 | 23.98 | 31.24 | 38.60 | 46.00 | 53.41 | 60.83 | 68.25 | 75.67 | 83.09 | 90.51 | 97.93 | 105.35 | 112.77 | 120.19 | 127.61 | 135.03 | 142.45 | 149.87 | 157.29 | 164.71 | 172.13 | 179.55 |
| | $W_{3,4}$ | 7.47 | 15.04 | 22.58 | 30.05 | 37.50 | 44.93 | 52.36 | 59.78 | 67.20 | 74.62 | 82.04 | 89.46 | 96.88 | 104.30 | 111.72 | 119.14 | 126.56 | 133.98 | 141.40 | 148.82 | 156.24 | 163.66 | 171.08 | 178.51 | 185.93 |
| | $J$ | 6.86 | 11.94 | 15.38 | 17.73 | 19.36 | 20.53 | 21.39 | 22.03 | 22.51 | 22.87 | 23.15 | 23.36 | 23.53 | 23.66 | 23.76 | 23.85 | 23.92 | 23.97 | 24.01 | 24.05 | 24.08 | 24.11 | 24.14 | 24.16 | 24.18 |
| | $K$ | 2.67 | 4.56 | 5.55 | 5.99 | 6.17 | 6.24 | 6.27 | 6.28 | 6.28 | 6.28 | 6.28 | 6.28 | 6.28 | 6.28 | 6.28 | 6.28 | 6.28 | 6.28 | 6.28 | 6.28 | 6.28 | 6.28 | 6.28 | 6.28 | 6.28 |
| 0.0 | $U_1$ | 12.26 | 22.50 | 29.77 | 35.39 | 39.65 | 42.97 | 45.60 | 47.73 | 49.48 | 50.94 | 52.18 | 53.25 | 54.17 | 54.97 | 55.67 | 56.30 | 56.85 | 57.35 | 57.81 | 58.21 | 58.58 | 58.93 | 59.24 | 59.52 | 59.79 |
| | $U_2$ | 0.18 | 0.39 | 0.69 | 1.10 | 1.64 | 2.26 | 2.92 | 3.60 | 4.26 | 4.90 | 5.51 | 6.08 | 6.61 | 7.10 | 7.56 | 7.99 | 8.38 | 8.75 | 9.09 | 9.40 | 9.70 | 9.97 | 10.22 | 10.46 | 10.69 |
| | $W_{1,2}$ | 4.67 | 10.37 | 16.86 | 23.88 | 31.14 | 38.49 | 45.89 | 53.30 | 60.72 | 68.14 | 75.56 | 82.98 | 90.40 | 97.82 | 105.24 | 112.66 | 120.08 | 127.50 | 134.92 | 142.34 | 149.76 | 157.18 | 164.60 | 172.02 | 179.44 |
| | $W_{3,4}$ | 7.45 | 15.01 | 22.54 | 30.01 | 37.46 | 44.88 | 52.31 | 59.73 | 67.15 | 74.57 | 81.99 | 89.42 | 96.84 | 104.26 | 111.68 | 119.10 | 126.52 | 133.94 | 141.36 | 148.78 | 156.20 | 163.62 | 171.04 | 178.46 | 185.88 |
| | $J$ | 6.88 | 11.97 | 15.42 | 17.77 | 19.41 | 20.58 | 21.44 | 22.07 | 22.55 | 22.90 | 23.18 | 23.39 | 23.56 | 23.69 | 23.80 | 23.88 | 23.95 | 24.01 | 24.05 | 24.09 | 24.12 | 24.15 | 24.17 | 24.19 | 24.21 |
| | $K$ | 2.69 | 4.59 | 5.59 | 6.03 | 6.21 | 6.28 | 6.31 | 6.32 | 6.33 | 6.33 | 6.33 | 6.33 | 6.33 | 6.33 | 6.33 | 6.33 | 6.33 | 6.33 | 6.33 | 6.33 | 6.33 | 6.33 | 6.33 | 6.33 | 6.33 |

TABLE S27. Parameters of the THF interaction Hamiltonian for different ratios $u_0/u_1$ and screening lengths $\xi$ of the double-gate interaction potential ($U_\xi = 24$ meV for $\xi = 10$nm). We use the same BM model parameters as in Table S26.

FIG. S30. Band structures of the BM and THF models near charge neutrality for $w_0/w_1 = 0.8$, depicted by lines and crosses, respectively. The BM bands are colored according to the weight of the $f$-electron wave function on them. We use the same BM parameters as in Table S28.

TABLE S28. Parameters of the $f$-electron Wannier functions for different values of the tunneling ratio $w_0/w_1$. $\mathcal{W}$ denotes the total weight of the THF $f$-electrons on the active bands. In computing the ratios $\gamma/U_1$ and $\gamma/U_2$, we employ the on-site and nearest-neighbor repulsion parameters $U_1$ and $U_2$ obtained numerically for $\xi = 10$ nm, as given in Table S29. We employ $v_F = 5.944$ eV·Å, $|\mathbf{K}| = 1.703\,\text{Å}^{-1}$, $w_1 = 110$ meV, and $\theta = 0.94°$ for the BM model.

| $w_0/w_1$ | $\alpha_1$ | $\alpha_2$ | $\lambda_1$ | $\lambda_2$ | $\lambda$ | $\mathcal{W}$ | $t_0$ | $\gamma$ | $M$ | $v_\star$ | $v'_\star$ | $v''_\star$ | $\gamma/U_1$ | $\gamma/U_2$ |
|---|---|---|---|---|---|---|---|---|---|---|---|---|---|---|
| 1.00 | 0.751 | 0.660 | 0.143 | 0.161 | 0.403 | 0.967 | 0.308 | 15.882 | 14.384 | -3.138 | 1.233 | 0.0581 | 0.27 | 8.25 |
| 0.90 | 0.772 | 0.636 | 0.155 | 0.167 | 0.339 | 0.981 | 0.379 | 3.479 | 14.816 | -3.527 | 1.389 | -0.0022 | 0.06 | 2.23 |
| 0.80 | 0.801 | 0.599 | 0.167 | 0.179 | 0.328 | 0.974 | 0.631 | -0.847 | 15.118 | -3.837 | 1.484 | -0.0385 | -0.18 | -6.50 |
| 0.70 | 0.834 | 0.551 | 0.179 | 0.185 | 0.340 | 0.954 | 0.959 | -23.46 | 15.322 | -4.086 | 1.525 | -0.0565 | -0.46 | -14.64 |
| 0.60 | 0.870 | 0.494 | 0.191 | 0.191 | 0.355 | 0.938 | 1.286 | -36.821 | 15.457 | -4.291 | 1.511 | -0.0609 | -0.78 | -21.52 |
| 0.50 | 0.904 | 0.428 | 0.209 | 0.197 | 0.363 | 0.922 | 1.559 | -49.467 | 15.543 | -4.463 | 1.439 | -0.0556 | -1.12 | -27.91 |
| 0.40 | 0.935 | 0.354 | 0.221 | 0.203 | 0.377 | 0.914 | 1.847 | -60.919 | 15.595 | -4.608 | 1.299 | -0.0439 | -1.46 | -32.78 |
| 0.30 | 0.962 | 0.274 | 0.239 | 0.209 | 0.377 | 0.903 | 2.024 | -70.671 | 15.627 | -4.727 | 1.082 | -0.0291 | -1.75 | -38.03 |
| 0.20 | 0.982 | 0.188 | 0.245 | 0.209 | 0.374 | 0.891 | 2.206 | -78.189 | 15.643 | -4.819 | 0.784 | -0.0146 | -1.95 | -42.85 |
| 0.10 | 0.995 | 0.097 | 0.257 | 0.215 | 0.372 | 0.885 | 2.325 | -82.961 | 15.652 | -4.878 | 0.414 | -0.0039 | -2.08 | -45.92 |
| 0.00 | 1.000 | 0.000 | 0.257 | 0.000 | 0.371 | 0.882 | 2.406 | -84.601 | 15.654 | -4.899 | -0.0000 | -0.0000 | -2.12 | -47.27 |

TABLE S29. Parameters of the $f$-electron Wannier functions and the THF single-particle Hamiltonian for different values of the tunneling ratio $w_0/w_1$. (Parameters $U_1$, $U_2$, $W_{1,2}$, $W_{3,4}$, $J$, and $K$ in meV for on-site and nearest-neighbor repulsion parameters $U_1$ and $U_2$ obtained numerically for $\xi = 10$ nm, as given in Table S29.)

| $w_0/w_1$ | $\xi$/nm | 2 | 4 | 6 | 8 | 10 | 12 | 14 | 16 | 18 | 20 | 22 | 24 | 26 | 28 | 30 | 32 | 34 | 36 | 38 | 40 | 42 | 44 | 46 | 48 | 50 |
|---|---|---|---|---|---|---|---|---|---|---|---|---|---|---|---|---|---|---|---|---|---|---|---|---|---|---|
| 1.0 | $U_1$ | 21.31 | 36.42 | 46.36 | 53.11 | 57.92 | 61.51 | 64.28 | 66.48 | 68.27 | 69.75 | 71.00 | 72.06 | 72.98 | 73.78 | 74.49 | 75.11 | 75.66 | 76.16 | 76.61 | 77.01 | 77.38 | 77.72 | 78.03 | 78.32 | 78.58 |
| | $U_2$ | 0.32 | 0.64 | 0.97 | 1.39 | 1.92 | 2.54 | 3.21 | 3.89 | 4.57 | 5.22 | 5.83 | 6.41 | 6.95 | 7.45 | 7.91 | 8.34 | 8.74 | 9.10 | 9.45 | 9.76 | 10.06 | 10.33 | 10.59 | 10.83 | 11.05 |
| | $W_{1,2}$ | 5.28 | 11.40 | 18.45 | 25.88 | 33.50 | 41.19 | 48.92 | 56.66 | 64.41 | 72.15 | 79.90 | 87.64 | 95.39 | 103.14 | 110.88 | 118.63 | 126.38 | 134.12 | 141.87 | 149.62 | 157.36 | 165.11 | 172.86 | 180.60 | 188.35 |
| | $W_{3,4}$ | 10.82 | 21.08 | 30.34 | 39.35 | 48.54 | 57.99 | 67.67 | 77.67 | 85.42 | 93.17 | 100.92 | 108.67 | 116.41 | 124.16 | 131.90 | 139.65 | 147.40 | 155.14 | 162.89 | 170.64 | 178.38 | 186.13 | 193.88 | 201.62 | 201.62 |
| | $J$ | 0.97 | 10.63 | 12.28 | 13.06 | 13.45 | 13.65 | 13.78 | 13.86 | 13.91 | 13.95 | 13.98 | 14.00 | 14.02 | 14.04 | 14.05 | 14.07 | 14.08 | 14.09 | 14.09 | 14.10 | 14.10 | 14.11 | 14.11 | 14.12 | 14.12 |
| | $K$ | 2.72 | 4.55 | 5.46 | 5.84 | 6.05 | 6.07 | 6.08 | 6.09 | 6.09 | 6.09 | 6.09 | 6.09 | 6.09 | 6.09 | 6.09 | 6.09 | 6.09 | 6.09 | 6.09 | 6.09 | 6.09 | 6.09 | 6.09 | 6.09 | 6.09 |
| 0.9 | $U_1$ | 21.08 | 34.66 | 43.72 | 50.61 | 55.00 | 59.34 | 62.24 | 64.55 | 66.42 | 67.96 | 69.26 | 70.37 | 71.32 | 72.14 | 72.87 | 73.51 | 74.08 | 74.59 | 75.05 | 75.46 | 75.84 | 76.18 | 76.50 | 76.79 | 77.05 |
| | $U_2$ | 0.20 | 0.41 | 0.67 | 1.00 | 1.56 | 2.17 | 2.84 | 3.53 | 4.21 | 4.87 | 5.50 | 6.09 | 6.63 | 7.14 | 7.61 | 8.05 | 8.45 | 8.83 | 9.17 | 9.49 | 9.79 | 10.07 | 10.33 | 10.56 | 10.79 |
| | $W_{1,2}$ | 5.11 | 11.81 | 18.84 | 26.29 | 33.92 | 41.62 | 49.35 | 57.09 | 64.83 | 72.58 | 80.32 | 88.07 | 95.82 | 103.56 | 111.31 | 119.06 | 126.80 | 134.55 | 142.30 | 150.04 | 157.79 | 165.54 | 173.28 | 181.03 | 188.78 |
| | $W_{3,4}$ | 9.99 | 19.64 | 28.45 | 36.66 | 44.60 | 52.42 | 60.20 | 67.95 | 75.70 | 83.45 | 91.20 | 98.95 | 106.69 | 114.44 | 122.19 | 129.93 | 137.68 | 145.43 | 153.17 | 160.92 | 168.67 | 176.41 | 184.16 | 191.91 | 199.65 |
| | $J$ | 6.68 | 10.34 | 12.06 | 12.89 | 13.31 | 13.53 | 13.65 | 13.72 | 13.76 | 13.79 | 13.81 | 13.82 | 13.83 | 13.84 | 13.84 | 13.84 | 13.85 | 13.85 | 13.85 | 13.85 | 13.85 | 13.85 | 13.85 | 13.85 | 13.85 |
| | $K$ | 2.56 | 4.31 | 5.19 | 5.46 | 5.71 | 5.77 | 5.79 | 5.80 | 5.80 | 5.80 | 5.80 | 5.80 | 5.80 | 5.80 | 5.80 | 5.80 | 5.80 | 5.80 | 5.80 | 5.80 | 5.80 | 5.80 | 5.80 | 5.80 | 5.80 |
| 0.8 | $U_1$ | 19.41 | 33.83 | 43.72 | 50.61 | 56.07 | 60.77 | 64.64 | 67.65 | 70.11 | 72.12 | 73.80 | 75.22 | 76.44 | 77.49 | 78.39 | 79.18 | 79.87 | 80.48 | 81.02 | 81.51 | 81.95 | 82.36 | 82.72 | 83.06 | 83.37 |
| | $U_2$ | 0.17 | 0.36 | 0.62 | 1.00 | 1.51 | 2.13 | 2.81 | 3.50 | 4.19 | 4.85 | 5.48 | 6.07 | 6.62 | 7.13 | 7.60 | 8.04 | 8.44 | 8.81 | 9.15 | 9.48 | 9.78 | 10.06 | 10.32 | 10.56 | 10.79 |
| | $W_{1,2}$ | 5.66 | 12.05 | 19.12 | 26.58 | 34.22 | 41.92 | 49.65 | 57.39 | 65.13 | 72.88 | 80.62 | 88.37 | 96.12 | 103.86 | 111.61 | 119.36 | 127.10 | 134.85 | 142.60 | 150.34 | 158.09 | 165.84 | 173.58 | 181.33 | 189.08 |
| | $W_{3,4}$ | 9.31 | 18.45 | 26.98 | 35.07 | 42.96 | 50.76 | 58.53 | 66.29 | 74.03 | 81.78 | 89.53 | 97.28 | 105.02 | 112.77 | 120.52 | 128.26 | 136.01 | 143.76 | 151.50 | 159.25 | 167.00 | 174.74 | 182.49 | 190.24 | 197.98 |
| | $J$ | 6.57 | 10.42 | 12.42 | 13.50 | 14.13 | 14.52 | 14.78 | 14.96 | 15.08 | 15.18 | 15.25 | 15.30 | 15.34 | 15.38 | 15.40 | 15.42 | 15.44 | 15.46 | 15.47 | 15.48 | 15.49 | 15.50 | 15.50 | 15.50 | 15.51 |
| | $K$ | 2.46 | 4.16 | 5.01 | 5.38 | 5.53 | 5.58 | 5.61 | 5.62 | 5.62 | 5.62 | 5.62 | 5.62 | 5.62 | 5.62 | 5.62 | 5.62 | 5.62 | 5.62 | 5.62 | 5.62 | 5.62 | 5.62 | 5.62 | 5.62 | 5.62 |
| 0.7 | $U_1$ | 17.41 | 30.75 | 40.35 | 45.23 | 54.86 | 57.69 | 59.95 | 61.79 | 63.31 | 64.60 | 65.69 | 66.63 | 67.45 | 68.17 | 68.81 | 69.37 | 69.88 | 70.34 | 70.75 | 71.13 | 71.47 | 71.78 | 72.07 | 72.07 | 72.34 |
| | $U_2$ | 0.18 | 0.39 | 0.67 | 1.07 | 1.60 | 2.23 | 2.91 | 3.61 | 4.30 | 4.96 | 5.59 | 6.17 | 6.72 | 7.23 | 7.70 | 8.14 | 8.54 | 8.91 | 9.26 | 9.58 | 9.88 | 10.15 | 10.41 | 10.65 | 10.88 |
| | $W_{1,2}$ | 5.73 | 12.14 | 19.23 | 26.60 | 34.32 | 42.03 | 49.70 | 57.23 | 64.98 | 72.73 | 80.48 | 88.23 | 95.97 | 103.72 | 111.72 | 119.47 | 127.21 | 134.71 | 142.45 | 150.20 | 157.95 | 165.69 | 173.44 | 181.19 | 188.93 |
| | $W_{3,4}$ | 8.80 | 17.52 | 25.83 | 33.83 | 41.68 | 49.47 | 57.23 | 64.98 | 72.73 | 80.48 | 88.23 | 95.97 | 103.72 | 111.47 | 119.21 | 126.96 | 134.71 | 142.45 | 150.20 | 157.95 | 165.69 | 173.44 | 181.19 | 188.93 | 196.68 |
| | $J$ | 6.55 | 12.09 | 14.37 | 15.25 | 16.27 | 15.85 | 16.27 | 16.58 | 16.80 | 16.98 | 17.11 | 17.21 | 17.30 | 17.36 | 17.42 | 17.49 | 17.52 | 17.57 | 17.58 | 17.60 | 17.62 | 17.63 | 17.63 | 17.63 | 17.63 |
| | $K$ | 2.41 | 4.08 | 4.96 | 5.30 | 5.44 | 5.50 | 5.52 | 5.53 | 5.53 | 5.53 | 5.53 | 5.53 | 5.53 | 5.53 | 5.53 | 5.53 | 5.53 | 5.53 | 5.53 | 5.53 | 5.53 | 5.53 | 5.53 | 5.53 | 5.53 |

Total weight of the heavy-fermions on the active bands after wannierisation: 97.4%

Wannier orbitals support: 0.0 – 1.0

| $u_0/u_1$ | $\xi$/nm | 2 | 4 | 6 | 8 | 10 | 12 | 14 | 16 | 18 | 20 | 22 | 24 | 26 | 28 | 30 | 32 | 34 | 36 | 38 | 40 | 42 | 44 | 46 | 48 | 50 |
|---|---|---|---|---|---|---|---|---|---|---|---|---|---|---|---|---|---|---|---|---|---|---|---|---|---|---|
| 0.6 | $U_1$ | 15.63 | 27.73 | 36.37 | 42.58 | 47.18 | 50.69 | 53.44 | 55.64 | 57.45 | 58.95 | 60.21 | 61.29 | 62.23 | 63.04 | 63.75 | 64.38 | 64.95 | 65.45 | 65.90 | 66.31 | 66.69 | 67.03 | 67.34 | 67.63 | 67.90 |
| | $U_2$ | 0.20 | 0.43 | 0.74 | 1.16 | 1.71 | 2.35 | 3.04 | 3.74 | 4.42 | 5.09 | 5.71 | 6.30 | 6.84 | 7.35 | 7.82 | 8.25 | 8.65 | 9.02 | 9.37 | 9.69 | 9.99 | 10.26 | 10.52 | 10.76 | 10.99 |
| | $W_{1,2}$ | 5.70 | 12.09 | 19.16 | 26.62 | 34.26 | 41.96 | 49.69 | 57.43 | 65.17 | 72.92 | 80.66 | 88.41 | 96.16 | 103.90 | 111.65 | 119.40 | 127.14 | 134.89 | 142.64 | 150.38 | 158.13 | 165.88 | 173.62 | 181.37 | 189.12 |
| | $W_{3,4}$ | 8.42 | 16.84 | 24.99 | 32.91 | 40.74 | 48.51 | 56.27 | 64.02 | 71.77 | 79.51 | 87.26 | 95.00 | 102.75 | 110.50 | 118.25 | 125.99 | 133.74 | 141.49 | 149.23 | 156.98 | 164.73 | 172.47 | 180.22 | 187.97 | 195.71 |
| | $J$ | 6.59 | 10.96 | 13.61 | 15.29 | 16.41 | 17.20 | 17.77 | 18.19 | 18.51 | 18.76 | 18.95 | 19.10 | 19.22 | 19.31 | 19.39 | 19.46 | 19.51 | 19.55 | 19.59 | 19.61 | 19.64 | 19.66 | 19.68 | 19.69 | 19.71 |
| | $K$ | 2.40 | 4.07 | 4.93 | 5.29 | 5.44 | 5.50 | 5.52 | 5.53 | 5.53 | 5.53 | 5.53 | 5.53 | 5.53 | 5.53 | 5.53 | 5.53 | 5.53 | 5.53 | 5.53 | 5.53 | 5.53 | 5.53 | 5.53 | 5.53 | 5.53 |
| 0.5 | $U_1$ | 14.39 | 25.70 | 33.90 | 39.88 | 44.34 | 47.77 | 50.47 | 52.64 | 54.43 | 55.91 | 57.16 | 58.24 | 59.16 | 59.97 | 60.68 | 61.31 | 61.87 | 62.37 | 62.82 | 63.23 | 63.60 | 63.95 | 64.26 | 64.55 | 64.81 |
| | $U_2$ | 0.21 | 0.45 | 0.77 | 1.21 | 1.77 | 2.42 | 3.11 | 3.81 | 4.50 | 5.16 | 5.78 | 6.37 | 6.91 | 7.42 | 7.89 | 8.32 | 8.72 | 9.09 | 9.44 | 9.75 | 10.05 | 10.33 | 10.59 | 10.82 | 11.05 |
| | $W_{1,2}$ | 5.60 | 11.92 | 18.95 | 26.40 | 34.02 | 41.72 | 49.45 | 57.19 | 64.94 | 72.68 | 80.43 | 88.17 | 95.92 | 103.67 | 111.41 | 119.16 | 126.91 | 134.65 | 142.40 | 150.15 | 157.89 | 165.64 | 173.39 | 181.13 | 188.88 |
| | $W_{3,4}$ | 8.15 | 16.36 | 24.38 | 32.26 | 40.06 | 47.83 | 55.58 | 63.33 | 71.08 | 78.83 | 86.57 | 94.32 | 102.07 | 109.81 | 117.56 | 125.31 | 133.05 | 140.80 | 148.55 | 156.29 | 164.04 | 171.79 | 179.53 | 187.28 | 195.03 |
| | $J$ | 6.67 | 11.28 | 14.19 | 16.10 | 17.41 | 18.34 | 19.02 | 19.53 | 19.91 | 20.20 | 20.43 | 20.61 | 20.75 | 20.87 | 20.96 | 21.04 | 21.10 | 21.15 | 21.19 | 21.23 | 21.26 | 21.28 | 21.30 | 21.32 | 21.34 |
| | $K$ | 2.43 | 4.13 | 4.99 | 5.37 | 5.52 | 5.57 | 5.60 | 5.61 | 5.61 | 5.61 | 5.61 | 5.61 | 5.61 | 5.61 | 5.61 | 5.61 | 5.61 | 5.61 | 5.61 | 5.61 | 5.61 | 5.61 | 5.61 | 5.61 | 5.61 |
| 0.4 | $U_1$ | 13.24 | 23.80 | 31.57 | 37.29 | 41.60 | 44.93 | 47.57 | 49.70 | 51.45 | 52.91 | 54.15 | 55.21 | 56.13 | 56.93 | 57.63 | 58.26 | 58.81 | 59.31 | 59.76 | 60.17 | 60.54 | 60.88 | 61.19 | 61.48 | 61.74 |
| | $U_2$ | 0.23 | 0.49 | 0.83 | 1.28 | 1.86 | 2.51 | 3.21 | 3.91 | 4.60 | 5.26 | 5.88 | 6.46 | 7.00 | 7.51 | 7.97 | 8.40 | 8.80 | 9.17 | 9.52 | 9.83 | 10.13 | 10.41 | 10.66 | 10.90 | 11.13 |
| | $W_{1,2}$ | 5.47 | 11.71 | 18.70 | 26.12 | 33.74 | 41.44 | 49.17 | 56.91 | 64.65 | 72.40 | 80.14 | 87.89 | 95.63 | 103.38 | 111.13 | 118.87 | 126.62 | 134.37 | 142.11 | 149.86 | 157.61 | 165.35 | 173.10 | 180.85 | 188.59 |
| | $W_{3,4}$ | 7.97 | 16.02 | 23.96 | 31.79 | 39.58 | 47.34 | 55.08 | 62.84 | 70.59 | 78.34 | 86.08 | 93.83 | 101.58 | 109.32 | 117.07 | 124.82 | 132.56 | 140.31 | 148.06 | 155.80 | 163.55 | 171.29 | 179.04 | 186.79 | 194.53 |
| | $J$ | 6.79 | 11.62 | 14.80 | 16.89 | 18.30 | 19.30 | 20.03 | 20.57 | 20.98 | 21.29 | 21.53 | 21.72 | 21.87 | 21.99 | 22.09 | 22.17 | 22.23 | 22.29 | 22.33 | 22.37 | 22.40 | 22.43 | 23.07 | 23.12 | 23.14 |
| | $K$ | 2.56 | 4.35 | 5.26 | 5.66 | 5.71 | 5.73 | 5.74 | 5.74 | 5.74 | 5.74 | 5.74 | 5.74 | 5.74 | 5.74 | 5.74 | 5.74 | 5.74 | 5.74 | 5.74 | 5.74 | 5.74 | 5.74 | 5.74 | 5.74 | 5.74 |
| 0.3 | $U_1$ | 12.50 | 22.62 | 30.18 | 35.80 | 40.07 | 43.38 | 46.00 | 48.13 | 49.88 | 51.34 | 52.58 | 53.65 | 54.56 | 55.36 | 56.06 | 56.69 | 57.24 | 57.74 | 58.19 | 58.60 | 58.97 | 59.31 | 59.63 | 60.28 | 60.55 |
| | $U_2$ | 0.22 | 0.45 | 0.79 | 1.25 | 1.82 | 2.48 | 3.18 | 3.88 | 4.57 | 5.23 | 5.86 | 6.44 | 6.98 | 7.49 | 7.95 | 8.38 | 8.78 | 9.15 | 9.50 | 9.81 | 10.11 | 10.39 | 10.64 | 10.88 | 11.11 |
| | $W_{1,2}$ | 5.29 | 11.11 | 17.98 | 25.35 | 32.95 | 40.64 | 48.36 | 56.09 | 63.84 | 71.59 | 79.34 | 87.08 | 94.83 | 102.58 | 110.32 | 118.07 | 125.82 | 133.57 | 141.31 | 149.06 | 156.80 | 164.55 | 172.30 | 180.04 | 187.79 |
| | $W_{3,4}$ | 7.84 | 15.78 | 23.67 | 31.48 | 39.25 | 47.01 | 54.76 | 62.51 | 70.26 | 78.00 | 85.75 | 93.50 | 101.24 | 108.99 | 116.74 | 124.48 | 132.23 | 139.98 | 147.72 | 155.47 | 163.22 | 170.96 | 178.71 | 186.46 | 194.20 |
| | $J$ | 6.90 | 11.91 | 15.26 | 17.54 | 19.15 | 20.31 | 21.16 | 21.80 | 22.29 | 22.66 | 22.95 | 23.18 | 23.36 | 23.51 | 23.63 | 23.72 | 23.80 | 23.86 | 23.92 | 23.96 | 24.00 | 24.03 | 24.06 | 24.08 | 24.10 |
| | $K$ | 2.56 | 4.35 | 5.26 | 5.66 | 5.71 | 5.73 | 5.74 | 5.74 | 5.91 | 5.91 | 5.91 | 5.91 | 5.91 | 5.91 | 5.91 | 5.91 | 5.91 | 5.91 | 5.91 | 5.91 | 5.91 | 5.91 | 5.91 | 5.91 | 5.91 |
| 0.2 | $U_1$ | 12.36 | 22.40 | 29.93 | 35.54 | 39.83 | 43.13 | 45.77 | 47.87 | 49.62 | 51.08 | 52.32 | 53.38 | 54.31 | 55.10 | 55.81 | 56.43 | 56.99 | 57.49 | 57.94 | 58.35 | 58.72 | 59.06 | 59.37 | 59.66 | 60.17 |
| | $U_2$ | 0.21 | 0.44 | 0.79 | 1.23 | 1.81 | 2.46 | 3.16 | 3.87 | 4.57 | 5.23 | 5.86 | 6.44 | 6.98 | 7.49 | 7.96 | 8.38 | 8.78 | 9.15 | 9.50 | 9.81 | 10.11 | 10.39 | 10.64 | 10.88 | 11.11 |
| | $W_{1,2}$ | 5.11 | 11.11 | 17.98 | 25.35 | 32.95 | 40.64 | 48.36 | 56.09 | 63.84 | 71.59 | 79.34 | 87.08 | 94.83 | 102.58 | 110.32 | 118.07 | 125.82 | 133.57 | 141.31 | 149.06 | 156.80 | 164.55 | 172.30 | 180.04 | 187.79 |
| | $W_{3,4}$ | 7.75 | 15.63 | 23.48 | 31.27 | 39.04 | 46.80 | 54.55 | 62.29 | 70.04 | 77.79 | 85.53 | 93.28 | 101.03 | 108.77 | 116.52 | 124.27 | 132.01 | 139.76 | 147.51 | 155.25 | 163.00 | 170.75 | 178.49 | 186.24 | 193.99 |
| | $J$ | 6.99 | 12.13 | 15.54 | 17.96 | 19.62 | 20.82 | 21.70 | 22.35 | 22.84 | 23.21 | 23.50 | 23.73 | 23.91 | 24.05 | 24.16 | 24.26 | 24.33 | 24.40 | 24.45 | 24.49 | 24.53 | 24.56 | 24.58 | 24.61 | 24.62 |
| | $K$ | 2.63 | 4.47 | 5.41 | 5.82 | 6.04 | 6.07 | 6.08 | 6.08 | 6.08 | 6.08 | 6.08 | 6.08 | 6.08 | 6.08 | 6.08 | 6.08 | 6.08 | 6.08 | 6.08 | 6.08 | 6.08 | 6.08 | 6.08 | 6.08 | 6.08 |
| 0.1 | $U_1$ | 12.35 | 22.40 | 29.93 | 35.56 | 39.83 | 43.14 | 45.75 | 47.90 | 49.66 | 51.12 | 52.36 | 53.42 | 54.33 | 55.14 | 55.85 | 56.47 | 57.03 | 57.53 | 57.98 | 58.39 | 58.76 | 59.10 | 59.42 | 59.70 | 59.92 |
| | $U_2$ | 0.20 | 0.44 | 0.76 | 1.21 | 1.79 | 2.45 | 3.14 | 3.85 | 4.54 | 5.20 | 5.82 | 6.41 | 6.95 | 7.45 | 7.92 | 8.35 | 8.75 | 9.12 | 9.47 | 9.79 | 10.08 | 10.36 | 10.61 | 10.85 | 11.09 |
| | $W_{1,2}$ | 4.99 | 10.81 | 17.62 | 24.97 | 32.56 | 40.24 | 47.96 | 55.70 | 63.45 | 71.19 | 78.94 | 86.68 | 94.43 | 102.18 | 109.92 | 117.67 | 125.42 | 133.16 | 140.91 | 148.66 | 156.40 | 164.15 | 171.90 | 179.64 | 187.39 |
| | $W_{3,4}$ | 7.71 | 15.54 | 23.37 | 31.16 | 38.92 | 46.67 | 54.42 | 62.17 | 69.92 | 77.66 | 85.41 | 93.16 | 100.90 | 108.65 | 116.36 | 124.10 | 131.89 | 139.64 | 147.38 | 155.13 | 162.88 | 170.62 | 178.37 | 186.12 | 193.86 |
| | $J$ | 7.07 | 12.30 | 15.84 | 18.26 | 19.97 | 21.19 | 22.09 | 22.75 | 23.25 | 23.63 | 23.92 | 24.15 | 24.33 | 24.47 | 24.58 | 24.67 | 24.75 | 24.81 | 24.86 | 24.90 | 24.94 | 24.97 | 24.99 | 25.02 | 25.04 |
| | $K$ | 2.69 | 4.56 | 5.52 | 5.94 | 6.11 | 6.17 | 6.19 | 6.20 | 6.21 | 6.21 | 6.21 | 6.21 | 6.21 | 6.21 | 6.21 | 6.21 | 6.21 | 6.21 | 6.21 | 6.21 | 6.21 | 6.21 | 6.21 | 6.21 | 6.21 |
| 0.0 | $U_1$ | 12.35 | 22.40 | 29.93 | 35.56 | 39.83 | 43.14 | 45.77 | 47.90 | 49.66 | 51.12 | 52.36 | 53.42 | 54.34 | 55.14 | 55.85 | 56.47 | 57.03 | 57.53 | 57.98 | 58.39 | 58.76 | 59.10 | 59.42 | 59.70 | 59.97 |
| | $U_2$ | 0.19 | 0.43 | 0.76 | 1.21 | 1.79 | 2.45 | 3.14 | 3.85 | 4.54 | 5.20 | 5.82 | 6.41 | 6.95 | 7.45 | 7.92 | 8.35 | 8.75 | 9.12 | 9.47 | 9.79 | 10.08 | 10.36 | 10.61 | 10.85 | 11.08 |
| | $W_{1,2}$ | 4.94 | 10.81 | 17.62 | 24.97 | 32.56 | 40.24 | 47.96 | 55.70 | 63.45 | 71.19 | 78.94 | 86.68 | 94.43 | 102.18 | 109.92 | 117.67 | 125.42 | 133.16 | 140.91 | 148.66 | 156.40 | 164.15 | 171.90 | 179.64 | 187.59 |
| | $W_{3,4}$ | 7.69 | 15.51 | 23.33 | 31.12 | 38.88 | 46.63 | 54.38 | 62.13 | 69.88 | 77.62 | 85.37 | 93.12 | 100.87 | 108.61 | 116.36 | 124.10 | 131.85 | 139.60 | 147.34 | 155.09 | 162.84 | 170.58 | 178.33 | 186.08 | 193.82 |
| | $J$ | 7.10 | 12.37 | 15.94 | 18.38 | 20.10 | 21.33 | 22.23 | 22.89 | 23.39 | 23.77 | 24.06 | 24.28 | 24.46 | 24.60 | 24.71 | 24.80 | 24.88 | 24.94 | 24.99 | 25.03 | 25.06 | 25.09 | 25.12 | 25.14 | 25.16 |
| | $K$ | 2.71 | 4.60 | 5.57 | 5.98 | 6.15 | 6.22 | 6.24 | 6.25 | 6.26 | 6.26 | 6.26 | 6.26 | 6.26 | 6.26 | 6.26 | 6.26 | 6.26 | 6.26 | 6.26 | 6.26 | 6.26 | 6.26 | 6.26 | 6.26 | 6.26 |

TABLE S29. Parameters of the THF interaction Hamiltonian for different ratios $u_0/u_1$ and screening lengths $\xi$ of the double-gate interaction potential ($U_\xi = 24\,$meV for $\xi = 10\,$nm). We use the same BM model parameters as in Table S28.

Total weight of the heavy-fermions on the active bands after wannierisation: 97.2%

Wannier orbitals support

FIG. S31. Band structures of the BM and THF models near charge neutrality for $w_0/w_1 = 0.8$, depicted by lines and crosses, respectively. The BM bands are colored according to the weight of the $f$-electron wave function on them. We use the same BM parameters as in Table S30.

TABLE S30. Parameters of the $f$-electron Wannier functions for different values of the tunneling ratio $w_0/w_1$. $\mathcal{W}$ denotes the total weight of the THF $f$-electrons on the active bands. In computing the ratios $\gamma/U_1$ and $\gamma/U_2$, we employ the on-site and nearest-neighbor repulsion parameters $U_1$ and $U_2$ obtained numerically for $\xi = 10$ nm, as given in Table S31. We employ $v_F = 5.944$ eV·Å, $|\mathbf{K}| = 1.703$ Å$^{-1}$, $w_1 = 110$ meV, and $\theta = 0.96°$ for the BM model.

| $w_0/w_1$ | $\alpha_1$ | $\alpha_2$ | $\lambda_1$ | $\lambda_2$ | $\lambda$ | $\mathcal{W}$ | $t_0$ | $\gamma$ | $M$ | $v_\star$ | $v_\star'$ | $v_\star''$ | $\gamma/U_1$ | $\gamma/U_2$ |
|---|---|---|---|---|---|---|---|---|---|---|---|---|---|---|
| 1.00 | 0.752 | 0.659 | 0.143 | 0.167 | 0.388 | 0.970 | 0.247 | 14.032 | 12.923 | -3.289 | 1.289 | 0.0455 | 0.24 | 7.30 |
| 0.90 | 0.775 | 0.633 | 0.155 | 0.173 | 0.335 | 0.982 | 0.345 | 1.192 | 13.141 | -3.650 | 1.430 | -0.0074 | 0.02 | 0.73 |
| 0.80 | 0.804 | 0.595 | 0.167 | 0.179 | 0.329 | 0.972 | 0.583 | -12.414 | 13.265 | -3.938 | 1.514 | -0.0388 | -0.22 | -7.68 |
| 0.70 | 0.838 | 0.546 | 0.185 | 0.185 | 0.342 | 0.952 | 0.868 | -26.197 | 13.323 | -4.172 | 1.546 | -0.0539 | -0.51 | -15.28 |
| 0.60 | 0.873 | 0.488 | 0.197 | 0.197 | 0.357 | 0.936 | 1.144 | -39.666 | 13.340 | -4.366 | 1.526 | -0.0571 | -0.83 | -21.77 |
| 0.50 | 0.906 | 0.422 | 0.209 | 0.203 | 0.368 | 0.922 | 1.396 | -52.366 | 13.332 | -4.530 | 1.448 | -0.0516 | -1.17 | -27.61 |
| 0.40 | 0.937 | 0.349 | 0.227 | 0.203 | 0.381 | 0.915 | 1.602 | -63.839 | 13.312 | -4.668 | 1.304 | -0.0405 | -1.52 | -32.16 |
| 0.30 | 0.963 | 0.270 | 0.239 | 0.209 | 0.386 | 0.908 | 1.790 | -73.590 | 13.288 | -4.784 | 1.084 | -0.0267 | -1.81 | -36.47 |
| 0.20 | 0.983 | 0.185 | 0.251 | 0.215 | 0.381 | 0.896 | 1.965 | -81.096 | 13.268 | -4.873 | 0.785 | -0.0133 | -2.01 | -41.05 |
| 0.10 | 0.995 | 0.096 | 0.257 | 0.215 | 0.378 | 0.888 | 2.103 | -85.854 | 13.254 | -4.930 | 0.414 | -0.0036 | -2.14 | -44.14 |
| 0.00 | 1.000 | 0.000 | 0.257 | 0.000 | 0.371 | 0.881 | 2.085 | -87.488 | 13.249 | -4.950 | -0.0000 | | -2.15 | -46.21 |

| $w_0/w_1$ | $\xi$/nm | 2 | 4 | 6 | 8 | 10 | 12 | 14 | 16 | 18 | 20 | 22 | 24 | 26 | 28 | 30 | 32 | 34 | 36 | 38 | 40 | 42 | 44 | 46 | 48 | 50 |
|---|---|---|---|---|---|---|---|---|---|---|---|---|---|---|---|---|---|---|---|---|---|---|---|---|---|---|
| 1.0 | $U_1$ | 21.97 | 37.52 | 47.73 | 54.66 | 59.59 | 63.26 | 66.09 | 68.33 | 70.16 | 71.66 | 72.93 | 74.01 | 74.94 | 75.75 | 76.46 | 77.08 | 77.64 | 78.16 | 78.60 | 79.01 | 79.38 | 79.72 | 80.03 | 80.32 | 80.58 |
| | $U_2$ | 0.59 | 0.93 | 1.36 | 1.92 | 2.57 | 3.28 | 3.99 | 4.69 | 5.36 | 6.00 | 6.60 | 7.15 | 7.66 | 8.13 | 8.57 | 8.98 | 9.35 | 9.70 | 10.02 | 10.32 | 10.60 | 10.85 | 11.10 | 11.32 | |
| | $W_{1,2}$ | 5.61 | 12.14 | 19.48 | 27.25 | 35.21 | 43.25 | 51.31 | 59.38 | 67.46 | 75.54 | 83.62 | 91.70 | 99.78 | 107.86 | 115.94 | 124.02 | 132.10 | 140.18 | 148.26 | 156.34 | 164.42 | 172.50 | 180.58 | 188.66 | 196.74 |
| | $W_{3,4}$ | 10.92 | 21.32 | 30.64 | 39.26 | 47.55 | 55.71 | 63.82 | 71.91 | 79.99 | 88.07 | 96.15 | 104.23 | 112.31 | 120.39 | 128.47 | 136.55 | 144.63 | 152.71 | 160.79 | 168.87 | 176.95 | 185.03 | 193.11 | 201.19 | 209.27 |
| | $J$ | 7.01 | 10.66 | 12.29 | 13.03 | 13.39 | 13.58 | 13.68 | 13.74 | 13.77 | 13.80 | 13.82 | 13.83 | 13.84 | 13.85 | 13.85 | 13.86 | 13.86 | 13.87 | 13.87 | 13.87 | 13.87 | 13.88 | 13.88 | 13.88 | 13.88 |
| | $K$ | 2.68 | 4.46 | 5.33 | 5.69 | 5.83 | 5.90 | 5.91 | 5.91 | 5.91 | 5.91 | 5.91 | 5.91 | 5.91 | 5.91 | 5.91 | 5.91 | 5.91 | 5.91 | 5.91 | 5.91 | 5.91 | 5.91 | 5.91 | 5.91 | 5.91 |
| 0.9 | $U_1$ | 21.40 | 36.92 | 47.31 | 54.44 | 59.56 | 63.37 | 66.30 | 68.63 | 70.51 | 72.07 | 73.37 | 74.48 | 75.43 | 76.26 | 76.98 | 77.63 | 78.20 | 78.71 | 79.16 | 79.58 | 79.96 | 80.30 | 80.62 | 80.91 | 81.17 |
| | $U_2$ | 0.19 | 0.40 | 0.68 | 1.08 | 1.63 | 2.27 | 2.98 | 3.70 | 4.41 | 5.09 | 5.74 | 6.34 | 6.90 | 7.42 | 7.90 | 8.34 | 8.75 | 9.13 | 9.48 | 9.81 | 10.11 | 10.39 | 10.65 | 10.90 | 11.12 |
| | $W_{1,2}$ | 5.84 | 12.48 | 19.88 | 27.67 | 35.64 | 43.68 | 51.74 | 59.81 | 67.89 | 75.97 | 84.05 | 92.13 | 100.21 | 108.29 | 116.37 | 124.45 | 132.53 | 140.61 | 148.69 | 156.77 | 164.85 | 172.93 | 181.01 | 189.09 | 197.17 |
| | $W_{3,4}$ | 10.11 | 19.93 | 28.96 | 37.45 | 45.69 | 53.83 | 61.93 | 70.02 | 78.10 | 86.18 | 94.26 | 102.34 | 110.42 | 118.50 | 126.58 | 134.66 | 142.74 | 150.82 | 158.90 | 166.98 | 175.06 | 183.14 | 191.22 | 199.30 | 207.38 |
| | $J$ | 6.78 | 10.50 | 12.26 | 13.12 | 13.56 | 13.81 | 13.95 | 14.04 | 14.13 | 14.18 | 14.20 | 14.21 | 14.22 | 14.22 | 14.23 | 14.23 | 14.24 | 14.24 | 14.24 | 14.24 | 14.24 | 14.24 | 14.25 | 14.25 | 14.25 |
| | $K$ | 2.53 | 4.24 | 5.08 | 5.43 | 5.56 | 5.62 | 5.64 | 5.65 | 5.65 | 5.65 | 5.65 | 5.65 | 5.65 | 5.65 | 5.65 | 5.65 | 5.65 | 5.65 | 5.65 | 5.65 | 5.65 | 5.65 | 5.65 | 5.65 | 5.65 |
| 0.8 | $U_1$ | 19.61 | 34.16 | 44.12 | 51.05 | 56.07 | 59.84 | 62.75 | 65.07 | 66.94 | 68.50 | 69.80 | 70.91 | 71.86 | 72.69 | 73.42 | 74.06 | 74.63 | 75.14 | 75.60 | 76.01 | 76.39 | 76.74 | 77.05 | 77.34 | 77.61 |
| | $U_2$ | 0.18 | 0.38 | 0.65 | 1.06 | 1.62 | 2.27 | 2.98 | 3.71 | 4.42 | 5.10 | 5.75 | 6.36 | 6.92 | 7.44 | 7.92 | 8.36 | 8.77 | 9.15 | 9.50 | 9.84 | 10.14 | 10.41 | 10.67 | 10.92 | 11.14 |
| | $W_{1,2}$ | 5.98 | 12.71 | 20.13 | 27.94 | 35.91 | 43.95 | 52.01 | 60.09 | 68.17 | 76.24 | 84.32 | 92.40 | 100.48 | 108.56 | 116.64 | 124.72 | 132.80 | 140.88 | 148.96 | 157.04 | 165.12 | 173.20 | 181.28 | 189.36 | 197.44 |
| | $W_{3,4}$ | 9.47 | 18.79 | 27.56 | 35.54 | 44.14 | 52.27 | 60.36 | 68.45 | 76.53 | 84.61 | 92.69 | 100.77 | 108.85 | 116.93 | 125.01 | 133.09 | 141.17 | 149.25 | 157.33 | 165.41 | 173.49 | 181.57 | 189.65 | 197.73 | 205.81 |
| | $J$ | 6.70 | 10.65 | 12.71 | 13.84 | 14.51 | 14.93 | 15.21 | 15.41 | 15.56 | 15.66 | 15.74 | 15.80 | 15.85 | 15.89 | 15.92 | 15.95 | 15.97 | 15.98 | 16.00 | 16.01 | 16.02 | 16.02 | 16.03 | 16.04 | 16.04 |
| | $K$ | 2.44 | 4.10 | 4.92 | 5.27 | 5.40 | 5.45 | 5.47 | 5.48 | 5.48 | 5.48 | 5.48 | 5.48 | 5.48 | 5.48 | 5.48 | 5.48 | 5.48 | 5.48 | 5.48 | 5.48 | 5.48 | 5.48 | 5.48 | 5.48 | 5.48 |
| 0.7 | $U_1$ | 17.60 | 30.93 | 40.05 | 46.84 | 51.67 | 55.32 | 58.16 | 60.43 | 62.28 | 63.81 | 65.10 | 66.19 | 67.14 | 67.96 | 68.68 | 69.32 | 69.88 | 70.35 | 70.85 | 71.26 | 71.64 | 71.98 | 72.30 | 72.59 | 72.85 |
| | $U_2$ | 0.19 | 0.41 | 0.71 | 1.15 | 1.71 | 2.38 | 3.10 | 3.82 | 4.54 | 5.22 | 5.87 | 6.47 | 7.03 | 7.55 | 8.03 | 8.47 | 8.88 | 9.26 | 9.61 | 9.93 | 10.23 | 10.51 | 10.77 | 11.02 | 11.24 |
| | $W_{1,2}$ | 6.03 | 12.77 | 20.20 | 28.01 | 35.99 | 44.03 | 52.09 | 60.17 | 68.25 | 76.32 | 84.40 | 92.48 | 100.56 | 108.64 | 116.72 | 124.80 | 132.88 | 140.96 | 149.04 | 157.12 | 165.20 | 173.28 | 181.36 | 189.44 | 197.52 |
| | $W_{3,4}$ | 8.98 | 17.92 | 26.49 | 34.78 | 42.55 | 51.06 | 59.15 | 67.24 | 75.32 | 83.40 | 91.48 | 99.56 | 107.64 | 115.72 | 123.80 | 131.88 | 139.96 | 148.04 | 156.12 | 164.20 | 172.28 | 180.36 | 188.44 | 196.52 | 204.60 |
| | $J$ | 6.71 | 10.93 | 13.33 | 14.77 | 15.70 | 16.33 | 16.77 | 17.10 | 17.34 | 17.52 | 17.66 | 17.77 | 17.86 | 17.93 | 17.99 | 18.03 | 18.07 | 18.10 | 18.13 | 18.15 | 18.16 | 18.18 | 18.18 | 18.20 | 18.21 |
| | $K$ | 2.40 | 4.04 | 4.86 | 5.20 | 5.33 | 5.39 | 5.40 | 5.41 | 5.41 | 5.41 | 5.42 | 5.42 | 5.42 | 5.42 | 5.42 | 5.42 | 5.42 | 5.42 | 5.42 | 5.42 | 5.42 | 5.42 | 5.42 | 5.42 | 5.42 |

TABLE S31: Parameters of the THF interaction Hamiltonian for different ratios $u_0/u_1$ and screening lengths $\xi$ of the double-gate interaction potential ($U_\xi = 24\,\mathrm{meV} = \xi$ for $\xi = 10\,\mathrm{nm}$). We use the same BM model parameters as in Table S30.

| $u_0/u_1$ | $\xi$/nm | 2 | 4 | 6 | 8 | 10 | 12 | 14 | 16 | 18 | 20 | 22 | 24 | 26 | 28 | 30 | 32 | 34 | 36 | 38 | 40 | 42 | 44 | 46 | 48 | 50 |
|---|---|---|---|---|---|---|---|---|---|---|---|---|---|---|---|---|---|---|---|---|---|---|---|---|---|---|
| 0.6 | $U_1$ | 15.86 | 28.10 | 36.82 | 43.09 | 47.72 | 51.26 | 54.02 | 56.24 | 58.06 | 59.56 | 60.83 | 61.92 | 62.85 | 63.67 | 64.38 | 65.01 | 65.58 | 66.08 | 66.54 | 66.95 | 67.32 | 67.66 | 67.98 | 68.27 | 68.53 |
| | $U_2$ | 0.21 | 0.45 | 0.78 | 1.24 | 1.82 | 2.50 | 3.22 | 3.95 | 4.66 | 5.34 | 5.99 | 6.59 | 7.15 | 7.66 | 8.14 | 8.58 | 8.99 | 9.36 | 9.71 | 10.04 | 10.34 | 10.62 | 10.88 | 11.12 | 11.35 |
| | $W_{1,2}$ | 5.99 | 12.70 | 20.12 | 27.92 | 35.89 | 43.03 | 52.00 | 60.07 | 68.15 | 76.23 | 84.01 | 91.85 | 100.47 | 108.55 | 116.63 | 124.71 | 132.79 | 140.87 | 148.95 | 157.03 | 165.11 | 173.19 | 181.26 | 189.34 | 197.42 |
| | $W_{3,4}$ | 8.63 | 17.28 | 25.70 | 33.93 | 42.07 | 50.17 | 58.26 | 66.34 | 74.42 | 82.50 | 90.58 | 98.66 | 106.74 | 114.82 | 122.90 | 130.98 | 139.06 | 147.14 | 155.22 | 163.30 | 171.38 | 179.46 | 187.54 | 195.62 | 203.70 |
| | $J$ | 6.77 | 11.26 | 13.99 | 15.73 | 16.89 | 17.71 | 18.30 | 18.73 | 19.06 | 19.32 | 19.51 | 19.67 | 19.79 | 19.89 | 19.97 | 20.03 | 20.08 | 20.13 | 20.16 | 20.19 | 20.22 | 20.24 | 20.26 | 20.27 | 20.29 |
| | $K$ | 2.40 | 4.05 | 4.87 | 5.22 | 5.35 | 5.40 | 5.42 | 5.43 | 5.43 | 5.43 | 5.43 | 5.43 | 5.43 | 5.43 | 5.43 | 5.43 | 5.43 | 5.43 | 5.43 | 5.43 | 5.43 | 5.43 | 5.43 | 5.43 | 5.43 |
| 0.5 | $U_1$ | 14.56 | 25.98 | 34.25 | 40.26 | 44.75 | 48.20 | 50.91 | 53.09 | 54.88 | 56.37 | 57.62 | 58.70 | 59.63 | 60.43 | 61.14 | 61.77 | 62.33 | 62.84 | 63.29 | 63.70 | 64.07 | 64.41 | 64.73 | 65.02 | 65.28 |
| | $U_2$ | 0.22 | 0.48 | 0.82 | 1.30 | 1.90 | 2.58 | 3.31 | 4.04 | 4.75 | 5.43 | 6.08 | 6.69 | 7.25 | 7.74 | 8.22 | 8.66 | 9.06 | 9.44 | 9.79 | 10.11 | 10.41 | 10.69 | 10.95 | 11.19 | 11.42 |
| | $W_{1,2}$ | 5.88 | 12.52 | 19.90 | 27.69 | 35.66 | 43.69 | 51.76 | 59.83 | 67.91 | 75.99 | 84.07 | 92.15 | 100.23 | 108.30 | 116.38 | 124.46 | 132.54 | 140.62 | 148.70 | 156.78 | 164.86 | 172.94 | 181.02 | 189.10 | 197.18 |
| | $W_{3,4}$ | 8.38 | 16.83 | 25.14 | 33.32 | 41.44 | 49.54 | 57.62 | 65.70 | 73.78 | 81.86 | 89.94 | 98.02 | 106.10 | 114.18 | 122.26 | 130.34 | 138.42 | 146.50 | 154.58 | 162.66 | 170.74 | 178.82 | 186.90 | 194.98 | 203.06 |
| | $J$ | 6.87 | 11.61 | 14.62 | 16.60 | 17.96 | 18.93 | 19.64 | 20.16 | 20.56 | 20.87 | 21.11 | 21.30 | 21.44 | 21.56 | 21.66 | 21.74 | 21.80 | 21.85 | 21.90 | 21.93 | 21.96 | 21.99 | 22.01 | 22.03 | 22.05 |
| | $K$ | 2.43 | 4.11 | 4.95 | 5.30 | 5.44 | 5.49 | 5.51 | 5.52 | 5.52 | 5.52 | 5.52 | 5.52 | 5.52 | 5.52 | 5.52 | 5.52 | 5.52 | 5.52 | 5.52 | 5.52 | 5.52 | 5.52 | 5.52 | 5.52 | 5.52 |
| 0.4 | $U_1$ | 13.44 | 24.13 | 31.98 | 37.75 | 42.09 | 45.44 | 48.09 | 50.24 | 51.99 | 53.46 | 54.70 | 55.77 | 56.69 | 57.49 | 58.19 | 58.82 | 59.38 | 59.88 | 60.33 | 60.74 | 61.11 | 61.45 | 61.76 | 62.05 | 62.31 |
| | $U_2$ | 0.24 | 0.52 | 0.88 | 1.37 | 1.99 | 2.68 | 3.40 | 4.14 | 4.85 | 5.53 | 6.17 | 6.76 | 7.32 | 7.83 | 8.31 | 8.74 | 9.15 | 9.52 | 9.87 | 10.19 | 10.49 | 10.77 | 11.03 | 11.27 | 11.49 |
| | $W_{1,2}$ | 5.75 | 12.30 | 19.63 | 27.40 | 35.36 | 43.09 | 51.45 | 59.53 | 67.61 | 75.68 | 83.76 | 91.84 | 99.56 | 108.00 | 116.08 | 124.16 | 132.24 | 140.32 | 148.40 | 156.48 | 164.56 | 172.64 | 180.72 | 188.80 | 196.88 |
| | $W_{3,4}$ | 8.20 | 16.51 | 24.74 | 32.89 | 41.00 | 49.09 | 57.17 | 65.26 | 73.34 | 81.42 | 89.50 | 97.58 | 105.66 | 113.74 | 121.81 | 129.89 | 137.97 | 146.05 | 154.13 | 162.21 | 170.29 | 178.37 | 186.45 | 194.53 | 202.61 |
| | $J$ | 6.99 | 11.98 | 15.26 | 17.48 | 19.04 | 20.16 | 20.99 | 21.61 | 22.09 | 22.45 | 22.74 | 22.96 | 23.14 | 23.28 | 23.40 | 23.49 | 23.57 | 23.64 | 23.69 | 23.74 | 23.77 | 23.81 | 23.83 | 23.86 | 23.88 |
| | $K$ | 2.50 | 4.22 | 5.08 | 5.44 | 5.58 | 5.63 | 5.65 | 5.66 | 5.66 | 5.66 | 5.66 | 5.66 | 5.66 | 5.66 | 5.66 | 5.66 | 5.66 | 5.66 | 5.66 | 5.66 | 5.66 | 5.66 | 5.66 | 5.66 | 5.66 |
| 0.3 | $U_1$ | 12.78 | 23.04 | 30.65 | 36.28 | 40.55 | 43.85 | 46.47 | 48.59 | 50.34 | 51.80 | 53.03 | 54.09 | 55.01 | 55.81 | 56.51 | 57.13 | 57.69 | 58.19 | 58.64 | 59.04 | 59.42 | 59.75 | 60.07 | 60.35 | 60.62 |
| | $U_2$ | 0.24 | 0.53 | 0.90 | 1.40 | 2.02 | 2.71 | 3.44 | 4.18 | 4.89 | 5.57 | 6.22 | 6.80 | 7.36 | 7.87 | 8.34 | 8.78 | 9.18 | 9.54 | 9.89 | 10.22 | 10.52 | 10.80 | 11.06 | 11.30 | 11.53 |
| | $W_{1,2}$ | 5.58 | 12.02 | 19.30 | 27.05 | 35.00 | 43.03 | 51.09 | 59.16 | 67.24 | 75.32 | 83.40 | 91.48 | 99.56 | 107.63 | 115.71 | 123.79 | 131.87 | 139.95 | 148.03 | 156.11 | 164.19 | 172.27 | 180.35 | 188.43 | 196.51 |
| | $W_{3,4}$ | 8.08 | 16.29 | 24.47 | 32.60 | 40.70 | 48.79 | 56.87 | 64.95 | 73.03 | 81.11 | 89.19 | 97.27 | 105.35 | 113.43 | 121.51 | 129.59 | 137.67 | 145.75 | 153.83 | 161.91 | 169.99 | 178.07 | 186.15 | 194.23 | 202.31 |
| | $J$ | 7.12 | 12.53 | 16.12 | 18.58 | 20.32 | 21.57 | 22.49 | 23.17 | 23.69 | 24.08 | 24.39 | 24.63 | 24.82 | 24.97 | 25.09 | 25.19 | 25.27 | 25.33 | 25.39 | 25.44 | 25.47 | 25.50 | 25.53 | 25.56 | 25.58 |
| | $K$ | 2.57 | 4.34 | 5.23 | 5.60 | 5.75 | 5.81 | 5.83 | 5.84 | 5.84 | 5.84 | 5.84 | 5.84 | 5.84 | 5.84 | 5.84 | 5.84 | 5.84 | 5.84 | 5.84 | 5.84 | 5.84 | 5.84 | 5.84 | 5.84 | 5.84 |
| 0.2 | $U_1$ | 12.61 | 22.79 | 30.38 | 36.02 | 40.29 | 43.61 | 46.24 | 48.36 | 50.12 | 51.58 | 52.74 | 53.80 | 54.74 | 55.53 | 56.23 | 56.86 | 57.41 | 57.91 | 58.36 | 58.77 | 59.15 | 59.41 | 59.80 | 60.08 | 60.35 |
| | $U_2$ | 0.23 | 0.50 | 0.86 | 1.36 | 1.98 | 2.67 | 3.40 | 4.14 | 4.85 | 5.53 | 6.17 | 6.77 | 7.32 | 7.83 | 8.28 | 8.72 | 9.12 | 9.50 | 9.84 | 10.17 | 10.46 | 10.74 | 11.00 | 11.24 | 11.47 |
| | $W_{1,2}$ | 5.39 | 11.71 | 18.93 | 26.65 | 34.59 | 42.62 | 50.68 | 58.75 | 66.83 | 74.90 | 82.98 | 91.06 | 99.14 | 107.22 | 115.30 | 123.38 | 131.46 | 139.54 | 147.62 | 155.70 | 163.78 | 171.86 | 179.94 | 188.02 | 196.10 |
| | $W_{3,4}$ | 8.00 | 16.15 | 24.30 | 32.41 | 40.50 | 48.59 | 56.67 | 64.75 | 72.83 | 80.91 | 88.99 | 97.07 | 105.15 | 113.23 | 121.31 | 129.39 | 137.47 | 145.53 | 153.63 | 161.71 | 169.79 | 177.87 | 185.95 | 194.03 | 202.11 |
| | $J$ | 7.22 | 12.63 | 16.24 | 18.72 | 20.47 | 21.74 | 22.65 | 23.33 | 23.84 | 24.23 | 24.53 | 24.76 | 24.95 | 25.10 | 25.22 | 25.32 | 25.40 | 25.46 | 25.52 | 25.56 | 25.60 | 25.63 | 25.66 | 25.68 | 25.70 |
| | $K$ | 2.65 | 4.47 | 5.39 | 5.77 | 5.92 | 5.98 | 6.00 | 6.01 | 6.01 | 6.01 | 6.01 | 6.01 | 6.01 | 6.01 | 6.01 | 6.01 | 6.01 | 6.01 | 6.01 | 6.01 | 6.01 | 6.01 | 6.01 | 6.01 | 6.01 |
| 0.1 | $U_1$ | 12.52 | 22.67 | 30.26 | 35.90 | 40.19 | 43.51 | 46.15 | 48.28 | 50.03 | 51.50 | 52.74 | 53.80 | 54.74 | 55.53 | 56.23 | 56.86 | 57.41 | 57.91 | 58.36 | 58.77 | 59.15 | 59.49 | 59.80 | 60.08 | 60.35 |
| | $U_2$ | 0.20 | 0.45 | 0.80 | 1.28 | 1.89 | 2.59 | 3.32 | 4.05 | 4.77 | 5.45 | 6.09 | 6.69 | 7.24 | 7.76 | 8.23 | 8.67 | 9.08 | 9.45 | 9.80 | 10.12 | 10.42 | 10.70 | 10.96 | 11.20 | 11.43 |
| | $W_{1,2}$ | 5.18 | 11.35 | 18.66 | 26.36 | 34.29 | 42.32 | 50.38 | 58.45 | 66.36 | 74.43 | 82.51 | 90.59 | 98.67 | 106.75 | 114.83 | 122.91 | 130.99 | 139.07 | 147.15 | 155.23 | 163.31 | 171.39 | 179.47 | 187.55 | 195.63 |
| | $W_{3,4}$ | 7.94 | 16.04 | 24.16 | 32.26 | 40.35 | 48.43 | 56.51 | 64.59 | 72.67 | 80.75 | 88.83 | 96.91 | 104.99 | 113.07 | 121.15 | 129.23 | 137.31 | 145.39 | 153.47 | 161.55 | 169.63 | 177.71 | 185.79 | 193.87 | 201.95 |
| | $J$ | 7.33 | 12.74 | 16.41 | 18.91 | 20.65 | 21.91 | 22.80 | 23.47 | 23.97 | 24.34 | 24.63 | 24.85 | 25.03 | 25.17 | 25.28 | 25.36 | 25.44 | 25.50 | 25.55 | 25.59 | 25.62 | 25.65 | 25.67 | 25.69 | 25.71 |
| | $K$ | 2.73 | 4.60 | 5.54 | 5.94 | 6.09 | 6.15 | 6.17 | 6.18 | 6.19 | 6.19 | 6.19 | 6.19 | 6.19 | 6.19 | 6.19 | 6.19 | 6.19 | 6.19 | 6.19 | 6.19 | 6.19 | 6.19 | 6.19 | 6.19 | 6.19 |
| 0.0 | $U_1$ | 12.69 | 22.97 | 30.66 | 36.37 | 40.70 | 44.06 | 46.72 | 48.87 | 50.64 | 52.11 | 53.36 | 54.43 | 55.36 | 56.16 | 56.86 | 57.49 | 58.05 | 58.55 | 59.00 | 59.41 | 59.79 | 60.13 | 60.44 | 60.73 | 60.99 |
| | $U_2$ | 0.20 | 0.45 | 0.80 | 1.28 | 1.89 | 2.59 | 3.32 | 4.05 | 4.77 | 5.45 | 6.09 | 6.69 | 7.24 | 7.76 | 8.23 | 8.67 | 9.08 | 9.45 | 9.80 | 10.12 | 10.42 | 10.70 | 10.96 | 11.20 | 11.43 |
| | $W_{1,2}$ | 5.18 | 11.48 | 18.66 | 26.36 | 34.29 | 42.32 | 50.38 | 58.45 | 66.53 | 74.61 | 82.69 | 90.76 | 98.84 | 106.92 | 115.00 | 123.08 | 131.16 | 139.24 | 147.32 | 155.40 | 163.48 | 171.56 | 179.64 | 187.72 | 195.80 |
| | $W_{3,4}$ | 7.94 | 16.07 | 24.19 | 32.30 | 40.39 | 48.47 | 56.55 | 64.63 | 72.71 | 80.79 | 88.87 | 96.95 | 105.03 | 113.11 | 121.19 | 129.27 | 137.35 | 145.43 | 153.51 | 161.59 | 169.67 | 177.75 | 185.83 | 193.91 | 201.99 |
| | $J$ | 7.29 | 12.69 | 16.34 | 18.85 | 20.62 | 21.88 | 22.81 | 23.50 | 24.01 | 24.41 | 24.71 | 24.94 | 25.13 | 25.28 | 25.39 | 25.49 | 25.57 | 25.63 | 25.69 | 25.73 | 25.77 | 25.80 | 25.83 | 25.85 | 25.87 |
| | $K$ | 2.71 | 4.57 | 5.50 | 5.89 | 6.05 | 6.10 | 6.13 | 6.14 | 6.14 | 6.14 | 6.14 | 6.14 | 6.14 | 6.14 | 6.14 | 6.14 | 6.14 | 6.14 | 6.14 | 6.14 | 6.14 | 6.14 | 6.14 | 6.14 | 6.14 |

Total weight of the heavy-fermions on the active bands after wannierisation: 96.8%

Wannier orbitals support

FIG. S32. Band structures of the BM and THF models near charge neutrality for $w_0/w_1 = 0.8$, depicted by lines and crosses, respectively. The BM bands are colored according to the weight of the $f$-electron wave function on them. We use the same BM parameters as in Table S32.

TABLE S32. Parameters of the $f$-electron Wannier functions and the THF single-particle Hamiltonian for different values of the tunneling ratio $w_0/w_1$. $\mathcal{W}$ denotes the total weight of the THF $f$-electrons on the active bands. In computing the ratios $\gamma/U_1$ and $\gamma/U_2$, we employ the on-site and nearest-neighbor repulsion parameters $U_1$ and $U_2$ obtained numerically for $\xi = 10$ nm, as given in Table S33. We employ $v_F = 5.944$ eV Å, $|\mathbf{K}| = 1.703$ Å$^{-1}$, $v_1 = 110$ meV, and $\theta = 0.98°$ for the BM model.

| $w_0/w_1$ | $\alpha_1$ | $\alpha_2$ | $\lambda_1$ | $\lambda_2$ | $\lambda$ | $\mathcal{W}$ | $t_0$ | $\gamma$ | $M$ | $v_\star$ | $v_\star'$ | $v_\star''$ | $\gamma/U_1$ | $\gamma/U_2$ |
|---|---|---|---|---|---|---|---|---|---|---|---|---|---|---|
| 1.00 | 0.754 | 0.657 | 0.149 | 0.167 | 0.372 | 0.972 | 0.185 | 12.035 | 11.319 | -3.428 | 1.340 | 0.0351 | 0.20 | 6.22 |
| 0.90 | 0.777 | 0.629 | 0.161 | 0.173 | 0.333 | 0.983 | 0.305 | -1.195 | 11.342 | -3.763 | 1.468 | -0.0113 | -0.02 | -0.70 |
| 0.80 | 0.807 | 0.590 | 0.173 | 0.185 | 0.330 | 0.968 | 0.516 | -15.051 | 11.303 | -4.031 | 1.541 | -0.0384 | -0.27 | -8.72 |
| 0.70 | 0.841 | 0.541 | 0.185 | 0.191 | 0.344 | 0.949 | 0.757 | -28.990 | 11.229 | -4.251 | 1.565 | -0.0511 | -0.56 | -15.84 |
| 0.60 | 0.876 | 0.483 | 0.197 | 0.197 | 0.359 | 0.934 | 0.983 | -42.549 | 11.138 | -4.435 | 1.539 | -0.0532 | -0.88 | -21.96 |
| 0.50 | 0.909 | 0.417 | 0.215 | 0.203 | 0.371 | 0.922 | 1.180 | -55.297 | 11.045 | -4.591 | 1.456 | -0.0477 | -1.22 | -27.27 |
| 0.40 | 0.939 | 0.345 | 0.227 | 0.209 | 0.381 | 0.914 | 1.351 | -66.787 | 10.959 | -4.725 | 1.309 | -0.0372 | -1.56 | -31.83 |
| 0.30 | 0.964 | 0.267 | 0.245 | 0.209 | 0.389 | 0.909 | 1.491 | -76.535 | 10.886 | -4.836 | 1.086 | -0.0244 | -1.86 | -35.57 |
| 0.20 | 0.983 | 0.183 | 0.251 | 0.215 | 0.382 | 0.895 | 1.663 | -84.028 | 10.831 | -4.923 | 0.785 | -0.0122 | -2.04 | -40.28 |
| 0.10 | 0.996 | 0.094 | 0.263 | 0.215 | 0.378 | 0.888 | 1.812 | -88.774 | 10.797 | -4.979 | 0.414 | -0.0033 | -2.16 | -43.23 |
| 0.00 | 1.000 | 0.000 | 0.263 | 0.000 | 0.380 | 0.888 | 1.863 | -90.402 | 10.786 | -4.998 | -0.0000 | -0.0000 | -2.22 | -43.74 |

| $w_0/w_1$ | $\xi$/nm | 2 | 4 | 6 | 8 | 10 | 12 | 14 | 16 | 18 | 20 | 22 | 24 | 26 | 28 | 30 | 32 | 34 | 36 | 38 | 40 | 42 | 44 | 46 | 48 | 50 |
|---|---|---|---|---|---|---|---|---|---|---|---|---|---|---|---|---|---|---|---|---|---|---|---|---|---|---|
| 1.0 | $U_1$ | 22.59 | 38.56 | 49.02 | 56.12 | 61.16 | 64.91 | 67.79 | 70.08 | 71.93 | 73.46 | 74.74 | 75.84 | 76.78 | 77.59 | 78.31 | 78.94 | 79.51 | 80.01 | 80.47 | 80.88 | 81.26 | 81.60 | 81.91 | 82.20 | 82.47 |
|  | $U_2$ | 2.63 | 4.37 | 5.20 | 5.54 | 5.66 | 5.71 | 5.73 | 5.74 | 5.74 | 5.74 | 5.74 | 5.74 | 5.74 | 5.74 | 5.74 | 5.74 | 5.74 | 5.74 | 5.74 | 5.74 | 5.74 | 5.74 | 5.74 | 5.74 | 5.74 |
|  | $W_{1,2}$ | 5.96 | 12.84 | 20.54 | 28.67 | 36.08 | 45.36 | 53.77 | 62.18 | 70.60 | 79.02 | 87.44 | 95.86 | 104.28 | 112.70 | 121.12 | 129.54 | 137.96 | 146.38 | 154.80 | 163.22 | 171.64 | 180.06 | 188.48 | 196.90 | 205.32 |
|  | $W_{3,4}$ | 11.03 | 21.50 | 31.13 | 40.02 | 48.62 | 57.11 | 65.55 | 73.98 | 82.41 | 90.83 | 99.25 | 107.67 | 116.00 | 124.51 | 132.93 | 141.35 | 149.77 | 158.19 | 166.61 | 175.03 | 183.45 | 191.87 | 200.29 | 208.71 | 217.13 |
|  | $J$ | 7.07 | 10.73 | 12.35 | 13.07 | 13.42 | 13.58 | 13.67 | 13.72 | 13.75 | 13.76 | 13.77 | 13.78 | 13.78 | 13.79 | 13.79 | 13.79 | 13.79 | 13.79 | 13.79 | 13.79 | 13.79 | 13.79 | 13.79 | 13.79 | 13.80 |
|  | $K$ | 2.50 | 4.17 | 4.97 | 5.30 | 5.42 | 5.47 | 5.49 | 5.49 | 5.49 | 5.49 | 5.49 | 5.49 | 5.49 | 5.49 | 5.49 | 5.49 | 5.49 | 5.49 | 5.49 | 5.49 | 5.49 | 5.49 | 5.49 | 5.49 | 5.49 |
| 0.9 | $U_1$ | 21.66 | 37.34 | 47.82 | 55.02 | 60.17 | 64.01 | 66.96 | 69.30 | 71.20 | 72.76 | 74.07 | 75.18 | 76.14 | 76.97 | 77.70 | 78.34 | 78.91 | 79.41 | 79.88 | 80.30 | 80.68 | 81.02 | 81.32 | 81.63 | 81.90 |
|  | $U_2$ | 0.19 | 0.41 | 0.70 | 1.13 | 1.71 | 2.40 | 3.14 | 3.89 | 4.63 | 5.33 | 6.00 | 6.62 | 7.19 | 7.72 | 8.21 | 8.66 | 9.07 | 9.46 | 9.81 | 10.14 | 10.45 | 10.73 | 10.99 | 11.24 | 11.47 |
|  | $W_{1,2}$ | 6.19 | 13.19 | 20.94 | 29.09 | 37.41 | 45.79 | 54.19 | 62.61 | 71.03 | 79.45 | 87.87 | 96.29 | 104.71 | 113.13 | 121.55 | 129.97 | 138.39 | 146.81 | 155.23 | 163.65 | 172.07 | 180.49 | 188.91 | 197.33 | 205.75 |
|  | $W_{3,4}$ | 10.25 | 20.24 | 29.50 | 38.28 | 46.84 | 55.31 | 63.75 | 72.17 | 80.60 | 89.02 | 97.44 | 105.86 | 114.28 | 122.70 | 131.12 | 139.54 | 147.96 | 156.38 | 164.80 | 173.22 | 181.64 | 190.06 | 198.48 | 206.90 | 215.32 |
|  | $J$ | 6.88 | 10.67 | 12.47 | 13.37 | 13.84 | 14.11 | 14.27 | 14.37 | 14.44 | 14.49 | 14.52 | 14.55 | 14.57 | 14.59 | 14.60 | 14.61 | 14.62 | 14.62 | 14.63 | 14.63 | 14.64 | 14.64 | 14.64 | 14.64 | 14.65 |
|  | $K$ | 2.50 | 4.17 | 4.97 | 5.30 | 5.42 | 5.47 | 5.49 | 5.49 | 5.49 | 5.49 | 5.49 | 5.49 | 5.49 | 5.49 | 5.49 | 5.49 | 5.49 | 5.49 | 5.49 | 5.49 | 5.49 | 5.49 | 5.49 | 5.49 | 5.49 |
| 0.8 | $U_1$ | 19.80 | 34.46 | 44.48 | 51.45 | 56.50 | 60.28 | 63.20 | 65.53 | 67.41 | 68.97 | 70.28 | 71.39 | 72.34 | 73.17 | 73.90 | 74.54 | 75.12 | 75.63 | 76.09 | 76.50 | 76.88 | 77.23 | 77.54 | 77.83 | 78.10 |
|  | $U_2$ | 0.18 | 0.40 | 0.69 | 1.14 | 1.73 | 2.42 | 3.17 | 3.92 | 4.66 | 5.37 | 6.03 | 6.65 | 7.22 | 7.75 | 8.24 | 8.69 | 9.11 | 9.49 | 9.85 | 10.18 | 10.48 | 10.77 | 11.03 | 11.28 | 11.50 |
|  | $W_{1,2}$ | 6.31 | 13.38 | 21.16 | 29.32 | 37.65 | 46.03 | 54.44 | 62.85 | 71.27 | 79.69 | 88.11 | 96.53 | 104.95 | 113.37 | 121.79 | 130.21 | 138.63 | 147.05 | 155.47 | 163.89 | 172.31 | 180.73 | 189.15 | 197.57 | 205.99 |
|  | $W_{3,4}$ | 9.64 | 19.16 | 28.19 | 36.86 | 45.38 | 53.84 | 62.27 | 70.70 | 79.12 | 87.54 | 95.96 | 104.38 | 112.80 | 121.22 | 129.64 | 138.06 | 146.48 | 154.90 | 163.32 | 171.74 | 180.16 | 188.58 | 197.00 | 205.42 | 213.84 |
|  | $J$ | 6.84 | 10.88 | 13.00 | 14.18 | 14.89 | 15.35 | 15.65 | 15.87 | 16.02 | 16.14 | 16.23 | 16.30 | 16.35 | 16.39 | 16.43 | 16.45 | 16.48 | 16.49 | 16.51 | 16.51 | 16.52 | 16.53 | 16.54 | 16.55 | 16.56 |
|  | $K$ | 2.42 | 4.05 | 4.84 | 5.16 | 5.28 | 5.33 | 5.35 | 5.35 | 5.35 | 5.35 | 5.35 | 5.35 | 5.35 | 5.35 | 5.35 | 5.35 | 5.35 | 5.35 | 5.35 | 5.35 | 5.35 | 5.35 | 5.35 | 5.35 | 5.35 |
| 0.7 | $U_1$ | 17.80 | 31.25 | 40.64 | 47.28 | 52.13 | 55.80 | 58.66 | 60.93 | 62.79 | 64.32 | 65.61 | 66.71 | 67.66 | 68.48 | 69.21 | 69.84 | 70.41 | 70.92 | 71.38 | 71.79 | 72.17 | 72.51 | 72.82 | 73.12 | 73.40 |
|  | $U_2$ | 0.20 | 0.43 | 0.76 | 1.22 | 1.83 | 2.53 | 3.28 | 4.04 | 4.78 | 5.49 | 6.15 | 6.77 | 7.34 | 7.87 | 8.35 | 8.80 | 9.22 | 9.60 | 9.95 | 10.28 | 10.59 | 10.87 | 11.13 | 11.38 | 11.61 |
|  | $W_{1,2}$ | 6.34 | 13.30 | 21.21 | 29.37 | 37.69 | 46.08 | 54.48 | 62.90 | 71.32 | 79.74 | 88.16 | 96.58 | 105.00 | 113.42 | 121.84 | 130.25 | 138.67 | 147.09 | 155.51 | 163.93 | 172.35 | 180.77 | 189.19 | 197.61 | 206.03 |
|  | $W_{3,4}$ | 9.18 | 18.34 | 27.18 | 35.78 | 44.27 | 52.72 | 61.15 | 69.57 | 77.99 | 86.41 | 94.83 | 103.25 | 111.67 | 120.09 | 128.51 | 136.93 | 145.35 | 153.77 | 162.19 | 170.61 | 179.03 | 187.45 | 195.87 | 204.29 | 212.71 |
|  | $J$ | 6.87 | 11.20 | 13.69 | 15.12 | 16.02 | 16.62 | 17.00 | 17.27 | 17.45 | 17.61 | 17.72 | 17.87 | 18.21 | 18.32 | 18.42 | 18.55 | 18.63 | 18.66 | 18.71 | 18.73 | 18.74 | 18.75 | 18.77 | 18.78 | 18.78 |
|  | $K$ | 2.39 | 4.00 | 4.79 | 5.11 | 5.23 | 5.28 | 5.29 | 5.30 | 5.30 | 5.30 | 5.30 | 5.30 | 5.30 | 5.30 | 5.30 | 5.30 | 5.30 | 5.30 | 5.30 | 5.30 | 5.30 | 5.30 | 5.30 | 5.30 | 5.30 |

| $u_0/u_1$ | $\xi$/nm | 2 | 4 | 6 | 8 | 10 | 12 | 14 | 16 | 18 | 20 | 22 | 24 | 26 | 28 | 30 | 32 | 34 | 36 | 38 | 40 | 42 | 44 | 46 | 48 | 50 |
|---|---|---|---|---|---|---|---|---|---|---|---|---|---|---|---|---|---|---|---|---|---|---|---|---|---|---|
| 0.6 | $U_1$ | 16.00 | 28.48 | 37.28 | 43.60 | 48.27 | 51.83 | 54.61 | 56.84 | 58.66 | 60.18 | 61.45 | 62.54 | 63.48 | 64.29 | 65.01 | 65.64 | 66.21 | 66.71 | 67.17 | 67.58 | 67.96 | 68.31 | 68.61 | 68.90 | 69.17 |
| | $U_2$ | 0.22 | 0.48 | 0.82 | 1.31 | 1.94 | 2.65 | 3.41 | 4.17 | 4.90 | 5.61 | 6.27 | 6.88 | 7.45 | 7.98 | 8.46 | 8.91 | 9.32 | 9.71 | 10.06 | 10.39 | 10.69 | 10.97 | 11.23 | 11.48 | 11.71 |
| | $W_{1,2}$ | 6.29 | 13.34 | 21.10 | 29.25 | 37.57 | 45.05 | 54.36 | 51.89 | 62.78 | 71.19 | 79.61 | 88.03 | 96.45 | 104.87 | 113.29 | 121.71 | 130.13 | 138.55 | 146.97 | 155.39 | 163.81 | 172.23 | 180.65 | 189.07 | 205.91 |
| | $W_{3,4}$ | 8.74 | 17.74 | 26.45 | 34.59 | 43.46 | 51.89 | 60.32 | 68.74 | 77.16 | 85.58 | 94.00 | 102.42 | 110.84 | 119.26 | 127.68 | 136.10 | 144.52 | 152.94 | 161.36 | 169.78 | 178.20 | 186.62 | 195.04 | 203.46 | 211.88 |
| | $J$ | 6.95 | 11.57 | 14.39 | 16.18 | 17.39 | 18.23 | 18.84 | 19.30 | 19.64 | 19.90 | 20.10 | 20.26 | 20.38 | 20.48 | 20.56 | 20.63 | 20.68 | 20.72 | 20.76 | 20.79 | 20.81 | 20.84 | 20.85 | 20.87 | 20.88 |
| | $K$ | 2.40 | 4.02 | 4.81 | 5.14 | 5.26 | 5.31 | 5.32 | 5.33 | 5.33 | 5.33 | 5.33 | 5.33 | 5.33 | 5.33 | 5.33 | 5.33 | 5.33 | 5.33 | 5.33 | 5.33 | 5.33 | 5.33 | 5.33 | 5.33 | 5.33 |
| 0.5 | $U_1$ | 14.74 | 26.28 | 34.62 | 40.67 | 45.18 | 48.65 | 51.37 | 53.56 | 55.35 | 56.84 | 58.10 | 59.18 | 60.11 | 60.92 | 61.63 | 62.26 | 62.82 | 63.33 | 63.78 | 64.19 | 64.57 | 64.91 | 65.22 | 65.51 | 65.77 |
| | $U_2$ | 0.24 | 0.51 | 0.88 | 1.39 | 2.03 | 2.75 | 3.51 | 4.27 | 5.01 | 5.71 | 6.37 | 6.98 | 7.55 | 8.07 | 8.55 | 9.00 | 9.41 | 9.79 | 10.14 | 10.47 | 10.77 | 11.05 | 11.32 | 11.56 | 11.79 |
| | $W_{1,2}$ | 6.18 | 13.15 | 20.87 | 29.01 | 37.33 | 45.71 | 54.11 | 62.53 | 70.94 | 79.36 | 87.78 | 96.20 | 104.62 | 113.04 | 121.46 | 129.88 | 138.30 | 146.72 | 155.14 | 163.56 | 171.98 | 180.40 | 188.82 | 197.24 | 205.66 |
| | $W_{3,4}$ | 8.61 | 17.32 | 25.92 | 34.42 | 42.87 | 51.31 | 59.73 | 68.15 | 76.57 | 84.99 | 93.41 | 101.83 | 110.25 | 118.67 | 127.09 | 135.51 | 143.93 | 152.35 | 160.77 | 169.19 | 177.61 | 186.03 | 194.45 | 202.87 | 211.29 |
| | $J$ | 7.07 | 11.96 | 15.07 | 17.12 | 18.53 | 19.54 | 20.27 | 20.82 | 21.24 | 21.55 | 21.80 | 21.99 | 22.15 | 22.27 | 22.37 | 22.45 | 22.51 | 22.57 | 22.61 | 22.65 | 22.68 | 22.71 | 22.73 | 22.75 | 22.77 |
| | $K$ | 2.44 | 4.09 | 4.90 | 5.23 | 5.36 | 5.41 | 5.42 | 5.43 | 5.43 | 5.43 | 5.43 | 5.43 | 5.43 | 5.43 | 5.43 | 5.43 | 5.43 | 5.43 | 5.43 | 5.43 | 5.43 | 5.43 | 5.43 | 5.43 | 5.43 |
| 0.4 | $U_1$ | 13.72 | 24.61 | 32.57 | 38.42 | 42.81 | 46.20 | 48.87 | 51.03 | 52.80 | 54.27 | 55.52 | 56.59 | 57.51 | 58.32 | 59.03 | 59.65 | 60.21 | 60.71 | 61.16 | 61.57 | 61.95 | 62.29 | 62.60 | 62.89 | 63.15 |
| | $U_2$ | 0.25 | 0.54 | 0.93 | 1.45 | 2.10 | 2.83 | 3.59 | 4.35 | 5.09 | 5.79 | 6.44 | 7.05 | 7.62 | 8.14 | 8.62 | 9.07 | 9.48 | 9.86 | 10.21 | 10.54 | 10.84 | 11.12 | 11.38 | 11.62 | 11.85 |
| | $W_{1,2}$ | 6.02 | 12.89 | 20.58 | 28.70 | 37.00 | 45.38 | 53.78 | 62.20 | 70.62 | 79.03 | 87.45 | 95.87 | 104.29 | 112.71 | 121.13 | 129.55 | 137.97 | 146.39 | 154.81 | 163.23 | 171.65 | 180.07 | 188.49 | 196.91 | 205.33 |
| | $W_{3,4}$ | 8.44 | 17.02 | 25.56 | 34.03 | 42.47 | 50.90 | 59.32 | 67.74 | 76.16 | 84.58 | 93.00 | 101.42 | 109.84 | 118.26 | 126.68 | 135.10 | 143.52 | 151.94 | 160.36 | 168.78 | 177.20 | 185.62 | 194.04 | 202.46 | 210.88 |
| | $J$ | 7.20 | 12.34 | 15.71 | 17.99 | 19.58 | 20.72 | 21.56 | 22.19 | 22.67 | 23.04 | 23.33 | 23.55 | 23.73 | 23.87 | 23.99 | 24.08 | 24.16 | 24.22 | 24.27 | 24.32 | 24.36 | 24.39 | 24.42 | 24.44 | 24.46 |
| | $K$ | 2.51 | 4.21 | 5.04 | 5.38 | 5.51 | 5.56 | 5.58 | 5.59 | 5.59 | 5.59 | 5.59 | 5.59 | 5.59 | 5.59 | 5.59 | 5.59 | 5.59 | 5.59 | 5.59 | 5.59 | 5.59 | 5.59 | 5.59 | 5.59 | 5.59 |
| 0.3 | $U_1$ | 12.98 | 23.38 | 31.07 | 36.76 | 41.00 | 44.38 | 47.02 | 49.15 | 50.90 | 52.37 | 53.61 | 54.67 | 55.58 | 56.39 | 57.09 | 57.71 | 58.27 | 58.77 | 59.22 | 59.63 | 60.00 | 60.34 | 60.65 | 60.94 | 61.20 |
| | $U_2$ | 0.26 | 0.56 | 0.96 | 1.50 | 2.15 | 2.88 | 3.65 | 4.41 | 5.15 | 5.84 | 6.49 | 7.11 | 7.67 | 8.20 | 8.68 | 9.12 | 9.53 | 9.91 | 10.26 | 10.58 | 10.89 | 11.16 | 11.43 | 11.67 | 11.90 |
| | $W_{1,2}$ | 5.86 | 12.62 | 20.25 | 28.35 | 36.65 | 45.02 | 53.43 | 61.84 | 70.26 | 78.68 | 87.10 | 95.52 | 103.94 | 112.36 | 120.78 | 129.20 | 137.62 | 146.04 | 154.46 | 162.88 | 171.30 | 179.72 | 188.14 | 196.56 | 204.98 |
| | $W_{3,4}$ | 8.34 | 16.82 | 25.31 | 33.76 | 42.19 | 50.62 | 59.04 | 67.46 | 75.88 | 84.30 | 92.72 | 101.14 | 109.56 | 117.98 | 126.40 | 134.82 | 143.24 | 151.66 | 160.08 | 168.50 | 176.92 | 185.34 | 193.76 | 202.18 | 210.60 |
| | $J$ | 7.34 | 12.60 | 16.28 | 18.74 | 20.48 | 21.75 | 22.68 | 23.36 | 23.91 | 24.32 | 24.64 | 24.89 | 25.09 | 25.25 | 25.38 | 25.48 | 25.57 | 25.64 | 25.70 | 25.75 | 25.79 | 25.83 | 25.86 | 25.89 | 25.91 |
| | $K$ | 2.59 | 4.34 | 5.20 | 5.55 | 5.69 | 5.74 | 5.76 | 5.77 | 5.77 | 5.77 | 5.77 | 5.77 | 5.77 | 5.77 | 5.77 | 5.77 | 5.77 | 5.77 | 5.77 | 5.77 | 5.77 | 5.77 | 5.77 | 5.77 | 5.77 |
| 0.2 | $U_1$ | 12.92 | 23.32 | 31.05 | 36.73 | 41.11 | 44.47 | 47.12 | 49.27 | 51.03 | 52.51 | 53.75 | 54.82 | 55.74 | 56.54 | 57.25 | 57.88 | 58.44 | 58.94 | 59.35 | 59.76 | 60.17 | 60.47 | 60.78 | 61.11 | 61.33 |
| | $U_2$ | 0.23 | 0.52 | 0.90 | 1.43 | 2.09 | 2.82 | 3.55 | 4.34 | 5.08 | 5.78 | 6.41 | 7.02 | 7.59 | 8.11 | 8.59 | 9.04 | 9.45 | 9.83 | 10.18 | 10.51 | 10.81 | 11.09 | 11.35 | 11.59 | 11.82 |
| | $W_{1,2}$ | 5.65 | 12.28 | 19.85 | 27.92 | 36.21 | 44.53 | 52.99 | 61.40 | 69.82 | 78.24 | 86.66 | 95.08 | 103.50 | 111.92 | 120.34 | 128.76 | 137.17 | 145.59 | 154.01 | 162.43 | 170.85 | 179.27 | 187.69 | 196.11 | 204.53 |
| | $W_{3,4}$ | 8.26 | 16.68 | 25.14 | 33.58 | 42.00 | 50.43 | 58.85 | 67.27 | 75.69 | 84.11 | 92.53 | 100.95 | 109.37 | 117.79 | 126.21 | 134.63 | 143.05 | 151.47 | 159.89 | 168.31 | 176.73 | 185.15 | 193.57 | 201.99 | 210.41 |
| | $J$ | 7.45 | 12.91 | 16.59 | 19.11 | 20.88 | 22.14 | 23.07 | 23.76 | 24.28 | 24.67 | 24.98 | 25.21 | 25.40 | 25.55 | 25.67 | 25.77 | 25.85 | 25.92 | 25.96 | 26.01 | 26.05 | 26.08 | 26.11 | 26.13 | 26.15 |
| | $K$ | 2.67 | 4.47 | 5.36 | 5.72 | 5.86 | 5.91 | 5.93 | 5.94 | 5.94 | 5.94 | 5.94 | 5.94 | 5.94 | 5.94 | 5.94 | 5.94 | 5.94 | 5.94 | 5.94 | 5.94 | 5.94 | 5.94 | 5.94 | 5.94 | 5.94 |
| 0.1 | $U_1$ | 12.85 | 23.22 | 30.95 | 36.69 | 41.03 | 44.39 | 47.00 | 49.21 | 50.98 | 52.45 | 53.70 | 54.77 | 55.69 | 56.50 | 57.21 | 57.83 | 58.39 | 58.89 | 59.35 | 59.76 | 60.13 | 60.47 | 60.78 | 61.07 | 61.33 |
| | $U_2$ | 0.22 | 0.51 | 0.89 | 1.42 | 2.08 | 2.81 | 3.57 | 4.34 | 5.08 | 5.78 | 6.43 | 7.04 | 7.61 | 8.13 | 8.61 | 9.05 | 9.46 | 9.84 | 10.19 | 10.51 | 10.82 | 11.10 | 11.35 | 11.59 | 11.82 |
| | $W_{1,2}$ | 5.51 | 12.05 | 19.58 | 27.63 | 35.92 | 44.28 | 52.68 | 61.10 | 69.52 | 77.93 | 86.35 | 94.77 | 103.19 | 111.61 | 120.03 | 128.45 | 136.87 | 145.29 | 153.71 | 162.13 | 170.55 | 178.97 | 187.39 | 195.81 | 204.23 |
| | $W_{3,4}$ | 8.21 | 16.60 | 25.04 | 33.47 | 41.90 | 50.32 | 58.74 | 67.16 | 75.58 | 84.00 | 92.42 | 100.84 | 109.26 | 117.68 | 126.10 | 134.52 | 142.94 | 151.36 | 159.78 | 168.20 | 176.62 | 185.04 | 193.46 | 201.88 | 210.30 |
| | $J$ | 7.52 | 13.09 | 16.83 | 19.38 | 21.18 | 22.46 | 23.39 | 24.08 | 24.60 | 24.99 | 25.29 | 25.52 | 25.71 | 25.86 | 25.97 | 26.06 | 26.14 | 26.20 | 26.24 | 26.29 | 26.33 | 26.36 | 26.39 | 26.41 | 26.43 |
| | $K$ | 2.72 | 4.57 | 5.48 | 5.85 | 5.99 | 6.04 | 6.06 | 6.07 | 6.07 | 6.07 | 6.07 | 6.07 | 6.07 | 6.07 | 6.07 | 6.07 | 6.07 | 6.07 | 6.07 | 6.07 | 6.07 | 6.07 | 6.07 | 6.07 | 6.07 |
| 0.0 | $U_1$ | 12.72 | 23.00 | 30.68 | 36.38 | 40.70 | 44.05 | 46.71 | 48.85 | 50.62 | 52.09 | 53.33 | 54.40 | 55.32 | 56.13 | 56.83 | 57.46 | 58.02 | 58.52 | 58.97 | 59.38 | 59.76 | 60.10 | 60.41 | 60.70 | 60.96 |
| | $U_2$ | 0.22 | 0.50 | 0.88 | 1.41 | 2.07 | 2.80 | 3.56 | 4.33 | 5.07 | 5.77 | 6.42 | 7.03 | 7.60 | 8.12 | 8.61 | 9.05 | 9.46 | 9.84 | 10.19 | 10.52 | 10.82 | 11.10 | 11.36 | 11.60 | 11.83 |
| | $W_{1,2}$ | 5.48 | 12.05 | 19.50 | 27.56 | 35.84 | 44.21 | 52.61 | 61.02 | 69.44 | 77.86 | 86.28 | 94.70 | 103.11 | 111.53 | 119.96 | 128.38 | 136.80 | 145.22 | 153.64 | 162.06 | 170.48 | 178.90 | 187.32 | 195.74 | 204.16 |
| | $W_{3,4}$ | 8.20 | 16.58 | 25.04 | 33.47 | 41.87 | 50.29 | 58.71 | 67.13 | 75.55 | 83.97 | 92.39 | 100.81 | 109.23 | 117.65 | 126.07 | 134.49 | 142.91 | 151.33 | 159.75 | 168.17 | 176.59 | 185.01 | 193.43 | 201.85 | 210.27 |
| | $J$ | 7.57 | 13.17 | 16.97 | 19.58 | 21.42 | 22.73 | 23.68 | 24.39 | 24.92 | 25.32 | 25.63 | 25.86 | 26.05 | 26.20 | 26.32 | 26.41 | 26.49 | 26.56 | 26.62 | 26.65 | 26.69 | 26.72 | 26.75 | 26.77 | 26.79 |
| | $K$ | 2.75 | 4.61 | 5.52 | 5.89 | 6.04 | 6.09 | 6.11 | 6.12 | 6.12 | 6.12 | 6.12 | 6.12 | 6.12 | 6.12 | 6.12 | 6.12 | 6.12 | 6.12 | 6.12 | 6.12 | 6.12 | 6.12 | 6.12 | 6.12 | 6.12 |

TABLE S33: Parameters of the THF interaction Hamiltonian for different ratios $u_0/u_1$ and screening lengths $\xi$ of the double-gate interaction potential ($U_\xi = 24$ meV for $\xi = 10$nm). We use the same BM model parameters as in Table S32.

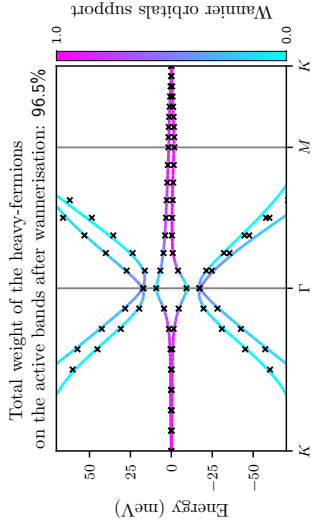

Total weight of the heavy-fermions on the active bands after wannierisation: 96.5%

Wannier orbitals support  0.0 — 1.0

FIG. S33. Band structures of the BM and THF models near charge neutrality for $w_0/w_1 = 0.8$, depicted by lines and crosses, respectively. The BM bands are colored according to the weight of the $f$-electron wave function on them. We use the same BM parameters as in Table S34.

TABLE S34. Parameters of the $f$-electron Wannier functions for different values of the tunneling ratio $w_0/w_1$. $\mathcal{W}$ denotes the total weight of the THF $f$-electrons on the active bands. In computing the ratios $\gamma'/U_1$ and $\gamma'/U_2$, we employ the on-site and nearest-neighbor repulsion parameters $U_1$ and $U_2$ obtained numerically for $\xi = 10$ nm, as given in Table S35. We employ $v_F = 5.944\,\mathrm{eV\,\mathring{A}}$, $|\mathbf{K}| = 1.703\,\mathring{A}^{-1}$, $w_1 = 110$ meV, and $\theta = 1.00°$ for the BM model.

| $w_0/w_1$ | $\alpha_1$ | $\alpha_2$ | $\lambda_1$ | $\lambda_2$ | $\lambda$ | $\mathcal{W}$ | $t_0$ | $\gamma$ | $M$ | $v_\star$ | $v_\star'$ | $\gamma'/U_1$ | $\gamma'/U_2$ |
|---|---|---|---|---|---|---|---|---|---|---|---|---|---|
| 1.00 | 0.756 | 0.655 | 0.149 | 0.167 | 0.360 | 0.975 | 0.134 | 9.906 | 9.583 | -3.555 | 1.388 | 0.0266 | 5.04 |
| 0.90 | 0.780 | 0.625 | 0.161 | 0.179 | 0.331 | 0.984 | 0.256 | -3.673 | 9.427 | -3.866 | 1.502 | -0.0140 | -2.03 |
| 0.80 | 0.811 | 0.585 | 0.173 | 0.185 | 0.332 | 0.965 | 0.434 | -17.752 | 9.241 | -4.117 | 1.567 | -0.0375 | -9.64 |
| 0.70 | 0.845 | 0.535 | 0.191 | 0.191 | 0.346 | 0.947 | 0.628 | -31.829 | 9.046 | -4.324 | 1.584 | -0.0481 | -16.32 |
| 0.60 | 0.879 | 0.477 | 0.203 | 0.197 | 0.361 | 0.932 | 0.804 | -45.469 | 8.858 | -4.498 | 1.552 | -0.0494 | -22.10 |
| 0.50 | 0.911 | 0.412 | 0.215 | 0.203 | 0.373 | 0.921 | 0.956 | -58.258 | 8.688 | -4.648 | 1.465 | -0.0439 | -27.13 |
| 0.40 | 0.940 | 0.341 | 0.233 | 0.209 | 0.383 | 0.913 | 1.086 | -69.761 | 8.541 | -4.777 | 1.313 | -0.0341 | -31.47 |
| 0.30 | 0.965 | 0.263 | 0.245 | 0.215 | 0.392 | 0.910 | 1.207 | -79.505 | 8.423 | -4.885 | 1.088 | -0.0223 | -34.85 |
| 0.20 | 0.984 | 0.180 | 0.257 | 0.215 | 0.385 | 0.897 | 1.380 | -86.985 | 8.337 | -4.970 | 0.785 | -0.0111 | -39.17 |
| 0.10 | 0.996 | 0.093 | 0.263 | 0.221 | 0.380 | 0.889 | 1.478 | -91.718 | 8.284 | -5.024 | 0.413 | -0.0030 | -41.99 |
| 0.00 | 1.000 | 0.000 | 0.263 | 0.000 | 0.380 | 0.888 | 1.516 | -93.341 | 8.267 | -5.043 | 0.0000 | -0.0000 | -42.80 |

| $w_0/w_1$ | $\xi$/nm | 2 | 4 | 6 | 8 | 10 | 12 | 14 | 16 | 18 | 20 | 22 | 24 | 26 | 28 | 30 | 32 | 34 | 36 | 38 | 40 | 42 | 44 | 46 | 48 | 50 |
|---|---|---|---|---|---|---|---|---|---|---|---|---|---|---|---|---|---|---|---|---|---|---|---|---|---|---|
| 1.0 | $U_1$ | 23.15 | 39.49 | 50.18 | 57.42 | 62.56 | 66.37 | 69.31 | 71.62 | 73.50 | 75.05 | 76.35 | 77.45 | 78.40 | 79.23 | 79.95 | 80.59 | 81.16 | 81.66 | 82.12 | 82.54 | 82.91 | 83.26 | 83.57 | 83.86 | 84.13 |
| | $U_2$ | 0.25 | 0.57 | 0.87 | 1.34 | 1.97 | 2.69 | 3.46 | 4.24 | 5.00 | 5.72 | 6.39 | 7.02 | 7.60 | 8.14 | 8.63 | 9.09 | 9.50 | 9.89 | 10.25 | 10.58 | 10.89 | 11.17 | 11.43 | 11.68 | 11.91 |
| | $W_{1,2}$ | 6.32 | 13.58 | 21.65 | 30.14 | 38.81 | 47.54 | 56.29 | 65.05 | 73.82 | 82.58 | 91.35 | 100.12 | 108.89 | 117.65 | 126.42 | 135.19 | 143.95 | 152.72 | 161.49 | 170.26 | 179.02 | 187.79 | 196.56 | 205.32 | 214.09 |
| | $W_{3,4}$ | 11.15 | 21.87 | 31.65 | 40.83 | 49.75 | 58.58 | 67.36 | 76.14 | 84.91 | 93.68 | 102.44 | 111.21 | 119.98 | 128.75 | 137.51 | 146.28 | 155.05 | 163.81 | 172.58 | 181.35 | 190.11 | 198.88 | 207.65 | 216.42 | 225.18 |
| | $J$ | 7.15 | 10.83 | 12.45 | 13.51 | 13.68 | 13.76 | 13.81 | 13.84 | 13.83 | 13.85 | 13.85 | 13.85 | 13.86 | 13.86 | 13.86 | 13.86 | 13.86 | 13.86 | 13.86 | 13.86 | 13.86 | 13.86 | 13.86 | 13.86 | 13.86 |
| | $K$ | 2.59 | 4.28 | 5.07 | 5.39 | 5.50 | 5.55 | 5.56 | 5.67 | 5.57 | 5.57 | 5.57 | 5.57 | 5.57 | 5.57 | 5.57 | 5.57 | 5.57 | 5.57 | 5.57 | 5.57 | 5.57 | 5.57 | 5.57 | 5.57 | 5.57 |
| 0.9 | $U_1$ | 21.92 | 37.75 | 48.33 | 55.58 | 60.77 | 64.63 | 67.61 | 69.96 | 71.86 | 73.43 | 74.75 | 75.87 | 76.83 | 77.66 | 78.39 | 79.04 | 79.61 | 80.12 | 80.59 | 81.00 | 81.38 | 81.72 | 82.04 | 82.33 | 82.60 |
| | $U_2$ | 0.20 | 0.42 | 0.73 | 1.19 | 1.81 | 2.53 | 3.31 | 4.09 | 4.86 | 5.58 | 6.26 | 6.90 | 7.48 | 8.02 | 8.52 | 8.98 | 9.40 | 9.79 | 10.14 | 10.48 | 10.78 | 11.07 | 11.34 | 11.58 | 11.81 |
| | $W_{1,2}$ | 6.54 | 13.91 | 22.03 | 30.54 | 39.22 | 47.95 | 56.70 | 65.47 | 74.23 | 83.00 | 91.76 | 100.53 | 109.30 | 118.07 | 126.83 | 135.60 | 144.37 | 153.13 | 161.90 | 170.67 | 179.44 | 188.20 | 196.97 | 205.74 | 214.50 |
| | $W_{3,4}$ | 10.40 | 20.58 | 30.09 | 39.17 | 48.05 | 56.86 | 65.64 | 74.41 | 83.18 | 91.95 | 100.72 | 109.49 | 118.25 | 127.02 | 135.79 | 144.55 | 153.32 | 162.09 | 170.85 | 179.62 | 188.39 | 197.16 | 205.92 | 214.69 | 223.46 |
| | $J$ | 7.00 | 10.86 | 12.71 | 13.65 | 14.15 | 14.44 | 14.62 | 14.74 | 14.82 | 14.88 | 14.92 | 14.96 | 14.98 | 15.00 | 15.01 | 15.03 | 15.04 | 15.05 | 15.06 | 15.06 | 15.07 | 15.07 | 15.07 | 15.07 | 15.07 |
| | $K$ | 2.47 | 4.10 | 4.86 | 5.17 | 5.28 | 5.33 | 5.34 | 5.35 | 5.35 | 5.35 | 5.35 | 5.35 | 5.35 | 5.35 | 5.35 | 5.35 | 5.35 | 5.35 | 5.35 | 5.35 | 5.35 | 5.35 | 5.35 | 5.35 | 5.35 |
| 0.8 | $U_1$ | 19.99 | 34.77 | 44.84 | 51.86 | 56.92 | 60.72 | 63.66 | 65.99 | 67.89 | 69.45 | 70.76 | 71.87 | 72.83 | 73.66 | 74.39 | 75.03 | 75.61 | 76.12 | 76.58 | 76.99 | 77.37 | 77.72 | 78.04 | 78.33 | 78.58 |
| | $U_2$ | 0.19 | 0.42 | 0.73 | 1.21 | 1.84 | 2.58 | 3.36 | 4.14 | 4.91 | 5.64 | 6.32 | 6.95 | 7.53 | 8.07 | 8.57 | 9.03 | 9.45 | 9.84 | 10.19 | 10.53 | 10.84 | 11.12 | 11.39 | 11.64 | 11.87 |
| | $W_{1,2}$ | 6.65 | 14.08 | 22.22 | 30.75 | 39.42 | 48.15 | 56.91 | 65.67 | 74.44 | 83.20 | 91.97 | 100.74 | 109.51 | 118.27 | 127.04 | 135.81 | 144.57 | 153.34 | 162.11 | 170.88 | 179.64 | 188.41 | 197.18 | 205.94 | 214.71 |
| | $W_{3,4}$ | 9.82 | 19.56 | 28.85 | 37.84 | 46.69 | 55.48 | 64.26 | 73.03 | 81.80 | 90.57 | 99.34 | 108.10 | 116.87 | 125.64 | 134.40 | 143.17 | 151.94 | 160.71 | 169.47 | 178.24 | 187.01 | 195.77 | 204.54 | 213.31 | 222.07 |
| | $J$ | 6.99 | 11.13 | 13.32 | 14.56 | 15.30 | 15.79 | 16.12 | 16.35 | 16.52 | 16.65 | 16.75 | 16.82 | 16.88 | 16.93 | 16.96 | 16.99 | 17.02 | 17.04 | 17.05 | 17.07 | 17.08 | 17.09 | 17.10 | 17.10 | 17.11 |
| | $K$ | 2.40 | 3.99 | 4.75 | 5.05 | 5.16 | 5.20 | 5.22 | 5.22 | 5.23 | 5.23 | 5.23 | 5.23 | 5.23 | 5.23 | 5.23 | 5.23 | 5.23 | 5.23 | 5.23 | 5.23 | 5.23 | 5.23 | 5.23 | 5.23 | 5.23 |
| 0.7 | $U_1$ | 18.00 | 31.58 | 41.03 | 47.71 | 52.59 | 56.28 | 59.15 | 61.44 | 63.30 | 64.84 | 66.13 | 67.24 | 68.18 | 69.01 | 69.73 | 70.37 | 70.94 | 71.45 | 71.91 | 72.33 | 72.70 | 73.05 | 73.36 | 73.65 | 73.92 |
| | $U_2$ | 0.21 | 0.46 | 0.80 | 1.30 | 1.95 | 2.69 | 3.48 | 4.27 | 5.03 | 5.76 | 6.44 | 7.07 | 7.65 | 8.19 | 8.68 | 9.14 | 9.56 | 9.95 | 10.30 | 10.63 | 10.94 | 11.23 | 11.49 | 11.74 | 11.97 |
| | $W_{1,2}$ | 6.66 | 14.10 | 22.24 | 30.76 | 39.44 | 48.17 | 56.93 | 65.69 | 74.45 | 83.22 | 91.99 | 100.75 | 109.52 | 118.29 | 127.06 | 135.82 | 144.59 | 153.36 | 162.13 | 170.89 | 179.66 | 188.43 | 197.19 | 205.96 | 214.73 |
| | $W_{3,4}$ | 9.38 | 18.79 | 27.91 | 36.83 | 45.65 | 54.44 | 63.21 | 71.98 | 80.75 | 89.52 | 98.29 | 107.05 | 115.82 | 124.59 | 133.35 | 142.12 | 150.89 | 159.65 | 168.42 | 177.19 | 185.96 | 194.72 | 203.49 | 212.26 | 221.02 |
| | $J$ | 7.04 | 11.49 | 14.00 | 15.61 | 16.62 | 17.31 | 17.80 | 18.15 | 18.42 | 18.62 | 18.77 | 18.90 | 19.00 | 19.06 | 19.11 | 19.17 | 19.21 | 19.25 | 19.27 | 19.30 | 19.31 | 19.33 | 19.34 | 19.36 | 19.37 |
| | $K$ | 2.38 | 3.96 | 4.72 | 5.03 | 5.13 | 5.17 | 5.19 | 5.19 | 5.19 | 5.19 | 5.19 | 5.19 | 5.19 | 5.19 | 5.19 | 5.19 | 5.19 | 5.19 | 5.19 | 5.19 | 5.19 | 5.19 | 5.19 | 5.19 | 5.19 |

| $u_0/u_1$ | $\xi/\mathrm{nm}$ | 2 | 4 | 6 | 8 | 10 | 12 | 14 | 16 | 18 | 20 | 22 | 24 | 26 | 28 | 30 | 32 | 34 | 36 | 38 | 40 | 42 | 44 | 46 | 48 | 50 |
|---|---|---|---|---|---|---|---|---|---|---|---|---|---|---|---|---|---|---|---|---|---|---|---|---|---|---|
| 0.6 | $U_1$ | 16.32 | 28.86 | 37.75 | 44.13 | 48.83 | 52.41 | 55.21 | 57.45 | 59.28 | 60.80 | 62.08 | 63.17 | 64.11 | 64.93 | 65.65 | 66.28 | 66.85 | 67.36 | 67.81 | 68.22 | 68.60 | 68.94 | 69.26 | 69.55 | 69.81 |
| | $U_2$ | 0.23 | 0.50 | 0.87 | 1.39 | 2.06 | 2.81 | 3.60 | 4.39 | 5.15 | 5.88 | 6.55 | 7.18 | 7.76 | 8.30 | 8.79 | 9.25 | 9.66 | 10.05 | 10.41 | 10.74 | 11.04 | 11.33 | 11.59 | 11.84 | 12.07 |
| | $W_{1,2}$ | 6.60 | 13.98 | 22.11 | 30.62 | 39.29 | 48.02 | 56.78 | 65.54 | 74.31 | 83.07 | 91.84 | 100.61 | 109.37 | 118.14 | 126.91 | 135.68 | 144.44 | 153.21 | 161.98 | 170.74 | 179.51 | 188.28 | 197.04 | 205.81 | 214.58 |
| | $W_{3,4}$ | 9.07 | 18.23 | 27.22 | 36.09 | 44.89 | 53.68 | 62.45 | 71.22 | 79.98 | 88.75 | 97.52 | 106.29 | 115.05 | 123.82 | 132.59 | 141.35 | 150.12 | 158.89 | 167.66 | 176.42 | 185.19 | 193.96 | 202.72 | 211.49 | 220.26 |
| | $J$ | 7.14 | 11.89 | 14.79 | 16.64 | 17.89 | 18.76 | 19.39 | 19.85 | 20.20 | 20.47 | 20.67 | 20.84 | 20.96 | 21.06 | 21.14 | 21.21 | 21.26 | 21.31 | 21.34 | 21.37 | 21.40 | 21.42 | 21.44 | 21.46 | 21.47 |
| | $K$ | 2.40 | 4.00 | 4.76 | 5.06 | 5.18 | 5.22 | 5.23 | 5.24 | 5.24 | 5.24 | 5.24 | 5.24 | 5.24 | 5.24 | 5.24 | 5.24 | 5.24 | 5.24 | 5.24 | 5.24 | 5.24 | 5.24 | 5.24 | 5.24 | 5.24 |
| 0.5 | $U_1$ | 15.00 | 26.71 | 35.15 | 41.27 | 45.82 | 49.31 | 52.05 | 54.26 | 56.06 | 57.56 | 58.82 | 59.90 | 60.84 | 61.65 | 62.36 | 63.00 | 63.56 | 64.06 | 64.52 | 64.93 | 65.30 | 65.64 | 65.96 | 66.25 | 66.51 |
| | $U_2$ | 0.23 | 0.53 | 0.93 | 1.47 | 2.15 | 2.91 | 3.70 | 4.49 | 5.25 | 5.97 | 6.65 | 7.28 | 7.85 | 8.39 | 8.88 | 9.33 | 9.75 | 10.10 | 10.50 | 10.82 | 11.12 | 11.41 | 11.67 | 11.92 | 12.15 |
| | $W_{1,2}$ | 6.47 | 13.78 | 21.86 | 30.36 | 39.03 | 47.76 | 56.51 | 65.27 | 74.04 | 82.80 | 91.57 | 100.34 | 109.11 | 117.87 | 126.64 | 135.41 | 144.17 | 152.94 | 161.71 | 170.48 | 179.24 | 188.01 | 196.78 | 205.54 | 214.31 |
| | $W_{3,4}$ | 8.85 | 17.83 | 26.74 | 35.57 | 44.36 | 53.13 | 61.90 | 70.67 | 79.44 | 88.21 | 96.97 | 105.74 | 114.51 | 123.27 | 132.04 | 140.81 | 149.58 | 158.34 | 167.11 | 175.88 | 184.64 | 193.41 | 202.18 | 210.95 | 219.71 |
| | $J$ | 7.27 | 12.30 | 15.51 | 17.62 | 19.07 | 20.10 | 20.85 | 21.41 | 21.83 | 22.15 | 22.40 | 22.60 | 22.75 | 22.87 | 22.97 | 23.05 | 23.12 | 23.17 | 23.22 | 23.26 | 23.29 | 23.32 | 23.34 | 23.36 | 23.38 |
| | $K$ | 2.44 | 4.08 | 4.86 | 5.17 | 5.29 | 5.33 | 5.34 | 5.35 | 5.35 | 5.35 | 5.35 | 5.35 | 5.35 | 5.35 | 5.35 | 5.35 | 5.35 | 5.35 | 5.35 | 5.35 | 5.35 | 5.35 | 5.35 | 5.35 | 5.35 |
| 0.4 | $U_1$ | 14.00 | 25.08 | 33.16 | 39.08 | 43.52 | 46.94 | 49.63 | 51.80 | 53.59 | 55.07 | 56.32 | 57.40 | 58.32 | 59.13 | 59.84 | 60.47 | 61.03 | 61.53 | 61.99 | 62.40 | 62.77 | 63.11 | 63.42 | 63.71 | 63.98 |
| | $U_2$ | 0.26 | 0.56 | 0.97 | 1.53 | 2.22 | 2.98 | 3.78 | 4.57 | 5.33 | 6.05 | 6.72 | 7.35 | 7.93 | 8.46 | 8.95 | 9.40 | 9.81 | 10.20 | 10.55 | 10.88 | 11.19 | 11.47 | 11.73 | 11.98 | 12.21 |
| | $W_{1,2}$ | 6.31 | 13.51 | 21.50 | 29.89 | 38.55 | 47.28 | 56.04 | 64.80 | 73.57 | 82.35 | 91.13 | 99.89 | 108.65 | 117.53 | 126.30 | 135.06 | 143.83 | 152.60 | 161.37 | 170.13 | 178.90 | 187.67 | 196.43 | 205.20 | 213.97 |
| | $W_{3,4}$ | 8.70 | 17.55 | 26.40 | 35.23 | 43.98 | 52.75 | 61.52 | 70.29 | 79.06 | 87.83 | 96.59 | 105.36 | 114.13 | 122.89 | 131.66 | 140.43 | 149.20 | 157.96 | 166.73 | 175.50 | 184.26 | 193.03 | 201.80 | 210.57 | 219.33 |
| | $J$ | 7.42 | 12.70 | 16.17 | 18.50 | 20.13 | 21.30 | 22.15 | 22.79 | 23.27 | 23.64 | 23.93 | 24.15 | 24.33 | 24.47 | 24.59 | 24.68 | 24.76 | 24.82 | 24.87 | 24.92 | 24.96 | 24.99 | 25.01 | 25.04 | 25.06 |
| | $K$ | 2.52 | 4.20 | 5.00 | 5.32 | 5.44 | 5.49 | 5.50 | 5.51 | 5.51 | 5.51 | 5.51 | 5.51 | 5.51 | 5.51 | 5.51 | 5.51 | 5.51 | 5.51 | 5.51 | 5.51 | 5.51 | 5.51 | 5.51 | 5.51 | 5.51 |
| 0.3 | $U_1$ | 13.22 | 23.79 | 31.58 | 37.33 | 41.66 | 45.02 | 47.67 | 49.82 | 51.58 | 53.05 | 54.29 | 55.36 | 56.28 | 57.08 | 57.79 | 58.41 | 58.97 | 59.47 | 59.92 | 60.33 | 60.71 | 61.05 | 61.36 | 61.64 | 61.91 |
| | $U_2$ | 0.27 | 0.59 | 1.02 | 1.59 | 2.28 | 3.05 | 3.85 | 4.64 | 5.40 | 6.12 | 6.79 | 7.41 | 7.99 | 8.52 | 9.01 | 9.46 | 9.87 | 10.25 | 10.61 | 10.94 | 11.24 | 11.53 | 11.79 | 12.03 | 12.26 |
| | $W_{1,2}$ | 6.14 | 13.24 | 21.23 | 29.69 | 38.34 | 47.06 | 55.82 | 64.58 | 73.34 | 82.11 | 90.88 | 99.64 | 108.41 | 117.18 | 125.95 | 134.71 | 143.48 | 152.25 | 161.01 | 169.78 | 178.55 | 187.31 | 196.08 | 204.85 | 213.62 |
| | $W_{3,4}$ | 8.60 | 17.37 | 26.17 | 34.96 | 43.73 | 52.50 | 61.27 | 70.04 | 78.80 | 87.57 | 96.34 | 105.11 | 113.87 | 122.64 | 131.41 | 140.17 | 148.94 | 157.71 | 166.47 | 175.24 | 184.01 | 192.78 | 201.54 | 210.31 | 219.08 |
| | $J$ | 7.58 | 13.19 | 16.79 | 19.33 | 21.12 | 22.42 | 23.37 | 24.08 | 24.62 | 25.02 | 25.32 | 25.56 | 25.74 | 25.89 | 26.01 | 26.11 | 26.30 | 26.21 | 26.53 | 26.48 | 26.56 | 26.53 | 26.56 | 26.62 | 26.56 |
| | $K$ | 2.60 | 4.34 | 5.17 | 5.50 | 5.63 | 5.67 | 5.69 | 5.69 | 5.70 | 5.70 | 5.70 | 5.70 | 5.70 | 5.70 | 5.70 | 5.70 | 5.70 | 5.70 | 5.70 | 5.70 | 5.70 | 5.70 | 5.70 | 5.70 | 5.70 |
| 0.2 | $U_1$ | 13.14 | 23.63 | 31.51 | 37.29 | 41.66 | 45.03 | 47.70 | 49.86 | 51.63 | 53.11 | 54.35 | 55.43 | 56.35 | 57.16 | 57.87 | 58.50 | 59.06 | 59.57 | 60.02 | 60.43 | 60.81 | 61.13 | 61.45 | 61.73 | 62.00 |
| | $U_2$ | 0.25 | 0.55 | 0.96 | 1.53 | 2.22 | 2.99 | 3.79 | 4.58 | 5.34 | 6.06 | 6.73 | 7.36 | 7.94 | 8.47 | 8.96 | 9.41 | 9.82 | 10.18 | 10.53 | 10.86 | 11.17 | 11.48 | 11.74 | 11.99 | 12.22 |
| | $W_{1,2}$ | 5.94 | 12.90 | 20.83 | 29.06 | 37.91 | 46.63 | 55.38 | 64.11 | 72.91 | 81.67 | 90.44 | 99.21 | 107.97 | 116.74 | 125.51 | 134.27 | 143.04 | 151.81 | 160.58 | 169.34 | 178.11 | 186.88 | 195.64 | 204.41 | 213.18 |
| | $W_{3,4}$ | 8.52 | 17.24 | 26.01 | 34.78 | 43.55 | 52.32 | 61.09 | 69.86 | 78.62 | 87.39 | 96.16 | 104.93 | 113.69 | 122.46 | 131.23 | 139.99 | 148.76 | 157.53 | 166.30 | 175.06 | 183.83 | 192.60 | 201.36 | 210.13 | 218.90 |
| | $J$ | 7.68 | 13.31 | 17.04 | 19.69 | 21.51 | 22.81 | 23.77 | 24.46 | 24.99 | 25.39 | 25.70 | 25.94 | 26.14 | 26.29 | 26.41 | 26.51 | 26.21 | 26.71 | 26.71 | 26.76 | 26.80 | 26.83 | 26.86 | 26.88 | 26.90 |
| | $K$ | 2.68 | 4.48 | 5.33 | 5.67 | 5.80 | 5.85 | 5.87 | 5.88 | 5.88 | 5.88 | 5.88 | 5.88 | 5.88 | 5.88 | 5.88 | 5.88 | 5.88 | 5.88 | 5.88 | 5.88 | 5.88 | 5.88 | 5.88 | 5.88 | 5.88 |
| 0.1 | $U_1$ | 13.10 | 23.66 | 31.49 | 37.30 | 41.68 | 45.07 | 47.75 | 49.92 | 51.70 | 53.18 | 54.43 | 55.50 | 56.43 | 57.24 | 57.95 | 58.58 | 59.14 | 59.64 | 60.09 | 60.50 | 60.88 | 61.22 | 61.53 | 61.81 | 62.08 |
| | $U_2$ | 0.25 | 0.55 | 0.96 | 1.53 | 2.22 | 2.99 | 3.79 | 4.58 | 5.34 | 6.06 | 6.73 | 7.36 | 7.94 | 8.47 | 8.96 | 9.38 | 9.80 | 10.18 | 10.53 | 10.86 | 11.17 | 11.45 | 11.71 | 11.96 | 12.19 |
| | $W_{1,2}$ | 5.79 | 12.66 | 20.54 | 29.06 | 37.59 | 46.31 | 55.06 | 63.83 | 72.59 | 81.36 | 90.12 | 98.89 | 107.66 | 116.42 | 125.19 | 133.96 | 142.73 | 151.49 | 160.26 | 169.03 | 177.79 | 186.56 | 195.33 | 204.10 | 212.86 |
| | $W_{3,4}$ | 8.48 | 17.16 | 25.92 | 34.69 | 43.45 | 52.22 | 60.99 | 69.76 | 78.52 | 87.29 | 96.06 | 104.82 | 113.59 | 122.36 | 131.13 | 139.89 | 148.66 | 157.43 | 166.19 | 174.96 | 183.73 | 192.50 | 201.26 | 210.03 | 218.80 |
| | $J$ | 7.77 | 13.48 | 17.34 | 19.97 | 21.81 | 23.12 | 24.07 | 24.77 | 25.29 | 25.69 | 26.01 | 26.24 | 26.44 | 26.63 | 26.77 | 26.83 | 26.84 | 26.66 | 26.71 | 27.00 | 27.04 | 27.07 | 27.10 | 27.12 | 27.14 |
| | $K$ | 2.74 | 4.58 | 5.45 | 5.80 | 5.93 | 5.98 | 6.00 | 6.01 | 6.01 | 6.01 | 6.01 | 6.01 | 6.01 | 6.01 | 6.01 | 6.01 | 6.01 | 6.01 | 6.01 | 6.01 | 6.01 | 6.01 | 6.01 | 6.01 | 6.01 |
| 0.0 | $U_1$ | 13.05 | 23.57 | 31.39 | 37.19 | 41.57 | 44.96 | 47.64 | 49.80 | 51.58 | 53.06 | 54.31 | 55.38 | 56.31 | 57.12 | 57.83 | 58.46 | 59.02 | 59.52 | 59.98 | 60.39 | 60.76 | 61.10 | 61.41 | 61.70 | 61.97 |
| | $U_2$ | 0.23 | 0.52 | 0.93 | 1.49 | 2.18 | 2.95 | 3.75 | 4.54 | 5.30 | 6.02 | 6.70 | 7.32 | 7.90 | 8.43 | 8.92 | 9.38 | 9.79 | 10.18 | 10.53 | 10.86 | 11.17 | 11.45 | 11.71 | 11.96 | 12.19 |
| | $W_{1,2}$ | 5.74 | 12.28 | 20.45 | 28.86 | 37.50 | 46.22 | 54.97 | 63.73 | 72.49 | 81.26 | 90.02 | 98.79 | 107.56 | 116.33 | 125.09 | 133.86 | 142.63 | 151.39 | 160.16 | 168.93 | 177.70 | 186.46 | 195.23 | 204.00 | 212.76 |
| | $W_{3,4}$ | 8.46 | 17.11 | 25.86 | 34.66 | 43.42 | 52.19 | 60.95 | 69.72 | 78.49 | 87.26 | 96.03 | 104.79 | 113.56 | 122.33 | 131.09 | 139.86 | 148.63 | 157.40 | 166.16 | 174.93 | 183.70 | 192.46 | 201.23 | 210.00 | 218.77 |
| | $J$ | 7.81 | 13.56 | 17.46 | 20.12 | 21.98 | 23.31 | 24.27 | 24.98 | 25.50 | 25.91 | 26.23 | 26.47 | 26.67 | 26.83 | 26.95 | 27.06 | 27.14 | 27.18 | 27.22 | 27.25 | 27.27 | 27.29 | 27.31 | 27.33 | 27.35 |
| | $K$ | 2.77 | 4.61 | 5.50 | 5.85 | 5.98 | 6.03 | 6.05 | 6.06 | 6.06 | 6.06 | 6.06 | 6.06 | 6.06 | 6.06 | 6.06 | 6.06 | 6.06 | 6.06 | 6.06 | 6.06 | 6.06 | 6.06 | 6.06 | 6.06 | 6.06 |

TABLE S35. Parameters of the THF interaction Hamiltonian for different ratios $u_0/u_1$ and screening lengths $\xi$ of the double-gate interaction potential ($U_\xi = 24\,\mathrm{meV}$ for $\xi = 10\,\mathrm{nm}$). We use the same BM model parameters as in Table S34.

| $w_0/w_1$ | $\alpha_1$ | $\alpha_2$ | $\lambda_1$ | $\lambda_2$ | $\lambda$ | $\mathcal{W}$ | $t_0$ | $\gamma$ | $M$ | $v_\star$ | $v''_\star$ | $\gamma/U_1$ | $\gamma/U_2$ |
|---|---|---|---|---|---|---|---|---|---|---|---|---|---|
| 1.00 | 0.758 | 0.652 | 0.155 | 0.173 | 0.349 | 0.979 | 0.078 | 7.658 | 7.724 | 1.432 | 0.0199 | 0.12 | 3.78 |
| 0.90 | 0.783 | 0.622 | 0.167 | 0.179 | 0.332 | 0.982 | 0.192 | -6.233 | 7.406 | 1.535 | -0.0157 | -0.10 | -3.25 |
| 0.80 | 0.814 | 0.581 | 0.179 | 0.185 | 0.334 | 0.961 | 0.333 | -20.511 | 7.087 | 1.590 | -0.0360 | -0.36 | -10.44 |
| 0.70 | 0.848 | 0.530 | 0.191 | 0.191 | 0.348 | 0.944 | 0.480 | -34.711 | 6.783 | 1.601 | -0.0450 | -0.65 | -16.73 |
| 0.60 | 0.881 | 0.472 | 0.203 | 0.203 | 0.362 | 0.930 | 0.608 | -48.423 | 6.506 | 1.564 | -0.0456 | -0.98 | -22.19 |
| 0.50 | 0.913 | 0.407 | 0.221 | 0.209 | 0.374 | 0.920 | 0.715 | -61.247 | 6.265 | 1.472 | -0.0403 | -1.32 | -26.97 |
| 0.40 | 0.942 | 0.336 | 0.233 | 0.209 | 0.384 | 0.913 | 0.804 | -72.761 | 6.063 | 1.317 | -0.0311 | -1.65 | -31.06 |
| 0.30 | 0.966 | 0.259 | 0.251 | 0.215 | 0.392 | 0.909 | 0.893 | -82.499 | 5.904 | 1.090 | -0.0203 | -1.95 | -34.31 |
| 0.20 | 0.984 | 0.178 | 0.257 | 0.221 | 0.391 | 0.901 | 1.036 | -89.965 | 5.790 | 0.785 | -0.0101 | -2.14 | -37.75 |
| 0.10 | 0.996 | 0.091 | 0.269 | 0.221 | 0.382 | 0.890 | 1.111 | -94.686 | 5.720 | 0.413 | -0.0027 | -2.24 | -40.83 |
| 0.00 | 1.000 | 0.000 | 0.269 | 0.000 | 0.380 | 0.887 | 1.158 | -96.304 | 5.697 | -5.085 | -0.0000 | -2.27 | -41.87 |

TABLE S36. Parameters of the $f$-electron Wannier functions for different values of the tunneling ratio $w_0/w_1$. $\mathcal{W}$ denotes the total weight of the THF $f$-electrons on the active bands. In computing the ratios $\gamma/U_1$ and $\gamma/U_2$, we employ the on-site and nearest-neighbor repulsion parameters $U_1$ and $U_2$ obtained numerically for $\xi = 10$ nm, as given in Table S37. We employ $v_F = 5.944$ eV·Å, $|\mathbf{K}| = 1.703$ Å$^{-1}$, $w_1 = 110$ meV, and $\theta = 1.02°$ for the BM model.



FIG. S34. Band structures of the BM and THF models near charge neutrality for $w_0/w_1 = 0.8$, depicted by lines and crosses, respectively. The BM bands are colored according to the weight of the $f$-electron wave function on them. We use the same BM parameters as in Table S36.

| $w_0/w_1$ | $\xi/\text{nm}$ | 2 | 4 | 6 | 8 | 10 | 12 | 14 | 16 | 18 | 20 | 22 | 24 | 26 | 28 | 30 | 32 | 34 | 36 | 38 | 40 | 42 | 44 | 46 | 48 | 50 |
|---|---|---|---|---|---|---|---|---|---|---|---|---|---|---|---|---|---|---|---|---|---|---|---|---|---|---|
| 1.0 | $U_1$ | 23.61 | 40.25 | 51.12 | 58.48 | 63.70 | 67.56 | 70.53 | 72.88 | 74.78 | 76.34 | 77.65 | 78.77 | 79.72 | 80.55 | 81.28 | 81.92 | 82.49 | 83.00 | 83.46 | 83.88 | 84.26 | 84.60 | 84.92 | 85.21 | 85.48 |
|  | $U_2$ | 0.25 | 0.52 | 0.86 | 1.37 | 2.02 | 2.78 | 3.59 | 4.40 | 5.19 | 5.93 | 6.62 | 7.27 | 7.86 | 8.41 | 8.91 | 9.37 | 9.80 | 10.19 | 10.55 | 10.89 | 11.20 | 11.48 | 11.75 | 12.00 | 12.23 |
|  | $W_{1,2}$ | 6.70 | 14.34 | 22.79 | 31.65 | 40.68 | 49.77 | 58.88 | 67.99 | 77.11 | 86.24 | 95.36 | 104.48 | 113.60 | 122.72 | 131.84 | 140.96 | 150.08 | 159.21 | 168.33 | 177.45 | 186.57 | 195.69 | 204.81 | 213.93 | 223.05 |
|  | $W_{3,4}$ | 11.28 | 22.18 | 32.21 | 41.69 | 50.94 | 60.11 | 69.25 | 78.38 | 87.50 | 96.62 | 105.74 | 114.87 | 123.99 | 133.11 | 142.23 | 151.35 | 160.47 | 169.59 | 178.71 | 187.84 | 196.96 | 206.08 | 215.20 | 224.32 | 233.44 |
|  | $J$ | 7.23 | 10.96 | 12.60 | 13.33 | 13.67 | 13.85 | 13.93 | 13.98 | 14.01 | 14.02 | 14.03 | 14.04 | 14.04 | 14.04 | 14.05 | 14.05 | 14.05 | 14.05 | 14.05 | 14.05 | 14.05 | 14.05 | 14.05 | 14.05 | 14.05 |
|  | $K$ | 2.55 | 4.19 | 4.95 | 5.24 | 5.35 | 5.39 | 5.40 | 5.40 | 5.40 | 5.41 | 5.41 | 5.41 | 5.41 | 5.41 | 5.41 | 5.41 | 5.41 | 5.41 | 5.41 | 5.41 | 5.41 | 5.41 | 5.41 | 5.41 | 5.41 |
| 0.9 | $U_1$ | 22.12 | 38.08 | 48.73 | 56.02 | 61.24 | 65.12 | 68.11 | 70.45 | 72.39 | 73.96 | 75.28 | 76.41 | 77.37 | 78.20 | 78.94 | 79.58 | 80.16 | 80.67 | 81.13 | 81.55 | 81.93 | 82.27 | 82.58 | 82.88 | 83.15 |
|  | $U_2$ | 0.20 | 0.43 | 0.76 | 1.26 | 1.92 | 2.68 | 3.49 | 4.31 | 5.10 | 5.85 | 6.54 | 7.19 | 7.79 | 8.34 | 8.84 | 9.30 | 9.73 | 10.12 | 10.49 | 10.82 | 11.13 | 11.42 | 11.69 | 11.94 | 12.17 |
|  | $W_{1,2}$ | 6.91 | 14.65 | 23.15 | 32.03 | 41.07 | 50.16 | 59.27 | 68.39 | 77.51 | 86.63 | 95.75 | 104.87 | 113.99 | 123.11 | 132.23 | 141.35 | 150.47 | 159.60 | 168.72 | 177.84 | 186.96 | 196.08 | 205.20 | 214.32 | 223.44 |
|  | $W_{3,4}$ | 10.56 | 20.94 | 30.72 | 40.10 | 49.32 | 58.48 | 67.61 | 76.74 | 85.86 | 94.98 | 104.10 | 113.23 | 122.35 | 131.47 | 140.59 | 149.71 | 158.83 | 167.95 | 177.07 | 186.20 | 195.32 | 204.44 | 213.56 | 222.68 | 231.80 |
|  | $J$ | 7.12 | 11.06 | 12.97 | 13.94 | 14.48 | 14.80 | 15.00 | 15.13 | 15.23 | 15.29 | 15.34 | 15.38 | 15.41 | 15.43 | 15.45 | 15.47 | 15.48 | 15.49 | 15.49 | 15.49 | 15.51 | 15.51 | 15.51 | 15.52 | 15.52 |
|  | $K$ | 2.44 | 4.03 | 4.76 | 5.05 | 5.15 | 5.19 | 5.20 | 5.21 | 5.21 | 5.21 | 5.21 | 5.21 | 5.21 | 5.21 | 5.21 | 5.21 | 5.21 | 5.21 | 5.21 | 5.21 | 5.21 | 5.21 | 5.21 | 5.21 | 5.21 |
| 0.8 | $U_1$ | 20.17 | 35.05 | 45.19 | 52.24 | 57.33 | 61.14 | 64.09 | 66.43 | 68.33 | 69.90 | 71.21 | 72.33 | 73.28 | 74.12 | 74.85 | 75.49 | 76.07 | 76.58 | 77.04 | 77.46 | 77.84 | 78.18 | 78.50 | 78.79 | 79.06 |
|  | $U_2$ | 0.20 | 0.44 | 0.78 | 1.29 | 1.96 | 2.74 | 3.55 | 4.37 | 5.16 | 5.91 | 6.61 | 7.25 | 7.85 | 8.40 | 8.90 | 9.37 | 9.79 | 10.19 | 10.55 | 10.89 | 11.19 | 11.48 | 11.75 | 12.00 | 12.23 |
|  | $W_{1,2}$ | 7.00 | 14.79 | 23.31 | 32.20 | 41.23 | 50.33 | 59.44 | 68.55 | 77.67 | 86.80 | 95.92 | 105.04 | 114.16 | 123.28 | 132.40 | 141.52 | 150.64 | 159.77 | 168.89 | 178.01 | 187.13 | 196.25 | 205.37 | 214.49 | 223.61 |
|  | $W_{3,4}$ | 10.01 | 19.98 | 29.55 | 38.86 | 48.05 | 57.19 | 66.32 | 75.44 | 84.57 | 93.69 | 102.81 | 111.93 | 121.05 | 130.18 | 139.30 | 148.42 | 157.54 | 166.66 | 175.78 | 184.90 | 194.02 | 203.15 | 212.27 | 221.39 | 230.51 |
|  | $J$ | 7.14 | 11.38 | 13.64 | 14.93 | 15.71 | 16.22 | 16.57 | 16.82 | 17.01 | 17.14 | 17.24 | 17.32 | 17.38 | 17.43 | 17.47 | 17.50 | 17.53 | 17.55 | 17.57 | 17.58 | 17.59 | 17.60 | 17.61 | 17.62 | 17.63 |
|  | $K$ | 2.38 | 3.94 | 4.66 | 4.95 | 5.05 | 5.09 | 5.10 | 5.11 | 5.11 | 5.11 | 5.11 | 5.11 | 5.11 | 5.11 | 5.11 | 5.11 | 5.11 | 5.11 | 5.11 | 5.11 | 5.11 | 5.11 | 5.11 | 5.11 | 5.11 |
| 0.7 | $U_1$ | 18.20 | 31.91 | 41.44 | 48.16 | 53.07 | 56.78 | 59.66 | 61.95 | 63.82 | 65.36 | 66.66 | 67.77 | 68.72 | 69.55 | 70.27 | 70.91 | 71.48 | 71.99 | 72.45 | 72.87 | 73.24 | 73.59 | 73.90 | 74.19 | 74.46 |
|  | $U_2$ | 0.22 | 0.48 | 0.85 | 1.39 | 2.07 | 2.86 | 3.68 | 4.50 | 5.28 | 6.03 | 6.73 | 7.37 | 7.97 | 8.51 | 9.01 | 9.48 | 9.90 | 10.29 | 10.65 | 10.99 | 11.30 | 11.59 | 11.85 | 12.10 | 12.33 |
|  | $W_{1,2}$ | 9.60 | 19.25 | 28.67 | 37.92 | 47.08 | 56.22 | 65.35 | 74.47 | 83.59 | 92.72 | 101.84 | 110.96 | 120.08 | 129.20 | 138.32 | 147.44 | 156.56 | 165.69 | 174.81 | 183.93 | 193.05 | 202.17 | 211.29 | 220.41 | 229.53 |
|  | $W_{3,4}$ | 7.22 | 11.78 | 14.43 | 16.04 | 17.08 | 17.80 | 18.31 | 18.68 | 18.95 | 19.16 | 19.32 | 19.45 | 19.54 | 19.62 | 19.68 | 19.73 | 19.77 | 19.80 | 19.83 | 19.85 | 19.87 | 19.89 | 19.90 | 19.92 | 19.93 |
|  | $K$ | 2.37 | 3.93 | 4.65 | 4.93 | 5.03 | 5.07 | 5.08 | 5.09 | 5.09 | 5.09 | 5.09 | 5.09 | 5.09 | 5.09 | 5.09 | 5.09 | 5.09 | 5.09 | 5.09 | 5.09 | 5.09 | 5.09 | 5.09 | 5.09 | 5.09 |

| $\xi$/nm | | 2 | 4 | 6 | 8 | 10 | 12 | 14 | 16 | 18 | 20 | 22 | 24 | 26 | 28 | 30 | 32 | 34 | 36 | 38 | 40 | 42 | 44 | 46 | 48 | 50 |
|---|---|---|---|---|---|---|---|---|---|---|---|---|---|---|---|---|---|---|---|---|---|---|---|---|---|---|
| $u_0/u_1$ | | | | | | | | | | | | | | | | | | | | | | | | | | |
| 0.6 | $U_1$ | 16.55 | 29.25 | 38.23 | 44.66 | 49.40 | 53.00 | 55.82 | 58.07 | 59.91 | 61.43 | 62.71 | 63.81 | 64.75 | 65.57 | 66.29 | 66.93 | 67.49 | 68.00 | 68.46 | 68.87 | 69.25 | 69.59 | 69.91 | 70.20 | 70.46 |
| | $U_2$ | 0.24 | 0.52 | 0.92 | 1.48 | 2.18 | 2.97 | 3.80 | 4.62 | 5.40 | 6.15 | 6.84 | 7.48 | 8.08 | 8.62 | 9.12 | 9.58 | 10.00 | 10.39 | 10.76 | 11.09 | 11.40 | 11.68 | 11.95 | 12.20 | 12.43 |
| | $W_{1,2}$ | 6.91 | 14.65 | 23.14 | 32.02 | 41.05 | 50.14 | 59.25 | 68.37 | 77.49 | 86.61 | 95.73 | 104.85 | 113.97 | 123.09 | 132.21 | 141.33 | 150.46 | 159.58 | 168.70 | 177.82 | 186.94 | 196.06 | 205.18 | 214.30 | 223.43 |
| | $W_{3,4}$ | 9.30 | 18.73 | 28.03 | 37.23 | 46.38 | 55.52 | 64.64 | 73.76 | 82.89 | 92.01 | 101.13 | 110.25 | 119.37 | 128.49 | 137.61 | 146.73 | 155.86 | 164.98 | 174.10 | 183.22 | 192.34 | 201.46 | 210.58 | 219.70 | 228.83 |
| | $J$ | 7.34 | 12.22 | 15.21 | 17.12 | 18.40 | 19.30 | 19.95 | 20.42 | 20.78 | 21.05 | 21.26 | 21.42 | 21.56 | 21.65 | 21.73 | 21.80 | 21.85 | 21.90 | 21.93 | 21.96 | 21.99 | 22.01 | 22.03 | 22.05 | 22.06 |
| | $K$ | 2.39 | 3.97 | 4.70 | 4.99 | 5.09 | 5.13 | 5.14 | 5.15 | 5.15 | 5.15 | 5.15 | 5.15 | 5.15 | 5.15 | 5.15 | 5.15 | 5.15 | 5.15 | 5.15 | 5.15 | 5.15 | 5.15 | 5.15 | 5.15 | 5.15 |
| 0.5 | $U_1$ | 15.26 | 27.15 | 35.69 | 41.87 | 46.47 | 49.99 | 52.75 | 54.97 | 56.78 | 58.29 | 59.56 | 60.64 | 61.58 | 62.39 | 63.11 | 63.74 | 64.30 | 64.81 | 65.27 | 65.68 | 66.05 | 66.39 | 66.71 | 67.00 | 67.26 |
| | $U_2$ | 0.25 | 0.56 | 0.98 | 1.56 | 2.27 | 3.07 | 3.90 | 4.71 | 5.50 | 6.24 | 6.93 | 7.57 | 8.16 | 8.71 | 9.21 | 9.67 | 10.09 | 10.48 | 10.84 | 11.17 | 11.48 | 11.76 | 12.03 | 12.28 | 12.51 |
| | $W_{1,2}$ | 6.78 | 14.43 | 22.88 | 31.74 | 40.76 | 49.85 | 58.96 | 68.08 | 77.20 | 86.32 | 95.44 | 104.56 | 113.68 | 122.80 | 131.93 | 141.05 | 150.17 | 159.29 | 168.41 | 177.53 | 186.65 | 195.77 | 204.90 | 214.02 | 223.14 |
| | $W_{3,4}$ | 9.10 | 18.36 | 27.58 | 36.75 | 45.89 | 55.01 | 64.14 | 73.26 | 82.38 | 91.50 | 100.62 | 109.75 | 118.87 | 127.99 | 137.11 | 146.23 | 155.35 | 164.47 | 173.59 | 182.72 | 191.84 | 200.96 | 210.08 | 219.20 | 228.32 |
| | $J$ | 7.49 | 12.66 | 15.95 | 18.12 | 19.61 | 20.66 | 21.43 | 21.99 | 22.42 | 22.74 | 22.99 | 23.19 | 23.35 | 23.47 | 23.57 | 23.65 | 23.71 | 23.77 | 23.81 | 23.85 | 23.88 | 23.91 | 23.93 | 23.95 | 23.97 |
| | $K$ | 2.45 | 4.06 | 4.81 | 5.11 | 5.21 | 5.25 | 5.27 | 5.27 | 5.27 | 5.27 | 5.27 | 5.27 | 5.27 | 5.27 | 5.27 | 5.27 | 5.27 | 5.27 | 5.27 | 5.27 | 5.27 | 5.27 | 5.27 | 5.27 | 5.27 |
| 0.4 | $U_1$ | 14.28 | 25.53 | 33.73 | 39.72 | 44.21 | 47.65 | 50.37 | 52.55 | 54.35 | 55.84 | 57.10 | 58.18 | 59.11 | 59.92 | 60.63 | 61.26 | 61.82 | 62.32 | 62.78 | 63.19 | 63.56 | 63.90 | 64.22 | 64.51 | 64.77 |
| | $U_2$ | 0.27 | 0.59 | 1.02 | 1.62 | 2.34 | 3.14 | 3.97 | 4.79 | 5.58 | 6.32 | 7.01 | 7.65 | 8.24 | 8.78 | 9.27 | 9.73 | 10.15 | 10.54 | 10.90 | 11.23 | 11.54 | 11.82 | 12.09 | 12.34 | 12.57 |
| | $W_{1,2}$ | 6.61 | 14.15 | 22.55 | 31.39 | 40.41 | 49.50 | 58.61 | 67.73 | 76.85 | 85.97 | 95.09 | 104.21 | 113.33 | 122.45 | 131.57 | 140.69 | 149.81 | 158.94 | 168.06 | 177.18 | 186.30 | 195.42 | 204.54 | 213.66 | 222.79 |
| | $W_{3,4}$ | 8.95 | 18.10 | 27.27 | 36.41 | 45.54 | 54.67 | 63.79 | 72.91 | 82.03 | 91.15 | 100.27 | 109.40 | 118.52 | 127.64 | 136.76 | 145.88 | 155.00 | 164.12 | 173.24 | 182.37 | 191.49 | 200.61 | 209.73 | 218.85 | 227.97 |
| | $J$ | 7.65 | 13.08 | 16.65 | 19.04 | 20.71 | 21.90 | 22.77 | 23.41 | 23.90 | 24.27 | 24.56 | 24.79 | 24.97 | 25.11 | 25.22 | 25.32 | 25.39 | 25.45 | 25.50 | 25.55 | 25.59 | 25.62 | 25.64 | 25.67 | 25.69 |
| | $K$ | 2.53 | 4.19 | 4.97 | 5.27 | 5.38 | 5.42 | 5.43 | 5.44 | 5.44 | 5.44 | 5.44 | 5.44 | 5.44 | 5.44 | 5.44 | 5.44 | 5.44 | 5.44 | 5.44 | 5.44 | 5.44 | 5.44 | 5.44 | 5.44 | 5.44 |
| 0.3 | $U_1$ | 13.52 | 24.28 | 32.20 | 38.03 | 42.42 | 45.80 | 48.48 | 50.64 | 52.42 | 53.89 | 55.15 | 56.22 | 57.14 | 57.95 | 58.65 | 59.28 | 59.84 | 60.34 | 60.80 | 61.21 | 61.58 | 61.92 | 62.23 | 62.52 | 62.79 |
| | $U_2$ | 0.28 | 0.61 | 1.06 | 1.67 | 2.40 | 3.21 | 4.04 | 4.86 | 5.65 | 6.38 | 7.07 | 7.71 | 8.30 | 8.84 | 9.33 | 9.79 | 10.21 | 10.60 | 10.96 | 11.29 | 11.59 | 11.88 | 12.14 | 12.39 | 12.62 |
| | $W_{1,2}$ | 6.43 | 13.87 | 22.23 | 31.04 | 40.06 | 49.14 | 58.25 | 67.36 | 76.48 | 85.60 | 94.72 | 103.85 | 112.97 | 122.09 | 131.21 | 140.33 | 149.45 | 158.57 | 167.69 | 176.82 | 185.94 | 195.06 | 204.18 | 213.30 | 222.42 |
| | $W_{3,4}$ | 8.86 | 17.93 | 27.06 | 36.19 | 45.31 | 54.43 | 63.56 | 72.68 | 81.80 | 90.92 | 100.04 | 109.16 | 118.28 | 127.40 | 136.52 | 145.64 | 154.77 | 163.89 | 173.01 | 182.13 | 191.25 | 200.37 | 209.49 | 218.62 | 227.74 |
| | $J$ | 7.81 | 13.74 | 17.66 | 20.34 | 22.23 | 23.58 | 24.70 | 25.46 | 25.98 | 26.33 | 26.59 | 26.82 | 27.01 | 27.13 | 27.24 | 27.33 | 27.41 | 27.48 | 27.53 | 27.60 | 27.65 | 27.70 | 27.74 | 27.83 | 27.86 |
| | $K$ | 2.62 | 4.34 | 5.14 | 5.45 | 5.57 | 5.61 | 5.62 | 5.63 | 5.63 | 5.63 | 5.63 | 5.63 | 5.63 | 5.63 | 5.63 | 5.63 | 5.63 | 5.63 | 5.63 | 5.63 | 5.63 | 5.63 | 5.63 | 5.63 | 5.63 |
| 0.2 | $U_1$ | 13.27 | 23.90 | 31.76 | 37.57 | 41.95 | 45.33 | 48.01 | 50.17 | 51.95 | 53.43 | 54.68 | 55.75 | 56.68 | 57.48 | 58.18 | 58.82 | 59.38 | 59.88 | 60.34 | 60.74 | 61.11 | 61.45 | 61.77 | 62.05 | 62.32 |
| | $U_2$ | 0.27 | 0.59 | 1.04 | 1.65 | 2.38 | 3.19 | 4.02 | 4.84 | 5.63 | 6.37 | 7.05 | 7.69 | 8.28 | 8.82 | 9.32 | 9.77 | 10.19 | 10.58 | 10.94 | 11.27 | 11.58 | 11.86 | 12.13 | 12.37 | 12.60 |
| | $W_{1,2}$ | 6.24 | 13.56 | 21.86 | 30.66 | 39.66 | 48.74 | 57.85 | 66.97 | 76.09 | 85.21 | 94.33 | 103.45 | 112.57 | 121.69 | 130.81 | 139.93 | 149.06 | 158.18 | 167.30 | 176.42 | 185.54 | 194.66 | 203.78 | 212.90 | 222.03 |
| | $W_{3,4}$ | 8.79 | 17.81 | 26.92 | 36.03 | 45.15 | 54.27 | 63.40 | 72.52 | 81.64 | 90.76 | 99.88 | 109.00 | 118.12 | 127.24 | 136.37 | 145.49 | 154.61 | 163.73 | 172.85 | 181.97 | 191.09 | 200.21 | 209.34 | 218.46 | 227.58 |
| | $J$ | 7.93 | 13.74 | 17.87 | 20.67 | 22.45 | 23.78 | 24.75 | 25.46 | 25.99 | 26.39 | 26.70 | 26.94 | 27.12 | 27.27 | 27.38 | 27.48 | 27.55 | 27.61 | 27.66 | 27.71 | 27.75 | 27.78 | 27.82 | 27.85 | 27.86 |
| | $K$ | 2.70 | 4.48 | 5.31 | 5.63 | 5.75 | 5.79 | 5.81 | 5.81 | 5.81 | 5.81 | 5.81 | 5.81 | 5.81 | 5.81 | 5.81 | 5.81 | 5.81 | 5.81 | 5.81 | 5.81 | 5.81 | 5.81 | 5.81 | 5.81 | 5.81 |
| 0.1 | $U_1$ | 13.37 | 24.11 | 32.04 | 37.92 | 42.43 | 45.85 | 48.56 | 50.74 | 52.53 | 54.02 | 55.28 | 56.36 | 57.29 | 58.10 | 58.81 | 59.44 | 60.01 | 60.51 | 60.96 | 61.38 | 61.75 | 62.09 | 62.41 | 62.69 | 62.96 |
| | $U_2$ | 0.25 | 0.55 | 0.99 | 1.59 | 2.32 | 3.13 | 3.96 | 4.78 | 5.56 | 6.31 | 7.00 | 7.64 | 8.22 | 8.77 | 9.26 | 9.72 | 10.14 | 10.53 | 10.89 | 11.22 | 11.53 | 11.81 | 12.08 | 12.33 | 12.56 |
| | $W_{1,2}$ | 6.07 | 13.28 | 21.53 | 30.31 | 39.31 | 48.39 | 57.50 | 66.61 | 75.73 | 84.85 | 93.97 | 103.09 | 112.22 | 121.34 | 130.46 | 139.58 | 148.70 | 157.82 | 166.94 | 176.06 | 185.19 | 194.31 | 203.43 | 212.55 | 221.67 |
| | $W_{3,4}$ | 8.74 | 17.71 | 26.82 | 35.94 | 45.06 | 54.18 | 63.30 | 72.42 | 81.54 | 90.66 | 99.78 | 108.91 | 118.03 | 127.15 | 136.27 | 145.39 | 154.51 | 163.63 | 172.76 | 181.88 | 191.00 | 200.12 | 209.24 | 218.36 | 227.48 |
| | $J$ | 8.01 | 13.90 | 17.87 | 20.57 | 22.45 | 23.78 | 24.75 | 25.46 | 25.99 | 26.39 | 26.70 | 26.94 | 27.12 | 27.27 | 27.38 | 27.48 | 27.55 | 27.61 | 27.66 | 27.71 | 27.75 | 27.78 | 27.82 | 27.85 | 27.87 |
| | $K$ | 2.76 | 4.58 | 5.43 | 5.76 | 5.88 | 5.92 | 5.94 | 5.95 | 5.95 | 5.95 | 5.95 | 5.95 | 5.95 | 5.95 | 5.95 | 5.95 | 5.95 | 5.95 | 5.95 | 5.95 | 5.95 | 5.95 | 5.95 | 5.95 | 5.95 |
| 0.0 | $U_1$ | 13.38 | 24.14 | 32.10 | 37.99 | 42.43 | 45.85 | 48.56 | 50.74 | 52.53 | 54.02 | 55.28 | 56.36 | 57.30 | 58.10 | 58.81 | 59.44 | 60.01 | 60.51 | 60.96 | 61.38 | 61.75 | 62.09 | 62.41 | 62.69 | 62.96 |
| | $U_2$ | 0.24 | 0.54 | 0.97 | 1.57 | 2.30 | 3.11 | 3.94 | 4.76 | 5.55 | 6.29 | 6.98 | 7.62 | 8.21 | 8.75 | 9.25 | 9.70 | 10.13 | 10.52 | 10.87 | 11.21 | 11.51 | 11.80 | 12.06 | 12.31 | 12.54 |
| | $W_{1,2}$ | 6.01 | 13.19 | 21.42 | 30.20 | 39.19 | 48.27 | 57.38 | 66.49 | 75.61 | 84.73 | 93.85 | 102.97 | 112.09 | 121.21 | 130.34 | 139.46 | 148.58 | 157.70 | 166.82 | 175.94 | 185.07 | 194.19 | 203.31 | 212.43 | 221.55 |
| | $W_{3,4}$ | 8.74 | 17.77 | 26.88 | 36.00 | 45.02 | 54.14 | 63.26 | 72.38 | 81.51 | 90.63 | 99.75 | 108.87 | 117.99 | 127.11 | 136.23 | 145.35 | 154.48 | 163.60 | 172.72 | 181.84 | 190.96 | 200.08 | 209.20 | 218.32 | 227.45 |
| | $J$ | 8.05 | 13.97 | 17.96 | 20.67 | 22.55 | 23.89 | 24.86 | 25.57 | 26.09 | 26.49 | 26.79 | 27.02 | 27.20 | 27.35 | 27.46 | 27.55 | 27.63 | 27.69 | 27.74 | 27.78 | 27.82 | 27.85 | 27.87 | 27.89 | 27.91 |
| | $K$ | 2.79 | 4.62 | 5.47 | 5.81 | 5.93 | 5.97 | 5.99 | 5.99 | 6.00 | 6.00 | 6.00 | 6.00 | 6.00 | 6.00 | 6.00 | 6.00 | 6.00 | 6.00 | 6.00 | 6.00 | 6.00 | 6.00 | 6.00 | 6.00 | 6.00 |

TABLE S37. Parameters of the THF interaction Hamiltonian for different ratios $u_0/u_1$ and screening lengths $\xi$ of the double-gate interaction potential ($U_\xi = 24$ meV for $\xi = 10$nm). We use the same BM model parameters as in Table S36.

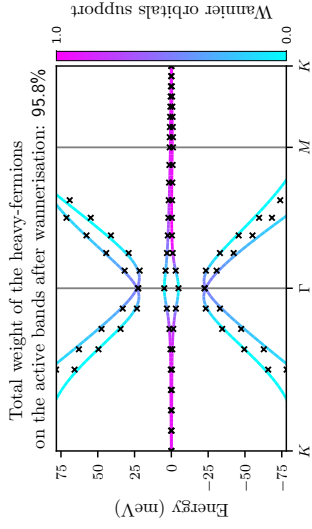

**Total weight of the heavy-fermions on the active bands after wannierisation: 95.8%**

FIG. S35. Band structures of the BM and THF models near charge neutrality for $w_0/w_1 = 0.8$, depicted by lines and crosses, respectively. The BM bands are colored according to the weight of the $f$-electron wave function on them. We use the same BM parameters as in Table S38.

TABLE S38. Parameters of the $f$-electron Wannier functions for different values of the tunneling ratio $w_0/w_1$. $\mathcal{W}$ denotes the total weight of the THF $f$-electrons on the active bands. In computing the ratios $\gamma/U_1$ and $\gamma/U_2$, we employ the on-site and nearest-neighbor repulsion parameters $U_1$ and $U_2$ obtained numerically for $\xi = 10$ nm, as given in Table S39. We employ $v_F = 5.944$ eV·Å, $|\mathbf{K}| = 1.703$ Å$^{-1}$, $w_1 = 110$ meV, and $\theta = 1.04°$ for the BM model.

| $w_0/w_1$ | $\alpha_1$ | $\alpha_2$ | $\lambda_1$ | $\lambda_2$ | $\lambda$ | $\mathcal{W}$ | $t_0$ | $\gamma$ | $M$ | $v_\star$ | $v'_\star$ | $\gamma/U_1$ | $\gamma/U_2$ |
|---|---|---|---|---|---|---|---|---|---|---|---|---|---|
| 1.00 | 0.760 | 0.649 | 0.155 | 0.173 | 0.342 | 0.982 | 0.018 | 5.302 | 5.752 | 1.472 | 0.0145 | 0.08 | 2.54 |
| 0.90 | 0.786 | 0.618 | 0.167 | 0.179 | 0.334 | 0.978 | 0.114 | -8.867 | 5.287 | 1.565 | -0.0166 | -0.14 | -4.38 |
| 0.80 | 0.818 | 0.576 | 0.179 | 0.191 | 0.336 | 0.958 | 0.215 | -23.324 | 4.847 | 1.612 | -0.0342 | -0.40 | -11.15 |
| 0.70 | 0.851 | 0.525 | 0.197 | 0.197 | 0.351 | 0.941 | 0.314 | -37.633 | 4.444 | 1.617 | -0.0418 | -0.70 | -17.08 |
| 0.60 | 0.884 | 0.467 | 0.209 | 0.203 | 0.364 | 0.928 | 0.396 | -51.407 | 4.087 | 1.575 | -0.0419 | -1.03 | -22.25 |
| 0.50 | 0.915 | 0.403 | 0.221 | 0.209 | 0.376 | 0.919 | 0.461 | -64.262 | 3.781 | 1.479 | -0.0368 | -1.36 | -26.77 |
| 0.40 | 0.943 | 0.332 | 0.230 | 0.215 | 0.386 | 0.913 | 0.517 | -75.783 | 3.530 | 1.321 | -0.0283 | -1.69 | -30.58 |
| 0.30 | 0.967 | 0.256 | 0.251 | 0.215 | 0.393 | 0.908 | 0.566 | -85.514 | 3.333 | 1.091 | -0.0184 | -1.98 | -33.77 |
| 0.20 | 0.985 | 0.175 | 0.263 | 0.221 | 0.393 | 0.901 | 0.696 | -92.968 | 3.193 | 0.785 | -0.0091 | -2.18 | -36.90 |
| 0.10 | 0.996 | 0.090 | 0.269 | 0.221 | 0.388 | 0.894 | 0.786 | -97.676 | 3.108 | 0.413 | -0.0024 | -2.29 | -39.33 |
| 0.00 | 1.000 | 0.000 | 0.269 | 0.000 | 0.381 | 0.887 | 0.790 | -99.289 | 3.080 | -5.125 | -0.0000 | -2.29 | -40.96 |

TABLE S38. Parameters of the $f$-electron Wannier functions and the THF single-particle Hamiltonian for different values of the tunneling ratio $w_0/w_1$. $\mathcal{W}$ denotes the total weight of the THF $f$-electrons on the active bands. In computing the ratios $\gamma/U_1$ and $\gamma/U_2$, we employ the on-site and nearest-neighbor repulsion parameters $U_1$ and $U_2$ obtained numerically for $\xi = 10$ nm, as given in Table S39.

| $w_0/w_1$ | | $\xi$/nm | 2 | 4 | 6 | 8 | 10 | 12 | 14 | 16 | 18 | 20 | 22 | 24 | 26 | 28 | 30 | 32 | 34 | 36 | 38 | 40 | 42 | 44 | 46 | 48 | 50 |
|---|---|---|---|---|---|---|---|---|---|---|---|---|---|---|---|---|---|---|---|---|---|---|---|---|---|---|---|---|
| 1.0 | | $U_1$ | 24.05 | 40.96 | 52.01 | 59.47 | 64.76 | 68.67 | 71.68 | 74.05 | 75.96 | 77.54 | 78.86 | 79.98 | 80.95 | 81.78 | 82.51 | 83.16 | 83.73 | 84.24 | 84.71 | 85.12 | 85.50 | 85.85 | 86.16 | 86.46 | 86.72 |
| | | $U_2$ | 0.50 | 0.86 | 1.39 | 2.09 | 2.89 | 3.73 | 4.57 | 5.38 | 6.15 | 6.86 | 7.52 | 8.13 | 8.68 | 9.19 | 9.66 | 10.09 | 10.49 | 10.86 | 11.19 | 11.51 | 11.80 | 12.07 | 12.32 | 12.55 | 12.77 |
| | | $W_{1,2}$ | 7.09 | 15.12 | 23.96 | 33.20 | 42.60 | 52.05 | 61.52 | 71.00 | 80.48 | 89.97 | 99.45 | 108.93 | 118.41 | 127.89 | 137.38 | 146.86 | 156.34 | 165.82 | 175.31 | 184.79 | 194.27 | 203.75 | 213.24 | 222.72 | 232.20 |
| | | $W_{3,4}$ | 11.42 | 22.52 | 32.80 | 42.60 | 52.19 | 61.72 | 71.21 | 80.70 | 90.18 | 99.66 | 109.15 | 118.63 | 128.11 | 137.60 | 147.08 | 156.56 | 166.04 | 175.53 | 185.01 | 194.49 | 203.97 | 213.46 | 222.94 | 232.42 | 241.90 |
| | | $J$ | 7.34 | 11.12 | 12.78 | 13.53 | 13.90 | 14.08 | 14.18 | 14.24 | 14.27 | 14.29 | 14.30 | 14.31 | 14.32 | 14.32 | 14.33 | 14.33 | 14.33 | 14.33 | 14.33 | 14.33 | 14.33 | 14.34 | 14.34 | 14.34 | 14.34 |
| | | $K$ | 2.51 | 4.11 | 4.83 | 5.10 | 5.24 | 5.23 | 5.24 | 5.24 | 5.25 | 5.25 | 5.25 | 5.25 | 5.25 | 5.25 | 5.25 | 5.25 | 5.25 | 5.25 | 5.25 | 5.25 | 5.25 | 5.25 | 5.25 | 5.25 | 5.25 |
| 0.9 | | $U_1$ | 22.36 | 38.46 | 49.19 | 56.53 | 61.78 | 65.69 | 68.69 | 71.06 | 72.98 | 74.57 | 75.89 | 77.01 | 77.98 | 78.82 | 79.55 | 80.20 | 80.77 | 81.29 | 81.75 | 82.17 | 82.55 | 82.89 | 83.21 | 83.50 | 83.77 |
| | | $U_2$ | 0.20 | 0.44 | 0.79 | 1.33 | 2.03 | 2.83 | 3.68 | 4.52 | 5.34 | 6.10 | 6.82 | 7.48 | 8.09 | 8.64 | 9.16 | 9.63 | 10.06 | 10.46 | 10.82 | 11.16 | 11.47 | 11.76 | 12.03 | 12.28 | 12.52 |
| | | $W_{1,2}$ | 7.28 | 15.41 | 24.29 | 33.55 | 42.95 | 52.41 | 61.88 | 71.36 | 80.84 | 90.32 | 99.81 | 109.30 | 118.77 | 128.25 | 137.74 | 147.22 | 156.70 | 166.18 | 175.66 | 185.15 | 194.63 | 204.11 | 213.59 | 223.08 | 232.56 |
| | | $W_{3,4}$ | 10.74 | 21.34 | 31.39 | 41.10 | 50.66 | 60.17 | 69.66 | 79.15 | 88.63 | 98.12 | 107.60 | 117.08 | 126.56 | 136.05 | 145.53 | 155.01 | 164.49 | 173.98 | 183.46 | 192.94 | 202.42 | 211.91 | 221.39 | 230.87 | 240.35 |
| | | $J$ | 7.25 | 11.26 | 13.22 | 14.22 | 14.78 | 15.11 | 15.31 | 15.45 | 15.55 | 15.62 | 15.67 | 15.70 | 15.73 | 15.75 | 15.76 | 15.78 | 15.79 | 15.80 | 15.80 | 15.81 | 15.81 | 15.81 | 15.81 | 15.81 | 15.82 |
| | | $K$ | 2.41 | 3.96 | 4.66 | 4.93 | 5.06 | 5.05 | 5.06 | 5.07 | 5.07 | 5.07 | 5.07 | 5.07 | 5.07 | 5.07 | 5.07 | 5.07 | 5.07 | 5.07 | 5.07 | 5.07 | 5.07 | 5.07 | 5.07 | 5.07 | 5.07 |
| 0.8 | | $U_1$ | 20.36 | 35.35 | 45.55 | 52.63 | 57.75 | 61.58 | 64.54 | 66.88 | 68.79 | 70.36 | 71.67 | 72.79 | 73.75 | 74.59 | 75.32 | 75.97 | 76.54 | 77.05 | 77.51 | 77.93 | 78.31 | 78.66 | 78.97 | 79.27 | 79.53 |
| | | $U_2$ | 0.21 | 0.46 | 0.90 | 1.47 | 2.20 | 3.03 | 3.76 | 4.60 | 5.42 | 6.19 | 6.90 | 7.56 | 8.17 | 8.73 | 9.24 | 9.71 | 10.14 | 10.54 | 10.90 | 11.24 | 11.55 | 11.84 | 12.11 | 12.36 | 12.60 |
| | | $W_{1,2}$ | 7.35 | 15.52 | 24.42 | 33.68 | 43.09 | 52.54 | 62.01 | 71.49 | 80.98 | 90.46 | 99.94 | 109.42 | 118.90 | 128.39 | 137.87 | 147.35 | 156.83 | 166.32 | 175.80 | 185.28 | 194.76 | 204.25 | 213.73 | 223.21 | 232.69 |
| | | $W_{3,4}$ | 10.27 | 20.48 | 30.30 | 39.93 | 49.47 | 58.97 | 68.46 | 77.94 | 87.43 | 96.91 | 106.39 | 115.88 | 125.36 | 134.84 | 144.32 | 153.80 | 163.29 | 172.77 | 182.25 | 191.73 | 201.22 | 210.70 | 220.18 | 229.66 | 239.15 |
| | | $J$ | 7.30 | 11.65 | 13.99 | 15.23 | 16.16 | 16.71 | 17.08 | 17.34 | 17.54 | 17.68 | 17.79 | 17.88 | 17.95 | 18.00 | 18.04 | 18.07 | 18.10 | 18.12 | 18.14 | 18.15 | 18.17 | 18.18 | 18.19 | 18.19 | 18.20 |
| | | $K$ | 2.36 | 3.89 | 4.58 | 4.85 | 4.94 | 4.97 | 4.98 | 4.99 | 4.99 | 4.99 | 4.99 | 4.99 | 4.99 | 4.99 | 4.99 | 4.99 | 4.99 | 4.99 | 4.99 | 4.99 | 4.99 | 4.99 | 4.99 | 4.99 | 4.99 |
| 0.7 | | $U_1$ | 18.41 | 32.25 | 41.85 | 48.63 | 53.55 | 57.28 | 60.17 | 62.47 | 64.35 | 65.90 | 67.20 | 68.31 | 69.26 | 70.09 | 70.82 | 71.46 | 72.03 | 72.53 | 72.98 | 73.42 | 73.79 | 74.14 | 74.45 | 74.75 | 75.01 |
| | | $U_2$ | 0.23 | 0.50 | 0.90 | 1.47 | 2.24 | 3.08 | 3.88 | 4.73 | 5.54 | 6.31 | 7.02 | 7.68 | 8.28 | 8.84 | 9.35 | 9.82 | 10.25 | 10.64 | 11.01 | 11.34 | 11.66 | 11.95 | 12.21 | 12.46 | 12.70 |
| | | $W_{1,2}$ | 7.33 | 15.49 | 24.38 | 33.64 | 43.04 | 52.49 | 61.97 | 71.45 | 80.93 | 90.41 | 99.89 | 109.37 | 118.86 | 128.34 | 137.82 | 147.30 | 156.79 | 166.27 | 175.75 | 185.23 | 194.71 | 204.20 | 213.68 | 223.16 | 232.65 |
| | | $W_{3,4}$ | 9.82 | 19.74 | 29.47 | 39.05 | 48.57 | 58.07 | 67.55 | 77.04 | 86.52 | 96.00 | 105.49 | 114.97 | 124.45 | 133.93 | 143.42 | 152.90 | 162.38 | 171.86 | 181.35 | 190.83 | 200.31 | 209.79 | 219.28 | 228.76 | 238.24 |
| | | $J$ | 7.40 | 12.04 | 14.81 | 16.48 | 17.57 | 18.31 | 18.83 | 19.20 | 19.47 | 19.68 | 19.83 | 19.96 | 20.06 | 20.14 | 20.20 | 20.24 | 20.30 | 20.34 | 20.37 | 20.42 | 20.44 | 20.46 | 20.48 | 20.49 | 20.49 |
| | | $K$ | 2.36 | 3.89 | 4.58 | 4.85 | 4.94 | 4.99 | 4.99 | 4.99 | 4.99 | 4.99 | 4.99 | 4.99 | 4.99 | 4.99 | 4.99 | 4.99 | 4.99 | 4.99 | 4.99 | 4.99 | 4.99 | 4.99 | 4.99 | 4.99 | 4.99 |

TABLE S39: Parameters of the THF interaction Hamiltonian for different ratios $u_0/u_1$ and screening lengths $\xi$ of the double-gate interaction potential ($U_\xi = 24$ meV for $\xi = 10$nm). We use the same BM model parameters as in Table S38.

| $u_0/u_1$ | | $\xi$/nm = 2 | 4 | 6 | 8 | 10 | 12 | 14 | 16 | 18 | 20 | 22 | 24 | 26 | 28 | 30 | 32 | 34 | 36 | 38 | 40 | 42 | 44 | 46 | 48 | 50 |
|---|---|---|---|---|---|---|---|---|---|---|---|---|---|---|---|---|---|---|---|---|---|---|---|---|---|---|
| 0.6 | $U_1$ | 16.79 | 29.64 | 38.72 | 45.20 | 49.98 | 53.60 | 56.43 | 58.70 | 60.54 | 62.07 | 63.36 | 64.46 | 65.40 | 66.23 | 66.95 | 67.59 | 68.15 | 68.66 | 69.12 | 69.53 | 69.91 | 70.25 | 70.57 | 70.86 | 71.13 |
|  | $U_2$ | 0.25 | 0.55 | 0.97 | 1.57 | 2.31 | 3.14 | 4.00 | 4.85 | 5.66 | 6.42 | 7.13 | 7.79 | 8.39 | 8.94 | 9.45 | 9.92 | 10.35 | 10.74 | 11.11 | 11.44 | 11.75 | 12.04 | 12.31 | 12.56 | 12.79 |
|  | $W_{1,2}$ | 7.24 | 15.33 | 24.20 | 33.44 | 42.84 | 52.30 | 61.77 | 71.25 | 80.73 | 90.21 | 99.70 | 109.18 | 118.66 | 128.14 | 137.63 | 147.11 | 156.59 | 166.07 | 175.56 | 185.04 | 194.52 | 204.00 | 213.48 | 222.97 | 232.45 |
|  | $W_{3,4}$ | 7.71 | 13.02 | 16.41 | 18.64 | 20.17 | 21.24 | 22.02 | 22.60 | 23.03 | 23.36 | 23.61 | 23.81 | 23.97 | 24.09 | 24.19 | 24.27 | 24.33 | 24.39 | 24.43 | 24.47 | 24.51 | 24.53 | 24.55 | 24.57 | 24.59 |
|  | $J$ | 2.39 | 3.94 | 4.65 | 4.92 | 5.01 | 5.04 | 5.06 | 5.06 | 5.07 | 5.07 | 5.07 | 5.07 | 5.07 | 5.07 | 5.07 | 5.07 | 5.07 | 5.07 | 5.07 | 5.07 | 5.07 | 5.07 | 5.07 | 5.07 | 5.07 |
|  | $K$ | 5.05 | 5.05 | 5.06 | 5.07 | 5.07 | 5.07 | 5.07 | 5.07 | 5.20 | 5.20 | 5.20 | 5.20 | 5.20 | 5.20 | 5.20 | 5.20 | 5.20 | 5.20 | 5.20 | 5.20 | 5.20 | 5.20 | 5.20 | 5.20 | 5.20 |
| 0.5 | $U_1$ | 15.52 | 25.95 | 36.22 | 42.47 | 47.10 | 50.65 | 53.43 | 55.65 | 57.48 | 59.02 | 60.27 | 61.35 | 62.29 | 63.11 | 63.83 | 64.46 | 65.03 | 65.53 | 65.99 | 66.40 | 66.78 | 67.12 | 67.43 | 67.72 | 67.99 |
|  | $U_2$ | 0.26 | 0.58 | 1.03 | 1.65 | 2.40 | 3.24 | 4.10 | 4.95 | 5.76 | 6.52 | 7.23 | 7.88 | 8.48 | 9.03 | 9.54 | 10.00 | 10.43 | 10.82 | 11.19 | 11.51 | 11.83 | 12.12 | 12.39 | 12.64 | 12.87 |
|  | $W_{1,2}$ | 7.09 | 15.09 | 23.92 | 33.15 | 42.55 | 52.00 | 61.47 | 70.95 | 80.43 | 89.99 | 99.39 | 108.88 | 118.36 | 127.84 | 137.32 | 146.81 | 156.29 | 165.77 | 175.25 | 184.74 | 194.22 | 203.70 | 213.18 | 222.67 | 232.15 |
|  | $W_{3,4}$ | 7.89 | 13.48 | 17.15 | 19.62 | 21.33 | 22.55 | 23.44 | 24.10 | 24.59 | 24.97 | 25.26 | 25.49 | 25.67 | 25.81 | 25.92 | 26.02 | 26.09 | 26.16 | 26.21 | 26.25 | 26.29 | 26.32 | 26.34 | 26.37 | 26.39 |
|  | $J$ | 2.45 | 4.05 | 4.77 | 5.05 | 5.15 | 5.18 | 5.19 | 5.20 | 5.20 | 5.20 | 5.20 | 5.20 | 5.20 | 5.20 | 5.20 | 5.20 | 5.20 | 5.20 | 5.20 | 5.20 | 5.20 | 5.20 | 5.20 | 5.20 | 5.20 |
|  | $K$ | 5.32 | 5.32 | 5.33 | 5.35 | 5.37 | 5.37 | 5.37 | 5.37 | 5.38 | 5.38 | 5.38 | 5.38 | 5.38 | 5.38 | 5.38 | 5.38 | 5.38 | 5.38 | 5.38 | 5.38 | 5.38 | 5.38 | 5.38 | 5.38 | 5.38 |
| 0.4 | $U_1$ | 14.52 | 24.78 | 32.83 | 38.73 | 43.17 | 46.59 | 49.29 | 51.47 | 53.19 | 54.74 | 56.00 | 57.07 | 58.01 | 58.81 | 59.52 | 60.15 | 60.71 | 61.21 | 61.67 | 62.08 | 62.45 | 62.79 | 63.11 | 63.39 | 63.66 |
|  | $U_2$ | 0.28 | 0.62 | 1.08 | 1.71 | 2.48 | 3.32 | 4.18 | 5.03 | 5.84 | 6.60 | 7.30 | 7.96 | 8.55 | 9.10 | 9.61 | 10.07 | 10.50 | 10.89 | 11.25 | 11.59 | 11.90 | 12.19 | 12.45 | 12.70 | 12.94 |
|  | $W_{1,2}$ | 6.92 | 14.81 | 23.59 | 32.80 | 42.19 | 51.64 | 61.11 | 70.60 | 80.07 | 89.53 | 99.03 | 108.54 | 118.00 | 127.48 | 136.96 | 146.45 | 155.93 | 165.41 | 174.89 | 184.38 | 193.86 | 203.34 | 212.82 | 222.31 | 231.79 |
|  | $W_{3,4}$ | 8.67 | 14.81 | 18.67 | 21.15 | 22.80 | 23.99 | 24.86 | 25.22 | 25.60 | 25.86 | 26.05 | 26.20 | 26.29 | 26.37 | 26.41 | 26.45 | 26.48 | 26.51 | 26.53 | 26.55 | 26.56 | 26.57 | 26.58 | 26.59 | 26.39 |
|  | $J$ | 2.63 | 4.33 | 5.11 | 5.22 | 5.31 | 5.34 | 5.36 | 5.37 | 5.37 | 5.37 | 5.37 | 5.37 | 5.37 | 5.37 | 5.37 | 5.37 | 5.37 | 5.37 | 5.37 | 5.37 | 5.37 | 5.37 | 5.37 | 5.37 | 5.37 |
|  | $K$ | 5.37 | 5.37 | 5.37 | 5.37 | 5.37 | 5.37 | 5.37 | 5.37 | 5.57 | 5.57 | 5.57 | 5.57 | 5.57 | 5.57 | 5.57 | 5.57 | 5.57 | 5.57 | 5.57 | 5.57 | 5.57 | 5.57 | 5.57 | 5.57 | 5.57 |
| 0.3 | $U_1$ | 13.81 | 24.34 | 32.33 | 38.23 | 42.63 | 46.05 | 48.74 | 50.92 | 52.70 | 54.19 | 55.43 | 56.50 | 57.44 | 58.26 | 58.97 | 59.60 | 60.16 | 60.66 | 61.12 | 61.53 | 61.90 | 62.24 | 62.55 | 62.84 | 63.11 |
|  | $U_2$ | 0.28 | 0.62 | 1.10 | 1.75 | 2.52 | 3.36 | 4.24 | 5.09 | 5.89 | 6.65 | 7.35 | 8.00 | 8.60 | 9.15 | 9.65 | 10.12 | 10.54 | 10.93 | 11.29 | 11.63 | 11.94 | 12.23 | 12.49 | 12.74 | 12.97 |
|  | $W_{1,2}$ | 6.73 | 14.51 | 23.24 | 32.43 | 41.81 | 51.26 | 60.73 | 70.21 | 79.69 | 89.17 | 98.65 | 108.14 | 117.62 | 127.10 | 136.58 | 146.07 | 155.55 | 165.03 | 174.51 | 184.00 | 193.48 | 202.96 | 212.44 | 221.93 | 231.41 |
|  | $W_{3,4}$ | 9.14 | 14.16 | 18.19 | 21.22 | 23.13 | 24.48 | 25.31 | 25.88 | 26.28 | 26.59 | 26.85 | 27.05 | 27.22 | 27.35 | 27.46 | 27.55 | 27.62 | 27.67 | 27.72 | 27.75 | 27.78 | 27.80 | 27.82 | 27.83 | 27.84 |
|  | $J$ | 2.72 | 4.48 | 5.28 | 5.41 | 5.50 | 5.54 | 5.56 | 5.57 | 5.57 | 5.57 | 5.57 | 5.57 | 5.57 | 5.57 | 5.57 | 5.57 | 5.57 | 5.57 | 5.57 | 5.57 | 5.57 | 5.57 | 5.57 | 5.57 | 5.57 |
|  | $K$ | 5.57 | 5.57 | 5.57 | 5.57 | 5.57 | 5.57 | 5.57 | 5.57 | 5.75 | 5.75 | 5.75 | 5.75 | 5.75 | 5.75 | 5.75 | 5.75 | 5.75 | 5.75 | 5.75 | 5.75 | 5.75 | 5.75 | 5.75 | 5.75 | 5.75 |
| 0.2 | $U_1$ | 13.34 | 24.36 | 32.32 | 38.23 | 42.63 | 46.05 | 48.74 | 50.99 | 52.78 | 54.27 | 55.53 | 56.52 | 57.44 | 58.35 | 59.06 | 59.69 | 60.25 | 60.76 | 61.21 | 61.62 | 62.00 | 62.34 | 62.65 | 62.94 | 63.21 |
|  | $U_2$ | 0.28 | 0.62 | 1.10 | 1.75 | 2.52 | 3.36 | 4.24 | 5.08 | 5.89 | 6.64 | 7.35 | 8.00 | 8.60 | 9.15 | 9.65 | 10.12 | 10.54 | 10.93 | 11.29 | 11.63 | 11.94 | 12.23 | 12.49 | 12.74 | 12.97 |
|  | $W_{1,2}$ | 6.54 | 14.20 | 22.87 | 32.05 | 41.42 | 50.87 | 60.34 | 69.83 | 79.30 | 88.78 | 98.26 | 107.74 | 116.91 | 126.39 | 135.88 | 145.36 | 151.10 | 160.58 | 170.06 | 179.54 | 189.03 | 198.51 | 207.99 | 217.62 | 227.10 |
|  | $W_{3,4}$ | 9.04 | 15.32 | 19.40 | 22.10 | 23.92 | 25.22 | 26.16 | 26.82 | 27.29 | 27.64 | 27.90 | 28.10 | 28.21 | 28.29 | 28.34 | 28.39 | 28.43 | 28.46 | 28.48 | 28.50 | 28.51 | 28.52 | 28.52 | 28.52 | 28.54 |
|  | $J$ | 2.78 | 4.58 | 5.40 | 5.72 | 5.75 | 5.75 | 5.75 | 5.75 | 5.89 | 5.89 | 5.89 | 5.89 | 5.89 | 5.89 | 5.89 | 5.89 | 5.89 | 5.89 | 5.89 | 5.89 | 5.89 | 5.89 | 5.89 | 5.89 | 5.89 |
|  | $K$ | 5.75 | 5.75 | 5.75 | 5.75 | 5.75 | 5.75 | 5.75 | 5.75 | 5.89 | 5.89 | 5.89 | 5.89 | 5.89 | 5.89 | 5.89 | 5.89 | 5.89 | 5.89 | 5.89 | 5.89 | 5.89 | 5.89 | 5.89 | 5.89 | 5.89 |
| 0.1 | $U_1$ | 13.51 | 24.34 | 32.33 | 38.23 | 42.63 | 46.10 | 48.81 | 50.99 | 52.78 | 54.27 | 55.53 | 56.52 | 57.44 | 58.35 | 59.06 | 60.25 | 60.76 | 61.21 | 61.62 | 62.34 | 62.73 | 63.07 | 63.39 | 63.67 | 63.94 |
|  | $U_2$ | 0.27 | 0.60 | 1.07 | 1.71 | 2.48 | 3.33 | 4.19 | 5.04 | 5.85 | 6.61 | 7.32 | 7.97 | 8.57 | 9.12 | 9.62 | 10.09 | 10.51 | 10.90 | 11.27 | 11.60 | 11.91 | 12.20 | 12.46 | 12.71 | 12.94 |
|  | $W_{1,2}$ | 6.38 | 13.95 | 22.58 | 31.74 | 41.11 | 50.56 | 60.02 | 69.50 | 78.98 | 88.46 | 97.95 | 107.43 | 116.91 | 126.39 | 135.88 | 145.36 | 154.84 | 164.32 | 173.81 | 183.29 | 192.77 | 202.25 | 211.74 | 221.22 | 230.70 |
|  | $W_{3,4}$ | 9.02 | 14.37 | 18.45 | 21.22 | 23.17 | 24.55 | 25.55 | 26.28 | 26.83 | 27.26 | 27.62 | 27.89 | 28.12 | 28.29 | 28.39 | 28.47 | 28.53 | 28.58 | 28.63 | 28.67 | 28.70 | 28.73 | 28.76 | 28.76 | 28.78 |
|  | $J$ | 2.81 | 4.62 | 5.45 | 5.76 | 5.92 | 5.89 | 5.89 | 5.89 | 5.89 | 5.89 | 5.89 | 5.89 | 5.89 | 5.89 | 5.89 | 5.89 | 5.89 | 5.89 | 5.89 | 5.89 | 5.89 | 5.89 | 5.89 | 5.89 | 5.89 |
|  | $K$ | 5.89 | 5.89 | 5.89 | 5.89 | 5.89 | 5.89 | 5.89 | 5.89 | 5.89 | 5.89 | 5.89 | 5.89 | 5.89 | 5.89 | 5.89 | 5.89 | 5.89 | 5.89 | 5.89 | 5.89 | 5.89 | 5.89 | 5.89 | 5.89 | 5.89 |
| 0.0 | $U_1$ | 13.72 | 24.70 | 32.81 | 38.79 | 43.28 | 46.74 | 49.47 | 51.67 | 53.47 | 54.97 | 56.24 | 57.24 | 58.35 | 59.07 | 59.79 | 60.42 | 60.98 | 61.49 | 61.94 | 62.36 | 62.73 | 63.07 | 63.39 | 63.67 | 63.94 |
|  | $U_2$ | 0.25 | 0.56 | 1.02 | 1.66 | 2.42 | 3.27 | 4.13 | 4.98 | 5.80 | 6.56 | 7.26 | 7.92 | 8.52 | 9.07 | 9.57 | 10.04 | 10.46 | 10.86 | 11.22 | 11.55 | 11.86 | 12.15 | 12.42 | 12.67 | 13.00 |
|  | $W_{1,2}$ | 6.30 | 13.81 | 22.42 | 31.57 | 40.93 | 50.37 | 59.84 | 69.32 | 78.80 | 88.28 | 97.76 | 107.25 | 116.73 | 126.21 | 135.69 | 145.18 | 150.97 | 160.45 | 169.93 | 179.42 | 188.90 | 198.38 | 207.86 | 217.34 | 226.83 |
|  | $W_{3,4}$ | 9.02 | 14.37 | 18.45 | 21.22 | 23.13 | 24.48 | 25.45 | 26.09 | 26.68 | 27.07 | 27.37 | 27.60 | 27.78 | 27.92 | 28.03 | 28.12 | 28.19 | 28.25 | 28.30 | 28.34 | 28.38 | 28.41 | 28.43 | 28.45 | 28.47 |
|  | $J$ | 2.81 | 4.62 | 5.45 | 5.76 | 5.93 | 5.94 | 5.94 | 5.94 | 5.94 | 5.94 | 5.94 | 5.94 | 5.94 | 5.94 | 5.94 | 5.94 | 5.94 | 5.94 | 5.94 | 5.94 | 5.94 | 5.94 | 5.94 | 5.94 | 5.94 |
|  | $K$ | 5.94 | 5.94 | 5.94 | 5.94 | 5.94 | 5.94 | 5.94 | 5.94 | 5.94 | 5.94 | 5.94 | 5.94 | 5.94 | 5.94 | 5.94 | 5.94 | 5.94 | 5.94 | 5.94 | 5.94 | 5.94 | 5.94 | 5.94 | 5.94 | 5.94 |

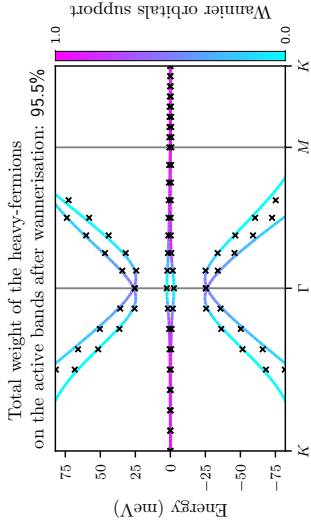

Total weight of the heavy-fermions
on the active bands after wannierisation: 95.5%

Wannier orbitals support

FIG. S36. Band structures of the BM and THF models near charge neutrality for $u_0/u_1 = 0.8$, depicted by lines and crosses, respectively. The BM bands are colored according to the weight of the $f$-electron wave function on them. We use the same BM parameters as in Table S40.

TABLE S40. Parameters of the $f$-electron Wannier functions and the THF single-particle Hamiltonian for different values of the tunneling ratio $u_0/u_1$. $\mathcal{W}$ denotes the total weight of the THF $f$-electrons on the active bands. In computing the ratios $\gamma/U_1$ and $\gamma/U_2$, we employ the on-site and nearest-neighbor repulsion parameters $U_1$ and $U_2$ obtained numerically for $\xi = 10$nm, as given in Table S41. We employ $v_F = 5.944$ eV·Å, $|\mathbf{K}| = 1.703$Å$^{-1}$, $w_1 = 110$meV, and $\theta = 1.06°$ for the BM model.

| $u_0/u_1$ | $\alpha_1$ | $\alpha_2$ | $\lambda_1$ | $\lambda_2$ | $\lambda$ | $\mathcal{W}$ | $t_0$ | $\gamma$ | $M$ | $v_\star$ | $v_\star'$ | $v_\star''$ | $\gamma/U_1$ | $\gamma/U_2$ |
|---|---|---|---|---|---|---|---|---|---|---|---|---|---|---|
| 1.00 | 0.763 | 0.647 | 0.161 | 0.179 | 0.340 | 0.984 | 0.052 | 2.848 | 3.676 | 1.510 | 0.0105 | 0.04 | 1.30 | |
| 0.90 | 0.790 | 0.613 | 0.173 | 0.185 | 0.332 | 0.975 | 0.022 | -11.570 | 3.078 | -4.127 | -0.0168 | -0.19 | -5.38 | |
| 0.80 | 0.821 | 0.571 | 0.185 | 0.191 | 0.339 | 0.955 | 0.080 | -26.184 | 2.528 | -4.335 | -0.0321 | -0.45 | -11.77 | |
| 0.70 | 0.854 | 0.520 | 0.197 | 0.191 | 0.353 | 0.939 | 0.131 | -40.591 | 2.035 | -4.510 | -0.0385 | -0.75 | -17.36 | |
| 0.60 | 0.887 | 0.462 | 0.209 | 0.203 | 0.366 | 0.927 | 0.166 | -54.421 | 1.606 | -4.663 | -0.0383 | -1.08 | -22.27 | |
| 0.50 | 0.917 | 0.398 | 0.227 | 0.209 | 0.377 | 0.918 | 0.190 | -67.301 | 1.242 | -4.796 | -0.0334 | -1.41 | -26.53 | |
| 0.40 | 0.945 | 0.327 | 0.239 | 0.215 | 0.394 | 0.917 | 0.244 | -78.828 | 0.944 | -4.914 | -0.0257 | -1.76 | -29.68 | |
| 0.30 | 0.968 | 0.252 | 0.257 | 0.221 | 0.402 | 0.913 | 0.269 | -88.551 | 0.714 | -5.014 | -0.0166 | -2.05 | -32.66 | |
| 0.20 | 0.985 | 0.173 | 0.263 | 0.221 | 0.399 | 0.906 | 0.305 | -95.991 | 0.549 | -5.092 | -0.0082 | -2.24 | -35.56 | |
| 0.10 | 0.996 | 0.089 | 0.275 | 0.221 | 0.393 | 0.897 | 0.459 | -100.687 | 0.451 | -5.144 | -0.0022 | -2.34 | -38.09 | |
| 0.00 | 1.000 | 0.000 | 0.275 | 0.000 | 0.390 | 0.893 | 0.497 | -102.295 | 0.418 | -5.162 | -0.0000 | -2.36 | -39.08 | |

| $\xi$/nm | | 2 | 4 | 6 | 8 | 10 | 12 | 14 | 16 | 18 | 20 | 22 | 24 | 26 | 28 | 30 | 32 | 34 | 36 | 38 | 40 | 42 | 44 | 46 | 48 | 50 |
|---|---|---|---|---|---|---|---|---|---|---|---|---|---|---|---|---|---|---|---|---|---|---|---|---|---|---|
| 1.0 | $U_1$ | 24.32 | 41.40 | 52.54 | 60.05 | 65.39 | 69.33 | 72.35 | 74.74 | 76.66 | 78.25 | 79.57 | 80.70 | 81.66 | 82.50 | 83.24 | 83.88 | 84.46 | 84.97 | 85.43 | 85.85 | 86.23 | 86.58 | 86.90 | 87.19 | 87.46 |
| | $U_2$ | 0.24 | 0.51 | 0.89 | 1.45 | 2.19 | 3.03 | 3.91 | 4.78 | 5.61 | 6.40 | 7.13 | 7.80 | 8.42 | 8.99 | 9.50 | 9.98 | 10.42 | 10.82 | 11.19 | 11.53 | 11.84 | 12.13 | 12.41 | 12.66 | 12.89 |
| | $W_{1,2}$ | 7.49 | 15.93 | 25.17 | 34.79 | 44.56 | 54.39 | 64.23 | 74.08 | 83.93 | 93.78 | 103.63 | 113.48 | 123.33 | 133.18 | 143.03 | 152.88 | 162.73 | 172.58 | 182.43 | 192.29 | 202.14 | 211.99 | 221.84 | 231.69 | 241.54 |
| | $W_{3,4}$ | 11.58 | 22.87 | 33.43 | 43.56 | 53.50 | 63.38 | 73.25 | 83.10 | 92.95 | 102.80 | 112.65 | 122.51 | 132.36 | 142.21 | 152.06 | 161.91 | 171.76 | 181.61 | 191.46 | 201.31 | 211.16 | 221.01 | 230.86 | 240.71 | 250.56 |
| | $J$ | 7.44 | 11.27 | 12.97 | 13.75 | 14.13 | 14.33 | 14.44 | 14.51 | 14.55 | 14.57 | 14.59 | 14.60 | 14.61 | 14.62 | 14.62 | 14.63 | 14.63 | 14.63 | 14.64 | 14.64 | 14.64 | 14.64 | 14.64 | 14.64 | 14.64 |
| | $K$ | 2.47 | 4.02 | 4.71 | 4.96 | 5.05 | 5.08 | 5.09 | 5.09 | 5.09 | 5.09 | 5.09 | 5.09 | 5.09 | 5.09 | 5.09 | 5.09 | 5.09 | 5.09 | 5.09 | 5.09 | 5.09 | 5.09 | 5.09 | 5.09 | 5.09 |
| 0.9 | $U_1$ | 22.36 | 38.77 | 49.57 | 56.95 | 62.23 | 66.15 | 69.17 | 71.55 | 73.48 | 75.07 | 76.40 | 77.53 | 78.49 | 79.33 | 80.07 | 80.72 | 81.29 | 81.81 | 82.27 | 82.69 | 83.07 | 83.43 | 83.75 | 84.03 | 84.30 |
| | $U_2$ | 0.21 | 0.46 | 0.84 | 1.41 | 2.15 | 3.00 | 3.88 | 4.75 | 5.59 | 6.38 | 7.11 | 7.79 | 8.41 | 8.97 | 9.49 | 9.97 | 10.40 | 10.81 | 11.18 | 11.52 | 11.83 | 12.13 | 12.40 | 12.65 | 12.89 |
| | $W_{1,2}$ | 7.66 | 16.19 | 25.46 | 35.10 | 44.88 | 54.71 | 64.55 | 74.40 | 84.25 | 94.10 | 103.95 | 113.80 | 123.65 | 133.50 | 143.35 | 153.20 | 163.05 | 172.90 | 182.75 | 192.60 | 202.45 | 212.30 | 222.15 | 232.01 | 241.86 |
| | $W_{3,4}$ | 10.92 | 21.75 | 32.10 | 42.14 | 52.06 | 61.93 | 71.79 | 81.64 | 91.49 | 101.35 | 111.20 | 121.05 | 130.90 | 140.75 | 150.60 | 160.45 | 170.30 | 180.15 | 190.00 | 199.85 | 209.70 | 219.55 | 229.40 | 239.26 | 249.11 |
| | $J$ | 7.41 | 11.54 | 13.58 | 14.65 | 15.26 | 15.64 | 15.88 | 16.06 | 16.18 | 16.27 | 16.34 | 16.39 | 16.43 | 16.46 | 16.49 | 16.51 | 16.52 | 16.54 | 16.55 | 16.55 | 16.56 | 16.56 | 16.56 | 16.58 | 16.58 |
| | $K$ | 2.38 | 3.89 | 4.56 | 4.81 | 4.90 | 4.93 | 4.94 | 4.94 | 4.94 | 4.94 | 4.94 | 4.94 | 4.94 | 4.94 | 4.94 | 4.94 | 4.94 | 4.94 | 4.94 | 4.94 | 4.94 | 4.94 | 4.94 | 4.94 | 4.94 |
| 0.8 | $U_1$ | 20.55 | 35.64 | 45.90 | 53.02 | 58.16 | 62.00 | 64.97 | 67.33 | 69.24 | 70.81 | 72.13 | 73.25 | 74.21 | 75.05 | 75.78 | 76.43 | 77.00 | 77.52 | 77.98 | 78.40 | 78.78 | 79.12 | 79.44 | 79.73 | 80.00 |
| | $U_2$ | 0.22 | 0.48 | 0.88 | 1.47 | 2.22 | 3.08 | 3.97 | 4.84 | 5.68 | 6.47 | 7.20 | 7.87 | 8.49 | 9.06 | 9.57 | 10.05 | 10.49 | 10.89 | 11.26 | 11.60 | 11.91 | 12.20 | 12.48 | 12.73 | 12.98 |
| | $W_{1,2}$ | 7.72 | 16.27 | 25.55 | 35.20 | 44.97 | 54.80 | 64.64 | 74.49 | 84.34 | 94.19 | 104.04 | 113.89 | 123.74 | 133.59 | 143.44 | 153.29 | 163.15 | 173.00 | 182.85 | 192.70 | 202.55 | 212.40 | 222.25 | 232.10 | 241.95 |
| | $W_{3,4}$ | 10.43 | 20.89 | 31.07 | 41.05 | 50.94 | 60.81 | 70.66 | 80.52 | 90.37 | 100.22 | 110.07 | 119.92 | 129.77 | 139.62 | 149.47 | 159.32 | 169.17 | 179.02 | 188.88 | 198.73 | 208.58 | 218.43 | 228.28 | 238.13 | 247.98 |
| | $J$ | 7.47 | 11.93 | 14.35 | 15.75 | 16.61 | 17.18 | 17.58 | 17.86 | 18.07 | 18.22 | 18.33 | 18.42 | 18.49 | 18.55 | 18.59 | 18.63 | 18.65 | 18.68 | 18.70 | 18.71 | 18.73 | 18.74 | 18.75 | 18.75 | 18.76 |
| | $K$ | 2.35 | 3.84 | 4.50 | 4.75 | 4.83 | 4.86 | 4.87 | 4.88 | 4.88 | 4.88 | 4.88 | 4.88 | 4.88 | 4.88 | 4.88 | 4.88 | 4.88 | 4.88 | 4.88 | 4.88 | 4.88 | 4.88 | 4.88 | 4.88 | 4.88 |
| 0.7 | $U_1$ | 18.62 | 32.59 | 42.26 | 49.08 | 54.04 | 57.78 | 60.69 | 63.00 | 64.88 | 66.44 | 67.74 | 68.85 | 69.81 | 70.64 | 71.37 | 72.01 | 72.58 | 73.09 | 73.55 | 73.97 | 74.35 | 74.69 | 75.01 | 75.30 | 75.57 |
| | $U_2$ | 0.24 | 0.53 | 0.95 | 1.57 | 2.34 | 3.20 | 4.09 | 4.97 | 5.80 | 6.59 | 7.32 | 7.99 | 8.60 | 9.17 | 9.68 | 10.16 | 10.59 | 10.99 | 11.36 | 11.70 | 12.02 | 12.31 | 12.58 | 12.83 | 13.06 |
| | $W_{1,2}$ | 7.68 | 16.22 | 25.46 | 34.90 | 44.72 | 54.72 | 64.53 | 74.26 | 83.99 | 93.72 | 103.96 | 113.43 | 123.67 | 133.51 | 143.37 | 153.22 | 163.07 | 172.92 | 182.77 | 192.62 | 202.47 | 212.32 | 222.17 | 232.02 | 241.87 |
| | $W_{3,4}$ | 10.06 | 20.26 | 30.30 | 40.23 | 50.11 | 59.97 | 69.83 | 79.68 | 89.53 | 99.38 | 109.23 | 119.08 | 128.93 | 138.78 | 148.63 | 158.48 | 168.33 | 178.19 | 188.04 | 197.89 | 207.74 | 217.59 | 227.44 | 237.29 | 247.14 |
| | $J$ | 7.59 | 12.41 | 15.22 | 16.94 | 18.07 | 18.84 | 19.38 | 19.78 | 20.07 | 20.29 | 20.46 | 20.59 | 20.69 | 20.77 | 20.84 | 20.89 | 20.93 | 20.96 | 20.99 | 21.03 | 21.05 | 21.07 | 21.08 | 21.09 | 21.09 |
| | $K$ | 2.35 | 3.85 | 4.52 | 4.77 | 4.85 | 4.88 | 4.89 | 4.90 | 4.90 | 4.90 | 4.90 | 4.90 | 4.90 | 4.90 | 4.90 | 4.90 | 4.90 | 4.90 | 4.90 | 4.90 | 4.90 | 4.90 | 4.90 | 4.90 | 4.90 |

TABLE S41: Parameters of the THIF interaction Hamiltonian for different ratios $u_0/u_1$ and screening lengths $\xi$ of the double-gate interaction potential ($U_\xi = 24$ meV for $\xi = 10$nm). We use the same BM model parameters as in Table S40.

| $u_0/u_1$ | $\xi$/nm | 2 | 4 | 6 | 8 | 10 | 12 | 14 | 16 | 18 | 20 | 22 | 24 | 26 | 28 | 30 | 32 | 34 | 36 | 38 | 40 | 42 | 44 | 46 | 48 | 50 |
|---|---|---|---|---|---|---|---|---|---|---|---|---|---|---|---|---|---|---|---|---|---|---|---|---|---|---|
| 0.6 | $U_1$ | 17.04 | 30.04 | 39.21 | 45.75 | 50.56 | 54.21 | 57.05 | 59.33 | 61.18 | 62.72 | 64.01 | 65.11 | 66.06 | 66.88 | 67.61 | 68.24 | 68.81 | 69.32 | 69.78 | 70.20 | 70.57 | 70.92 | 71.23 | 71.52 | 71.79 |
| | $U_2$ | 0.26 | 0.57 | 1.02 | 1.66 | 2.44 | 3.31 | 4.21 | 5.08 | 5.92 | 6.70 | 7.43 | 8.10 | 8.71 | 9.27 | 9.79 | 10.26 | 10.69 | 11.09 | 11.46 | 11.80 | 12.11 | 12.40 | 12.67 | 12.92 | 13.16 |
| | $W_{1,2}$ | 7.57 | 16.03 | 25.28 | 34.91 | 44.68 | 54.50 | 64.19 | 73.64 | 83.89 | 93.89 | 103.74 | 113.59 | 123.44 | 133.30 | 143.15 | 153.00 | 162.85 | 172.70 | 182.55 | 192.40 | 202.25 | 212.10 | 221.95 | 231.80 | 241.65 |
| | $W_{3,4}$ | 9.80 | 19.80 | 29.75 | 39.65 | 49.51 | 59.37 | 69.22 | 79.07 | 88.92 | 98.77 | 108.63 | 118.48 | 128.33 | 138.18 | 148.03 | 157.88 | 167.73 | 177.58 | 187.43 | 197.28 | 207.13 | 216.98 | 226.83 | 236.68 | 246.53 |
| | $J$ | 7.75 | 12.90 | 16.07 | 18.00 | 19.44 | 20.39 | 21.06 | 21.56 | 21.93 | 22.21 | 22.42 | 22.59 | 22.72 | 22.82 | 22.90 | 22.97 | 23.02 | 23.07 | 23.10 | 23.13 | 23.16 | 23.18 | 23.20 | 23.22 | 23.23 |
| | $K$ | 2.39 | 3.92 | 4.60 | 4.85 | 4.97 | 4.98 | 4.98 | 4.98 | 4.98 | 4.98 | 4.98 | 4.98 | 4.98 | 4.98 | 4.98 | 4.98 | 4.98 | 4.98 | 4.98 | 4.98 | 4.98 | 4.98 | 4.98 | 4.98 | 4.98 |
| 0.5 | $U_1$ | 15.78 | 26.02 | 34.32 | 40.37 | 44.89 | 48.36 | 51.10 | 53.29 | 55.09 | 56.58 | 57.81 | 58.84 | 59.72 | 60.48 | 61.14 | 61.72 | 62.23 | 62.69 | 63.10 | 63.47 | 63.80 | 64.11 | 64.39 | 64.64 | 65.54 |
| | $U_2$ | 0.27 | 0.61 | 1.01 | 1.74 | 2.54 | 3.41 | 4.31 | 5.18 | 6.02 | 6.80 | 7.52 | 8.19 | 8.80 | 9.36 | 9.87 | 10.34 | 10.78 | 11.17 | 11.54 | 11.88 | 12.19 | 12.48 | 12.75 | 13.00 | 13.24 |
| | $W_{1,2}$ | 7.41 | 15.78 | 24.99 | 34.60 | 44.37 | 54.19 | 64.03 | 73.88 | 83.73 | 93.58 | 103.43 | 113.28 | 123.13 | 132.98 | 142.83 | 152.68 | 162.53 | 172.38 | 182.23 | 192.08 | 201.93 | 211.78 | 221.63 | 231.49 | 241.34 |
| | $W_{3,4}$ | 9.62 | 19.48 | 29.37 | 39.23 | 49.09 | 58.94 | 68.79 | 78.65 | 88.50 | 98.35 | 108.20 | 118.05 | 127.90 | 137.75 | 147.60 | 157.45 | 167.30 | 177.15 | 187.00 | 196.85 | 206.70 | 216.56 | 226.41 | 236.26 | 246.11 |
| | $J$ | 7.93 | 13.40 | 16.89 | 19.17 | 20.74 | 21.84 | 22.63 | 23.22 | 23.66 | 23.99 | 24.24 | 24.44 | 24.60 | 24.72 | 24.82 | 24.90 | 24.97 | 25.02 | 25.06 | 25.10 | 25.13 | 25.16 | 25.18 | 25.20 | 25.22 |
| | $K$ | 2.46 | 4.03 | 4.73 | 4.99 | 5.08 | 5.11 | 5.12 | 5.12 | 5.12 | 5.12 | 5.12 | 5.12 | 5.12 | 5.12 | 5.12 | 5.12 | 5.12 | 5.12 | 5.12 | 5.12 | 5.12 | 5.12 | 5.12 | 5.12 | 5.12 |
| 0.4 | $U_1$ | 14.58 | 24.88 | 32.93 | 38.84 | 43.28 | 46.70 | 49.40 | 51.58 | 53.36 | 54.85 | 56.10 | 57.11 | 57.99 | 58.75 | 59.41 | 59.99 | 60.50 | 60.96 | 61.37 | 61.74 | 62.08 | 62.39 | 62.67 | 62.93 | 63.17 |
| | $U_2$ | 0.30 | 0.67 | 1.17 | 1.85 | 2.66 | 3.54 | 4.43 | 5.31 | 6.14 | 6.92 | 7.64 | 8.30 | 8.91 | 9.46 | 9.98 | 10.44 | 10.87 | 11.27 | 11.64 | 11.97 | 12.28 | 12.57 | 12.84 | 13.09 | 13.33 |
| | $W_{1,2}$ | 7.25 | 15.52 | 24.69 | 34.28 | 44.05 | 53.87 | 63.71 | 73.56 | 83.41 | 93.26 | 103.11 | 112.96 | 122.81 | 132.66 | 142.51 | 152.36 | 162.21 | 172.06 | 181.91 | 191.76 | 201.61 | 211.46 | 221.31 | 231.17 | 241.02 |
| | $W_{3,4}$ | 9.51 | 19.27 | 29.11 | 38.96 | 48.81 | 58.66 | 68.51 | 78.36 | 88.21 | 98.06 | 107.91 | 117.76 | 127.61 | 137.46 | 147.31 | 157.16 | 167.02 | 176.87 | 186.72 | 196.57 | 206.42 | 216.27 | 226.12 | 235.97 | 245.82 |
| | $J$ | 8.14 | 13.92 | 17.73 | 20.30 | 22.10 | 23.38 | 24.31 | 25.01 | 25.54 | 25.94 | 26.25 | 26.49 | 26.68 | 26.84 | 26.96 | 27.06 | 27.15 | 27.21 | 27.27 | 27.32 | 27.36 | 27.39 | 27.42 | 27.45 | 27.47 |
| | $K$ | 2.55 | 4.17 | 4.90 | 5.16 | 5.26 | 5.29 | 5.31 | 5.31 | 5.31 | 5.31 | 5.31 | 5.31 | 5.31 | 5.31 | 5.31 | 5.31 | 5.31 | 5.31 | 5.31 | 5.31 | 5.31 | 5.31 | 5.31 | 5.31 | 5.31 |
| 0.3 | $U_1$ | 13.88 | 24.51 | 32.51 | 38.40 | 42.83 | 46.25 | 48.93 | 51.12 | 52.91 | 54.40 | 55.65 | 56.70 | 57.56 | 58.34 | 59.00 | 59.59 | 60.11 | 60.58 | 60.99 | 61.37 | 61.71 | 62.01 | 62.30 | 62.56 | 62.80 |
| | $U_2$ | 0.31 | 0.69 | 1.21 | 1.90 | 2.71 | 3.59 | 4.49 | 5.36 | 6.19 | 6.97 | 7.69 | 8.35 | 8.95 | 9.51 | 10.02 | 10.49 | 10.92 | 11.32 | 11.68 | 12.02 | 12.33 | 12.62 | 12.88 | 13.14 | 13.37 |
| | $W_{1,2}$ | 7.06 | 15.22 | 24.34 | 33.91 | 43.67 | 53.49 | 63.33 | 73.18 | 83.03 | 92.88 | 102.73 | 112.58 | 122.43 | 132.28 | 142.13 | 151.98 | 161.83 | 171.68 | 181.53 | 191.38 | 201.23 | 211.08 | 220.93 | 230.79 | 240.64 |
| | $W_{3,4}$ | 9.43 | 19.12 | 28.93 | 38.77 | 48.61 | 58.46 | 68.31 | 78.16 | 88.02 | 97.87 | 107.72 | 117.57 | 127.42 | 137.27 | 147.12 | 156.97 | 166.82 | 176.67 | 186.53 | 196.37 | 206.22 | 216.07 | 225.92 | 235.78 | 245.63 |
| | $J$ | 8.44 | 14.62 | 18.80 | 21.65 | 23.66 | 25.09 | 26.13 | 26.87 | 27.43 | 27.86 | 28.18 | 28.43 | 28.63 | 28.79 | 28.92 | 29.02 | 29.11 | 29.16 | 29.23 | 29.28 | 29.32 | 29.39 | 29.44 | 29.48 | 29.50 |
| | $K$ | 2.74 | 4.48 | 5.25 | 5.54 | 5.64 | 5.68 | 5.69 | 5.69 | 5.69 | 5.69 | 5.69 | 5.69 | 5.69 | 5.69 | 5.69 | 5.69 | 5.69 | 5.69 | 5.69 | 5.69 | 5.69 | 5.69 | 5.69 | 5.69 | 5.69 |
| 0.2 | $U_1$ | 13.64 | 23.96 | 31.85 | 37.71 | 42.13 | 45.55 | 48.22 | 50.41 | 52.19 | 53.68 | 54.93 | 55.98 | 56.84 | 57.62 | 58.28 | 58.87 | 59.38 | 59.84 | 60.25 | 60.62 | 60.96 | 61.26 | 61.54 | 61.80 | 62.04 |
| | $U_2$ | 0.30 | 0.68 | 1.19 | 1.88 | 2.70 | 3.58 | 4.48 | 5.36 | 6.19 | 6.97 | 7.68 | 8.35 | 8.95 | 9.51 | 10.02 | 10.49 | 10.92 | 11.31 | 11.68 | 12.02 | 12.33 | 12.62 | 12.88 | 13.13 | 13.37 |
| | $W_{1,2}$ | 6.87 | 14.90 | 23.96 | 33.52 | 43.27 | 53.09 | 62.93 | 72.77 | 82.62 | 92.47 | 102.32 | 112.17 | 122.03 | 131.88 | 141.73 | 151.58 | 161.43 | 171.28 | 181.13 | 190.98 | 200.83 | 210.68 | 220.53 | 230.39 | 240.23 |
| | $W_{3,4}$ | 9.37 | 19.02 | 28.64 | 38.44 | 48.18 | 58.03 | 67.88 | 77.73 | 87.58 | 97.43 | 107.28 | 117.13 | 126.98 | 136.84 | 146.68 | 156.53 | 166.38 | 176.23 | 186.08 | 195.93 | 205.78 | 215.63 | 225.49 | 235.34 | 245.19 |
| | $J$ | 8.53 | 14.78 | 19.02 | 21.88 | 23.88 | 25.30 | 26.33 | 27.07 | 27.62 | 28.05 | 28.36 | 28.61 | 28.80 | 28.96 | 29.09 | 29.16 | 29.25 | 29.29 | 29.33 | 29.39 | 29.40 | 29.45 | 29.50 | 29.53 | 29.55 |
| | $K$ | 2.80 | 4.59 | 5.38 | 5.67 | 5.78 | 5.83 | 5.83 | 5.83 | 5.83 | 5.83 | 5.83 | 5.83 | 5.83 | 5.83 | 5.83 | 5.83 | 5.83 | 5.83 | 5.83 | 5.83 | 5.83 | 5.83 | 5.83 | 5.83 | 5.83 |
| 0.1 | $U_1$ | 13.69 | 24.63 | 32.69 | 38.63 | 43.09 | 46.53 | 49.25 | 51.44 | 53.23 | 54.73 | 55.99 | 57.04 | 57.91 | 58.69 | 59.35 | 59.93 | 60.45 | 60.91 | 61.32 | 61.70 | 62.04 | 62.34 | 62.62 | 62.88 | 63.12 |
| | $U_2$ | 0.28 | 0.64 | 1.15 | 1.83 | 2.64 | 3.53 | 4.43 | 5.30 | 6.13 | 6.91 | 7.63 | 8.30 | 8.90 | 9.46 | 9.97 | 10.44 | 10.87 | 11.27 | 11.63 | 11.97 | 12.28 | 12.57 | 12.84 | 13.09 | 13.33 |
| | $W_{1,2}$ | 6.70 | 14.64 | 23.66 | 33.20 | 42.94 | 52.76 | 62.60 | 72.44 | 82.29 | 92.14 | 101.99 | 111.84 | 121.70 | 131.55 | 141.40 | 151.25 | 161.10 | 170.95 | 180.80 | 190.65 | 200.50 | 210.35 | 220.20 | 230.05 | 239.90 |
| | $W_{3,4}$ | 9.33 | 18.95 | 28.73 | 38.55 | 48.33 | 58.23 | 68.08 | 77.91 | 87.76 | 97.61 | 107.46 | 117.31 | 127.16 | 137.02 | 146.87 | 156.72 | 166.57 | 176.42 | 186.27 | 196.12 | 206.00 | 215.82 | 225.72 | 235.52 | 245.40 |
| | $J$ | 8.56 | 14.85 | 19.08 | 21.96 | 23.96 | 25.38 | 26.40 | 27.15 | 27.70 | 28.13 | 28.44 | 28.69 | 28.88 | 29.04 | 29.16 | 29.25 | 29.33 | 29.39 | 29.45 | 29.49 | 29.53 | 29.57 | 29.59 | 29.60 | 29.62 |
| | $K$ | 2.83 | 4.62 | 5.42 | 5.72 | 5.83 | 5.87 | 5.88 | 5.88 | 5.88 | 5.88 | 5.88 | 5.88 | 5.88 | 5.88 | 5.88 | 5.88 | 5.88 | 5.88 | 5.88 | 5.88 | 5.88 | 5.88 | 5.88 | 5.88 | 5.88 |
| 0.0 | $U_1$ | 13.74 | 24.73 | 32.83 | 38.79 | 43.27 | 46.73 | 49.45 | 51.65 | 53.45 | 54.95 | 56.21 | 57.26 | 58.13 | 58.91 | 59.57 | 60.16 | 60.68 | 61.15 | 61.56 | 61.94 | 62.28 | 62.59 | 62.87 | 63.13 | 63.35 |
| | $U_2$ | 0.27 | 0.62 | 1.13 | 1.80 | 2.62 | 3.50 | 4.40 | 5.28 | 6.11 | 6.89 | 7.61 | 8.27 | 8.88 | 9.44 | 9.95 | 10.42 | 10.85 | 11.25 | 11.61 | 11.95 | 12.26 | 12.55 | 12.82 | 13.07 | 13.31 |
| | $W_{1,2}$ | 6.64 | 14.53 | 23.54 | 33.07 | 42.81 | 52.63 | 62.47 | 72.31 | 82.16 | 92.01 | 101.86 | 111.71 | 121.56 | 131.41 | 141.27 | 151.12 | 160.97 | 170.82 | 180.67 | 190.52 | 200.37 | 210.22 | 220.07 | 229.92 | 239.77 |
| | $W_{3,4}$ | 9.31 | 18.93 | 28.70 | 38.52 | 48.36 | 58.21 | 68.06 | 77.91 | 87.76 | 97.61 | 107.46 | 117.31 | 127.17 | 137.02 | 146.87 | 156.72 | 166.57 | 176.42 | 186.27 | 196.12 | 206.02 | 215.82 | 225.67 | 235.52 | 245.37 |
| | $J$ | 8.60 | 14.92 | 19.16 | 22.04 | 24.05 | 25.48 | 26.51 | 27.26 | 27.81 | 28.24 | 28.56 | 28.81 | 29.00 | 29.16 | 29.29 | 29.39 | 29.40 | 29.45 | 29.50 | 29.55 | 29.57 | 29.57 | 29.59 | 29.62 | 29.64 |
| | $K$ | 2.85 | 4.66 | 5.47 | 5.78 | 5.88 | 5.88 | 5.88 | 5.88 | 5.88 | 5.88 | 5.88 | 5.88 | 5.88 | 5.88 | 5.88 | 5.88 | 5.88 | 5.88 | 5.88 | 5.88 | 5.88 | 5.88 | 5.88 | 5.88 | 5.88 |

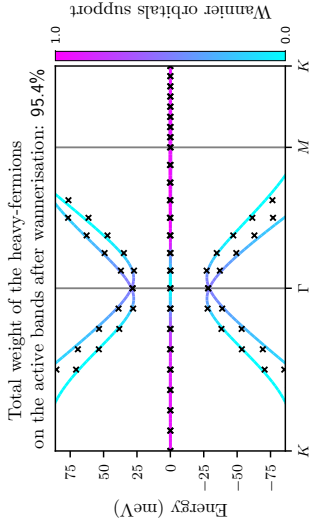

Total weight of the heavy-fermions on the active bands after wannierisation: 95.4%

Wannier orbitals support

FIG. S37. Band structures of the BM and THF models near charge neutrality for $v_0/v_1 = 0.8$, depicted by lines and crosses, respectively. The BM bands are colored according to the weight of the $f$-electron wave function on them. We use the same BM parameters as in Table S42.

TABLE S42. Parameters of the $f$-electron Wannier functions and the THF single-particle Hamiltonian for different values of the tunneling ratio $v_0/v_1$. $W$ denotes the total weight of the THF $f$-electrons on the active bands. In computing the ratios $\gamma/U_1$ and $\gamma/U_2$, we employ the on-site and nearest-neighbor repulsion parameters $U_1$ and $U_2$ obtained numerically for $\xi = 10$ nm, as given in Table S43. We employ $v_F = 5.944$ eV Å, $|\mathbf{K}| = 1.703$ Å$^{-1}$, $v_1 = 110$ meV, and $\theta = 1.08°$ for the BM model.

| $v_0/v_1$ | $\alpha_1$ | $\alpha_2$ | $\lambda_1$ | $\lambda_2$ | $\lambda$ | $W$ | $t_0$ | $\gamma$ | $M$ | $v_*$ | $v'_*$ | $\gamma/U_1$ | $\gamma/U_2$ |
|---|---|---|---|---|---|---|---|---|---|---|---|---|---|
| 1.00 | 0.765 | 0.643 | 0.161 | 0.179 | 0.339 | 0.985 | 0.129 | 0.307 | 1.504 | -3.965 | 1.544 | 0.0075 | 0.00 | 0.13 |
| 0.90 | 0.793 | 0.609 | 0.173 | 0.185 | 0.333 | 0.972 | 0.084 | -14.333 | 0.786 | -4.200 | 1.618 | -0.0164 | -0.23 | -6.29 |
| 0.80 | 0.824 | 0.566 | 0.185 | 0.191 | 0.346 | 0.954 | 0.057 | -29.090 | 0.137 | -4.396 | 1.652 | -0.0297 | -0.50 | -12.20 |
| 0.70 | 0.857 | 0.515 | 0.197 | 0.197 | 0.355 | 0.937 | 0.069 | -43.583 | -0.437 | -4.564 | 1.645 | -0.0351 | -0.80 | -17.60 |
| 0.60 | 0.889 | 0.457 | 0.215 | 0.203 | 0.368 | 0.925 | 0.080 | -57.461 | -0.933 | -4.710 | 1.595 | -0.0347 | -1.12 | -22.24 |
| 0.50 | 0.920 | 0.393 | 0.227 | 0.209 | 0.386 | 0.921 | 0.063 | -70.363 | -1.350 | -4.839 | 1.492 | -0.0302 | -1.47 | -25.91 |
| 0.40 | 0.946 | 0.323 | 0.245 | 0.215 | 0.395 | 0.916 | 0.080 | -81.893 | -1.689 | -4.953 | 1.329 | -0.0231 | -1.80 | -29.31 |
| 0.30 | 0.968 | 0.249 | 0.257 | 0.221 | 0.402 | 0.913 | 0.083 | -91.608 | -1.951 | -5.051 | 1.094 | -0.0149 | -2.08 | -32.16 |
| 0.20 | 0.985 | 0.171 | 0.269 | 0.221 | 0.407 | 0.907 | 0.093 | -99.034 | -2.137 | -5.128 | 0.786 | -0.0074 | -2.24 | -34.69 |
| 0.10 | 0.996 | 0.087 | 0.275 | 0.227 | 0.393 | 0.897 | 0.079 | -103.718 | -2.248 | -5.179 | 0.412 | -0.0020 | -2.36 | -37.34 |
| 0.00 | 1.000 | 0.000 | 0.275 | 0.000 | 0.390 | 0.893 | 0.107 | -105.322 | -2.285 | -5.196 | -0.0000 | -0.0000 | -2.39 | -38.27 |

| $v_0/v_1$ | $\xi$/nm | 2 | 4 | 6 | 8 | 10 | 12 | 14 | 16 | 18 | 20 | 22 | 24 | 26 | 28 | 30 | 32 | 34 | 36 | 38 | 40 | 42 | 44 | 46 | 48 | 50 |
|---|---|---|---|---|---|---|---|---|---|---|---|---|---|---|---|---|---|---|---|---|---|---|---|---|---|---|
| 1.0 | $U_1$ | 24.57 | 41.79 | 53.01 | 60.58 | 65.95 | 69.92 | 72.96 | 75.35 | 77.29 | 78.88 | 80.21 | 81.34 | 82.30 | 83.15 | 83.88 | 84.53 | 85.10 | 85.62 | 86.08 | 86.50 | 86.88 | 87.23 | 87.55 | 87.84 | 88.11 |
| | $U_2$ | 0.64 | 0.51 | 0.91 | 1.51 | 2.29 | 3.17 | 4.09 | 4.99 | 5.85 | 6.66 | 7.40 | 8.11 | 8.71 | 9.30 | 9.81 | 10.30 | 10.74 | 11.14 | 11.52 | 11.86 | 12.18 | 12.47 | 12.74 | 13.00 | 13.23 |
| | $W_{1,2}$ | 7.91 | 16.76 | 26.40 | 36.42 | 44.57 | 56.77 | 66.99 | 77.21 | 87.44 | 97.67 | 107.89 | 118.12 | 128.34 | 138.57 | 148.79 | 159.02 | 169.25 | 179.47 | 189.70 | 199.92 | 210.15 | 220.38 | 230.60 | 240.83 | 251.05 |
| | $W_{3,4}$ | 11.74 | 23.06 | 34.11 | 44.57 | 54.87 | 65.12 | 75.36 | 85.59 | 95.81 | 106.04 | 116.27 | 126.49 | 136.72 | 146.94 | 157.17 | 167.40 | 177.62 | 187.85 | 198.07 | 208.30 | 218.53 | 228.75 | 238.98 | 249.20 | 259.43 |
| | $J$ | 7.55 | 11.46 | 13.20 | 14.02 | 14.43 | 14.65 | 14.78 | 14.87 | 14.92 | 14.95 | 14.98 | 15.00 | 15.01 | 15.02 | 15.03 | 15.04 | 15.04 | 15.05 | 15.05 | 15.05 | 15.06 | 15.06 | 15.06 | 15.06 | 15.06 |
| | $K$ | 2.43 | 3.94 | 4.59 | 4.83 | 4.91 | 4.93 | 4.94 | 4.94 | 4.94 | 4.95 | 4.95 | 4.95 | 4.95 | 4.95 | 4.95 | 4.95 | 4.95 | 4.95 | 4.95 | 4.95 | 4.95 | 4.95 | 4.95 | 4.95 | 4.95 |
| 0.9 | $U_1$ | 22.74 | 39.06 | 49.91 | 57.34 | 62.64 | 66.58 | 69.60 | 72.00 | 73.93 | 75.52 | 76.85 | 77.98 | 78.95 | 79.79 | 80.53 | 81.18 | 81.76 | 82.27 | 82.73 | 83.15 | 83.53 | 83.88 | 84.20 | 84.49 | 84.76 |
| | $U_2$ | 0.22 | 0.48 | 0.88 | 1.49 | 2.28 | 3.17 | 4.09 | 4.99 | 5.85 | 6.66 | 7.41 | 8.10 | 8.72 | 9.30 | 9.83 | 10.31 | 10.75 | 11.15 | 11.53 | 11.87 | 12.19 | 12.48 | 12.76 | 13.01 | 13.25 |
| | $W_{1,2}$ | 8.05 | 16.98 | 26.66 | 36.68 | 46.84 | 57.04 | 67.26 | 77.49 | 87.71 | 97.94 | 108.16 | 118.39 | 128.62 | 138.84 | 149.07 | 159.29 | 169.52 | 179.74 | 189.97 | 200.20 | 210.42 | 220.65 | 230.87 | 241.10 | 251.33 |
| | $W_{3,4}$ | 11.12 | 22.20 | 32.85 | 43.23 | 53.52 | 63.76 | 73.99 | 84.22 | 94.45 | 104.67 | 114.90 | 125.12 | 135.35 | 145.58 | 155.80 | 166.03 | 176.25 | 186.48 | 196.71 | 206.93 | 217.16 | 227.38 | 237.61 | 247.84 | 258.06 |
| | $J$ | 7.56 | 11.79 | 13.90 | 15.03 | 15.68 | 16.09 | 16.36 | 16.55 | 16.69 | 16.79 | 16.86 | 16.92 | 16.96 | 17.00 | 17.03 | 17.05 | 17.07 | 17.08 | 17.09 | 17.10 | 17.11 | 17.12 | 17.13 | 17.13 | 17.13 |
| | $K$ | 2.35 | 3.83 | 4.47 | 4.70 | 4.78 | 4.80 | 4.81 | 4.81 | 4.81 | 4.82 | 4.82 | 4.82 | 4.82 | 4.82 | 4.82 | 4.82 | 4.82 | 4.82 | 4.82 | 4.82 | 4.82 | 4.82 | 4.82 | 4.82 | 4.82 |
| 0.8 | $U_1$ | 20.59 | 35.70 | 45.95 | 53.06 | 58.20 | 62.04 | 65.01 | 67.36 | 69.27 | 70.85 | 72.17 | 73.29 | 74.25 | 75.09 | 75.82 | 76.46 | 77.04 | 77.55 | 78.01 | 78.43 | 78.81 | 79.16 | 79.47 | 79.77 | 80.04 |
| | $U_2$ | 0.23 | 0.52 | 0.95 | 1.58 | 2.38 | 3.28 | 4.20 | 5.11 | 5.97 | 6.77 | 7.52 | 8.20 | 8.83 | 9.40 | 9.93 | 10.41 | 10.85 | 11.25 | 11.63 | 11.97 | 12.29 | 12.58 | 12.85 | 13.11 | 13.34 |
| | $W_{1,2}$ | 8.09 | 17.04 | 26.73 | 36.76 | 46.91 | 57.12 | 67.34 | 77.56 | 87.79 | 98.01 | 108.24 | 118.46 | 128.69 | 138.91 | 149.14 | 159.37 | 169.59 | 179.82 | 190.04 | 200.27 | 210.50 | 220.72 | 230.95 | 241.17 | 251.40 |
| | $W_{3,4}$ | 10.65 | 21.39 | 31.88 | 42.21 | 52.47 | 62.71 | 72.94 | 83.17 | 93.39 | 103.62 | 113.85 | 124.07 | 134.30 | 144.52 | 154.75 | 164.98 | 175.20 | 185.43 | 195.65 | 205.88 | 216.10 | 226.33 | 236.56 | 246.78 | 257.01 |
| | $J$ | 7.65 | 12.25 | 14.77 | 16.24 | 17.17 | 17.79 | 18.22 | 18.53 | 18.76 | 18.94 | 19.07 | 19.17 | 19.25 | 19.31 | 19.36 | 19.40 | 19.43 | 19.46 | 19.48 | 19.50 | 19.51 | 19.53 | 19.54 | 19.55 | 19.56 |
| | $K$ | 2.33 | 3.79 | 4.42 | 4.65 | 4.73 | 4.76 | 4.77 | 4.77 | 4.77 | 4.77 | 4.77 | 4.77 | 4.77 | 4.77 | 4.77 | 4.77 | 4.77 | 4.77 | 4.77 | 4.77 | 4.77 | 4.77 | 4.77 | 4.77 | 4.77 |
| 0.7 | $U_1$ | 18.84 | 32.94 | 42.69 | 49.55 | 54.54 | 58.50 | 61.22 | 63.54 | 65.43 | 66.99 | 68.30 | 69.41 | 70.37 | 71.20 | 71.93 | 72.57 | 73.14 | 73.66 | 74.12 | 74.53 | 74.91 | 75.26 | 75.57 | 75.87 | 76.13 |
| | $U_2$ | 0.25 | 0.55 | 1.01 | 1.66 | 2.48 | 3.38 | 4.30 | 5.21 | 6.07 | 6.88 | 7.62 | 8.30 | 8.93 | 9.50 | 10.02 | 10.50 | 10.94 | 11.35 | 11.72 | 12.06 | 12.38 | 12.67 | 12.94 | 13.20 | 13.43 |
| | $W_{1,2}$ | 8.03 | 16.94 | 26.61 | 36.63 | 46.79 | 56.99 | 67.21 | 77.44 | 87.66 | 97.89 | 108.11 | 118.34 | 128.56 | 138.79 | 149.02 | 159.24 | 169.47 | 179.69 | 189.92 | 200.15 | 210.37 | 220.60 | 230.82 | 241.05 | 251.27 |
| | $W_{3,4}$ | 10.31 | 20.79 | 31.17 | 41.46 | 51.70 | 61.94 | 72.17 | 82.39 | 92.62 | 102.84 | 113.07 | 123.30 | 133.52 | 143.75 | 153.97 | 164.20 | 174.43 | 184.65 | 194.88 | 205.10 | 215.33 | 225.56 | 235.78 | 246.01 | 256.23 |
| | $J$ | 7.79 | 12.74 | 15.63 | 17.41 | 18.57 | 19.36 | 19.92 | 20.33 | 20.63 | 20.85 | 21.02 | 21.16 | 21.26 | 21.34 | 21.41 | 21.46 | 21.50 | 21.53 | 21.56 | 21.59 | 21.61 | 21.63 | 21.64 | 21.65 | 21.66 |
| | $K$ | 2.34 | 3.82 | 4.45 | 4.69 | 4.77 | 4.79 | 4.80 | 4.80 | 4.80 | 4.80 | 4.81 | 4.81 | 4.81 | 4.81 | 4.81 | 4.81 | 4.81 | 4.81 | 4.81 | 4.81 | 4.81 | 4.81 | 4.81 | 4.81 | 4.81 |

| $u_0/u_1$ | $\xi$/nm | 2 | 4 | 6 | 8 | 10 | 12 | 14 | 16 | 18 | 20 | 22 | 24 | 26 | 28 | 30 | 32 | 34 | 36 | 38 | 40 | 42 | 44 | 46 | 48 | 50 |
|---|---|---|---|---|---|---|---|---|---|---|---|---|---|---|---|---|---|---|---|---|---|---|---|---|---|---|
| 0.6 | $U_1$ | 17.28 | 30.44 | 39.70 | 46.29 | 51.13 | 54.80 | 57.66 | 59.50 | 61.81 | 63.35 | 64.65 | 65.75 | 66.70 | 67.53 | 68.25 | 68.89 | 69.46 | 69.97 | 70.43 | 70.85 | 71.22 | 71.57 | 71.89 | 72.18 | 72.44 |
| | $U_2$ | 0.27 | 0.60 | 1.08 | 1.76 | 2.58 | 3.49 | 4.42 | 5.32 | 6.18 | 6.99 | 7.73 | 8.41 | 9.03 | 9.60 | 10.12 | 10.60 | 11.04 | 11.44 | 11.81 | 12.15 | 12.47 | 12.76 | 13.03 | 13.29 | 13.52 |
| | $W_{1,2}$ | 7.91 | 16.75 | 26.39 | 36.40 | 46.55 | 56.75 | 66.97 | 77.19 | 87.42 | 97.65 | 107.87 | 118.10 | 128.32 | 138.55 | 148.77 | 159.00 | 169.23 | 179.45 | 189.68 | 199.90 | 210.13 | 220.36 | 230.58 | 240.81 | 251.03 |
| | $W_{3,4}$ | 10.07 | 20.37 | 30.60 | 40.91 | 51.15 | 61.38 | 71.61 | 81.83 | 92.06 | 102.29 | 112.51 | 122.74 | 132.96 | 143.19 | 153.42 | 163.64 | 173.87 | 184.09 | 194.32 | 204.54 | 214.77 | 225.00 | 235.22 | 245.45 | 255.67 |
| | $J$ | 7.96 | 13.26 | 16.61 | 18.50 | 19.98 | 20.95 | 21.64 | 22.15 | 22.52 | 22.80 | 23.02 | 23.19 | 23.32 | 23.42 | 23.50 | 23.57 | 23.62 | 23.67 | 23.70 | 23.74 | 23.76 | 23.78 | 23.80 | 23.82 | 23.83 |
| | $K$ | 2.39 | 3.89 | 4.54 | 4.78 | 4.86 | 4.89 | 4.90 | 4.90 | 4.90 | 4.90 | 4.90 | 4.90 | 4.90 | 4.90 | 4.90 | 4.90 | 4.90 | 4.90 | 4.90 | 4.90 | 4.90 | 4.90 | 4.90 | 4.90 | 4.90 |
| 0.5 | $U_1$ | 15.83 | 28.07 | 36.82 | 43.13 | 47.80 | 51.37 | 54.16 | 56.40 | 58.23 | 59.75 | 61.03 | 62.12 | 63.05 | 63.88 | 64.60 | 65.23 | 65.80 | 66.31 | 66.76 | 67.18 | 67.55 | 67.89 | 68.21 | 68.50 | 68.76 |
| | $U_2$ | 0.30 | 0.66 | 1.18 | 1.87 | 2.72 | 3.63 | 4.56 | 5.46 | 6.32 | 7.12 | 7.85 | 8.53 | 9.15 | 9.72 | 10.24 | 10.71 | 11.15 | 11.55 | 11.92 | 12.26 | 12.58 | 12.87 | 13.14 | 13.39 | 13.63 |
| | $W_{1,2}$ | 7.76 | 16.51 | 26.11 | 36.11 | 46.26 | 56.46 | 66.68 | 76.90 | 87.13 | 97.35 | 107.58 | 117.81 | 128.03 | 138.26 | 148.48 | 158.71 | 168.94 | 179.16 | 189.39 | 199.61 | 209.84 | 220.06 | 230.29 | 240.52 | 250.74 |
| | $W_{3,4}$ | 9.91 | 20.08 | 30.31 | 40.54 | 50.77 | 60.99 | 71.22 | 81.45 | 91.67 | 101.90 | 112.12 | 122.35 | 132.58 | 142.80 | 153.03 | 163.26 | 173.48 | 183.71 | 193.93 | 204.16 | 214.38 | 224.61 | 234.84 | 245.06 | 255.29 |
| | $J$ | 8.17 | 13.83 | 17.44 | 19.83 | 21.47 | 22.63 | 23.47 | 24.09 | 24.56 | 24.91 | 25.19 | 25.40 | 25.57 | 25.70 | 25.81 | 25.90 | 25.97 | 26.03 | 26.07 | 26.11 | 26.15 | 26.18 | 26.20 | 26.22 | 26.24 |
| | $K$ | 2.46 | 4.01 | 4.68 | 4.93 | 5.01 | 5.04 | 5.05 | 5.05 | 5.05 | 5.05 | 5.05 | 5.05 | 5.05 | 5.05 | 5.05 | 5.05 | 5.05 | 5.05 | 5.05 | 5.05 | 5.05 | 5.05 | 5.05 | 5.05 | 5.05 |
| 0.4 | $U_1$ | 14.86 | 26.50 | 34.92 | 41.04 | 45.61 | 49.11 | 51.86 | 54.07 | 55.88 | 57.39 | 58.66 | 59.74 | 60.68 | 61.49 | 62.21 | 62.84 | 63.40 | 63.91 | 64.36 | 64.78 | 65.15 | 65.49 | 65.81 | 66.10 | 66.36 |
| | $U_2$ | 0.31 | 0.70 | 1.23 | 1.94 | 2.79 | 3.71 | 4.64 | 5.54 | 6.40 | 7.20 | 7.93 | 8.61 | 9.22 | 9.79 | 10.31 | 10.78 | 11.22 | 11.62 | 11.99 | 12.33 | 12.64 | 12.93 | 13.20 | 13.45 | 13.69 |
| | $W_{1,2}$ | 7.58 | 16.21 | 25.77 | 35.75 | 45.89 | 56.09 | 66.31 | 76.53 | 86.76 | 96.98 | 107.21 | 117.43 | 127.66 | 137.88 | 148.11 | 158.34 | 168.56 | 178.79 | 189.01 | 199.24 | 209.47 | 219.69 | 229.92 | 240.14 | 250.37 |
| | $W_{3,4}$ | 9.79 | 19.88 | 30.06 | 40.28 | 50.50 | 60.73 | 70.95 | 81.18 | 91.41 | 101.63 | 111.86 | 122.08 | 132.31 | 142.54 | 152.76 | 162.99 | 173.21 | 183.44 | 193.67 | 203.89 | 214.12 | 224.34 | 234.57 | 244.79 | 255.02 |
| | $J$ | 8.38 | 14.32 | 18.23 | 20.86 | 22.68 | 23.98 | 24.93 | 25.63 | 26.10 | 26.56 | 26.87 | 27.11 | 27.30 | 27.45 | 27.58 | 27.68 | 27.76 | 27.83 | 27.88 | 27.93 | 27.97 | 28.00 | 28.03 | 28.05 | 28.08 |
| | $K$ | 2.56 | 4.16 | 4.86 | 5.11 | 5.20 | 5.23 | 5.24 | 5.24 | 5.24 | 5.24 | 5.24 | 5.24 | 5.24 | 5.24 | 5.24 | 5.24 | 5.24 | 5.24 | 5.24 | 5.24 | 5.24 | 5.24 | 5.24 | 5.24 | 5.24 |
| 0.3 | $U_1$ | 14.18 | 25.38 | 33.56 | 39.55 | 44.04 | 47.49 | 50.21 | 52.40 | 54.20 | 55.69 | 56.96 | 58.04 | 58.97 | 59.78 | 60.49 | 61.12 | 61.68 | 62.19 | 62.64 | 63.05 | 63.43 | 63.77 | 64.08 | 64.37 | 64.64 |
| | $U_2$ | 0.32 | 0.72 | 1.27 | 1.99 | 2.85 | 3.77 | 4.70 | 5.60 | 6.46 | 7.25 | 7.98 | 8.66 | 9.27 | 9.84 | 10.36 | 10.83 | 11.26 | 11.66 | 12.03 | 12.37 | 12.68 | 12.98 | 13.26 | 13.51 | 13.74 |
| | $W_{1,2}$ | 7.38 | 15.90 | 25.40 | 35.37 | 45.51 | 55.70 | 65.92 | 76.14 | 86.37 | 96.59 | 106.82 | 117.04 | 127.27 | 137.50 | 147.72 | 157.95 | 168.17 | 178.40 | 188.63 | 198.85 | 209.08 | 219.30 | 229.53 | 239.76 | 249.98 |
| | $W_{3,4}$ | 9.72 | 19.74 | 29.90 | 40.11 | 50.33 | 60.55 | 70.78 | 81.00 | 91.23 | 101.45 | 111.68 | 121.91 | 132.13 | 142.36 | 152.58 | 162.81 | 173.04 | 183.26 | 193.49 | 203.71 | 213.94 | 224.17 | 234.39 | 244.62 | 254.84 |
| | $J$ | 8.70 | 15.06 | 19.35 | 22.28 | 24.32 | 25.79 | 26.86 | 27.64 | 28.22 | 28.65 | 28.98 | 29.23 | 29.42 | 29.58 | 29.71 | 29.81 | 29.90 | 29.97 | 30.03 | 30.08 | 30.12 | 30.16 | 30.19 | 30.22 | 30.24 |
| | $K$ | 2.66 | 4.33 | 5.05 | 5.31 | 5.40 | 5.43 | 5.45 | 5.45 | 5.45 | 5.45 | 5.45 | 5.45 | 5.45 | 5.45 | 5.45 | 5.45 | 5.45 | 5.45 | 5.45 | 5.45 | 5.45 | 5.45 | 5.45 | 5.45 | 5.45 |
| 0.2 | $U_1$ | 13.87 | 24.89 | 32.99 | 38.93 | 43.39 | 46.83 | 49.55 | 51.73 | 53.53 | 55.02 | 56.30 | 57.36 | 58.29 | 59.10 | 59.81 | 60.45 | 61.01 | 61.51 | 61.97 | 62.38 | 62.75 | 63.09 | 63.41 | 63.69 | 63.96 |
| | $U_2$ | 0.32 | 0.72 | 1.27 | 1.99 | 2.85 | 3.78 | 4.71 | 5.61 | 6.46 | 7.26 | 7.99 | 8.67 | 9.28 | 9.85 | 10.37 | 10.84 | 11.27 | 11.67 | 12.04 | 12.38 | 12.70 | 12.99 | 13.26 | 13.51 | 13.74 |
| | $W_{1,2}$ | 7.18 | 15.59 | 25.04 | 34.99 | 45.13 | 55.32 | 65.54 | 75.75 | 85.98 | 96.20 | 106.43 | 116.66 | 126.88 | 137.11 | 147.33 | 157.56 | 167.79 | 178.01 | 188.24 | 198.46 | 208.69 | 218.92 | 229.14 | 239.37 | 249.59 |
| | $W_{3,4}$ | 9.67 | 19.69 | 29.79 | 39.90 | 50.21 | 60.43 | 70.66 | 80.88 | 91.11 | 101.33 | 111.56 | 121.79 | 132.01 | 142.24 | 152.46 | 162.69 | 172.91 | 183.14 | 193.36 | 203.59 | 213.82 | 224.04 | 234.27 | 244.50 | 254.72 |
| | $J$ | 8.78 | 15.20 | 19.52 | 22.44 | 24.50 | 25.97 | 27.05 | 27.82 | 28.40 | 28.84 | 29.17 | 29.42 | 29.62 | 29.77 | 29.89 | 30.00 | 30.08 | 30.15 | 30.21 | 30.27 | 30.30 | 30.33 | 30.35 | 30.33 | 30.35 |
| | $K$ | 2.75 | 4.48 | 5.23 | 5.50 | 5.59 | 5.62 | 5.64 | 5.64 | 5.64 | 5.64 | 5.64 | 5.64 | 5.64 | 5.64 | 5.64 | 5.64 | 5.64 | 5.64 | 5.64 | 5.64 | 5.64 | 5.64 | 5.64 | 5.64 | 5.64 |
| 0.1 | $U_1$ | 14.02 | 25.19 | 33.39 | 39.41 | 43.93 | 47.41 | 50.15 | 52.35 | 54.16 | 55.66 | 56.93 | 58.01 | 58.95 | 59.76 | 60.48 | 61.11 | 61.68 | 62.18 | 62.64 | 63.05 | 63.43 | 63.77 | 64.08 | 64.37 | 64.64 |
| | $U_2$ | 0.29 | 0.67 | 1.20 | 1.92 | 2.78 | 3.70 | 4.63 | 5.54 | 6.39 | 7.19 | 7.92 | 8.60 | 9.22 | 9.79 | 10.30 | 10.78 | 11.22 | 11.61 | 11.98 | 12.32 | 12.64 | 12.93 | 13.20 | 13.45 | 13.68 |
| | $W_{1,2}$ | 7.00 | 15.30 | 24.71 | 34.64 | 44.77 | 54.96 | 65.17 | 75.40 | 85.62 | 95.85 | 106.07 | 116.30 | 126.52 | 136.75 | 146.98 | 157.20 | 167.43 | 177.65 | 187.88 | 198.10 | 208.33 | 218.56 | 228.78 | 239.01 | 249.23 |
| | $W_{3,4}$ | 9.63 | 19.58 | 29.71 | 39.90 | 50.12 | 60.34 | 70.57 | 80.79 | 91.02 | 101.24 | 111.47 | 121.70 | 131.92 | 142.15 | 152.37 | 162.60 | 172.83 | 183.05 | 193.28 | 203.50 | 213.73 | 223.95 | 234.18 | 244.41 | 254.63 |
| | $J$ | 8.78 | 15.20 | 19.52 | 22.46 | 24.50 | 25.90 | 26.94 | 27.69 | 28.26 | 28.68 | 29.00 | 29.25 | 29.45 | 29.60 | 29.73 | 29.83 | 29.92 | 29.99 | 30.05 | 30.10 | 30.15 | 30.17 | 30.20 | 30.22 | 30.24 |
| | $K$ | 2.82 | 4.59 | 5.35 | 5.63 | 5.73 | 5.76 | 5.77 | 5.77 | 5.77 | 5.77 | 5.77 | 5.77 | 5.77 | 5.77 | 5.77 | 5.77 | 5.77 | 5.77 | 5.77 | 5.77 | 5.77 | 5.77 | 5.77 | 5.77 | 5.77 |
| 0.0 | $U_1$ | 14.08 | 25.29 | 33.53 | 39.58 | 44.11 | 47.61 | 50.35 | 52.57 | 54.38 | 55.88 | 57.15 | 58.24 | 59.18 | 59.99 | 60.71 | 61.34 | 61.91 | 62.41 | 62.87 | 63.28 | 63.66 | 64.00 | 64.31 | 64.60 | 64.87 |
| | $U_2$ | 0.28 | 0.65 | 1.18 | 1.90 | 2.75 | 3.67 | 4.61 | 5.51 | 6.37 | 7.17 | 7.90 | 8.58 | 9.20 | 9.76 | 10.28 | 10.76 | 11.19 | 11.59 | 11.96 | 12.30 | 12.62 | 12.91 | 13.18 | 13.43 | 13.67 |
| | $W_{1,2}$ | 6.94 | 15.20 | 24.59 | 34.51 | 44.63 | 54.83 | 65.04 | 75.26 | 85.49 | 95.71 | 105.94 | 116.16 | 126.39 | 136.62 | 146.84 | 157.07 | 167.29 | 177.52 | 187.75 | 197.97 | 208.20 | 218.42 | 228.65 | 238.88 | 249.10 |
| | $W_{3,4}$ | 9.61 | 19.56 | 29.68 | 39.88 | 50.09 | 60.31 | 70.54 | 80.76 | 90.99 | 101.22 | 111.44 | 121.67 | 131.89 | 142.12 | 152.35 | 162.57 | 172.80 | 183.02 | 193.25 | 203.47 | 213.70 | 223.93 | 234.15 | 244.38 | 254.60 |
| | $J$ | 8.82 | 15.27 | 19.60 | 22.52 | 24.55 | 25.97 | 27.00 | 27.75 | 28.30 | 28.71 | 29.03 | 29.27 | 29.46 | 29.61 | 29.73 | 29.84 | 29.91 | 29.97 | 30.02 | 30.07 | 30.10 | 30.13 | 30.16 | 30.18 | 30.20 |
| | $K$ | 2.84 | 4.63 | 5.40 | 5.68 | 5.78 | 5.81 | 5.82 | 5.82 | 5.82 | 5.82 | 5.82 | 5.82 | 5.82 | 5.82 | 5.82 | 5.82 | 5.82 | 5.82 | 5.82 | 5.82 | 5.82 | 5.82 | 5.82 | 5.82 | 5.82 |

TABLE S43: Parameters of the THF interaction Hamiltonian for different ratios $u_0/u_1$ and screening lengths $\xi$ of the double-gate interaction potential ($U_\xi = 24$ meV for $\xi = 10$nm). We use the same BM model parameters as in Table S42.

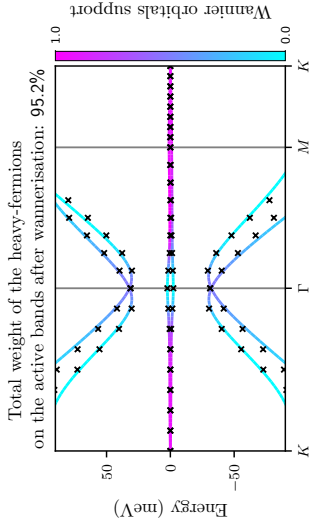

Total weight of the heavy-fermions on the active bands after wannierisation: 95.2%

Wannier orbitals support

Energy (meV) — axis labels: $K$ $\Gamma$ $M$ $K$

FIG. S38. Band structures of the BM and THF models near charge neutrality for $u_0/u_1 = 0.8$, depicted by lines and crosses, respectively. The BM bands are colored according to the weight of the $f$-electron wave function on them. We use the same BM parameters as in Table S44.

TABLE S44. Parameters of the $f$-electron Wannier functions for different values of the tunneling ratio $u_0/u_1$. $W$ denotes the total weight of the THF $f$-electrons on the active bands. In computing the ratios $\gamma/U_1$ and $\gamma/U_2$, we employ the on-site and nearest-neighbor repulsion parameters $U_1$ and $U_2$ obtained numerically for $\xi = 10$ nm, as given in Table S45. We employ $v_F = 5.944$ eV Å, $|\mathbf{K}| = 1.703$ Å$^{-1}$, $w_1 = 110$ meV, and $\theta = 1.10°$ for the BM model.

| $u_0/u_1$ | $\alpha_1$ | $\alpha_2$ | $\lambda_1$ | $\lambda_2$ | $\lambda$ | $W$ | $t_0$ | $\gamma$ | $M$ | $v_*$ | $v_*''$ | $\gamma/U_1$ | $\gamma/U_2$ |
|---|---|---|---|---|---|---|---|---|---|---|---|---|---|
| 1.00 | 0.768 | 0.640 | 0.161 | 0.179 | 0.337 | 0.983 | 0.218 | -2.314 | -0.757 | 1.576 | 0.0055 | -0.03 | -0.96 |
| 0.90 | 0.796 | 0.605 | 0.173 | 0.185 | 0.335 | 0.969 | 0.209 | -17.153 | -1.583 | 1.642 | 0.0156 | -0.27 | -7.09 |
| 0.80 | 0.828 | 0.561 | 0.191 | 0.197 | 0.348 | 0.952 | 0.228 | -32.036 | -2.322 | 1.670 | -0.0272 | -0.55 | -12.67 |
| 0.70 | 0.860 | 0.510 | 0.203 | 0.203 | 0.357 | 0.935 | 0.286 | -46.606 | -2.970 | 1.659 | -0.0318 | -0.85 | -17.79 |
| 0.60 | 0.892 | 0.452 | 0.215 | 0.209 | 0.376 | 0.928 | 0.313 | -60.527 | -3.526 | 1.604 | -0.0313 | -1.18 | -21.90 |
| 0.50 | 0.921 | 0.389 | 0.233 | 0.215 | 0.387 | 0.920 | 0.364 | -73.447 | -3.991 | 1.498 | -0.0271 | -1.52 | -25.70 |
| 0.40 | 0.948 | 0.320 | 0.245 | 0.215 | 0.396 | 0.915 | 0.417 | -84.978 | -4.367 | 1.332 | -0.0207 | -1.83 | -28.94 |
| 0.30 | 0.969 | 0.246 | 0.263 | 0.221 | 0.403 | 0.912 | 0.445 | -94.683 | -4.658 | 1.096 | -0.0134 | -2.11 | -31.67 |
| 0.20 | 0.986 | 0.168 | 0.260 | 0.227 | 0.410 | 0.911 | 0.440 | -102.095 | -4.864 | 0.786 | -0.0066 | -2.34 | -33.56 |
| 0.10 | 0.996 | 0.087 | 0.275 | 0.227 | 0.399 | 0.900 | 0.321 | -106.768 | -4.987 | 0.412 | -0.0017 | -2.41 | -36.10 |
| 0.00 | 1.000 | 0.000 | 0.281 | 0.000 | 0.390 | 0.893 | 0.291 | -108.367 | -5.027 | -5.229 | -0.0000 | -2.41 | -37.48 |

| $u_0/u_1$ | $\xi$/nm | 2 | 4 | 6 | 8 | 10 | 12 | 14 | 16 | 18 | 20 | 22 | 24 | 26 | 28 | 30 | 32 | 34 | 36 | 38 | 40 | 42 | 44 | 46 | 48 | 50 |
|---|---|---|---|---|---|---|---|---|---|---|---|---|---|---|---|---|---|---|---|---|---|---|---|---|---|---|
| 1.0 | $U_1$ | 24.83 | 42.20 | 53.52 | 61.14 | 66.55 | 70.54 | 73.60 | 76.00 | 77.55 | 79.55 | 80.88 | 82.01 | 82.99 | 83.83 | 84.57 | 85.22 | 85.79 | 86.31 | 86.77 | 87.19 | 87.57 | 87.92 | 88.24 | 88.53 | 88.80 |
| | $U_2$ | 0.24 | 0.52 | 0.95 | 1.59 | 2.41 | 3.33 | 4.28 | 5.21 | 6.10 | 6.92 | 7.69 | 8.41 | 9.02 | 9.61 | 10.14 | 10.63 | 11.07 | 11.48 | 11.86 | 12.20 | 12.52 | 12.82 | 13.09 | 13.35 | 13.59 |
| | $W_{1,2}$ | 8.32 | 17.60 | 27.65 | 38.06 | 48.61 | 59.19 | 69.80 | 80.40 | 91.01 | 101.62 | 112.23 | 122.83 | 133.44 | 144.00 | 154.66 | 165.27 | 175.87 | 186.48 | 197.00 | 207.70 | 218.31 | 228.92 | 239.52 | 250.13 | 260.74 |
| | $W_{3,4}$ | 11.92 | 23.68 | 34.83 | 45.63 | 56.31 | 66.94 | 77.55 | 88.16 | 98.77 | 109.39 | 119.99 | 130.60 | 141.20 | 151.81 | 162.42 | 173.03 | 183.64 | 194.25 | 204.85 | 215.46 | 226.07 | 236.68 | 247.29 | 257.89 | 268.50 |
| | $J$ | 7.68 | 11.67 | 13.46 | 14.32 | 14.76 | 15.01 | 15.16 | 15.26 | 15.32 | 15.40 | 15.40 | 15.42 | 15.44 | 15.45 | 15.46 | 15.47 | 15.48 | 15.49 | 15.49 | 15.50 | 15.50 | 15.50 | 15.50 | 15.51 | 15.51 |
| | $K$ | 2.39 | 3.86 | 4.48 | 4.70 | 4.77 | 4.79 | 4.80 | 4.80 | 4.80 | 4.81 | 4.81 | 4.81 | 4.81 | 4.81 | 4.81 | 4.81 | 4.81 | 4.81 | 4.81 | 4.81 | 4.81 | 4.81 | 4.81 | 4.81 | 4.81 |
| 0.9 | $U_1$ | 22.91 | 39.32 | 50.22 | 57.67 | 62.99 | 66.85 | 69.98 | 72.38 | 74.32 | 75.91 | 77.25 | 78.38 | 79.35 | 80.19 | 80.93 | 81.58 | 82.16 | 82.67 | 83.14 | 83.56 | 83.94 | 84.29 | 84.61 | 84.90 | 85.17 |
| | $U_2$ | 0.23 | 0.51 | 0.94 | 1.59 | 2.42 | 3.35 | 4.30 | 5.24 | 6.12 | 6.95 | 7.71 | 8.41 | 9.05 | 9.63 | 10.17 | 10.65 | 11.10 | 11.51 | 11.89 | 12.23 | 12.55 | 12.85 | 13.12 | 13.38 | 13.62 |
| | $W_{1,2}$ | 8.45 | 17.79 | 27.87 | 38.29 | 48.84 | 59.43 | 70.03 | 80.63 | 91.24 | 101.85 | 112.46 | 123.07 | 133.67 | 144.28 | 154.89 | 165.50 | 176.11 | 186.71 | 197.32 | 207.93 | 218.54 | 229.15 | 239.76 | 250.36 | 260.97 |
| | $W_{3,4}$ | 11.33 | 22.67 | 33.64 | 44.38 | 55.03 | 65.66 | 76.27 | 86.88 | 97.49 | 108.10 | 118.70 | 129.31 | 139.92 | 150.53 | 161.14 | 171.74 | 182.35 | 192.96 | 203.57 | 214.18 | 224.78 | 235.39 | 246.00 | 256.61 | 267.22 |
| | $J$ | 7.72 | 12.05 | 14.23 | 15.16 | 16.53 | 16.82 | 17.13 | 17.28 | 17.36 | 17.42 | 17.47 | 17.51 | 17.54 | 17.56 | 17.58 | 17.59 | 17.61 | 17.62 | 17.63 | 17.63 | 17.64 | 17.64 | 17.65 | 17.65 | 17.65 |
| | $K$ | 2.33 | 3.77 | 4.37 | 4.59 | 4.66 | 4.68 | 4.69 | 4.69 | 4.69 | 4.70 | 4.70 | 4.70 | 4.70 | 4.70 | 4.70 | 4.70 | 4.70 | 4.70 | 4.70 | 4.70 | 4.70 | 4.70 | 4.70 | 4.70 | 4.70 |
| 0.8 | $U_1$ | 20.77 | 35.98 | 46.30 | 53.45 | 58.60 | 62.46 | 65.44 | 67.80 | 69.71 | 71.29 | 72.61 | 73.73 | 74.70 | 75.54 | 76.27 | 76.92 | 77.49 | 78.00 | 78.47 | 78.89 | 79.27 | 79.61 | 79.93 | 80.22 | 80.49 |
| | $U_2$ | 0.24 | 0.55 | 1.01 | 1.68 | 2.53 | 3.46 | 4.42 | 5.36 | 6.24 | 7.06 | 7.83 | 8.52 | 9.16 | 9.74 | 10.27 | 10.76 | 11.20 | 11.61 | 11.99 | 12.33 | 12.65 | 12.95 | 13.22 | 13.48 | 13.71 |
| | $W_{1,2}$ | 8.47 | 17.82 | 27.90 | 38.33 | 48.87 | 59.40 | 70.06 | 80.67 | 91.28 | 101.89 | 112.49 | 123.10 | 133.71 | 144.32 | 154.93 | 165.53 | 176.14 | 186.75 | 197.36 | 207.97 | 218.57 | 229.18 | 239.79 | 250.40 | 261.01 |
| | $W_{3,4}$ | 10.89 | 21.91 | 32.73 | 43.42 | 54.06 | 64.68 | 75.29 | 85.90 | 96.51 | 107.12 | 117.72 | 128.33 | 138.94 | 149.55 | 160.16 | 170.76 | 181.37 | 191.98 | 202.59 | 213.20 | 223.80 | 234.41 | 245.02 | 255.63 | 266.24 |
| | $J$ | 7.84 | 12.55 | 15.16 | 16.69 | 17.66 | 18.30 | 18.76 | 19.08 | 19.32 | 19.50 | 19.63 | 19.74 | 19.82 | 19.89 | 19.94 | 19.98 | 20.01 | 20.04 | 20.06 | 20.08 | 20.10 | 20.11 | 20.12 | 20.13 | 20.14 |
| | $K$ | 2.31 | 3.74 | 4.35 | 4.56 | 4.64 | 4.66 | 4.67 | 4.67 | 4.67 | 4.67 | 4.67 | 4.67 | 4.67 | 4.67 | 4.67 | 4.67 | 4.67 | 4.67 | 4.67 | 4.67 | 4.67 | 4.67 | 4.67 | 4.67 | 4.67 |
| 0.7 | $U_1$ | 19.06 | 33.30 | 43.12 | 50.03 | 55.05 | 58.73 | 61.76 | 64.09 | 65.98 | 67.55 | 68.86 | 69.97 | 70.93 | 71.77 | 72.50 | 73.14 | 73.72 | 74.23 | 74.69 | 75.11 | 75.49 | 75.83 | 76.15 | 76.44 | 76.71 |
| | $U_2$ | 0.26 | 0.58 | 1.07 | 1.76 | 2.62 | 3.56 | 4.52 | 5.46 | 6.34 | 7.16 | 7.92 | 8.62 | 9.25 | 9.83 | 10.36 | 10.85 | 11.29 | 11.70 | 12.07 | 12.42 | 12.74 | 13.03 | 13.31 | 13.56 | 13.80 |
| | $W_{1,2}$ | 8.40 | 17.70 | 27.76 | 38.18 | 48.72 | 59.31 | 69.91 | 80.52 | 91.12 | 101.73 | 112.34 | 122.95 | 133.56 | 144.17 | 154.77 | 165.38 | 175.99 | 186.60 | 197.21 | 207.81 | 218.42 | 229.03 | 239.64 | 250.25 | 260.85 |
| | $W_{3,4}$ | 7.99 | 13.08 | 16.05 | 17.89 | 19.09 | 19.90 | 20.48 | 20.90 | 21.21 | 21.44 | 21.61 | 21.74 | 21.85 | 21.93 | 22.00 | 22.05 | 22.09 | 22.13 | 22.16 | 22.18 | 22.20 | 22.22 | 22.23 | 22.24 | 22.25 |
| | $J$ | 21.35 | 32.07 | 42.72 | 53.35 | 63.96 | 74.57 | 85.18 | 95.79 | 106.40 | 117.00 | 127.61 | 138.22 | 148.83 | 159.44 | 170.05 | 180.65 | 191.26 | 201.87 | 212.48 | 223.09 | 233.69 | 244.30 | 254.91 | 265.52 | 276.13 |
| | $K$ | 2.33 | 3.78 | 4.39 | 4.61 | 4.68 | 4.71 | 4.72 | 4.72 | 4.72 | 4.72 | 4.72 | 4.72 | 4.72 | 4.72 | 4.72 | 4.72 | 4.72 | 4.72 | 4.72 | 4.72 | 4.72 | 4.72 | 4.72 | 4.72 | 4.72 |

| $u_0/u_1$ | $\xi$/nm | 2 | 4 | 6 | 8 | 10 | 12 | 14 | 16 | 18 | 20 | 22 | 24 | 26 | 28 | 30 | 32 | 34 | 36 | 38 | 40 | 42 | 44 | 46 | 48 | 50 |
|---|---|---|---|---|---|---|---|---|---|---|---|---|---|---|---|---|---|---|---|---|---|---|---|---|---|---|
| 0.6 | $U_1$ | 17.33 | 30.50 | 39.75 | 46.34 | 51.18 | 54.85 | 57.71 | 59.99 | 61.85 | 63.39 | 64.69 | 65.79 | 66.74 | 67.57 | 68.29 | 68.93 | 69.50 | 70.01 | 70.47 | 70.88 | 71.26 | 71.61 | 71.92 | 72.21 | 72.48 |
|  | $U_2$ | 0.29 | 0.65 | 1.17 | 1.89 | 2.76 | 3.71 | 4.67 | 5.61 | 6.49 | 7.31 | 8.06 | 8.75 | 9.38 | 9.96 | 10.49 | 10.97 | 11.41 | 11.82 | 12.19 | 12.54 | 12.85 | 13.15 | 13.42 | 13.68 | 13.91 |
|  | $W_{1,2}$ | 8.28 | 17.51 | 27.55 | 37.95 | 48.49 | 59.08 | 69.68 | 80.29 | 90.89 | 101.50 | 112.11 | 122.72 | 133.33 | 143.93 | 154.54 | 165.15 | 175.76 | 186.37 | 196.98 | 207.58 | 218.19 | 228.80 | 239.41 | 250.02 | 260.62 |
|  | $W_{3,4}$ | 10.35 | 20.96 | 31.60 | 42.23 | 52.84 | 63.45 | 74.06 | 84.67 | 95.28 | 105.88 | 116.49 | 127.10 | 137.71 | 148.32 | 158.93 | 169.53 | 180.14 | 190.75 | 201.36 | 211.97 | 222.57 | 233.18 | 243.79 | 254.40 | 265.01 |
|  | $J$ | 8.20 | 13.67 | 17.05 | 19.22 | 20.68 | 21.71 | 22.44 | 22.98 | 23.38 | 23.68 | 23.92 | 24.10 | 24.24 | 24.35 | 24.44 | 24.52 | 24.57 | 24.62 | 24.66 | 24.70 | 24.73 | 24.75 | 24.77 | 24.79 | 24.80 |
|  | $K$ | 2.39 | 3.87 | 4.49 | 4.72 | 4.79 | 4.82 | 4.83 | 4.83 | 4.83 | 4.83 | 4.83 | 4.83 | 4.83 | 4.83 | 4.83 | 4.83 | 4.83 | 4.83 | 4.83 | 4.83 | 4.83 | 4.83 | 4.83 | 4.83 | 4.83 |
| 0.5 | $U_1$ | 16.10 | 28.52 | 37.38 | 43.75 | 48.46 | 52.06 | 54.87 | 57.12 | 58.96 | 60.48 | 61.77 | 62.86 | 63.81 | 64.63 | 65.35 | 65.99 | 66.55 | 67.06 | 67.52 | 67.93 | 68.31 | 68.65 | 68.97 | 69.26 | 69.52 |
|  | $U_2$ | 0.31 | 0.69 | 1.24 | 1.97 | 2.86 | 3.81 | 4.77 | 5.70 | 6.58 | 7.40 | 8.15 | 8.84 | 9.47 | 10.05 | 10.57 | 11.06 | 11.50 | 11.90 | 12.27 | 12.62 | 12.93 | 13.23 | 13.50 | 13.75 | 13.99 |
|  | $W_{1,2}$ | 8.10 | 17.23 | 27.23 | 37.62 | 48.16 | 58.74 | 69.34 | 79.95 | 90.56 | 101.16 | 111.77 | 122.39 | 132.99 | 143.60 | 154.20 | 164.81 | 175.42 | 186.03 | 196.64 | 207.24 | 217.85 | 228.46 | 239.07 | 249.68 | 260.29 |
|  | $W_{3,4}$ | 10.19 | 20.69 | 31.27 | 41.88 | 52.49 | 63.09 | 73.70 | 84.31 | 94.92 | 105.53 | 116.13 | 126.74 | 137.35 | 147.96 | 158.57 | 169.18 | 179.79 | 190.39 | 201.00 | 211.61 | 222.22 | 232.82 | 243.43 | 254.04 | 264.65 |
|  | $J$ | 8.41 | 14.22 | 17.93 | 20.37 | 22.04 | 23.22 | 24.07 | 24.70 | 25.17 | 25.52 | 25.80 | 26.01 | 26.18 | 26.31 | 26.42 | 26.50 | 26.57 | 26.63 | 26.68 | 26.72 | 26.75 | 26.78 | 26.80 | 26.83 | 26.84 |
|  | $K$ | 2.47 | 4.00 | 4.64 | 4.87 | 4.95 | 4.98 | 4.99 | 4.99 | 4.99 | 4.99 | 4.99 | 4.99 | 4.99 | 4.99 | 4.99 | 4.99 | 4.99 | 4.99 | 4.99 | 4.99 | 4.99 | 4.99 | 4.99 | 4.99 | 4.99 |
| 0.4 | $U_1$ | 15.15 | 26.99 | 35.52 | 41.71 | 46.32 | 49.85 | 52.63 | 54.85 | 56.67 | 58.18 | 59.46 | 60.55 | 61.48 | 62.30 | 63.02 | 63.65 | 64.22 | 64.72 | 65.18 | 65.59 | 65.97 | 66.31 | 66.62 | 66.92 | 67.18 |
|  | $U_2$ | 0.32 | 0.73 | 1.29 | 2.04 | 2.94 | 3.89 | 4.86 | 5.79 | 6.66 | 7.48 | 8.23 | 8.92 | 9.59 | 10.12 | 10.64 | 11.12 | 11.56 | 11.97 | 12.34 | 12.68 | 13.00 | 13.29 | 13.56 | 13.82 | 14.05 |
|  | $W_{1,2}$ | 7.91 | 16.92 | 26.87 | 37.25 | 47.78 | 58.36 | 68.96 | 79.56 | 90.17 | 100.77 | 111.38 | 121.99 | 132.60 | 143.21 | 153.82 | 164.43 | 175.04 | 185.64 | 196.25 | 206.86 | 217.47 | 228.08 | 238.69 | 249.29 | 259.90 |
|  | $W_{3,4}$ | 10.09 | 20.50 | 31.05 | 41.64 | 52.24 | 62.85 | 73.46 | 84.07 | 94.67 | 105.28 | 115.89 | 126.50 | 137.11 | 147.71 | 158.32 | 168.93 | 179.54 | 190.15 | 200.75 | 211.36 | 221.97 | 232.58 | 243.19 | 253.80 | 264.40 |
|  | $J$ | 8.62 | 14.73 | 18.73 | 21.42 | 23.27 | 24.59 | 25.54 | 26.25 | 26.76 | 27.15 | 27.45 | 27.73 | 27.92 | 28.07 | 28.19 | 28.29 | 28.37 | 28.44 | 28.49 | 28.54 | 28.58 | 28.61 | 28.64 | 28.66 | 28.68 |
|  | $K$ | 2.57 | 4.15 | 4.83 | 5.06 | 5.14 | 5.17 | 5.18 | 5.18 | 5.18 | 5.18 | 5.18 | 5.18 | 5.18 | 5.18 | 5.18 | 5.18 | 5.18 | 5.18 | 5.18 | 5.18 | 5.18 | 5.18 | 5.18 | 5.18 | 5.18 |
| 0.3 | $U_1$ | 14.48 | 25.89 | 34.20 | 40.26 | 44.80 | 48.28 | 51.03 | 53.23 | 55.04 | 56.54 | 57.81 | 58.89 | 59.83 | 60.64 | 61.36 | 61.99 | 62.55 | 63.06 | 63.51 | 63.92 | 64.30 | 64.64 | 64.96 | 65.24 | 65.51 |
|  | $U_2$ | 0.33 | 0.75 | 1.33 | 2.09 | 3.05 | 4.01 | 4.91 | 5.84 | 6.72 | 7.53 | 8.28 | 8.97 | 9.59 | 10.12 | 10.64 | 11.17 | 11.61 | 12.01 | 12.38 | 12.72 | 13.04 | 13.33 | 13.60 | 13.86 | 14.09 |
|  | $W_{1,2}$ | 7.70 | 16.60 | 26.50 | 36.87 | 47.38 | 57.96 | 68.56 | 79.17 | 89.78 | 100.38 | 110.99 | 121.60 | 132.21 | 142.82 | 153.43 | 164.03 | 174.64 | 185.25 | 195.86 | 206.46 | 217.07 | 227.68 | 238.29 | 248.90 | 259.51 |
|  | $W_{3,4}$ | 9.99 | 20.38 | 30.90 | 41.48 | 52.08 | 62.69 | 73.30 | 83.90 | 94.51 | 105.12 | 115.73 | 126.34 | 136.94 | 147.55 | 158.16 | 168.77 | 179.38 | 189.99 | 200.59 | 211.20 | 221.81 | 232.42 | 243.03 | 253.63 | 264.24 |
|  | $J$ | 8.81 | 15.16 | 19.40 | 22.27 | 24.26 | 25.69 | 26.72 | 27.49 | 28.06 | 28.50 | 28.83 | 29.10 | 29.30 | 29.47 | 29.60 | 29.71 | 29.80 | 29.87 | 29.93 | 29.98 | 30.02 | 30.06 | 30.09 | 30.10 | 30.12 |
|  | $K$ | 2.67 | 4.31 | 5.01 | 5.25 | 5.33 | 5.36 | 5.38 | 5.38 | 5.39 | 5.39 | 5.39 | 5.39 | 5.39 | 5.39 | 5.39 | 5.39 | 5.39 | 5.39 | 5.39 | 5.39 | 5.39 | 5.39 | 5.39 | 5.39 | 5.39 |
| 0.2 | $U_1$ | 13.95 | 25.02 | 33.13 | 39.08 | 43.55 | 46.99 | 49.70 | 51.89 | 53.68 | 55.18 | 56.44 | 57.51 | 58.45 | 59.26 | 59.97 | 60.60 | 61.16 | 61.67 | 62.12 | 62.53 | 62.91 | 63.25 | 63.56 | 63.85 | 64.12 |
|  | $U_2$ | 0.34 | 0.77 | 1.38 | 2.14 | 3.10 | 4.07 | 4.97 | 5.90 | 6.77 | 7.58 | 8.33 | 9.01 | 9.64 | 10.21 | 10.73 | 11.21 | 11.65 | 12.05 | 12.42 | 12.77 | 13.08 | 13.37 | 13.65 | 13.90 | 14.13 |
|  | $W_{1,2}$ | 7.54 | 16.34 | 26.22 | 36.54 | 47.06 | 57.64 | 68.24 | 78.84 | 89.45 | 100.06 | 110.67 | 121.27 | 131.88 | 142.49 | 153.10 | 163.71 | 174.32 | 184.92 | 195.53 | 206.14 | 216.75 | 227.36 | 237.96 | 248.57 | 259.18 |
|  | $W_{3,4}$ | 9.98 | 20.31 | 30.82 | 41.39 | 51.99 | 62.59 | 73.20 | 83.81 | 94.42 | 105.02 | 115.63 | 126.24 | 136.85 | 147.46 | 158.06 | 168.67 | 179.28 | 189.89 | 200.50 | 211.11 | 221.71 | 232.32 | 242.93 | 253.54 | 264.15 |
|  | $J$ | 8.98 | 15.54 | 19.98 | 23.01 | 25.13 | 26.63 | 27.78 | 28.59 | 29.21 | 29.68 | 30.03 | 30.30 | 30.55 | 30.73 | 30.87 | 30.99 | 31.09 | 31.17 | 31.23 | 31.29 | 31.33 | 31.37 | 31.41 | 31.44 | 31.46 |
|  | $K$ | 2.77 | 4.48 | 5.20 | 5.44 | 5.53 | 5.56 | 5.57 | 5.58 | 5.58 | 5.58 | 5.58 | 5.58 | 5.58 | 5.58 | 5.58 | 5.58 | 5.58 | 5.58 | 5.58 | 5.58 | 5.58 | 5.58 | 5.58 | 5.58 | 5.58 |
| 0.1 | $U_1$ | 14.16 | 25.41 | 33.66 | 39.71 | 44.24 | 47.73 | 50.47 | 52.68 | 54.49 | 56.00 | 57.27 | 58.35 | 59.30 | 60.12 | 60.84 | 61.48 | 62.04 | 62.55 | 63.00 | 63.40 | 63.76 | 64.11 | 64.42 | 64.71 | 64.98 |
|  | $U_2$ | 0.32 | 0.72 | 1.29 | 2.06 | 2.96 | 3.92 | 4.88 | 5.81 | 6.69 | 7.51 | 8.26 | 8.94 | 9.57 | 10.14 | 10.67 | 11.15 | 11.59 | 11.99 | 12.36 | 12.71 | 13.02 | 13.31 | 13.59 | 13.84 | 14.08 |
|  | $W_{1,2}$ | 7.34 | 16.03 | 25.84 | 36.17 | 46.69 | 57.26 | 67.86 | 78.47 | 89.07 | 99.68 | 110.29 | 120.90 | 131.51 | 142.11 | 152.72 | 163.33 | 173.94 | 184.55 | 195.15 | 205.76 | 216.37 | 226.98 | 237.59 | 248.20 | 258.80 |
|  | $W_{3,4}$ | 9.94 | 20.24 | 30.73 | 41.30 | 51.90 | 62.51 | 73.11 | 83.72 | 94.33 | 104.94 | 115.54 | 126.15 | 136.76 | 147.37 | 157.97 | 168.58 | 179.19 | 189.80 | 200.41 | 211.02 | 221.63 | 232.23 | 242.84 | 253.45 | 264.06 |
|  | $J$ | 9.06 | 15.68 | 20.12 | 23.14 | 25.23 | 26.71 | 27.78 | 28.56 | 29.16 | 29.59 | 29.92 | 30.18 | 30.38 | 30.55 | 30.68 | 30.78 | 30.86 | 30.93 | 30.98 | 31.02 | 31.06 | 31.09 | 31.11 | 31.14 | 31.17 |
|  | $K$ | 2.84 | 4.59 | 5.33 | 5.59 | 5.68 | 5.71 | 5.72 | 5.72 | 5.72 | 5.72 | 5.72 | 5.72 | 5.72 | 5.72 | 5.72 | 5.72 | 5.72 | 5.72 | 5.72 | 5.72 | 5.72 | 5.72 | 5.72 | 5.72 | 5.72 |
| 0.0 | $U_1$ | 14.41 | 25.85 | 34.23 | 40.36 | 44.95 | 48.48 | 51.25 | 53.48 | 55.30 | 56.81 | 58.09 | 59.18 | 60.12 | 60.94 | 61.66 | 62.29 | 62.86 | 63.37 | 63.82 | 64.24 | 64.61 | 64.96 | 65.27 | 65.56 | 65.83 |
|  | $U_2$ | 0.29 | 0.68 | 1.24 | 2.00 | 2.96 | 3.92 | 4.88 | 5.81 | 6.69 | 7.51 | 8.20 | 8.89 | 9.52 | 10.09 | 10.62 | 11.10 | 11.54 | 11.94 | 12.31 | 12.66 | 12.97 | 13.27 | 13.54 | 13.79 | 14.03 |
|  | $W_{1,2}$ | 7.34 | 16.03 | 25.66 | 35.98 | 46.49 | 57.07 | 67.67 | 78.27 | 88.88 | 99.49 | 110.10 | 120.70 | 131.31 | 141.92 | 152.53 | 163.14 | 173.74 | 184.35 | 194.96 | 205.56 | 216.18 | 226.79 | 237.39 | 248.00 | 258.61 |
|  | $W_{3,4}$ | 9.92 | 20.20 | 30.69 | 41.26 | 51.86 | 62.46 | 73.07 | 83.68 | 94.29 | 104.89 | 115.50 | 126.11 | 136.72 | 147.33 | 157.93 | 168.54 | 179.15 | 189.76 | 200.37 | 210.98 | 221.58 | 232.19 | 242.80 | 253.41 | 264.02 |
|  | $J$ | 9.08 | 15.70 | 20.12 | 23.09 | 25.14 | 26.57 | 27.60 | 28.34 | 28.89 | 29.30 | 29.62 | 29.86 | 30.04 | 30.19 | 30.31 | 30.40 | 30.48 | 30.53 | 30.59 | 30.63 | 30.67 | 30.70 | 30.72 | 30.75 | 30.76 |
|  | $K$ | 2.86 | 4.63 | 5.38 | 5.64 | 5.73 | 5.76 | 5.77 | 5.77 | 5.77 | 5.77 | 5.77 | 5.77 | 5.77 | 5.77 | 5.77 | 5.77 | 5.77 | 5.77 | 5.77 | 5.77 | 5.77 | 5.77 | 5.77 | 5.77 | 5.77 |

TABLE S45. Parameters of the THF interaction Hamiltonian for different ratios $u_0/u_1$ and screening lengths $\xi$ of the double-gate interaction potential ($U_\xi = 24$ meV nm for $\xi = 10$nm). We use the same BM model parameters as in Table S44.

Total weight of the heavy-fermions on the active bands after wannierisation: 94.9%

FIG. S39. Band structures of the BM and THF models near charge neutrality for $v_0/v_1 = 0.8$, depicted by lines and crosses, respectively. The BM bands are colored according to the weight of the $f$-electron wave function on them. We use the same BM parameters as in Table S46.

TABLE S46. Parameters of the $f$-electron Wannier functions and the THF single-particle Hamiltonian for different values of the tunneling ratio $v_0/v_1$. $W$ denotes the total weight of the THF $f$-electrons on the active bands. In computing the ratios $\gamma/U_1$ and $\gamma/U_2$, we employ the on-site and nearest-neighbor repulsion parameters $U_1$ and $U_2$ obtained numerically for $\xi = 10$ nm, as given in Table S47. We employ $v_F = 5.944$ eV·Å, $|\mathbf{K}| = 1.703$ Å$^{-1}$, $v_1 = 110$ meV, and $\theta = 1.12°$ for the BM model.

| $v_0/v_1$ | $\alpha_1$ | $\alpha_2$ | $\lambda_1$ | $\lambda_2$ | $\lambda$ | $W$ | $t_0$ | $\gamma$ | $M$ | $v_\star$ | $v_\star'$ | $v_\star''$ | $\gamma/U_1$ | $\gamma/U_2$ |
|---|---|---|---|---|---|---|---|---|---|---|---|---|---|---|
| 1.00 | 0.771 | 0.637 | 0.167 | 0.185 | 0.337 | 0.982 | 0.319 | -5.007 | -3.099 | 1.606 | 0.0043 | -0.07 | -1.98 | |
| 0.90 | 0.799 | 0.601 | 0.179 | 0.191 | 0.337 | 0.965 | 0.350 | -20.023 | -4.024 | -4.329 | 1.664 | 0.0143 | -0.32 | -7.81 |
| 0.80 | 0.831 | 0.557 | 0.191 | 0.197 | 0.351 | 0.949 | 0.415 | -35.019 | -4.843 | -4.505 | 1.686 | 0.0245 | -0.59 | -13.09 |
| 0.70 | 0.863 | 0.505 | 0.203 | 0.203 | 0.359 | 0.933 | 0.518 | -49.658 | -5.557 | -4.658 | 1.671 | 0.0285 | -0.89 | -17.94 |
| 0.60 | 0.894 | 0.448 | 0.221 | 0.209 | 0.378 | 0.926 | 0.589 | -63.616 | -6.168 | -4.794 | 1.613 | -0.0279 | -1.23 | -21.85 |
| 0.50 | 0.923 | 0.384 | 0.233 | 0.215 | 0.388 | 0.919 | 0.678 | -76.551 | -6.677 | -4.916 | 1.504 | -0.0241 | -1.56 | -25.47 |
| 0.40 | 0.949 | 0.316 | 0.251 | 0.221 | 0.397 | 0.915 | 0.765 | -88.081 | -7.088 | -5.025 | 1.335 | -0.0184 | -1.87 | -28.57 |
| 0.30 | 0.970 | 0.243 | 0.263 | 0.221 | 0.405 | 0.912 | 0.815 | -97.776 | -7.404 | -5.119 | 1.097 | -0.0118 | -2.15 | -31.07 |
| 0.20 | 0.986 | 0.166 | 0.275 | 0.227 | 0.410 | 0.911 | 0.829 | -105.175 | -7.628 | -5.193 | 0.786 | -0.0058 | -2.37 | -33.00 |
| 0.10 | 0.996 | 0.085 | 0.281 | 0.227 | 0.403 | 0.903 | 0.717 | -109.836 | -7.762 | -5.242 | 0.412 | -0.0015 | -2.46 | -35.09 |
| 0.00 | 1.000 | 0.000 | 0.281 | 0.000 | 0.400 | 0.899 | 0.651 | -111.430 | -7.806 | -5.259 | -0.0000 | 0.0000 | -2.48 | -35.90 |

| $v_0/v_1$ | $\xi/$nm | 2 | 4 | 6 | 8 | 10 | 12 | 14 | 16 | 18 | 20 | 22 | 24 | 26 | 28 | 30 | 32 | 34 | 36 | 38 | 40 | 42 | 44 | 46 | 48 | 50 |
|---|---|---|---|---|---|---|---|---|---|---|---|---|---|---|---|---|---|---|---|---|---|---|---|---|---|---|
| 1.0 | $U_1$ | 25.06 | 42.57 | 53.96 | 61.63 | 67.00 | 71.08 | 74.15 | 76.57 | 78.52 | 80.12 | 81.46 | 82.60 | 83.57 | 84.42 | 85.16 | 85.81 | 86.39 | 86.90 | 87.37 | 87.79 | 88.17 | 88.52 | 88.84 | 89.13 | 89.40 |
| | $U_2$ | 0.24 | 0.54 | 0.99 | 1.67 | 2.53 | 3.50 | 4.48 | 5.44 | 6.35 | 7.20 | 7.98 | 8.69 | 9.33 | 9.93 | 10.47 | 10.96 | 11.41 | 11.82 | 12.20 | 12.55 | 12.87 | 13.17 | 13.45 | 13.70 | 13.94 |
| | $W_{1,2}$ | 8.75 | 18.45 | 28.03 | 39.74 | 50.68 | 61.66 | 72.65 | 83.65 | 94.64 | 105.64 | 116.64 | 127.63 | 138.63 | 149.63 | 160.63 | 171.62 | 182.62 | 193.62 | 204.62 | 215.61 | 226.61 | 237.61 | 248.61 | 259.60 | 270.60 |
| | $W_{3,4}$ | 12.12 | 24.12 | 35.58 | 46.75 | 57.80 | 68.82 | 79.82 | 90.82 | 101.82 | 112.82 | 123.81 | 134.81 | 145.81 | 156.81 | 167.80 | 178.80 | 189.80 | 200.80 | 211.79 | 222.79 | 233.79 | 244.78 | 255.78 | 266.78 | 277.78 |
| | $J$ | 7.82 | 11.89 | 13.75 | 14.65 | 15.13 | 15.40 | 15.58 | 15.69 | 15.76 | 15.82 | 15.86 | 15.89 | 15.91 | 15.93 | 15.94 | 15.95 | 15.96 | 15.97 | 15.98 | 15.98 | 15.98 | 15.99 | 15.99 | 15.99 | 16.00 |
| | $K$ | 2.35 | 3.78 | 4.37 | 4.57 | 4.64 | 4.66 | 4.67 | 4.67 | 4.67 | 4.67 | 4.67 | 4.67 | 4.67 | 4.67 | 4.67 | 4.67 | 4.67 | 4.67 | 4.67 | 4.67 | 4.67 | 4.67 | 4.67 | 4.67 | 4.67 |
| 0.9 | $U_1$ | 23.08 | 39.59 | 50.54 | 58.02 | 63.36 | 67.33 | 70.38 | 72.78 | 74.72 | 76.36 | 77.66 | 78.77 | 79.77 | 80.61 | 81.35 | 82.00 | 82.58 | 83.09 | 83.56 | 83.98 | 84.36 | 84.71 | 85.03 | 85.32 | 85.59 |
| | $U_2$ | 0.24 | 0.53 | 0.99 | 1.69 | 2.56 | 3.53 | 4.52 | 5.49 | 6.40 | 7.24 | 8.02 | 8.73 | 9.38 | 9.97 | 10.51 | 11.00 | 11.45 | 11.86 | 12.24 | 12.59 | 12.92 | 13.21 | 13.49 | 13.75 | 13.99 |
| | $W_{1,2}$ | 8.86 | 18.61 | 29.10 | 39.93 | 50.87 | 61.85 | 72.84 | 83.84 | 94.83 | 105.83 | 116.83 | 127.82 | 138.82 | 149.82 | 160.82 | 171.81 | 182.81 | 193.81 | 204.81 | 215.80 | 226.80 | 237.80 | 248.80 | 259.79 | 270.79 |
| | $W_{3,4}$ | 11.56 | 23.17 | 34.47 | 45.57 | 56.61 | 67.62 | 78.62 | 89.62 | 100.61 | 111.61 | 122.61 | 133.61 | 144.60 | 155.60 | 166.60 | 177.60 | 188.59 | 199.59 | 210.59 | 221.59 | 232.58 | 243.58 | 254.58 | 265.58 | 276.57 |
| | $J$ | 7.89 | 12.32 | 14.58 | 15.81 | 16.54 | 17.00 | 17.31 | 17.53 | 17.69 | 17.80 | 17.89 | 17.96 | 18.01 | 18.05 | 18.08 | 18.11 | 18.13 | 18.15 | 18.16 | 18.17 | 18.18 | 18.18 | 18.19 | 18.19 | 18.20 |
| | $K$ | 2.30 | 3.71 | 4.28 | 4.48 | 4.55 | 4.57 | 4.58 | 4.58 | 4.58 | 4.58 | 4.58 | 4.58 | 4.58 | 4.58 | 4.58 | 4.58 | 4.58 | 4.58 | 4.58 | 4.58 | 4.58 | 4.58 | 4.58 | 4.58 | 4.58 |
| 0.8 | $U_1$ | 20.97 | 36.29 | 46.67 | 53.86 | 59.04 | 62.91 | 65.90 | 68.27 | 70.19 | 71.77 | 73.09 | 74.22 | 75.18 | 76.02 | 76.76 | 77.40 | 77.98 | 78.49 | 78.96 | 79.38 | 79.76 | 80.10 | 80.42 | 80.71 | 80.98 |
| | $U_2$ | 0.25 | 1.07 | 1.79 | 2.68 | 3.65 | 4.64 | 5.61 | 6.51 | 7.36 | 8.13 | 8.84 | 9.49 | 10.08 | 10.61 | 11.11 | 11.55 | 11.95 | 12.32 | 12.69 | 13.01 | 13.31 | 13.59 | 13.84 | 14.08 | |
| | $W_{1,2}$ | 8.86 | 18.61 | 29.10 | 39.93 | 50.87 | 61.85 | 72.84 | 83.84 | 94.83 | 105.83 | 116.83 | 127.82 | 138.82 | 149.82 | 160.82 | 171.81 | 182.81 | 193.81 | 204.81 | 215.80 | 226.80 | 237.80 | 248.80 | 259.79 | 270.79 |
| | $W_{3,4}$ | 11.14 | 22.44 | 33.62 | 44.68 | 55.70 | 66.77 | 77.77 | 88.71 | 99.70 | 110.70 | 121.70 | 132.70 | 143.69 | 154.69 | 165.69 | 176.68 | 187.68 | 198.68 | 209.68 | 220.67 | 231.67 | 242.67 | 253.67 | 264.66 | 275.66 |
| | $J$ | 8.03 | 12.87 | 15.56 | 17.15 | 18.15 | 18.83 | 19.30 | 19.64 | 19.88 | 20.07 | 20.21 | 20.32 | 20.40 | 20.47 | 20.52 | 20.56 | 20.60 | 20.63 | 20.65 | 20.67 | 20.69 | 20.70 | 20.71 | 20.72 | 20.73 |
| | $K$ | 2.30 | 3.70 | 4.28 | 4.48 | 4.54 | 4.56 | 4.57 | 4.57 | 4.57 | 4.57 | 4.57 | 4.57 | 4.57 | 4.57 | 4.57 | 4.57 | 4.57 | 4.57 | 4.57 | 4.57 | 4.57 | 4.57 | 4.57 | 4.57 | 4.57 |
| 0.7 | $U_1$ | 19.29 | 33.66 | 43.57 | 50.52 | 55.57 | 59.37 | 62.31 | 64.65 | 66.55 | 68.12 | 69.43 | 70.55 | 71.51 | 72.35 | 73.08 | 73.72 | 74.30 | 74.81 | 75.27 | 75.69 | 76.07 | 76.42 | 76.73 | 77.02 | 77.29 |
| | $U_2$ | 0.27 | 0.61 | 1.13 | 1.87 | 2.77 | 3.75 | 4.74 | 5.71 | 6.61 | 7.46 | 8.23 | 8.94 | 9.58 | 10.17 | 10.70 | 11.19 | 11.64 | 12.05 | 12.43 | 12.78 | 13.10 | 13.40 | 13.67 | 13.93 | 14.17 |
| | $W_{1,2}$ | 8.77 | 18.46 | 28.93 | 39.75 | 50.69 | 61.67 | 72.66 | 83.65 | 94.65 | 105.65 | 116.64 | 127.64 | 138.64 | 149.64 | 160.63 | 171.63 | 182.63 | 193.63 | 204.63 | 215.62 | 226.62 | 237.62 | 248.61 | 259.61 | 270.61 |
| | $W_{3,4}$ | 10.84 | 21.93 | 33.01 | 44.03 | 55.04 | 66.04 | 77.04 | 88.04 | 99.04 | 110.03 | 121.03 | 132.03 | 143.03 | 154.02 | 165.02 | 176.02 | 187.02 | 198.01 | 209.01 | 220.01 | 231.01 | 242.00 | 253.00 | 264.00 | 274.99 |
| | $J$ | 8.20 | 13.42 | 16.48 | 18.37 | 19.60 | 20.44 | 21.03 | 21.46 | 21.77 | 22.01 | 22.18 | 22.32 | 22.42 | 22.51 | 22.57 | 22.63 | 22.67 | 22.71 | 22.73 | 22.76 | 22.78 | 22.79 | 22.81 | 22.82 | 22.83 |
| | $K$ | 2.33 | 3.75 | 4.33 | 4.54 | 4.60 | 4.62 | 4.63 | 4.63 | 4.63 | 4.63 | 4.63 | 4.63 | 4.63 | 4.63 | 4.63 | 4.63 | 4.63 | 4.63 | 4.63 | 4.63 | 4.63 | 4.63 | 4.63 | 4.63 | 4.63 |

TABLE S47. Parameters of the THF interaction Hamiltonian for different ratios $u_0/u_1$ and screening lengths $\xi$ of the double-gate interaction potential ($U_\xi = 24\,\mathrm{meV}$ for $\xi = 10\,\mathrm{nm}$). We use the same BM model parameters as in Table S46.

| $u_0/u_1$ | | 2 | 4 | 6 | 8 | 10 | 12 | 14 | 16 | 18 | 20 | 22 | 24 | 26 | 28 | 30 | 32 | 34 | 36 | 38 | 40 | 42 | 44 | 46 | 48 | 50 |
|---|---|---|---|---|---|---|---|---|---|---|---|---|---|---|---|---|---|---|---|---|---|---|---|---|---|---|
| 0.6 | $U_1$ | 17.58 | 30.01 | 40.26 | 46.91 | 51.78 | 55.47 | 58.35 | 60.64 | 62.51 | 64.05 | 65.35 | 66.46 | 67.41 | 68.24 | 68.96 | 69.61 | 70.18 | 70.69 | 71.15 | 71.56 | 71.94 | 72.28 | 72.60 | 72.89 | 73.16 |
| | $U_2$ | 0.30 | 0.68 | 1.23 | 2.00 | 2.91 | 3.90 | 4.89 | 5.85 | 6.76 | 7.60 | 8.37 | 9.07 | 9.71 | 10.30 | 10.83 | 11.32 | 11.76 | 12.17 | 12.55 | 12.90 | 13.22 | 13.51 | 13.79 | 14.04 | 14.28 |
| | $W_{1,2}$ | 8.64 | 18.26 | 28.70 | 39.51 | 50.44 | 61.42 | 72.41 | 83.40 | 94.40 | 105.40 | 116.40 | 127.39 | 138.39 | 149.39 | 160.39 | 171.38 | 182.38 | 193.38 | 204.38 | 215.37 | 226.37 | 237.37 | 248.36 | 259.36 | 270.36 |
| | $W_{3,4}$ | 10.63 | 21.57 | 32.57 | 43.58 | 54.58 | 65.57 | 76.57 | 87.57 | 98.57 | 109.56 | 120.56 | 131.56 | 142.56 | 153.55 | 164.55 | 175.55 | 186.55 | 197.54 | 208.54 | 219.54 | 230.53 | 241.53 | 252.53 | 263.53 | 274.52 |
| | $J$ | 8.43 | 14.04 | 17.51 | 19.74 | 21.24 | 22.28 | 23.03 | 23.57 | 23.98 | 24.28 | 24.52 | 24.70 | 24.84 | 24.95 | 25.04 | 25.12 | 25.17 | 25.22 | 25.26 | 25.30 | 25.32 | 25.35 | 25.37 | 25.39 | 25.40 |
| | $K$ | 2.39 | 3.84 | 4.45 | 4.65 | 4.72 | 4.75 | 4.75 | 4.75 | 4.75 | 4.75 | 4.75 | 4.75 | 4.75 | 4.75 | 4.75 | 4.75 | 4.75 | 4.75 | 4.75 | 4.75 | 4.75 | 4.75 | 4.75 | 4.75 | 4.75 |
| 0.5 | $U_1$ | 16.38 | 28.98 | 37.05 | 44.38 | 49.13 | 52.75 | 55.58 | 57.85 | 59.69 | 61.23 | 62.51 | 63.61 | 64.56 | 65.38 | 66.10 | 66.74 | 67.31 | 67.82 | 68.28 | 68.69 | 69.07 | 69.41 | 69.73 | 70.02 | 70.29 |
| | $U_2$ | 0.32 | 0.72 | 1.30 | 2.08 | 3.01 | 4.00 | 4.99 | 5.95 | 6.85 | 7.69 | 8.46 | 9.16 | 9.80 | 10.38 | 10.91 | 11.40 | 11.84 | 12.25 | 12.63 | 12.97 | 13.29 | 13.59 | 13.86 | 14.12 | 14.36 |
| | $W_{1,2}$ | 8.45 | 17.97 | 28.37 | 39.16 | 50.09 | 61.07 | 72.06 | 83.05 | 94.05 | 105.05 | 116.04 | 127.04 | 138.04 | 149.03 | 160.03 | 171.03 | 182.03 | 193.02 | 204.02 | 215.02 | 226.02 | 237.01 | 248.01 | 259.01 | 270.01 |
| | $W_{3,4}$ | 10.49 | 21.32 | 32.27 | 43.26 | 54.25 | 65.25 | 76.24 | 87.24 | 98.24 | 109.23 | 120.23 | 131.23 | 142.23 | 153.22 | 164.22 | 175.22 | 186.22 | 197.21 | 208.21 | 219.21 | 230.21 | 241.20 | 252.20 | 263.20 | 274.20 |
| | $J$ | 8.65 | 14.61 | 18.42 | 20.91 | 22.62 | 23.81 | 24.67 | 25.30 | 25.77 | 26.13 | 26.40 | 26.62 | 26.78 | 26.91 | 27.02 | 27.11 | 27.18 | 27.23 | 27.28 | 27.32 | 27.35 | 27.38 | 27.41 | 27.43 | 27.44 |
| | $K$ | 2.47 | 3.98 | 4.60 | 4.82 | 4.89 | 4.91 | 4.92 | 4.92 | 4.92 | 4.92 | 4.92 | 4.92 | 4.92 | 4.92 | 4.92 | 4.92 | 4.92 | 4.92 | 4.92 | 4.92 | 4.92 | 4.92 | 4.92 | 4.92 | 4.92 |
| 0.4 | $U_1$ | 15.45 | 27.47 | 36.13 | 42.39 | 47.04 | 50.60 | 53.39 | 55.63 | 57.46 | 58.98 | 60.26 | 61.35 | 62.29 | 63.11 | 63.83 | 64.47 | 65.04 | 65.54 | 66.00 | 66.41 | 66.79 | 67.13 | 67.45 | 67.74 | 68.00 |
| | $U_2$ | 0.34 | 0.76 | 1.35 | 2.15 | 3.08 | 4.08 | 5.07 | 6.03 | 6.93 | 7.77 | 8.53 | 9.23 | 9.87 | 10.45 | 10.98 | 11.47 | 11.91 | 12.32 | 12.69 | 13.04 | 13.36 | 13.65 | 13.93 | 14.18 | 14.42 |
| | $W_{1,2}$ | 8.24 | 17.65 | 27.63 | 38.39 | 49.31 | 60.28 | 71.27 | 82.27 | 93.26 | 104.26 | 115.26 | 126.25 | 137.25 | 148.25 | 159.25 | 170.25 | 181.24 | 192.24 | 203.24 | 214.24 | 225.23 | 236.23 | 247.23 | 258.23 | 269.61 |
| | $W_{3,4}$ | 10.39 | 21.15 | 32.07 | 43.04 | 54.03 | 65.02 | 76.02 | 87.02 | 98.01 | 109.01 | 120.01 | 131.01 | 142.00 | 153.00 | 164.00 | 175.00 | 185.99 | 196.99 | 207.99 | 218.99 | 229.98 | 240.98 | 251.98 | 262.97 | 273.97 |
| | $J$ | 9.08 | 15.62 | 19.96 | 22.72 | 24.64 | 25.80 | 26.62 | 27.20 | 27.62 | 27.93 | 28.17 | 28.35 | 28.49 | 28.60 | 28.69 | 28.77 | 28.83 | 28.88 | 28.93 | 28.97 | 29.01 | 29.14 | 29.21 | 29.26 | 29.28 |
| | $K$ | 2.58 | 4.14 | 4.79 | 5.02 | 5.11 | 5.12 | 5.12 | 5.12 | 5.12 | 5.12 | 5.12 | 5.12 | 5.12 | 5.12 | 5.12 | 5.12 | 5.12 | 5.12 | 5.12 | 5.12 | 5.12 | 5.12 | 5.12 | 5.12 | 5.12 |
| 0.3 | $U_1$ | 14.74 | 26.32 | 33.79 | 39.82 | 44.34 | 47.82 | 50.55 | 52.79 | 54.62 | 56.06 | 57.33 | 58.50 | 59.59 | 60.53 | 61.34 | 62.02 | 62.69 | 63.26 | 63.76 | 64.22 | 64.63 | 65.03 | 65.38 | 65.70 | 66.22 |
| | $U_2$ | 0.34 | 0.79 | 1.40 | 2.21 | 3.15 | 4.14 | 5.14 | 6.10 | 7.00 | 7.83 | 8.59 | 9.28 | 9.93 | 10.51 | 11.04 | 11.52 | 11.96 | 12.36 | 12.72 | 13.06 | 13.38 | 13.68 | 13.95 | 14.20 | 14.47 |
| | $W_{1,2}$ | 8.04 | 17.33 | 27.31 | 38.08 | 49.03 | 60.00 | 70.99 | 81.99 | 92.98 | 103.98 | 114.98 | 125.97 | 136.97 | 147.97 | 158.97 | 169.96 | 180.96 | 191.96 | 202.96 | 213.95 | 224.95 | 235.95 | 246.95 | 258.22 | 269.22 |
| | $W_{3,4}$ | 10.29 | 20.93 | 31.93 | 42.78 | 53.76 | 64.76 | 75.76 | 86.76 | 97.75 | 108.75 | 119.75 | 130.74 | 141.74 | 152.74 | 163.73 | 174.73 | 185.73 | 196.73 | 207.73 | 218.72 | 229.72 | 240.72 | 251.72 | 262.72 | 273.83 |
| | $J$ | 9.25 | 15.98 | 20.59 | 23.47 | 25.55 | 27.00 | 28.01 | 28.72 | 29.24 | 29.61 | 29.89 | 30.09 | 30.25 | 30.37 | 30.46 | 30.54 | 30.59 | 30.64 | 30.68 | 30.71 | 30.74 | 30.78 | 30.81 | 30.84 | 30.87 |
| | $K$ | 2.69 | 4.32 | 4.99 | 5.23 | 5.30 | 5.33 | 5.34 | 5.34 | 5.34 | 5.34 | 5.34 | 5.34 | 5.34 | 5.34 | 5.34 | 5.34 | 5.34 | 5.34 | 5.34 | 5.34 | 5.34 | 5.34 | 5.34 | 5.34 | 5.34 |
| 0.2 | $U_1$ | 14.36 | 25.73 | 34.05 | 40.14 | 44.69 | 48.20 | 50.95 | 53.17 | 54.98 | 56.49 | 57.77 | 58.85 | 59.79 | 60.61 | 61.34 | 61.96 | 62.52 | 63.03 | 63.48 | 63.90 | 64.27 | 64.62 | 64.93 | 65.22 | 65.49 |
| | $U_2$ | 0.36 | 0.76 | 1.38 | 2.19 | 3.13 | 4.13 | 5.13 | 6.09 | 6.99 | 7.82 | 8.58 | 9.28 | 9.91 | 10.50 | 11.03 | 11.51 | 11.95 | 12.36 | 12.74 | 13.08 | 13.40 | 13.69 | 13.97 | 14.26 | 14.50 |
| | $W_{1,2}$ | 7.69 | 16.77 | 26.99 | 37.72 | 48.67 | 59.60 | 70.59 | 81.58 | 92.58 | 103.58 | 114.57 | 125.57 | 136.57 | 147.57 | 158.56 | 169.56 | 180.56 | 191.56 | 202.55 | 213.55 | 224.55 | 235.55 | 246.54 | 257.54 | 268.88 |
| | $W_{3,4}$ | 10.26 | 20.87 | 31.78 | 42.74 | 53.72 | 64.71 | 75.71 | 86.71 | 97.70 | 108.70 | 119.70 | 130.69 | 141.69 | 152.69 | 163.68 | 174.68 | 185.68 | 196.68 | 207.68 | 218.68 | 229.67 | 240.67 | 251.67 | 262.67 | 273.74 |
| | $J$ | 9.30 | 16.15 | 20.73 | 23.82 | 25.97 | 27.49 | 28.58 | 29.39 | 29.99 | 30.44 | 30.79 | 31.06 | 31.27 | 31.43 | 31.55 | 31.65 | 31.73 | 31.80 | 31.85 | 31.90 | 31.93 | 31.97 | 32.00 | 32.02 | 32.05 |
| | $K$ | 2.86 | 4.59 | 5.17 | 5.42 | 5.50 | 5.52 | 5.53 | 5.53 | 5.53 | 5.53 | 5.53 | 5.53 | 5.53 | 5.53 | 5.53 | 5.53 | 5.53 | 5.53 | 5.53 | 5.53 | 5.53 | 5.53 | 5.53 | 5.53 | 5.53 |
| 0.1 | $U_1$ | 14.41 | 25.78 | 34.12 | 40.26 | 44.83 | 48.35 | 51.11 | 53.32 | 55.14 | 56.65 | 57.93 | 59.02 | 59.96 | 60.78 | 61.50 | 62.13 | 62.70 | 63.20 | 63.66 | 64.08 | 64.45 | 64.79 | 65.11 | 65.40 | 65.44 |
| | $U_2$ | 0.33 | 0.75 | 1.37 | 2.17 | 3.11 | 4.11 | 5.11 | 6.07 | 6.97 | 7.80 | 8.57 | 9.27 | 9.90 | 10.48 | 11.01 | 11.50 | 11.94 | 12.35 | 12.72 | 13.07 | 13.39 | 13.68 | 13.96 | 14.22 | 14.46 |
| | $W_{1,2}$ | 7.66 | 16.71 | 26.93 | 37.66 | 48.61 | 59.54 | 70.53 | 81.52 | 92.52 | 103.52 | 114.51 | 125.51 | 136.51 | 147.51 | 158.50 | 169.50 | 180.50 | 191.50 | 202.49 | 213.49 | 224.49 | 235.49 | 246.48 | 257.48 | 268.64 |
| | $W_{3,4}$ | 10.25 | 20.88 | 31.77 | 42.72 | 53.70 | 64.70 | 75.69 | 86.69 | 97.69 | 108.69 | 119.68 | 130.68 | 141.67 | 152.67 | 163.67 | 174.67 | 185.66 | 196.66 | 207.66 | 218.66 | 229.65 | 240.65 | 251.65 | 262.64 | 273.70 |
| | $J$ | 9.34 | 16.18 | 20.76 | 23.91 | 26.00 | 27.52 | 28.61 | 29.40 | 30.00 | 30.45 | 30.79 | 31.05 | 31.26 | 31.42 | 31.54 | 31.64 | 31.72 | 31.79 | 31.85 | 31.90 | 31.93 | 31.97 | 32.00 | 32.03 | 32.06 |
| | $K$ | 2.87 | 4.48 | 5.17 | 5.42 | 5.52 | 5.60 | 5.63 | 5.66 | 5.67 | 5.67 | 5.67 | 5.67 | 5.67 | 5.67 | 5.67 | 5.67 | 5.67 | 5.67 | 5.67 | 5.67 | 5.67 | 5.67 | 5.67 | 5.67 | 5.67 |
| 0.0 | $U_1$ | 14.41 | 25.83 | 34.19 | 40.30 | 44.88 | 48.39 | 51.16 | 53.38 | 55.20 | 56.71 | 57.98 | 59.07 | 60.01 | 60.83 | 61.55 | 62.18 | 62.75 | 63.25 | 63.71 | 64.12 | 64.50 | 64.84 | 65.16 | 65.44 | 65.71 |
| | $U_2$ | 0.33 | 0.75 | 1.36 | 2.16 | 3.10 | 4.10 | 5.10 | 6.06 | 6.96 | 7.79 | 8.56 | 9.26 | 9.89 | 10.47 | 11.00 | 11.49 | 11.93 | 12.34 | 12.71 | 13.06 | 13.38 | 13.67 | 13.95 | 14.20 | 14.44 |
| | $W_{1,2}$ | 7.62 | 16.66 | 26.87 | 37.59 | 48.50 | 59.47 | 70.46 | 81.46 | 92.45 | 103.45 | 114.45 | 125.44 | 136.44 | 147.44 | 158.43 | 169.43 | 180.43 | 191.43 | 202.43 | 213.42 | 224.42 | 235.42 | 246.41 | 257.41 | 268.41 |
| | $W_{3,4}$ | 10.24 | 20.89 | 31.76 | 42.71 | 53.69 | 64.69 | 75.68 | 86.68 | 97.68 | 108.68 | 119.67 | 130.67 | 141.66 | 152.66 | 163.66 | 174.66 | 185.65 | 196.65 | 207.65 | 218.65 | 229.64 | 240.64 | 251.64 | 262.62 | 273.66 |
| | $J$ | 9.37 | 16.21 | 20.80 | 23.90 | 26.04 | 27.55 | 28.63 | 29.42 | 30.01 | 30.45 | 30.79 | 31.04 | 31.25 | 31.41 | 31.53 | 31.64 | 31.72 | 31.79 | 31.84 | 31.89 | 31.93 | 31.96 | 31.99 | 32.02 | 32.04 |
| | $K$ | 2.88 | 4.63 | 5.35 | 5.58 | 5.65 | 5.68 | 5.71 | 5.72 | 5.72 | 5.72 | 5.72 | 5.72 | 5.72 | 5.72 | 5.72 | 5.72 | 5.72 | 5.72 | 5.72 | 5.72 | 5.72 | 5.72 | 5.72 | 5.72 | 5.72 |

Total weight of the heavy-fermions on the active bands after wannierisation: 94.6%

Energy (meV): 50, 0, −50 — axis labels $K$, $\Gamma$, $M$, $K$.

Wannier orbitals support: 0.0 — 1.0

FIG. S40. Band structures of the BM and THF models near charge neutrality for $w_0/w_1 = 0.8$, depicted by lines and crosses, respectively. The BM bands are colored according to the weight of the $f$-electron wave function on them. We use the same BM parameters as in Table S48.

TABLE S48. Parameters of the $f$-electron Wannier functions and the THF single-particle Hamiltonian for different values of the tunneling ratio $w_0/w_1$. $W$ denotes the total weight of the THF $f$-electrons on the active bands. In computing the ratios $\gamma/U_1$ and $\gamma/U_2$, we employ the on-site and nearest-neighbor repulsion parameters $U_1$ and $U_2$ obtained numerically for $\xi = 10$ nm, as given in Table S49. We employ $v_F = 5.944$ eV·Å, $|K| = 1.703$ Å$^{-1}$, $w_1 = 110$ meV, and $\theta = 1.14°$ for the BM model.

| $w_0/w_1$ | $\alpha_1$ | $\alpha_2$ | $\lambda_1$ | $\lambda_2$ | $\lambda$ | $W$ | $t_0$ | $\gamma$ | $M$ | $v_*$ | $v'_*$ | $v''_*$ | $\gamma/U_1$ | $\gamma/U_2$ |
|---|---|---|---|---|---|---|---|---|---|---|---|---|---|---|
| 1.00 | 0.774 | 0.633 | 0.167 | 0.185 | 0.336 | 0.979 | 0.438 | −7.765 | −5.517 |  | 1.633 | 0.0038 | −0.11 | −2.92 |
| 0.90 | 0.803 | 0.596 | 0.179 | 0.191 | 0.339 | 0.962 | 0.509 | −22.941 | −6.529 | −4.386 | 1.085 | −0.0127 | −0.36 | −8.48 |
| 0.80 | 0.834 | 0.552 | 0.197 | 0.197 | 0.353 | 0.946 | 0.619 | −38.037 | −7.422 | −4.554 | 1.702 | −0.0217 | −0.64 | −13.45 |
| 0.70 | 0.866 | 0.500 | 0.209 | 0.203 | 0.367 | 0.934 | 0.745 | −52.737 | −8.196 | −4.701 | 1.682 | −0.0252 | −0.95 | −17.86 |
| 0.60 | 0.897 | 0.443 | 0.221 | 0.209 | 0.379 | 0.925 | 0.878 | −66.726 | −8.856 | −4.832 | 1.621 | −0.0246 | −1.27 | −21.78 |
| 0.50 | 0.925 | 0.380 | 0.239 | 0.215 | 0.390 | 0.919 | 1.010 | −79.673 | −9.405 | −4.951 | 1.509 | −0.0213 | −1.60 | −25.22 |
| 0.40 | 0.950 | 0.312 | 0.251 | 0.221 | 0.398 | 0.914 | 1.124 | −91.201 | −9.847 | −5.057 | 1.338 | −0.0162 | −1.91 | −28.20 |
| 0.30 | 0.971 | 0.240 | 0.269 | 0.227 | 0.406 | 0.913 | 1.214 | −100.886 | −10.187 | −5.149 | 1.099 | −0.0104 | −2.19 | −30.51 |
| 0.20 | 0.986 | 0.164 | 0.275 | 0.227 | 0.411 | 0.911 | 1.221 | −108.271 | −10.428 | −5.223 | 0.787 | −0.0051 | −2.40 | −32.39 |
| 0.10 | 0.996 | 0.084 | 0.281 | 0.227 | 0.406 | 0.904 | 1.137 | −112.921 | −10.571 | −5.271 | 0.412 | −0.0014 | −2.50 | −34.23 |
| 0.00 | 1.000 | 0.000 | 0.286 | 0.000 | 0.000 | 0.901 | 1.064 | −114.511 | −10.619 | −5.288 | −0.000 | −0.0000 | −2.52 | −34.99 |

| $v_0/v_1$ | $\xi/\mathrm{nm}$ | 2 | 4 | 6 | 8 | 10 | 12 | 14 | 16 | 18 | 20 | 22 | 24 | 26 | 28 | 30 | 32 | 34 | 36 | 38 | 40 | 42 | 44 | 46 | 48 | 50 |
|---|---|---|---|---|---|---|---|---|---|---|---|---|---|---|---|---|---|---|---|---|---|---|---|---|---|---|
| 1.0 | $U_1$ | 25.29 | 42.95 | 54.41 | 62.14 | 67.60 | 71.64 | 74.72 | 77.15 | 79.11 | 80.72 | 82.07 | 83.21 | 84.18 | 85.03 | 85.77 | 86.42 | 87.00 | 87.52 | 87.99 | 88.41 | 88.79 | 89.14 | 89.45 | 89.75 | 90.02 |
|  | $U_2$ | 23.29 | 39.92 | 50.94 | 58.46 | 63.83 | 67.81 | 70.87 | 73.28 | 75.23 | 76.83 | 78.17 | 79.31 | 80.28 | 81.13 | 81.87 | 82.52 | 83.10 | 83.62 | 84.08 | 84.50 | 84.84 | 85.16 | 85.45 | 85.70 | 86.11 |
|  | $W_{1,2}$ | 9.19 | 19.32 | 30.22 | 41.44 | 52.78 | 64.16 | 75.55 | 86.94 | 98.34 | 109.73 | 121.12 | 132.52 | 143.91 | 155.30 | 166.70 | 178.09 | 189.48 | 200.88 | 212.27 | 223.67 | 235.06 | 246.45 | 257.85 | 269.24 | 280.63 |
|  | $W_{3,4}$ | 12.33 | 24.59 | 36.39 | 47.92 | 59.30 | 70.77 | 82.17 | 93.56 | 104.95 | 116.35 | 127.74 | 139.14 | 150.53 | 161.92 | 173.32 | 184.71 | 196.10 | 207.50 | 218.89 | 230.29 | 241.68 | 253.07 | 264.47 | 275.86 | 287.25 |
|  | $J$ | 7.98 | 12.15 | 14.07 | 15.03 | 15.54 | 15.85 | 16.04 | 16.17 | 16.26 | 16.33 | 16.37 | 16.41 | 16.44 | 16.46 | 16.48 | 16.49 | 16.50 | 16.51 | 16.52 | 16.52 | 16.53 | 16.53 | 16.54 | 16.54 | 16.54 |
|  | $K$ | 2.32 | 3.71 | 4.26 | 4.45 | 4.51 | 4.53 | 4.54 | 4.54 | 4.54 | 4.54 | 4.54 | 4.54 | 4.54 | 4.54 | 4.54 | 4.54 | 4.54 | 4.54 | 4.54 | 4.54 | 4.54 | 4.54 | 4.54 | 4.54 | 4.54 |
| 0.9 | $U_1$ | 21.16 | 36.60 | 47.04 | 54.26 | 59.47 | 63.35 | 66.35 | 68.73 | 70.65 | 72.24 | 73.56 | 74.69 | 75.66 | 76.50 | 77.23 | 77.88 | 78.46 | 78.97 | 79.46 | 79.86 | 80.24 | 80.58 | 80.89 | 81.19 | 81.46 |
|  | $U_2$ | 19.33 | 33.71 | 43.62 | 50.56 | 55.61 | 59.40 | 62.34 | 64.68 | 66.58 | 68.15 | 69.46 | 70.58 | 71.54 | 72.37 | 73.10 | 73.75 | 74.32 | 74.84 | 75.30 | 75.72 | 76.10 | 76.44 | 76.76 | 77.06 | 77.32 |
|  | $W_{1,2}$ | 9.19 | 19.32 | 30.22 | 41.59 | 52.93 | 64.31 | 75.70 | 87.09 | 98.48 | 109.88 | 121.27 | 132.66 | 144.06 | 155.45 | 166.84 | 178.24 | 189.63 | 201.02 | 212.42 | 223.81 | 235.21 | 246.60 | 257.99 | 269.39 | 280.78 |
|  | $W_{3,4}$ | 11.80 | 23.69 | 35.34 | 46.82 | 58.24 | 69.64 | 81.04 | 92.43 | 103.83 | 115.22 | 126.61 | 138.01 | 149.40 | 160.80 | 172.19 | 183.58 | 194.98 | 206.37 | 217.76 | 229.16 | 240.55 | 251.95 | 263.34 | 274.73 | 286.13 |
|  | $J$ | 8.07 | 12.63 | 14.97 | 16.25 | 17.02 | 17.52 | 17.85 | 18.09 | 18.26 | 18.39 | 18.47 | 18.54 | 18.59 | 18.64 | 18.68 | 18.70 | 18.73 | 18.75 | 18.77 | 18.78 | 18.79 | 18.80 | 18.80 | 18.81 | 18.81 |
|  | $K$ | 2.28 | 3.65 | 4.20 | 4.38 | 4.44 | 4.46 | 4.47 | 4.47 | 4.47 | 4.47 | 4.47 | 4.47 | 4.47 | 4.47 | 4.47 | 4.47 | 4.47 | 4.47 | 4.47 | 4.47 | 4.47 | 4.47 | 4.47 | 4.47 | 4.47 |
| 0.8 | $U_1$ |  |  |  |  |  |  |  |  |  |  |  |  |  |  |  |  |  |  |  |  |  |  |  |  |  |
|  | $U_2$ |  |  |  |  |  |  |  |  |  |  |  |  |  |  |  |  |  |  |  |  |  |  |  |  |  |
|  | $W_{1,2}$ | 9.25 | 19.42 | 30.33 | 41.56 | 52.90 | 64.27 | 75.66 | 87.06 | 98.45 | 109.84 | 121.24 | 132.63 | 144.02 | 155.42 | 166.81 | 178.20 | 189.60 | 200.99 | 212.39 | 223.78 | 235.17 | 246.57 | 257.96 | 269.35 | 280.75 |
|  | $W_{3,4}$ | 11.40 | 23.02 | 34.55 | 45.99 | 57.40 | 68.80 | 80.19 | 91.59 | 102.98 | 114.37 | 125.77 | 137.16 | 148.56 | 159.95 | 171.34 | 182.74 | 194.13 | 205.52 | 216.92 | 228.31 | 239.70 | 251.10 | 262.49 | 273.89 | 285.28 |
|  | $J$ | 8.23 | 13.21 | 15.08 | 17.62 | 18.67 | 19.37 | 19.86 | 20.21 | 20.47 | 20.66 | 20.80 | 20.92 | 21.00 | 21.07 | 21.12 | 21.17 | 21.20 | 21.23 | 21.26 | 21.28 | 21.29 | 21.31 | 21.32 | 21.33 | 21.34 |
|  | $K$ | 2.28 | 3.65 | 4.20 | 4.39 | 4.45 | 4.47 | 4.48 | 4.48 | 4.48 | 4.48 | 4.48 | 4.48 | 4.48 | 4.48 | 4.48 | 4.48 | 4.48 | 4.48 | 4.48 | 4.48 | 4.48 | 4.48 | 4.48 | 4.48 | 4.48 |
| 0.7 | $U_1$ |  |  |  |  |  |  |  |  |  |  |  |  |  |  |  |  |  |  |  |  |  |  |  |  |  |
|  | $U_2$ |  |  |  |  |  |  |  |  |  |  |  |  |  |  |  |  |  |  |  |  |  |  |  |  |  |
|  | $W_{1,2}$ | 9.16 | 19.27 | 30.15 | 41.38 | 52.72 | 64.09 | 75.48 | 86.87 | 98.27 | 109.66 | 121.05 | 132.45 | 143.84 | 155.23 | 166.63 | 178.02 | 189.41 | 200.81 | 212.20 | 223.60 | 234.99 | 246.38 | 257.78 | 269.17 | 280.56 |
|  | $W_{3,4}$ | 11.12 | 22.54 | 33.08 | 45.39 | 56.79 | 68.18 | 79.58 | 90.97 | 102.37 | 113.76 | 125.15 | 136.55 | 147.94 | 159.33 | 170.73 | 182.12 | 193.52 | 204.91 | 216.30 | 227.70 | 239.09 | 250.48 | 261.88 | 273.27 | 284.66 |
|  | $J$ | 8.82 | 14.82 | 17.00 | 18.98 | 20.28 | 21.17 | 21.80 | 22.26 | 22.60 | 22.85 | 23.04 | 23.19 | 23.30 | 23.40 | 23.47 | 23.53 | 23.58 | 23.63 | 23.65 | 23.70 | 23.71 | 23.73 | 23.74 | 23.75 | 23.76 |
|  | $K$ | 2.32 | 3.71 | 4.27 | 4.48 | 4.53 | 4.55 | 4.55 | 4.55 | 4.55 | 4.55 | 4.55 | 4.55 | 4.55 | 4.55 | 4.55 | 4.55 | 4.55 | 4.55 | 4.55 | 4.55 | 4.55 | 4.55 | 4.55 | 4.55 | 4.55 |

TABLE S49. Parameters of the THF interaction Hamiltonian for different ratios $u_0/u_1$ and screening lengths $\xi$ of the double-gate interaction potential ($U_\xi = 24$ meV for $\xi = 10$ nm). We use the same BM model parameters as in Table S48.

| $u_0/u_1$ | $\xi$/nm | 2 | 4 | 6 | 8 | 10 | 12 | 14 | 16 | 18 | 20 | 22 | 24 | 26 | 28 | 30 | 32 | 34 | 36 | 38 | 40 | 42 | 44 | 46 | 48 | 50 |
|---|---|---|---|---|---|---|---|---|---|---|---|---|---|---|---|---|---|---|---|---|---|---|---|---|---|---|
| 0.6 | $U_1$ | 17.84 | 31.33 | 40.78 | 47.48 | 52.39 | 56.10 | 58.99 | 61.29 | 63.17 | 64.72 | 66.02 | 67.13 | 68.09 | 68.92 | 69.64 | 70.29 | 70.86 | 71.37 | 71.83 | 72.25 | 72.62 | 72.97 | 73.29 | 73.58 | 73.84 |
| | $U_2$ | 0.31 | 0.71 | 1.30 | 2.10 | 3.06 | 4.09 | 5.12 | 6.11 | 7.03 | 7.89 | 8.67 | 9.39 | 10.04 | 10.63 | 11.17 | 11.66 | 12.12 | 12.53 | 12.91 | 13.26 | 13.58 | 13.88 | 14.15 | 14.41 | 14.65 |
| | $W_{1,2}$ | 9.00 | 19.03 | 29.88 | 41.09 | 52.42 | 63.80 | 75.19 | 86.58 | 97.97 | 109.37 | 120.76 | 132.15 | 143.55 | 154.94 | 166.33 | 177.73 | 189.12 | 200.52 | 211.91 | 223.30 | 234.70 | 246.09 | 257.48 | 268.88 | 280.27 |
| | $W_{3,4}$ | 10.93 | 22.20 | 33.57 | 44.96 | 56.36 | 67.75 | 79.14 | 90.54 | 101.93 | 113.32 | 124.72 | 136.11 | 147.51 | 158.90 | 170.29 | 181.69 | 193.08 | 204.47 | 215.87 | 227.26 | 238.65 | 250.05 | 261.44 | 272.84 | 284.23 |
| | $J$ | 8.66 | 14.43 | 17.99 | 20.27 | 21.80 | 22.86 | 23.62 | 24.17 | 24.58 | 24.88 | 25.12 | 25.30 | 25.44 | 25.55 | 25.64 | 25.72 | 25.77 | 25.82 | 25.86 | 25.90 | 25.92 | 25.95 | 25.97 | 25.98 | 26.00 |
| | $K$ | 2.39 | 3.82 | 4.40 | 4.60 | 4.66 | 4.68 | 4.68 | 4.68 | 4.68 | 4.68 | 4.68 | 4.68 | 4.68 | 4.68 | 4.68 | 4.68 | 4.68 | 4.68 | 4.68 | 4.68 | 4.68 | 4.68 | 4.68 | 4.68 | 4.68 |
| 0.5 | $U_1$ | 16.65 | 29.43 | 38.49 | 44.99 | 49.78 | 53.43 | 56.27 | 58.55 | 60.40 | 61.94 | 63.23 | 64.33 | 65.28 | 66.11 | 66.83 | 67.47 | 68.04 | 68.55 | 69.01 | 69.42 | 69.80 | 70.15 | 70.46 | 70.75 | 71.02 |
| | $U_2$ | 0.35 | 0.75 | 1.37 | 2.19 | 3.16 | 4.19 | 5.22 | 6.21 | 7.13 | 7.98 | 8.76 | 9.48 | 10.13 | 10.72 | 11.26 | 11.75 | 12.20 | 12.60 | 12.99 | 13.34 | 13.66 | 13.95 | 14.23 | 14.49 | 14.73 |
| | $W_{1,2}$ | 8.81 | 18.72 | 29.53 | 40.73 | 52.06 | 63.44 | 74.82 | 86.21 | 97.61 | 109.00 | 120.39 | 131.79 | 143.18 | 154.58 | 165.97 | 177.36 | 188.76 | 200.15 | 211.54 | 222.94 | 234.33 | 245.72 | 257.12 | 268.51 | 279.91 |
| | $W_{3,4}$ | 10.79 | 21.97 | 33.29 | 44.67 | 56.06 | 67.45 | 78.84 | 90.24 | 101.63 | 113.02 | 124.42 | 135.81 | 147.20 | 158.60 | 169.99 | 181.38 | 192.78 | 204.17 | 215.57 | 226.96 | 238.35 | 249.75 | 261.14 | 272.53 | 283.93 |
| | $J$ | 8.90 | 15.02 | 18.92 | 21.48 | 23.21 | 24.43 | 25.30 | 25.94 | 26.41 | 26.77 | 27.04 | 27.25 | 27.42 | 27.55 | 27.65 | 27.74 | 27.81 | 27.87 | 27.91 | 27.95 | 27.98 | 28.01 | 28.04 | 28.06 | 28.07 |
| | $K$ | 2.48 | 3.96 | 4.56 | 4.77 | 4.83 | 4.85 | 4.86 | 4.86 | 4.86 | 4.86 | 4.86 | 4.86 | 4.86 | 4.86 | 4.86 | 4.86 | 4.86 | 4.86 | 4.86 | 4.86 | 4.86 | 4.86 | 4.86 | 4.86 | 4.86 |
| 0.4 | $U_1$ | 15.74 | 27.97 | 36.74 | 43.07 | 47.77 | 51.36 | 54.17 | 56.42 | 58.25 | 59.78 | 61.07 | 62.16 | 63.11 | 63.93 | 64.65 | 65.29 | 65.85 | 66.36 | 66.82 | 67.23 | 67.61 | 67.95 | 68.27 | 68.56 | 68.83 |
| | $U_2$ | 0.34 | 0.79 | 1.42 | 2.26 | 3.23 | 4.27 | 5.30 | 6.28 | 7.21 | 8.06 | 8.84 | 9.55 | 10.19 | 10.78 | 11.32 | 11.81 | 12.26 | 12.67 | 13.05 | 13.40 | 13.72 | 14.01 | 14.29 | 14.54 | 14.78 |
| | $W_{1,2}$ | 8.59 | 18.59 | 29.15 | 40.33 | 51.66 | 63.03 | 74.42 | 85.81 | 97.20 | 108.60 | 119.99 | 131.38 | 142.78 | 154.17 | 165.56 | 176.96 | 188.35 | 199.74 | 211.14 | 222.53 | 233.93 | 245.32 | 256.71 | 268.11 | 279.50 |
| | $W_{3,4}$ | 10.70 | 21.81 | 33.11 | 44.47 | 55.73 | 67.25 | 78.64 | 90.03 | 101.43 | 112.82 | 124.21 | 135.61 | 147.00 | 158.39 | 169.79 | 181.18 | 192.57 | 203.97 | 215.36 | 226.76 | 238.15 | 249.54 | 260.94 | 272.33 | 283.72 |
| | $J$ | 9.13 | 15.57 | 19.76 | 22.55 | 24.46 | 25.80 | 26.77 | 27.48 | 28.01 | 28.41 | 28.72 | 28.96 | 29.14 | 29.29 | 29.41 | 29.50 | 29.58 | 29.65 | 29.70 | 29.74 | 29.78 | 29.81 | 29.84 | 29.86 | 29.88 |
| | $K$ | 2.58 | 4.13 | 4.76 | 4.97 | 5.04 | 5.06 | 5.07 | 5.07 | 5.07 | 5.07 | 5.07 | 5.07 | 5.07 | 5.07 | 5.07 | 5.07 | 5.07 | 5.07 | 5.07 | 5.07 | 5.07 | 5.07 | 5.07 | 5.07 | 5.07 |
| 0.3 | $U_1$ | 15.01 | 26.76 | 35.27 | 41.46 | 46.08 | 49.61 | 52.39 | 54.62 | 56.44 | 57.95 | 59.23 | 60.32 | 61.26 | 62.08 | 62.80 | 63.43 | 64.00 | 64.50 | 64.96 | 65.37 | 65.75 | 66.09 | 66.41 | 66.70 | 66.96 |
| | $U_2$ | 0.36 | 0.82 | 1.48 | 2.33 | 3.32 | 4.36 | 5.39 | 6.37 | 7.30 | 8.15 | 8.93 | 9.64 | 10.28 | 10.88 | 11.42 | 11.91 | 12.36 | 12.77 | 13.14 | 13.49 | 13.81 | 14.11 | 14.38 | 14.64 | 14.86 |
| | $W_{1,2}$ | 8.39 | 18.07 | 28.79 | 39.95 | 51.27 | 62.64 | 74.03 | 85.42 | 96.82 | 108.21 | 119.60 | 131.00 | 142.39 | 153.78 | 165.18 | 176.57 | 187.96 | 199.36 | 210.75 | 222.14 | 233.54 | 244.93 | 256.33 | 267.72 | 279.11 |
| | $W_{3,4}$ | 10.62 | 21.65 | 32.91 | 44.27 | 55.65 | 67.04 | 78.43 | 89.83 | 101.22 | 112.61 | 124.01 | 135.40 | 146.79 | 158.19 | 169.58 | 180.97 | 192.37 | 203.76 | 215.15 | 226.55 | 237.94 | 249.33 | 260.73 | 272.12 | 283.51 |
| | $J$ | 9.35 | 16.04 | 20.45 | 23.42 | 25.42 | 26.83 | 27.85 | 28.59 | 29.15 | 29.57 | 29.89 | 30.14 | 30.33 | 30.48 | 30.60 | 30.70 | 30.78 | 30.84 | 30.90 | 30.94 | 30.98 | 31.01 | 31.04 | 31.06 | 31.31 |
| | $K$ | 2.70 | 4.31 | 4.96 | 5.18 | 5.25 | 5.28 | 5.29 | 5.29 | 5.29 | 5.29 | 5.29 | 5.29 | 5.29 | 5.29 | 5.29 | 5.29 | 5.29 | 5.29 | 5.29 | 5.29 | 5.29 | 5.29 | 5.29 | 5.29 | 5.29 |
| 0.2 | $U_1$ | 14.59 | 26.12 | 34.52 | 40.66 | 45.25 | 48.77 | 51.54 | 53.77 | 55.59 | 57.10 | 58.38 | 59.47 | 60.41 | 61.22 | 61.94 | 62.58 | 63.14 | 63.65 | 64.11 | 64.52 | 64.90 | 65.24 | 65.55 | 65.84 | 66.11 |
| | $U_2$ | 0.37 | 0.84 | 1.50 | 2.35 | 3.33 | 4.37 | 5.40 | 6.38 | 7.30 | 8.14 | 8.92 | 9.63 | 10.27 | 10.86 | 11.40 | 11.89 | 12.34 | 12.75 | 13.12 | 13.47 | 13.79 | 14.09 | 14.36 | 14.62 | 14.86 |
| | $W_{1,2}$ | 8.03 | 17.51 | 28.15 | 39.29 | 50.60 | 61.97 | 73.36 | 84.75 | 96.14 | 107.53 | 118.93 | 130.32 | 141.71 | 153.11 | 164.50 | 175.89 | 187.29 | 198.68 | 210.07 | 221.47 | 232.86 | 244.26 | 255.65 | 267.04 | 278.44 |
| | $W_{3,4}$ | 10.58 | 21.59 | 32.85 | 44.20 | 55.58 | 66.97 | 78.36 | 89.76 | 101.15 | 112.54 | 123.94 | 135.33 | 146.72 | 158.12 | 169.51 | 180.90 | 192.30 | 203.69 | 215.08 | 226.48 | 237.87 | 249.27 | 260.66 | 272.06 | 283.45 |
| | $J$ | 9.52 | 16.44 | 21.07 | 24.22 | 26.32 | 27.75 | 28.78 | 29.52 | 30.09 | 30.52 | 30.85 | 31.11 | 31.31 | 31.46 | 31.59 | 31.69 | 31.77 | 31.84 | 31.90 | 31.94 | 31.98 | 32.01 | 32.04 | 32.06 | 32.73 |
| | $K$ | 2.82 | 4.47 | 5.15 | 5.38 | 5.45 | 5.47 | 5.48 | 5.48 | 5.48 | 5.48 | 5.48 | 5.48 | 5.48 | 5.48 | 5.48 | 5.48 | 5.48 | 5.48 | 5.48 | 5.48 | 5.48 | 5.48 | 5.48 | 5.48 | 5.48 |
| 0.1 | $U_1$ | 14.64 | 26.21 | 34.65 | 40.82 | 45.43 | 48.96 | 51.74 | 53.97 | 55.79 | 57.31 | 58.59 | 59.68 | 60.62 | 61.44 | 62.16 | 62.79 | 63.36 | 63.86 | 64.32 | 64.73 | 65.11 | 65.45 | 65.77 | 66.06 | 66.33 |
| | $U_2$ | 0.34 | 0.81 | 1.46 | 2.31 | 3.30 | 4.34 | 5.37 | 6.35 | 7.27 | 8.12 | 8.90 | 9.61 | 10.25 | 10.84 | 11.38 | 11.87 | 12.32 | 12.73 | 13.10 | 13.45 | 13.77 | 14.07 | 14.34 | 14.60 | 14.84 |
| | $W_{1,2}$ | 7.96 | 17.41 | 28.03 | 39.16 | 50.48 | 61.84 | 73.23 | 84.62 | 96.02 | 107.41 | 118.80 | 130.19 | 141.59 | 152.98 | 164.37 | 175.77 | 187.16 | 198.55 | 209.95 | 221.34 | 232.73 | 244.13 | 255.52 | 266.92 | 278.31 |
| | $W_{3,4}$ | 10.57 | 21.57 | 32.83 | 44.18 | 55.56 | 66.95 | 78.34 | 89.73 | 101.13 | 112.52 | 123.91 | 135.30 | 146.70 | 158.09 | 169.48 | 180.88 | 192.27 | 203.66 | 215.06 | 226.45 | 237.84 | 249.24 | 260.63 | 272.03 | 283.42 |
| | $J$ | 9.66 | 16.62 | 21.31 | 24.47 | 26.58 | 28.02 | 29.05 | 29.80 | 30.37 | 30.80 | 31.14 | 31.40 | 31.60 | 31.76 | 31.89 | 31.99 | 32.07 | 32.14 | 32.20 | 32.24 | 32.28 | 32.32 | 32.35 | 32.37 | 32.79 |
| | $K$ | 2.88 | 4.59 | 5.28 | 5.51 | 5.59 | 5.61 | 5.62 | 5.62 | 5.62 | 5.62 | 5.62 | 5.62 | 5.62 | 5.62 | 5.62 | 5.62 | 5.62 | 5.62 | 5.62 | 5.62 | 5.62 | 5.62 | 5.62 | 5.62 | 5.62 |
| 0.0 | $U_1$ | 14.64 | 26.21 | 34.65 | 40.82 | 45.43 | 48.96 | 51.74 | 53.97 | 55.79 | 57.31 | 58.59 | 59.68 | 60.62 | 61.44 | 62.16 | 62.79 | 63.36 | 63.86 | 64.32 | 64.73 | 65.11 | 65.45 | 65.77 | 66.06 | 66.33 |
| | $U_2$ | 0.34 | 0.79 | 1.44 | 2.29 | 3.27 | 4.31 | 5.34 | 6.35 | 7.25 | 8.10 | 8.88 | 9.59 | 10.23 | 10.82 | 11.36 | 11.85 | 12.29 | 12.70 | 13.08 | 13.43 | 13.75 | 14.05 | 14.32 | 14.58 | 14.82 |
| | $W_{1,2}$ | 7.96 | 17.41 | 28.03 | 39.16 | 50.48 | 61.84 | 73.23 | 84.62 | 96.02 | 107.41 | 118.80 | 130.19 | 141.59 | 152.98 | 164.37 | 175.77 | 187.16 | 198.55 | 209.95 | 221.34 | 232.73 | 244.13 | 255.52 | 266.92 | 278.31 |
| | $W_{3,4}$ | 10.57 | 21.57 | 32.83 | 44.18 | 55.56 | 66.95 | 78.34 | 89.73 | 101.13 | 112.52 | 123.91 | 135.30 | 146.70 | 158.09 | 169.48 | 180.88 | 192.27 | 203.66 | 215.06 | 226.45 | 237.84 | 249.24 | 260.63 | 272.03 | 283.42 |
| | $J$ | 9.66 | 16.69 | 21.39 | 24.56 | 26.74 | 28.19 | 29.23 | 29.99 | 30.56 | 31.00 | 31.34 | 31.60 | 31.81 | 31.97 | 32.10 | 32.20 | 32.28 | 32.35 | 32.41 | 32.45 | 32.49 | 32.53 | 32.56 | 32.58 | 32.83 |
| | $K$ | 2.90 | 4.63 | 5.33 | 5.56 | 5.64 | 5.67 | 5.67 | 5.67 | 5.67 | 5.67 | 5.67 | 5.67 | 5.67 | 5.67 | 5.67 | 5.67 | 5.67 | 5.67 | 5.67 | 5.67 | 5.67 | 5.67 | 5.67 | 5.67 | 5.67 |

**Total weight of the heavy-fermions on the active bands after wannierisation: 94.4%**

Energy (meV) axis: 100, 50, 0, −50, −100 — horizontal axis: K, Γ, M, K

Wannier orbitals support (color scale 1.0 – 0.0)

| $w_0/w_1$ | $\alpha_1$ | $\alpha_2$ | $\lambda_1$ | $\lambda_2$ | $\lambda$ | $W$ | $t_0$ | $\gamma$ | $M$ | $v_\star$ | $v'_\star$ | $v''_\star$ | $\gamma/U_1$ | $\gamma/U_2$ |
|---|---|---|---|---|---|---|---|---|---|---|---|---|---|---|
| 1.00 | 0.777 | 0.630 | 0.173 | 0.185 | 0.338 | 0.977 | 0.570 | −10.582 | −8.003 | −4.253 | 1.659 | 0.0038 | | −3.76 |
| 0.90 | 0.806 | 0.592 | 0.185 | 0.191 | 0.341 | 0.959 | 0.683 | −25.900 | −9.095 | −4.439 | 1.704 | −0.0108 | −0.40 | −9.05 |
| 0.80 | 0.837 | 0.547 | 0.197 | 0.203 | 0.355 | 0.944 | 0.838 | −41.087 | −10.054 | −4.599 | 1.717 | −0.0188 | −0.69 | −13.76 |
| 0.70 | 0.869 | 0.496 | 0.209 | 0.209 | 0.369 | 0.932 | 1.008 | −55.840 | −10.882 | −4.740 | 1.693 | −0.0219 | −0.99 | −17.96 |
| 0.60 | 0.899 | 0.439 | 0.227 | 0.215 | 0.381 | 0.924 | 1.180 | −69.858 | −11.587 | −4.867 | 1.629 | −0.0215 | −1.32 | −21.69 |
| 0.50 | 0.927 | 0.376 | 0.239 | 0.221 | 0.391 | 0.918 | 1.350 | −82.814 | −12.172 | −4.983 | 1.514 | −0.0185 | −1.64 | −24.98 |
| 0.40 | 0.951 | 0.308 | 0.257 | 0.221 | 0.408 | 0.920 | 1.512 | −94.338 | −12.643 | −5.088 | 1.341 | −0.0141 | −1.98 | −27.32 |
| 0.30 | 0.972 | 0.237 | 0.269 | 0.227 | 0.416 | 0.918 | 1.627 | −104.012 | −13.005 | −5.178 | 1.100 | −0.0091 | −2.26 | −29.57 |
| 0.20 | 0.987 | 0.162 | 0.281 | 0.227 | 0.412 | 0.911 | 1.627 | −111.383 | −13.261 | −5.251 | 0.787 | −0.0044 | −2.43 | −31.86 |
| 0.10 | 0.997 | 0.083 | 0.286 | 0.233 | 0.414 | 0.910 | 1.587 | −116.022 | −13.413 | −5.298 | 0.412 | −0.0012 | −2.56 | −33.02 |
| 0.00 | 1.000 | 0.000 | 0.286 | 0.000 | 0.000 | 0.904 | 1.475 | −117.608 | −13.464 | −5.315 | −0.000 | 0.0000 | −2.57 | −33.98 |

TABLE S50. Parameters of the $f$-electron Wannier functions and the THF single-particle Hamiltonian for different values of the tunneling ratio $w_0/w_1$. $W$ denotes the total weight of the THF $f$-electrons on the active bands. In computing the ratios $\gamma/U_1$ and $\gamma/U_2$, we employ the on-site and nearest-neighbor repulsion parameters $U_1$ and $U_2$ obtained numerically for $\xi = 10$ nm, as given in Table S51. We employ $v_F = 5.944$ eV·Å, $|\mathbf{K}| = 1.703$ Å$^{-1}$, $w_1 = 110$ meV, and $\theta = 1.16°$ for the BM model.

FIG. S41. Band structures of the BM and THF models near charge neutrality for $w_0/w_1 = 0.8$, depicted by lines and crosses, respectively. The BM bands are colored according to the weight of the $f$-electron wave function on them. We use the same BM parameters as in Table S50.

| $w_0/w_1$ | | 2 | 4 | 6 | 8 | 10 | 12 | 14 | 16 | 18 | 20 | 22 | 24 | 26 | 28 | 30 | 32 | 34 | 36 | 38 | 40 | 42 | 44 | 46 | 48 | 50 |
|---|---|---|---|---|---|---|---|---|---|---|---|---|---|---|---|---|---|---|---|---|---|---|---|---|---|---|
| 1.0 | $U_1$ | 25.46 | 43.20 | 54.72 | 62.47 | 67.95 | 72.06 | 75.10 | 77.53 | 79.50 | 81.11 | 82.46 | 83.60 | 84.58 | 85.42 | 86.16 | 86.82 | 87.40 | 87.92 | 88.38 | 88.80 | 89.19 | 89.53 | 89.85 | 90.15 | 90.42 |
| | $U_2$ | 0.26 | 0.58 | 1.09 | 1.85 | 2.81 | 3.86 | 4.91 | 5.93 | 6.89 | 7.77 | 8.58 | 9.34 | 9.98 | 10.59 | 11.14 | 11.66 | 12.15 | 12.52 | 12.90 | 13.26 | 13.58 | 13.89 | 14.17 | 14.42 | 14.67 |
| | $W_{1,4}$ | 9.63 | 20.21 | 31.53 | 43.18 | 54.92 | 66.71 | 78.50 | 90.29 | 102.09 | 113.89 | 125.69 | 137.48 | 149.28 | 161.08 | 172.87 | 184.67 | 196.47 | 208.26 | 220.06 | 231.86 | 243.65 | 255.45 | 267.25 | 279.04 | 290.84 |
| | $W_{2,3}$ | 12.55 | 25.09 | 37.23 | 49.14 | 60.98 | 72.78 | 84.58 | 96.38 | 108.18 | 119.98 | 131.77 | 143.57 | 155.37 | 167.16 | 178.96 | 190.76 | 202.55 | 214.35 | 226.15 | 237.95 | 249.74 | 261.54 | 273.34 | 285.13 | 296.93 |
| | $J$ | 8.13 | 12.39 | 14.38 | 15.38 | 15.93 | 16.27 | 16.48 | 16.62 | 16.72 | 16.80 | 16.85 | 16.89 | 16.92 | 16.95 | 16.96 | 16.98 | 16.99 | 17.00 | 17.01 | 17.02 | 17.02 | 17.03 | 17.03 | 17.04 | 17.04 |
| | $K$ | 2.29 | 3.63 | 4.16 | 4.34 | 4.39 | 4.41 | 4.42 | 4.42 | 4.42 | 4.42 | 4.42 | 4.42 | 4.42 | 4.42 | 4.42 | 4.42 | 4.42 | 4.42 | 4.42 | 4.42 | 4.42 | 4.42 | 4.42 | 4.42 | 4.42 |
| 0.9 | $U_1$ | 23.47 | 40.20 | 51.27 | 58.82 | 64.21 | 68.07 | 71.27 | 73.69 | 75.64 | 77.25 | 78.50 | 79.73 | 80.71 | 81.55 | 82.29 | 82.94 | 83.52 | 84.04 | 84.51 | 84.93 | 85.31 | 85.66 | 85.98 | 86.27 | 86.54 |
| | $U_2$ | 0.25 | 0.58 | 1.11 | 1.89 | 2.86 | 3.91 | 4.97 | 5.99 | 6.95 | 7.83 | 8.64 | 9.37 | 10.04 | 10.64 | 11.19 | 11.70 | 12.15 | 12.57 | 12.96 | 13.31 | 13.64 | 13.94 | 14.22 | 14.48 | 14.72 |
| | $W_{1,4}$ | 9.69 | 20.30 | 31.63 | 43.28 | 55.03 | 66.81 | 78.60 | 90.40 | 102.19 | 113.99 | 125.79 | 137.58 | 149.38 | 161.18 | 172.97 | 184.77 | 196.57 | 208.37 | 220.16 | 231.96 | 243.76 | 255.55 | 267.35 | 279.15 | 290.94 |
| | $W_{2,3}$ | 12.04 | 24.24 | 36.24 | 48.11 | 59.93 | 71.73 | 83.53 | 95.33 | 107.13 | 118.92 | 130.72 | 142.52 | 154.31 | 166.11 | 177.91 | 189.70 | 201.50 | 213.30 | 225.09 | 236.89 | 248.69 | 260.49 | 272.28 | 284.08 | 295.88 |
| | $J$ | 8.26 | 12.94 | 15.35 | 16.69 | 17.51 | 18.03 | 18.38 | 18.63 | 18.81 | 18.94 | 19.04 | 19.12 | 19.18 | 19.22 | 19.26 | 19.29 | 19.31 | 19.33 | 19.35 | 19.37 | 19.39 | 19.39 | 19.39 | 19.39 | 19.40 |
| | $K$ | 2.26 | 3.59 | 4.11 | 4.29 | 4.34 | 4.36 | 4.37 | 4.37 | 4.37 | 4.37 | 4.37 | 4.37 | 4.37 | 4.37 | 4.37 | 4.37 | 4.37 | 4.37 | 4.37 | 4.37 | 4.37 | 4.37 | 4.37 | 4.37 | 4.37 |
| 0.8 | $U_1$ | 21.36 | 36.02 | 47.42 | 54.68 | 59.90 | 63.81 | 66.82 | 69.20 | 71.12 | 72.71 | 74.04 | 75.17 | 76.14 | 76.98 | 77.72 | 78.37 | 78.95 | 79.46 | 79.93 | 80.34 | 80.73 | 81.07 | 81.41 | 81.68 | 81.95 |
| | $U_2$ | 0.28 | 0.64 | 1.20 | 2.01 | 2.99 | 4.04 | 5.11 | 6.12 | 7.08 | 7.96 | 8.76 | 9.49 | 10.15 | 10.76 | 11.31 | 11.81 | 12.27 | 12.68 | 13.07 | 13.42 | 13.75 | 14.05 | 14.33 | 14.59 | 14.83 |
| | $W_{1,4}$ | 9.65 | 20.24 | 31.57 | 43.21 | 54.96 | 66.74 | 78.54 | 90.33 | 102.13 | 113.92 | 125.72 | 137.52 | 149.32 | 161.11 | 172.91 | 184.71 | 196.50 | 208.30 | 220.10 | 231.89 | 243.69 | 255.49 | 267.28 | 279.08 | 290.88 |
| | $W_{2,3}$ | 11.68 | 23.62 | 35.51 | 47.34 | 59.15 | 70.95 | 82.74 | 94.54 | 106.34 | 118.14 | 129.93 | 141.73 | 153.53 | 165.32 | 177.12 | 188.92 | 200.71 | 212.51 | 224.31 | 236.10 | 247.90 | 259.70 | 271.50 | 283.29 | 295.09 |
| | $J$ | 8.43 | 13.54 | 16.40 | 18.09 | 19.18 | 19.86 | 20.32 | 20.65 | 20.77 | 21.03 | 21.23 | 21.37 | 21.49 | 21.58 | 21.65 | 21.70 | 21.75 | 21.78 | 21.81 | 21.84 | 21.85 | 21.87 | 21.89 | 21.91 | 21.92 |
| | $K$ | 3.27 | 3.61 | 4.14 | 4.31 | 4.37 | 4.38 | 4.39 | 4.39 | 4.39 | 4.39 | 4.39 | 4.39 | 4.39 | 4.39 | 4.39 | 4.39 | 4.39 | 4.39 | 4.39 | 4.39 | 4.39 | 4.39 | 4.39 | 4.39 | 4.39 |
| 0.7 | $U_1$ | 19.57 | 34.09 | 44.07 | 51.06 | 56.14 | 59.95 | 63.02 | 65.25 | 67.16 | 68.73 | 70.05 | 71.17 | 72.13 | 72.97 | 73.70 | 74.35 | 74.92 | 75.43 | 75.90 | 76.31 | 76.69 | 77.04 | 77.36 | 77.65 | 77.92 |
| | $U_2$ | 0.30 | 0.69 | 1.29 | 2.12 | 3.11 | 4.17 | 5.23 | 6.25 | 7.20 | 8.08 | 8.88 | 9.61 | 10.27 | 10.87 | 11.42 | 11.92 | 12.37 | 12.79 | 13.17 | 13.53 | 13.85 | 14.15 | 14.43 | 14.69 | 14.93 |
| | $W_{1,4}$ | 9.54 | 20.07 | 31.37 | 42.95 | 54.75 | 66.54 | 78.33 | 90.12 | 101.92 | 113.71 | 125.51 | 137.31 | 149.11 | 160.90 | 172.70 | 184.50 | 196.29 | 208.09 | 219.89 | 231.69 | 243.48 | 255.28 | 267.08 | 278.87 | 290.67 |
| | $W_{2,3}$ | 11.41 | 23.17 | 34.98 | 46.78 | 58.58 | 70.38 | 82.18 | 93.97 | 105.77 | 117.57 | 129.36 | 141.16 | 152.96 | 164.75 | 176.55 | 188.35 | 200.15 | 211.94 | 223.74 | 235.54 | 247.33 | 259.13 | 270.93 | 282.72 | 294.52 |
| | $J$ | 8.66 | 14.18 | 17.45 | 19.48 | 20.82 | 21.73 | 22.37 | 22.84 | 23.18 | 23.44 | 23.63 | 23.78 | 23.90 | 23.99 | 24.06 | 24.12 | 24.17 | 24.21 | 24.24 | 24.26 | 24.28 | 24.29 | 24.31 | 24.33 | 24.35 |
| | $K$ | 2.31 | 3.68 | 4.22 | 4.40 | 4.45 | 4.47 | 4.47 | 4.48 | 4.48 | 4.48 | 4.48 | 4.48 | 4.48 | 4.48 | 4.48 | 4.48 | 4.48 | 4.48 | 4.48 | 4.48 | 4.48 | 4.48 | 4.48 | 4.48 | 4.48 |

TABLE S51. Parameters of the THF interaction Hamiltonian for different ratios $u_0/u_1$ and screening lengths $\xi$ of the double-gate interaction potential ($U_\xi = 24$ meV for $\xi = 10$nm). We use the same BM model parameters as in Table S50.

| $u_0/u_1$ | $\xi$/nm | 2 | 4 | 6 | 8 | 10 | 12 | 14 | 16 | 18 | 20 | 22 | 24 | 26 | 28 | 30 | 32 | 34 | 36 | 38 | 40 | 42 | 44 | 46 | 48 | 50 |
|---|---|---|---|---|---|---|---|---|---|---|---|---|---|---|---|---|---|---|---|---|---|---|---|---|---|---|
| 0.6 | $U_1$ | 18.10 | 31.77 | 41.31 | 48.06 | 53.01 | 56.74 | 59.65 | 61.96 | 63.85 | 65.40 | 66.71 | 67.82 | 68.78 | 69.61 | 70.34 | 70.98 | 71.55 | 72.07 | 72.53 | 72.94 | 73.32 | 73.67 | 73.98 | 74.28 | 74.54 |
|  | $U_2$ | 0.32 | 0.74 | 1.37 | 2.22 | 3.22 | 4.29 | 5.35 | 6.36 | 7.31 | 8.19 | 8.98 | 9.71 | 10.37 | 10.97 | 11.52 | 12.01 | 12.47 | 12.88 | 13.27 | 13.62 | 13.94 | 14.24 | 14.52 | 14.78 | 15.02 |
|  | $W_{1,2}$ | 9.37 | 19.81 | 31.08 | 42.70 | 54.44 | 66.22 | 78.02 | 89.81 | 101.61 | 113.41 | 125.20 | 137.00 | 148.80 | 160.59 | 172.39 | 184.19 | 195.98 | 207.78 | 219.58 | 231.37 | 243.17 | 254.97 | 266.77 | 278.56 | 290.36 |
|  | $W_{3,4}$ | 11.33 | 22.85 | 34.60 | 46.39 | 58.18 | 69.98 | 81.78 | 93.57 | 105.37 | 117.17 | 128.96 | 140.76 | 152.56 | 164.35 | 176.15 | 187.95 | 199.74 | 211.54 | 223.34 | 235.13 | 246.93 | 258.73 | 270.53 | 282.32 | 294.12 |
|  | $J$ | 8.90 | 14.82 | 18.46 | 20.79 | 22.35 | 23.43 | 24.20 | 24.76 | 25.17 | 25.48 | 25.71 | 25.89 | 26.03 | 26.15 | 26.24 | 26.31 | 26.37 | 26.41 | 26.45 | 26.49 | 26.51 | 26.54 | 26.56 | 26.57 | 26.59 |
|  | $K$ | 2.39 | 3.79 | 4.35 | 4.59 | 4.61 | 4.62 | 4.62 | 4.62 | 4.62 | 4.62 | 4.62 | 4.62 | 4.62 | 4.62 | 4.62 | 4.62 | 4.62 | 4.62 | 4.62 | 4.62 | 4.62 | 4.62 | 4.62 | 4.62 | 4.62 |
| 0.5 | $U_1$ | 16.93 | 29.00 | 39.07 | 45.63 | 50.46 | 54.13 | 56.99 | 59.28 | 61.14 | 62.68 | 63.98 | 65.09 | 66.04 | 66.86 | 67.59 | 68.20 | 68.80 | 69.31 | 69.77 | 70.19 | 70.56 | 70.91 | 71.23 | 71.52 | 71.78 |
|  | $U_2$ | 0.41 | 0.93 | 1.43 | 2.30 | 3.32 | 4.39 | 5.45 | 6.46 | 7.41 | 8.28 | 9.07 | 9.80 | 10.46 | 11.05 | 11.60 | 12.10 | 12.55 | 12.96 | 13.35 | 13.70 | 14.02 | 14.32 | 14.60 | 14.85 | 15.09 |
|  | $W_{1,2}$ | 9.17 | 19.49 | 30.72 | 42.33 | 54.07 | 65.85 | 77.64 | 89.43 | 101.23 | 113.03 | 124.82 | 136.62 | 148.42 | 160.22 | 172.01 | 183.81 | 195.61 | 207.40 | 219.20 | 231.00 | 242.79 | 254.59 | 266.39 | 278.18 | 289.98 |
|  | $W_{3,4}$ | 11.11 | 22.63 | 34.35 | 46.12 | 57.91 | 69.70 | 81.50 | 93.30 | 105.09 | 116.89 | 128.69 | 140.48 | 152.28 | 164.08 | 175.87 | 187.67 | 199.47 | 211.26 | 223.06 | 234.86 | 246.66 | 258.45 | 270.25 | 282.05 | 293.84 |
|  | $J$ | 9.15 | 15.44 | 19.43 | 22.03 | 23.83 | 25.03 | 25.90 | 26.55 | 27.02 | 27.38 | 27.65 | 27.86 | 28.03 | 28.16 | 28.26 | 28.34 | 28.41 | 28.47 | 28.51 | 28.55 | 28.59 | 28.61 | 28.64 | 28.66 | 28.67 |
|  | $K$ | 2.48 | 3.93 | 4.52 | 4.72 | 4.77 | 4.79 | 4.80 | 4.80 | 4.80 | 4.80 | 4.80 | 4.80 | 4.80 | 4.80 | 4.80 | 4.80 | 4.80 | 4.80 | 4.80 | 4.80 | 4.80 | 4.80 | 4.80 | 4.80 | 4.80 |
| 0.4 | $U_1$ | 15.72 | 27.01 | 36.64 | 42.95 | 47.64 | 51.21 | 54.02 | 56.26 | 58.10 | 59.62 | 60.90 | 62.00 | 62.94 | 63.76 | 64.48 | 65.12 | 65.68 | 66.19 | 66.65 | 67.06 | 67.44 | 67.78 | 68.10 | 68.39 | 68.65 |
|  | $U_2$ | 0.38 | 0.86 | 1.54 | 2.43 | 3.45 | 4.53 | 5.59 | 6.60 | 7.54 | 8.40 | 9.20 | 9.92 | 10.57 | 11.17 | 11.71 | 12.20 | 12.65 | 13.07 | 13.45 | 13.80 | 14.12 | 14.42 | 14.69 | 14.95 | 15.19 |
|  | $W_{1,2}$ | 8.99 | 19.21 | 30.39 | 41.99 | 53.73 | 65.50 | 77.30 | 89.09 | 100.89 | 112.68 | 124.48 | 136.28 | 148.07 | 159.87 | 171.67 | 183.47 | 195.26 | 207.06 | 218.86 | 230.66 | 242.45 | 254.25 | 266.05 | 277.84 | 289.64 |
|  | $W_{3,4}$ | 11.04 | 22.51 | 34.20 | 45.96 | 57.75 | 69.54 | 81.34 | 93.13 | 104.93 | 116.73 | 128.52 | 140.32 | 152.12 | 163.91 | 175.71 | 187.51 | 199.30 | 211.10 | 222.90 | 234.69 | 246.49 | 258.29 | 270.09 | 281.88 | 293.68 |
|  | $J$ | 9.43 | 16.09 | 20.46 | 23.38 | 25.39 | 26.82 | 27.85 | 28.61 | 29.18 | 29.61 | 29.94 | 30.20 | 30.41 | 30.57 | 30.70 | 30.81 | 30.89 | 30.96 | 31.02 | 31.07 | 31.12 | 31.15 | 31.18 | 31.21 | 31.23 |
|  | $K$ | 2.57 | 4.12 | 4.73 | 4.92 | 4.99 | 5.01 | 5.01 | 5.01 | 5.01 | 5.01 | 5.01 | 5.01 | 5.01 | 5.01 | 5.01 | 5.01 | 5.01 | 5.01 | 5.01 | 5.01 | 5.01 | 5.01 | 5.01 | 5.01 | 5.01 |
| 0.3 | $U_1$ | 15.03 | 26.79 | 35.24 | 41.47 | 46.08 | 49.61 | 52.18 | 54.61 | 56.23 | 57.94 | 59.22 | 60.11 | 61.05 | 61.87 | 62.59 | 63.22 | 63.78 | 64.29 | 64.75 | 65.17 | 65.55 | 65.88 | 66.20 | 66.40 | 66.24 |
|  | $U_2$ | 0.38 | 0.89 | 1.59 | 2.49 | 3.52 | 4.59 | 5.65 | 6.66 | 7.60 | 8.45 | 9.25 | 9.97 | 10.63 | 11.22 | 11.76 | 12.26 | 12.69 | 13.11 | 13.49 | 13.84 | 14.16 | 14.46 | 14.73 | 14.99 | 15.24 |
|  | $W_{1,2}$ | 8.77 | 18.88 | 30.02 | 41.61 | 53.34 | 65.11 | 76.90 | 88.70 | 100.49 | 112.29 | 124.09 | 135.88 | 147.68 | 159.48 | 171.28 | 183.07 | 194.87 | 206.67 | 218.46 | 230.26 | 242.06 | 253.85 | 265.65 | 277.45 | 289.24 |
|  | $W_{3,4}$ | 10.95 | 22.21 | 34.29 | 45.85 | 57.63 | 69.42 | 81.22 | 93.02 | 104.81 | 116.61 | 128.41 | 140.20 | 152.00 | 163.80 | 175.60 | 187.39 | 199.19 | 210.99 | 222.78 | 234.58 | 246.38 | 258.18 | 269.97 | 281.77 | 293.47 |
|  | $J$ | 9.65 | 16.69 | 21.61 | 24.80 | 27.01 | 28.57 | 29.61 | 30.36 | 30.89 | 31.26 | 31.54 | 31.75 | 31.92 | 32.05 | 32.15 | 32.25 | 32.32 | 32.39 | 32.44 | 32.48 | 32.53 | 32.56 | 32.59 | 32.81 | 32.83 |
|  | $K$ | 2.71 | 4.31 | 4.93 | 5.12 | 5.20 | 5.21 | 5.23 | 5.23 | 5.24 | 5.24 | 5.24 | 5.24 | 5.24 | 5.24 | 5.24 | 5.24 | 5.24 | 5.24 | 5.24 | 5.24 | 5.24 | 5.24 | 5.24 | 5.24 | 5.24 |
| 0.2 | $U_1$ | 14.88 | 26.58 | 35.00 | 41.18 | 45.27 | 48.79 | 51.36 | 53.78 | 55.60 | 57.11 | 58.38 | 59.47 | 60.41 | 61.23 | 61.95 | 62.59 | 63.15 | 63.65 | 64.11 | 64.52 | 64.90 | 65.24 | 65.56 | 65.84 | 66.11 |
|  | $U_2$ | 0.38 | 0.88 | 1.58 | 2.48 | 3.51 | 4.59 | 5.65 | 6.66 | 7.58 | 8.45 | 9.24 | 9.97 | 10.63 | 11.21 | 11.76 | 12.26 | 12.69 | 13.11 | 13.49 | 13.84 | 14.16 | 14.46 | 14.73 | 14.99 | 15.23 |
|  | $W_{1,2}$ | 8.52 | 18.52 | 29.62 | 41.18 | 52.91 | 64.68 | 76.47 | 88.27 | 100.07 | 111.86 | 123.66 | 135.46 | 147.25 | 159.05 | 170.85 | 182.64 | 194.44 | 206.24 | 218.03 | 229.83 | 241.63 | 253.42 | 265.22 | 277.02 | 288.82 |
|  | $W_{3,4}$ | 10.95 | 22.34 | 34.00 | 45.72 | 57.50 | 69.29 | 81.09 | 92.88 | 104.68 | 116.48 | 128.28 | 140.07 | 151.87 | 163.67 | 175.47 | 187.26 | 199.06 | 210.86 | 222.66 | 234.45 | 246.25 | 258.05 | 269.84 | 281.64 | 293.43 |
|  | $J$ | 9.80 | 17.16 | 22.01 | 25.28 | 26.99 | 28.46 | 29.50 | 30.26 | 30.80 | 31.17 | 31.45 | 31.66 | 31.83 | 31.96 | 32.06 | 32.14 | 32.21 | 32.27 | 32.32 | 32.36 | 32.40 | 32.43 | 32.46 | 33.28 | 33.31 |
|  | $K$ | 2.89 | 4.59 | 5.26 | 5.48 | 5.43 | 5.43 | 5.44 | 5.44 | 5.44 | 5.44 | 5.44 | 5.44 | 5.44 | 5.44 | 5.44 | 5.44 | 5.44 | 5.44 | 5.44 | 5.44 | 5.44 | 5.44 | 5.44 | 5.44 | 5.44 |
| 0.1 | $U_1$ | 14.62 | 26.16 | 34.56 | 40.69 | 45.27 | 48.79 | 51.36 | 53.78 | 55.60 | 57.11 | 58.38 | 59.47 | 60.41 | 61.23 | 61.95 | 62.59 | 63.15 | 63.65 | 64.11 | 64.52 | 64.90 | 65.24 | 65.56 | 65.84 | 66.11 |
|  | $U_2$ | 0.38 | 0.88 | 1.58 | 2.48 | 3.51 | 4.59 | 5.65 | 6.66 | 7.58 | 8.47 | 9.26 | 9.97 | 10.63 | 11.21 | 11.76 | 12.26 | 12.71 | 13.13 | 13.50 | 13.85 | 14.17 | 14.47 | 14.74 | 14.99 | 15.20 |
|  | $W_{1,2}$ | 8.42 | 18.33 | 29.39 | 40.95 | 52.67 | 64.45 | 76.24 | 88.03 | 99.83 | 111.63 | 123.42 | 135.22 | 147.02 | 158.81 | 170.61 | 182.41 | 194.20 | 206.00 | 217.80 | 229.59 | 241.39 | 253.19 | 264.99 | 276.78 | 288.58 |
|  | $W_{3,4}$ | 10.91 | 22.29 | 33.93 | 45.72 | 57.50 | 69.29 | 81.05 | 92.85 | 104.65 | 116.44 | 128.24 | 140.04 | 151.83 | 163.63 | 175.43 | 187.22 | 199.02 | 210.82 | 222.62 | 234.41 | 246.21 | 258.01 | 269.80 | 281.60 | 293.43 |
|  | $J$ | 9.95 | 17.18 | 22.02 | 25.28 | 27.52 | 29.10 | 30.24 | 31.06 | 31.66 | 32.13 | 32.48 | 32.76 | 32.97 | 33.13 | 33.27 | 33.37 | 33.46 | 33.53 | 33.59 | 33.64 | 33.68 | 33.72 | 33.75 | 33.78 | 33.80 |
|  | $K$ | 2.89 | 4.50 | 5.26 | 5.48 | 5.55 | 5.57 | 5.58 | 5.58 | 5.58 | 5.58 | 5.58 | 5.58 | 5.58 | 5.58 | 5.58 | 5.58 | 5.58 | 5.58 | 5.58 | 5.58 | 5.58 | 5.58 | 5.58 | 5.58 | 5.58 |
| 0.0 | $U_1$ | 14.80 | 26.47 | 34.97 | 41.16 | 45.78 | 49.33 | 52.17 | 54.35 | 57.09 | 57.69 | 58.97 | 60.06 | 61.01 | 61.83 | 62.55 | 63.18 | 63.75 | 64.26 | 64.71 | 65.13 | 65.50 | 65.85 | 66.16 | 66.45 | 66.72 |
|  | $U_2$ | 0.37 | 0.85 | 1.53 | 2.43 | 3.46 | 4.54 | 5.60 | 6.61 | 7.55 | 8.42 | 9.21 | 9.93 | 10.58 | 11.18 | 11.72 | 12.22 | 12.67 | 13.08 | 13.46 | 13.81 | 14.13 | 14.43 | 14.71 | 14.96 | 15.20 |
|  | $W_{1,2}$ | 8.18 | 18.19 | 29.24 | 40.79 | 52.51 | 64.29 | 76.08 | 87.87 | 99.67 | 111.46 | 123.26 | 135.06 | 146.85 | 158.65 | 170.45 | 182.24 | 194.04 | 205.84 | 217.63 | 229.43 | 241.23 | 253.03 | 264.82 | 276.62 | 288.42 |
|  | $W_{3,4}$ | 9.95 | 22.23 | 33.93 | 45.68 | 57.46 | 69.25 | 81.05 | 92.85 | 104.65 | 116.44 | 128.24 | 140.04 | 151.83 | 163.63 | 175.43 | 187.22 | 199.02 | 210.82 | 222.62 | 234.41 | 246.21 | 258.01 | 269.80 | 281.60 | 293.80 |
|  | $J$ | 9.05 | 17.18 | 22.02 | 25.28 | 27.52 | 29.10 | 30.24 | 31.06 | 31.66 | 32.13 | 32.48 | 32.76 | 32.97 | 33.13 | 33.27 | 33.37 | 33.46 | 33.53 | 33.59 | 33.64 | 33.68 | 33.72 | 33.75 | 33.78 | 33.80 |
|  | $K$ | 2.92 | 4.63 | 5.26 | 5.44 | 5.62 | 5.63 | 5.63 | 5.63 | 5.63 | 5.63 | 5.63 | 5.63 | 5.63 | 5.63 | 5.63 | 5.63 | 5.63 | 5.63 | 5.63 | 5.63 | 5.63 | 5.63 | 5.63 | 5.63 | 5.63 |

Total weight of the heavy-fermions on the active bands after wannierisation: 94.2%

Wannier orbitals support  0.0 — 1.0

Energy (meV): 100, 50, 0, -50, -100 along $K$, $\Gamma$, $M$, $K$

FIG. S42. Band structures of the BM and THF models near charge neutrality for $w_0/w_1 = 0.8$, depicted by lines and crosses, respectively. The BM bands are colored according to the weight of the $f$-electron wave function on them. We use the same BM parameters as in Table S52.

| $w_0/w_1$ | $\alpha_1$ | $\alpha_2$ | $\lambda_1$ | $\lambda_2$ | $\lambda$ | $W$ | $t_0$ | $\gamma$ | $M$ | $v_\star$ | $v'_\star$ | $v''_\star$ | $\gamma/U_1$ | $\gamma/U_2$ |
|---|---|---|---|---|---|---|---|---|---|---|---|---|---|---|
| 1.00 | 0.780 | 0.626 | 0.173 | 0.191 | 0.338 | 0.973 | 0.708 | -13.453 | -10.554 | -4.311 | 1.683 | 0.0044 | -0.20 | -4.56 |
| 0.90 | 0.809 | 0.588 | 0.185 | 0.197 | 0.343 | 0.956 | 0.873 | -28.899 | -11.717 | -4.487 | 1.723 | -0.0087 | -0.45 | -9.56 |
| 0.80 | 0.840 | 0.543 | 0.197 | 0.203 | 0.358 | 0.942 | 1.073 | -44.166 | -12.735 | -4.641 | 1.730 | -0.0158 | -0.73 | -14.02 |
| 0.70 | 0.871 | 0.491 | 0.215 | 0.209 | 0.371 | 0.931 | 1.290 | -58.966 | -13.613 | -4.777 | 1.703 | -0.0187 | -1.04 | -18.01 |
| 0.60 | 0.901 | 0.434 | 0.227 | 0.215 | 0.382 | 0.922 | 1.496 | -73.008 | -14.358 | -4.901 | 1.636 | -0.0184 | -1.36 | -21.59 |
| 0.50 | 0.929 | 0.371 | 0.245 | 0.221 | 0.401 | 0.922 | 1.717 | -85.971 | -14.976 | -5.014 | 1.519 | -0.0159 | -1.71 | -24.34 |
| 0.40 | 0.953 | 0.305 | 0.257 | 0.227 | 0.409 | 0.919 | 1.895 | -97.491 | -15.473 | -5.116 | 1.344 | -0.0121 | -2.02 | -26.98 |
| 0.30 | 0.972 | 0.234 | 0.275 | 0.227 | 0.416 | 0.918 | 2.033 | -107.153 | -15.855 | -5.205 | 1.101 | -0.0078 | -2.29 | -29.14 |
| 0.20 | 0.987 | 0.160 | 0.281 | 0.233 | 0.420 | 0.916 | 2.064 | -114.510 | -16.125 | -5.277 | 0.787 | -0.0038 | -2.49 | -30.88 |
| 0.10 | 0.997 | 0.082 | 0.286 | 0.233 | 0.415 | 0.909 | 2.015 | -119.138 | -16.285 | -5.324 | 0.412 | -0.0010 | -2.58 | -32.46 |
| 0.00 | 1.000 | 0.000 | 0.292 | 0.000 | 0.409 | 0.904 | 1.911 | -120.720 | -16.338 | -5.341 | 0.000 | 0.0000 | -2.59 | -33.36 |

TABLE S52. Parameters of the $f$-electron Wannier functions for different values of the tunneling ratio $w_0/w_1$. $W$ denotes the total weight of the THF $f$-electrons on the active bands. In computing the ratios $\gamma/U_1$ and $\gamma/U_2$, we employ the on-site and nearest-neighbor repulsion parameters $U_1$ and $U_2$ obtained numerically for $\xi = 10$ nm, as given in Table S53. We employ $v_F = 5.944$ eV·Å, $|K| = 1.703$ Å$^{-1}$, $w_1 = 110$ meV, and $\theta = 1.18°$ for the BM model.

| $w_0/w_1$ | | 2 | 4 | 6 | 8 | 10 | 12 | 14 | 16 | 18 | 20 | 22 | 24 | 26 | 28 | 30 | 32 | 34 | 36 | 38 | 40 | 42 | 44 | 46 | 48 | 50 |
|---|---|---|---|---|---|---|---|---|---|---|---|---|---|---|---|---|---|---|---|---|---|---|---|---|---|---|
| **1.0** | $U_1$ | 25.71 | 43.61 | 55.21 | 63.01 | 68.53 | 72.60 | 75.71 | 78.15 | 80.12 | 81.74 | 83.09 | 84.24 | 85.22 | 86.07 | 86.81 | 87.47 | 88.05 | 88.57 | 89.03 | 89.45 | 89.84 | 90.19 | 90.51 | 90.80 | 91.07 |
| | $U_2$ | 0.26 | 0.59 | 1.14 | 1.95 | 2.95 | 4.04 | 5.13 | 6.18 | 7.16 | 8.06 | 8.88 | 9.62 | 10.30 | 10.91 | 11.47 | 11.98 | 12.44 | 12.86 | 13.25 | 13.61 | 13.94 | 14.24 | 14.52 | 14.78 | 15.02 |
| | $W_{1,2}$ | 10.08 | 21.10 | 32.86 | 44.03 | 57.09 | 69.28 | 81.49 | 93.69 | 105.90 | 118.11 | 130.31 | 142.52 | 154.73 | 166.94 | 179.14 | 191.35 | 203.56 | 215.76 | 227.97 | 240.18 | 252.39 | 264.59 | 276.80 | 289.01 | 301.22 |
| | $W_{1,2}$ | 12.78 | 25.62 | 38.11 | 50.41 | 62.65 | 74.87 | 87.08 | 99.28 | 111.49 | 123.70 | 135.91 | 148.11 | 160.32 | 172.53 | 184.73 | 196.94 | 209.15 | 221.35 | 233.56 | 245.77 | 257.98 | 270.19 | 282.39 | 294.60 | 306.81 |
| | $J$ | 8.30 | 12.67 | 14.73 | 15.78 | 16.37 | 16.72 | 16.95 | 17.11 | 17.22 | 17.30 | 17.36 | 17.41 | 17.44 | 17.47 | 17.49 | 17.51 | 17.52 | 17.53 | 17.54 | 17.55 | 17.55 | 17.56 | 17.56 | 17.57 | 17.57 |
| | $K$ | 2.25 | 3.56 | 4.06 | 4.23 | 4.28 | 4.29 | 4.30 | 4.30 | 4.30 | 4.30 | 4.30 | 4.30 | 4.30 | 4.30 | 4.30 | 4.30 | 4.30 | 4.30 | 4.30 | 4.30 | 4.30 | 4.30 | 4.30 | 4.30 | 4.30 |
| **0.9** | $U_1$ | 23.64 | 40.46 | 51.59 | 59.17 | 64.57 | 68.58 | 71.66 | 74.08 | 76.04 | 77.65 | 78.99 | 80.13 | 81.11 | 81.96 | 82.70 | 83.35 | 83.93 | 84.45 | 84.91 | 85.33 | 85.72 | 86.07 | 86.38 | 86.68 | 86.95 |
| | $U_2$ | 0.26 | 0.61 | 1.18 | 2.01 | 3.02 | 4.12 | 5.21 | 6.26 | 7.24 | 8.13 | 8.95 | 9.70 | 10.37 | 10.99 | 11.54 | 12.05 | 12.51 | 12.94 | 13.32 | 13.68 | 14.01 | 14.31 | 14.59 | 14.85 | 15.09 |
| | $W_{1,2}$ | 10.11 | 21.16 | 32.93 | 44.59 | 57.16 | 69.35 | 81.55 | 93.76 | 105.97 | 118.17 | 130.38 | 142.59 | 154.79 | 167.00 | 179.21 | 191.42 | 203.62 | 215.83 | 228.04 | 240.25 | 252.45 | 264.66 | 276.87 | 289.07 | 301.28 |
| | $W_{1,2}$ | 12.31 | 24.82 | 37.18 | 49.45 | 61.67 | 73.88 | 86.09 | 98.30 | 110.51 | 122.71 | 134.92 | 147.13 | 159.33 | 171.54 | 183.75 | 195.96 | 208.16 | 220.37 | 232.58 | 244.79 | 256.99 | 269.20 | 281.41 | 293.61 | 305.82 |
| | $J$ | 8.45 | 13.24 | 15.74 | 17.13 | 17.98 | 18.52 | 18.90 | 19.16 | 19.35 | 19.48 | 19.59 | 19.67 | 19.73 | 19.77 | 19.81 | 19.84 | 19.87 | 19.89 | 19.90 | 19.92 | 19.93 | 19.94 | 19.95 | 19.96 | 19.96 |
| | $K$ | 2.23 | 3.53 | 4.03 | 4.20 | 4.24 | 4.26 | 4.27 | 4.27 | 4.27 | 4.27 | 4.27 | 4.27 | 4.27 | 4.27 | 4.27 | 4.27 | 4.27 | 4.27 | 4.27 | 4.27 | 4.27 | 4.27 | 4.27 | 4.27 | 4.27 |
| **0.8** | $U_1$ | 21.56 | 37.24 | 47.80 | 55.10 | 60.35 | 64.27 | 67.29 | 69.68 | 71.61 | 73.20 | 74.53 | 75.66 | 76.64 | 77.48 | 78.22 | 78.87 | 79.44 | 79.96 | 80.42 | 80.84 | 81.22 | 81.57 | 81.89 | 82.18 | 82.45 |
| | $U_2$ | 0.27 | 0.67 | 1.27 | 2.12 | 3.15 | 4.25 | 5.34 | 6.39 | 7.36 | 8.26 | 9.08 | 9.82 | 10.49 | 11.10 | 11.66 | 12.16 | 12.62 | 13.05 | 13.43 | 13.79 | 14.12 | 14.42 | 14.70 | 14.96 | 15.20 |
| | $W_{1,2}$ | 10.06 | 21.08 | 32.83 | 44.50 | 57.00 | 69.25 | 81.46 | 93.66 | 105.87 | 118.08 | 130.28 | 142.49 | 154.70 | 166.90 | 179.11 | 191.32 | 203.53 | 215.73 | 227.94 | 240.15 | 252.35 | 264.56 | 276.77 | 288.98 | 301.18 |
| | $W_{1,2}$ | 11.96 | 24.23 | 36.50 | 48.73 | 60.94 | 73.15 | 85.36 | 97.57 | 109.78 | 121.98 | 134.19 | 146.40 | 158.60 | 170.81 | 183.02 | 195.23 | 207.43 | 219.64 | 231.85 | 244.05 | 256.26 | 268.47 | 280.68 | 292.88 | 305.09 |
| | $J$ | 8.65 | 13.89 | 16.82 | 18.58 | 19.69 | 20.44 | 20.96 | 21.33 | 21.60 | 21.80 | 21.95 | 22.07 | 22.16 | 22.23 | 22.28 | 22.33 | 22.37 | 22.39 | 22.42 | 22.44 | 22.46 | 22.47 | 22.48 | 22.49 | 22.50 |
| | $K$ | 2.25 | 3.56 | 4.07 | 4.23 | 4.28 | 4.29 | 4.30 | 4.30 | 4.30 | 4.30 | 4.30 | 4.30 | 4.30 | 4.30 | 4.30 | 4.30 | 4.30 | 4.30 | 4.30 | 4.30 | 4.30 | 4.30 | 4.30 | 4.30 | 4.30 |
| **0.7** | $U_1$ | 19.80 | 34.46 | 44.52 | 51.55 | 56.66 | 60.49 | 63.45 | 65.81 | 67.72 | 69.29 | 70.61 | 71.74 | 72.70 | 73.54 | 74.27 | 74.92 | 75.49 | 76.01 | 76.47 | 76.89 | 77.27 | 77.62 | 77.93 | 78.22 | 78.50 |
| | $U_2$ | 0.29 | 0.73 | 1.36 | 2.23 | 3.27 | 4.38 | 5.47 | 6.52 | 7.49 | 8.38 | 9.20 | 9.94 | 10.61 | 11.21 | 11.77 | 12.27 | 12.73 | 13.15 | 13.54 | 13.89 | 14.22 | 14.52 | 14.80 | 15.06 | 15.30 |
| | $W_{1,2}$ | 9.94 | 20.88 | 32.61 | 44.67 | 56.83 | 69.02 | 81.23 | 93.43 | 105.64 | 117.84 | 130.05 | 142.26 | 154.47 | 166.67 | 178.88 | 191.09 | 203.30 | 215.50 | 227.71 | 239.92 | 252.12 | 264.33 | 276.54 | 288.75 | 300.95 |
| | $W_{1,2}$ | 11.71 | 23.81 | 36.01 | 48.21 | 60.42 | 72.63 | 84.84 | 97.04 | 109.25 | 121.46 | 133.66 | 145.87 | 158.08 | 170.29 | 182.49 | 194.70 | 206.91 | 219.12 | 231.32 | 243.53 | 255.74 | 267.94 | 280.15 | 292.36 | 304.57 |
| | $J$ | 8.89 | 14.56 | 17.92 | 20.01 | 21.38 | 22.31 | 22.96 | 23.44 | 23.79 | 24.04 | 24.24 | 24.39 | 24.51 | 24.60 | 24.67 | 24.73 | 24.78 | 24.82 | 24.85 | 24.88 | 24.90 | 24.92 | 24.93 | 24.95 | 24.96 |
| | $K$ | 2.31 | 3.65 | 4.16 | 4.33 | 4.38 | 4.40 | 4.40 | 4.40 | 4.40 | 4.40 | 4.40 | 4.40 | 4.40 | 4.40 | 4.40 | 4.40 | 4.40 | 4.40 | 4.40 | 4.40 | 4.40 | 4.40 | 4.40 | 4.40 | 4.40 |

TABLE S53. Band structures of the BM and THF single-particle Hamiltonian for different values of the tunneling ratio $w_0/w_1$. $W$ denotes the total weight of the THF $f$-electrons on the active bands. In computing the ratios $\gamma/U_1$ and $\gamma/U_2$, we employ the on-site and nearest-neighbor repulsion parameters $U_1$ and $U_2$ obtained numerically for $\xi = 10$ nm, as given in Table S53. We employ $v_F = 5.944$ eV·Å, $|K| = 1.703$ Å$^{-1}$, $w_1 = 110$ meV, and $\theta = 1.18°$ for the BM model.

| $u_0/u_1$ | $\xi$/nm | 2 | 4 | 6 | 8 | 10 | 12 | 14 | 16 | 18 | 20 | 22 | 24 | 26 | 28 | 30 | 32 | 34 | 36 | 38 | 40 | 42 | 44 | 46 | 48 | 50 |
|---|---|---|---|---|---|---|---|---|---|---|---|---|---|---|---|---|---|---|---|---|---|---|---|---|---|---|
| 0.6 | $U_1$ | 18.37 | 33.22 | 41.84 | 48.65 | 53.63 | 57.39 | 60.31 | 62.64 | 64.52 | 66.09 | 67.40 | 68.51 | 69.47 | 70.30 | 71.03 | 71.68 | 72.25 | 72.76 | 73.23 | 73.64 | 74.02 | 74.37 | 74.68 | 74.98 | 75.24 |
| | $U_2$ | 0.34 | 0.78 | 1.44 | 2.33 | 3.38 | 4.49 | 5.58 | 6.63 | 7.60 | 8.49 | 9.30 | 10.03 | 10.70 | 11.31 | 11.86 | 12.36 | 12.82 | 13.24 | 13.63 | 13.98 | 14.31 | 14.61 | 14.89 | 15.15 | 15.39 |
| | $W_{1,2}$ | 9.76 | 20.61 | 32.30 | 44.34 | 56.50 | 68.69 | 80.90 | 93.10 | 105.31 | 117.52 | 129.72 | 141.93 | 154.14 | 166.34 | 178.55 | 190.76 | 202.97 | 215.17 | 227.38 | 239.59 | 251.79 | 264.00 | 276.21 | 288.42 | 300.62 |
| | $W_{3,4}$ | 11.54 | 23.52 | 35.66 | 47.85 | 60.00 | 72.26 | 84.47 | 96.67 | 108.88 | 121.09 | 133.29 | 145.50 | 157.71 | 169.92 | 182.12 | 194.33 | 206.96 | 218.75 | 230.05 | 243.16 | 255.37 | 267.57 | 279.78 | 291.99 | 304.20 |
| | $J$ | 9.14 | 15.21 | 18.05 | 21.33 | 22.92 | 24.01 | 24.79 | 25.35 | 25.76 | 26.07 | 26.31 | 26.49 | 26.63 | 26.74 | 26.83 | 26.90 | 26.96 | 27.01 | 27.05 | 27.08 | 27.11 | 27.13 | 27.15 | 27.16 | 27.18 |
| | $K$ | 2.39 | 3.77 | 4.30 | 4.48 | 4.53 | 4.55 | 4.55 | 4.55 | 4.55 | 4.55 | 4.55 | 4.55 | 4.55 | 4.55 | 4.55 | 4.55 | 4.55 | 4.55 | 4.55 | 4.55 | 4.55 | 4.55 | 4.55 | 4.55 | 4.55 |
| 0.5 | $U_1$ | 16.92 | 29.86 | 37.26 | 45.54 | 50.36 | 54.02 | 56.88 | 59.16 | 61.02 | 62.56 | 63.85 | 64.96 | 65.91 | 66.73 | 67.46 | 68.10 | 68.67 | 69.18 | 69.64 | 70.05 | 70.43 | 70.78 | 71.09 | 71.38 | 71.65 |
| | $U_2$ | 0.28 | 0.86 | 1.55 | 2.47 | 3.53 | 4.64 | 5.74 | 6.77 | 7.74 | 8.62 | 9.43 | 10.16 | 10.83 | 11.43 | 11.98 | 12.48 | 12.94 | 13.36 | 13.74 | 14.09 | 14.42 | 14.72 | 15.00 | 15.26 | 15.50 |
| | $W_{1,2}$ | 9.57 | 20.33 | 31.98 | 44.01 | 56.17 | 68.36 | 80.56 | 92.77 | 104.97 | 117.18 | 129.39 | 141.60 | 153.80 | 166.01 | 178.22 | 190.42 | 202.63 | 214.84 | 227.05 | 239.25 | 251.46 | 263.67 | 275.87 | 288.08 | 300.29 |
| | $W_{3,4}$ | 11.44 | 23.34 | 35.45 | 47.62 | 59.82 | 72.03 | 84.23 | 96.44 | 108.65 | 120.85 | 133.06 | 145.27 | 157.47 | 169.68 | 181.89 | 194.10 | 206.30 | 218.51 | 230.72 | 242.93 | 255.13 | 267.34 | 279.55 | 291.75 | 303.96 |
| | $J$ | 9.44 | 15.04 | 20.09 | 22.82 | 24.68 | 25.98 | 26.92 | 27.61 | 28.12 | 28.50 | 28.80 | 29.03 | 29.21 | 29.35 | 29.47 | 29.56 | 29.64 | 29.70 | 29.75 | 29.80 | 29.83 | 29.86 | 29.89 | 29.91 | 29.93 |
| | $K$ | 2.49 | 3.93 | 4.49 | 4.67 | 4.72 | 4.74 | 4.74 | 4.74 | 4.74 | 4.74 | 4.74 | 4.74 | 4.74 | 4.74 | 4.74 | 4.74 | 4.74 | 4.74 | 4.74 | 4.74 | 4.74 | 4.74 | 4.74 | 4.74 | 4.74 |
| 0.4 | $U_1$ | 15.34 | 27.31 | 35.04 | 42.19 | 46.84 | 50.40 | 53.20 | 55.43 | 57.27 | 58.79 | 60.07 | 61.16 | 62.10 | 62.92 | 63.64 | 64.28 | 64.84 | 65.35 | 65.81 | 66.22 | 66.60 | 66.94 | 67.26 | 67.55 | 67.81 |
| | $U_2$ | 0.39 | 0.90 | 1.61 | 2.55 | 3.61 | 4.73 | 5.82 | 6.86 | 7.82 | 8.70 | 9.51 | 10.24 | 10.90 | 11.50 | 12.05 | 12.55 | 13.01 | 13.42 | 13.81 | 14.16 | 14.48 | 14.78 | 15.06 | 15.32 | 15.56 |
| | $W_{1,2}$ | 9.35 | 19.98 | 31.60 | 43.61 | 55.76 | 67.95 | 80.15 | 92.36 | 104.57 | 116.77 | 128.98 | 141.19 | 153.39 | 165.60 | 177.81 | 190.02 | 202.22 | 214.43 | 226.64 | 238.85 | 251.05 | 263.26 | 275.47 | 287.67 | 299.88 |
| | $W_{3,4}$ | 11.37 | 23.21 | 35.29 | 47.36 | 59.56 | 71.76 | 83.97 | 96.17 | 108.38 | 120.59 | 132.79 | 145.00 | 157.21 | 169.42 | 181.62 | 193.83 | 206.04 | 218.24 | 230.45 | 242.66 | 254.87 | 267.07 | 279.28 | 291.49 | 303.69 |
| | $J$ | 9.70 | 16.53 | 20.99 | 23.96 | 26.00 | 27.44 | 28.47 | 29.24 | 29.80 | 30.23 | 30.57 | 30.82 | 31.02 | 31.18 | 31.31 | 31.42 | 31.50 | 31.57 | 31.63 | 31.68 | 31.72 | 31.76 | 31.79 | 31.81 | 31.83 |
| | $K$ | 2.61 | 4.11 | 4.69 | 4.88 | 4.94 | 4.96 | 4.96 | 4.96 | 4.96 | 4.96 | 4.96 | 4.96 | 4.96 | 4.96 | 4.96 | 4.96 | 4.96 | 4.96 | 4.96 | 4.96 | 4.96 | 4.96 | 4.96 | 4.96 | 4.96 |
| 0.3 | $U_1$ | 14.93 | 26.64 | 35.12 | 43.22 | 49.45 | 52.72 | 55.45 | 57.77 | 59.76 | 61.16 | 62.10 | 62.92 | 63.64 | 64.28 | 64.84 | 65.35 | 65.81 | 66.22 | 66.60 | 66.94 | 67.26 | 67.55 | 66.78 | 67.05 | 67.05 |
| | $U_2$ | 0.41 | 0.94 | 1.68 | 2.63 | 3.71 | 4.82 | 5.92 | 6.92 | 7.88 | 8.70 | 9.51 | 10.32 | 10.98 | 11.56 | 12.10 | 12.63 | 13.08 | 13.50 | 13.88 | 14.23 | 14.55 | 14.85 | 15.13 | 15.39 | 15.63 |
| | $W_{1,2}$ | 9.13 | 19.65 | 31.22 | 43.26 | 55.36 | 67.55 | 79.75 | 91.96 | 104.17 | 116.37 | 128.58 | 140.79 | 153.00 | 165.20 | 177.41 | 189.62 | 201.83 | 214.03 | 226.24 | 238.45 | 250.65 | 262.86 | 275.07 | 287.27 | 299.48 |
| | $W_{3,4}$ | 11.33 | 23.13 | 35.29 | 47.46 | 59.56 | 71.76 | 83.90 | 96.17 | 108.38 | 120.59 | 132.73 | 144.94 | 157.14 | 169.35 | 181.56 | 193.77 | 205.97 | 218.18 | 230.38 | 242.59 | 254.80 | 267.01 | 279.22 | 291.42 | 303.63 |
| | $J$ | 9.92 | 17.04 | 21.74 | 24.91 | 27.01 | 29.53 | 30.71 | 31.58 | 32.22 | 32.72 | 33.09 | 33.39 | 33.62 | 33.81 | 33.98 | 34.12 | 34.24 | 34.34 | 34.42 | 34.38 | 34.43 | 34.47 | 34.50 | 34.53 | 34.56 |
| | $K$ | 2.73 | 4.47 | 4.91 | 5.10 | 5.18 | 5.18 | 5.19 | 5.19 | 5.19 | 5.19 | 5.19 | 5.19 | 5.19 | 5.19 | 5.19 | 5.19 | 5.19 | 5.19 | 5.19 | 5.19 | 5.19 | 5.19 | 5.19 | 5.19 | 5.19 |
| 0.2 | $U_1$ | 14.96 | 26.72 | 35.25 | 42.87 | 49.65 | 52.72 | 55.12 | 54.07 | 56.50 | 58.02 | 59.30 | 60.40 | 61.34 | 62.16 | 62.88 | 63.51 | 64.08 | 64.59 | 65.05 | 65.46 | 65.84 | 66.18 | 66.49 | 66.78 | 67.05 |
| | $U_2$ | 0.41 | 0.91 | 1.65 | 2.59 | 3.67 | 4.79 | 5.88 | 6.92 | 7.88 | 8.70 | 9.59 | 10.29 | 10.95 | 11.56 | 12.10 | 12.60 | 13.06 | 13.50 | 13.85 | 14.21 | 14.53 | 14.83 | 15.11 | 15.37 | 15.61 |
| | $W_{1,2}$ | 8.94 | 19.35 | 30.88 | 42.87 | 55.01 | 67.09 | 79.40 | 91.61 | 103.81 | 116.02 | 128.23 | 140.44 | 152.64 | 164.85 | 177.06 | 189.26 | 201.47 | 213.68 | 225.88 | 238.09 | 250.30 | 262.51 | 274.71 | 286.92 | 299.13 |
| | $W_{3,4}$ | 11.30 | 23.08 | 35.14 | 47.30 | 59.43 | 71.69 | 83.90 | 96.11 | 108.31 | 120.52 | 132.73 | 144.94 | 157.14 | 169.35 | 181.56 | 193.77 | 205.97 | 218.18 | 230.38 | 242.59 | 254.80 | 267.01 | 279.22 | 291.42 | 303.63 |
| | $J$ | 10.10 | 17.41 | 22.30 | 25.61 | 27.91 | 29.53 | 30.71 | 31.58 | 32.22 | 32.72 | 33.09 | 33.39 | 33.62 | 33.81 | 33.98 | 34.07 | 34.17 | 34.25 | 34.32 | 34.38 | 34.43 | 34.47 | 34.50 | 34.53 | 34.56 |
| | $K$ | 2.73 | 4.51 | 5.10 | 5.30 | 5.38 | 5.39 | 5.39 | 5.39 | 5.39 | 5.39 | 5.39 | 5.39 | 5.39 | 5.39 | 5.39 | 5.39 | 5.39 | 5.39 | 5.39 | 5.39 | 5.39 | 5.39 | 5.39 | 5.39 | 5.39 |
| 0.1 | $U_1$ | 15.14 | 27.03 | 35.66 | 42.55 | 46.10 | 49.65 | 52.44 | 54.67 | 56.50 | 58.02 | 59.30 | 60.40 | 61.34 | 62.16 | 62.88 | 63.51 | 64.08 | 64.59 | 65.05 | 65.46 | 65.84 | 66.18 | 66.49 | 66.78 | 67.05 |
| | $U_2$ | 0.40 | 0.91 | 1.65 | 2.59 | 3.67 | 4.79 | 5.88 | 6.92 | 7.88 | 8.70 | 9.56 | 10.29 | 10.95 | 11.56 | 12.10 | 12.60 | 13.06 | 13.47 | 13.85 | 14.21 | 14.53 | 14.83 | 15.11 | 15.37 | 15.61 |
| | $W_{1,2}$ | 8.77 | 19.08 | 30.57 | 42.55 | 54.69 | 66.88 | 79.08 | 91.28 | 103.49 | 115.70 | 127.90 | 140.11 | 152.32 | 164.53 | 176.73 | 188.94 | 201.15 | 213.36 | 225.56 | 237.77 | 249.98 | 262.19 | 274.39 | 286.60 | 298.81 |
| | $W_{3,4}$ | 11.27 | 23.03 | 35.08 | 47.24 | 59.43 | 71.64 | 83.84 | 96.05 | 108.26 | 120.46 | 132.67 | 144.88 | 157.08 | 169.29 | 181.50 | 193.71 | 205.91 | 218.12 | 230.33 | 242.53 | 254.71 | 266.92 | 279.12 | 291.33 | 303.54 |
| | $J$ | 10.21 | 17.61 | 22.56 | 25.87 | 28.13 | 29.72 | 30.85 | 31.68 | 32.28 | 32.74 | 33.08 | 33.35 | 33.56 | 33.72 | 33.88 | 33.98 | 34.12 | 34.24 | 34.34 | 34.41 | 34.43 | 34.56 | 34.63 | 34.66 | 34.68 |
| | $K$ | 2.91 | 4.59 | 5.19 | 5.38 | 5.44 | 5.52 | 5.53 | 5.53 | 5.53 | 5.53 | 5.53 | 5.53 | 5.53 | 5.53 | 5.53 | 5.53 | 5.53 | 5.53 | 5.53 | 5.53 | 5.53 | 5.53 | 5.53 | 5.53 | 5.53 |
| 0.0 | $U_1$ | 15.14 | 27.03 | 35.66 | 41.94 | 46.61 | 50.19 | 52.99 | 55.23 | 57.08 | 58.61 | 59.89 | 60.99 | 61.93 | 62.75 | 63.48 | 64.11 | 64.68 | 65.19 | 65.65 | 66.06 | 66.44 | 66.78 | 67.10 | 67.39 | 67.66 |
| | $U_2$ | 0.38 | 0.88 | 1.60 | 2.55 | 3.62 | 4.74 | 5.83 | 6.87 | 7.83 | 8.71 | 9.52 | 10.25 | 10.91 | 11.52 | 12.06 | 12.56 | 13.02 | 13.43 | 13.82 | 14.17 | 14.49 | 14.79 | 15.07 | 15.33 | 15.57 |
| | $W_{1,2}$ | 8.68 | 18.94 | 30.42 | 42.39 | 54.53 | 66.71 | 78.91 | 91.12 | 103.32 | 115.53 | 127.74 | 139.95 | 152.15 | 164.36 | 176.57 | 188.78 | 200.98 | 213.19 | 225.40 | 237.60 | 249.81 | 262.02 | 274.22 | 286.43 | 298.64 |
| | $W_{3,4}$ | 11.25 | 23.00 | 35.05 | 47.21 | 59.34 | 71.60 | 83.81 | 96.01 | 108.22 | 120.43 | 132.64 | 144.84 | 157.05 | 169.26 | 181.46 | 193.67 | 205.88 | 218.09 | 230.29 | 242.50 | 254.71 | 266.92 | 279.12 | 291.33 | 303.54 |
| | $J$ | 10.94 | 17.64 | 22.57 | 25.87 | 28.13 | 29.72 | 30.85 | 31.68 | 32.28 | 32.74 | 33.08 | 33.35 | 33.56 | 33.72 | 33.85 | 33.96 | 34.04 | 34.11 | 34.17 | 34.22 | 34.26 | 34.30 | 34.32 | 34.35 | 34.37 |
| | $K$ | 2.94 | 4.63 | 5.23 | 5.44 | 5.50 | 5.52 | 5.53 | 5.58 | 5.58 | 5.58 | 5.58 | 5.58 | 5.58 | 5.58 | 5.58 | 5.58 | 5.58 | 5.58 | 5.58 | 5.58 | 5.58 | 5.58 | 5.58 | 5.58 | 5.58 |

TABLE S53: Parameters of the THF interaction Hamiltonian for different ratios $u_0/u_1$ and screening lengths $\xi$ of the double-gate interaction potential ($U_\xi = 24$ meV for $\xi = 10$ nm). We use the same BM model parameters as in Table S52.

Total weight of the heavy-fermions on the active bands after wannierisation: 94.0%

Wannier orbitals support  0.0 — 1.0

*Energy (meV)* — 100, 50, 0, −50, −100 — axis labels K, Γ, M, K

FIG. S43. Band structures of the BM and THF models near charge neutrality for $w_0/w_1 = 0.8$, depicted by lines and crosses, respectively. The BM bands are colored according to the weight of the $f$-electron wave function on them. We use the same BM parameters as in Table S54.

TABLE S54. Parameters of the $f$-electron Wannier functions for different values of the tunneling ratio $w_0/w_1$. $W$ denotes the total weight of the THF $f$-electrons on the active bands. In computing the ratios $\gamma/U_1$ and $\gamma/U_2$, we employ the on-site and nearest-neighbor repulsion parameters $U_1$ and $U_2$ obtained numerically for $\xi = 10$ nm, as given in Table S55. We employ $v_F = 5.944$ eV·Å, $|\mathbf{K}| = 1.703$ Å$^{-1}$, $w_1 = 110$ meV, and $\theta = 1.20°$ for the BM model.

| $w_0/w_1$ | $\alpha_1$ | $\alpha_2$ | $\lambda_1$ | $\lambda_2$ | $\lambda$ | $W$ | $t_0$ | $\gamma$ | $M$ | $v_\star$ | $v'_\star$ | $v''_\star$ | $\gamma/U_1$ | $\gamma/U_2$ |
|---|---|---|---|---|---|---|---|---|---|---|---|---|---|---|
| 1.00 | 0.783 | 0.622 | 0.179 | 0.191 | 0.339 | 0.970 | 0.872 | -16.373 | -13.163 | -4.365 | 1.705 | 0.0053 | -2.24 | -5.27 |
| 0.90 | 0.812 | 0.584 | 0.191 | 0.197 | 0.346 | 0.953 | 1.078 | -31.933 | -14.390 | -4.533 | 1.740 | -0.0064 | -0.49 | -10.01 |
| 0.80 | 0.843 | 0.538 | 0.203 | 0.203 | 0.360 | 0.940 | 1.325 | -47.272 | -15.462 | -4.679 | 1.743 | -0.0129 | -0.78 | -14.26 |
| 0.70 | 0.874 | 0.486 | 0.215 | 0.209 | 0.373 | 0.929 | 1.581 | -62.114 | -16.384 | -4.811 | 1.713 | -0.0155 | -1.09 | -18.05 |
| 0.60 | 0.903 | 0.430 | 0.233 | 0.215 | 0.384 | 0.922 | 1.829 | -76.176 | -17.166 | -4.931 | 1.643 | -0.0154 | -1.41 | -21.45 |
| 0.50 | 0.930 | 0.367 | 0.245 | 0.221 | 0.402 | 0.922 | 2.081 | -89.144 | -17.815 | -5.042 | 1.524 | -0.0134 | -1.75 | -24.11 |
| 0.40 | 0.954 | 0.301 | 0.263 | 0.227 | 0.418 | 0.924 | 2.340 | -100.658 | -18.336 | -5.143 | 1.347 | -0.0102 | -2.08 | -26.27 |
| 0.30 | 0.973 | 0.231 | 0.275 | 0.227 | 0.422 | 0.920 | 2.437 | -110.308 | -18.735 | -5.231 | 1.102 | -0.0066 | -2.34 | -28.49 |
| 0.20 | 0.987 | 0.158 | 0.286 | 0.233 | 0.422 | 0.917 | 2.529 | -117.652 | -19.017 | -5.302 | 0.788 | -0.0032 | -2.53 | -30.27 |
| 0.10 | 0.997 | 0.081 | 0.292 | 0.233 | 0.415 | 0.909 | 2.452 | -122.269 | -19.186 | -5.348 | 0.412 | -0.0009 | -2.61 | -31.90 |
| 0.00 | 1.000 | 0.000 | 0.292 | 0.000 | 0.409 | 0.903 | 2.353 | -123.847 | -19.241 | -5.365 | 0.000 | -0.0000 | -2.61 | -32.75 |

| $w_0/w_1$ | $\xi/$nm | 2 | 4 | 6 | 8 | 10 | 12 | 14 | 16 | 18 | 20 | 22 | 24 | 26 | 28 | 30 | 32 | 34 | 36 | 38 | 40 | 42 | 44 | 46 | 48 | 50 |
|---|---|---|---|---|---|---|---|---|---|---|---|---|---|---|---|---|---|---|---|---|---|---|---|---|---|---|
| 1.0 | $U_1$ | 25.88 | 43.86 | 55.51 | 63.34 | 68.88 | 72.96 | 76.08 | 78.53 | 80.50 | 82.13 | 83.48 | 84.63 | 85.61 | 86.46 | 87.20 | 87.86 | 88.44 | 88.96 | 89.43 | 89.85 | 90.23 | 90.58 | 90.90 | 91.19 | 91.46 |
|  | $U_2$ | 0.27 | 0.62 | 1.20 | 2.06 | 3.11 | 4.24 | 5.37 | 6.44 | 7.44 | 8.36 | 9.19 | 9.95 | 10.63 | 11.26 | 11.82 | 12.33 | 12.80 | 13.22 | 13.61 | 13.97 | 14.30 | 14.61 | 14.89 | 15.15 | 15.39 |
|  | $W_{1,2}$ | 10.55 | 22.01 | 34.22 | 46.71 | 59.29 | 71.90 | 84.53 | 97.15 | 109.77 | 122.40 | 135.02 | 147.65 | 160.27 | 172.90 | 185.52 | 198.14 | 210.77 | 223.39 | 236.02 | 248.64 | 261.27 | 273.89 | 286.52 | 299.14 | 311.76 |
|  | $W_{3,4}$ | 13.03 | 26.17 | 39.03 | 51.73 | 64.38 | 77.01 | 89.64 | 102.26 | 114.89 | 127.51 | 140.14 | 152.76 | 165.39 | 178.01 | 190.64 | 203.26 | 215.89 | 228.51 | 241.13 | 253.76 | 266.38 | 279.01 | 291.63 | 304.26 | 316.88 |
|  | $J$ | 8.48 | 12.96 | 15.10 | 16.20 | 16.83 | 17.21 | 17.46 | 17.64 | 17.76 | 17.85 | 17.92 | 17.97 | 18.00 | 18.03 | 18.06 | 18.08 | 18.10 | 18.12 | 18.12 | 18.14 | 18.14 | 18.14 | 18.15 | 18.15 | 18.15 |
|  | $K$ | 2.22 | 3.49 | 3.97 | 4.12 | 4.17 | 4.18 | 4.18 | 4.18 | 4.18 | 4.18 | 4.18 | 4.18 | 4.18 | 4.18 | 4.18 | 4.18 | 4.18 | 4.18 | 4.18 | 4.18 | 4.18 | 4.18 | 4.18 | 4.18 | 4.18 |
| 0.9 | $U_1$ | 23.82 | 40.74 | 51.91 | 59.52 | 64.95 | 68.97 | 72.05 | 74.48 | 76.44 | 78.05 | 79.40 | 80.54 | 81.52 | 82.37 | 83.11 | 83.76 | 84.34 | 84.86 | 85.33 | 85.75 | 86.13 | 86.48 | 86.80 | 87.09 | 87.36 |
|  | $U_2$ | 0.28 | 0.65 | 1.25 | 2.13 | 3.19 | 4.32 | 5.45 | 6.53 | 7.53 | 8.44 | 9.28 | 10.03 | 10.71 | 11.33 | 11.90 | 12.41 | 12.87 | 13.30 | 13.69 | 14.05 | 14.38 | 14.68 | 14.96 | 15.23 | 15.47 |
|  | $W_{1,2}$ | 10.55 | 22.04 | 34.24 | 46.73 | 59.32 | 71.93 | 84.55 | 97.17 | 109.80 | 122.42 | 135.05 | 147.67 | 160.30 | 172.92 | 185.55 | 198.17 | 210.79 | 223.42 | 236.04 | 248.67 | 261.29 | 273.92 | 286.54 | 299.17 | 311.79 |
|  | $W_{3,4}$ | 12.58 | 25.42 | 38.16 | 50.83 | 63.47 | 76.09 | 88.72 | 101.34 | 113.97 | 126.59 | 139.22 | 151.84 | 164.47 | 177.09 | 189.72 | 202.34 | 214.96 | 227.59 | 240.21 | 252.84 | 265.46 | 278.09 | 290.71 | 303.34 | 315.96 |
|  | $J$ | 8.64 | 13.56 | 16.13 | 17.58 | 18.46 | 19.03 | 19.42 | 19.69 | 19.89 | 20.03 | 20.14 | 20.22 | 20.30 | 20.34 | 20.37 | 20.41 | 20.43 | 20.45 | 20.47 | 20.48 | 20.49 | 20.50 | 20.51 | 20.52 | 20.52 |
|  | $K$ | 2.21 | 3.48 | 3.96 | 4.11 | 4.15 | 4.16 | 4.17 | 4.17 | 4.17 | 4.17 | 4.17 | 4.17 | 4.17 | 4.17 | 4.17 | 4.17 | 4.17 | 4.17 | 4.17 | 4.17 | 4.17 | 4.17 | 4.17 | 4.17 | 4.17 |
| 0.8 | $U_1$ | 21.78 | 37.58 | 48.21 | 55.55 | 60.83 | 64.76 | 67.79 | 70.19 | 72.12 | 73.72 | 75.05 | 76.19 | 77.16 | 78.00 | 78.74 | 79.39 | 79.97 | 80.49 | 80.95 | 81.37 | 81.75 | 82.10 | 82.42 | 82.71 | 82.98 |
|  | $U_2$ | 0.30 | 0.70 | 1.34 | 2.24 | 3.31 | 4.46 | 5.58 | 6.66 | 7.65 | 8.57 | 9.40 | 10.15 | 10.83 | 11.45 | 12.01 | 12.52 | 12.98 | 13.41 | 13.80 | 14.15 | 14.48 | 14.79 | 15.07 | 15.33 | 15.57 |
|  | $W_{1,2}$ | 10.48 | 21.93 | 34.12 | 46.61 | 59.19 | 71.80 | 84.42 | 97.05 | 109.67 | 122.29 | 134.92 | 147.54 | 160.17 | 172.79 | 185.42 | 198.04 | 210.67 | 223.29 | 235.92 | 248.54 | 261.16 | 273.79 | 286.41 | 299.04 | 311.66 |
|  | $W_{3,4}$ | 12.25 | 24.87 | 37.53 | 50.16 | 62.79 | 75.42 | 88.04 | 100.67 | 113.29 | 125.91 | 138.54 | 151.16 | 163.79 | 176.41 | 189.04 | 201.66 | 214.29 | 226.91 | 239.54 | 252.16 | 264.78 | 277.41 | 290.03 | 302.66 | 315.28 |
|  | $J$ | 8.87 | 14.25 | 17.27 | 19.08 | 20.23 | 21.00 | 21.53 | 21.91 | 22.19 | 22.40 | 22.55 | 22.67 | 22.76 | 22.83 | 22.89 | 22.93 | 22.97 | 23.00 | 23.02 | 23.04 | 23.06 | 23.08 | 23.09 | 23.10 | 23.11 |
|  | $K$ | 2.24 | 4.01 | 4.16 | 4.20 | 4.22 | 4.22 | 4.22 | 4.22 | 4.22 | 4.22 | 4.22 | 4.22 | 4.22 | 4.22 | 4.22 | 4.22 | 4.22 | 4.22 | 4.22 | 4.22 | 4.22 | 4.22 | 4.22 | 4.22 | 4.22 |
| 0.7 | $U_1$ | 20.04 | 34.85 | 44.09 | 52.07 | 57.21 | 61.06 | 64.03 | 66.40 | 68.31 | 69.89 | 71.22 | 72.34 | 73.31 | 74.15 | 74.88 | 75.53 | 76.10 | 76.62 | 77.08 | 77.50 | 77.88 | 78.23 | 78.55 | 78.84 | 79.11 |
|  | $U_2$ | 0.33 | 0.76 | 1.43 | 2.35 | 3.44 | 4.58 | 5.71 | 6.78 | 7.78 | 8.69 | 9.51 | 10.26 | 10.94 | 11.56 | 12.12 | 12.63 | 13.09 | 13.51 | 13.90 | 14.26 | 14.59 | 14.89 | 15.17 | 15.43 | 15.67 |
|  | $W_{1,2}$ | 10.33 | 21.71 | 33.88 | 46.36 | 58.94 | 71.55 | 84.17 | 96.79 | 109.42 | 122.04 | 134.67 | 147.29 | 159.92 | 172.54 | 185.16 | 197.79 | 210.41 | 223.04 | 235.66 | 248.29 | 260.91 | 273.54 | 286.16 | 298.79 | 311.41 |
|  | $W_{3,4}$ | 12.02 | 24.49 | 37.07 | 49.69 | 62.31 | 74.93 | 87.56 | 100.18 | 112.81 | 125.43 | 138.06 | 150.68 | 163.30 | 175.93 | 188.55 | 201.18 | 213.80 | 226.43 | 239.05 | 251.68 | 264.30 | 276.92 | 289.55 | 302.17 | 314.80 |
|  | $J$ | 9.12 | 14.95 | 18.39 | 20.53 | 21.93 | 22.88 | 23.54 | 24.02 | 24.37 | 24.64 | 24.83 | 24.98 | 25.10 | 25.19 | 25.27 | 25.37 | 25.37 | 25.27 | 25.27 | 25.27 | 25.27 | 25.27 | 25.27 | 25.54 | 25.55 |
|  | $K$ | 2.30 | 3.61 | 4.11 | 4.27 | 4.31 | 4.33 | 4.33 | 4.33 | 4.33 | 4.33 | 4.33 | 4.33 | 4.33 | 4.33 | 4.33 | 4.33 | 4.33 | 4.33 | 4.33 | 4.33 | 4.33 | 4.33 | 4.33 | 4.33 | 4.33 |

| $u_0/u_1$ | $\xi$/nm | 2 | 4 | 6 | 8 | 10 | 12 | 14 | 16 | 18 | 20 | 22 | 24 | 26 | 28 | 30 | 32 | 34 | 36 | 38 | 40 | 42 | 44 | 46 | 48 | 50 |
|---|---|---|---|---|---|---|---|---|---|---|---|---|---|---|---|---|---|---|---|---|---|---|---|---|---|---|
| 0.6 | $U_1$ | 18.62 | 32.61 | 42.34 | 49.20 | 54.21 | 57.99 | 60.92 | 63.26 | 65.15 | 66.72 | 68.03 | 69.15 | 70.11 | 70.94 | 71.68 | 72.32 | 72.89 | 73.41 | 73.87 | 74.29 | 74.67 | 75.01 | 75.33 | 75.62 | 75.89 |
| | $U_2$ | 0.35 | 0.82 | 1.51 | 2.46 | 3.55 | 4.70 | 5.83 | 6.90 | 7.89 | 8.79 | 9.62 | 10.36 | 11.04 | 11.66 | 12.21 | 12.72 | 13.18 | 13.60 | 13.99 | 14.35 | 14.68 | 14.98 | 15.26 | 15.52 | 15.76 |
| | $W_{1,2}$ | 10.15 | 21.42 | 33.55 | 46.02 | 58.60 | 71.21 | 83.83 | 96.45 | 109.08 | 121.70 | 134.32 | 146.95 | 159.57 | 172.20 | 184.82 | 197.45 | 210.07 | 222.70 | 235.32 | 247.94 | 260.57 | 273.19 | 285.82 | 298.44 | 311.07 |
| | $W_{3,4}$ | 11.87 | 24.21 | 36.75 | 49.35 | 61.97 | 74.59 | 87.22 | 99.84 | 112.47 | 125.09 | 137.71 | 150.34 | 162.96 | 175.59 | 188.21 | 200.84 | 213.46 | 226.09 | 238.71 | 251.34 | 263.96 | 276.58 | 289.21 | 301.83 | 314.46 |
| | $J$ | 9.40 | 15.63 | 19.46 | 21.90 | 23.52 | 24.64 | 25.42 | 25.99 | 26.41 | 26.72 | 26.96 | 27.14 | 27.28 | 27.39 | 27.48 | 27.55 | 27.61 | 27.66 | 27.70 | 27.73 | 27.75 | 27.78 | 27.80 | 27.81 | 27.83 |
| | $K$ | 2.39 | 3.75 | 4.26 | 4.42 | 4.47 | 4.48 | 4.49 | 4.49 | 4.49 | 4.49 | 4.49 | 4.49 | 4.49 | 4.49 | 4.49 | 4.49 | 4.49 | 4.49 | 4.49 | 4.49 | 4.49 | 4.49 | 4.49 | 4.49 | 4.49 |
| 0.5 | $U_1$ | 17.21 | 30.33 | 39.58 | 46.19 | 51.04 | 54.73 | 57.60 | 59.89 | 61.76 | 63.31 | 64.61 | 65.72 | 66.67 | 67.50 | 68.22 | 68.86 | 69.43 | 69.95 | 70.41 | 70.82 | 71.20 | 71.54 | 71.86 | 72.15 | 72.42 |
| | $U_2$ | 0.39 | 0.89 | 1.63 | 2.59 | 3.70 | 4.85 | 5.97 | 7.04 | 8.02 | 8.93 | 9.75 | 10.49 | 11.16 | 11.77 | 12.33 | 12.83 | 13.28 | 13.70 | 14.10 | 14.46 | 14.78 | 15.09 | 15.37 | 15.63 | 15.87 |
| | $W_{1,2}$ | 9.05 | 21.13 | 33.22 | 45.67 | 58.25 | 70.86 | 83.48 | 96.10 | 108.73 | 121.35 | 133.98 | 146.60 | 159.22 | 171.85 | 184.47 | 197.10 | 209.72 | 222.35 | 234.97 | 247.60 | 260.22 | 272.85 | 285.47 | 298.09 | 310.72 |
| | $W_{3,4}$ | 11.77 | 24.04 | 36.55 | 49.15 | 61.76 | 74.38 | 87.01 | 99.63 | 112.25 | 124.88 | 137.50 | 150.13 | 162.75 | 175.38 | 188.00 | 200.63 | 213.25 | 225.87 | 238.50 | 251.12 | 263.75 | 276.37 | 289.00 | 301.62 | 314.25 |
| | $J$ | 9.70 | 16.37 | 20.61 | 23.39 | 25.28 | 26.59 | 27.54 | 28.23 | 28.74 | 29.12 | 29.42 | 29.65 | 29.83 | 29.97 | 30.08 | 30.17 | 30.25 | 30.31 | 30.36 | 30.40 | 30.44 | 30.47 | 30.49 | 30.52 | 30.53 |
| | $K$ | 2.60 | 3.91 | 4.45 | 4.62 | 4.67 | 4.68 | 4.69 | 4.69 | 4.69 | 4.69 | 4.69 | 4.69 | 4.69 | 4.69 | 4.69 | 4.69 | 4.69 | 4.69 | 4.69 | 4.69 | 4.69 | 4.69 | 4.69 | 4.69 | 4.69 |
| 0.4 | $U_1$ | 16.00 | 28.43 | 37.27 | 43.64 | 48.36 | 51.96 | 54.78 | 57.03 | 58.87 | 60.40 | 61.69 | 62.79 | 63.73 | 64.55 | 65.27 | 65.91 | 66.48 | 66.99 | 67.45 | 67.86 | 68.24 | 68.58 | 68.90 | 69.19 | 69.45 |
| | $U_2$ | 0.43 | 0.97 | 1.73 | 2.72 | 3.83 | 4.98 | 6.11 | 7.17 | 8.15 | 9.05 | 9.86 | 10.60 | 11.27 | 11.88 | 12.43 | 12.94 | 13.40 | 13.81 | 14.20 | 14.55 | 14.88 | 15.18 | 15.46 | 15.72 | 15.96 |
| | $W_{1,2}$ | 9.29 | 20.80 | 32.85 | 45.29 | 57.87 | 70.47 | 83.09 | 95.72 | 108.34 | 120.97 | 133.59 | 146.21 | 158.84 | 171.47 | 184.09 | 196.71 | 209.34 | 221.96 | 234.59 | 247.21 | 259.84 | 272.46 | 285.08 | 297.73 | 310.36 |
| | $W_{3,4}$ | 11.72 | 23.95 | 36.44 | 49.02 | 61.63 | 74.26 | 86.88 | 99.50 | 112.13 | 124.75 | 137.38 | 150.00 | 162.63 | 175.25 | 187.87 | 200.50 | 213.12 | 225.75 | 238.37 | 251.00 | 263.62 | 276.25 | 288.87 | 301.49 | 314.12 |
| | $J$ | 10.00 | 17.10 | 21.67 | 24.76 | 26.89 | 28.40 | 29.48 | 30.29 | 30.89 | 31.35 | 31.70 | 31.97 | 32.19 | 32.36 | 32.50 | 32.61 | 32.71 | 32.78 | 32.85 | 32.90 | 32.94 | 32.98 | 33.00 | 33.04 | 33.07 |
| | $K$ | 2.62 | 4.10 | 4.66 | 4.84 | 4.89 | 4.91 | 4.91 | 4.91 | 4.91 | 4.91 | 4.91 | 4.91 | 4.91 | 4.91 | 4.91 | 4.91 | 4.91 | 4.91 | 4.91 | 4.91 | 4.91 | 4.91 | 4.91 | 4.91 | 4.91 |
| 0.3 | $U_1$ | 15.48 | 27.53 | 36.20 | 42.48 | 47.15 | 50.72 | 53.52 | 55.76 | 57.59 | 59.11 | 60.40 | 61.49 | 62.43 | 63.25 | 63.97 | 64.61 | 65.18 | 65.68 | 66.14 | 66.55 | 66.93 | 67.27 | 67.59 | 67.88 | 68.15 |
| | $U_2$ | 0.43 | 0.99 | 1.77 | 2.77 | 3.87 | 5.02 | 6.15 | 7.21 | 8.19 | 9.10 | 9.92 | 10.66 | 11.32 | 11.92 | 12.47 | 12.97 | 13.43 | 13.85 | 14.23 | 14.60 | 14.93 | 15.21 | 15.49 | 15.75 | 15.99 |
| | $W_{1,2}$ | 9.52 | 20.47 | 32.48 | 44.91 | 57.48 | 70.08 | 82.70 | 95.33 | 107.95 | 120.58 | 133.20 | 145.82 | 158.45 | 171.08 | 183.70 | 196.32 | 208.95 | 221.57 | 234.20 | 246.82 | 259.45 | 272.07 | 284.69 | 297.32 | 309.94 |
| | $W_{3,4}$ | 11.68 | 23.82 | 36.28 | 48.86 | 61.47 | 74.09 | 86.71 | 99.34 | 111.96 | 124.59 | 137.21 | 149.84 | 162.46 | 175.08 | 187.71 | 200.33 | 212.96 | 225.58 | 238.21 | 250.83 | 263.46 | 276.08 | 288.71 | 301.33 | 313.96 |
| | $J$ | 10.23 | 17.55 | 22.40 | 25.66 | 27.92 | 29.51 | 30.68 | 31.56 | 32.22 | 32.72 | 33.12 | 33.43 | 33.68 | 33.86 | 34.02 | 34.14 | 34.24 | 34.33 | 34.41 | 34.46 | 34.50 | 34.54 | 34.57 | 34.64 | 34.66 |
| | $K$ | 2.74 | 4.30 | 4.88 | 5.06 | 5.12 | 5.13 | 5.14 | 5.14 | 5.14 | 5.14 | 5.14 | 5.14 | 5.14 | 5.14 | 5.14 | 5.14 | 5.14 | 5.14 | 5.14 | 5.14 | 5.14 | 5.14 | 5.14 | 5.14 | 5.14 |
| 0.2 | $U_1$ | 15.29 | 27.28 | 35.05 | 42.24 | 46.92 | 50.51 | 53.31 | 55.57 | 57.40 | 58.93 | 60.22 | 61.32 | 62.26 | 63.08 | 63.80 | 64.44 | 65.01 | 65.52 | 65.98 | 66.39 | 66.77 | 67.11 | 67.43 | 67.72 | 67.98 |
| | $U_2$ | 0.41 | 0.95 | 1.72 | 2.71 | 3.83 | 4.99 | 6.11 | 7.18 | 8.16 | 9.06 | 9.87 | 10.61 | 11.28 | 11.89 | 12.45 | 12.95 | 13.41 | 13.83 | 14.21 | 14.57 | 14.89 | 15.17 | 15.47 | 15.73 | 15.97 |
| | $W_{1,2}$ | 9.12 | 19.85 | 31.78 | 44.19 | 56.75 | 69.35 | 81.97 | 94.59 | 107.22 | 119.84 | 132.47 | 145.09 | 157.72 | 170.34 | 182.96 | 195.59 | 208.21 | 220.84 | 233.46 | 246.09 | 258.71 | 271.34 | 283.96 | 296.59 | 309.21 |
| | $W_{3,4}$ | 11.62 | 23.76 | 36.19 | 48.76 | 61.37 | 73.99 | 86.62 | 99.24 | 111.87 | 124.49 | 137.11 | 149.74 | 162.36 | 174.99 | 187.61 | 200.23 | 212.86 | 225.48 | 238.11 | 250.73 | 263.36 | 275.98 | 288.61 | 301.23 | 313.86 |
| | $J$ | 10.50 | 18.08 | 23.12 | 26.49 | 28.81 | 30.44 | 31.63 | 32.45 | 33.08 | 33.55 | 33.89 | 34.15 | 34.34 | 34.56 | 34.71 | 34.82 | 34.93 | 35.01 | 35.07 | 35.13 | 35.18 | 35.22 | 35.25 | 35.28 | 35.31 |
| | $K$ | 2.85 | 4.47 | 5.07 | 5.26 | 5.32 | 5.33 | 5.34 | 5.34 | 5.34 | 5.34 | 5.34 | 5.34 | 5.34 | 5.34 | 5.34 | 5.34 | 5.34 | 5.34 | 5.34 | 5.34 | 5.34 | 5.34 | 5.34 | 5.34 | 5.34 |
| 0.1 | $U_1$ | 15.39 | 27.43 | 36.19 | 42.53 | 47.29 | 50.92 | 53.77 | 56.05 | 57.91 | 59.46 | 60.77 | 61.89 | 62.85 | 63.68 | 64.41 | 65.04 | 65.61 | 66.12 | 66.58 | 66.99 | 67.37 | 67.71 | 68.03 | 68.18 | 68.32 |
| | $U_2$ | 0.39 | 0.92 | 1.68 | 2.66 | 3.78 | 4.94 | 6.06 | 7.13 | 8.11 | 9.01 | 9.83 | 10.57 | 11.24 | 11.85 | 12.41 | 12.91 | 13.37 | 13.79 | 14.18 | 14.53 | 14.86 | 15.16 | 15.44 | 15.70 | 15.94 |
| | $W_{1,2}$ | 9.03 | 19.75 | 31.68 | 44.09 | 56.64 | 69.25 | 81.86 | 94.49 | 107.11 | 119.74 | 132.36 | 144.99 | 157.61 | 170.24 | 182.86 | 195.48 | 208.11 | 220.73 | 233.36 | 245.98 | 258.61 | 271.23 | 283.85 | 296.48 | 309.21 |
| | $W_{3,4}$ | 11.60 | 23.73 | 36.19 | 48.78 | 61.39 | 74.01 | 86.63 | 99.26 | 111.89 | 124.51 | 137.14 | 149.76 | 162.38 | 175.01 | 187.63 | 200.25 | 212.88 | 225.50 | 238.13 | 250.75 | 263.38 | 276.00 | 288.63 | 301.25 | 313.88 |
| | $J$ | 10.52 | 18.14 | 23.13 | 26.46 | 28.75 | 30.34 | 31.47 | 32.29 | 32.93 | 33.34 | 33.69 | 33.98 | 34.14 | 34.41 | 34.55 | 34.65 | 34.73 | 34.82 | 34.88 | 34.93 | 34.98 | 35.02 | 35.06 | 35.14 | 35.25 |
| | $K$ | 2.96 | 4.63 | 5.26 | 5.46 | 5.49 | 5.49 | 5.49 | 5.49 | 5.49 | 5.49 | 5.49 | 5.49 | 5.49 | 5.49 | 5.49 | 5.49 | 5.49 | 5.49 | 5.49 | 5.49 | 5.49 | 5.49 | 5.49 | 5.49 | 5.49 |
| 0.0 | $U_1$ | 15.48 | 27.60 | 36.36 | 42.71 | 47.44 | 51.04 | 53.87 | 56.13 | 57.98 | 59.51 | 60.80 | 61.91 | 62.85 | 63.68 | 64.40 | 65.04 | 65.61 | 66.12 | 66.58 | 66.99 | 67.37 | 67.71 | 68.03 | 68.32 | 68.59 |
| | $U_2$ | 0.39 | 0.92 | 1.68 | 2.66 | 3.78 | 4.94 | 6.06 | 7.13 | 8.11 | 9.01 | 9.83 | 10.57 | 11.24 | 11.85 | 12.41 | 12.91 | 13.37 | 13.79 | 14.18 | 14.53 | 14.86 | 15.16 | 15.44 | 15.70 | 15.94 |
| | $W_{1,2}$ | 9.03 | 19.73 | 31.66 | 44.06 | 56.62 | 69.23 | 81.84 | 94.46 | 107.09 | 119.71 | 132.33 | 144.96 | 157.58 | 170.21 | 182.83 | 195.45 | 208.08 | 220.70 | 233.33 | 245.95 | 258.58 | 271.20 | 283.82 | 296.45 | 309.04 |
| | $W_{3,4}$ | 11.59 | 23.72 | 36.18 | 48.77 | 61.38 | 74.00 | 86.62 | 99.25 | 111.87 | 124.50 | 137.12 | 149.75 | 162.37 | 175.00 | 187.62 | 200.24 | 212.87 | 225.49 | 238.12 | 250.74 | 263.37 | 275.99 | 288.62 | 301.24 | 313.89 |
| | $J$ | 10.52 | 18.14 | 23.12 | 26.46 | 28.75 | 30.30 | 31.47 | 32.29 | 32.93 | 33.34 | 33.69 | 33.98 | 34.15 | 34.34 | 34.49 | 34.59 | 34.67 | 34.74 | 34.79 | 34.83 | 34.87 | 34.90 | 34.92 | 34.94 | 34.94 |
| | $K$ | 2.96 | 4.62 | 5.26 | 5.46 | 5.52 | 5.53 | 5.54 | 5.54 | 5.54 | 5.54 | 5.54 | 5.54 | 5.54 | 5.54 | 5.54 | 5.54 | 5.54 | 5.54 | 5.54 | 5.54 | 5.54 | 5.54 | 5.54 | 5.54 | 5.54 |

TABLE S55. Parameters of the THF interaction Hamiltonian for different ratios $u_0/u_1$ and screening lengths $\xi$ of the double-gate interaction potential ($U_\xi = 24\,$meV for $\xi = 10\,$nm). We use the same BM model parameters as in Table S54.

Total weight of the heavy-fermions
on the active bands after wannierisation: 93.8%

Wannier orbitals support
0.0 ——— 1.0

Energy (meV): 100, 50, 0, −50, −100

K  Γ  M  K

FIG. S44. Band structures of the BM and THF models near charge neutrality for $w_0/w_1 = 0.8$, depicted by lines and crosses, respectively. The BM bands are colored according to the weight of the $f$-electron wave function on them. We use the same BM parameters as in Table S56.

TABLE S56. Parameters of the $f$-electron Wannier functions and the THF single-particle Hamiltonian for different values of the tunneling ratio $w_0/w_1$. $W$ denotes the total weight of the THF $f$-electrons on the active bands. In computing the ratios $\gamma/U_1$ and $\gamma/U_2$, we employ the on-site and nearest-neighbor repulsion parameters $U_1$ and $U_2$ obtained numerically for $\xi = 10$ nm, as given in Table S57. We employ $v_F = 5.944$ eV·Å, $|\mathbf{K}| = 1.703$ Å$^{-1}$, $w_1 = 110$ meV, and $\theta = 1.22°$ for the BM model.

| $w_0/w_1$ | $\alpha_1$ | $\alpha_2$ | $\lambda_1$ | $\lambda_2$ | $\lambda$ | $W$ | $t_0$ | $\gamma$ | $M$ | $v_\star$ | $v'_\star$ | $v''_\star$ | $\gamma/U_1$ | $\gamma/U_2$ |
|---|---|---|---|---|---|---|---|---|---|---|---|---|---|---|
| 1.00 | 0.786 | 0.618 | 0.179 | 0.191 | 0.342 | 0.966 | 1.050 | −19.338 | −15.826 | −4.415 | 1.726 | 0.0065 | −0.28 | −5.90 |
| 0.90 | 0.815 | 0.579 | 0.197 | 0.348 | 0.362 | 0.950 | 1.300 | −35.000 | −17.111 | −4.574 | 1.756 | −0.0040 | −0.54 | −10.42 |
| 0.80 | 0.846 | 0.533 | 0.209 | 0.362 | 0.374 | 0.938 | 1.590 | −50.403 | −18.231 | −4.716 | 1.756 | −0.0099 | −0.82 | −14.46 |
| 0.70 | 0.876 | 0.482 | 0.221 | 0.215 | 0.374 | 0.928 | 1.886 | −65.281 | −19.194 | −4.843 | 1.722 | −0.0125 | −1.13 | −18.08 |
| 0.60 | 0.905 | 0.425 | 0.233 | 0.215 | 0.393 | 0.925 | 2.187 | −79.360 | −20.010 | −4.960 | 1.649 | −0.0126 | −1.47 | −21.07 |
| 0.50 | 0.932 | 0.363 | 0.251 | 0.221 | 0.410 | 0.926 | 2.509 | −92.331 | −20.685 | −5.069 | 1.528 | −0.0110 | −1.81 | −23.56 |
| 0.40 | 0.955 | 0.297 | 0.263 | 0.227 | 0.427 | 0.929 | 2.828 | −103.839 | −21.228 | −5.168 | 1.349 | −0.0084 | −2.15 | −25.56 |
| 0.30 | 0.974 | 0.228 | 0.275 | 0.233 | 0.422 | 0.920 | 2.861 | −113.477 | −21.644 | −5.255 | 1.103 | −0.0054 | −2.37 | −28.09 |
| 0.20 | 0.988 | 0.156 | 0.286 | 0.233 | 0.429 | 0.921 | 3.001 | −120.807 | −21.938 | −5.325 | 0.788 | −0.0026 | −2.59 | −29.47 |
| 0.10 | 0.997 | 0.080 | 0.292 | 0.233 | 0.418 | 0.911 | 2.972 | −125.414 | −22.113 | −5.371 | 0.412 | −0.0007 | −2.64 | −31.14 |
| 0.00 | 1.000 | 0.000 | 0.292 | 0.000 | 0.409 | 0.903 | 2.803 | −126.988 | −22.171 | −5.387 | 0.000 | 0.0000 | −2.63 | −32.16 |

| $w_0/w_1$ | $\xi$/nm | 2 | 4 | 6 | 8 | 10 | 12 | 14 | 16 | 18 | 20 | 22 | 24 | 26 | 28 | 30 | 32 | 34 | 36 | 38 | 40 | 42 | 44 | 46 | 48 | 50 |
|---|---|---|---|---|---|---|---|---|---|---|---|---|---|---|---|---|---|---|---|---|---|---|---|---|---|---|
| 1.0 | $U_1$ | 26.05 | 44.12 | 55.82 | 63.68 | 69.23 | 73.32 | 76.45 | 78.91 | 80.89 | 82.51 | 83.87 | 85.02 | 86.00 | 86.85 | 87.59 | 88.25 | 88.83 | 89.35 | 89.82 | 90.24 | 90.62 | 90.97 | 91.29 | 91.59 | 91.86 |
| | $U_2$ | 0.28 | 0.65 | 1.27 | 2.18 | 3.28 | 4.45 | 5.61 | 6.71 | 7.73 | 8.67 | 9.51 | 10.28 | 10.97 | 11.60 | 12.17 | 12.68 | 13.15 | 13.58 | 13.98 | 14.34 | 14.67 | 14.98 | 15.26 | 15.52 | 15.77 |
| | $W_{1,2}$ | 10.99 | 22.94 | 35.58 | 48.51 | 61.52 | 74.56 | 87.61 | 100.66 | 113.70 | 126.75 | 139.80 | 152.85 | 165.90 | 178.95 | 192.00 | 205.05 | 218.09 | 231.14 | 244.19 | 257.24 | 270.29 | 283.34 | 296.39 | 309.44 | 322.48 |
| | $W_{3,4}$ | 13.29 | 26.75 | 39.99 | 53.10 | 66.17 | 79.22 | 92.27 | 105.32 | 118.37 | 131.42 | 144.47 | 157.52 | 170.57 | 183.62 | 196.66 | 209.71 | 222.76 | 235.81 | 248.86 | 261.91 | 274.96 | 288.01 | 301.05 | 314.10 | 327.15 |
| | $J$ | 8.66 | 13.26 | 15.47 | 16.62 | 17.29 | 17.70 | 17.97 | 18.16 | 18.29 | 18.39 | 18.46 | 18.51 | 18.56 | 18.59 | 18.61 | 18.64 | 18.65 | 18.66 | 18.68 | 18.68 | 18.69 | 18.70 | 18.70 | 18.71 | 18.71 |
| | $K$ | 2.19 | 3.43 | 3.88 | 4.02 | 4.06 | 4.07 | 4.07 | 4.08 | 4.08 | 4.08 | 4.08 | 4.08 | 4.08 | 4.08 | 4.08 | 4.08 | 4.08 | 4.08 | 4.08 | 4.08 | 4.08 | 4.08 | 4.08 | 4.08 | 4.08 |
| 0.9 | $U_1$ | 24.00 | 41.02 | 52.25 | 59.89 | 65.33 | 69.37 | 72.46 | 74.89 | 76.86 | 78.47 | 79.82 | 80.96 | 81.94 | 82.79 | 83.54 | 84.19 | 84.77 | 85.29 | 85.76 | 86.18 | 86.56 | 86.91 | 87.23 | 87.52 | 87.79 |
| | $U_2$ | 0.29 | 0.68 | 1.32 | 2.25 | 3.36 | 4.54 | 5.70 | 6.80 | 7.82 | 8.75 | 9.60 | 10.36 | 11.06 | 11.68 | 12.25 | 12.77 | 13.23 | 13.66 | 14.06 | 14.42 | 14.75 | 15.05 | 15.34 | 15.60 | 15.84 |
| | $W_{1,2}$ | 10.99 | 22.93 | 35.57 | 48.50 | 61.51 | 74.55 | 87.60 | 100.64 | 113.69 | 126.74 | 139.79 | 152.84 | 165.89 | 178.94 | 191.98 | 205.03 | 218.08 | 231.13 | 244.18 | 257.23 | 270.28 | 283.33 | 296.37 | 309.42 | 322.47 |
| | $W_{3,4}$ | 12.87 | 26.05 | 39.18 | 52.26 | 65.31 | 78.36 | 91.41 | 104.46 | 117.51 | 130.56 | 143.61 | 156.66 | 169.71 | 182.76 | 195.80 | 208.85 | 221.90 | 234.95 | 248.00 | 261.05 | 274.10 | 287.15 | 300.19 | 313.24 | 326.29 |
| | $J$ | 8.85 | 13.90 | 16.56 | 18.06 | 18.98 | 19.58 | 19.98 | 20.27 | 20.47 | 20.62 | 20.73 | 20.82 | 20.89 | 20.94 | 20.98 | 21.01 | 21.04 | 21.07 | 21.09 | 21.10 | 21.11 | 21.12 | 21.12 | 21.12 | 21.13 |
| | $K$ | 2.19 | 3.43 | 3.88 | 4.02 | 4.06 | 4.07 | 4.08 | 4.08 | 4.08 | 4.08 | 4.08 | 4.08 | 4.08 | 4.08 | 4.08 | 4.08 | 4.08 | 4.08 | 4.08 | 4.08 | 4.08 | 4.08 | 4.08 | 4.08 | 4.08 |
| 0.8 | $U_1$ | 21.99 | 37.92 | 48.62 | 56.00 | 61.30 | 65.24 | 68.28 | 70.69 | 72.63 | 74.23 | 75.57 | 76.70 | 77.68 | 78.52 | 79.26 | 79.91 | 80.49 | 81.01 | 81.47 | 81.89 | 82.27 | 82.62 | 82.94 | 83.23 | 83.50 |
| | $U_2$ | 0.31 | 0.74 | 1.42 | 2.36 | 3.49 | 4.67 | 5.83 | 6.93 | 7.95 | 8.88 | 9.72 | 10.48 | 11.17 | 11.80 | 12.36 | 12.88 | 13.34 | 13.77 | 14.16 | 14.52 | 14.85 | 15.16 | 15.44 | 15.70 | 15.95 |
| | $W_{1,2}$ | 10.90 | 22.79 | 35.42 | 48.34 | 61.35 | 74.39 | 87.44 | 100.49 | 113.53 | 126.58 | 139.63 | 152.68 | 165.73 | 178.78 | 191.83 | 204.88 | 217.92 | 230.97 | 244.02 | 257.07 | 270.12 | 283.17 | 296.22 | 309.27 | 322.32 |
| | $W_{3,4}$ | 12.35 | 25.17 | 38.16 | 51.20 | 64.24 | 77.29 | 90.34 | 103.39 | 116.43 | 129.48 | 142.53 | 155.58 | 168.63 | 181.68 | 194.73 | 207.78 | 220.82 | 233.87 | 246.92 | 259.97 | 273.02 | 286.07 | 299.12 | 312.17 | 325.22 |
| | $J$ | 9.09 | 14.62 | 17.72 | 19.58 | 20.77 | 21.56 | 22.10 | 22.49 | 22.78 | 22.99 | 23.14 | 23.26 | 23.36 | 23.43 | 23.48 | 23.53 | 23.57 | 23.60 | 23.62 | 23.64 | 23.66 | 23.67 | 23.68 | 23.69 | 23.70 |
| | $K$ | 2.23 | 3.94 | 4.09 | 4.13 | 4.14 | 4.14 | 4.14 | 4.14 | 4.14 | 4.14 | 4.14 | 4.14 | 4.14 | 4.14 | 4.14 | 4.14 | 4.14 | 4.14 | 4.14 | 4.14 | 4.14 | 4.14 | 4.14 | 4.14 | 4.14 |
| 0.7 | $U_1$ | 20.29 | 35.25 | 45.47 | 52.60 | 57.76 | 61.63 | 64.62 | 67.00 | 68.92 | 70.50 | 71.83 | 72.96 | 73.92 | 74.76 | 75.50 | 76.15 | 76.72 | 77.24 | 77.70 | 78.12 | 78.50 | 78.85 | 79.17 | 79.46 | 79.73 |
| | $U_2$ | 0.34 | 0.80 | 1.51 | 2.48 | 3.61 | 4.80 | 5.96 | 7.06 | 8.07 | 9.00 | 9.84 | 10.60 | 11.28 | 11.91 | 12.47 | 12.98 | 13.45 | 13.88 | 14.27 | 14.62 | 14.96 | 15.26 | 15.54 | 15.80 | 16.05 |
| | $W_{1,2}$ | 10.72 | 22.56 | 35.16 | 48.07 | 61.08 | 74.12 | 87.16 | 100.21 | 113.26 | 126.31 | 139.36 | 152.41 | 165.46 | 178.50 | 191.55 | 204.60 | 217.65 | 230.70 | 243.75 | 256.80 | 269.85 | 282.90 | 295.94 | 308.99 | 322.04 |
| | $W_{3,4}$ | 12.35 | 25.17 | 38.16 | 51.20 | 64.24 | 77.29 | 90.34 | 103.39 | 116.43 | 129.48 | 142.53 | 155.58 | 168.63 | 181.68 | 194.73 | 207.78 | 220.82 | 233.87 | 246.92 | 259.97 | 273.02 | 286.07 | 299.12 | 312.17 | 325.22 |
| | $J$ | 9.29 | 15.34 | 18.67 | 20.77 | 22.23 | 23.26 | 23.94 | 24.37 | 24.87 | 25.17 | 25.58 | 25.70 | 25.79 | 25.86 | 25.97 | 26.04 | 26.10 | 26.13 | 26.14 | 26.26 | 26.06 | 26.10 | 26.13 | 26.14 | 26.14 |
| | $K$ | 2.29 | 3.58 | 4.06 | 4.20 | 4.25 | 4.26 | 4.26 | 4.26 | 4.26 | 4.26 | 4.26 | 4.26 | 4.26 | 4.26 | 4.26 | 4.26 | 4.26 | 4.26 | 4.26 | 4.26 | 4.26 | 4.26 | 4.26 | 4.26 | 4.26 |

TABLE S57. Parameters of the THF interaction Hamiltonian for different ratios $u_0/u_1$ and screening lengths $\xi$ of the double-gate interaction potential ($U_\xi = 24\,\text{meV}$ for $\xi = 10\,\text{nm}$). We use the same BM model parameters as in Table S56.

| $u_0/u_1$ | | $\xi/\text{nm}$ → | 2 | 4 | 6 | 8 | 10 | 12 | 14 | 16 | 18 | 20 | 22 | 24 | 26 | 28 | 30 | 32 | 34 | 36 | 38 | 40 | 42 | 44 | 46 | 48 | 50 |
|---|---|---|---|---|---|---|---|---|---|---|---|---|---|---|---|---|---|---|---|---|---|---|---|---|---|---|---|
| 0.6 | $U_1$ | | 18.63 | 32.61 | 42.32 | 49.17 | 54.17 | 57.94 | 60.87 | 63.20 | 65.09 | 66.65 | 67.97 | 69.08 | 70.04 | 70.88 | 71.61 | 72.25 | 72.83 | 73.34 | 73.80 | 74.22 | 74.60 | 74.94 | 75.26 | 75.55 | 75.82 |
| | $U_2$ | | 0.38 | 0.88 | 1.63 | 2.62 | 3.77 | 4.95 | 6.11 | 7.21 | 8.22 | 9.14 | 9.97 | 10.73 | 11.41 | 12.03 | 12.59 | 13.11 | 13.57 | 14.00 | 14.38 | 14.74 | 15.07 | 15.38 | 15.66 | 15.92 | 16.16 |
| | $W_{1,2}$ | | 10.57 | 22.29 | 34.86 | 47.76 | 60.77 | 73.80 | 86.85 | 99.80 | 112.94 | 125.99 | 139.04 | 152.09 | 165.14 | 178.19 | 191.24 | 204.29 | 217.33 | 230.38 | 243.43 | 256.48 | 269.53 | 282.58 | 295.63 | 308.68 | 321.72 |
| | $W_{3,4}$ | | 12.21 | 24.93 | 37.88 | 50.91 | 63.95 | 76.99 | 90.04 | 103.09 | 116.14 | 129.19 | 142.24 | 155.28 | 168.33 | 181.38 | 194.43 | 207.48 | 220.53 | 233.58 | 246.63 | 259.67 | 272.72 | 285.77 | 298.82 | 311.87 | 324.92 |
| | $J$ | | 9.68 | 16.12 | 20.10 | 22.65 | 24.35 | 25.53 | 26.37 | 26.97 | 27.42 | 27.76 | 28.02 | 28.21 | 28.37 | 28.49 | 28.59 | 28.67 | 28.73 | 28.78 | 28.83 | 28.86 | 28.89 | 28.92 | 28.94 | 28.96 | 28.97 |
| | $K$ | | 2.39 | 3.73 | 4.22 | 4.37 | 4.41 | 4.43 | 4.43 | 4.43 | 4.43 | 4.43 | 4.43 | 4.43 | 4.43 | 4.43 | 4.43 | 4.43 | 4.43 | 4.43 | 4.43 | 4.43 | 4.43 | 4.43 | 4.43 | 4.43 | 4.43 |
| 0.5 | $U_1$ | | 17.22 | 30.33 | 39.57 | 46.16 | 51.01 | 54.69 | 57.56 | 59.85 | 61.71 | 63.26 | 64.56 | 65.66 | 66.62 | 67.44 | 68.17 | 68.81 | 69.38 | 69.89 | 70.35 | 70.77 | 71.14 | 71.49 | 71.80 | 72.10 | 72.36 |
| | $U_2$ | | 0.42 | 0.97 | 1.75 | 2.77 | 3.92 | 5.11 | 6.26 | 7.35 | 8.36 | 9.27 | 10.10 | 10.86 | 11.54 | 12.15 | 12.71 | 13.22 | 13.69 | 14.11 | 14.49 | 14.85 | 15.18 | 15.48 | 15.77 | 16.03 | 16.27 |
| | $W_{1,2}$ | | 10.37 | 21.99 | 34.52 | 47.41 | 60.42 | 73.45 | 86.50 | 99.55 | 112.59 | 125.64 | 138.69 | 151.74 | 164.79 | 177.84 | 190.89 | 203.94 | 216.98 | 230.03 | 243.08 | 256.13 | 269.18 | 282.23 | 295.28 | 308.33 | 321.37 |
| | $W_{3,4}$ | | 12.13 | 24.79 | 37.71 | 50.72 | 63.76 | 76.81 | 89.85 | 102.90 | 115.95 | 129.00 | 142.05 | 155.10 | 168.15 | 181.20 | 194.25 | 207.29 | 220.34 | 233.39 | 246.44 | 259.49 | 272.54 | 285.59 | 298.63 | 311.68 | 324.73 |
| | $J$ | | 10.00 | 16.89 | 21.29 | 24.18 | 26.15 | 27.54 | 28.53 | 29.26 | 29.80 | 30.22 | 30.53 | 30.78 | 30.97 | 31.12 | 31.25 | 31.35 | 31.43 | 31.49 | 31.55 | 31.60 | 31.64 | 31.67 | 31.70 | 31.72 | 31.74 |
| | $K$ | | 2.50 | 3.90 | 4.41 | 4.57 | 4.62 | 4.63 | 4.64 | 4.64 | 4.64 | 4.64 | 4.64 | 4.64 | 4.64 | 4.64 | 4.64 | 4.64 | 4.64 | 4.64 | 4.64 | 4.64 | 4.64 | 4.64 | 4.64 | 4.64 | 4.64 |
| 0.4 | $U_1$ | | 16.03 | 28.39 | 37.21 | 43.56 | 48.27 | 51.86 | 54.67 | 56.92 | 58.76 | 60.28 | 61.57 | 62.66 | 63.61 | 64.43 | 65.15 | 65.79 | 66.35 | 66.86 | 67.32 | 67.73 | 68.11 | 68.46 | 68.77 | 69.06 | 69.33 |
| | $U_2$ | | 0.46 | 1.05 | 1.87 | 2.90 | 4.06 | 5.25 | 6.40 | 7.49 | 8.49 | 9.40 | 10.23 | 10.98 | 11.65 | 12.27 | 12.82 | 13.33 | 13.79 | 14.21 | 14.60 | 14.95 | 15.28 | 15.58 | 15.86 | 16.12 | 16.37 |
| | $W_{1,2}$ | | 10.17 | 21.69 | 34.19 | 47.07 | 60.07 | 73.10 | 86.15 | 99.20 | 112.24 | 125.29 | 138.34 | 151.39 | 164.44 | 177.49 | 190.54 | 203.59 | 216.63 | 229.68 | 242.73 | 255.78 | 268.83 | 281.88 | 294.93 | 307.98 | 321.02 |
| | $W_{3,4}$ | | 12.09 | 24.71 | 37.61 | 50.62 | 63.66 | 76.70 | 89.75 | 102.80 | 115.85 | 128.90 | 141.94 | 154.99 | 168.04 | 181.09 | 194.14 | 207.19 | 220.24 | 233.29 | 246.33 | 259.38 | 272.43 | 285.48 | 298.53 | 311.58 | 324.63 |
| | $J$ | | 10.32 | 17.61 | 22.41 | 25.62 | 27.85 | 29.43 | 30.58 | 31.43 | 32.07 | 32.55 | 32.92 | 33.22 | 33.46 | 33.64 | 33.79 | 33.91 | 34.01 | 34.10 | 34.17 | 34.22 | 34.27 | 34.31 | 34.34 | 34.38 | 34.41 |
| | $K$ | | 2.63 | 4.09 | 4.63 | 4.80 | 4.84 | 4.86 | 4.86 | 4.86 | 4.86 | 4.86 | 4.86 | 4.86 | 4.86 | 4.86 | 4.86 | 4.86 | 4.86 | 4.86 | 4.86 | 4.86 | 4.86 | 4.86 | 4.86 | 4.86 | 4.86 |
| 0.3 | $U_1$ | | 15.80 | 27.20 | 36.86 | 42.06 | 46.71 | 50.27 | 53.09 | 55.30 | 57.13 | 58.65 | 59.93 | 61.03 | 61.97 | 62.79 | 63.51 | 64.15 | 64.72 | 65.23 | 65.68 | 66.09 | 66.47 | 66.81 | 67.13 | 67.42 | 67.68 |
| | $U_2$ | | 0.46 | 1.06 | 1.89 | 2.93 | 4.10 | 5.29 | 6.45 | 7.53 | 8.53 | 9.44 | 10.27 | 11.01 | 11.69 | 12.30 | 12.86 | 13.37 | 13.83 | 14.25 | 14.64 | 14.99 | 15.32 | 15.62 | 15.90 | 16.16 | 16.40 |
| | $W_{1,2}$ | | 9.90 | 21.27 | 33.72 | 46.58 | 59.58 | 72.61 | 85.66 | 98.71 | 111.75 | 124.80 | 137.85 | 150.90 | 163.95 | 177.00 | 190.04 | 203.10 | 216.14 | 229.19 | 242.24 | 255.29 | 268.34 | 281.39 | 294.44 | 307.49 | 320.53 |
| | $W_{3,4}$ | | 12.03 | 24.61 | 37.51 | 50.51 | 63.54 | 76.59 | 89.64 | 102.68 | 115.73 | 128.78 | 141.79 | 154.84 | 167.89 | 180.93 | 194.00 | 207.04 | 220.08 | 233.13 | 246.18 | 259.23 | 272.28 | 285.33 | 298.38 | 311.43 | 324.51 |
| | $J$ | | 10.51 | 18.05 | 22.95 | 26.26 | 28.54 | 30.14 | 31.30 | 32.15 | 32.78 | 33.26 | 33.63 | 33.92 | 34.14 | 34.32 | 34.47 | 34.58 | 34.68 | 34.76 | 34.82 | 34.88 | 34.93 | 34.96 | 35.00 | 35.03 | 35.05 |
| | $K$ | | 2.75 | 4.29 | 4.85 | 5.03 | 5.08 | 5.09 | 5.10 | 5.10 | 5.10 | 5.10 | 5.10 | 5.10 | 5.10 | 5.10 | 5.10 | 5.10 | 5.10 | 5.10 | 5.10 | 5.10 | 5.10 | 5.10 | 5.10 | 5.10 | 5.10 |
| 0.2 | $U_1$ | | 15.51 | 26.37 | 35.84 | 42.71 | 47.42 | 51.02 | 53.83 | 56.09 | 57.94 | 59.47 | 60.76 | 61.86 | 62.80 | 63.63 | 64.35 | 64.99 | 65.56 | 66.07 | 66.52 | 66.94 | 67.32 | 67.67 | 67.98 | 68.27 | 68.53 |
| | $U_2$ | | 0.46 | 1.06 | 1.82 | 2.86 | 4.03 | 5.22 | 6.38 | 7.47 | 8.47 | 9.38 | 10.21 | 10.96 | 11.64 | 12.26 | 12.81 | 13.32 | 13.78 | 14.21 | 14.59 | 14.94 | 15.28 | 15.58 | 15.86 | 16.12 | 16.36 |
| | $W_{1,2}$ | | 9.72 | 21.00 | 33.41 | 46.26 | 59.26 | 72.29 | 85.33 | 98.38 | 111.43 | 124.48 | 137.53 | 150.57 | 163.62 | 176.67 | 189.72 | 202.77 | 215.82 | 228.87 | 241.92 | 254.96 | 268.01 | 281.06 | 294.11 | 307.16 | 320.21 |
| | $W_{3,4}$ | | 10.72 | 24.58 | 37.47 | 50.40 | 63.51 | 76.55 | 89.60 | 102.53 | 115.58 | 128.62 | 141.79 | 154.84 | 167.88 | 180.93 | 193.97 | 207.04 | 220.07 | 233.13 | 246.10 | 259.23 | 272.21 | 285.33 | 298.31 | 311.43 | 324.40 |
| | $J$ | | 10.72 | 18.45 | 23.61 | 27.09 | 29.50 | 31.21 | 32.44 | 33.35 | 34.02 | 34.54 | 34.93 | 35.24 | 35.48 | 35.68 | 35.83 | 35.96 | 36.06 | 36.15 | 36.22 | 36.28 | 36.33 | 36.37 | 36.41 | 36.44 | 36.46 |
| | $K$ | | 2.85 | 4.46 | 5.05 | 5.23 | 5.28 | 5.30 | 5.30 | 5.30 | 5.30 | 5.30 | 5.30 | 5.30 | 5.30 | 5.30 | 5.30 | 5.30 | 5.30 | 5.30 | 5.30 | 5.30 | 5.30 | 5.30 | 5.30 | 5.30 | 5.30 |
| 0.1 | $U_1$ | | 15.51 | 27.63 | 36.37 | 42.71 | 47.42 | 51.02 | 53.81 | 56.09 | 57.94 | 59.47 | 60.76 | 61.86 | 62.80 | 63.63 | 64.35 | 64.99 | 65.56 | 66.07 | 66.52 | 66.94 | 67.32 | 67.63 | 67.98 | 68.27 | 68.53 |
| | $U_2$ | | 0.43 | 1.01 | 1.82 | 2.79 | 4.03 | 5.22 | 6.38 | 7.47 | 8.47 | 9.38 | 10.21 | 10.96 | 11.64 | 12.26 | 12.81 | 13.32 | 13.78 | 14.21 | 14.59 | 14.94 | 15.28 | 15.58 | 15.86 | 16.12 | 16.36 |
| | $W_{1,2}$ | | 11.98 | 24.52 | 37.40 | 50.40 | 63.43 | 76.48 | 89.53 | 102.57 | 115.62 | 128.67 | 141.72 | 154.77 | 167.82 | 180.87 | 193.92 | 206.96 | 220.01 | 233.06 | 246.11 | 259.16 | 272.21 | 285.26 | 298.31 | 311.35 | 324.40 |
| | $W_{3,4}$ | | 11.95 | 24.48 | 37.36 | 50.35 | 63.43 | 76.48 | 89.48 | 102.53 | 115.58 | 128.62 | 141.67 | 154.72 | 167.77 | 180.82 | 193.87 | 206.92 | 219.97 | 233.02 | 246.06 | 259.11 | 272.16 | 285.21 | 298.26 | 311.31 | 324.36 |
| | $J$ | | 10.80 | 18.58 | 23.75 | 27.20 | 29.56 | 31.22 | 32.40 | 33.26 | 33.90 | 34.37 | 34.73 | 35.01 | 35.24 | 35.41 | 35.54 | 35.65 | 35.74 | 35.81 | 35.87 | 35.92 | 35.97 | 36.00 | 36.03 | 36.06 | 36.08 |
| | $K$ | | 2.95 | 4.59 | 5.19 | 5.37 | 5.42 | 5.44 | 5.45 | 5.45 | 5.45 | 5.45 | 5.45 | 5.45 | 5.45 | 5.45 | 5.45 | 5.45 | 5.45 | 5.45 | 5.45 | 5.45 | 5.45 | 5.45 | 5.45 | 5.45 | 5.45 |
| 0.0 | $U_1$ | | 15.82 | 28.17 | 37.06 | 43.49 | 48.26 | 51.90 | 54.75 | 57.02 | 58.88 | 60.42 | 61.72 | 62.82 | 63.77 | 64.60 | 65.32 | 65.96 | 66.53 | 67.04 | 67.50 | 67.92 | 68.30 | 68.64 | 68.96 | 69.25 | 69.51 |
| | $U_2$ | | 0.41 | 0.96 | 1.75 | 2.79 | 3.95 | 5.14 | 6.30 | 7.39 | 8.40 | 9.31 | 10.14 | 10.90 | 11.58 | 12.19 | 12.75 | 13.26 | 13.72 | 14.15 | 14.54 | 14.89 | 15.22 | 15.52 | 15.80 | 16.06 | 16.31 |
| | $W_{1,2}$ | | 11.05 | 24.48 | 37.36 | 50.35 | 63.35 | 76.39 | 89.48 | 102.53 | 115.58 | 128.62 | 141.67 | 154.72 | 167.77 | 180.82 | 193.87 | 206.92 | 219.97 | 233.02 | 246.06 | 259.11 | 272.16 | 285.21 | 298.26 | 311.31 | 324.36 |
| | $W_{3,4}$ | | 10.81 | 18.57 | 23.67 | 27.06 | 29.35 | 30.95 | 32.08 | 32.89 | 33.49 | 33.93 | 34.27 | 34.53 | 34.73 | 34.89 | 35.01 | 35.11 | 35.20 | 35.26 | 35.32 | 35.36 | 35.40 | 35.43 | 35.46 | 35.49 | 35.51 |
| | $J$ | | 10.81 | 18.57 | 23.67 | 27.06 | 29.35 | 30.95 | 32.08 | 32.89 | 33.49 | 33.93 | 34.27 | 34.53 | 34.73 | 34.89 | 35.01 | 35.11 | 35.20 | 35.26 | 35.32 | 35.36 | 35.40 | 35.43 | 35.46 | 35.49 | 35.51 |
| | $K$ | | 2.98 | 4.63 | 5.24 | 5.42 | 5.49 | 5.50 | 5.50 | 5.50 | 5.50 | 5.50 | 5.50 | 5.50 | 5.50 | 5.50 | 5.50 | 5.50 | 5.50 | 5.50 | 5.50 | 5.50 | 5.50 | 5.50 | 5.50 | 5.50 | 5.50 |

The band structure figure (near charge neutrality) shows Energy (meV) on the vertical axis with tick marks at 100, 50, 0, −50, −100, and horizontal axis labels K, Γ, M, K. The panel is annotated "Total weight of the heavy-fermions on the active bands after wannierisation: 93.6%" with a colorbar labeled "Wannier orbitals support" ranging from 0.0 to 1.0.

TABLE S58. Parameters of the $f$-electron Wannier functions for different values of the tunneling ratio $w_0/w_1$. $W$ denotes the total weight of the THF $f$-electrons on the active bands. In computing the ratios $\gamma'/U_1$ and $\gamma'/U_2$, we employ the on-site and nearest-neighbor repulsion parameters $U_1$ and $U_2$ obtained numerically for $\xi = 10$ nm, as given in Table S59. We employ $v_F = 5.944\,\mathrm{eV\cdot\mathring{A}}$, $|\mathbf{K}| = 1.703\,\mathring{A}^{-1}$, $w_1 = 110$ meV, and $\theta = 1.24°$ for the BM model.

| $w_0/w_1$ | $\alpha_1$ | $\alpha_2$ | $\lambda_1$ | $\lambda_2$ | $\lambda$ | $W$ | $t_0$ | $\gamma$ | $M$ | $v_*$ | $v_*'$ | $\gamma'/v_*$ | $\gamma'/U_1$ | $\gamma'/U_2$ |
|---|---|---|---|---|---|---|---|---|---|---|---|---|---|---|
| 1.00 | 0.789 | 0.614 | 0.179 | 0.197 | 0.343 | 0.963 | 1.246 | -22.343 | -18.540 | -14.185 | 1.745 | 0.0081 | -0.32 | -6.49 |
| 0.90 | 0.818 | 0.575 | 0.191 | 0.203 | 0.351 | 0.948 | 1.537 | -38.097 | -19.877 | -4.613 | 1.771 | -0.0015 | -0.58 | -10.78 |
| 0.80 | 0.849 | 0.529 | 0.209 | 0.209 | 0.364 | 0.936 | 1.869 | -53.557 | -21.040 | -4.749 | 1.767 | -0.0069 | -0.87 | -14.62 |
| 0.70 | 0.879 | 0.478 | 0.221 | 0.215 | 0.376 | 0.926 | 2.203 | -68.468 | -22.040 | -4.873 | 1.731 | -0.0094 | -1.17 | -18.08 |
| 0.60 | 0.907 | 0.421 | 0.239 | 0.221 | 0.394 | 0.924 | 2.541 | -82.561 | -22.886 | -4.987 | 1.655 | -0.0098 | -1.51 | -20.95 |
| 0.50 | 0.933 | 0.359 | 0.251 | 0.227 | 0.411 | 0.925 | 2.900 | -95.533 | -23.586 | -5.094 | 1.532 | -0.0087 | -1.85 | -23.34 |
| 0.40 | 0.956 | 0.293 | 0.269 | 0.227 | 0.427 | 0.929 | 3.252 | -107.033 | -24.148 | -5.191 | 1.352 | -0.0067 | -2.19 | -25.26 |
| 0.30 | 0.974 | 0.225 | 0.281 | 0.233 | 0.434 | 0.927 | 3.444 | -116.659 | -24.580 | -5.278 | 1.105 | -0.0043 | -2.45 | -27.15 |
| 0.20 | 0.988 | 0.154 | 0.292 | 0.233 | 0.429 | 0.920 | 3.448 | -123.976 | -24.884 | -5.347 | 0.788 | -0.0021 | -2.61 | -29.03 |
| 0.10 | 0.997 | 0.079 | 0.298 | 0.239 | 0.426 | 0.916 | 3.455 | -128.572 | -25.065 | -5.393 | 0.412 | -0.0006 | -2.71 | -30.23 |
| 0.00 | 1.000 | 0.000 | 0.298 | 0.000 | 0.422 | 0.912 | 3.266 | -130.142 | -25.126 | -5.409 | -0.000 | 0.0000 | -2.72 | -30.95 |

FIG. S45. Band structures of the BM and THF models near charge neutrality for $w_0/w_1 = 0.8$, depicted by lines and crosses, respectively. The BM bands are colored according to the weight of the $f$-electron wave function on them. We use the same BM parameters as in Table S58.

| $w_0/w_1$ | $\xi$/nm | 2 | 4 | 6 | 8 | 10 | 12 | 14 | 16 | 18 | 20 | 22 | 24 | 26 | 28 | 30 | 32 | 34 | 36 | 38 | 40 | 42 | 44 | 46 | 48 | 50 |
|---|---|---|---|---|---|---|---|---|---|---|---|---|---|---|---|---|---|---|---|---|---|---|---|---|---|---|
| 1.0 | $U_1$ | 26.23 | 44.40 | 56.16 | 64.05 | 69.62 | 73.73 | 76.87 | 79.33 | 81.31 | 82.94 | 84.30 | 85.45 | 86.43 | 87.29 | 88.03 | 88.69 | 89.27 | 89.79 | 90.26 | 90.68 | 91.06 | 91.41 | 91.73 | 92.03 | 92.30 |
| | $U_2$ | 0.29 | 0.68 | 1.34 | 2.30 | 3.44 | 4.66 | 5.85 | 6.98 | | 8.97 | 9.83 | 10.61 | 11.31 | 11.95 | 12.52 | 13.04 | 13.51 | 13.94 | 14.34 | 14.70 | 15.04 | 15.34 | 15.63 | 15.89 | 16.14 |
| | $W_{1,2}$ | 11.46 | 23.87 | 36.97 | 50.34 | 63.78 | 77.26 | 90.73 | 104.21 | 117.69 | 131.17 | 144.65 | 158.13 | 171.61 | 185.09 | 198.57 | 212.05 | 225.53 | 239.01 | 252.49 | 265.97 | 279.45 | 292.93 | 306.41 | 319.89 | 333.37 |
| | $W_{3,4}$ | | 27.36 | 40.09 | 54.52 | 68.01 | 81.49 | 94.98 | 108.46 | 121.94 | 135.42 | 148.90 | 162.38 | 175.86 | 189.34 | 202.82 | 216.30 | 229.78 | 243.26 | 256.74 | 270.22 | 283.70 | 297.18 | 310.66 | 324.14 | 337.62 |
| | $J$ | 8.85 | 13.57 | 17.08 | 17.78 | 18.32 | 18.51 | 18.72 | 18.86 | 19.04 | 19.06 | 19.10 | 19.15 | 19.18 | 19.21 | 19.23 | 19.25 | 19.26 | 19.27 | 19.28 | 19.29 | 19.30 | 19.30 | 19.31 | 19.31 | |
| | $K$ | 2.16 | 3.36 | 3.70 | 3.92 | 3.96 | 3.97 | 3.97 | 3.97 | 3.97 | 3.97 | 3.97 | 3.97 | 3.97 | 3.97 | 3.97 | 3.97 | 3.97 | 3.97 | 3.97 | 3.97 | 3.97 | 3.97 | 3.97 | 3.97 | |
| 0.9 | $U_1$ | 24.18 | 41.31 | 52.59 | 60.26 | 65.72 | 69.77 | 72.86 | 75.31 | 77.27 | 78.89 | 80.24 | 81.39 | 82.37 | 83.22 | 83.96 | 84.62 | 85.20 | 85.72 | 86.60 | 86.99 | 87.34 | 87.66 | 87.95 | 88.22 | |
| | $U_2$ | 0.30 | 0.72 | 1.40 | 2.38 | 3.54 | 4.75 | 5.95 | 7.08 | 8.12 | 9.07 | 9.93 | 10.70 | 11.40 | 12.03 | 12.61 | 13.13 | 13.60 | 14.03 | 14.42 | 14.79 | 15.12 | 15.43 | 15.71 | 15.98 | 16.22 |
| | $W_{1,2}$ | 11.43 | 23.83 | 36.92 | 50.29 | 63.74 | 77.21 | 90.69 | 104.16 | 117.64 | 131.12 | 144.60 | 158.08 | 171.57 | 185.05 | 198.53 | 212.01 | 225.49 | 238.97 | 252.45 | 265.93 | 279.41 | 292.89 | 306.37 | 319.85 | 333.33 |
| | $W_{3,4}$ | 13.16 | 26.70 | 40.23 | 53.73 | 67.21 | 80.69 | 94.17 | 107.65 | 121.13 | 134.61 | 148.09 | 161.57 | 175.05 | 188.53 | 202.02 | 215.50 | 228.98 | 242.46 | 255.94 | 269.42 | 282.90 | 296.38 | 309.86 | 323.34 | 336.82 |
| | $J$ | 8.96 | 14.24 | 16.98 | 18.53 | 19.48 | 20.10 | 20.53 | 20.82 | 21.19 | 21.30 | 21.39 | 21.46 | 21.51 | 21.55 | 21.58 | 21.61 | 21.63 | 21.65 | 21.66 | 21.67 | 21.68 | 21.69 | 21.70 | 21.71 | |
| | $K$ | 2.17 | 3.38 | 3.81 | 3.64 | 3.88 | 3.96 | 3.98 | 3.98 | 3.99 | 3.99 | 3.99 | 3.99 | 3.99 | 3.99 | 3.99 | 3.99 | 3.99 | 3.99 | 3.99 | 3.99 | 3.99 | 3.99 | 3.99 | 3.99 | |
| 0.8 | $U_1$ | 22.22 | 38.27 | 49.04 | 56.46 | 61.78 | 65.74 | 68.79 | 71.20 | 73.15 | 74.75 | 76.09 | 77.23 | 78.20 | 79.05 | 79.79 | 80.44 | 81.02 | 81.54 | 82.00 | 82.42 | 82.81 | 83.16 | 83.47 | 83.77 | 84.04 |
| | $U_2$ | 0.30 | 0.72 | 1.49 | 2.49 | 3.79 | 5.01 | 6.21 | 7.33 | 8.36 | 9.31 | 10.16 | 10.93 | 11.62 | 12.25 | 12.82 | 13.34 | 13.81 | 14.24 | 14.63 | 14.99 | 15.32 | 15.63 | 15.91 | 16.18 | 16.42 |
| | $W_{1,2}$ | 11.33 | 23.67 | 36.75 | 50.11 | 63.55 | 77.02 | 90.50 | 103.98 | 117.46 | 130.94 | 144.42 | 157.90 | 171.38 | 184.86 | 198.34 | 211.82 | 225.30 | 238.78 | 252.26 | 265.74 | 279.22 | 292.70 | 306.18 | 319.66 | 333.14 |
| | $W_{3,4}$ | 12.88 | 26.22 | 39.68 | 53.15 | 66.63 | 80.11 | 93.59 | 107.07 | 120.55 | 134.03 | 147.51 | 160.99 | 174.47 | 187.95 | 201.43 | 214.91 | 228.39 | 241.87 | 255.35 | 268.83 | 282.31 | 295.79 | 309.27 | 322.75 | 336.23 |
| | $J$ | 9.32 | 14.99 | 18.18 | 20.00 | 21.30 | 22.11 | 22.67 | 23.07 | 23.36 | 23.57 | 23.73 | 23.85 | 23.94 | 24.01 | 24.07 | 24.12 | 24.15 | 24.18 | 24.21 | 24.23 | 24.24 | 24.25 | 24.27 | 24.28 | 24.29 |
| | $K$ | 3.44 | 3.88 | 4.02 | 4.05 | 4.06 | 4.07 | 4.07 | 4.07 | 4.07 | 4.07 | 4.07 | 4.07 | 4.07 | 4.07 | 4.07 | 4.07 | 4.07 | 4.07 | 4.07 | 4.07 | 4.07 | 4.07 | 4.07 | 4.07 | |
| 0.7 | $U_1$ | 20.54 | 35.65 | 45.96 | 53.14 | 58.33 | 62.22 | 65.22 | 67.60 | 69.53 | 71.12 | 72.45 | 73.58 | 74.55 | 75.39 | 76.12 | 76.77 | 77.35 | 77.87 | 78.33 | 78.75 | 79.13 | 79.46 | 79.80 | 80.09 | 80.36 |
| | $U_2$ | 0.36 | 0.84 | 1.59 | 2.61 | 3.79 | 5.01 | 6.21 | 7.33 | 8.36 | 9.31 | 10.16 | 10.93 | 11.62 | 12.25 | 12.82 | 13.34 | 13.81 | 14.24 | 14.63 | 14.99 | 15.32 | 15.63 | 15.91 | 16.18 | 16.42 |
| | $W_{1,2}$ | 11.26 | 23.42 | 36.47 | 49.82 | 63.26 | 76.73 | 90.21 | 103.69 | 117.17 | 130.65 | 144.13 | 157.61 | 171.09 | 184.57 | 198.05 | 211.53 | 225.01 | 238.49 | 251.97 | 265.45 | 278.93 | 292.41 | 305.89 | 319.37 | 332.85 |
| | $W_{3,4}$ | 12.68 | 25.88 | 39.28 | 52.74 | 66.22 | 79.70 | 93.18 | 106.66 | 120.14 | 133.62 | 147.10 | 160.58 | 174.06 | 187.54 | 201.02 | 214.50 | 227.98 | 241.46 | 254.94 | 268.42 | 281.90 | 295.38 | 308.86 | 322.34 | 335.82 |
| | $J$ | 9.61 | 15.73 | 19.35 | 21.58 | 23.04 | 24.02 | 24.71 | 25.20 | 25.56 | 25.82 | 26.02 | 26.17 | 26.29 | 26.38 | 26.46 | 26.53 | 26.58 | 26.63 | 26.68 | 26.70 | 26.71 | 26.72 | 26.73 | 26.74 | |
| | $K$ | 2.29 | 3.55 | 4.01 | 4.14 | 4.18 | 4.19 | 4.20 | 4.20 | 4.20 | 4.20 | 4.20 | 4.20 | 4.20 | 4.20 | 4.20 | 4.20 | 4.20 | 4.20 | 4.20 | 4.20 | 4.20 | 4.20 | 4.20 | 4.20 | |

TABLE S59: Parameters of the THF interaction Hamiltonian for different ratios $u_0/u_1$ and screening lengths $\xi$ of the double-gate interaction potential ($U_\xi = 24\,\mathrm{meV}$ for $\xi = 10\,\mathrm{nm}$). We use the same BM model parameters as in Table S58.

| $u_0/u_1$ | $\xi$/nm | 2 | 4 | 6 | 8 | 10 | 12 | 14 | 16 | 18 | 20 | 22 | 24 | 26 | 28 | 30 | 32 | 34 | 36 | 38 | 40 | 42 | 44 | 46 | 48 | 50 |
|---|---|---|---|---|---|---|---|---|---|---|---|---|---|---|---|---|---|---|---|---|---|---|---|---|---|---|
| 0.6 | $U_1$ | 18.91 | 33.06 | 42.86 | 49.77 | 54.80 | 58.00 | 61.54 | 63.88 | 65.78 | 67.35 | 68.67 | 69.79 | 70.75 | 71.58 | 72.31 | 72.96 | 73.53 | 74.05 | 74.51 | 74.93 | 75.31 | 75.65 | 75.97 | 76.26 | 76.53 |
| | $U_2$ | 0.40 | 0.92 | 1.71 | 2.75 | 3.94 | 5.17 | 6.36 | 7.48 | 8.51 | 9.45 | 10.29 | 11.06 | 11.75 | 12.38 | 12.95 | 13.46 | 13.93 | 14.36 | 14.75 | 15.11 | 15.44 | 15.75 | 16.03 | 16.29 | 16.53 |
| | $W_{1,2}$ | 10.97 | 23.13 | 36.15 | 49.49 | 62.93 | 76.40 | 89.88 | 103.36 | 116.84 | 130.32 | 143.80 | 157.28 | 170.76 | 184.24 | 197.72 | 211.20 | 224.68 | 238.16 | 251.64 | 265.12 | 278.60 | 292.08 | 305.56 | 319.04 | 332.52 |
| | $W_{3,4}$ | 12.55 | 25.66 | 39.03 | 52.48 | 65.95 | 79.43 | 92.91 | 106.39 | 119.87 | 133.35 | 146.83 | 160.31 | 173.79 | 187.27 | 200.75 | 214.23 | 227.71 | 241.19 | 254.67 | 268.15 | 281.63 | 295.11 | 308.59 | 322.07 | 335.55 |
| | $J$ | 9.94 | 16.54 | 20.61 | 23.21 | 24.94 | 26.13 | 26.97 | 27.58 | 28.03 | 28.37 | 28.63 | 28.83 | 28.98 | 29.10 | 29.20 | 29.27 | 29.34 | 29.39 | 29.43 | 29.47 | 29.50 | 29.52 | 29.54 | 29.56 | 29.57 |
| | $K$ | 2.39 | 3.70 | 4.17 | 4.32 | 4.36 | 4.37 | 4.37 | 4.37 | 4.37 | 4.37 | 4.37 | 4.37 | 4.37 | 4.37 | 4.37 | 4.37 | 4.37 | 4.37 | 4.37 | 4.37 | 4.37 | 4.37 | 4.37 | 4.37 | 4.37 |
| 0.5 | $U_1$ | 17.51 | 30.81 | 40.15 | 46.80 | 51.69 | 55.39 | 58.28 | 60.58 | 62.45 | 64.00 | 65.30 | 66.41 | 67.37 | 68.19 | 68.92 | 69.56 | 70.14 | 70.65 | 71.11 | 71.52 | 71.90 | 72.25 | 72.56 | 72.85 | 73.12 |
| | $U_2$ | 0.44 | 1.01 | 1.83 | 2.90 | 4.09 | 5.32 | 6.51 | 7.62 | 8.65 | 9.58 | 10.43 | 11.19 | 11.88 | 12.50 | 13.06 | 13.58 | 14.04 | 14.47 | 14.86 | 15.22 | 15.55 | 15.85 | 16.14 | 16.40 | 16.64 |
| | $W_{1,2}$ | 10.77 | 22.82 | 35.80 | 49.13 | 62.57 | 76.04 | 89.52 | 103.00 | 116.48 | 129.96 | 143.44 | 156.92 | 170.40 | 183.88 | 197.36 | 210.84 | 224.32 | 237.80 | 251.28 | 264.76 | 278.24 | 291.72 | 305.20 | 318.68 | 332.16 |
| | $W_{3,4}$ | 12.48 | 25.55 | 38.88 | 52.32 | 65.79 | 79.26 | 92.74 | 106.22 | 119.70 | 133.18 | 146.66 | 160.14 | 173.62 | 187.10 | 200.58 | 214.06 | 227.54 | 241.02 | 254.50 | 267.98 | 281.46 | 294.94 | 308.42 | 321.90 | 335.38 |
| | $J$ | 10.28 | 17.33 | 21.83 | 24.77 | 26.77 | 28.16 | 29.16 | 29.89 | 30.44 | 30.85 | 31.16 | 31.41 | 31.60 | 31.75 | 31.87 | 31.97 | 32.05 | 32.12 | 32.17 | 32.22 | 32.25 | 32.29 | 32.32 | 32.34 | 32.36 |
| | $K$ | 2.51 | 3.88 | 4.38 | 4.53 | 4.57 | 4.58 | 4.58 | 4.58 | 4.58 | 4.58 | 4.58 | 4.58 | 4.58 | 4.58 | 4.58 | 4.58 | 4.58 | 4.58 | 4.58 | 4.58 | 4.58 | 4.58 | 4.58 | 4.58 | 4.58 |
| 0.4 | $U_1$ | 16.33 | 28.88 | 37.81 | 44.23 | 48.97 | 52.50 | 55.42 | 57.68 | 59.53 | 61.00 | 62.35 | 63.45 | 64.39 | 65.22 | 65.94 | 66.58 | 67.15 | 67.66 | 68.11 | 68.53 | 68.91 | 69.25 | 69.57 | 69.86 | 70.12 |
| | $U_2$ | 0.48 | 1.09 | 1.95 | 3.03 | 4.24 | 5.46 | 6.65 | 7.76 | 8.78 | 9.71 | 10.55 | 11.30 | 11.99 | 12.61 | 13.17 | 13.68 | 14.15 | 14.57 | 14.96 | 15.32 | 15.65 | 15.95 | 16.23 | 16.49 | 16.74 |
| | $W_{1,2}$ | 10.34 | 22.18 | 35.08 | 48.39 | 61.82 | 75.29 | 88.76 | 102.24 | 115.72 | 129.20 | 142.68 | 156.16 | 169.64 | 183.12 | 196.60 | 210.08 | 223.56 | 237.04 | 250.52 | 264.00 | 277.48 | 290.96 | 304.44 | 317.92 | 331.40 |
| | $W_{3,4}$ | 12.44 | 25.46 | 38.79 | 52.23 | 65.71 | 79.18 | 92.66 | 106.13 | 119.61 | 133.09 | 146.57 | 160.05 | 173.53 | 187.01 | 200.49 | 213.97 | 227.45 | 240.93 | 254.41 | 267.89 | 281.37 | 294.85 | 308.33 | 321.81 | 335.29 |
| | $J$ | 10.60 | 18.07 | 22.96 | 26.23 | 28.49 | 30.08 | 31.23 | 32.08 | 32.72 | 33.20 | 33.58 | 33.87 | 34.10 | 34.28 | 34.43 | 34.55 | 34.65 | 34.73 | 34.80 | 34.85 | 34.90 | 34.94 | 34.98 | 35.01 | 35.03 |
| | $K$ | 2.63 | 4.06 | 4.60 | 4.76 | 4.80 | 4.81 | 4.82 | 4.82 | 4.82 | 4.82 | 4.82 | 4.82 | 4.82 | 4.82 | 4.82 | 4.82 | 4.82 | 4.82 | 4.82 | 4.82 | 4.82 | 4.82 | 4.82 | 4.82 | 4.82 |
| 0.3 | $U_1$ | 15.69 | 27.85 | 36.58 | 42.88 | 47.51 | 51.14 | 53.94 | 56.19 | 58.00 | 59.55 | 60.83 | 61.93 | 62.87 | 63.69 | 64.40 | 65.04 | 65.62 | 66.12 | 66.58 | 67.00 | 67.37 | 67.72 | 68.03 | 68.32 | 68.58 |
| | $U_2$ | 0.48 | 1.10 | 1.97 | 3.06 | 4.27 | 5.50 | 6.69 | 7.80 | 8.82 | 9.74 | 10.58 | 11.34 | 12.02 | 12.64 | 13.22 | 13.73 | 14.18 | 14.60 | 15.00 | 15.35 | 15.68 | 15.99 | 16.28 | 16.54 | 16.77 |
| | $W_{1,2}$ | 10.10 | 21.81 | 34.67 | 47.96 | 61.39 | 74.86 | 88.33 | 101.81 | 115.29 | 128.77 | 142.25 | 155.73 | 169.21 | 182.69 | 196.17 | 209.65 | 223.13 | 236.61 | 250.09 | 263.57 | 277.05 | 290.53 | 304.01 | 317.49 | 330.97 |
| | $W_{3,4}$ | 12.42 | 25.41 | 38.73 | 52.16 | 65.63 | 79.11 | 92.58 | 106.06 | 119.54 | 133.02 | 146.50 | 159.98 | 173.46 | 186.94 | 200.42 | 213.91 | 227.38 | 240.86 | 254.34 | 267.82 | 281.30 | 294.78 | 308.26 | 321.74 | 335.22 |
| | $J$ | 10.85 | 18.61 | 23.75 | 27.22 | 29.62 | 31.33 | 32.56 | 33.47 | 34.15 | 34.67 | 35.07 | 35.39 | 35.64 | 35.85 | 36.01 | 36.13 | 36.24 | 36.32 | 36.40 | 36.46 | 36.51 | 36.56 | 36.60 | 36.63 | 36.65 |
| | $K$ | 2.77 | 4.28 | 4.82 | 4.99 | 5.03 | 5.05 | 5.05 | 5.05 | 5.05 | 5.05 | 5.05 | 5.05 | 5.05 | 5.05 | 5.05 | 5.05 | 5.05 | 5.05 | 5.05 | 5.05 | 5.05 | 5.05 | 5.05 | 5.05 | 5.05 |
| 0.2 | $U_1$ | 15.56 | 27.70 | 36.44 | 42.78 | 47.49 | 51.08 | 53.90 | 56.16 | 58.00 | 59.53 | 60.82 | 61.92 | 62.86 | 63.69 | 64.40 | 65.04 | 65.62 | 66.13 | 66.59 | 67.00 | 67.38 | 67.71 | 68.03 | 68.32 | 68.59 |
| | $U_2$ | 0.47 | 1.08 | 1.93 | 3.04 | 4.25 | 5.48 | 6.67 | 7.78 | 8.80 | 9.73 | 10.57 | 11.33 | 12.01 | 12.63 | 13.21 | 13.71 | 14.17 | 14.60 | 14.99 | 15.34 | 15.67 | 15.98 | 16.26 | 16.52 | 16.76 |
| | $W_{1,2}$ | 9.03 | 21.55 | 34.38 | 47.67 | 61.09 | 74.56 | 88.03 | 101.51 | 114.99 | 128.47 | 141.95 | 155.43 | 168.91 | 182.39 | 195.87 | 209.35 | 222.83 | 236.31 | 249.79 | 263.27 | 276.75 | 290.23 | 303.71 | 317.19 | 330.67 |
| | $W_{3,4}$ | 12.36 | 25.32 | 38.62 | 52.05 | 65.52 | 78.99 | 92.47 | 105.95 | 119.43 | 132.91 | 146.39 | 159.87 | 173.35 | 186.83 | 200.31 | 213.79 | 227.27 | 240.75 | 254.23 | 267.71 | 281.19 | 294.67 | 308.15 | 321.63 | 335.11 |
| | $J$ | 11.13 | 19.15 | 24.47 | 28.04 | 30.49 | 32.21 | 33.44 | 34.34 | 35.00 | 35.51 | 35.90 | 36.19 | 36.43 | 36.61 | 36.76 | 36.88 | 36.98 | 37.06 | 37.12 | 37.18 | 37.23 | 37.27 | 37.30 | 37.33 | 37.35 |
| | $K$ | 2.96 | 4.58 | 5.16 | 5.34 | 5.39 | 5.40 | 5.41 | 5.41 | 5.41 | 5.41 | 5.41 | 5.41 | 5.41 | 5.41 | 5.41 | 5.41 | 5.41 | 5.41 | 5.41 | 5.41 | 5.41 | 5.41 | 5.41 | 5.41 | 5.41 |
| 0.1 | $U_1$ | 15.69 | 27.94 | 36.75 | 43.13 | 47.87 | 51.49 | 54.32 | 56.58 | 58.43 | 59.96 | 61.26 | 62.36 | 63.30 | 64.13 | 64.85 | 65.49 | 66.06 | 66.57 | 67.03 | 67.44 | 67.82 | 68.17 | 68.48 | 68.77 | 69.04 |
| | $U_2$ | 0.47 | 1.08 | 1.93 | 3.03 | 4.25 | 5.48 | 6.67 | 7.78 | 8.80 | 9.73 | 10.53 | 11.29 | 11.97 | 12.59 | 13.16 | 13.67 | 14.13 | 14.56 | 14.95 | 15.31 | 15.64 | 15.94 | 16.22 | 16.48 | 16.72 |
| | $W_{1,2}$ | 9.85 | 21.61 | 34.25 | 47.53 | 60.95 | 74.42 | 87.89 | 101.37 | 114.85 | 128.33 | 141.81 | 155.29 | 168.77 | 182.25 | 195.73 | 209.21 | 222.69 | 236.17 | 249.65 | 263.13 | 276.61 | 290.09 | 303.57 | 317.05 | 330.53 |
| | $W_{3,4}$ | 12.35 | 25.29 | 38.60 | 52.03 | 65.49 | 78.97 | 92.45 | 105.93 | 119.41 | 132.89 | 146.37 | 159.85 | 173.33 | 186.81 | 200.29 | 213.77 | 227.25 | 240.73 | 254.21 | 267.69 | 281.17 | 294.65 | 308.13 | 321.61 | 335.09 |
| | $J$ | 11.15 | 19.19 | 24.50 | 28.05 | 30.48 | 32.18 | 33.39 | 34.27 | 34.93 | 35.42 | 35.79 | 36.08 | 36.31 | 36.49 | 36.64 | 36.74 | 36.84 | 36.92 | 36.98 | 37.03 | 37.08 | 37.12 | 37.15 | 37.18 | 37.20 |
| | $K$ | 2.99 | 4.63 | 5.21 | 5.39 | 5.44 | 5.45 | 5.46 | 5.46 | 5.46 | 5.46 | 5.46 | 5.46 | 5.46 | 5.46 | 5.46 | 5.46 | 5.46 | 5.46 | 5.46 | 5.46 | 5.46 | 5.46 | 5.46 | 5.46 | 5.46 |
| 0.0 | $U_1$ | 15.69 | 27.94 | 36.75 | 43.13 | 47.87 | 51.49 | 54.38 | 56.43 | 58.43 | 59.96 | 61.26 | 62.36 | 63.30 | 64.13 | 64.85 | 65.49 | 66.06 | 66.57 | 67.03 | 67.44 | 67.82 | 68.17 | 68.48 | 68.77 | 69.04 |
| | $U_2$ | 0.45 | 1.05 | 1.91 | 3.00 | 4.21 | 5.44 | 6.62 | 7.74 | 8.76 | 9.69 | 10.53 | 11.29 | 11.97 | 12.59 | 13.16 | 13.67 | 14.13 | 14.56 | 14.95 | 15.31 | 15.64 | 15.94 | 16.22 | 16.48 | 16.72 |
| | $W_{1,2}$ | 9.85 | 21.55 | 34.25 | 47.53 | 60.95 | 74.42 | 87.89 | 101.37 | 114.85 | 128.33 | 141.81 | 155.29 | 168.77 | 182.25 | 195.73 | 209.21 | 222.69 | 236.17 | 249.65 | 263.13 | 276.61 | 290.09 | 303.57 | 317.05 | 330.53 |
| | $W_{3,4}$ | 12.35 | 25.29 | 38.60 | 52.03 | 65.49 | 78.97 | 92.45 | 105.93 | 119.41 | 132.89 | 146.37 | 159.85 | 173.33 | 186.81 | 200.29 | 213.77 | 227.25 | 240.73 | 254.21 | 267.69 | 281.17 | 294.65 | 308.13 | 321.61 | 335.09 |
| | $J$ | 11.15 | 19.19 | 24.50 | 28.05 | 30.48 | 32.18 | 33.39 | 34.27 | 34.93 | 35.42 | 35.79 | 36.08 | 36.31 | 36.49 | 36.64 | 36.74 | 36.84 | 36.92 | 36.98 | 37.03 | 37.08 | 37.12 | 37.15 | 37.18 | 37.20 |
| | $K$ | 2.99 | 4.63 | 5.21 | 5.39 | 5.44 | 5.45 | 5.46 | 5.46 | 5.46 | 5.46 | 5.46 | 5.46 | 5.46 | 5.46 | 5.46 | 5.46 | 5.46 | 5.46 | 5.46 | 5.46 | 5.46 | 5.46 | 5.46 | 5.46 | 5.46 |

Total weight of the heavy-fermions on the active bands after wannierisation: 93.4%

Wannier orbitals support 0.0 — 1.0

FIG. S46. Band structures of the BM and THF models near charge neutrality for $w_0/w_1 = 0.8$, depicted by lines and crosses, respectively. The BM bands are colored according to the weight of the $f$-electron wave function on them. We use the same BM parameters as in Table S60.

TABLE S60. Parameters of the $f$-electron Wannier functions and the THF single-particle Hamiltonian for different values of the tunneling ratio $w_0/w_1$. $W$ denotes the total weight of the THF $f$-electrons on the active bands. In computing the ratios $\gamma'/U_1$ and $\gamma'/U_2$, we employ the on-site and nearest-neighbor repulsion parameters $U_1$ and $U_2$ obtained numerically for $\xi = 10$ nm, as given in Table S61. We employ $v_F = 5.944$ eV·Å, $|\mathbf{K}| = 1.703$ Å$^{-1}$, $w_1 = 110$ meV, and $\theta = 1.26°$ for the BM model.

| $w_0/w_1$ | $\alpha_1$ | $\alpha_2$ | $\lambda_1$ | $\lambda_2$ | $\lambda$ | $W$ | $t_0$ | $\gamma$ | $M$ | $v_\star$ | $v'_\star$ | $v''_\star$ | $\gamma'/U_1$ | $\gamma'/U_2$ |
|---|---|---|---|---|---|---|---|---|---|---|---|---|---|---|
| 1.00 | 0.792 | 0.610 | 0.185 | 0.197 | 0.346 | 0.960 | 1.456 | -25.386 | -21.300 | -4.503 | 1.764 | 0.0098 | -0.36 | -7.01 |
| 0.90 | 0.821 | 0.571 | 0.197 | 0.203 | 0.353 | 0.945 | 1.787 | -41.222 | -22.683 | -4.649 | 1.785 | 0.0011 | -0.62 | -11.11 |
| 0.80 | 0.851 | 0.525 | 0.209 | 0.209 | 0.366 | 0.934 | 2.163 | -56.732 | -23.886 | -4.781 | 1.778 | -0.0040 | -0.91 | -14.77 |
| 0.70 | 0.881 | 0.473 | 0.227 | 0.215 | 0.385 | 0.929 | 2.550 | -71.671 | -24.919 | -4.901 | 1.757 | -0.0065 | -1.23 | -17.89 |
| 0.60 | 0.909 | 0.416 | 0.239 | 0.221 | 0.400 | 0.926 | 2.908 | -85.776 | -25.792 | -5.013 | 1.661 | -0.0071 | -1.56 | -20.70 |
| 0.50 | 0.935 | 0.355 | 0.257 | 0.227 | 0.420 | 0.929 | 3.368 | -98.748 | -26.515 | -5.117 | 1.536 | -0.0065 | -1.91 | -22.84 |
| 0.40 | 0.957 | 0.290 | 0.269 | 0.233 | 0.429 | 0.928 | 3.683 | -110.240 | -27.096 | -5.214 | 1.354 | -0.0050 | -2.22 | -24.95 |
| 0.30 | 0.975 | 0.223 | 0.281 | 0.233 | 0.434 | 0.927 | 3.893 | -119.853 | -27.540 | -5.299 | 1.106 | -0.0033 | -2.48 | -26.79 |
| 0.20 | 0.988 | 0.152 | 0.292 | 0.239 | 0.437 | 0.925 | 3.990 | -127.156 | -27.855 | -5.368 | 0.789 | -0.0016 | -2.68 | -28.23 |
| 0.10 | 0.997 | 0.078 | 0.298 | 0.239 | 0.429 | 0.917 | 3.920 | -131.743 | -28.042 | -5.413 | 0.412 | -0.0004 | -2.74 | -29.65 |
| 0.00 | 1.000 | 0.000 | 0.298 | 0.000 | 0.423 | 0.911 | 3.732 | -133.309 | -28.104 | -5.429 | 0.000 | 0.0000 | -2.74 | -30.43 |

| $w_0/w_1$ | $\xi$/nm | 2 | 4 | 6 | 8 | 10 | 12 | 14 | 16 | 18 | 20 | 22 | 24 | 26 | 28 | 30 | 32 | 34 | 36 | 38 | 40 | 42 | 44 | 46 | 48 | 50 |
|---|---|---|---|---|---|---|---|---|---|---|---|---|---|---|---|---|---|---|---|---|---|---|---|---|---|---|
| 1.0 | $U_1$ | 26.40 | 44.67 | 54.47 | 64.39 | 68.98 | 74.10 | 77.24 | 79.71 | 81.70 | 83.33 | 84.69 | 85.84 | 86.82 | 87.68 | 88.42 | 89.08 | 89.66 | 90.18 | 90.65 | 91.08 | 91.46 | 91.81 | 92.13 | 92.42 | 92.69 |
| | $U_2$ | 0.30 | 0.72 | 1.42 | 2.42 | 3.62 | 4.87 | 6.10 | 7.26 | 8.32 | 9.29 | 10.16 | 10.95 | 11.66 | 12.29 | 12.87 | 13.40 | 13.87 | 14.31 | 14.71 | 15.07 | 15.41 | 15.72 | 16.00 | 16.27 | 16.51 |
| | $W_{1,2}$ | 11.03 | 24.81 | 38.37 | 52.19 | 66.08 | 79.99 | 93.90 | 107.82 | 121.74 | 135.66 | 149.58 | 163.50 | 177.41 | 191.33 | 205.25 | 219.17 | 233.09 | 247.01 | 260.92 | 274.84 | 288.76 | 302.68 | 316.60 | 330.52 | 344.43 |
| | $W_{3,4}$ | 13.85 | 27.99 | 42.02 | 55.98 | 69.91 | 83.83 | 97.75 | 111.66 | 125.58 | 139.50 | 153.42 | 167.33 | 181.25 | 195.18 | 209.09 | 223.01 | 236.93 | 250.85 | 264.77 | 278.69 | 292.60 | 306.52 | 320.44 | 334.36 | 348.28 |
| | $J$ | 9.06 | 13.90 | 16.27 | 17.54 | 18.29 | 18.75 | 19.07 | 19.28 | 19.43 | 19.55 | 19.63 | 19.69 | 19.74 | 19.78 | 19.81 | 19.83 | 19.85 | 19.86 | 19.88 | 19.89 | 19.90 | 19.90 | 19.91 | 19.91 | 19.92 |
| | $K$ | 2.13 | 3.30 | 3.71 | 3.83 | 3.86 | 3.87 | 3.87 | 3.87 | 3.87 | 3.87 | 3.87 | 3.87 | 3.87 | 3.87 | 3.87 | 3.87 | 3.87 | 3.87 | 3.87 | 3.87 | 3.87 | 3.87 | 3.87 | 3.87 | 3.87 |
| 0.9 | $U_1$ | 24.40 | 41.65 | 53.00 | 60.71 | 66.19 | 70.25 | 73.36 | 75.81 | 77.78 | 79.40 | 80.76 | 81.90 | 82.88 | 83.74 | 84.48 | 85.14 | 85.72 | 86.24 | 86.70 | 87.13 | 87.51 | 87.86 | 88.18 | 88.47 | 88.74 |
| | $U_2$ | 0.31 | 0.75 | 1.48 | 2.50 | 3.71 | 4.97 | 6.20 | 7.35 | 8.41 | 9.38 | 10.25 | 11.03 | 11.74 | 12.38 | 12.96 | 13.48 | 13.96 | 14.39 | 14.79 | 15.15 | 15.49 | 15.80 | 16.08 | 16.35 | 16.59 |
| | $W_{1,2}$ | 11.88 | 24.74 | 38.29 | 52.10 | 65.99 | 79.90 | 93.82 | 107.74 | 121.66 | 135.57 | 149.49 | 163.41 | 177.33 | 191.25 | 205.17 | 219.08 | 233.00 | 246.92 | 260.84 | 274.76 | 288.68 | 302.60 | 316.51 | 330.43 | 344.35 |
| | $W_{3,4}$ | 13.47 | 27.37 | 41.31 | 55.24 | 69.16 | 83.08 | 97.00 | 110.92 | 124.84 | 138.75 | 152.67 | 166.59 | 180.51 | 194.43 | 208.35 | 222.26 | 236.18 | 250.10 | 264.02 | 277.94 | 291.86 | 305.77 | 319.69 | 333.61 | 347.53 |
| | $J$ | 9.28 | 14.11 | 16.41 | 17.41 | 19.02 | 20.00 | 20.64 | 21.08 | 21.38 | 21.60 | 21.76 | 21.88 | 21.97 | 22.04 | 22.09 | 22.13 | 22.18 | 22.22 | 22.25 | 22.26 | 22.28 | 22.28 | 22.29 | 22.29 | 22.29 |
| | $K$ | 2.15 | 3.33 | 3.74 | 3.86 | 3.89 | 3.90 | 3.91 | 3.91 | 3.91 | 3.91 | 3.91 | 3.91 | 3.91 | 3.91 | 3.91 | 3.91 | 3.91 | 3.91 | 3.91 | 3.91 | 3.91 | 3.91 | 3.91 | 3.91 | 3.91 |
| 0.8 | $U_1$ | 22.44 | 38.63 | 49.47 | 56.93 | 62.27 | 66.25 | 69.31 | 71.73 | 73.68 | 75.29 | 76.63 | 77.77 | 78.75 | 79.59 | 80.33 | 80.99 | 81.57 | 82.09 | 82.55 | 82.97 | 83.35 | 83.70 | 84.02 | 84.32 | 84.59 |
| | $U_2$ | 0.34 | 0.82 | 1.58 | 2.62 | 3.84 | 5.10 | 6.33 | 7.48 | 8.54 | 9.50 | 10.37 | 11.15 | 11.86 | 12.50 | 13.07 | 13.59 | 14.07 | 14.50 | 14.90 | 15.26 | 15.60 | 15.90 | 16.19 | 16.45 | 16.70 |
| | $W_{1,2}$ | 11.76 | 24.56 | 38.09 | 51.90 | 65.79 | 79.69 | 93.61 | 107.53 | 121.45 | 135.37 | 149.28 | 163.20 | 177.12 | 191.04 | 204.96 | 218.88 | 232.79 | 246.71 | 260.63 | 274.55 | 288.47 | 302.39 | 316.31 | 330.22 | 344.14 |
| | $W_{3,4}$ | 13.24 | 26.92 | 40.80 | 54.71 | 68.62 | 82.54 | 96.46 | 110.37 | 124.29 | 138.21 | 152.13 | 166.05 | 179.97 | 193.89 | 207.80 | 221.72 | 235.64 | 249.56 | 263.48 | 277.40 | 291.31 | 305.23 | 319.15 | 333.07 | 346.99 |
| | $J$ | 9.56 | 15.37 | 18.65 | 20.61 | 21.85 | 22.68 | 23.25 | 23.66 | 23.95 | 24.16 | 24.33 | 24.45 | 24.54 | 24.62 | 24.67 | 24.72 | 24.76 | 24.79 | 24.81 | 24.83 | 24.85 | 24.87 | 24.88 | 24.89 | 24.89 |
| | $K$ | 2.20 | 3.40 | 3.82 | 3.95 | 3.98 | 3.99 | 4.00 | 4.00 | 4.00 | 4.00 | 4.00 | 4.00 | 4.00 | 4.00 | 4.00 | 4.00 | 4.00 | 4.00 | 4.00 | 4.00 | 4.00 | 4.00 | 4.00 | 4.00 | 4.00 |
| 0.7 | $U_1$ | 20.55 | 35.65 | 45.94 | 53.11 | 58.29 | 62.17 | 65.17 | 67.55 | 69.47 | 71.06 | 72.39 | 73.52 | 74.49 | 75.33 | 76.06 | 76.71 | 77.27 | 77.80 | 78.27 | 78.69 | 79.07 | 79.42 | 79.73 | 80.03 | 80.30 |
| | $U_2$ | 0.38 | 0.91 | 1.71 | 2.78 | 4.01 | 5.27 | 6.50 | 7.64 | 8.70 | 9.65 | 10.52 | 11.30 | 12.00 | 12.63 | 13.20 | 13.72 | 14.20 | 14.63 | 15.02 | 15.39 | 15.72 | 16.03 | 16.31 | 16.58 | 16.82 |
| | $W_{1,2}$ | 11.64 | 24.32 | 37.82 | 51.62 | 65.51 | 79.41 | 93.33 | 107.25 | 121.17 | 135.09 | 149.00 | 162.92 | 176.84 | 190.75 | 204.67 | 218.60 | 232.52 | 246.43 | 260.35 | 274.27 | 288.19 | 302.11 | 316.02 | 329.94 | 343.86 |
| | $W_{3,4}$ | 13.03 | 26.02 | 40.45 | 54.34 | 68.25 | 82.17 | 96.09 | 110.01 | 123.93 | 137.84 | 151.76 | 165.68 | 179.60 | 193.52 | 207.44 | 221.35 | 235.27 | 249.19 | 263.11 | 277.03 | 290.95 | 304.86 | 318.78 | 332.70 | 346.62 |
| | $J$ | 16.21 | 19.97 | 22.31 | 23.84 | 24.89 | 25.62 | 26.14 | 26.53 | 26.81 | 27.03 | 27.20 | 27.33 | 27.43 | 27.51 | 27.57 | 27.62 | 27.67 | 27.70 | 27.73 | 27.76 | 27.78 | 27.79 | 27.81 | 27.82 | 27.82 |
| | $K$ | 2.28 | 3.52 | 3.96 | 4.09 | 4.12 | 4.13 | 4.13 | 4.14 | 4.14 | 4.14 | 4.14 | 4.14 | 4.14 | 4.14 | 4.14 | 4.14 | 4.14 | 4.14 | 4.14 | 4.14 | 4.14 | 4.14 | 4.14 | 4.14 | 4.14 |

| $u_0/u_1$ | | 2 | 4 | 6 | 8 | 10 | 12 | 14 | 16 | 18 | 20 | 22 | 24 | 26 | 28 | 30 | 32 | 34 | 36 | 38 | 40 | 42 | 44 | 46 | 48 | 50 |
|---|---|---|---|---|---|---|---|---|---|---|---|---|---|---|---|---|---|---|---|---|---|---|---|---|---|---|
| 0.6 | $U_1$ | 19.04 | 33.26 | 43.10 | 50.03 | 55.07 | 58.87 | 61.82 | 64.16 | 66.06 | 67.64 | 68.95 | 70.07 | 71.03 | 71.87 | 72.60 | 73.25 | 73.82 | 74.34 | 74.80 | 75.22 | 75.60 | 75.94 | 76.26 | 76.55 | 76.82 |
| | $U_2$ | 0.42 | 0.98 | 1.81 | 2.91 | 4.14 | 5.41 | 6.63 | 7.78 | 8.83 | 9.78 | 10.64 | 11.41 | 12.11 | 12.74 | 13.31 | 13.83 | 14.30 | 14.73 | 15.13 | 15.49 | 15.82 | 16.13 | 16.41 | 16.68 | 16.92 |
| | $W_{1,2}$ | 11.40 | 24.01 | 37.48 | 51.27 | 65.15 | 79.06 | 92.98 | 106.90 | 120.81 | 134.73 | 148.65 | 162.57 | 176.49 | 190.41 | 204.33 | 218.24 | 232.16 | 246.08 | 260.00 | 273.92 | 287.84 | 301.75 | 315.67 | 329.59 | 343.51 |
| | $W_{3,4}$ | 12.91 | 26.42 | 40.22 | 54.10 | 68.01 | 81.92 | 95.84 | 109.76 | 123.68 | 137.60 | 151.51 | 165.43 | 179.35 | 193.27 | 207.19 | 221.11 | 235.03 | 248.95 | 262.86 | 276.78 | 290.70 | 304.62 | 318.54 | 332.45 | 346.37 |
| | $J$ | 10.23 | 17.02 | 21.21 | 23.89 | 25.68 | 26.92 | 27.79 | 28.42 | 28.89 | 29.24 | 29.51 | 29.71 | 29.88 | 30.00 | 30.10 | 30.19 | 30.25 | 30.31 | 30.35 | 30.39 | 30.42 | 30.45 | 30.47 | 30.49 | 30.50 |
| | $K$ | 2.39 | 3.68 | 4.13 | 4.27 | 4.30 | 4.31 | 4.32 | 4.32 | 4.32 | 4.32 | 4.32 | 4.32 | 4.32 | 4.32 | 4.32 | 4.32 | 4.32 | 4.32 | 4.32 | 4.32 | 4.32 | 4.32 | 4.32 | 4.32 | 4.32 |
| 0.5 | $U_1$ | 17.52 | 30.81 | 40.14 | 46.78 | 51.66 | 55.35 | 58.24 | 60.53 | 62.40 | 63.95 | 65.25 | 66.36 | 67.32 | 68.14 | 68.87 | 69.51 | 70.08 | 70.59 | 71.05 | 71.47 | 71.85 | 72.19 | 72.51 | 72.80 | 73.07 |
| | $U_2$ | 0.47 | 1.09 | 1.96 | 3.08 | 4.32 | 5.59 | 6.81 | 7.94 | 8.99 | 9.93 | 10.79 | 11.56 | 12.25 | 12.88 | 13.45 | 13.96 | 14.43 | 14.86 | 15.25 | 15.62 | 15.95 | 16.25 | 16.54 | 16.80 | 17.04 |
| | $W_{1,2}$ | 11.20 | 23.72 | 37.16 | 50.93 | 64.81 | 78.72 | 92.64 | 106.55 | 120.47 | 134.39 | 148.31 | 162.23 | 176.15 | 190.06 | 203.98 | 217.90 | 231.82 | 245.74 | 259.66 | 273.57 | 287.49 | 301.41 | 315.33 | 329.25 | 343.17 |
| | $W_{3,4}$ | 12.85 | 26.31 | 40.09 | 53.97 | 67.87 | 81.79 | 95.71 | 109.62 | 123.54 | 137.46 | 151.38 | 165.30 | 179.22 | 193.14 | 207.06 | 220.97 | 234.89 | 248.81 | 262.73 | 276.65 | 290.57 | 304.48 | 318.40 | 332.32 | 346.24 |
| | $J$ | 10.59 | 17.87 | 22.53 | 25.59 | 27.68 | 29.14 | 30.19 | 30.96 | 31.54 | 31.97 | 32.31 | 32.57 | 32.77 | 32.94 | 33.07 | 33.17 | 33.26 | 33.33 | 33.39 | 33.44 | 33.48 | 33.52 | 33.55 | 33.57 | 33.60 |
| | $K$ | 2.51 | 3.87 | 4.34 | 4.48 | 4.52 | 4.53 | 4.53 | 4.53 | 4.53 | 4.53 | 4.53 | 4.53 | 4.53 | 4.53 | 4.53 | 4.53 | 4.53 | 4.53 | 4.53 | 4.53 | 4.53 | 4.53 | 4.53 | 4.53 | 4.53 |
| 0.4 | $U_1$ | 16.00 | 29.33 | 38.87 | 44.85 | 49.63 | 53.27 | 56.12 | 58.39 | 60.24 | 61.78 | 63.07 | 64.18 | 65.13 | 65.95 | 66.67 | 67.31 | 67.88 | 68.39 | 68.85 | 69.27 | 69.64 | 69.99 | 70.31 | 70.60 | 70.86 |
| | $U_2$ | 0.50 | 1.14 | 2.04 | 3.17 | 4.42 | 5.69 | 6.90 | 8.04 | 9.08 | 10.02 | 10.87 | 11.64 | 12.33 | 12.96 | 13.53 | 14.04 | 14.51 | 14.94 | 15.33 | 15.69 | 16.02 | 16.32 | 16.61 | 16.87 | 17.11 |
| | $W_{1,2}$ | 10.99 | 23.37 | 36.77 | 50.53 | 64.41 | 78.32 | 92.23 | 106.15 | 120.07 | 133.99 | 147.90 | 161.82 | 175.74 | 189.66 | 203.57 | 217.49 | 231.41 | 245.33 | 259.25 | 273.17 | 287.09 | 301.00 | 314.93 | 328.84 | 342.76 |
| | $W_{3,4}$ | 12.81 | 26.24 | 40.00 | 53.87 | 67.68 | 81.60 | 95.61 | 109.53 | 123.45 | 137.37 | 151.29 | 165.20 | 179.13 | 193.04 | 206.96 | 220.88 | 234.80 | 248.71 | 262.63 | 276.55 | 290.47 | 304.39 | 318.31 | 332.23 | 346.14 |
| | $J$ | 10.90 | 18.56 | 23.55 | 26.88 | 29.16 | 30.77 | 31.93 | 32.78 | 33.42 | 33.90 | 34.25 | 34.51 | 34.79 | 34.98 | 35.12 | 35.24 | 35.34 | 35.42 | 35.49 | 35.54 | 35.59 | 35.63 | 35.67 | 35.69 | 35.72 |
| | $K$ | 2.65 | 4.07 | 4.57 | 4.72 | 4.76 | 4.77 | 4.77 | 4.77 | 4.77 | 4.77 | 4.77 | 4.77 | 4.77 | 4.77 | 4.77 | 4.77 | 4.77 | 4.77 | 4.77 | 4.77 | 4.77 | 4.77 | 4.77 | 4.77 | 4.77 |
| 0.3 | $U_1$ | 15.63 | 27.78 | 37.21 | 43.59 | 48.32 | 51.92 | 54.74 | 57.00 | 58.85 | 60.38 | 61.67 | 62.76 | 63.71 | 64.53 | 65.26 | 65.89 | 66.46 | 66.97 | 67.43 | 67.84 | 68.22 | 68.56 | 68.88 | 69.17 | 69.44 |
| | $U_2$ | 0.51 | 1.17 | 2.08 | 3.23 | 4.48 | 5.74 | 6.96 | 8.09 | 9.13 | 10.07 | 10.93 | 11.69 | 12.38 | 13.01 | 13.57 | 14.08 | 14.55 | 14.98 | 15.37 | 15.73 | 16.06 | 16.36 | 16.65 | 16.91 | 17.15 |
| | $W_{1,2}$ | 10.73 | 23.01 | 36.37 | 50.13 | 64.00 | 77.91 | 91.82 | 105.74 | 119.66 | 133.58 | 147.49 | 161.41 | 175.33 | 189.25 | 203.17 | 217.09 | 231.00 | 244.92 | 258.84 | 272.76 | 286.68 | 300.60 | 314.52 | 328.44 | 342.35 |
| | $W_{3,4}$ | 12.79 | 26.20 | 39.94 | 53.81 | 67.68 | 81.60 | 95.51 | 109.43 | 123.35 | 137.27 | 151.19 | 165.11 | 179.02 | 192.94 | 206.86 | 220.78 | 234.70 | 248.62 | 262.53 | 276.45 | 290.37 | 304.29 | 318.21 | 332.13 | 346.04 |
| | $J$ | 11.14 | 19.09 | 24.33 | 27.84 | 30.26 | 31.97 | 33.21 | 34.12 | 34.80 | 35.31 | 35.71 | 36.02 | 36.26 | 36.46 | 36.62 | 36.75 | 36.85 | 36.94 | 37.01 | 37.07 | 37.12 | 37.16 | 37.20 | 37.24 | 37.26 |
| | $K$ | 2.78 | 4.28 | 4.80 | 4.95 | 4.99 | 5.01 | 5.01 | 5.01 | 5.01 | 5.01 | 5.01 | 5.01 | 5.01 | 5.01 | 5.01 | 5.01 | 5.01 | 5.01 | 5.01 | 5.01 | 5.01 | 5.01 | 5.01 | 5.01 | 5.01 |
| 0.2 | $U_1$ | 15.81 | 28.11 | 36.05 | 42.83 | 47.52 | 51.11 | 53.92 | 56.17 | 58.01 | 59.53 | 60.82 | 61.91 | 62.86 | 63.68 | 64.40 | 65.04 | 65.70 | 66.27 | 66.78 | 67.24 | 67.65 | 68.01 | 68.37 | 68.69 | 68.98 |
| | $U_2$ | 0.51 | 1.14 | 2.05 | 3.19 | 4.44 | 5.71 | 6.93 | 8.07 | 9.11 | 10.05 | 10.91 | 11.67 | 12.36 | 12.99 | 13.56 | 14.07 | 14.57 | 15.00 | 15.39 | 15.75 | 16.08 | 16.39 | 16.67 | 16.94 | 17.17 |
| | $W_{1,2}$ | 10.53 | 22.71 | 36.03 | 49.78 | 63.65 | 77.55 | 91.47 | 105.38 | 119.30 | 133.22 | 147.14 | 161.05 | 174.97 | 188.89 | 202.81 | 216.73 | 230.65 | 244.57 | 258.49 | 272.40 | 286.32 | 300.24 | 314.16 | 328.08 | 342.00 |
| | $W_{3,4}$ | 12.77 | 26.18 | 39.91 | 53.78 | 67.68 | 81.60 | 95.51 | 109.43 | 123.35 | 137.27 | 151.19 | 165.10 | 179.02 | 192.94 | 206.86 | 220.78 | 234.70 | 248.62 | 262.53 | 276.45 | 290.37 | 304.29 | 318.21 | 332.13 | 345.97 |
| | $J$ | 11.34 | 19.60 | 25.10 | 28.74 | 31.23 | 32.97 | 34.22 | 35.12 | 35.80 | 36.31 | 36.69 | 36.99 | 37.23 | 37.43 | 37.58 | 37.71 | 37.84 | 37.95 | 38.04 | 38.10 | 38.18 | 38.23 | 38.28 | 38.32 | 38.38 |
| | $K$ | 2.90 | 4.46 | 5.00 | 5.16 | 5.20 | 5.22 | 5.22 | 5.22 | 5.22 | 5.22 | 5.22 | 5.22 | 5.22 | 5.22 | 5.22 | 5.22 | 5.22 | 5.22 | 5.22 | 5.22 | 5.22 | 5.22 | 5.22 | 5.22 | 5.22 |
| 0.1 | $U_1$ | 16.03 | 28.50 | 37.45 | 43.91 | 48.69 | 52.33 | 55.18 | 57.46 | 59.32 | 60.86 | 62.16 | 63.26 | 64.21 | 65.04 | 65.77 | 66.41 | 66.98 | 67.50 | 67.95 | 68.37 | 68.74 | 69.08 | 69.40 | 69.69 | 69.96 |
| | $U_2$ | 0.47 | 1.09 | 1.99 | 3.13 | 4.38 | 5.65 | 6.87 | 8.01 | 9.05 | 9.99 | 10.85 | 11.62 | 12.31 | 12.94 | 13.51 | 14.02 | 14.49 | 14.92 | 15.31 | 15.67 | 16.00 | 16.31 | 16.59 | 16.85 | 17.10 |
| | $W_{1,2}$ | 10.23 | 22.42 | 35.53 | 49.25 | 63.12 | 77.02 | 90.94 | 104.86 | 118.78 | 132.69 | 146.61 | 160.53 | 174.45 | 188.37 | 202.28 | 216.20 | 230.12 | 244.04 | 257.96 | 271.88 | 285.80 | 299.71 | 313.63 | 327.55 | 341.47 |
| | $W_{3,4}$ | 12.72 | 26.08 | 39.82 | 53.69 | 67.57 | 81.50 | 95.42 | 109.33 | 123.25 | 137.17 | 151.09 | 165.00 | 178.92 | 192.84 | 206.76 | 220.67 | 234.59 | 248.52 | 262.44 | 276.36 | 290.28 | 304.19 | 318.11 | 332.03 | 345.95 |
| | $J$ | 11.46 | 19.67 | 25.07 | 28.71 | 31.19 | 32.93 | 34.18 | 35.09 | 35.77 | 36.28 | 36.67 | 36.96 | 37.22 | 37.37 | 37.50 | 37.56 | 37.61 | 37.66 | 37.69 | 37.71 | 37.73 | 37.75 | 37.76 | 37.77 | 37.78 |
| | $K$ | 2.98 | 4.58 | 5.14 | 5.30 | 5.35 | 5.37 | 5.37 | 5.37 | 5.37 | 5.37 | 5.37 | 5.37 | 5.37 | 5.37 | 5.37 | 5.37 | 5.37 | 5.37 | 5.37 | 5.37 | 5.37 | 5.37 | 5.37 | 5.37 | 5.37 |
| 0.0 | $U_1$ | 15.85 | 28.13 | 36.65 | 43.33 | 48.07 | 51.69 | 54.52 | 56.79 | 58.64 | 60.17 | 61.46 | 62.56 | 63.51 | 64.34 | 65.06 | 65.70 | 66.27 | 66.78 | 67.24 | 67.65 | 68.06 | 68.37 | 68.74 | 69.08 | 69.35 |
| | $U_2$ | 0.49 | 1.11 | 2.00 | 3.14 | 4.38 | 5.65 | 6.87 | 8.01 | 9.05 | 9.99 | 10.85 | 11.62 | 12.31 | 12.94 | 13.51 | 14.02 | 14.49 | 14.92 | 15.31 | 15.67 | 16.00 | 16.31 | 16.59 | 16.85 | 17.10 |
| | $W_{1,2}$ | 10.23 | 22.68 | 35.80 | 49.52 | 63.39 | 77.30 | 91.22 | 105.13 | 119.05 | 132.97 | 146.89 | 160.80 | 174.72 | 188.64 | 202.56 | 216.47 | 230.39 | 244.31 | 258.23 | 272.15 | 286.06 | 299.98 | 313.90 | 327.82 | 341.73 |
| | $W_{3,4}$ | 12.72 | 26.08 | 39.79 | 53.62 | 67.44 | 81.35 | 95.26 | 109.18 | 123.10 | 137.02 | 150.93 | 164.85 | 178.77 | 192.69 | 206.61 | 220.52 | 234.44 | 248.36 | 262.28 | 276.19 | 290.11 | 304.03 | 317.95 | 331.87 | 345.78 |
| | $J$ | 11.46 | 19.67 | 25.07 | 28.66 | 31.10 | 32.80 | 34.02 | 34.89 | 35.54 | 36.03 | 36.39 | 36.68 | 36.90 | 37.08 | 37.22 | 37.33 | 37.42 | 37.50 | 37.56 | 37.61 | 37.66 | 37.69 | 37.73 | 37.75 | 37.78 |
| | $K$ | 3.01 | 4.63 | 5.19 | 5.36 | 5.41 | 5.42 | 5.42 | 5.42 | 5.42 | 5.42 | 5.42 | 5.42 | 5.42 | 5.42 | 5.42 | 5.42 | 5.42 | 5.42 | 5.42 | 5.42 | 5.42 | 5.42 | 5.42 | 5.42 | 5.42 |

TABLE S61: Parameters of the THF interaction Hamiltonian for different ratios $u_0/u_1$ and screening lengths $\xi$ of the double-gate interaction potential ($U'_\xi = 24$ meV for $\xi = 10$nm). We use the same BM model parameters as in Table S60.

FIG. S47. Band structures of the BM and THF models near charge neutrality for $w_0/w_1 = 0.8$, depicted by lines and crosses, respectively. The BM bands are colored according to the weight of the $f$-electron wave function on them. We use the same BM parameters as in Table S62.

Figure annotations: Total weight of the heavy-fermions on the active bands after wannierisation: 93.2%. Colorbar: Wannier orbitals support (0.6 – 1.0). $y$-axis: Energy (meV). $x$-axis: $K$ $\Gamma$ $M$ $K$.

TABLE S62. Parameters of the $f$-electron Wannier functions for different values of the tunneling ratio $w_0/w_1$. $W$ denotes the total weight of the THF $f$-electrons on the active bands. In computing the ratios $\gamma/U_1$ and $\gamma/U_2$, we employ the on-site and nearest-neighbor repulsion parameters $U_1$ and $U_2$ obtained numerically for $\xi = 10$ nm, as given in Table S63. We employ $v_F = 5.944\,\text{eV\,Å}$, $|\mathbf{K}| = 1.703\,\text{Å}^{-1}$, $w_1 = 110$ meV, and $\theta = 1.28°$ for the BM model.

| $w_0/w_1$ | $\alpha_1$ | $\alpha_2$ | $\lambda_1$ | $\lambda_2$ | $\lambda$ | $W$ | $t_0$ | $\gamma$ | $M$ | $v_\star$ | $v'_\star$ | $v''_\star$ | $\gamma/U_1$ | $\gamma/U_2$ |
|---|---|---|---|---|---|---|---|---|---|---|---|---|---|---|
| 1.00 | 0.795 | 0.606 | 0.185 | 0.197 | 0.346 | 0.957 | 1.676 | -28.463 | -24.104 | -4.581 | 1.781 | 0.0118 | -0.40 | -7.51 |
| 0.90 | 0.824 | 0.566 | 0.197 | 0.203 | 0.355 | 0.942 | 2.052 | -44.373 | -25.528 | -4.683 | 1.798 | 0.0038 | -0.67 | -11.39 |
| 0.80 | 0.854 | 0.520 | 0.215 | 0.209 | 0.368 | 0.932 | 2.467 | -59.928 | -26.766 | -4.810 | 1.789 | -0.0011 | -0.95 | -14.89 |
| 0.70 | 0.883 | 0.469 | 0.227 | 0.215 | 0.386 | 0.928 | 2.893 | -74.891 | -27.829 | -4.927 | 1.740 | -0.0037 | -1.27 | -17.87 |
| 0.60 | 0.911 | 0.412 | 0.245 | 0.221 | 0.405 | 0.928 | 3.356 | -89.005 | -28.727 | -5.037 | 1.667 | -0.0046 | -1.61 | -20.42 |
| 0.50 | 0.936 | 0.351 | 0.257 | 0.227 | 0.421 | 0.929 | 3.800 | -101.976 | -29.470 | -5.140 | 1.540 | -0.0044 | -1.95 | -22.59 |
| 0.40 | 0.958 | 0.287 | 0.275 | 0.233 | 0.434 | 0.931 | 4.188 | -113.459 | -30.067 | -5.235 | 1.356 | -0.0035 | -2.28 | -24.47 |
| 0.30 | 0.976 | 0.220 | 0.286 | 0.233 | 0.441 | 0.931 | 4.445 | -123.058 | -30.525 | -5.319 | 1.107 | -0.0023 | -2.54 | -26.14 |
| 0.20 | 0.989 | 0.157 | 0.295 | 0.239 | 0.438 | 0.925 | 4.453 | -130.349 | -30.848 | -5.387 | 0.789 | -0.0011 | -2.70 | -27.84 |
| 0.10 | 0.997 | 0.077 | 0.304 | 0.239 | 0.433 | 0.919 | 4.398 | -134.925 | -31.040 | -5.432 | 0.412 | -0.0003 | -2.78 | -29.03 |
| 0.00 | 1.000 | 0.000 | 0.304 | 0.000 | 0.423 | 0.911 | 4.203 | -136.488 | -31.104 | -5.448 | -0.000 | 0.0000 | -2.76 | -29.93 |

TABLE S63. $f$-electron single-particle Hamiltonian and repulsion parameters (in meV) as a function of $\xi$.

| $w_0/w_1$ | $\xi/\text{nm}$ | 2 | 4 | 6 | 8 | 10 | 12 | 14 | 16 | 18 | 20 | 22 | 24 | 26 | 28 | 30 | 32 | 34 | 36 | 38 | 40 | 42 | 44 | 46 | 48 | 50 |
|---|---|---|---|---|---|---|---|---|---|---|---|---|---|---|---|---|---|---|---|---|---|---|---|---|---|---|
| 1.0 | $U_1$ | 26.63 | 45.03 | 56.90 | 64.86 | 70.48 | 74.61 | 77.77 | 80.24 | 82.23 | 83.87 | 85.23 | 86.38 | 87.37 | 88.23 | 88.97 | 89.63 | 90.21 | 90.74 | 91.20 | 91.63 | 92.01 | 92.36 | 92.68 | 92.98 | 93.25 |
| | $U_2$ | 0.31 | 0.75 | 1.49 | 2.54 | 3.79 | 5.09 | 6.35 | 7.53 | 8.61 | 9.60 | 10.48 | 11.28 | 11.99 | 12.64 | 13.22 | 13.75 | 14.23 | 14.67 | 15.07 | 15.43 | 15.77 | 16.08 | 16.37 | 16.63 | 16.88 |
| | $W_{1,2}$ | 12.41 | 25.77 | 39.79 | 54.06 | 68.40 | 82.75 | 97.11 | 111.48 | 125.84 | 140.21 | 154.57 | 168.93 | 183.30 | 197.66 | 212.02 | 226.39 | 240.75 | 255.12 | 269.48 | 283.84 | 298.21 | 312.57 | 326.93 | 341.30 | 355.66 |
| | $W_{3,4}$ | 14.15 | 28.65 | 43.09 | 57.48 | 71.85 | 86.22 | 100.58 | 114.95 | 129.31 | 143.67 | 158.04 | 172.40 | 186.76 | 201.13 | 215.49 | 229.86 | 244.22 | 258.59 | 272.95 | 287.31 | 301.68 | 316.04 | 330.40 | 344.77 | 359.13 |
| | $J$ | 9.26 | 14.23 | 16.68 | 18.00 | 18.77 | 19.26 | 19.59 | 19.81 | 19.97 | 20.09 | 20.18 | 20.24 | 20.29 | 20.33 | 20.36 | 20.39 | 20.40 | 20.42 | 20.43 | 20.44 | 20.45 | 20.46 | 20.47 | 20.47 | 20.48 |
| | $K$ | 2.11 | 3.24 | 3.63 | 3.74 | 3.77 | 3.78 | 3.78 | 3.78 | 3.78 | 3.78 | 3.78 | 3.78 | 3.78 | 3.78 | 3.78 | 3.78 | 3.78 | 3.78 | 3.78 | 3.78 | 3.78 | 3.78 | 3.78 | 3.78 | 3.78 |
| 0.9 | $U_1$ | 24.60 | 41.96 | 53.36 | 61.11 | 66.91 | 70.68 | 73.79 | 76.25 | 78.22 | 79.85 | 81.20 | 82.35 | 83.33 | 84.19 | 84.93 | 85.59 | 86.17 | 86.69 | 87.16 | 87.58 | 87.96 | 88.31 | 88.63 | 88.93 | 89.20 |
| | $U_2$ | 0.33 | 0.79 | 1.56 | 2.64 | 3.89 | 5.20 | 6.46 | 7.64 | 8.72 | 9.70 | 10.58 | 11.38 | 12.09 | 12.74 | 13.32 | 13.85 | 14.32 | 14.76 | 15.16 | 15.53 | 15.86 | 16.17 | 16.46 | 16.73 | 16.97 |
| | $W_{1,2}$ | 12.34 | 25.67 | 39.68 | 53.95 | 68.28 | 82.64 | 97.00 | 111.36 | 125.73 | 140.09 | 154.46 | 168.82 | 183.18 | 197.55 | 211.91 | 226.28 | 240.64 | 255.00 | 269.37 | 283.73 | 298.09 | 312.46 | 326.82 | 341.19 | 355.55 |
| | $W_{3,4}$ | 13.79 | 28.07 | 42.43 | 56.79 | 71.16 | 85.52 | 99.88 | 114.25 | 128.61 | 142.98 | 157.34 | 171.70 | 186.07 | 200.43 | 214.79 | 229.16 | 243.52 | 257.89 | 272.25 | 286.61 | 300.98 | 315.34 | 329.71 | 344.07 | 358.43 |
| | $J$ | 9.50 | 14.96 | 17.85 | 19.51 | 20.52 | 21.19 | 21.63 | 21.95 | 22.19 | 22.33 | 22.45 | 22.55 | 22.62 | 22.67 | 22.71 | 22.75 | 22.77 | 22.80 | 22.83 | 22.84 | 22.85 | 22.86 | 22.87 | 22.87 | 22.87 |
| | $K$ | 2.13 | 3.28 | 3.67 | 3.78 | 3.81 | 3.82 | 3.83 | 3.83 | 3.83 | 3.83 | 3.83 | 3.83 | 3.83 | 3.83 | 3.83 | 3.83 | 3.83 | 3.83 | 3.83 | 3.83 | 3.83 | 3.83 | 3.83 | 3.83 | 3.83 |
| 0.8 | $U_1$ | 22.68 | 39.00 | 49.91 | 57.41 | 62.78 | 66.78 | 69.85 | 72.27 | 74.23 | 75.84 | 77.18 | 78.32 | 79.30 | 80.15 | 80.89 | 81.54 | 82.12 | 82.64 | 83.11 | 83.53 | 83.91 | 84.26 | 84.58 | 84.87 | 85.14 |
| | $U_2$ | 0.36 | 0.86 | 1.66 | 2.76 | 4.03 | 5.33 | 6.59 | 7.77 | 8.85 | 9.82 | 10.70 | 11.49 | 12.20 | 12.85 | 13.43 | 13.96 | 14.43 | 14.87 | 15.27 | 15.63 | 15.97 | 16.28 | 16.56 | 16.83 | 17.08 |
| | $W_{1,2}$ | 12.20 | 25.47 | 39.46 | 53.72 | 68.05 | 82.41 | 96.77 | 111.13 | 125.50 | 139.86 | 154.22 | 168.59 | 182.95 | 197.31 | 211.68 | 226.04 | 240.41 | 254.77 | 269.13 | 283.50 | 297.86 | 312.22 | 326.59 | 340.95 | 355.32 |
| | $W_{3,4}$ | 13.58 | 27.72 | 42.08 | 56.44 | 70.80 | 85.16 | 99.52 | 113.88 | 128.24 | 142.61 | 156.97 | 171.33 | 185.69 | 200.05 | 214.42 | 228.78 | 243.14 | 257.50 | 271.86 | 286.23 | 300.59 | 314.95 | 329.31 | 343.68 | 358.01 |
| | $J$ | 9.80 | 15.76 | 19.12 | 21.12 | 22.40 | 23.24 | 23.82 | 24.23 | 24.53 | 24.75 | 24.91 | 25.03 | 25.12 | 25.20 | 25.26 | 25.30 | 25.34 | 25.37 | 25.39 | 25.41 | 25.43 | 25.44 | 25.46 | 25.47 | 25.48 |
| | $K$ | 2.19 | 3.37 | 3.77 | 3.88 | 3.91 | 3.92 | 3.93 | 3.93 | 3.93 | 3.93 | 3.93 | 3.93 | 3.93 | 3.93 | 3.93 | 3.93 | 3.93 | 3.93 | 3.93 | 3.93 | 3.93 | 3.93 | 3.93 | 3.93 | 3.93 |
| 0.7 | $U_1$ | 20.81 | 36.07 | 46.44 | 53.66 | 58.87 | 62.77 | 65.78 | 68.17 | 70.10 | 71.69 | 73.02 | 74.15 | 75.12 | 75.96 | 76.70 | 77.35 | 77.93 | 78.44 | 78.91 | 79.33 | 79.71 | 80.06 | 80.38 | 80.67 | 80.94 |
| | $U_2$ | 0.40 | 0.95 | 1.79 | 2.92 | 4.19 | 5.50 | 6.75 | 7.93 | 9.00 | 9.97 | 10.85 | 11.63 | 12.34 | 12.98 | 13.56 | 14.08 | 14.56 | 15.00 | 15.39 | 15.76 | 16.09 | 16.40 | 16.69 | 16.95 | 17.20 |
| | $W_{1,2}$ | 12.03 | 25.27 | 39.17 | 53.42 | 67.75 | 82.11 | 96.47 | 110.83 | 125.19 | 139.56 | 153.93 | 168.29 | 182.65 | 197.01 | 211.38 | 225.74 | 240.10 | 254.47 | 268.84 | 283.20 | 297.56 | 311.93 | 326.29 | 340.66 | 355.02 |
| | $W_{3,4}$ | 13.38 | 27.37 | 41.63 | 55.97 | 70.32 | 84.68 | 99.05 | 113.41 | 127.77 | 142.14 | 156.50 | 170.87 | 185.23 | 199.60 | 213.96 | 228.32 | 242.68 | 257.05 | 271.41 | 285.78 | 300.14 | 314.50 | 328.87 | 343.23 | 357.59 |
| | $J$ | 10.15 | 16.56 | 20.66 | 23.16 | 24.80 | 25.90 | 26.64 | 27.13 | 27.46 | 27.68 | 27.83 | 27.94 | 28.03 | 28.11 | 28.19 | 28.23 | 28.27 | 28.28 | 28.33 | 28.36 | 28.38 | 28.39 | 28.40 | 28.41 | 28.42 |
| | $K$ | 2.28 | 3.50 | 3.91 | 4.03 | 4.06 | 4.07 | 4.08 | 4.08 | 4.08 | 4.08 | 4.08 | 4.08 | 4.08 | 4.08 | 4.08 | 4.08 | 4.08 | 4.08 | 4.08 | 4.08 | 4.08 | 4.08 | 4.08 | 4.08 | 4.08 |

| $u_0/u_1$ | $\xi$/nm | 2 | 4 | 6 | 8 | 10 | 12 | 14 | 16 | 18 | 20 | 22 | 24 | 26 | 28 | 30 | 32 | 34 | 36 | 38 | 40 | 42 | 44 | 46 | 48 | 50 |
|---|---|---|---|---|---|---|---|---|---|---|---|---|---|---|---|---|---|---|---|---|---|---|---|---|---|---|
| 0.6 | $U_1$ | 19.16 | 33.45 | 43.32 | 50.26 | 55.31 | 59.12 | 62.07 | 64.41 | 66.31 | 67.89 | 69.20 | 70.32 | 71.29 | 72.12 | 72.86 | 73.50 | 74.08 | 74.59 | 75.05 | 75.47 | 75.85 | 76.20 | 76.52 | 76.81 | 77.08 |
| | $U_2$ | 0.45 | 1.05 | 1.93 | 3.08 | 4.36 | 5.66 | 6.92 | 8.08 | 9.15 | 10.12 | 10.99 | 11.77 | 12.48 | 13.12 | 13.69 | 14.21 | 14.69 | 15.12 | 15.52 | 15.88 | 16.21 | 16.52 | 16.81 | 17.07 | 17.32 |
| | $W_{3,4}$ | 11.83 | 24.91 | 38.84 | 53.08 | 67.41 | 81.77 | 96.13 | 110.49 | 124.86 | 139.22 | 153.58 | 167.95 | 182.31 | 196.67 | 211.04 | 225.40 | 239.77 | 254.13 | 268.49 | 282.86 | 297.22 | 311.59 | 325.95 | 340.31 | 354.68 |
| | $W_{3,4}$ | 13.28 | 27.20 | 41.43 | 55.76 | 70.11 | 84.47 | 98.83 | 113.20 | 127.56 | 141.92 | 156.29 | 170.65 | 185.02 | 199.38 | 213.74 | 228.11 | 242.47 | 256.83 | 271.20 | 285.56 | 299.93 | 314.29 | 328.65 | 343.02 | 357.38 |
| | $J$ | 10.52 | 17.50 | 21.82 | 24.59 | 26.44 | 27.71 | 28.61 | 29.27 | 29.75 | 30.11 | 30.39 | 30.60 | 30.77 | 30.90 | 31.00 | 31.09 | 31.16 | 31.21 | 31.26 | 31.30 | 31.33 | 31.36 | 31.38 | 31.40 | 31.42 |
| | $K$ | 2.39 | 3.66 | 4.09 | 4.22 | 4.25 | 4.26 | 4.26 | 4.26 | 4.26 | 4.26 | 4.26 | 4.26 | 4.26 | 4.26 | 4.26 | 4.26 | 4.26 | 4.26 | 4.26 | 4.26 | 4.26 | 4.26 | 4.26 | 4.26 | 4.26 |
| 0.5 | $U_1$ | 17.78 | 31.22 | 40.64 | 47.33 | 52.24 | 55.96 | 58.85 | 61.16 | 63.03 | 64.59 | 65.89 | 67.00 | 67.96 | 68.79 | 69.52 | 70.16 | 70.73 | 71.24 | 71.70 | 72.12 | 72.50 | 72.84 | 73.16 | 73.45 | 73.72 |
| | $U_2$ | 0.49 | 1.14 | 2.06 | 3.22 | 4.51 | 5.82 | 7.07 | 8.23 | 9.29 | 10.25 | 11.12 | 11.90 | 12.60 | 13.24 | 13.81 | 14.33 | 14.81 | 15.25 | 15.63 | 15.99 | 16.32 | 16.63 | 16.91 | 17.18 | 17.42 |
| | $W_{1,2}$ | 11.62 | 24.59 | 38.49 | 52.72 | 67.05 | 81.40 | 95.76 | 110.13 | 124.49 | 138.85 | 153.22 | 167.58 | 181.94 | 196.31 | 210.67 | 225.03 | 239.40 | 253.76 | 268.13 | 282.49 | 296.86 | 311.22 | 325.58 | 339.95 | 354.31 |
| | $W_{3,4}$ | 13.22 | 27.10 | 41.31 | 55.63 | 69.99 | 84.35 | 98.71 | 113.07 | 127.44 | 141.80 | 156.16 | 170.53 | 184.89 | 199.26 | 213.62 | 227.98 | 242.35 | 256.71 | 271.08 | 285.44 | 299.80 | 314.17 | 328.53 | 342.89 | 357.26 |
| | $J$ | 10.88 | 18.35 | 23.11 | 26.23 | 28.35 | 29.83 | 30.89 | 31.67 | 32.24 | 32.68 | 33.02 | 33.28 | 33.48 | 33.64 | 33.77 | 33.88 | 33.96 | 34.03 | 34.09 | 34.14 | 34.18 | 34.22 | 34.25 | 34.27 | 34.30 |
| | $K$ | 2.52 | 3.85 | 4.31 | 4.44 | 4.47 | 4.48 | 4.49 | 4.49 | 4.49 | 4.49 | 4.49 | 4.49 | 4.49 | 4.49 | 4.49 | 4.49 | 4.49 | 4.49 | 4.49 | 4.49 | 4.49 | 4.49 | 4.49 | 4.49 | 4.49 |
| 0.4 | $U_1$ | 16.72 | 29.51 | 38.58 | 45.07 | 49.87 | 53.51 | 56.36 | 58.64 | 60.49 | 62.03 | 63.33 | 64.43 | 65.38 | 66.20 | 66.93 | 67.57 | 68.14 | 68.65 | 69.11 | 69.52 | 69.90 | 70.24 | 70.56 | 70.85 | 71.12 |
| | $U_2$ | 0.53 | 1.21 | 2.16 | 3.34 | 4.64 | 5.94 | 7.19 | 8.34 | 9.40 | 10.36 | 11.22 | 12.00 | 12.70 | 13.33 | 13.90 | 14.42 | 14.89 | 15.32 | 15.71 | 16.07 | 16.41 | 16.71 | 17.00 | 17.26 | 17.50 |
| | $W_{1,2}$ | 11.40 | 24.27 | 38.07 | 52.34 | 66.67 | 81.02 | 95.38 | 109.75 | 124.11 | 138.47 | 152.84 | 167.20 | 181.56 | 195.93 | 210.29 | 224.66 | 239.02 | 253.38 | 267.75 | 282.11 | 296.47 | 310.84 | 325.20 | 339.57 | 353.93 |
| | $W_{3,4}$ | 13.20 | 27.05 | 41.25 | 55.57 | 69.92 | 84.28 | 98.64 | 113.00 | 127.37 | 141.73 | 156.10 | 170.46 | 184.82 | 199.19 | 213.55 | 227.91 | 242.28 | 256.64 | 271.01 | 285.37 | 299.73 | 314.10 | 328.46 | 342.82 | 357.19 |
| | $J$ | 11.21 | 19.09 | 24.22 | 27.64 | 29.99 | 31.64 | 32.83 | 33.71 | 34.37 | 34.87 | 35.25 | 35.55 | 35.79 | 35.98 | 36.13 | 36.25 | 36.35 | 36.43 | 36.50 | 36.56 | 36.61 | 36.65 | 36.69 | 36.72 | 36.74 |
| | $K$ | 2.66 | 4.04 | 4.54 | 4.68 | 4.72 | 4.73 | 4.73 | 4.73 | 4.73 | 4.73 | 4.73 | 4.73 | 4.73 | 4.73 | 4.73 | 4.73 | 4.73 | 4.73 | 4.73 | 4.73 | 4.73 | 4.73 | 4.73 | 4.73 | 4.73 |
| 0.3 | $U_1$ | 16.05 | 28.44 | 37.29 | 43.66 | 48.39 | 51.99 | 54.81 | 57.07 | 58.91 | 60.44 | 61.73 | 62.83 | 63.78 | 64.60 | 65.32 | 65.96 | 66.53 | 67.04 | 67.49 | 67.91 | 68.29 | 68.63 | 68.94 | 69.23 | 69.50 |
| | $U_2$ | 0.56 | 1.25 | 2.22 | 3.41 | 4.71 | 6.01 | 7.26 | 8.41 | 9.46 | 10.42 | 11.28 | 12.05 | 12.75 | 13.38 | 13.95 | 14.47 | 14.94 | 15.37 | 15.76 | 16.11 | 16.45 | 16.75 | 17.05 | 17.31 | 17.55 |
| | $W_{1,2}$ | 11.17 | 23.92 | 37.75 | 51.96 | 66.28 | 80.63 | 94.99 | 109.36 | 123.72 | 138.09 | 152.45 | 166.81 | 181.17 | 195.54 | 209.90 | 224.27 | 238.63 | 252.99 | 267.36 | 281.72 | 296.09 | 310.45 | 324.81 | 339.18 | 353.54 |
| | $W_{3,4}$ | 13.18 | 27.01 | 41.21 | 55.52 | 69.87 | 84.23 | 98.59 | 112.96 | 127.32 | 141.69 | 156.05 | 170.41 | 184.78 | 199.14 | 213.50 | 227.87 | 242.23 | 256.59 | 270.96 | 285.32 | 299.69 | 314.05 | 328.41 | 342.78 | 357.14 |
| | $J$ | 11.65 | 19.99 | 25.60 | 29.20 | 31.73 | 33.52 | 34.80 | 35.74 | 36.46 | 36.98 | 37.39 | 37.70 | 37.95 | 38.15 | 38.31 | 38.44 | 38.55 | 38.64 | 38.71 | 38.77 | 38.82 | 38.87 | 38.90 | 38.94 | 38.96 |
| | $K$ | 2.79 | 4.27 | 4.77 | 4.92 | 4.96 | 4.97 | 4.97 | 4.97 | 4.97 | 4.97 | 4.97 | 4.97 | 4.97 | 4.97 | 4.97 | 4.97 | 4.97 | 4.97 | 4.97 | 4.97 | 4.97 | 4.97 | 4.97 | 4.97 | 4.97 |
| 0.2 | $U_1$ | 16.00 | 28.42 | 37.32 | 43.75 | 48.52 | 52.14 | 54.76 | 57.03 | 58.87 | 60.40 | 61.70 | 62.80 | 63.74 | 64.57 | 65.29 | 65.93 | 66.50 | 67.01 | 67.47 | 67.88 | 68.26 | 68.60 | 68.92 | 69.21 | 69.48 |
| | $U_2$ | 0.53 | 1.23 | 2.19 | 3.38 | 4.68 | 5.99 | 7.24 | 8.39 | 9.45 | 10.40 | 11.26 | 12.04 | 12.74 | 13.37 | 13.95 | 14.46 | 14.91 | 15.34 | 15.73 | 16.09 | 16.43 | 16.73 | 17.02 | 17.28 | 17.52 |
| | $W_{1,2}$ | 11.13 | 23.72 | 37.54 | 51.74 | 66.06 | 80.41 | 94.78 | 109.14 | 123.50 | 137.86 | 152.23 | 166.59 | 180.95 | 195.32 | 209.68 | 224.04 | 238.41 | 252.77 | 267.13 | 281.50 | 295.86 | 310.22 | 324.59 | 338.95 | 353.12 |
| | $W_{3,4}$ | 13.15 | 26.96 | 41.15 | 55.46 | 69.81 | 84.17 | 98.54 | 112.90 | 127.26 | 141.63 | 155.99 | 170.36 | 184.72 | 199.08 | 213.45 | 227.81 | 242.17 | 256.54 | 270.90 | 285.27 | 299.63 | 313.99 | 328.36 | 342.72 | 357.09 |
| | $J$ | 11.76 | 20.25 | 25.87 | 29.57 | 32.11 | 33.90 | 35.18 | 36.12 | 36.84 | 37.37 | 37.77 | 38.08 | 38.33 | 38.52 | 38.69 | 38.82 | 38.93 | 39.02 | 39.10 | 39.16 | 39.22 | 39.26 | 39.30 | 39.33 | 39.36 |
| | $K$ | 2.91 | 4.45 | 4.98 | 5.13 | 5.17 | 5.18 | 5.18 | 5.18 | 5.18 | 5.18 | 5.18 | 5.18 | 5.18 | 5.18 | 5.18 | 5.18 | 5.18 | 5.18 | 5.18 | 5.18 | 5.18 | 5.18 | 5.18 | 5.18 | 5.18 |
| 0.1 | $U_1$ | 16.16 | 28.66 | 37.69 | 44.43 | 49.32 | 53.03 | 55.94 | 58.25 | 60.13 | 61.68 | 62.99 | 64.11 | 65.07 | 65.91 | 66.65 | 67.30 | 67.89 | 68.41 | 68.89 | 69.32 | 69.71 | 70.07 | 70.39 | 70.70 | 70.98 |
| | $U_2$ | 0.51 | 1.20 | 2.15 | 3.33 | 4.62 | 5.93 | 7.18 | 8.33 | 9.39 | 10.35 | 11.21 | 11.99 | 12.69 | 13.32 | 13.90 | 14.42 | 14.89 | 15.32 | 15.71 | 16.07 | 16.41 | 16.72 | 17.01 | 17.27 | 17.52 |
| | $W_{1,2}$ | 10.74 | 23.28 | 37.04 | 51.23 | 65.54 | 79.89 | 94.25 | 108.62 | 122.98 | 137.35 | 151.71 | 166.07 | 180.44 | 194.80 | 209.16 | 223.53 | 237.89 | 252.26 | 266.62 | 280.98 | 295.35 | 309.71 | 324.07 | 338.44 | 352.80 |
| | $W_{3,4}$ | 13.13 | 26.93 | 41.11 | 55.42 | 69.77 | 84.13 | 98.50 | 112.86 | 127.22 | 141.59 | 155.95 | 170.31 | 184.68 | 199.04 | 213.40 | 227.77 | 242.13 | 256.50 | 270.86 | 285.22 | 299.59 | 313.95 | 328.31 | 342.68 | 357.04 |
| | $J$ | 11.76 | 20.19 | 25.73 | 29.39 | 31.89 | 33.64 | 34.88 | 35.79 | 36.47 | 36.98 | 37.37 | 37.67 | 37.91 | 38.09 | 38.25 | 38.37 | 38.48 | 38.56 | 38.63 | 38.69 | 38.74 | 38.78 | 38.82 | 38.85 | 38.88 |
| | $K$ | 3.00 | 4.58 | 5.12 | 5.27 | 5.31 | 5.33 | 5.33 | 5.33 | 5.33 | 5.33 | 5.33 | 5.33 | 5.33 | 5.33 | 5.33 | 5.33 | 5.33 | 5.33 | 5.33 | 5.33 | 5.33 | 5.33 | 5.33 | 5.33 | 5.33 |
| 0.0 | $U_1$ | 16.38 | 29.07 | 38.15 | 44.68 | 49.51 | 53.18 | 56.05 | 58.34 | 60.21 | 61.76 | 63.06 | 64.17 | 65.12 | 65.95 | 66.68 | 67.32 | 67.89 | 68.40 | 68.86 | 69.28 | 69.66 | 70.00 | 70.32 | 70.61 | 70.88 |
| | $U_2$ | 0.48 | 1.14 | 2.08 | 3.26 | 4.56 | 5.87 | 7.12 | 8.28 | 9.34 | 10.30 | 11.17 | 11.95 | 12.65 | 13.28 | 13.86 | 14.38 | 14.85 | 15.28 | 15.67 | 16.04 | 16.37 | 16.68 | 16.96 | 17.22 | 17.47 |
| | $W_{1,2}$ | 10.62 | 23.08 | 36.43 | 51.01 | 65.32 | 79.68 | 94.04 | 108.40 | 122.76 | 137.13 | 151.49 | 165.85 | 180.22 | 194.58 | 208.94 | 223.31 | 237.67 | 252.04 | 266.40 | 280.76 | 295.13 | 309.49 | 323.85 | 338.22 | 352.58 |
| | $W_{3,4}$ | 13.11 | 26.88 | 41.06 | 55.37 | 69.72 | 84.08 | 98.44 | 112.81 | 127.17 | 141.53 | 155.90 | 170.26 | 184.62 | 198.99 | 213.35 | 227.72 | 242.08 | 256.44 | 270.81 | 285.17 | 299.53 | 313.90 | 328.26 | 342.63 | 356.99 |
| | $J$ | 11.76 | 20.15 | 25.64 | 29.27 | 31.72 | 33.43 | 34.64 | 35.51 | 36.15 | 36.63 | 36.99 | 37.27 | 37.49 | 37.67 | 37.80 | 37.91 | 38.00 | 38.08 | 38.14 | 38.19 | 38.23 | 38.27 | 38.30 | 38.32 | 38.35 |
| | $K$ | 3.03 | 4.62 | 5.17 | 5.32 | 5.37 | 5.38 | 5.38 | 5.38 | 5.38 | 5.38 | 5.38 | 5.38 | 5.38 | 5.38 | 5.38 | 5.38 | 5.38 | 5.38 | 5.38 | 5.38 | 5.38 | 5.38 | 5.38 | 5.38 | 5.38 |

TABLE S63: Parameters of the THF interaction Hamiltonian for different ratios $u_0/u_1$ and screening lengths $\xi$ of the double-gate interaction potential ($U_\xi = 24$ meV for $\xi = 10$ nm). We use the same BM model parameters as in Table S62.

Total weight of the heavy-fermions on the active bands after wannierisation: 93.0%

Wannier orbitals support 0.0 — 1.0

FIG. S48. Band structures of the BM and THF models near charge neutrality for $v_0/v_1 = 0.8$, depicted by lines and crosses, respectively. The BM bands are colored according to the weight of the $f$-electron wave function on them. We use the same BM parameters as in Table S64.

TABLE S64. Parameters of the $f$-electron Wannier functions and the THF single-particle Hamiltonian for different values of the tunneling ratio $v_0/v_1$. $W$ denotes the total weight of the THF $f$-electrons on the active bands. In computing the ratios $\gamma/U_1$ and $\gamma/U_2$, we employ the on-site and nearest-neighbor repulsion parameters $U_1$ and $U_2$ obtained numerically for $\xi = 10$ nm, as given in Table S65. We employ $v_F = 5.944$ eV·Å, $|\mathbf{K}| = 1.703$ Å$^{-1}$, $w_1 = 110$ meV, and $\theta = 1.30°$ for the BM model.

| $v_0/v_1$ | $\alpha_1$ | $\alpha_2$ | $\lambda_1$ | $\lambda_2$ | $\lambda$ | $W$ | $t_0$ | $\gamma$ | $M$ | $v_\star$ | $v_\star'$ | $v_\star''$ | $\gamma/U_1$ | $\gamma/U_2$ |
|---|---|---|---|---|---|---|---|---|---|---|---|---|---|---|
| 1.00 | 0.798 | 0.602 | 0.191 | 0.203 | 0.349 | 0.954 | 1.913 | -31.571 | -26.947 | -4.579 | 1.797 | 0.0139 | -0.45 | -7.94 |
| 0.90 | 0.827 | 0.562 | 0.203 | 0.209 | 0.357 | 0.940 | 2.335 | -47.547 | -28.409 | -4.714 | 1.811 | 0.0064 | -0.71 | -11.65 |
| 0.80 | 0.857 | 0.516 | 0.215 | 0.215 | 0.370 | 0.930 | 2.785 | -63.142 | -29.678 | -4.838 | 1.798 | 0.0018 | -1.00 | -14.98 |
| 0.70 | 0.886 | 0.464 | 0.227 | 0.221 | 0.388 | 0.927 | 3.247 | -78.126 | -30.768 | -4.952 | 1.754 | -0.0009 | -1.31 | -17.84 |
| 0.60 | 0.913 | 0.407 | 0.245 | 0.221 | 0.413 | 0.931 | 3.813 | -92.247 | -31.688 | -5.059 | 1.673 | -0.0021 | -1.67 | -20.08 |
| 0.50 | 0.938 | 0.347 | 0.263 | 0.227 | 0.427 | 0.932 | 4.291 | -105.215 | -32.450 | -5.161 | 1.544 | -0.0023 | -2.01 | -22.21 |
| 0.40 | 0.959 | 0.284 | 0.275 | 0.233 | 0.435 | 0.931 | 4.638 | -116.689 | -33.062 | -5.255 | 1.358 | -0.0020 | -2.31 | -24.20 |
| 0.30 | 0.976 | 0.218 | 0.286 | 0.239 | 0.442 | 0.931 | 4.913 | -126.275 | -33.531 | -5.338 | 1.108 | -0.0013 | -2.57 | -25.81 |
| 0.20 | 0.989 | 0.149 | 0.296 | 0.239 | 0.438 | 0.924 | 4.923 | -133.552 | -33.862 | -5.406 | 0.789 | -0.0007 | -2.72 | -27.45 |
| 0.10 | 0.997 | 0.076 | 0.304 | 0.239 | 0.440 | 0.923 | 4.934 | -138.119 | -34.060 | -5.451 | 0.412 | -0.0002 | -2.84 | -28.31 |
| 0.00 | 1.000 | 0.000 | 0.304 | 0.000 | 0.434 | 0.918 | 4.729 | -139.678 | -34.125 | -5.466 | -0.0000 | -0.0000 | -2.83 | -29.03 |

| $v_0/v_1$ | $\xi$/nm | 2 | 4 | 6 | 8 | 10 | 12 | 14 | 16 | 18 | 20 | 22 | 24 | 26 | 28 | 30 | 32 | 34 | 36 | 38 | 40 | 42 | 44 | 46 | 48 | 50 |
|---|---|---|---|---|---|---|---|---|---|---|---|---|---|---|---|---|---|---|---|---|---|---|---|---|---|---|
| 1.0 | $U_1$ | 26.80 | 45.28 | 57.20 | 65.19 | 70.82 | 74.96 | 78.12 | 80.60 | 82.60 | 84.23 | 85.60 | 86.75 | 87.74 | 88.60 | 89.34 | 90.00 | 90.59 | 91.11 | 91.58 | 92.00 | 92.38 | 92.73 | 93.05 | 93.35 | 93.62 |
| | $U_2$ | 0.32 | 0.79 | 1.57 | 2.68 | 3.98 | 5.42 | 6.61 | 7.82 | 8.92 | 9.92 | 10.81 | 11.62 | 12.34 | 13.00 | 13.58 | 14.12 | 14.60 | 15.04 | 15.44 | 15.81 | 16.15 | 16.46 | 16.75 | 17.01 | 17.26 |
| | $W_{1,2}$ | 12.89 | 26.74 | 41.23 | 55.96 | 70.75 | 85.56 | 100.37 | 115.19 | 130.01 | 144.82 | 159.64 | 174.45 | 189.27 | 204.09 | 218.90 | 233.72 | 248.54 | 263.35 | 278.17 | 292.98 | 307.80 | 322.62 | 337.43 | 352.25 | 367.06 |
| | $W_{3,4}$ | 14.46 | 29.33 | 44.20 | 59.03 | 73.85 | 88.67 | 103.48 | 118.30 | 133.12 | 147.93 | 162.75 | 177.57 | 192.38 | 207.20 | 222.01 | 236.83 | 251.64 | 266.46 | 281.28 | 296.09 | 310.91 | 325.73 | 340.54 | 355.36 | 370.18 |
| | $J$ | 9.47 | 14.57 | 17.09 | 18.46 | 19.27 | 19.78 | 20.12 | 20.36 | 20.53 | 20.65 | 20.74 | 20.81 | 20.86 | 20.90 | 20.93 | 20.96 | 20.98 | 21.00 | 21.01 | 21.02 | 21.03 | 21.04 | 21.04 | 21.05 | 21.05 |
| | $K$ | 2.08 | 3.18 | 3.55 | 3.45 | 3.68 | 3.69 | 3.69 | 3.69 | 3.69 | 3.69 | 3.69 | 3.69 | 3.69 | 3.69 | 3.69 | 3.69 | 3.75 | 3.75 | 3.75 | 3.75 | 3.75 | 3.75 | 3.69 | 3.69 | 3.69 |
| 0.9 | $U_1$ | 24.82 | 43.30 | 53.77 | 61.55 | 67.08 | 71.16 | 74.29 | 76.75 | 78.73 | 80.35 | 81.71 | 82.86 | 83.85 | 84.70 | 85.44 | 86.10 | 86.68 | 87.21 | 87.67 | 88.10 | 88.48 | 88.83 | 89.15 | 89.44 | 89.71 |
| | $U_2$ | 0.34 | 0.83 | 1.65 | 2.78 | 4.08 | 5.42 | 6.72 | 7.92 | 9.02 | 10.02 | 10.91 | 11.72 | 12.44 | 13.09 | 13.68 | 14.21 | 14.69 | 15.13 | 15.53 | 15.90 | 16.23 | 16.55 | 16.83 | 17.10 | 17.35 |
| | $W_{1,2}$ | 12.80 | 26.61 | 41.09 | 55.81 | 70.60 | 85.41 | 100.23 | 115.04 | 129.86 | 144.68 | 159.49 | 174.31 | 189.13 | 203.94 | 218.76 | 233.57 | 248.39 | 263.21 | 278.02 | 292.84 | 307.65 | 322.47 | 337.29 | 352.10 | 366.92 |
| | $W_{3,4}$ | 14.12 | 28.79 | 43.58 | 58.39 | 73.20 | 88.02 | 102.83 | 117.65 | 132.47 | 147.28 | 162.10 | 176.91 | 191.73 | 206.55 | 221.36 | 236.18 | 250.99 | 265.81 | 280.63 | 295.45 | 310.26 | 325.08 | 339.89 | 354.71 | 369.52 |
| | $J$ | 9.74 | 15.33 | 18.31 | 20.02 | 21.07 | 21.75 | 22.21 | 22.53 | 22.76 | 22.93 | 23.05 | 23.14 | 23.22 | 23.27 | 23.31 | 23.35 | 23.38 | 23.40 | 23.42 | 23.43 | 23.44 | 23.46 | 23.46 | 23.47 | 23.48 |
| | $K$ | 2.12 | 3.23 | 3.61 | 3.71 | 3.74 | 3.75 | 3.75 | 3.75 | 3.75 | 3.75 | 3.75 | 3.75 | 3.75 | 3.75 | 3.75 | 3.75 | 3.75 | 3.75 | 3.75 | 3.75 | 3.75 | 3.75 | 3.75 | 3.75 | 3.75 |
| 0.8 | $U_1$ | 22.92 | 39.37 | 50.36 | 57.00 | 63.30 | 67.31 | 70.38 | 72.81 | 74.78 | 76.39 | 77.74 | 78.91 | 79.91 | 80.71 | 81.45 | 82.10 | 82.69 | 83.20 | 83.67 | 84.09 | 84.48 | 84.82 | 85.14 | 85.44 | 85.71 |
| | $U_2$ | 0.37 | 0.94 | 1.75 | 2.90 | 4.21 | 5.56 | 6.85 | 8.05 | 9.15 | 10.14 | 11.03 | 11.83 | 12.55 | 13.20 | 13.79 | 14.32 | 14.80 | 15.24 | 15.64 | 16.01 | 16.34 | 16.65 | 16.93 | 17.21 | 17.45 |
| | $W_{1,2}$ | 12.65 | 26.38 | 40.84 | 55.56 | 70.35 | 85.16 | 99.97 | 114.79 | 129.61 | 144.42 | 159.24 | 174.05 | 188.87 | 203.69 | 218.50 | 233.32 | 248.14 | 262.95 | 277.77 | 292.58 | 307.40 | 322.22 | 337.03 | 351.85 | 366.66 |
| | $W_{3,4}$ | 13.89 | 28.40 | 43.13 | 57.93 | 72.74 | 87.55 | 102.37 | 117.19 | 132.00 | 146.82 | 161.63 | 176.45 | 191.27 | 206.08 | 220.90 | 235.71 | 250.53 | 265.35 | 280.16 | 294.98 | 309.80 | 324.61 | 339.43 | 354.24 | 369.06 |
| | $J$ | 10.05 | 16.15 | 19.59 | 21.65 | 22.95 | 23.81 | 24.40 | 24.81 | 25.11 | 25.33 | 25.50 | 25.62 | 25.71 | 25.79 | 25.85 | 25.89 | 25.93 | 25.96 | 25.98 | 26.00 | 26.02 | 26.03 | 26.05 | 26.06 | 26.06 |
| | $K$ | 2.18 | 3.33 | 3.73 | 3.82 | 3.85 | 3.86 | 3.86 | 3.86 | 3.86 | 3.86 | 3.86 | 3.86 | 3.86 | 3.86 | 3.86 | 3.86 | 3.86 | 3.86 | 3.86 | 3.86 | 3.86 | 3.86 | 3.86 | 3.86 | 3.86 |
| 0.7 | $U_1$ | 21.07 | 36.49 | 46.95 | 54.22 | 59.40 | 63.38 | 66.40 | 68.79 | 70.73 | 72.32 | 73.66 | 74.79 | 75.76 | 76.61 | 77.35 | 78.00 | 78.59 | 79.06 | 79.56 | 79.98 | 80.36 | 80.71 | 81.02 | 81.32 | 81.59 |
| | $U_2$ | 0.42 | 0.99 | 1.88 | 3.06 | 4.38 | 5.72 | 7.01 | 8.21 | 9.30 | 10.29 | 11.18 | 11.97 | 12.69 | 13.34 | 13.92 | 14.45 | 14.93 | 15.36 | 15.76 | 16.13 | 16.46 | 16.77 | 17.06 | 17.33 | 17.57 |
| | $W_{1,2}$ | 12.47 | 26.11 | 40.55 | 55.25 | 70.04 | 84.85 | 99.66 | 114.47 | 129.29 | 144.11 | 158.92 | 173.74 | 188.56 | 203.37 | 218.19 | 233.01 | 247.82 | 262.64 | 277.45 | 292.27 | 307.09 | 321.90 | 336.72 | 351.53 | 366.35 |
| | $W_{3,4}$ | 13.74 | 28.14 | 42.84 | 57.62 | 72.43 | 87.25 | 102.06 | 116.88 | 131.69 | 146.51 | 161.33 | 176.14 | 190.96 | 205.77 | 220.59 | 235.41 | 250.22 | 265.04 | 279.86 | 294.67 | 309.49 | 324.30 | 339.12 | 353.94 | 368.75 |
| | $J$ | 10.41 | 17.05 | 20.98 | 23.41 | 25.00 | 26.07 | 26.82 | 27.35 | 27.74 | 28.03 | 28.24 | 28.41 | 28.54 | 28.64 | 28.72 | 28.78 | 28.83 | 28.87 | 28.91 | 28.94 | 28.98 | 29.00 | 29.00 | 29.02 | 29.02 |
| | $K$ | 2.27 | 3.47 | 3.86 | 3.98 | 4.01 | 4.01 | 4.02 | 4.02 | 4.02 | 4.02 | 4.02 | 4.02 | 4.02 | 4.02 | 4.02 | 4.02 | 4.02 | 4.02 | 4.02 | 4.02 | 4.02 | 4.02 | 4.02 | 4.02 | 4.02 |

| $u_0/u_1$ | $\xi$/nm | 2 | 4 | 6 | 8 | 10 | 12 | 14 | 16 | 18 | 20 | 22 | 24 | 26 | 28 | 30 | 32 | 34 | 36 | 38 | 40 | 42 | 44 | 46 | 48 | 50 |
|---|---|---|---|---|---|---|---|---|---|---|---|---|---|---|---|---|---|---|---|---|---|---|---|---|---|---|
| 0.6 | $U_1$ | 19.17 | 33.44 | 43.29 | 50.22 | 55.26 | 59.06 | 62.01 | 64.35 | 66.25 | 67.82 | 69.13 | 70.25 | 71.22 | 72.05 | 72.78 | 73.43 | 74.00 | 74.52 | 74.98 | 75.40 | 75.78 | 76.12 | 76.44 | 76.73 | 77.00 |
| | $U_2$ | 0.48 | 1.12 | 2.06 | 3.26 | 4.59 | 5.94 | 7.22 | 8.41 | 9.49 | 10.47 | 11.35 | 12.15 | 12.86 | 13.50 | 14.08 | 14.60 | 15.08 | 15.52 | 15.91 | 16.28 | 16.61 | 16.92 | 17.21 | 17.47 | 17.72 |
| | $W_{1,2}$ | 12.28 | 25.84 | 40.24 | 54.04 | 69.73 | 84.53 | 99.35 | 114.16 | 128.98 | 143.79 | 158.61 | 173.43 | 187.89 | 203.06 | 217.88 | 232.69 | 247.51 | 262.32 | 277.14 | 291.06 | 306.77 | 321.59 | 336.40 | 351.22 | 366.04 |
| | $W_{3,4}$ | 13.66 | 28.00 | 42.68 | 57.46 | 72.26 | 87.08 | 101.89 | 116.71 | 131.52 | 146.34 | 161.16 | 175.97 | 190.78 | 205.60 | 220.42 | 235.24 | 250.05 | 264.87 | 279.68 | 294.50 | 309.32 | 324.13 | 338.95 | 353.77 | 368.58 |
| | $J$ | 10.83 | 18.04 | 22.52 | 25.40 | 27.33 | 28.67 | 29.62 | 30.32 | 30.83 | 31.21 | 31.51 | 31.74 | 31.92 | 32.06 | 32.17 | 32.26 | 32.33 | 32.40 | 32.45 | 32.49 | 32.52 | 32.55 | 32.58 | 32.60 | 32.62 |
| | $K$ | 2.39 | 3.64 | 4.05 | 4.17 | 4.20 | 4.21 | 4.21 | 4.21 | 4.21 | 4.21 | 4.21 | 4.21 | 4.21 | 4.21 | 4.21 | 4.21 | 4.21 | 4.21 | 4.21 | 4.21 | 4.21 | 4.21 | 4.21 | 4.21 | 4.21 |
| 0.5 | $U_1$ | 17.86 | 31.35 | 40.79 | 47.49 | 52.40 | 56.12 | 59.01 | 61.32 | 63.20 | 64.75 | 66.06 | 67.17 | 68.12 | 68.95 | 69.68 | 70.32 | 70.90 | 71.41 | 71.87 | 72.28 | 72.66 | 73.01 | 73.32 | 73.62 | 73.88 |
| | $U_2$ | 0.52 | 1.21 | 2.18 | 3.40 | 4.74 | 6.08 | 7.36 | 8.54 | 9.62 | 10.60 | 11.47 | 12.26 | 12.97 | 13.61 | 14.19 | 14.71 | 15.19 | 15.62 | 16.02 | 16.38 | 16.71 | 17.02 | 17.30 | 17.57 | 17.82 |
| | $W_{1,2}$ | 12.06 | 25.51 | 39.87 | 54.56 | 69.35 | 84.16 | 98.97 | 113.78 | 128.60 | 143.42 | 158.23 | 173.05 | 187.86 | 202.68 | 217.50 | 232.31 | 247.13 | 261.95 | 276.76 | 291.58 | 306.39 | 321.21 | 336.03 | 350.84 | 365.66 |
| | $W_{3,4}$ | 13.61 | 27.92 | 42.58 | 57.35 | 72.15 | 86.97 | 101.78 | 116.60 | 131.41 | 146.23 | 161.05 | 175.86 | 190.68 | 205.50 | 220.31 | 235.13 | 249.94 | 264.76 | 279.57 | 294.39 | 309.21 | 324.02 | 338.84 | 353.66 | 368.47 |
| | $J$ | 11.20 | 18.88 | 23.79 | 27.01 | 29.20 | 30.73 | 31.83 | 32.63 | 33.23 | 33.68 | 34.03 | 34.30 | 34.51 | 34.68 | 34.81 | 34.92 | 35.01 | 35.09 | 35.15 | 35.20 | 35.24 | 35.28 | 35.31 | 35.34 | 35.36 |
| | $K$ | 2.52 | 3.84 | 4.27 | 4.40 | 4.43 | 4.44 | 4.44 | 4.44 | 4.44 | 4.44 | 4.44 | 4.44 | 4.44 | 4.44 | 4.44 | 4.44 | 4.44 | 4.44 | 4.44 | 4.44 | 4.44 | 4.44 | 4.44 | 4.44 | 4.44 |
| 0.4 | $U_1$ | 17.02 | 30.01 | 39.18 | 45.75 | 50.58 | 54.25 | 57.11 | 59.40 | 61.27 | 62.81 | 64.11 | 65.21 | 66.17 | 66.99 | 67.72 | 68.36 | 68.93 | 69.44 | 69.90 | 70.32 | 70.69 | 71.04 | 71.35 | 71.64 | 71.91 |
| | $U_2$ | 0.54 | 1.26 | 2.25 | 3.48 | 4.82 | 6.16 | 7.44 | 8.62 | 9.70 | 10.67 | 11.55 | 12.33 | 13.04 | 13.68 | 14.25 | 14.78 | 15.25 | 15.68 | 16.08 | 16.44 | 16.78 | 17.08 | 17.37 | 17.63 | 17.88 |
| | $W_{1,2}$ | 11.81 | 25.14 | 39.47 | 54.15 | 68.93 | 83.73 | 98.55 | 113.36 | 128.18 | 143.00 | 157.81 | 172.63 | 187.44 | 202.26 | 217.08 | 231.89 | 246.71 | 261.52 | 276.34 | 291.16 | 305.97 | 320.79 | 335.61 | 350.42 | 365.24 |
| | $W_{3,4}$ | 13.58 | 27.86 | 42.51 | 57.28 | 72.08 | 86.89 | 101.71 | 116.52 | 131.34 | 146.16 | 160.97 | 175.79 | 190.61 | 205.42 | 220.24 | 235.05 | 249.87 | 264.69 | 279.50 | 294.32 | 309.13 | 323.95 | 338.77 | 353.58 | 368.40 |
| | $J$ | 11.51 | 19.57 | 24.77 | 28.26 | 30.63 | 32.29 | 33.49 | 34.37 | 35.02 | 35.52 | 35.90 | 36.19 | 36.43 | 36.61 | 36.76 | 36.88 | 36.98 | 37.06 | 37.13 | 37.19 | 37.24 | 37.28 | 37.31 | 37.34 | 37.37 |
| | $K$ | 2.67 | 4.05 | 4.51 | 4.64 | 4.68 | 4.69 | 4.69 | 4.69 | 4.69 | 4.69 | 4.69 | 4.69 | 4.69 | 4.69 | 4.69 | 4.69 | 4.69 | 4.69 | 4.69 | 4.69 | 4.69 | 4.69 | 4.69 | 4.69 | 4.69 |
| 0.3 | $U_1$ | 16.36 | 28.95 | 37.92 | 44.37 | 49.13 | 52.76 | 55.60 | 57.87 | 59.72 | 61.26 | 62.55 | 63.65 | 64.60 | 65.43 | 66.15 | 66.79 | 67.36 | 67.87 | 68.33 | 68.74 | 69.12 | 69.46 | 69.78 | 70.07 | 70.34 |
| | $U_2$ | 0.56 | 1.30 | 2.31 | 3.55 | 4.89 | 6.23 | 7.51 | 8.69 | 9.76 | 10.73 | 11.59 | 12.37 | 13.08 | 13.73 | 14.31 | 14.83 | 15.30 | 15.72 | 16.13 | 16.49 | 16.82 | 17.13 | 17.42 | 17.68 | 17.93 |
| | $W_{1,2}$ | 11.58 | 24.79 | 39.00 | 53.76 | 68.53 | 83.34 | 98.15 | 112.97 | 127.78 | 142.60 | 157.42 | 172.23 | 187.05 | 201.87 | 216.68 | 231.50 | 246.32 | 261.13 | 275.95 | 290.76 | 305.58 | 320.40 | 335.21 | 350.03 | 364.84 |
| | $W_{3,4}$ | 13.57 | 27.83 | 42.49 | 57.24 | 72.04 | 86.85 | 101.61 | 116.48 | 131.30 | 146.11 | 160.93 | 175.75 | 190.56 | 205.38 | 220.19 | 235.01 | 249.83 | 264.64 | 279.46 | 294.28 | 309.09 | 323.91 | 338.72 | 353.54 | 368.36 |
| | $J$ | 11.79 | 20.15 | 25.65 | 29.32 | 31.85 | 33.63 | 34.91 | 35.85 | 36.55 | 37.09 | 37.50 | 37.82 | 38.07 | 38.27 | 38.43 | 38.56 | 38.67 | 38.76 | 38.84 | 38.90 | 38.95 | 39.00 | 39.03 | 39.06 | 39.09 |
| | $K$ | 2.81 | 4.26 | 4.75 | 4.88 | 4.92 | 4.93 | 4.93 | 4.93 | 4.93 | 4.93 | 4.93 | 4.93 | 4.93 | 4.93 | 4.93 | 4.93 | 4.93 | 4.93 | 4.93 | 4.93 | 4.93 | 4.93 | 4.93 | 4.93 | 4.93 |
| 0.2 | $U_1$ | 16.08 | 28.54 | 37.37 | 43.50 | 48.66 | 52.29 | 55.13 | 57.40 | 59.26 | 60.79 | 62.09 | 63.19 | 64.14 | 64.96 | 65.69 | 66.33 | 66.90 | 67.41 | 67.87 | 68.28 | 68.66 | 69.01 | 69.32 | 69.61 | 69.88 |
| | $U_2$ | 0.55 | 1.28 | 2.29 | 3.53 | 4.88 | 6.22 | 7.50 | 8.68 | 9.75 | 10.72 | 11.60 | 12.38 | 13.09 | 13.72 | 14.30 | 14.82 | 15.29 | 15.73 | 16.12 | 16.49 | 16.82 | 17.12 | 17.41 | 17.67 | 17.92 |
| | $W_{1,2}$ | 11.19 | 24.21 | 38.44 | 53.09 | 67.86 | 82.67 | 97.48 | 112.30 | 127.11 | 141.93 | 156.75 | 171.56 | 186.38 | 201.19 | 216.01 | 230.83 | 245.64 | 260.46 | 275.27 | 290.09 | 304.91 | 319.72 | 334.54 | 349.36 | 364.17 |
| | $W_{3,4}$ | 13.54 | 27.77 | 42.42 | 57.17 | 71.98 | 86.79 | 101.60 | 116.42 | 131.24 | 146.05 | 160.87 | 175.68 | 190.50 | 205.32 | 220.13 | 234.95 | 249.76 | 264.58 | 279.40 | 294.21 | 309.03 | 323.85 | 338.66 | 353.48 | 368.29 |
| | $J$ | 12.10 | 20.77 | 26.50 | 30.35 | 33.05 | 34.79 | 36.11 | 37.07 | 37.78 | 38.32 | 38.74 | 39.06 | 39.31 | 39.51 | 39.68 | 39.81 | 39.92 | 40.01 | 40.08 | 40.14 | 40.19 | 40.23 | 40.27 | 40.30 | 40.33 |
| | $K$ | 3.02 | 4.58 | 5.09 | 5.24 | 5.28 | 5.29 | 5.29 | 5.29 | 5.29 | 5.29 | 5.29 | 5.29 | 5.29 | 5.29 | 5.29 | 5.29 | 5.29 | 5.29 | 5.29 | 5.29 | 5.29 | 5.29 | 5.29 | 5.29 | 5.29 |
| 0.1 | $U_1$ | 16.33 | 28.97 | 38.01 | 43.80 | 48.68 | 52.98 | 55.84 | 58.13 | 59.99 | 61.53 | 62.83 | 63.94 | 64.89 | 65.72 | 66.44 | 67.08 | 67.65 | 68.16 | 68.62 | 69.04 | 69.42 | 69.76 | 70.08 | 70.37 | 70.64 |
| | $U_2$ | 0.52 | 1.23 | 2.23 | 3.47 | 4.81 | 6.16 | 7.43 | 8.62 | 9.69 | 10.67 | 11.54 | 12.33 | 13.03 | 13.67 | 14.25 | 14.77 | 15.24 | 15.68 | 16.07 | 16.44 | 16.77 | 17.08 | 17.36 | 17.63 | 17.87 |
| | $W_{1,2}$ | 11.09 | 24.06 | 38.27 | 52.92 | 67.69 | 82.50 | 97.31 | 112.12 | 126.94 | 141.76 | 156.57 | 171.39 | 186.20 | 201.02 | 215.84 | 230.65 | 245.47 | 260.29 | 275.10 | 289.92 | 304.73 | 319.55 | 334.37 | 349.18 | 364.00 |
| | $W_{3,4}$ | 13.51 | 27.74 | 42.37 | 57.14 | 71.94 | 86.75 | 101.57 | 116.38 | 131.20 | 146.01 | 160.83 | 175.65 | 190.46 | 205.28 | 220.09 | 234.91 | 249.73 | 264.54 | 279.36 | 294.18 | 308.99 | 323.81 | 338.62 | 353.44 | 368.26 |
| | $J$ | 12.12 | 20.78 | 26.46 | 30.24 | 32.81 | 34.60 | 35.88 | 36.81 | 37.50 | 38.02 | 38.42 | 38.72 | 38.96 | 39.14 | 39.30 | 39.42 | 39.52 | 39.60 | 39.67 | 39.73 | 39.78 | 39.82 | 39.85 | 39.88 | 39.91 |
| | $K$ | 3.05 | 4.62 | 5.15 | 5.29 | 5.34 | 5.35 | 5.35 | 5.35 | 5.35 | 5.35 | 5.35 | 5.35 | 5.35 | 5.35 | 5.35 | 5.35 | 5.35 | 5.35 | 5.35 | 5.35 | 5.35 | 5.35 | 5.35 | 5.35 | 5.35 |
| 0.0 | $U_1$ | 16.33 | 28.97 | 38.01 | 43.80 | 48.68 | 52.98 | 55.84 | 58.13 | 59.99 | 61.53 | 62.83 | 63.94 | 64.89 | 65.72 | 66.44 | 67.08 | 67.65 | 68.16 | 68.62 | 69.04 | 69.42 | 69.76 | 70.08 | 70.37 | 70.64 |
| | $U_2$ | 0.52 | 1.23 | 2.23 | 3.47 | 4.81 | 6.16 | 7.43 | 8.62 | 9.69 | 10.67 | 11.54 | 12.33 | 13.03 | 13.67 | 14.25 | 14.77 | 15.24 | 15.68 | 16.07 | 16.44 | 16.77 | 17.08 | 17.36 | 17.63 | 17.87 |
| | $W_{1,2}$ | 11.09 | 24.06 | 38.27 | 52.92 | 67.69 | 82.50 | 97.31 | 112.12 | 126.94 | 141.76 | 156.57 | 171.39 | 186.20 | 201.02 | 215.84 | 230.65 | 245.47 | 260.29 | 275.10 | 289.92 | 304.73 | 319.55 | 334.37 | 349.18 | 364.00 |
| | $W_{3,4}$ | 13.51 | 27.74 | 42.37 | 57.14 | 71.94 | 86.75 | 101.57 | 116.38 | 131.20 | 146.01 | 160.83 | 175.65 | 190.46 | 205.28 | 220.09 | 234.91 | 249.73 | 264.54 | 279.36 | 294.18 | 308.99 | 323.81 | 338.62 | 353.44 | 368.26 |
| | $J$ | 12.12 | 20.78 | 26.46 | 30.24 | 32.81 | 34.60 | 35.88 | 36.81 | 37.50 | 38.02 | 38.42 | 38.72 | 38.96 | 39.14 | 39.30 | 39.42 | 39.52 | 39.60 | 39.67 | 39.73 | 39.78 | 39.82 | 39.85 | 39.88 | 39.91 |
| | $K$ | 3.05 | 4.62 | 5.15 | 5.29 | 5.34 | 5.35 | 5.35 | 5.35 | 5.35 | 5.35 | 5.35 | 5.35 | 5.35 | 5.35 | 5.35 | 5.35 | 5.35 | 5.35 | 5.35 | 5.35 | 5.35 | 5.35 | 5.35 | 5.35 | 5.35 |

TABLE S65. Parameters of the THIF interaction Hamiltonian for different ratios $u_0/u_1$ and screening lengths $\xi$ of the double-gate interaction potential ($U_\xi = 24$ meV for $\xi = 10$ nm). We use the same BM model parameters as in Table S64.

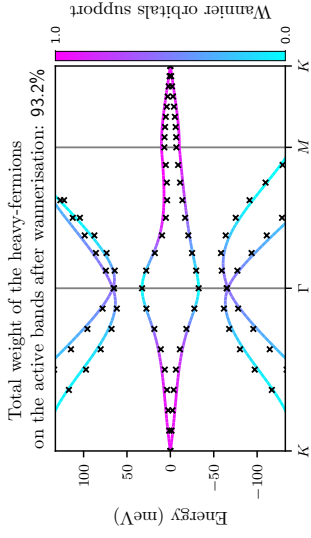

Total weight of the heavy-fermions on the active bands after wannierisation: 93.2%

Wannier orbitals support  0.6 — 1.0

Energy (meV) axis: -100 … 0 … 100 along path K – M – Γ – M – K

FIG. S49. Band structures of the BM and THF models near charge neutrality for $w_0/w_1 = 0.8$, depicted by lines and crosses, respectively. The BM bands are colored according to the weight of the $f$-electron wave function on them. We use the same BM parameters as in Table S66.

TABLE S66. Parameters of the $f$-electron Wannier functions for different values of the tunneling ratio $w_0/w_1$. $W$ denotes the total weight of the THF $f$-electrons on the active bands. In computing the ratios $\gamma/U_1$ and $\gamma/U_2$, we employ the on-site and nearest-neighbor repulsion parameters $U_1$ and $U_2$ obtained numerically for $\xi = 10$ nm, as given in Table S67. We employ $v_F = 5.944$ eV·Å, $|\mathbf{K}| = 1.703$ Å$^{-1}$, $w_1 = 110$ meV, and $\theta = 1.32°$ for the BM model.

| $w_0/w_1$ | $\alpha_1$ | $\alpha_2$ | $\lambda_1$ | $\lambda_2$ | $\lambda$ | $W$ | $t_0$ | $\gamma$ | $M$ | $v'_\star$ | $v''_\star$ | $\gamma/U_1$ | $\gamma/U_2$ |
|---|---|---|---|---|---|---|---|---|---|---|---|---|---|
| 1.00 | 0.801 | 0.598 | 0.191 | 0.203 | 0.351 | 0.951 | 2.167 | -34.708 | -29.827 | 1.812 | 0.0161 | -0.49 | -8.32 |
| 0.90 | 0.830 | 0.558 | 0.209 | 0.215 | 0.359 | 0.938 | 2.627 | -50.743 | -31.322 | 1.823 | 0.0091 | -0.75 | -11.88 |
| 0.80 | 0.859 | 0.511 | 0.215 | 0.221 | 0.378 | 0.932 | 3.131 | -66.373 | -32.620 | 1.808 | 0.0046 | -1.05 | -14.94 |
| 0.70 | 0.888 | 0.460 | 0.221 | 0.225 | 0.393 | 0.928 | 3.609 | -81.375 | -33.734 | 1.762 | 0.0018 | -1.36 | -17.73 |
| 0.60 | 0.915 | 0.404 | 0.245 | 0.227 | 0.414 | 0.931 | 4.220 | -95.501 | -34.675 | 1.678 | 0.0003 | -1.71 | -19.95 |
| 0.50 | 0.939 | 0.344 | 0.263 | 0.233 | 0.428 | 0.932 | 4.727 | -108.466 | -35.454 | 1.547 | -0.0004 | -2.04 | -22.00 |
| 0.40 | 0.960 | 0.280 | 0.281 | 0.233 | 0.446 | 0.936 | 5.119 | -119.929 | -36.080 | 1.360 | -0.0006 | -2.38 | -23.65 |
| 0.30 | 0.977 | 0.214 | 0.292 | 0.239 | 0.454 | 0.937 | 5.443 | -129.503 | -36.559 | 1.109 | -0.0004 | -2.65 | -25.12 |
| 0.20 | 0.989 | 0.147 | 0.289 | 0.239 | 0.447 | 0.930 | 5.554 | -136.767 | -36.898 | 0.790 | -0.0002 | -2.80 | -26.67 |
| 0.10 | 0.997 | 0.075 | 0.304 | 0.239 | 0.443 | 0.925 | 5.430 | -141.324 | -37.099 | 0.412 | -0.0001 | -2.87 | -27.79 |
| 0.00 | 1.000 | 0.000 | 0.310 | 0.000 | 0.434 | 0.917 | 5.214 | -142.879 | -37.166 | -5.484 | 0.0000 | -2.85 | -28.59 |

TABLE S67.

| $w_0/w_1$ | $\xi$/nm | 2 | 4 | 6 | 8 | 10 | 12 | 14 | 16 | 18 | 20 | 22 | 24 | 26 | 28 | 30 | 32 | 34 | 36 | 38 | 40 | 42 | 44 | 46 | 48 | 50 |
|---|---|---|---|---|---|---|---|---|---|---|---|---|---|---|---|---|---|---|---|---|---|---|---|---|---|---|
| **1.0** | $U_1$ | 26.97 | 45.55 | 57.51 | 65.53 | 71.18 | 75.33 | 78.50 | 80.99 | 82.98 | 84.62 | 85.99 | 87.14 | 88.13 | 88.99 | 89.74 | 90.40 | 90.98 | 91.50 | 91.97 | 92.40 | 92.78 | 93.13 | 93.45 | 93.74 | 94.02 |
| | $U_2$ | 25.03 | 42.63 | 54.16 | 61.97 | 67.52 | 71.61 | 74.75 | 77.21 | 79.20 | 80.83 | 82.19 | 83.34 | 84.32 | 85.18 | 85.92 | 86.58 | 87.16 | 87.68 | 88.15 | 88.57 | 88.96 | 89.31 | 89.63 | 89.92 | 90.19 |
| | $W_{1,2}$ | 13.38 | 27.72 | 42.69 | 57.88 | 73.13 | 88.40 | 103.68 | 118.95 | 134.23 | 149.50 | 164.78 | 180.06 | 195.33 | 210.61 | 225.88 | 241.16 | 256.43 | 271.71 | 286.98 | 302.26 | 317.53 | 332.81 | 348.09 | 363.36 | 378.64 |
| | $W_{3,4}$ | 14.78 | 30.04 | 45.34 | 60.62 | 75.90 | 91.18 | 106.45 | 121.73 | 137.00 | 152.28 | 167.55 | 182.83 | 198.10 | 213.38 | 228.66 | 243.93 | 259.21 | 273.48 | 289.76 | 305.03 | 320.31 | 335.58 | 350.86 | 366.13 | 381.41 |
| | $J$ | 9.69 | 14.92 | 17.53 | 18.96 | 19.80 | 20.34 | 20.70 | 20.95 | 21.12 | 21.25 | 21.35 | 21.42 | 21.47 | 21.51 | 21.55 | 21.57 | 21.60 | 21.61 | 21.63 | 21.64 | 21.65 | 21.65 | 21.66 | 21.67 | 21.67 |
| | $K$ | 2.05 | 3.12 | 3.47 | 3.57 | 3.59 | 3.60 | 3.60 | 3.60 | 3.60 | 3.60 | 3.60 | 3.60 | 3.60 | 3.60 | 3.60 | 3.60 | 3.60 | 3.60 | 3.60 | 3.60 | 3.60 | 3.60 | 3.60 | 3.60 | 3.60 |
| **0.9** | $U_1$ | 23.25 | 39.31 | 49.71 | 56.95 | 62.15 | 66.02 | 69.02 | 71.38 | 73.29 | 74.83 | 76.10 | 77.18 | 78.10 | 78.88 | 79.58 | 80.18 | 80.73 | 81.20 | 81.63 | 82.02 | 82.38 | 82.70 | 82.99 | 83.27 | 83.52 |
| | $U_2$ | 23.25 | 39.31 | 49.71 | 56.95 | 62.15 | 66.02 | 69.02 | 71.38 | 73.29 | 74.83 | 76.10 | 77.18 | 78.10 | 78.88 | 79.58 | 80.18 | 80.73 | 81.20 | 81.63 | 82.02 | 82.38 | 82.70 | 82.99 | 83.27 | 83.52 |
| | $W_{1,2}$ | 13.28 | 27.56 | 42.52 | 57.71 | 72.96 | 88.23 | 103.50 | 118.78 | 134.05 | 149.33 | 164.60 | 179.88 | 195.16 | 210.43 | 225.71 | 240.98 | 256.26 | 271.53 | 286.81 | 302.08 | 317.36 | 332.64 | 347.91 | 363.19 | 378.46 |
| | $W_{3,4}$ | 14.46 | 29.53 | 44.76 | 60.02 | 75.29 | 90.57 | 105.84 | 121.12 | 136.40 | 151.67 | 166.95 | 182.22 | 197.50 | 212.77 | 228.05 | 243.32 | 258.60 | 273.87 | 289.15 | 304.43 | 319.70 | 334.98 | 350.25 | 365.53 | 380.80 |
| | $J$ | 9.97 | 15.71 | 18.77 | 20.53 | 21.61 | 22.31 | 22.78 | 23.11 | 23.35 | 23.52 | 23.64 | 23.74 | 23.81 | 23.87 | 23.91 | 23.94 | 23.97 | 23.99 | 24.01 | 24.03 | 24.04 | 24.05 | 24.06 | 24.07 | 24.07 |
| | $K$ | 2.10 | 3.19 | 3.54 | 3.64 | 3.66 | 3.67 | 3.67 | 3.67 | 3.67 | 3.67 | 3.67 | 3.67 | 3.67 | 3.67 | 3.67 | 3.67 | 3.67 | 3.67 | 3.67 | 3.67 | 3.67 | 3.67 | 3.67 | 3.67 | 3.67 |
| **0.8** | $U_1$ | 22.93 | 38.38 | 50.06 | 57.89 | 63.27 | 67.28 | 70.36 | 72.78 | 74.74 | 76.35 | 77.70 | 78.84 | 79.82 | 80.67 | 81.41 | 82.07 | 82.65 | 83.17 | 83.63 | 84.05 | 84.44 | 84.79 | 85.10 | 85.40 | 85.67 |
| | $U_2$ | 22.93 | 38.38 | 50.06 | 57.89 | 63.27 | 67.28 | 70.36 | 72.78 | 74.74 | 76.35 | 77.70 | 78.84 | 79.82 | 80.67 | 81.41 | 82.07 | 82.65 | 83.17 | 83.63 | 84.05 | 84.44 | 84.79 | 85.10 | 85.40 | 85.67 |
| | $W_{1,2}$ | 13.12 | 27.34 | 42.27 | 57.46 | 72.71 | 87.98 | 103.25 | 118.53 | 133.80 | 149.08 | 164.35 | 179.63 | 194.90 | 210.18 | 225.45 | 240.73 | 256.01 | 271.28 | 286.56 | 301.83 | 317.11 | 332.38 | 347.66 | 362.93 | 378.21 |
| | $W_{3,4}$ | 14.35 | 29.18 | 44.36 | 59.61 | 74.88 | 90.15 | 105.43 | 120.70 | 135.98 | 151.25 | 166.53 | 181.80 | 197.08 | 212.36 | 227.63 | 242.91 | 258.18 | 273.46 | 288.73 | 304.01 | 319.28 | 334.56 | 349.83 | 365.11 | 380.39 |
| | $J$ | 10.33 | 16.62 | 20.20 | 22.35 | 23.72 | 24.63 | 25.27 | 25.71 | 26.03 | 26.27 | 26.45 | 26.59 | 26.69 | 26.77 | 26.84 | 26.89 | 26.93 | 26.96 | 26.99 | 27.01 | 27.03 | 27.05 | 27.06 | 27.07 | 27.08 |
| | $K$ | 2.17 | 3.29 | 3.66 | 3.76 | 3.79 | 3.80 | 3.80 | 3.80 | 3.80 | 3.80 | 3.80 | 3.80 | 3.80 | 3.80 | 3.80 | 3.80 | 3.80 | 3.80 | 3.80 | 3.80 | 3.80 | 3.80 | 3.80 | 3.80 | 3.80 |
| **0.7** | $U_1$ | 21.22 | 36.71 | 47.21 | 54.50 | 59.75 | 63.67 | 66.70 | 69.09 | 71.03 | 72.63 | 73.96 | 75.10 | 76.07 | 76.91 | 77.65 | 78.30 | 78.88 | 79.40 | 79.86 | 80.28 | 80.66 | 81.01 | 81.32 | 81.62 | 81.89 |
| | $U_2$ | 21.22 | 36.71 | 47.21 | 54.50 | 59.75 | 63.67 | 66.70 | 69.09 | 71.03 | 72.63 | 73.96 | 75.10 | 76.07 | 76.91 | 77.65 | 78.30 | 78.88 | 79.40 | 79.86 | 80.28 | 80.66 | 81.01 | 81.32 | 81.62 | 81.89 |
| | $W_{1,2}$ | 12.92 | 27.14 | 42.37 | 57.64 | 72.88 | 88.10 | 103.35 | 118.60 | 133.85 | 149.10 | 166.24 | 181.52 | 196.80 | 212.07 | 227.35 | 242.62 | 257.90 | 273.17 | 288.45 | 303.72 | 319.00 | 334.28 | 349.55 | 364.83 | 380.10 |
| | $W_{3,4}$ | 14.11 | 28.94 | 44.09 | 59.33 | 74.59 | 89.87 | 105.14 | 120.42 | 135.69 | 150.97 | 166.24 | 181.52 | 196.80 | 212.07 | 227.35 | 242.62 | 257.90 | 273.17 | 288.45 | 303.72 | 319.00 | 334.28 | 349.55 | 364.83 | 380.10 |
| | $J$ | 10.70 | 17.52 | 21.57 | 24.08 | 25.72 | 26.83 | 27.60 | 28.16 | 28.56 | 28.86 | 29.09 | 29.26 | 29.39 | 29.50 | 29.58 | 29.65 | 29.70 | 29.74 | 29.77 | 29.80 | 29.81 | 29.84 | 29.86 | 29.88 | 29.91 |
| | $K$ | 2.27 | 3.44 | 3.82 | 3.92 | 3.95 | 3.96 | 3.96 | 3.96 | 3.96 | 3.96 | 3.96 | 3.96 | 3.96 | 3.96 | 3.96 | 3.96 | 3.96 | 3.96 | 3.96 | 3.96 | 3.96 | 3.96 | 3.96 | 3.96 | 3.96 |

TABLE S67. Parameters of the THF interaction Hamiltonian for different ratios $u_0/u_1$ and screening lengths $\xi$ of the double-gate interaction potential ($U_\xi = 24$ meV for $\xi = 10$ nm). We use the same BM model parameters as in Table S66.

| $u_0/u_1$ | | 2 | 4 | 6 | 8 | 10 | 12 | 14 | 16 | 18 | 20 | 22 | 24 | 26 | 28 | 30 | 32 | 34 | 36 | 38 | 40 | 42 | 44 | 46 | 48 | 50 |
|---|---|---|---|---|---|---|---|---|---|---|---|---|---|---|---|---|---|---|---|---|---|---|---|---|---|---|
| $\xi$/nm | | | | | | | | | | | | | | | | | | | | | | | | | | |
| 0.6 | $U_1$ | 19.44 | 33.87 | 43.82 | 50.80 | 55.88 | 59.70 | 62.65 | 65.00 | 66.91 | 68.49 | 69.81 | 70.93 | 71.89 | 72.73 | 73.46 | 74.11 | 74.68 | 75.20 | 75.66 | 76.08 | 76.46 | 76.81 | 77.12 | 77.42 | 77.68 |
| | $U_2$ | 0.50 | 1.17 | 2.16 | 3.41 | 4.79 | 6.17 | 7.48 | 8.70 | 9.80 | 10.79 | 11.69 | 12.49 | 13.21 | 13.85 | 14.44 | 14.97 | 15.45 | 15.88 | 16.28 | 16.65 | 16.99 | 17.30 | 17.58 | 17.85 | 18.10 |
| | $W_{1,2}$ | 12.72 | 26.75 | 41.62 | 56.79 | 72.04 | 87.30 | 102.58 | 117.85 | 133.13 | 148.40 | 163.68 | 178.95 | 194.23 | 209.51 | 224.78 | 240.06 | 255.33 | 270.61 | 285.88 | 301.16 | 316.43 | 331.71 | 346.98 | 362.26 | 377.54 |
| | $W_{3,4}$ | 11.49 | 19.35 | 24.35 | 27.62 | 29.83 | 31.37 | 32.47 | 33.28 | 33.88 | 34.33 | 34.67 | 34.94 | 35.15 | 35.32 | 35.45 | 35.55 | 35.65 | 35.72 | 35.78 | 35.83 | 35.87 | 35.91 | 35.94 | 35.97 | 35.99 |
| | $J$ | 11.12 | 18.50 | 23.07 | 26.01 | 27.97 | 29.32 | 30.28 | 30.97 | 31.49 | 31.87 | 32.17 | 32.40 | 32.57 | 32.71 | 32.83 | 32.92 | 33.00 | 33.05 | 33.10 | 33.14 | 33.18 | 33.21 | 33.23 | 33.25 | 33.27 |
| | $K$ | 2.39 | 3.62 | 4.02 | 4.15 | 4.16 | 4.16 | 4.16 | 4.16 | 4.16 | 4.16 | 4.16 | 4.16 | 4.16 | 4.16 | 4.16 | 4.16 | 4.16 | 4.16 | 4.16 | 4.16 | 4.16 | 4.16 | 4.16 | 4.16 | 4.16 |
| 0.5 | $U_1$ | 18.15 | 31.82 | 41.36 | 48.12 | 53.07 | 56.82 | 59.72 | 62.04 | 63.93 | 65.49 | 66.79 | 67.91 | 68.87 | 69.70 | 70.43 | 71.07 | 71.64 | 72.16 | 72.62 | 73.03 | 73.41 | 73.76 | 74.08 | 74.37 | 74.64 |
| | $U_2$ | 0.54 | 1.26 | 2.28 | 3.55 | 4.93 | 6.31 | 7.62 | 8.83 | 9.93 | 10.91 | 11.80 | 12.60 | 13.32 | 13.96 | 14.54 | 15.07 | 15.54 | 15.98 | 16.38 | 16.75 | 17.08 | 17.39 | 17.68 | 17.95 | 18.19 |
| | $W_{1,2}$ | 12.49 | 26.41 | 41.25 | 56.40 | 71.65 | 86.92 | 102.19 | 117.47 | 132.74 | 148.02 | 163.29 | 178.57 | 193.84 | 209.12 | 224.39 | 239.67 | 254.94 | 270.22 | 285.50 | 300.77 | 316.05 | 331.32 | 346.60 | 361.87 | 377.15 |
| | $W_{3,4}$ | 14.00 | 28.74 | 43.85 | 59.08 | 74.34 | 89.62 | 104.89 | 120.17 | 135.44 | 150.72 | 165.99 | 181.27 | 196.54 | 211.82 | 227.09 | 242.37 | 257.65 | 272.92 | 288.20 | 303.47 | 318.75 | 334.02 | 349.30 | 364.58 | 379.85 |
| | $J$ | 13.49 | 21.35 | 26.35 | 29.62 | 31.72 | 33.14 | 34.15 | 34.90 | 35.47 | 35.91 | 36.26 | 36.53 | 36.75 | 36.92 | 37.06 | 37.18 | 37.27 | 37.35 | 37.41 | 37.46 | 37.51 | 37.54 | 37.58 | 37.60 | 37.62 |
| | $K$ | 2.53 | 3.82 | 4.24 | 4.36 | 4.39 | 4.40 | 4.40 | 4.40 | 4.40 | 4.40 | 4.40 | 4.40 | 4.40 | 4.40 | 4.40 | 4.40 | 4.40 | 4.40 | 4.40 | 4.40 | 4.40 | 4.40 | 4.40 | 4.40 | 4.40 |
| 0.4 | $U_1$ | 16.96 | 29.89 | 39.02 | 45.55 | 50.37 | 54.02 | 56.88 | 59.16 | 61.02 | 62.56 | 63.85 | 64.96 | 65.91 | 66.73 | 67.46 | 68.10 | 68.67 | 69.18 | 69.64 | 70.05 | 70.43 | 70.77 | 71.09 | 71.38 | 71.65 |
| | $U_2$ | 0.59 | 1.35 | 2.40 | 3.69 | 5.07 | 6.45 | 7.75 | 8.95 | 10.05 | 11.03 | 11.91 | 12.70 | 13.42 | 14.06 | 14.64 | 15.16 | 15.64 | 16.07 | 16.47 | 16.84 | 17.17 | 17.48 | 17.76 | 18.03 | 18.27 |
| | $W_{1,2}$ | 12.28 | 26.10 | 40.55 | 55.69 | 70.93 | 86.07 | 101.84 | 117.17 | 132.40 | 147.67 | 162.95 | 178.22 | 193.50 | 208.77 | 224.05 | 239.32 | 254.60 | 269.87 | 285.15 | 300.42 | 315.70 | 330.98 | 346.25 | 361.53 | 376.80 |
| | $W_{3,4}$ | 14.00 | 28.72 | 43.82 | 59.05 | 74.32 | 89.59 | 104.87 | 120.14 | 135.42 | 150.69 | 165.97 | 181.24 | 196.52 | 211.79 | 227.07 | 242.35 | 257.62 | 272.90 | 288.17 | 303.45 | 318.72 | 334.00 | 349.27 | 364.55 | 379.85 |
| | $J$ | 11.87 | 20.19 | 25.29 | 29.23 | 31.72 | 33.45 | 34.75 | 35.69 | 36.39 | 36.93 | 37.35 | 37.67 | 37.93 | 38.14 | 38.31 | 38.45 | 38.56 | 38.65 | 38.73 | 38.79 | 38.85 | 38.90 | 38.94 | 38.97 | 39.00 |
| | $K$ | 2.67 | 4.04 | 4.48 | 4.60 | 4.64 | 4.65 | 4.65 | 4.65 | 4.65 | 4.65 | 4.65 | 4.65 | 4.65 | 4.65 | 4.65 | 4.65 | 4.65 | 4.65 | 4.65 | 4.65 | 4.65 | 4.65 | 4.65 | 4.65 | 4.65 |
| 0.3 | $U_1$ | 16.24 | 28.77 | 37.71 | 44.15 | 48.91 | 52.54 | 55.37 | 57.65 | 59.50 | 61.04 | 62.33 | 63.43 | 64.38 | 65.21 | 65.93 | 66.57 | 67.14 | 67.65 | 68.11 | 68.53 | 68.91 | 69.25 | 69.56 | 69.86 | 70.12 |
| | $U_2$ | 0.59 | 1.40 | 2.47 | 3.77 | 5.15 | 6.53 | 7.83 | 9.03 | 10.12 | 11.09 | 11.97 | 12.76 | 13.47 | 14.12 | 14.70 | 15.22 | 15.70 | 16.13 | 16.53 | 16.89 | 17.23 | 17.54 | 17.82 | 18.09 | 18.33 |
| | $W_{1,2}$ | 12.06 | 25.78 | 40.05 | 55.09 | 70.49 | 85.76 | 101.03 | 116.30 | 131.58 | 146.85 | 162.13 | 177.41 | 192.68 | 207.96 | 223.23 | 238.51 | 253.78 | 269.06 | 284.34 | 299.01 | 314.89 | 330.16 | 345.44 | 360.71 | 375.99 |
| | $W_{3,4}$ | 14.00 | 28.71 | 43.81 | 59.04 | 74.30 | 89.57 | 104.85 | 120.12 | 135.40 | 150.67 | 165.95 | 181.22 | 196.50 | 211.77 | 227.05 | 242.32 | 257.60 | 272.88 | 288.15 | 303.43 | 318.70 | 333.98 | 349.25 | 364.53 | 379.80 |
| | $J$ | 12.16 | 20.91 | 25.75 | 28.94 | 31.04 | 32.45 | 33.46 | 34.21 | 34.78 | 35.22 | 35.57 | 35.84 | 36.06 | 36.24 | 36.39 | 36.51 | 36.62 | 36.71 | 36.79 | 36.85 | 36.91 | 36.96 | 37.01 | 37.05 | 41.04 |
| | $K$ | 2.82 | 4.26 | 4.73 | 4.85 | 4.88 | 4.89 | 4.89 | 4.89 | 4.89 | 4.89 | 4.89 | 4.89 | 4.89 | 4.89 | 4.89 | 4.89 | 4.89 | 4.89 | 4.89 | 4.89 | 4.89 | 4.89 | 4.89 | 4.89 | 4.89 |
| 0.2 | $U_1$ | 16.30 | 28.90 | 37.90 | 44.37 | 49.16 | 52.81 | 55.66 | 57.95 | 59.80 | 61.34 | 62.64 | 63.74 | 64.69 | 65.52 | 66.24 | 66.88 | 67.45 | 67.96 | 68.42 | 68.84 | 69.21 | 69.56 | 69.88 | 70.17 | 70.43 |
| | $U_2$ | 0.57 | 1.37 | 2.44 | 3.74 | 5.13 | 6.53 | 7.83 | 9.03 | 10.12 | 11.05 | 11.94 | 12.73 | 13.44 | 14.08 | 14.66 | 15.19 | 15.67 | 16.10 | 16.50 | 16.86 | 17.20 | 17.51 | 17.79 | 18.06 | 18.35 |
| | $W_{1,2}$ | 11.80 | 25.39 | 40.12 | 55.25 | 70.49 | 85.76 | 101.03 | 116.34 | 131.58 | 146.89 | 162.13 | 177.43 | 192.67 | 207.99 | 223.23 | 238.20 | 253.78 | 269.06 | 284.34 | 299.01 | 314.29 | 329.85 | 345.13 | 360.45 | 375.99 |
| | $W_{3,4}$ | 13.97 | 28.62 | 43.75 | 58.98 | 74.24 | 89.51 | 104.78 | 120.08 | 135.33 | 150.63 | 165.89 | 181.16 | 196.43 | 211.73 | 227.00 | 242.28 | 257.55 | 272.78 | 288.09 | 303.37 | 318.64 | 333.92 | 349.19 | 364.44 | 379.71 |
| | $J$ | 12.16 | 21.11 | 26.01 | 29.32 | 31.75 | 33.49 | 34.80 | 35.77 | 36.49 | 37.04 | 37.47 | 37.81 | 38.08 | 38.30 | 38.48 | 38.62 | 38.74 | 38.84 | 38.93 | 39.00 | 39.21 | 39.70 | 40.13 | 40.51 | 41.23 |
| | $K$ | 2.82 | 4.26 | 4.72 | 4.85 | 4.89 | 4.89 | 4.89 | 4.89 | 4.89 | 4.89 | 4.89 | 4.89 | 4.89 | 4.89 | 4.89 | 4.89 | 4.89 | 4.89 | 4.89 | 4.89 | 4.89 | 5.11 | 5.11 | 5.11 | 5.11 |
| 0.1 | $U_1$ | 16.68 | 29.54 | 38.71 | 45.28 | 50.14 | 53.83 | 56.71 | 59.00 | 60.87 | 62.42 | 63.73 | 64.84 | 65.79 | 66.62 | 67.35 | 67.99 | 68.56 | 69.07 | 69.53 | 69.95 | 70.33 | 70.67 | 70.99 | 71.28 | 71.55 |
| | $U_2$ | 0.54 | 1.28 | 2.32 | 3.61 | 5.00 | 6.38 | 7.69 | 8.90 | 10.01 | 10.98 | 11.87 | 12.66 | 13.38 | 14.02 | 14.60 | 15.13 | 15.61 | 16.04 | 16.44 | 16.80 | 17.14 | 17.45 | 17.73 | 18.00 | 18.25 |
| | $W_{1,2}$ | 11.41 | 25.12 | 39.83 | 54.95 | 70.18 | 85.45 | 100.72 | 116.00 | 131.27 | 146.55 | 161.82 | 177.10 | 192.37 | 207.65 | 222.92 | 238.20 | 253.47 | 268.75 | 284.03 | 299.30 | 314.58 | 329.85 | 345.13 | 360.40 | 375.68 |
| | $W_{3,4}$ | 13.91 | 28.58 | 43.66 | 58.89 | 74.15 | 89.42 | 104.70 | 119.97 | 135.25 | 150.52 | 165.80 | 181.07 | 196.35 | 211.62 | 226.90 | 242.17 | 257.45 | 272.73 | 288.00 | 303.28 | 318.55 | 333.83 | 349.11 | 364.38 | 379.71 |
| | $J$ | 12.43 | 21.32 | 27.17 | 31.07 | 33.74 | 35.60 | 36.94 | 37.92 | 38.64 | 39.19 | 39.61 | 39.94 | 40.19 | 40.40 | 40.56 | 40.69 | 40.77 | 40.88 | 40.77 | 40.83 | 40.89 | 40.93 | 40.97 | 40.84 | 41.20 |
| | $K$ | 2.95 | 4.44 | 4.93 | 5.06 | 5.10 | 5.11 | 5.11 | 5.11 | 5.11 | 5.11 | 5.11 | 5.11 | 5.11 | 5.11 | 5.11 | 5.11 | 5.11 | 5.11 | 5.11 | 5.26 | 5.26 | 5.11 | 4.89 | 5.26 | 5.26 |
| 0.0 | $U_1$ | 16.43 | 28.38 | 37.00 | 43.28 | 47.94 | 51.48 | 54.27 | 56.50 | 58.35 | 59.89 | 61.20 | 62.33 | 63.29 | 64.12 | 64.84 | 65.49 | 66.06 | 66.58 | 67.05 | 67.47 | 67.87 | 70.67 | 70.99 | 71.28 | 71.55 |
| | $U_2$ | 0.54 | 1.28 | 2.32 | 3.61 | 5.00 | 6.38 | 7.69 | 8.98 | 10.07 | 10.98 | 11.87 | 12.66 | 13.38 | 14.02 | 14.60 | 15.13 | 15.61 | 16.04 | 16.44 | 16.80 | 17.14 | 17.45 | 17.73 | 18.00 | 18.25 |
| | $W_{1,2}$ | 11.41 | 25.12 | 39.40 | 54.74 | 70.08 | 85.34 | 100.61 | 115.89 | 131.16 | 146.44 | 161.71 | 176.99 | 192.26 | 207.54 | 222.81 | 238.09 | 253.36 | 268.64 | 283.91 | 299.19 | 314.46 | 329.74 | 345.01 | 360.19 | 375.47 |
| | $W_{3,4}$ | 13.91 | 28.56 | 43.66 | 58.89 | 74.15 | 89.42 | 104.69 | 119.97 | 135.24 | 150.52 | 165.80 | 181.07 | 196.35 | 211.62 | 226.90 | 242.17 | 257.45 | 272.73 | 288.00 | 303.28 | 318.55 | 333.83 | 349.11 | 364.38 | 379.66 |
| | $J$ | 12.43 | 21.27 | 27.05 | 30.86 | 33.34 | 35.04 | 36.26 | 37.13 | 37.79 | 38.30 | 38.68 | 38.99 | 39.25 | 39.45 | 39.61 | 39.75 | 39.86 | 39.96 | 40.04 | 40.11 | 40.17 | 40.22 | 40.26 | 40.31 | 40.49 |
| | $K$ | 3.06 | 4.62 | 5.12 | 5.26 | 5.30 | 5.31 | 5.31 | 5.31 | 5.31 | 5.31 | 5.31 | 5.31 | 5.31 | 5.31 | 5.31 | 5.31 | 5.31 | 5.31 | 5.31 | 5.31 | 5.31 | 5.31 | 5.31 | 5.31 | 5.31 |

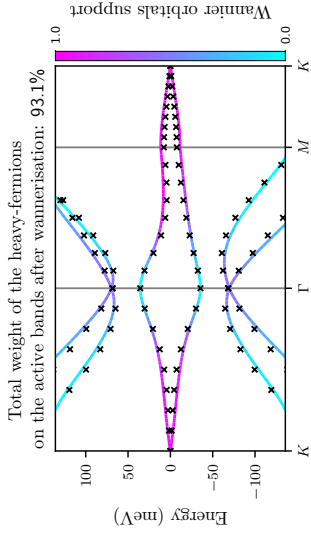

Total weight of the heavy-fermions on the active bands after wannierisation: 93.1%

Wannier orbitals support

FIG. S50. Band structures of the BM and THF models near charge neutrality for $v/v_1 = 0.8$, depicted by lines and crosses, respectively. The BM bands are colored according to the weight of the $f$-electron wave function on them. We use the same BM parameters as in Table S68.

TABLE S68. Parameters of the $f$-electron Wannier functions and the THF single-particle Hamiltonian for different values of the tunneling ratio $v_0/v_1$. $W$ denotes the total weight of the THF $f$-electrons on the active bands. In computing the ratios $\gamma^*/U_1$ and $v_\star'/U_2$, we employ the on-site and nearest-neighbor repulsion parameters $U_1$ and $U_2$ obtained numerically for $\xi = 10$ nm, as given in Table S69. We employ $v_F = 5.944$ eV·Å, $|\mathbf{K}| = 1.703\,\text{Å}^{-1}$, $w_1 = 110$ meV, and $\theta = 1.34°$ for the BM model.

| $v_0/v_1$ | $\alpha_1$ | $\alpha_2$ | $\lambda_1$ | $\lambda_2$ | $\lambda$ | $W$ | $t_0$ | $\gamma$ | $M$ | $v_\star$ | $v_\star'$ | $v_\star'/U_2$ | $\gamma^*/U_1$ | $\gamma^*/U_2$ |
|---|---|---|---|---|---|---|---|---|---|---|---|---|---|---|
| 1.00 | 0.804 | 0.594 | 0.191 | 0.203 | 0.354 | 0.949 | 2.433 | -37.871 | -32.742 | -4.645 | 1.827 | 0.0183 | -0.53 | -8.67 |
| 0.90 | 0.833 | 0.554 | 0.203 | 0.209 | 0.362 | 0.936 | 2.932 | -53.959 | -34.267 | -4.771 | 1.835 | 0.0117 | -0.79 | -12.07 |
| 0.80 | 0.862 | 0.507 | 0.221 | 0.215 | 0.380 | 0.931 | 3.474 | -69.620 | -35.590 | -4.887 | 1.817 | 0.0073 | -1.09 | -15.01 |
| 0.70 | 0.890 | 0.456 | 0.233 | 0.221 | 0.394 | 0.927 | 3.984 | -84.637 | -36.726 | -4.937 | 1.768 | 0.0044 | -1.40 | -17.69 |
| 0.60 | 0.917 | 0.399 | 0.251 | 0.227 | 0.420 | 0.933 | 4.720 | -98.767 | -37.685 | -5.100 | 1.683 | 0.0026 | -1.77 | -19.65 |
| 0.50 | 0.940 | 0.341 | 0.263 | 0.233 | 0.429 | 0.931 | 5.171 | -111.727 | -38.480 | -5.199 | 1.551 | 0.0015 | -2.08 | -21.80 |
| 0.40 | 0.961 | 0.277 | 0.281 | 0.233 | 0.447 | 0.936 | 5.587 | -123.180 | -39.118 | -5.291 | 1.362 | 0.0008 | -2.41 | -23.39 |
| 0.30 | 0.977 | 0.212 | 0.292 | 0.239 | 0.455 | 0.937 | 5.928 | -132.740 | -39.607 | -5.373 | 1.110 | 0.0004 | -2.68 | -24.82 |
| 0.20 | 0.989 | 0.145 | 0.304 | 0.239 | 0.459 | 0.936 | 6.082 | -139.991 | -39.952 | -5.440 | 0.790 | 0.0002 | -2.88 | -26.00 |
| 0.10 | 0.997 | 0.075 | 0.310 | 0.245 | 0.443 | 0.925 | 5.919 | -144.539 | -40.157 | -5.484 | 0.412 | 0.0000 | -2.89 | -27.40 |
| 0.00 | 1.000 | 0.000 | 0.310 | 0.000 | 0.434 | 0.917 | 5.704 | -146.091 | -40.226 | -5.500 | -0.000 | -0.0000 | -2.87 | -28.15 |

| $v_0/v_1$ | $\xi$/nm | 2 | 4 | 6 | 8 | 10 | 12 | 14 | 16 | 18 | 20 | 22 | 24 | 26 | 28 | 30 | 32 | 34 | 36 | 38 | 40 | 42 | 44 | 46 | 48 | 50 |
|---|---|---|---|---|---|---|---|---|---|---|---|---|---|---|---|---|---|---|---|---|---|---|---|---|---|---|
| 1.0 | $U_1$ | 27.15 | 45.82 | 57.83 | 65.87 | 71.53 | 75.69 | 78.87 | 81.36 | 83.36 | 85.00 | 86.37 | 87.52 | 88.52 | 89.37 | 90.12 | 90.78 | 91.36 | 91.89 | 92.36 | 92.78 | 93.16 | 93.51 | 93.84 | 94.13 | 94.40 |
| | $U_2$ | 0.35 | 0.87 | 1.75 | 2.97 | 4.47 | 5.89 | 7.15 | 8.40 | 9.54 | 10.57 | 11.49 | 12.31 | 13.05 | 13.71 | 14.31 | 14.85 | 15.34 | 15.78 | 16.19 | 16.56 | 16.90 | 17.21 | 17.50 | 17.77 | 18.02 |
| | $W_{1,2}$ | 13.88 | 28.71 | 44.16 | 59.83 | 75.55 | 91.29 | 107.03 | 122.77 | 138.51 | 154.25 | 169.99 | 185.74 | 201.48 | 217.22 | 232.96 | 248.70 | 264.45 | 280.19 | 295.93 | 311.67 | 327.41 | 343.15 | 358.90 | 374.64 | 390.38 |
| | $W_{3,4}$ | 15.12 | 30.77 | 46.15 | 62.26 | 78.00 | 93.74 | 109.48 | 125.22 | 140.97 | 156.71 | 172.45 | 188.19 | 203.93 | 219.67 | 235.42 | 251.16 | 266.90 | 282.64 | 298.38 | 314.13 | 329.87 | 345.61 | 361.35 | 377.09 | 392.83 |
| | $J$ | 9.91 | 15.28 | 17.97 | 19.44 | 20.32 | 20.88 | 21.25 | 21.51 | 21.69 | 21.83 | 21.92 | 22.00 | 22.05 | 22.10 | 22.13 | 22.16 | 22.18 | 22.20 | 22.21 | 22.22 | 22.23 | 22.24 | 22.25 | 22.25 | 22.26 |
| | $K$ | 2.73 | 3.07 | 3.40 | 3.49 | 3.51 | 3.52 | 3.52 | 3.52 | 3.52 | 3.52 | 3.52 | 3.52 | 3.52 | 3.52 | 3.52 | 3.52 | 3.52 | 3.52 | 3.52 | 3.52 | 3.52 | 3.52 | 3.52 | 3.52 | 3.52 |
| 0.9 | $U_1$ | 25.24 | 42.90 | 54.55 | 62.40 | 67.96 | 72.07 | 75.22 | 77.69 | 79.67 | 81.31 | 82.67 | 83.82 | 84.81 | 85.66 | 86.41 | 87.07 | 87.65 | 88.17 | 88.64 | 89.06 | 89.45 | 89.80 | 90.12 | 90.41 | 90.68 |
| | $U_2$ | 0.37 | 0.92 | 1.83 | 3.07 | 4.47 | 5.89 | 7.25 | 8.51 | 9.64 | 10.67 | 11.59 | 12.41 | 13.14 | 13.80 | 14.40 | 14.94 | 15.43 | 15.87 | 16.28 | 16.65 | 16.99 | 17.30 | 17.59 | 17.86 | 18.11 |
| | $W_{1,2}$ | 13.75 | 28.53 | 43.96 | 59.63 | 75.35 | 91.08 | 106.82 | 122.57 | 138.31 | 154.05 | 169.79 | 185.53 | 201.27 | 217.02 | 232.76 | 248.50 | 264.24 | 279.98 | 295.73 | 311.47 | 327.21 | 342.95 | 358.69 | 374.44 | 390.18 |
| | $W_{3,4}$ | 14.82 | 30.29 | 45.97 | 61.69 | 77.43 | 93.17 | 108.92 | 124.66 | 140.40 | 156.14 | 171.88 | 187.62 | 203.37 | 219.11 | 234.85 | 250.59 | 266.33 | 282.08 | 297.82 | 313.56 | 329.30 | 345.04 | 360.79 | 376.53 | 392.27 |
| | $J$ | 10.22 | 16.09 | 19.04 | 20.69 | 21.66 | 22.28 | 22.69 | 22.99 | 23.22 | 23.39 | 23.52 | 23.63 | 23.72 | 23.79 | 23.85 | 23.91 | 23.95 | 23.99 | 24.24 | 24.42 | 24.55 | 24.58 | 24.63 | 24.65 | 24.66 |
| | $K$ | 2.08 | 3.14 | 3.48 | 3.57 | 3.60 | 3.60 | 3.60 | 3.60 | 3.60 | 3.60 | 3.60 | 3.60 | 3.60 | 3.60 | 3.60 | 3.60 | 3.60 | 3.60 | 3.60 | 3.60 | 3.60 | 3.60 | 3.60 | 3.60 | 3.60 |
| 0.8 | $U_1$ | 23.18 | 39.77 | 50.82 | 58.39 | 63.80 | 67.82 | 70.91 | 73.34 | 75.31 | 76.92 | 78.27 | 79.42 | 80.40 | 81.25 | 81.99 | 82.64 | 83.22 | 83.74 | 84.21 | 84.63 | 85.01 | 85.35 | 85.68 | 85.98 | 86.25 |
| | $U_2$ | 0.42 | 1.02 | 1.97 | 3.23 | 4.64 | 6.06 | 7.42 | 8.67 | 9.80 | 10.82 | 11.73 | 12.55 | 13.28 | 13.94 | 14.53 | 15.07 | 15.56 | 16.00 | 16.40 | 16.77 | 17.11 | 17.43 | 17.72 | 17.98 | 18.23 |
| | $W_{1,2}$ | 13.59 | 28.29 | 43.70 | 59.36 | 75.07 | 90.81 | 106.55 | 122.29 | 138.03 | 153.78 | 169.52 | 185.26 | 201.00 | 216.74 | 232.49 | 248.23 | 263.97 | 279.71 | 295.45 | 311.20 | 326.94 | 342.68 | 358.42 | 374.16 | 389.91 |
| | $W_{3,4}$ | 14.62 | 29.96 | 45.60 | 61.33 | 77.05 | 92.79 | 108.53 | 124.27 | 140.01 | 155.76 | 171.50 | 187.24 | 202.98 | 218.72 | 234.47 | 250.21 | 265.95 | 281.69 | 297.43 | 313.18 | 328.92 | 344.66 | 360.40 | 376.14 | 391.89 |
| | $J$ | 10.59 | 17.03 | 20.20 | 21.89 | 22.89 | 23.52 | 24.24 | 25.22 | 25.86 | 26.31 | 26.63 | 26.87 | 27.05 | 27.19 | 27.29 | 27.37 | 27.44 | 27.49 | 27.53 | 27.56 | 27.59 | 27.61 | 27.63 | 27.65 | 27.68 |
| | $K$ | 2.16 | 3.26 | 3.61 | 3.70 | 3.73 | 3.73 | 3.73 | 3.73 | 3.73 | 3.73 | 3.73 | 3.73 | 3.73 | 3.73 | 3.73 | 3.73 | 3.73 | 3.73 | 3.73 | 3.73 | 3.73 | 3.73 | 3.73 | 3.73 | 3.73 |
| 0.7 | $U_1$ | 21.49 | 37.14 | 47.74 | 55.07 | 60.35 | 64.29 | 67.33 | 69.73 | 71.68 | 73.28 | 74.62 | 75.75 | 76.73 | 77.57 | 78.31 | 78.96 | 79.54 | 80.05 | 80.52 | 80.95 | 81.33 | 81.68 | 81.99 | 82.29 | 82.56 |
| | $U_2$ | 0.46 | 1.10 | 2.09 | 3.37 | 4.78 | 6.21 | 7.56 | 8.80 | 9.93 | 10.95 | 11.86 | 12.67 | 13.40 | 14.06 | 14.65 | 15.18 | 15.67 | 16.11 | 16.51 | 16.88 | 17.22 | 17.53 | 17.82 | 18.09 | 18.34 |
| | $W_{1,2}$ | 13.49 | 28.18 | 43.60 | 59.25 | 74.93 | 90.63 | 106.35 | 122.08 | 137.82 | 153.57 | 169.31 | 185.05 | 200.79 | 216.53 | 232.28 | 248.02 | 263.76 | 279.50 | 295.24 | 310.99 | 326.73 | 342.47 | 358.21 | 373.95 | 389.70 |
| | $W_{3,4}$ | 14.49 | 29.75 | 45.35 | 61.06 | 76.79 | 92.53 | 108.27 | 124.01 | 139.76 | 155.50 | 171.24 | 186.98 | 202.72 | 218.47 | 234.21 | 249.95 | 265.69 | 281.43 | 297.17 | 312.92 | 328.66 | 344.40 | 360.14 | 375.88 | 391.63 |
| | $J$ | 10.91 | 17.60 | 21.39 | 23.75 | 25.26 | 26.38 | 27.19 | 27.63 | 28.36 | 28.76 | 29.09 | 29.30 | 29.47 | 29.70 | 29.87 | 30.11 | 30.19 | 30.25 | 30.30 | 30.35 | 30.39 | 30.42 | 30.44 | 30.46 | 30.51 |
| | $K$ | 2.27 | 3.41 | 3.78 | 3.87 | 3.90 | 3.91 | 3.91 | 3.91 | 3.91 | 3.91 | 3.91 | 3.91 | 3.91 | 3.91 | 3.91 | 3.91 | 3.91 | 3.91 | 3.91 | 3.91 | 3.91 | 3.91 | 3.91 | 3.91 | 3.91 |

Band structures of the BM and THF models near charge neutrality for $v/v_1 = 0.8$, depicted by lines and crosses, respectively. The BM bands are colored according to the weight of the $f$-electron wave function on them. We use the same BM parameters as in Table S68.

TABLE S69. Parameters of the TIHF interaction Hamiltonian for different ratios $u_0/u_1$ and screening lengths $\xi$ of the double-gate interaction potential ($U_\xi = 24$ meV for $\xi = 10$ nm). We use the same BM model parameters as in Table S68.

| $u_0/u_1$ | $\xi$/nm | 2 | 4 | 6 | 8 | 10 | 12 | 14 | 16 | 18 | 20 | 22 | 24 | 26 | 28 | 30 | 32 | 34 | 36 | 38 | 40 | 42 | 44 | 46 | 48 | 50 |
|---|---|---|---|---|---|---|---|---|---|---|---|---|---|---|---|---|---|---|---|---|---|---|---|---|---|---|
| 0.6 | $U_1$ | 19.48 | 33.93 | 43.88 | 50.86 | 55.93 | 59.75 | 62.70 | 65.00 | 66.96 | 68.53 | 69.85 | 70.97 | 71.94 | 72.77 | 73.51 | 74.15 | 74.73 | 75.24 | 75.70 | 76.12 | 76.50 | 76.85 | 77.17 | 77.46 | 77.73 |
| | $U_2$ | 0.53 | 1.25 | 2.29 | 3.60 | 5.03 | 6.44 | 7.79 | 9.02 | 10.14 | 11.15 | 12.05 | 12.86 | 13.58 | 14.24 | 14.82 | 15.36 | 15.84 | 16.28 | 16.68 | 17.05 | 17.38 | 17.69 | 17.98 | 18.25 | 18.50 |
| | $W_{1,2}$ | 11.96 | 27.70 | 43.00 | 58.70 | 74.41 | 90.15 | 105.89 | 121.63 | 137.37 | 153.11 | 168.85 | 184.60 | 200.34 | 216.08 | 231.82 | 247.56 | 263.31 | 279.05 | 294.79 | 310.53 | 326.27 | 342.02 | 357.76 | 373.50 | 389.24 |
| | $W_{3,4}$ | 14.44 | 29.65 | 45.24 | 60.94 | 76.68 | 92.41 | 108.16 | 123.90 | 139.64 | 155.38 | 171.12 | 186.86 | 202.61 | 218.35 | 234.09 | 249.83 | 265.57 | 281.32 | 297.06 | 312.80 | 328.54 | 344.28 | 360.03 | 375.77 | 391.51 |
| | $J$ | 11.43 | 19.04 | 23.76 | 26.79 | 28.82 | 30.22 | 31.22 | 31.94 | 32.48 | 32.88 | 33.19 | 33.43 | 33.61 | 33.76 | 33.88 | 33.97 | 34.05 | 34.11 | 34.16 | 34.21 | 34.24 | 34.27 | 34.30 | 34.32 | 34.34 |
| | $K$ | 2.39 | 3.60 | 3.98 | 4.08 | 4.11 | 4.12 | 4.12 | 4.12 | 4.12 | 4.12 | 4.12 | 4.12 | 4.12 | 4.12 | 4.12 | 4.12 | 4.12 | 4.12 | 4.12 | 4.12 | 4.12 | 4.12 | 4.12 | 4.12 | 4.12 |
| 0.5 | $U_1$ | 18.45 | 32.30 | 41.95 | 48.77 | 53.75 | 57.52 | 60.44 | 62.77 | 64.66 | 66.22 | 67.54 | 68.65 | 69.61 | 70.45 | 71.18 | 71.82 | 72.40 | 72.91 | 73.37 | 73.79 | 74.17 | 74.51 | 74.83 | 75.12 | 75.39 |
| | $U_2$ | 0.56 | 1.31 | 2.38 | 3.70 | 5.12 | 6.54 | 7.88 | 9.12 | 10.23 | 11.24 | 12.13 | 12.94 | 13.66 | 14.31 | 14.90 | 15.43 | 15.91 | 16.35 | 16.75 | 17.12 | 17.46 | 17.77 | 18.05 | 18.32 | 18.57 |
| | $W_{1,2}$ | 11.27 | 27.01 | 42.64 | 57.93 | 73.64 | 89.37 | 105.11 | 120.85 | 136.59 | 152.34 | 168.08 | 183.83 | 199.57 | 215.31 | 231.06 | 246.80 | 262.53 | 278.27 | 294.01 | 309.75 | 325.50 | 341.24 | 356.98 | 372.72 | 388.81 |
| | $W_{3,4}$ | 14.40 | 29.57 | 45.15 | 60.85 | 76.58 | 92.32 | 108.06 | 123.80 | 139.54 | 155.28 | 171.02 | 186.76 | 202.50 | 218.25 | 233.99 | 249.72 | 265.46 | 281.20 | 296.96 | 312.70 | 328.44 | 344.18 | 359.91 | 375.65 | 391.41 |
| | $J$ | 11.79 | 19.83 | 24.92 | 28.23 | 30.40 | 32.02 | 33.12 | 33.92 | 34.57 | 34.97 | 35.31 | 35.58 | 35.79 | 35.96 | 36.08 | 36.19 | 36.28 | 36.35 | 36.41 | 36.46 | 36.50 | 36.53 | 36.56 | 36.59 | 36.61 |
| | $K$ | 2.53 | 3.81 | 4.21 | 4.32 | 4.35 | 4.36 | 4.36 | 4.36 | 4.36 | 4.36 | 4.36 | 4.36 | 4.36 | 4.36 | 4.36 | 4.36 | 4.36 | 4.36 | 4.36 | 4.36 | 4.36 | 4.36 | 4.36 | 4.36 | 4.36 |
| 0.4 | $U_1$ | 17.27 | 30.39 | 39.63 | 46.23 | 51.08 | 54.77 | 57.64 | 59.93 | 61.80 | 63.34 | 64.64 | 65.75 | 66.70 | 67.53 | 68.25 | 68.89 | 69.47 | 69.98 | 70.44 | 70.85 | 71.23 | 71.58 | 71.89 | 72.18 | 72.45 |
| | $U_2$ | 0.61 | 1.40 | 2.50 | 3.83 | 5.27 | 6.68 | 8.01 | 9.24 | 10.35 | 11.35 | 12.24 | 13.04 | 13.76 | 14.41 | 14.99 | 15.52 | 16.00 | 16.44 | 16.84 | 17.20 | 17.54 | 17.85 | 18.14 | 18.40 | 18.65 |
| | $W_{1,2}$ | 11.18 | 27.01 | 42.64 | 57.55 | 73.26 | 89.04 | 104.73 | 120.58 | 136.22 | 151.96 | 167.53 | 183.44 | 199.03 | 214.83 | 230.52 | 246.31 | 262.10 | 277.94 | 293.58 | 309.42 | 325.12 | 340.86 | 356.60 | 372.34 | 388.46 |
| | $W_{3,4}$ | 14.40 | 29.57 | 45.15 | 60.82 | 76.56 | 92.30 | 108.04 | 123.77 | 139.51 | 155.25 | 170.99 | 186.73 | 202.47 | 218.21 | 233.95 | 249.69 | 265.43 | 281.17 | 296.91 | 312.65 | 328.39 | 344.13 | 359.87 | 375.61 | 391.41 |
| | $J$ | 12.18 | 20.68 | 26.00 | 29.86 | 32.37 | 34.14 | 35.41 | 36.35 | 37.05 | 37.57 | 37.97 | 38.28 | 38.52 | 38.72 | 38.88 | 39.01 | 39.12 | 39.21 | 39.29 | 39.36 | 39.42 | 39.48 | 39.53 | 39.58 | 39.63 |
| | $K$ | 2.68 | 4.03 | 4.46 | 4.57 | 4.60 | 4.61 | 4.61 | 4.61 | 4.61 | 4.61 | 4.61 | 4.61 | 4.61 | 4.61 | 4.61 | 4.61 | 4.61 | 4.61 | 4.61 | 4.61 | 4.61 | 4.61 | 4.61 | 4.61 | 4.61 |
| 0.3 | $U_1$ | 16.57 | 29.27 | 38.24 | 44.76 | 49.54 | 53.17 | 56.02 | 58.29 | 60.15 | 61.68 | 62.98 | 64.09 | 65.03 | 65.85 | 66.58 | 67.21 | 67.78 | 68.29 | 68.75 | 69.17 | 69.55 | 69.89 | 70.21 | 70.50 | 70.76 |
| | $U_2$ | 0.64 | 1.47 | 2.60 | 3.95 | 5.38 | 6.80 | 8.13 | 9.26 | 10.37 | 11.37 | 12.26 | 13.06 | 13.79 | 14.43 | 15.02 | 15.55 | 16.03 | 16.46 | 16.86 | 17.23 | 17.57 | 17.88 | 18.16 | 18.43 | 18.70 |
| | $W_{1,2}$ | 10.72 | 26.46 | 42.20 | 57.94 | 73.26 | 89.04 | 104.73 | 120.47 | 136.22 | 151.96 | 167.53 | 183.27 | 199.03 | 214.83 | 230.52 | 246.31 | 262.10 | 277.78 | 293.42 | 309.16 | 324.90 | 340.64 | 356.38 | 372.12 | 388.00 |
| | $W_{3,4}$ | 14.36 | 29.53 | 45.11 | 60.78 | 76.52 | 92.26 | 108.00 | 123.73 | 139.47 | 155.21 | 170.95 | 186.69 | 202.43 | 218.17 | 233.91 | 249.65 | 265.39 | 281.13 | 296.87 | 312.61 | 328.35 | 344.09 | 359.83 | 375.57 | 391.38 |
| | $J$ | 12.47 | 21.27 | 26.84 | 30.83 | 33.55 | 35.45 | 36.83 | 37.85 | 38.62 | 39.21 | 39.66 | 40.02 | 40.30 | 40.53 | 40.71 | 40.86 | 40.98 | 41.08 | 41.20 | 41.27 | 41.33 | 41.38 | 41.42 | 41.46 | 41.80 |
| | $K$ | 2.83 | 4.25 | 4.70 | 4.82 | 4.85 | 4.86 | 4.86 | 4.86 | 4.86 | 4.86 | 4.86 | 4.86 | 4.86 | 4.86 | 4.86 | 4.86 | 4.86 | 4.86 | 4.86 | 4.86 | 4.86 | 4.86 | 4.86 | 4.86 | 4.86 |
| 0.2 | $U_1$ | 16.18 | 28.64 | 37.54 | 43.94 | 48.69 | 52.30 | 55.13 | 57.39 | 59.24 | 60.77 | 62.06 | 63.16 | 64.11 | 64.93 | 65.66 | 66.30 | 66.87 | 67.38 | 67.83 | 68.25 | 68.62 | 68.97 | 69.28 | 69.57 | 69.84 |
| | $U_2$ | 0.69 | 1.57 | 2.75 | 4.13 | 5.58 | 6.99 | 8.32 | 9.45 | 10.56 | 11.56 | 12.44 | 13.24 | 13.96 | 14.60 | 15.18 | 15.71 | 16.19 | 16.62 | 17.02 | 17.39 | 17.73 | 18.04 | 18.32 | 18.59 | 18.73 |
| | $W_{1,2}$ | 10.47 | 26.01 | 41.20 | 56.79 | 72.50 | 88.23 | 103.97 | 119.71 | 135.45 | 151.20 | 166.94 | 182.68 | 198.42 | 214.16 | 229.90 | 245.65 | 261.39 | 277.13 | 292.87 | 308.61 | 324.36 | 340.10 | 355.84 | 371.58 | 387.75 |
| | $W_{3,4}$ | 14.36 | 29.48 | 45.11 | 60.73 | 76.45 | 92.19 | 107.93 | 123.68 | 139.42 | 155.16 | 170.90 | 186.64 | 202.38 | 218.12 | 233.87 | 249.61 | 265.35 | 281.09 | 296.84 | 312.58 | 328.32 | 344.06 | 359.80 | 375.55 | 391.38 |
| | $J$ | 12.69 | 21.77 | 27.63 | 31.48 | 34.07 | 35.86 | 37.13 | 38.05 | 38.75 | 39.23 | 39.62 | 39.90 | 40.15 | 40.33 | 40.48 | 40.59 | 40.69 | 40.77 | 40.83 | 40.89 | 40.93 | 40.97 | 41.01 | 41.04 | 41.06 |
| | $K$ | 2.96 | 4.44 | 4.91 | 5.03 | 5.06 | 5.08 | 5.08 | 5.08 | 5.08 | 5.08 | 5.08 | 5.08 | 5.08 | 5.08 | 5.08 | 5.08 | 5.08 | 5.08 | 5.08 | 5.08 | 5.08 | 5.08 | 5.08 | 5.08 | 5.08 |
| 0.1 | $U_1$ | 16.65 | 29.47 | 38.59 | 45.15 | 49.98 | 53.66 | 56.53 | 58.82 | 60.69 | 62.23 | 63.53 | 64.64 | 65.59 | 66.42 | 67.15 | 67.79 | 68.37 | 68.87 | 69.33 | 69.75 | 70.13 | 70.47 | 70.79 | 71.08 | 71.35 |
| | $U_2$ | 0.56 | 1.89 | 2.49 | 3.75 | 5.19 | 6.61 | 7.95 | 9.18 | 10.30 | 11.30 | 12.19 | 13.00 | 13.72 | 14.37 | 14.96 | 15.49 | 15.97 | 16.41 | 16.81 | 17.17 | 17.51 | 17.82 | 18.11 | 18.37 | 18.67 |
| | $W_{1,2}$ | 10.04 | 25.53 | 41.20 | 56.79 | 72.50 | 88.23 | 103.97 | 119.71 | 135.45 | 151.20 | 166.94 | 182.68 | 198.42 | 214.16 | 229.90 | 245.65 | 261.39 | 277.13 | 292.87 | 308.61 | 324.36 | 340.10 | 355.84 | 371.58 | 387.32 |
| | $W_{3,4}$ | 14.32 | 29.43 | 44.98 | 60.67 | 76.40 | 92.13 | 107.88 | 123.62 | 139.37 | 155.11 | 170.85 | 186.59 | 202.33 | 218.08 | 233.82 | 249.56 | 265.30 | 281.04 | 296.79 | 312.53 | 328.27 | 344.01 | 359.75 | 375.49 | 391.29 |
| | $J$ | 12.75 | 21.77 | 27.63 | 34.37 | 36.23 | 37.57 | 38.65 | 39.40 | 39.80 | 40.22 | 40.54 | 40.79 | 41.15 | 41.28 | 41.41 | 41.50 | 41.58 | 41.63 | 41.66 | 41.70 | 41.72 | 41.74 | 41.76 | 41.77 | 42.77 |
| | $K$ | 3.01 | 4.57 | 5.05 | 5.18 | 5.21 | 5.22 | 5.22 | 5.22 | 5.22 | 5.22 | 5.22 | 5.22 | 5.22 | 5.22 | 5.22 | 5.22 | 5.22 | 5.22 | 5.22 | 5.22 | 5.22 | 5.22 | 5.22 | 5.22 | 5.22 |
| 0.0 | $U_1$ | 17.05 | 30.11 | 39.40 | 46.06 | 50.95 | 54.67 | 57.57 | 59.88 | 61.76 | 63.31 | 64.62 | 65.73 | 66.69 | 67.52 | 68.25 | 68.89 | 69.47 | 69.98 | 70.44 | 70.86 | 71.24 | 71.58 | 71.90 | 72.19 | 72.46 |
| | $U_2$ | 0.59 | 1.39 | 2.49 | 3.84 | 5.28 | 6.70 | 8.03 | 9.26 | 10.37 | 11.37 | 12.26 | 13.06 | 13.79 | 14.43 | 15.02 | 15.55 | 16.03 | 16.46 | 16.86 | 17.23 | 17.57 | 17.88 | 18.16 | 18.37 | 18.62 |
| | $W_{1,2}$ | 9.83 | 25.48 | 41.06 | 56.58 | 72.29 | 88.02 | 103.76 | 119.50 | 125.62 | 150.80 | 166.55 | 182.30 | 198.07 | 213.95 | 229.56 | 245.65 | 261.30 | 276.92 | 292.92 | 308.60 | 324.16 | 340.10 | 355.55 | 371.33 | 387.11 |
| | $W_{3,4}$ | 14.32 | 29.43 | 44.98 | 60.64 | 76.36 | 92.10 | 107.83 | 123.57 | 139.31 | 155.04 | 170.78 | 186.52 | 202.26 | 218.00 | 233.73 | 249.47 | 265.21 | 280.95 | 296.69 | 312.42 | 328.16 | 343.90 | 359.64 | 375.38 | 391.01 |
| | $J$ | 12.75 | 21.77 | 27.63 | 31.48 | 34.04 | 35.86 | 37.13 | 38.05 | 38.74 | 39.23 | 39.62 | 39.90 | 40.15 | 40.33 | 40.48 | 40.59 | 40.69 | 40.77 | 40.83 | 40.89 | 40.93 | 40.97 | 41.01 | 41.04 | 41.06 |
| | $K$ | 3.07 | 4.62 | 5.10 | 5.23 | 5.26 | 5.28 | 5.28 | 5.28 | 5.28 | 5.28 | 5.28 | 5.28 | 5.28 | 5.28 | 5.28 | 5.28 | 5.28 | 5.28 | 5.28 | 5.28 | 5.28 | 5.28 | 5.28 | 5.28 | 5.28 |

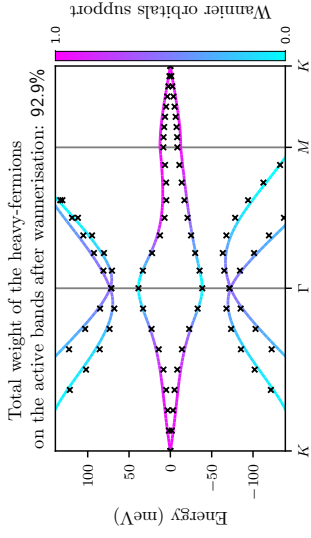

FIG. S51. Band structures of the BM and THF models near charge neutrality for $w_0/w_1 = 0.8$, depicted by lines and crosses, respectively. The BM bands are colored according to the weight of the $f$-electron wave function on them. We use the same BM parameters as in Table S70.

TABLE S70. Parameters of the $f$-electron Wannier functions and the THF single-particle Hamiltonian for different values of the tunneling ratio $w_0/w_1$. $W$ denotes the total weight of the THF $f$-electrons on the active bands. In computing the ratios $\gamma/U_1$ and $\gamma/U_2$, we employ the on-site and nearest-neighbor repulsion parameters $U_1$ and $U_2$ obtained numerically for $\xi = 10$ nm, as given in Table S71. We employ $v_F = 5.944$ eV Å, $|\mathbf{K}| = 1.703$ Å$^{-1}$, $w_1 = 110$ meV, and $\theta = 1.36°$ for the BM model.

| $w_0/w_1$ | $\alpha_1$ | $\alpha_2$ | $\lambda_1$ | $\lambda_2$ | $\lambda$ | $W$ | $t_0$ | $\gamma$ | $M$ | $v_\star$ | $v'_\star$ | $v''_\star$ | $\gamma/U_1$ | $\gamma/U_2$ |
|---|---|---|---|---|---|---|---|---|---|---|---|---|---|---|
| 1.00 | 0.807 | 0.591 | 0.197 | 0.203 | 0.357 | 0.947 | 2.712 | -41.059 | -35.689 | -4.674 | 1.840 | 0.0207 | -0.57 | -8.99 |
| 0.90 | 0.836 | 0.549 | 0.209 | 0.368 | 0.381 | 0.937 | 3.267 | -57.194 | -37.240 | -4.797 | 1.845 | 0.0144 | -0.84 | -12.17 |
| 0.80 | 0.864 | 0.503 | 0.221 | 0.215 | 0.381 | 0.929 | 3.827 | -72.882 | -38.586 | -4.910 | 1.825 | 0.0100 | -1.13 | -15.06 |
| 0.70 | 0.893 | 0.451 | 0.239 | 0.221 | 0.407 | 0.933 | 4.531 | -87.912 | -39.741 | -1.47 | 1.775 | 0.0069 | -1.47 | -17.35 |
| 0.60 | 0.918 | 0.396 | 0.251 | 0.227 | 0.422 | 0.933 | 5.149 | -102.044 | -40.718 | -5.119 | 1.688 | 0.0048 | -1.80 | -19.52 |
| 0.50 | 0.942 | 0.336 | 0.269 | 0.233 | 0.441 | 0.937 | 5.658 | -114.998 | -41.526 | -5.217 | 1.554 | 0.0032 | -2.15 | -21.36 |
| 0.40 | 0.962 | 0.274 | 0.281 | 0.239 | 0.449 | 0.937 | 6.092 | -126.440 | -42.175 | -5.308 | 1.364 | 0.0021 | -2.45 | -23.08 |
| 0.30 | 0.978 | 0.210 | 0.298 | 0.239 | 0.457 | 0.938 | 6.443 | -135.987 | -42.673 | -5.390 | 1.111 | 0.0012 | -2.79 | -24.46 |
| 0.20 | 0.990 | 0.143 | 0.304 | 0.245 | 0.459 | 0.936 | 6.577 | -143.226 | -43.024 | -5.456 | 0.790 | 0.0006 | -2.89 | -25.68 |
| 0.10 | 0.997 | 0.074 | 0.310 | 0.245 | 0.447 | 0.928 | 6.554 | -147.764 | -43.234 | -5.500 | 0.412 | 0.0001 | -2.94 | -26.78 |
| 0.00 | 1.000 | 0.000 | 0.316 | 0.000 | 0.442 | 0.922 | 6.241 | -149.313 | -43.303 | -5.515 | -0.000 | -0.0000 | -2.93 | -27.49 |

| $w_0/w_1$ | $\xi$/nm | 2 | 4 | 6 | 8 | 10 | 12 | 14 | 16 | 18 | 20 | 22 | 24 | 26 | 28 | 30 | 32 | 34 | 36 | 38 | 40 | 42 | 44 | 46 | 48 | 50 |
|---|---|---|---|---|---|---|---|---|---|---|---|---|---|---|---|---|---|---|---|---|---|---|---|---|---|---|
| 1.0 | $U_1$ | 27.33 | 46.09 | 58.15 | 66.21 | 71.89 | 76.07 | 79.25 | 81.74 | 83.75 | 85.39 | 86.76 | 87.92 | 88.91 | 89.76 | 90.51 | 91.17 | 91.76 | 92.28 | 92.75 | 93.17 | 93.56 | 93.91 | 94.23 | 94.52 | 94.80 |
| | $U_2$ | 0.37 | 0.92 | 1.85 | 3.13 | 4.57 | 6.03 | 7.42 | 8.70 | 9.86 | 10.99 | 11.83 | 12.66 | 13.40 | 14.07 | 14.68 | 15.22 | 15.71 | 16.15 | 16.56 | 16.93 | 17.28 | 17.59 | 17.88 | 18.15 | 18.40 |
| | $W_{1,2}$ | 14.38 | 29.71 | 45.65 | 61.80 | 77.99 | 94.21 | 110.42 | 126.63 | 142.85 | 159.07 | 175.28 | 191.50 | 207.71 | 223.93 | 240.14 | 256.36 | 272.57 | 288.79 | 305.00 | 321.22 | 337.43 | 353.65 | 369.86 | 386.08 | 402.29 |
| | $W_{3,4}$ | 15.46 | 31.53 | 47.72 | 63.93 | 80.14 | 96.36 | 112.57 | 128.79 | 145.01 | 161.22 | 177.44 | 193.65 | 209.87 | 226.08 | 242.30 | 258.51 | 274.73 | 290.94 | 307.16 | 323.37 | 339.59 | 355.80 | 372.02 | 388.23 | 404.45 |
| | $J$ | 10.14 | 15.64 | 18.41 | 19.93 | 20.85 | 21.42 | 21.81 | 22.08 | 22.27 | 22.40 | 22.50 | 22.58 | 22.64 | 22.68 | 22.72 | 22.75 | 22.77 | 22.79 | 22.80 | 22.81 | 22.82 | 22.83 | 22.84 | 22.84 | 22.85 |
| | $K$ | 2.01 | 3.02 | 3.33 | 3.41 | 3.43 | 3.44 | 3.44 | 3.44 | 3.44 | 3.44 | 3.44 | 3.44 | 3.44 | 3.44 | 3.44 | 3.44 | 3.44 | 3.44 | 3.44 | 3.44 | 3.44 | 3.44 | 3.44 | 3.44 | 3.44 |
| 0.9 | $U_1$ | 25.28 | 42.99 | 54.58 | 62.42 | 67.98 | 72.09 | 75.23 | 77.70 | 79.68 | 81.31 | 82.68 | 83.83 | 84.81 | 85.67 | 86.42 | 87.07 | 87.66 | 88.18 | 88.65 | 89.07 | 89.45 | 89.80 | 90.12 | 90.42 | 90.69 |
| | $U_2$ | 0.40 | 0.99 | 1.95 | 3.25 | 4.70 | 6.16 | 7.55 | 8.83 | 9.98 | 11.02 | 11.95 | 12.77 | 13.52 | 14.18 | 14.78 | 15.32 | 15.81 | 16.26 | 16.67 | 17.04 | 17.38 | 17.69 | 17.99 | 18.25 | 18.50 |
| | $W_{1,2}$ | 14.24 | 29.52 | 45.45 | 61.59 | 77.78 | 93.99 | 110.21 | 126.43 | 142.64 | 158.85 | 175.07 | 191.28 | 207.50 | 223.71 | 239.93 | 256.14 | 272.36 | 288.57 | 304.79 | 321.01 | 337.22 | 353.44 | 369.65 | 385.87 | 402.08 |
| | $W_{3,4}$ | 15.19 | 31.08 | 47.22 | 63.42 | 79.63 | 95.84 | 112.06 | 128.27 | 144.49 | 160.70 | 176.92 | 193.13 | 209.35 | 225.56 | 241.78 | 257.99 | 274.21 | 290.42 | 306.64 | 322.85 | 339.07 | 355.28 | 371.50 | 387.71 | 403.93 |
| | $J$ | 10.49 | 16.55 | 19.71 | 21.73 | 22.88 | 23.65 | 24.17 | 24.53 | 24.79 | 24.98 | 25.12 | 25.23 | 25.31 | 25.37 | 25.42 | 25.46 | 25.50 | 25.52 | 25.54 | 25.55 | 25.57 | 25.58 | 25.59 | 25.60 | 25.61 |
| | $K$ | 2.07 | 3.10 | 3.42 | 3.51 | 3.53 | 3.53 | 3.53 | 3.54 | 3.54 | 3.54 | 3.54 | 3.54 | 3.54 | 3.54 | 3.54 | 3.54 | 3.54 | 3.54 | 3.54 | 3.54 | 3.54 | 3.54 | 3.54 | 3.54 | 3.54 |
| 0.8 | $U_1$ | 23.43 | 40.16 | 51.29 | 58.90 | 64.34 | 68.38 | 71.47 | 73.91 | 75.88 | 77.50 | 78.85 | 80.00 | 80.98 | 81.83 | 82.57 | 83.23 | 83.81 | 84.33 | 84.80 | 85.22 | 85.60 | 85.95 | 86.27 | 86.56 | 86.84 |
| | $U_2$ | 0.44 | 1.07 | 2.07 | 3.38 | 4.84 | 6.30 | 7.69 | 8.96 | 10.11 | 11.15 | 12.07 | 12.90 | 13.64 | 14.30 | 14.90 | 15.44 | 15.93 | 16.37 | 16.78 | 17.15 | 17.49 | 17.80 | 18.09 | 18.36 | 18.60 |
| | $W_{1,2}$ | 14.06 | 29.25 | 45.15 | 61.28 | 77.48 | 93.69 | 109.90 | 126.12 | 142.33 | 158.55 | 174.76 | 190.98 | 207.19 | 223.41 | 239.62 | 255.84 | 272.05 | 288.27 | 304.48 | 320.70 | 336.91 | 353.13 | 369.34 | 385.56 | 401.77 |
| | $W_{3,4}$ | 15.00 | 30.77 | 46.87 | 63.06 | 79.26 | 95.47 | 111.69 | 127.91 | 144.12 | 160.34 | 176.55 | 192.77 | 208.98 | 225.20 | 241.42 | 257.63 | 273.85 | 290.06 | 306.28 | 322.49 | 338.71 | 354.92 | 371.14 | 387.35 | 403.57 |
| | $J$ | 10.85 | 17.49 | 21.19 | 23.44 | 24.86 | 25.80 | 26.45 | 26.91 | 27.23 | 27.48 | 27.65 | 27.79 | 27.88 | 27.98 | 28.04 | 28.09 | 28.13 | 28.17 | 28.19 | 28.21 | 28.23 | 28.25 | 28.26 | 28.27 | 28.28 |
| | $K$ | 2.15 | 3.23 | 3.55 | 3.65 | 3.67 | 3.68 | 3.68 | 3.68 | 3.68 | 3.68 | 3.68 | 3.68 | 3.68 | 3.68 | 3.68 | 3.68 | 3.68 | 3.68 | 3.68 | 3.68 | 3.68 | 3.68 | 3.68 | 3.68 | 3.68 |
| 0.7 | $U_1$ | 21.29 | 36.79 | 47.28 | 54.55 | 59.80 | 63.73 | 66.77 | 69.13 | 71.06 | 72.66 | 73.99 | 75.12 | 76.10 | 76.94 | 77.68 | 78.33 | 78.91 | 79.43 | 79.89 | 80.31 | 80.69 | 81.04 | 81.36 | 81.65 | 81.92 |
| | $U_2$ | 0.51 | 1.21 | 2.26 | 3.60 | 5.07 | 6.53 | 7.90 | 9.17 | 10.31 | 11.34 | 12.25 | 13.07 | 13.81 | 14.47 | 15.06 | 15.60 | 16.09 | 16.53 | 16.94 | 17.31 | 17.65 | 17.96 | 18.25 | 18.52 | 18.76 |
| | $W_{1,2}$ | 13.87 | 28.97 | 44.85 | 60.98 | 77.17 | 93.38 | 109.59 | 125.80 | 142.02 | 158.23 | 174.45 | 190.67 | 206.88 | 223.10 | 239.31 | 255.53 | 271.74 | 287.96 | 304.18 | 320.39 | 336.61 | 352.82 | 369.04 | 385.25 | 401.47 |
| | $W_{3,4}$ | 14.90 | 30.61 | 46.68 | 62.86 | 79.06 | 95.28 | 111.49 | 127.71 | 143.92 | 160.14 | 176.35 | 192.57 | 208.78 | 225.00 | 241.21 | 257.43 | 273.64 | 289.86 | 306.07 | 322.29 | 338.51 | 354.72 | 370.94 | 387.15 | 403.37 |
| | $J$ | 11.31 | 18.55 | 22.88 | 25.59 | 27.37 | 28.58 | 29.44 | 30.05 | 30.50 | 30.83 | 31.08 | 31.28 | 31.43 | 31.55 | 31.64 | 31.72 | 31.78 | 31.83 | 31.88 | 31.91 | 31.94 | 31.96 | 31.98 | 31.99 | 32.01 |
| | $K$ | 2.26 | 3.39 | 3.73 | 3.83 | 3.85 | 3.86 | 3.86 | 3.86 | 3.86 | 3.86 | 3.86 | 3.86 | 3.86 | 3.86 | 3.86 | 3.86 | 3.86 | 3.86 | 3.86 | 3.86 | 3.86 | 3.86 | 3.86 | 3.86 | 3.86 |

TABLE S7I. Parameters of the THIF interaction Hamiltonian for different ratios $u_0/u_1$ and screening lengths $\xi$ of the double-gate interaction potential ($U_\xi = 24$ meV for $\xi = 10$nm). We use the same BM model parameters as in Table S7O.

| $u_0/u_1$ | | $\xi$/nm = 2 | 4 | 6 | 8 | 10 | 12 | 14 | 16 | 18 | 20 | 22 | 24 | 26 | 28 | 30 | 32 | 34 | 36 | 38 | 40 | 42 | 44 | 46 | 48 | 50 |
|---|---|---|---|---|---|---|---|---|---|---|---|---|---|---|---|---|---|---|---|---|---|---|---|---|---|---|
| 0.6 | $U_1$ | 19.76 | 34.38 | 44.42 | 51.45 | 56.56 | 60.40 | 63.37 | 65.72 | 67.64 | 69.21 | 70.54 | 71.66 | 72.63 | 73.47 | 74.20 | 74.85 | 75.42 | 75.94 | 76.40 | 76.82 | 77.20 | 77.55 | 77.86 | 78.16 | 78.43 |
| | $U_2$ | 0.55 | 1.31 | 2.40 | 3.76 | 5.23 | 6.68 | 8.06 | 9.32 | 10.45 | 11.47 | 12.39 | 13.20 | 13.93 | 14.59 | 15.18 | 15.72 | 16.20 | 16.65 | 17.05 | 17.42 | 17.76 | 18.07 | 18.36 | 18.63 | 18.87 |
| | $W_{1,2}$ | 13.64 | 28.64 | 44.48 | 60.60 | 76.79 | 93.00 | 109.29 | 125.43 | 141.45 | 157.86 | 174.08 | 190.29 | 206.51 | 222.72 | 238.94 | 255.15 | 271.37 | 287.58 | 303.80 | 320.01 | 336.23 | 352.44 | 368.66 | 384.88 | 401.09 |
| | $W_{3,4}$ | 14.84 | 30.50 | 46.55 | 62.73 | 78.93 | 95.15 | 111.36 | 127.58 | 143.79 | 160.01 | 176.22 | 192.44 | 208.65 | 224.87 | 241.08 | 257.30 | 273.51 | 289.73 | 305.94 | 322.16 | 338.38 | 354.59 | 370.81 | 387.02 | 403.24 |
| | $J$ | 11.73 | 19.51 | 24.32 | 27.39 | 29.45 | 30.86 | 31.87 | 32.59 | 33.13 | 33.53 | 33.83 | 34.07 | 34.25 | 34.34 | 34.58 | 34.61 | 34.68 | 34.75 | 34.80 | 34.84 | 34.88 | 34.91 | 34.93 | 34.95 | 34.97 |
| | $K$ | 2.39 | 3.58 | 3.94 | 4.04 | 4.06 | 4.07 | 4.07 | 4.07 | 4.07 | 4.07 | 4.07 | 4.07 | 4.07 | 4.07 | 4.07 | 4.07 | 4.07 | 4.07 | 4.07 | 4.07 | 4.07 | 4.07 | 4.07 | 4.07 | 4.07 |
| 0.5 | $U_1$ | 18.37 | 32.15 | 41.75 | 48.53 | 53.50 | 57.25 | 60.16 | 62.48 | 64.37 | 65.93 | 67.24 | 68.35 | 69.31 | 70.14 | 70.87 | 71.52 | 72.09 | 72.60 | 73.06 | 73.48 | 73.86 | 74.21 | 74.52 | 74.81 | 75.08 |
| | $U_2$ | 0.60 | 1.41 | 2.53 | 3.91 | 5.38 | 6.84 | 8.20 | 9.45 | 10.58 | 11.60 | 12.51 | 13.32 | 14.05 | 14.70 | 15.29 | 15.82 | 16.31 | 16.76 | 17.15 | 17.52 | 17.86 | 18.17 | 18.45 | 18.72 | 18.97 |
| | $W_{1,2}$ | 13.42 | 28.32 | 44.13 | 60.24 | 76.43 | 92.64 | 108.85 | 125.07 | 141.28 | 157.50 | 173.71 | 189.93 | 206.14 | 222.36 | 238.57 | 254.79 | 271.00 | 287.22 | 303.44 | 319.65 | 335.87 | 352.08 | 368.30 | 384.51 | 400.73 |
| | $W_{3,4}$ | 14.83 | 30.44 | 46.51 | 62.68 | 78.88 | 95.10 | 111.31 | 127.53 | 143.74 | 159.96 | 176.17 | 192.39 | 208.60 | 224.82 | 241.03 | 257.25 | 273.46 | 289.68 | 305.89 | 322.11 | 338.32 | 354.54 | 370.76 | 386.97 | 403.19 |
| | $J$ | 12.16 | 20.47 | 25.76 | 29.22 | 31.57 | 33.22 | 34.40 | 35.26 | 35.90 | 36.40 | 36.78 | 37.07 | 37.30 | 37.49 | 37.64 | 37.76 | 37.86 | 37.94 | 38.01 | 38.06 | 38.11 | 38.15 | 38.19 | 38.22 | 38.25 |
| | $K$ | 2.54 | 3.79 | 4.18 | 4.28 | 4.31 | 4.31 | 4.31 | 4.31 | 4.31 | 4.31 | 4.31 | 4.31 | 4.31 | 4.31 | 4.31 | 4.31 | 4.31 | 4.31 | 4.31 | 4.31 | 4.31 | 4.31 | 4.31 | 4.31 | 4.31 |
| 0.4 | $U_1$ | 17.50 | 30.77 | 40.10 | 46.74 | 51.63 | 55.33 | 58.21 | 60.51 | 62.38 | 63.93 | 65.23 | 66.34 | 67.30 | 68.13 | 68.85 | 69.50 | 70.07 | 70.58 | 71.04 | 71.46 | 71.83 | 72.18 | 72.50 | 72.79 | 73.05 |
| | $U_2$ | 0.63 | 1.46 | 2.61 | 4.00 | 5.48 | 6.93 | 8.29 | 9.54 | 10.67 | 11.68 | 12.58 | 13.39 | 14.12 | 14.77 | 15.36 | 15.89 | 16.38 | 16.82 | 17.22 | 17.58 | 17.92 | 18.23 | 18.52 | 18.79 | 19.03 |
| | $W_{1,2}$ | 13.17 | 27.95 | 43.73 | 59.83 | 76.02 | 92.23 | 108.44 | 124.66 | 140.87 | 157.09 | 173.30 | 189.52 | 205.73 | 221.95 | 238.16 | 254.38 | 270.59 | 286.81 | 303.02 | 319.24 | 335.45 | 351.67 | 367.89 | 384.10 | 400.32 |
| | $W_{3,4}$ | 14.82 | 30.41 | 46.48 | 62.65 | 78.85 | 95.06 | 111.28 | 127.49 | 143.70 | 159.92 | 176.13 | 192.35 | 208.56 | 224.78 | 241.00 | 257.21 | 273.43 | 289.65 | 305.86 | 322.08 | 338.29 | 354.51 | 370.72 | 386.94 | 403.15 |
| | $J$ | 12.50 | 21.21 | 26.84 | 30.57 | 33.11 | 34.90 | 36.18 | 37.12 | 37.83 | 38.37 | 38.78 | 39.10 | 39.36 | 39.53 | 39.73 | 39.79 | 39.86 | 39.93 | 40.09 | 40.14 | 40.20 | 40.25 | 40.30 | 40.37 | 40.40 |
| | $K$ | 2.69 | 4.02 | 4.43 | 4.56 | 4.57 | 4.57 | 4.57 | 4.57 | 4.57 | 4.57 | 4.57 | 4.57 | 4.57 | 4.57 | 4.57 | 4.57 | 4.57 | 4.57 | 4.57 | 4.57 | 4.57 | 4.57 | 4.57 | 4.57 | 4.57 |
| 0.3 | $U_1$ | 16.80 | 29.65 | 38.75 | 45.27 | 50.08 | 53.74 | 56.59 | 58.87 | 60.73 | 62.27 | 63.57 | 64.67 | 65.62 | 66.45 | 67.18 | 67.82 | 68.39 | 68.90 | 69.36 | 69.77 | 70.15 | 70.49 | 70.81 | 71.10 | 71.37 |
| | $U_2$ | 0.66 | 1.52 | 2.68 | 4.08 | 5.56 | 7.01 | 8.37 | 9.62 | 10.74 | 11.75 | 12.65 | 13.46 | 14.18 | 14.83 | 15.42 | 15.95 | 16.43 | 16.87 | 17.27 | 17.64 | 17.98 | 18.29 | 18.58 | 18.84 | 19.10 |
| | $W_{1,2}$ | 12.94 | 27.62 | 43.37 | 59.46 | 75.65 | 91.85 | 108.07 | 124.28 | 140.50 | 156.71 | 172.93 | 189.14 | 205.36 | 221.57 | 237.79 | 254.00 | 270.22 | 286.43 | 302.65 | 318.86 | 335.08 | 351.30 | 367.51 | 383.73 | 399.94 |
| | $W_{3,4}$ | 14.82 | 30.43 | 46.42 | 62.64 | 78.83 | 95.04 | 111.26 | 127.47 | 143.69 | 159.91 | 176.12 | 192.33 | 208.55 | 224.76 | 240.98 | 257.19 | 273.41 | 289.63 | 305.84 | 322.06 | 338.27 | 354.49 | 370.70 | 386.92 | 403.15 |
| | $J$ | 12.81 | 21.86 | 27.78 | 31.74 | 34.45 | 36.38 | 37.74 | 38.75 | 39.51 | 40.09 | 40.54 | 40.89 | 41.17 | 41.39 | 41.57 | 41.71 | 41.83 | 41.93 | 42.02 | 42.09 | 42.15 | 42.20 | 42.24 | 42.28 | 42.32 |
| | $K$ | 2.84 | 4.24 | 4.67 | 4.81 | 4.82 | 4.82 | 4.82 | 4.82 | 4.82 | 4.82 | 4.82 | 4.82 | 4.82 | 4.82 | 4.82 | 4.82 | 4.82 | 4.82 | 4.82 | 4.82 | 4.82 | 4.82 | 4.82 | 4.82 | 4.82 |
| 0.2 | $U_1$ | 16.51 | 29.01 | 37.93 | 44.69 | 49.48 | 53.13 | 55.97 | 58.25 | 60.11 | 61.65 | 62.94 | 64.04 | 64.99 | 65.82 | 66.54 | 67.18 | 67.75 | 68.27 | 68.72 | 69.14 | 69.52 | 69.86 | 70.18 | 70.47 | 70.73 |
| | $U_2$ | 0.66 | 1.52 | 2.68 | 4.10 | 5.58 | 7.03 | 8.39 | 9.63 | 10.76 | 11.76 | 12.67 | 13.44 | 14.14 | 14.85 | 15.41 | 15.93 | 16.42 | 16.86 | 17.29 | 17.65 | 17.99 | 18.30 | 18.59 | 18.85 | 19.10 |
| | $W_{1,2}$ | 12.72 | 27.30 | 43.02 | 59.10 | 75.28 | 91.49 | 107.70 | 123.92 | 140.13 | 156.35 | 172.57 | 188.78 | 205.00 | 221.21 | 237.43 | 253.64 | 269.86 | 286.07 | 302.29 | 318.50 | 334.72 | 350.93 | 367.15 | 383.36 | 399.58 |
| | $W_{3,4}$ | 14.82 | 30.42 | 46.46 | 62.63 | 78.83 | 95.04 | 111.24 | 127.47 | 143.68 | 159.90 | 176.12 | 192.33 | 208.54 | 224.76 | 240.97 | 257.19 | 273.41 | 289.62 | 305.84 | 322.05 | 338.27 | 354.49 | 370.70 | 386.92 | 403.13 |
| | $J$ | 13.02 | 22.28 | 28.37 | 32.45 | 35.25 | 37.22 | 38.64 | 39.69 | 40.47 | 41.07 | 41.53 | 41.89 | 42.17 | 42.40 | 42.59 | 42.74 | 42.86 | 42.96 | 43.05 | 43.12 | 43.18 | 43.23 | 43.28 | 43.32 | 43.35 |
| | $K$ | 2.98 | 4.43 | 4.88 | 5.00 | 5.04 | 5.04 | 5.04 | 5.04 | 5.04 | 5.04 | 5.04 | 5.04 | 5.04 | 5.04 | 5.04 | 5.04 | 5.04 | 5.04 | 5.04 | 5.04 | 5.04 | 5.04 | 5.04 | 5.04 | 5.04 |
| 0.1 | $U_1$ | 16.77 | 29.66 | 38.82 | 45.39 | 50.23 | 53.91 | 56.78 | 59.08 | 60.95 | 62.50 | 63.80 | 64.91 | 65.86 | 66.69 | 67.42 | 68.06 | 68.63 | 69.14 | 69.60 | 70.02 | 70.40 | 70.74 | 71.06 | 71.35 | 71.62 |
| | $U_2$ | 0.63 | 1.47 | 2.64 | 4.02 | 5.51 | 6.97 | 8.34 | 9.59 | 10.71 | 11.73 | 12.63 | 13.44 | 14.17 | 14.82 | 15.41 | 15.94 | 16.42 | 16.86 | 17.25 | 17.63 | 17.97 | 18.28 | 18.57 | 18.83 | 19.08 |
| | $W_{1,2}$ | 12.50 | 26.97 | 42.66 | 58.73 | 74.91 | 91.12 | 107.33 | 123.55 | 139.76 | 155.98 | 172.19 | 188.41 | 204.62 | 220.84 | 237.05 | 253.27 | 269.49 | 285.70 | 301.92 | 318.13 | 334.35 | 350.56 | 366.78 | 382.99 | 399.21 |
| | $W_{3,4}$ | 14.78 | 30.37 | 46.57 | 62.57 | 78.77 | 94.98 | 111.20 | 127.41 | 143.63 | 159.84 | 176.06 | 192.27 | 208.49 | 224.70 | 240.88 | 257.13 | 273.35 | 289.57 | 305.78 | 322.00 | 338.21 | 354.43 | 370.64 | 386.86 | 403.07 |
| | $J$ | 13.13 | 22.39 | 28.41 | 32.41 | 35.26 | 37.16 | 38.52 | 39.51 | 40.47 | 40.80 | 41.22 | 41.54 | 41.81 | 42.00 | 42.17 | 42.32 | 42.45 | 42.57 | 42.63 | 42.68 | 42.72 | 42.76 | 42.80 | 42.82 | 42.85 |
| | $K$ | 3.07 | 4.57 | 5.03 | 5.15 | 5.18 | 5.19 | 5.19 | 5.19 | 5.19 | 5.19 | 5.19 | 5.19 | 5.19 | 5.19 | 5.19 | 5.19 | 5.19 | 5.19 | 5.19 | 5.19 | 5.19 | 5.19 | 5.19 | 5.19 | 5.19 |
| 0.0 | $U_1$ | 17.08 | 30.19 | 39.49 | 46.15 | 51.04 | 54.76 | 57.65 | 59.96 | 61.84 | 63.39 | 64.70 | 65.81 | 66.77 | 67.60 | 68.33 | 68.97 | 69.54 | 70.06 | 70.52 | 70.93 | 71.31 | 71.66 | 71.97 | 72.27 | 72.53 |
| | $U_2$ | 0.60 | 1.41 | 2.56 | 3.95 | 5.43 | 6.89 | 8.25 | 9.51 | 10.63 | 11.65 | 12.55 | 13.37 | 14.09 | 14.75 | 15.34 | 15.87 | 16.35 | 16.79 | 17.19 | 17.56 | 17.90 | 18.21 | 18.50 | 18.77 | 19.01 |
| | $W_{1,2}$ | 12.38 | 26.81 | 42.48 | 58.55 | 74.73 | 90.94 | 107.15 | 123.36 | 139.58 | 155.79 | 172.01 | 188.22 | 204.44 | 220.66 | 236.87 | 253.09 | 269.30 | 285.52 | 301.73 | 317.95 | 334.16 | 350.38 | 366.59 | 382.81 | 399.02 |
| | $W_{3,4}$ | 14.75 | 30.33 | 46.36 | 62.52 | 78.73 | 94.94 | 111.15 | 127.37 | 143.58 | 159.80 | 176.02 | 192.23 | 208.44 | 224.66 | 240.88 | 257.09 | 273.31 | 289.52 | 305.74 | 321.95 | 338.17 | 354.38 | 370.60 | 386.81 | 403.03 |
| | $J$ | 13.11 | 22.39 | 28.38 | 32.38 | 35.06 | 36.92 | 38.24 | 39.20 | 39.92 | 40.45 | 40.86 | 41.22 | 41.48 | 41.71 | 41.90 | 42.05 | 42.18 | 42.30 | 42.39 | 42.46 | 42.53 | 42.59 | 42.64 | 42.69 | 42.73 |
| | $K$ | 3.10 | 4.61 | 5.08 | 5.21 | 5.24 | 5.25 | 5.25 | 5.25 | 5.25 | 5.25 | 5.25 | 5.25 | 5.25 | 5.25 | 5.25 | 5.25 | 5.25 | 5.25 | 5.25 | 5.25 | 5.25 | 5.25 | 5.25 | 5.25 | 5.25 |

138

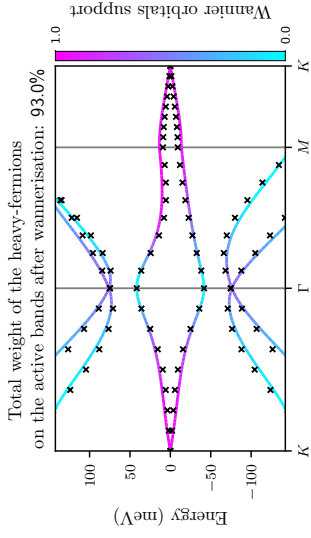

Total weight of the heavy-fermions on the active bands after wannierisation: 93.0%

Wannier orbitals support

*Band structure figure: Energy (meV) vs. K–Γ–M–K path, BM bands (lines) and THF bands (crosses), colorbar 0.0–1.0.*

**FIG. S52.** Band structures of the BM and THF models near charge neutrality for $w_0/w_1 = 0.8$, depicted by lines and crosses, respectively. The BM bands are colored according to the weight of the $f$-electron wave function on them. We use the same BM parameters as in Table S72.

**TABLE S72.** Parameters of the $f$-electron Wannier functions and the THF single-particle Hamiltonian for different values of the tunneling ratio $w_0/w_1$. $W$ denotes the total weight of the THF $f$-electrons on the active bands. In computing the ratios $\gamma/U_1$ and $\gamma/U_2$, we employ the on-site and nearest-neighbor repulsion parameters $U_1$ and $U_2$ obtained numerically for $\xi = 10$ nm, as given in Table S73. We employ $v_F = 5.944$ eV·Å, $|\mathbf{K}| = 1.703$ Å$^{-1}$, $w_1 = 110$ meV, and $\theta = 1.38°$ for the BM model.

| $w_0/w_1$ | $\alpha_1$ | $\alpha_2$ | $\lambda_1$ | $\lambda_2$ | $\lambda$ | $W$ | $t_0$ | $\gamma$ | $M$ | $v_*$ | $v_*'$ | $v_*''$ | $\gamma/U_1$ | $\gamma/U_2$ |
|---|---|---|---|---|---|---|---|---|---|---|---|---|---|---|
| 1.00 | 0.810 | 0.587 | 0.197 | 0.209 | 0.359 | 0.945 | 3.006 | -44.270 | -38.665 | -4.702 | 1.835 | 0.0230 | -0.61 | -9.27 |
| 0.90 | 0.838 | 0.545 | 0.209 | 0.215 | 0.371 | 0.935 | 3.597 | -60.447 | -40.240 | -4.821 | 1.856 | 0.0170 | -0.88 | -12.32 |
| 0.80 | 0.867 | 0.499 | 0.227 | 0.221 | 0.386 | 0.930 | 4.191 | -76.159 | -41.607 | -4.931 | 1.833 | 0.0126 | -1.18 | -15.05 |
| 0.70 | 0.895 | 0.447 | 0.239 | 0.227 | 0.409 | 0.933 | 4.954 | -91.199 | -42.779 | -5.036 | 1.781 | 0.0094 | -1.51 | -17.27 |
| 0.60 | 0.920 | 0.392 | 0.257 | 0.227 | 0.423 | 0.932 | 5.587 | -105.332 | -43.771 | -5.137 | 1.692 | 0.0069 | -1.84 | -19.40 |
| 0.50 | 0.943 | 0.333 | 0.269 | 0.233 | 0.441 | 0.936 | 6.121 | -118.279 | -44.592 | -5.233 | 1.557 | 0.0049 | -2.18 | -21.18 |
| 0.40 | 0.962 | 0.272 | 0.286 | 0.239 | 0.450 | 0.936 | 6.574 | -129.710 | -45.251 | -5.324 | 1.366 | 0.0033 | -2.48 | -22.84 |
| 0.30 | 0.978 | 0.208 | 0.298 | 0.239 | 0.457 | 0.937 | 6.940 | -139.244 | -45.757 | -5.405 | 1.111 | 0.0020 | -2.74 | -24.17 |
| 0.20 | 0.990 | 0.142 | 0.310 | 0.245 | 0.460 | 0.936 | 7.112 | -146.469 | -46.114 | -5.471 | 0.790 | 0.0009 | -2.92 | -25.30 |
| 0.10 | 0.997 | 0.072 | 0.316 | 0.245 | 0.459 | 0.934 | 7.157 | -150.999 | -46.327 | -5.515 | 0.412 | 0.0002 | -3.02 | -26.09 |
| 0.00 | 1.000 | 0.000 | 0.316 | 0.000 | 0.451 | 0.928 | 7.068 | -152.544 | -46.397 | -5.530 | -0.000 | -0.0000 | -3.01 | -26.63 |

**TABLE S73.** Band structures of the BM and THF models near charge neutrality for $w_0/w_1 = 0.8$, depicted by lines and crosses, respectively. The BM bands are colored according to the weight of the $f$-electron wave function on them. We use the same BM parameters as in Table S72.

| $w_0/w_1$ | $\xi/\mathrm{nm}$ | 2 | 4 | 6 | 8 | 10 | 12 | 14 | 16 | 18 | 20 | 22 | 24 | 26 | 28 | 30 | 32 | 34 | 36 | 38 | 40 | 42 | 44 | 46 | 48 | 50 |
|---|---|---|---|---|---|---|---|---|---|---|---|---|---|---|---|---|---|---|---|---|---|---|---|---|---|---|
| 1.0 | $U_1$ | 27.51 | 36.37 | 58.47 | 66.57 | 72.26 | 76.45 | 79.63 | 82.13 | 84.14 | 85.79 | 87.16 | 88.32 | 89.31 | 90.17 | 90.92 | 91.58 | 92.16 | 92.68 | 93.15 | 93.58 | 93.96 | 94.31 | 94.63 | 94.93 | 95.20 |
| | $U_2$ | 0.39 | 0.97 | 1.95 | 3.28 | 4.78 | 6.28 | 7.70 | 9.00 | 10.18 | 11.23 | 12.17 | 13.02 | 13.76 | 14.44 | 15.04 | 15.59 | 16.08 | 16.53 | 16.94 | 17.31 | 17.66 | 17.97 | 18.26 | 18.53 | 18.79 |
| | $W_{1,2}$ | 14.88 | 30.72 | 47.16 | 63.79 | 80.47 | 97.16 | 113.86 | 130.55 | 147.25 | 163.94 | 180.64 | 197.34 | 214.03 | 230.73 | 247.42 | 264.12 | 280.81 | 297.51 | 314.21 | 330.90 | 347.60 | 364.29 | 380.99 | 397.68 | 414.38 |
| | $W_{3,4}$ | 15.82 | 32.30 | 48.96 | 65.64 | 82.34 | 99.03 | 115.73 | 132.42 | 149.12 | 165.82 | 182.51 | 199.21 | 215.90 | 232.60 | 249.29 | 265.99 | 282.69 | 299.38 | 316.08 | 332.77 | 349.47 | 366.16 | 382.86 | 399.56 | 416.25 |
| | $J$ | 10.38 | 16.01 | 18.87 | 20.44 | 21.38 | 21.98 | 22.38 | 22.65 | 22.85 | 22.99 | 23.09 | 23.17 | 23.23 | 23.27 | 23.31 | 23.34 | 23.36 | 23.38 | 23.39 | 23.41 | 23.42 | 23.43 | 23.43 | 23.44 | 23.44 |
| | $K$ | 1.08 | 2.96 | 3.34 | 3.36 | 3.36 | 3.36 | 3.36 | 3.36 | 3.36 | 3.36 | 3.36 | 3.36 | 3.36 | 3.36 | 3.36 | 3.36 | 3.36 | 3.36 | 3.36 | 3.36 | 3.36 | 3.36 | 3.36 | 3.36 | 3.36 |
| 0.9 | $U_1$ | 25.50 | 43.34 | 54.99 | 62.87 | 68.45 | 72.57 | 75.71 | 78.19 | 80.18 | 81.81 | 83.18 | 84.33 | 85.32 | 86.17 | 86.92 | 87.58 | 88.16 | 88.68 | 89.15 | 89.57 | 89.96 | 90.31 | 90.63 | 90.92 | 91.19 |
| | $U_2$ | 0.42 | 1.04 | 2.05 | 3.40 | 4.90 | 6.41 | 7.83 | 9.13 | 10.30 | 11.35 | 12.29 | 13.12 | 13.87 | 14.54 | 15.15 | 15.69 | 16.18 | 16.63 | 17.04 | 17.42 | 17.76 | 18.07 | 18.37 | 18.64 | 18.89 |
| | $W_{1,2}$ | 14.73 | 30.51 | 46.93 | 63.56 | 80.23 | 96.93 | 113.62 | 130.32 | 147.01 | 163.71 | 180.40 | 197.10 | 213.79 | 230.49 | 247.19 | 263.88 | 280.58 | 297.27 | 313.97 | 330.66 | 347.36 | 364.06 | 380.75 | 397.45 | 414.14 |
| | $W_{3,4}$ | 15.56 | 31.89 | 48.49 | 65.17 | 81.86 | 98.55 | 115.25 | 131.94 | 148.64 | 165.33 | 182.03 | 198.72 | 215.42 | 232.12 | 248.81 | 265.51 | 282.20 | 298.90 | 315.59 | 332.29 | 348.99 | 365.68 | 382.38 | 399.07 | 415.77 |
| | $J$ | 10.74 | 16.95 | 20.30 | 22.23 | 23.44 | 24.22 | 24.75 | 25.12 | 25.38 | 25.57 | 25.71 | 25.82 | 25.90 | 25.97 | 26.02 | 26.06 | 26.09 | 26.11 | 26.13 | 26.15 | 26.17 | 26.18 | 26.19 | 26.20 | 26.21 |
| | $K$ | 2.05 | 3.06 | 3.37 | 3.44 | 3.46 | 3.47 | 3.47 | 3.47 | 3.47 | 3.47 | 3.47 | 3.47 | 3.47 | 3.47 | 3.47 | 3.47 | 3.47 | 3.47 | 3.47 | 3.47 | 3.47 | 3.47 | 3.47 | 3.47 | 3.47 |
| 0.8 | $U_1$ | 23.57 | 40.38 | 51.54 | 59.17 | 64.62 | 68.66 | 71.76 | 74.20 | 76.17 | 77.79 | 79.15 | 80.29 | 81.27 | 82.13 | 82.87 | 83.52 | 84.11 | 84.63 | 85.09 | 85.52 | 85.90 | 86.25 | 86.57 | 86.86 | 87.13 |
| | $U_2$ | 0.46 | 1.13 | 2.18 | 3.55 | 5.06 | 6.56 | 7.98 | 9.27 | 10.44 | 11.49 | 12.42 | 13.25 | 14.00 | 14.67 | 15.27 | 15.81 | 16.30 | 16.75 | 17.16 | 17.53 | 17.87 | 18.18 | 18.48 | 18.75 | 19.00 |
| | $W_{1,2}$ | 14.54 | 30.23 | 46.63 | 63.25 | 79.92 | 96.61 | 113.31 | 130.00 | 146.70 | 163.39 | 180.09 | 196.79 | 213.48 | 230.18 | 246.87 | 263.57 | 280.26 | 296.96 | 313.66 | 330.35 | 347.05 | 363.74 | 380.44 | 397.13 | 413.83 |
| | $W_{3,4}$ | 15.31 | 31.61 | 48.18 | 64.84 | 81.53 | 98.22 | 114.92 | 131.61 | 148.31 | 165.00 | 181.70 | 198.40 | 215.09 | 231.79 | 248.48 | 265.18 | 281.88 | 298.57 | 315.27 | 331.96 | 348.66 | 365.35 | 382.05 | 398.75 | 415.44 |
| | $J$ | 11.14 | 17.92 | 21.77 | 24.08 | 25.55 | 26.53 | 27.20 | 27.68 | 28.02 | 28.27 | 28.46 | 28.60 | 28.71 | 28.80 | 28.86 | 28.92 | 28.96 | 29.00 | 29.02 | 29.05 | 29.07 | 29.08 | 29.10 | 29.11 | 29.12 |
| | $K$ | 2.14 | 3.19 | 3.51 | 3.59 | 3.61 | 3.62 | 3.62 | 3.62 | 3.62 | 3.62 | 3.62 | 3.62 | 3.62 | 3.62 | 3.62 | 3.62 | 3.62 | 3.62 | 3.62 | 3.62 | 3.62 | 3.62 | 3.62 | 3.62 | 3.62 |
| 0.7 | $U_1$ | 21.52 | 37.15 | 47.71 | 55.02 | 60.28 | 64.17 | 67.28 | 69.64 | 71.58 | 73.18 | 74.52 | 75.65 | 76.63 | 77.47 | 78.20 | 78.86 | 79.44 | 79.96 | 80.42 | 80.84 | 81.22 | 81.57 | 81.89 | 82.18 | 82.45 |
| | $U_2$ | 0.53 | 1.27 | 2.37 | 3.77 | 5.28 | 6.78 | 8.19 | 9.47 | 10.63 | 11.67 | 12.60 | 13.43 | 14.17 | 14.84 | 15.43 | 15.97 | 16.46 | 16.91 | 17.32 | 17.69 | 18.03 | 18.34 | 18.63 | 18.90 | 19.15 |
| | $W_{1,2}$ | 14.34 | 29.94 | 46.31 | 62.92 | 79.60 | 96.29 | 112.98 | 129.68 | 146.37 | 163.07 | 179.76 | 196.46 | 213.16 | 229.85 | 246.55 | 263.24 | 279.94 | 296.63 | 313.33 | 330.03 | 346.72 | 363.42 | 380.11 | 396.81 | 413.50 |
| | $W_{3,4}$ | 15.05 | 31.41 | 48.00 | 64.66 | 81.35 | 98.04 | 114.74 | 131.43 | 148.13 | 164.82 | 181.52 | 198.22 | 214.91 | 231.61 | 248.30 | 265.00 | 281.69 | 298.39 | 315.09 | 331.78 | 348.48 | 365.17 | 381.87 | 398.56 | 415.26 |
| | $J$ | 11.60 | 19.10 | 23.44 | 26.20 | 28.01 | 29.24 | 30.10 | 30.72 | 31.17 | 31.50 | 31.76 | 31.95 | 32.10 | 32.22 | 32.32 | 32.39 | 32.45 | 32.50 | 32.54 | 32.58 | 32.61 | 32.63 | 32.65 | 32.66 | 32.68 |
| | $K$ | 2.26 | 3.36 | 3.69 | 3.78 | 3.80 | 3.80 | 3.81 | 3.81 | 3.81 | 3.81 | 3.81 | 3.81 | 3.81 | 3.81 | 3.81 | 3.81 | 3.81 | 3.81 | 3.81 | 3.81 | 3.81 | 3.81 | 3.81 | 3.81 | 3.81 |

| $u_0/u_1$ | $\xi$/nm | 2 | 4 | 6 | 8 | 10 | 12 | 14 | 16 | 18 | 20 | 22 | 24 | 26 | 28 | 30 | 32 | 34 | 36 | 38 | 40 | 42 | 44 | 46 | 48 | 50 |
|---|---|---|---|---|---|---|---|---|---|---|---|---|---|---|---|---|---|---|---|---|---|---|---|---|---|---|
| 0.6 | $U_1$ | 20.05 | 34.84 | 44.97 | 52.06 | 57.20 | 61.05 | 64.04 | 66.40 | 68.32 | 69.90 | 71.23 | 72.36 | 73.32 | 74.16 | 74.90 | 75.55 | 76.12 | 76.64 | 77.10 | 77.52 | 77.90 | 78.25 | 78.57 | 78.86 | 79.13 |
| | $U_2$ | 0.57 | 1.36 | 2.50 | 3.91 | 5.43 | 6.93 | 8.33 | 9.61 | 10.77 | 11.80 | 12.72 | 13.55 | 14.29 | 14.95 | 15.55 | 16.08 | 16.57 | 17.02 | 17.42 | 17.79 | 18.13 | 18.45 | 18.74 | 19.00 | 19.25 |
| | $W_{1,2}$ | 14.10 | 29.59 | 45.93 | 62.53 | 79.21 | 95.90 | 112.59 | 129.29 | 145.98 | 162.68 | 179.37 | 196.07 | 212.77 | 229.46 | 246.16 | 262.85 | 279.55 | 296.24 | 312.94 | 329.64 | 346.33 | 363.03 | 379.72 | 396.42 | 413.11 |
| | $W_{3,4}$ | 15.25 | 31.36 | 47.89 | 64.55 | 81.23 | 97.93 | 114.62 | 131.32 | 148.01 | 164.71 | 181.40 | 198.10 | 214.80 | 231.49 | 248.19 | 264.88 | 281.58 | 298.27 | 314.97 | 331.67 | 348.36 | 365.06 | 381.75 | 398.45 | 415.14 |
| | $J$ | 12.03 | 19.98 | 24.88 | 28.00 | 30.08 | 31.50 | 32.51 | 33.23 | 33.77 | 34.17 | 34.47 | 34.71 | 34.89 | 35.03 | 35.15 | 35.24 | 35.31 | 35.38 | 35.43 | 35.47 | 35.50 | 35.53 | 35.56 | 35.58 | 35.60 |
| | $K$ | 2.39 | 3.56 | 3.91 | 4.00 | 4.02 | 4.03 | 4.03 | 4.03 | 4.03 | 4.03 | 4.03 | 4.03 | 4.03 | 4.03 | 4.03 | 4.03 | 4.03 | 4.03 | 4.03 | 4.03 | 4.03 | 4.03 | 4.03 | 4.03 | 4.03 |
| 0.5 | $U_1$ | 18.67 | 32.63 | 42.33 | 49.18 | 54.18 | 57.95 | 60.88 | 63.21 | 65.10 | 66.62 | 67.98 | 69.10 | 70.06 | 70.89 | 71.62 | 72.27 | 72.84 | 73.36 | 73.82 | 74.24 | 74.62 | 74.96 | 75.28 | 75.57 | 75.84 |
| | $U_2$ | 0.62 | 1.46 | 2.64 | 4.06 | 5.67 | 7.08 | 8.47 | 9.75 | 10.92 | 11.99 | 12.84 | 13.66 | 14.40 | 15.06 | 15.65 | 16.19 | 16.67 | 17.12 | 17.52 | 17.89 | 18.23 | 18.54 | 18.83 | 19.10 | 19.35 |
| | $W_{1,2}$ | 13.87 | 29.26 | 45.57 | 62.17 | 78.84 | 95.53 | 112.22 | 128.92 | 145.61 | 162.31 | 179.00 | 195.70 | 212.40 | 229.09 | 245.79 | 262.48 | 279.18 | 295.88 | 312.57 | 329.27 | 345.96 | 362.66 | 379.35 | 396.05 | 412.75 |
| | $W_{3,4}$ | 15.24 | 31.34 | 47.86 | 64.51 | 81.20 | 97.89 | 114.59 | 131.28 | 147.98 | 164.67 | 181.37 | 198.06 | 214.76 | 231.46 | 248.15 | 264.85 | 281.54 | 298.24 | 314.93 | 331.63 | 348.33 | 365.02 | 381.72 | 398.41 | 415.11 |
| | $J$ | 12.47 | 20.96 | 26.34 | 29.85 | 32.22 | 33.87 | 35.06 | 35.92 | 36.57 | 37.05 | 37.43 | 37.72 | 37.95 | 38.13 | 38.28 | 38.40 | 38.49 | 38.57 | 38.64 | 38.70 | 38.75 | 38.79 | 38.82 | 38.85 | 38.87 |
| | $K$ | 2.54 | 3.78 | 4.15 | 4.24 | 4.27 | 4.27 | 4.27 | 4.27 | 4.27 | 4.27 | 4.27 | 4.27 | 4.27 | 4.27 | 4.27 | 4.27 | 4.27 | 4.27 | 4.27 | 4.27 | 4.27 | 4.27 | 4.27 | 4.27 | 4.27 |
| 0.4 | $U_1$ | 17.81 | 31.28 | 40.72 | 47.42 | 52.34 | 56.07 | 58.97 | 61.28 | 63.16 | 64.71 | 66.02 | 67.13 | 68.09 | 68.92 | 69.65 | 70.29 | 70.86 | 71.38 | 71.84 | 72.25 | 72.63 | 72.98 | 73.29 | 73.59 | 73.85 |
| | $U_2$ | 0.65 | 1.52 | 2.72 | 4.16 | 5.68 | 7.17 | 8.56 | 9.83 | 10.98 | 12.00 | 12.92 | 13.74 | 14.47 | 15.13 | 15.72 | 16.26 | 16.74 | 17.18 | 17.59 | 17.96 | 18.29 | 18.61 | 18.90 | 19.16 | 19.41 |
| | $W_{1,2}$ | 13.61 | 28.80 | 44.57 | 61.75 | 78.42 | 95.11 | 111.81 | 128.50 | 145.20 | 161.89 | 178.59 | 195.28 | 211.98 | 228.68 | 245.37 | 262.07 | 278.76 | 295.46 | 312.15 | 328.85 | 345.55 | 362.24 | 378.94 | 395.63 | 412.33 |
| | $W_{3,4}$ | 15.24 | 31.32 | 47.83 | 64.49 | 81.17 | 97.86 | 114.56 | 131.25 | 147.95 | 164.64 | 181.34 | 198.03 | 214.73 | 231.43 | 248.12 | 264.82 | 281.52 | 298.21 | 314.91 | 331.60 | 348.30 | 365.00 | 381.69 | 398.39 | 415.08 |
| | $J$ | 12.82 | 21.71 | 27.44 | 31.20 | 33.76 | 35.55 | 36.84 | 37.78 | 38.48 | 39.02 | 39.44 | 39.74 | 40.00 | 40.20 | 40.36 | 40.49 | 40.59 | 40.69 | 40.76 | 40.83 | 40.88 | 40.92 | 40.96 | 40.99 | 41.02 |
| | $K$ | 2.70 | 4.01 | 4.40 | 4.50 | 4.53 | 4.53 | 4.54 | 4.54 | 4.54 | 4.54 | 4.54 | 4.54 | 4.54 | 4.54 | 4.54 | 4.54 | 4.54 | 4.54 | 4.54 | 4.54 | 4.54 | 4.54 | 4.54 | 4.54 | 4.54 |
| 0.3 | $U_1$ | 17.12 | 30.17 | 39.39 | 45.98 | 50.83 | 54.51 | 57.38 | 59.67 | 61.54 | 63.07 | 64.39 | 65.50 | 66.45 | 67.28 | 68.01 | 68.65 | 69.22 | 69.73 | 70.19 | 70.60 | 70.98 | 71.33 | 71.64 | 71.93 | 72.20 |
| | $U_2$ | 0.68 | 1.57 | 2.79 | 4.23 | 5.76 | 7.25 | 8.64 | 9.91 | 11.05 | 12.08 | 12.98 | 13.81 | 14.55 | 15.19 | 15.78 | 16.33 | 16.80 | 17.24 | 17.66 | 18.01 | 18.35 | 18.66 | 18.95 | 19.22 | 19.48 |
| | $W_{1,2}$ | 13.38 | 28.51 | 44.80 | 61.38 | 78.05 | 94.73 | 111.42 | 128.12 | 144.82 | 161.51 | 178.21 | 194.91 | 211.60 | 228.30 | 245.00 | 261.69 | 278.39 | 295.08 | 311.78 | 328.47 | 345.17 | 361.86 | 378.56 | 395.25 | 411.95 |
| | $W_{3,4}$ | 15.24 | 31.32 | 47.83 | 64.49 | 81.17 | 97.86 | 114.55 | 131.25 | 147.94 | 164.64 | 181.33 | 198.03 | 214.73 | 231.42 | 248.12 | 264.82 | 281.51 | 298.21 | 314.90 | 331.60 | 348.29 | 364.99 | 381.69 | 398.38 | 415.08 |
| | $J$ | 13.13 | 22.37 | 28.38 | 32.37 | 35.10 | 37.02 | 38.40 | 39.41 | 40.16 | 40.74 | 41.18 | 41.51 | 41.80 | 42.00 | 42.16 | 42.30 | 42.42 | 42.50 | 42.57 | 42.63 | 42.68 | 42.72 | 42.76 | 42.79 | 42.82 |
| | $K$ | 2.86 | 4.23 | 4.65 | 4.76 | 4.79 | 4.79 | 4.79 | 4.79 | 4.79 | 4.79 | 4.79 | 4.79 | 4.79 | 4.79 | 4.79 | 4.79 | 4.79 | 4.79 | 4.79 | 4.79 | 4.79 | 4.79 | 4.79 | 4.79 | 4.79 |
| 0.2 | $U_1$ | 16.77 | 29.03 | 38.15 | 44.46 | 50.10 | 53.76 | 56.62 | 58.91 | 60.77 | 62.31 | 63.61 | 64.72 | 65.67 | 66.50 | 67.22 | 67.86 | 68.43 | 68.95 | 69.40 | 69.82 | 70.20 | 70.54 | 70.86 | 71.15 | 71.42 |
| | $U_2$ | 0.69 | 1.59 | 2.81 | 4.26 | 5.79 | 7.28 | 8.67 | 9.93 | 11.07 | 12.09 | 13.01 | 13.82 | 14.55 | 15.21 | 15.80 | 16.33 | 16.82 | 17.26 | 17.66 | 18.03 | 18.37 | 18.68 | 18.97 | 19.23 | 19.48 |
| | $W_{1,2}$ | 13.17 | 28.24 | 44.46 | 61.03 | 77.70 | 94.38 | 111.08 | 127.77 | 144.47 | 161.16 | 177.86 | 194.56 | 211.25 | 227.95 | 244.64 | 261.34 | 278.03 | 294.73 | 311.43 | 328.12 | 344.82 | 361.51 | 378.21 | 394.90 | 411.60 |
| | $W_{3,4}$ | 15.24 | 31.31 | 47.82 | 64.47 | 81.17 | 97.86 | 114.55 | 131.25 | 147.93 | 164.63 | 181.32 | 198.02 | 214.72 | 231.41 | 248.11 | 264.80 | 281.50 | 298.20 | 314.89 | 331.59 | 348.29 | 364.98 | 381.68 | 398.38 | 415.06 |
| | $J$ | 13.35 | 22.81 | 29.01 | 33.14 | 35.97 | 37.96 | 39.38 | 40.43 | 41.22 | 41.82 | 42.27 | 42.63 | 42.92 | 43.14 | 43.31 | 43.45 | 43.57 | 43.66 | 43.70 | 43.79 | 43.86 | 43.91 | 43.95 | 43.98 | 44.01 |
| | $K$ | 2.99 | 4.43 | 4.86 | 4.97 | 5.00 | 5.01 | 5.01 | 5.01 | 5.01 | 5.01 | 5.01 | 5.01 | 5.01 | 5.01 | 5.01 | 5.01 | 5.01 | 5.01 | 5.01 | 5.01 | 5.01 | 5.01 | 5.01 | 5.01 | 5.01 |
| 0.1 | $U_1$ | 16.69 | 29.50 | 38.60 | 45.14 | 49.96 | 53.52 | 56.48 | 58.77 | 60.63 | 62.18 | 63.48 | 64.58 | 65.54 | 66.36 | 67.09 | 67.73 | 68.30 | 68.81 | 69.27 | 69.69 | 70.07 | 70.41 | 70.73 | 71.02 | 71.29 |
| | $U_2$ | 0.68 | 1.58 | 2.81 | 4.26 | 5.79 | 7.28 | 8.67 | 9.93 | 11.07 | 12.09 | 13.01 | 13.82 | 14.55 | 15.21 | 15.80 | 16.32 | 16.80 | 17.26 | 17.66 | 18.03 | 18.37 | 18.68 | 18.97 | 19.24 | 19.48 |
| | $W_{1,2}$ | 13.01 | 28.01 | 44.20 | 60.77 | 77.43 | 94.12 | 110.81 | 127.51 | 144.21 | 160.90 | 177.60 | 194.29 | 210.99 | 227.69 | 244.38 | 261.08 | 277.77 | 294.47 | 311.16 | 327.86 | 344.56 | 361.25 | 377.95 | 394.64 | 411.34 |
| | $W_{3,4}$ | 15.24 | 31.31 | 47.82 | 64.47 | 81.15 | 97.85 | 114.54 | 131.24 | 147.92 | 164.62 | 181.31 | 198.01 | 214.71 | 231.40 | 248.10 | 264.79 | 281.49 | 298.18 | 314.89 | 331.58 | 348.28 | 364.98 | 381.67 | 398.37 | 415.06 |
| | $J$ | 13.49 | 23.07 | 29.31 | 33.43 | 36.23 | 38.17 | 39.56 | 40.56 | 41.28 | 41.87 | 42.32 | 42.67 | 42.96 | 43.19 | 43.37 | 43.51 | 43.63 | 43.68 | 43.79 | 43.86 | 43.92 | 43.97 | 44.01 | 44.04 | 44.08 |
| | $K$ | 3.08 | 4.56 | 5.01 | 5.12 | 5.15 | 5.16 | 5.16 | 5.16 | 5.16 | 5.16 | 5.16 | 5.16 | 5.16 | 5.16 | 5.16 | 5.16 | 5.16 | 5.16 | 5.16 | 5.16 | 5.16 | 5.16 | 5.16 | 5.16 | 5.16 |
| 0.0 | $U_1$ | 16.98 | 30.00 | 39.23 | 45.84 | 50.71 | 54.41 | 57.30 | 59.60 | 61.47 | 63.02 | 64.33 | 65.45 | 66.39 | 67.22 | 67.95 | 68.59 | 69.17 | 69.68 | 70.14 | 70.55 | 70.93 | 71.28 | 71.60 | 71.89 | 72.16 |
| | $U_2$ | 0.66 | 1.54 | 2.75 | 4.20 | 5.73 | 7.22 | 8.61 | 9.89 | 11.03 | 12.05 | 12.97 | 13.79 | 14.52 | 15.18 | 15.77 | 16.30 | 16.79 | 17.23 | 17.63 | 18.00 | 18.34 | 18.65 | 18.94 | 19.21 | 19.46 |
| | $W_{1,2}$ | 12.89 | 27.81 | 44.02 | 60.58 | 77.25 | 93.93 | 110.63 | 127.32 | 144.02 | 160.71 | 177.41 | 194.11 | 210.80 | 227.50 | 244.20 | 260.89 | 277.58 | 294.28 | 310.98 | 327.67 | 344.37 | 361.06 | 377.76 | 394.45 | 411.15 |
| | $W_{3,4}$ | 15.23 | 31.27 | 47.78 | 64.43 | 81.11 | 97.81 | 114.50 | 131.19 | 147.89 | 164.59 | 181.28 | 197.98 | 214.67 | 231.37 | 248.06 | 264.76 | 281.46 | 298.15 | 314.85 | 331.54 | 348.24 | 364.94 | 381.63 | 398.33 | 415.03 |
| | $J$ | 13.51 | 23.03 | 29.33 | 33.43 | 36.23 | 38.17 | 39.56 | 40.56 | 41.28 | 41.87 | 42.30 | 42.61 | 42.89 | 43.10 | 43.27 | 43.40 | 43.51 | 43.60 | 43.68 | 43.74 | 43.79 | 43.84 | 43.88 | 43.91 | 43.94 |
| | $K$ | 3.11 | 4.61 | 5.06 | 5.18 | 5.21 | 5.22 | 5.22 | 5.22 | 5.22 | 5.22 | 5.22 | 5.22 | 5.22 | 5.22 | 5.22 | 5.22 | 5.22 | 5.22 | 5.22 | 5.22 | 5.22 | 5.22 | 5.22 | 5.22 | 5.22 |

TABLE S73. Parameters of the THF interaction Hamiltonian for different ratios $u_0/u_1$ and screening lengths $\xi$ of the double-gate interaction potential ($U_\xi = 24$ meV for $\xi = 10$ nm). We use the same BM model parameters as in Table S72.

Total weight of the heavy-fermions on the active bands after wannierisation: 92.8%

Wannier orbitals support

FIG. S53. Band structures of the BM and THF models near charge neutrality for $w_0/w_1 = 0.8$, depicted by lines and crosses, respectively. The BM bands are colored according to the weight of the $f$-electron wave function on them. We use the same BM parameters as in Table S74.

TABLE S74. Parameters of the $f$-electron Wannier functions and the THF single-particle Hamiltonian for different values of the tunneling ratio $w_0/w_1$. $\mathcal{W}$ denotes the total weight of the THF $f$-electrons on the active bands. In computing the ratios $\gamma/U_1$ and $\gamma/U_2$, we employ the on-site and nearest-neighbor repulsion parameters $U_1$ and $U_2$ obtained numerically for $\xi = 10$ nm, as given in Table S75. We employ $v_F = 5.944$ eV·Å, $|\mathbf{K}| = 1.703$ Å$^{-1}$, $w_1 = 110$ meV, and $\theta = 1.40°$ for the BM model.

| $w_0/w_1$ | $\alpha_1$ | $\alpha_2$ | $\lambda_1$ | $\lambda_2$ | $\lambda$ | $\mathcal{W}$ | $t_0$ | $\gamma$ | $M$ | $v'_\star$ | $v''_\star$ | $\gamma/U_1$ | $\gamma/U_2$ |
|---|---|---|---|---|---|---|---|---|---|---|---|---|---|
| 1.00 | 0.813 | 0.583 | 0.203 | 0.209 | 0.361 | 0.942 | 3.316 | -47.501 | -41.670 | -4.728 | 0.0255 | -0.65 | -9.53 |
| 0.90 | 0.841 | 0.541 | 0.215 | 0.373 | 0.373 | 0.933 | 3.938 | -63.715 | -43.265 | -4.843 | 0.0196 | -0.92 | -12.46 |
| 0.80 | 0.869 | 0.495 | 0.227 | 0.388 | 0.388 | 0.928 | 4.565 | -79.448 | -44.650 | -4.951 | 0.0152 | -1.22 | -15.08 |
| 0.70 | 0.897 | 0.443 | 0.245 | 0.227 | 0.415 | 0.935 | 5.439 | -94.497 | -45.839 | -5.054 | 0.0117 | -1.56 | -17.11 |
| 0.60 | 0.922 | 0.387 | 0.257 | 0.233 | 0.433 | 0.937 | 6.057 | -108.629 | -46.844 | -5.154 | 0.0089 | -1.91 | -19.10 |
| 0.50 | 0.944 | 0.330 | 0.275 | 0.233 | 0.443 | 0.937 | 6.622 | -121.569 | -47.676 | -5.249 | 0.0065 | -2.22 | -20.94 |
| 0.40 | 0.963 | 0.269 | 0.286 | 0.239 | 0.450 | 0.936 | 7.061 | -132.988 | -48.345 | -5.339 | 0.0045 | -2.51 | -22.60 |
| 0.30 | 0.979 | 0.206 | 0.298 | 0.245 | 0.458 | 0.937 | 7.442 | -142.509 | -48.858 | -5.420 | 0.0027 | -2.76 | -23.89 |
| 0.20 | 0.990 | 0.140 | 0.310 | 0.245 | 0.459 | 0.937 | 7.669 | -149.722 | -49.220 | -5.485 | 0.0013 | -2.95 | -24.92 |
| 0.10 | 0.997 | 0.072 | 0.316 | 0.245 | 0.460 | 0.934 | 7.673 | -154.242 | -49.436 | -5.529 | 0.0003 | -3.04 | -25.75 |
| 0.00 | 1.000 | 0.000 | 0.316 | 0.000 | 0.451 | 0.928 | 7.585 | -155.784 | -49.507 | -5.544 | 0.0000 | -3.02 | -26.26 |

| $w_0/w_1$ | $\xi/$nm | 2 | 4 | 6 | 8 | 10 | 12 | 14 | 16 | 18 | 20 | 22 | 24 | 26 | 28 | 30 | 32 | 34 | 36 | 38 | 40 | 42 | 44 | 46 | 48 | 50 |
|---|---|---|---|---|---|---|---|---|---|---|---|---|---|---|---|---|---|---|---|---|---|---|---|---|---|---|
| 1.0 | $U_1$ | 27.71 | 46.67 | 58.83 | 66.96 | 72.67 | 76.86 | 80.06 | 82.56 | 84.57 | 86.22 | 87.59 | 88.75 | 89.74 | 90.60 | 91.35 | 92.01 | 92.60 | 93.12 | 93.59 | 94.02 | 94.40 | 94.75 | 95.07 | 95.37 | 95.64 |
| | $U_2$ | 0.41 | 1.52 | 2.05 | 3.44 | 4.98 | 6.53 | 7.98 | 9.31 | 10.50 | 11.57 | 12.52 | 13.36 | 14.12 | 14.80 | 15.41 | 15.96 | 16.45 | 16.90 | 17.31 | 17.69 | 18.04 | 18.35 | 18.65 | 18.92 | 19.17 |
| | $W_{1,2}$ | 15.39 | 31.75 | 48.69 | 65.81 | 82.98 | 100.16 | 117.34 | 134.52 | 151.71 | 168.89 | 186.07 | 203.26 | 220.44 | 237.62 | 254.81 | 271.99 | 289.17 | 306.35 | 323.54 | 340.72 | 357.90 | 375.09 | 392.27 | 409.45 | 426.64 |
| | $W_{3,4}$ | 16.19 | 33.10 | 50.23 | 67.40 | 84.58 | 101.76 | 118.94 | 136.13 | 153.31 | 170.49 | 187.67 | 204.86 | 222.04 | 239.23 | 256.41 | 273.59 | 290.77 | 307.96 | 325.14 | 342.32 | 359.51 | 376.69 | 393.87 | 411.06 | 428.24 |
| | $J$ | 10.62 | 16.41 | 19.34 | 20.96 | 21.94 | 22.56 | 22.97 | 23.25 | 23.45 | 23.60 | 23.70 | 23.78 | 23.85 | 23.90 | 23.93 | 23.96 | 23.98 | 24.00 | 24.01 | 24.03 | 24.04 | 24.05 | 24.05 | 24.06 | 24.06 |
| | $K$ | 1.96 | 2.91 | 3.20 | 3.27 | 3.29 | 3.29 | 3.29 | 3.29 | 3.29 | 3.29 | 3.29 | 3.29 | 3.29 | 3.29 | 3.29 | 3.29 | 3.29 | 3.29 | 3.29 | 3.29 | 3.29 | 3.29 | 3.29 | 3.29 | 3.29 |
| 0.9 | $U_1$ | 25.73 | 43.69 | 55.41 | 63.32 | 68.92 | 73.05 | 76.21 | 78.69 | 80.68 | 82.32 | 83.69 | 84.84 | 85.83 | 86.69 | 87.43 | 88.08 | 88.68 | 89.20 | 89.67 | 90.09 | 90.47 | 90.82 | 91.15 | 91.44 | 91.71 |
| | $U_2$ | 0.44 | 1.09 | 2.16 | 3.56 | 5.11 | 6.66 | 8.11 | 9.43 | 10.62 | 11.68 | 12.63 | 13.48 | 14.23 | 14.91 | 15.51 | 16.06 | 16.56 | 17.01 | 17.42 | 17.79 | 18.14 | 18.45 | 18.75 | 19.02 | 19.27 |
| | $W_{1,2}$ | 15.23 | 31.51 | 48.43 | 65.55 | 82.72 | 99.90 | 117.08 | 134.26 | 151.45 | 168.63 | 185.81 | 203.00 | 220.18 | 237.36 | 254.55 | 271.73 | 288.91 | 306.09 | 323.28 | 340.46 | 357.64 | 374.83 | 392.01 | 409.19 | 426.38 |
| | $W_{3,4}$ | 15.94 | 32.72 | 49.79 | 66.95 | 84.13 | 101.31 | 118.49 | 135.68 | 152.86 | 170.04 | 187.23 | 204.41 | 221.59 | 238.78 | 255.96 | 273.14 | 290.33 | 307.51 | 324.69 | 341.88 | 359.06 | 376.24 | 393.42 | 410.61 | 427.79 |
| | $J$ | 11.00 | 17.36 | 20.79 | 22.77 | 23.99 | 24.79 | 25.33 | 25.71 | 25.97 | 26.17 | 26.31 | 26.42 | 26.50 | 26.57 | 26.61 | 26.65 | 26.69 | 26.71 | 26.73 | 26.75 | 26.76 | 26.78 | 26.79 | 26.80 | 26.80 |
| | $K$ | 2.04 | 3.02 | 3.31 | 3.38 | 3.40 | 3.41 | 3.41 | 3.41 | 3.41 | 3.41 | 3.41 | 3.41 | 3.41 | 3.41 | 3.41 | 3.41 | 3.41 | 3.41 | 3.41 | 3.41 | 3.41 | 3.41 | 3.41 | 3.41 | 3.41 |
| 0.8 | $U_1$ | 23.83 | 40.79 | 52.03 | 59.70 | 65.17 | 69.23 | 72.34 | 74.79 | 76.77 | 78.39 | 79.74 | 80.89 | 81.87 | 82.73 | 83.47 | 84.13 | 84.71 | 85.23 | 85.70 | 86.12 | 86.50 | 86.85 | 87.17 | 87.47 | 87.74 |
| | $U_2$ | 0.48 | 1.18 | 2.29 | 3.71 | 5.27 | 6.81 | 8.25 | 9.57 | 10.76 | 11.82 | 12.76 | 13.60 | 14.36 | 15.03 | 15.64 | 16.18 | 16.68 | 17.13 | 17.53 | 17.91 | 18.25 | 18.57 | 18.86 | 19.13 | 19.38 |
| | $W_{1,2}$ | 15.02 | 31.22 | 48.11 | 65.23 | 82.39 | 99.57 | 116.75 | 133.94 | 151.12 | 168.30 | 185.48 | 202.67 | 219.85 | 237.03 | 254.22 | 271.40 | 288.58 | 305.77 | 322.95 | 340.13 | 357.32 | 374.50 | 391.68 | 408.86 | 426.05 |
| | $W_{3,4}$ | 15.68 | 32.46 | 49.51 | 66.66 | 83.83 | 101.01 | 118.20 | 135.38 | 152.56 | 169.74 | 186.93 | 204.11 | 221.29 | 238.48 | 255.66 | 272.84 | 290.03 | 307.21 | 324.39 | 341.58 | 358.76 | 375.94 | 393.12 | 410.31 | 427.49 |
| | $J$ | 11.41 | 18.35 | 22.28 | 24.64 | 26.13 | 27.13 | 27.80 | 28.28 | 28.63 | 28.88 | 29.07 | 29.21 | 29.32 | 29.40 | 29.47 | 29.52 | 29.57 | 29.60 | 29.63 | 29.65 | 29.67 | 29.69 | 29.70 | 29.72 | 29.73 |
| | $K$ | 2.13 | 3.16 | 3.46 | 3.54 | 3.56 | 3.56 | 3.56 | 3.56 | 3.56 | 3.56 | 3.56 | 3.56 | 3.56 | 3.56 | 3.56 | 3.56 | 3.56 | 3.56 | 3.56 | 3.56 | 3.56 | 3.56 | 3.56 | 3.56 | 3.56 |
| 0.7 | $U_1$ | 21.58 | 37.24 | 47.81 | 55.12 | 60.38 | 64.32 | 67.35 | 69.74 | 71.68 | 73.28 | 74.62 | 75.75 | 76.72 | 77.57 | 78.31 | 78.96 | 79.54 | 80.05 | 80.52 | 80.94 | 81.32 | 81.67 | 81.99 | 82.28 | 82.55 |
| | $U_2$ | 0.56 | 1.35 | 2.51 | 3.96 | 5.52 | 7.06 | 8.49 | 9.80 | 10.98 | 12.03 | 12.96 | 13.80 | 14.55 | 15.22 | 15.82 | 16.36 | 16.85 | 17.30 | 17.71 | 18.08 | 18.42 | 18.74 | 19.03 | 19.30 | 19.55 |
| | $W_{1,2}$ | 14.83 | 30.94 | 47.81 | 64.91 | 82.08 | 99.26 | 116.45 | 133.63 | 150.80 | 167.99 | 185.17 | 202.35 | 219.54 | 236.72 | 253.90 | 271.09 | 288.27 | 305.45 | 322.64 | 339.82 | 357.00 | 374.19 | 391.37 | 408.55 | 425.73 |
| | $W_{3,4}$ | 15.72 | 32.34 | 49.37 | 66.51 | 83.69 | 100.87 | 118.05 | 135.23 | 152.42 | 169.60 | 186.78 | 203.97 | 221.15 | 238.33 | 255.52 | 272.70 | 289.88 | 307.07 | 324.25 | 341.43 | 358.62 | 375.80 | 392.98 | 410.16 | 427.35 |
| | $J$ | 11.94 | 19.04 | 24.10 | 26.95 | 28.82 | 30.10 | 30.99 | 31.63 | 32.10 | 32.45 | 32.72 | 32.93 | 33.08 | 33.20 | 33.30 | 33.38 | 33.44 | 33.49 | 33.53 | 33.57 | 33.60 | 33.63 | 33.65 | 33.67 | 33.68 |
| | $K$ | 2.25 | 3.33 | 3.65 | 3.73 | 3.75 | 3.76 | 3.76 | 3.76 | 3.76 | 3.76 | 3.76 | 3.76 | 3.76 | 3.76 | 3.76 | 3.76 | 3.76 | 3.76 | 3.76 | 3.76 | 3.76 | 3.76 | 3.76 | 3.76 | 3.76 |


| $u_0/u_1$ | $\xi/$nm | 2 | 4 | 6 | 8 | 10 | 12 | 14 | 16 | 18 | 20 | 22 | 24 | 26 | 28 | 30 | 32 | 34 | 36 | 38 | 40 | 42 | 44 | 46 | 48 | 50 |
|---|---|---|---|---|---|---|---|---|---|---|---|---|---|---|---|---|---|---|---|---|---|---|---|---|---|---|
| 0.6 | $U_1$ | 19.99 | 34.73 | 44.83 | 51.89 | 57.01 | 60.85 | 63.83 | 66.19 | 68.10 | 69.68 | 71.00 | 72.13 | 73.09 | 73.93 | 74.67 | 75.32 | 75.89 | 76.41 | 76.87 | 77.29 | 77.67 | 78.02 | 78.33 | 78.63 | 78.89 |
| | $U_2$ | 0.62 | 1.40 | 2.66 | 4.12 | 5.69 | 7.22 | 8.65 | 10.05 | 11.12 | 12.16 | 13.09 | 13.92 | 14.67 | 15.33 | 15.93 | 16.47 | 16.96 | 17.41 | 17.82 | 18.19 | 18.53 | 18.84 | 19.13 | 19.40 | 19.65 |
| | $W_{1,2}$ | 14.60 | 30.01 | 47.45 | 64.55 | 81.71 | 98.89 | 116.07 | 133.25 | 150.44 | 167.62 | 184.80 | 201.99 | 219.17 | 236.35 | 253.54 | 270.72 | 287.90 | 305.09 | 322.27 | 339.45 | 356.63 | 373.82 | 391.00 | 408.18 | 425.37 |
| | $W_{3,4}$ | 15.69 | 32.27 | 49.29 | 66.43 | 83.61 | 100.79 | 117.97 | 135.15 | 152.34 | 169.52 | 186.70 | 203.88 | 221.07 | 238.23 | 255.43 | 272.62 | 289.80 | 306.98 | 324.17 | 341.35 | 358.53 | 375.72 | 392.90 | 410.08 | 427.26 |
| | $J$ | 12.39 | 20.60 | 25.69 | 28.95 | 31.14 | 32.65 | 33.72 | 34.51 | 35.09 | 35.52 | 35.86 | 36.12 | 36.32 | 36.48 | 36.61 | 36.72 | 36.80 | 36.87 | 36.93 | 36.98 | 37.02 | 37.05 | 37.08 | 37.11 | 37.13 |
| | $K$ | 2.39 | 3.54 | 3.87 | 3.96 | 3.98 | 3.98 | 3.98 | 3.98 | 3.98 | 3.98 | 3.98 | 3.98 | 3.98 | 3.98 | 3.98 | 3.98 | 3.98 | 3.98 | 3.98 | 3.98 | 3.98 | 3.98 | 3.98 | 3.98 | 3.98 |
| 0.5 | $U_1$ | 18.00 | 33.01 | 42.79 | 49.68 | 54.70 | 58.49 | 61.43 | 63.77 | 65.67 | 67.23 | 68.55 | 69.68 | 70.63 | 71.47 | 72.20 | 72.84 | 73.42 | 73.93 | 74.39 | 74.81 | 75.19 | 75.54 | 75.85 | 76.15 | 76.42 |
| | $U_2$ | 0.65 | 1.53 | 2.70 | 4.24 | 5.80 | 7.33 | 8.76 | 10.05 | 11.22 | 12.26 | 13.19 | 14.02 | 14.76 | 15.42 | 16.02 | 16.56 | 17.05 | 17.49 | 17.90 | 18.27 | 18.61 | 18.93 | 19.22 | 19.48 | 19.73 |
| | $W_{1,2}$ | 14.34 | 30.23 | 47.04 | 64.13 | 81.29 | 98.47 | 115.65 | 132.84 | 150.02 | 167.20 | 184.39 | 201.57 | 218.75 | 235.94 | 253.12 | 270.30 | 287.49 | 304.67 | 321.85 | 339.03 | 356.22 | 373.40 | 390.58 | 407.77 | 424.95 |
| | $W_{3,4}$ | 15.67 | 32.23 | 49.24 | 66.38 | 83.56 | 100.74 | 117.92 | 135.10 | 152.28 | 169.47 | 186.65 | 203.83 | 221.01 | 238.20 | 255.38 | 272.57 | 289.75 | 306.93 | 324.12 | 341.30 | 358.48 | 375.67 | 392.85 | 410.03 | 427.21 |
| | $J$ | 12.79 | 21.48 | 26.97 | 30.54 | 32.94 | 34.62 | 35.81 | 36.68 | 37.33 | 37.82 | 38.19 | 38.48 | 38.71 | 38.89 | 39.04 | 39.16 | 39.25 | 39.33 | 39.40 | 39.46 | 39.50 | 39.54 | 39.58 | 39.61 | 39.63 |
| | $K$ | 2.55 | 3.76 | 4.12 | 4.21 | 4.23 | 4.24 | 4.24 | 4.24 | 4.24 | 4.24 | 4.24 | 4.24 | 4.24 | 4.24 | 4.24 | 4.24 | 4.24 | 4.24 | 4.24 | 4.24 | 4.24 | 4.24 | 4.24 | 4.24 | 4.24 |
| 0.4 | $U_1$ | 18.12 | 31.79 | 41.34 | 48.11 | 53.07 | 56.81 | 59.73 | 62.05 | 63.94 | 65.50 | 66.81 | 67.92 | 68.88 | 69.72 | 70.45 | 71.09 | 71.66 | 72.18 | 72.64 | 73.05 | 73.43 | 73.78 | 74.10 | 74.39 | 74.66 |
| | $U_2$ | 0.67 | 1.58 | 2.82 | 4.31 | 5.88 | 7.41 | 8.83 | 10.13 | 11.29 | 12.33 | 13.25 | 14.08 | 14.82 | 15.48 | 16.08 | 16.62 | 17.11 | 17.55 | 17.96 | 18.33 | 18.67 | 18.98 | 19.27 | 19.54 | 19.79 |
| | $W_{1,2}$ | 14.07 | 29.85 | 46.62 | 63.70 | 80.86 | 98.04 | 115.22 | 132.40 | 149.59 | 166.77 | 183.95 | 201.14 | 218.32 | 235.50 | 252.69 | 269.87 | 287.05 | 304.24 | 321.42 | 338.60 | 355.78 | 372.97 | 390.15 | 407.33 | 424.52 |
| | $W_{3,4}$ | 15.66 | 32.21 | 49.22 | 66.36 | 83.53 | 100.71 | 117.89 | 135.08 | 152.26 | 169.44 | 186.63 | 203.81 | 221.00 | 238.18 | 255.36 | 272.54 | 289.72 | 306.91 | 324.09 | 341.27 | 358.46 | 375.64 | 392.82 | 410.01 | 427.20 |
| | $J$ | 13.14 | 22.22 | 28.03 | 31.84 | 34.41 | 36.21 | 37.49 | 38.43 | 39.13 | 39.66 | 40.07 | 40.38 | 40.63 | 40.83 | 40.99 | 41.12 | 41.22 | 41.31 | 41.38 | 41.45 | 41.50 | 41.54 | 41.58 | 41.61 | 41.64 |
| | $K$ | 2.71 | 4.00 | 4.38 | 4.47 | 4.49 | 4.50 | 4.50 | 4.50 | 4.50 | 4.50 | 4.50 | 4.50 | 4.50 | 4.50 | 4.50 | 4.50 | 4.50 | 4.50 | 4.50 | 4.50 | 4.50 | 4.50 | 4.50 | 4.50 | 4.50 |
| 0.3 | $U_1$ | 17.44 | 30.70 | 40.03 | 46.69 | 51.58 | 55.28 | 58.17 | 60.48 | 62.35 | 63.91 | 65.21 | 66.32 | 67.28 | 68.11 | 68.83 | 69.48 | 70.05 | 70.56 | 71.02 | 71.44 | 71.82 | 72.16 | 72.48 | 72.77 | 73.04 |
| | $U_2$ | 0.70 | 1.63 | 2.89 | 4.39 | 5.96 | 7.49 | 8.91 | 10.20 | 11.36 | 12.39 | 13.32 | 14.14 | 14.88 | 15.54 | 16.14 | 16.68 | 17.17 | 17.61 | 18.01 | 18.38 | 18.72 | 19.03 | 19.32 | 19.59 | 19.84 |
| | $W_{1,2}$ | 13.83 | 29.50 | 46.25 | 63.32 | 80.48 | 97.66 | 114.84 | 132.02 | 149.20 | 166.39 | 183.57 | 200.75 | 217.94 | 235.12 | 252.30 | 269.49 | 286.67 | 303.85 | 321.04 | 338.22 | 355.40 | 372.59 | 389.77 | 406.95 | 424.14 |
| | $W_{3,4}$ | 15.67 | 32.22 | 49.23 | 66.36 | 83.54 | 100.72 | 117.90 | 135.08 | 152.27 | 169.45 | 186.63 | 203.82 | 221.00 | 238.18 | 255.37 | 272.55 | 289.73 | 306.91 | 324.10 | 341.28 | 358.46 | 375.65 | 392.83 | 410.01 | 427.20 |
| | $J$ | 13.60 | 23.36 | 29.68 | 33.86 | 36.72 | 38.72 | 40.15 | 41.21 | 41.99 | 42.59 | 43.04 | 43.40 | 43.68 | 43.91 | 44.09 | 44.24 | 44.36 | 44.46 | 44.54 | 44.61 | 44.67 | 44.72 | 44.77 | 44.80 | 43.52 |
| | $K$ | 2.87 | 4.23 | 4.63 | 4.73 | 4.75 | 4.76 | 4.76 | 4.76 | 4.76 | 4.76 | 4.76 | 4.76 | 4.76 | 4.76 | 4.76 | 4.76 | 4.76 | 4.76 | 4.76 | 4.76 | 4.76 | 4.76 | 4.76 | 4.76 | 4.76 |
| 0.2 | $U_1$ | 17.02 | 30.04 | 39.28 | 45.88 | 50.75 | 54.44 | 57.32 | 59.62 | 61.49 | 63.04 | 64.34 | 65.45 | 66.41 | 67.24 | 67.96 | 68.60 | 69.18 | 69.69 | 70.15 | 70.56 | 70.94 | 71.29 | 71.60 | 71.90 | 72.16 |
| | $U_2$ | 0.71 | 1.66 | 2.93 | 4.44 | 6.01 | 7.53 | 8.95 | 10.24 | 11.40 | 12.43 | 13.35 | 14.16 | 14.91 | 15.57 | 16.16 | 16.70 | 17.19 | 17.63 | 18.03 | 18.41 | 18.74 | 19.05 | 19.34 | 19.61 | 19.87 |
| | $W_{1,2}$ | 13.63 | 29.21 | 45.93 | 62.99 | 80.15 | 97.32 | 114.51 | 131.69 | 148.87 | 166.06 | 183.24 | 200.42 | 217.60 | 234.79 | 251.97 | 269.15 | 286.34 | 303.52 | 320.70 | 337.89 | 355.07 | 372.25 | 389.44 | 406.62 | 423.80 |
| | $W_{3,4}$ | 15.67 | 32.23 | 49.23 | 66.37 | 83.53 | 100.72 | 117.90 | 135.03 | 152.28 | 169.46 | 186.64 | 203.82 | 221.00 | 238.18 | 255.36 | 272.55 | 289.73 | 306.91 | 324.09 | 341.28 | 358.46 | 375.64 | 392.82 | 410.01 | 427.20 |
| | $J$ | 13.69 | 23.36 | 29.68 | 33.86 | 36.72 | 38.72 | 40.15 | 41.21 | 41.99 | 42.59 | 43.04 | 43.40 | 43.68 | 43.91 | 44.09 | 44.25 | 44.38 | 44.48 | 44.57 | 44.61 | 44.67 | 44.72 | 44.77 | 44.80 | 44.84 |
| | $K$ | 3.00 | 4.42 | 4.84 | 4.95 | 4.97 | 4.98 | 4.98 | 4.98 | 4.98 | 4.98 | 4.98 | 4.98 | 4.98 | 4.98 | 4.98 | 4.98 | 4.98 | 4.98 | 4.98 | 4.98 | 4.98 | 4.98 | 4.98 | 4.98 | 4.98 |
| 0.1 | $U_1$ | 17.02 | 30.05 | 39.34 | 45.88 | 50.77 | 54.44 | 57.32 | 59.62 | 61.49 | 63.04 | 64.34 | 65.45 | 66.41 | 67.24 | 67.96 | 68.60 | 69.18 | 69.69 | 70.15 | 70.56 | 70.94 | 71.29 | 71.60 | 71.90 | 72.16 |
| | $U_2$ | 0.70 | 1.64 | 2.91 | 4.42 | 5.99 | 7.52 | 8.93 | 10.22 | 11.38 | 12.42 | 13.34 | 14.16 | 14.90 | 15.56 | 16.16 | 16.70 | 17.17 | 17.63 | 18.03 | 18.40 | 18.74 | 19.05 | 19.34 | 19.61 | 19.86 |
| | $W_{1,2}$ | 13.45 | 28.95 | 45.65 | 62.71 | 79.86 | 97.04 | 114.22 | 131.40 | 148.58 | 165.77 | 182.95 | 200.14 | 217.32 | 234.50 | 251.69 | 268.87 | 286.05 | 303.23 | 320.42 | 337.60 | 354.78 | 371.97 | 389.15 | 406.33 | 423.47 |
| | $W_{3,4}$ | 15.64 | 32.21 | 49.21 | 66.35 | 83.52 | 100.70 | 117.88 | 135.07 | 152.25 | 169.43 | 186.61 | 203.80 | 220.98 | 238.16 | 255.35 | 272.53 | 289.71 | 306.90 | 324.08 | 341.26 | 358.44 | 375.63 | 392.81 | 409.99 | 427.18 |
| | $J$ | 13.82 | 23.60 | 29.98 | 34.21 | 37.08 | 39.00 | 40.52 | 41.57 | 42.35 | 42.94 | 43.40 | 43.75 | 44.03 | 44.25 | 44.43 | 44.57 | 44.69 | 44.78 | 44.87 | 44.94 | 45.00 | 45.05 | 45.09 | 45.13 | 45.16 |
| | $K$ | 3.10 | 4.56 | 4.99 | 5.10 | 5.12 | 5.13 | 5.13 | 5.13 | 5.13 | 5.13 | 5.13 | 5.13 | 5.13 | 5.13 | 5.13 | 5.13 | 5.13 | 5.13 | 5.13 | 5.13 | 5.13 | 5.13 | 5.13 | 5.13 | 5.13 |
| 0.0 | $U_1$ | 17.31 | 30.55 | 39.91 | 46.59 | 51.50 | 55.23 | 58.18 | 60.44 | 62.32 | 63.88 | 65.19 | 66.30 | 67.26 | 68.09 | 68.82 | 69.47 | 70.04 | 70.55 | 71.01 | 71.43 | 71.81 | 72.16 | 72.47 | 72.76 | 73.03 |
| | $U_2$ | 0.68 | 1.59 | 2.86 | 4.36 | 5.93 | 7.46 | 8.89 | 10.18 | 11.34 | 12.38 | 13.30 | 14.13 | 14.87 | 15.53 | 16.13 | 16.67 | 17.16 | 17.60 | 18.00 | 18.37 | 18.71 | 19.03 | 19.32 | 19.59 | 19.83 |
| | $W_{1,2}$ | 13.34 | 28.78 | 45.64 | 62.71 | 79.67 | 96.85 | 114.03 | 131.21 | 148.40 | 165.58 | 182.76 | 199.95 | 217.13 | 234.31 | 251.50 | 268.68 | 285.86 | 303.04 | 320.23 | 337.41 | 354.59 | 371.78 | 388.96 | 406.14 | 423.33 |
| | $W_{3,4}$ | 15.64 | 32.17 | 49.17 | 66.31 | 83.48 | 100.66 | 117.84 | 135.03 | 152.21 | 169.39 | 186.57 | 203.76 | 220.94 | 238.12 | 255.31 | 272.49 | 289.67 | 306.86 | 324.04 | 341.22 | 358.41 | 375.59 | 392.77 | 409.95 | 427.14 |
| | $J$ | 13.84 | 23.59 | 29.91 | 34.07 | 36.87 | 38.82 | 40.20 | 41.24 | 42.00 | 42.59 | 43.04 | 43.40 | 43.70 | 43.87 | 44.00 | 44.11 | 44.19 | 44.27 | 44.33 | 44.38 | 44.43 | 44.46 | 44.48 | 44.50 | 44.52 |
| | $K$ | 3.13 | 4.61 | 5.04 | 5.15 | 5.18 | 5.19 | 5.19 | 5.19 | 5.19 | 5.19 | 5.19 | 5.19 | 5.19 | 5.19 | 5.19 | 5.19 | 5.19 | 5.19 | 5.19 | 5.19 | 5.19 | 5.19 | 5.19 | 5.19 | 5.19 |

TABLE S75. Parameters of the THF interaction Hamiltonian for different ratios $u_0/u_1$ and screening lengths $\xi$ of the double-gate interaction potential ($U_\xi = 24$ meV for $\xi = 10$ nm). We use the same BM model parameters as in Table S74.

Total weight of the heavy-fermions on the active bands after wannierisation: 93.2%

Wannier orbitals support 0.0 — 1.0

FIG. S54. Band structures of the BM and THF models near charge neutrality for $w_0/w_1 = 0.8$, depicted by lines and crosses, respectively. The BM bands are colored according to the weight of the $f$-electron wave function on them. We use the same BM parameters as in Table S76.

| $w_0/w_1$ | $\alpha_1$ | $\alpha_2$ | $\lambda_1$ | $\lambda_2$ | $\lambda$ | $w_0$ | $W$ | $t_0$ | $\gamma$ | $M$ | $v_\star$ | $v_\star'$ | $v_\star''$ | $\gamma/U_1$ | $\gamma/U_2$ |
|---|---|---|---|---|---|---|---|---|---|---|---|---|---|---|---|
| 1.00 | 0.816 | 0.579 | 0.203 | 0.209 | 0.362 | 0.939 | 3.633 | -50.752 | -44.700 | -4.752 | 1.877 | 0.0279 | | -0.69 | -9.79 |
| 0.90 | 0.844 | 0.537 | 0.215 | 0.215 | 0.377 | 0.933 | 4.291 | -66.999 | -46.314 | -4.864 | 1.875 | 0.0222 | | -0.97 | -12.55 |
| 0.80 | 0.872 | 0.490 | 0.227 | 0.221 | 0.397 | 0.932 | 5.049 | -82.749 | -47.715 | -4.970 | 1.849 | 0.0177 | | -1.27 | -14.95 |
| 0.70 | 0.899 | 0.439 | 0.245 | 0.227 | 0.416 | 0.934 | 5.867 | -97.805 | -48.918 | -5.072 | 1.793 | 0.0140 | | -1.60 | -17.06 |
| 0.60 | 0.923 | 0.384 | 0.263 | 0.233 | 0.435 | 0.936 | 6.512 | -111.935 | -49.935 | -5.170 | 1.701 | 0.0109 | | -1.94 | -18.98 |
| 0.50 | 0.945 | 0.327 | 0.275 | 0.239 | 0.444 | 0.936 | 7.100 | -124.867 | -50.778 | -5.264 | 1.563 | 0.0081 | | -2.25 | -20.76 |
| 0.40 | 0.964 | 0.265 | 0.292 | 0.239 | 0.465 | 0.944 | 7.692 | -136.275 | -51.455 | -5.353 | 1.370 | 0.0056 | | -2.60 | -22.07 |
| 0.30 | 0.979 | 0.203 | 0.304 | 0.245 | 0.472 | 0.944 | 8.065 | -145.782 | -51.974 | -5.433 | 1.113 | 0.0034 | | -2.85 | -23.34 |
| 0.20 | 0.990 | 0.138 | 0.316 | 0.245 | 0.477 | 0.945 | 8.352 | -152.983 | -52.341 | -5.499 | 0.791 | 0.0016 | | -3.06 | -24.27 |
| 0.10 | 0.997 | 0.071 | 0.322 | 0.245 | 0.460 | 0.934 | 8.193 | -157.494 | -52.560 | -5.542 | 0.412 | 0.0004 | | -3.06 | -25.42 |
| 0.00 | 1.000 | 0.000 | 0.322 | 0.000 | 0.451 | 0.928 | 8.107 | -159.033 | -52.652 | -5.557 | 0.000 | 0.0000 | | -3.04 | -25.90 |

TABLE S76. Parameters of the $f$-electron Wannier functions and the THF single-particle Hamiltonian for different values of the tunneling ratio $w_0/w_1$. $W$ denotes the total weight of the THF $f$-electrons on the active bands. In computing the ratios $\gamma/U_1$ and $\gamma/U_2$, we employ the on-site and nearest-neighbor repulsion parameters $U_1$ and $U_2$ obtained numerically for $\xi = 10$ nm, as given in Table S77. We employ $v_F = 5.944$ eV·Å, $|\mathbf{K}| = 1.703$ Å$^{-1}$, $w_1 = 110$ meV, and $\theta = 1.42°$ for the BM model.

| $w_0/w_1$ | $\xi/$nm | 2 | 4 | 6 | 8 | 10 | 12 | 14 | 16 | 18 | 20 | 22 | 24 | 26 | 28 | 30 | 32 | 34 | 36 | 38 | 40 | 42 | 44 | 46 | 48 | 50 |
|---|---|---|---|---|---|---|---|---|---|---|---|---|---|---|---|---|---|---|---|---|---|---|---|---|---|---|
| 1.0 | $U_1$ | 28.00 | 47.12 | 59.36 | 67.54 | 73.28 | 77.49 | 80.69 | 83.20 | 85.22 | 86.87 | 88.24 | 89.41 | 90.40 | 91.26 | 92.01 | 92.67 | 93.26 | 93.78 | 94.25 | 94.68 | 95.06 | 95.41 | 95.74 | 96.03 | 96.30 |
| | $U_2$ | 0.42 | 1.06 | 2.15 | 3.59 | 5.18 | 6.76 | 8.25 | 9.60 | 10.81 | 11.89 | 12.85 | 13.71 | 14.47 | 15.15 | 15.77 | 16.32 | 16.82 | 17.27 | 17.68 | 18.06 | 18.41 | 18.72 | 19.02 | 19.29 | 19.54 |
| | $W_{1,2}$ | 15.91 | 32.78 | 50.23 | 67.85 | 85.51 | 103.19 | 120.86 | 138.54 | 156.22 | 173.90 | 191.57 | 209.25 | 226.93 | 244.61 | 262.28 | 279.96 | 297.64 | 315.32 | 332.99 | 350.67 | 368.35 | 386.03 | 403.70 | 421.38 | 439.06 |
| | $W_{3,4}$ | 16.56 | 33.92 | 51.53 | 69.19 | 86.86 | 104.54 | 122.22 | 139.89 | 157.57 | 175.25 | 192.93 | 210.60 | 228.28 | 245.96 | 263.64 | 281.31 | 298.99 | 316.67 | 334.35 | 352.02 | 369.70 | 387.38 | 405.06 | 422.73 | 440.41 |
| | $J$ | 10.88 | 16.81 | 19.82 | 21.49 | 22.48 | 23.12 | 23.54 | 23.83 | 24.04 | 24.19 | 24.29 | 24.37 | 24.44 | 24.48 | 24.52 | 24.55 | 24.57 | 24.59 | 24.61 | 24.62 | 24.63 | 24.64 | 24.64 | 24.65 | 24.66 |
| | $K$ | 1.94 | 2.87 | 3.14 | 3.20 | 3.22 | 3.22 | 3.22 | 3.22 | 3.22 | 3.22 | 3.22 | 3.22 | 3.22 | 3.22 | 3.22 | 3.22 | 3.22 | 3.22 | 3.22 | 3.22 | 3.22 | 3.22 | 3.22 | 3.22 | 3.22 |
| 0.9 | $U_1$ | 25.87 | 43.90 | 55.64 | 63.57 | 69.18 | 73.32 | 76.48 | 78.96 | 80.95 | 82.59 | 83.96 | 85.11 | 86.10 | 86.96 | 87.70 | 88.36 | 88.95 | 89.47 | 89.94 | 90.36 | 90.75 | 91.10 | 91.43 | 91.71 | 91.98 |
| | $U_2$ | 0.46 | 1.15 | 2.28 | 3.74 | 5.34 | 6.92 | 8.40 | 9.74 | 10.95 | 12.03 | 12.99 | 13.84 | 14.60 | 15.28 | 15.89 | 16.44 | 16.94 | 17.39 | 17.80 | 18.18 | 18.52 | 18.84 | 19.13 | 19.41 | 19.66 |
| | $W_{1,2}$ | 15.74 | 32.54 | 49.97 | 67.58 | 85.25 | 102.92 | 120.60 | 138.27 | 155.95 | 173.63 | 191.31 | 208.98 | 226.66 | 244.34 | 262.02 | 279.69 | 297.37 | 315.05 | 332.73 | 350.40 | 368.08 | 385.76 | 403.44 | 421.11 | 438.79 |
| | $W_{3,4}$ | 16.34 | 33.57 | 51.13 | 68.78 | 86.45 | 104.13 | 121.81 | 139.49 | 157.16 | 174.84 | 192.52 | 210.20 | 227.87 | 245.55 | 263.23 | 280.91 | 298.58 | 316.26 | 333.94 | 351.62 | 369.29 | 386.97 | 404.65 | 422.33 | 440.00 |
| | $J$ | 11.28 | 17.80 | 21.34 | 23.39 | 24.66 | 25.49 | 26.05 | 26.45 | 26.73 | 26.93 | 27.08 | 27.19 | 27.28 | 27.35 | 27.40 | 27.44 | 27.48 | 27.50 | 27.53 | 27.55 | 27.56 | 27.57 | 27.58 | 27.59 | 27.60 |
| | $K$ | 2.02 | 2.98 | 3.26 | 3.33 | 3.34 | 3.35 | 3.35 | 3.35 | 3.35 | 3.35 | 3.35 | 3.35 | 3.35 | 3.35 | 3.35 | 3.35 | 3.35 | 3.35 | 3.35 | 3.35 | 3.35 | 3.35 | 3.35 | 3.35 | 3.35 |
| 0.8 | $U_1$ | 23.77 | 40.66 | 51.85 | 59.50 | 64.95 | 69.00 | 72.10 | 74.55 | 76.52 | 78.14 | 79.49 | 80.64 | 81.62 | 82.47 | 83.21 | 83.87 | 84.45 | 84.97 | 85.44 | 85.86 | 86.24 | 86.59 | 86.91 | 87.21 | 87.48 |
| | $U_2$ | 0.52 | 1.28 | 2.44 | 3.93 | 5.53 | 7.11 | 8.58 | 9.92 | 11.12 | 12.19 | 13.15 | 13.99 | 14.75 | 15.43 | 16.04 | 16.59 | 17.09 | 17.53 | 17.94 | 18.32 | 18.66 | 18.98 | 19.27 | 19.54 | 19.79 |
| | $W_{1,2}$ | 15.53 | 32.25 | 49.65 | 67.26 | 84.92 | 102.60 | 120.27 | 137.95 | 155.63 | 173.31 | 190.98 | 208.66 | 226.34 | 244.02 | 261.69 | 279.37 | 297.05 | 314.73 | 332.41 | 350.08 | 367.76 | 385.44 | 403.12 | 420.79 | 438.47 |
| | $W_{3,4}$ | 16.21 | 33.35 | 50.68 | 68.53 | 86.20 | 103.87 | 121.55 | 139.23 | 156.91 | 174.58 | 192.26 | 209.94 | 227.62 | 245.29 | 262.97 | 280.65 | 298.33 | 316.00 | 333.68 | 351.36 | 369.04 | 386.71 | 404.39 | 422.07 | 439.75 |
| | $J$ | 11.73 | 18.89 | 22.98 | 25.45 | 27.03 | 28.09 | 28.81 | 29.32 | 29.70 | 29.97 | 30.17 | 30.33 | 30.45 | 30.54 | 30.62 | 30.67 | 30.72 | 30.76 | 30.79 | 30.82 | 30.84 | 30.86 | 30.87 | 30.89 | 30.90 |
| | $K$ | 2.12 | 3.13 | 3.42 | 3.49 | 3.51 | 3.51 | 3.51 | 3.51 | 3.51 | 3.51 | 3.51 | 3.51 | 3.51 | 3.51 | 3.51 | 3.51 | 3.51 | 3.51 | 3.51 | 3.51 | 3.51 | 3.51 | 3.51 | 3.51 | 3.51 |
| 0.7 | $U_1$ | 21.85 | 37.67 | 48.32 | 55.68 | 60.97 | 64.92 | 67.96 | 70.37 | 72.31 | 73.92 | 75.25 | 76.39 | 77.37 | 78.21 | 78.95 | 79.60 | 80.18 | 80.70 | 81.16 | 81.58 | 81.97 | 82.31 | 82.63 | 82.93 | 83.20 |
| | $U_2$ | 0.58 | 1.41 | 2.62 | 4.13 | 5.73 | 7.31 | 8.77 | 10.10 | 11.30 | 12.36 | 13.31 | 14.15 | 14.90 | 15.58 | 16.18 | 16.73 | 17.22 | 17.67 | 18.08 | 18.46 | 18.80 | 19.12 | 19.41 | 19.68 | 19.93 |
| | $W_{1,2}$ | 15.31 | 31.93 | 49.31 | 66.91 | 84.57 | 102.25 | 119.92 | 137.60 | 155.28 | 172.95 | 190.63 | 208.31 | 225.99 | 243.66 | 261.34 | 279.02 | 296.70 | 314.37 | 332.05 | 349.73 | 367.41 | 385.08 | 402.76 | 420.44 | 438.12 |
| | $W_{3,4}$ | 16.14 | 33.23 | 50.75 | 68.39 | 86.05 | 103.73 | 121.41 | 139.08 | 156.76 | 174.44 | 192.12 | 209.79 | 227.47 | 245.15 | 262.83 | 280.50 | 298.18 | 315.86 | 333.54 | 351.21 | 368.89 | 386.57 | 404.25 | 421.92 | 439.60 |
| | $J$ | 12.21 | 20.01 | 24.65 | 27.55 | 29.45 | 30.73 | 31.63 | 32.27 | 32.74 | 33.09 | 33.36 | 33.56 | 33.71 | 33.84 | 33.93 | 34.01 | 34.07 | 34.13 | 34.17 | 34.21 | 34.24 | 34.26 | 34.28 | 34.30 | 34.31 |
| | $K$ | 2.25 | 3.31 | 3.61 | 3.69 | 3.71 | 3.71 | 3.71 | 3.71 | 3.71 | 3.71 | 3.71 | 3.71 | 3.71 | 3.71 | 3.71 | 3.71 | 3.71 | 3.71 | 3.71 | 3.71 | 3.71 | 3.71 | 3.71 | 3.71 | 3.71 |

| $u_0/u_1$ | $\xi$/nm | 2 | 4 | 6 | 8 | 10 | 12 | 14 | 16 | 18 | 20 | 22 | 24 | 26 | 28 | 30 | 32 | 34 | 36 | 38 | 40 | 42 | 44 | 46 | 48 | 50 |
|---|---|---|---|---|---|---|---|---|---|---|---|---|---|---|---|---|---|---|---|---|---|---|---|---|---|---|
| 0.6 | $U_1$ | 20.28 | 35.20 | 45.39 | 52.50 | 57.65 | 61.52 | 64.51 | 66.88 | 68.80 | 70.38 | 71.71 | 72.83 | 73.80 | 74.64 | 75.38 | 76.02 | 76.60 | 77.12 | 77.58 | 78.00 | 78.38 | 78.73 | 79.05 | 79.34 | 79.61 |
|  | $U_2$ | 0.64 | 1.51 | 2.77 | 4.29 | 5.90 | 7.47 | 8.92 | 10.25 | 11.43 | 12.49 | 13.43 | 14.27 | 15.02 | 15.69 | 16.30 | 16.84 | 17.33 | 17.78 | 18.19 | 18.56 | 18.91 | 19.22 | 19.51 | 19.78 | 20.03 |
|  | $W_{1,2}$ | 15.08 | 31.59 | 48.94 | 66.54 | 84.20 | 101.87 | 119.55 | 137.22 | 154.90 | 172.58 | 190.26 | 207.93 | 225.61 | 243.29 | 260.97 | 278.64 | 296.32 | 314.00 | 331.68 | 349.35 | 367.03 | 384.71 | 402.39 | 420.06 | 437.74 |
|  | $W_{3,4}$ | 16.11 | 33.17 | 50.68 | 68.32 | 85.99 | 103.66 | 121.34 | 139.02 | 156.69 | 174.37 | 192.05 | 209.73 | 227.40 | 245.08 | 262.76 | 280.44 | 298.11 | 315.79 | 333.47 | 351.15 | 368.82 | 386.50 | 404.18 | 421.86 | 439.53 |
|  | $J$ | 12.70 | 21.09 | 26.26 | 29.57 | 31.78 | 33.30 | 34.38 | 35.16 | 35.74 | 36.17 | 36.50 | 36.76 | 36.96 | 37.12 | 37.25 | 37.35 | 37.44 | 37.51 | 37.57 | 37.61 | 37.65 | 37.69 | 37.72 | 37.74 | 37.76 |
|  | $K$ | 2.40 | 3.52 | 3.84 | 3.92 | 3.94 | 3.94 | 3.94 | 3.94 | 3.94 | 3.94 | 3.94 | 3.94 | 3.94 | 3.94 | 3.94 | 3.94 | 3.94 | 3.94 | 3.94 | 3.94 | 3.94 | 3.94 | 3.94 | 3.94 | 3.94 |
| 0.5 | $U_1$ | 19.20 | 33.50 | 43.38 | 50.33 | 55.39 | 59.20 | 62.15 | 64.50 | 66.40 | 67.98 | 69.30 | 70.42 | 71.38 | 72.22 | 72.95 | 73.60 | 74.17 | 74.69 | 75.15 | 75.57 | 75.95 | 76.29 | 76.61 | 76.90 | 77.17 |
|  | $U_2$ | 0.67 | 1.59 | 2.87 | 4.40 | 6.01 | 7.58 | 9.03 | 10.35 | 11.54 | 12.59 | 13.53 | 14.36 | 15.11 | 15.78 | 16.38 | 16.93 | 17.42 | 17.86 | 18.27 | 18.64 | 18.99 | 19.30 | 19.59 | 19.86 | 20.11 |
|  | $W_{1,2}$ | 14.81 | 31.21 | 48.53 | 66.12 | 83.77 | 101.45 | 119.12 | 136.80 | 154.48 | 172.16 | 189.83 | 207.51 | 225.19 | 242.87 | 260.54 | 278.22 | 295.90 | 313.58 | 331.25 | 348.93 | 366.61 | 384.29 | 401.96 | 419.64 | 437.32 |
|  | $W_{3,4}$ | 16.10 | 33.14 | 50.64 | 68.28 | 85.95 | 103.62 | 121.30 | 138.98 | 156.65 | 174.33 | 192.01 | 209.69 | 227.36 | 245.04 | 262.72 | 280.40 | 298.07 | 315.75 | 333.43 | 351.11 | 368.78 | 386.46 | 404.14 | 421.82 | 439.49 |
|  | $J$ | 13.11 | 21.98 | 27.56 | 31.17 | 33.59 | 35.27 | 36.46 | 37.33 | 37.98 | 38.46 | 38.83 | 39.12 | 39.35 | 39.53 | 39.67 | 39.79 | 39.89 | 39.96 | 40.03 | 40.08 | 40.13 | 40.17 | 40.20 | 40.23 | 40.26 |
|  | $K$ | 2.72 | 3.99 | 4.35 | 4.44 | 4.46 | 4.47 | 4.47 | 4.47 | 4.47 | 4.47 | 4.47 | 4.47 | 4.47 | 4.47 | 4.47 | 4.47 | 4.47 | 4.47 | 4.47 | 4.47 | 4.47 | 4.47 | 4.47 | 4.47 | 4.47 |
| 0.4 | $U_1$ | 17.50 | 31.40 | 40.84 | 47.55 | 52.46 | 56.18 | 59.08 | 61.39 | 63.26 | 64.82 | 66.12 | 67.23 | 68.19 | 69.02 | 69.75 | 70.39 | 70.96 | 71.47 | 71.94 | 72.35 | 72.73 | 73.08 | 73.39 | 73.68 | 73.95 |
|  | $U_2$ | 0.73 | 1.70 | 3.01 | 4.56 | 6.17 | 7.73 | 9.18 | 10.49 | 11.66 | 12.71 | 13.64 | 14.47 | 15.21 | 15.88 | 16.48 | 17.02 | 17.51 | 17.95 | 18.36 | 18.73 | 19.07 | 19.39 | 19.68 | 19.95 | 20.19 |
|  | $W_{1,2}$ | 14.61 | 30.92 | 48.21 | 65.80 | 83.45 | 101.12 | 118.80 | 136.48 | 154.15 | 171.83 | 189.51 | 207.19 | 224.86 | 242.54 | 260.22 | 277.90 | 295.58 | 313.25 | 330.93 | 348.61 | 366.29 | 383.96 | 401.64 | 419.32 | 437.00 |
|  | $W_{3,4}$ | 16.14 | 33.18 | 50.69 | 68.32 | 85.99 | 103.66 | 121.34 | 139.02 | 156.69 | 174.37 | 192.05 | 209.73 | 227.40 | 245.08 | 262.76 | 280.44 | 298.11 | 315.79 | 333.47 | 351.15 | 368.83 | 386.50 | 404.18 | 421.86 | 439.53 |
|  | $J$ | 13.56 | 22.96 | 29.03 | 33.04 | 35.77 | 37.70 | 39.09 | 40.12 | 40.89 | 41.47 | 41.93 | 42.29 | 42.57 | 42.80 | 42.98 | 43.13 | 43.25 | 43.36 | 43.44 | 43.52 | 43.58 | 43.63 | 43.68 | 43.71 | 43.75 |
|  | $K$ | 2.99 | 4.35 | 4.44 | 4.46 | 4.47 | 4.47 | 4.47 | 4.47 | 4.47 | 4.47 | 4.47 | 4.47 | 4.47 | 4.47 | 4.47 | 4.47 | 4.47 | 4.47 | 4.47 | 4.47 | 4.47 | 4.47 | 4.47 | 4.47 | 4.47 |
| 0.3 | $U_1$ | 17.28 | 30.40 | 39.65 | 46.25 | 51.11 | 54.79 | 57.66 | 59.96 | 61.82 | 63.37 | 64.67 | 65.78 | 66.73 | 67.56 | 68.29 | 68.93 | 69.50 | 70.01 | 70.47 | 70.89 | 71.26 | 71.61 | 71.92 | 72.22 | 72.48 |
|  | $U_2$ | 0.78 | 1.75 | 3.08 | 4.63 | 6.25 | 7.80 | 9.24 | 10.55 | 11.72 | 12.77 | 13.70 | 14.53 | 15.27 | 15.93 | 16.53 | 17.07 | 17.56 | 18.00 | 18.41 | 18.78 | 19.12 | 19.43 | 19.72 | 19.99 | 20.24 |
|  | $W_{1,2}$ | 14.37 | 30.68 | 47.84 | 65.42 | 83.07 | 100.74 | 118.42 | 136.09 | 153.77 | 171.45 | 189.12 | 206.81 | 224.49 | 242.16 | 259.84 | 277.52 | 295.20 | 312.87 | 330.55 | 348.23 | 365.91 | 383.58 | 401.26 | 418.94 | 436.62 |
|  | $W_{3,4}$ | 16.15 | 33.19 | 50.70 | 68.35 | 86.02 | 103.70 | 121.35 | 139.05 | 156.73 | 174.40 | 192.08 | 209.76 | 227.43 | 245.11 | 262.79 | 280.47 | 298.14 | 315.82 | 333.50 | 351.18 | 368.85 | 386.53 | 404.21 | 421.89 | 439.55 |
|  | $J$ | 13.87 | 23.61 | 29.96 | 34.18 | 37.07 | 39.11 | 40.59 | 41.67 | 42.49 | 43.11 | 43.59 | 43.99 | 44.29 | 44.51 | 44.69 | 44.84 | 44.95 | 45.03 | 45.11 | 45.18 | 45.23 | 45.28 | 45.38 | 45.43 | 45.45 |
|  | $K$ | 2.88 | 4.22 | 4.60 | 4.70 | 4.73 | 4.73 | 4.73 | 4.73 | 4.73 | 4.73 | 4.73 | 4.73 | 4.73 | 4.73 | 4.73 | 4.73 | 4.73 | 4.73 | 4.73 | 4.73 | 4.73 | 4.73 | 4.73 | 4.73 | 4.73 |
| 0.2 | $U_1$ | 16.83 | 29.65 | 38.74 | 45.26 | 50.07 | 53.72 | 56.56 | 58.85 | 60.72 | 62.26 | 63.56 | 64.66 | 65.62 | 66.44 | 67.16 | 67.80 | 68.37 | 68.88 | 69.34 | 69.76 | 70.13 | 70.47 | 70.79 | 71.09 | 71.35 |
|  | $U_2$ | 0.78 | 1.79 | 3.13 | 4.69 | 6.30 | 7.86 | 9.30 | 10.60 | 11.77 | 12.81 | 13.74 | 14.57 | 15.31 | 15.97 | 16.57 | 17.11 | 17.60 | 18.04 | 18.45 | 18.82 | 19.16 | 19.47 | 19.76 | 20.03 | 20.28 |
|  | $W_{1,2}$ | 14.18 | 30.31 | 47.54 | 65.12 | 82.77 | 100.44 | 118.12 | 135.79 | 153.47 | 171.15 | 188.83 | 206.50 | 224.18 | 241.86 | 259.54 | 277.21 | 294.89 | 312.57 | 330.25 | 347.92 | 365.60 | 383.28 | 400.96 | 418.63 | 436.31 |
|  | $W_{3,4}$ | 16.17 | 33.21 | 50.72 | 68.35 | 86.02 | 103.70 | 121.37 | 139.05 | 156.73 | 174.40 | 192.08 | 209.76 | 227.43 | 245.11 | 262.79 | 280.47 | 298.15 | 315.83 | 333.50 | 351.18 | 368.86 | 386.54 | 404.21 | 421.89 | 439.57 |
|  | $J$ | 14.11 | 24.12 | 30.59 | 34.85 | 37.73 | 39.74 | 41.17 | 42.28 | 43.12 | 43.76 | 44.22 | 44.63 | 44.93 | 45.15 | 45.33 | 45.44 | 45.53 | 45.59 | 45.64 | 45.68 | 45.72 | 45.75 | 45.78 | 45.80 | 45.82 |
|  | $K$ | 3.05 | 4.42 | 4.82 | 4.92 | 4.94 | 4.95 | 4.95 | 4.95 | 4.95 | 4.95 | 4.95 | 4.95 | 4.95 | 4.95 | 4.95 | 4.95 | 4.95 | 4.95 | 4.95 | 4.95 | 4.95 | 4.95 | 4.95 | 4.95 | 4.95 |
| 0.1 | $U_1$ | 17.65 | 31.11 | 40.59 | 47.34 | 52.29 | 56.04 | 58.96 | 61.29 | 63.18 | 64.74 | 66.05 | 67.17 | 68.13 | 68.96 | 69.69 | 70.34 | 70.91 | 71.42 | 71.89 | 72.30 | 72.68 | 73.03 | 73.35 | 73.64 | 73.91 |
|  | $U_2$ | 0.72 | 1.70 | 3.02 | 4.58 | 6.20 | 7.76 | 9.21 | 10.52 | 11.69 | 12.74 | 13.68 | 14.51 | 15.25 | 15.92 | 16.52 | 17.06 | 17.55 | 17.99 | 18.40 | 18.77 | 19.11 | 19.43 | 19.72 | 19.99 | 20.24 |
|  | $W_{1,2}$ | 13.79 | 29.74 | 46.93 | 64.49 | 82.14 | 99.81 | 117.49 | 135.16 | 152.84 | 170.52 | 188.19 | 205.87 | 223.55 | 241.23 | 258.90 | 276.58 | 294.26 | 311.94 | 329.62 | 347.29 | 364.97 | 382.65 | 400.33 | 418.00 | 435.68 |
|  | $W_{3,4}$ | 16.11 | 33.13 | 50.63 | 68.26 | 85.93 | 103.61 | 121.24 | 138.96 | 156.64 | 174.31 | 191.99 | 209.67 | 227.34 | 245.02 | 262.70 | 280.38 | 298.05 | 315.73 | 333.41 | 351.09 | 368.77 | 386.45 | 404.12 | 421.80 | 439.48 |
|  | $J$ | 14.16 | 24.12 | 30.50 | 34.75 | 37.73 | 39.74 | 41.16 | 42.26 | 43.10 | 43.74 | 44.20 | 44.60 | 44.90 | 45.12 | 45.30 | 45.43 | 45.53 | 45.59 | 45.65 | 45.68 | 45.71 | 45.74 | 45.68 | 45.71 | 45.16 |
|  | $K$ | 3.11 | 4.55 | 4.97 | 5.07 | 5.09 | 5.10 | 5.10 | 5.10 | 5.10 | 5.10 | 5.10 | 5.10 | 5.10 | 5.10 | 5.10 | 5.10 | 5.10 | 5.10 | 5.10 | 5.10 | 5.10 | 5.10 | 5.10 | 5.10 | 5.10 |
| 0.0 | $U_1$ | 17.36 | 30.60 | 39.95 | 46.63 | 51.53 | 55.25 | 58.15 | 60.46 | 62.34 | 63.90 | 65.20 | 66.32 | 67.27 | 68.10 | 68.83 | 69.48 | 70.05 | 70.56 | 71.02 | 71.44 | 71.81 | 72.16 | 72.48 | 72.77 | 73.04 |
|  | $U_2$ | 0.70 | 1.65 | 2.97 | 4.52 | 6.14 | 7.71 | 9.16 | 10.47 | 11.65 | 12.70 | 13.64 | 14.47 | 15.22 | 15.89 | 16.49 | 17.03 | 17.52 | 17.97 | 18.37 | 18.75 | 19.09 | 19.40 | 19.69 | 19.96 | 20.21 |
|  | $W_{1,2}$ | 13.79 | 29.74 | 46.93 | 64.49 | 82.14 | 99.81 | 117.49 | 135.16 | 152.84 | 170.52 | 188.19 | 205.87 | 223.55 | 241.23 | 258.90 | 276.58 | 294.26 | 311.94 | 329.62 | 347.29 | 364.97 | 382.65 | 400.33 | 418.00 | 435.68 |
|  | $W_{3,4}$ | 16.14 | 33.16 | 50.66 | 68.29 | 85.96 | 103.63 | 121.31 | 138.99 | 156.66 | 174.34 | 192.02 | 209.69 | 227.37 | 245.05 | 262.73 | 280.40 | 298.08 | 315.76 | 333.43 | 351.11 | 368.79 | 386.47 | 404.14 | 421.82 | 439.50 |
|  | $J$ | 14.17 | 24.11 | 30.51 | 34.70 | 37.52 | 39.46 | 40.83 | 41.83 | 42.56 | 43.11 | 43.53 | 43.85 | 44.10 | 44.30 | 44.46 | 44.59 | 44.70 | 44.78 | 44.86 | 44.92 | 44.97 | 45.01 | 45.05 | 45.08 | 45.10 |
|  | $K$ | 3.15 | 4.60 | 5.02 | 5.12 | 5.15 | 5.16 | 5.16 | 5.16 | 5.16 | 5.16 | 5.16 | 5.16 | 5.16 | 5.16 | 5.16 | 5.16 | 5.16 | 5.16 | 5.16 | 5.16 | 5.16 | 5.16 | 5.16 | 5.16 | 5.16 |

TABLE S77. Parameters of the THF interaction Hamiltonian for different ratios $u_0/u_1$ and screening lengths $\xi$ of the double-gate interaction potential ($U_\xi = 24$ meV for $\xi = 10$ nm). We use the same BM model parameters as in Table S76.

**Total weight of the heavy-fermions on the active bands after wannierisation: 93.1%**

Energy (meV) axis: 150, 100, 50, 0, −50, −100, −150
Horizontal axis labels: Γ   K   M   K

Wannier orbitals support: 0.6 — 1.0

FIG. S55. Band structures of the BM and THF models near charge neutrality for $w_0/w_1 = 0.8$, depicted by lines and crosses, respectively. The BM bands are colored according to the weight of the $f$-electron wave function on them. We use the same BM parameters as in Table S78.

TABLE S78. Parameters of the $f$-electron Wannier functions and the THF single-particle Hamiltonian for different values of the tunneling ratio $w_0/w_1$. $W$ denotes the total weight of the THF $f$-electrons on the active bands. In computing the ratios $\gamma/U_1$ and $\gamma/U_2$, we employ the on-site and nearest-neighbor repulsion parameters $U_1$ and $U_2$ obtained numerically for $\xi = 10\,\mathrm{nm}$, as given in Table S79. We employ $v_F = 5.944\,\mathrm{eV\cdot\mathring{A}}$, $|\mathbf{K}| = 1.703\,\mathring{A}^{-1}$, $w_1 = 110\,\mathrm{meV}$, and $\theta = 1.44°$ for the BM model.

| $w_0/w_1$ | $\alpha_1$ | $\alpha_2$ | $\lambda_1$ | $\lambda_2$ | $\lambda$ | $W$ | $t_0$ | $\gamma$ | $M$ | $v_\star$ | $v'_\star$ | $v''_\star$ | $\gamma/U_1$ | $\gamma/U_2$ |
|---|---|---|---|---|---|---|---|---|---|---|---|---|---|---|
| 1.00 | 0.818 | 0.575 | 0.203 | 0.209 | 0.364 | 0.938 | 3.966 | −54.020 | −47.755 | −4.775 | 1.884 | 0.0303 | −0.73 | −10.01 |
| 0.90 | 0.846 | 0.533 | 0.215 | 0.379 | 0.932 | 4.654 | −70.296 | −49.385 | −4.884 | 1.884 | 0.0247 | −1.01 | −12.65 |
| 0.80 | 0.874 | 0.486 | 0.233 | 0.221 | 0.399 | 0.931 | 5.454 | −86.063 | −50.801 | −4.988 | 1.856 | 0.0201 | −1.31 | −14.95 |
| 0.70 | 0.900 | 0.435 | 0.245 | 0.227 | 0.418 | 0.933 | 6.304 | −101.123 | −52.016 | −5.088 | 1.798 | 0.0162 | −1.64 | −16.99 |
| 0.60 | 0.925 | 0.381 | 0.263 | 0.233 | 0.436 | 0.936 | 6.973 | −115.250 | −53.044 | −5.184 | 1.705 | 0.0128 | −1.98 | −18.85 |
| 0.50 | 0.946 | 0.324 | 0.281 | 0.239 | 0.445 | 0.936 | 7.583 | −128.173 | −53.896 | −5.278 | 1.566 | 0.0096 | −2.29 | −20.58 |
| 0.40 | 0.965 | 0.262 | 0.292 | 0.239 | 0.467 | 0.944 | 8.225 | −139.569 | −54.581 | −5.367 | 1.371 | 0.0067 | −2.63 | −21.82 |
| 0.30 | 0.980 | 0.200 | 0.310 | 0.245 | 0.480 | 0.948 | 8.611 | −149.063 | −55.106 | −5.446 | 1.114 | 0.0041 | −2.91 | −22.98 |
| 0.20 | 0.991 | 0.136 | 0.245 | 0.477 | 0.945 | 8.878 | −156.252 | −55.477 | −5.511 | 0.793 | 0.0019 | −3.07 | −23.99 |
| 0.10 | 0.998 | 0.070 | 0.322 | 0.251 | 0.460 | 0.933 | 8.717 | −160.755 | −55.698 | −5.555 | 0.412 | 0.0005 | −3.07 | −25.10 |
| 0.00 | 1.000 | 0.000 | 0.328 | 0.000 | 0.475 | 0.941 | 8.909 | −162.291 | −55.772 | −5.570 | −0.000 | −0.0000 | −3.20 | −24.97 |

| $w_0/w_1$ | $\xi/\mathrm{nm}$ | 2 | 4 | 6 | 8 | 10 | 12 | 14 | 16 | 18 | 20 | 22 | 24 | 26 | 28 | 30 | 32 | 34 | 36 | 38 | 40 | 42 | 44 | 46 | 48 | 50 |
|---|---|---|---|---|---|---|---|---|---|---|---|---|---|---|---|---|---|---|---|---|---|---|---|---|---|---|
| 1.0 | $U_1$ | 28.21 | 47.45 | 59.75 | 67.96 | 73.71 | 77.94 | 81.15 | 83.66 | 85.68 | 87.34 | 88.71 | 89.88 | 90.87 | 91.73 | 92.48 | 93.14 | 93.73 | 94.26 | 94.73 | 95.15 | 95.54 | 95.89 | 96.21 | 96.50 | 96.78 |
| | $U_2$ | 0.44 | 1.12 | 2.25 | 3.76 | 5.40 | 7.02 | 8.53 | 9.91 | 11.13 | 12.37 | 13.20 | 14.06 | 14.83 | 15.52 | 16.13 | 16.69 | 17.19 | 17.65 | 18.06 | 18.44 | 18.79 | 19.11 | 19.40 | 19.67 | 19.92 |
| | $W_{1,2}$ | 16.43 | 33.83 | 51.79 | 69.92 | 88.08 | 106.26 | 124.44 | 142.62 | 160.80 | 178.97 | 197.15 | 215.33 | 233.51 | 251.69 | 269.87 | 288.05 | 306.23 | 324.41 | 342.59 | 360.76 | 378.94 | 397.12 | 415.30 | 433.48 | 451.66 |
| | $W_{3,4}$ | 16.43 | 34.76 | 52.86 | 71.02 | 89.19 | 107.37 | 125.55 | 143.73 | 161.91 | 180.09 | 198.26 | 216.44 | 234.62 | 252.80 | 270.98 | 289.16 | 307.34 | 325.51 | 343.69 | 361.87 | 380.05 | 398.23 | 416.41 | 434.58 | 452.76 |
| | $J$ | 11.14 | 17.21 | 20.30 | 22.02 | 23.05 | 23.70 | 24.13 | 24.43 | 24.64 | 24.79 | 24.90 | 24.98 | 25.05 | 25.09 | 25.13 | 25.16 | 25.18 | 25.20 | 25.22 | 25.23 | 25.24 | 25.25 | 25.26 | 25.27 |
| | $K$ | 1.92 | 2.82 | 3.07 | 3.14 | 3.15 | 3.15 | 3.15 | 3.15 | 3.15 | 3.15 | 3.15 | 3.15 | 3.15 | 3.15 | 3.15 | 3.15 | 3.15 | 3.15 | 3.15 | 3.15 | 3.15 | 3.15 | 3.15 | 3.15 |
| 0.9 | $U_1$ | 26.11 | 44.27 | 56.08 | 64.04 | 69.67 | 73.82 | 76.99 | 79.48 | 81.48 | 83.12 | 84.49 | 85.64 | 86.63 | 87.49 | 88.24 | 88.90 | 89.48 | 90.00 | 90.47 | 90.90 | 91.28 | 91.63 | 91.95 | 92.25 | 92.52 |
| | $U_2$ | 0.48 | 1.21 | 2.39 | 3.91 | 5.56 | 7.18 | 8.68 | 10.05 | 11.28 | 12.37 | 13.33 | 14.19 | 14.96 | 15.64 | 16.26 | 16.81 | 17.31 | 17.77 | 18.18 | 18.56 | 18.90 | 19.22 | 19.52 | 19.79 | 20.04 |
| | $W_{1,2}$ | 16.25 | 33.57 | 51.51 | 69.63 | 87.80 | 105.97 | 124.15 | 142.33 | 160.51 | 178.69 | 196.87 | 215.04 | 233.22 | 251.40 | 269.58 | 287.76 | 305.94 | 324.12 | 342.30 | 360.48 | 378.66 | 396.83 | 415.01 | 433.19 | 451.37 |
| | $W_{3,4}$ | 16.75 | 34.44 | 52.49 | 70.64 | 88.82 | 106.99 | 125.17 | 143.35 | 161.53 | 179.71 | 197.89 | 216.07 | 234.25 | 252.42 | 270.60 | 288.78 | 306.96 | 325.14 | 343.32 | 361.50 | 379.68 | 397.86 | 416.04 | 434.21 | 452.39 |
| | $J$ | 11.55 | 18.23 | 21.84 | 23.94 | 25.23 | 26.08 | 26.65 | 27.04 | 27.33 | 27.53 | 27.68 | 27.80 | 27.89 | 27.95 | 28.01 | 28.05 | 28.08 | 28.11 | 28.13 | 28.16 | 28.18 | 28.19 | 28.20 | 28.21 |
| | $K$ | 2.01 | 2.94 | 3.21 | 3.28 | 3.29 | 3.29 | 3.29 | 3.29 | 3.29 | 3.29 | 3.29 | 3.29 | 3.29 | 3.29 | 3.29 | 3.29 | 3.29 | 3.29 | 3.29 | 3.29 | 3.29 | 3.29 | 3.29 | 3.29 |
| 0.8 | $U_1$ | 24.00 | 41.03 | 52.29 | 59.97 | 65.45 | 69.51 | 72.62 | 75.07 | 77.04 | 78.67 | 80.02 | 81.17 | 82.15 | 83.00 | 83.75 | 84.40 | 84.99 | 85.51 | 85.97 | 86.40 | 86.78 | 87.13 | 87.45 | 87.74 | 88.01 |
| | $U_2$ | 0.54 | 1.34 | 2.56 | 4.10 | 5.76 | 7.38 | 8.87 | 10.23 | 11.45 | 12.53 | 13.50 | 14.35 | 15.11 | 15.80 | 16.41 | 16.96 | 17.46 | 17.91 | 18.32 | 18.70 | 19.05 | 19.36 | 19.66 | 19.93 | 20.18 |
| | $W_{1,2}$ | 16.03 | 33.27 | 51.18 | 69.30 | 87.46 | 105.64 | 123.82 | 142.00 | 160.18 | 178.35 | 196.53 | 214.71 | 232.89 | 251.07 | 269.25 | 287.43 | 305.61 | 323.79 | 341.97 | 360.14 | 378.32 | 396.50 | 414.68 | 432.86 | 451.04 |
| | $W_{3,4}$ | 16.63 | 34.24 | 52.27 | 70.41 | 88.59 | 106.76 | 124.94 | 143.12 | 161.30 | 179.48 | 197.66 | 215.84 | 234.01 | 252.19 | 270.37 | 288.55 | 306.73 | 324.91 | 343.09 | 361.27 | 379.45 | 397.63 | 415.80 | 433.98 | 452.16 |
| | $J$ | 12.02 | 19.34 | 23.52 | 26.04 | 27.64 | 28.71 | 29.46 | 30.00 | 30.38 | 30.60 | 30.81 | 30.97 | 31.08 | 31.18 | 31.25 | 31.31 | 31.36 | 31.40 | 31.43 | 31.45 | 31.47 | 31.49 | 31.51 | 31.52 | 31.53 |
| | $K$ | 2.12 | 3.10 | 3.37 | 3.44 | 3.46 | 3.46 | 3.46 | 3.46 | 3.46 | 3.46 | 3.46 | 3.46 | 3.46 | 3.46 | 3.46 | 3.46 | 3.46 | 3.46 | 3.46 | 3.46 | 3.46 | 3.46 | 3.46 | 3.46 |
| 0.7 | $U_1$ | 22.13 | 38.10 | 48.84 | 56.25 | 61.57 | 65.54 | 68.59 | 71.00 | 72.95 | 74.56 | 75.90 | 77.04 | 78.01 | 78.86 | 79.60 | 80.25 | 80.83 | 81.35 | 81.82 | 82.24 | 82.62 | 82.97 | 83.29 | 83.58 | 83.85 |
| | $U_2$ | 0.61 | 1.47 | 2.73 | 4.29 | 5.95 | 7.56 | 9.06 | 10.41 | 11.62 | 12.69 | 13.65 | 14.50 | 15.26 | 15.94 | 16.55 | 17.10 | 17.60 | 18.05 | 18.46 | 18.83 | 19.18 | 19.50 | 19.79 | 20.06 | 20.31 |
| | $W_{1,2}$ | 16.57 | 32.94 | 50.83 | 68.94 | 87.10 | 105.27 | 123.45 | 141.63 | 159.81 | 177.99 | 196.16 | 214.34 | 232.52 | 250.71 | 268.88 | 287.06 | 305.24 | 324.75 | 341.60 | 359.78 | 377.96 | 396.14 | 414.32 | 432.50 | 450.67 |
| | $W_{3,4}$ | 16.57 | 34.13 | 52.15 | 70.29 | 88.46 | 106.64 | 124.81 | 142.99 | 161.17 | 179.35 | 197.53 | 215.71 | 233.89 | 252.07 | 270.25 | 288.43 | 306.60 | 324.78 | 342.96 | 361.14 | 379.32 | 397.50 | 415.68 | 433.86 | 452.04 |
| | $J$ | 12.52 | 20.48 | 25.21 | 28.15 | 30.07 | 31.37 | 32.27 | 32.91 | 33.38 | 33.73 | 33.99 | 34.19 | 34.35 | 34.47 | 34.57 | 34.64 | 34.71 | 34.76 | 34.80 | 34.84 | 34.86 | 34.89 | 34.91 | 34.93 | 34.94 |
| | $K$ | 2.25 | 3.28 | 3.58 | 3.65 | 3.66 | 3.67 | 3.67 | 3.67 | 3.67 | 3.67 | 3.67 | 3.67 | 3.67 | 3.67 | 3.67 | 3.67 | 3.67 | 3.67 | 3.67 | 3.67 | 3.67 | 3.67 | 3.67 | 3.67 | 3.67 |

| $u_0/u_1$ | ξ/nm | 2 | 4 | 6 | 8 | 10 | 12 | 14 | 16 | 18 | 20 | 22 | 24 | 26 | 28 | 30 | 32 | 34 | 36 | 38 | 40 | 42 | 44 | 46 | 48 | 50 |
|---|---|---|---|---|---|---|---|---|---|---|---|---|---|---|---|---|---|---|---|---|---|---|---|---|---|---|
| 0.6 | $U_1$ | 20.58 | 35.67 | 45.95 | 53.12 | 58.31 | 62.19 | 65.19 | 67.57 | 69.50 | 71.08 | 72.41 | 73.54 | 74.51 | 75.35 | 76.09 | 76.74 | 77.32 | 77.83 | 78.30 | 78.72 | 79.10 | 79.45 | 79.76 | 80.06 | 80.33 |
| | $U_2$ | 0.66 | 1.58 | 2.88 | 4.46 | 6.11 | 7.72 | 9.21 | 10.55 | 11.75 | 12.82 | 13.78 | 14.62 | 15.38 | 16.05 | 16.66 | 17.21 | 17.70 | 18.15 | 18.56 | 18.94 | 19.28 | 19.60 | 19.89 | 20.16 | 20.41 |
| | $W_{1,2}$ | 15.56 | 32.59 | 50.45 | 68.56 | 86.72 | 104.89 | 123.07 | 141.25 | 159.43 | 177.61 | 195.79 | 213.97 | 232.14 | 250.32 | 268.50 | 286.68 | 304.86 | 323.04 | 341.22 | 359.40 | 377.58 | 395.75 | 413.93 | 432.11 | 450.29 |
| | $W_{3,4}$ | 16.55 | 34.09 | 52.10 | 70.24 | 88.41 | 106.58 | 124.70 | 142.94 | 161.12 | 179.30 | 197.48 | 215.66 | 233.83 | 252.01 | 270.19 | 288.37 | 306.55 | 324.73 | 342.91 | 361.09 | 379.27 | 397.45 | 415.62 | 433.80 | 451.98 |
| | $J$ | 13.01 | 21.58 | 26.84 | 30.19 | 32.42 | 33.95 | 35.05 | 35.81 | 36.39 | 36.82 | 37.15 | 37.40 | 37.60 | 37.76 | 37.89 | 37.99 | 38.07 | 38.14 | 38.20 | 38.25 | 38.28 | 38.32 | 38.35 | 38.37 | 38.39 |
| | $K$ | 2.40 | 3.50 | 3.81 | 3.88 | 3.90 | 3.90 | 3.90 | 3.90 | 3.90 | 3.90 | 3.90 | 3.90 | 3.90 | 3.90 | 3.90 | 3.90 | 3.90 | 3.90 | 3.90 | 3.90 | 3.90 | 3.90 | 3.90 | 3.90 | 3.90 |
| 0.5 | $U_1$ | 19.51 | 33.99 | 43.98 | 50.98 | 56.08 | 59.91 | 62.88 | 65.23 | 67.15 | 68.72 | 70.05 | 71.17 | 72.14 | 72.97 | 73.71 | 74.36 | 74.93 | 75.45 | 75.91 | 76.33 | 76.71 | 77.06 | 77.37 | 77.67 | 77.93 |
| | $U_2$ | 1.65 | 2.98 | 4.57 | 6.23 | 7.83 | 9.31 | 10.65 | 11.85 | 12.92 | 13.87 | 14.71 | 15.47 | 16.14 | 16.75 | 17.29 | 17.78 | 18.24 | 18.65 | 19.02 | 19.36 | 19.68 | 19.97 | 20.24 | 20.49 | 20.49 |
| | $W_{1,2}$ | 15.29 | 32.20 | 50.03 | 68.13 | 86.29 | 104.46 | 122.64 | 140.82 | 159.00 | 177.18 | 195.36 | 213.54 | 231.72 | 249.90 | 268.07 | 286.25 | 304.43 | 322.61 | 340.79 | 358.97 | 377.15 | 395.33 | 413.51 | 431.69 | 449.86 |
| | $W_{3,4}$ | 16.54 | 34.07 | 52.07 | 70.21 | 88.38 | 106.55 | 124.73 | 142.91 | 161.09 | 179.27 | 197.45 | 215.63 | 233.80 | 251.98 | 270.16 | 288.34 | 306.52 | 324.70 | 342.88 | 361.06 | 379.24 | 397.42 | 415.59 | 433.77 | 451.95 |
| | $J$ | 13.43 | 22.48 | 28.15 | 31.80 | 34.24 | 35.92 | 37.12 | 37.98 | 38.62 | 39.11 | 39.47 | 39.76 | 39.98 | 40.16 | 40.30 | 40.42 | 40.51 | 40.59 | 40.65 | 40.71 | 40.75 | 40.79 | 40.82 | 40.85 | 40.88 |
| | $K$ | 2.56 | 3.73 | 4.06 | 4.14 | 4.16 | 4.16 | 4.16 | 4.16 | 4.16 | 4.16 | 4.16 | 4.16 | 4.16 | 4.16 | 4.16 | 4.16 | 4.16 | 4.16 | 4.16 | 4.16 | 4.16 | 4.16 | 4.16 | 4.16 | 4.16 |
| 0.4 | $U_1$ | 18.16 | 31.81 | 41.34 | 48.09 | 53.04 | 56.78 | 59.69 | 62.00 | 63.89 | 65.44 | 66.75 | 67.87 | 68.82 | 69.66 | 70.38 | 71.03 | 71.60 | 72.11 | 72.57 | 72.99 | 73.37 | 73.72 | 74.03 | 74.32 | 74.59 |
| | $U_2$ | 0.76 | 1.77 | 3.14 | 4.74 | 6.40 | 7.99 | 9.47 | 10.80 | 11.99 | 13.05 | 13.99 | 14.83 | 15.58 | 16.25 | 16.85 | 17.39 | 17.89 | 18.33 | 18.74 | 19.11 | 19.46 | 19.77 | 20.06 | 20.33 | 20.58 |
| | $W_{1,2}$ | 15.09 | 31.92 | 49.73 | 67.82 | 85.98 | 104.15 | 122.33 | 140.51 | 158.68 | 176.86 | 195.04 | 213.22 | 231.40 | 249.58 | 267.76 | 285.94 | 304.12 | 322.30 | 340.47 | 358.65 | 376.83 | 395.01 | 413.19 | 431.37 | 449.55 |
| | $W_{3,4}$ | 16.59 | 34.12 | 52.13 | 70.26 | 88.43 | 106.61 | 124.79 | 142.97 | 161.15 | 179.33 | 197.51 | 215.68 | 233.86 | 252.04 | 270.22 | 288.40 | 306.58 | 324.76 | 342.94 | 361.12 | 379.30 | 397.47 | 415.65 | 433.83 | 452.01 |
| | $J$ | 13.90 | 23.50 | 29.67 | 33.74 | 36.50 | 38.44 | 39.84 | 40.86 | 41.63 | 42.22 | 42.67 | 43.03 | 43.31 | 43.53 | 43.72 | 43.87 | 43.99 | 44.09 | 44.17 | 44.25 | 44.31 | 44.36 | 44.40 | 44.44 | 44.47 |
| | $K$ | 2.73 | 3.98 | 4.33 | 4.41 | 4.43 | 4.43 | 4.43 | 4.43 | 4.43 | 4.43 | 4.43 | 4.43 | 4.43 | 4.43 | 4.43 | 4.43 | 4.43 | 4.43 | 4.43 | 4.43 | 4.43 | 4.43 | 4.43 | 4.43 | 4.43 |
| 0.3 | $U_1$ | 17.36 | 30.53 | 39.80 | 46.41 | 51.27 | 54.95 | 57.83 | 60.12 | 61.99 | 63.53 | 64.84 | 65.94 | 66.89 | 67.72 | 68.45 | 69.09 | 69.66 | 70.17 | 70.63 | 71.05 | 71.42 | 71.77 | 72.08 | 72.38 | 72.64 |
| | $U_2$ | 0.84 | 1.84 | 3.22 | 4.83 | 6.51 | 8.08 | 9.57 | 10.90 | 12.09 | 13.16 | 14.10 | 14.94 | 15.69 | 16.30 | 16.90 | 17.47 | 17.96 | 18.38 | 18.78 | 19.16 | 19.50 | 19.81 | 20.10 | 20.37 | 20.62 |
| | $W_{1,2}$ | 14.88 | 31.62 | 49.40 | 67.48 | 85.64 | 103.82 | 121.99 | 140.17 | 158.35 | 176.53 | 194.71 | 212.89 | 231.06 | 249.25 | 267.42 | 285.60 | 303.78 | 321.96 | 340.14 | 358.32 | 376.50 | 394.68 | 412.86 | 431.04 | 449.21 |
| | $W_{3,4}$ | 16.62 | 34.16 | 52.17 | 70.30 | 88.47 | 106.64 | 124.82 | 143.00 | 161.18 | 179.36 | 197.54 | 215.72 | 233.89 | 252.07 | 270.25 | 288.43 | 306.61 | 324.79 | 342.97 | 361.15 | 379.33 | 397.51 | 415.68 | 433.86 | 452.05 |
| | $J$ | 14.24 | 24.24 | 30.75 | 35.05 | 38.01 | 40.17 | 41.70 | 42.83 | 43.68 | 44.31 | 44.85 | 45.25 | 45.58 | 45.84 | 46.05 | 46.23 | 46.36 | 46.48 | 46.59 | 46.67 | 46.74 | 46.81 | 46.86 | 46.91 | 46.95 |
| | $K$ | 2.89 | 4.21 | 4.58 | 4.67 | 4.70 | 4.70 | 4.70 | 4.70 | 4.70 | 4.70 | 4.70 | 4.70 | 4.70 | 4.70 | 4.70 | 4.70 | 4.70 | 4.70 | 4.70 | 4.70 | 4.70 | 4.70 | 4.70 | 4.70 | 4.70 |
| 0.2 | $U_1$ | 17.13 | 30.14 | 39.40 | 45.90 | 50.84 | 54.52 | 57.39 | 59.68 | 61.55 | 63.10 | 64.40 | 65.50 | 66.46 | 67.28 | 68.03 | 68.65 | 69.22 | 69.73 | 70.19 | 70.61 | 70.99 | 71.34 | 71.65 | 71.94 | 72.21 |
| | $U_2$ | 0.80 | 1.85 | 3.24 | 4.85 | 6.51 | 8.08 | 9.57 | 10.90 | 12.09 | 13.14 | 14.08 | 14.92 | 15.66 | 16.33 | 16.93 | 17.47 | 17.96 | 18.41 | 18.82 | 19.19 | 19.53 | 19.85 | 20.14 | 20.41 | 20.65 |
| | $W_{1,2}$ | 14.64 | 31.29 | 49.04 | 67.12 | 85.27 | 103.45 | 121.62 | 139.80 | 157.98 | 176.16 | 194.34 | 212.52 | 230.70 | 248.88 | 267.06 | 285.24 | 303.41 | 321.59 | 339.77 | 357.95 | 376.13 | 394.31 | 412.49 | 430.67 | 448.85 |
| | $W_{3,4}$ | 16.62 | 34.15 | 52.16 | 70.30 | 88.47 | 106.64 | 124.82 | 143.00 | 161.18 | 179.36 | 197.54 | 215.72 | 233.89 | 252.07 | 270.25 | 288.43 | 306.61 | 324.79 | 342.97 | 361.15 | 379.33 | 397.51 | 415.68 | 433.86 | 452.04 |
| | $J$ | 14.45 | 24.65 | 31.20 | 35.49 | 38.46 | 40.63 | 42.17 | 43.31 | 44.17 | 44.82 | 45.36 | 45.77 | 46.10 | 46.36 | 46.58 | 46.73 | 46.89 | 47.03 | 47.15 | 47.24 | 47.32 | 47.39 | 47.45 | 47.50 | 47.54 |
| | $K$ | 3.03 | 4.41 | 4.80 | 4.89 | 4.92 | 4.92 | 4.92 | 4.92 | 4.92 | 4.92 | 4.92 | 4.92 | 4.92 | 4.92 | 4.92 | 4.92 | 4.92 | 4.92 | 4.92 | 4.92 | 4.92 | 4.92 | 4.92 | 4.92 | 4.92 |
| 0.1 | $U_1$ | 17.70 | 31.16 | 40.63 | 47.38 | 52.32 | 56.00 | 58.88 | 61.30 | 63.19 | 64.75 | 66.06 | 67.24 | 67.96 | 68.87 | 69.70 | 70.35 | 70.92 | 71.43 | 71.89 | 72.31 | 72.69 | 73.04 | 73.35 | 73.65 | 73.91 |
| | $U_2$ | 0.75 | 1.76 | 3.13 | 4.74 | 6.50 | 8.01 | 9.48 | 10.82 | 12.01 | 13.07 | 14.01 | 14.85 | 15.60 | 16.28 | 16.88 | 17.42 | 17.92 | 18.36 | 18.77 | 19.15 | 19.49 | 19.80 | 20.09 | 20.36 | 20.61 |
| | $W_{1,2}$ | 14.36 | 31.02 | 48.61 | 66.68 | 84.83 | 103.00 | 121.18 | 139.36 | 157.54 | 175.72 | 193.90 | 212.08 | 230.25 | 248.43 | 266.61 | 284.79 | 302.97 | 321.15 | 339.33 | 357.51 | 375.69 | 393.87 | 412.04 | 430.22 | 448.40 |
| | $W_{3,4}$ | 16.56 | 34.07 | 52.07 | 70.21 | 88.38 | 106.55 | 124.73 | 142.91 | 161.09 | 179.27 | 197.45 | 215.62 | 233.80 | 251.98 | 270.16 | 288.34 | 306.52 | 324.70 | 342.88 | 361.06 | 379.24 | 397.41 | 415.59 | 433.77 | 451.95 |
| | $J$ | 14.56 | 24.65 | 31.20 | 35.49 | 38.38 | 40.38 | 41.81 | 42.84 | 43.61 | 44.19 | 44.63 | 45.03 | 45.24 | 45.45 | 45.63 | 45.76 | 45.88 | 45.98 | 46.06 | 46.12 | 46.17 | 46.22 | 46.26 | 46.30 | 46.33 |
| | $K$ | 3.13 | 4.55 | 4.95 | 5.04 | 5.07 | 5.07 | 5.07 | 5.07 | 5.07 | 5.07 | 5.07 | 5.07 | 5.07 | 5.07 | 5.07 | 5.07 | 5.07 | 5.07 | 5.07 | 5.07 | 5.07 | 5.07 | 5.07 | 5.07 | 5.07 |
| 0.0 | $U_1$ | 17.07 | 30.03 | 39.32 | 45.91 | 50.77 | 54.46 | 57.33 | 59.63 | 61.50 | 63.05 | 64.35 | 65.45 | 66.41 | 67.24 | 67.96 | 68.61 | 69.18 | 69.69 | 70.15 | 70.56 | 70.94 | 71.29 | 71.60 | 71.89 | 72.16 |
| | $U_2$ | 0.79 | 1.83 | 3.23 | 4.84 | 6.50 | 8.09 | 9.56 | 10.89 | 12.08 | 13.13 | 14.07 | 14.91 | 15.65 | 16.32 | 16.92 | 17.47 | 17.96 | 18.40 | 18.81 | 19.18 | 19.53 | 19.84 | 20.13 | 20.40 | 20.65 |
| | $W_{1,2}$ | 14.42 | 30.97 | 48.70 | 66.77 | 84.92 | 103.10 | 121.27 | 139.45 | 157.63 | 175.81 | 193.99 | 212.15 | 230.35 | 248.51 | 266.70 | 284.88 | 303.06 | 321.24 | 339.42 | 357.60 | 375.78 | 393.96 | 412.14 | 430.32 | 448.50 |
| | $W_{3,4}$ | 16.61 | 34.15 | 52.15 | 70.29 | 88.46 | 106.63 | 124.81 | 142.99 | 161.17 | 179.35 | 197.53 | 215.71 | 233.88 | 252.06 | 270.24 | 288.42 | 306.60 | 324.78 | 342.96 | 361.14 | 379.32 | 397.50 | 415.67 | 433.85 | 452.03 |
| | $J$ | 14.65 | 25.02 | 31.79 | 36.09 | 39.37 | 41.52 | 43.08 | 44.22 | 45.07 | 45.72 | 46.22 | 46.61 | 46.92 | 47.17 | 47.38 | 47.54 | 47.68 | 47.79 | 47.88 | 47.96 | 48.03 | 48.09 | 48.14 | 48.18 | 48.22 |
| | $K$ | 3.36 | 4.60 | 5.00 | 5.10 | 5.12 | 5.13 | 5.13 | 5.13 | 5.13 | 5.13 | 5.13 | 5.13 | 5.13 | 5.13 | 5.13 | 5.13 | 5.13 | 5.13 | 5.13 | 5.13 | 5.13 | 5.13 | 5.13 | 5.13 | 5.13 |

TABLE S79. Parameters of the THF interaction Hamiltonian for different ratios $u_0/u_1$ and screening lengths $\xi$ of the double-gate interaction potential ($U_\xi = 24\,\text{meV}$ for $\xi = 10\,\text{nm}$). We use the same BM model parameters as in Table S78.

Total weight of the heavy-fermions on the active bands after wannierisation: 93.3%

**Energy (meV)** — y-axis ticks: 150, 100, 50, 0, −50, −100, −150; x-axis labels: $K$, $M$, $K$

Legend: **Wannier orbitals support** (colorbar 0.0 – 1.0)

FIG. S56. Band structures of the BM and THF models near charge neutrality for $w_0/w_1 = 0.8$, depicted by lines and crosses, respectively. The BM bands are colored according to the weight of the $f$-electron wave function on them. We use the same BM parameters as in Table S80.

TABLE S80. Parameters of the $f$-electron Wannier functions and the THF single-particle Hamiltonian for different values of the tunneling ratio $w_0/w_1$. $\mathcal{W}$ denotes the total weight of the THF $f$-electrons on the active bands. In computing the ratios $\gamma'/U_1$ and $\gamma''/U_2$, we employ the on-site and nearest-neighbor repulsion parameters $U_1$ and $U_2$ obtained numerically for $\xi = 10$ nm, as given in Table S81. We employ $v_F = 5.944$ eV·Å, $|\mathbf{K}| = 1.703$ Å$^{-1}$, $w_1 = 110$ meV, and $\theta = 1.46°$ for the BM model.

| $w_0/w_1$ | $\alpha_1$ | $\alpha_2$ | $\lambda_1$ | $\lambda_2$ | $\lambda$ | $\mathcal{W}$ | $t_0$ | $\gamma$ | $M$ | $v'_\star$ | $v''_\star$ | $v''_\star/U_2$ | $\gamma'/U_1$ | $\gamma''/U_2$ |
|---|---|---|---|---|---|---|---|---|---|---|---|---|---|---|
| 1.00 | 0.821 | 0.571 | 0.209 | 0.215 | 0.366 | 0.936 | 4.306 | −57.305 | −50.832 | −4.796 | 1.899 | 0.0327 | −0.77 | −10.20 |
| 0.90 | 0.849 | 0.529 | 0.221 | 0.221 | 0.381 | 0.930 | 5.028 | −73.607 | −52.477 | −4.903 | 1.892 | 0.0271 | −1.05 | −12.74 |
| 0.80 | 0.876 | 0.482 | 0.233 | 0.227 | 0.404 | 0.933 | 5.956 | −89.386 | −53.905 | −5.005 | 1.863 | 0.0225 | −1.36 | −14.87 |
| 0.70 | 0.903 | 0.431 | 0.251 | 0.227 | 0.427 | 0.937 | 6.776 | −104.450 | −55.132 | −5.103 | 1.804 | 0.0183 | −1.70 | −16.81 |
| 0.60 | 0.926 | 0.377 | 0.263 | 0.233 | 0.438 | 0.937 | 7.478 | −118.573 | −56.170 | −5.199 | 1.709 | 0.0146 | −2.02 | −18.69 |
| 0.50 | 0.948 | 0.320 | 0.281 | 0.239 | 0.459 | 0.943 | 8.184 | −131.488 | −57.030 | −5.292 | 1.568 | 0.0110 | −2.36 | −20.20 |
| 0.40 | 0.966 | 0.259 | 0.298 | 0.245 | 0.475 | 0.947 | 8.763 | −142.872 | −57.722 | −5.380 | 1.373 | 0.0077 | −2.69 | −21.51 |
| 0.30 | 0.980 | 0.199 | 0.310 | 0.245 | 0.480 | 0.947 | 9.132 | −152.352 | −58.252 | −5.459 | 1.115 | 0.0047 | −2.93 | −22.74 |
| 0.20 | 0.991 | 0.135 | 0.322 | 0.251 | 0.485 | 0.948 | 9.443 | −159.529 | −58.627 | −5.524 | 0.792 | 0.0022 | −3.13 | −23.61 |
| 0.10 | 0.998 | 0.069 | 0.328 | 0.251 | 0.479 | 0.944 | 9.472 | −164.024 | −58.850 | −5.567 | 0.412 | 0.0006 | −3.20 | −24.37 |
| 0.00 | 1.000 | 0.000 | 0.328 | 0.000 | 0.475 | 0.941 | 9.448 | −165.556 | −58.925 | −5.582 | 0.000 | 0.0000 | −3.21 | −24.66 |

| $w_0/w_1$ | $\xi$/nm | 2 | 4 | 6 | 8 | 10 | 12 | 14 | 16 | 18 | 20 | 22 | 24 | 26 | 28 | 30 | 32 | 34 | 36 | 38 | 40 | 42 | 44 | 46 | 48 | 50 |
|---|---|---|---|---|---|---|---|---|---|---|---|---|---|---|---|---|---|---|---|---|---|---|---|---|---|---|
| 1.0 | $U_1$ | 28.43 | 47.78 | 60.13 | 68.37 | 74.14 | 78.37 | 81.59 | 84.11 | 86.13 | 87.79 | 89.17 | 90.33 | 91.33 | 92.19 | 92.94 | 93.61 | 94.19 | 94.72 | 95.19 | 95.61 | 96.00 | 96.35 | 96.67 | 96.97 | 97.24 |
| | $U_2$ | 0.46 | 1.17 | 2.37 | 3.93 | 5.62 | 7.28 | 8.82 | 10.22 | 11.46 | 12.57 | 13.55 | 14.42 | 15.19 | 15.88 | 16.50 | 17.06 | 17.57 | 18.02 | 18.44 | 18.82 | 19.17 | 19.49 | 19.78 | 20.06 | 20.31 |
| | $W_{1,2}$ | 16.96 | 34.89 | 53.37 | 72.01 | 90.69 | 109.37 | 128.06 | 146.74 | 165.43 | 184.12 | 202.81 | 221.49 | 240.18 | 258.87 | 277.56 | 296.24 | 314.93 | 333.62 | 352.31 | 370.99 | 389.68 | 408.37 | 427.06 | 445.74 | 464.43 |
| | $W_{3,4}$ | 17.35 | 35.63 | 54.22 | 72.88 | 91.57 | 110.25 | 128.94 | 147.63 | 166.32 | 185.00 | 203.69 | 222.38 | 241.07 | 259.75 | 278.44 | 297.13 | 315.82 | 334.50 | 353.19 | 371.88 | 390.56 | 409.25 | 427.94 | 446.63 | 465.31 |
| | $J$ | 11.40 | 17.61 | 20.78 | 22.54 | 23.60 | 24.27 | 24.71 | 25.01 | 25.22 | 25.38 | 25.49 | 25.57 | 25.64 | 25.69 | 25.72 | 25.75 | 25.78 | 25.80 | 25.81 | 25.82 | 25.83 | 25.84 | 25.85 | 25.86 | 25.86 |
| | $K$ | 1.90 | 2.77 | 3.02 | 3.07 | 3.09 | 3.09 | 3.09 | 3.09 | 3.09 | 3.09 | 3.09 | 3.09 | 3.09 | 3.09 | 3.09 | 3.09 | 3.09 | 3.09 | 3.09 | 3.09 | 3.09 | 3.09 | 3.09 | 3.09 | 3.09 |
| 0.9 | $U_1$ | 26.35 | 44.65 | 56.53 | 64.53 | 70.18 | 74.34 | 77.52 | 80.01 | 82.01 | 83.66 | 85.03 | 86.18 | 87.17 | 88.03 | 88.78 | 89.44 | 90.03 | 90.55 | 91.02 | 91.44 | 91.83 | 92.18 | 92.50 | 92.79 | 93.07 |
| | $U_2$ | 0.50 | 1.27 | 2.50 | 4.08 | 5.78 | 7.43 | 8.97 | 10.36 | 11.60 | 12.71 | 13.68 | 14.55 | 15.32 | 16.01 | 16.63 | 17.18 | 17.69 | 18.14 | 18.56 | 18.94 | 19.29 | 19.61 | 19.90 | 20.17 | 20.43 |
| | $W_{1,2}$ | 16.76 | 34.61 | 53.07 | 71.70 | 90.38 | 109.06 | 127.75 | 146.44 | 165.13 | 183.81 | 202.50 | 221.19 | 239.88 | 258.56 | 277.25 | 295.94 | 314.63 | 333.31 | 352.00 | 370.69 | 389.38 | 408.06 | 426.75 | 445.44 | 464.12 |
| | $W_{3,4}$ | 17.16 | 35.32 | 53.88 | 72.54 | 91.22 | 109.91 | 128.59 | 147.28 | 165.97 | 184.65 | 203.34 | 222.03 | 240.72 | 259.40 | 278.09 | 296.78 | 315.47 | 334.15 | 352.84 | 371.53 | 390.22 | 408.90 | 427.59 | 446.28 | 464.97 |
| | $J$ | 11.83 | 18.66 | 22.35 | 24.49 | 25.81 | 26.66 | 27.24 | 27.64 | 27.93 | 28.13 | 28.29 | 28.40 | 28.49 | 28.56 | 28.61 | 28.65 | 28.69 | 28.71 | 28.74 | 28.77 | 28.78 | 28.79 | 28.80 | 28.81 | 28.81 |
| | $K$ | 1.99 | 2.90 | 3.16 | 3.22 | 3.23 | 3.23 | 3.23 | 3.23 | 3.23 | 3.23 | 3.23 | 3.23 | 3.23 | 3.23 | 3.23 | 3.23 | 3.23 | 3.23 | 3.23 | 3.23 | 3.23 | 3.23 | 3.23 | 3.23 | 3.23 |
| 0.8 | $U_1$ | 24.06 | 41.11 | 52.38 | 60.06 | 65.53 | 69.59 | 72.70 | 75.15 | 77.12 | 78.75 | 80.10 | 81.25 | 82.23 | 83.08 | 83.83 | 84.49 | 85.07 | 85.59 | 86.06 | 86.48 | 86.86 | 87.21 | 87.53 | 87.82 | 88.10 |
| | $U_2$ | 0.58 | 1.42 | 2.71 | 4.31 | 6.01 | 7.66 | 9.19 | 10.57 | 11.80 | 12.90 | 13.87 | 14.73 | 15.50 | 16.18 | 16.80 | 17.35 | 17.85 | 18.30 | 18.72 | 19.10 | 19.45 | 19.77 | 20.06 | 20.33 | 20.58 |
| | $W_{1,2}$ | 16.55 | 34.31 | 52.75 | 71.38 | 90.05 | 108.74 | 127.42 | 146.11 | 164.80 | 183.49 | 202.17 | 220.86 | 239.55 | 258.24 | 276.92 | 295.61 | 314.30 | 332.99 | 351.67 | 370.36 | 389.05 | 407.74 | 426.42 | 445.11 | 463.80 |
| | $W_{3,4}$ | 16.95 | 35.16 | 53.69 | 72.35 | 91.03 | 109.71 | 128.40 | 147.09 | 165.78 | 184.46 | 203.15 | 221.84 | 240.52 | 259.21 | 277.90 | 296.59 | 315.27 | 333.96 | 352.65 | 371.34 | 390.02 | 408.71 | 427.40 | 446.09 | 464.77 |
| | $J$ | 12.33 | 19.86 | 24.16 | 26.76 | 28.42 | 29.53 | 30.29 | 30.83 | 31.21 | 31.50 | 31.71 | 31.87 | 32.00 | 32.09 | 32.17 | 32.23 | 32.28 | 32.32 | 32.35 | 32.38 | 32.40 | 32.42 | 32.44 | 32.45 | 32.46 |
| | $K$ | 2.11 | 3.07 | 3.33 | 3.39 | 3.41 | 3.41 | 3.41 | 3.41 | 3.41 | 3.41 | 3.41 | 3.41 | 3.41 | 3.41 | 3.41 | 3.41 | 3.41 | 3.41 | 3.41 | 3.41 | 3.41 | 3.41 | 3.41 | 3.41 | 3.41 |
| 0.7 | $U_1$ | 22.09 | 38.02 | 48.72 | 56.11 | 61.41 | 65.37 | 68.42 | 70.82 | 72.77 | 74.37 | 75.71 | 76.85 | 77.82 | 78.67 | 79.41 | 80.06 | 80.64 | 81.16 | 81.62 | 82.04 | 82.43 | 82.77 | 83.09 | 83.39 | 83.66 |
| | $U_2$ | 0.65 | 1.56 | 2.89 | 4.51 | 6.21 | 7.86 | 9.38 | 10.75 | 11.97 | 13.06 | 14.02 | 14.88 | 15.64 | 16.33 | 16.94 | 17.49 | 17.99 | 18.44 | 18.85 | 19.23 | 19.58 | 19.89 | 20.19 | 20.46 | 20.71 |
| | $W_{1,2}$ | 16.33 | 33.99 | 52.41 | 71.03 | 89.70 | 108.39 | 127.08 | 145.76 | 164.44 | 183.13 | 201.82 | 220.51 | 239.20 | 257.89 | 276.57 | 295.26 | 313.95 | 332.64 | 351.32 | 370.01 | 388.70 | 407.38 | 426.07 | 444.76 | 463.45 |
| | $W_{3,4}$ | 17.02 | 35.08 | 53.61 | 72.26 | 90.93 | 109.62 | 128.31 | 146.99 | 165.68 | 184.37 | 203.06 | 221.74 | 240.43 | 259.12 | 277.81 | 296.49 | 315.18 | 333.87 | 352.56 | 371.24 | 389.93 | 408.62 | 427.31 | 445.99 | 464.68 |
| | $J$ | 12.81 | 21.26 | 26.23 | 29.22 | 31.08 | 32.32 | 33.18 | 33.80 | 34.25 | 34.59 | 34.85 | 35.04 | 35.32 | 35.55 | 35.75 | 35.90 | 36.05 | 36.18 | 36.27 | 36.33 | 36.36 | 36.38 | 36.40 | 36.62 | 36.40 |
| | $K$ | 2.24 | 3.26 | 3.54 | 3.61 | 3.62 | 3.62 | 3.62 | 3.62 | 3.62 | 3.62 | 3.62 | 3.62 | 3.62 | 3.62 | 3.62 | 3.62 | 3.62 | 3.62 | 3.62 | 3.62 | 3.62 | 3.62 | 3.62 | 3.62 | 3.62 |

TABLE S81: Parameters of the THF interaction Hamiltonian for different ratios $u_0/u_1$ and screening lengths $\xi$ of the double-gate interaction potential ($U_\xi = 24$ meV for $\xi = 10$ nm). We use the same BM model parameters as in Table S80.

| $u_0/u_1$ | | 2 | 4 | 6 | 8 | 10 | 12 | 14 | 16 | 18 | 20 | 22 | 24 | 26 | 28 | 30 | 32 | 34 | 36 | 38 | 40 | 42 | 44 | 46 | 48 | 50 |
|---|---|---|---|---|---|---|---|---|---|---|---|---|---|---|---|---|---|---|---|---|---|---|---|---|---|---|
| 0.6 | $U_1$ | 20.80 | 36.03 | 46.38 | 53.59 | 58.79 | 62.69 | 65.70 | 68.00 | 70.02 | 71.61 | 72.94 | 74.07 | 75.04 | 75.89 | 76.62 | 77.27 | 77.85 | 78.37 | 78.83 | 79.25 | 79.63 | 79.98 | 80.30 | 80.59 | 80.86 |
| | $U_2$ | 0.69 | 1.65 | 3.01 | 4.64 | 6.34 | 7.99 | 9.50 | 10.87 | 12.09 | 13.17 | 14.13 | 14.98 | 15.74 | 16.43 | 17.04 | 17.59 | 18.08 | 18.54 | 18.95 | 19.32 | 19.67 | 19.99 | 20.28 | 20.55 | 20.80 |
| | $W_{1,2}$ | 16.06 | 35.03 | 52.00 | 70.61 | 89.28 | 107.97 | 126.65 | 145.34 | 164.03 | 182.72 | 201.40 | 220.09 | 238.78 | 257.46 | 276.15 | 294.84 | 313.53 | 332.21 | 350.90 | 369.59 | 388.28 | 406.96 | 425.65 | 444.34 | 463.03 |
| | $W_{3,4}$ | 17.00 | 33.63 | 53.54 | 72.19 | 90.87 | 109.56 | 128.24 | 146.93 | 165.62 | 184.31 | 202.90 | 221.68 | 240.37 | 259.06 | 277.74 | 296.43 | 315.12 | 333.81 | 352.49 | 371.18 | 389.87 | 408.55 | 427.24 | 445.93 | 464.62 |
| | $J$ | 13.34 | 22.10 | 27.46 | 30.87 | 33.12 | 34.67 | 35.76 | 36.55 | 37.13 | 37.56 | 37.89 | 38.15 | 38.35 | 38.50 | 38.63 | 38.73 | 38.81 | 38.88 | 38.94 | 38.98 | 39.02 | 39.06 | 39.09 | 39.11 | 39.13 |
| | $K$ | 3.87 | 3.78 | 3.85 | 3.86 | 3.87 | 3.87 | 3.87 | 3.87 | 3.87 | 3.87 | 3.87 | 3.87 | 3.87 | 3.87 | 3.87 | 3.87 | 3.87 | 3.87 | 3.87 | 3.87 | 3.87 | 3.87 | 3.87 | 3.87 | 3.87 |
| 0.5 | $U_1$ | 19.36 | 33.72 | 43.63 | 50.58 | 55.65 | 59.46 | 62.41 | 64.76 | 66.66 | 68.24 | 69.56 | 70.68 | 71.64 | 72.48 | 73.21 | 73.86 | 74.43 | 74.94 | 75.41 | 75.83 | 76.21 | 76.55 | 76.87 | 77.16 | 77.43 |
| | $U_2$ | 0.75 | 1.77 | 3.16 | 4.81 | 6.51 | 8.14 | 9.65 | 11.01 | 12.22 | 13.29 | 14.25 | 15.10 | 15.85 | 16.53 | 17.14 | 17.69 | 18.18 | 18.63 | 19.05 | 19.42 | 19.76 | 20.08 | 20.37 | 20.64 | 20.89 |
| | $W_{1,2}$ | 15.83 | 33.29 | 51.64 | 70.25 | 88.92 | 107.61 | 126.29 | 144.98 | 163.67 | 182.36 | 201.04 | 219.73 | 238.42 | 257.11 | 275.79 | 294.48 | 313.17 | 331.85 | 350.54 | 369.23 | 387.92 | 406.60 | 425.29 | 443.98 | 462.67 |
| | $W_{3,4}$ | 17.02 | 35.05 | 53.57 | 72.22 | 90.90 | 109.58 | 128.27 | 146.96 | 165.64 | 184.33 | 202.92 | 221.70 | 240.39 | 259.08 | 277.76 | 296.45 | 315.14 | 333.83 | 352.51 | 371.20 | 389.89 | 408.58 | 427.27 | 445.95 | 464.64 |
| | $J$ | 13.84 | 23.20 | 29.09 | 32.50 | 35.50 | 37.29 | 38.58 | 39.52 | 40.22 | 40.75 | 41.16 | 41.48 | 41.74 | 41.94 | 42.10 | 42.24 | 42.34 | 42.43 | 42.51 | 42.57 | 42.63 | 42.67 | 42.71 | 42.74 | 42.77 |
| | $K$ | 2.56 | 3.72 | 4.03 | 4.11 | 4.12 | 4.13 | 4.13 | 4.13 | 4.13 | 4.13 | 4.13 | 4.13 | 4.13 | 4.13 | 4.13 | 4.13 | 4.13 | 4.13 | 4.13 | 4.13 | 4.13 | 4.13 | 4.13 | 4.13 | 4.13 |
| 0.4 | $U_1$ | 18.24 | 31.93 | 41.47 | 48.23 | 53.18 | 56.92 | 59.82 | 62.14 | 64.02 | 65.58 | 66.89 | 68.00 | 68.96 | 69.79 | 70.52 | 71.16 | 71.74 | 72.25 | 72.71 | 73.13 | 73.50 | 73.85 | 74.17 | 74.46 | 74.73 |
| | $U_2$ | 0.80 | 1.86 | 3.29 | 4.94 | 6.64 | 8.27 | 9.77 | 11.12 | 12.32 | 13.39 | 14.34 | 15.19 | 15.94 | 16.61 | 17.22 | 17.77 | 18.26 | 18.71 | 19.12 | 19.49 | 19.83 | 20.15 | 20.44 | 20.71 | 20.96 |
| | $W_{1,2}$ | 15.60 | 32.97 | 51.30 | 69.91 | 88.58 | 107.26 | 125.94 | 144.63 | 163.32 | 182.00 | 200.69 | 219.38 | 238.07 | 256.76 | 275.44 | 294.13 | 312.82 | 331.51 | 350.19 | 368.88 | 387.57 | 406.26 | 424.94 | 443.63 | 462.32 |
| | $W_{3,4}$ | 17.08 | 35.11 | 53.61 | 72.26 | 90.94 | 109.63 | 128.31 | 147.00 | 165.68 | 184.37 | 202.96 | 221.73 | 240.42 | 259.11 | 277.79 | 296.48 | 315.17 | 333.85 | 352.54 | 371.22 | 389.91 | 408.62 | 427.31 | 445.96 | 464.66 |
| | $J$ | 14.28 | 24.14 | 30.47 | 34.45 | 37.50 | 39.52 | 40.83 | 41.95 | 42.83 | 43.49 | 44.00 | 44.30 | 44.60 | 44.84 | 45.03 | 45.18 | 45.30 | 45.45 | 45.53 | 45.61 | 45.68 | 45.73 | 45.78 | 45.82 | 45.86 |
| | $K$ | 2.74 | 3.96 | 4.30 | 4.38 | 4.40 | 4.40 | 4.40 | 4.40 | 4.40 | 4.40 | 4.40 | 4.40 | 4.40 | 4.40 | 4.40 | 4.40 | 4.40 | 4.40 | 4.40 | 4.40 | 4.40 | 4.40 | 4.40 | 4.40 | 4.40 |
| 0.3 | $U_1$ | 17.69 | 31.06 | 40.45 | 47.12 | 52.02 | 55.74 | 58.63 | 60.93 | 62.81 | 64.34 | 65.66 | 66.77 | 67.73 | 68.56 | 69.28 | 69.93 | 70.50 | 71.01 | 71.47 | 71.89 | 72.27 | 72.61 | 72.93 | 73.22 | 73.49 |
| | $U_2$ | 0.84 | 1.90 | 3.34 | 5.00 | 6.70 | 8.33 | 9.82 | 11.17 | 12.37 | 13.44 | 14.40 | 15.27 | 16.02 | 16.70 | 17.30 | 17.84 | 18.34 | 18.78 | 19.19 | 19.57 | 19.91 | 20.22 | 20.51 | 20.78 | 21.03 |
| | $W_{1,2}$ | 15.35 | 32.62 | 50.92 | 69.52 | 88.18 | 106.87 | 125.55 | 144.24 | 162.93 | 181.62 | 200.30 | 219.00 | 237.68 | 256.36 | 275.05 | 293.74 | 312.13 | 330.81 | 349.50 | 368.19 | 386.88 | 405.56 | 424.25 | 442.94 | 461.63 |
| | $W_{3,4}$ | 17.08 | 35.11 | 53.63 | 72.28 | 90.95 | 109.64 | 128.33 | 147.01 | 165.70 | 184.38 | 203.10 | 221.79 | 240.48 | 259.17 | 277.85 | 296.54 | 315.23 | 333.89 | 352.58 | 371.25 | 389.98 | 408.67 | 427.35 | 445.89 | 464.73 |
| | $J$ | 14.33 | 24.77 | 31.37 | 35.71 | 38.73 | 40.83 | 42.36 | 43.54 | 44.51 | 45.17 | 45.87 | 46.48 | 46.87 | 47.23 | 47.48 | 47.60 | 47.88 | 48.00 | 48.19 | 48.36 | 48.49 | 48.64 | 48.76 | 48.88 | 48.97 |
| | $K$ | 2.00 | 3.96 | 4.30 | 4.38 | 4.46 | 4.66 | 4.67 | 4.67 | 4.67 | 4.67 | 4.67 | 4.67 | 4.67 | 4.67 | 4.67 | 4.67 | 4.67 | 4.67 | 4.67 | 4.67 | 4.67 | 4.67 | 4.67 | 4.67 | 4.67 |
| 0.2 | $U_1$ | 17.23 | 30.32 | 39.56 | 46.16 | 51.01 | 54.69 | 57.57 | 59.88 | 61.75 | 63.28 | 64.58 | 65.68 | 66.63 | 67.46 | 68.19 | 68.85 | 69.40 | 69.91 | 70.37 | 70.79 | 71.17 | 71.51 | 71.83 | 72.50 | 72.77 |
| | $U_2$ | 0.82 | 1.94 | 3.39 | 5.06 | 6.76 | 8.38 | 9.87 | 11.22 | 12.42 | 13.48 | 14.43 | 15.27 | 16.02 | 16.70 | 17.30 | 17.84 | 18.33 | 18.78 | 19.19 | 19.57 | 19.91 | 20.22 | 20.51 | 20.78 | 21.03 |
| | $W_{1,2}$ | 15.16 | 32.34 | 50.60 | 69.22 | 87.88 | 106.57 | 125.25 | 143.94 | 162.63 | 181.31 | 200.00 | 218.69 | 237.38 | 256.06 | 274.75 | 293.44 | 312.13 | 330.81 | 349.50 | 368.19 | 386.88 | 405.56 | 424.25 | 442.94 | 461.63 |
| | $W_{3,4}$ | 16.93 | 34.98 | 53.59 | 72.28 | 90.95 | 109.64 | 128.33 | 147.01 | 165.70 | 184.39 | 203.07 | 221.76 | 240.45 | 259.13 | 277.82 | 296.51 | 315.19 | 333.88 | 352.57 | 371.25 | 389.94 | 408.63 | 427.31 | 446.00 | 464.74 |
| | $J$ | 14.84 | 25.29 | 32.11 | 36.64 | 39.75 | 41.95 | 43.54 | 44.72 | 45.60 | 46.28 | 46.81 | 47.23 | 47.47 | 47.56 | 47.86 | 48.06 | 48.23 | 48.39 | 48.48 | 48.58 | 48.66 | 48.73 | 48.79 | 48.83 | 48.93 |
| | $K$ | 2.04 | 3.41 | 4.78 | 4.87 | 4.89 | 4.89 | 4.89 | 4.89 | 4.89 | 4.89 | 4.89 | 4.89 | 4.89 | 4.89 | 4.89 | 4.89 | 4.89 | 4.89 | 4.89 | 4.89 | 4.89 | 4.89 | 4.89 | 4.89 | 4.89 |
| 0.1 | $U_1$ | 17.32 | 30.04 | 39.08 | 45.46 | 50.26 | 53.83 | 56.63 | 58.86 | 60.66 | 62.09 | 63.64 | 64.95 | 66.02 | 67.06 | 67.84 | 68.42 | 69.19 | 69.78 | 70.20 | 70.75 | 71.11 | 71.42 | 71.89 | 72.46 | 72.75 |
| | $U_2$ | 0.82 | 1.91 | 3.36 | 5.03 | 6.73 | 8.36 | 9.86 | 11.20 | 12.40 | 13.47 | 14.42 | 15.26 | 16.01 | 16.69 | 17.29 | 17.84 | 18.33 | 18.73 | 19.19 | 19.56 | 19.91 | 20.22 | 20.51 | 20.78 | 21.03 |
| | $W_{1,2}$ | 14.97 | 32.07 | 50.33 | 68.91 | 87.58 | 106.26 | 124.95 | 143.63 | 162.32 | 181.00 | 199.69 | 218.38 | 237.00 | 255.76 | 274.45 | 293.13 | 311.82 | 330.51 | 349.20 | 367.88 | 386.57 | 405.26 | 423.95 | 442.64 | 461.32 |
| | $W_{3,4}$ | 16.93 | 34.97 | 53.63 | 72.28 | 90.95 | 109.64 | 128.33 | 147.01 | 165.70 | 184.39 | 203.08 | 221.77 | 240.45 | 259.14 | 277.83 | 296.52 | 315.20 | 333.89 | 352.58 | 371.27 | 389.95 | 408.64 | 427.33 | 446.02 | 464.69 |
| | $J$ | 14.96 | 25.55 | 32.35 | 36.89 | 39.98 | 42.16 | 43.74 | 44.84 | 45.72 | 46.39 | 46.92 | 47.45 | 47.55 | 47.80 | 47.86 | 48.06 | 48.21 | 48.39 | 48.49 | 48.65 | 48.73 | 48.79 | 48.84 | 48.88 | 48.92 |
| | $K$ | 3.14 | 4.55 | 4.93 | 5.02 | 5.04 | 5.05 | 5.05 | 5.05 | 5.05 | 5.05 | 5.05 | 5.05 | 5.05 | 5.05 | 5.05 | 5.05 | 5.05 | 5.05 | 5.05 | 5.05 | 5.05 | 5.05 | 5.05 | 5.05 | 5.05 |
| 0.0 | $U_1$ | 17.40 | 30.64 | 39.98 | 46.65 | 51.54 | 55.26 | 58.15 | 60.45 | 62.33 | 63.89 | 65.19 | 66.30 | 67.26 | 68.09 | 68.82 | 69.46 | 70.03 | 70.54 | 71.00 | 71.42 | 71.80 | 72.15 | 72.46 | 72.75 | 73.02 |
| | $U_2$ | 0.81 | 1.90 | 3.34 | 5.01 | 6.71 | 8.34 | 9.84 | 11.19 | 12.39 | 13.46 | 14.41 | 15.25 | 16.01 | 16.68 | 17.29 | 17.83 | 18.33 | 18.77 | 19.18 | 19.56 | 19.90 | 20.22 | 20.51 | 20.78 | 21.03 |
| | $W_{1,2}$ | 14.89 | 31.96 | 50.21 | 68.80 | 87.46 | 106.14 | 124.83 | 143.51 | 162.20 | 180.89 | 199.57 | 218.26 | 236.95 | 255.64 | 274.33 | 293.02 | 311.70 | 330.39 | 349.08 | 367.77 | 386.45 | 405.14 | 423.83 | 442.52 | 461.20 |
| | $W_{3,4}$ | 16.93 | 34.97 | 53.62 | 72.26 | 90.94 | 109.63 | 128.31 | 147.00 | 165.69 | 184.38 | 203.06 | 221.75 | 240.44 | 259.13 | 277.81 | 296.50 | 315.19 | 333.88 | 352.56 | 371.25 | 389.94 | 408.63 | 427.32 | 446.00 | 464.69 |
| | $J$ | 15.00 | 25.64 | 32.43 | 36.95 | 40.03 | 42.18 | 43.75 | 44.84 | 45.72 | 46.39 | 46.92 | 47.45 | 47.80 | 47.86 | 48.16 | 48.29 | 48.40 | 48.49 | 48.58 | 48.65 | 48.73 | 48.78 | 48.84 | 48.78 | 48.82 |
| | $K$ | 3.18 | 4.60 | 4.98 | 5.07 | 5.09 | 5.10 | 5.10 | 5.10 | 5.10 | 5.10 | 5.10 | 5.10 | 5.10 | 5.10 | 5.10 | 5.10 | 5.10 | 5.10 | 5.10 | 5.10 | 5.10 | 5.10 | 5.10 | 5.10 | 5.10 |

FIG. SS7. Band structures of the BM and THF models near charge neutrality for $w_0/w_1 = 0.8$, depicted by lines and crosses, respectively. The BM bands are colored according to the weight of the $f$-electron wave function on them. We use the same BM parameters as in Table SS2.

TABLE SS2. Parameters of the $f$-electron Wannier functions and the THF single-particle Hamiltonian for different values of the tunneling ratio $w_0/w_1$. $\mathcal{W}$ denotes the total weight of the THF $f$-electrons on the active bands. In computing the ratios $\gamma/U_1$ and $\gamma/U_2$, we employ the on-site and nearest-neighbor repulsion parameters $U_1$ and $U_2$ obtained numerically for $\xi = 10$ nm, as given in Table SS3. We employ $v_F = 5.944$ eV·Å, $|\mathbf{K}| = 1.703$ Å$^{-1}$, $w_1 = 110$ meV, and $\theta = 1.48°$ for the BM model.

| $w_0/w_1$ | $\alpha_1$ | $\alpha_2$ | $\lambda_1$ | $\lambda_2$ | $\lambda$ | $\mathcal{W}$ | $t_0$ | $\gamma$ | $M$ | $v_\star$ | $v_\star'$ | $v_\star''$ | $\gamma/U_1$ | $\gamma/U_2$ |
|---|---|---|---|---|---|---|---|---|---|---|---|---|---|---|
| 1.00 | 0.824 | 0.567 | 0.209 | 0.215 | 0.371 | 0.935 | 4.658 | -60.605 | -53.931 | -4.816 | 1.909 | 0.0351 | -0.81 | -10.35 |
| 0.90 | 0.852 | 0.524 | 0.221 | 0.221 | 0.389 | 0.933 | 5.499 | -76.930 | -55.588 | -4.921 | 1.901 | 0.0295 | -1.10 | -12.72 |
| 0.80 | 0.879 | 0.477 | 0.239 | 0.227 | 0.410 | 0.935 | 6.448 | -92.720 | -57.027 | -5.021 | 1.869 | 0.0247 | -1.41 | -14.81 |
| 0.70 | 0.904 | 0.427 | 0.251 | 0.233 | 0.428 | 0.937 | 7.230 | -107.786 | -58.264 | -5.117 | 1.809 | 0.0204 | -1.74 | -16.75 |
| 0.60 | 0.927 | 0.374 | 0.269 | 0.233 | 0.439 | 0.936 | 7.954 | -121.905 | -59.311 | -5.212 | 1.712 | 0.0163 | -2.05 | -18.57 |
| 0.50 | 0.949 | 0.316 | 0.286 | 0.239 | 0.460 | 0.943 | 8.709 | -134.810 | -60.178 | -5.304 | 1.571 | 0.0124 | -2.40 | -20.00 |
| 0.40 | 0.966 | 0.257 | 0.298 | 0.245 | 0.476 | 0.947 | 9.284 | -146.181 | -60.876 | -5.392 | 1.375 | 0.0087 | -2.72 | -21.30 |
| 0.30 | 0.980 | 0.197 | 0.310 | 0.245 | 0.480 | 0.947 | 9.657 | -155.649 | -61.412 | -5.471 | 1.116 | 0.0053 | -2.95 | -22.51 |
| 0.20 | 0.991 | 0.134 | 0.322 | 0.251 | 0.485 | 0.948 | 9.979 | -162.814 | -61.790 | -5.535 | 0.792 | 0.0025 | -3.14 | -23.35 |
| 0.10 | 0.998 | 0.069 | 0.328 | 0.251 | 0.479 | 0.943 | 10.011 | -167.300 | -62.016 | -5.578 | 0.413 | 0.0007 | -3.21 | -24.08 |
| 0.00 | 1.000 | 0.000 | 0.328 | 0.000 | 0.482 | 0.944 | 10.040 | -168.829 | -62.090 | -5.594 | -0.0000 | -0.0000 | -3.26 | -24.25 |

| $w_0/w_1$ | $\xi/\text{nm}$ | 2 | 4 | 6 | 8 | 10 | 12 | 14 | 16 | 18 | 20 | 22 | 24 | 26 | 28 | 30 | 32 | 34 | 36 | 38 | 40 | 42 | 44 | 46 | 48 | 50 |
|---|---|---|---|---|---|---|---|---|---|---|---|---|---|---|---|---|---|---|---|---|---|---|---|---|---|---|
| 1.0 | $U_1$ | 28.56 | 47.07 | 60.36 | 68.61 | 74.39 | 78.63 | 81.85 | 84.37 | 86.39 | 88.05 | 89.43 | 90.59 | 91.59 | 92.45 | 93.20 | 93.87 | 94.45 | 94.98 | 95.45 | 95.87 | 96.26 | 96.61 | 96.93 | 97.23 | 97.50 |
| | $U_2$ | 0.48 | 1.24 | 2.49 | 4.11 | 5.85 | 7.505 | 9.12 | 10.54 | 11.80 | 12.92 | 13.91 | 14.78 | 15.56 | 16.26 | 16.88 | 17.44 | 17.95 | 18.41 | 18.83 | 19.21 | 19.56 | 19.88 | 20.17 | 20.45 | 20.70 |
| | $W_{1,2}$ | 17.50 | 35.97 | 54.97 | 74.13 | 93.02 | 112.53 | 131.73 | 150.93 | 170.16 | 189.34 | 208.54 | 227.74 | 246.94 | 266.15 | 285.35 | 304.55 | 323.76 | 342.96 | 362.16 | 381.37 | 400.57 | 419.77 | 438.97 | 458.18 | 477.38 |
| | $W_{3,4}$ | 1.88 | 36.52 | 55.61 | 74.79 | 93.99 | 113.19 | 132.40 | 151.60 | 170.80 | 190.00 | 209.21 | 228.41 | 247.61 | 266.82 | 286.02 | 305.22 | 324.42 | 343.63 | 362.83 | 382.03 | 401.24 | 420.44 | 439.64 | 458.84 | 478.05 |
| | $J$ | 11.67 | 18.05 | 21.32 | 23.14 | 24.24 | 24.94 | 25.40 | 25.72 | 25.94 | 26.11 | 26.22 | 26.31 | 26.38 | 26.43 | 26.47 | 26.50 | 26.53 | 26.55 | 26.57 | 26.58 | 26.59 | 26.60 | 26.61 | 26.62 | 26.62 |
| | $K$ | 1.88 | 2.73 | 2.96 | 3.01 | 3.02 | 3.03 | 3.03 | 3.03 | 3.03 | 3.03 | 3.03 | 3.03 | 3.03 | 3.03 | 3.03 | 3.03 | 3.03 | 3.03 | 3.03 | 3.03 | 3.03 | 3.03 | 3.03 | 3.03 | 3.03 |
| 0.9 | $U_1$ | 26.29 | 44.53 | 56.37 | 64.34 | 69.98 | 74.13 | 77.30 | 79.79 | 81.79 | 83.43 | 84.79 | 85.95 | 86.94 | 87.80 | 88.55 | 89.21 | 89.79 | 90.31 | 90.78 | 91.21 | 91.59 | 91.94 | 92.26 | 92.56 | 92.83 |
| | $U_2$ | 0.54 | 1.37 | 2.66 | 4.30 | 6.05 | 7.74 | 9.31 | 10.72 | 11.97 | 13.08 | 14.07 | 14.94 | 15.72 | 16.41 | 17.03 | 17.59 | 18.09 | 18.55 | 18.97 | 19.35 | 19.70 | 20.02 | 20.31 | 20.58 | 20.84 |
| | $W_{1,2}$ | 17.30 | 35.69 | 54.67 | 73.83 | 93.02 | 112.22 | 131.42 | 150.62 | 169.83 | 189.03 | 208.23 | 227.44 | 246.64 | 265.84 | 285.04 | 304.25 | 323.45 | 342.65 | 361.86 | 381.06 | 400.26 | 419.47 | 438.67 | 457.87 | 477.07 |
| | $W_{3,4}$ | 17.60 | 36.25 | 55.32 | 74.49 | 93.69 | 112.89 | 132.09 | 151.29 | 170.50 | 189.70 | 208.90 | 228.10 | 247.31 | 266.51 | 285.71 | 304.92 | 324.12 | 343.32 | 362.53 | 381.73 | 400.93 | 420.13 | 439.34 | 458.54 | 477.74 |
| | $J$ | 12.14 | 19.18 | 23.01 | 25.25 | 26.65 | 27.56 | 28.18 | 28.61 | 28.91 | 29.14 | 29.31 | 29.43 | 29.53 | 29.60 | 29.66 | 29.70 | 29.74 | 29.77 | 29.80 | 29.82 | 29.83 | 29.85 | 29.86 | 29.87 | 29.88 |
| | $K$ | 1.98 | 2.87 | 3.11 | 3.16 | 3.17 | 3.18 | 3.18 | 3.18 | 3.18 | 3.18 | 3.18 | 3.18 | 3.18 | 3.18 | 3.18 | 3.18 | 3.18 | 3.18 | 3.18 | 3.18 | 3.18 | 3.18 | 3.18 | 3.18 | 3.18 |
| 0.8 | $U_1$ | 24.13 | 41.20 | 52.47 | 60.15 | 65.63 | 69.68 | 72.79 | 75.24 | 77.22 | 78.84 | 80.19 | 81.34 | 82.32 | 83.17 | 83.92 | 84.57 | 85.16 | 85.68 | 86.14 | 86.57 | 86.95 | 87.30 | 87.62 | 87.91 | 88.18 |
| | $U_2$ | 0.61 | 1.51 | 2.86 | 4.51 | 6.26 | 7.95 | 9.51 | 10.92 | 12.15 | 13.26 | 14.24 | 15.10 | 15.88 | 16.57 | 17.18 | 17.74 | 18.24 | 18.70 | 19.11 | 19.49 | 19.84 | 20.16 | 20.45 | 20.73 | 20.98 |
| | $W_{1,2}$ | 17.08 | 35.37 | 54.34 | 73.49 | 92.68 | 111.88 | 131.08 | 150.28 | 169.48 | 188.69 | 207.89 | 227.09 | 246.30 | 265.50 | 284.70 | 303.90 | 323.11 | 342.31 | 361.51 | 380.72 | 399.92 | 419.12 | 438.32 | 457.53 | 476.73 |
| | $W_{3,4}$ | 17.17 | 36.10 | 55.15 | 74.32 | 93.51 | 112.71 | 131.92 | 151.12 | 170.32 | 189.52 | 208.73 | 227.93 | 247.13 | 266.34 | 285.54 | 304.74 | 323.94 | 343.15 | 362.35 | 381.55 | 400.76 | 419.96 | 439.16 | 458.36 | 477.57 |
| | $J$ | 12.65 | 20.39 | 24.82 | 27.50 | 29.22 | 30.37 | 31.16 | 31.72 | 32.12 | 32.42 | 32.64 | 32.81 | 32.94 | 33.04 | 33.12 | 33.18 | 33.23 | 33.28 | 33.31 | 33.34 | 33.36 | 33.38 | 33.40 | 33.41 | 33.43 |
| | $K$ | 2.10 | 3.04 | 3.29 | 3.35 | 3.36 | 3.36 | 3.36 | 3.36 | 3.36 | 3.36 | 3.36 | 3.36 | 3.36 | 3.36 | 3.36 | 3.36 | 3.36 | 3.36 | 3.36 | 3.36 | 3.36 | 3.36 | 3.36 | 3.36 | 3.36 |
| 0.7 | $U_1$ | 22.37 | 38.46 | 49.25 | 56.69 | 62.02 | 66.00 | 69.05 | 71.47 | 73.42 | 75.03 | 76.37 | 77.51 | 78.48 | 79.33 | 80.07 | 80.72 | 81.30 | 81.82 | 82.29 | 82.71 | 83.09 | 83.44 | 83.76 | 84.05 | 84.32 |
| | $U_2$ | 0.67 | 1.63 | 3.01 | 4.69 | 6.44 | 8.12 | 9.67 | 11.06 | 12.30 | 13.40 | 14.37 | 15.23 | 16.00 | 16.69 | 17.31 | 17.86 | 18.36 | 18.82 | 19.23 | 19.61 | 19.96 | 20.27 | 20.57 | 20.84 | 21.09 |
| | $W_{1,2}$ | 16.83 | 35.03 | 53.97 | 73.11 | 92.30 | 111.50 | 130.70 | 149.90 | 169.11 | 188.31 | 207.51 | 226.72 | 245.92 | 265.12 | 284.32 | 303.53 | 322.73 | 341.93 | 361.14 | 380.34 | 399.54 | 418.75 | 437.95 | 457.15 | 476.35 |
| | $W_{3,4}$ | 17.47 | 36.02 | 55.06 | 74.23 | 93.42 | 112.62 | 131.83 | 151.03 | 170.23 | 189.43 | 208.64 | 227.84 | 247.04 | 266.25 | 285.45 | 304.65 | 323.85 | 343.06 | 362.26 | 381.46 | 400.67 | 419.87 | 439.07 | 458.27 | 477.48 |
| | $J$ | 13.19 | 21.58 | 26.58 | 29.70 | 31.74 | 33.13 | 34.10 | 34.79 | 35.30 | 35.68 | 35.97 | 36.19 | 36.36 | 36.50 | 36.61 | 36.70 | 36.76 | 36.82 | 36.87 | 36.91 | 36.94 | 36.97 | 36.99 | 37.02 | 37.03 |
| | $K$ | 2.24 | 3.24 | 3.50 | 3.57 | 3.58 | 3.58 | 3.58 | 3.58 | 3.58 | 3.58 | 3.58 | 3.58 | 3.58 | 3.58 | 3.58 | 3.58 | 3.58 | 3.58 | 3.58 | 3.58 | 3.58 | 3.58 | 3.58 | 3.58 | 3.58 |

| $\xi/\text{nm}$ | | 2 | 4 | 6 | 8 | 10 | 12 | 14 | 16 | 18 | 20 | 22 | 24 | 26 | 28 | 30 | 32 | 34 | 36 | 38 | 40 | 42 | 44 | 46 | 48 | 50 |
|---|---|---|---|---|---|---|---|---|---|---|---|---|---|---|---|---|---|---|---|---|---|---|---|---|---|---|
| $u_0/u_1$ | | | | | | | | | | | | | | | | | | | | | | | | | | |
| 0.6 | $U_1$ | 21.10 | 36.50 | 46.95 | 54.21 | 59.45 | 63.37 | 66.39 | 68.78 | 70.72 | 72.32 | 73.65 | 74.78 | 75.76 | 76.60 | 77.34 | 77.99 | 78.57 | 79.08 | 79.55 | 79.97 | 80.35 | 80.70 | 81.02 | 81.31 | 81.58 |
| | $U_2$ | 0.71 | 1.71 | 3.13 | 4.81 | 6.56 | 8.24 | 9.79 | 11.17 | 12.41 | 13.50 | 14.48 | 15.34 | 16.10 | 16.79 | 17.40 | 17.96 | 18.46 | 18.91 | 19.32 | 19.70 | 20.05 | 20.36 | 20.66 | 20.93 | 21.18 |
| | $W_{1,2}$ | 16.55 | 34.64 | 53.55 | 72.69 | 91.87 | 111.07 | 130.28 | 149.48 | 168.68 | 187.88 | 207.09 | 226.29 | 245.49 | 264.70 | 283.90 | 303.10 | 322.30 | 341.51 | 360.71 | 379.91 | 399.12 | 418.32 | 437.52 | 456.72 | 475.93 |
| | $W_{3,4}$ | 17.45 | 35.08 | 55.01 | 74.17 | 93.37 | 112.57 | 131.77 | 150.98 | 170.18 | 189.38 | 208.59 | 227.79 | 246.99 | 266.19 | 285.40 | 304.60 | 323.80 | 343.00 | 362.21 | 381.41 | 400.61 | 419.82 | 439.02 | 458.22 | 477.42 |
| | $J$ | 13.66 | 22.60 | 28.04 | 31.49 | 33.76 | 35.32 | 36.41 | 37.19 | 37.77 | 38.20 | 38.53 | 38.78 | 38.98 | 39.14 | 39.26 | 39.36 | 39.44 | 39.51 | 39.57 | 39.61 | 39.65 | 39.68 | 39.71 | 39.73 | 39.76 |
| | $K$ | 2.40 | 3.46 | 3.74 | 3.83 | 3.83 | 3.83 | 3.83 | 3.83 | 3.83 | 3.83 | 3.83 | 3.83 | 3.83 | 3.83 | 3.83 | 3.83 | 3.83 | 3.83 | 3.83 | 3.83 | 3.83 | 3.83 | 3.83 | 3.83 | 3.83 |
| 0.5 | $U_1$ | 19.60 | 34.12 | 44.10 | 51.10 | 56.19 | 60.02 | 62.98 | 65.34 | 67.25 | 68.82 | 70.14 | 71.27 | 72.23 | 73.07 | 73.80 | 74.45 | 75.03 | 75.54 | 76.00 | 76.42 | 76.80 | 77.15 | 77.47 | 77.76 | 78.03 |
| | $U_2$ | 0.78 | 1.84 | 3.29 | 4.99 | 6.74 | 8.41 | 9.94 | 11.32 | 12.55 | 13.63 | 14.60 | 15.45 | 16.22 | 16.90 | 17.51 | 18.05 | 18.53 | 18.97 | 19.42 | 19.80 | 20.14 | 20.46 | 20.75 | 21.03 | 21.28 |
| | $W_{1,2}$ | 16.33 | 34.32 | 53.20 | 72.33 | 91.52 | 110.72 | 129.92 | 149.12 | 168.33 | 187.53 | 206.73 | 225.94 | 245.14 | 264.34 | 283.54 | 302.75 | 321.95 | 341.15 | 360.36 | 379.56 | 398.76 | 417.97 | 437.17 | 456.37 | 475.57 |
| | $W_{3,4}$ | 17.49 | 36.02 | 55.05 | 74.22 | 93.41 | 112.61 | 131.81 | 151.01 | 170.22 | 189.42 | 208.62 | 227.83 | 247.03 | 266.23 | 285.44 | 304.64 | 323.84 | 343.05 | 362.25 | 381.45 | 400.65 | 419.86 | 439.06 | 458.26 | 477.47 |
| | $J$ | 14.18 | 23.74 | 29.74 | 33.62 | 36.23 | 38.05 | 39.34 | 40.29 | 41.01 | 41.56 | 41.98 | 42.31 | 42.57 | 42.78 | 42.95 | 43.11 | 43.20 | 43.27 | 43.34 | 43.39 | 43.44 | 43.47 | 43.51 | 43.51 | 43.53 |
| | $K$ | 2.57 | 3.70 | 4.01 | 4.08 | 4.09 | 4.09 | 4.09 | 4.10 | 4.10 | 4.10 | 4.10 | 4.10 | 4.10 | 4.10 | 4.10 | 4.10 | 4.10 | 4.10 | 4.10 | 4.10 | 4.10 | 4.10 | 4.10 | 4.10 | 4.10 |
| 0.4 | $U_1$ | 18.53 | 32.40 | 42.04 | 48.86 | 53.84 | 57.60 | 60.52 | 62.85 | 64.74 | 66.30 | 67.61 | 68.73 | 69.69 | 70.52 | 71.25 | 71.89 | 72.47 | 72.98 | 73.44 | 73.86 | 74.24 | 74.58 | 74.90 | 75.19 | 75.46 |
| | $U_2$ | 0.83 | 1.93 | 3.41 | 5.12 | 6.86 | 8.53 | 10.05 | 11.42 | 12.64 | 13.73 | 14.69 | 15.54 | 16.30 | 16.98 | 17.58 | 18.13 | 18.63 | 19.08 | 19.49 | 19.87 | 20.21 | 20.53 | 20.82 | 21.09 | 21.34 |
| | $W_{1,2}$ | 16.10 | 33.99 | 52.85 | 71.98 | 91.16 | 110.36 | 129.57 | 148.77 | 167.97 | 187.17 | 206.38 | 225.58 | 244.78 | 263.99 | 283.19 | 302.39 | 321.59 | 340.80 | 360.00 | 379.20 | 398.41 | 417.61 | 436.81 | 456.01 | 475.22 |
| | $W_{3,4}$ | 17.53 | 36.07 | 55.10 | 74.26 | 93.46 | 112.66 | 131.86 | 151.06 | 170.27 | 189.47 | 208.67 | 227.87 | 247.08 | 266.28 | 285.48 | 304.69 | 323.89 | 343.09 | 362.29 | 381.50 | 400.70 | 419.90 | 439.11 | 458.31 | 477.51 |
| | $J$ | 14.02 | 24.68 | 31.11 | 35.34 | 38.21 | 40.22 | 41.68 | 42.75 | 43.55 | 44.14 | 44.64 | 45.02 | 45.31 | 45.55 | 45.75 | 45.91 | 46.04 | 46.15 | 46.24 | 46.32 | 46.38 | 46.44 | 46.48 | 46.53 | 46.56 |
| | $K$ | 2.75 | 3.95 | 4.28 | 4.35 | 4.37 | 4.37 | 4.37 | 4.37 | 4.37 | 4.37 | 4.37 | 4.37 | 4.37 | 4.37 | 4.37 | 4.37 | 4.37 | 4.37 | 4.37 | 4.37 | 4.37 | 4.37 | 4.37 | 4.37 | 4.37 |
| 0.3 | $U_1$ | 18.02 | 31.60 | 40.92 | 47.84 | 52.78 | 56.52 | 59.43 | 61.74 | 63.63 | 65.18 | 66.49 | 67.61 | 68.56 | 69.39 | 70.12 | 70.77 | 71.34 | 71.85 | 72.31 | 72.73 | 73.11 | 73.45 | 73.77 | 74.06 | 74.33 |
| | $U_2$ | 0.86 | 1.96 | 3.46 | 5.17 | 6.92 | 8.58 | 10.11 | 11.47 | 12.69 | 13.78 | 14.74 | 15.59 | 16.33 | 17.01 | 17.62 | 18.17 | 18.66 | 19.11 | 19.53 | 19.91 | 20.24 | 20.56 | 20.85 | 21.12 | 21.37 |
| | $W_{1,2}$ | 15.84 | 33.63 | 52.46 | 71.58 | 90.76 | 109.96 | 129.16 | 148.37 | 167.57 | 186.77 | 205.98 | 225.18 | 244.38 | 263.58 | 282.79 | 301.99 | 321.19 | 340.40 | 359.60 | 378.80 | 398.00 | 417.21 | 436.41 | 455.61 | 474.82 |
| | $W_{3,4}$ | 17.54 | 36.08 | 55.11 | 74.28 | 93.47 | 112.68 | 131.88 | 151.08 | 170.28 | 189.48 | 208.69 | 227.89 | 247.09 | 266.30 | 285.50 | 304.70 | 323.90 | 343.10 | 362.31 | 381.51 | 400.71 | 419.92 | 439.12 | 458.32 | 477.53 |
| | $J$ | 14.93 | 25.30 | 31.98 | 36.39 | 39.39 | 41.50 | 43.02 | 44.14 | 44.98 | 45.60 | 46.13 | 46.52 | 46.81 | 47.08 | 47.29 | 47.46 | 47.60 | 47.72 | 47.81 | 47.89 | 47.96 | 48.02 | 48.08 | 48.11 | 48.15 |
| | $K$ | 3.06 | 4.40 | 4.76 | 4.84 | 4.86 | 4.86 | 4.86 | 4.86 | 4.86 | 4.86 | 4.86 | 4.86 | 4.86 | 4.86 | 4.86 | 4.86 | 4.86 | 4.86 | 4.86 | 4.86 | 4.86 | 4.86 | 4.86 | 4.86 | 4.86 |
| 0.2 | $U_1$ | 17.66 | 30.87 | 40.22 | 46.88 | 51.78 | 55.49 | 58.39 | 60.71 | 62.60 | 64.11 | 65.42 | 66.53 | 67.48 | 68.31 | 69.04 | 69.68 | 70.25 | 70.76 | 71.22 | 71.64 | 72.02 | 72.36 | 72.68 | 72.97 | 73.63 |
| | $U_2$ | 0.90 | 1.98 | 3.48 | 5.20 | 6.95 | 8.61 | 10.14 | 11.52 | 12.74 | 13.83 | 14.78 | 15.63 | 16.39 | 17.07 | 17.66 | 18.21 | 18.70 | 19.15 | 19.56 | 19.94 | 20.28 | 20.60 | 20.89 | 21.16 | 21.41 |
| | $W_{1,2}$ | 15.64 | 33.35 | 52.16 | 71.28 | 90.46 | 109.66 | 128.86 | 148.06 | 167.27 | 186.47 | 205.67 | 224.87 | 244.08 | 263.28 | 282.48 | 301.69 | 320.89 | 340.09 | 359.29 | 378.50 | 397.70 | 416.90 | 436.11 | 455.31 | 474.51 |
| | $W_{3,4}$ | 17.56 | 36.11 | 55.14 | 74.31 | 93.50 | 112.70 | 131.90 | 151.10 | 170.30 | 189.51 | 208.71 | 227.91 | 247.12 | 266.32 | 285.52 | 304.72 | 323.93 | 343.13 | 362.33 | 381.54 | 400.74 | 419.94 | 439.14 | 458.35 | 477.53 |
| | $J$ | 15.18 | 26.03 | 32.97 | 37.54 | 40.65 | 42.82 | 44.38 | 45.52 | 46.37 | 47.02 | 47.52 | 47.91 | 48.22 | 48.47 | 48.67 | 48.84 | 48.98 | 49.11 | 49.21 | 49.30 | 49.37 | 49.43 | 49.49 | 49.53 | 49.57 |
| | $K$ | 3.06 | 4.40 | 4.76 | 4.84 | 4.86 | 4.86 | 4.86 | 4.86 | 4.86 | 4.86 | 4.86 | 4.86 | 4.86 | 4.86 | 4.86 | 4.86 | 4.86 | 4.86 | 4.86 | 4.86 | 4.86 | 4.86 | 4.86 | 4.86 | 4.86 |
| 0.1 | $U_1$ | 17.60 | 30.81 | 40.18 | 46.85 | 51.75 | 55.47 | 58.36 | 60.67 | 62.55 | 64.10 | 65.41 | 66.52 | 67.48 | 68.31 | 69.04 | 69.68 | 70.25 | 70.76 | 71.22 | 71.64 | 72.03 | 72.41 | 72.76 | 73.07 | 73.37 |
| | $U_2$ | 0.85 | 1.98 | 3.48 | 5.21 | 6.96 | 8.62 | 10.15 | 11.51 | 12.73 | 13.81 | 14.76 | 15.61 | 16.37 | 17.05 | 17.66 | 18.20 | 18.70 | 19.15 | 19.56 | 19.94 | 20.28 | 20.60 | 20.89 | 21.16 | 21.41 |
| | $W_{1,2}$ | 15.41 | 33.03 | 51.82 | 70.93 | 90.11 | 109.31 | 128.51 | 147.71 | 166.91 | 186.12 | 205.32 | 224.52 | 243.72 | 262.93 | 282.13 | 301.33 | 320.54 | 339.74 | 358.94 | 378.14 | 397.35 | 416.55 | 435.75 | 454.96 | 474.16 |
| | $W_{3,4}$ | 17.56 | 36.10 | 55.14 | 74.30 | 93.50 | 112.70 | 131.90 | 151.11 | 170.31 | 189.51 | 208.71 | 227.91 | 247.12 | 266.32 | 285.52 | 304.72 | 323.93 | 343.13 | 362.33 | 381.54 | 400.74 | 419.94 | 439.14 | 458.35 | 477.55 |
| | $J$ | 15.39 | 26.20 | 33.22 | 37.86 | 41.02 | 43.20 | 44.82 | 45.99 | 46.87 | 47.52 | 48.06 | 48.46 | 48.78 | 49.04 | 49.25 | 49.42 | 49.56 | 49.68 | 49.78 | 49.86 | 49.93 | 49.99 | 50.04 | 50.09 | 50.13 |
| | $K$ | 3.16 | 4.54 | 4.91 | 4.99 | 5.01 | 5.02 | 5.02 | 5.02 | 5.02 | 5.02 | 5.02 | 5.02 | 5.02 | 5.02 | 5.02 | 5.02 | 5.02 | 5.02 | 5.02 | 5.02 | 5.02 | 5.02 | 5.02 | 5.02 | 5.02 |
| 0.0 | $U_1$ | 17.51 | 30.81 | 40.18 | 46.85 | 51.75 | 55.47 | 58.36 | 60.67 | 62.55 | 64.10 | 65.41 | 66.52 | 67.48 | 68.31 | 69.04 | 69.68 | 70.25 | 70.76 | 71.22 | 71.64 | 72.02 | 72.36 | 72.68 | 72.97 | 73.24 |
| | $U_2$ | 0.85 | 1.99 | 3.49 | 5.21 | 6.96 | 8.62 | 10.15 | 11.51 | 12.73 | 13.81 | 14.76 | 15.61 | 16.37 | 17.05 | 17.66 | 18.21 | 18.70 | 19.15 | 19.56 | 19.94 | 20.28 | 20.60 | 20.89 | 21.16 | 21.41 |
| | $W_{1,2}$ | 15.39 | 33.03 | 51.82 | 70.93 | 90.11 | 109.31 | 128.51 | 147.71 | 166.91 | 186.12 | 205.32 | 224.52 | 243.72 | 262.92 | 282.13 | 301.33 | 320.53 | 339.74 | 358.94 | 378.14 | 397.35 | 416.55 | 435.75 | 454.96 | 474.16 |
| | $W_{3,4}$ | 17.56 | 36.11 | 55.14 | 74.30 | 93.50 | 112.70 | 131.90 | 151.11 | 170.31 | 189.51 | 208.71 | 227.92 | 247.12 | 266.32 | 285.52 | 304.72 | 323.93 | 343.13 | 362.33 | 381.54 | 400.74 | 419.94 | 439.14 | 458.35 | 477.55 |
| | $J$ | 15.39 | 26.20 | 33.22 | 37.86 | 41.02 | 43.23 | 44.82 | 45.99 | 46.87 | 47.54 | 48.06 | 48.46 | 48.78 | 49.04 | 49.25 | 49.42 | 49.56 | 49.68 | 49.78 | 49.86 | 49.93 | 49.99 | 50.04 | 50.09 | 50.13 |
| | $K$ | 3.19 | 4.59 | 4.96 | 5.05 | 5.07 | 5.07 | 5.07 | 5.07 | 5.07 | 5.07 | 5.07 | 5.07 | 5.07 | 5.07 | 5.07 | 5.07 | 5.07 | 5.07 | 5.07 | 5.07 | 5.07 | 5.07 | 5.07 | 5.07 | 5.07 |

TABLE S83: Parameters of the THF interaction Hamiltonian for different ratios $u_0/u_1$ and screening lengths $\xi$ of the double-gate interaction potential ($U_\xi = 24$ meV for $\xi = 10$nm). We use the same BM model parameters as in Table S82.

Total weight of the heavy-fermions on the active bands after wannierisation: 93.4%

FIG. S58. Band structures of the BM and THF models near charge neutrality for $w_0/w_1 = 0.8$, depicted by lines and crosses, respectively. The BM bands are colored according to the weight of the $f$-electron wave function on them. We use the same BM parameters as in Table S84.

TABLE S84. Parameters of the $f$-electron Wannier functions and the THF single-particle Hamiltonian for different values of the tunneling ratio $w_0/w_1$. $W$ denotes the total weight of the THF $f$-electrons on the active bands. In computing the ratios $\gamma/U_1$ and $\gamma/U_2$, we employ the on-site and nearest-neighbor repulsion parameters $U_1$ and $U_2$ obtained numerically for $\xi = 10$ nm, as given in Table S85. We employ $v_F = 5.944$ eV·Å, $|\mathbf{K}| = 1.703\,\text{Å}^{-1}$, $w_1 = 110$ meV, and $\theta = 1.50°$ for the BM model.

| $w_0/w_1$ | $\alpha_1$ | $\alpha_2$ | $\lambda_1$ | $\lambda_2$ | $\lambda$ | $W$ | $t_0$ | $\gamma$ | $M$ | $v_\star$ | $v_\star'$ | $\gamma/U_1$ | $\gamma/U_2$ |
|---|---|---|---|---|---|---|---|---|---|---|---|---|---|
| 1.00 | 0.827 | 0.563 | 0.209 | 0.215 | 0.373 | 0.933 | 5.019 | -63.919 | -57.050 | -4.835 | 0.0374 | -0.85 | -10.51 |
| 0.90 | 0.854 | 0.521 | 0.227 | 0.221 | 0.391 | 0.932 | 5.898 | -80.265 | -58.717 | -4.937 | 0.0319 | -1.14 | -12.79 |
| 0.80 | 0.881 | 0.474 | 0.239 | 0.227 | 0.412 | 0.934 | 6.884 | -96.064 | -60.167 | -5.035 | 0.0269 | -1.45 | -14.80 |
| 0.70 | 0.906 | 0.423 | 0.257 | 0.233 | 0.431 | 0.937 | 7.726 | -111.130 | -61.412 | -5.131 | 0.0224 | -1.78 | -16.65 |
| 0.60 | 0.929 | 0.371 | 0.269 | 0.239 | 0.440 | 0.936 | 8.449 | -125.243 | -62.467 | -5.225 | 0.0180 | -2.09 | -18.43 |
| 0.50 | 0.950 | 0.313 | 0.286 | 0.239 | 0.468 | 0.946 | 9.242 | -138.139 | -63.341 | -5.316 | 0.0137 | -2.45 | -19.76 |
| 0.40 | 0.967 | 0.253 | 0.304 | 0.245 | 0.491 | 0.955 | 10.016 | -149.498 | -64.044 | -5.403 | 0.0096 | -2.81 | -20.86 |
| 0.30 | 0.981 | 0.194 | 0.316 | 0.245 | 0.493 | 0.953 | 10.367 | -158.952 | -64.584 | -5.482 | 0.0059 | -3.04 | -22.07 |
| 0.20 | 0.991 | 0.132 | 0.322 | 0.251 | 0.485 | 0.948 | 10.518 | -166.105 | -64.966 | -5.547 | 0.0028 | -3.16 | -23.09 |
| 0.10 | 0.998 | 0.067 | 0.334 | 0.251 | 0.495 | 0.951 | 10.745 | -170.583 | -65.193 | -5.589 | 0.0007 | -3.32 | -23.51 |
| 0.00 | 1.000 | 0.000 | 0.334 | 0.000 | 0.482 | 0.944 | 10.587 | -172.110 | -65.269 | -5.604 | 0.0000 | -3.28 | -23.97 |

Large parameter table (values versus $\xi$/nm; all energies in meV). Rows grouped by $w_0/w_1$ with quantities $U_1$, $U_2$, $W_{1,2}$, $W_{3,4}$, $J$, $K$.

| $w_0/w_1$ | $\xi$/nm → | 2 | 4 | 6 | 8 | 10 | 12 | 14 | 16 | 18 | 20 | 22 | 24 | 26 | 28 | 30 | 32 | 34 | 36 | 38 | 40 | 42 | 44 | 46 | 48 | 50 |
|---|---|---|---|---|---|---|---|---|---|---|---|---|---|---|---|---|---|---|---|---|---|---|---|---|---|---|
| **1.0** | $U_1$ | 28.79 | 48.32 | 60.76 | 69.04 | 74.84 | 79.09 | 82.31 | 84.84 | 86.87 | 88.52 | 89.90 | 91.07 | 92.07 | 92.93 | 93.68 | 94.35 | 94.93 | 95.46 | 95.93 | 96.35 | 96.74 | 97.09 | 97.41 | 97.71 | 97.98 |
| | $U_2$ | 1.30 | 2.61 | 4.29 | 6.08 | 7.82 | 9.42 | 10.86 | 12.13 | 13.26 | 14.26 | 15.14 | 15.93 | 16.63 | 17.26 | 17.82 | 18.33 | 18.79 | 19.21 | 19.59 | 19.94 | 20.26 | 20.56 | 20.83 | 21.09 | 21.37 |
| | $W_{1,2}$ | 18.04 | 37.05 | 56.59 | 75.99 | 95.07 | 115.71 | 135.44 | 155.16 | 174.89 | 194.61 | 214.34 | 234.07 | 253.79 | 273.52 | 293.24 | 312.97 | 332.69 | 352.42 | 372.14 | 391.87 | 411.59 | 431.32 | 451.04 | 470.77 | 490.50 |
| | $W_{3,4}$ | 11.94 | 18.47 | 21.81 | 23.68 | 24.80 | 25.51 | 25.99 | 26.31 | 26.54 | 26.70 | 26.82 | 26.91 | 26.98 | 27.03 | 27.07 | 27.10 | 27.13 | 27.15 | 27.17 | 27.18 | 27.19 | 27.20 | 27.21 | 27.22 | 27.22 |
| | $J$ | 1.86 | 2.69 | 2.91 | 2.95 | 2.96 | 2.97 | 2.97 | 2.97 | 2.97 | 2.97 | 2.97 | 2.97 | 2.97 | 2.97 | 2.97 | 2.97 | 2.97 | 2.97 | 2.97 | 2.97 | 2.97 | 2.97 | 2.97 | 2.97 | 2.97 |
| | $K$ | 2.24 | 3.21 | 3.47 | 3.53 | 3.54 | 3.54 | 3.54 | 3.54 | 3.54 | 3.54 | 3.54 | 3.54 | 3.54 | 3.54 | 3.54 | 3.54 | 3.54 | 3.54 | 3.54 | 3.54 | 3.54 | 3.54 | 3.54 | 3.54 | 3.54 |
| **0.9** | $U_1$ | 20.65 | 34.41 | 44.91 | 56.82 | 64.63 | 70.49 | 74.65 | 77.83 | 80.32 | 82.97 | 85.34 | 86.50 | 87.49 | 88.34 | 89.09 | 89.75 | 90.33 | 90.86 | 91.33 | 91.75 | 92.14 | 92.49 | 92.81 | 93.11 | 93.38 |
| | $U_2$ | 0.56 | 1.43 | 2.79 | 4.48 | 6.28 | 8.01 | 9.60 | 11.03 | 12.30 | 13.43 | 14.42 | 15.30 | 16.08 | 16.78 | 17.40 | 17.96 | 18.47 | 18.93 | 19.35 | 19.73 | 20.08 | 20.40 | 20.70 | 20.97 | 21.22 |
| | $W_{1,2}$ | 17.83 | 36.76 | 56.27 | 75.95 | 95.67 | 115.39 | 135.12 | 154.84 | 174.57 | 194.29 | 214.02 | 233.74 | 253.47 | 273.20 | 292.92 | 312.65 | 332.37 | 352.10 | 371.82 | 391.55 | 411.27 | 431.00 | 450.72 | 470.45 | 490.17 |
| | $W_{3,4}$ | 12.42 | 19.62 | 23.54 | 25.82 | 27.24 | 28.16 | 28.79 | 29.22 | 29.53 | 29.76 | 29.93 | 30.05 | 30.14 | 30.22 | 30.28 | 30.32 | 30.36 | 30.39 | 30.41 | 30.43 | 30.45 | 30.46 | 30.48 | 30.49 | 30.49 |
| | $J$ | 1.97 | 2.83 | 3.06 | 3.11 | 3.12 | 3.13 | 3.13 | 3.13 | 3.13 | 3.13 | 3.13 | 3.13 | 3.13 | 3.13 | 3.13 | 3.13 | 3.13 | 3.13 | 3.13 | 3.13 | 3.13 | 3.13 | 3.13 | 3.13 | 3.13 |
| | $K$ | 2.09 | 2.95 | 3.08 | 3.12 | 3.13 | 3.13 | 3.13 | 3.13 | 3.13 | 3.13 | 3.13 | 3.13 | 3.13 | 3.13 | 3.13 | 3.13 | 3.13 | 3.13 | 3.13 | 3.13 | 3.13 | 3.13 | 3.13 | 3.13 | 3.13 |
| **0.8** | $U_1$ | 17.60 | 36.43 | 55.93 | 73.60 | 76.30 | 79.70 | 81.04 | 82.92 | 83.77 | 84.52 | 85.18 | 85.76 | 86.28 | 86.75 | 87.17 | 87.55 | 87.90 | 88.22 | 88.52 | 88.79 | — | — | — | — | 88.79 |
| | $U_2$ | 0.64 | 1.57 | 2.98 | 4.69 | 6.49 | 8.21 | 9.80 | 11.22 | 12.48 | 13.60 | 14.59 | 15.46 | 16.24 | 16.93 | 17.56 | 18.12 | 18.62 | 19.08 | 19.49 | 19.87 | 20.22 | 20.54 | 20.84 | 21.11 | 21.37 |
| | $W_{1,2}$ | 18.19 | 37.42 | 57.03 | 76.73 | 96.45 | 116.18 | 135.90 | 155.63 | 175.35 | 195.08 | 214.80 | 234.53 | 254.25 | 273.98 | 293.70 | 313.43 | 333.15 | 352.88 | 372.60 | 392.33 | 412.05 | 431.78 | 451.50 | 471.23 | 490.95 |
| | $W_{3,4}$ | 12.96 | 20.86 | 25.37 | 28.10 | 29.84 | 31.00 | 31.79 | 32.35 | 32.76 | 33.05 | 33.27 | 33.44 | 33.57 | 33.67 | 33.75 | 33.81 | 33.87 | 33.91 | 33.93 | 33.97 | 33.99 | 34.01 | 34.03 | 34.04 | 34.06 |
| | $J$ | 2.09 | 3.01 | 3.25 | 3.30 | 3.32 | 3.32 | 3.32 | 3.32 | 3.32 | 3.32 | 3.32 | 3.32 | 3.32 | 3.32 | 3.32 | 3.32 | 3.32 | 3.32 | 3.32 | 3.32 | 3.32 | 3.32 | 3.32 | 3.32 | 3.32 |
| | $K$ | 2.09 | 3.01 | 3.25 | 3.30 | 3.32 | 3.32 | 3.32 | 3.32 | 3.32 | 3.32 | 3.32 | 3.32 | 3.32 | 3.32 | 3.32 | 3.32 | 3.32 | 3.32 | 3.32 | 3.32 | 3.32 | 3.32 | 3.32 | 3.32 | 3.32 |
| **0.7** | $U_1$ | 22.58 | 38.79 | 49.65 | 57.11 | 62.47 | 66.45 | 69.52 | 71.94 | 73.89 | 75.50 | 76.84 | 77.98 | 78.96 | 79.81 | 80.55 | 81.20 | 81.78 | 82.30 | 82.77 | 83.19 | 83.57 | 83.92 | 84.24 | 84.54 | 84.80 |
| | $U_2$ | 0.70 | 1.70 | 3.15 | 4.88 | 6.68 | 8.39 | 9.97 | 11.38 | 12.64 | 13.75 | 14.73 | 15.59 | 16.37 | 17.07 | 17.68 | 18.24 | 18.75 | 19.20 | 19.62 | 20.00 | 20.34 | 20.66 | 20.96 | 21.23 | 21.48 |
| | $W_{1,2}$ | 17.03 | 36.09 | 55.55 | 75.23 | 94.94 | 114.66 | 134.39 | 154.11 | 173.84 | 193.56 | 213.29 | 233.01 | 252.74 | 272.46 | 292.19 | 311.91 | 331.64 | 351.36 | 371.09 | 390.82 | 410.54 | 430.27 | 449.99 | 469.72 | 490.46 |
| | $W_{3,4}$ | 13.51 | 22.09 | 27.19 | 30.36 | 32.43 | 33.81 | 34.81 | 35.51 | 36.02 | 36.40 | 36.69 | 36.91 | 37.09 | 37.22 | 37.33 | 37.41 | 37.47 | 37.53 | 37.57 | 37.63 | 37.66 | 37.69 | 37.71 | 37.73 | 37.75 |
| | $J$ | 2.24 | 3.21 | 3.47 | 3.53 | 3.54 | 3.54 | 3.54 | 3.54 | 3.54 | 3.54 | 3.54 | 3.54 | 3.54 | 3.54 | 3.54 | 3.54 | 3.54 | 3.54 | 3.54 | 3.54 | 3.54 | 3.54 | 3.54 | 3.54 | 3.54 |
| | $K$ | 2.24 | 3.21 | 3.47 | 3.53 | 3.54 | 3.54 | 3.54 | 3.54 | 3.54 | 3.54 | 3.54 | 3.54 | 3.54 | 3.54 | 3.54 | 3.54 | 3.54 | 3.54 | 3.54 | 3.54 | 3.54 | 3.54 | 3.54 | 3.54 | 3.54 |

| $u_0/u_1$ | $\xi$/nm | 2 | 4 | 6 | 8 | 10 | 12 | 14 | 16 | 18 | 20 | 22 | 24 | 26 | 28 | 30 | 32 | 34 | 36 | 38 | 40 | 42 | 44 | 46 | 48 | 50 |
|---|---|---|---|---|---|---|---|---|---|---|---|---|---|---|---|---|---|---|---|---|---|---|---|---|---|---|
| 0.6 | $U_1$ | 21.38 | 36.94 | 47.48 | 54.79 | 60.05 | 63.09 | 67.02 | 69.42 | 71.36 | 72.96 | 74.30 | 75.43 | 76.40 | 77.25 | 77.99 | 78.64 | 79.22 | 79.74 | 80.20 | 80.62 | 81.00 | 81.35 | 81.67 | 81.96 | 82.23 |
|  | $U_2$ | 0.74 | 1.78 | 3.25 | 5.00 | 6.79 | 8.51 | 10.08 | 11.49 | 12.74 | 13.85 | 14.83 | 15.69 | 16.47 | 17.16 | 17.77 | 18.33 | 18.83 | 19.29 | 19.70 | 20.08 | 20.43 | 20.75 | 21.04 | 21.31 | 21.57 |
|  | $W_{1,2}$ | 35.69 | 55.13 | 74.94 | 94.50 | 114.23 | 133.95 | 153.68 | 173.40 | 193.13 | 212.85 | 232.58 | 252.30 | 272.03 | 291.75 | 311.48 | 331.20 | 350.93 | 370.65 | 390.38 | 410.11 | 429.83 | 449.56 | 469.28 | 489.01 | 508.73 |
|  | $W_{3,4}$ | 17.91 | 36.95 | 56.50 | 76.19 | 95.91 | 115.63 | 135.35 | 155.08 | 174.80 | 194.53 | 214.26 | 233.98 | 253.71 | 273.43 | 293.16 | 312.89 | 332.61 | 352.34 | 372.06 | 391.79 | 411.51 | 431.24 | 450.96 | 470.69 | 490.41 |
|  | $J$ | 13.09 | 23.11 | 28.65 | 32.14 | 34.43 | 35.99 | 37.09 | 37.87 | 38.45 | 38.88 | 39.21 | 39.46 | 39.65 | 39.81 | 39.93 | 40.03 | 40.11 | 40.18 | 40.23 | 40.28 | 40.31 | 40.35 | 40.37 | 40.40 | 40.42 |
|  | $K$ | 3.44 | 3.71 | 3.78 | 3.78 | 3.79 | 3.79 | 3.79 | 3.79 | 3.79 | 3.79 | 3.79 | 3.79 | 3.79 | 3.79 | 3.79 | 3.79 | 3.79 | 3.79 | 3.79 | 3.79 | 3.79 | 3.79 | 3.79 | 3.79 | 3.79 |
| 0.5 | $U_1$ | 19.68 | 34.23 | 44.23 | 51.24 | 56.33 | 60.16 | 63.12 | 65.48 | 67.38 | 68.96 | 70.28 | 71.40 | 72.37 | 73.21 | 73.94 | 74.59 | 75.16 | 75.68 | 76.14 | 76.56 | 76.94 | 77.28 | 77.60 | 77.89 | 78.16 |
|  | $U_2$ | 0.82 | 1.93 | 3.44 | 5.20 | 6.99 | 8.69 | 10.25 | 11.65 | 12.89 | 13.98 | 14.95 | 15.82 | 16.58 | 17.27 | 17.88 | 18.44 | 18.94 | 19.40 | 19.80 | 20.18 | 20.52 | 20.84 | 21.13 | 21.41 | 21.66 |
|  | $W_{1,2}$ | 35.40 | 54.82 | 74.47 | 94.19 | 113.91 | 133.63 | 153.36 | 173.08 | 192.81 | 212.53 | 232.26 | 251.98 | 271.71 | 291.43 | 311.16 | 330.89 | 350.61 | 370.34 | 390.06 | 409.79 | 429.51 | 449.24 | 468.96 | 488.69 | 508.41 |
|  | $W_{3,4}$ | 17.97 | 37.02 | 56.58 | 76.27 | 95.98 | 115.71 | 135.43 | 155.16 | 174.88 | 194.61 | 214.33 | 234.06 | 253.79 | 273.51 | 293.24 | 312.96 | 332.69 | 352.41 | 372.14 | 391.86 | 411.59 | 431.31 | 451.04 | 470.76 | 490.49 |
|  | $J$ | 14.56 | 24.37 | 30.53 | 34.53 | 37.22 | 39.10 | 40.44 | 41.43 | 42.17 | 42.73 | 43.16 | 43.50 | 43.77 | 43.99 | 44.16 | 44.31 | 44.42 | 44.52 | 44.60 | 44.67 | 44.73 | 44.78 | 44.82 | 44.85 | 44.89 |
|  | $K$ | 2.57 | 3.69 | 3.98 | 4.05 | 4.06 | 4.06 | 4.06 | 4.06 | 4.06 | 4.06 | 4.06 | 4.06 | 4.06 | 4.06 | 4.06 | 4.06 | 4.06 | 4.06 | 4.06 | 4.06 | 4.06 | 4.06 | 4.06 | 4.06 | 4.06 |
| 0.4 | $U_1$ | 18.20 | 31.93 | 41.45 | 48.19 | 53.13 | 56.86 | 59.76 | 62.07 | 63.95 | 65.50 | 66.81 | 67.92 | 68.88 | 69.71 | 70.44 | 71.08 | 71.65 | 72.16 | 72.62 | 73.04 | 73.42 | 73.76 | 74.08 | 74.37 | 74.64 |
|  | $U_2$ | 0.89 | 2.07 | 3.62 | 5.38 | 7.17 | 8.86 | 10.40 | 11.79 | 13.02 | 14.10 | 15.07 | 15.92 | 16.69 | 17.37 | 17.98 | 18.53 | 19.03 | 19.48 | 19.89 | 20.26 | 20.61 | 20.93 | 21.22 | 21.49 | 21.74 |
|  | $W_{1,2}$ | 35.35 | 54.54 | 74.19 | 93.90 | 113.60 | 133.25 | 153.07 | 172.80 | 192.52 | 212.25 | 231.97 | 251.70 | 271.42 | 291.15 | 310.87 | 330.60 | 350.32 | 370.05 | 389.78 | 409.50 | 429.23 | 448.95 | 468.68 | 488.40 | 508.13 |
|  | $W_{3,4}$ | 18.03 | 37.12 | 56.68 | 76.36 | 96.06 | 115.80 | 135.53 | 155.25 | 174.98 | 194.70 | 214.43 | 234.16 | 253.88 | 273.61 | 293.33 | 313.06 | 332.79 | 352.51 | 372.23 | 391.96 | 411.68 | 431.41 | 451.14 | 470.86 | 490.59 |
|  | $J$ | 15.00 | 25.46 | 32.15 | 36.58 | 39.61 | 41.76 | 43.31 | 44.47 | 45.35 | 46.02 | 46.54 | 46.96 | 47.29 | 47.56 | 47.78 | 47.96 | 48.11 | 48.24 | 48.34 | 48.42 | 48.50 | 48.56 | 48.62 | 48.67 | 48.71 |
|  | $K$ | 2.75 | 3.94 | 4.25 | 4.32 | 4.34 | 4.34 | 4.34 | 4.34 | 4.34 | 4.34 | 4.34 | 4.34 | 4.34 | 4.34 | 4.34 | 4.34 | 4.34 | 4.34 | 4.34 | 4.34 | 4.34 | 4.34 | 4.34 | 4.34 | 4.34 |
| 0.3 | $U_1$ | 17.87 | 31.33 | 40.75 | 47.45 | 52.36 | 56.07 | 58.97 | 61.27 | 63.15 | 64.70 | 66.01 | 67.12 | 68.07 | 68.90 | 69.63 | 70.27 | 70.84 | 71.35 | 71.81 | 72.23 | 72.61 | 72.95 | 73.27 | 73.56 | 73.83 |
|  | $U_2$ | 0.90 | 2.09 | 3.65 | 5.42 | 7.20 | 8.89 | 10.44 | 11.82 | 13.05 | 14.14 | 15.10 | 15.97 | 16.73 | 17.40 | 18.01 | 18.57 | 19.07 | 19.52 | 19.92 | 20.26 | 20.64 | 20.96 | 21.25 | 21.52 | 21.77 |
|  | $W_{1,2}$ | 34.76 | 54.13 | 73.78 | 93.49 | 113.21 | 132.94 | 152.66 | 172.39 | 192.11 | 211.84 | 231.56 | 251.29 | 271.01 | 290.74 | 310.46 | 330.19 | 349.91 | 369.64 | 389.37 | 409.09 | 428.82 | 448.54 | 468.27 | 487.99 | 507.72 |
|  | $W_{3,4}$ | 18.11 | 37.17 | 56.67 | 76.36 | 96.08 | 115.80 | 135.53 | 155.25 | 174.98 | 194.69 | 214.41 | 234.14 | 253.86 | 273.59 | 293.31 | 313.04 | 332.76 | 352.49 | 372.21 | 391.94 | 411.66 | 431.39 | 451.11 | 470.84 | 490.56 |
|  | $J$ | 15.35 | 26.36 | 33.25 | 37.96 | 41.09 | 43.28 | 44.83 | 46.03 | 46.90 | 47.57 | 48.09 | 48.49 | 48.81 | 49.07 | 49.28 | 49.45 | 49.60 | 49.72 | 49.82 | 49.90 | 49.97 | 50.03 | 50.09 | 50.13 | 50.17 |
|  | $K$ | 2.03 | 4.19 | 4.52 | 4.59 | 4.61 | 4.61 | 4.61 | 4.61 | 4.61 | 4.61 | 4.61 | 4.61 | 4.61 | 4.61 | 4.61 | 4.61 | 4.61 | 4.61 | 4.61 | 4.61 | 4.61 | 4.61 | 4.61 | 4.61 | 4.61 |
| 0.2 | $U_1$ | 17.38 | 30.57 | 39.87 | 46.49 | 51.37 | 55.06 | 57.94 | 60.24 | 62.11 | 63.66 | 64.95 | 66.07 | 67.02 | 67.85 | 68.58 | 69.22 | 69.79 | 70.30 | 70.76 | 71.18 | 71.56 | 71.90 | 72.22 | 72.51 | 72.78 |
|  | $U_2$ | 0.92 | 2.12 | 3.69 | 5.47 | 7.26 | 8.95 | 10.49 | 11.87 | 13.10 | 14.19 | 15.15 | 16.01 | 16.76 | 17.44 | 18.05 | 18.60 | 19.10 | 19.55 | 19.96 | 20.31 | 20.66 | 20.97 | 21.27 | 21.54 | 21.81 |
|  | $W_{1,2}$ | 34.26 | 53.60 | 73.24 | 92.95 | 112.67 | 132.39 | 152.12 | 171.84 | 191.57 | 211.29 | 231.02 | 250.74 | 270.47 | 290.19 | 309.92 | 329.64 | 349.37 | 369.10 | 388.82 | 408.55 | 428.27 | 448.00 | 467.72 | 487.45 | 507.45 |
|  | $W_{3,4}$ | 18.08 | 37.15 | 56.71 | 76.40 | 96.12 | 115.78 | 135.50 | 155.23 | 174.95 | 194.68 | 214.41 | 234.13 | 253.86 | 273.58 | 293.31 | 313.03 | 332.76 | 352.48 | 372.21 | 391.93 | 411.66 | 431.38 | 451.11 | 470.84 | 490.63 |
|  | $J$ | 15.54 | 26.84 | 34.08 | 38.86 | 42.12 | 44.34 | 45.88 | 47.04 | 47.91 | 48.57 | 49.08 | 49.47 | 49.78 | 50.02 | 50.20 | 50.32 | 50.44 | 50.52 | 50.58 | 50.63 | 50.68 | 50.72 | 50.76 | 50.79 | 50.13 |
|  | $K$ | 4.30 | 4.74 | 4.82 | 4.83 | 4.84 | 4.84 | 4.84 | 4.84 | 4.84 | 4.84 | 4.84 | 4.84 | 4.84 | 4.84 | 4.84 | 4.84 | 4.84 | 4.84 | 4.84 | 4.84 | 4.84 | 4.84 | 4.84 | 4.84 | 4.84 |
| 0.1 | $U_1$ | 17.84 | 30.40 | 39.50 | 46.05 | 50.88 | 54.55 | 57.42 | 59.72 | 61.59 | 63.14 | 64.43 | 65.55 | 66.51 | 67.34 | 68.07 | 68.71 | 69.28 | 69.79 | 70.25 | 70.67 | 71.05 | 71.39 | 71.70 | 71.98 | 72.24 |
|  | $U_2$ | 0.92 | 2.14 | 3.70 | 5.48 | 7.26 | 8.95 | 10.49 | 11.87 | 13.09 | 14.18 | 15.15 | 16.01 | 16.76 | 17.44 | 18.02 | 18.57 | 19.06 | 19.52 | 19.93 | 20.31 | 20.64 | 20.97 | 21.27 | 21.56 | 21.81 |
|  | $W_{1,2}$ | 34.20 | 53.53 | 73.18 | 92.89 | 112.62 | 132.34 | 152.07 | 171.79 | 191.52 | 211.24 | 230.97 | 250.69 | 270.42 | 290.14 | 309.87 | 329.59 | 349.32 | 369.04 | 388.77 | 408.49 | 428.22 | 447.94 | 467.67 | 487.39 | 507.45 |
|  | $W_{3,4}$ | 18.18 | 37.15 | 56.74 | 76.34 | 96.06 | 115.79 | 135.51 | 155.24 | 174.96 | 194.69 | 214.41 | 234.14 | 253.86 | 273.59 | 293.31 | 313.04 | 332.76 | 352.49 | 372.21 | 391.94 | 411.66 | 431.39 | 451.11 | 470.84 | 490.56 |
|  | $J$ | 15.76 | 26.84 | 34.00 | 38.86 | 42.15 | 44.47 | 46.15 | 47.47 | 48.43 | 49.14 | 49.68 | 50.08 | 50.40 | 50.63 | 50.80 | 50.92 | 51.01 | 51.08 | 51.13 | 51.18 | 51.22 | 51.25 | 51.28 | 51.51 | 51.91 |
|  | $K$ | 3.17 | 4.54 | 4.89 | 4.97 | 4.99 | 4.99 | 4.99 | 4.99 | 4.99 | 4.99 | 4.99 | 4.99 | 4.99 | 4.99 | 4.99 | 4.99 | 4.99 | 4.99 | 4.99 | 4.99 | 4.99 | 4.99 | 4.99 | 4.99 | 4.99 |
| 0.0 | $U_1$ | 17.36 | 30.84 | 40.48 | 47.58 | 53.01 | 56.27 | 59.18 | 61.50 | 63.38 | 64.94 | 66.25 | 67.37 | 68.33 | 69.16 | 69.89 | 70.53 | 71.10 | 71.62 | 72.08 | 72.50 | 72.87 | 73.22 | 73.54 | 73.84 | 74.10 |
|  | $U_2$ | 0.88 | 2.06 | 3.62 | 5.39 | 7.18 | 8.88 | 10.43 | 11.82 | 13.05 | 14.14 | 15.11 | 15.96 | 16.73 | 17.41 | 18.02 | 18.57 | 19.03 | 19.52 | 19.93 | 20.31 | 20.66 | 20.97 | 21.27 | 21.54 | 21.79 |
|  | $W_{1,2}$ | 34.04 | 53.38 | 73.01 | 92.72 | 112.44 | 132.16 | 151.89 | 171.61 | 191.34 | 211.07 | 230.79 | 250.52 | 270.24 | 289.97 | 309.69 | 329.42 | 349.14 | 368.87 | 388.59 | 408.32 | 428.04 | 447.77 | 467.49 | 487.22 | 506.94 |
|  | $W_{3,4}$ | 18.03 | 37.09 | 56.65 | 76.34 | 96.06 | 115.78 | 135.55 | 155.23 | 174.94 | 194.68 | 214.41 | 234.13 | 253.86 | 273.58 | 293.31 | 313.03 | 332.76 | 352.48 | 372.21 | 391.93 | 411.66 | 431.38 | 451.11 | 470.83 | 490.56 |
|  | $J$ | 15.74 | 26.74 | 33.85 | 38.52 | 41.69 | 43.80 | 45.49 | 46.65 | 47.52 | 48.18 | 48.69 | 49.08 | 49.41 | 49.66 | 49.87 | 50.04 | 50.20 | 50.32 | 50.39 | 50.47 | 50.54 | 50.60 | 50.65 | 50.69 | 50.73 |
|  | $K$ | 3.21 | 4.59 | 4.95 | 5.03 | 5.04 | 5.05 | 5.05 | 5.05 | 5.05 | 5.05 | 5.05 | 5.05 | 5.05 | 5.05 | 5.05 | 5.05 | 5.05 | 5.05 | 5.05 | 5.05 | 5.05 | 5.05 | 5.05 | 5.05 | 5.05 |

TABLE S85. Parameters of the THF interaction Hamiltonian for different ratios $u_0/u_1$ and screening lengths $\xi$ of the double-gate interaction potential ($U_\xi = 24$ meV for $\xi = 10$ nm). We use the same BM model parameters as in Table S84.

FIG. S59. Band structures of the BM and THF models near charge neutrality for $w_0/w_1 = 0.8$, depicted by lines and crosses, respectively. The BM bands are colored according to the weight of the $f$-electron wave function on them. We use the same BM parameters as in Table S86.

TABLE S86. Parameters of the $f$-electron single-particle Hamiltonian for different values of the tunneling ratio $w_0/w_1$. $W$ denotes the total weight of the THF $f$-electrons on the active bands. In computing the ratios $\gamma/U_1$ and $\gamma/U_2$, we employ the on-site and nearest-neighbor repulsion parameters $U_1$ and $U_2$ obtained numerically for $\xi = 10$ nm, as given in Table S87. We employ $v_F = 5.944$ eV·Å, $|\mathbf{K}| = 1.703$ Å$^{-1}$, $w_1 = 110$ meV, and $\theta = 1.52°$ for the BM model.

| $w_0/w_1$ | $\alpha_1$ | $\alpha_2$ | $\lambda_1$ | $\lambda_2$ | $\lambda$ | $W$ | $t_0$ | $\gamma$ | $M$ | $v'_\star$ | $v''_\star$ | $\gamma/U_1$ | $\gamma/U_2$ |
|---|---|---|---|---|---|---|---|---|---|---|---|---|---|
| 1.00 | 0.829 | 0.559 | 0.215 | 0.215 | 0.380 | 0.935 | 5.457 | -67.247 | -60.187 | -4.853 | 1.928 | 0.0397 | -2.09 | -10.58 |
| 0.90 | 0.856 | 0.517 | 0.227 | 0.221 | 0.394 | 0.931 | 6.308 | -83.610 | -61.864 | -4.953 | 1.916 | 0.0342 | -1.18 | -12.83 |
| 0.80 | 0.883 | 0.469 | 0.245 | 0.227 | 0.420 | 0.937 | 7.353 | -99.416 | -63.322 | -5.050 | 1.881 | 0.0291 | -1.50 | -14.71 |
| 0.70 | 0.908 | 0.420 | 0.257 | 0.233 | 0.432 | 0.936 | 8.197 | -114.482 | -64.575 | -5.144 | 1.818 | 0.0243 | -1.81 | -16.58 |
| 0.60 | 0.931 | 0.366 | 0.275 | 0.239 | 0.454 | 0.943 | 9.076 | -128.589 | -65.636 | -5.237 | 1.720 | 0.0195 | -2.16 | -18.14 |
| 0.50 | 0.951 | 0.310 | 0.286 | 0.245 | 0.475 | 0.949 | 9.870 | -141.474 | -66.517 | -5.328 | 1.576 | 0.0149 | -2.51 | -19.50 |
| 0.40 | 0.968 | 0.251 | 0.304 | 0.245 | 0.491 | 0.955 | 10.551 | -152.821 | -67.225 | -5.415 | 1.378 | 0.0105 | -2.84 | -20.68 |
| 0.30 | 0.981 | 0.192 | 0.316 | 0.251 | 0.498 | 0.956 | 11.012 | -162.262 | -67.769 | -5.493 | 1.117 | 0.0064 | -3.09 | -21.74 |
| 0.20 | 0.991 | 0.131 | 0.328 | 0.251 | 0.496 | 0.953 | 11.212 | -169.404 | -68.153 | -5.557 | 0.793 | 0.0030 | -3.24 | -22.66 |
| 0.10 | 0.998 | 0.067 | 0.334 | 0.251 | 0.495 | 0.951 | 11.294 | -173.873 | -68.382 | -5.600 | 0.413 | 0.0008 | -3.33 | -23.25 |
| 0.00 | 1.000 | 0.000 | 0.334 | 0.000 | 0.482 | 0.944 | 11.137 | -175.397 | -68.459 | -5.615 | -0.000 | 0.0000 | -3.29 | -23.69 |

| $w_0/w_1$ | $\xi$/nm | 2 | 4 | 6 | 8 | 10 | 12 | 14 | 16 | 18 | 20 | 22 | 24 | 26 | 28 | 30 | 32 | 34 | 36 | 38 | 40 | 42 | 44 | 46 | 48 | 50 |
|---|---|---|---|---|---|---|---|---|---|---|---|---|---|---|---|---|---|---|---|---|---|---|---|---|---|---|
| 1.0 | $U_1$ | 28.74 | 48.22 | 60.63 | 68.89 | 74.67 | 78.91 | 82.14 | 84.66 | 86.68 | 88.34 | 89.72 | 90.88 | 91.88 | 92.74 | 93.49 | 94.16 | 94.74 | 95.27 | 95.74 | 96.16 | 96.55 | 96.90 | 97.22 | 97.52 | 97.79 |
| | $U_2$ | 0.54 | 1.84 | 2.78 | 4.52 | 6.38 | 8.13 | 9.75 | 11.21 | 12.50 | 13.64 | 14.65 | 15.53 | 16.32 | 17.03 | 17.66 | 18.22 | 18.73 | 19.19 | 19.61 | 20.00 | 20.35 | 20.67 | 20.97 | 21.24 | 21.50 |
| | $W_{1,2}$ | 18.60 | 38.16 | 58.24 | 78.46 | 98.70 | 118.96 | 139.21 | 159.47 | 179.72 | 199.98 | 220.23 | 240.49 | 260.74 | 281.00 | 301.25 | 321.51 | 341.76 | 362.01 | 382.27 | 402.52 | 422.78 | 443.03 | 463.29 | 483.54 | 503.80 |
| | $W_{3,4}$ | 18.63 | 38.36 | 58.50 | 78.73 | 98.98 | 119.23 | 139.48 | 159.74 | 179.99 | 200.25 | 220.50 | 240.75 | 261.01 | 281.27 | 301.52 | 321.77 | 342.03 | 362.28 | 382.54 | 402.79 | 423.05 | 443.30 | 463.56 | 483.82 | 504.07 |
| | $J$ | 12.23 | 18.95 | 22.42 | 24.38 | 25.56 | 26.32 | 26.82 | 27.17 | 27.42 | 27.59 | 27.72 | 27.82 | 27.89 | 27.95 | 28.00 | 28.03 | 28.06 | 28.08 | 28.10 | 28.11 | 28.13 | 28.14 | 28.15 | 28.15 | 28.16 |
| | $K$ | 2.84 | 2.65 | 2.85 | 2.90 | 2.91 | 2.91 | 2.91 | 2.91 | 2.91 | 2.91 | 2.91 | 2.91 | 2.91 | 2.91 | 2.91 | 2.91 | 2.91 | 2.91 | 2.91 | 2.91 | 2.91 | 2.91 | 2.91 | 2.91 | 2.91 |
| 0.9 | $U_1$ | 26.76 | 45.25 | 57.21 | 65.25 | 70.92 | 75.10 | 78.28 | 80.78 | 82.79 | 84.43 | 85.80 | 86.96 | 87.95 | 88.81 | 89.56 | 90.22 | 90.81 | 91.33 | 91.80 | 92.22 | 92.61 | 92.96 | 93.28 | 93.58 | 93.85 |
| | $U_2$ | 0.59 | 1.50 | 2.92 | 4.67 | 6.51 | 8.28 | 9.90 | 11.35 | 12.64 | 13.78 | 14.78 | 15.66 | 16.45 | 17.15 | 17.78 | 18.34 | 18.85 | 19.31 | 19.73 | 20.12 | 20.47 | 20.79 | 21.08 | 21.36 | 21.61 |
| | $W_{1,2}$ | 18.36 | 37.84 | 57.90 | 78.11 | 98.36 | 118.61 | 138.86 | 159.12 | 179.37 | 199.63 | 219.88 | 240.14 | 260.39 | 280.65 | 300.90 | 321.16 | 341.41 | 361.67 | 381.92 | 402.18 | 422.43 | 442.69 | 462.94 | 483.20 | 503.45 |
| | $W_{3,4}$ | 18.48 | 38.12 | 58.24 | 78.48 | 98.71 | 118.97 | 139.22 | 159.47 | 179.73 | 199.98 | 220.24 | 240.49 | 260.75 | 281.00 | 301.26 | 321.51 | 341.77 | 362.02 | 382.28 | 402.53 | 422.78 | 443.04 | 463.29 | 483.55 | 503.80 |
| | $J$ | 12.71 | 20.07 | 24.07 | 26.40 | 27.84 | 28.78 | 29.41 | 29.85 | 30.16 | 30.38 | 30.55 | 30.67 | 30.77 | 30.85 | 30.91 | 30.95 | 30.99 | 31.03 | 31.06 | 31.08 | 31.09 | 31.11 | 31.11 | 31.12 | 31.12 |
| | $K$ | 1.95 | 2.80 | 3.02 | 3.06 | 3.07 | 3.08 | 3.08 | 3.08 | 3.08 | 3.08 | 3.08 | 3.08 | 3.08 | 3.08 | 3.08 | 3.08 | 3.08 | 3.08 | 3.08 | 3.08 | 3.08 | 3.08 | 3.08 | 3.08 | 3.08 |
| 0.8 | $U_1$ | 24.37 | 41.55 | 52.87 | 60.58 | 66.07 | 70.13 | 73.24 | 75.70 | 77.67 | 79.29 | 80.65 | 81.80 | 82.78 | 83.63 | 84.38 | 85.03 | 85.62 | 86.14 | 86.60 | 87.03 | 87.41 | 87.76 | 88.08 | 88.37 | 88.64 |
| | $U_2$ | 0.68 | 1.67 | 3.14 | 4.92 | 6.76 | 8.52 | 10.13 | 11.58 | 12.87 | 14.01 | 14.96 | 15.84 | 16.62 | 17.32 | 17.95 | 18.51 | 19.01 | 19.47 | 19.89 | 20.27 | 20.62 | 20.94 | 21.24 | 21.51 | 21.77 |
| | $W_{1,2}$ | 18.14 | 37.53 | 57.57 | 77.77 | 98.02 | 118.27 | 138.53 | 158.78 | 179.04 | 199.29 | 219.55 | 239.80 | 260.06 | 280.31 | 300.56 | 320.82 | 341.07 | 361.33 | 381.58 | 401.84 | 422.09 | 442.35 | 462.60 | 482.86 | 503.11 |
| | $W_{3,4}$ | 18.18 | 38.03 | 58.13 | 78.35 | 98.60 | 118.85 | 139.11 | 159.36 | 179.62 | 199.87 | 220.13 | 240.38 | 260.64 | 280.89 | 301.15 | 321.40 | 341.66 | 361.91 | 382.17 | 402.42 | 422.67 | 442.93 | 463.18 | 483.44 | 503.69 |
| | $J$ | 13.32 | 21.46 | 26.15 | 29.00 | 30.84 | 32.07 | 32.92 | 33.52 | 33.96 | 34.29 | 34.54 | 34.72 | 34.87 | 34.98 | 35.07 | 35.14 | 35.20 | 35.25 | 35.29 | 35.32 | 35.35 | 35.37 | 35.39 | 35.41 | 35.42 |
| | $K$ | 2.08 | 2.98 | 3.21 | 3.26 | 3.27 | 3.27 | 3.27 | 3.27 | 3.27 | 3.27 | 3.27 | 3.27 | 3.27 | 3.27 | 3.27 | 3.27 | 3.27 | 3.27 | 3.27 | 3.27 | 3.27 | 3.27 | 3.27 | 3.27 | 3.27 |
| 0.7 | $U_1$ | 22.87 | 39.25 | 50.19 | 57.70 | 63.08 | 67.08 | 70.16 | 72.58 | 74.54 | 76.16 | 77.50 | 78.64 | 79.62 | 80.47 | 81.21 | 81.87 | 82.45 | 82.97 | 83.43 | 83.86 | 84.24 | 84.59 | 84.91 | 85.20 | 85.47 |
| | $U_2$ | 0.73 | 1.77 | 3.27 | 5.06 | 6.90 | 8.66 | 10.26 | 11.70 | 12.97 | 14.09 | 15.08 | 15.95 | 16.74 | 17.43 | 18.05 | 18.61 | 19.12 | 19.58 | 20.00 | 20.38 | 20.72 | 21.05 | 21.34 | 21.61 | 21.87 |
| | $W_{1,2}$ | 17.87 | 37.15 | 57.16 | 77.36 | 97.61 | 117.86 | 138.12 | 158.37 | 178.62 | 198.88 | 219.13 | 239.39 | 259.64 | 279.90 | 300.15 | 320.41 | 340.66 | 360.91 | 381.17 | 401.43 | 421.68 | 441.94 | 462.19 | 482.44 | 502.70 |
| | $W_{3,4}$ | 18.39 | 37.96 | 58.05 | 78.27 | 98.52 | 118.77 | 139.03 | 159.28 | 179.54 | 199.79 | 220.05 | 240.30 | 260.56 | 280.81 | 301.07 | 321.32 | 341.58 | 361.83 | 382.08 | 402.34 | 422.59 | 442.85 | 463.10 | 483.36 | 503.61 |
| | $J$ | 18.34 | 22.59 | 27.77 | 30.98 | 33.06 | 34.47 | 35.46 | 36.16 | 36.67 | 37.05 | 37.33 | 37.55 | 37.72 | 37.86 | 37.97 | 38.05 | 38.12 | 38.18 | 38.23 | 38.26 | 38.29 | 38.32 | 38.35 | 38.37 | 38.38 |
| | $K$ | 2.23 | 3.19 | 3.44 | 3.49 | 3.50 | 3.50 | 3.50 | 3.50 | 3.50 | 3.50 | 3.50 | 3.50 | 3.50 | 3.50 | 3.50 | 3.50 | 3.50 | 3.50 | 3.50 | 3.50 | 3.50 | 3.50 | 3.50 | 3.50 | 3.50 |

| $u_0/u_1$ | $\xi$/nm | 2 | 4 | 6 | 8 | 10 | 12 | 14 | 16 | 18 | 20 | 22 | 24 | 26 | 28 | 30 | 32 | 34 | 36 | 38 | 40 | 42 | 44 | 46 | 48 | 50 |
|---|---|---|---|---|---|---|---|---|---|---|---|---|---|---|---|---|---|---|---|---|---|---|---|---|---|---|
| 0.6 | $U_1$ | 21.18 | 36.59 | 47.04 | 54.28 | 59.51 | 63.43 | 66.44 | 68.83 | 70.77 | 72.36 | 73.69 | 74.82 | 75.80 | 76.64 | 77.38 | 78.03 | 78.61 | 79.12 | 79.59 | 80.01 | 80.39 | 80.74 | 81.05 | 81.35 | 81.62 |
| | $U_2$ | 0.80 | 1.91 | 3.45 | 5.25 | 7.09 | 8.84 | 10.43 | 11.85 | 13.11 | 14.23 | 15.21 | 16.09 | 16.86 | 17.55 | 18.17 | 18.73 | 19.23 | 19.69 | 20.11 | 20.48 | 20.83 | 21.15 | 21.45 | 21.72 | 21.97 |
| | $W_{1,2}$ | 17.62 | 36.81 | 56.80 | 77.00 | 97.24 | 117.49 | 137.74 | 158.00 | 178.25 | 198.51 | 218.76 | 239.02 | 259.27 | 279.53 | 299.78 | 320.04 | 340.29 | 360.55 | 380.80 | 401.06 | 421.31 | 441.57 | 461.82 | 482.08 | 502.33 |
| | $W_{3,4}$ | 18.42 | 37.99 | 58.07 | 78.29 | 98.54 | 118.79 | 139.05 | 159.30 | 179.56 | 199.81 | 220.07 | 240.32 | 260.58 | 280.83 | 301.09 | 321.34 | 341.60 | 361.85 | 382.10 | 402.36 | 422.61 | 442.87 | 463.12 | 483.38 | 503.63 |
| | $J$ | 14.41 | 23.85 | 29.63 | 33.30 | 35.74 | 37.41 | 38.60 | 39.46 | 40.10 | 40.58 | 40.94 | 41.23 | 41.45 | 41.63 | 41.77 | 41.89 | 41.98 | 42.06 | 42.12 | 42.18 | 42.22 | 42.26 | 42.29 | 42.32 | 42.34 |
| | $K$ | 2.40 | 3.12 | 3.62 | 3.69 | 3.74 | 3.75 | 3.76 | 3.76 | 3.76 | 3.76 | 3.76 | 3.76 | 3.76 | 3.76 | 3.76 | 3.76 | 3.76 | 3.76 | 3.76 | 3.76 | 3.76 | 3.76 | 3.76 | 3.76 | 3.76 |
| 0.5 | $U_1$ | 19.72 | 34.28 | 44.27 | 51.28 | 56.36 | 60.19 | 63.15 | 65.50 | 67.41 | 68.99 | 70.31 | 71.43 | 72.39 | 73.23 | 73.96 | 74.61 | 75.19 | 75.70 | 76.16 | 76.58 | 76.96 | 77.31 | 77.63 | 77.92 | 78.19 |
| | $U_2$ | 0.87 | 2.04 | 3.61 | 5.42 | 7.25 | 8.99 | 10.57 | 11.98 | 13.24 | 14.34 | 15.32 | 16.19 | 16.96 | 17.65 | 18.27 | 18.82 | 19.32 | 19.78 | 20.19 | 20.57 | 20.92 | 21.23 | 21.53 | 21.80 | 22.05 |
| | $W_{1,2}$ | 17.39 | 36.49 | 56.46 | 76.65 | 96.89 | 117.14 | 137.39 | 157.65 | 177.90 | 198.16 | 218.41 | 238.67 | 258.92 | 279.18 | 299.43 | 319.69 | 339.94 | 360.20 | 380.45 | 400.71 | 420.96 | 441.22 | 461.47 | 481.73 | 501.98 |
| | $W_{3,4}$ | 18.47 | 38.05 | 58.13 | 78.35 | 98.60 | 118.85 | 139.11 | 159.36 | 179.62 | 199.87 | 220.13 | 240.38 | 260.64 | 280.89 | 301.15 | 321.40 | 341.66 | 361.91 | 382.17 | 402.42 | 422.68 | 442.93 | 463.19 | 483.44 | 503.70 |
| | $J$ | 14.95 | 25.01 | 31.34 | 35.44 | 38.21 | 40.15 | 41.54 | 42.56 | 43.33 | 43.91 | 44.36 | 44.72 | 45.00 | 45.23 | 45.41 | 45.56 | 45.68 | 45.78 | 45.87 | 45.94 | 46.00 | 46.05 | 46.10 | 46.13 | 46.17 |
| | $K$ | 2.58 | 3.67 | 3.95 | 4.02 | 4.03 | 4.03 | 4.03 | 4.03 | 4.03 | 4.03 | 4.03 | 4.03 | 4.03 | 4.03 | 4.03 | 4.03 | 4.03 | 4.03 | 4.03 | 4.03 | 4.03 | 4.03 | 4.03 | 4.03 | 4.03 |
| 0.4 | $U_1$ | 18.57 | 32.44 | 42.06 | 48.86 | 53.83 | 57.59 | 60.50 | 62.82 | 64.71 | 66.27 | 67.58 | 68.69 | 69.65 | 70.48 | 71.21 | 71.86 | 72.43 | 72.94 | 73.40 | 73.82 | 74.20 | 74.54 | 74.86 | 75.15 | 75.42 |
| | $U_2$ | 0.92 | 2.14 | 3.74 | 5.56 | 7.39 | 9.12 | 10.69 | 12.09 | 13.34 | 14.44 | 15.41 | 16.28 | 17.04 | 17.73 | 18.34 | 18.90 | 19.40 | 19.85 | 20.26 | 20.64 | 20.98 | 21.30 | 21.60 | 21.87 | 22.12 |
| | $W_{1,2}$ | 17.17 | 36.19 | 56.13 | 76.31 | 96.56 | 116.81 | 137.06 | 157.32 | 177.57 | 197.83 | 218.08 | 238.34 | 258.59 | 278.85 | 299.10 | 319.36 | 339.61 | 359.86 | 380.12 | 400.37 | 420.63 | 440.88 | 461.14 | 481.39 | 501.65 |
| | $W_{3,4}$ | 18.52 | 38.12 | 58.21 | 78.43 | 98.67 | 118.93 | 139.18 | 159.44 | 179.69 | 199.95 | 220.20 | 240.46 | 260.71 | 280.97 | 301.22 | 321.48 | 341.73 | 361.99 | 382.24 | 402.49 | 422.75 | 443.01 | 463.26 | 483.51 | 503.77 |
| | $J$ | 15.41 | 26.00 | 32.78 | 37.25 | 40.29 | 42.44 | 44.00 | 45.15 | 46.02 | 46.69 | 47.21 | 47.62 | 47.95 | 48.22 | 48.43 | 48.61 | 48.76 | 48.88 | 48.98 | 49.07 | 49.14 | 49.21 | 49.26 | 49.31 | 49.35 |
| | $K$ | 2.76 | 3.93 | 4.23 | 4.30 | 4.31 | 4.31 | 4.31 | 4.31 | 4.31 | 4.31 | 4.31 | 4.31 | 4.31 | 4.31 | 4.31 | 4.31 | 4.31 | 4.31 | 4.31 | 4.31 | 4.31 | 4.31 | 4.31 | 4.31 | 4.31 |
| 0.3 | $U_1$ | 17.95 | 31.45 | 40.89 | 47.59 | 52.50 | 56.22 | 59.11 | 61.42 | 63.30 | 64.85 | 66.15 | 67.26 | 68.22 | 69.05 | 69.78 | 70.42 | 70.99 | 71.50 | 71.96 | 72.38 | 72.76 | 73.10 | 73.42 | 73.71 | 73.98 |
| | $U_2$ | 0.95 | 2.19 | 3.81 | 5.63 | 7.46 | 9.19 | 10.76 | 12.17 | 13.42 | 14.51 | 15.48 | 16.34 | 17.11 | 17.79 | 18.39 | 18.94 | 19.44 | 19.90 | 20.31 | 20.69 | 21.03 | 21.35 | 21.64 | 21.91 | 22.16 |
| | $W_{1,2}$ | 16.94 | 35.86 | 55.78 | 75.96 | 96.20 | 116.45 | 136.70 | 156.96 | 177.21 | 197.47 | 217.72 | 237.98 | 258.23 | 278.49 | 298.74 | 319.00 | 339.25 | 359.51 | 379.76 | 400.02 | 420.27 | 440.53 | 460.78 | 481.04 | 501.29 |
| | $W_{3,4}$ | 18.55 | 38.15 | 58.25 | 78.47 | 98.72 | 118.97 | 139.23 | 159.48 | 179.74 | 199.99 | 220.25 | 240.50 | 260.76 | 281.01 | 301.26 | 321.52 | 341.77 | 362.03 | 382.28 | 402.54 | 422.79 | 443.05 | 463.30 | 483.56 | 503.81 |
| | $J$ | 15.66 | 26.68 | 33.76 | 38.44 | 41.64 | 43.91 | 45.57 | 46.77 | 47.69 | 48.40 | 48.95 | 49.39 | 49.74 | 50.02 | 50.25 | 50.44 | 50.59 | 50.73 | 50.84 | 50.94 | 51.02 | 51.09 | 51.15 | 51.20 | 51.24 |
| | $K$ | 2.93 | 4.23 | 4.53 | 4.57 | 4.58 | 4.58 | 4.58 | 4.58 | 4.58 | 4.58 | 4.58 | 4.58 | 4.58 | 4.58 | 4.58 | 4.58 | 4.58 | 4.58 | 4.58 | 4.58 | 4.58 | 4.58 | 4.58 | 4.58 | 4.58 |
| 0.2 | $U_1$ | 17.72 | 31.13 | 40.54 | 47.23 | 52.15 | 55.88 | 58.77 | 61.07 | 62.95 | 64.51 | 65.81 | 66.92 | 67.88 | 68.71 | 69.44 | 70.08 | 70.65 | 71.17 | 71.63 | 72.04 | 72.42 | 72.77 | 73.09 | 73.38 | 73.65 |
| | $U_2$ | 0.95 | 2.20 | 3.82 | 5.64 | 7.48 | 9.20 | 10.77 | 12.18 | 13.43 | 14.52 | 15.49 | 16.35 | 17.12 | 17.80 | 18.41 | 18.96 | 19.46 | 19.92 | 20.33 | 20.71 | 21.06 | 21.37 | 21.67 | 21.94 | 22.19 |
| | $W_{1,2}$ | 16.70 | 35.52 | 55.42 | 75.60 | 95.83 | 116.09 | 136.34 | 156.59 | 176.85 | 197.10 | 217.36 | 237.61 | 257.87 | 278.12 | 298.38 | 318.63 | 338.89 | 359.14 | 379.40 | 399.65 | 419.91 | 440.16 | 460.42 | 480.67 | 500.93 |
| | $W_{3,4}$ | 18.56 | 38.16 | 58.25 | 78.47 | 98.72 | 118.97 | 139.23 | 159.48 | 179.74 | 199.99 | 220.24 | 240.50 | 260.75 | 281.01 | 301.26 | 321.52 | 341.77 | 362.03 | 382.28 | 402.54 | 422.79 | 443.05 | 463.30 | 483.56 | 503.81 |
| | $J$ | 15.96 | 27.10 | 34.31 | 39.08 | 42.34 | 44.64 | 46.30 | 47.53 | 48.46 | 49.18 | 49.74 | 50.17 | 50.52 | 50.80 | 51.03 | 51.22 | 51.38 | 51.51 | 51.62 | 51.71 | 51.79 | 51.86 | 51.92 | 51.97 | 52.02 |
| | $K$ | 3.10 | 4.40 | 4.74 | 4.80 | 4.81 | 4.81 | 4.81 | 4.81 | 4.81 | 4.81 | 4.81 | 4.81 | 4.81 | 4.81 | 4.81 | 4.81 | 4.81 | 4.81 | 4.81 | 4.81 | 4.81 | 4.81 | 4.81 | 4.81 | 4.81 |
| 0.1 | $U_1$ | 17.82 | 31.34 | 40.87 | 47.57 | 52.48 | 56.20 | 59.09 | 61.39 | 63.27 | 64.82 | 66.12 | 67.23 | 68.19 | 69.02 | 69.75 | 70.39 | 70.96 | 71.47 | 71.93 | 72.35 | 72.72 | 73.07 | 73.38 | 73.68 | 73.95 |
| | $U_2$ | 0.93 | 2.17 | 3.78 | 5.60 | 7.43 | 9.16 | 10.74 | 12.15 | 13.40 | 14.50 | 15.47 | 16.33 | 17.10 | 17.78 | 18.39 | 18.94 | 19.44 | 19.90 | 20.31 | 20.69 | 21.04 | 21.35 | 21.65 | 21.92 | 22.19 |
| | $W_{1,2}$ | 16.54 | 35.30 | 55.18 | 75.35 | 95.59 | 115.84 | 136.10 | 156.35 | 176.61 | 196.86 | 217.12 | 237.37 | 257.63 | 277.88 | 298.14 | 318.39 | 338.65 | 358.90 | 379.16 | 399.41 | 419.67 | 439.92 | 460.18 | 480.43 | 500.69 |
| | $W_{3,4}$ | 18.58 | 38.18 | 58.28 | 78.50 | 98.74 | 119.00 | 139.25 | 159.50 | 179.76 | 200.01 | 220.27 | 240.52 | 260.78 | 281.03 | 301.29 | 321.54 | 341.80 | 362.05 | 382.31 | 402.56 | 422.82 | 443.07 | 463.33 | 483.58 | 503.84 |
| | $J$ | 16.04 | 27.20 | 34.40 | 39.16 | 42.42 | 44.72 | 46.38 | 47.61 | 48.54 | 49.26 | 49.81 | 50.25 | 50.60 | 50.88 | 51.11 | 51.30 | 51.46 | 51.59 | 51.70 | 51.79 | 51.87 | 51.94 | 52.00 | 52.05 | 52.09 |
| | $K$ | 3.19 | 4.53 | 4.88 | 4.96 | 4.97 | 4.97 | 4.97 | 4.97 | 4.97 | 4.97 | 4.97 | 4.97 | 4.97 | 4.97 | 4.97 | 4.97 | 4.97 | 4.97 | 4.97 | 4.97 | 4.97 | 4.97 | 4.97 | 4.97 | 4.97 |
| 0.0 | $U_1$ | 18.18 | 31.91 | 41.51 | 48.32 | 53.30 | 57.07 | 59.99 | 62.32 | 64.22 | 65.78 | 67.10 | 68.21 | 69.17 | 70.01 | 70.74 | 71.38 | 71.96 | 72.47 | 72.93 | 73.35 | 73.73 | 74.07 | 74.39 | 74.68 | 74.95 |
| | $U_2$ | 0.91 | 2.13 | 3.74 | 5.56 | 7.40 | 9.14 | 10.71 | 12.12 | 13.37 | 14.47 | 15.45 | 16.32 | 17.08 | 17.77 | 18.39 | 18.94 | 19.44 | 19.89 | 20.31 | 20.69 | 21.03 | 21.35 | 21.64 | 21.92 | 22.17 |
| | $W_{1,2}$ | 16.39 | 35.09 | 54.90 | 75.15 | 95.36 | 115.62 | 135.87 | 156.12 | 176.38 | 196.63 | 216.89 | 237.14 | 257.40 | 277.65 | 297.91 | 318.16 | 338.42 | 358.67 | 378.93 | 399.18 | 419.44 | 439.69 | 459.95 | 480.20 | 500.46 |
| | $W_{3,4}$ | 18.60 | 38.21 | 58.30 | 78.52 | 98.77 | 119.02 | 139.28 | 159.53 | 179.79 | 200.04 | 220.30 | 240.55 | 260.81 | 281.06 | 301.32 | 321.57 | 341.83 | 362.08 | 382.34 | 402.59 | 422.85 | 443.10 | 463.36 | 483.61 | 503.87 |
| | $J$ | 16.10 | 27.29 | 34.47 | 39.18 | 42.35 | 44.56 | 46.15 | 47.30 | 48.16 | 48.84 | 49.32 | 49.72 | 50.03 | 50.28 | 50.48 | 50.65 | 50.78 | 50.90 | 50.99 | 51.07 | 51.14 | 51.19 | 51.24 | 51.29 | 51.32 |
| | $K$ | 3.23 | 4.58 | 4.93 | 5.00 | 5.02 | 5.02 | 5.02 | 5.02 | 5.02 | 5.02 | 5.02 | 5.02 | 5.02 | 5.02 | 5.02 | 5.02 | 5.02 | 5.02 | 5.02 | 5.02 | 5.02 | 5.02 | 5.02 | 5.02 | 5.02 |

TABLE S87. Parameters of the THF interaction Hamiltonian for different ratios $u_0/u_1$ and screening lengths $\xi$ of the double-gate interaction potential ($U_\xi = 24$ meV for $\xi = 10$ nm). We use the same BM model parameters as in Table S86.

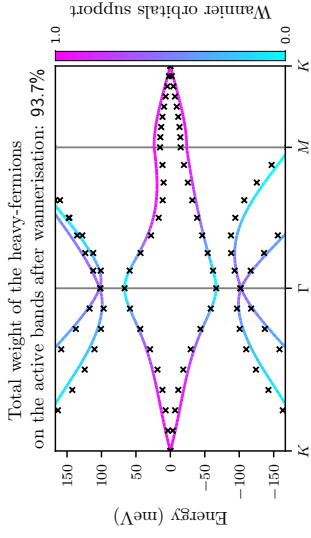

FIG. S60. Band structures of the BM and THF models near charge neutrality for $w_0/w_1 = 0.8$, depicted by lines and crosses, respectively. The BM bands are colored according to the weight of the $f$-electron wave function on them. We use the same BM parameters as in Table S88.

TABLE S88. Parameters of the $f$-electron Wannier functions and the THF single-particle Hamiltonian for different values of the tunneling ratio $w_0/w_1$. $W$ denotes the total weight of the THF $f$-electrons on the active bands. In computing the ratios $\gamma/U_1$ and $\gamma/U_2$, we employ the on-site and nearest-neighbor repulsion parameters $U_1$ and $U_2$ obtained numerically for $\xi = 10$ nm, as given in Table S89. We employ $v_F = 5.944$ eV·Å, $|\mathbf{K}| = 1.703$ Å$^{-1}$, $w_1 = 110$ meV, and $\theta = 1.54°$ for the BM model.

| $w_0/w_1$ | $\alpha_1$ | $\alpha_2$ | $\lambda_1$ | $\lambda_2$ | $\lambda$ | $W$ | $t_0$ | $\gamma$ | $M$ | $v'_\star$ | $v''_\star$ | $v'''_\star$ | $\gamma/U_1$ | $\gamma/U_2$ |
|---|---|---|---|---|---|---|---|---|---|---|---|---|---|---|
| 1.00 | 0.832 | 0.555 | 0.215 | 0.221 | 0.383 | 0.555 | 5.844 | -70.587 | -63.342 | -4.870 | 1.936 | 0.0420 | -0.94 | -10.71 |
| 0.90 | 0.859 | 0.512 | 0.227 | 0.227 | 0.401 | 0.934 | 6.837 | -86.965 | -65.027 | -4.968 | 1.923 | 0.0364 | -1.23 | -12.81 |
| 0.80 | 0.885 | 0.465 | 0.245 | 0.227 | 0.423 | 0.937 | 7.840 | -102.777 | -66.492 | -5.063 | 1.887 | 0.0312 | -1.55 | -14.67 |
| 0.70 | 0.909 | 0.416 | 0.257 | 0.233 | 0.435 | 0.937 | 8.702 | -117.841 | -67.752 | -5.157 | 1.823 | 0.0261 | -1.85 | -16.49 |
| 0.60 | 0.932 | 0.363 | 0.275 | 0.239 | 0.455 | 0.942 | 9.576 | -131.942 | -68.820 | -5.249 | 1.723 | 0.0211 | -2.19 | -18.02 |
| 0.50 | 0.952 | 0.306 | 0.292 | 0.245 | 0.485 | 0.954 | 10.524 | -144.816 | -69.705 | -5.339 | 1.578 | 0.0161 | -2.58 | -19.21 |
| 0.40 | 0.969 | 0.249 | 0.304 | 0.245 | 0.492 | 0.954 | 11.089 | -156.150 | -70.418 | -5.425 | 1.379 | 0.0113 | -2.86 | -20.50 |
| 0.30 | 0.982 | 0.190 | 0.322 | 0.251 | 0.498 | 0.956 | 11.560 | -165.579 | -70.965 | -5.503 | 1.118 | 0.0069 | -3.11 | -21.53 |
| 0.20 | 0.992 | 0.129 | 0.328 | 0.251 | 0.501 | 0.956 | 11.853 | -172.709 | -71.352 | -5.567 | 0.793 | 0.0033 | -3.29 | -22.32 |
| 0.10 | 0.998 | 0.066 | 0.334 | 0.257 | 0.499 | 0.953 | 11.948 | -177.171 | -71.583 | -5.610 | 0.413 | 0.0009 | -3.38 | -22.90 |
| 0.00 | 1.000 | 0.000 | 0.334 | 0.000 | 0.483 | 0.944 | 11.689 | -178.691 | -71.659 | -5.625 | -0.000 | -0.0000 | -3.30 | -23.42 |

| $w_0/w_1$ | $\xi$/nm | 2 | 4 | 6 | 8 | 10 | 12 | 14 | 16 | 18 | 20 | 22 | 24 | 26 | 28 | 30 | 32 | 34 | 36 | 38 | 40 | 42 | 44 | 46 | 48 | 50 |
|---|---|---|---|---|---|---|---|---|---|---|---|---|---|---|---|---|---|---|---|---|---|---|---|---|---|---|
| 1.0 | $U_1$ | 28.67 | 48.57 | 61.03 | 69.32 | 75.13 | 79.38 | 82.61 | 85.13 | 87.16 | 88.82 | 90.20 | 91.36 | 92.36 | 93.22 | 93.98 | 94.64 | 95.23 | 95.75 | 96.22 | 96.65 | 97.03 | 97.38 | 97.71 | 98.00 | 98.27 |
| | $U_2$ | 0.57 | 1.46 | 2.90 | 4.70 | 6.59 | 8.40 | 10.05 | 11.53 | 12.84 | 13.99 | 15.00 | 15.90 | 16.69 | 17.40 | 18.03 | 18.60 | 19.11 | 19.58 | 20.00 | 20.38 | 20.73 | 21.06 | 21.35 | 21.63 | 21.88 |
| | $W_{1,2}$ | 19.15 | 39.27 | 59.90 | 80.65 | 101.43 | 122.22 | 143.12 | 163.81 | 184.00 | 205.39 | 226.18 | 246.27 | 267.76 | 288.56 | 309.35 | 330.14 | 350.93 | 371.72 | 392.51 | 413.30 | 434.09 | 454.89 | 475.68 | 496.47 | 517.26 |
| | $W_{3,4}$ | 19.07 | 39.31 | 59.97 | 80.74 | 101.52 | 122.31 | 143.11 | 163.60 | 184.00 | 205.28 | 226.07 | 247.06 | 267.85 | 288.65 | 309.44 | 330.23 | 351.02 | 371.81 | 392.60 | 413.39 | 434.18 | 454.54 | 475.77 | 496.56 | 517.35 |
| | $J$ | 12.52 | 19.38 | 22.94 | 24.93 | 26.14 | 26.91 | 27.42 | 27.78 | 28.08 | 28.20 | 28.34 | 28.43 | 28.51 | 28.57 | 28.61 | 28.65 | 28.68 | 28.70 | 28.72 | 28.74 | 28.76 | 28.77 | 28.78 | 28.78 | 28.78 |
| | $K$ | 1.82 | 2.61 | 2.80 | 2.84 | 2.85 | 2.85 | 2.85 | 2.85 | 2.85 | 2.85 | 2.85 | 2.85 | 2.85 | 2.85 | 2.85 | 2.85 | 2.85 | 2.85 | 2.85 | 2.85 | 2.85 | 2.85 | 2.85 | 2.85 | 2.85 |
| 0.9 | $U_1$ | 26.74 | 45.20 | 57.14 | 65.17 | 70.83 | 75.00 | 78.18 | 80.67 | 82.68 | 84.32 | 85.69 | 86.85 | 87.84 | 88.70 | 89.45 | 90.11 | 90.74 | 91.22 | 91.69 | 92.11 | 92.50 | 92.85 | 93.17 | 93.46 | 93.73 |
| | $U_2$ | 0.63 | 1.60 | 3.08 | 4.90 | 6.79 | 8.59 | 10.24 | 11.71 | 13.01 | 14.15 | 15.16 | 16.05 | 16.84 | 17.55 | 18.18 | 18.74 | 19.25 | 19.72 | 20.14 | 20.52 | 20.87 | 21.19 | 21.49 | 21.77 | 22.02 |
| | $W_{1,2}$ | 18.92 | 38.95 | 59.56 | 80.31 | 101.09 | 121.88 | 142.67 | 163.46 | 184.25 | 205.05 | 225.84 | 246.63 | 267.42 | 288.21 | 309.00 | 329.79 | 350.59 | 371.38 | 392.17 | 412.96 | 433.75 | 454.54 | 475.33 | 496.13 | 516.92 |
| | $W_{3,4}$ | 18.94 | 39.11 | 59.75 | 80.51 | 101.30 | 122.09 | 142.88 | 163.67 | 184.46 | 205.26 | 226.04 | 246.84 | 267.63 | 288.42 | 309.21 | 330.00 | 350.79 | 371.59 | 392.38 | 413.17 | 433.96 | 454.75 | 475.54 | 496.33 | 517.13 |
| | $J$ | 13.04 | 20.62 | 24.76 | 27.18 | 28.68 | 29.67 | 30.34 | 30.86 | 31.31 | 31.55 | 31.87 | 31.93 | 31.98 | 32.02 | 32.08 | 32.10 | 32.12 | 32.13 | 32.13 | 32.14 | 32.15 | 32.15 | 32.16 | 32.16 | 32.16 |
| | $K$ | 1.94 | 2.76 | 2.97 | 3.02 | 3.03 | 3.03 | 3.03 | 3.03 | 3.03 | 3.03 | 3.03 | 3.03 | 3.03 | 3.03 | 3.03 | 3.03 | 3.03 | 3.03 | 3.03 | 3.03 | 3.03 | 3.03 | 3.03 | 3.03 | 3.03 |
| 0.8 | $U_1$ | 24.56 | 41.84 | 53.21 | 60.95 | 66.45 | 70.52 | 73.64 | 76.09 | 78.07 | 79.70 | 81.05 | 82.20 | 83.19 | 84.04 | 84.78 | 85.44 | 86.02 | 86.54 | 87.01 | 87.43 | 87.82 | 88.17 | 88.49 | 88.78 | 89.05 |
| | $U_2$ | 0.71 | 1.75 | 3.28 | 5.12 | 7.01 | 8.80 | 10.44 | 11.90 | 13.19 | 14.32 | 15.33 | 16.21 | 17.00 | 17.70 | 18.33 | 18.89 | 19.40 | 19.86 | 20.28 | 20.66 | 21.01 | 21.33 | 21.63 | 21.90 | 22.16 |
| | $W_{1,2}$ | 18.68 | 38.62 | 59.20 | 79.95 | 100.73 | 121.52 | 142.31 | 163.11 | 183.90 | 204.69 | 225.48 | 246.27 | 267.06 | 287.85 | 308.64 | 329.44 | 350.23 | 371.02 | 391.81 | 412.60 | 433.39 | 454.18 | 474.98 | 495.77 | 516.56 |
| | $W_{3,4}$ | 18.50 | 39.02 | 59.65 | 80.41 | 101.20 | 121.99 | 142.78 | 163.57 | 184.36 | 205.15 | 225.95 | 246.73 | 267.53 | 288.32 | 309.11 | 329.90 | 350.69 | 371.48 | 392.27 | 413.07 | 433.86 | 454.65 | 475.44 | 496.23 | 517.02 |
| | $J$ | 13.63 | 21.96 | 26.74 | 29.64 | 31.51 | 32.75 | 33.61 | 34.22 | 34.99 | 35.43 | 35.57 | 35.78 | 35.85 | 35.91 | 35.95 | 35.99 | 36.02 | 36.05 | 36.07 | 36.09 | 36.11 | 36.12 | 36.12 | 36.13 | 36.13 |
| | $K$ | 2.08 | 2.95 | 3.17 | 3.22 | 3.23 | 3.23 | 3.23 | 3.23 | 3.23 | 3.23 | 3.23 | 3.23 | 3.23 | 3.23 | 3.23 | 3.23 | 3.23 | 3.23 | 3.23 | 3.23 | 3.23 | 3.23 | 3.23 | 3.23 | 3.23 |
| 0.7 | $U_1$ | 23.10 | 39.61 | 50.61 | 58.16 | 63.56 | 67.58 | 70.66 | 73.09 | 75.06 | 76.67 | 78.02 | 79.16 | 80.14 | 80.99 | 81.73 | 82.39 | 82.97 | 83.49 | 83.96 | 84.38 | 84.76 | 85.11 | 85.43 | 85.72 | 85.99 |
| | $U_2$ | 0.76 | 1.85 | 3.41 | 5.26 | 7.15 | 8.94 | 10.57 | 12.02 | 13.31 | 14.44 | 15.44 | 16.32 | 17.11 | 17.81 | 18.43 | 18.99 | 19.50 | 19.97 | 20.38 | 20.76 | 21.11 | 21.43 | 21.73 | 22.00 | 22.26 |
| | $W_{1,2}$ | 18.40 | 38.23 | 58.79 | 79.54 | 100.31 | 121.10 | 141.89 | 162.68 | 183.48 | 204.27 | 225.06 | 245.85 | 266.64 | 287.43 | 308.22 | 329.02 | 349.81 | 370.60 | 391.39 | 412.18 | 432.97 | 453.76 | 474.55 | 495.35 | 516.14 |
| | $W_{3,4}$ | 18.87 | 38.96 | 59.59 | 80.34 | 101.13 | 121.92 | 142.71 | 163.50 | 184.29 | 205.08 | 225.87 | 246.67 | 267.46 | 288.25 | 309.04 | 329.83 | 350.62 | 371.41 | 392.21 | 413.00 | 433.79 | 454.58 | 475.37 | 496.16 | 516.95 |
| | $J$ | 14.16 | 23.10 | 28.38 | 31.64 | 33.75 | 35.18 | 36.07 | 37.39 | 37.77 | 38.06 | 38.28 | 38.45 | 38.58 | 38.69 | 38.77 | 38.84 | 38.90 | 38.97 | 39.01 | 39.04 | 39.08 | 39.10 | 39.10 | 39.11 | 39.11 |
| | $K$ | 2.23 | 3.17 | 3.40 | 3.45 | 3.46 | 3.47 | 3.47 | 3.47 | 3.47 | 3.47 | 3.47 | 3.47 | 3.47 | 3.47 | 3.47 | 3.47 | 3.47 | 3.47 | 3.47 | 3.47 | 3.47 | 3.47 | 3.47 | 3.47 | 3.47 |

| $u_0/u_1$ | $\xi$/nm | 2 | 4 | 6 | 8 | 10 | 12 | 14 | 16 | 18 | 20 | 22 | 24 | 26 | 28 | 30 | 32 | 34 | 36 | 38 | 40 | 42 | 44 | 46 | 48 | 50 |
|---|---|---|---|---|---|---|---|---|---|---|---|---|---|---|---|---|---|---|---|---|---|---|---|---|---|---|
| 0.6 | $U_1$ | 21.48 | 37.07 | 47.61 | 54.91 | 60.17 | 64.10 | 67.13 | 69.53 | 71.47 | 73.07 | 74.40 | 75.54 | 76.51 | 77.35 | 78.09 | 78.74 | 79.32 | 79.84 | 80.31 | 80.73 | 81.11 | 81.46 | 81.77 | 82.07 | 82.34 |
| | $U_2$ | 0.83 | 1.98 | 3.58 | 5.43 | 7.32 | 9.10 | 10.72 | 12.17 | 13.44 | 14.57 | 15.56 | 16.44 | 17.22 | 17.92 | 18.54 | 19.10 | 19.61 | 20.07 | 20.48 | 20.86 | 21.21 | 21.53 | 21.83 | 22.10 | 22.36 |
| | $W_{1,2}$ | 17.69 | 37.89 | 58.42 | 79.15 | 99.93 | 120.72 | 141.51 | 162.31 | 183.10 | 203.89 | 224.68 | 245.47 | 266.26 | 287.05 | 307.84 | 328.64 | 349.43 | 370.22 | 391.01 | 411.80 | 432.59 | 453.38 | 474.18 | 494.97 | 515.76 |
| | $W_{3,4}$ | 18.90 | 38.99 | 59.61 | 80.37 | 101.15 | 121.94 | 142.74 | 163.53 | 184.32 | 205.11 | 225.90 | 246.69 | 267.48 | 288.27 | 309.07 | 329.86 | 350.65 | 371.44 | 392.23 | 413.02 | 433.81 | 454.61 | 475.40 | 496.19 | 516.98 |
| | $J$ | 14.75 | 24.37 | 30.22 | 33.93 | 36.39 | 38.07 | 39.26 | 40.12 | 40.75 | 41.23 | 41.59 | 41.87 | 42.09 | 42.27 | 42.41 | 42.52 | 42.62 | 42.69 | 42.76 | 42.81 | 42.85 | 42.89 | 42.92 | 42.95 | 42.97 |
| | $K$ | 2.40 | 3.41 | 3.66 | 3.71 | 3.72 | 3.72 | 3.72 | 3.72 | 3.72 | 3.72 | 3.72 | 3.72 | 3.72 | 3.72 | 3.72 | 3.72 | 3.72 | 3.72 | 3.72 | 3.72 | 3.72 | 3.72 | 3.72 | 3.72 | 3.72 |
| 0.5 | $U_1$ | 19.03 | 34.10 | 44.04 | 51.01 | 56.08 | 59.89 | 62.84 | 65.19 | 67.09 | 68.66 | 69.98 | 71.10 | 72.07 | 72.90 | 73.63 | 74.28 | 74.85 | 75.37 | 75.83 | 76.25 | 76.63 | 76.97 | 77.29 | 77.58 | 77.85 |
| | $U_2$ | 0.92 | 2.16 | 3.80 | 5.66 | 7.54 | 9.30 | 10.91 | 12.33 | 13.60 | 14.71 | 15.70 | 16.57 | 17.34 | 18.03 | 18.65 | 19.21 | 19.71 | 20.17 | 20.58 | 20.96 | 21.31 | 21.63 | 21.92 | 22.20 | 22.45 |
| | $W_{1,2}$ | 17.45 | 37.62 | 58.14 | 78.89 | 99.66 | 120.44 | 141.23 | 162.02 | 182.81 | 203.60 | 224.40 | 245.19 | 265.98 | 286.77 | 307.56 | 328.35 | 349.14 | 369.94 | 390.73 | 411.52 | 432.31 | 453.10 | 473.89 | 494.68 | 515.47 |
| | $W_{3,4}$ | 19.10 | 39.10 | 59.73 | 80.49 | 101.27 | 122.06 | 142.85 | 163.64 | 184.43 | 205.23 | 226.02 | 246.81 | 267.60 | 288.39 | 309.18 | 329.97 | 350.77 | 371.56 | 392.35 | 413.14 | 433.93 | 454.72 | 475.51 | 496.31 | 517.10 |
| | $J$ | 15.36 | 25.72 | 32.25 | 36.50 | 39.39 | 41.41 | 42.87 | 43.95 | 44.76 | 45.38 | 45.86 | 46.23 | 46.54 | 46.78 | 46.97 | 47.13 | 47.27 | 47.38 | 47.47 | 47.55 | 47.62 | 47.67 | 47.72 | 47.76 | 47.80 |
| | $K$ | 2.58 | 3.66 | 3.93 | 3.99 | 4.00 | 4.00 | 4.00 | 4.00 | 4.00 | 4.00 | 4.00 | 4.00 | 4.00 | 4.00 | 4.00 | 4.00 | 4.00 | 4.00 | 4.00 | 4.00 | 4.00 | 4.00 | 4.00 | 4.00 | 4.00 |
| 0.4 | $U_1$ | 18.88 | 32.94 | 42.68 | 49.54 | 54.54 | 58.32 | 61.25 | 63.58 | 65.47 | 67.04 | 68.35 | 69.47 | 70.43 | 71.26 | 71.99 | 72.64 | 73.21 | 73.72 | 74.18 | 74.60 | 74.98 | 75.33 | 75.64 | 75.93 | 76.21 |
| | $U_2$ | 0.95 | 2.21 | 3.87 | 5.74 | 7.62 | 9.38 | 10.98 | 12.40 | 13.66 | 14.78 | 15.76 | 16.63 | 17.40 | 18.09 | 18.71 | 19.26 | 19.77 | 20.23 | 20.64 | 21.01 | 21.36 | 21.68 | 21.97 | 22.25 | 22.50 |
| | $W_{1,2}$ | 17.69 | 37.25 | 57.74 | 78.47 | 99.25 | 120.03 | 140.83 | 161.62 | 182.41 | 203.20 | 223.99 | 244.78 | 265.57 | 286.36 | 307.16 | 327.95 | 348.74 | 369.53 | 390.32 | 411.11 | 431.90 | 452.70 | 473.49 | 494.28 | 515.07 |
| | $W_{3,4}$ | 19.13 | 39.13 | 59.76 | 80.52 | 101.30 | 122.09 | 142.88 | 163.68 | 184.47 | 205.26 | 226.05 | 246.84 | 267.63 | 288.42 | 309.22 | 330.01 | 350.80 | 371.59 | 392.38 | 413.17 | 433.96 | 454.75 | 475.55 | 496.34 | 517.13 |
| | $J$ | 15.76 | 26.54 | 33.41 | 37.91 | 40.97 | 43.13 | 44.68 | 45.83 | 46.70 | 47.36 | 47.88 | 48.28 | 48.61 | 48.87 | 49.08 | 49.26 | 49.40 | 49.52 | 49.62 | 49.71 | 49.78 | 49.84 | 49.90 | 49.94 | 49.98 |
| | $K$ | 2.77 | 3.92 | 4.21 | 4.27 | 4.28 | 4.29 | 4.29 | 4.29 | 4.29 | 4.29 | 4.29 | 4.29 | 4.29 | 4.29 | 4.29 | 4.29 | 4.29 | 4.29 | 4.29 | 4.29 | 4.29 | 4.29 | 4.29 | 4.29 | 4.29 |
| 0.3 | $U_1$ | 18.27 | 31.97 | 41.52 | 48.28 | 53.23 | 56.97 | 59.88 | 62.20 | 64.08 | 65.64 | 66.95 | 68.00 | 69.02 | 69.85 | 70.58 | 71.23 | 71.80 | 72.31 | 72.77 | 73.19 | 73.57 | 73.91 | 74.23 | 74.52 | 74.79 |
| | $U_2$ | 0.98 | 2.27 | 3.94 | 5.82 | 7.69 | 9.45 | 11.04 | 12.45 | 13.72 | 14.83 | 15.81 | 16.68 | 17.45 | 18.14 | 18.76 | 19.31 | 19.81 | 20.27 | 20.68 | 21.06 | 21.41 | 21.73 | 22.02 | 22.29 | 22.54 |
| | $W_{1,2}$ | 17.45 | 36.92 | 57.39 | 78.11 | 98.89 | 119.68 | 140.47 | 161.26 | 182.05 | 202.84 | 223.63 | 244.42 | 265.22 | 286.01 | 306.80 | 327.59 | 348.38 | 369.17 | 389.96 | 410.75 | 431.54 | 452.33 | 473.13 | 493.92 | 514.71 |
| | $W_{3,4}$ | 19.11 | 39.11 | 59.81 | 80.55 | 101.35 | 122.14 | 142.93 | 163.72 | 184.51 | 205.31 | 226.10 | 246.89 | 267.68 | 288.47 | 309.26 | 330.05 | 350.84 | 371.64 | 392.43 | 413.22 | 434.01 | 454.80 | 475.59 | 496.38 | 517.17 |
| | $J$ | 16.10 | 27.23 | 34.39 | 39.11 | 42.33 | 44.59 | 46.23 | 47.45 | 48.36 | 49.06 | 49.61 | 50.04 | 50.39 | 50.67 | 50.89 | 51.08 | 51.24 | 51.37 | 51.47 | 51.57 | 51.65 | 51.71 | 51.77 | 51.82 | 51.86 |
| | $K$ | 2.95 | 4.17 | 4.47 | 4.54 | 4.56 | 4.56 | 4.56 | 4.56 | 4.56 | 4.56 | 4.56 | 4.56 | 4.56 | 4.56 | 4.56 | 4.56 | 4.56 | 4.56 | 4.56 | 4.56 | 4.56 | 4.56 | 4.56 | 4.56 | 4.56 |
| 0.2 | $U_1$ | 17.85 | 31.37 | 40.81 | 47.51 | 52.43 | 56.15 | 59.05 | 61.36 | 63.24 | 64.79 | 66.10 | 67.22 | 68.17 | 69.01 | 69.73 | 70.38 | 70.95 | 71.46 | 71.92 | 72.34 | 72.71 | 73.05 | 73.37 | 73.66 | 73.93 |
| | $U_2$ | 0.99 | 2.30 | 3.98 | 5.86 | 7.74 | 9.50 | 11.09 | 12.51 | 13.76 | 14.87 | 15.86 | 16.72 | 17.50 | 18.18 | 18.80 | 19.36 | 19.86 | 20.31 | 20.73 | 21.10 | 21.45 | 21.77 | 22.06 | 22.33 | 22.58 |
| | $W_{1,2}$ | 17.08 | 36.41 | 56.84 | 77.56 | 98.33 | 119.12 | 139.91 | 160.70 | 181.49 | 202.28 | 223.08 | 243.87 | 264.66 | 285.45 | 306.24 | 327.03 | 347.82 | 368.61 | 389.41 | 410.20 | 430.99 | 451.78 | 472.57 | 493.36 | 514.41 |
| | $W_{3,4}$ | 19.07 | 39.21 | 59.84 | 80.60 | 101.38 | 122.17 | 142.96 | 163.75 | 184.54 | 205.34 | 226.13 | 246.92 | 267.71 | 288.50 | 309.29 | 330.08 | 350.88 | 371.67 | 392.46 | 413.25 | 434.04 | 454.83 | 475.63 | 496.42 | 517.21 |
| | $J$ | 16.37 | 27.77 | 35.14 | 40.01 | 43.34 | 45.69 | 47.38 | 48.64 | 49.59 | 50.32 | 50.89 | 51.33 | 51.69 | 51.98 | 52.22 | 52.41 | 52.57 | 52.70 | 52.82 | 52.91 | 53.00 | 53.07 | 53.13 | 53.18 | 53.22 |
| | $K$ | 3.10 | 4.38 | 4.70 | 4.77 | 4.78 | 4.79 | 4.79 | 4.79 | 4.79 | 4.79 | 4.79 | 4.79 | 4.79 | 4.79 | 4.79 | 4.79 | 4.79 | 4.79 | 4.79 | 4.79 | 4.79 | 4.79 | 4.79 | 4.79 | 4.79 |
| 0.1 | $U_1$ | 17.85 | 31.33 | 40.78 | 47.49 | 52.42 | 56.14 | 59.05 | 61.36 | 63.24 | 64.80 | 66.12 | 67.24 | 68.20 | 69.04 | 69.77 | 70.42 | 71.00 | 71.51 | 71.98 | 72.40 | 72.78 | 73.12 | 73.45 | 73.73 | 73.94 |
| | $U_2$ | 0.93 | 2.29 | 3.98 | 5.86 | 7.77 | 9.53 | 11.12 | 12.54 | 13.80 | 14.92 | 15.90 | 16.77 | 17.55 | 18.24 | 18.86 | 19.42 | 19.92 | 20.38 | 20.79 | 21.16 | 21.52 | 21.84 | 22.13 | 22.41 | 22.59 |
| | $W_{1,2}$ | 16.90 | 36.15 | 56.56 | 77.27 | 98.05 | 118.83 | 139.63 | 160.42 | 181.21 | 202.00 | 222.79 | 243.58 | 264.37 | 285.16 | 305.96 | 326.75 | 347.54 | 368.33 | 389.12 | 409.91 | 430.70 | 451.50 | 472.29 | 493.08 | 514.15 |
| | $W_{3,4}$ | 19.12 | 39.12 | 59.75 | 80.60 | 101.29 | 122.08 | 142.87 | 163.66 | 184.45 | 205.24 | 226.04 | 246.83 | 267.62 | 288.41 | 309.20 | 329.99 | 350.78 | 371.57 | 392.36 | 413.15 | 433.95 | 454.74 | 475.53 | 496.57 | 517.22 |
| | $J$ | 16.46 | 28.02 | 35.47 | 40.38 | 43.73 | 46.08 | 47.78 | 49.03 | 49.97 | 50.69 | 51.25 | 51.69 | 52.05 | 52.33 | 52.56 | 52.75 | 52.90 | 53.03 | 53.14 | 53.23 | 53.31 | 53.39 | 53.44 | 53.49 | 53.54 |
| | $K$ | 3.20 | 4.53 | 4.86 | 4.93 | 4.94 | 4.94 | 4.94 | 4.94 | 4.94 | 4.94 | 4.94 | 4.94 | 4.94 | 4.94 | 4.94 | 4.94 | 4.94 | 4.94 | 4.94 | 4.94 | 4.94 | 4.94 | 4.94 | 4.94 | 4.94 |
| 0.0 | $U_1$ | 18.52 | 32.46 | 42.18 | 49.06 | 54.08 | 57.87 | 60.87 | 63.13 | 65.05 | 66.62 | 67.94 | 69.06 | 70.02 | 70.85 | 71.59 | 72.23 | 72.81 | 73.32 | 73.78 | 74.20 | 74.58 | 74.93 | 75.25 | 75.54 | 75.81 |
| | $U_2$ | 0.93 | 2.20 | 3.87 | 5.75 | 7.63 | 9.40 | 11.00 | 12.43 | 13.70 | 14.81 | 15.80 | 16.67 | 17.44 | 18.13 | 18.75 | 19.31 | 19.81 | 20.27 | 20.68 | 21.06 | 21.41 | 21.73 | 22.02 | 22.30 | 22.55 |
| | $W_{1,2}$ | 16.90 | 36.16 | 56.56 | 77.27 | 98.05 | 118.84 | 139.63 | 160.42 | 181.21 | 202.00 | 222.79 | 243.58 | 264.37 | 285.17 | 305.96 | 326.75 | 347.54 | 368.33 | 389.12 | 409.91 | 430.70 | 451.49 | 472.29 | 493.08 | 513.87 |
| | $W_{3,4}$ | 19.39 | 39.43 | 59.75 | 80.48 | 101.25 | 122.04 | 142.83 | 163.62 | 184.41 | 205.20 | 225.99 | 246.78 | 267.58 | 288.37 | 309.16 | 329.95 | 350.74 | 371.53 | 392.32 | 413.11 | 433.90 | 454.69 | 475.49 | 496.28 | 517.11 |
| | $J$ | 16.46 | 27.84 | 35.10 | 39.83 | 43.02 | 45.22 | 46.80 | 47.96 | 48.80 | 49.45 | 49.95 | 50.34 | 50.65 | 50.89 | 51.09 | 51.25 | 51.39 | 51.51 | 51.61 | 51.67 | 51.74 | 51.79 | 51.84 | 51.88 | 51.92 |
| | $K$ | 3.24 | 4.58 | 4.91 | 4.98 | 5.00 | 5.00 | 5.00 | 5.00 | 5.00 | 5.00 | 5.00 | 5.00 | 5.00 | 5.00 | 5.00 | 5.00 | 5.00 | 5.00 | 5.00 | 5.00 | 5.00 | 5.00 | 5.00 | 5.00 | 5.00 |

TABLE S89. Parameters of the THF interaction Hamiltonian for different ratios $u_0/u_1$ and screening lengths $\xi$ of the double-gate interaction potential ($U_\xi = 24$ meV for $\xi = 10$ nm). We use the same BM model parameters as in Table S88.

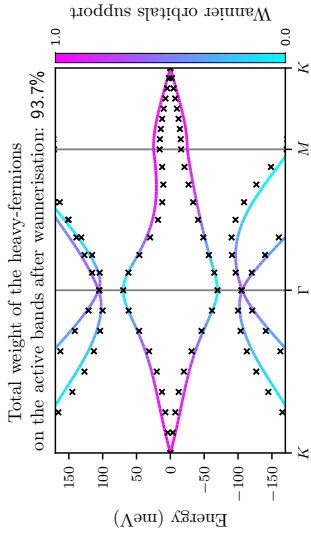

Total weight of the heavy-fermions on the active bands after wannierisation: 93.7%

Wannier orbitals support 0.0 — 1.0

Energy (meV) vs Γ M K Γ K

FIG. S61. Band structures of the BM and THF models near charge neutrality for $w_0/w_1 = 0.8$, depicted by lines and crosses, respectively. The BM bands are colored according to the weight of the $f$-electron wave function on them. We use the same BM parameters as in Table S90.

TABLE S90. Parameters of the $f$-electron Wannier functions and the THF single-particle Hamiltonian for different values of the tunneling ratio $w_0/w_1$. $W$ denotes the total weight of the THF $f$-electrons on the active bands. In computing the ratios $\gamma/U_1$ and $\gamma/U_2$, we employ the on-site and nearest-neighbor repulsion parameters $U_1$ and $U_2$ obtained numerically for $\xi = 10$ nm, as given in Table S91. We employ $v_F = 5.944$ eV·Å, $|\mathbf{K}| = 1.703 \text{Å}^{-1}$, $w_1 = 110$ meV, and $\theta = 1.56°$ for the BM model.

| $w_0/w_1$ | $\alpha_1$ | $\alpha_2$ | $\lambda_1$ | $\lambda_2$ | $\lambda$ | $W$ | $t_0$ | $\gamma$ | $M$ | $v_\star$ | $v'_\star$ | $\gamma/U_1$ | $\gamma/U_2$ |
|---|---|---|---|---|---|---|---|---|---|---|---|---|---|
| 1.00 | 0.834 | 0.551 | 0.221 | 0.385 | 0.932 | 6.240 | -73.939 | -66.514 | -4.886 | 1.945 | 0.0443 | -0.98 | -10.82 |
| 0.90 | 0.861 | 0.508 | 0.227 | 0.406 | 0.935 | 7.333 | -90.329 | -68.205 | -4.982 | 1.930 | 0.0386 | -1.27 | -12.82 |
| 0.80 | 0.887 | 0.462 | 0.233 | 0.425 | 0.937 | 8.303 | -106.145 | -69.677 | -5.076 | 1.893 | 0.0332 | -1.58 | -14.65 |
| 0.70 | 0.911 | 0.413 | 0.263 | 0.436 | 0.936 | 9.186 | -121.207 | -70.942 | -5.168 | 1.827 | 0.0279 | -1.89 | -16.42 |
| 0.60 | 0.933 | 0.359 | 0.281 | 0.462 | 0.945 | 10.109 | -135.301 | -72.015 | -5.260 | 1.726 | 0.0226 | -2.24 | -17.85 |
| 0.50 | 0.953 | 0.303 | 0.292 | 0.485 | 0.954 | 11.056 | -148.165 | -72.906 | -5.349 | 1.580 | 0.0173 | -2.61 | -19.07 |
| 0.40 | 0.969 | 0.247 | 0.310 | 0.492 | 0.954 | 11.631 | -159.486 | -73.622 | -5.435 | 1.381 | 0.0121 | -2.89 | -20.32 |
| 0.30 | 0.982 | 0.188 | 0.322 | 0.498 | 0.955 | 12.112 | -168.901 | -74.173 | -5.513 | 1.119 | 0.0074 | -3.13 | -21.33 |
| 0.20 | 0.992 | 0.128 | 0.334 | 0.501 | 0.955 | 12.406 | -176.020 | -74.562 | -5.577 | 0.793 | 0.0035 | -3.31 | -22.10 |
| 0.10 | 0.998 | 0.065 | 0.340 | 0.499 | 0.953 | 12.508 | -180.474 | -74.794 | -5.620 | 0.413 | 0.0009 | -3.39 | -22.66 |
| 0.00 | 1.000 | 0.000 | 0.340 | 0.498 | 0.951 | 12.363 | -181.992 | -74.871 | -5.635 | 0.000 | -0.0000 | -3.39 | -22.99 |

FIG. S61. Band structures of the BM and THF models near charge neutrality for $w_0/w_1 = 0.8$, depicted by lines and crosses, respectively. The BM bands are colored according to the weight of the $f$-electron wave function on them. We use the same BM parameters as in Table S90.

| $w_0/w_1$ | | $\xi/$nm 2 | 4 | 6 | 8 | 10 | 12 | 14 | 16 | 18 | 20 | 22 | 24 | 26 | 28 | 30 | 32 | 34 | 36 | 38 | 40 | 42 | 44 | 46 | 48 | 50 |
|---|---|---|---|---|---|---|---|---|---|---|---|---|---|---|---|---|---|---|---|---|---|---|---|---|---|---|
| 1.0 | $U_1$ | 29.20 | 48.92 | 61.45 | 69.77 | 75.59 | 79.85 | 83.08 | 85.62 | 87.65 | 89.31 | 90.69 | 91.80 | 92.85 | 93.72 | 94.47 | 95.13 | 95.72 | 96.25 | 96.72 | 97.14 | 97.53 | 97.88 | 98.20 | 98.50 | 98.77 |
| | $U_2$ | 0.59 | 1.53 | 3.03 | 4.89 | 6.83 | 8.68 | 10.38 | 12.05 | 13.18 | 14.34 | 15.36 | 16.26 | 17.06 | 17.77 | 18.41 | 18.98 | 19.49 | 19.96 | 20.38 | 20.77 | 21.12 | 21.44 | 21.74 | 22.02 | 22.27 |
| | $W_{1,2}$ | 19.71 | 40.39 | 61.57 | 82.87 | 104.20 | 125.53 | 146.87 | 168.20 | 189.54 | 210.87 | 232.21 | 253.46 | 274.88 | 296.21 | 317.55 | 338.88 | 360.21 | 381.55 | 402.88 | 424.21 | 445.55 | 466.89 | 488.22 | 509.56 | 530.89 |
| | $W_{3,4}$ | 19.52 | 40.27 | 61.48 | 82.79 | 104.12 | 125.45 | 146.78 | 168.11 | 189.45 | 210.79 | 232.12 | 253.46 | 274.79 | 296.13 | 317.46 | 338.80 | 360.13 | 381.47 | 402.80 | 424.14 | 445.47 | 466.81 | 488.14 | 509.48 | 530.81 |
| | $J$ | 12.80 | 19.82 | 23.45 | 25.49 | 26.72 | 27.51 | 28.03 | 28.38 | 28.64 | 28.82 | 28.95 | 29.05 | 29.12 | 29.18 | 29.23 | 29.26 | 29.29 | 29.31 | 29.33 | 29.35 | 29.36 | 29.37 | 29.38 | 29.39 | 29.39 |
| | $K$ | 1.81 | 2.57 | 2.75 | 2.79 | 2.80 | 2.80 | 2.80 | 2.80 | 2.80 | 2.80 | 2.80 | 2.80 | 2.80 | 2.80 | 2.80 | 2.80 | 2.80 | 2.80 | 2.80 | 2.80 | 2.80 | 2.80 | 2.80 | 2.80 | 2.80 |
| 0.9 | $U_1$ | 26.85 | 45.35 | 57.31 | 65.35 | 71.01 | 75.18 | 78.36 | 80.86 | 82.86 | 84.51 | 85.88 | 87.04 | 88.03 | 88.89 | 89.64 | 90.30 | 90.88 | 91.40 | 91.87 | 92.30 | 92.68 | 93.03 | 93.35 | 93.65 | 93.92 |
| | $U_2$ | 0.67 | 1.69 | 3.23 | 5.11 | 7.05 | 8.89 | 10.56 | 12.05 | 13.36 | 14.51 | 15.53 | 16.43 | 17.22 | 17.93 | 18.57 | 19.13 | 19.65 | 20.11 | 20.53 | 20.92 | 21.27 | 21.59 | 21.89 | 22.16 | 22.42 |
| | $W_{1,2}$ | 19.67 | 40.07 | 61.23 | 82.53 | 103.85 | 125.19 | 146.52 | 167.86 | 189.19 | 210.53 | 231.86 | 253.19 | 274.53 | 295.86 | 317.20 | 338.53 | 359.87 | 381.20 | 402.54 | 423.87 | 445.21 | 466.54 | 487.88 | 509.21 | 530.55 |
| | $W_{3,4}$ | 19.42 | 40.10 | 61.29 | 82.60 | 103.92 | 125.26 | 146.59 | 167.93 | 189.26 | 210.60 | 231.93 | 253.27 | 274.60 | 295.94 | 317.27 | 338.61 | 359.94 | 381.28 | 402.61 | 423.95 | 445.28 | 466.61 | 487.95 | 509.28 | 530.62 |
| | $J$ | 13.36 | 21.12 | 25.37 | 27.86 | 29.41 | 30.42 | 31.11 | 31.59 | 31.92 | 32.17 | 32.35 | 32.49 | 32.59 | 32.68 | 32.74 | 32.79 | 32.83 | 32.89 | 32.93 | 32.95 | 32.96 | 32.97 | 32.98 | 32.98 | — |
| | $K$ | 1.93 | 2.73 | 2.93 | 2.97 | 2.98 | 2.98 | 2.98 | 2.98 | 2.98 | 2.98 | 2.98 | 2.98 | 2.98 | 2.98 | 2.98 | 2.98 | 2.98 | 2.98 | 2.98 | 2.98 | 2.98 | 2.98 | 2.98 | 2.98 | 2.98 |
| 0.8 | $U_1$ | 24.83 | 42.27 | 53.72 | 61.49 | 67.01 | 71.10 | 74.23 | 76.69 | 78.67 | 80.30 | 81.66 | 82.81 | 83.80 | 84.65 | 85.39 | 86.05 | 86.63 | 87.16 | 87.62 | 88.05 | 88.43 | 88.78 | 89.10 | 89.39 | 89.67 |
| | $U_2$ | 0.74 | 1.83 | 3.42 | 5.31 | 7.24 | 9.07 | 10.74 | 12.22 | 13.52 | 14.67 | 15.68 | 16.57 | 17.37 | 18.07 | 18.70 | 19.27 | 19.78 | 20.24 | 20.66 | 21.04 | 21.40 | 21.72 | 22.02 | 22.29 | 22.55 |
| | $W_{1,2}$ | 19.22 | 39.72 | 60.86 | 82.15 | 103.48 | 124.81 | 146.14 | 167.48 | 188.81 | 210.15 | 231.48 | 252.82 | 274.15 | 295.49 | 316.82 | 338.16 | 359.49 | 380.83 | 402.16 | 423.50 | 444.83 | 466.17 | 487.50 | 508.84 | 530.17 |
| | $W_{3,4}$ | 19.42 | 40.02 | 61.20 | 82.50 | 103.84 | 125.16 | 146.50 | 167.83 | 189.17 | 210.50 | 231.84 | 253.17 | 274.51 | 295.84 | 317.17 | 338.51 | 359.84 | 381.18 | 402.51 | 423.85 | 445.18 | 466.52 | 487.85 | 509.19 | 530.52 |
| | $J$ | 13.05 | 22.45 | 27.31 | 30.25 | 32.14 | 33.39 | 34.26 | 34.87 | 35.31 | 35.64 | 35.88 | 36.07 | 36.22 | 36.33 | 36.42 | 36.49 | 36.55 | 36.59 | 36.63 | 36.66 | 36.69 | 36.71 | 36.73 | 36.75 | 36.76 |
| | $K$ | 2.07 | 2.93 | 3.14 | 3.18 | 3.19 | 3.19 | 3.19 | 3.19 | 3.19 | 3.19 | 3.19 | 3.19 | 3.19 | 3.19 | 3.19 | 3.19 | 3.19 | 3.19 | 3.19 | 3.19 | 3.19 | 3.19 | 3.19 | 3.19 | 3.19 |
| 0.7 | $U_1$ | 23.39 | 40.07 | 51.16 | 58.76 | 64.19 | 68.22 | 71.31 | 73.75 | 75.72 | 77.34 | 78.69 | 79.83 | 80.80 | 81.66 | 82.41 | 83.06 | 83.64 | 84.16 | 84.63 | 85.05 | 85.44 | 85.78 | 86.10 | 86.40 | 86.67 |
| | $U_2$ | 0.79 | 1.92 | 3.54 | 5.44 | 7.38 | 9.21 | 10.86 | 12.34 | 13.64 | 14.78 | 15.79 | 16.68 | 17.47 | 18.17 | 18.80 | 19.37 | 19.88 | 20.34 | 20.76 | 21.14 | 21.49 | 21.81 | 22.11 | 22.39 | 22.64 |
| | $W_{1,2}$ | 18.93 | 39.33 | 60.44 | 81.72 | 103.05 | 124.38 | 145.72 | 167.05 | 188.39 | 209.72 | 231.06 | 252.39 | 273.73 | 295.06 | 316.39 | 337.73 | 359.06 | 380.40 | 401.73 | 423.07 | 444.40 | 465.74 | 487.07 | 508.41 | 529.74 |
| | $W_{3,4}$ | 19.35 | 39.97 | 61.14 | 82.44 | 103.77 | 125.11 | 146.44 | 167.77 | 189.11 | 210.44 | 231.77 | 253.11 | 274.45 | 295.78 | 317.12 | 338.45 | 359.79 | 381.12 | 402.46 | 423.79 | 445.13 | 466.46 | 487.80 | 509.13 | 530.47 |
| | $J$ | 14.49 | 23.60 | 28.96 | 32.26 | 34.39 | 35.82 | 36.81 | 37.52 | 38.03 | 38.41 | 38.68 | 38.91 | 39.08 | 39.21 | 39.32 | 39.40 | 39.47 | 39.53 | 39.57 | 39.61 | 39.64 | 39.67 | 39.69 | 39.71 | 39.73 |
| | $K$ | 2.23 | 3.15 | 3.37 | 3.42 | 3.43 | 3.43 | 3.43 | 3.43 | 3.43 | 3.43 | 3.43 | 3.43 | 3.43 | 3.43 | 3.43 | 3.43 | 3.43 | 3.43 | 3.43 | 3.43 | 3.43 | 3.43 | 3.43 | 3.43 | 3.43 |

| $v_0/v_1$ | $\xi$/nm | 2 | 4 | 6 | 8 | 10 | 12 | 14 | 16 | 18 | 20 | 22 | 24 | 26 | 28 | 30 | 32 | 34 | 36 | 38 | 40 | 42 | 44 | 46 | 48 | 50 |
|---|---|---|---|---|---|---|---|---|---|---|---|---|---|---|---|---|---|---|---|---|---|---|---|---|---|---|
| 0.6 | $U_1$ | 21.56 | 37.19 | 47.75 | 55.06 | 60.32 | 64.25 | 67.28 | 69.68 | 71.61 | 73.21 | 74.55 | 75.68 | 76.65 | 77.50 | 78.24 | 78.89 | 79.47 | 79.98 | 80.45 | 80.87 | 81.25 | 81.60 | 81.92 | 82.21 | 82.48 |
|  | $U_2$ | 0.87 | 2.08 | 3.74 | 5.65 | 7.58 | 9.39 | 11.04 | 12.50 | 13.79 | 14.92 | 15.92 | 16.81 | 17.59 | 18.29 | 18.92 | 19.48 | 19.99 | 20.46 | 20.86 | 21.24 | 21.59 | 21.92 | 22.21 | 22.49 | 22.74 |
|  | $W_{1,2}$ | 18.70 | 39.00 | 60.09 | 81.37 | 102.70 | 124.03 | 145.37 | 166.70 | 188.04 | 209.37 | 230.71 | 252.04 | 273.38 | 294.71 | 316.04 | 337.38 | 358.71 | 380.05 | 401.38 | 422.72 | 444.05 | 465.39 | 486.72 | 508.06 | 529.39 |
|  | $W_{3,4}$ |  | 40.03 | 61.20 | 82.51 | 103.83 | 125.17 | 146.50 | 167.84 | 189.17 | 210.51 | 231.84 | 253.18 | 274.51 | 295.85 | 317.18 | 338.52 | 359.85 | 381.19 | 402.52 | 423.85 | 445.19 | 466.52 | 487.86 | 509.19 | 530.53 |
|  | $J$ | 15.13 | 25.00 | 31.01 | 34.83 | 37.36 | 39.11 | 40.34 | 41.24 | 41.91 | 42.41 | 42.79 | 43.09 | 43.33 | 43.52 | 43.67 | 43.79 | 43.89 | 43.98 | 44.05 | 44.10 | 44.15 | 44.19 | 44.23 | 44.26 | 44.29 |
|  | $K$ | 2.40 | 3.39 | 3.63 | 3.68 | 3.69 | 3.69 | 3.69 | 3.69 | 3.69 | 3.69 | 3.69 | 3.69 | 3.69 | 3.69 | 3.69 | 3.69 | 3.69 | 3.69 | 3.69 | 3.69 | 3.69 | 3.69 | 3.69 | 3.69 | 3.69 |
| 0.5 | $U_1$ | 19.93 | 34.59 | 44.63 | 51.66 | 56.76 | 60.59 | 63.56 | 65.91 | 67.82 | 69.40 | 70.72 | 71.84 | 72.80 | 73.64 | 74.38 | 75.02 | 75.60 | 76.11 | 76.57 | 76.99 | 77.37 | 77.72 | 78.04 | 78.33 | 78.60 |
|  | $U_2$ | 0.95 | 2.23 | 3.93 | 5.85 | 7.77 | 9.57 | 11.20 | 12.65 | 13.93 | 15.06 | 16.04 | 16.92 | 17.70 | 18.40 | 19.02 | 19.58 | 20.09 | 20.56 | 20.96 | 21.34 | 21.69 | 22.02 | 22.30 | 22.58 | 22.83 |
|  | $W_{1,2}$ | 18.48 | 38.71 | 59.78 | 81.06 | 102.38 | 123.71 | 145.05 | 166.38 | 187.72 | 209.05 | 230.39 | 251.72 | 273.06 | 294.39 | 315.73 | 337.06 | 358.40 | 379.73 | 401.07 | 422.40 | 443.73 | 465.07 | 486.40 | 507.74 | 529.07 |
|  | $W_{3,4}$ |  | 40.13 | 61.30 | 82.61 | 103.93 | 125.27 | 146.60 | 167.94 | 189.27 | 210.61 | 231.94 | 253.28 | 274.61 | 295.95 | 317.28 | 338.62 | 359.95 | 381.29 | 402.62 | 423.96 | 445.29 | 466.63 | 487.96 | 509.29 | 530.63 |
|  | $J$ | 15.71 | 26.26 | 32.87 | 37.17 | 40.07 | 42.10 | 43.55 | 44.63 | 45.43 | 46.05 | 46.53 | 46.90 | 47.20 | 47.44 | 47.63 | 47.79 | 47.92 | 48.03 | 48.12 | 48.20 | 48.26 | 48.32 | 48.37 | 48.41 | 48.44 |
|  | $K$ | 2.59 | 3.65 | 3.90 | 3.96 | 3.97 | 3.97 | 3.97 | 3.97 | 3.97 | 3.97 | 3.97 | 3.97 | 3.97 | 3.97 | 3.97 | 3.97 | 3.97 | 3.97 | 3.97 | 3.97 | 3.97 | 3.97 | 3.97 | 3.97 | 3.97 |
| 0.4 | $U_1$ | 18.50 | 32.49 | 42.15 | 48.98 | 53.97 | 57.73 | 60.60 | 62.98 | 64.87 | 66.44 | 67.75 | 68.87 | 69.83 | 70.66 | 71.39 | 72.03 | 72.61 | 73.12 | 73.58 | 74.00 | 74.38 | 74.72 | 75.04 | 75.33 | 75.60 |
|  | $U_2$ | 0.98 | 2.29 | 4.00 | 5.93 | 7.85 | 9.65 | 11.27 | 12.71 | 13.99 | 15.11 | 16.10 | 16.98 | 17.76 | 18.45 | 19.07 | 19.63 | 20.14 | 20.59 | 21.01 | 21.39 | 21.74 | 22.06 | 22.35 | 22.63 | 22.88 |
|  | $W_{1,2}$ | 17.97 | 38.00 | 59.02 | 80.29 | 101.61 | 122.94 | 144.28 | 165.61 | 186.95 | 208.28 | 229.62 | 250.95 | 272.29 | 293.62 | 314.96 | 336.29 | 357.63 | 378.96 | 400.30 | 421.63 | 442.96 | 464.30 | 485.63 | 506.97 | 528.30 |
|  | $W_{3,4}$ |  | 40.21 | 61.39 | 82.69 | 104.02 | 125.30 | 146.69 | 168.02 | 189.36 | 210.69 | 232.03 | 253.36 | 274.70 | 296.03 | 317.37 | 338.70 | 360.04 | 381.37 | 402.71 | 424.04 | 445.37 | 466.71 | 488.04 | 509.38 | 530.71 |
|  | $J$ | 16.12 | 27.09 | 34.04 | 38.58 | 41.65 | 43.81 | 45.36 | 46.50 | 47.37 | 48.02 | 48.54 | 48.94 | 49.26 | 49.52 | 49.73 | 49.90 | 50.04 | 50.16 | 50.26 | 50.34 | 50.41 | 50.47 | 50.53 | 50.57 | 50.61 |
|  | $K$ | 2.78 | 3.91 | 4.19 | 4.24 | 4.26 | 4.26 | 4.26 | 4.26 | 4.26 | 4.26 | 4.26 | 4.26 | 4.26 | 4.26 | 4.26 | 4.26 | 4.26 | 4.26 | 4.26 | 4.26 | 4.26 | 4.26 | 4.26 | 4.26 | 4.26 |
| 0.3 | $U_1$ | 18.19 | 31.88 | 41.44 | 48.22 | 53.19 | 56.95 | 59.87 | 62.19 | 64.07 | 65.63 | 66.94 | 68.06 | 69.02 | 69.85 | 70.58 | 71.22 | 71.80 | 72.31 | 72.77 | 73.19 | 73.57 | 73.91 | 74.23 | 74.52 | 74.79 |
|  | $U_2$ | 1.02 | 2.37 | 4.11 | 6.04 | 7.96 | 9.76 | 11.38 | 12.82 | 14.09 | 15.21 | 16.19 | 17.07 | 17.84 | 18.54 | 19.16 | 19.71 | 20.22 | 20.68 | 21.09 | 21.47 | 21.82 | 22.13 | 22.43 | 22.70 | 22.96 |
|  | $W_{1,2}$ | 17.76 | 37.71 | 58.72 | 79.98 | 101.30 | 122.64 | 143.97 | 165.30 | 186.64 | 207.97 | 229.31 | 250.64 | 271.98 | 293.31 | 314.65 | 335.98 | 357.32 | 378.65 | 399.99 | 421.32 | 442.66 | 463.99 | 485.33 | 506.66 | 528.00 |
|  | $W_{3,4}$ |  | 40.24 | 61.42 | 82.73 | 104.05 | 125.39 | 146.72 | 168.06 | 189.39 | 210.73 | 232.06 | 253.40 | 274.73 | 296.07 | 317.40 | 338.74 | 360.07 | 381.41 | 402.74 | 424.08 | 445.41 | 466.74 | 488.08 | 509.41 | 530.75 |
|  | $J$ | 16.73 | 28.32 | 35.77 | 40.67 | 44.01 | 46.35 | 48.04 | 49.29 | 50.23 | 50.95 | 51.51 | 51.95 | 52.31 | 52.59 | 52.82 | 53.01 | 53.17 | 53.30 | 53.41 | 53.50 | 53.58 | 53.65 | 53.71 | 53.76 | 53.81 |
|  | $K$ | 3.11 | 4.38 | 4.68 | 4.75 | 4.76 | 4.76 | 4.76 | 4.76 | 4.76 | 4.76 | 4.76 | 4.76 | 4.76 | 4.76 | 4.76 | 4.76 | 4.76 | 4.76 | 4.76 | 4.76 | 4.76 | 4.76 | 4.76 | 4.76 | 4.76 |
| 0.2 | $U_1$ | 18.07 | 31.88 | 41.44 | 48.22 | 53.12 | 56.88 | 59.78 | 62.09 | 63.97 | 65.53 | 66.84 | 67.96 | 68.92 | 69.75 | 70.48 | 71.12 | 71.69 | 72.20 | 72.66 | 73.08 | 73.46 | 73.80 | 74.12 | 74.40 | 74.52 |
|  | $U_2$ | 1.02 | 2.37 | 4.11 | 6.04 | 7.97 | 9.76 | 11.38 | 12.82 | 14.09 | 15.21 | 16.20 | 17.08 | 17.85 | 18.55 | 19.17 | 19.72 | 20.23 | 20.68 | 21.08 | 21.47 | 21.82 | 22.13 | 22.44 | 22.71 | 22.97 |
|  | $W_{1,2}$ | 17.60 | 37.48 | 58.47 | 79.73 | 101.05 | 122.38 | 143.72 | 165.05 | 186.39 | 207.72 | 229.06 | 250.39 | 271.74 | 293.06 | 314.40 | 335.73 | 357.06 | 378.40 | 399.73 | 421.07 | 442.40 | 463.74 | 485.07 | 506.41 | 527.74 |
|  | $W_{3,4}$ |  | 40.24 | 61.42 | 82.73 | 104.05 | 125.38 | 146.72 | 168.06 | 189.40 | 210.73 | 232.07 | 253.40 | 274.73 | 296.07 | 317.40 | 338.74 | 360.07 | 381.41 | 402.74 | 424.08 | 445.41 | 466.74 | 488.08 | 509.41 | 530.74 |
|  | $J$ | 16.91 | 28.62 | 36.14 | 41.06 | 44.40 | 46.75 | 48.44 | 49.67 | 50.60 | 51.32 | 51.87 | 52.31 | 52.66 | 52.95 | 53.18 | 53.37 | 53.53 | 53.65 | 53.76 | 53.85 | 53.92 | 53.98 | 54.04 | 54.09 | 54.13 |
|  | $K$ | 3.22 | 4.52 | 4.84 | 4.90 | 4.92 | 4.92 | 4.92 | 4.92 | 4.92 | 4.92 | 4.92 | 4.92 | 4.92 | 4.92 | 4.92 | 4.92 | 4.92 | 4.92 | 4.92 | 4.92 | 4.92 | 4.92 | 4.92 | 4.92 | 4.92 |
| 0.1 | $U_1$ | 18.22 | 32.07 | 41.65 | 48.44 | 53.43 | 57.20 | 60.13 | 62.46 | 64.35 | 65.92 | 67.24 | 68.37 | 69.33 | 70.17 | 70.91 | 71.56 | 72.14 | 72.66 | 73.13 | 73.55 | 73.93 | 74.28 | 74.60 | 74.89 | 75.06 |
|  | $U_2$ | 1.00 | 2.37 | 4.11 | 6.04 | 7.96 | 9.76 | 11.38 | 12.82 | 14.08 | 15.20 | 16.19 | 17.07 | 17.84 | 18.54 | 19.16 | 19.71 | 20.22 | 20.68 | 21.08 | 21.46 | 21.81 | 22.13 | 22.43 | 22.70 | 22.96 |
|  | $W_{1,2}$ | 17.55 | 37.48 | 58.47 | 79.73 | 101.05 | 122.38 | 143.72 | 165.05 | 186.39 | 207.72 | 229.06 | 250.39 | 271.74 | 293.06 | 314.40 | 335.73 | 357.06 | 378.40 | 399.73 | 421.07 | 442.40 | 463.74 | 485.07 | 506.41 | 527.67 |
|  | $W_{3,4}$ |  | 40.23 | 61.41 | 82.71 | 104.04 | 125.36 | 146.70 | 168.04 | 189.37 | 210.71 | 232.04 | 253.38 | 274.71 | 296.05 | 317.38 | 338.72 | 360.06 | 381.39 | 402.73 | 424.06 | 445.40 | 466.73 | 488.07 | 509.40 | 530.72 |
|  | $J$ | 16.91 | 28.62 | 36.14 | 41.06 | 44.40 | 46.74 | 48.43 | 49.67 | 50.60 | 51.32 | 51.88 | 52.32 | 52.66 | 52.95 | 53.18 | 53.37 | 53.53 | 53.65 | 53.76 | 53.85 | 53.92 | 53.98 | 54.04 | 54.09 | 54.09 |
|  | $K$ | 3.22 | 4.52 | 4.84 | 4.90 | 4.92 | 4.92 | 4.92 | 4.92 | 4.92 | 4.92 | 4.92 | 4.92 | 4.92 | 4.92 | 4.92 | 4.92 | 4.92 | 4.92 | 4.92 | 4.92 | 4.92 | 4.92 | 4.92 | 4.92 | 4.92 |
| 0.0 | $U_1$ | 18.38 | 32.21 | 41.85 | 48.68 | 53.67 | 57.44 | 60.37 | 62.70 | 64.59 | 66.16 | 67.47 | 68.59 | 69.55 | 70.38 | 71.11 | 71.76 | 72.33 | 72.85 | 73.31 | 73.72 | 74.10 | 74.45 | 74.77 | 75.06 | 75.33 |
|  | $U_2$ | 1.00 | 2.38 | 4.06 | 5.99 | 7.91 | 9.71 | 11.33 | 12.78 | 14.05 | 15.17 | 16.16 | 17.03 | 17.81 | 18.50 | 19.12 | 19.68 | 20.19 | 20.64 | 21.06 | 21.44 | 21.79 | 22.11 | 22.40 | 22.67 | 22.93 |
|  | $W_{1,2}$ | 17.50 | 37.35 | 58.33 | 79.59 | 100.91 | 122.24 | 143.58 | 164.91 | 186.25 | 207.58 | 228.92 | 250.25 | 271.59 | 292.92 | 314.26 | 335.59 | 356.93 | 378.26 | 399.60 | 420.93 | 442.27 | 463.60 | 484.94 | 506.27 | 527.61 |
|  | $W_{3,4}$ |  | 40.22 | 61.40 | 82.70 | 104.03 | 125.35 | 146.69 | 168.03 | 189.36 | 210.70 | 232.03 | 253.37 | 274.70 | 296.04 | 317.37 | 338.71 | 360.05 | 381.38 | 402.72 | 424.05 | 445.39 | 466.72 | 488.06 | 509.39 | 530.72 |
|  | $J$ | 16.91 | 28.62 | 36.14 | 41.06 | 44.40 | 46.74 | 48.43 | 49.67 | 50.60 | 51.32 | 51.88 | 52.32 | 52.66 | 52.95 | 53.18 | 53.37 | 53.53 | 53.65 | 53.76 | 53.85 | 53.92 | 53.98 | 54.04 | 54.09 | 54.16 |
|  | $K$ | 3.26 | 4.57 | 4.89 | 4.96 | 4.98 | 4.98 | 4.98 | 4.98 | 4.98 | 4.98 | 4.98 | 4.98 | 4.98 | 4.98 | 4.98 | 4.98 | 4.98 | 4.98 | 4.98 | 4.98 | 4.98 | 4.98 | 4.98 | 4.98 | 4.98 |

TABLE S91. Parameters of the THF interaction Hamiltonian for different ratios $v_0/v_1$ and screening lengths $\xi$ of the double-gate interaction potential ($U_\xi = 24$ meV for $\xi = 10$nm). We use the same BM model parameters as in Table S90.

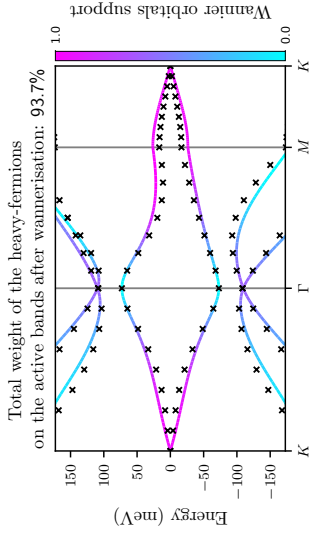

FIG. S62. Band structures of the BM and THF models near charge neutrality for $w_0/w_1 = 0.8$, depicted by lines and crosses, respectively. The BM bands are colored according to the weight of the $f$-electron wave function on them. We use the same BM parameters as in Table S92.

TABLE S92. Parameters of the $f$-electron Wannier functions and the THF single-particle Hamiltonian for different values of the tunneling ratio $w_0/w_1$. $W$ denotes the total weight of the THF $f$-electrons on the active bands. In computing the ratios $\gamma/U_1$ and $\gamma/U_2$, we employ the on-site and nearest-neighbor repulsion parameters $U_1$ and $U_2$ obtained numerically for $\xi = 10$ nm, as given in Table S93. We employ $v_F = 5.944$ eV·Å, $|\mathbf{K}| = 1.703$ Å$^{-1}$, $w_1 = 110$ meV, and $\theta = 1.58°$ for the BM model.

| $w_0/w_1$ | $\alpha_1$ | $\alpha_2$ | $\lambda_1$ | $\lambda_2$ | $\lambda$ | $W$ | $t_0$ | $\gamma$ | $M$ | $v_\star'$ | $v_\star$ | $v_\star''$ | $\gamma/U_1$ | $\gamma/U_2$ |
|---|---|---|---|---|---|---|---|---|---|---|---|---|---|---|
| 1.00 | 0.837 | 0.547 | 0.221 | 0.221 | 0.387 | 0.931 | 6.646 | -77.302 | -69.701 | -4.901 | 1.953 | 0.0464 | -1.02 | -10.93 |
| 0.90 | 0.864 | 0.504 | 0.227 | 0.233 | 0.415 | 0.938 | 7.832 | -93.703 | -71.398 | -4.995 | 1.937 | 0.0407 | -1.32 | -12.77 |
| 0.80 | 0.889 | 0.458 | 0.251 | 0.251 | 0.427 | 0.937 | 8.801 | -109.522 | -72.874 | -5.088 | 1.898 | 0.0351 | -1.62 | -14.62 |
| 0.70 | 0.913 | 0.408 | 0.263 | 0.239 | 0.447 | 0.941 | 9.779 | -124.580 | -74.145 | -5.179 | 1.831 | 0.0296 | -1.95 | -16.24 |
| 0.60 | 0.935 | 0.355 | 0.281 | 0.239 | 0.469 | 0.948 | 10.752 | -138.666 | -75.223 | -5.270 | 1.729 | 0.0240 | -2.30 | -17.65 |
| 0.50 | 0.954 | 0.300 | 0.298 | 0.245 | 0.486 | 0.953 | 11.591 | -151.518 | -76.117 | -5.359 | 1.583 | 0.0184 | -2.64 | -18.93 |
| 0.40 | 0.970 | 0.245 | 0.310 | 0.251 | 0.493 | 0.954 | 12.177 | -162.827 | -76.838 | -5.445 | 1.382 | 0.0129 | -2.91 | -20.15 |
| 0.30 | 0.983 | 0.185 | 0.328 | 0.251 | 0.521 | 0.964 | 12.807 | -172.230 | -77.391 | -5.523 | 1.120 | 0.0079 | -3.25 | -20.95 |
| 0.20 | 0.992 | 0.125 | 0.334 | 0.257 | 0.528 | 0.966 | 13.187 | -179.338 | -77.782 | -5.586 | 0.794 | 0.0037 | -3.46 | -21.63 |
| 0.10 | 0.998 | 0.065 | 0.340 | 0.257 | 0.499 | 0.953 | 13.070 | -183.784 | -78.015 | -5.629 | 0.413 | 0.0010 | -3.41 | -22.42 |
| 0.00 | 1.000 | 0.000 | 0.340 | 0.000 | 0.500 | 0.952 | 12.995 | -185.299 | -78.093 | -5.644 | -0.000 | -0.0000 | -3.42 | -22.68 |

| $w_0/w_1$ | | $\xi$/nm | 2 | 4 | 6 | 8 | 10 | 12 | 14 | 16 | 18 | 20 | 22 | 24 | 26 | 28 | 30 | 32 | 34 | 36 | 38 | 40 | 42 | 44 | 46 | 48 | 50 |
|---|---|---|---|---|---|---|---|---|---|---|---|---|---|---|---|---|---|---|---|---|---|---|---|---|---|---|---|
| 1.0 | | $U_1$ | 29.44 | 49.28 | 61.87 | 70.22 | 76.06 | 80.33 | 83.58 | 86.11 | 88.14 | 89.81 | 91.19 | 92.36 | 93.36 | 94.22 | 94.97 | 95.64 | 96.23 | 96.75 | 97.22 | 97.65 | 98.03 | 98.39 | 98.71 | 99.00 | 99.28 |
| | | $U_2$ | 0.62 | 1.60 | 3.17 | 5.09 | 7.07 | 8.96 | 10.66 | 12.18 | 13.52 | 14.69 | 15.72 | 16.63 | 17.43 | 18.15 | 18.78 | 19.36 | 19.87 | 20.34 | 20.76 | 21.15 | 21.50 | 21.83 | 22.13 | 22.41 | 22.66 |
| | | $W_{1,2}$ | 20.27 | 41.52 | 63.26 | 85.11 | 106.39 | 128.88 | 150.76 | 172.05 | 194.53 | 216.42 | 238.30 | 260.19 | 282.08 | 303.96 | 325.85 | 347.73 | 369.62 | 391.50 | 413.39 | 435.27 | 457.16 | 479.04 | 500.93 | 522.81 | 544.70 |
| | | $W_{3,4}$ | 19.98 | 41.26 | 63.01 | 84.87 | 106.75 | 128.63 | 150.51 | 172.40 | 194.29 | 216.17 | 238.06 | 259.94 | 281.83 | 303.71 | 325.60 | 347.49 | 369.37 | 391.26 | 413.14 | 435.03 | 456.91 | 478.80 | 500.68 | 522.57 | 544.70 |
| | | $J$ | 13.09 | 20.27 | 23.98 | 26.05 | 27.31 | 28.10 | 28.63 | 28.99 | 29.25 | 29.43 | 29.56 | 29.66 | 29.74 | 29.79 | 29.84 | 29.87 | 29.90 | 29.93 | 29.94 | 29.96 | 29.97 | 29.98 | 29.99 | 30.00 | 30.01 |
| | | $K$ | 1.79 | 2.63 | 2.70 | 2.74 | 2.75 | 2.75 | 2.75 | 2.75 | 2.75 | 2.75 | 2.75 | 2.75 | 2.75 | 2.75 | 2.75 | 2.75 | 2.75 | 2.75 | 2.75 | 2.75 | 2.75 | 2.75 | 2.75 | 2.75 | 2.75 |
| 0.9 | | $U_1$ | 26.75 | 45.17 | 57.07 | 65.08 | 70.72 | 74.88 | 78.05 | 80.54 | 82.54 | 84.19 | 85.55 | 86.71 | 87.70 | 88.56 | 89.31 | 89.97 | 90.55 | 91.07 | 91.54 | 91.97 | 92.35 | 92.70 | 93.02 | 93.32 | 93.59 |
| | | $U_2$ | 0.72 | 1.80 | 3.42 | 5.35 | 7.34 | 9.21 | 10.90 | 12.41 | 13.73 | 14.90 | 15.92 | 16.82 | 17.62 | 18.33 | 18.97 | 19.54 | 20.05 | 20.51 | 20.94 | 21.32 | 21.67 | 22.00 | 22.30 | 22.57 | 22.83 |
| | | $W_{1,2}$ | 20.05 | 41.22 | 62.93 | 84.78 | 106.66 | 128.55 | 150.43 | 172.32 | 194.20 | 216.09 | 237.97 | 259.86 | 281.69 | 303.63 | 325.52 | 347.40 | 369.29 | 391.17 | 413.06 | 434.94 | 456.83 | 478.71 | 500.60 | 522.48 | 544.37 |
| | | $W_{3,4}$ | 19.91 | 41.14 | 62.87 | 84.73 | 106.61 | 128.49 | 150.38 | 172.26 | 194.15 | 216.03 | 237.92 | 259.80 | 281.69 | 303.57 | 325.46 | 347.35 | 369.23 | 391.12 | 413.00 | 434.89 | 456.77 | 478.66 | 500.54 | 522.43 | 544.31 |
| | | $J$ | 13.71 | 21.72 | 26.14 | 28.75 | 30.39 | 31.47 | 32.21 | 32.73 | 33.10 | 33.38 | 33.58 | 33.74 | 33.85 | 33.95 | 34.02 | 34.08 | 34.13 | 34.16 | 34.19 | 34.22 | 34.24 | 34.26 | 34.27 | 34.28 | 34.30 |
| | | $K$ | 1.92 | 2.70 | 2.89 | 2.93 | 2.94 | 2.94 | 2.94 | 2.94 | 2.94 | 2.94 | 2.94 | 2.94 | 2.94 | 2.94 | 2.94 | 2.94 | 2.94 | 2.94 | 2.94 | 2.94 | 2.94 | 2.94 | 2.94 | 2.94 | 2.94 |
| 0.8 | | $U_1$ | 25.05 | 42.60 | 54.10 | 61.91 | 67.45 | 71.54 | 74.68 | 77.14 | 79.13 | 80.76 | 82.12 | 83.27 | 84.26 | 85.11 | 85.86 | 86.51 | 87.10 | 87.62 | 88.09 | 88.51 | 88.89 | 89.24 | 89.56 | 89.86 | 90.13 |
| | | $U_2$ | 0.77 | 1.91 | 3.56 | 5.51 | 7.49 | 9.36 | 11.05 | 12.55 | 13.86 | 15.02 | 16.04 | 16.94 | 17.74 | 18.45 | 19.08 | 19.65 | 20.16 | 20.63 | 21.05 | 21.43 | 21.78 | 22.11 | 22.41 | 22.68 | 22.94 |
| | | $W_{1,2}$ | 19.77 | 40.84 | 62.53 | 84.38 | 106.26 | 128.14 | 150.03 | 171.91 | 193.80 | 215.68 | 237.57 | 259.45 | 281.34 | 303.22 | 325.11 | 346.99 | 368.88 | 390.77 | 412.65 | 434.54 | 456.42 | 478.31 | 500.19 | 522.08 | 543.96 |
| | | $W_{3,4}$ | 19.86 | 41.04 | 62.77 | 84.62 | 106.51 | 128.39 | 150.27 | 172.16 | 194.04 | 215.93 | 237.81 | 259.70 | 281.58 | 303.47 | 325.36 | 347.24 | 369.13 | 391.01 | 412.90 | 434.78 | 456.67 | 478.55 | 500.44 | 522.32 | 544.21 |
| | | $J$ | 14.28 | 22.05 | 27.90 | 30.89 | 32.80 | 34.07 | 34.95 | 35.56 | 36.01 | 36.34 | 36.58 | 36.77 | 36.91 | 37.03 | 37.11 | 37.19 | 37.24 | 37.29 | 37.33 | 37.36 | 37.39 | 37.41 | 37.43 | 37.44 | 37.46 |
| | | $K$ | 2.06 | 2.90 | 3.10 | 3.14 | 3.15 | 3.15 | 3.15 | 3.15 | 3.15 | 3.15 | 3.15 | 3.15 | 3.15 | 3.15 | 3.15 | 3.15 | 3.15 | 3.15 | 3.15 | 3.15 | 3.15 | 3.15 | 3.15 | 3.15 | 3.15 |
| 0.7 | | $U_1$ | 23.27 | 39.85 | 50.88 | 58.44 | 63.84 | 67.86 | 70.94 | 73.37 | 75.34 | 76.95 | 78.30 | 79.44 | 80.42 | 81.27 | 82.01 | 82.67 | 83.25 | 83.77 | 84.23 | 84.65 | 85.04 | 85.39 | 85.71 | 86.00 | 86.27 |
| | | $U_2$ | 0.84 | 2.04 | 3.73 | 5.69 | 7.67 | 9.53 | 11.21 | 12.69 | 14.01 | 15.16 | 16.17 | 17.07 | 17.86 | 18.56 | 19.20 | 19.76 | 20.27 | 20.74 | 21.16 | 21.54 | 21.89 | 22.21 | 22.51 | 22.79 | 23.04 |
| | | $W_{1,2}$ | 19.48 | 40.48 | 62.15 | 83.99 | 105.87 | 127.75 | 149.64 | 171.52 | 193.41 | 215.29 | 237.18 | 259.06 | 280.95 | 302.83 | 324.72 | 346.61 | 368.49 | 390.38 | 412.26 | 434.15 | 456.03 | 477.92 | 499.80 | 521.69 | 543.57 |
| | | $W_{3,4}$ | 19.87 | 41.04 | 62.77 | 84.62 | 106.50 | 128.38 | 150.27 | 172.15 | 194.04 | 215.92 | 237.81 | 259.69 | 281.58 | 303.46 | 325.35 | 347.24 | 369.12 | 391.01 | 412.89 | 434.78 | 456.66 | 478.55 | 500.43 | 522.32 | 544.20 |
| | | $J$ | 15.07 | 22.99 | 28.18 | 33.33 | 35.59 | 37.12 | 38.18 | 38.95 | 39.51 | 39.93 | 40.25 | 40.50 | 40.69 | 40.84 | 40.96 | 41.06 | 41.13 | 41.20 | 41.25 | 41.31 | 41.33 | 41.37 | 41.39 | 41.42 | 41.44 |
| | | $K$ | 2.22 | 3.12 | 3.34 | 3.38 | 3.39 | 3.39 | 3.39 | 3.39 | 3.39 | 3.39 | 3.39 | 3.39 | 3.39 | 3.39 | 3.39 | 3.39 | 3.39 | 3.39 | 3.39 | 3.39 | 3.39 | 3.39 | 3.39 | 3.39 | 3.39 |

TABLE S93: Parameters of the THF interaction Hamiltonian for different ratios $u_0/u_1$ and screening lengths $\xi$ of the double-gate interaction potential ($U_\xi = 24\ \text{meV}$ for $\xi = 10\ \text{nm}$). We use the same BM model parameters as in Table S92.

| $u_0/u_1$ | $\xi/\text{nm}$ | 2 | 4 | 6 | 8 | 10 | 12 | 14 | 16 | 18 | 20 | 22 | 24 | 26 | 28 | 30 | 32 | 34 | 36 | 38 | 40 | 42 | 44 | 46 | 48 | 50 |
|---|---|---|---|---|---|---|---|---|---|---|---|---|---|---|---|---|---|---|---|---|---|---|---|---|---|---|
| 0.6 | $U_1$ | 21.57 | 37.18 | 47.72 | 55.01 | 60.27 | 64.19 | 67.22 | 69.61 | 71.55 | 73.14 | 74.48 | 75.61 | 76.58 | 77.43 | 78.17 | 78.82 | 79.41 | 79.91 | 80.38 | 80.80 | 81.18 | 81.53 | 81.85 | 82.14 | 82.41 |
| | $U_2$ | 0.92 | 2.19 | 3.92 | 5.88 | 7.86 | 9.70 | 11.37 | 12.84 | 14.14 | 15.29 | 16.30 | 17.19 | 17.97 | 18.68 | 19.30 | 19.87 | 20.38 | 20.84 | 21.26 | 21.64 | 21.99 | 22.31 | 22.61 | 22.88 | 23.14 |
| | $W_{1,2}$ | 19.26 | 40.14 | 61.79 | 83.63 | 105.51 | 127.39 | 149.28 | 171.16 | 193.05 | 214.93 | 236.82 | 258.70 | 280.59 | 302.47 | 324.36 | 346.24 | 368.13 | 390.01 | 411.90 | 433.79 | 455.67 | 477.56 | 499.44 | 521.33 | 543.21 |
| | $W_{3,4}$ | 19.92 | 41.10 | 62.83 | 84.68 | 106.56 | 128.44 | 150.33 | 172.21 | 194.10 | 215.98 | 237.87 | 259.76 | 281.64 | 303.53 | 325.41 | 347.30 | 369.19 | 391.07 | 412.96 | 434.84 | 456.73 | 478.61 | 500.49 | 522.38 | 544.27 |
| | $J$ | 15.52 | 25.65 | 31.84 | 35.77 | 38.39 | 40.20 | 41.48 | 42.41 | 43.11 | 43.64 | 44.04 | 44.35 | 44.60 | 44.80 | 44.96 | 45.09 | 45.19 | 45.28 | 45.36 | 45.42 | 45.47 | 45.51 | 45.55 | 45.58 | 45.61 |
| | $K$ | 2.40 | 3.37 | 3.60 | 3.65 | 3.66 | 3.66 | 3.66 | 3.66 | 3.66 | 3.66 | 3.66 | 3.66 | 3.66 | 3.66 | 3.66 | 3.66 | 3.66 | 3.66 | 3.66 | 3.66 | 3.66 | 3.66 | 3.66 | 3.66 | 3.66 |
| 0.5 | $U_1$ | 20.24 | 35.08 | 45.23 | 52.31 | 57.44 | 61.29 | 64.27 | 66.63 | 68.55 | 70.13 | 71.46 | 72.58 | 73.55 | 74.39 | 75.12 | 75.77 | 76.34 | 76.86 | 77.32 | 77.74 | 78.12 | 78.47 | 78.79 | 79.08 | 79.35 |
| | $U_2$ | 0.96 | 2.31 | 4.06 | 6.04 | 8.00 | 9.84 | 11.49 | 12.96 | 14.26 | 15.40 | 16.39 | 17.28 | 18.06 | 18.76 | 19.39 | 19.95 | 20.46 | 20.92 | 21.33 | 21.72 | 22.07 | 22.39 | 22.68 | 22.96 | 23.21 |
| | $W_{1,2}$ | 19.02 | 39.81 | 61.44 | 83.27 | 105.14 | 127.03 | 148.91 | 170.80 | 192.68 | 214.57 | 236.45 | 258.34 | 280.22 | 302.11 | 324.00 | 345.88 | 367.77 | 389.65 | 411.54 | 433.42 | 455.31 | 477.19 | 499.08 | 520.96 | 542.85 |
| | $W_{3,4}$ | 19.98 | 41.18 | 62.90 | 84.76 | 106.63 | 128.52 | 150.40 | 172.29 | 194.17 | 216.06 | 237.95 | 259.83 | 281.72 | 303.60 | 325.49 | 347.37 | 369.26 | 391.14 | 413.03 | 434.91 | 456.80 | 478.68 | 500.57 | 522.46 | 544.34 |
| | $J$ | 16.06 | 26.80 | 33.50 | 37.83 | 40.74 | 42.78 | 44.23 | 45.30 | 46.11 | 46.72 | 47.19 | 47.56 | 47.86 | 48.09 | 48.28 | 48.44 | 48.57 | 48.68 | 48.77 | 48.84 | 48.91 | 48.96 | 49.01 | 49.05 | 49.08 |
| | $K$ | 2.59 | 3.63 | 3.88 | 3.93 | 3.94 | 3.94 | 3.94 | 3.94 | 3.94 | 3.94 | 3.94 | 3.94 | 3.94 | 3.94 | 3.94 | 3.94 | 3.94 | 3.94 | 3.94 | 3.94 | 3.94 | 3.94 | 3.94 | 3.94 | 3.94 |
| 0.4 | $U_1$ | 18.95 | 33.97 | 43.91 | 50.80 | 55.97 | 59.79 | 62.75 | 65.10 | 67.00 | 68.58 | 69.90 | 71.02 | 71.98 | 72.82 | 73.55 | 74.20 | 74.78 | 75.29 | 75.75 | 76.17 | 76.55 | 76.90 | 77.22 | 77.51 | 77.78 |
| | $U_2$ | 1.01 | 2.36 | 4.14 | 6.11 | 8.08 | 9.91 | 11.56 | 13.03 | 14.32 | 15.45 | 16.45 | 17.34 | 18.12 | 18.82 | 19.44 | 20.00 | 20.51 | 20.97 | 21.39 | 21.77 | 22.12 | 22.44 | 22.73 | 23.01 | 23.26 |
| | $W_{1,2}$ | 18.74 | 39.43 | 61.03 | 82.86 | 104.74 | 126.62 | 148.50 | 170.39 | 192.27 | 214.16 | 236.04 | 257.93 | 279.81 | 301.70 | 323.58 | 345.47 | 367.35 | 389.24 | 411.13 | 433.01 | 454.90 | 476.78 | 498.67 | 520.55 | 542.44 |
| | $W_{3,4}$ | 20.01 | 41.22 | 62.94 | 84.80 | 106.68 | 128.56 | 150.44 | 172.33 | 194.21 | 216.10 | 237.99 | 259.87 | 281.76 | 303.64 | 325.53 | 347.41 | 369.30 | 391.18 | 413.07 | 434.95 | 456.84 | 478.73 | 500.61 | 522.50 | 544.38 |
| | $J$ | 16.48 | 27.63 | 34.67 | 39.24 | 42.33 | 44.48 | 46.03 | 47.17 | 48.03 | 48.68 | 49.19 | 49.59 | 49.90 | 50.16 | 50.37 | 50.53 | 50.67 | 50.79 | 50.89 | 50.97 | 51.04 | 51.10 | 51.15 | 51.20 | 51.24 |
| | $K$ | 2.79 | 3.90 | 4.16 | 4.22 | 4.23 | 4.23 | 4.23 | 4.23 | 4.23 | 4.23 | 4.23 | 4.23 | 4.23 | 4.23 | 4.23 | 4.23 | 4.23 | 4.23 | 4.23 | 4.23 | 4.23 | 4.23 | 4.23 | 4.23 | 4.23 |
| 0.3 | $U_1$ | 18.22 | 31.86 | 41.37 | 48.00 | 53.02 | 56.75 | 59.64 | 61.95 | 63.83 | 65.38 | 66.60 | 67.80 | 68.75 | 69.58 | 70.31 | 70.95 | 71.52 | 72.04 | 72.50 | 72.91 | 73.29 | 73.64 | 73.95 | 74.24 | 74.51 |
| | $U_2$ | 1.07 | 2.49 | 4.29 | 6.27 | 8.22 | 10.04 | 11.67 | 13.13 | 14.42 | 15.55 | 16.56 | 17.44 | 18.21 | 18.86 | 19.49 | 20.04 | 20.55 | 21.00 | 21.42 | 21.80 | 22.15 | 22.47 | 22.78 | 23.03 | 23.29 |
| | $W_{1,2}$ | 18.61 | 39.25 | 60.84 | 82.67 | 104.54 | 126.42 | 148.31 | 170.19 | 192.08 | 213.96 | 235.85 | 257.74 | 279.62 | 301.51 | 323.39 | 345.28 | 367.16 | 389.05 | 410.93 | 432.82 | 454.70 | 476.59 | 498.47 | 520.36 | 542.25 |
| | $W_{3,4}$ | 20.12 | 41.35 | 63.09 | 84.95 | 106.82 | 128.71 | 150.59 | 172.48 | 194.36 | 216.25 | 238.14 | 260.02 | 281.91 | 303.79 | 325.68 | 347.56 | 369.45 | 391.33 | 413.22 | 435.10 | 456.99 | 478.87 | 500.76 | 522.64 | 544.53 |
| | $J$ | 17.25 | 29.29 | 36.74 | 41.71 | 45.07 | 47.41 | 49.00 | 50.15 | 50.99 | 51.64 | 52.14 | 52.52 | 52.82 | 53.05 | 53.24 | 53.37 | 53.48 | 53.57 | 53.64 | 53.70 | 53.74 | 53.78 | 53.81 | 53.84 | 53.86 |
| | $K$ | 2.97 | 4.15 | 4.43 | 4.49 | 4.51 | 4.51 | 4.51 | 4.51 | 4.51 | 4.51 | 4.51 | 4.51 | 4.51 | 4.51 | 4.51 | 4.51 | 4.51 | 4.51 | 4.51 | 4.51 | 4.51 | 4.51 | 4.51 | 4.51 | 4.51 |
| 0.2 | $U_1$ | 17.69 | 30.34 | 40.46 | 46.98 | 51.85 | 55.54 | 58.42 | 60.71 | 62.58 | 64.13 | 65.38 | 66.55 | 67.48 | 68.31 | 69.04 | 69.68 | 70.25 | 70.76 | 71.22 | 71.64 | 72.01 | 72.36 | 72.68 | 72.97 | 73.23 |
| | $U_2$ | 1.11 | 2.55 | 4.38 | 6.34 | 8.29 | 10.10 | 11.67 | 13.18 | 14.45 | 15.57 | 16.56 | 17.44 | 18.21 | 18.86 | 19.49 | 20.04 | 20.55 | 21.00 | 21.42 | 21.81 | 22.16 | 22.48 | 22.77 | 23.09 | 23.32 |
| | $W_{1,2}$ | 18.44 | 39.02 | 60.60 | 82.42 | 104.29 | 126.18 | 148.06 | 169.95 | 191.83 | 213.72 | 235.60 | 257.49 | 279.37 | 301.26 | 323.15 | 345.03 | 366.92 | 388.80 | 410.69 | 432.57 | 454.46 | 476.34 | 498.23 | 520.11 | 542.00 |
| | $W_{3,4}$ | 20.12 | 41.35 | 63.09 | 84.95 | 106.82 | 128.71 | 150.59 | 172.48 | 194.36 | 216.25 | 238.14 | 260.02 | 281.91 | 303.79 | 325.68 | 347.56 | 369.45 | 391.33 | 413.22 | 435.10 | 456.99 | 478.87 | 500.76 | 522.64 | 544.59 |
| | $J$ | 17.24 | 29.22 | 36.74 | 41.71 | 45.07 | 47.41 | 49.11 | 50.35 | 51.26 | 51.96 | 52.51 | 52.94 | 53.28 | 53.56 | 53.78 | 53.98 | 54.13 | 54.27 | 54.38 | 54.47 | 54.55 | 54.61 | 54.67 | 54.72 | 57.35 |
| | $K$ | 3.13 | 4.37 | 4.66 | 4.72 | 4.74 | 4.74 | 4.74 | 4.74 | 4.74 | 4.74 | 4.74 | 4.74 | 4.74 | 4.74 | 4.74 | 4.74 | 4.74 | 4.74 | 4.74 | 4.74 | 4.74 | 4.74 | 4.74 | 4.74 | 4.74 |
| 0.1 | $U_1$ | 18.52 | 32.43 | 42.25 | 49.11 | 54.13 | 57.91 | 60.85 | 63.18 | 65.08 | 66.65 | 67.96 | 69.08 | 70.04 | 70.88 | 71.61 | 72.25 | 72.83 | 73.34 | 73.80 | 74.22 | 74.60 | 74.95 | 75.26 | 75.56 | 75.82 |
| | $U_2$ | 1.05 | 2.44 | 4.24 | 6.23 | 8.20 | 10.03 | 11.67 | 13.13 | 14.42 | 15.53 | 16.55 | 17.43 | 18.21 | 18.88 | 19.51 | 20.07 | 20.57 | 21.03 | 21.45 | 21.83 | 22.18 | 22.50 | 22.79 | 23.07 | 23.35 |
| | $W_{1,2}$ | 18.12 | 38.57 | 60.12 | 81.93 | 103.81 | 125.69 | 147.57 | 169.46 | 191.34 | 213.23 | 235.11 | 257.00 | 278.88 | 300.77 | 322.66 | 344.54 | 366.43 | 388.31 | 410.20 | 432.08 | 453.97 | 475.85 | 497.74 | 519.62 | 541.51 |
| | $W_{3,4}$ | 20.08 | 41.30 | 63.03 | 84.88 | 106.76 | 128.65 | 150.53 | 172.42 | 194.30 | 216.19 | 238.07 | 259.96 | 281.84 | 303.72 | 325.61 | 347.50 | 369.38 | 391.27 | 413.15 | 435.04 | 456.92 | 478.81 | 500.69 | 522.58 | 544.46 |
| | $J$ | 17.24 | 29.23 | 36.74 | 41.71 | 45.07 | 47.41 | 49.02 | 50.18 | 51.05 | 51.71 | 52.22 | 52.62 | 52.93 | 53.16 | 53.36 | 53.52 | 53.66 | 53.78 | 53.88 | 53.96 | 54.11 | 54.30 | 54.56 | 54.68 | 54.95 |
| | $K$ | 3.23 | 4.52 | 4.82 | 4.88 | 4.89 | 4.90 | 4.90 | 4.90 | 4.90 | 4.90 | 4.90 | 4.90 | 4.90 | 4.90 | 4.90 | 4.90 | 4.90 | 4.90 | 4.90 | 4.90 | 4.90 | 4.90 | 4.90 | 4.90 | 4.90 |
| 0.0 | $U_1$ | 18.59 | 32.54 | 42.25 | 49.01 | 54.13 | 57.91 | 60.85 | 63.18 | 65.08 | 66.65 | 67.96 | 69.08 | 70.04 | 70.88 | 71.61 | 72.25 | 72.83 | 73.34 | 73.80 | 74.22 | 74.60 | 74.95 | 75.26 | 75.37 | 75.56 |
| | $U_2$ | 1.04 | 2.42 | 4.22 | 6.20 | 8.17 | 10.00 | 11.65 | 13.11 | 14.40 | 15.53 | 16.52 | 17.41 | 18.19 | 18.88 | 19.51 | 20.07 | 20.57 | 21.03 | 21.45 | 21.83 | 22.18 | 22.50 | 22.79 | 23.07 | 23.32 |
| | $W_{1,2}$ | 18.05 | 38.47 | 60.02 | 81.83 | 103.70 | 125.59 | 147.47 | 169.36 | 191.24 | 213.13 | 235.01 | 256.90 | 278.78 | 300.67 | 322.55 | 344.44 | 366.33 | 388.21 | 410.10 | 431.98 | 453.87 | 475.75 | 497.64 | 519.52 | 541.41 |
| | $W_{3,4}$ | 20.13 | 41.29 | 63.02 | 84.88 | 106.76 | 128.65 | 150.53 | 172.41 | 194.30 | 216.18 | 238.07 | 259.95 | 281.84 | 303.72 | 325.61 | 347.49 | 369.38 | 391.27 | 413.15 | 435.04 | 456.92 | 478.81 | 500.69 | 522.58 | 544.46 |
| | $J$ | 17.29 | 29.23 | 36.74 | 41.71 | 45.07 | 45.21 | 46.14 | 46.94 | 47.67 | 48.31 | 48.88 | 49.26 | 49.56 | 49.79 | 49.99 | 50.15 | 50.29 | 50.41 | 50.51 | 50.59 | 50.66 | 50.72 | 50.77 | 50.81 | 54.99 |
| | $K$ | 3.27 | 4.57 | 4.88 | 4.94 | 4.95 | 4.95 | 4.95 | 4.95 | 4.95 | 4.95 | 4.95 | 4.95 | 4.95 | 4.95 | 4.95 | 4.95 | 4.95 | 4.95 | 4.95 | 4.95 | 4.95 | 4.95 | 4.95 | 4.95 | 4.95 |

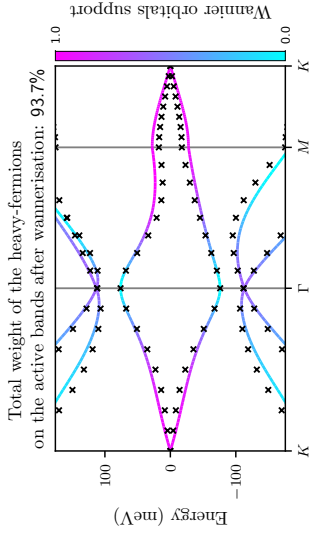

Total weight of the heavy-fermions on the active bands after wannierisation: 93.7%

Wannier orbitals support  1.0 — 0.0

FIG. S63. Band structures of the BM and THF models near charge neutrality for $w_0/w_1 = 0.8$, depicted by lines and crosses, respectively. The BM bands are colored according to the weight of the $f$-electron wave function on them. We use the same BM parameters as in Table S94.

TABLE S94. Parameters of the $f$-electron Wannier functions and the THF single-particle Hamiltonian for different values of the tunneling ratio $w_0/w_1$. $\mathcal{W}$ denotes the total weight of the THF $f$-electrons on the active bands. In computing the ratios $\gamma/U_1$ and $\gamma/U_2$, we employ the on-site and nearest-neighbor repulsion parameters $U_1$ and $U_2$ obtained numerically for $\xi = 10$ nm, as given in Table S95. We employ $v_F = 5.944\,\mathrm{eV\,Å}$, $|\mathbf{K}| = 1.703\,\mathrm{Å}^{-1}$, $w_1 = 110$ meV, and $\theta = 1.60°$ for the BM model.

| $w_0/w_1$ | $\alpha_1$ | $\alpha_2$ | $\lambda_1$ | $\lambda_2$ | $\lambda$ | $\mathcal{W}$ | $t_0$ | $\gamma$ | $M$ | $v_\star$ | $v_\star'$ | $v_\star''$ | $\gamma/U_1$ | $\gamma/U_2$ |
|---|---|---|---|---|---|---|---|---|---|---|---|---|---|---|
| 1.00 | 0.839 | 0.543 | 0.221 | 0.221 | 0.393 | 0.933 | 7.141 | -80.674 | -72.903 | -4.915 | 1.961 | 0.0486 | -1.06 | -10.97 |
| 0.90 | 0.866 | 0.500 | 0.239 | 0.237 | 0.418 | 0.938 | 8.313 | -97.085 | -74.604 | -5.008 | 1.943 | 0.0427 | -1.37 | -12.79 |
| 0.80 | 0.891 | 0.455 | 0.251 | 0.233 | 0.429 | 0.937 | 9.278 | -112.905 | -76.085 | -5.099 | 1.903 | 0.0370 | -1.66 | -14.59 |
| 0.70 | 0.914 | 0.405 | 0.269 | 0.239 | 0.448 | 0.941 | 10.278 | -127.959 | -77.361 | -5.190 | 1.836 | 0.0312 | -1.98 | -16.17 |
| 0.60 | 0.936 | 0.352 | 0.281 | 0.245 | 0.478 | 0.952 | 11.382 | -142.036 | -78.442 | -5.280 | 1.732 | 0.0253 | -2.36 | -17.46 |
| 0.50 | 0.955 | 0.298 | 0.298 | 0.245 | 0.487 | 0.953 | 12.131 | -154.878 | -79.340 | -5.369 | 1.585 | 0.0194 | -2.66 | -18.79 |
| 0.40 | 0.971 | 0.241 | 0.316 | 0.251 | 0.514 | 0.963 | 12.865 | -166.174 | -80.063 | -5.454 | 1.383 | 0.0136 | -3.02 | -19.82 |
| 0.30 | 0.983 | 0.182 | 0.328 | 0.251 | 0.525 | 0.966 | 13.447 | -175.564 | -80.619 | -5.532 | 1.120 | 0.0083 | -3.29 | -20.68 |
| 0.20 | 0.992 | 0.124 | 0.340 | 0.257 | 0.531 | 0.967 | 13.796 | -182.661 | -81.012 | -5.595 | 0.794 | 0.0040 | -3.49 | -21.39 |
| 0.10 | 0.998 | 0.064 | 0.346 | 0.257 | 0.522 | 0.963 | 13.838 | -187.100 | -81.246 | -5.638 | 0.413 | 0.0010 | -3.54 | -21.98 |
| 0.00 | 1.000 | 0.000 | 0.346 | 0.000 | 0.500 | 0.951 | 13.560 | -188.612 | -81.324 | -5.652 | -0.0000 | -0.0000 | -3.44 | -22.44 |

TABLE S95 (continued): same BM parameters as in Table S94.

| $w_0/w_1$ | $\xi$/nm | 2 | 4 | 6 | 8 | 10 | 12 | 14 | 16 | 18 | 20 | 22 | 24 | 26 | 28 | 30 | 32 | 34 | 36 | 38 | 40 | 42 | 44 | 46 | 48 | 50 |
|---|---|---|---|---|---|---|---|---|---|---|---|---|---|---|---|---|---|---|---|---|---|---|---|---|---|---|
| 1.0 | $U_1$ | 29.44 | 49.27 | 61.84 | 70.18 | 76.01 | 80.28 | 83.52 | 86.05 | 88.08 | 89.74 | 91.12 | 92.29 | 93.29 | 94.15 | 94.91 | 95.57 | 96.16 | 96.68 | 97.15 | 97.58 | 97.97 | 98.32 | 98.64 | 98.94 | 99.21 |
| | $U_2$ | 0.66 | 1.70 | 3.34 | 5.32 | 7.35 | 9.27 | 11.00 | 12.54 | 13.88 | 15.06 | 16.10 | 17.02 | 17.82 | 18.54 | 19.18 | 19.76 | 20.27 | 20.74 | 21.17 | 21.55 | 21.91 | 22.23 | 22.53 | 22.81 | 23.07 |
| | $W_{1,2}$ | 20.85 | 42.68 | 64.98 | 87.40 | 109.84 | 132.28 | 154.72 | 177.16 | 199.61 | 222.05 | 244.49 | 266.93 | 289.38 | 311.82 | 334.26 | 356.71 | 379.15 | 401.59 | 424.04 | 446.48 | 468.92 | 491.36 | 513.81 | 536.25 | 558.69 |
| | $W_{3,4}$ | 20.46 | 42.28 | 64.58 | 87.00 | 109.44 | 131.88 | 154.32 | 176.77 | 199.21 | 221.65 | 244.09 | 266.54 | 288.98 | 311.42 | 333.87 | 356.31 | 378.75 | 401.19 | 423.64 | 446.08 | 468.52 | 490.97 | 513.41 | 535.85 | 558.29 |
| | $J$ | 13.41 | 20.77 | 24.60 | 26.76 | 28.06 | 28.89 | 29.45 | 29.83 | 30.09 | 30.29 | 30.43 | 30.53 | 30.61 | 30.67 | 30.72 | 30.76 | 30.79 | 30.81 | 30.83 | 30.85 | 30.86 | 30.87 | 30.88 | 30.89 | 30.90 |
| | $K$ | 1.77 | 2.49 | 2.66 | 2.69 | 2.70 | 2.70 | 2.70 | 2.70 | 2.70 | 2.70 | 2.70 | 2.70 | 2.70 | 2.70 | 2.70 | 2.70 | 2.70 | 2.70 | 2.70 | 2.70 | 2.70 | 2.70 | 2.70 | 2.70 | 2.70 |
| 0.9 | $U_1$ | 26.93 | 45.44 | 57.39 | 65.42 | 71.07 | 75.24 | 78.42 | 80.91 | 82.91 | 84.55 | 85.92 | 87.08 | 88.07 | 88.93 | 89.68 | 90.34 | 90.92 | 91.45 | 91.91 | 92.34 | 92.72 | 93.07 | 93.40 | 93.70 | 93.96 |
| | $U_2$ | 0.75 | 1.88 | 3.57 | 5.56 | 7.59 | 9.50 | 11.22 | 12.74 | 14.08 | 15.25 | 16.28 | 17.19 | 18.00 | 18.71 | 19.35 | 19.92 | 20.44 | 20.90 | 21.33 | 21.71 | 22.07 | 22.39 | 22.69 | 22.97 | 23.22 |
| | $W_{1,2}$ | 20.61 | 42.36 | 64.64 | 87.05 | 109.49 | 131.93 | 154.37 | 176.81 | 199.26 | 221.70 | 244.14 | 266.59 | 289.03 | 311.47 | 333.91 | 356.36 | 378.80 | 401.24 | 423.69 | 446.13 | 468.57 | 491.01 | 513.46 | 535.90 | 558.34 |
| | $W_{3,4}$ | 20.40 | 42.16 | 64.46 | 86.87 | 109.31 | 131.75 | 154.19 | 176.64 | 199.08 | 221.52 | 243.96 | 266.41 | 288.85 | 311.29 | 333.74 | 356.18 | 378.62 | 401.06 | 423.51 | 445.95 | 468.39 | 490.84 | 513.28 | 535.72 | 558.16 |
| | $J$ | 14.03 | 22.21 | 26.71 | 29.37 | 31.04 | 32.14 | 32.89 | 33.41 | 33.79 | 34.06 | 34.27 | 34.42 | 34.54 | 34.64 | 34.71 | 34.78 | 34.83 | 34.88 | 34.91 | 34.93 | 34.95 | 34.97 | 34.98 | 34.99 | 35.00 |
| | $K$ | 1.90 | 2.67 | 2.85 | 2.88 | 2.89 | 2.89 | 2.89 | 2.89 | 2.89 | 2.89 | 2.89 | 2.89 | 2.89 | 2.89 | 2.89 | 2.89 | 2.89 | 2.89 | 2.89 | 2.89 | 2.89 | 2.89 | 2.89 | 2.89 | 2.89 |
| 0.8 | $U_1$ | 25.32 | 43.03 | 54.61 | 62.46 | 68.02 | 72.14 | 75.28 | 77.75 | 79.74 | 81.37 | 82.73 | 83.89 | 84.87 | 85.73 | 86.48 | 87.13 | 87.72 | 88.24 | 88.71 | 89.13 | 89.52 | 89.87 | 90.19 | 90.48 | 90.75 |
| | $U_2$ | 0.80 | 1.98 | 3.70 | 5.71 | 7.74 | 9.64 | 11.35 | 12.87 | 14.20 | 15.37 | 16.40 | 17.31 | 18.11 | 18.82 | 19.46 | 20.03 | 20.54 | 21.00 | 21.43 | 21.82 | 22.17 | 22.49 | 22.79 | 23.07 | 23.32 |
| | $W_{1,2}$ | 20.33 | 41.97 | 64.23 | 86.63 | 109.07 | 131.51 | 153.95 | 176.40 | 198.84 | 221.28 | 243.72 | 266.17 | 288.61 | 311.05 | 333.50 | 355.94 | 378.38 | 400.82 | 423.27 | 445.71 | 468.15 | 490.60 | 513.04 | 535.48 | 557.92 |
| | $W_{3,4}$ | 20.35 | 42.08 | 64.36 | 86.78 | 109.22 | 131.66 | 154.10 | 176.54 | 198.99 | 221.43 | 243.87 | 266.31 | 288.75 | 311.20 | 333.64 | 356.08 | 378.53 | 400.97 | 423.41 | 445.85 | 468.30 | 490.74 | 513.18 | 535.63 | 558.07 |
| | $J$ | 14.60 | 23.45 | 28.48 | 31.51 | 33.43 | 34.71 | 35.59 | 36.21 | 36.65 | 36.98 | 37.22 | 37.41 | 37.55 | 37.66 | 37.75 | 37.82 | 37.88 | 37.93 | 37.96 | 38.00 | 38.02 | 38.04 | 38.06 | 38.08 | 38.09 |
| | $K$ | 2.05 | 2.87 | 3.06 | 3.10 | 3.11 | 3.11 | 3.11 | 3.11 | 3.11 | 3.11 | 3.11 | 3.11 | 3.11 | 3.11 | 3.11 | 3.11 | 3.11 | 3.11 | 3.11 | 3.11 | 3.11 | 3.11 | 3.11 | 3.11 | 3.11 |
| 0.7 | $U_1$ | 23.56 | 40.32 | 51.43 | 59.04 | 64.47 | 68.50 | 71.60 | 74.04 | 76.00 | 77.62 | 78.97 | 80.12 | 81.10 | 81.95 | 82.69 | 83.33 | 83.93 | 84.45 | 84.91 | 85.34 | 85.72 | 86.07 | 86.39 | 86.68 | 86.95 |
| | $U_2$ | 0.87 | 2.12 | 3.87 | 5.89 | 7.91 | 9.80 | 11.51 | 13.02 | 14.34 | 15.50 | 16.53 | 17.43 | 18.22 | 18.93 | 19.57 | 20.14 | 20.65 | 21.11 | 21.54 | 21.92 | 22.27 | 22.60 | 22.89 | 23.17 | 23.43 |
| | $W_{1,2}$ | 21.60 | 43.60 | 69.84 | 86.21 | 108.67 | 131.12 | 153.56 | 176.00 | 198.45 | 220.89 | 243.33 | 265.77 | 288.22 | 310.66 | 333.10 | 355.54 | 377.99 | 400.43 | 422.87 | 445.32 | 467.76 | 490.20 | 512.64 | 535.09 | 557.53 |
| | $W_{3,4}$ | 20.37 | 42.09 | 64.37 | 86.78 | 109.22 | 131.66 | 154.10 | 176.55 | 198.99 | 221.43 | 243.87 | 266.32 | 288.76 | 311.20 | 333.65 | 356.09 | 378.53 | 400.98 | 423.42 | 445.86 | 468.30 | 490.75 | 513.19 | 535.63 | 558.08 |
| | $J$ | 15.24 | 24.82 | 30.47 | 33.96 | 36.23 | 37.77 | 38.84 | 39.60 | 40.16 | 40.58 | 40.90 | 41.14 | 41.33 | 41.48 | 41.60 | 41.70 | 41.77 | 41.84 | 41.89 | 41.93 | 41.97 | 42.00 | 42.03 | 42.05 | 42.07 |
| | $K$ | 2.22 | 3.10 | 3.31 | 3.35 | 3.36 | 3.36 | 3.36 | 3.36 | 3.36 | 3.36 | 3.36 | 3.36 | 3.36 | 3.36 | 3.36 | 3.36 | 3.36 | 3.36 | 3.36 | 3.36 | 3.36 | 3.36 | 3.36 | 3.36 | 3.36 |

| $u_0/u_1$ | $\xi$/nm | 2 | 4 | 6 | 8 | 10 | 12 | 14 | 16 | 18 | 20 | 22 | 24 | 26 | 28 | 30 | 32 | 34 | 36 | 38 | 40 | 42 | 44 | 46 | 48 | 50 |
|---|---|---|---|---|---|---|---|---|---|---|---|---|---|---|---|---|---|---|---|---|---|---|---|---|---|---|
| 0.6 | $U_1$ | 21.54 | 37.11 | 47.62 | 54.90 | 60.14 | 64.06 | 67.08 | 69.47 | 71.40 | 72.99 | 74.33 | 75.46 | 76.43 | 77.27 | 78.01 | 78.66 | 79.24 | 79.76 | 80.22 | 80.64 | 81.02 | 81.37 | 81.69 | 81.98 | 82.25 |
| | $U_2$ | 0.97 | 2.50 | 4.10 | 6.12 | 8.13 | 10.01 | 11.69 | 13.19 | 14.50 | 15.65 | 16.66 | 17.56 | 18.35 | 19.05 | 19.68 | 20.25 | 20.76 | 21.22 | 21.64 | 22.02 | 22.37 | 22.70 | 22.99 | 23.27 | 23.52 |
| | $W_{1,2}$ | 19.84 | 41.30 | 63.52 | 85.92 | 108.36 | 130.80 | 153.24 | 175.68 | 198.12 | 220.57 | 243.01 | 265.45 | 287.90 | 310.34 | 332.78 | 355.23 | 377.67 | 400.11 | 422.55 | 445.00 | 467.44 | 489.88 | 512.33 | 534.77 | 557.21 |
| | $W_{3,4}$ | 20.45 | 42.19 | 64.48 | 86.89 | 109.33 | 131.77 | 154.21 | 176.65 | 199.10 | 221.54 | 243.98 | 266.43 | 288.87 | 311.31 | 333.76 | 356.20 | 378.64 | 401.08 | 423.53 | 445.97 | 468.41 | 490.86 | 513.30 | 535.74 | 558.18 |
| | $J$ | 15.93 | 26.33 | 32.70 | 36.77 | 39.48 | 41.36 | 42.70 | 43.68 | 44.41 | 44.97 | 45.40 | 45.73 | 45.99 | 46.21 | 46.38 | 46.52 | 46.63 | 46.72 | 46.80 | 46.87 | 46.93 | 46.97 | 47.01 | 47.05 | 47.08 |
| | $K$ | 2.40 | 3.35 | 3.57 | 3.62 | 3.63 | 3.63 | 3.63 | 3.63 | 3.63 | 3.63 | 3.63 | 3.63 | 3.63 | 3.63 | 3.63 | 3.63 | 3.63 | 3.63 | 3.63 | 3.63 | 3.63 | 3.63 | 3.63 | 3.63 | 3.63 |
| 0.5 | $U_1$ | 20.05 | 35.58 | 45.82 | 52.96 | 58.13 | 62.00 | 64.99 | 67.36 | 69.28 | 70.87 | 72.20 | 73.32 | 74.29 | 75.11 | 75.87 | 76.52 | 77.09 | 77.61 | 78.07 | 78.49 | 78.87 | 79.22 | 79.54 | 79.83 | 80.10 |
| | $U_2$ | 1.01 | 2.38 | 4.20 | 6.23 | 8.24 | 10.11 | 11.79 | 13.28 | 14.59 | 15.73 | 16.74 | 17.64 | 18.43 | 19.13 | 19.76 | 20.32 | 20.83 | 21.29 | 21.71 | 22.09 | 22.44 | 22.77 | 23.06 | 23.33 | 23.59 |
| | $W_{1,2}$ | 19.56 | 40.92 | 63.12 | 85.51 | 107.94 | 130.38 | 152.83 | 175.27 | 197.71 | 220.16 | 242.60 | 265.04 | 287.48 | 309.93 | 332.37 | 354.81 | 377.26 | 399.70 | 422.14 | 444.58 | 467.03 | 489.47 | 511.91 | 534.36 | 556.80 |
| | $W_{3,4}$ | 20.49 | 42.24 | 64.52 | 86.94 | 109.37 | 131.81 | 154.25 | 176.70 | 199.14 | 221.58 | 244.03 | 266.47 | 288.91 | 311.35 | 333.80 | 356.24 | 378.69 | 401.13 | 423.57 | 446.01 | 468.45 | 490.90 | 513.34 | 535.79 | 558.23 |
| | $J$ | 16.42 | 27.34 | 34.13 | 38.49 | 41.42 | 43.45 | 44.91 | 45.98 | 46.77 | 47.38 | 47.85 | 48.22 | 48.51 | 48.74 | 48.93 | 49.09 | 49.21 | 49.32 | 49.40 | 49.48 | 49.54 | 49.60 | 49.64 | 49.68 | 49.72 |
| | $K$ | 2.60 | 3.63 | 3.86 | 3.90 | 3.91 | 3.92 | 3.92 | 3.92 | 3.92 | 3.92 | 3.92 | 3.92 | 3.92 | 3.92 | 3.92 | 3.92 | 3.92 | 3.92 | 3.92 | 3.92 | 3.92 | 3.92 | 3.92 | 3.92 | 3.92 |
| 0.4 | $U_1$ | 18.35 | 33.35 | 43.15 | 50.03 | 55.05 | 58.83 | 61.77 | 64.10 | 66.02 | 67.56 | 68.87 | 69.99 | 70.95 | 71.78 | 72.51 | 73.15 | 73.73 | 74.24 | 74.70 | 75.12 | 75.50 | 75.84 | 76.16 | 76.45 | 76.72 |
| | $U_2$ | 1.08 | 2.52 | 4.36 | 6.38 | 8.38 | 10.24 | 11.90 | 13.38 | 14.67 | 15.81 | 16.81 | 17.70 | 18.48 | 19.18 | 19.81 | 20.37 | 20.87 | 21.33 | 21.75 | 22.13 | 22.48 | 22.80 | 23.10 | 23.37 | 23.63 |
| | $W_{1,2}$ | 19.28 | 40.67 | 62.85 | 85.24 | 107.68 | 130.12 | 152.56 | 175.00 | 197.44 | 219.89 | 242.33 | 264.77 | 287.22 | 309.66 | 332.10 | 354.55 | 376.99 | 399.43 | 421.87 | 444.32 | 466.76 | 489.20 | 511.65 | 534.09 | 556.53 |
| | $W_{3,4}$ | 20.59 | 42.37 | 64.60 | 87.07 | 109.51 | 131.95 | 154.39 | 176.84 | 199.28 | 221.72 | 244.17 | 266.61 | 289.05 | 311.50 | 333.94 | 356.38 | 378.82 | 401.27 | 423.71 | 446.15 | 468.60 | 491.04 | 513.48 | 535.92 | 558.37 |
| | $J$ | 16.98 | 28.55 | 35.92 | 40.76 | 44.07 | 46.42 | 48.16 | 49.41 | 50.38 | 51.14 | 51.73 | 52.21 | 52.59 | 52.90 | 53.16 | 53.37 | 53.55 | 53.70 | 53.82 | 53.93 | 54.02 | 54.10 | 54.17 | 54.23 | 54.29 |
| | $K$ | 2.80 | 3.89 | 4.14 | 4.19 | 4.20 | 4.21 | 4.21 | 4.21 | 4.21 | 4.21 | 4.21 | 4.21 | 4.21 | 4.21 | 4.21 | 4.21 | 4.21 | 4.21 | 4.21 | 4.21 | 4.21 | 4.21 | 4.21 | 4.21 | 4.21 |
| 0.3 | $U_1$ | 18.35 | 32.06 | 41.60 | 48.35 | 53.29 | 57.02 | 59.92 | 62.23 | 64.11 | 65.67 | 66.97 | 68.08 | 69.04 | 69.87 | 70.60 | 71.24 | 71.81 | 72.32 | 72.78 | 73.20 | 73.58 | 73.92 | 74.24 | 74.53 | 74.80 |
| | $U_2$ | 1.15 | 2.60 | 4.46 | 6.49 | 8.49 | 10.33 | 11.99 | 13.46 | 14.75 | 15.89 | 16.91 | 17.79 | 18.58 | 19.24 | 19.89 | 20.45 | 20.96 | 21.42 | 21.83 | 22.21 | 22.56 | 22.88 | 23.18 | 23.45 | 23.70 |
| | $W_{1,2}$ | 19.18 | 40.40 | 62.50 | 84.95 | 107.38 | 129.56 | 152.00 | 174.44 | 196.89 | 219.33 | 241.64 | 264.22 | 286.06 | 309.10 | 331.54 | 353.99 | 376.43 | 398.87 | 421.31 | 443.76 | 466.20 | 488.64 | 511.09 | 533.53 | 555.97 |
| | $W_{3,4}$ | 20.65 | 42.45 | 64.75 | 87.16 | 109.60 | 132.04 | 154.48 | 176.92 | 199.37 | 221.81 | 244.25 | 266.70 | 289.14 | 311.58 | 334.02 | 356.47 | 378.91 | 401.35 | 423.80 | 446.24 | 468.68 | 491.13 | 513.57 | 536.01 | 558.45 |
| | $J$ | 17.36 | 29.34 | 37.05 | 42.16 | 45.67 | 48.16 | 49.91 | 51.35 | 52.40 | 53.18 | 53.79 | 54.25 | 54.61 | 54.90 | 55.12 | 55.26 | 55.35 | 55.44 | 56.13 | 56.25 | 56.35 | 56.44 | 56.52 | 56.58 | 56.64 |
| | $K$ | 2.98 | 4.15 | 4.42 | 4.47 | 4.48 | 4.48 | 4.48 | 4.48 | 4.48 | 4.48 | 4.48 | 4.48 | 4.48 | 4.48 | 4.48 | 4.48 | 4.48 | 4.48 | 4.48 | 4.48 | 4.48 | 4.48 | 4.48 | 4.48 | 4.48 |
| 0.2 | $U_1$ | 17.89 | 31.34 | 40.74 | 47.41 | 52.31 | 56.01 | 58.90 | 61.20 | 63.07 | 64.62 | 65.92 | 67.02 | 67.69 | 68.81 | 69.54 | 70.18 | 70.75 | 71.26 | 71.72 | 72.14 | 72.52 | 72.86 | 73.18 | 73.47 | 73.74 |
| | $U_2$ | 1.15 | 2.61 | 4.48 | 6.51 | 8.51 | 10.36 | 12.02 | 13.49 | 14.78 | 15.92 | 16.92 | 17.80 | 18.58 | 19.28 | 19.90 | 20.45 | 20.97 | 21.43 | 21.84 | 22.22 | 22.57 | 22.89 | 23.19 | 23.46 | 23.72 |
| | $W_{1,2}$ | 19.00 | 40.15 | 62.30 | 84.60 | 107.12 | 129.50 | 151.94 | 174.38 | 196.83 | 219.27 | 241.71 | 264.15 | 286.60 | 309.04 | 331.48 | 353.93 | 376.37 | 398.81 | 421.25 | 443.70 | 466.14 | 488.58 | 510.99 | 533.39 | 555.83 |
| | $W_{3,4}$ | 20.69 | 42.60 | 64.80 | 87.21 | 109.65 | 132.09 | 154.54 | 176.98 | 199.42 | 221.86 | 244.31 | 266.75 | 289.19 | 311.64 | 334.08 | 356.52 | 378.97 | 401.41 | 423.85 | 446.29 | 468.74 | 491.18 | 513.62 | 536.07 | 558.51 |
| | $J$ | 17.64 | 29.90 | 37.85 | 43.13 | 46.77 | 49.37 | 51.27 | 52.70 | 53.79 | 54.56 | 55.15 | 55.66 | 56.05 | 56.36 | 56.60 | 56.80 | 56.96 | 57.09 | 57.20 | 57.29 | 57.37 | 57.43 | 57.50 | 57.57 | 57.94 |
| | $K$ | 3.14 | 4.36 | 4.65 | 4.70 | 4.71 | 4.72 | 4.72 | 4.72 | 4.72 | 4.72 | 4.72 | 4.72 | 4.72 | 4.72 | 4.72 | 4.72 | 4.72 | 4.72 | 4.72 | 4.72 | 4.72 | 4.72 | 4.72 | 4.72 | 4.72 |
| 0.1 | $U_1$ | 18.10 | 31.71 | 41.21 | 47.95 | 52.89 | 56.62 | 59.52 | 61.84 | 63.72 | 65.27 | 66.58 | 67.69 | 68.65 | 69.48 | 70.21 | 70.85 | 71.42 | 71.93 | 72.40 | 72.81 | 73.19 | 73.54 | 73.85 | 74.14 | 74.41 |
| | $U_2$ | 1.13 | 2.61 | 4.48 | 6.51 | 8.51 | 10.36 | 12.03 | 13.49 | 14.79 | 15.92 | 16.92 | 17.80 | 18.58 | 19.28 | 19.90 | 20.46 | 20.97 | 21.43 | 21.84 | 22.22 | 22.57 | 22.89 | 23.19 | 23.46 | 23.72 |
| | $W_{1,2}$ | 18.79 | 39.87 | 62.00 | 84.38 | 106.82 | 129.26 | 151.70 | 174.14 | 196.59 | 219.03 | 241.47 | 263.91 | 286.36 | 308.80 | 331.24 | 353.68 | 376.13 | 398.57 | 421.01 | 443.46 | 465.90 | 488.34 | 510.78 | 533.23 | 555.67 |
| | $W_{3,4}$ | 20.67 | 42.47 | 64.77 | 87.18 | 109.62 | 132.06 | 154.51 | 176.95 | 199.39 | 221.83 | 244.28 | 266.72 | 289.16 | 311.61 | 334.05 | 356.49 | 378.94 | 401.38 | 423.82 | 446.26 | 468.71 | 491.15 | 513.59 | 536.04 | 558.48 |
| | $J$ | 17.74 | 30.00 | 38.00 | 43.25 | 46.85 | 49.45 | 51.16 | 52.52 | 53.70 | 54.53 | 55.15 | 55.63 | 56.03 | 56.34 | 56.56 | 56.76 | 56.91 | 57.05 | 57.13 | 57.25 | 57.43 | 57.55 | 57.65 | 57.81 | 57.88 |
| | $K$ | 3.25 | 4.51 | 4.80 | 4.86 | 4.86 | 4.87 | 4.88 | 4.88 | 4.88 | 4.88 | 4.88 | 4.88 | 4.88 | 4.88 | 4.88 | 4.88 | 4.88 | 4.88 | 4.88 | 4.88 | 4.88 | 4.88 | 4.88 | 4.88 | 4.88 |
| 0.0 | $U_1$ | 18.93 | 33.09 | 42.92 | 49.84 | 54.89 | 58.70 | 61.65 | 64.00 | 65.90 | 67.47 | 68.79 | 69.91 | 70.88 | 71.71 | 72.45 | 73.09 | 73.67 | 74.18 | 74.65 | 75.06 | 75.44 | 75.79 | 76.11 | 76.40 | 76.67 |
| | $U_2$ | 1.07 | 2.50 | 4.35 | 6.39 | 8.41 | 10.27 | 11.94 | 13.42 | 14.72 | 15.87 | 16.87 | 17.76 | 18.55 | 19.25 | 19.88 | 20.44 | 20.95 | 21.41 | 21.82 | 22.21 | 22.56 | 22.88 | 23.17 | 23.45 | 23.70 |
| | $W_{1,2}$ | 18.57 | 39.57 | 61.69 | 84.06 | 106.49 | 128.93 | 151.38 | 173.83 | 196.26 | 218.71 | 241.15 | 263.59 | 286.03 | 308.48 | 330.92 | 353.36 | 375.81 | 398.25 | 420.69 | 443.13 | 465.58 | 488.02 | 510.46 | 532.91 | 555.35 |
| | $W_{3,4}$ | 20.50 | 42.36 | 64.65 | 87.07 | 109.50 | 131.94 | 154.38 | 176.82 | 199.27 | 221.71 | 244.16 | 266.60 | 289.04 | 311.49 | 333.93 | 356.37 | 378.82 | 401.26 | 423.70 | 446.14 | 468.59 | 491.03 | 513.47 | 535.92 | 558.36 |
| | $J$ | 17.66 | 29.79 | 37.50 | 42.51 | 45.88 | 48.24 | 49.93 | 51.16 | 52.09 | 52.80 | 53.29 | 53.79 | 54.13 | 54.43 | 54.63 | 54.82 | 54.97 | 55.10 | 55.21 | 55.30 | 55.37 | 55.44 | 55.50 | 55.55 | 55.59 |
| | $K$ | 3.28 | 4.56 | 4.86 | 4.92 | 4.93 | 4.93 | 4.93 | 4.93 | 4.93 | 4.93 | 4.93 | 4.93 | 4.93 | 4.93 | 4.93 | 4.93 | 4.93 | 4.93 | 4.93 | 4.93 | 4.93 | 4.93 | 4.93 | 4.93 | 4.93 |

TABLE S95. Parameters of the THF interaction Hamiltonian for different ratios $u_0/u_1$ and screening lengths $\xi$ of the double-gate interaction potential ($U_\xi = 24\,\mathrm{meV}$ for $\xi = 10\,\mathrm{nm}$). We use the same BM model parameters as in Table S94.



\end{document}